\documentclass[]{cernyrep}
\usepackage{ifthen}
\newcommand{\BLKP}{% add blank page when on even page after PDF file
  \ifthenelse{\isodd{\value{page}}}{\relax}{\mbox{}\thispagestyle{empty}\newpage}}
\def    \nn             {\nonumber} 
\def    \=              {\;=\;} 
\def    \lsim           {\raisebox{-3pt}{$\>\stackrel{<}{\scriptstyle\sim}\>$}} 
\def    \gsim           {\raisebox{-3pt}{$\>\stackrel{>}{\scriptstyle\sim}\>$}} 
\def    \gtrsim         {\raisebox{-3pt}{$\>\stackrel{>}{\scriptstyle\sim}\>$}} 
 
\newcommand     \be     {\begin{equation}} 
\newcommand     \ee     {\end{equation}} 
\newcommand     \ba     {\begin{eqnarray}} 
\newcommand     \ea     {\end{eqnarray}} 
 
\newcommand     \sss            {\scriptscriptstyle}

\newcommand     \lambdamsb     {\ifmmode 
          \Lambda_4^{\rm \scriptscriptstyle \overline{MS}} \else 
         $\Lambda_4^{\rm \scriptscriptstyle \overline{MS}}$ \fi} 
\newcommand     \MSB            {\ifmmode {\overline{\rm MS}} \else 
                                 $\overline{\rm MS}$  \fi}

\newcommand     \ptmin     {\ifmmode p_{\scriptscriptstyle T}^{\sss min} \else 
                           $p_{\scriptscriptstyle T}^{\sss min}$ \fi}

\newcommand\as{\alpha_{\rm s}}

\newcommand\aem{\alpha_{\rm em}} 
 
\newcommand\qqb{{q\overline{q}}} 
 
\newcommand\bb{\overline{b}}

\def \ppbar {\mbox{$p \bar p$}} 
\def \ttbar {\mbox{$t \bar t$}} 
\def \bbbar {\mbox{$b \bar b$}} 
\def \ccbar {\mbox{$c \bar c$}} 
 
\newcommand{\pbarp}{\ensuremath{\bar{p} p}}
\newcommand{\ep}{\ensuremath{e^+ e^-}}
\newcommand{\gaga}{\ensuremath{\gamma \gamma}}

\def \pt   {\mbox{$p_{\scriptscriptstyle T}$}} 
 
\def \et   {\mbox{$E_{\scriptscriptstyle T}$}}

\def \to   {\mbox{$\rightarrow$}} 
\def    \mb             {\mbox{$m_b$}} 
\def    \mc             {\mbox{$m_c$}} 

\newcommand \jpsi{\ifmmode{J/\psi}\else{$J/\psi$}\fi} 
\newcommand \psip{\ifmmode{\psi(2S)}\else{$\psi(2S)$}\fi} 
\newcommand \psipp{\ifmmode{\psi(3770)}\else{$\psi(3770)$}\fi} 
\def\chicj #1{\chi_{c{#1}}}
\newcommand \chic[1]{\ifmmode{\chicj{#1}}\else{$\chicj{#1}$}\fi} 
\def\chibj #1{\chi_{b{#1}}}
\newcommand \chib[1]{\ifmmode{\chibj{#1}}\else{$\chibj{#1}$}\fi} 
\newcommand \etac{\ifmmode{\eta_c(1S)}\else{$\eta_c(1S)$}\fi}
\newcommand \etacp{\ifmmode{\eta_c(2S)}\else{$\eta_c(2S)$}\fi}
\newcommand \hc{\ifmmode{h_c(1P)}\else{$h_c(1P)$}\fi}
\newcommand{\ups}{\ensuremath{\Upsilon(1S)}}

\newcommand{\Jpsi}{\ensuremath{J/\psi}}
\newcommand{\un}[1]{\ensuremath{\; \mathrm{#1}}}

\def\calL{{\cal L}}

\newcommand{\pp}{\pi^+\pi^-}

\newcommand{\ra}{\rightarrow}

\newcommand{\beq}{\begin{equation}}
\newcommand{\eeq}{\end{equation}}
\newcommand{\bfg}{\begin{figure}}
\newcommand{\efg}{\end{figure}}
\newcommand{\bitm}{\begin{itemize}}
\newcommand{\eitm}{\end{itemize}}
\newcommand{\bnum}{\begin{enumerate}}
\newcommand{\enum}{\end{enumerate}}
\newcommand{\btbl}{\begin{table}}
\newcommand{\etbl}{\end{table}}
\newcommand{\btbu}{\begin{tabular}}
\newcommand{\etbu}{\end{tabular}}
\newcommand{\dedx}  {{\mathrm{d}}E/{\mathrm{d}}x}
\newcommand{\epem}  {\mathrm{e}^+\mathrm{e}^-}

\newcommand{\om}  {\ensuremath{\omega}}

\newcommand{\zz}  {\mathrm{Z}}
\newcommand{\unit}[1]{\,{\mathrm{#1}}}
\def\pep2{PEP-II}
\def\KS    {\ensuremath{K^0_{\scriptscriptstyle S}}}
\def\gevc {\ensuremath{{\mathrm{\,Ge\kern -0.1em V\!/}c}}}
\def\mevc {\ensuremath{{\mathrm{\,Me\kern -0.1em V\!/}c}}}
\def\gevcc {\ensuremath{{\mathrm{\,Ge\kern -0.1em V\!/}c^2}}}
\def\mevcc {\ensuremath{{\mathrm{\,Me\kern -0.1em V\!/}c^2}}}
\def\invfb   {\ensuremath{\mbox{\,fb}^{-1}}}
\def\babar {\mbox{BaBar}}

\def\pip   {\ensuremath{\pi^+}}
\def\pim   {\ensuremath{\pi^-}}

\def\Kp    {\ensuremath{K^+}}

\def\PM     {\ensuremath{\pm}}
\def\fsec    {\rm{\it{fs}}}                    % define how to write fsec
\def\MeV     {{\mbox{$\mathrm{MeV}$}}}
\def\GeV     {{\mbox{$\mathrm{GeV}$}}}
\def\mvc     {{\mbox{$\mathrm{MeV/}c^2$}}}

\def\Ntot    {162}                              % events in signal + sidebands
\def\Nevt    {22}                               % events in signal region
\def\Nback   {6.1}                              % fit background events
\def\dBack   {0.5}                              % fit backround uncertainty
\def\Nsig    {15.9}                             % excess
\def\Mass    {3519}                             % Mass
\def\dMass   {2}                                % Mass uncertainty
\def\BackD   {1.6}                              % fit background events
\def\dBackD  {0.35}                             % fit backround uncertainty
\def\NsigD   {5.4}                              % excess
\def\MassD   {3518}                             % Fit Mass
\def\dMassD  {3 }                               % Fit Mass uncertainty

\def\amass   {3518.7}                           % Average Mass
\def\damass  {1.7}                              % Average Mass uncertainty
\def\Pois    {{\mbox{$1.5 \times 10^{-5}$}}}

\def\lc      {{\mbox{$\Lambda_{c}^{+} $}}}

\def\kpi     {{\mbox{$K^- \pi^+$}}}

\def\ccs     {{\mbox{$\Omega_{cc}^{+}$}}}
\def\ccu     {{\mbox{$\Xi_{cc}^{++}$}}}
\def\ccd     {{\mbox{$\Xi_{cc}^{+}$}}}
\def\ccdlcki {{\mbox{$\ccd \rightarrow \lc K^-\pi^+ $}}}

\def\lcki    {{\mbox{$\ccd \rightarrow \lc K^-\pi^+ \ $}}}
\def\dpk     {{\mbox{$\ccd \rightarrow p D^+ K^- \ $}}}
\def\dpkn    {{\mbox{$\ccd \rightarrow p D^+ K^- $}}}

\def\bea     {\begin{eqnarray}}
\def\eea     {\end{eqnarray}}

\newcommand{\RRS}{\rm RS}

\newcommand{\OS}{\rm OS}

\newcommand{\slj}[3]{\mbox{$^{#1}${\ifcase#2\or S\or
         P\or D\or F\or G\fi}$_{#3}$}}
\newcommand{\sLj}[3]{{}^{#1}\!#2_{#3}}
\newcommand{\kev}{\hbox{ keV}}
\newcommand{\mev}{\hbox{ MeV}}
\newcommand{\gev}{\hbox{ GeV}}
\newcommand{\kst}[1]{\ifmmode |#1> \else%
$|#1\rangle$ %
\fi}

\def\kst#1{\ifmmode |#1\rangle \else%
$|#1>$ %
\fi}
\def\psip{\ifmmode \psi^{~\prime} \else%
$\psi^{~\prime}$ %
\fi}
\def\chic{\ifmmode \chi_c \else%
$\chi_c$ %
\fi}
\def\etac{\ifmmode \eta_c \else%
$\eta_c$ %
\fi}
\def\Y{\ifmmode \Upsilon \else%
$\Upsilon$ %
\fi}
\def\chib{\ifmmode \chi_b \else%
$\chi_b$ %
\fi}
\def\chibp{\ifmmode \chi_b' \else%
$\chi_b$ %
\fi}
\def\Q#1#2#3#4{\ifmmode
 \,#1\,{^{#2}#3}_{#4}
\else%
$#1\,{^{#2}#3}_{#4}$ %
\fi}
\def\bb{\bar bb}
\def\cc{\bar cc}
\def\tt{\bar tt}
\def\mm{\ifmmode \mu^+\mu^- \else $\mu^+\mu^-$  \fi}
\def\LL{\ifmmode l^+l^- \else $l^+l^-$  \fi}

\def\etal{{\it et al.}}
\def\B{{\cal B}}

\def\bfsigma{\mbox{\boldmath $\sigma$}}

\def\vo{V^{(0)}_o}

\def\bfnabla{\mbox{\boldmath $\nabla$}}

\def\bfsigma{\mbox{\boldmath $\sigma$}}

\def\lQ{\Lambda_{\rm QCD}}

\def\al{\alpha}
\def\als{\alpha_{\rm s}}
\def\siml{{\ \lower-1.2pt\vbox{\hbox{\rlap{$<$}\lower6pt\vbox{\hbox{$\sim$}}}}\ }} 
\def\simg{{\ \lower-1.2pt\vbox{\hbox{\rlap{$>$}\lower6pt\vbox{\hbox{$\sim$}}}}\ }} 
\newcommand{\MS}{\overline{\rm MS}}
\def\lla{\langle\!\langle}
\def\rra{\rangle\!\rangle}
\def\ecalif#1#2{|{\cal E}_{\rm #1,#2}|}
\def    \journal#1#2#3#4{{\it #1 } {\bf #2}(#3)#4}
\def    \ap     #1#2#3{{Ann. Phys.} {\bf #1}, #3 (#2}
\def    \app    #1#2#3{{Acta Phys. Polon. }{\bf #1}, #3 (#2)}
\def    \np     #1#2#3{{Nucl. Phys.} {\bf #1}, #3 (#2)}
\def    \npb    #1#2#3{{Nucl. Phys. B} {\bf #1}, #3 (#2)}
\def    \pl     #1#2#3{{Phys. Lett.} {\bf #1}, #3 (#2)}
\def    \plb    #1#2#3{{Phys. Lett. B} {\bf #1}, #3 (#2)}
\def    \prl    #1#2#3{{Phys. Rev. Lett.} {\bf #1}, #3 (#2)}
\def    \pr     #1#2#3{{Phys. Rev.} {\bf #1}, #3 (#2)}
\def    \prd    #1#2#3{{Phys. Rev.} D {\bf #1}, #3 (#2)}
\def    \prep   #1#2#3{{Phys. Rep.} {\bf #1}, #3 (#2)}
\def    \sjnp   #1#2#3{{Sov. J. Nucl. Phys. }{\bf #1}, #3 (#2)}
\def    \zeit   #1#2#3{{Z. Phys.} {\bf #1}, #3 (#2)}
\def    \zp     #1#2#3{{Z. Phys.} {\bf #1}, #3 (#2)}
\def    \zpc    #1#2#3{{Z. Phys. C} {\bf #1}, #3 (#2)}
\def    \ibid   #1#2#3{{ibid.} {\bf #1}, #3 (#2)}
\def    \hepph  #1 {{\tt hep-ph/#1}}
\def    \hepex  #1 {{\tt hep-ex/#1}}
\newcommand{\dd}{\displaystyle}
\def\HTslash#1{\setbox0=\hbox{$#1$}#1\hskip-\wd0\dimen0=5pt\advance
       \dimen0 by-\ht0\advance\dimen0 by\dp0\lower0.5\dimen0\hbox
	 to\wd0{\hss\sl/\/\hss}}
\newcommand{\J}{J/\psi}
\newcommand{\Jpi}{\pi^0 J/\psi}
\newcommand{\Jeta}{\eta J/\psi}
\newcommand{\Jpipi}{\pi^0\pi^0 J/\psi}
\newcommand{\gx}{\gamma\chi_{c1,c2}}
\newcommand{\ggee}{\gamma\gamma e^+e^-}

\newcommand{\ggll}{\gamma\gamma l^+l^-}
\newcommand{\MgJ}{M_{\gamma_h,J/\psi}}
\newcommand{\Mgg}{M_{\gamma\gamma}}
\newcommand{\ptochic}{\psi(2S)\ra \gamma\chi_{c1,2}}
\newcommand{\gguu}{\gamma\gamma\mu^+\mu^-}
\newcommand{\pppp}{\psi(2S) \rt J/\psi \pi^+ \pi^- }
\newcommand{\rt}{\rightarrow}

\newcommand{\ppll}{\psi(2S) \rightarrow
 \pi^+ \pi^- J/\psi$, $J/\psi \rightarrow l^+ l^-}

\newcommand{\jp}{J/\psi}
\newcommand{\pipi}{\pi^{+}\pi^{-}}

\newcommand{\Rsub}{\rm\scriptscriptstyle}
\def\dipmtx#1#2{\ifmmode
{\cal E}_{\rm #1,#2}
\else%
${\cal E}_{\rm #1,#2}$ %
\fi}

\def\centeron#1#2{{\setbox0=\hbox{#1}\setbox1=\hbox{#2}\ifdim
\wd1>\wd0\kern.5\wd1\kern-.5\wd0\fi
\copy0\kern-.5\wd0\kern-.5\wd1\copy1\ifdim\wd0>\wd1
\kern.5\wd0\kern-.5\wd1\fi}}
\def\centerover#1#2{\centeron{#1}{\setbox0=\hbox{#1}\setbox
1=\hbox{#2}\raise\ht0\hbox{\raise\dp1\hbox{\copy1}}}}
\def\centerunder#1#2{\centeron{#1}{\setbox0=\hbox{#1}\setbox
1=\hbox{#2}\lower\dp0\hbox{\lower\ht1\hbox{\copy1}}}}

\def\lsim{\;\centeron{\raise.35ex\hbox{$<$}}{\lower.65ex\hbox
{$\sim$}}\;}
\def\gsim{\;\centeron{\raise.35ex\hbox{$>$}}{\lower.65ex\hbox
{$\sim$}}\;}

\let\holdvec=\vec
\def\vec#1{\hbox{$\holdvec #1\,$}}

\def\O{{\cal O}}

\def\MS{{\overline{{\rm MS}}}}

\def\ra{\rightarrow}

\def\al{\alpha}

\def\O#1#2{\mbox{\boldmath $O$}_{\mbox{\scriptsize\boldmath $#1$},#2}}

\def\sitbf#1{\mbox{\scriptsize\boldmath $#1$}}

\def\bsigma{\mbox{\boldmath $\sigma$}}

\def\ms{$\overline{\rm MS}$ }
\def\chip#1{\chi_{\mbox{\scriptsize\boldmath $#1$}}}

\def\OMIT#1{}

\def\Dsl{\hbox{/\kern-.6000em D}} %roman D

\def\dsl{\,\raise.15ex\hbox{/}\mkern-11.5mu D}
\def\bsigma{\mbox{\boldmath $\sigma$}}

\def\psixp#1{\psi_{\mathbf{#1}}}
\def\chip#1{\chi_{\mathbf{#1}}}
\def\bsigma{\mbox{\boldmath $\sigma$}}

\def\ltap{\ \raise.3ex\hbox{$<$\kern-.75em\lower1ex\hbox{$\sim$}}\ }
\def\gtap{\ \raise.3ex\hbox{$>$\kern-.75em\lower1ex\hbox{$\sim$}}\ }

\def\OMIT#1{}

\def\ms{$\overline{\rm MS}$ }
\def\O#1#2{\mbox{\boldmath $O$}_{\mbox{\scriptsize $\mathbf #1$},#2}}
\newcommand{\bmk}{\mathbf k}
\newcommand{\bmp}{\mathbf p}

\newcommand{\bmr}{\mathbf r}

\newcommand{\bmS}{\mathbf S}
\newcommand{\bmsigma}{\mathbf \bsigma}

\def\lsim{\mathrel{\raise.3ex\hbox{$<$\kern-.75em\lower1ex\hbox{$\sim$}}}}
\def\gsim{\mathrel{\raise.3ex\hbox{$>$\kern-.75em\lower1ex\hbox{$\sim$}}}}
\def\Li2{{\rm Li}_2}

\def\bfsigma{\mbox{\boldmath $\sigma$}}
\def\bfnabla{\mbox{\boldmath $\nabla$}}

\newcommand{\cO}{{\cal O}}
\newcommand{\Slash}{\not\!}

\newcommand{\smvs}{\vbox{\vskip 8mm}}

\def\simg{{\ \lower-1.2pt\vbox{\hbox{\rlap{$>$}\lower6pt\vbox{\hbox{$\sim$}}}}\ }}
\def\siml{{\ \lower-1.2pt\vbox{\hbox{\rlap{$<$}\lower6pt\vbox{\hbox{$\sim$}}}}\ }}
\def\bfnabla{\mbox{\boldmath $\nabla$}}

\def\bfsigma{\mbox{\boldmath $\sigma$}}
\def\al{\alpha}
\def\lQ{\Lambda_{\rm QCD}}
\def\vs{V^{(0)}_s}

\newcommand{\vx}{\vec{x}}
\newcommand{\vr}{\vec{r}}
\newcommand{\tr}{{\rm Tr}} 
\newcommand{\sgh}{\sigma_H(p_0,\vec{p})}
\newcommand{\sg}{\sigma(\omega,\vec{p})}
\newcommand{\sgt}{\ensuremath{\sigma(\omega,T)}}
\newcommand{\tc}{\ensuremath{T_c}}
\newcommand{\grecon}{\ensuremath{G_{\rm recon}(\tau, T)}}
\newcommand{\ec}{\ensuremath{\eta_c}}
\newcommand{\ass}{\ensuremath{\chi_{c_0}}}
\newcommand{\asx}{\ensuremath{\chi_{c_1}}}
\def \ef    {\mbox{$E_{\scriptscriptstyle F}$}}
\def \nch   {\mbox{$N_{\scriptscriptstyle ch}$}}

\newcommand{\sabs}{$\sigma_{\rm abs}$}
\newcommand{\xf}{$x_{\rm F}$} 
\newcommand     \ber     {\begin{eqnarray}}
\newcommand     \eer    {\end{eqnarray}}

\begin{document}

\thispagestyle{empty}
\setlength{\unitlength}{1mm}
\begin{picture}(0.001,0.001)
%\graphpaper(0,-280)(210,290)
\put(125,8){CERN--2005--005}
\put(125,3){20 June 2005}
\put(-10,-50){\includegraphics[width=15cm]{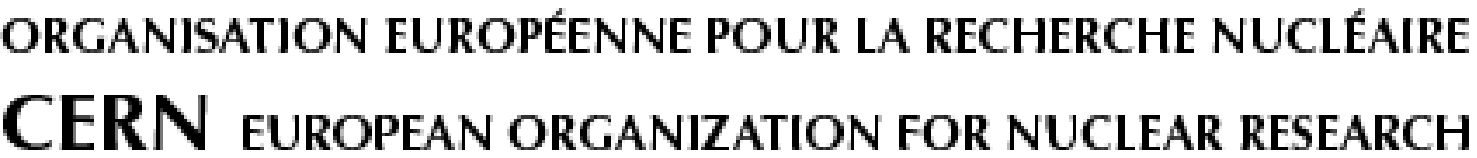}}
\put(-10,-100){\huge\bfseries HEAVY QUARKONIUM PHYSICS}
\put(115,-165){\Large\bfseries QWG Conveners}
\put(115,-173){\Large N.~Brambilla}
\put(115,-179){\Large M.~Kr\"amer}
\put(115,-185){\Large R.~Mussa}
\put(115,-191){\Large A.~Vairo}
\put(65,-250){\makebox(0,0){GENEVA}}
\put(65,-255){\makebox(0,0){2005}}
\end{picture}
\newpage
\thispagestyle{empty}
\mbox{}
\vfill
\begin{picture}(0.001,0.001)
%\put(80,-16){\makebox(0,0){CERN--XXX copies printed--July 2005}}
\end{picture}
\begin{center}
\end{center}

\newpage
\pagestyle{plain}
\pagenumbering{roman}
\setcounter{page}{3}
\begin{center}
\mbox{}\\[2mm]
\Huge  HEAVY QUARKONIUM PHYSICS\\[17mm]
\includegraphics*[scale=0.8]{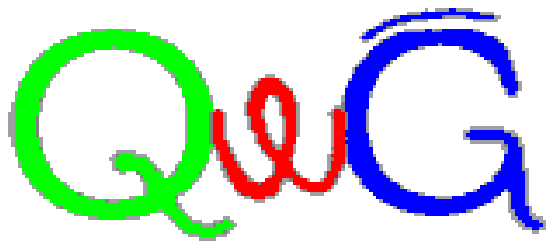}\\[13mm]
\end{center}
\begin{flushleft}
{\bf Authors:} 
\emph{QWG \& Topic conveners}:
N.~Brambilla$^{\,42}$, %Milano
M.~Kr\"amer${}^{16}$, %Edinburgh
R.~Mussa${}^{26}$, %INFN  Torino
A.~Vairo$^{\,42,36}$; \\ %Milano 
\emph{Topic Conveners}:
G.~Bali$^{\,19}$, %Glasgow;
G.~T.~Bodwin$^{\,1}$, %Argonne
E.~Braaten$^{\,45}$, %Ohio State
E.~Eichten$^{\,17}$, %Fermilab
S.~Eidelman$^{\,6}$, % novosibirsk 
S.~Godfrey$^{\,7}$, % Carleton
A.~Hoang$^{\,43}$, %Munich
M.~Jamin$^{\,44}$, %TU Munich
D.~Kharzeev$^{\,5}$, %bnl
M.~P.~Lombardo$^{\,24}$, %frascati
C.~Louren\c{c}o$^{\,11}$, %cern
A.~B.~Meyer$^{\,20}$, %IEP Hamburg 
V.~Papadimitriou$^{\,17,58}$, %Fermilab
C.~Patrignani$^{\,25}$, %INFN Genoa 
M.~Rosati$^{\,28}$, %iowastate
M.~A.~Sanchis-Lozano$^{\,62}$, %Valencia
H.~Satz$^{\,4}$, %bielefeld
J.~Soto$^{\,2}$; \\ %Barcelona
\emph{Contributors}:
D.~Z.~Besson$^{\,30}$, % U. Kansas 
D.~Bettoni$^{\,23}$, % INFN - Sezione di Ferrara
A.~B\"ohrer$^{\,55}$,  %Siegen
S.~Boogert$^{\,37}$, %U. College London
C.-H.~Chang$^{\,9,29}$, %CCAST Beijing and ITP Beijing
P.~Cooper$^{\,17}$, %Fermilab
P.~Crochet$^{\,13}$, %CNRS
S.~Datta$^{\,4}$, %bielefeld
C.~Davies$^{\,19}$, %Glasgow
A.~Deandrea$^{\,39}$, %U. de Lyon 
R.~Faustov$^{\,53}$, %Moscow  
T.~Ferguson$^{\,8}$, % carnegie mellon 
R.~Galik$^{\,14}$, % Cornell 
F.~A.~Harris$^{\,21}$, % U.Hawaii
O.~Iouchtchenko$^{\,11}$, %
O.~Kaczmarek$^{\,4}$, %bielefeld
F.~Karsch$^{\,4}$, %bielefeld
M.~Kienzle$^{\,18}$, %U. Geneva 
V.~V.~Kiselev$^{\,54}$, % Serpukhov, IHEP 
S.~R.~Klein$^{\,33}$, %lbnl
P.~Kroll$^{\,64}$, %U. Wuppertal 
A.~Kronfeld$^{\,17}$, %Fermilab 
Y.-P.~Kuang$^{\,61}$, % Tsinghua University
V.~Laporta$^{\,3}$, %bari
J.~Lee$^{\,32}$, %Argonne and KAIS Seoul
A.~Leibovich$^{\,49}$, %U. of Pittsburgh  
J.~P.~Ma$^{\,29}$, %itp beijing
P.~Mackenzie$^{\,17}$, %${}^{6}$ %Fermilab
L.~Maiani$^{\,50}$, %roma
M.~L.~Mangano$^{\,11}$, %CERN 
A.~Meyer$^{\,17}$, % Fermilab
X.~H.~Mo$^{\,22}$, %IHEP Beijing
C.~Morningstar$^{\,8}$, %Carnegie Mellon
A.~Nairz$^{\,11}$, % CERN
J.~Napolitano$^{\,51}$, %RPI
S.~Olsen$^{\,21}$, %U.Hawaii
A.~Penin$^{\,31}$, %U. Karlsruhe
P.~Petreczky$^{\,52}$, %rikenbnl
F.~Piccinini$^{\,47}$, %pavia
A.~Pineda$^{\,2}$, %Barcelona
A.~D.~Polosa$^{\,3,10}$, %bari,fermiroma
L.~Ramello$^{\,48}$, %alessandria
R.~Rapp$^{\,57}$, %Texas
J.~-M.~Richard$^{\,12}$, %Grenoble
V.~Riquer$^{\,11}$, %cern
S.~Ricciardi$^{\,38}$, % London Royal Holloway
E.~Robutti$^{\,25}$, %INFN Genova 
O.~Schneider$^{\,34}$, %Lausanne
E.~Scomparin$^{\,60}$, %torino
J.~Simone$^{\,17}$, %Fermilab,
T.~Skwarnicki$^{\,56}$ % Syracuse
G.~Stancari$^{\,17,23}$, %Fermilab+INFN Ferrara 
I.~W.~Stewart$^{\,41}$, %MIT
Yu.~Sumino$^{\,59}$, %Tohoku
T.~Teubner$^{\,35}$, %Liverpool
J.~Tseng$^{\,46}$, %Oxford
R.~Vogt$^{\,15,33}$, %Berkeley and Davies
P.~Wang$^{\,22}$, %IHEP Beijing
B.~Yabsley$^{\,63}$, % Virginia Tech
C.~Z.~Yuan$^{\,22}$, %IHEP Beijing
F.~Zantow$^{\,4}$, %bielefeld
Z.~G.~Zhao$^{\,40}$,  % U.of Michigan
A.~Zieminski$^{\,27}$\\[12mm]  %Indiana
${}^{1}$ 
HEP Division, Argonne National Laboratory, Argonne, Illinois, USA \\
${}^{2}$
Universitat de Barcelona, Barcelona, Catalonia, Spain\\
${}^{3}$
Universit\`{a} di Bari and INFN Bari, Italy \\
${}^{4}$
Facult\"at f\"ur Physik, Universit\"at Bielefeld, Germany\\
${}^{5}$
Brookhaven National Laboratory, Upton, New York, USA \\
${}^{6}$
Budker Institute, Novosibirsk, Russia \\
${}^{7}$
Carleton University, Ottawa,  Canada \\
${}^{8}$
Carnegie Mellon University, Pittsburgh, Pennsylvania, USA \\
${}^{9}$ 
CCAST (World Laboratory), Beijing, China \\
${}^{10}$
Centro Studi e Ricerche ``E. Fermi'', Roma, Italy \\
${}^{11}$
CERN, Geneva, Switzerland \\
${}^{12}$
CNRS-IN2P3-Universite Joseph Fourier, Grenoble, France \\
${}^{13}$
CNRS-IN2P3, France \\
${}^{14}$
Cornell University, Ithaca, New York, USA \\
${}^{15}$ 
University of California, Davis, California, USA\\
${}^{16}$ 
School of Physics, The University of Edinburgh, Edinburgh, United Kingdom; \\
\hspace*{3mm} now at Institut f\"ur Theoretische Physik E, RWTH Aachen, 
Aachen, Germany\\
${}^{17}$
Fermi National Accelerator Laboratory, Batavia, Illinois, USA\\
${}^{18}$
Geneva University, Geneva, Switzerland \\
${}^{19}$
Glasgow University, Glasgow, United Kingdom \\
${}^{20}$ 
Institut f\"ur Experimentalphysik, Universit\"at Hamburg, Hamburg, Germany\\
${}^{21}$
Hawaii University, Honolulu, USA \\
${}^{22}$ 
Institute of High Energy Physics, Chinese Academy of Sciences, Beijing, China \\
${}^{23}$
INFN, Ferrara, Italy \\
${}^{24}$
INFN, Frascati, Italy \\
${}^{25}$
INFN, Genova, Italy \\
${}^{26}$
INFN, Torino, Italy \\
${}^{27}$ 
Indiana University, Bloomington, Indiana, USA\\
${}^{28}$
Iowa State University, Ames, Iowa, USA\\
${}^{29}$ 
Institute of Theoretical Physics, Chinese Academy of Sciences, Beijing, China\\
${}^{30}$ 
University of Kansas, Lawrence, USA \\
${}^{31}$ 
Institut f\"ur Theoretische Teilchenphysik,
Universit\"at Karlsruhe,  Karlsruhe, Germany\\
${}^{32}$ 
Department of Physics, Korea University, Seoul, Korea\\
${}^{33}$ 
Laurence Berkeley National Laboratory, Berkeley, California, USA\\
${}^{34}$ 
Ecole Polytechnique F\'ed\'erale de Lausanne, Lausanne, Switzerland\\
${}^{35}$ 
University of Liverpool,
United Kingdom \\
${}^{36}$ 
CFIF, Instituto Superior Tecnico, Lisbon, Portugal\\
${}^{37}$ 
University College London, London, 
United Kingdom \\
${}^{38}$
Univ. of London, Royal Holloway, United Kingdom \\
${}^{39}$
Universit\'e de Lyon 1, 
Villeurbanne, France\\
${}^{40}$
Univ. of Michigan, Ann Arbor, Michigan, USA \\
${}^{41}$ 
Massachusetts Institute of Technology, Cambridge, USA \\
${}^{42}$
Universit\`a di Milano \& INFN Milano, Italy \\
${}^{43}$ 
Max-Planck-Institut f\"ur Physik, Munich, Germany \\
${}^{44}$ 
Physik-Department, TU M\"unchen, Garching, Germany \\
${}^{45}$ 
Ohio State University, Columbus, Ohio, USA\\
${}^{46}$
Oxford University, Oxford, United Kingdom \\
${}^{47}$
Universit\`{a} di Pavia \& INFN Pavia, Italy \\
${}^{48}$
Universit\`{a} del Piemonte Orientale, Alessandria, Italy \\
${}^{49}$
University of Pittsburgh, Pittsburgh, Pennsylvania, USA\\
${}^{50}$
Universit\`{a} di Roma ``La Sapienza'' \& INFN Roma, Italy \\
${}^{51}$
RPI, Troy, New York, USA \\
${}^{52}$
Physics Department and RIKEN-BNL, Brookhaven National Laboratory, 
Upton, New York, USA \\
${}^{53}$
Russian Academy of Science, Scientific Council for Cybernetics, Moscow, Russia\\
${}^{54}$
Russian State Research Center ``Institute for High Energy Physics'', 
Protvino, \\\hspace*{3mm} Moscow Region, Russia \\
${}^{55}$
Siegen University, Siegen, Germany \\
${}^{56}$
Syracuse University, Syracuse, New York, USA \\
${}^{57}$
Texas A\& M University, Texas, USA \\
${}^{58}$
Texas Tech University, Lubbock, Texas, USA \\
${}^{59}$
Tohoku University, Tohoku, Japan \\
${}^{60}$
Universit\`{a} di Torino \& INFN, Torino, Italy \\
${}^{61}$
Tsinghua University, Beijing, People's Republic of China\\
${}^{62}$
Instituto de F\'{\i}sica Corpuscular (IFIC) and Departamento de F\'{i}sica Te\'orica,
Centro Mixto \\\hspace*{3mm} Universidad de Valencia-CSIC, Valencia, Spain\\
${}^{63}$
Virginia Tech University, Virginia, USA \\
${}^{64}$
Universit\"at Wuppertal, Wuppertal, Germany
\end{flushleft}

\newpage 
%10/12/2004
%30 May 2005 MG
%%%%%%%%%%%%%%%%%%%%%%%%%%%%%%%%%%%%%%%%%%%%%%%%%%%%%%%%%%%%
\begin{center}
\mbox{}\\[4cm]
\large\bfseries\textsc{Abstract}\\[10mm]
\end{center}

%%%%%%%%%%%%%%%%%%%%%%%%%%%%%%%%%%%%%%%%%%%%%%%%%%%%%%%%%%%%

\noindent 
This report is the result of the collaboration and research effort of
the Quarkonium Working Group over the last three years. It provides a
comprehensive overview of the state of the art in heavy-quarkonium
theory and experiment, covering quarkonium spectroscopy, decay, and
production; the determination of QCD parameters from quarkonium
observables; quarkonia in media; and the effects on quarkonia of
physics beyond the Standard Model. An introduction to common
theoretical and experimental tools is included. Future opportunities
for research in quarkonium physics are also discussed.

\newpage
%9/12/2004
%30 May 2005 MG
%%%%%%%%%%%%%%%%%%%%%%%%%%%%%%%%%%%%%%%%%%%%%%%%%%%%%%%%%%%%
\begin{center}
\large\bfseries\textsc{Foreword}\\[3mm]
\end{center}

%%%%%%%%%%%%%%%%%%%%%%%%%%%%%%%%%%%%%%%%%%%%%%%%%%%%%%%%%%%%
\noindent
As the community of high-energy physicists impatiently awaits the
startup of the LHC and the opening of the new energy frontier, it is
very welcome news that so much challenging and exciting data are
constantly being produced in the field of quarkonium physics. The
proliferation of puzzling measurements has led over the past several
years to new challenges for the theorists, requiring the introduction
of new ideas, and providing new probes for the understanding of QCD at
its deeper levels.

Ten years ago, reports by the CDF Collaboration signalled the end of an
era in quarkonium physics, but at the same time opened new windows on
this field, which contributed so much to the development of QCD. The
observation of the top quark with a mass of about 175~GeV closed all
hopes of including toponium in the family of clean and useful
quarkonium states.  In parallel, the observation of an excess in
charmonium production by orders of magnitude over what was predicted
in the then available theoretical models gave birth to the modern
theoretical understanding of charmonium production. Since then, in
addition to successful explanations, a large set of puzzles kept being
generated by data obtained at the Tevatron, at HERA, and in low-energy
$\mathrm{e^+e^-}$ colliders: the apparent violation of universality
emerging when comparing data from the hadron and the $\mathrm{ep}$
colliders, the poor agreement (at the limit of inconsistency!) between
the predictions for the polarization of the $\mathrm{J/\psi}$ produced
in hadronic collisions and the actual data, the excess of double
charmonium production first observed by Belle. The solution to these
puzzles still remains to be found, as new data keep pouring in.

But the surprises and advances have not been limited to the complex
issue of the production mechanisms. The spectroscopy of quarkonium has
also received challenging inputs from the observation of new narrow
states, whose understanding requires an added dose of sophistication
in the theory, together perhaps with the need for inclusion of more
exotic patterns of bound states (hybrids, molecules,
tetraquarks). Progress in lattice calculations and effective field
theories has turned quarkonium physics into a powerful tool to measure
the mass of the heavy quarks and the strength of the QCD coupling,
providing accuracies comparable to or better than those allowed by any
other technique. The properties of production and absorption of
quarkonium in a nuclear medium are beginning to provide quantitative
inputs for the study of QCD at high density and temperature, giving a
unique experimental test bed for analytical and lattice studies.

The interplay of solid theoretical work and of accurate and versatile
experimental techniques has brought quarkonium physics to a
renaissance, with a flourishing of activity second only to the golden
age which followed the discovery of charmonium almost 30 years ago.
The appearance of this CERN Report, which documents the state of the
art through the contributions of the leaders in the field, represents
therefore a timely and much needed publication. The inclusion of both
the theoretical and experimental perspectives leads to a precious
resource for the active researcher, as well as for the young newcomers
to the field.

I am happy  to praise the organizers of  the Quarkonium Working Group,
the conveners and  all the participants, who have  worked so hard over
the past couple of years to produce this Report, which will provide an
essential guide  to this ever-exciting  area of research for  years to
come.

\vskip 2truecm
\centerline{Michelangelo Mangano}
\centerline{CERN PH Department}

\newpage 
%10/12/2004
%30 May 2005 MG
%%%%%%%%%%%%%%%%%%%%%%%%%%%%%%%%%%%%%%%%%%%%%%%%%%%%%%%%%%%%
\begin{center}
\large\bfseries\textsc{Preface}\\[3mm]
\end{center}

%%%%%%%%%%%%%%%%%%%%%%%%%%%%%%%%%%%%%%%%%%%%%%%%%%%%%%%%%%%%
\noindent
On the eve of the startup of the LHC and the search for new physics
beyond the Standard Model at energy scales of several TeV, there is
still a sector of the Standard Model that evades our control: the
sector of strongly interacting particles, i.e. quarks and gluons.  We
believe we have the field theory that describes strong interaction,
QCD, but we are not yet able to extract from it in a controlled way a
great part of the hadron properties. These same hadron properties
obviously play a relevant role in many searches for new physics and
new phenomena, CP violation being a strong case in hand.  At the LHC
hadron processes will again take the stage.  It is, therefore,
relevant to get hold of the strong sector of the Standard Model.  For
several reasons heavy quarkonium offers a unique opportunity in this
direction. Quarkonium systems may be crucially important to improve
our understanding of QCD. They probe all the energy regimes of QCD,
from the hard region, where an expansion in the coupling constant is
possible, to the low-energy region, where nonperturbative effects
dominate. Heavy-quark--antiquark bound states are thus an ideal, and
to some extent, unique laboratory where our understanding of
nonperturbative QCD and its interplay with perturbative QCD may be
tested in a controlled framework.

Moreover, in the last few years a wealth of new experimental results
have become available.  The diversity, quantity and accuracy of the
data currently being collected is impressive and includes:

\begin{itemize}
\item data on quarkonium formation  from BES at BEPC, E835 at Fermilab,
      KEDR (upgraded) at VEPP-4M, and CLEO~III at CESR;

\item clean samples of charmonia produced in B-decays, in photon--photon
      fusion and in initial-state radiation from the B-meson factory
      experiments BaBar at SLAC and Belle at KEK, including the
      unexpected observation of associated
      $\mathrm{(c\overline{c})(c\overline{c})}$ production;

\item heavy quarkonia production from gluon--gluon fusion in 
      $\mathrm{p\bar{p}}$ annihilations at 2~TeV from the CDF and D0
      experiments at Fermilab, including the first observation of
      $\mathrm{B_c}$ candidates;

\item charmonia production in photon--gluon fusion from the ZEUS and H1
      experiments at DESY;

\item charmonia production in heavy-ion collisions from the PHENIX and STAR
      experiments at RHIC, and the NA60 experiment at CERN.

\end{itemize}

These experiments may operate as heavy quarkonium factories, producing
quarkonium states in large amounts. If properly analysed and
interpreted, the data can lead to surprising results and major
progress in our understanding of QCD. This is exemplified by the very
recent discovery of a new unexpected narrow charmonium state,
temporarily labelled $X(3872)$, which was announced by the Belle
Collaboration at the Lepton--Photon Conference 2003 and confirmed
within a month by the CDF Collaboration at Fermilab, during the 2nd
QWG Workshop.

In the near future, even larger data samples are expected from the
CLEO-c and BES~III upgraded experiments, while the B factories, the
Fermilab Tevatron, and the DESY experiments will continue to supply
valuable data for several years.  New facilities will become
operational (LHC at CERN, Panda at GSI, much-higher-luminosity B
factories at KEK and SLAC, a Linear Collider, etc.) offering fantastic
challenges and opportunities, which we must start facing today.
Considerable efforts are also being made to study deconfined quark
matter, at SPS, RHIC and LHC energies, for which heavy quarkonium is
among the most crucial probes. The complexity of these studies
requires a close communication and the exchange of ideas between
experts in quarkonium physics and heavy-ion collisions.

Effective field theories, such as Nonrelativistic QCD (NRQCD), provide
new tools and definite predictions concerning, for instance,
heavy-quarkonium production and decays.  New effective field theories
for heavy quarkonium, as potential NRQCD (pNRQCD) and velocity NRQCD
(vNRQCD), have recently been developed and are producing a wealth of
new results.  The lattice implementation of such effective theories
has been partially carried out and many more results with drastically
reduced systematic uncertainties are expected in the near future.  The
progress in the understanding of non-relativistic effective field
theories makes it possible to go beyond phenomenological models and,
for the first time, face the possibility of providing a unified
description of all aspects of heavy-quarkonium physics.  This allows
us to use quarkonium as a benchmark for our understanding of QCD, for
the precise determination of relevant Standard Model parameters
(\eg heavy quark masses, $\alpha_{\rm s}$), and for new physics
searches.

It is crucial, now, to ensure an efficient communication between
experimentalists and theorists, within the broad quarkonium physics
community. This has been the main motivation for the creation of an
international research collaboration, the Quarkonium Working Group,
which constitutes the support platform of this CERN Report (see also
\url{http://www.qwg.to.infn.it})

The aim of the QWG is essentially twofold.  First, to guarantee an
intense and efficient exchange of results and ideas between
experimentalists and theorists, now that many new measurements are
becoming available.  Second, to overcome the dispersal of the research
in this field and jointly study the different approaches and
techniques, by establishing new collaborations and improving existing
ones. The concrete goals are:

\begin{itemize}
\item to achieve a better understanding of the dynamics of the 
      strong interaction and of strongly coupled theories, using
      quarkonium systems;

\item to gain detailed knowledge of the physics of 
      confinement/deconfinement;

\item to improve the determination of the fundamental parameters of
      the Standard Model and constrain the allowed parameter space for
      new physics;

\item to identify missing experimental information required to 
      improve our understanding of QCD, and to identify theoretical
      calculations needed for the interpretation of current and future
      experiments;

\item to make this information available to people working in  
      related fields.

\end{itemize}

This CERN Report presents the state of the art in heavy-quarkonium
physics at the end of 2004 and is a first step to achieving the goals
of the QWG.  The Report includes experimental and theoretical results
by different approaches and different communities (high-energy,
perturbative, lattice, nuclear, \etc) in a common language.  The
progress in the field and the impact of such progress on other areas
are presented, open problems and outstanding puzzles are discussed,
and the future opportunities of this field are outlined.

Given the richness of the physics involved in the project, the
research goals have been pursued by specifying seven main topics 
organized by theoretical and experimental topic conveners: 

\begin{itemize}
\item Quarkonium spectroscopy 
      [Conveners: G.~Bali, N.~Brambilla, J.~Soto (TH); R.~Mussa (EXP)];

\item Quarkonium decays 
      [Conveners: E.~Eichten, A.~Vairo (TH); C.~Patrignani (EXP)];

\item Quarkonium production 
      [Conveners: G.~Bodwin, E.~Braaten, M.~Kr\"amer (TH);
      A.~B.~Meyer, V.~Papadimitriou (EXP)];

\item Precision determination of Standard Model parameters 
      [Conveners: A.~Hoang, M.~Jamin (TH); S.~Eidelman (EXP)];

\item Quarkonium in media 
      [Conveners: D.~Kharzeev, M.~P.~Lombardo, H.~Satz (TH);
      C.~Louren\c{c}o, M.~Rosati (EXP)];

\item Beyond the Standard Model [Convener: M.~A.~Sanchis-Lozano (TH)];

\item Future opportunities [Conveners: S.~Godfrey, M.~A.~Sanchis-Lozano (TH)].
\end{itemize}

The Quarkonium Working Group was initiated in 2002 by Nora Brambilla,
Roberto Mussa and Antonio Vairo, who were, shortly afterwards, joined
by Armin B\"ohrer and Michael Kr\"amer as the QWG conveners team.
Most of the topic conveners listed above belong to the initial group
of people who supported the QWG and contributed to its research
programme. The CERN TH Division and CERN, and especially Michelangelo
Mangano, have played an important role in the history of the QWG, by
hosting the first QWG meeting and by supporting the enterprise of
compiling the CERN Report.

The QWG has organized three international meetings, which were held at
CERN (2002), Fermilab (2003) and IHEP Beijing (2004). Approximately
250 theoretical and experimental physicists participated in the
meetings.  The organizers, participants, and supporting institutions
are listed below. The third meeting was preceded by the first QWG
graduate school organized at the ITP Beijing with about 100
participating graduate students.

We would like to express here our sincerest thanks to all those who
have contributed to this enterprise and made this document possible,
in particular the topic conveners and the organizers and participants
of the three QWG meetings. We also gratefully acknowledge the support
from the institutions that hosted the QWG meetings.  Finally, we would
like to express our deepest thanks to Armin B\"ohrer who was of key
relevance at the start of the QWG by producing and hosting in
Siegen the first QWG Web page, designing the QWG logo, participating
in the organization of the first two QWG workshops, and supporting in
all ways the development of the QWG.  We also thank E.~Berger,
D.~Kharzeev and A.~Zieminski for having been topical conveners of a
topical section later absorbed by other ones.

As of September 2004, Vaia Papadimitriou joined the QWG conveners
team.  As of December 2004, Aldo Deandrea and Xiaoyan Shen agreed to
join the topical conveners team.

The Quarkonium Working Group has very quickly coalesced into an
active, international community of physicists working and
collaborating on quarkonium physics, QCD, and the related impact on
the Standard Model and physics beyond the Standard Model. Given the
continuous flux of data and the order-of-magnitude(s) improvement in
the statistical analysis coming and expected to come from present and
future accelerator experiments, this promises to remain a very rich
research area for several years to come.  To fully benefit from it, we
believe it is important that the community of physicists working in
the field maintains a common area of discussion, transcending
individual experimental and theoretical collaborations.  It is our
hope that this CERN Report will provide a basis for such future
developments.

\vskip 2truecm
\centerline{The QWG Conveners}
\centerline{Nora Brambilla, Michael Kr\"amer, Roberto Mussa, Antonio Vairo}

\newpage

\begin{flushleft}
\textbf{1st International Workshop on Heavy Quarkonium}\\[1mm]
CERN, 8--10~November 2002\\[1mm]
\url{http://www.qwg.to.infn.it/WS-nov02/index.html}\\
\emph{Organizing Committee: A.~B\"ohrer, N.~Brambilla, O.~Iouchtchenko,
M.~Kienzle, M.~Kr\"amer, C.~Louren\c{c}o, M.~L.~Mangano, R.~Mussa,
T.~Teubner, A.~Vairo}\\[1mm] 
\emph{Speakers:} G.~Bali, E.~L.~Berger,
G.~Bodwin, A.~B\"ohrer, E.~Braaten, N.~Brambilla, H.~Castilia-Valdez,
C.~Davies, A.~Deandrea, E.~Eichten, S.~Eidelman, R.~Faustov,
T.~Ferguson, M.~Graham, S.~Godfrey, F.~A.~Harris, A.~Hoang,
O.~Iouchtchenko, F.~Karsch, V.~V.~Kiselev, B.~Kniehl, M.~Kr\"amer,
M.-P.~Lombardo, C.~Louren\c{c}o, J.~P.~Ma, M.~L.~Mangano, A.~Meyer,
R.~Mussa, S.~Necco, V.~Papadimitriou, A.~Penin, K.~Peters,
P.~Petreczky, A.~Pich, A.~Pineda, A.~Polleri, M.~Rosati,
M.~A.~Sanchis-Lozano, H.~Satz, J.~Simone, G.~Stancari, I.~W.~Stewart,
Y.~Sumino, T.~Teubner, J.~Tseng, A.~Vairo, R.~Waldi, B.~Yabsley\\[1mm]
{\it Supported by the CERN TH Division}\\[1mm] 
Agenda and slides
available at \url{http://www.qwg.to.infn.it/WS-nov02/WSagenda.html}\\
List of participants available on website.\\[5mm]
\textbf{2nd International Workshop on Heavy Quarkonium}\\[1mm]
Fermilab, 20--22~September 2003\\[1mm]
\url{http://www.qwg.to.infn.it/WS-sep03/index.html}\\
\emph{Organizing Committee: G.~Bodwin, A.~B\"ohrer,
N.~Brambilla, E.~Eichten, V.~Jain, M.~Kr\"amer, R.~Mussa,
V.~Papadimitriou, S.~Pordes, A.~Vairo}\\[1mm] 
{\it Speakers:}
J.~Appel, G.~Bali, G.~Bauer, D.~Bettoni, E.~Braaten,
N.~Brambilla, K.~T.~Chao, P.~Cooper, A.~Deandrea, E.~Eichten, A.~El-Khadra,
J.~Erler, R.~Faustov, T.~Ferguson, S.~Fleming, R.~Galik, S.~Godfrey, 
Y.~Gotra, Z.~J.~Guo, F.~A.~Harris, T.~Hatsuda, A.~Hoang, J.~Huang, U.~Husemann,
V.~Jain, M.~Jamin, S.~Kelly, D.~Kharzeev, S.~R.~Klein, 
J.~Lee, A.~Leibovich, P.~Mackenzie, A.~B.~Meyer,
C.~Morningstar, H.~Muramatsu, R.~Mussa, A.~Nairz, J.~Napolitano,
V.~Papadimitriou, C.~Patrignani, J.~C.~Peng, P.~Petreczky, S.~Pordes,
J.~W.~Qiu, R.~Rapp, S.~Ricciardi, J-M.~Richard, E.~Robutti, M.~Rosati,
M.~A.~Sanchis-Lozano, H.~Sato, X.~Shen, J.~Simone, R.~Spighi,
L.~Stanco, R.~Thews, T.~Umeda, A.~Vairo, R.~Vogt, M.~Voloshin,
C.~Y.~Wong, W.~Xie, B.~Yabsley, C.~Z.~Yuan, G.~Zanderighi, Z.G.~Zhao\\[1mm]
\emph{Supported by Fermilab}\\[1mm]
Agenda and slides available at 
\url{http://www.qwg.to.infn.it/WS-sep03/WSagenda.html}\\
List of participants available on website.\\
\newpage
\textbf{3rd International Workshop on Heavy Quarkonium}\\[1mm] 
IHEP Beijing, 12--15~October 2004\\[1mm] 
\url{http://www.qwg.to.infn.it/WS-oct04/index.html}\\
\emph{Organizing Committee: N.~Brambilla, K.~T.~Chao, A.~Deandrea, 
M.~Kr\"amer, Y.~P.~Kuang, W.~G.~Li, J.~P.~Ma, R.~Mussa, V.~Papadimitriou, 
C.~F.~Qiao, X.~Y.~Shen, A.~Vairo, C.~Z.~Yuan}\\[1mm] 
\emph{Speakers:} G.~Bali, S.~Baranov, D.~Bernard, G.~Bodwin,
E.~Braaten, N.~Brambilla, R.~Brugnera, C.-H.~Chang, K.-T.~Chao,
H.-S.~Chen, Y.-Q.~Chen, C.-H.~Ching, A.~Deandrea, E.~Eichten,
G.~Feild, C.~Gao, M.~Garcia-Perez, Z.~Guo, L.~Han, B.~Heltsley, P.~Ko,
J.~Lee, T.~Lee, Z.~Liu, J.-P.~Ma, T.~Mehen, R.~Mussa, S.~L.~Olsen,
C.~Patrignani, K.~Peters, P.~Petreczky, O.~Philipsen, A.~Pineda,
C.-F.~Qiao, M.~Rosati, G.~Rong, M.~A.~Sanchis-Lozano, X.~Y.~Shen,
J.~Soto, Y.~Sumino, A.~Tomaradze, A.~Vairo, J.~X.~Wang, P.~Wang,
Y.~L.~Wang, Z.~Wang, C.~Y.~Wong, J.~L.~Wu, B.~Zhang, D.~Zhang,
S.-L.~Zhu, N.~Zhou\\[1mm]
\emph{Supported by the Chinese Center for Advanced Science and
Technology (CCAST), the Institute of Theoretical Physics (ITP), Beijing, the
Institute of High Energy Physics (IHEP), Beijing, the Nature Science
Foundation of China (NSFC), Peking University, and Tshinghua University}\\[1mm]
Agenda and slides available at 
\url{http://www.qwg.to.infn.it/WS-oct04/WSagenda.html}\\
List of participants available on the website.\\[5mm]
\textbf{1st International School  of the QWG}\\[1mm] 
\emph{Topical Seminar School on Heavy Quarkonia at Accelerators: 
New Theoretical Tools and Experimental Techniques}\\
ITP, Beijing, 8--11~October 2004\\[1mm]
\url{http://www.qwg.to.infn.it/TS-oct04/index.html}\\
\emph{Organizing Committee: N.~Brambilla, K.~T.~Chao, 
M.~Kr\"amer, Y.~P.~Kuang, W.~G.~Li, J.~P.~Ma, M.~L.~Mangano, 
R.~Mussa, V.~Papadimitriou, 
C.~F.~Qiao, X.~Y.~Shen, A.~Vairo, C.~Z.~Yuan}\\[1mm]
\emph{Lecturers:} \\[1mm]
~~~\begin{tabularx}{.92\linewidth}{@{}lX}
G.~Bodwin:& \emph{Introduction to NRQCD and quarkonium production};\\
G.~Bali:  & \emph{Quarkonium and Exotics};\\ 
N.~Brambilla: &
\emph{Introduction to pNRQCD, quarkonium spectroscopy and decays};\\
B.~Heltsley: & \emph{Crystal Calorimetry in modern particle physics
detectors}; \\
M.~Laine: & \emph{Introduction to quarkonium at high temperatures}; \\
R.~Mussa: & \emph{Angular Distributions in Helicity Formalism}; \\
S.~Nikitin: & \emph{Resonant depolarization technique for
high precision mass measurement at $e^+e^-$ storage rings};\\
C.~Patrignani: & \emph{How an experimental measure (quarkonium) enters
the data particle}; \\
M.~Rosati: & \emph{Quarkonium and Heavy Ion Experiments}; \\
J.~Soto: & \emph{Introduction to EFTs. Renormalization
Group in NRQCD and pNRQCD; SCET and heavy quarkonium}.
\end{tabularx}\\[1mm]
\emph{Supported by the Chinese Center for Advanced Science and
Technology (CCAST), the Institute of Theoretical Physics (ITP), Beijing, the
Institute of High Energy Physics (IHEP), Beijing, the Nature Science
Foundation of China (NSFC), Peking University, and Tshinghua University}\\[1mm]
\emph{Sponsored by CERN and Fermilab}\\[1mm]
Agenda and slides available at 
\url{http://www.qwg.to.infn.it/TS-oct04/TSAgenda.html}
\end{flushleft}

\cleardoublepage
\begingroup\baselineskip.98\baselineskip
\tableofcontents\endgroup
\cleardoublepage
\setcounter{page}{0}
\pagenumbering{arabic}
\pagestyle{fancy}

%6/12/2004

\chapter {COMMON THEORETICAL TOOLS}
\label{chapter:commontheoreticaltools}
{\it Authors:} G. Bali, N.~Brambilla, J. Soto, A.Vairo

\section[QCD]{QCD
$\!$\footnote{Author: J.~Soto}} 
\label{sec:unoqcd}

Quantum Chromodynamics (QCD) \cite{Fritzsch:1973pi} is the sector of
the Standard Model (SM) which is relevant for the strong
interactions. It is obtained from the full SM by setting the weak and
electromagnetic coupling constants to zero and freezing the scalar
doublet to its vacuum expectation value. What remains is a Yang--Mills
(YM) theory with local gauge group $SU(3)$ (colour) vectorially coupled
to six Dirac fields (quarks) of different masses (flavours). The vector
fields in the YM Lagrangian (gluons) live in the adjoint
representation and transform like connections under the local gauge
group whereas the quark fields live in the fundamental representation
and transform covariantly. The QCD Lagrangian reads
\begin{equation}
{\cal L}_{\rm  QCD}=
-\frac{1}{4}F^a_{\mu\nu}F^{a\;\mu\nu}
+\sum_{\{q\}}\bar q\left( i\gamma^\mu D_\mu -m_q\right) q\,,
\end{equation}
where $\{q\}=u,d,s,c,b,t$, $\; F^a_{\mu\nu}=\partial_\mu A_\nu^a -
\partial_\nu A_\mu^a + g f^{abc}A_\mu^b A_\nu^c$, $\;
D_\mu=\partial_\mu-iT^aA_\mu^a$.  $\; f^{abc}$ are the $SU(3)$
structure constants and $T^a$ form a basis of the fundamental
representation of the $SU(3)$ algebra.  When coupled to
electromagnetism, gluons behave as neutral particles whereas $u$, $c$
and $t$ quarks have charges $+2/3$ and $d$, $s$ and $b$ quarks have
charges $-1/3$.

The main properties of QCD follow:

\begin{itemize}
\item It is Poincar\'e, parity, time reversal and (hence) charge
      conjugation invariant. It is in addition invariant under
      $U(1)^6$ which implies individual flavour conservation.

\item Being a non-Abelian gauge theory, the physical spectrum consists
      of colour singlet states only. The simplest of these states have
      the quantum numbers of quark--antiquark pairs (mesons) or of
      three quarks (baryons), although other possibilities are not
      excluded.

\item The QCD effective coupling constant $\als (q)$ decreases as the
      momentum transfer scale $q$ increases (asymptotic freedom)
      \cite{Gross:1973id,Politzer:1973fx}.  This allows to make
      perturbative calculations in $\als$ at high energies.

\item At low energies it develops an intrinsic scale (mass gap),
      usually referred as $\Lambda_{\rm QCD}$, which provides the main
      contribution to the masses of most hadrons.  At scales $q\sim
      \Lambda_{\rm QCD}$, $\als (q) \sim 1$ and perturbation theory
      cannot be used.  Investigations must be carried out using
      nonperturbative techniques, the best established of which is
      lattice QCD.

      Quarks are conventionally divided into light $m_q \ll
      \Lambda_{\rm QCD}$, $\; q=u,d,s$ and heavy $m_Q \gg \Lambda_{\rm
      QCD}$, $\; Q=c,b,t$
\begin{eqnarray}
&& m_u=1.5 \div 4.0\, {\rm MeV} \,  ,\quad m_d=4 \div 8\, {\rm MeV}
 \, ,\quad m_s=80 \div 130\, {\rm MeV}\,,
\nonumber
\\  && \\ &&  m_c=1.15 \div 1.35\, {\rm GeV}  \, ,\quad  m_b=4.1 \div 4.4\,
 {\rm GeV}  \, ,\quad m_t=174.3\pm 5.1\, {\rm GeV} \,.
\nonumber
\end{eqnarray}
      These are $\overline{\rm MS}$ masses at scale $2$~GeV, $m_c$ and
$m_b$ for the light quarks, charm and bottom respectively.  All values
are taken from \cite{Eidelman:2004wy}.  The extraction of the values
of the heavy quark masses will be discussed in
\Chapter~\ref{chapter:precisiondeterminations}.

\item If light quark masses are neglected, the $U(1)^ 3$ flavour
      conservation symmetry of the QCD Lagrangian in this sector is
      enlarged to a $U(3)\otimes U(3)$ group. The axial $U(1)$
      subgroup is explicitly broken by quantum effects (axial
      anomaly). The vector $U(1)$ subgroup provides light flavour
      conservation. The remaining $SU(3)\otimes SU(3)$ subgroup, known
      as chiral symmetry group, turns out to be spontaneously broken
      down to the diagonal $SU(3)$ (flavour symmetry).  This produces
      eight Goldstone bosons, which, upon taking into account the
      explicit breaking of the symmetry due to the non-zero quark
      masses, acquire masses that are much smaller than $\Lambda_{\rm
      QCD}$.

\item Hadrons containing heavy quarks have masses of the order of
      $m_Q$ rather than of the order $\Lambda_{\rm QCD}$. They enjoy
      particular kinematical features that allow for specific
      theoretical treatments.  The study of hadrons containing two
      heavy quarks is the aim of this report.
\end{itemize}

\section[Effective field theories]
        {Effective field theories
        $\!$\footnote{Authors:  N.~Brambilla, A.~Vairo}} 
\label{sec:unoeft}

From the point of view of QCD the description of hadrons containing
two heavy quarks is a rather challenging problem, which adds to the
complications of the bound state in field theory those coming from a
nonperturbative low-energy dynamics.  A proper relativistic treatment
of the bound state based on the Bethe--Salpeter equation
\cite{Salpeter:1951sz} has proved difficult.  Perturbative
calculations have turned out unpractical at higher order and the
method has been abandoned in recent QED calculations.  Moreover, the
entanglement of all energy modes in a fully relativistic treatment is
more an obstacle than an advantage for the factorization of physical
quantities into high-energy perturbative and low energy
nonperturbative contributions. Partial semirelativistic reductions and
models have been often adopted to overcome these difficulties at the
price to introduce uncontrolled approximations and lose contact with
QCD.  The fully relativistic dynamics can, in principle, be treated
without approximations in lattice gauge theories. This is in
perspective the best founded and most promising approach. As we will
detail in the following, it is not without difficulties at the present
for heavy quarkonium.

A nonrelativistic treatment of the heavy quarkonium dynamics, which is
suggested by the large mass of the heavy quarks, has clear advantages.
The velocity of the quarks in the bound state provides a small
parameter in which the dynamical scales may be hierarchically ordered
and the QCD amplitudes systematically expanded. Factorization formulas
become easier to achieve.  A priori we do not know if a
nonrelativistic description will work well enough for all heavy
quarkonium systems in nature. For instance, the charm quark may not be
heavy enough. The fact that most of the theoretical predictions
presented in the report are based on such a nonrelativistic assumption
and the success of most of them may be seen as a support to the
assumption.

\longpage

On the example of positronium in QED, a nonrelativistic bound state is
characterized by at least three scales: the scale of the mass $m$
(called hard), the scale of the momentum transfer $p \sim mv$ (soft)
and the scale of the kinetic energy of the quark and antiquark in the
centre-of-mass frame $E \sim p^2/m \sim mv^2$ (ultrasoft).  The scales
$mv$ and $mv^2$ are dynamically generated, $v$ is the heavy-quark
velocity in the centre-of-mass frame.  In a nonrelativistic system: $v
\ll 1$, and the above scales are hierarchically ordered: $m \gg m v
\gg mv^2$.  In perturbation theory $v\sim \als$. Feynman diagrams will
get contributions from all momentum regions associated with the
scales.  Since these momentum regions depend on $\als$ each Feynman
diagram contributes to a given observable with a series in $\als$ and
a non trivial counting.  For energy scales close to $\lQ$ perturbation
theory breaks down and one has to rely on nonperturbative methods.
The wide span of energy scales involved makes also a lattice
calculation in full QCD extremely challenging since one needs a
space--time grid that is large compared to the largest length of the
problem, $1/mv^2$, and a lattice spacing that is small compared to the
smallest one, $1/m$.  To simulate, for instance, a $b\bar{b}$ state
where $m/mv^2\sim 10$, one needs lattices as large as $100^4$, which
are beyond present computing capabilities \cite{Thacker:1990bm} (see
also the next sections of the chapter).

We may, however, also take advantage of the existence of a hierarchy
of scales by substituting QCD with simpler but equivalent Effective
Field Theories (EFTs).  EFTs have become increasingly popular in
particle physics during the last decades.  They provide a realization
of Wilson renormalization group ideas and fully exploit the properties
of local quantum field theories.  An EFT is a quantum field theory
with the following properties: a) it contains the relevant degrees of
freedom to describe phenomena that occur in certain limited range of
energies and momenta and b) it contains an intrinsic energy scale
$\Lambda$ that sets the limit of applicability of the EFT. The
Lagrangian of an EFT is organized in operators of increasing
dimension, hence, an EFT is in general non-renormalizable in the usual
sense. In spite of this, it can be made finite to any finite order in
$1/\Lambda$ by renormalizing (matching) the constants (matching
coefficients) in front of the operators in the Lagrangian until that
order. This means that one needs more renormalization conditions when
the order in $1/\Lambda$ is increased.  However, even if the only way
of fixing the constants would be by means of experimental data, this
would reduce but not spoil the predictive power of the EFT. If the
data are abundant, the constants can be fit once for ever and used
later on to make predictions on new experiments.

The prototype of EFT for heavy quarks is the Heavy Quark Effective
Theory (HQET), which is the EFT of QCD suitable to describe systems
with only one heavy quark \cite{hqet}. These systems are characterized
by two energy scales: $m$ and $\lQ$. HQET is obtained by integrating
out the scale $m$ and built as a systematic expansion in powers of
$\lQ/m$. As discussed above, bound states made of two heavy quarks are
characterized by more scales.  Integrating out only the scale $m$,
which for heavy quarks can be done perturbatively, leads to an EFT,
Nonrelativistic QCD (NRQCD)
\cite{Caswell:1985ui,Thacker:1990bm,Bodwin:1994jh}, that still
contains the lower scales as dynamical degrees of freedom.
Disentangling the remaining scales is relevant both technically, since
it enables perturbative calculations otherwise quite complicate, and
more fundamentally, since it allows to factorize nonperturbative
contributions into the expectation values or matrix elements of few
operators. These may be eventually evaluated on the lattice, extracted
from the data or calculated in QCD vacuum models.  In the last few
years, the problem of systematically treating these remaining
dynamical scales in an effective theory framework has been addressed
by several groups and has now reached a solid level of understanding
(a list of references to the original literature can be found in
\cite{Brambilla:2004jw,Hoang:2002ae,reveft}).  In one approach an
additional effective theory (pNRQCD) very close to a
quantum-mechanical description of the bound system, containing only
the heavy quarkonium field and ultrasoft degrees of freedom, is
matched to NRQCD \cite{Pineda:1997bj,Brambilla:1999xf,Pineda:2000sz}.
An alternative approach, formulated only for the weak coupling case
$mv^2 \gg \lQ$, does not involve matching from NRQCD, but instead
matches a different effective theory (vNRQCD) to full QCD directly at
the hard scale \cite{Luke:1999kz,Manohar:1999xd,Hoang:2002yy}.
\longpage

In the next section we will give a brief general introduction to
NRQCD, since this is the framework for many applications reviewed in
this report. More specific presentations of NRQCD can be found in
\Chapter~\ref{chapter:spectroscopy}, \Section~\ref{sec:nrqcd}, 
\Chapter~\ref{chapter:decay}, \Section~\ref{sec:secnrqcd} and 
\Chapter~\ref{chapter:production}, \Section~\ref{sec:prodsec-nrqcdfact}.
NRQCD on the lattice will be presented mainly in the following
\Section~\ref{sec:HQactions} and in \Chapter~\ref{chapter:spectroscopy}, 
\Section~\ref{sec:spdqlc}. In \Chapter~\ref{chapter:decay}, 
\Section~\ref{sec:resumminglargez} a short presentation of SCET, an 
EFT suited to describe collinear fields interacting with soft degrees
of freedom, in combination with NRQCD may be found.

\subsection[Nonrelativistic QCD]{Nonrelativistic QCD} 
\label{sec:unonrqcd}

NRQCD is obtained by integrating out modes of energy and momentum $m$
from QCD Green functions describing heavy quark--antiquark pairs.  It
is characterized by an ultraviolet (UV) cut-off
$\nu_{NR}=\{\nu_p,\nu_s\}$ that satisfies $E, p, \lQ \ll \nu_{NR} \ll
m$; $\nu_p$ is the UV cut-off of the relative three-momentum of the
heavy quark and antiquark; $\nu_s$ is the UV cut-off of the energy of
the heavy quark and antiquark, and of the four-momenta of the gluons
and light quarks. NRQCD is, therefore, designated to describe the
dynamics of heavy quark--antiquark pairs (not necessarily of the same
flavour) at energy scales in the centre-of-mass frame much smaller than
their masses. At these energies quark--antiquark pairs cannot be
created so it is enough to use Pauli spinors for both the heavy quark
and the heavy antiquark degrees of freedom. Other degrees of freedom
of the theory are gluons and light quarks of four momentum smaller
than $\nu_s$.

The high-energy modes that have been integrated out have a relevant
effect on the low-energy physics. This effect is not lost, but encoded
into the matching coefficients $c$ and new local interactions of the
NRQCD Lagrangian.  In principle, there are infinite such terms to be
included, in practice only few of them are needed.  Each operator can
be counted in $v$.  The velocity $v$ and $\als$ (in the matching
coefficients) are the two small expansion parameters of NRQCD.  If we
aim at an accuracy of order $(\als^k\; v^n)$ we have to keep in the
Lagrangian only terms and matching coefficients that contribute up to
that order to the physical observable under study.  The couplings $m$,
$g$, $c$ are determined by the requirement that NRQCD reproduces the
results of QCD up to order $(\als^k\;v^n)$.

If the quark and antiquark have the same flavour, they can annihilate
into hard gluons. In NRQCD their effect is encoded in the imaginary
parts of the four-fermion matching coefficients (denoted by $f$ in the
following).  Their role in the description of heavy quarkonium
annihilations in NRQCD will be discussed in \Chapter~\ref{chapter:decay}.

In general, at each matching step the non-analytic behaviour in the
scale that is integrated out becomes explicit in the matching
coefficients.  Since in this case we are integrating out the mass, it
becomes an explicit parameter in the expansion in powers of $1/m$ in
the Lagrangian, while the dependence in $\ln(m/\nu)$ is encoded into
the matching coefficients.

Up to field redefinitions the NRQCD Lagrangian for one heavy flavour of mass $m$ and 
$n_f$ massless quarks at ${\cal O}(1/m^2)$, but including the kinetic 
energy term ${\bf D}^4 / (8 m^3)$, reads    
\cite{Caswell:1985ui,Bodwin:1994jh,Manohar:1997qy,Bauer:1997gs,Blok:1996iz}:
\begingroup
\allowdisplaybreaks
\begin{eqnarray}
{\cal L}_{\rm NRQCD}&=&
{\cal L}_g+{\cal L}_l+{\cal L}_{\psi}+{\cal L}_{\chi}+{\cal L}_{\psi\chi},
\label{eq:LagNRQCD}
\\
\nonumber
\\
{\cal L}_g&=&
-\frac{1}{4}F^{\mu\nu a}F_{\mu \nu}^a 
+ c_1^{g} \frac{1}{4 m^2}g f_{abc} F_{\mu\nu}^a F^{\mu\,b}{}_\alpha F^{\nu\alpha\,c},
\label{eq:Lg}
\\
\nonumber
\\
{\cal L}_l&=&\sum_{i=1}^{n_f} \bar q_i i\, \dsl q_i 
+ c_1^{ll}\displaystyle \frac{g^2}{8m^2}\sum_{i,j=1}^{n_f} 
\bar{q_i} T^a \gamma^\mu q_i \ \bar{q}_j T^a \gamma_\mu q_j  
+ c_2^{ll}\displaystyle \frac{g^2}{8m^2}\sum_{i,j=1}^{n_f} 
\bar{q_i} T^a \gamma^\mu \gamma_5 q_i \ \bar{q}_j T^a \gamma_\mu \gamma_5 q_j 
\nonumber
\\ 
&&
+ c_3^{ll}\displaystyle \frac{g^2}{8m^2}\sum_{i,j=1}^{n_f} 
\bar{q_i}  \gamma^\mu q_i \ \bar{q}_j \gamma_\mu q_j 
+ c_4^{ll}\displaystyle \frac{g^2}{8m^2}\sum_{i,j=1}^{n_f} 
\bar{q_i} \gamma^\mu \gamma_5 q_i \ \bar{q}_j \gamma_\mu \gamma_5 q_j,
\label{eq:Ll}
\\
\nonumber
\\
{\cal L}_{\psi}&=&
\psi^{\dagger} \Biggl\{ i D_0
+ \, c_2\frac{\mathbf{D}^2}{2 m} + \, c_4\frac{\mathbf{D}^4}{8 m^3}
+ c_F\, g \frac{\mathbf{\bfsigma \cdot B}}{2 m}
\nonumber
\\ 
&& \qquad
+ c_D \, g \frac{\mathbf{D \cdot E} - \mathbf{E \cdot D}}{8 m^2}
+ i c_S \, g \frac{\mathbf{\bfsigma \cdot \left(D \times E -E \times D\right) }}{8 m^2} \Biggr\} \psi
\nonumber
\\
&&
+c_1^{hl}\displaystyle\frac{g^2}{8m^2}\sum_{i=1}^{n_f} \psi^{\dagger} T^a \psi \ \bar{q}_i\gamma_0 T^a q_i 
+c_2^{hl}\displaystyle\frac{g^2}{8m^2}\sum_{i=1}^{n_f} \psi^{\dagger}\gamma^\mu\gamma_5
T^a \psi \ \bar{q}_i\gamma_\mu\gamma_5 T^a q_i 
\nonumber
\\
&&
+c_3^{hl}\displaystyle\frac{g^2}{8m^2}\sum_{i=1}^{n_f} \psi^{\dagger} \psi \ \bar{q}_i\gamma_0 q_i
+c_4^{hl}\displaystyle\frac{g^2}{8m^2}\sum_{i=1}^{n_f} \psi^{\dagger}\gamma^\mu\gamma_5
\psi \ \bar{q}_i\gamma_\mu\gamma_5 q_i,
\label{eq:Lhl}
\\
\nonumber
\\
{\cal L}_{\chi} &=&  \hbox{c.c. of }  {\cal L}_{\psi}, 
\\
\nonumber
\\
{\cal L}_{\psi\chi} &=&
  \frac{f_1(^1S_0)}{m^2} O_1(^1S_0) 
+ \frac{f_1(^3S_1)}{m^2} O_1(^3S_1)
+ \frac{f_8(^1S_0)}{m^2} O_8(^1S_0)
+ \frac{f_8(^3S_1)}{m^2} O_8(^3S_1),
\label{eq:4fermiondim6}
\\
&& O_1(^1S_0) = \psi^{\dag} \chi \, \chi^{\dag} \psi,\qquad\qquad~~
   O_1(^3S_1) = \psi^{\dag} {\bfsigma} \chi \, \chi^{\dag} {\bfsigma} \psi, 
\nonumber
\\
&& O_8(^1S_0) = \psi^{\dag} {\rm T}^a \chi \, \chi^{\dag} {\rm T}^a \psi,\qquad
   O_8(^3S_1) = \psi^{\dag} {\rm T}^a {\bfsigma} \chi \, \chi^{\dag} {\rm T}^a {\bfsigma} \psi,
\nonumber
\end{eqnarray}
\endgroup
where $\psi$ is the Pauli spinor that annihilates the quark, $\chi$ is
the Pauli spinor that creates the antiquark, $i D_0=i\partial_0
-gA_0^a\,T^a$, $i{\bf D}=i\bfnabla+g{\bf A}^a\,T^a$, ${\bf E}^i =
F^{i0\,a}\,T^a$, ${\bf B}^i = -\epsilon_{ijk}F^{jk\,a}\,T^a/2$ and
c.c. stands for charge conjugate.  The allowed operators in the
Lagrangian are constrained by the symmetries of QCD. However, due to
the particular kinematical region we are focusing, Lorentz invariance
is not linearly realized in the heavy quark sector. In practice,
Lorentz invariance is realized through the existence of relations
between the matching coefficients, \eg $ c_2 = c_4 = 1$, $ c_S = 2c_F
-1$
\cite{Luke:1992cs,Chen:1993sx,Sundrum:1997ut,Manohar:1997qy,Finkemeier:1997re,Brambilla:2003nt}.

The matching coefficients may be calculated in perturbation theory.
For the heavy quark (antiquark) bilinear sector as well as for the
purely gluonic sector up to ${\cal O}(1/m^2)$ the matching
coefficients have been obtained at one loop in
\cite{Manohar:1997qy}. The complete LL running of these coefficients
in the basis of operators (\ref{eq:Lg})--(\ref{eq:Lhl}) has been
calculated in \cite{Bauer:1997gs,Blok:1996iz}\footnote{%
                            After correcting a few misprints in the 
                            anomalous dimension matrix~\cite{privPir}, the
                            results of \cite{Blok:1996iz} agree with those 
                            of Ref.~\cite{Bauer:1997gs}.}.
For $c_F$ a NLL evaluation can be found in \cite{Amoros:1997rx}.  In
the four heavy fermion sector the matching coefficients $f$ of the
$1/m^2$ operators have been obtained at one loop in
\cite{Pineda:1998kj}.  As discussed above, in this sector the matching
coefficients have a non-zero imaginary part.  Due to their relevance
in heavy quarkonium decay processes, the calculation of corrections of
higher order in $\als$ has a long history
\cite{Bodwin:1994jh,Petrelli:1997ge,BCGRpwave,Mackenzie:1981sf,
Maltoniphd,Czarnecki:1997vz,Beneke:1997jm,BCGRswave}.  We summarize it
in \Section~\ref{sec:perturexpan} of \Chapter~\ref{chapter:decay}.  An
updated list of imaginary parts of four fermion matching coefficients
may be found in \cite{Vairo:2003gh}.

Since several scales remain dynamical in NRQCD, it is not possible to
give a homogeneous power counting for each operator without extra
assumptions, \ie the power counting in $v$ is not unambiguous. To
obtain a better defined power counting one should go to EFTs of lower
energy. It should be noticed that the importance of a given operator
for a practical calculation does not depend only on its size, but also
on the leading power of $\als$ of the corresponding matching
coefficient.

Finally, since modes of energy $m$ have been removed from the
Lagrangian, NRQCD lattice simulations may use lattices that are
coarser by about a factor $1/v^4$ ($\sim 100$ in the $b\bar{b}$ case)
than those needed by full QCD \cite{Thacker:1990bm}. We will come back
to this in \Section~\ref{sec:HQactions}.

\subsection{Lower energy EFTs} 

Effective field theories suited to describe the low energy modes of
the heavy quarkonium dynamics that will be used in this report are
pNRQCD and vNRQCD.  Here we will not give details on these EFTs since
specific introductions to pNRQCD can be found in
\Chapter~\ref{chapter:spectroscopy}, \Section~\ref{sec:spnrqcdlc} and
\Chapter~\ref{chapter:decay}, \Section~\ref{sec:pnrqcd}, and to vNRQCD in 
\Chapter~\ref{chapter:precisiondeterminations}, \Section~\ref{sec:vnrqcd}.
For detailed recent reviews on effective field theories for heavy
quarkonium we refer the reader to \cite{Brambilla:2004jw} and
\cite{Hoang:2002ae}, which are mainly devoted to pNRQCD and vNRQCD
respectively.

What we want to point out here is that in all these EFTs objects like
potentials show up.  For short range (or weakly coupled) quarkonia the
potentials may be built order by order in perturbation theory. At
higher order the pure potential picture breaks down and the
interaction of the heavy quark fields with the low-energy gluons has
to be taken into account (see the pNRQCD Lagrangian
of \Chapter~\ref{chapter:spectroscopy}, \Eq~(\ref{eq:wc}) and the
vNRQCD Lagrangian of \Chapter~\ref{chapter:precisiondeterminations},
\Eqs~(\ref{eq:Lus}) and (\ref{eq:Lp})).  For long range (or strongly
coupled) quarkonia the potentials are nonperturbative objects that may
be expressed in terms of gluon fields expectation values. Noteworthy,
the pNRQCD Lagrangian in the strong coupling regime reduces exactly,
under some circumstances, to the simple case of a heavy quarkonium
field interacting with a potential (see
\Chapter~\ref{chapter:spectroscopy}, \Eq(\ref{eq:scr})).

The potential picture that emerges from these EFTs is quite different
from the one of traditional potential models and superior. Not only
the potential is derived from QCD, but higher-order corrections can be
systematically included without being plagued by divergences or
\emph{ad hoc} cut-offs; these are absorbed in the renormalization
procedure of the EFT. Nevertheless, traditional potential models,
which so much have contributed to the early understanding of the heavy
quarkonium properties, may be still useful and will often appear in
the report.  First, a potential model can be seen, in absence of
competitive lattice data, as a specific \emph{ansatz} on the form of
the low-energy QCD dynamics encoded in the potential defined by an
EFT.  Second, potential models still provide the only available tool
to describe physical systems for which a suitable EFT has not been
built yet.  This is, for instance, the case of systems coupled to open
flavour channels.

\section[Lattice introduction]
        {Lattice introduction$\!$\footnote{Author:  G.~Bali}}

Low energy nonperturbative QCD can either be modelled or simulated on
the Lattice. Lattice gauge theory methods are particularly powerful in
heavy quark physics when combined with effective field theories
(EFTs). Lattice QCD input significantly increases the predictive power
of EFTs as more and more low energy parameters can be calculated
reliably directly from QCD and less fits to experimental data are
required for this purpose. Past lattice QCD results were often
obtained within the quenched approximation (neglecting sea quarks) or
with unrealistically heavy up and down quarks and $n_f=2$, rather than
$n_f=2+1$.  At present these limitations are gradually being removed.

We shall only describe general aspects of lattice gauge theory
simulations.  Recent reviews of different aspects of Lattice QCD can
for instance be found in
Refs.~\cite{Bali:2000gf,Creutz:2003qy,Davies:2002cx,DeGrand:2003xu,DiPierro:2000nt,
Gupta:1997nd,Kenway:kv,Kronfeld:2002pi,Luscher:2002pz,McNeile:2002uy,Munster:2000ez}. Several
books~\cite{Seiler:pw,Creutz:mg,Itzykson:sx,Rothe:kp,Montvay:cy,Smit:ug}
on the subject have been written and the summary talks of the yearly
proceedings of lattice conferences (see Ref.~\cite{Edwards:mp} for the
most recent ones) provide an overview of the
field. Ref.~\cite{Rebbi:tx} contains collections of early papers.

Obviously there are infinitely many gauge invariant ways to discretize
the continuum QCD action.  We will summarise and define the actions
most commonly used and address limitations of the method, before we
discuss extrapolations and sources of systematic errors.

\subsection{General aspects}

Lattice simulations rely on stochastic (Monte Carlo) methods.  Hence
all results inevitably carry statistical errors which however are no
problem of principle as they can be made arbitrarily small on
(arbitrarily) big computers or by means of algorithmic and
methodological improvements. In order to carry out path integral
quantisation in a mathematically sound ways, the discretisation of
space--time appears necessary. This also enables us to map continuous
problems onto a finite computer. Discretisation, \ie for instance
replacing derivatives $\partial_t\phi(t)$ by
$[\phi(t+a)-\phi(t-a)]/(2a)$ with ``lattice spacing'' $a$ and, in this
example, lattice ``errors'' of ${\cal O}(a^2)$, inevitably carries the
smell of {\em inexactness}. We stress however that the very nature of
QCD itself requires us to introduce an ultra-violet regulator and, as
we shall see below, lattice discretisation is one possible choice.
Continuum results are then obtained by removing the regulator,
$a\rightarrow 0$.

Observables are calculated (``measured'') taking their expectation
values in the path integral approach: this amounts to calculating
averages over all possible ``configurations'' of gauge fields on the
lattice, weighted with the respective exponent of the action. In
simulations with sea quarks, producing these configurations is costly
and the ILDG~\cite{Irving:2003uk} (International Lattice Data Grid) is
due to be set up, with the aim of standardising formats of organising
and labelling such lattice data, in a way that allows for easy
distributed storage, retrieval and sharing of such deposits among
different lattice groups.

The typical observables are $n$-point Green functions. In order to
determine a hadronic rest mass one has to construct an operator with
the respective quantum numbers: spin $J$, parity $P$, charge
conjugation $C$, isospin, flavour content \etc This is then projected
onto zero momentum and the 2-point Green function calculated, creating
the particle at time $0$ and destroying it at time $t$. For large $t$
this will then decay exponentially, $\propto \exp(-mt)$, with $m$
being the ground state mass within the channel in question. There
exist numerous ``wave functions'' with the right quantum numbers, some
with better and some with inferior overlap to the physical ground
state. It is a refined art to identify spatial ``smearing'' or
``fuzzing'' functions that maximise this overlap and allow to extract
the mass at moderate $t$-values, where the signal still dominates over
the statistical noise. The multi-exponential $t$-dependence of Green
functions complicates the identification of excited states, \ie
sub-leading or sub-sub-leading exponents. By working with very precise
data, realising a variational multi-state basis of test wave
functions~\cite{Michael:1988jr}, and employing sophisticated fitting
techniques~\cite{Michael:1994sz,Lepage:2001ym}, it has however in some
cases become possible to calculate moderately low lying radial
excitations.

Lattice QCD is formulated in Euclidean space time: in the continuum,
this amounts to replacing Lorentz boosts and
$O(3)$ rotational symmetry by
$O(4)$ rotations. The reason for this is that a real (and bounded)
action is required to allow for a probabilistic interpretation of the
path integral measure and computer simulation.
As an analytical
continuation to Minkowski space time of a finite number of finite-precision
data points is impossible, the predictive power is confined to quantities
that have a Euclidean space time interpretation such as masses and
matrix elements. 

Lattice discretisation unavoidably breaks rotational $O(4)$ invariance, on the
scale of the lattice spacing $a$. As the continuum limit $a\rightarrow 0$
is approached, any fixed physical correlation length $\xi$ will become
much larger than the lattice spacing. Provided the interaction ranges that appear within
the action are localised in space time, all physics will
become independent of the underlying discretisation and a universal
continuum limit will be reached, in which $O(4)$ invariance is restored.
Asymptotic freedom implies that such a continuum limit is approached as
the lattice coupling constant, $g\rightarrow 0$.

Replacing $O(4)$ invariance by its hypercubic subgroup means that in
particular higher spin states are hard to identify. For instance $J=4$
cannot easily be distinguished from $J=0$ on a hypercubic lattice. At
finite lattice spacing $a$, only discrete translations in space and
imaginary time are possible. This results in the maximum modulus of
Euclidean four-momentum components of $\pi/a$, providing the required
ultraviolet regularisation. Although an infrared cut-off is not
necessary in principle, on a finite computer only a finite number of
lattice points can be realised. Typically toroidal boundary conditions
are taken in all directions for the gauge fields while fermions, being
Grassmann-valued fields, are antiperiodic in time. This results in
quantisation of momentum components in steps of $2\pi/(La)$ where $L$
denotes the number of lattice points along the dimension in question:
not all momenta can be realised and this leads to kinematic
constraints when it comes to calculating decay matrix elements or to
extracting a particle mass from a dispersion relation.

The temporal extent $aL_{\tau}$ of the lattice can also be interpreted as
an inverse temperature (see \eg Ref.~\cite{Kapusta:tk})
and in this case QCD matter at
high temperature can be simulated. There are some subtleties related to
this approach. For instance the limit of infinite Euclidean time cannot
be taken anymore. Details of thermal field theory are discussed in
\Chapter~\ref{chapter:charm-beauty-in-media}.

While the lattice regulator inevitably violates Poincar\'e invariance it 
preserves gauge invariance and most global
symmetries of QCD. The exception was chiral symmetry which,
one had to hope, would become restored in the continuum limit.
However, within the past 10 years, formulations of chiral lattice
fermions~\cite{Neuberger:1997fp,Kaplan:1992bt}
have evolved that implement an exact lattice chiral
symmetry, which in the continuum limit corresponds to the continuum
chiral symmetry. These are known as overlap fermions or domain wall
fermions (which in some sense are a special case of the former)
and in some literature (somewhat inaccurately) as Ginsparg--Wilson fermions 
since the lattice-Dirac operator used obeys the so-called
Ginsparg--Wilson relation~\cite{Ginsparg:1981bj}.
We shall refer to these implementations as chiral fermions.

At presently available light quark masses chiral fermions are typically two
orders of magnitude more expensive to simulate than traditional formulations.
As the quark mass is decreased chiral fermions become more competitive.
Obvious advantages of chiral formulations are the applicability of
chiral perturbation theory also at finite lattice spacing
and a more continuum-like mixing between many lattice operators.
With respect to quarkonia in which both valence quarks are heavy
these new developments are at present of limited
significance as light quark mass effects are usually sub-leading.

Lattice QCD is a {\em first principles} approach. No parameters apart
from those that are inherent to QCD, \ie strong coupling constant at a
certain scale and quark masses, have to be introduced. In order to fix
these $n_f+1$ parameters $n_f+1$ low energy quantities are matched to
their experimental values. In simulations of quarkonia the lattice
spacing $a(\beta,m_i)$, that corresponds to given values of the
inverse lattice strong coupling, $\beta=6/g^2$ and lattice quark
masses $m_i$, is frequently obtained by matching to spin-averaged
experimental level splittings. In simulations with un-realistic sea
quark content one might hope that this increases the reliability of
other predictions as the systematics are partly correlated.  With
realistic sea quark content the predictive power with respect to
quarkonium physics can be enhanced by using independent input such as
the experimental proton mass $m_p$ or the pion decay constant,
$f_{\pi}$, instead.  A scale that cannot directly be accessed by
experiment but which owes its popularity to the accuracy and ease with
which it can be calculated is the Sommer scale
$r_0$~\cite{Sommer:1993ce}, implicitly defined through,
\begin{equation}
r^2 \left.\frac{dV(r)}{dr}\right|_{r=r_0}=1.65\,,
\end{equation}
where $V(r)$ denotes the static quark--antiquark potential and the
numerical value on the right hand side is adjusted such that fits of
the bottomonium spectrum to phenomenological or lattice potentials
yield $r_0\approx 0.5$~fm.  $r_0$ is also well-defined in the theory
with sea quarks and its model dependence is much smaller than that of
the string tension. Within the quenched approximation scale
uncertainties cannot be avoided anyway and hence such model dependence
is admissible.  In simulations with sea quarks this is different but
$r_0$ still provides a convenient reference scale, that can be used to
relate different lattice results with each other.

\subsection{Actions and finite $a$ effects}
We shall discuss the gauge and heavy quark actions that are
usually employed. In simulations with sea quarks, in addition a light
quark action needs to be specified.

Results from lattice simulations are inevitably obtained at a finite
lattice spacing $a$. Ideally, they are then extrapolated to the
physically relevant (and universal) continuum limit $a\rightarrow 0$.
Within the quenched approximation, such extrapolations have become the
standard while in simulations with light sea quarks a sufficient
variation of the lattice spacing is often still prohibitively
expensive in terms of computer time.  The leading order $a$ behaviour
depends on the choice of the discretisation.

One can follow Symanzik~\cite{Symanzik:1983dc} and use a continuum
effective field theory to show that the cutoff effects have the form
$a^n(\ln\Lambda a)^{m}$, where $\Lambda$ denotes a low energy scale of
the order of a few hundred MeV and $m\geq 0$.  The leading power is
usually (see below) $n=1$ or~$2$ and within this leading term, $m=0$.
By changing the discretisation, the leading terms can be reduced or
eliminated. This strategy is called ``improvement'', and it is used to
hasten the approach to the continuum limit.

In a classical mechanical system improvement is straightforward.
However, even in this case there exists a break-even point at which
further improvement becomes computationally more expensive than the
equivalent reduction of the lattice spacing, due to the exploding
number of terms and interaction range. Typically this point is reached
around $n\approx 5$.  In a quantum field theory the situation is more
complex. In QCD the (Wilson) coefficients of improvement terms obtain
quantum corrections which can be obtained perturbatively as a power
series in the strong coupling constant $g^2$, in a suitable scheme.
Following an effective field theory philosophy, such calculations can
be done and the size of next order corrections estimated. However, at
sufficiently small $a$ any $c_1g^{2n}(a)a+c_2a^2$ expression will be
dominated by the first term that, in this example, is proportional to
$a$. To eliminate such terms the coefficient has to be determined
nonperturbatively. Otherwise little is gained in a continuum limit
extrapolation, other than a reduction of the slope of the leading
order term. At a given finite $a$ value there is however still some
gain from improvement as the results will be more continuum-like.
Examples for a systematic nonperturbative improvement programme
exist~\cite{Luscher:1996ug}.

In the lattice literature often the word ``scaling'' is meant to imply
that an effective continuum limit is reached: within the ``scaling
region'' mass ratios appear to be independent of $a$, within
statistical errors. If $a$ is reduced even further, eventually one
will encounter ``asymptotic scaling'', \ie lattice masses $a(g)m$ will
depend on the coupling $g^2$ in the way expected from the perturbative
two-loop $\beta$ function. It is quite clear by now that ``asymptotic
scaling'' in terms of the bare lattice coupling might never be
achieved on large lattices.  However, asymptotic scaling has been
verified for a particular choice of the coupling, as a function of the
linear extent of tiny lattices, see \eg Ref.~\cite{Capitani:1998mq}.

\subsubsection{Gauge actions}

In lattice simulations, $SU(3)$ group elements $U_{x,\mu}$ are
typically represented as complex $3\times 3$ matrices that live on
directed links connecting a lattice site $x$ with the neighbouring
site $x+a\hat{\mu}$. Traces of products of such ``link variables'' or
``links'' along closed paths (Wilson loops) are gauge invariant.  The
simplest non-trivial such example is a $1\times 1$ square, an
elementary ``plaquette''. The lattice action should preserve gauge
invariance which means that it can be expressed as a sum over such
loops. Fermion fields $\psi_x$ and $\bar{\psi}_x$ are living on the
lattice sites and a quark can be ``transported'' from site
$x+a\hat{\mu}$ to site $x$ by means of a left multiplication with
$U_{x,\mu}$: the combination
$\bar{\psi}_xU_{x,\mu}\psi_{x+a\hat{\mu}}$ is gauge invariant.

The simplest gauge action is the so-called Wilson
action~\cite{Wilson:1974sk}, which is proportional to the trace of the
sum over all elementary plaquettes:
\begin{equation}
S_W=-\beta\sum_{x,\mu>\nu}\mbox{Re}\,\mbox{tr}\,\Pi_{x,\mu,\nu},
\end{equation}
where $x$ runs over all lattice sites and
$\Pi_{x,\mu\nu}=U_{x,\mu}U_{x+a\hat{\mu},\nu}U_{x+a\hat{\nu},\mu}^{\dagger}
U_{x,\nu}^{\dagger}$.
Up to an irrelevant constant the Wilson action
agrees with the Euclidean continuum action to ${\cal O}(a^2)$:
\begin{equation}
\label{eq:wilac}
S_{\rm YM}=\int\!d^4\!x\,
\frac{1}{4g^2}\sum_{a=1}^8 F_{\mu\nu}^a(x)F_{\mu\nu}^a(x)=S_W+\mbox{const.}+
{\cal O}(a^2)\,,
\end{equation}
where we identify $\beta=6/g^2$. Asymptotic freedom tells us that
$a\rightarrow 0$ as $\beta\rightarrow\infty$.  In simulations without
sea quarks it has been established that $\beta=6$ corresponds to a
lattice spacing $a\approx 0.1\,\mbox{fm} \approx
(2\,\mbox{GeV})^{-1}$.  With sea quarks (using the same gluonic
action) the same lattice spacing will be obtained at a somewhat
smaller $\beta$-value as the running of $a(g)$ with the coupling $g$
will be somewhat slower.  As mentioned above, perturbation theory in
terms of the lattice coupling $g^2$ is not yet reliable around
$g^2\approx 1$, to describe the running of $a(g^2)$ (asymptotic
scaling).

The ${\cal O}(a^2)$ artifacts within \Eq~(\ref{eq:wilac}) can be
replaced by ${\cal O}(a^4)$ lattice corrections, by adding two paths
consisting of six links, for instance a $1\times 2$ rectangle and a
``chair''. The result is known as the Symanzik--Weisz
action~\cite{Weisz:1982zw} and the coefficients of the individual
terms have been calculated to one loop [${\cal O}(g^2)$]
accuracy~\cite{Weisz:1983bn}. At tree level, only the coefficient of
the rectangle assumes a non-trivial value.  One (somewhat arbitrary)
choice in the space of actions is the
Iwasaki-action~\cite{Iwasaki:1984cj}, again the sum of plaquette and
rectangle, but with the relative weight fixed to a constant,
originally motivated by demanding invariance of physical mass ratios
under numerical renormalisation group transformations, within a
certain $\beta$ window. In addition to simulations with these gauge
actions~\cite{Necco:2003vh,Michael:nd,AliKhan:2001tx}, there have also
been simulations employing a combination of the plaquette in the
fundamental and in its adjoint representation~\cite{Michael:nd} as
well as simulations on anisotropic lattices, using an anisotropic
Wilson action~\cite{Chen:2000ej,Bali:2000un,Okamoto:2001jb} or
anisotropic variants of actions including Symanzik--Weisz style
terms~\cite{Morningstar:1999rf}.

The main motivation for adding such extra terms to the action is to
achieve a more continuum-like behaviour already at finite lattice
spacing. It also turns out that simulations with chiral fermions
benefit from such a choice which implies a ``smoother'' gauge field
background.

In order to achieve full ${\cal O}(a^2)$ improvement the coefficients
of the extra terms would have to be determined nonperturbatively, for
instance by imposing continuum relations: in the pure gauge theory
example above one could impose rotational invariance of the static
quark potential at two distances, \eg $V(3,0,0)=V(2,2,1),
V(5,0,0)=V(3,4,0)$ to fix the two coefficients, or use dispersion
relations of glueballs or torelons. This is laboursome and in
general the fermions will not be nonperturbatively improved beyond
${\cal O}(a^2)$ anyway. So in practice, only approximate improvement
has been implemented, either by using the perturbative coefficients at
a given order or by employing a so-called ``tadpole'' improvement
prescription.

The latter is motivated by two observations. The first one is that
short-distance lattice quantities differ considerably from their
continuum counterparts, even at lattice spacings at which one would,
based on the $\MS$ scheme continuum experience, assume perturbation
theory to be valid. For instance around $a^{-1}=2\,\mbox{GeV}$ the
numerical value for the plaquette with Wilson action reads
$\Box=\frac{1}{3}\langle \mbox{tr}\, \Pi\rangle\approx 0.6$ while at
$g=0$ this should obviously be normalised to {\em one}.  This is
closely related to the breaking of continuum rotational symmetry on
the scale of a lattice spacing $a$. Parisi~\cite{Parisi:1980pe}
hypothesized that such ultra-violet effects could largely be factored
out and put into commuting pre-factors. This mean-field improvement
amounts to dividing links that appear within lattice operators by
constant factors, \eg $u_0=\Box^{1/4}$.  An independent observation is
that lattice perturbation theory, whose convergence behaviour in terms
of the lattice coupling $g^2$ is well known to be quite bad, differs
from continuum perturbation theory largely by a class of
lattice-specific tadpole diagrams which are numerically large.  By
normalising everything with respect to other ``measured'' observables
like $u_0$ these contributions cancel at one loop order and one might
hope that tadpole dominance and cancellation approximately generalises
to higher orders as well~\cite{Lepage:1992xa}.

Finally, there is the idea of (classically) ``perfect'' actions~\cite{Hasenfratz:1993sp}. 
If one found an action that lies right on top of a renormalisation
group trajectory then, independent of the lattice spacing $a$,
one would obtain continuum results. Such actions can be identified by
demanding independence of physical results under a change of the underlying
scale. An action that contains a finite set of couplings is suggested and
these are then optimised with respect to such constraints.
In practice, one can of course at
best construct an action that is close to such a trajectory in which case
decreasing the lattice spacing still helps to reduce deviations of
the nearly perfect action from a real renormalisation group trajectory
which one attempts to approximate. An example of such an (approximately)
perfect action and its construction can be found in
Ref.~\cite{Niedermayer:2000yx}.

\shortpage
\subsubsection{Light quark actions}

The Dirac action is bi-linear in the quark fields. In the language of
perturbation theory this amounts to the non-existence of vertices
containing an odd number of quark fields. This means that the quark
part of a lattice calculation can to some extent be separated from the
gauge field evaluation: the gluon fields contain all information of
the QCD vacuum, including sea quark loops, provided these are
unquenched (see below). Hadronic $n$-point functions can be
obtained from contractions of colour fields, $\Gamma$-matrices and
quark-propagators, calculated on this gluonic background.

We denote a discretisation of the continuum Euclidean Dirac operator
$[D_{\mu}\gamma_{\mu}+m_i]$ as $M_i[U]$. Each quark flavour $i$ now contributes
a factor,
\begin{equation}
S_{f_i}=(\bar{\psi},M_i[U]\psi),
\end{equation}
to the action, where the scalar product $(\cdot,\cdot)$ is over all
$V=L^3l_{\tau}$ sites of Euclidean space time, colour and
Dirac-spinor index. Note that $M_i$ depends on the gauge fields
$U$. Components of $M_i^{-1}$ correspond to quark propagators. Often
it is sufficient to calculate propagators that originate from only one
source point. In this case only one space--time row of the otherwise
$12V\times 12V$ matrix $M_i^{-1}$ needs to be calculated.  As the
non-diagonal contributions to the Dirac operator all originate from a
first order co-variant derivative, $M_i$ will be a sparse matrix with
non-vanishing elements only in the vicinity of the (space time)
diagonal.  This tremendously helps to reduce the computational task.
Quark propagators can be contracted into hadronic Green functions,
expectation values (over gauge configurations) of which will decay
with the mass in question in the limit of large Euclidean times.

One complication arises from the fermions as these are represented by
anti-commuting Grassmann numbers. Realising these directly on a
computer implies a factorial (with the number of lattice points)
complexity~\cite{Creutz:1998ee} but fortunately they can be integrated
out analytically as,
\begin{equation}
\label{eq:det}
\int[d\psi][d\bar{\psi}]e^{(\bar{\psi},M_i[U]\psi)}=\det M_i[U]
=\int [d\phi][d\phi^+]e^{(\phi^+,M_i^{-1}[U]\phi)},
\end{equation}
where $\phi$ and $\phi^+$ are auxiliary Boson (pseudo-fermion)
fields.  The price one pays is that calculating $\det M_i[U]$ (or
$M_i^{-1}$) involves effective interactions over several lattice
sites. This renders simulations containing sea quark effects two to
three orders of magnitude more expensive than using the quenched (or
valence quark) approximation, $\det M_i[U]=\mbox{const.}$.

As one would expect
ratios of light hadron masses from lattice simulations of
quenched QCD
have been found to be inconsistent with
the observed spectrum~\cite{Aoki:2002fd}. However,
the differences are typically smaller than 10~\%, suggesting
that the quenched approximation has some predictive power
if cautiously consumed. Apart from the obvious shortcomings
like a stable $\Upsilon(4S)$,
the consequences of violating unitarity
at light quark mass can become dramatic in some
channels~\cite{Bardeen:2001jm}. Roughly speaking as
the axial anomaly does not exist in quenched QCD
the $\eta'$ will be a surplus light Goldstone Boson or, more
precisely, a ghost particle. The impact
of this can be investigated in quenched chiral perturbation theory.

Ultimately, one needs to include sea quarks and there are three
classes of light quark actions: staggered, Wilson-type, and
chiral.

After trivially rescaling the quark fields,
$\psi_x \rightarrow a^{-3/2}\psi_x,
\bar{\psi}\rightarrow a^{-3/2}\bar{\psi}_x$, to allow for a
representation as dimensionless numbers, a na\"{\i}ve discretisation of the 
Dirac action would read,
\begin{equation}
\label{eq:naive}
S_N=\sum_x\left\{ ma\bar{\psi}_x\psi_x+\frac{1}{2}\sum_{\mu}
  \gamma_{\mu}\bar{\psi}_x\left[U_{x,\mu}\psi_{x+a\hat{\mu}}-
    U_{x-a\hat{\mu},\mu}^{\dagger}\psi_{x-a\hat{\mu}}\right]\right\}.
\end{equation}
This action corresponds to the continuum action, up to ${\cal O}(a^2)$
terms, however, it turns out that it corresponds to 16
mass-degenerate species of Dirac fermions in the continuum limit,
rather than to one: the famous fermion-doubling
problem~\cite{Nielsen:1980rz,Wilsonf}.  In the lattice literature
these species are now often called tastes, instead of flavours, to
emphasize that they are unphysical.
\shortpage

It has been noted however that by means of a unitary transformation,
the na\"{\i}ve action can be diagonalised in spinor-space, into four
identical non-interacting terms, each corresponding to four continuum
tastes. The result is the so-called Kogut--Susskind (KS)
action~\cite{Susskind:pi}, in which 16 spin-taste components are
distributed within a $2^4$ hypercube, a construction that is known in
the continuum as K\"ahler fermions. The advantage is that one taste of
KS fermions corresponds to $n_f=4$ continuum tastes rather than
$n_f=16$. Another nice feature is that even at finite lattice spacing
one of the 15 ($n_f^2-1$) pions will become exactly massless as
$m\rightarrow 0$.  The price that one pays is strong spin-taste
mixing at finite lattice spacing and large coefficients accompanying
the leading ${\cal O}(a^2)$ lattice artifacts. KS-type fermions are
referred to as ``staggered'' and there are improved versions of them,
most notably the Naik action~\cite{Naik:1986bn}, the
AsqTad~\cite{Lepage:1998vj} ($a$ squared tadpole improved) action and
HYP actions~\cite{Hasenfratz:2001hp,Follana:2003fe} (in which parallel
transporters are smeared ``iteratively'' within hypercubes). The
latter two choices notably reduce the tastes mixing interactions.

In order to bring down $n_f=4$ to $n_f=1$, as required to achieve
$n_f=2+1$, sometimes the determinant within \Eq~(\ref{eq:det}) is
replaced by its fourth positive
root~\cite{Bitar:1992dk,Davies:2003ik}.  It can be shown that within
perturbation theory this indeed corresponds to replacing the
$n_f$-factors accompanying sea quark loops by $n_f/4$.  However, some
caution is in place. The operator $\sqrt{M}$ is
non-local~\cite{Bunk:2004br} and if its non-locality altered
$\sqrt{\det M} = \det\sqrt{M}$, universality could be lost in the
continuum limit.  One might argue that $A$ is not the only operator
with the property $\det A =\sqrt{\det M}$ but also in the Schwinger
model there exist some discouraging results for the behaviour of the
topological winding number at small quark masses~\cite{Durr:2003xs}.
Moreover, the valence quark action automatically differs from the one
used for the sea quarks as each taste of sea quarks will correspond to
4 tastes of valence quarks~\cite{Aubin:2003mg}.

Nonetheless, large scale simulations with this action are pursued at
present as the computational costs of going to light sea quark masses
appear much smaller than with other actions. Moreover, as long as the
sea quark masses are not too small, this approximation to QCD is not
completely wrong and in fact likely to be more realistic than quenched
QCD. Indeed, in quarkonium physics where light quark mass effects are
sub-leading, first results appear very
encouraging~\cite{Davies:2003ik}. There also exist first theoretical
attempts of constructing a local representation of the $n_F<4$
staggered action~\cite{Maresca:2004me,Adams:2004mf}.

Another ``solution'' to the fermion doubling problem are Wilson
fermions~\cite{Wilsonf}: the lattice analogue of the term,
$-\frac{1}{2}aD_{\mu}D_{\mu}$, is added to the $M$ of
\Eq~(\ref{eq:naive}). This increases the masses of the 15
doublers by amounts
that are proportional to $a^{-1}$, removing the unwanted modes.
Like in the case of staggered fermions the chiral symmetry that
QCD classically enjoys at $m=0$ is explicitly broken at any finite
lattice spacing $a$. In addition, one encounters additive mass renormalisation
and a rather awkward eigenvalue spectrum of the lattice Dirac operator
as well as ${\cal O}(a)$ lattice terms. The latter can be removed by adding
yet another counterterm to $M$: $\propto -ic_{sw}\sigma_{\mu\nu}F_{\mu\nu}$.
The resulting action is known as the Sheikholeslami--Wohlert (SW) or clover
action~\cite{Sheikholeslami:1985ij}.
The $c_{sw}$ coefficient
is known to one loop [${\cal O}(g^2)$] in perturbation
theory~\cite{Sheikholeslami:1985ij,Aoki:1998qd} but has also
been determined nonperturbatively in quenched QCD with Wilson
gauge action~\cite{Luscher:1996ug}, in $n_f=2$ QCD with
Wilson~\cite{Jansen:1998mx}
and $n_f=3$ QCD with Iwasaki gauge actions~\cite{Aoki:2002vh}.
Another variant is the FLIC (fat link irrelevant clover)
action~\cite{Zanotti:2001yb}.
Finally, there exists twisted mass QCD~\cite{Frezzotti:2000nk}, in which
an imaginary mass term is introduced into the Wilson action. 
Unfortunately, in this case there will be mixing between parity partners
within Green functions, something that one also encounters in staggered
formulations. However, the changed eigenvalue spectrum of $M$
renders smaller quark masses accessible. Moreover,
in the case of a purely imaginary renormalized mass parameter,
${\cal O}(a)$ improvement holds.

Finally, formulations of chiral lattice fermions exist. These are
automatically ${\cal O}(a)$ improved and do not suffer from the
fermion doubling problem.  Realisations of these fall into three
categories: overlap fermions, based on the Neuberger
action~\cite{Neuberger:1997fp}, domain wall fermions, which live on a
five-dimensional lattice and become chiral as the size of the fifth
dimension is sent to infinity~\cite{Kaplan:1992bt} and perfect
actions~\cite{Hasenfratz:1993sp,Hasenfratz:2000xz,Gattringer:2003qx}.
As always there is no free lunch and at presently accessible sea quark
masses these formulations are around two orders of magnitude more
expensive than the ``traditional'' quark actions, described above.
For this reason, these formulations have not yet been applied to
quarkonia (although one quenched study with ``chiral'' charm quarks
exists~\cite{Tamhankar:2004ci}) but in the future as algorithmic and
hardware development will reduce costs, gauge configurations with
chiral sea quarks will become increasingly available, in particular
also because at lighter quark masses chiral fermions will become more
competitive.

\subsubsection{Heavy quark actions}
\label{sec:HQactions}

To a very good approximation bottom quarks can be neglected from the
sea as their presence will only affect the theory at very short
distances. This is also true for charm quarks but, depending on the
phenomenology one is interested in, to a somewhat lesser extent.  In
principle nothing speaks against employing the same quark actions as
above to the heavy quark sector as well. With a naive treatment of
cutoff effects, lattice corrections $\propto(ma)^n$ arise.  This
suggests that to make contact with the continuum limit, the condition
$m<a^{-1}$ has to apply: as $m$ becomes large the lattice spacing has
to be made finer and finer, the number of lattice points larger and
larger and computational costs will explode, if not for charm then
certainly for bottom.
 
One possible way out would be to introduce an anisotropy,
$\xi=a_{\sigma}/a_{\tau}$ with a temporal lattice spacing
$a_{\tau}\ll m^{-1}$ while the spatial lattice spacing can be kept
coarser. An obvious application of anisotropic actions is finite
temperature physics~\cite{Datta:2003ww}
but an anisotropy has also been employed
successfully in investigations of pure gauge
theories~\cite{Morningstar:1999rf,Juge:2004xr} as well as
in charmonium physics~\cite{Chen:2000ej,Okamoto:2001jb}.
Obviously, the anisotropy of the gauge action has to
be matched to that of the light quark and heavy quark actions, in order
to obtain a sensible continuum limit. This matching certainly becomes very
expensive when sea quarks are included and even more so in the presence
of improvement terms.

Another starting point are effective field theories, in particular NRQCD
which relies on a power counting in terms of the relative heavy
quark velocity, $v$. In addition, EFTs automatically provide the framework for
factorisation of physical processes into nonperturbative low energy
QCD and perturbative high energy QCD contributions.
The fermionic part of the
${\cal O}(v^4)$ Euclidean continuum NRQCD Lagrange density with quark
fields $\psi$ and antiquark fields $\chi$
reads~\cite{Caswell:1985ui,Thacker:1990bm},
\begin{equation}
\label{eq:lagnrqcd}
{\mathcal L}=-
\psi^{\dagger}\left[D_4+H\right]\psi
-
\chi^{\dagger}[D_4-H^{\dagger}]\chi+{\mathcal L}_{\psi\chi},
\end{equation}
with
\begin{eqnarray}
\label{eq:ef1quark}
H&=&m+\delta m-c_2\frac{{\mathbf D}^2}{2m}-
c_F\frac{g{\bfsigma}\cdot{\mathbf B}}{2m}
-c_4\frac{({\mathbf D}^2)^2}{8m^3}\\\nonumber
&-&ic_D\frac{g({\mathbf D}\cdot{\mathbf E}
-{\mathbf E}\cdot{\mathbf D})}{8m^2}
+c_S\frac{g{\bfsigma}\cdot
({\mathbf D}\times  {\mathbf E}
-{\mathbf E}\times {\mathbf D})}{8m^2}
+\cdots,
\end{eqnarray}
where the matching coefficients $c_i(m/\mu,g^2)=1+{\cal O}(g^2)$,
$\delta m={\cal O}(g^2)$ are functions of the matching scale $\mu$ and
coupling $g^2$.  In the continuum $c_2=c_4=1$, however, this is in
general different on the lattice, where rotational invariance is
broken and to ${\cal O}(v^4)$ an additional term $\propto
a^2\sum_iD_i^4/m$ appears.  There are many obvious ways of
discretising the above equation on the lattice and often the published
expressions involve ``tadpole'' improvement factors $u_0=1+{\cal
O}(g^2)$. On a lattice with infinite temporal extent it is possible to
use a discretisation of the above Hamiltonian within the kernel of a
time-symmetric evolution equation~\cite{Lepage:1992tx} such that
fields at time $t+a$ can be computed entirely from fields at time $t$
(and vice versa). This turns the computation of propagators
particularly economical.  In reality, computations are performed on a
finite torus but as long as propagators fall off sufficiently fast in
Euclidean time, the resulting error of this approximation will be
small.

In addition there are the four-fermion interaction terms ${\cal
L}_{\psi\chi}$ which (in the case of flavour singlet quarkonia) are
accompanied by factors $\propto \als$ and have to be considered at
${\cal O}(v^4)$.  In principle it is known how to do this in lattice
simulations~\cite{Lepage:1992tx}.  For the $B_c$ system, where
annihilation is not possible, there will be further suppression of
these terms by an additional factor $\als$.  Finally, due to
integrating out heavy quark loops, two new purely gluonic operators
are encountered~\cite{Manohar:1997qy,Pineda:1998kj}, accompanied by
factors $1/m^2$. This ``unquenching'' of the heavy quark can in
principle easily be implemented in lattice simulations too.  However,
this is obviously an effect, less important than achieving a realistic
light flavour sea quark content.

Starting from a latticized NRQCD action there are in principle
different ways to calculate quark propagators. Usually the full
fermionic matrix that appears within a lattice discretisation of
\Eq~(\ref{eq:ef1quark}) is inverted, as described above, exploiting a
Hamiltonian evolution equation. As an alternative one could also
analytically expand the Green functions of interest in powers of $1/m$
and calculate the resulting coefficients individually.  It is
worthwhile to mention that in the continuum the expression ``HQET''
refers to heavy-light systems and ``NRQCD'' to quarkonia. In the
lattice literature however, NRQCD is used for both, heavy-heavy and
heavy-light system, indicating that the propagator is obtained as the
inverse of the lattice NRQCD quark matrix. The term HQET implies an
expansion of heavy quark propagators about the static limit. As these
are somewhat smeared out in space, NRQCD propagators can be determined
more accurately than HQET ones, however, with the invention of new
``fat'' static quark actions~\cite{Hasenfratz:2001hp} that reduce
$\delta m$ within \Eq~(\ref{eq:ef1quark}) above this has recently
changed.

The $m/\mu$ dependence of the matching coefficients $c_i$ has been
calculated in the $\MS$ scheme to various orders in perturbation
theory but so far no result on the $ma$ dependence exists in lattice
schemes. This seems to be changing, however~\cite{Trottier:2003bw}.
Such corrections are important as in the Coulomb-limit, in which
NRQCD power counting rules are formulated, $\als={\cal O}(v)$.  The
difference $\delta m$ between kinetic and rest mass can be determined
nonperturbatively from the $\Upsilon$ dispersion relation.

The Fermilab method~\cite{El-Khadra:1996mp} constitutes a hybrid
between heavy quark and light quark methods. It is based on an
expansion in terms of the lattice spacing, starting from the Wilson
quark action that encompasses the correct heavy quark symmetry. For
$ma\ll 1$ this is equivalent to the Symanzik-improvement programme,
the lowest order correction resembling the SW/clover term. However, at
$ma> 1$ the result is interpreted in terms of the heavy quark terms
that one obtains from a $1/m$ expansion.  Evidently, the light-quark
clover term has the same structure as the ${\mathbf \sigma}\cdot
{\mathbf B}$ fine structure interaction, in particular on anisotropic
lattices, where the difference can be attributed to the matching
coefficients.

An extension of the Fermilab method is an effective field theory
framework for describing discretization
effects~\cite{Kronfeld:2000ck}.  This theory lumps all discretization
effects into short-distance coefficients of the NRQCD/HQET effective
Lagrangian.  Compared to the continuum HQET or NRQCD, the coefficients
now depend on both short distances, $m_Q^{-1}$ and~$a$.  This theory
is also a natural extension of Symanzik's theory of cutoff effects
into the regime $m_Qa\not\ll1$~\cite{Harada:2001fi}.  The theory of
heavy-quark cutoff effects is not limited to the Fermilab method and
can be used to compare the relative size of cutoff effects in various
ways of discretising the heavy-quark action~\cite{Kronfeld:2003sd}.

Finally, it is possible to solve NRQCD on the lattice by computing
static propagators with field strength insertions, in the spirit of
the $1/m$ HQET expansion.  This can either be done on the level of
quarkonium Green functions (an approach that so far has never been
attempted) or within the framework of static potentials and
relativistic corrections derived from
NRQCD~\cite{Pineda:2000sz,Bali:1997am}.  When constructing Green
functions one has to keep the power counting in mind as well as the
fact that the lowest order NRQCD Lagrangian goes beyond the static
limit as the kinetic term is required.  It is also possible to put
pNRQCD~\cite{Brambilla:1999xf} onto the lattice. In the limit
$\Lambda< mv^2$ quarkonia are represented as colour singlet or colour
octet states, propagating in the QCD vacuum~\cite{Bali:2003jq}.  This
condition is only met for would-be toponium and to some extent for
the lowest lying bottomonium states.  However, this approach is
conceptionally interesting and reduces the number of relevant decay
matrix elements.

\subsection{Extrapolations}

In lattice simulations there are in general three kinds of effects:
finite volume effects, lattice artifacts and errors due to wrong light
quark masses. Within NRQCD there are additional error sources due to
the truncation of the effective field theory at a fixed order in the
velocity $v$ and determination of matching coefficients to a given
accuracy in $\als$. In addition to these controlled errors there are
error sources that are not controlled by a small parameter like
quenching or the use of ill-defined light quark actions. The
statistical analysis of lattice data is not trivial but we shall not
discuss the possible errors, caveats and pitfalls here as this would
be too technical.

Due to the confinement phenomenon and screening of colour,
finite size effects are usually quite benign and\,---\,once the lattice is
sufficiently large\,---\,fall off at least like $1/(La)^3$. Because of this
it is often sufficient to repeat simulations on 2-3 different volumes
to check if finite size effects can be resolved within statistical errors,
rather than to attempt proper infinite volume extrapolations.
Obviously, higher lying states and charmonia are spatially more extended
than lower lying states and bottomonia. In simulations with sea quarks
the lattice size has to be large, compared to the pion mass.
For instance the condition $La>4/m_{\pi}$ yields $La>5.7$~fm
at physical pion mass. There are no large-volume
lattice results as yet obtained at such light quark masses.
To disentangle possible finite volume from other systematic effects,
sequences of lattice simulations at different lattice spacings
are often obtained at a volume that is fixed in physical units.

The power $n$ of the dominant finite lattice spacing effect ${\cal
O}(a^n)$ is in general known and can be fitted to lattice data if
sufficient leverage in $a$ is provided. In the context of
``improvement'' (to a given order of perturbation theory or\/ {\em ad
hoc}) sometimes the coefficient of the leading order term is small
since it is suppressed by powers of $g^2$ such that the sub-leading
term has to be accounted for as well. Within the context of effective
field theories one cannot extrapolate to the continuum limit as the
lattice spacing provides the cut-off scale but one can check
independence of the results with respect to variations of $a$. Once
the $ma$ dependences of the short range matching coefficients are
determined, the scaling should improve. A notable exception is the
Fermilab action which has a continuum limit. However, the functional
form in the cross-over region between $ma>1$ and $ma<1$ is not as
simple as $a^n(\ln a)^m$.

As computer power is limited, lattice sea quark masses are typically
not much smaller than the strange quark mass but with the so-called
AsqTad ``$n_f=2+1$'' action values $m\approx 0.2\, m_s$ have been
reported~\cite{Davies:2003ik}.  Lattice results have to be chirally
extrapolated to the physical limit.  Chiral corrections to quarkonium
mass splittings are to leading order proportional to
$m_{\pi}^2$~\cite{Grinstein:1996gm}.  While within present-day
lattice calculations of light hadronic quantities as well as of $B$
and $D$ physics, such finite mass effects are frequently the dominant
source of systematic error, in the case of quarkonia, the dependence
appears to be much milder, due to the absence of a light valence quark
content.

If effective field theories are realised or simulations are only
available at very few lattice spacings cut-off effects can be
estimated by power counting rules and/or by varying the action(s). In
the absence of fully unquenched results, some experience can be
gained by comparing to experiment, on the likely effect of
implementing a wrong number of sea quarks but this error source is not
controllable from\/ {\em first principles.}  A real\/ {\em ab initio}
study must go beyond the valence quark approximation.

\BLKP
%10/12/2004

%\documentclass[11pt,twoside]{cernrep}
%\usepackage{graphicx,epsfig}
%\usepackage{here,cite,url}
%\input{newcommand.tex}

%\begin{document}
\chapter{COMMON EXPERIMENTAL TOOLS}
\label{chapter:commonexperimenttools}

{\it Convener:} R.~Mussa\par\noindent
{{\it Authors:} A.~B\"ohrer, S.~Eidelman, T.~Ferguson, R.~Galik, F.~A.~Harris,
M.~Kienzle, A.~B.~Meyer, A.~Meyer, X.~H.~Mo, V.~Papadimitriou, E.~Robutti, G.~Stancari, 
P.~Wang, B.~Yabsley, C.~Z.~Yuan}

\section[Overview]{Overview$\!$\footnote{Author: R.~Mussa}}
\label{sec:chap2-intro}

This chapter aims to provide an overview of the experimental facilities 
which are contributing to provide the wealth of data on heavy quarkonia
during the current decade. 
The experiments can be sorted in 7 broad classes, according to the accelerator 
which is being used. The world laboratory on heavy quarkonium can count on 
dedicated experiments working in the most important  HEP facilities, such as:

\begin{itemize}
\item Three $\tau$-charm factories, described in
      \Section~\ref{sec:taucharm}: \textbf{BES}, which provided record
      samples of $\jpsi$'s and $\psip$'s in the last years, and will
      run a new intensive program at these energies from 2006 on
      (BES~III), {\bf CLEO}, which after 25 years of running at
      $\Upsilon(nS)$ energies is presently involved in a 3 years
      program (CLEO-c) across open charm threshold, but also {\bf
      KEDR} which, exploiting the polarimeter in the VEPP-4 collider,
      has recently provided high precision measurements of $\jpsi$ and
      $\psip$ masses;
\item Three  B-factories:  after  CLEO,  {\bf  BaBar}  and  {\bf  Belle},
      described  in \Section~\ref{sec:Bfact},  have proved  to  have a
      large physics potential also  as charmonium factories, through a
      rich   variety   of   reactions   (B   decays   to   charmonium,
      $\gamma\gamma$  , ISR,  double  $c\bar{c}$), and  can easily  be
      exploited to study bottomonium physics;
\item One $\bar{p}p$ charmonium factory: the Antiproton Accumulator of
      the Tevatron, at Fermilab, was exploited by the {\bf E835}
      experiment, described in \Section~\ref{sec:ppbar}, to scan all
      known narrow charmonium states in formation from $\ppbar$
      annihilation.
\end{itemize}

In these last years, clean record samples of all the narrow vector
resonances have been accumulated.  \Table~\ref{tab:ccbars} shows the
record samples of charmonia produced (or formed) in:
\begin{itemize}
\item one B-factory (via B decays, $\gamma\gamma$, radiative return)
      with 250 $\mathrm{fb}^{-1}$ (such quantity is continuously
      increasing at present);
\item the highest statistics runs recently done by the $\tau$-charm
      factory BES (58~M $J\psi$'s and 14~M $\psi(2S)$)
\item the data samples formed in the $\ppbar$ charmonium experiment E835
\end{itemize}

In 2003, CLEO~III accumulated the largest data samples of
$\Upsilon(1,2,3S)$ states: 29~M, 9~M, 6~M respectively.  If the
production of $\Upsilon$ states may now stop for a while, the
available samples on charmonium are expected to boost in the future
years, not only as a result of the steady growth in data from
B-factories, but mainly from the dedicated efforts of the CLEO-c
project, which aims to take 1 billion $\jpsi$'s in 2006, and the
BES~III upgrade, from 2007 on.  Another $\ppbar$ charmonium factory is
going to start data taking at GSI in the next decade.

\newpage

\begin{table}[htb]
\caption{This table summarizes the numbers of charmonium states
         produced or formed , not necessarily detected, in the
         B-factories, $\tau$-charm factories and in $\ppbar$.}
\label{tab:ccbars}
\begin{center}
\begin{tabular}{|l|lllllll|}
\hline
Particle & $\psi(2S)$ & $\eta_c(2S)$ & $\chi_{c2}$ & 
$\chi_{c1}$ & $\chi_{c0}$ & $J/\psi$ & $\eta_c(1s)$\\
\hline 
 B decays      & 0.8M & 0.4M & 0.3M & 0.9M & 0.75M & 2.5M & 0.75M \\
$\gamma\gamma$ &  --   & 1.6M & 1M &   --   & 1.2M &  --   & 8.0 M  \\
 ISR           &  4M  &    --  &   --  & --     &  --    & 9M & --    \\
 $\psi(2S)$ decays & 14M & ?  & 0.9M & 1.2M & 1.2M & 8.1M & 39K      \\
 $J/\psi$ decays &   --  &   --   &   --  &   --   &   --   & 58M  & 0.14 M \\
 $\ppbar$      & 2.8M &   ?   &  1M   &  1M   &  1.2M & 0.8M & 7M \\
\hline 
\end{tabular}
\end{center}
\end{table}

Beside the dedicated experiments, many other facilities  provide not just 
valuable information on the mechanisms of heavy quarkonium production, 
but have nonetheless a high chance to discover new states:

\begin{itemize}
\item A Z-factory: the four LEP experiments, described in
      \Section~\ref{sec:LEPexp}, studied heavy quarkonium production
      in $\gaga$ fusion.
\item 2 Hadron Colliders, described in \Section~\ref{sec:Hcoll}:
      Tevatron, where {\bf CDF} and {\bf D0} can investigate the
      production mechanisms of prompt heavy quarkonium at high energy
      and RHIC, where Star and Phenix can search in heavy quarkonium
      suppression the signature of deconfined quark--gluon plasma.
\item 1 ep Collider: HERA, described in \Section~\ref{sec:HERA}, where the
      experiments {\bf H1} and {\bf ZEUS} can study charmonium
      photoproduction, and {\bf HERA-B} studies charmonium production
      in pA interactions.

\end{itemize}

The list of available sources of new data is far from complete: other
Fixed Target Experiments, such as NA50, NA60, study charmonium
production in pN, NN interactions.

At the end of the chapter, a set of appendices give details on some of
the experimental techniques which are widely employed in this field,
for the determination of narrow resonance parameters such as masses,
widths and branching ratios. These appendices aim to focus on some of
systematic limits that the present generation of high statistics
experiments is likely to reach, and give insights on the future
challenges in this field:

\begin{itemize}
\item Appendix \ref{sec:redepol} explains the physical principle of
      resonant depolarization, which provides the absolute energy
      calibration of the narrow vector states of charmonium and
      bottomonium.
\item Results from all \ep scanning experiments crucially depend on
      the subtraction of radiative corrections on the initial state: a
      detailed and comprehensive review of the analytical expression
      which connects the experimental excitation curve to physical
      quantities such as partial widths and branching ratios is given
      in Appendix \ref{sec:radcorr}.
\item Scanning techniques using $\ppbar$ annihilations are less
      affected by radiative corrections; the physical limits of the
      stochastic cooling on antiproton beams are reviewed in
      Appendix~\ref{sec:ppcool}.
\item Appendix~\ref{sec:gagalum} reviews the available software tools
      to calculate the luminosity in $\gaga$ fusion experiments, an
      issue which may become relevant as we hope to measure $\gaga$
      widths with accuracies better than 10\% with the current high
      statistics samples from B-factories.
\item Recent evidences both in $\ep$ and $\ppbar$ formation
      experiments have shown that the interference between continuum
      and resonant amplitudes can be observed in the charmonium system
      and may soon lead to a better understanding of some experimental
      puzzles, and therefore to a substantial reduction on systematic
      errors on branching fractions.  This issue is addressed in
      Appendix~\ref{sec:interf}.
\end{itemize}

\newpage
\section[$\tau$-charm factories]
        {$\tau$-charm factories}
\label{sec:taucharm}

\subsection[BES]{BES $\!$\footnote{Author: F.~A.~Harris}}
\label{sec:bes}

\begin{figure}[!htb]
\centerline{\epsfysize 4.0 truein
\epsfbox{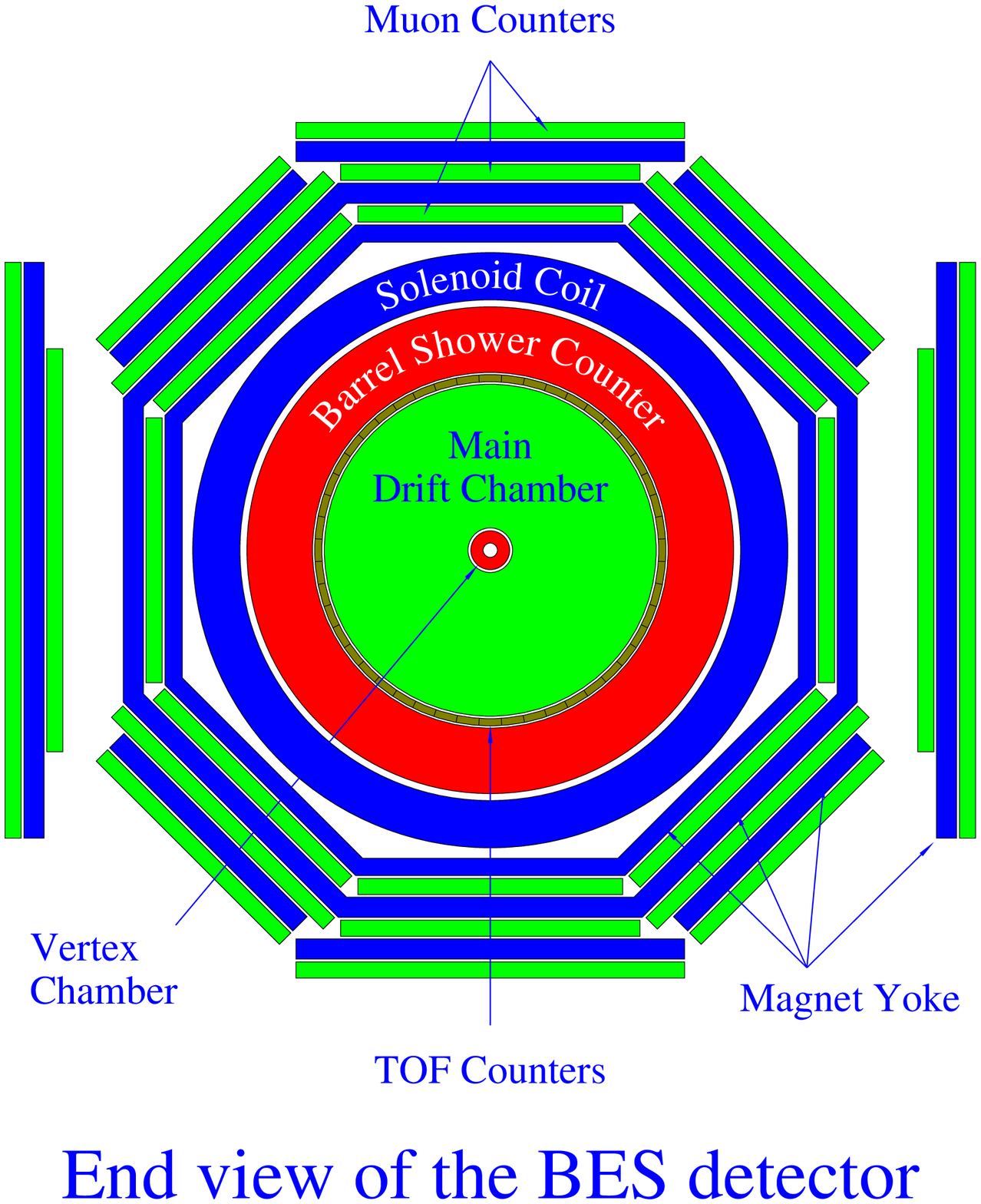}}
\caption{End view of BES (BES~II) detector.}\label{fig:besii}
\end{figure}

BES is a conventional solenoidal magnet detector that is described in
detail in Ref.~\cite{bes}; BES~II is the upgraded version of the BES
detector, which is described in Ref.~\cite{bes2} and shown in
\Figure~\ref{fig:besii}. In BES~II, a 12-layer vertex chamber (VC)
surrounding the 1.2~mm thick beryllium beam pipe provides trigger and
track coordinate information. A forty-layer main drift chamber (MDC),
located radially outside the VC, provides trajectory and energy loss
($dE/dx$) information for charged tracks over $85\%$ of the total
solid angle.  The momentum resolution is $\sigma _p/p = 0.017
\sqrt{1+p^2}$ ($p$ in $\hbox{\rm GeV}/c$), and the $dE/dx$ resolution
for hadron tracks is $\sim 8\%$.  An array of 48 scintillation
counters surrounding the MDC measures the time-of-flight (TOF) of
charged tracks with a resolution of $\sim 200$ ps for hadrons.
Radially outside the TOF system is a 12 radiation length, lead--gas
barrel shower counter (BSC), operating in self-quenching streamer
mode. This measures the energies of electrons and photons over $\sim
80\%$ of the total solid angle with an energy resolution of
$\sigma_E/E=22\%/\sqrt{E}$ ($E$ in GeV).  Surrounding the BSC is a
solenoidal magnet that provides a 0.4 Tesla magnetic field over the
tracking volume of the detector. Outside of the solenoidal coil is an
iron flux return that is instrumented with three double layers of
counters that identify muons of momentum greater than 0.5~GeV/$c$.
The BES~II parameters are summarized in \Table~\ref{tab:parameters}, and
a summary of the BES data sets is given in \Table~\ref{tab:datasets}.

\begin{table}[!htb]
\caption{Summary of BES~II detector parameters.}\label{tab:parameters} 
\begin{center}
\vspace{0.1in}
\begin{tabular}{|c|c|c|}
\hline
Detector  & Major parameter            & BES~II \\\hline
VC        & $\sigma_{xy}$($\mu m$)     & 100  \\\hline
          & $\sigma_{xy}$($\mu m$)     & 190--220 \\
MDC       & $\Delta p/p$ ($\%$)        & $1.7\sqrt{1+p^2}$ \\
          & $\sigma_{dE/dx}$ ($\%$)    & 8.4                \\\hline
TOF       & $\sigma_T$ (ps)            & 200    \\
          & $L_{atten}$ (m)            & 3.5 -- 5.5  \\\hline
BSC       & $\Delta E/\sqrt{E}$ ($\%$) & $22\%$ \\
          & $\sigma_{z}$(cm)           & 2.3 \\\hline
$\mu$ counter & $\sigma_{z}$(cm)       & 5.5 \\\hline
DAQ       & dead time (ms)             &  8  \\\hline
\end{tabular}
\end{center}
\end{table}

\begin{table}[!htb]
\caption{Summary of BES data sets.}\label{tab:datasets}
\begin{center}
%\centering
\vspace{0.1in}
\begin{tabular}{|c|c|c|c|}
\hline
Detector  & Physics  & $E_{CM}$ (GeV)  &  Sample              \\\hline
         & $J/\psi$  &  3.097          & $7.8 \times 10^6$   \\
 BES~I    &  $m_{\tau}$ &  3.55 scan & 5 pb$^{-1}$  \\ 
          & $D_s$, $D$ &  4.03           & 22.3 pb$^{-1}$         \\
          & { $\psi(2S)$} & { 3.686} & { $3.8 \times
10^6$ } \\\hline
          & $R$--scan &  2--5 scan       & 6 + 85 points       \\
 BES~II   & { $\psi(2S)$-scan} &  { $\sim$ 3.686}  & { 24 points}     \\
          & {$J/\psi$} & {3.097 }  & {$ 58 \times
10^6$}   \\ 
          & $\psi(3770)$ para. & $\psi(3770)$ scan &  \\ 
          & {$\psi(2S)$} & {3.686} & {14 $\times
   10^6$}   \\ 
          & {$\psi(3770)$} & $\sim$3.770 & $\sim$ 27 pb$^{-1}$ \\ 
          & continuum & 3.65 &  6.4 pb$^{-1}$ \\ \hline
\end{tabular}
\end{center}
\end{table}

\subsection[The CLEO  detector]
           {The CLEO detector
            $\!$\footnote{Author: R.~Galik}}

\subsubsection[The CLEO III and CLEO-c detectors]
              {The CLEO III and CLEO-c detectors}

In the twenty-five year history of the CLEO Collaboration there had
been a succession of detector upgrades that led from CLEO~I to CLEO
I.5 to CLEO~II\cite{CLEOII} to CLEO~II.V.  In preparation for its last
running at the $\Upsilon$(4S)\footnote{The last such $B\overline{B}$
running was in June 2001.}  there was a large scale modification,
primarily aimed at bringing the hadron identification capabilities up
to the same high level as the tracking and electromagnetic
calorimetry.  This configuration, described below, was called CLEO
III.  All of the $b\overline{b}$ resonance data ($\Upsilon$(1S),
$\Upsilon$(2S), $\Upsilon$(3S)) and the $\Upsilon$(5S) running of
2002--3 were taken with this CLEO~III configuration. A conference
proceeding on the commissioning and initial performance evaluation of
CLEO~III has been published\cite{Viehhauser}.

\begin{figure} 
\begin{center} 
\includegraphics[width=0.6\textwidth,angle=270]{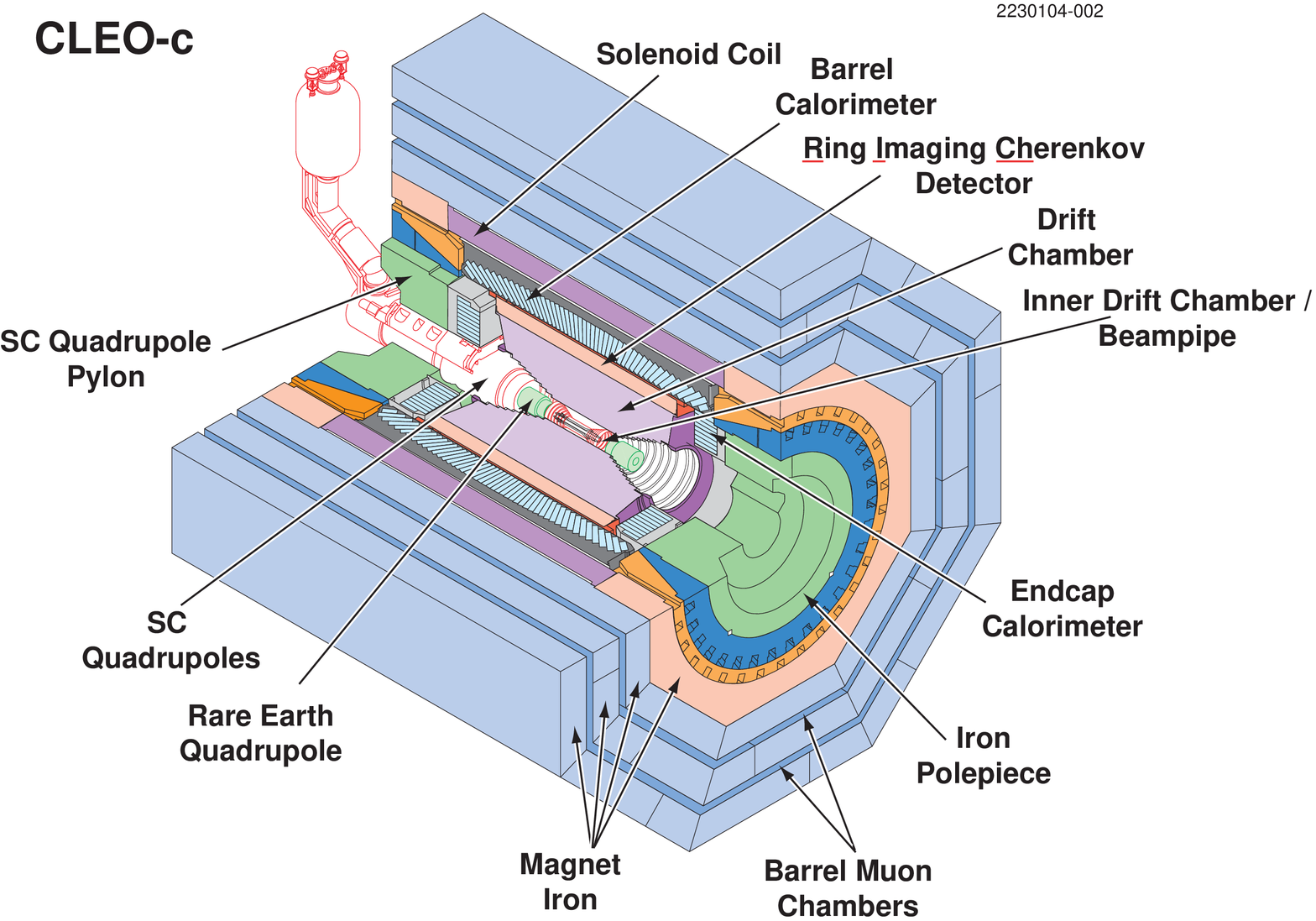} 
\end{center} 
\caption{CLEO-c.}
\label{fig:CLEOc}
\end{figure}

The transition to running in the $c\overline{c}$ region called for
rethinking the optimization of various components, particularly
tracking, in that the magnetic field would be lowered from 1.5~T to
1.0T to accommodate CESR having to handle the solenoid compensation of
such ``soft'' beams.  A thorough study was completed and available as
a Laboratory preprint\cite{cleocycz}, often referred to as the
``CLEO-c Yellow Book''.  The modifications are described below and the
cut-away view of the detector is shown in \Figure~\ref{fig:CLEOc}
  
%%%%%%%%%%%%%%%%%%%%%% FIGURE %%%%%%%%%%%%%%%%%%%%%%%%%%%%%%%% 

\subsubsection[CLEO~III]{CLEO~III}

As noted above the thrust of the upgrade to CLEO~III was to greatly
enhance hadronic particle identification without sacrificing the
excellent charged particle tracking and electromagnetic calorimetry of
CLEO~II\cite{CLEOII}. CLEO chose to accomplish this with a {\bf
ring-imaging Cherenkov (RICH) detector} which has an active region of
81\% of $4\pi$, matching that of the barrel calorimeter.  Details of
the construction and performance have been published\cite{RICH}; a
summary follows.

The RICH construction has LiF radiators of thickness 1cm followed by a
nitrogen-filled expansion volume of 16~cm.  The Cherenkov photons then
pass through a CaF$_{2}$ window into the photo-sensitive gas, for
which a mixture of triethylamine (TEA) and methane (CH$_{4}$) is used.
Readout is done on the 250,000 cathode pads that sense the avalanche
of electrons liberated in the TEA--CH$_{4}$ and accelerated to anode
wires.  To minimize effects of total internal reflection the LiF
radiators in the central region, \ie nearest the interaction
region (IR), have a sawtooth pattern cut on their outer surface.

From Bhabha scattering calibrations, the single photon angular
resolution ranges from 13 (nearest the IR) to 19 mrad (furtherest from
the IR).  The number of {\it detected} photons averages 12 in the
central, sawtooth region and 11 in the outer, flat radiator regions.
This leads to a Cherenkov angle determination of resolution better
than 5 mrad, except at the very outer edges along the beam direction,
in good agreement with simulations of the device.

As always, the performance is a trade-off between fake-rate (mis-ID)
and efficiency.  Charged kaons and pions in the decay of a $D$ meson
in the chain $D^{*} \to D\pi \to (K\pi)\pi$ can be identified using
kinematics.  Such a sample shows that below $p = 2 {\rm GeV/}c$ even
90\% efficiency for kaon identification has less than a 2\% fake-rate
for pions.  At $p = 2.6 {\rm GeV/}c$, the kinematic limit for $B$
decay, 80\% efficiency still has only a fake-rate of 8\%.  These
identification capabilities are enhanced by using dE/dx in the drift
chamber (described next). The ultimate efficiency/fake performance is
very specific to the decay channel of interest.

The RICH takes up some 15~cm in radius more than the previous
scintillator system from CLEO~II.  This meant a new {\bf drift
chamber} was to be built that would have the same momentum resolution
as that of CLEO~II but with reduced radius, spanning 12--82~cm from the
beam line.  Again, a detailed document has been published\cite{DR3},
of which a summary follows.

Accomplishing this meant minimizing mass (use of a helium based gas,
namely 60\%He--40\%C$_{3}$H$_{8}$; thin inner support cylinder, 0.12\%
radiation lengths; aluminum field wires with gold-plating), carefully
monitoring hole and wire positions, and paying close attention to
field wire geometry.  The innermost 16 layers are axial while the
outer 31 layers are stereo with sequential superlayers (of four layers
each) alternating in the sign of the stereo angle.  Both the axial and
stereo sections participate in the CLEO~III trigger.  The end plates
consist of a highly tapered assembly for the axial layers (allowing
full tracking coverage over 93\% of the solid angle) and a slightly
conical outer section that minimizes end plate mass (greatly improving
the energy resolution of the end cap CsI electromagnetic calorimeter).
The outer cylinder is instrumented with cathode strips for additional
$z$ measurements.

Spatial resolution within the cells is parametrized by two Gaussians
with the narrower constrained to have 80\% of the fitted area.
Averaged over the full cell this narrow component is 88 $\mu$m with
the middle of the cell being as good as 65 $\mu$m.  Some figures of
merit from 5~GeV/$c$ Bhabha tracks are a momentum resolution of 55
MeV/$c$ , a $z$ resolution of 1.2~mm from the cathodes and of 1.5~mm
from the stereo anodes, and dE/dx resolution of 5.0\%, which means
$K/\pi$ separation to 700~MeV/$c$ of hadron momentum.  All measures of
performance are beyond the design specifications.

To provide extremely accurate track position measurements in both the
azimuthal and $z$ coordinates, CLEO had installed a three layer,
double sided {\bf silicon vertex detector}\cite{Hill} which was the
distinguishing feature of CLEO~II.V.  For CLEO~III this was upgraded
to a four layer device \cite{Fast} with the smallest radius being 2.5
cm.  While the $z$ readout sides performed well throughout the
lifetime of CLEO~III the $r-\phi$ side quickly showed declining
efficiency, in unusual patterns, that has never been explained.  This
led us to rethink this innermost tracker with the advent of CESR-c
(see below).

The other hardware components of CLEO~III were the same as for CLEO
II.  The $\sim$8000 CsI crystals of the {\bf electromagnetic
calorimeter} still perform very well; the endcap regions were
re-stacked to allow for better focusing quadrupoles and greatly
benefited from the reduced material in the drift chamber endplate.
The {\bf muon system} was unchanged as was the {\bf superconducting
solenoid} with the exception of some reshaping of the endcap pole
pieces.  The magnetic field for all the $\Upsilon$ region running was
1.5~T.  The {\bf trigger} and {\bf data acquisition} systems were
totally revamped for CLEO~III, allowing CLEO to be extremely efficient
and redundant for even low multiplicity events and have minimal dead
time up to read out rates of 1 kHz.

\subsubsection[CLEO-c]{CLEO-c}

Very few changes were needed in preparing for the transition to CLEO-c
data collection in the $c\overline{c}$ region.  Both the average
multiplicity and average momenta of charged tracks are lower, so
particle identification via the RICH and dE/dx becomes even better
than at CLEO~III energies.  The lowered magnetic field strength of
1.0T means recalibration of the drift chamber, but actually improves
the ability to trigger on and find low momentum tracks.  The muon
chambers become less useful for identifying leptons from the
interaction region in that such muons range out in the iron; however,
the chambers are still a useful veto of cosmic rays.  The CsI
calorimeter routinely identifies showers down to 70~MeV, so it needed
no modification, other than changing the thresholds in its trigger
hardware to accommodate lowered energies of Bhabha scattering events.

The premature aging of the CLEO~III silicon meant that we had to
either replace it or substitute a small wire chamber.  The CLEO-c
program does not have the stringent vertexing requirements of CLEO~III
(the $D$ mesons are at rest in CLEO-c!). Further, track reconstruction
is optimized by having fewer scattering surfaces.  After detailed
studies of mass reconstruction and other figures of merit, it was
decided to build a six-layer stereo chamber with similar design as the
main drift chamber.  In this case the {\it outer} skin is very thin
($\sim$ 0.1\% of a radiation length), so that this small chamber and
the larger one look as much as possible like a single volume of gas.
The stereo wires (strung at 10--15 degrees) are needed to get $z$
information for low-momentum tracks that do not reach beyond the axial
layers of the main chamber.

This new wire chamber has been installed, calibrated, commissioned and
fully integrated into CLEO hardware and software; it is highly
efficient and has a very low noise occupancy.  The first CLEO-c data
uses this new device in its track fitting algorithms, although work
continues in areas such as calibration and alignment to optimize its
contributions to tracking.
 
\subsection[KEDR]{KEDR $\!$\footnote{Author: S.~Eidelman}}
\label{sec:kedr} 

\begin{figure}[h]
\begin{center}
\includegraphics[width=13cm]{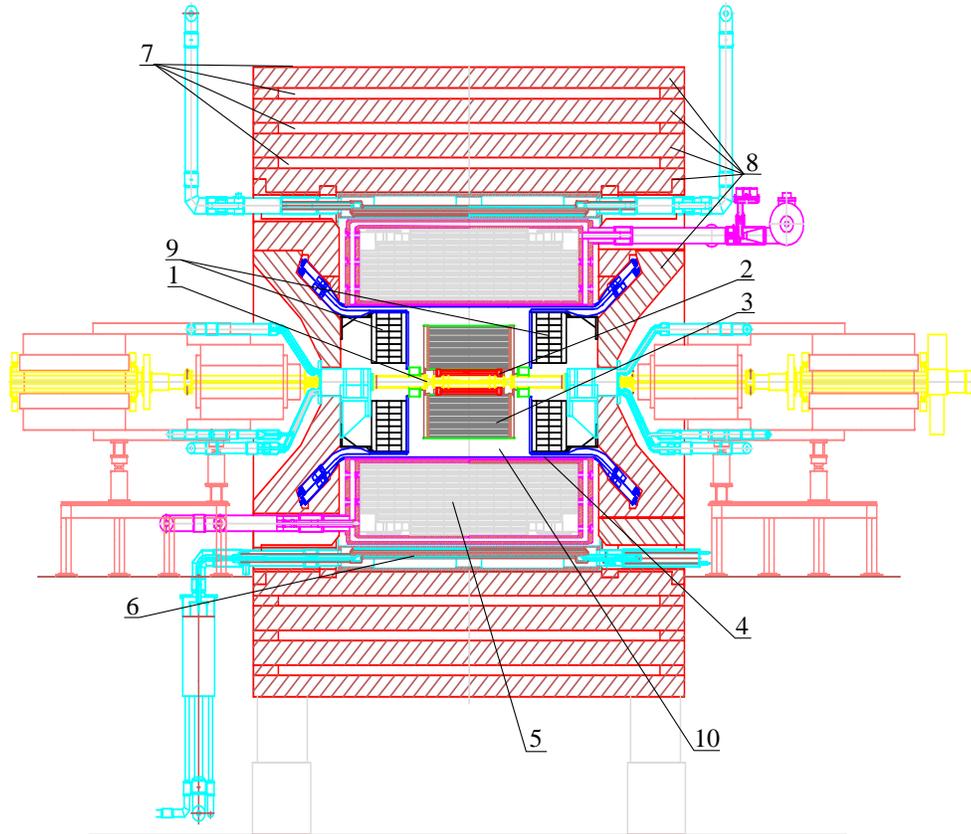}
\end{center}
\caption{%
Layout of the KEDR detector: 
1~--~beam pipe,
2~--~vertex detector,
3~--~drift chamber,
4~--~TOF scintillation counters,
5~--~LKr barrel calorimeter, 
6~--~superconducting coil, 
7~--~muon tubes,
8~--~magnet yoke,
9~--~CsI endcap calorimeter,
10~--~Aerogel Cherenkov counters}
\label{figkedr}
\end{figure}

The KEDR detector described in detail elsewhere~\cite{KEDR} is shown
in \Figure~\ref{figkedr}.

It consists of the vertex detector, the drift chamber, the
time-of-flight system of scintillation counters, the particle
identification system based on the aerogel Cherenkov counters, the
calorimeter (the liquid krypton in the barrel part and the CsI
crystals in the end caps) and the muon tube system inside and outside
of the magnet yoke.  In this experiment the magnetic field was off and
the liquid krypton calorimeter as well as aerogel counters were out of
operation.

The detection efficiency, determined by the visible peak height and
the table value of the leptonic width, is about 0.25 for the $J/\psi$
(\mbox{$\sim 20\cdot 10^{3}$} events) and about 0.28 for the $\psi'$
(\mbox{$\sim 6\cdot 10^{3}$} events).

Luminosity was measured by events of Bhabha scattering detected in the
end-cap CsI calorimeter.

\newpage
\section[B-factories]{B-factories}
\label{sec:Bfact}

\subsection[BaBar]{BaBar $\!$\footnote{Author: E.~Robutti}}

\babar\ is a general-purpose detector, located at the only interaction
point of the electron and positron beams of the \pep2\ asymmetric
collider at the Stanford Linear Accelerator Center. Although its
design has been optimized for the study of time-dependent $CP$
asymmetries in the decay of neutral $B$ mesons, it is well suited for
the study of a broad range of physics channels of interest, taking
profit from the large samples of data made available by the high
luminosity.

\begin{figure}[h!]
  \begin{center}
    \includegraphics*[width=16cm]{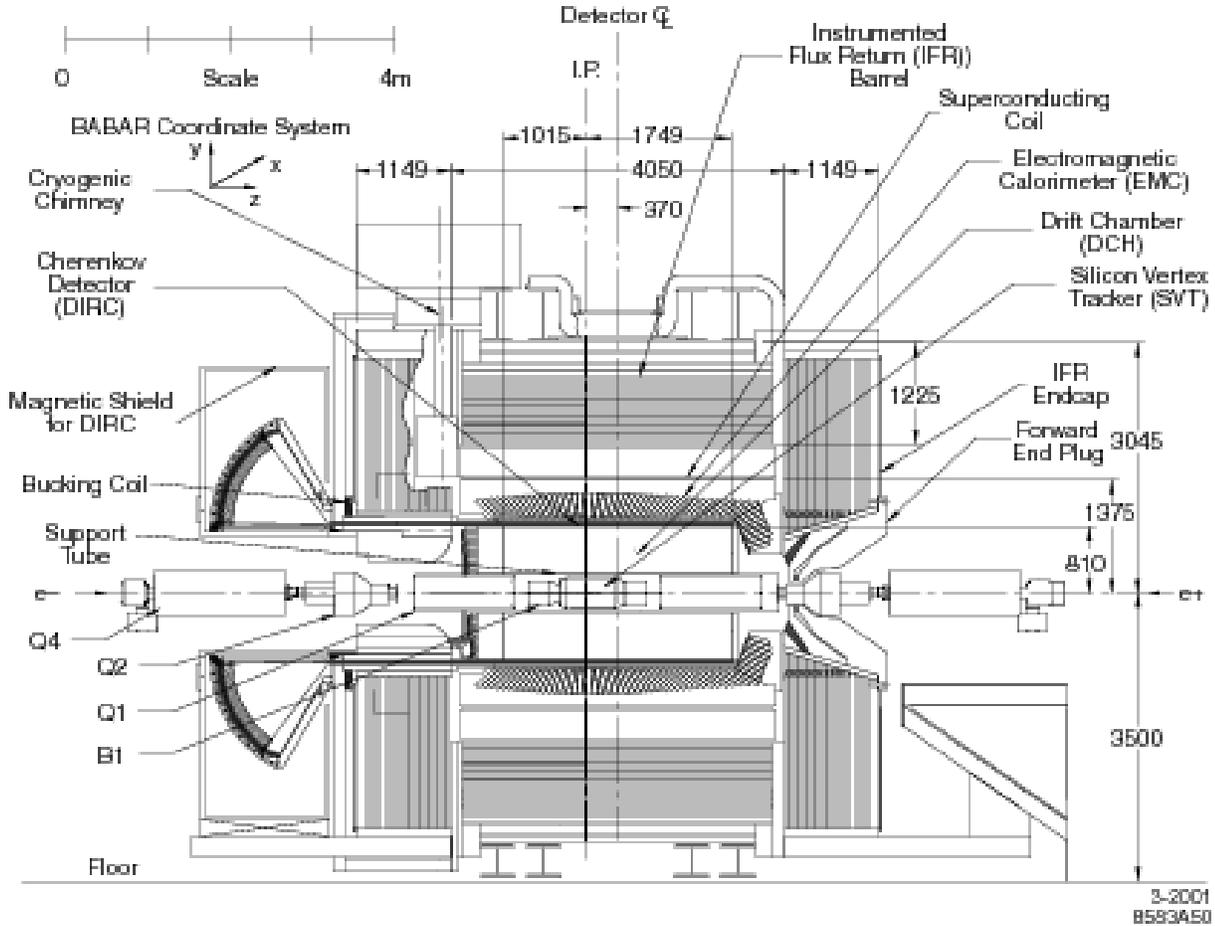}
    \caption{\babar\ detector longitudinal section.}
    \label{fig:babarDet_elev}
  \end{center}
\end{figure}

The \pep2\ $B$-factory operates at an energy of 10.58~GeV, equal to
the mass of the $\Upsilon(4S)$ meson; the colliding electron and
positron beams have an energy of 9 and 3.1~GeV, respectively,
corresponding to a Lorentz boost of the centre of mass of
$\beta=0.55$. The maximum instantaneous luminosity now exceeds
$9\times10^{33}~\mathrm{cm}^{-2}\mathrm{s}^{-1}$, well above the
design value of $3\times10^{33}~\mathrm{cm}^{-2}\mathrm{s}^{-1}$.  The
peak cross-section for formation of the $\Upsilon(4S)$ (which then
decays exclusively to $B^+B^-$ or $B^0\overline{B}^0$) is about 1~nb;
at the same energy, the total cross-section for $\ep\rightarrow \qqb$
($q=u, d, s, c$) is about 3~nb: in particular,
$\sigma(\ep\rightarrow\ccbar) \approx 1.3~\mathrm{nb}$. Of particular
interest for the study of charmonium states are also events where the
effective \ep\ energy is lowered by the initial emission of a photon
({\it Initial State Radiation}, or ISR), and $\gamma \gamma$ fusion
processes, where the two photons are radiated by the colliding beams:
both of them occur at substantial rates in the energy range of the
charmonium spectrum.

\begin{figure}[h!]
  \begin{center}
    \includegraphics*[width=16cm]{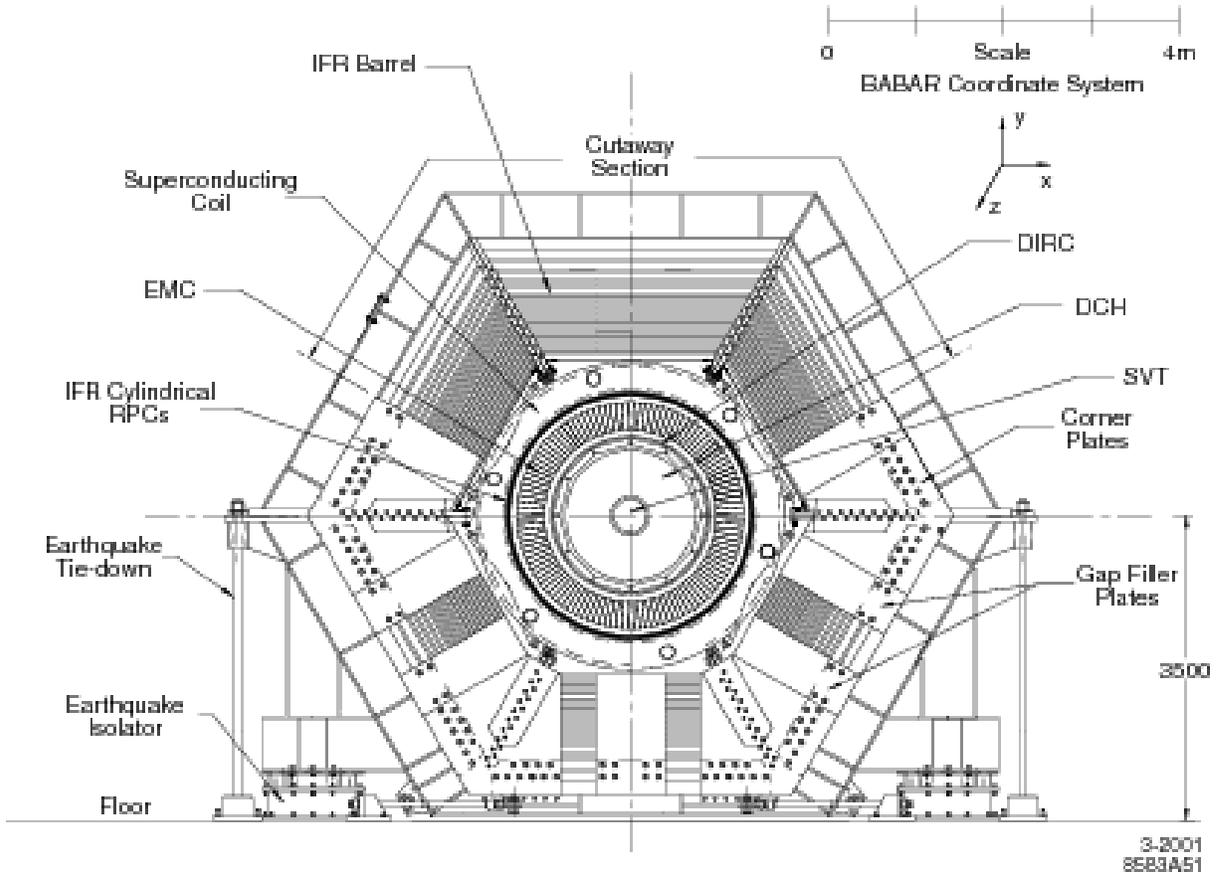}
    \caption{\babar\ detector end view.}
    \label{fig:babarDet_end}
  \end{center}
\end{figure}

A longitudinal section and an end view of the \babar\ detector are
shown in \Figure~\ref{fig:babarDet_elev} and
\Figure~\ref{fig:babarDet_end}. respectively. The structure is that
typical of full-coverage detectors at collider machines, except for a
slight asymmetry in $z$, with a larger acceptance in the positive
direction of the electron beam (``forward''), which reflects the
asymmetry in the beam energies.

The inner part of the apparatus is surrounded by a superconducting
solenoid providing the 1.5~T magnetic field used for the measurement
of particle charges and momenta.  It includes the tracking, particle
identification and electromagnetic calorimetry systems.

The tracking system is composed of a {\it Silicon Vertex Tracker}
(SVT) and a {\it Drift CHamber} (DCH). The SVT is a five-layer,
double-sided silicon strip detector, which is used for precision
measurements of the primary and secondary decay vertices, as well as a
stand-alone tracking device for particles with low transverse momentum
($50-120~\mevc$). The DCH is a 40-layer cylindrical drift chamber with
a helium--isobutane mixture as the sensitive gas, and is the primary
device used for the measurement of particle momenta; it is also used
for the reconstruction of secondary vertices outside the outer radius
of the SVT. Both detectors provide redundant $\mathrm{d}E/\mathrm{d}x$
samplings for particle identification of charged hadrons with momenta
below $\sim700~\mevc$.

The tracking reconstruction efficiency exceeds 95\% for tracks with
transverse momentum above $200$ $\mevc$. The resolution for the track
impact parameters is about 25 and 40 $\mu\mathrm{m}$ in the transverse
plane and along the detector axis, respectively. The momentum
resolution is well described by the linear relation:
$\sigma_{p_\mathrm{t}}/p_\mathrm{t} \simeq 0.45\% + 0.13\% \cdot
p_\mathrm{t} (\gevc)$. The $\mathrm{d}E/\mathrm{d}x$ resolution at
$1~\gevc$ is about 7.5\%.

Separation of pions and kaons at momenta above $500~\mevc$ is provided
by the DIRC ({\it Detector of Internally Reflected Cherenkov
light}). This is a novel kind of ring-imaging Cherenkov detector, in
which Cherenkov light is produced in bars of fused silica and
transported by total internal reflection, preserving the angle of
emission, to a water tank viewed by an array of photomultipliers
tubes. The pion--kaon separation obtained after association of signals
to the tracks ranges from about $10 \sigma$ at $1~\gevc$ to about $3
\sigma$ at $4~\gevc$.

The {\it ElectroMagnetic Calorimeter} (EMC) is a finely segmented
array of CsI(Tl) crystals with projective geometry. Its energy
resolution is well described by the relation $\sigma_E/E \simeq 2.3\%
\cdot E({\rm GeV})^{-1/4} \oplus 1.9\%$; the angular resolution ranges
from about 12~mrad at low energies to about 3~mrad at high energies;
the width of the reconstructed $\pi^0$ mass peak is about
$7~\mevc^2$.

Outside the superconducting coil is the detector for muons and neutral
hadrons, called {\it Instrumented Flux Return} (IFR): the iron return
yoke of the magnet is segmented into layers of increasing thickness
from the inside to the outside, interspersed with Resistive Plate
Chambers as the active elements. Muons are identified by criteria
exploiting the deepest penetration of their tracks into the iron: a
typical efficiency for a selector was about 90\% in the momentum range
$1.5 < p < 3~\gevc$\, with a pion fake rate for pions of about
6--8\%. The RPC have suffered a loss of efficiency since the beginning
of operation, causing a small degradation in the performance of muon
selectors. In the barrel section of the IFR, they will be substituted
by Limited Streamer Tubes, starting from Summer 2004.

\Table~\ref{tab:babarPerf} summarizes parameters and performances of
the different subsystems composing the \babar\ detector.

\begin{table}
\caption{Overview of the coverage, segmentation, and performance of
         the \babar\ detector systems.  The notation (C), (F), and (B)
         refers to the central barrel, forward and backward components
         of the system, respectively.  The detector coverage in the
         laboratory frame is specified in terms of the polar angles
         $\theta_1$ (forward) and $\theta_2$ (backward). Performance
         numbers are quoted for $1~\gevc^2$ particles, except where
         noted.}
\label{tab:babarPerf}
\begin{tabular}{@{}lcccccccc@{}}
\hline\hline
System &$\theta_1$ & & No.  & ADC & TDC&No.  & Segmentation & Performance \\ &($\theta_2)$& & Channels  & (bits)&(ns)&Layers & & \\
\hline\hline
 SVT & 20.1$^{\circ}$ & & 150K& 4&--&5 & 50--100 $\mu m~r-\phi$& $\sigma_{d_0}=55~\mu\mathrm{m}$ \\ & (--29.8$^{\circ}$) & & & & & & 100--200 $~\mu\mathrm{m}~z$ & $\sigma_{z_0}=65~\mu\mathrm{m}$ \\
\cline{1-8}
 DCH & 17.2$^{\circ}$ & & 7,104 &8&2& 40 & 6--8~mm & $\sigma_{\phi}=1~\mathrm{mrad}$\\ & (--27.4$^{\circ}$) & & & & & & drift distance & $\sigma_{tan\lambda}=0.001$\\ & & & & & & & & $\sigma_{\pt}/\pt = 0.47\%$ \\ & & & & & & & & $\sigma(dE/dx)=7.5$\%\\
\hline
  DIRC & 25.5$^{\circ}$ & & 10,752&--&0.5& &35 $\times$ 17~mm$^2$& $\sigma_{\theta_C} =2.5\mathrm{mrad}$\\
   & (--38.6$^{\circ}$) & & & & & &  ($r\Delta\phi\times\Delta r$)& per track \\
   & & & & & & & 144 bars & \\
    \hline
     EMC(C) & 27.1$^{\circ}$ & & $2 \times 5760$ &17--18&--& & 47 $\times$ 47~mm$^2$ & $\sigma_E/E=3.0\%$ \\
    & (--39.2$^{\circ}$) & & & & & & 5760 cystals &$\sigma_{\phi}= 3.9~\mathrm{mrad}$ \\
   EMC(F) & 15.8$^{\circ}$ & &$2 \times 820$ & & & & 820 crystals &$\sigma_{\theta}=3.9~\mathrm{mrad}$ \\
  & (27.1$^{\circ}$) & & & & & & & \\
\hline
 IFR(C) & 47$^{\circ}$ & & 22K+2K &1& 0.5 & 19+2 & 20--38~mm & 90\% $\mu^{\pm}$ eff.  \\
  & (--57$^{\circ}$) & & & & & & & 6--8\% $\pi^{\pm}$ mis-id \\
  IFR(F) & 20$^{\circ}$ & & 14.5K & & & 18 & 28--38~mm & (loose selection, \\ & (47$^{\circ}$) & & & & && &
 $1.5$--$3.0~\gevc^2$) \\
  IFR(B) & --57$^{\circ}$ & & 14.5K & & & 18 & 28--38~mm \\
   & (--26$^{\circ}$) & & & & & & \\
\hline\hline
\end{tabular}
\end{table}

The trigger system includes a first hardware level, L1, collecting
information from the DCH, EMC and IFR, and a software level, L3,
selecting events for different classes of processes of physics
interest. Output rates are currently around 3~kHz for L1 and 120 Hz
for L3. The combined efficiency exceeds 99.9\% for $B\overline{B}$
events, and is about 99\%, 96\% and 92\% for \ccbar, $uds$ and
$\tau\tau$ events.

\subsection[Belle]{Belle $\!$\footnote{Author: B.~Yabsley}}

The purpose of the Belle experiment is to study time-dependent CP
asymmetries in the decay of $B$-mesons, such as $B^0 \to
J/\psi\,K_S^0$, $\pi^+\pi^-$, and $\phi K_S^0$.  The experiment is
therefore designed to provide boosted $B^0\overline{B}{}^0$ pairs,
allowing decay-time differences to be measured as differences in
$B$-meson decay position; vertex resolution of order
$50\,\mu\mathrm{m}$, to measure those decay positions; and
high-acceptance tracking and electromagnetic calorimetry, to measure
the decay products.  Efficient electron and muon identification are
required to reconstruct the $J/\psi$, and kaon/pion separation is
required to distinguish kaons (\eg for $B$-meson flavour
tagging) and pions (\eg for separation of $B^0 \to \pi^+
\pi^-$ from $K^+ \pi^-$ decays).  Detection of $K_L^0$ mesons is also
desirable, to allow measurement of $B^0 \to J/\psi\,K_L^0$ and $\phi
K_L^0$ modes as a complement to $J/\psi\,K_S^0$ and $\phi K_S^0$.

\begin{figure}[ht]
  \begin{center}
    \includegraphics[angle=-90,width=\textwidth]{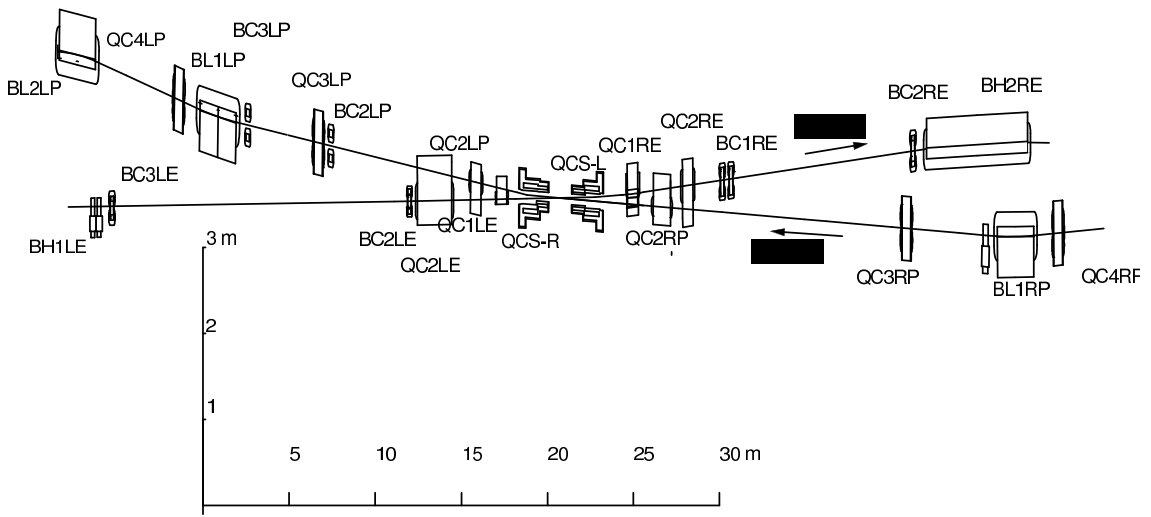}
  \end{center}
  \caption{Layout of the KEKB interaction region.}
  \label{fig:belle-ir}
\end{figure}
Belle is therefore suited to a wide range of other physics analyses,
particularly in the $e^+ e^- \to c \bar{c}$ continuum, and in the production and
decay of charmonium states. The experiment has an active programme of study 
in both of these fields.

\begin{figure}[ht]
  \begin{center}
    \includegraphics[width=14cm]{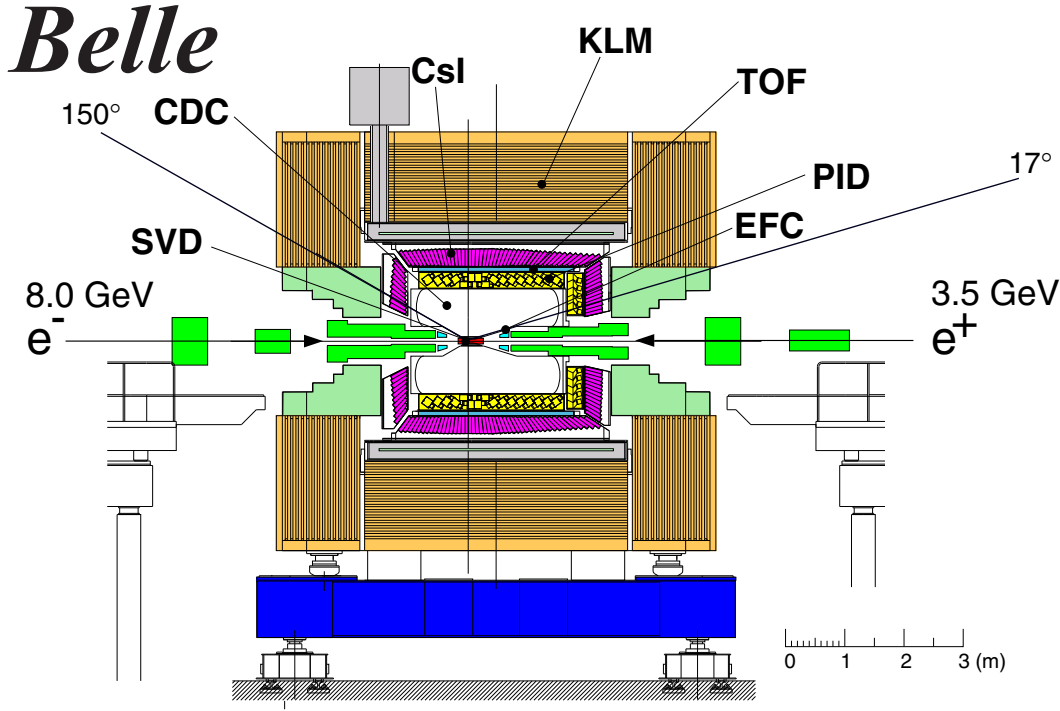}
  \end{center}
  \caption{Side view of the Belle detector.}
  \label{fig:belle-side}
\end{figure}

The detector is located at the interaction point of the KEKB $e^+ e^-$
collider~\cite{KEKB} at K.E.K. in Tsukuba, Japan. 
KEKB consists of an injection linear accelerator and 
two storage rings 3~km in circumference, with asymmetric energies:
8~GeV for electrons and 3.5~GeV for positrons.
The $e^+ e^-$ centre-of-mass system has an energy at the $\Upsilon(4S)$ resonance
and a Lorentz boost of $\beta \gamma = 0.425$. 
The interaction region is shown in \Figure~\ref{fig:belle-ir}: 
the lower-energy positron beam is aligned with the axis of the Belle detector,
and the higher-energy electron beam crosses it at an angle of 22~mrad.
This arrangement allows a dense fill pattern without parasitic collisions,
and also eliminates the need for separation bend magnets. 
KEKB's luminosity is the world's highest,
exceeding the $10^{34}\,\mathrm{cm}^{-2}\mathrm{s}^{-1}$ design value:
with the introduction of continuous beam injection, a record luminosity of
$13.9 \times 10^{33}\,\mathrm{cm}^{-2}\mathrm{s}^{-1}$ was achieved in June 2004;
further improvements are foreseen with the introduction of crab cavities.

The Belle detector~\cite{belle-detector}, shown in side view in
\Figure~\ref{fig:belle-side}, is built into a 1.5~Tesla
superconducting solenoid magnet of 1.7 metre radius.  (Compensating
solenoids and final-focus quadrupole magnets can also be seen on the
beamline, inside the main solenoid volume.)  The design is that of a
classic barrel spectrometer, but with an asymmetry along the beam axis
to provide roughly uniform acceptance in the $e^+ e^-$ centre-of-mass.

\begin{figure}[ht]
  \begin{center}
    \includegraphics[width=\textwidth]{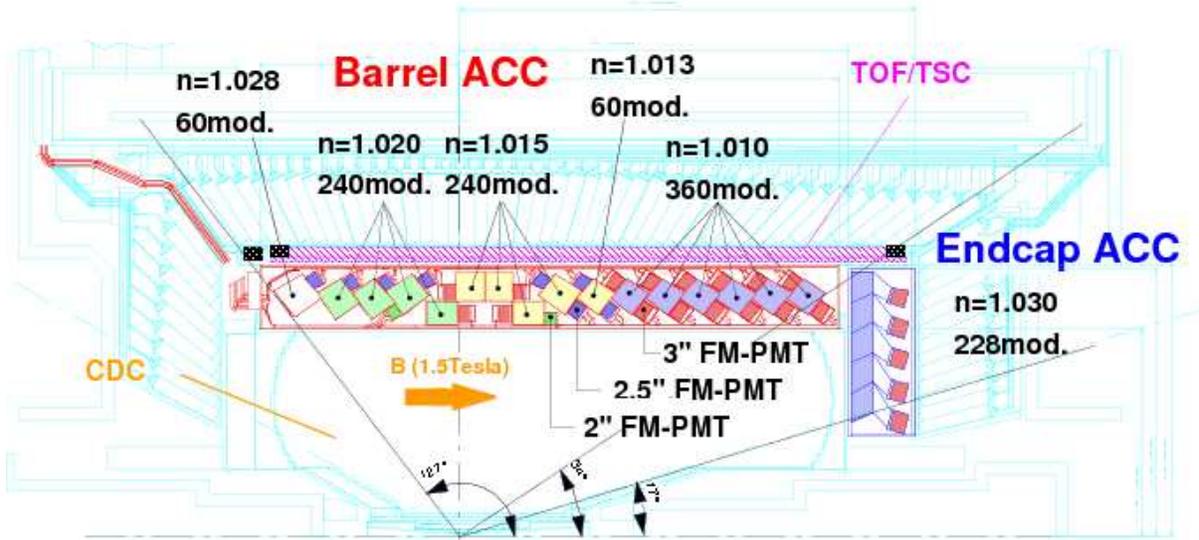}
  \end{center}
  \caption{Half-section of the inner region of the Belle detector,
        showing the layout of the subdetectors used in 
        kaon/pion separation.}
  \label{fig:belle-kid}
\end{figure}

Precision tracking and vertex measurements are provided
by a central drift chamber (CDC) and a silicon vertex detector (SVD).
The CDC is a small-cell cylindrical drift chamber
with 50 layers of anode wires including 18 layers of stereo wires.
A low-$Z$ gas mixture (He (50\%) and $\mathrm{C}_2\mathrm{H}_6$ (50\%))
is used to minimize multiple Coulomb scattering,
ensuring good momentum resolution for low momentum particles.
The tracking acceptance is $17^\circ < \theta < 150^\circ$,
where the laboratory polar angle $\theta$
is measured with respect to the (negative of the) positron beam axis.
The SVD consists of double-sided silicon strip detectors arranged in a barrel,
covering 86\% of the solid angle.
Three layers at radii of 3.0, 4.5 and 6.0~cm surround the beam-pipe,
a double-wall beryllium cylinder of 2.3~cm radius and 1~mm  thickness.
The strip pitches are $42\,\mu\mathrm{m}$ in the the $z$ (beam-axis) coordinate
and $25\,\mu\mathrm{m}$ for the azimuthal coordinate $r\phi$;
in each view, a pair of neighbouring strips is ganged together for readout.
The impact parameter resolution for reconstructed tracks is measured as
a function of the track momentum $p$ (measured in $\mathrm{GeV}/c$) to be 
$\sigma_{xy} = [19 \oplus 50/(p\beta\sin^{3/2}\theta)]\,\mu\mathrm{m}$ and
$\sigma_{z} = [36 \oplus 42/(p\beta\sin^{5/2}\theta)]\,\mu\mathrm{m}$.
The momentum resolution of the combined tracking system is 
$\sigma_{p_{\mathrm{t}}}/p_{\mathrm{t}} = (0.30/\beta \oplus 0.19p_{\mathrm{t}})$\%, 
where $p_{\mathrm{t}}$ is the transverse momentum in $\mathrm{GeV}/c$.

The subdetectors used in kaon/pion separation are shown in
\Figure~\ref{fig:belle-kid}: the CDC, a barrel arrangement of
time-of-flight counters (TOF), and an array of aerogel Cherenkov
counters (ACC).  The CDC measures energy loss for charged particles
with a resolution of $\sigma(dE/dx)$ = 6.9\% for minimum-ionizing
pions.  The TOF consists of 128 plastic scintillators viewed on both
ends by fine-mesh photo-multipliers that operate stably in the 1.5~T
magnetic field.  Their time resolution is 95~ps ($rms$) for
minimum-ionizing particles, providing three standard deviation
(3$\sigma$) $K^\pm/\pi^\pm$ separation below 1.0~GeV/$c$, and
2$\sigma$ up to 1.5~GeV/$c$.  The ACC consists of 1188 aerogel blocks
with refractive indices between 1.01 and 1.03 (see
\Figure~\ref{fig:belle-kid}) depending on the polar angle.  Fine-mesh
photo-multipliers detect the Cherenkov light: the effective number of
photoelectrons is $\sim 6$ for $\beta =1$ particles.  Information from
the three subdetectors is combined into likelihoods $\mathcal{L}_K$,
$\mathcal{L}_\pi$ \etc for various particle identification hypotheses,
and likelihood ratios such as $\mathcal{R}_{K/\pi} =
\mathcal{L}_K/(\mathcal{L}_K +\mathcal{L}_\pi)$ are used as
discriminators.  A typical selection with $\mathcal{R}_{K/\pi} > 0.6$
retains about 90\% of the charged kaons with a charged pion
misidentification rate of about 6\%.

Photons and other neutrals are reconstructed in a CsI(Tl) calorimeter
consisting of 8736 crystal blocks in a projective geometry,
16.1 radiation lengths deep, covering the same angular region as the CDC.
The energy resolution is 1.8\% for photons above 3~GeV.
Electron identification is based on a combination of
$dE/dx$ measurements in the CDC,
the response of the ACC,
the position and shape of the electromagnetic shower,
and the ratio of the cluster energy to the particle momentum.
The electron identification efficiency is determined 
from two-photon $e^+e^-\rightarrow e^+e^-e^+e^-$ 
processes to be more than 90\% for $p >1.0\,\mathrm{GeV}/c$.
The hadron misidentification probability,
determined using tagged pions from
inclusive $K_S^0\rightarrow \pi^+\pi^-$ decays, is below $0.5\%$.  

Outside the solenoid, the flux return is instrumented
to provide a $K_L^0$ and muon detector (KLM).
The active volume consists of 14 layers of iron absorber (4.7~cm thick)
alternating with resistive plate counters (RPCs),
covering polar angles $20^\circ < \theta < 155^\circ$.
The overall muon identification efficiency,
determined by using a two-photon process 
$e^+e^-\rightarrow e^+e^-\mu^+\mu^-$  
and simulated muons embedded in $B\overline{B}$ candidate events,
is greater than 90\% for tracks with $p > 1\,\mathrm{GeV}/c$ detected
in the CDC.  The corresponding
pion misidentification probability, determined using
$K_S^0\rightarrow \pi^+\pi^-$  decays, is less than 2\%. 

The Belle trigger and event selection are essentially open for hadronic events,
with over 99\% efficiency for $B\overline{B}$ and somewhat less for
$e^+e^- \to c\bar{c}$ and light-quark continuum processes. 
Analysis of such events is performed using a common hadronic event skim;
special provision is made to retain events with a $J/\psi$ or $\psi(2S)$
candidate but otherwise low multiplicity.
Tau-pair and two-photon
($e^+ e^- \to e^+ e^- \gamma \gamma \to e^+ e^- X$)
events are studied using dedicated triggers and data skims.

$154\,\mathrm{fb}^{-1}$ of data were taken in the configuration
described above.  An upgrade in summer 2003 replaced the SVD and the
innermost drift-chamber layers with a four-layer silicon detector
covering the same range in polar angle as the CDC.  The beam-pipe
radius was reduced to 1.5~cm and the inner SVD layer to 2.0~cm,
placing the first reconstructed hit of each track closer to the
interaction point.  Position resolution is similar to that of the
original SVD, with strip pitches of $75\,\mu\mathrm{m}$ ($z$) and
$50\,\mu\mathrm{m}$ ($r\phi$); every strip is read out.  A further
$124\,\mathrm{fb}^{-1}$ has been collected in this configuration
through the middle of June 2004.  Possible future upgrades to the
particle identification system, and further upgrades to the vertexing,
are currently under study.

\section[$\bar{p}p$ charm factories]{$\bar{p}p$ charm factories
$\!$\footnote{Author: R.~Mussa}
}
\label{sec:ppbar}
\subsection[E835]{E835}

The E835 experiment was located in the Fermilab Antiproton
Accumulator, where a stochastically cooled ($\Delta p/p \sim 10^{-4}$)
beam intersects an internal jet target of molecular hydrogen. The
$\bar{p}$ beam was injected in the Accumulator with an energy of
8.9~GeV and decelerated to the 3.7--6.4~GeV energy range, to form the
charmonium states. Stochastic cooling allowed to reduce RMS spreads on
$\sqrt{s}$ to less than 250~keV.  The E835 experiment was the
continuation of the E760 experiment, that took data in years 1990--91,
at a typical instantaneous luminosity ${\cal L}\sim 0.5 \cdot
10^{31}$.  The E760/E835 detector, described in detail in
\cite{e835NIM}, was a non-magnetic cylindrical spectrometer with full
azimuthal coverage and polar angle acceptance from 2 to 70 degrees in
the lab frame. It consisted of a lead-glass EM calorimeter divided
into a barrel and a forward section.  The inner part of the barrel was
instrumented with a multicell threshold \v{C}erenkov counter,
triggering hodoscopes and charged tracking chambers.  The plastic
scintillator hodoscopes and the \v Cerenkov were used for triggering:
pulse heights from these devices allow to identify electrons/positrons
and to distinguish them singly from electron--positron pairs due to
$\gamma$ conversions and to $\pi^0$ Dalitz decays.

The E835 detector was a major upgrade of the E760 detector:
\begin{itemize}
\item  
The variable target density allowed to keep a constant 
instantaneous luminosity (${\cal L}\sim 2 \cdot 10^{31}$) throughout each
stack. 
\item
In order to withstand the $\sim$3 MHz interaction rate,
all detector channels were instrumented with multi-hit TDCs.
\item
The inner tracking detector, a proportional multiwire drift chamber, 
was replaced by an increased number of straw tubes and scintillating fibers,
which were used for measuring the polar angle $\theta$ and providing
trigger information based on this coordinate.
\end{itemize}
\begin{figure}[ht]
\begin{center}
\includegraphics[angle=270,width=.9\textwidth]{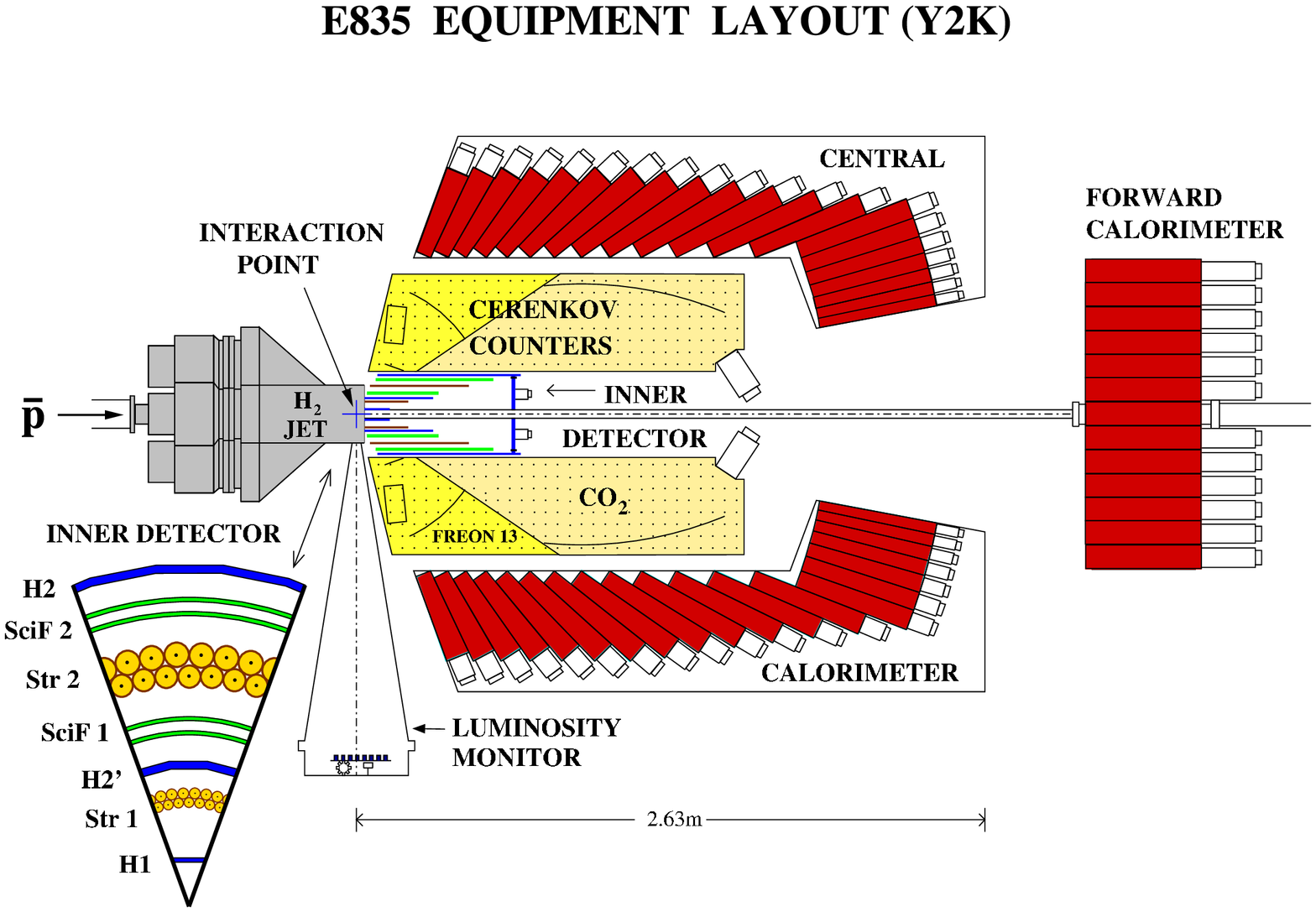}
\vspace*{-0.2cm}
\end{center}
\caption{The E835  detector in year 2000.}
\label{fig:e835}
\end{figure}

The calorimeter had an energy resolution $\sigma_E/E = 0.014 
+ 0.06 / \sqrt{E(\mbox{GeV})}$ and an
angular resolution (r.m.s.) of 11~mrad in $\phi$ and 6~mrad in $\theta$.
The angular resolution of the inner tracking system was 11~mrad in 
$\phi$, whereas in $\theta$ it varies from 3~mrad at small angles to
11~mrad at large angles, dominated by size of the interaction region,
and by multiple scattering at lower momenta.

\begin{table}[ht]
\caption{Integrated luminosities $\calL dt$ (in pb$^{-1}$) 
         taken by E760, E835-I, E835-II}
\label{tab:lumE760+E835}
\begin{center}
\begin{tabular}{|l|l|ccc|}
\hline
State & Decay Channels & E760   &  E835-I& E835-II \\
\hline
$\etac$   & $\gaga$  &  2.76  &  17.7  &  -- \\
$\jpsi$   & $\ep$  & 0.63  &   1.69 &  -- \\
$\chicj{0}$  & $\jpsi\gamma, \gaga, 2\pi^0, 2\eta$ & 
  --    &   2.57 & 32.8 \\
$\chicj{1}$  &  $\jpsi\gamma $ &  1.03  &   7.26 & 6.3 \\
$\hc$ search & $\jpsi\pi^0, \etac\gamma$ &  15.9  &   46.9 & 50.5 \\
$\chicj{2}$  & $\jpsi\gamma, \gaga$ & 1.16  &   12.4 & 1.1 \\
$\etacp$ search & $\gaga$ & 6.36  &   35.0 & -- \\
$\psip$   & $\ep,\chi_{cJ}\gamma, \jpsi\pi^0,$ & & & \\
& $\jpsi\pi^+\pi^-,
\jpsi\pi^0\pi^0,\jpsi\eta$ & 1.47  &   11.8 &15.0 \\
above     & \jpsi+X  &  --   &    2.6 & 7.5 \\
\hline
\end{tabular}
\end{center}
\end{table}

\Table~\ref{tab:lumE760+E835} summarizes the data taken by the two
experiments, subdivided in energy regions. The $h_c$ search region
extends from 3523 to 3529~MeV/c$^2$, \ie 6~MeV/c$^2$ around the
centre of gravity of P states (located at 3525.3).  The $\etacp$ was
searched between 3575 and 3660~MeV/c$^2$.  Experiment E835 took data
in 1996--7 (phase I) and 2000 (phase II).  During the long shutdown
between the two runs, substantial changes in the Antiproton Source
allowed to smoothly scan the $\chicj{0}$ region but prevented to take
new data down to $\jpsi$ and $\etac$ energies.

\section[Experiments at LEP]
        {Experiments at LEP
         $\!$\footnote{Authors: A.~B\"ohrer, M.~Kienzle}}
\label{sec:LEPexp}

At four of the eight straight sections 
of the LEP $\epem$-collider at CERN \cite{LEPbk} 
four collaborations have installed their 
detectors: ALEPH \cite{ALEPHDET}, DELPHI \cite{DELPHIDET}, L3 \cite{L3DET}, 
and OPAL \cite{OPALDET}. The design of the detectors is guided 
by the physics of interest. The detectors consist of several subdetectors 
each dedicated to special aspects of the final state under investigation. 

The main physics goal at LEP is the test of the Standard Model. The
mass and width of the $\zz$ boson are being measured to a high
precision. The couplings of the leptons and quarks to $\gamma / \zz$
are investigated. Special emphasis is put on the study of
$\tau$-decays. The $\tau$-polarization gives a good insight into the
couplings. The high production probability of the heavy flavours,
charm and bottom, allows for investigations of effects, such as
branching ratios, hadron masses, time dependent mixing \etc Indirect
information on the top mass is extracted and the influence from the
Higgs mass is studied. Direct Higgs-search is one of the most
important topics in the new physics area. Supersymmetric particles, if
they exist in the accessible range, should not be able to escape
detection. The strong interaction, with confinement and asymptotic
freedom still not understood, is to be investigated. The perturbative
part (\eg $\as$-determination) and the non-perturbative part,
fragmentation and particle production, guided the design of the
detectors as well.

In addition, the general features of the detectors have to keep the 
systematic uncertainties for their measurements very small to profit from 
the excellent energy calibration of LEP and to efficiently use the high 
event statistics.

All LEP detectors have therefore in common, a good hermiticity as well
as a good efficiency. The total (hadronic) energy has to be measured
as completely as possible. The total absorption guarantees that all
particles except neutrinos are seen. Muons also deposit only a small
fraction of their energy, but are detected in special muon chambers
and by their characteristic signature in the hadron calorimeter. Care
for efficient detection and identification of leptons is taken. In
general particle identification is provided. Good two-track resolution
is possible inside jets of hadrons; energy loss measurements on more
than hundred samplings, high granularity of the calorimeters are
needed. High precision tracking and vertexing of secondary vertices
guaranties good detection and momentum resolution for charged
particles, even in the case when they do not come from the primary
interaction point.

The trigger system ensures that all events of interest are seen with 
low background. The triggers of the four LEP detectors have a high 
redundancy. For example, hadronic events are found when the energy 
exceeds a few GeV in the electromagnetic calorimeter (total energy trigger), 
or two tracks are seen together with energy deposition in 
the hadron calorimeter, which 
exceeds the energy expected for a minimum ionizing particle ($\mu$-trigger). 
The efficiency for hadronic events is $\ge 99.99\%$ with an uncertainty of 
$0.01\%$.

These requirements lead to four LEP detector designs with a similar
general outline, while the detectors differ in their details (see
\Table~\ref{tab:detectors}, \cite{ARMINHAB}). The detectors show a
cylindrical symmetry around the beam pipe. In the forward direction,
calorimeters are installed for the measurement of the luminosity with
high precision. The main body has closest to the beam pipe a vertex
detector mounted, with precision measurements of the hits from tracks
crossing; a general tracking system, which may consist of separate
tracking devices; an electromagnetic calorimeter for measuring
electrons and photons; a coil of a magnet in order to bend charged
particles for the momentum measurement in the tracking devices; a
hadron calorimeter for hadronic showers absorbing strong interacting
particle, but passed by muons; the latter are detected in the muon
chambers, surrounding the experiments.

\begin{table}[htbp]
\caption[Characteristics of the four LEP experiments]
        {Characteristics of the four LEP 
         experiments \protect\cite{ALEPHDET}--\protect\cite{ARMINHAB}.}
\label{tab:detectors}
\begin{center}
\begin{tabular}{|c|c|c|c|c|}
\hline
       & ALEPH         & DELPHI  & L3      & OPAL    \\
 B-field & 1.5T          & 1.2T    & 0.5T    & 0.435T  \\
\hline
Si VTX & 2 layers      & 3 layers& 2 layers& 2 layers\\
       & $R\phi\,z$      & $R\phi\,z$& $R\phi\,z$& $R\phi\,z$\\
r=0.1m & $12\mu,12\mu$ & $9\mu,7.6\mu$ &           & $5\mu,13\mu$   \\
\hline
inner tr. & 8pts,150$\mu$,5cm & 24pts,100$\mu$ & TEC+z.chb & 159pts        \\ 
r=0.3m & drift ch.           & jet ch. $R\phi$ &          & $135\mu$,6cm   \\ 
\cline{1-3}
main tr.& TPC, 1atm     & TPC, 1atm  &37 pts, 30 to 70 $\mu$& JET 4 atm\\
detector &             &         &        &         \\ \cline{4-4}
$\dedx$& 4.6$\%$       & 5.5$\%$ & BGO e.m. cal & 3.5$\%$ \\
       &               &         & 4\% at 200 MeV  &         \\ \cline{3-3}
r=1.1m &               &RICH,1cmC$_6$F$_{14}$  & HCAL 60 U plates &\\
       &               &gas C$_5$F$_{12}$ & 55$\%/\sqrt{E}$  &         \\
\hline
r=1.8  & ECAL 21.5$X_0$  &OD 5pts,150$\mu$  & filter 1$\lambda$,5pts&$z\,$chb 
6x300$\mu$ \\ \cline{3-3}
       & 18$\%/\sqrt{E}$,3sp&HPC 18X$_0$& support pipe & coil 1.7X$_0$\\ 
\cline{1-2} \cline{4-5}
r=2.2  & coil 1.6X$_0$   & 33$\%/\sqrt{E}$,9sp  &muon chb&lead glass 20X$_0$\\
       &               &coil 2X$_0$& 3sets   & 5$\%/\sqrt{E}$ \\ 
\cline{1-3} \cline{5-5}
r=2.9  & HCAL 1.2mFe   &HCAL 1.2mFe&lever arm2.7m&HCAL 1mFe \\ 
\cline{1-3} \cline{5-5}
       & muon chb      &muon chb &      &muon chb \\
       & 2 layers      &2 layers &         &4 layers \\ \cline{3-3}
r=5.7  & lever arm 0.5m&lever arm .3/.6m & coil & lever arm .7m \\
\hline
\hline
Lumi.   & calorimeter      & calorimeter & wire ch.  & calorimeter \\  
forward & tungsten/silicon & lead/scint. & BGO+prop. & tube ch.    \\
        & 24--58mrad        & 29--185mrad  & 25--70mrad & 58--120mrad \\
\hline
\end{tabular}
\end{center}
\end{table}

In the following all four detectors will be described. The ALEPH detector 
will be presented in some detail. For the other three detectors, special 
aspects relevant for the subject of this paper are discussed.

\subsection[ALEPH detector]{ALEPH detector}

\begin{figure}[htbp]
\begin{center}
\includegraphics[width=14.22cm,height=7.74cm]{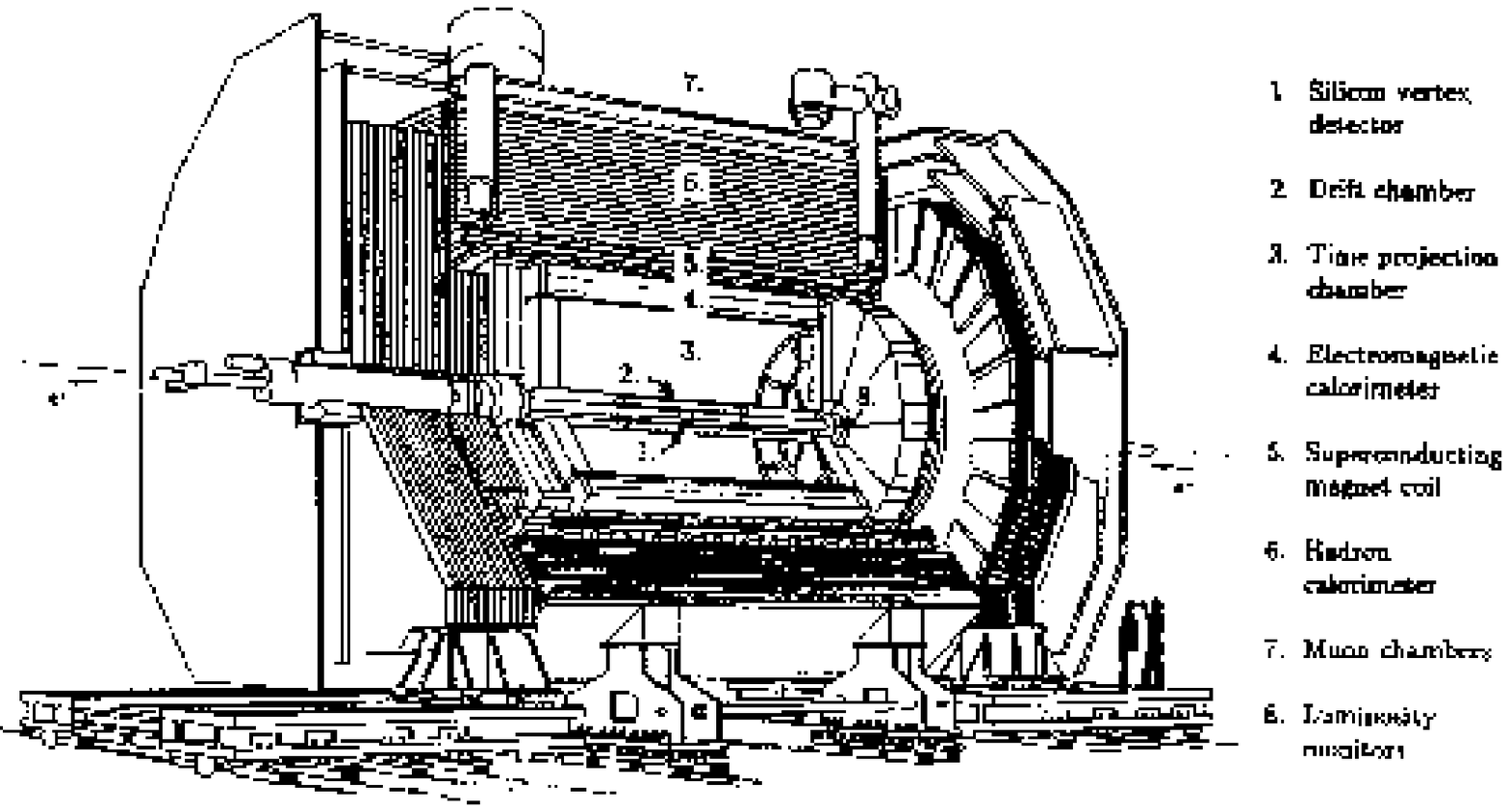}
\end{center}
\caption[ALEPH detector]
        {ALEPH detector \protect \cite{ALEPHDET}.}\label{fig:alephdetector}
\end{figure}

The ALEPH detector (\Figure~\ref{fig:alephdetector}) \cite{ALEPHDET}
shows the typical cylindrical symmetry around the beam pipe. The
interaction point of the electron and positron beams is at the centre
of the detector. The tracking chambers and the electromagnetic
calorimeter are immersed in a solenoidal magnetic field of
$1.5\unit{T}$ produced by the superconducting coil (with a length of
$6.4\unit{m}$ and a diameter of $5.3\unit{m}$). Outside the coil the
hadron calorimeter is used as return yoke.
\longpage

The beam pipe inside ALEPH, with a length of $5.5\unit{m}$ extends
between the two `low-$\beta$' qua\-dru\-po\-les, which focuses the
electron and positron beams onto the interaction point. The tube is
made of $1.5\unit{mm}$ thick aluminium, with an inner diameter of
$106\unit{mm}$. The central part ($760\unit{mm}$ length), however, is
made of beryllium, $1.1\unit{mm}$ thick.

Closest to the interaction point, the silicon vertex detector (VDET) is 
installed. It consists of two concentric rings with average radius 
$6.5\unit{cm}$ and $11.3\unit{cm}$. The inner layers has 9 silicon wafers 
in azimuth, the outer layer has 15 wafers; both layers are 
four wafers ($5.12 \times 5.12 \times 0.03 \unit{cm}^3$) long 
in $z$-direction. The arrangement in azimuth is such that the 
wafers overlap by $5\%$. This allows an internal relative alignment with 
tracks passing through adjacent wafers. The point resolution in the $r-\phi$ 
and $r-z$ view is $12\unit{\mu m}$. The hit association of VDET hits to 
tracks extrapolated from the TPC is found by Monte Carlo 
to be $98\%$ for tracks in hadronic events with two vertex hits in the 
acceptance of the vertex detector: $|\mathrm{cos} \, \theta| < 0.85$.

Around the vertex detector the inner tracking chamber (ITC) is built
with the same polar geometrical acceptance as the vertex
detector. This conventional cylindrical drift chamber is filled with
$80\%$ argon and $20\%$ carbon dioxide with ethanol. The chamber
provides eight measurements in $r-\phi$ in a radial range between
$16\unit{cm}$ and $26\unit{cm}$, with the wires stretched in
$z$-direction and arranged in eight concentric layers of hexagonal
drift cells. In $r-\phi$ the position of hits is measured to
$150\unit{\mu m}$; in $z$ the position is obtained by the measurements
of the difference of the arrival time of the pulses at both ends of
the $2\unit{m}$ long wires. The precision reached is
$5\unit{cm}$. However, only the $r-\phi$ measurements are used for the
tracking; the information of $z$ can be used for track association
with the tracks reconstructed in the TPC. An important aspect of the
ITC is that it is the only tracker used for the trigger.

The time projection chamber (TPC) serves as the main tracking chamber in 
ALEPH. In a volume extending in radius from $0.3\unit{m}$ to $1.8\unit{m}$, 
with a length of $4.4\unit{m}$ up to 21 space points are measured. The 
ionization charge is recorded in proportional wire chambers at both ends of 
the drift volume, reading out cathode pads arranged in 21 concentric 
circles; up to 338 $\dedx$ samples are used for particle identification. 
The $z$ coordinate of the hits in the TPC is calculated from the 
drift time of the electrons collected. For this, the magnet field, electric 
field both pointing in horizontal direction 
(and their distortions), and the drift velocity must be known perfectly. 
These quantities are determined from a measured magnetic field map, by laser 
calibration and study of reconstructed tracks and their vertices. The 
resolution is found in $r-\phi$ as $173\unit{\mu m}$ and in $z$ as 
$740\unit{\mu m}$.

In hadronic events, $98.6\%$ of the tracks are reconstructed, when
they cross at least four out of 21 pad rows, $|\mathrm{cos} \, \theta|
< 0.966$. The momentum resolution has been determined with di-muon
events. The transverse momentum resolution $\sigma(1/p_t)$ is $1.2
\times 10^{-3}$ ($p_t$ in ${\mathrm{GeV}}/c$) for the TPC alone;
including ITC an VDET the resolution is $\sigma(p_t)/p_t = 0.0006
\cdot p_t \oplus 0.005$ ($p_t$ in ${\mathrm{GeV}}/c$); $\oplus$
implies that the two errors are added in quadrature.

The TPC is surrounded by the electromagnetic calorimeter (ECAL), which 
consists of a barrel part and two endcaps, in order to measure 
electromagnetic energy in an angular range $|\mathrm{cos} \, \theta| < 0.98$. 
With its fine segmentation in projective towers of 
approximately $3\unit{cm}$ by $3\unit{cm}$, \ie 
$0.9^{\circ}$ by $0.9^{\circ}$, the angular resolution is
$\sigma_{\theta,\phi} = 2.5/\sqrt{E} + 0.25$ ($E$ in $\mathrm{GeV}$;
$\sigma_{\theta,\phi}$ in $\mathrm{mrad}$). The towers are read out in three 
segments in depth called storeys of 4, 9, and 9 radiation lengths. 
This lead-proportional tube chamber
has an energy resolution for electromagnetic showers of $\sigma_{E}/E =
0.18/\sqrt{E} + 0.009$ ($E$ in $\mathrm{GeV}$).

The outer shell used as return yoke, is the hadron calorimeter (HCAL). 
It is made from iron plates of $5\unit{cm}$ thickness, interleaved 
with 22 layers of plastic streamer tubes and one layer of tubes in front. The 
towers are arranged in projective direction to the primary vertex with a 
solid angle of $3.7^{\circ}$ by $3.7^{\circ}$, corresponding to $4 \times 4$ 
of the electromagnetic calorimeter towers. Both the cathode pads defining the 
towers (pads of different tubes forming one tower are connected galvanically 
within one storey) and wires in the 
$1\unit{cm}$ wide tubes are read out. The latter are used for muon 
identification and as a trigger. The energy resolution can be parameterized 
$\sigma_{E}/E = 0.85/\sqrt{E}$ ($E$ in $\mathrm{GeV}$). In addition, two 
double layers of streamer tubes are installed around the hadron calorimeter 
outside the magnetic field and serve as muon detectors.

\subsection[DELPHI detector]{DELPHI detector}

\begin{figure}[ht]
\begin{center}
\includegraphics[width=\textwidth]{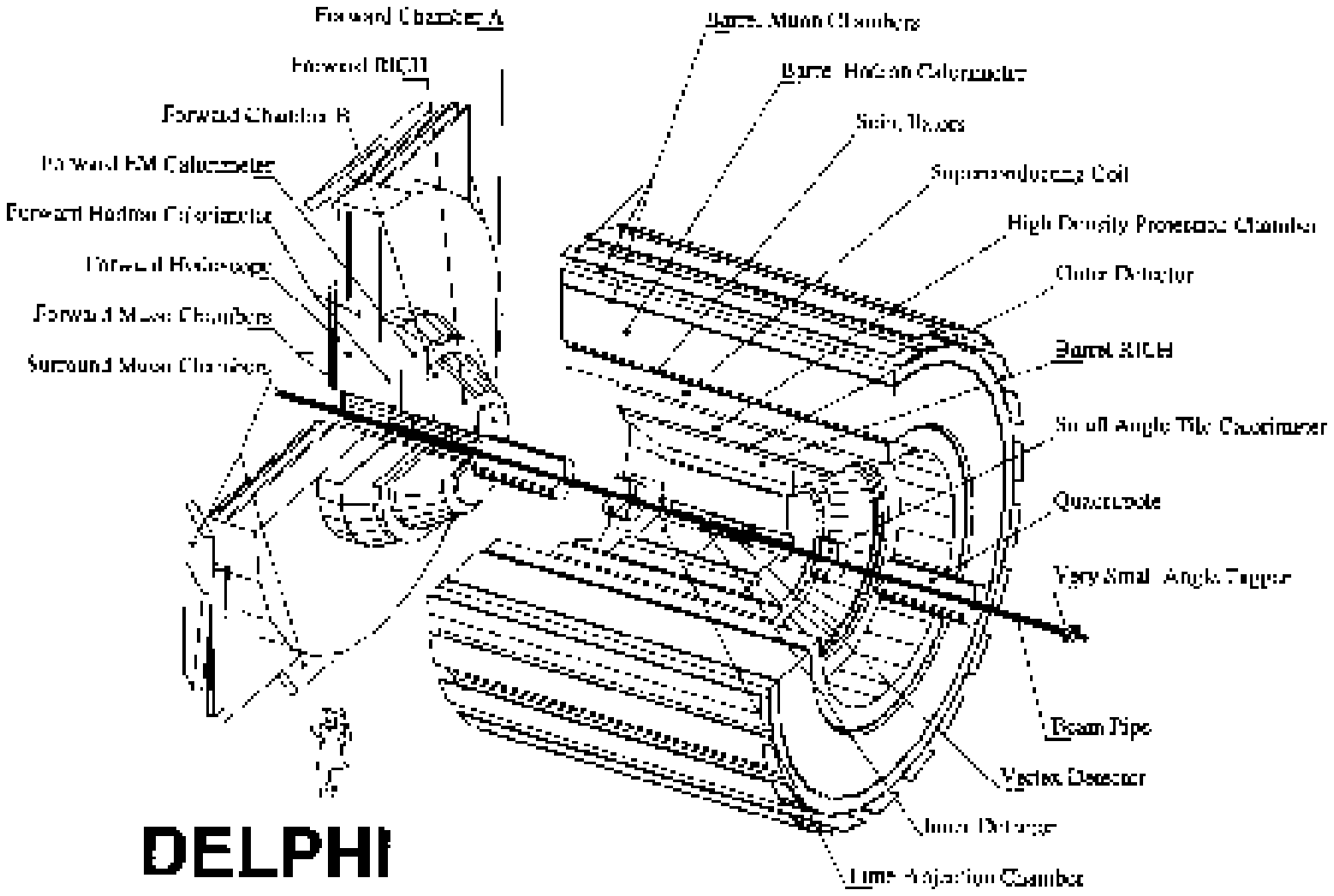}
\end{center}
\caption[DELPHI detector]
        {DELPHI detector~\cite{DELPHIDET}.}
\label{fig:delphidetector}
\end{figure}

The layout of the DELPHI detector \cite{DELPHIDET} is shown in
\Figure~\ref{fig:delphidetector}. The subdetectors are arranged in a
cylinder symmetrical arrangement with only the hadron calorimeter and
the muon chambers being outside the superconducting coil. The vertex
detector closest to the beam pipe is made of silicon wafers. It
provides measurements in three layers with information in both $z$-
and $r-\phi$-direction. The single hit resolution is found to be
$9\unit{\mu m}$ and $7.6\unit{\mu m}$.  The vertex detector is
surrounded by the inner detector (ID) of a jet-chamber geometry with
five multi wire proportional chambers (MWPC) layers.  The main
tracking device is a Time Projection Chamber (TPC) measuring up to 16
space points per track. Together with the outer detector (OD) with 5
layers of drift tubes the four tracking chambers provide a momentum
resolution of $\sigma(p)/p = 0.0006 \cdot p$ ($p$ in
$\mathrm{GeV}/c$).

A specialty of the DELPHI detector is the Ring Imaging Cherenkov detector 
(RICH) enclosed by the outer detector. The particle identification in the 
RICH complements the identification with $\dedx$ in the TPC. The DELPHI 
collaboration has chosen to use a gas and a liquid RICH (C$_5$F$_{12}$ and 
C$_6$F$_{14}$), having two different 
refractive indices. While the $\dedx$ measurement is most powerful in the 
momentum range below $1\unit{GeV}/c$, the liquid radiator allows for 
particle identification from $0.7\unit{GeV}/c$ to $8\unit{GeV}/c$ and the 
gaseous radiator from $2.5\unit{GeV}/c$ to $25\unit{GeV}/c$, with angular 
resolution between $1.2\unit{mrad}$ and $5.2\unit{mrad}$.

The high density projection chamber (HPC) consists of layers of TPCs,
which are separated by lead wires. These wires separate the drift
cells and provide the drift field, but also serve as converter
material for the electromagnetically interacting particles. The energy
deposits on the pads are monitored with $\pi^0$'s, where one decay
photon converted in the material {\em in front of} the HPC and the
momentum is precisely measured: with the $\pi^0$ mass as a constraint,
the energy resolution is measured to $\sigma(E)/E = 0.33/\sqrt{E}
\oplus 0.043$ ($E$ in GeV).

Outside the magnet coil a layer of scintillators is installed, mainly for 
trigger purposes. The hadron calorimeter (HCAL) made from iron interleaved 
with limited streamer tubes, serves as return yoke and muon filter, as well. 
Muon identification is supported by additional muon chambers. The resolution 
of the HCAL is $\sigma(E)/E = 1.12/\sqrt{E} \oplus 0.21$ ($E$ in GeV).

\clearpage

\subsection[L3 detector]{L3 detector}

\begin{figure}[h]
\begin{center}
\includegraphics[width=.9\textwidth]{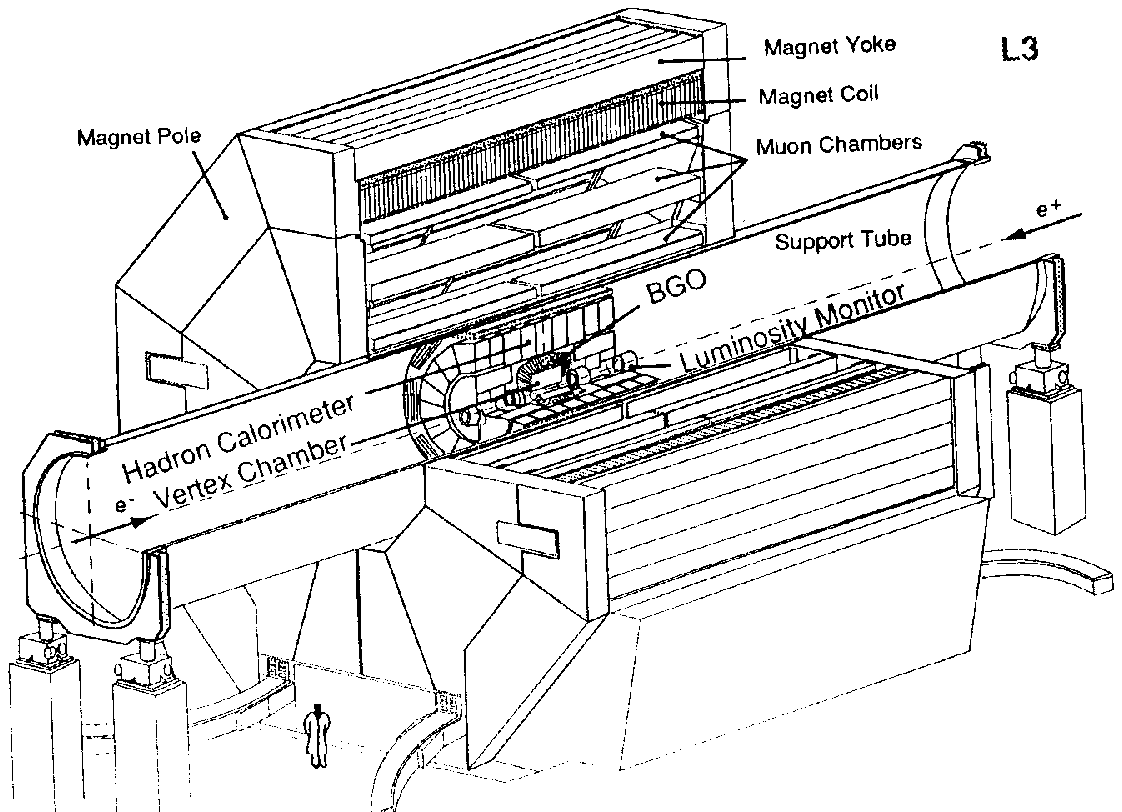}
\end{center}
\caption[L3 detector]
        {L3 detector~\cite{L3DET}.}
\label{fig:L3detector}
\end{figure}

The subdetectors in the detector of the L3 collaboration
(\Figure~\ref{fig:L3detector}) \cite{L3DET} are mounted inside a
support tube with a diameter of $4.45\unit{m}$ with the exception of
the muon detection system. The muon chambers are only surrounded by a
very large low field air magnet ($0.5\unit{T}$).  The coil has an
inner diameter of $11.9\unit{m}$. The size of the magnet allows a long
lever arm for the muon momentum measurement. This requires a high
precision alignment and monitoring of these chambers.

The tracking system consists of a silicon vertex detector and a central 
track detector. The latter is a Time Expansion Chamber (TEC) providing 
37~points on standard wires for the $r-\phi$ measurement; in addition 
14~wires resolving left-right ambiguities. The $z$ coordinate is measured 
on 11 wires by charge division. The surrounding two cylindrical proportional 
chambers are designed to provide a good $z$~measurement. With a total lever 
arm of $0.32\unit{m}$ the momentum resolution is $\sigma(p_t)/p^2_t = 0.0206 
\pm 0.0006$ ($p_t$ in ${\mathrm{GeV}}/c$).

Muons in $\epem \rightarrow \mu^+ \mu^-$ are measured with the high precision
of $\sigma(p)/p \approx 2.5\%$, with the long lever arm to the muon chambers. 
Apart from the muon detection, special emphasis was put on a high precision 
measurement for electromagnetic showers. They are measured in a crystal 
calorimeter read out by photomultipliers. The crystals of bismuth 
germanium oxide (BGO) have a shape of a truncated pyramid, $24\unit{cm}$ long 
and of $2 \times 2 \unit{cm}^2$ at the inner and $3 \times 3 \unit{cm}^2$ at 
the outer end. The energy resolution varies from $5\%$ at $100 \unit{MeV}$ 
to $1.4\%$ at high energy.

A layer of scintillation counters is used for time-of-flight measurement. 
Besides its trigger task, it efficiently rejects cosmic shower events. A 
uranium calorimeter with proportional wire chambers measures hadronic 
showers and absorbs most particles except muons. Around this calorimeter 
a muon filter is mounted, made of brass plates interleaved with five layers 
of proportional tubes.
 
\clearpage

\subsection[OPAL detector]{OPAL detector}

\begin{figure}[h]
\begin{center}
\includegraphics[width=\textwidth]{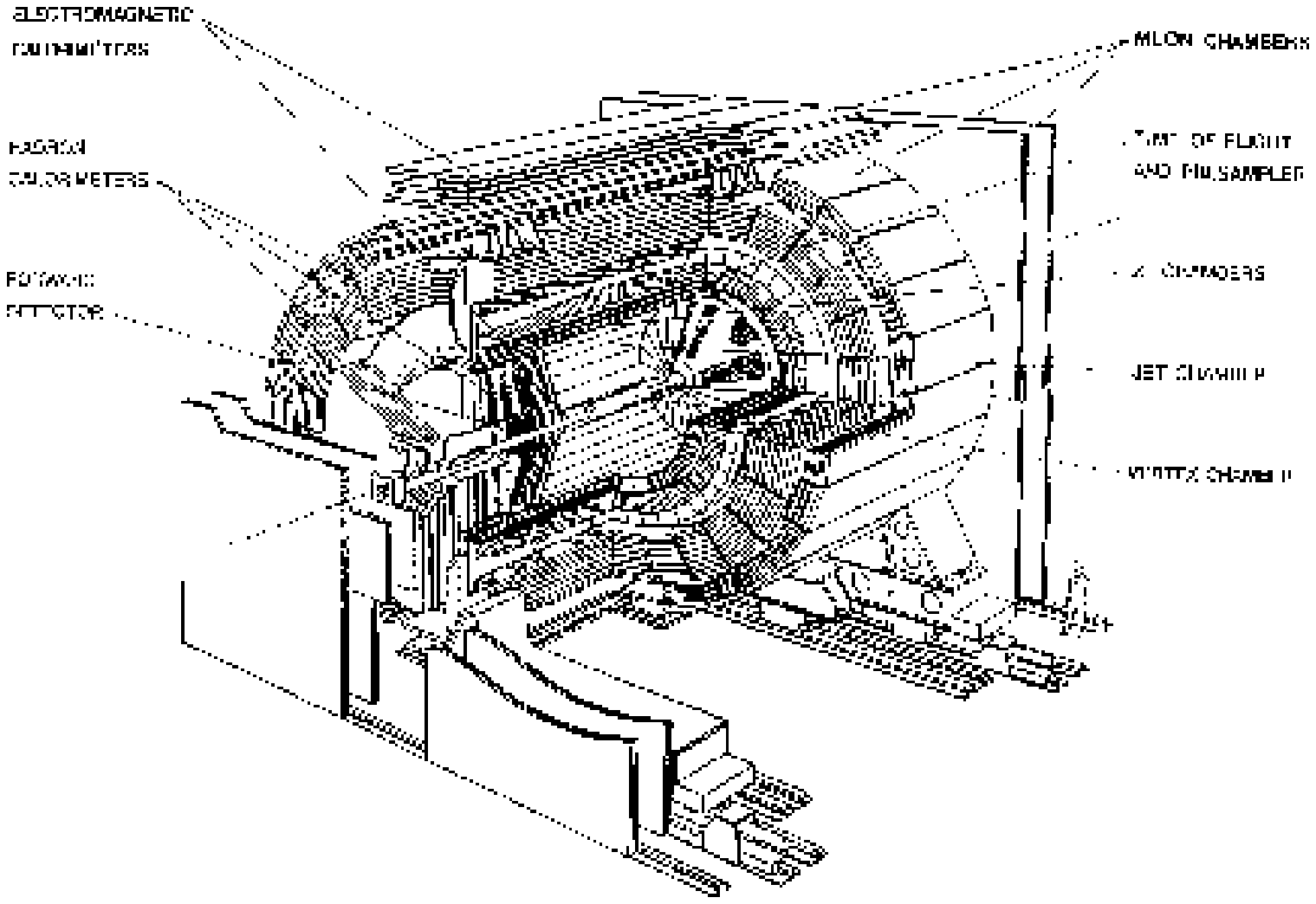}
\end{center}
\caption[OPAL detector]
        {OPAL detector \protect \cite{OPALDET}.}
\label{fig:opaldetector}
\end{figure}

The OPAL detector (\Figure~\ref{fig:opaldetector}) \cite{OPALDET}
comprises a tracking system inside a solenoidal magnet of
$0.435\unit{T}$, which consists of a vertex detector a jet-chamber and
a $z$-chamber. The new vertex detector of OPAL with two concentric
layers of silicon wafers is placed at radii of $6.1\unit{cm}$ and
$7.5\unit{cm}$. The single hit resolution in $r - \phi$ is $5\unit{\mu
m}$, in $z$ $13\unit{\mu m}$. The main tracking with the jet-chamber
provides up to 159 space points ($\sigma_{r\phi} = 135 \unit{\mu m}$,
$\sigma_z = 6 \unit{cm}$) per track. It allows good particle
identification with the energy ionization loss $\dedx$. The
$z$-direction of tracks is substantially improved with information
from the $z$-chambers, which are made of modules of drift chambers
with 6 staggered anodes strung in $\phi$-direction. The momentum
resolution is measured to $\sigma_p/p^2 = 0.0022\unit{GeV^{-1}}$.

A time-of-flight system, consisting of scintillation counters, allows 
particle identification in the momentum range from $0.6\unit{GeV}/c$ to 
$2.5\unit{GeV}/c$. It is used for triggering and for cosmic shower rejection.

Electromagnetic showers are measured with an assembly of lead glass blocks, 
with $10 \times 10 \unit{cm}^2$ and $37 \unit{cm}$ in depth, 
read out with photomultipliers. The energy resolution is about 
$\sigma(E)/E = 0.05/\sqrt{E}$ ($E$ in GeV), 
when combined with a presampler mounted in 
front of the calorimeter. Hadrons are measured with nine chambers, 
limited streamer tubes, interleaved with eight layers of iron plates, where 
the hadrons may shower. Muons are detected in addition in four layers of 
drift chambers, the muon chambers.

\newpage

\section[Experiments at hadron colliders]{Experiments at hadron colliders}
\label{sec:Hcoll}

\subsection[CDF in Run II]
           {CDF in Run II $\!$\footnote{Author: V.~Papadimitriou}}

The CDF Run II detector\cite{runIITDR}, in operation since 2001, is an
azimuthally and forward-backward symmetric apparatus designed to study
$p\bar{p}$ collisions at the Tevatron. It is a general purpose
solenoidal detector which combines precision charged particle tracking
with fast projective calorimetry and fine grained muon detection.  
Tracking systems are contained in a superconducting
solenoid, 1.5~m in radius and 4.8~m in length, which generates a 1.4 T
magnetic field parallel to the beam axis. Calorimetry and muon systems
are all outside the solenoid.  The main features of the detector
systems are summarized below.

The tracking system consists of a silicon microstrip
system\cite{runIISi} and of an open-cell wire drift
chamber\cite{runIICOT} that surrounds the silicon.  The silicon
microstrip detector consists of seven layers (eight layers for
$1.0<|\eta|< 2.0$) in a barrel geometry that extends from a radius of
$r = 1.5$~cm from the beam line to $r = 28$~cm. The layer closest to
the beam pipe is a radiation-hard, single sided detector called Layer
00 which employs LHC designs for sensors supporting high-bias
voltages. This enables signal-to-noise performance even after extreme
radiation doses. The remaining seven layers are radiation-hard, double
sided detectors.  The first five layers after Layer 00 comprise the
SVXII system and the two outer layers comprise the ISL system. This
entire system allows track reconstruction in three dimensions.  The
impact parameter resolution of the combination of SVXII and ISL is 40
$\mu$m including a 30 $\mu$m contribution from the beamline.  The
$z_0$ resolution of the SVXII and ISL is 70 $\mu$m.  The 3.1~m long
cylindrical drift chamber (COT) covers the radial range from 40 to
137~cm and provides 96 measurement layers, organized into alternating
axial and $\pm 2^{\circ}$ stereo superlayers.  The COT provides
coverage for $|\eta| \leq $1. The hit position resolution is
approximately 140 $\mu$m and the momentum resolution
$\sigma(p_T)/p_T^2 =$0.0015 (GeV/c)$^{-1}$.  The COT provides in
addition $dE/dx$ information for the tracks.

A Time-of-Flight (TOF) detector\cite{runIITOF}, based on plastic
scintillators and fine-mesh photomultipliers is installed in a few
centimeters clearance just outside the COT. The TOF resolution is
$\approx 100$ ps and it provides at least two standard deviation
separation between $K^{\pm}$ and $\pi^{\pm}$ for momenta $p <$ 1.6
GeV/c.

\shortpage

Segmented electromagnetic and hadronic sampling calorimeters surround the
tracking system and measure the energy flow of interacting particles in the 
pseudo-rapidity range $|\eta|<$ 3.64.
The central calorimeters (and the endwall hadronic calorimeter) cover the 
pseudorapidity range $|\eta|<$ 1.1(1.3). 
The central electromagnetic calorimeter\cite{runIICEM} (CEM) uses lead sheets 
interspersed with polystyrene
scintillator as the active medium and employs phototube readout.
Its energy resolution is $13.5\%/\sqrt{E_T} \oplus 2\%$.
The central hadronic calorimeter\cite{runIICHA} (CHA) uses steel absorber 
interspersed with acrylic scintillator as the active medium.
Its energy resolution is $75\%/\sqrt{E_T} \oplus 3\%$.
The plug calorimeters cover the pseudorapidity region  1.1 $<|\eta|<$ 3.64.
They are sampling scintillator calorimeters which are read out with plastic 
fibers and phototubes. The energy resolution of the plug electromagnetic 
calorimeter\cite{runIIPEM} is $16\%/\sqrt{E} \oplus 1\%$.
 The energy resolution of the plug hadronic 
calorimeter is $74\%/\sqrt{E} \oplus 4\%$. 

The muon system resides beyond the calorimetry. Four layers of planar
drift chambers (CMU) detect muons with $p_T >$ 1.4~GeV/c which
penetrate the five absorption lengths of calorimeter steel. An
additional four layers of planar drift chambers (CMP) instrument 0.6 m
of steel outside the magnet return yoke and detect muons with $p_T >$
2.0~GeV/c.  The CMU and CMP chambers each provide coverage in the
pseudo-rapidity range $|\eta|<0.6$. The Intermediate MUon detectors
(IMU) are covering the region 1.0 $<|\eta|<$1.5.

The beam luminosity is determined by using gas Cherenkov counters located in 
the $3.7 < |\eta| < 4.7$ region which measure the average number of inelastic
$p\bar{p}$ collisions per bunch crossing\cite{runIIlum}.

The trigger and data acquisition systems are designed to accommodate
the high rates and large data volume of Run II. Based on preliminary
information from tracking, calorimetry, and muon systems, the output
of the first level of the trigger is used to limit the rate for
accepted events to $\approx$ 18~kHz at the luminosity range of 3--7
10$^{31}~cm^{-2}s^{-1}$.  At the next trigger stage, with more refined
information and additional tracking information from the silicon
detector, the rate is reduced further to $\approx$ 300~Hz.  The third
and final level of the trigger, with access to the complete event
information, uses software algorithms and a computing farm, and
reduces the output rate to $\approx$ 75~Hz, which is written to
permanent storage.

The CDF Run I and Run 0 detector, which operated in the time period
1987--1996, is described elsewhere~\cite{runIdet}. Major differences
for Run II include: the replacement of the central tracking system;
the replacement of a gas sampling calorimeter in the plug-forward
region with a scintillating tile calorimeter; preshower detectors;
extension of the muon coverage, a TOF detector and upgrades of
trigger, readout electronics, and data acquisition systems.

\subsection[DO detector in Run II]{DO detector in Run II
$\!$\footnote{Author: Arnd Meyer}
}
\label{sec:D0}

The DO Run II detector, in operation since 2001, is made of
the following main elements.
The central tracking system consists of a silicon  microstrip tracker (SMT)
and a central fiber tracker (CFT),  both located within a 2~T
superconducting solenoidal  magnet~\cite{run2det}. The SMT has $\approx
800,000$ individual strips, with typical pitch of $50-80$ $\mu$m, and a
design  optimized for tracking and vertexing capability at $|\eta|<3$.  The
system has a six-barrel longitudinal structure, each with  a set of four
layers arranged axially around the beam pipe,  and interspersed with 16
radial disks. The CFT has eight thin  coaxial barrels, each supporting two
doublets of overlapping  scintillating fibers of 0.835~mm diameter, one
doublet being  parallel to the collision axis, and the other alternating
by  $\pm 3^{\circ}$ relative to the axis. Light signals are transferred 
via clear light fibers to solid-state photon counters (VLPC) that  have
$\approx 80\%$ quantum efficiency.

Central and forward preshower detectors located just outside  of the
superconducting coil (in front of the calorimetry) are  constructed of
several layers of extruded triangular scintillator  strips that are read
out using wavelength-shifting fibers and  VLPCs. The next layer of
detection involves three  liquid-argon/uranium calorimeters: a central
section (CC) covering  $|\eta|$ up to $\approx 1$, and two end calorimeters
(EC)  extending coverage to $|\eta|\approx 4$, all housed in separate 
cryostats~\cite{run1det}. In addition to the preshower detectors, 
scintillators between the CC and EC cryostats provide sampling  of
developing showers at $1.1<|\eta|<1.4$.

The muon system resides beyond the calorimetry, and consists of  a layer of
tracking detectors and scintillation trigger counters  before 1.8~T
toroids, followed by two more similar layers after  the toroids. Tracking
at $|\eta|<1$ relies on 10~cm wide drift  tubes~\cite{run1det}, while 1~cm
mini drift tubes are used at  $1<|\eta|<2$.

Luminosity is measured using plastic scintillator arrays located  in front
of the EC cryostats, covering $2.7 < |\eta| < 4.4$.  A forward-proton
detector, situated in the Tevatron tunnel on  either side of the
interaction region, consists of a total of  18 Roman pots used for
measuring high-momentum charged-particle  trajectories close to the
incident beam directions.
\shortpage

The trigger and data acquisition systems are designed to accommodate  the
large luminosity of Run II. Based on preliminary information  from
tracking, calorimetry, and muon systems, the output of the first  level of
the trigger is used to limit the rate for accepted events  to $\approx$
1.5~kHz. At the next trigger stage, with more refined  information, the
rate is reduced further to $\approx$ 800~Hz.  The third and final level of
the trigger, with access to the complete event information, uses software
algorithms and a computing farm, and reduces the output rate to $\approx$
50~Hz, which is written to permanent storage.

The DO Run I detector is described elsewhere~\cite{run1det}. Major
differences for Run II include: the replacement of the central tracking
system, optimized for the absence of a central magnetic field, by a magnetic
tracking system; preshower detectors; and upgrades of trigger, readout
electronics, and data acquisition systems.

\clearpage

\section[Experiments at HERA]
        {Experiments at HERA
         $\!$\footnote{Author: Andreas B.\ Meyer}}
\label{sec:HERA}

\begin{figure}[htb]
\unitlength1.0cm
\begin{picture}(16,7)
\put(1,0){\includegraphics[width=14.0cm]{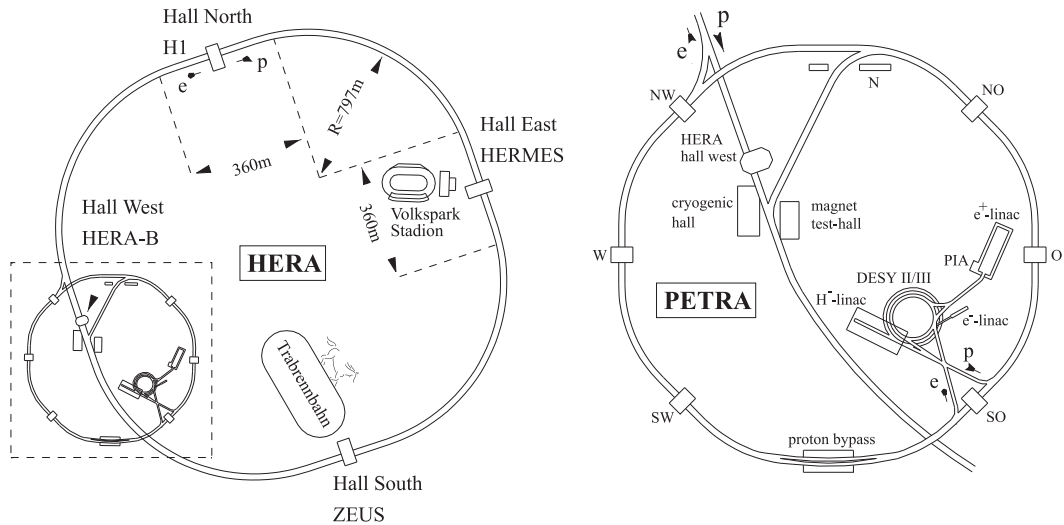}}
\end{picture}
\caption{The HERA collider with the four experiments H1, ZEUS, HERMES
         and HERA-B on the left and its pre-accelerators on the
         right.}
\label{fig:heracollider}
\end{figure}

The electron positron storage ring HERA
(\Figure~\ref{fig:heracollider}) at the DESY laboratory in Hamburg
collides 27.5~GeV electrons or positrons with 920~GeV
protons\footnote{Until 1998 the proton energy was 820~GeV.}.  The
storage ring has a circumference of $6.4\,{\rm km}$ and consists of
two separate accelerators with a maximum of 180 colliding bunches
each, providing a bunch crossing rate of 10~MHz.  Four experiments are
situated at HERA. The two collider experiments H1 and ZEUS have been
in operation since 1992. In 1995 the HERMES experiment started data
taking using the polarized electron beam on a fixed polarized gas
target \cite{Ackerstaff:1998av}.  The HERA-B proton proton fixed
target experiment was operated between 1998 and 2003.  HERA-B makes
use of the proton beam halo using a wire target and is described in
\Section~\ref{sec:FTexp-HERAB}.

\begin{figure}[htb]
\unitlength1.0cm
\begin{picture}(16,7)
\put(2.5,0){\includegraphics[width=10cm]{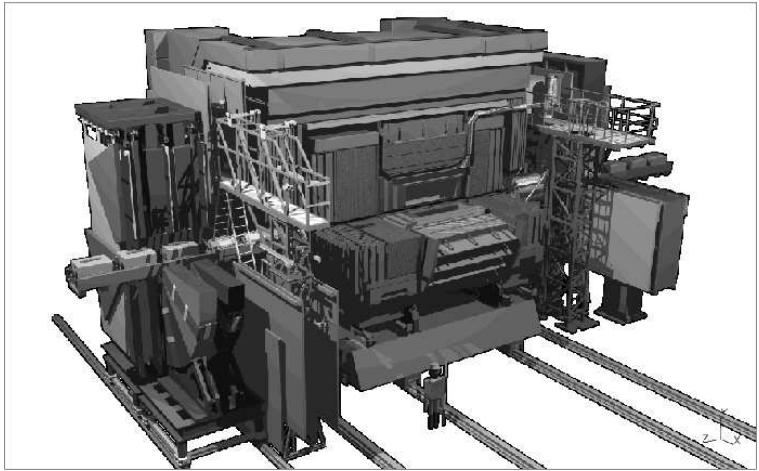}}
\end{picture}
\caption{Schematic view of the ZEUS Detector.}
\label{fig:zeus_experiment}
\end{figure}

The H1 and ZEUS detectors are typical multi-purpose collider
experiments. A schematic view of the ZEUS detector is shown in
\Figure~\ref{fig:zeus_experiment}.  The physics programs comprise the
full spectrum of QCD studies, measurements of the proton structure
functions and exclusive hadronic final states, as well as electroweak
physics and searches for new physics
phenomena~\cite{h1zeus_publications}.  With an $ep$ centre-of-mass
energy of 320~GeV the HERA collider experiments H1 and ZEUS are close
to the present energy frontier for accelerator based experiments.
Only the Tevatron experiments CDF and D0 (described in
section~\ref{sec:Hcoll}) have access to higher centre-of-mass
energies.  Events in deep inelastic $ep$ scattering have been measured
down to values of $x$ as low as $\sim 10^{-6}$ and up to values of
$Q^2$ of $30,000$~GeV$^2$.  In QCD, measurements of exclusive final
states comprise jet physics, heavy flavour production, processes in
hard and soft diffraction and hadron spectroscopy.

In the years between 1992 and 2000 the collider experiments H1 and ZEUS
collected an integrated luminosity of 100 pb$^{-1}$ each. 
The bulk data were taken in the years 1996 through 2000.
In the years 2001/2 a major luminosity upgrade was put in place.
The interaction points were equipped with new focusing magnets
which allow for substantially increased specific luminosities.
Since 2003/4 the HERA collider is running and
an integrated luminosity of 700 pb$^{-1}$ is expected to be produced
for each of the two experiments~\cite{hera_upgrade}.

\begin{figure}[htb]
\begin{center}
\includegraphics[width=11cm]{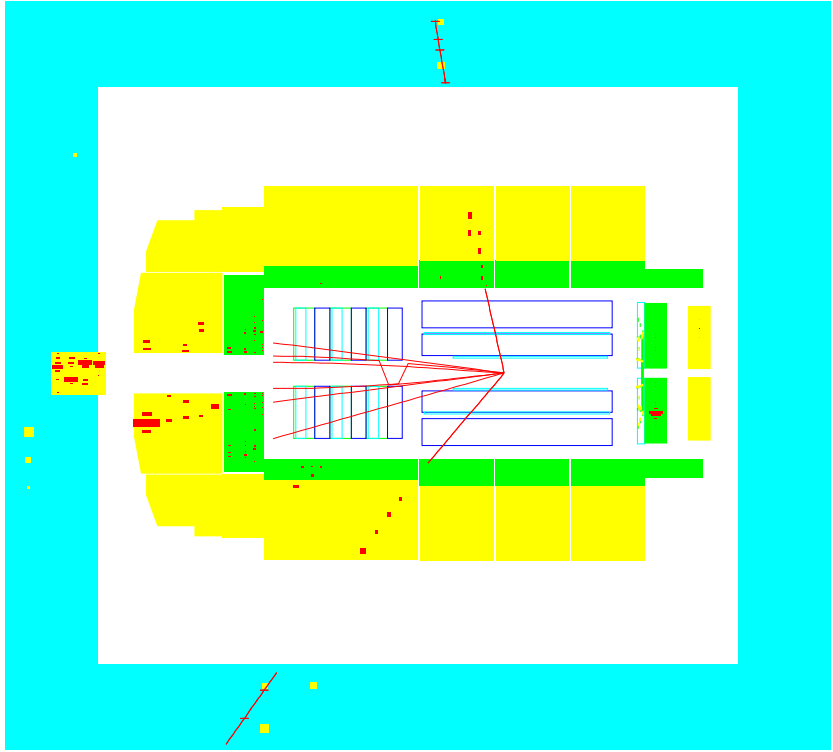}
\end{center}
\caption{Display of a charmonium event candidate in the H1 Detector.}
\label{fig:h1event}
\end{figure}

The designs of the H1 and ZEUS detectors were chosen to be somewhat
complementary, with emphasis on the reconstruction
of the scattered electron in the case of H1 and on the 
precise calorimetric measurement
of the hadronic final states in the case of ZEUS.
Both experiments are capable of the triggering and 
reconstruction of charmonium and bottomonium events down to very low 
transverse momenta $p_{t,\psi} \sim 0$. A candidate charmonium
event is displayed in figure~\ref{fig:h1event}.
In the following the experiments are described in detail, emphasizing 
those components that are most relevant for the triggering 
and reconstruction of quarkonium events with 
two decay leptons in the final state.

\subsection[H1]{H1}

The design of the 2800 ton H1 detector~\cite{Abt:1997xv},
schematically shown in \Figure~\ref{fig:h1event}, emphasizes charged
particle tracking in the central region as well as high calorimetric
re\-solution for electromagnetic energy depositions.

The primary components of the H1 tracking system are two coaxial
cylindrical jet-type drift chambers (CJC) covering the polar angle
region between 15$^\circ$ and 165$^\circ$.  The two chambers consist
of 30 (60) drift cells respectively with 24 (32) sense wires strung
parallel to the beam axis.  The sense wires are read out at both ends,
and the $z$-coordinate is measured by charge division with a
resolution of $\sigma_z=22\,{\rm mm}$.  The spatial resolution of the
CJC in the ${r \varphi}$ plane is $\sigma_{r \varphi}=170\,\mu{\rm
m}$.  The momentum re\-solution in the coordinate transverse to the
1.2 Tesla solenoidal field of \mbox{$\sigma(p_T)/p_T \, = \, 0.01 \,
p_T [{\rm GeV}] \, \oplus \, 0.015$}.  The magnetic field is produced
by a 5~m~long superconducting solenoid of 5.8~m in diameter which
encloses the calorimeter.  Two further inner drift chambers and two
multiwire proportional chambers (MWPC), serve to measure the
longitudinal track coordinates and to provide trigger information.
The Forward Tracking Detectors cover a polar angular range between
$5^\circ $ and $ 30^\circ$. The system consists of three supermodules
composed of three planar drift chambers, a multiwire proportional
chamber, a transition radiator and a radial drift chamber.  The MWPCs
serve for trigger purposes and complement the polar angular coverage
of the central proportional chambers.

The H1 main calorimeter employs a fine-grain liquid argon (LAr)
sandwich structure in the barrel and forward (proton-beam) region
(with angular range from 4$^\circ$ to 155$^\circ$ in polar angle).  In
the backward region (with angular range from 155$^\circ$ to
177.5$^\circ$) a lead/scintillating-fiber
calorimeter~\cite{Nicholls:1996di} provides an excellent energy
re\-solution of \mbox{$\sigma(E)/E \, = \, 0.07/\sqrt{E {\rm [GeV]}}
\, \oplus \, 0.01$,} and a time re\-solution better than 1~ns.  The
electromagnetic section of the liquid argon calorimeter uses lead
plates as absorber material. In the hadronic section (which provide a
depth of 4 to 6 nuclear interaction lengths) steel plates are
used. Both sections are segmented transversely in cells of $4 \times
4$~cm$^2$ in cross-section and are further segmented in longitudinal
shower direction.  In total there are 31,000 electromagnetic and
14,000 hadronic readout channels.  The electromagnetic LAr calorimeter
achieves an energy re\-solution of \mbox{$\sigma(E)/E \, = \,
0.12/\sqrt{E {\rm [GeV]}} \, \oplus \, 0.01$.}  The high degree of
segmentation allows for a distinction between hadronic and
electromagnetic energy depositions in the offline reconstruction,
resulting in a hadronic energy re\-solution of $\sigma(E)/E \, = \,
0.55/\sqrt{E {\rm [GeV]}} \, \oplus \, 0.01$.

Muons are identified as minimum ionizing particles in both the
calorimeters and in the magnetic field iron return yoke surrounding
the magnetic coil.  The iron system is instrumented with 16 layers of
limited-streamer tubes of 1~cm$^2$ cell size. Altogether the muon
system consists of 100k channels.  Up to five out of 16 layers are
used for triggering.  In order to provide a two-dimensional
measurement five of the 16 layers are equipped in addition with strip
electrodes glued perpendicular to the sense wire direction.

The H1 trigger and readout system consists of four levels of hardware
and software filtering.  The triggering of charmonium event candidates
relies on track pattern recognition in the central jet chambers and
timing information in the MWPC.  For the detection of the scattered
electron calorimeter triggers are used.  For the muon decay channel
coincidences of hits in the same sector of the instrumented iron (in
different layers) are required at the first trigger level.

\subsection[ZEUS]{ZEUS}

The ZEUS detector~\cite{Derrick:1992kk,zeusdet} makes use of a
700~ton compensating uranium sampling calorimeter, with equal
sampling fractions for electromagnetic and hadronic shower components.
The calorimeter is made up of layers of 2.6~mm SCSN-38 scintillator
and 3.3~mm stainless-steel-clad depleted-uranium plates. One layer
corresponds to 1.0 radiation length ($X_0$) and $0.04$ interaction
lengths. This choice of layer thicknesses results in a sampling
fraction of $4\%$ for electromagnetic and hadronic shower components,
and hence compensation, and $7\%$ for minimum-ionizing particles.
The compensation results in a very good hadronic energy re\-solution
of \mbox{$\sigma(E)/E \, = \, 0.35/\sqrt{E {\rm [GeV]}} \, \oplus \,
0.02$.}  The re\-solution for electromagnetic showers is
\mbox{$\sigma(E)/E \, = \, 0.18/\sqrt{E {\rm [GeV]}} \, \oplus \,
0.01$.}

The ZEUS solenoidal coil of diameter 1.9~m and length 2.6~m provides a
1.43 T magnetic field for the charged-particle tracking volume.  The
tracking system consists of a central wire chamber covering the polar
angular region from 15$^\circ$ to 164$^\circ$ , a forward planar
tracking detector from 8$^\circ$ to 28$^\circ$ and a second planar
tracking chamber in the backward direction, covering the region from
158$^\circ$ to 170$^\circ$.  The momentum re\-sol\-u\-tion attained is
\mbox{$\sigma(p_T)/p_T\,=\,0.005 \, p_T \, \oplus \, 0.015$} and a
track is extrapolated to the calorimeter face with a transverse
re\-solution of about 3~mm.  Ionization measurements from the central
tracking chamber also serve to identify electron--positron pairs from
$J/\psi$ decays.

The muon system is constructed of limited streamer tubes inside and
outside of the magnetic return yoke, covering the region in polar
angle from 10$^\circ$ to 171$^\circ$. Hits in the inner chambers
provide muon triggers for $J/\psi$ decays.

The ZEUS trigger algorithm is primarily calorimeter-based, exploiting
the excellent time resolution of the calorimeter, while that of H1
emphasizes tracking algorithms for reconstruction of the interaction
vertex.  The shaping, sampling, and pipelining algorithm of the
readout developed for the ZEUS calorimeter and used in modified form
for the silicon and presampler systems permits the reconstruction of
shower times with respect to the bunch crossings with a re\-solution
of better than 1~ns, providing essential rejection against upstream
beam--gas interactions, as well as allowing 5~$\mu$s for the
calculations of the calorimeter trigger processor.  For the triggering
of the charmonium production channels a muon track candidate in the
central drift chamber with one or more hits in the muon chambers can
be validated by energy in the calorimeter above a threshold of 460
MeV.

\subsection[HERA-B]{HERA-B}
\label{sec:FTexp-HERAB}

\begin{figure}
\begin{center}
  \includegraphics[width=\textwidth]{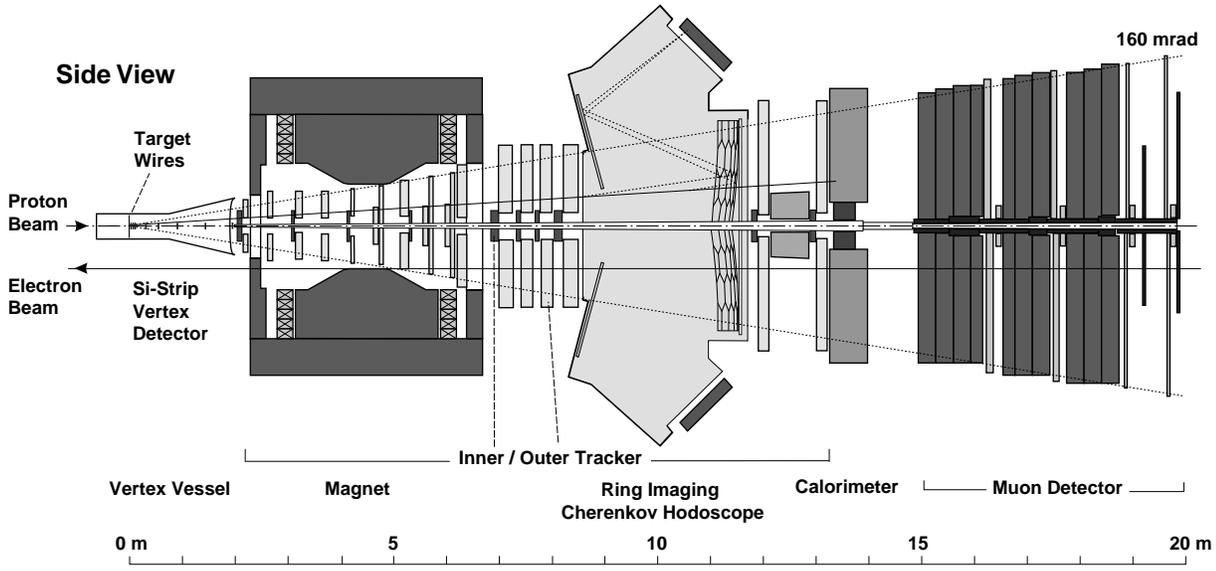}
\end{center}
\caption{Side view of the HERA-B Spectrometer.}
\label{fig:herab_experiment}       % Give a unique label
\end{figure}

The fixed target experiment HERA-B is located at the HERA storage ring
at DESY (see \Section~\ref{sec:HERA}), The data taking took place in
the years between 2000 and 2003.  At HERA-B, wire targets are inserted
into the halo of the 920~GeV HERA proton beam to spawn inelastic $pA$
collisions in which charmonium and other heavy flavour states are
produced.  The $pN$ ($N=p,n$) centre-of-mass energy is $\sqrt{s} =
41.6$~GeV.  A side view of the HERA-B spectrometer is shown in
\Figure~\ref{fig:herab_experiment}.  A detailed description of the
apparatus is given in Ref.~\cite{hb_tdr,Padilla:de,hb_oct}.

The wire target~\cite{Ehret:df} consists of two wire stations, each
containing four target wires of different materials.  A servo system
automatically steers the target wires during the data taking in order
to maintain a constant interaction rate.  The spectrometer has a
geometrical coverage from 15 mrad to 220 mrad in the horizontal plane
and from 15 mrad to 160 mrad in the vertical plane.  The
instrumentation emphasises vertexing, tracking and particle
identification.  The silicon vertex detector system~\cite{Bauer:tp} is
realized by a system of 20 Roman pots containing seven planar stations
(four stereo views) of double-sided silicon micro-strip detectors
which are operated in a vacuum vessel at 10 to 15~mm distance from the
proton beam. An additional station is mounted immediately behind the 3
mm thick Aluminium window of the vacuum vessel.  The tracker is
divided into a fine grained inner tracker using micro-strip gas
chambers with gas electron multipliers and a large area outer tracker
consisting of honeycomb drift cells with wire pitches between 5~mm
near and 10~mm~\cite{Zeuner:dx,Pruneau:2002yf,Capeans:dw}.  Particle
identification is performed by a Ring Imaging Cherenkov
hodoscope~\cite{Pyrlik:du,Arino:ur}, an electromagnetic
calori\-meter~\cite{Zoccoli:dn} and a muon
detector~\cite{Buchler:1998rh,Arefev:cu}.  The calorimeter is divided
into three radial parts with decreasing granularities.  The muon
system consists of four tracking stations. It is built from gas-pixel
chambers in the radially innermost region and from proportional tube
chambers, some with segmented cathodes (pads), everywhere else.

The detector components used for charmonium analyses include
the tracking and vertex detectors, the calorimeter and the muon system.
A complex trigger and read-out chain ~\cite{hb_daq} allows for a 
reduction of an initial interaction rate of several MHz to a 
final output rate of order 100 Hz. A dedicated $J/\psi$-trigger 
is based on the selection of $\mu^+\mu^-$ and $e^+e^-$ pairs
and subsequent reconstruction of invariant masses.

\section[Appendices]{Appendices}

\subsection[Resonant depolarization for absolute mass measurements]
           {Resonant depolarization for absolute mass measurements 
            $\!$\footnote{Author: S.~Eidelman}}
\label{sec:redepol}

Electrons and positrons in storage rings can become polarized due to
emission of synchrotron radiation according to the Sokolov--Ternov
effect \cite{SokTern}. The spins of the polarized electrons precess
around the vertical guiding magnetic field with the precession frequency
$\Omega$, which in the plane orbit approximation is
directly related to the particle energy $E$ and the
beam revolution frequency $\omega$:
\begin{equation}
\Omega/\omega = 1 + \gamma \cdot \mu^{\prime}/\mu_0 = 1 + \nu \:,
\label{eqn:Omega}
\end{equation}
where $\gamma=E/m_{\rm e}$, $m_{\rm e}$ is the electron mass, 
$\mu^{\prime}$ and $\mu_0$ are the anomalous and normal parts
of the electron magnetic moment. The $\nu$ is a spin tune, which
represents the spin precession frequency in the coordinate basis
related to the particle velocity vector.

The precession frequency can be determined using the {\it
resonant depolarization}. To this end one needs a polarized
beam in the storage ring which is affected by the external
electromagnetic field with the frequency $\Omega_D$ given by the
relation
\begin{equation}
     \Omega \pm \Omega_D = \omega \cdot n 
\label{eqn:OmegaD}
\end{equation}
with any integer $n$ (for VEPP-4M in the $J/\psi(1S)$ region $n=3$).

The precession frequency is measured
at the moment of the polarization destruction detected by the
{\it polarimeter},
while the {\it depolarizer} frequency is being scanned.
The process of
forced depolarization is relatively slow compared to the period of the
synchrotron oscillations of the particle energy. This allows to
determine the average spin tune $\left<\nu\right>$ and corresponding
average energy of the particles $\left<E\right>$ 
with higher accuracy than the beam energy spread $\sigma_{\rm E}$.

Formula (\ref{eqn:Omega}) gives the value of $\gamma$
averaged over the beam revolution time. Thus, 
for a symmetric machine, it corresponds  to the energy in the
interaction point.

The method described has been developed in Novosibirsk and first 
applied to the
$\phi$ meson mass measurement at the VEPP-2M storage ring~\cite{BUK}.
Later it was successfully used to measure masses of the 
$\psi$-~\cite{Psi12} 
and $\Upsilon$-meson family~\cite{UPS1,UPS23,UPSM},
see also Ref.~\cite{REVISIT}, in which the values of the masses were
rescaled to take into account the change of the electron mass value.
The relative mass accuracy achieved in these experiments was 
$1\cdot 10^{-5}$ for the $\Upsilon$-family
and $3\cdot 10^{-5}$ for the $\psi$-family. The resonant
depolarization experiments on bottomonium masses  were also performed
with the CUSB detector at CESR \cite{mCUSB} ($\Upsilon(1S)$) and
with the ARGUS detector at DORIS \cite{mARGUS} ($\Upsilon(2S)$).
The accuracy of the $J/\psi(1S)$ meson mass measurement was improved in 
the Fermilab $p\bar{p}$-experiment E760~\cite{E760} to 
$1.2\cdot 10^{-5}$ using the $\psi(2S)$ mass value from Ref.~\cite{Psi12}.
The resonant depolarization method was also successfully
applied to the high precision measurement of the Z-boson mass at 
LEP~\cite{LEP}.
The comprehensive review of this technique and
its application to particle mass measurements can be found 
in~\cite{review}.

\subsection[\ensuremath{\mathrm{e^+e^-}} scanning, radiative corrections]
           {\ensuremath{\mathrm{e^+e^-}} scanning, radiative corrections
            $\!$\footnote{Author: P.~Wang}}
\label{sec:radcorr}

\subsubsection[Introduction]{Introduction}
    The measurement of the  mass and total width  of $1^{--}$ resonances
by $e^{+}e^{-}$ colliding experiments is done by scanning the
resonance curve and fitting the data with the 
theoretical cross-section as a function of these parameters. 
The $e^+e^-$ partial width is also determined from this fitting,
\ie $\Gamma_{ee}$ is measured as the coupling of the resonance to
the incoming $e^{+}e^{-}$, instead of decaying process; while most
other decay modes are measured as branching ratios by dividing
the number of the observed events decaying into this mode by the 
total number of resonance events. 
Such fitting requires precise calculation of initial  
state radiative corrections. This is done by 
the structure function  approach~\cite{rad.1,rad.2,rad.3}. 
It yields the accuracy of $0.1\%$. In this scheme, 
the radiatively corrected cross-section is expressed as 
\begin{eqnarray}
\sigma(s)\;=\int^{1-s_m/s}_0 dx \; \tilde{\sigma}(s(1-x))F(x,s) ,
\label{eq:rad}
\end{eqnarray}
where $\sqrt{s}$ is the C.M. energy of the colliding beam, $\sqrt{s_m}$ is
the cut-off of the invariant mass in the event selection, and 
\begin{equation}
\tilde{\sigma}(s)=\frac{\sigma_{B}(s)}{|1-\Pi(s)|^2}.
\end{equation}
with $\sigma _{B} (s)$ the 
Born order cross-section and $\Pi(s)$ the vacuum polarization. 
In \Eq~(\ref{eq:rad})
\begin{equation}
F(x,s)\;=\;\beta x^{\beta-1}\delta^{V+S}+\delta^{H} ,
\label{eq:Fexp}
\end{equation}
with
\begin{equation}
\beta\;=\;\frac{2\alpha}{\pi} \left(\ln \frac{s}{m^{2}_{e}}-1\right) ,
\label{eq:beta}
\end {equation}
\begin{equation}
\delta^{V+S}\;=\;1+\frac{3}{4}\beta+\frac{\alpha}{\pi}
\left(\frac{\pi^{2}}{3}-\frac{1}{2}\right)+\beta^{2}
\left(\frac{9}{32}-\frac{\pi^{2}}{12}\right) ,
\label{eq:V+S}
\end{equation}
\begin{eqnarray}
\delta^{H}\; & = & \;-\beta\left(1-\frac{x}{2}\right) \nonumber \\
              &   & \nonumber \\
& & +\frac{1}{8}\beta^{2}\left[4(2-x)\ln\frac{1}{x}-
\frac{1+3(1-x)^{2}}{x}\ln(1-x)-6+x\right] .
\end{eqnarray}
Here the conversion of  bremsstrahlung photons to 
real $e^{+}e^{-}$ pairs is 
included in the cross-section which is the usual 
experimental situation.  Thus there is cancellation between the 
contributions of virtual and real $e^{+}e^{-}$ pairs in the leading
term~\cite{rad.3}.

    Since this note discusses resonances, like $\psi'$ and
$\psi''$. so $\sigma _{B} (s)$ is expressed by Breit--Wigner formula. 
For narrow resonances, like $J/\psi$ and $\psi'$, it is 
\begin{equation}
\sigma_{B}(s)\;=\frac{12\pi \Gamma^0_{ee} \Gamma_{f}}{(s-M^{2})^{2}
+\Gamma^{2} M^{2}} ,
\end{equation}                                       
where $M$ and $\Gamma$  are the mass and total width of the resonance; 
$\Gamma^0_{ee}$ and $\Gamma _{f}$   are the partial  widths 
to the $e^{+}e^{-}$  mode  and to  the final 
state  respectively. Usually  to
measure $M$, $\Gamma$ and $\Gamma_{ee}$, $f$ is inclusive hadrons.
Here $\Gamma^0_{ee}$ describes the coupling strength of the resonance 
to  $e^{+}e^{-}$ through a virtual photon. For example, in potential
model, $\Gamma^0_{ee}$ is related to the wave function at the
origin $\psi(0)$ in the way
\begin{equation}
\Gamma^0_{ee} = \frac{4\alpha^2 Q_q^2 |\psi(0)|^2}{M^2}
\end{equation}
where $Q_q$ is the charge carried by the quark in the quarkonium
and $\alpha$ is the QED fine structure constant. 
Since the decay of a
quarkonium $1^{--}$ state to $e^{+}e^{-}$ pair is through a
virtual photon, there is always vacuum polarization associated with
this process. So the experimentally measured $e^+e^-$ partial width, 
denoted explicitly as $\Gamma_{ee}^{exp}$, is related to 
$\Gamma^0_{ee}$ by the expression  
\begin{equation}
\Gamma _{ee}^{exp}=\frac{\Gamma^0_{ee}}{|1-\Pi(M^2)|^2}.
\label{eq:gee}
\end{equation}
Particle Data Group adopts the convention of Ref~\cite{Luth}
that $\Gamma_{ee}$ means $\Gamma_{ee}^{exp}$. The radiatively
corrected resonance cross-section is
\begin{equation}
\sigma_{res}(s) = \;\int^1_0 dx \; F(x,s)
\frac{12\pi \Gamma^{exp}_{ee} \Gamma_{f}}{(s-M^{2})^{2}
+\Gamma^{2} M^{2}}.
\label{eq:xres}
\end{equation} 
For resonances, as long as $\sqrt{s}-\sqrt{s_m} \gg \Gamma$, the
integral of \Eq~(\ref{eq:xres}) is insensitive to $\sqrt{s_m}$, because
the Breit-Wigner formula behaves like a $\delta$ function. One can put
the upper limit of integral to $1$.

\subsubsection[Analytical expression]{Analytical expression}

For the practical purpose of fitting, the expression of radiative
corrected resonance cross-section in \Eq~(\ref{eq:xres}) is integrated
analytically~\cite{Cahn}. To get the best accuracy, one rewrites
$F(x,s)$ in \Eq~(\ref{eq:Fexp}) in terms of a series expansion:
\begin{eqnarray}
\begin{array}{lll}
\displaystyle
F(x,s) &\displaystyle = \; \beta x^{\beta -1} 
\left[ 1+ \frac{3}{4} \beta
  + \frac{\alpha}{\pi} \left( \frac{\pi ^2}{3}-\frac{1}{2}\right)
 + \beta ^2 \left( \frac{9}{32}- \frac{\pi ^2}{12} \right) \right] \\ 
\noalign{\vskip7pt}
\displaystyle{}&\displaystyle + x^{\beta} 
\left(- \beta- \frac{\beta^2}{4}\right)
+ x^{\beta+1} \left(\frac{\beta}{2}- \frac{3}{8} \beta^2 \right)
 +O(x^{\beta+2} \beta^2)   \\
\noalign{\vskip7pt}
\displaystyle{}&\displaystyle =  
\beta x^{\beta -1} \delta^{V+S} + \delta^H ,
%       (2.12)
\end{array}
\label{eq:Four}
\end{eqnarray}
with
\begin{equation}
\delta^{H}\;=\;x^{\beta}\left(-\beta-\frac{\beta^{2}}{4}\right)
+x^{\beta+1}\left(\frac{\beta}{2}-\frac{3}{8}
\beta^{2}\right) .
%       (2.13)
\end{equation}

Notice that here the omitted terms start from $x^{\beta +2}\beta
^{2}$, while the three terms which are kept all have $\beta$ term in
their coefficients.  \Eq[b]~(\ref{eq:Four}) differs from \Eq~(\ref{eq:Fexp}) in
the $\delta^{H}$ term.  Their equivalence can be verified if one
writes $x^{\beta} = 1 + \beta \ln x$, $x^{\beta +1} = x + \beta x\ln
x$ and $\ln (1-x) = -x - x^{2}/2 +\ldots $.
\shortpage[2]

With $F(x,s)$ in the form of \Eq(~\ref{eq:Four}), the radiatively
corrected resonance cross-section of \Eq~(\ref{eq:xres}) can be expressed
as
\begin{eqnarray}
\begin{array}{llllll}
\displaystyle
\sigma_{res}(s)\; &\displaystyle =\;\frac{12\pi\Gamma_{ee}\Gamma_{f}}{s^{2}}
\{\delta^{V+S} [
a^{\beta-2}\Phi(\cos\theta,\beta) \\
\noalign{\vskip7pt}
\displaystyle{}&\displaystyle+\beta ( \frac{1}{\beta-2}+
\frac{2(s-M^{2})}{(\beta-3)s}
+\frac{3(s-M^{2})^{2}-M^{2}\Gamma^{2}}{(\beta-4)s^2} ) ] \\
\noalign{\vskip7pt}
\displaystyle{}&\displaystyle -\beta(1+\frac{\beta}{4})[
\frac{1}{1+\beta}a^{\beta-1}\Phi(\cos\theta,\beta+1) \\
\noalign{\vskip7pt}
\displaystyle{}&\displaystyle+\frac{1}{\beta-1}
+\frac{2(s-M^{2})}{(\beta-2)s}
+\frac{3(s-M^{2})^{2}-M^{2}\Gamma^{2}}{(\beta-3)s^{2}}] \\
\noalign{\vskip7pt}
\displaystyle{}&\displaystyle +(\frac{\beta}{2}
-\frac{3}{8}\beta^{2})[\frac{1}{2}
\ln\frac{1+2a\cos\theta+a^{2}}{a^{2}} \\
\noalign{\vskip7pt}
\displaystyle{}&\displaystyle-\cot\theta(\arctan
\frac{1+a\cos\theta}{a\sin\theta}
-\frac{\pi}{2}+\theta)]\} ,
\end{array}
\label{eq:bwa}
\end{eqnarray}
with 
\begin{equation}
\Phi (\cos \theta ,\nu)\; \equiv \;
\frac{\pi \nu \sin[(1-\nu )\theta]}
{\sin\theta \;\sin \pi \nu } ;
\label{eq:phi}
\end{equation}
\begin{equation}    
   a^{2}\;=\;\left(1-\frac{M^{2}}{s}\right)^{2}
+\frac{M^{2}\Gamma^{2}}{s^{2}} ;
\label{eq:aexp}
\end{equation}
\begin{equation}
  \cos\theta\;=\;\frac{1}{a}\left(\frac{M^{2}}{s}-1\right) .
\label{eq:theta}
\end{equation}

\subsubsection[Narrow resonances]{Narrow resonances}
Below the open charm or bottom threshold, the resonances are narrow
with widths from tens to hundreds KeV, while the beam energy
resolution of $e^+e^-$ colliders is of the order of MeV. If the
resonance width is comparable or smaller than the beam energy
resolution, the observed resonance cross-section is the cross-section
by \Eq~(\ref{eq:bwa}) folded with the beam energy resolution function
$G(W,W')$. Also in the observed cross-section, there is always a
continuum part from direct virtual photon annihilation which is
usually treated as $1/s$ dependence. So the experimentally observed
cross-section is
\begin{equation}
\sigma_{obs}(W)\;=\;\frac{R}{s}+
\int_{0}^{\infty}G(W,W')\sigma_{res}(W')dW' .
\label{eq:xobs}
\end{equation}
In the above, $R$ is a fitting parameter and
$G(W,W')$ is usually taken as a Gaussian function:
\begin{equation}
    G(W,W')\;=\;\frac{1}{\sqrt{2\pi}\Delta}
    \exp\left[-\frac{(W-W')^{2}}
    {2\Delta^{2}}\right] ,
\label{eq:gww}
\end{equation}
with $\Delta$ the standard deviation of the Gaussian distribution.

\shortpage[2]

In the fitting of the experimental data with the theoretical curve,  
$M$, $\Gamma$, $\Gamma_{ee}$, $R$ and $\Delta$ are
obtained. 

\subsubsection[$\mu^+\mu^-$ final state]{$\mu^+\mu^-$ final state}

Usually the $\mu^+\mu^-$ curve is also fitted, to extract
$\Gamma_{\mu\mu}$. The fitting of inclusive hadron can be combined
with the $\mu^+\mu^-$ curve to obtain $M$, $\Gamma$, $\Gamma_{ee}$,
$\Gamma_{\mu\mu}$, $R$ and $\Delta$.  Here unlike for an inclusive
hadronic final 
state, the continuum $\mu^+\mu^-$ cross-section is
calculated by QED~\cite{Bere80}, and the interference between virtual
photon and the resonance must be included.  The cross-section of
$e^+e^-\rightarrow\mu^+\mu^-$ at Born order is
\begin{equation}
\sigma_B(s) = \frac{4\pi\alpha^2}{3s} [ 1 + 2 ReB(s) + |B(s)|^2 ]
\end{equation}
where
\begin{equation}
B(s) = \frac{3s\Gamma_{ee}/M\alpha}{(s-M^2)+iM\Gamma}.
\end{equation}
With radiative correction, it can be expressed as
\begin{eqnarray}
&& \sigma_{\mu+\mu^-}(s) =  
      \frac{1}{|1-\Pi(s)|^2} \left\{\frac{4\pi\alpha^2}{3s}
                \left[1+\frac{\beta}{2}\left(2\ln x_f -\ln (1-x_f) +\frac{3}{2}
              -x_f\right) + \frac{\alpha}{\pi}\left(\frac{\pi^2}{3}-\frac{1}{2}\right)\right]\right.
              \nonumber \\
&&      \nonumber                          \\
&& \>   +C_1 \delta^{V+S} \left[a^{\beta-2}\Phi (\cos\theta,\beta)
                       +\beta\left(\frac{x_f^{\beta-2}}{\beta-2}
                             +\frac{x_f^{\beta-3}}{\beta-3}R_2
                             +\frac{x_f^{\beta-4}}{\beta-4}R_3\right)\right] 
\nonumber \\
&&               \nonumber               \\
&& \>  +\left[-\beta\delta^{V+S}C_2 + 
               \left(-\beta-\frac{\beta^2}{4}\right)C_1 \right]
              \left[\frac{a^{\beta-1}}{1+\beta}\Phi(\cos\theta,\beta+1)
               +\frac{x_f^{\beta-1}}{\beta-1}
               +\frac{x_f^{\beta-2}}{\beta-2}R_2
               +\frac{x_f^{\beta-3}}{\beta-3}R_3\right]  \nonumber   \\
&&                \nonumber  \\
&& \>   +\left[\left(\beta +\frac{\beta^2}{4}\right)C_2 +
                 \left(\frac{\beta}{2}-\frac{3}{8}\beta^2\right)C_1\right]    
                \left[\frac{1}{2}\ln\frac{x_f^2+2ax_f\cos\theta+a^2}{a^2} \right.
\nonumber\\    
&& \qquad\qquad\qquad\qquad\qquad\qquad\qquad\qquad
\left.\left.
      -\cot\theta\left(\tan^{-1}\frac{x_f+a\cos\theta}{a\sin\theta}
                           -\frac{\pi}{2}+\theta\right)\right]\right\}   
\label{eq:xmu}
\end{eqnarray}
where
\begin{eqnarray}
C_1 & = & [8\pi\alpha\frac{\sqrt{\Gamma^0_e\Gamma^0_{\mu}}}{M}(s-M^2)
      +12\pi(\frac{\Gamma^0_e\Gamma^0_{\mu}}{M^2})s]/s^2,
        \nonumber \\
     & &   \nonumber \\
C_2 & = & [8\pi\alpha\frac{\sqrt{\Gamma^0_e\Gamma^0_{\mu}}}{M}
      +12\pi(\frac{\Gamma^0_e\Gamma^0_{\mu}}{M^2})]/s,
         \nonumber \\
     & &    \nonumber \\
R_2 & = & \frac{2(s-M^2)}{s} = -2a\cos\theta    \\
     & &    \nonumber \\
R_3 & = & a^2 (4\cos^2\theta - 1 )     \nonumber \\
     & &    \nonumber \\
x_f & = & 1 - \frac{s_m}{s}.     \nonumber
\end{eqnarray}
$\Phi (\cos \theta ,\nu)$, $a$ and $\theta$ are defined previously in 
\Eq~(\ref{eq:phi}),(\ref{eq:aexp}) and (\ref{eq:theta}).
Here unlike the resonance term, the continuum term depends on 
the invariant mass cut $\sqrt{s_m}$ in the 
$\mu^+\mu^-$ event selection. 
Also $\Gamma_{\mu\mu}$ is defined similar to $\Gamma_{ee}$, in the way
\begin{equation}
\Gamma_{\mu\mu} \equiv \Gamma^{exp}_{\mu\mu} = 
\frac{\Gamma_{\mu\mu}^0}{|1-\Pi(M_{res}^2)|^2}.
\end{equation}
For $\mu$ pair final state, 
unlike the inclusive hadrons, the vacuum polarization 
cannot be absorbed into the definition
of $\Gamma_{ee}$ in all terms, 
so it must be calculated explicitly. The leptonic
part of $\Pi(s)$ is well known. (For example, in
Ref~\cite{Bere73}, although there the definition of $\Pi(s)$ has a
minus sign relative to the more common convention used here.)  
The hadronic part was first calculated
in Ref~\cite{vac}, and a more recent treatment is found in
Ref~\cite{Pich}. 

For narrow resonances, the $\mu$ pair cross-section also need to be
folded with the beam energy resolution function.

\subsubsection[Very narrow resonances]
              {Very narrow resonances}

For the narrow resonances with $\Gamma \ll \Delta$, 
(\eg $\Gamma$ is an order of magnitude smaller than $\Delta$), 
then \Eq~(\ref{eq:xobs}) is insensitive to
$\Gamma$, the fitting becomes impractical. In such case, the area
method~\cite{JS,Tsai} can be used to extract 
\begin{equation}
(Area)_0 = \frac{6\pi^2\Gamma_{ee}\Gamma_f}{M^2\Gamma},
\end{equation} 
together with $M$, $R$ and $\Delta$ from the fitting.
Here the final state $f$ is inclusive hadrons.
This method requires additional information on the leptonic branching
ratio ${\cal B}(l^+l^-)$, 
which is obtained from counting the $\mu$ pair events on top of 
the resonance. With the  
assumption that ${\cal B}(e^+e^-)={\cal B}(\mu^+\mu^-)$ and 
$\Gamma_{had}=\Gamma(1-2 {\cal B}(l^+l^-))$(If the resonance is above 
the $\tau$ production threshold, $\Gamma_{had}=\Gamma(1-3 {\cal B}(l^+l^-))$.
A phase space correction is needed for $\psi(2S)$, for it is close to
the $\tau$ threshold.), $\Gamma$ and $\Gamma_{ee}$ can be solved from 
$(Area)_0$ and ${\cal B}(l^+l^-)$.  

Both of the original papers on the area method in Ref.~\cite{Tsai} 
and\cite{JS} mistreated 
radiative correction. This was pointed out later on by
Ref~\cite{Luth}. For the convenience of the readers, here the 
complete formulae of area method are presented.

The experimentally observed cross-section is 
\begin{equation}
\sigma_{obs}(W)=\frac{R}{s}+(Area)_0 \delta^{V+S} Gr(W-M),
\label{eq:area}
\end{equation}
where $\delta^{V+S}$ is defined in \Eq~(\ref{eq:V+S}), and
the radiatively corrected Gaussian function 
\begin{equation}
Gr(w)=\left(\frac{2\Delta}{W}\right)^\beta\frac{1}{\Delta}
F(\frac{w}{\Delta},\beta).
\end{equation} 
The function $F(z,\beta)$ is approximated~\cite{JS} as
\begin{eqnarray}
F(z,\beta)& \approx &\frac{\Gamma(1+\beta)}{\sqrt{2\pi}}e^{-z^2/2}
\left[0.31-\frac{0.73z}{\sqrt{1+\left(\frac{z}{1+1.37\beta}\right)^2 }}
+z^2\right]^{-\beta/2} \nonumber \\
                  \nonumber         \\
& & +\theta(z)\beta z^{\beta}\left(\frac{z^{2.18}}{1+z^{3.18}}\right)
                  \nonumber\\
 \nonumber\\
& & \times
\left\{1+\frac{(1-\beta)(2-\beta)/2}{\left[\left(z-\frac{46}{z^2+21}\right)^2
+2.44+1.5\beta\right]}\right\},
\label{eq:Fzb}
\end{eqnarray}
where $\Gamma(1+\beta)$ is the Gamma function and $\theta(z)$ is the
step function. Notice that to derive \Eq~(\ref{eq:area}), $\delta^H$
term in \Eq~(\ref{eq:Fexp}) is neglected, and the Breit--Wigner is
approximated as a $\delta$ function compared with $\Delta$.  These
limit the accuracy of the results.

For large positive or negative $z$, there are asymptotic expansions
of $F(z,\beta)$~\cite{JS}, which are useful to calculate the
resonance cross-sections away from the peak, \eg radiative tails. 

\subsubsection[Resonance near threshold]{Resonance near threshold}

$\psi(3770)$ and $\Upsilon(4S)$ are near the threshold of $D\bar{D}$
or $B\bar{B}$ production. They decay predominately into $D\bar{D}$ or
$B\bar{B}$.  The line shape is cut off at the threshold. In the
radiative correction expressed by the integral of \Eq~(\ref{eq:rad}),
the cut off $\sqrt{s_m}=2m_P$, with $m_P$ the mass of the pseudoscalar
meson ($D^0$ or $B^\pm$) produced at the threshold.  (If the resonance
is well above the threshold, \ie  $\sqrt{s}-2m_P \gg \Gamma$, the
integral is not sensitive to the upper limit.)  The line shape of the
resonances is
\begin{eqnarray}
\sigma_{res}(s)\;= \frac{R}{s} + 
\int^{1-4M_P^2/s}_0 dx \; F(x,s) 
\frac{12\pi \Gamma^{exp}_{ee} \Gamma_{f}}{(s-M^{2})^{2}
+\Gamma^{2} M^{2}}.
\label{eq:x3770}
\end{eqnarray}
In the above, the first term is the continuum cross-section due to 
direct virtual photon annihilation. This term could be greater than 
the resonance itself. For example, for $\psi(3770)$, the continuum
cross-section is $13$nb; while the radiatively corrected resonance
cross-section is about 8nb.

\subsubsection[The energy dependent width]{The energy dependent width}

Above the open charm or bottom threshold, the resonances are wide,
usually over 10MeV.  For such wide resonances, the energy dependence
of its width need to be considered. Such dependence cannot be derived
from first principle, and the formula is model dependent.  For
example, in the MARK III scan of $\psi(3770)$, it is in the form:
\begin{equation}
\Gamma(E) \propto \frac{p^3_{D^0}}{1+(rp_{D^0})^2}
 +\frac{p^3_{D^\pm}}{1+(rp_{D^\pm})^2}
\end{equation}
and the width listed by PDG is defined as
\begin{equation}
 \Gamma_{\psi''}=\Gamma(E=M_{\psi''})
\end{equation}
In the above, $p_{D^0}$ and  $p_{D^\pm}$ are the momentum of $D^0$ and
$D^\pm$ produced. $r$ is a free parameter. Usually the fitting is not
sensitive to $r$.   
So in \Eq~(\ref{eq:x3770}),
\begin{equation}
\Gamma(E) = \frac{\Gamma_{\psi''}\left(
\frac{p^3_{D^0}}{1+(rp_{D^0})^2}
 +\frac{p^3_{D^\pm}}{1+(rp_{D^\pm})^2}\right)}
{\frac{(p^0)^3_{D^0}}{1+(rp^0_{D^0})^2}
 +\frac{(p^0)^3_{D^\pm}}{1+(rp^0_{D^\pm})^2}}
\end{equation}
where 
\begin{equation}
p_{D^0} = \frac{1}{2}\sqrt{E^2-4m_{D^0}^2};
\end{equation}
\begin{equation}
p_{D^\pm} = \frac{1}{2}\sqrt{E^2-4m_{D^\pm}^2};
\end{equation}
\begin{equation}
p^0_{D^0} = \frac{1}{2}\sqrt{M^2-4m_{D^0}^2};
\end{equation}
and
\begin{equation}
p^0_{D^\pm} = \frac{1}{2}\sqrt{M^2-4m_{D^\pm}^2}
\end{equation}

The width of $\Upsilon(4S)$ and the states above are expressed
similarly.
 
The Breit--Wigner with the energy dependent width cannot be integrated
analytically with $F(x,s)$. In the fitting, the cross-section is
numerically integrated.  On the other hand, for these wide resonances,
usually the finite beam energy spread can be neglected.

\subsubsection[The shift and scale down of the maximum height]
              {The shift and scale down of the maximum height} 

With the radiative correction, the maximum height of the resonance is
shifted from the mass of the resonance $M$ to~\cite{Zline} 
\begin{equation}
 M + \Delta\sqrt{s}_{max} 
\end{equation}
where
\begin{equation}
 \Delta\sqrt{s}_{max} \approx \frac{\beta\pi}{8}\Gamma
\end{equation}
and the maximum height of the resonant peak is reduced by a
factor~\cite{Zline} 
\begin{equation}
 \rho \approx \left(\frac{\Gamma}{M}\right)^\beta \delta^{V+S};
\end{equation}
where $\beta$ and $\delta^{V+S}$ are defined in \Eq~(\ref{eq:beta})
and (\ref{eq:V+S}).  For the narrow resonances, the shift of the
maximum height due to radiative correction is small, due to the narrow
widths. On the other hand, the finite beam resolution also shifts the
maximum height of the observed resonance shape. It is roughly at the
order of one-tenth of $\Delta$. This needs to be taken into account in
the precision measurements of the branching ratios.

\subsection[$\bar{p}p$ scanning techniques and limits]
           {$\bar{p}p$ scanning techniques and limits
            $\!$\footnote{Author: G.~Stancari}}
\label{sec:ppcool}

\subsubsection[Introduction]{Introduction}

The \pbarp\ formation technique, where the antiproton beam annihilates
with a hydrogen target, has been thoroughly exploited to scan all
known charmonium states, overcoming the limitations of the $e^+e^-$,
which can actually form only vector states.  A successful program was
carried out at CERN's ISR by R704~\cite{baglin:1986,baglin:1987} and
at the Fermilab Antiproton Source by
E760~\cite{armstrong:1992,armstrong:1992b,armstrong:1993} and
E835~\cite{ambrogiani:1999,bagnasco:2002,ambrogiani:2003,andreotti:2003}.
All detectors used so far were non-magnetic.  Experiment PANDA at the
future GSI facility also includes a program of charmonium
studies~\cite{GSIcdr:2001} and will be the first provided with a
magnetic field.

Many aspects of antiproton beam conditioning for charmonium studies
are discussed in Ref.~\cite{mcginnis:2003}.  The antiproton beam
energy is scanned across the resonance in steps appropriate for the
width of the resonance under study.  The observed cross-section is
given by $\sigma_{obs}(W=\sqrt{s}) = \sigma_{cont}+
\int_{0}^{\infty}G(W,W')$ $\sigma_{BW}(W')$ $dW'$.  The mass, width
and peak cross-section $\sigma_{BW}(M_R)$ are determined by the number
of observed events, after deconvoluting the beam energy spectrum
$G(W,W')$ and subtracting the continuum cross-section $\sigma_{cont}$
from the observed cross-section $\sigma_{obs}(W=\sqrt{s}) =
\sigma_{cont}$ $+$ $\int_{0}^{\infty}G(W,W')$ $\sigma_{BW}(W')$ $dW'$.
They are not directly influenced by the detector resolution.

\begin{figure}[htb]
\begin{center}
  \includegraphics[width=.85\textwidth]{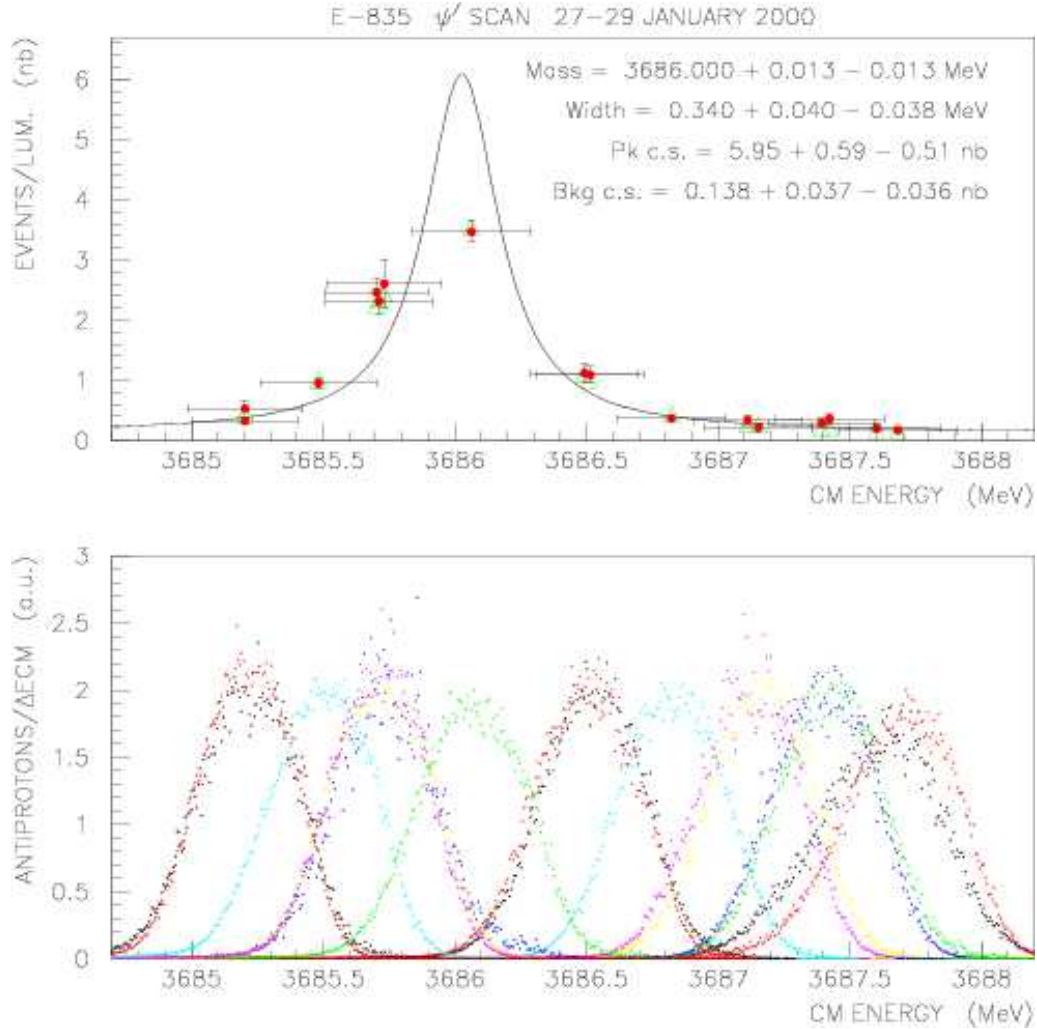}
\end{center}
\caption{(colour) Resonance scan at the $\psi(2S)$.}
\label{fig:scan}
\end{figure}

For instance, \Figure~\ref{fig:scan} shows a 16-point scan of the
\psip\ resonance.  The bottom plot shows the normalized beam energy
distributions as the beam was decelerated.  The top plot indicates the
measured cross-sections (red circles) compared with the best
predictions (green triangles) from a maximum-likelihood fit to the
convolution of the beam distributions (bottom plot) with a
Breit--Wigner resonance curve (solid line).

An important role is played by the beam energy distribution.  This
function can be measured from the Schottky revolution frequency
spectrum, the bunching radiofrequency, the orbit length of particles
in the rf bucket and the slip factor of the
machine~\cite{armstrong:1993,werkema:2000}.  The only quantity that
needs external input is the orbit length.  It needs to be calibrated
with the scan of a narrow resonance whose mass is well known.
Typically, the \psip\ is chosen, because the absolute value of its
mass is measured with extreme accuracy (25~keV) by the resonant
depolarization method in $e^+e^-$
~\cite{artamonov:2000,aulchenko:2003}, described in Section
(\ref{sec:redepol}).  Using the \psip\ for calibration, the other
masses are determined with an uncertainty \( \leq 200\un{keV} \).  The
main contribution to the uncertainty comes from the orbit length: its
value is obtained by comparison of the reference orbit with the
beam-position monitor readings during the scan of the resonance under
study.

\subsubsection[Signal extraction in hadronic annihilations]
              {Signal extraction in hadronic annihilations}

The rate of charmonium formation formed depends on the coupling
between the initial state and the resonance.  In $e^+e^-$\
annihilation, the couplings of \ep\ to both \jpsi\ and \ups\ are of
the order of~$10^{-2}$. The branching fractions \( \pbarp \to \ccbar
\) are of the order of $10^{-4}$--$10^{-3}$ for charmonium, but
probably much smaller for bottomonium: \( \approx 10^{-7} \) is the
theoretical prediction~\cite{chernyak:1984}, and \( < 5 \times 10^{-4}
\) is the experimental upper limit for \( \pbarp \to \ups
\)~\cite{baru:1996}.  The relatively low intensity of antiproton beams
is partially compensated by the availability of jet targets; typical
luminosities were \( 2 \times 10^{31}\un{cm^{-2} s^{-1}} \) for E835;
at GSI, an increase of a factor~10 is expected.

Formation cross-sections for charmonium states in $\ppbar$
annihilations range between 10 and 10$^3$~nb, but only a small
fraction can be detected.  In antiproton--proton annihilations, the
limiting factor is the large total cross-section (70~mb).  This
implies that a clean charmonium signal (pb--nb) can be extracted only
by identifying its inclusive or exclusive electromagnetic decays to a
high-invariant-mass \ep\ or \gaga\ pair, such as \( \pbarp \to \chi_c
\to \jpsi + X \to \ep + X \).

Hadronic channels such as $\pi^0\pi^0$ and $\eta\eta$ have recently
been investigated. Using data taken in 2000, E835 has provided the
first evidence \cite{Andreotti:2003sk} of a charmonium signal
exploiting the interference between resonance and continuum at the
$\chi_{c0}$ energy.

\subsubsection[Limits on energy and width resolution]
              {Limits on energy and width resolution}

A small beam energy spread is desirable because it reduces the
uncertainty on the mass by sharpening the resonance peak.  However,
efforts to make the beam much narrower than the resonance are
obviously not necessary.  For antiproton beams, the minimum attainable
momentum spread is determined by longitudinal stability (Keil--Schnell
criterion~\cite{keil:1969}) rather than stochastic cooling
power. Typically, with a beam current \( I = 50\un{mA} \), one can
achieve a momentum spread \( \sigma_p / p = 10^{-4} \), which
translates to \( \sigma_{\sqrt{s}} = 0.2\un{MeV} \) in the
centre-of-momentum frame.  As the beam intensity increases, the
minimum attainable momentum spread increases as~$\sqrt{I}$.  In
$e^+e^-$ machines there is no need for stochastic cooling, but the
energy spread is dominated by initial state radiation: \(
\sigma_{\sqrt{s}} = 1\un{MeV} \) at the \jpsi, \( \sigma_{\sqrt{s}} =
4\un{MeV} \) at the \ups.

\subsection[Luminosity of photon photon scattering]
           {Luminosity of photon photon scattering 
            $\!$\footnote{Author: M.~Kienzle}}
\label{sec:gagalum}

The cross-section for a $\gamma \gamma$ process is related to the cross-section
at the $e^+ e^-$ level, which is measured in the laboratory, by the formula
\begin{equation}
d\sigma (e^+ e^- \rightarrow e^+ e^- X) = \sigma (\gamma_1\gamma_2
\rightarrow X) \frac{d^2n_1}{dz_1 dP_1^2} \frac{d^2n_2}{dz_2 dP_2^2} dz_1dz_2
dP_1^2dP_2^2
\label{eq:epa} 
\end{equation}
where $z_i$ is the scaled photon energy in the laboratory frame and
$P_i^2$ is the photon mass. This is the equivalent photon
approximation (EPA)\cite{Budnev} where the longitudinal polarization
component as well as the mass of the incoming photons are neglected in
$\sigma (\gamma\gamma \rightarrow X)$. The $P_i^2$ integration can be
carried out to give the photon "density" in the $e^\pm$ (the photon
flux)
\begin{equation}
\nonumber
f_{\gamma/e}(z, P_{min}, P_{max}) = \int_{P^2_{min}}^{P^2_{max}} \frac{d^2n}{dz dP^2} dP^2 =
\end{equation}
\begin{equation}
=\frac{\alpha}{2\pi} \left [\frac{1 + (1-z)^2}{z} \ln \frac{P^2_{max}}{P^2_{min}} - 2m^2_e z \left (
\frac{1}{P^2_{min}}-\frac{1}{P^2_{max}} \right ) \right ] .
\end{equation}

 For untagged experiments (the scattered $e^\pm$ are undetected ) $P_{min}$ is the kinematic limit:
\begin{equation}
P^2_{min} = \frac{m_e^2z^2}{1-z}
\end{equation}
and $P_{max} \simeq E_{beam}$.  \par For resonance production,
\Eq(\ref{eq:epa}) simplifies since one of the $z_i$ integrations can
be performed with the constraint $z_1z_2 = \tau = M^2 /s_{e^+ e^- }$
where $M$ is the resonance mass. It is then customary to define
luminosity functions:
\begin{equation}
\frac{d\calL }{dM} = \frac{2 \tau} {M} \int dz_1dz_2 f_{\gamma/e}(z_1)f_{\gamma/e}(z_2) 
\delta (z_1z_2 - \tau )
\label{eq:lgg} 
\end{equation}
so that
\begin{equation}
d\sigma (e^+ e^- \rightarrow e^+ e^- X) = \int M \frac{d\calL }{dM}\sigma (\gamma\gamma
 \rightarrow X).
\end{equation}
This luminosity curve makes it easy to determine the counting rate for
resonance production knowing the width of the resonance in the $\gamma
\gamma $ channel.  \par The most accurate Monte Carlo computation of
two-photon production in $e^+ e^-$ collisions is the program
GALUGA\cite{galuga} widely used to extract the luminosity function and
the photon structure function in various kinematical regions.

\subsection[Interference with continuum in \ensuremath{\mathrm{e^+e^-}}
            experiments]
           {Interference with continuum in \ensuremath{\mathrm{e^+e^-}}
            experiments
            $\!$\footnote{Authors: C.Z.~Yuan, P.~Wang, X.~H.~Mo}}
\label{sec:interf}

\subsubsection[Introduction]{Introduction}
\label{sec:interference-intro}

It is well known that the $e^+e^-$ experiments have lots of advantages
in the study of the charmonium physics: large cross-section, small
background, and well-determined initial state (both four-momentum and
quantum numbers). However, there is an inevitable amplitude\,---\,the
continuum amplitude
\[
      e^+e^- \rightarrow \gamma^* \rightarrow hadrons
\]
accompanied with the production of the resonances. This amplitude
does not go through the resonance, but in general it can produce
the same final hadronic states as charmonia do. This amplitude has
been overlooked in many previous studies.

\subsubsection[Experimentally observed cross-section]
              {Experimentally observed cross-section}

\begin{figure}[hbt]
\begin{center}
\includegraphics[width=3.cm,height=2cm]{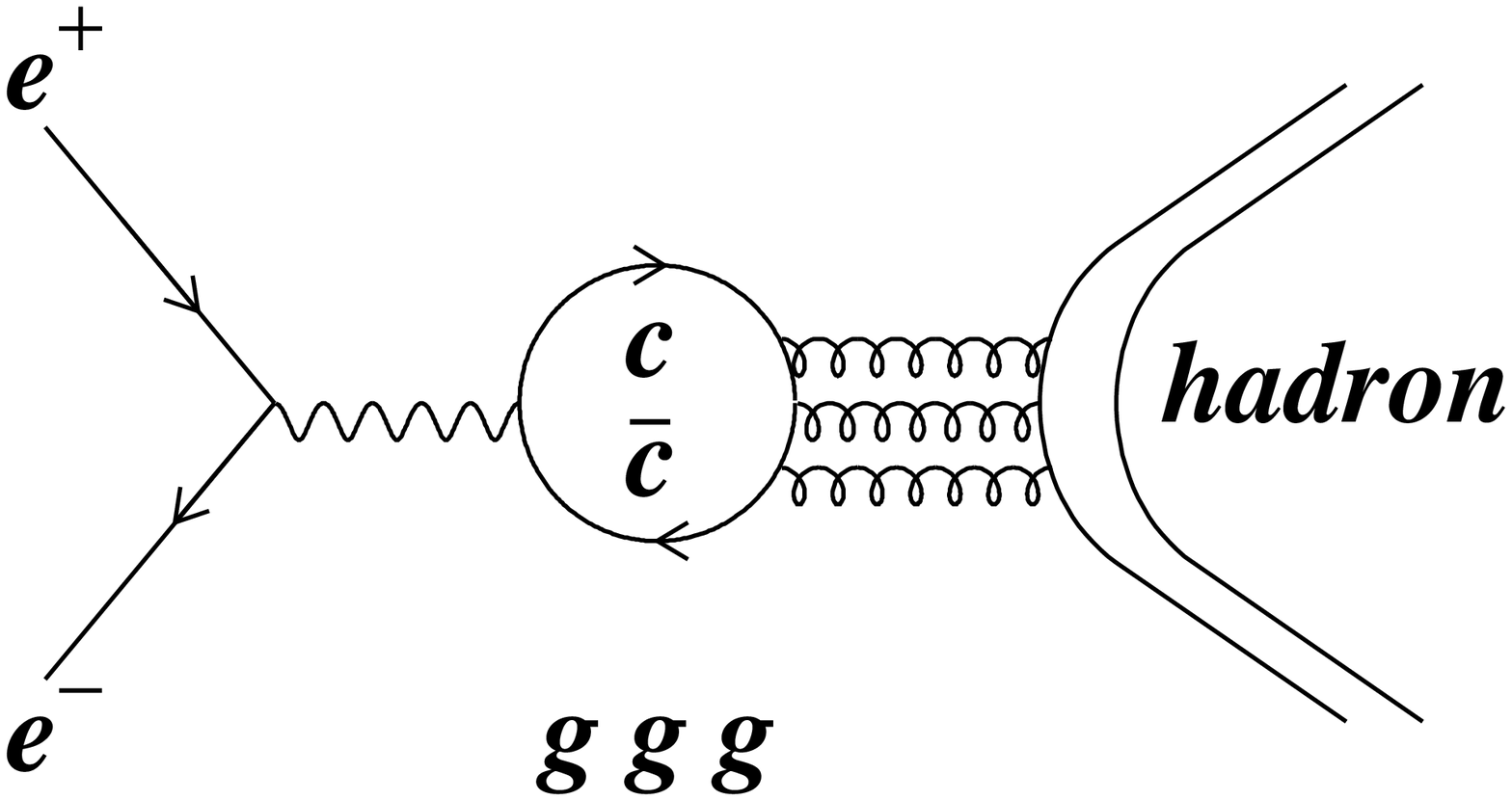}
\qquad
\includegraphics[width=3.cm,height=2cm]{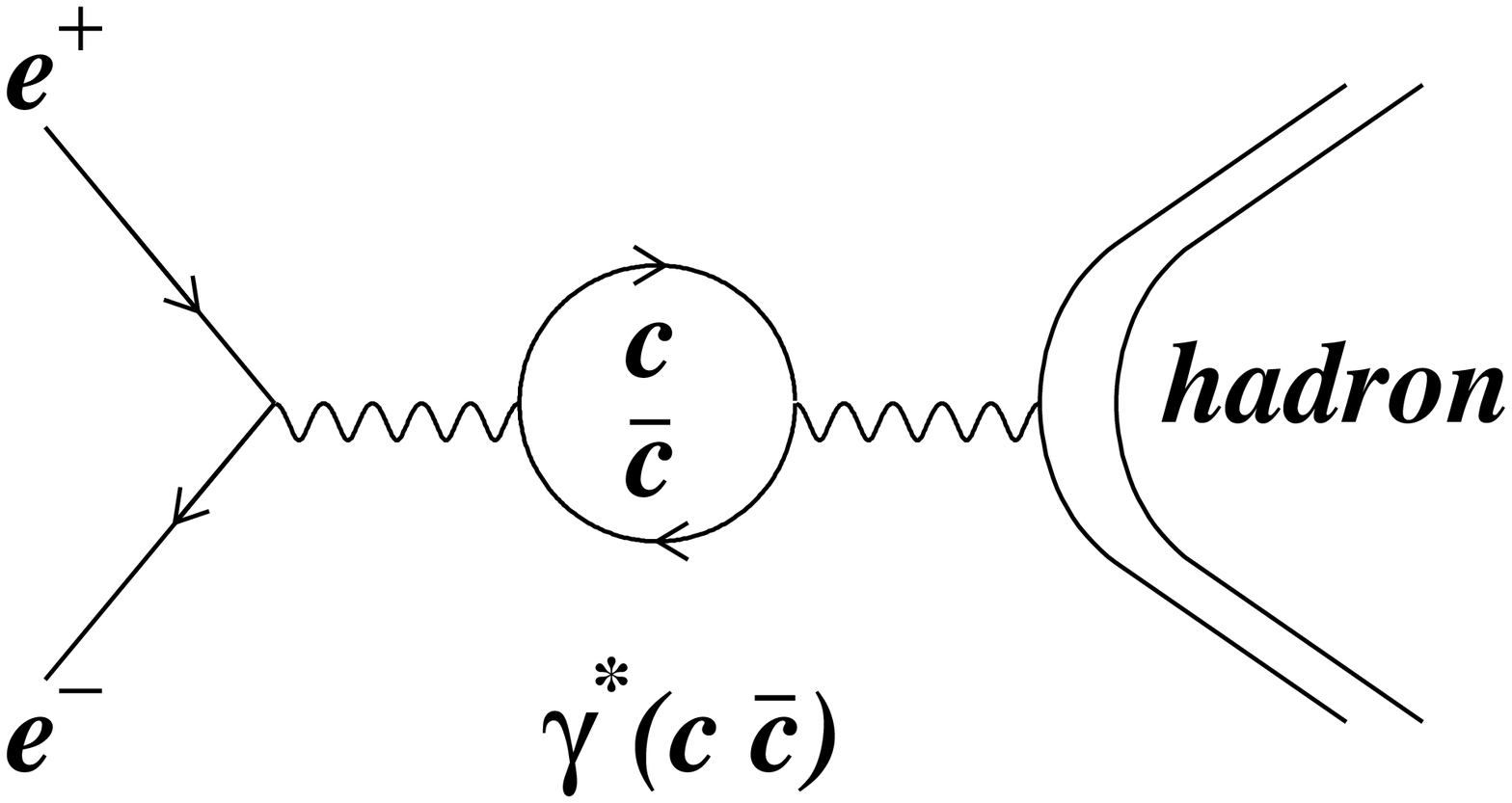}
\qquad
\includegraphics[width=3.cm,height=2cm]{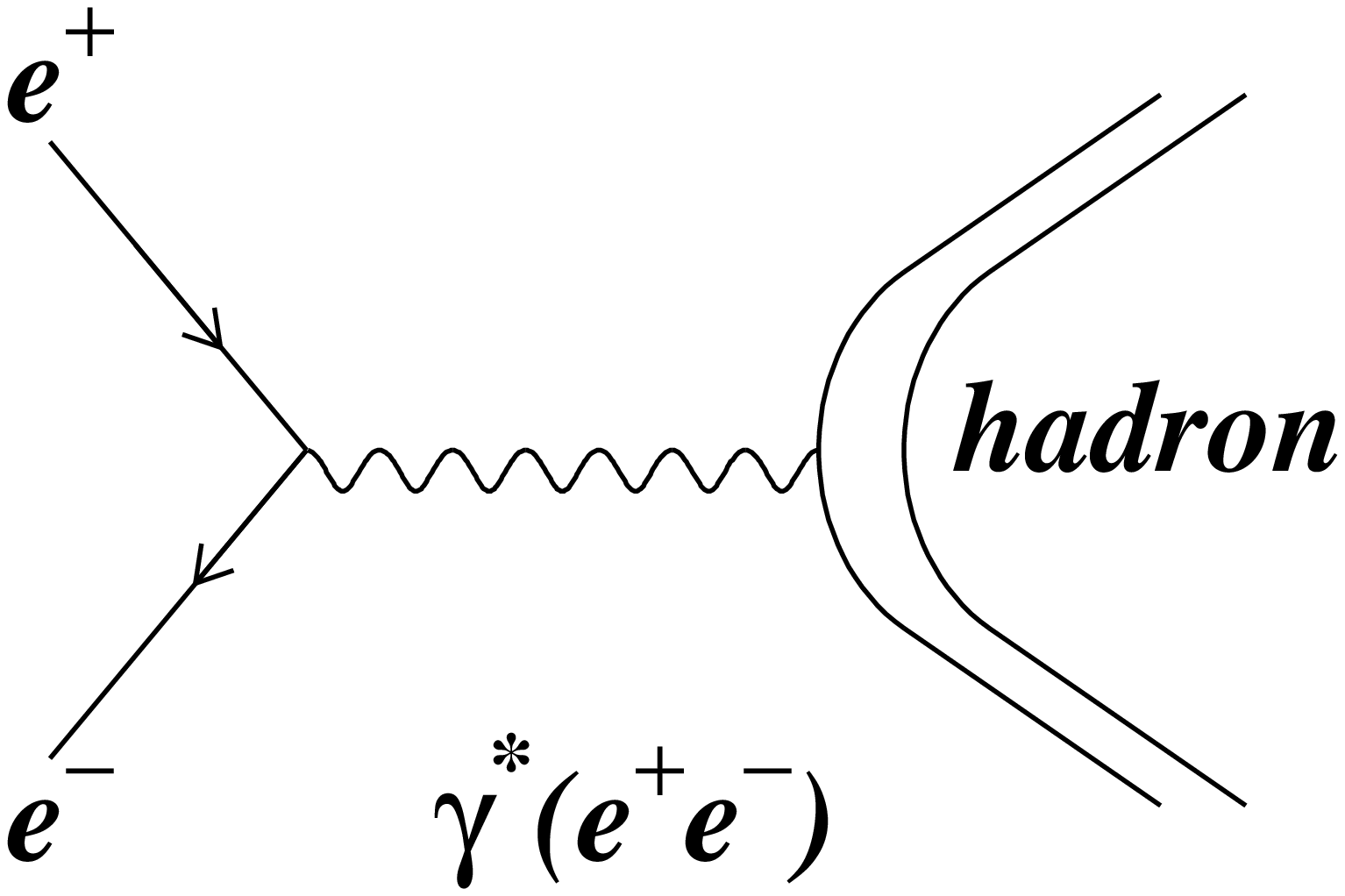}
\end{center}
\caption{The Feynman diagrams of $e^+e^-\rightarrow light\, \,
         hadrons$ at charmonium resonance. From left to right are of
         three-gluon annihilation, of one-photon annihilation and of
         one-photon continuum.}\label{fig:threefymn} 
\end{figure}

We know that $J/\psi$ and $\psi(3686)$ (shortened as $\psi^\prime$)
decay into light hadrons through three-gluon and one-photon
annihilation of which the amplitudes are denoted by $a_{3g}$ and
$a_{\gamma}$ respectively. This is also true for $\psi(3770)$
(shortened as $\psi^{\prime\prime}$) in its OZI suppressed decay into
light hadrons. In general, for the resonance ${\cal R}$ (${\cal
R}=J/\psi$, $\psi^\prime$ or $\psi^{\prime\prime}$), the cross-section
at the Born order is expressed as
\begin{equation} 
\sigma_{B}(s)= \frac{4\pi s\alpha^2}{3}|a_{3g}+a_{\gamma}|^2~,
\label{eq:born} 
\end{equation} 
where $\sqrt{s}$ is the C.M. energy,
$\alpha$ is the fine structure constant. If the $J/\psi$,
$\psi^\prime$ or $\psi^{\prime\prime}$ is produced in $e^+e^-$
collision, the process
\begin{equation}
e^+e^- \rightarrow \gamma^* \rightarrow hadrons
\end{equation}
could produce the same final hadronic states as charmonium decays
do~\cite{rudaz}. We denote its amplitude by $a_c$, then the
cross-section becomes
\begin{equation} 
\sigma^{\prime}_{B}(s) = \frac{4\pi s \alpha^2}{3}|a_{3g}+a_{\gamma}+a_c|^2~.
\label{eq:bornp} 
\end{equation} 
So what truly contribute to the experimentally measured cross-section
are three classes of Feynman diagrams, \ie the three-gluon decays,
the one-photon decays, and the one-photon continuum process, as
illustrated in \Figure~\ref{fig:threefymn}. To analyze the
experimental results, we must take into account three amplitudes and
two relative phases.

For an exclusive mode, $a_c$ can be expressed by
\begin{equation}
a_c(s) = \frac{{\cal F}(s)}{s} e^{i \phi^{\prime}}~,
\label{eq:agee}
\end{equation}
where $\phi^\prime$ is the phase relative to $a_{3g}$; ${\cal
F}(s)$ depends on the individual mode, and for simplicity, the
phase space factor is incorporated into $|{\cal F}(s)|^2$. The
one-photon annihilation amplitude can be written as
\begin{equation}
a_{\gamma}(s)=\frac{3\Gamma_{ee}{\cal F}(s)/(\alpha \sqrt{s})}
{s-m_{\cal R}^2+i m_{\cal R} \Gamma_t}e^{i\phi}~, \label{eq:agcc}
\end{equation}
where $m_{\cal R}$ and $\Gamma_t$ are the mass and the total width
of ${\cal R}$, $\Gamma_{ee}$ is the partial width to $e^+e^-$,
$\phi$ is the phase relative to $a_{3g}$. The strong decay
amplitude $a_{3g}$ is defined by ${\cal C}\equiv
|a_{3g}/a_{\gamma}|$, which is the relative strength to
$a_{\gamma}$, so
\begin{equation}
a_{3g}(s) = {\cal C} \cdot \frac{3\Gamma_{ee}{\cal F}(s)/(\alpha
\sqrt{s})} {s-m_{\cal R}^2+i m_{\cal R} \Gamma_t}~. \label{eq:aggg}
\end{equation}
For resonances, ${\cal C}$ can be taken as a constant.

In principle, $a_{3g}$, $a_{\gamma}$ and $a_c$ depend on individual
exclusive mode both in absolute values and in relative strength. In
this note, for illustrative purpose, following assumptions are used
for an exclusive hadronic mode: ${\cal F}(s)$ is replaced by
$\sqrt{R(s)}$, where $R(s)$ is the ratio of the inclusive hadronic
cross-section to the $\mu^+\mu^-$ cross-section measured at nearby
energy \cite{besR}; in \Eq~(\ref{eq:aggg}),
\begin{equation} 
{\cal C}= \sqrt{\frac{B({\cal R}\rightarrow
ggg\rightarrow hadrons)} {B({\cal R}\rightarrow \gamma^*
\rightarrow hadrons) }}~. 
\label{eq:cratio} 
\end{equation} 
Here $B({\cal R}\rightarrow \gamma^* \rightarrow hadrons)
=B_{\mu^+\mu^-}R(s)$, where $B_{\mu^+\mu^-}$ is the $\mu^+\mu^-$
branching ratio; while $B({\cal R}\rightarrow ggg \rightarrow
hadrons)$ is calculated as following: we first estimate the branching
ratio of $B({\cal R}\rightarrow \gamma gg)+B({\cal R}\rightarrow ggg)$
by subtracting the lepton pairs, $\gamma^* \rightarrow hadrons$, and
the modes with charmonium production from the total branching ratio
(100\%). Then using pQCD result~\cite{guali} $B({\cal R}\rightarrow
\gamma gg)/B({\cal R}\rightarrow ggg)\approx 6 \%$ we obtain $B({\cal
R}\rightarrow ggg \rightarrow hadrons)$. \Table~\ref{tab:estimation}
lists all the estimations used as inputs in the calculations, where
$\sigma^{\cal R}_B$ is the total resonance cross-section of Born order
at $s=m_{\cal R}^2$ obtained from 
\begin{equation} 
\sigma_0^{\cal
R}(s)=\frac{12\pi\Gamma_{ee}\Gamma_t} {(s-m_{\cal R}^2)^2+m_{\cal
R}^2\Gamma_t^2}~. 
\label{eq:sgmzeror} 
\end{equation}

\begin{table} 
\caption{Estimated amplitudes at $J/\psi$, $\psi^\prime$ and
         $\psi^{\prime\prime}$ peaks}\label{tab:estimation} 
\begin{center}
\begin{tabular}{r||ccc} 
\hline \hline $\sqrt{s}$~~~~~~~~
& $m_{J/\psi}$   &   $m_{\psi^\prime}$ & $m_{\psi^{\prime\prime}}$
\\ \hline \hline
$| a_{3g}(m^2_{\cal R}) |^2 \propto $& 70\% $\sigma^{J/\psi}_B$  &
19\% $\sigma^{\psi^\prime}_B$
      & $\sim$ 1\% $\sigma^{\psi^{\prime\prime}}_B$  \\
$| a_{\gamma}(m^2_{\cal R}) |^2 \propto $& 13\% $\sigma^{J/\psi}_B$  &
1.6\% $\sigma^{\psi^\prime}_B$
& $2.5\times 10^{-5} \sigma^{\psi^{\prime\prime}}_B$   \\
$| a_c(m^2_{\cal R}) |^2 \propto $& 20~nb & 14~nb & 14~nb  \\
\hline \hline 
\end{tabular} 
\end{center}
\end{table}

The cross-section by $e^+e^-$ collision incorporating radiative
correction on the Born order is expressed by~\cite{radycz}
\begin{equation}
\sigma_{r.c.} (s)=\int \limits_{0}^{x_m} dx F(x,s)
\frac{\sigma_{0}(s(1-x))}{|1-\Pi (s(1-x))|^2}~, \label{eq:radsec}
\end{equation}
where $\sigma_{0}$ is $\sigma_{B}$ or $\sigma^{\prime}_{B}$ by
\Eq~(\ref{eq:born}) or (\ref{eq:bornp}), $F(x,s)$ has been calculated in
Ref.~\cite{radycz} and $\Pi (s)$ is the vacuum polarization
factor~\cite{vacuumycz}; the upper limit of the integration
$x_m=1-s_m/s$ where $\sqrt{s_m}$ is the experimentally required
minimum invariant mass of the final state $f$ after losing energy
to multi-photon emission. In this note, we assume that
$\sqrt{s_m}$ equals to $90\%$ of the resonance mass, \ie
$x_m=0.2$.

For narrow resonances like $J/\psi$ and $\psi^\prime$, one should
consider the energy spread function of $e^+e^-$ colliders:
\begin{equation}
G(\sqrt{s},\sqrt{s'})=\frac{1}{\sqrt{2 \pi} \Delta}
          e^{ -\frac{(\sqrt{s}-\sqrt{s'})^2}{2 {\Delta}^2} },
\label{eq:spread}
\end{equation}
where $\Delta$ describes the C.M. energy spread of the accelerator,
$\sqrt{s}$ and $\sqrt{s'}$ are the nominal and actual C. M. energy
respectively. Then the experimentally measured cross-section
\begin{equation}
\sigma_{exp} (s)=\int \limits_{0}^{\infty}
   \sigma_{r.c.} (s') G(\sqrt{s},\sqrt{s'}) d\sqrt{s'}~.
\label{eq:expsec}
\end{equation}

The radiative correction reduces the maximum cross-sections of
$J/\psi$, $\psi^\prime$ and $\psi^{\prime\prime}$ by $52\%$,
$49\%$ and $29\%$ respectively. The energy spread further reduces
the cross-sections of $J/\psi$ and $\psi^\prime$ by an order of
magnitude. The radiative correction and energy spread also shift
the maximum height of the resonance peak to above the resonance
mass. Take $\psi^\prime$ as an example, from \Eq~(\ref{eq:sgmzeror}),
$\sigma_B^{\psi^\prime}=7887$~nb at $\psi^\prime$ mass; substitute
$\sigma_0(s)$ in \Eq~(\ref{eq:radsec}) by $\sigma_0^{\cal R}(s)$ in
\Eq~(\ref{eq:sgmzeror}), $\sigma_{r.c.}$ reaches the maximum of
$4046$~nb at $\sqrt{s}=m_{\psi^\prime}+9$~keV; with the energy
spread $\Delta=1.3$~MeV at BES/BEPC, combining
\Eqs~(\ref{eq:sgmzeror}$-$\ref{eq:expsec}), $\sigma_{exp}$ reaches the
maximum of $640$~nb at $\sqrt{s}=m_{\psi^\prime}+0.14$~MeV.
Similarly, at $J/\psi$, with BES/BEPC energy spread
$\Delta=1.0$~MeV, the maximum of $\sigma_{exp}$ is $2988$~nb. At
CESRc~\cite{cleocycz}, the maximum of $\sigma_{exp}$ at $J/\psi$
is $1270$~nb ($\Delta$= 2.5~MeV), and at $\psi^\prime$, it is
$250$~nb ($\Delta$= 3.6~MeV). In this note, we calculate
$\sigma_{exp}$ at the energies which yield the maximum inclusive
hadronic cross-sections.

To measure an exclusive mode in $e^+e^-$ experiment, the
contribution of the continuum part should be subtracted from the
experimentally measured $\sigma^{\prime}_{exp}$ to get the
physical quantity $\sigma_{exp}$, where $\sigma_{exp}$ and
$\sigma^{\prime}_{exp}$ indicate the experimental cross-sections
calculated from \Eqs(\ref{eq:radsec}$-$\ref{eq:expsec}) with the
substitution of $\sigma_{B}$ and $\sigma^{\prime}_{B}$ from
\Eqs~(\ref{eq:born}) and (\ref{eq:bornp}) respectively for $\sigma_{0}$
in \Eq~(\ref{eq:radsec}). Up to now, most of the measurements did not
include this contribution and $\sigma^{\prime}_{exp}=\sigma_{exp}$
is assumed at least at $J/\psi$ and $\psi^\prime$. As a
consequence, the theoretical analyses are based on pure
contribution from the resonance; on the other hand, the
experiments actually measure quantities with the contribution of
the continuum amplitude included.

\begin{figure}[hbt]
\begin{center}
\includegraphics[width=5.5cm,height=5cm]{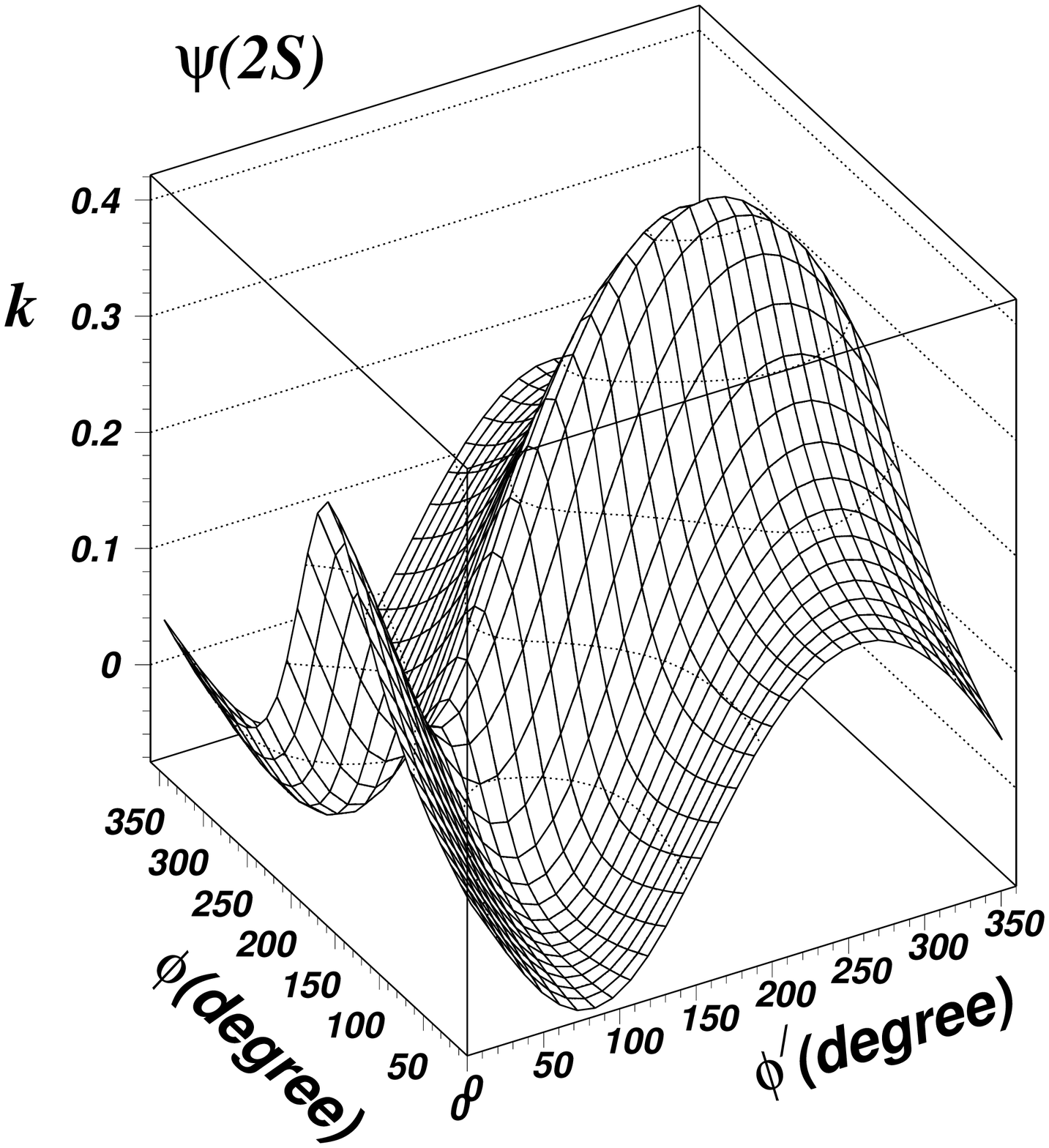}
\qquad
\includegraphics[width=5.5cm,height=5cm]{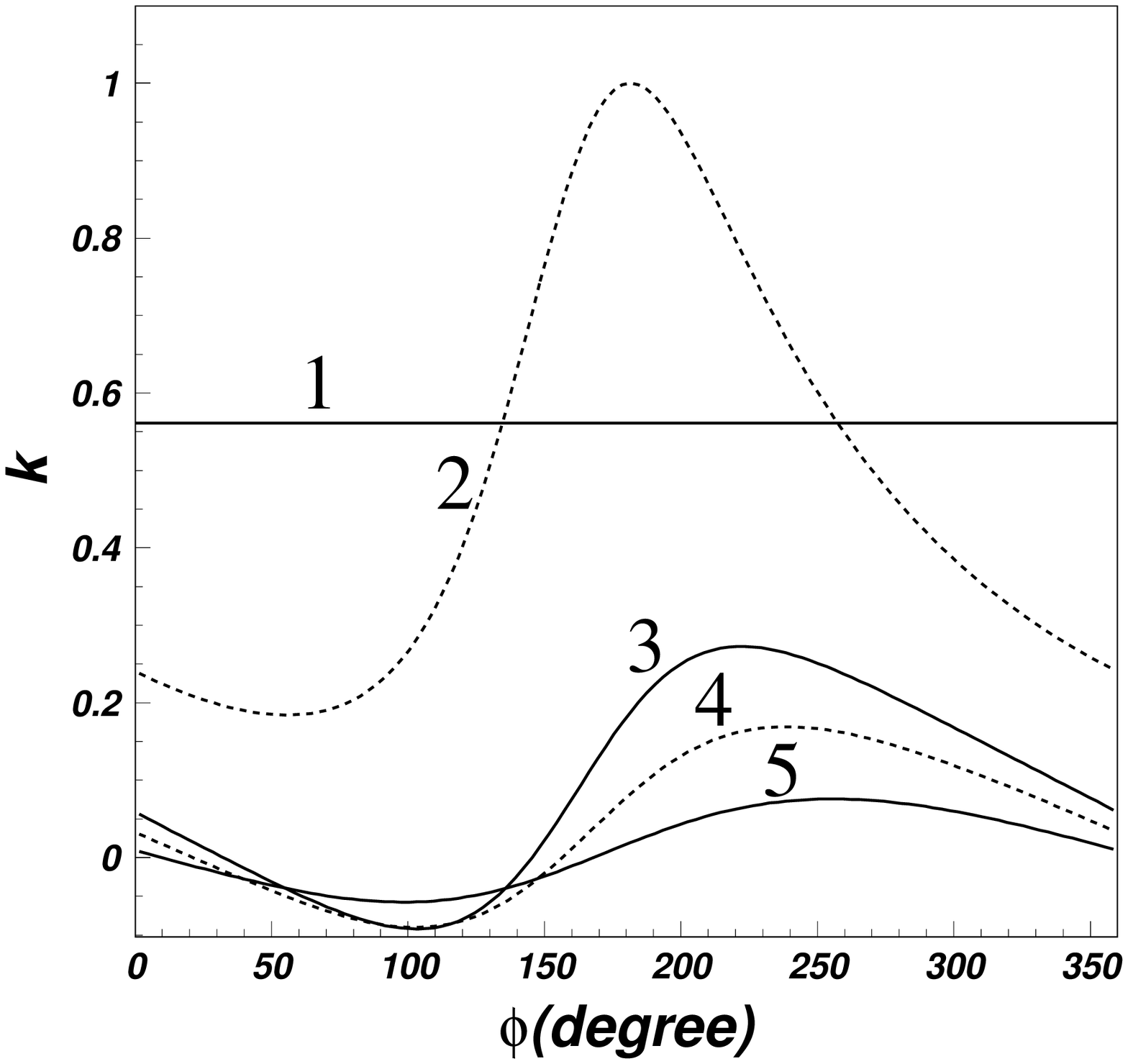}
\end{center}
\caption{Left: $k$ as a function of $\phi$ and $\phi^{\prime}$ for
         $\psi^\prime$, with input from
         \Table~\ref{tab:estimation}. Right: $k$ as a function of
         $\phi$ ($\phi = \phi^{\prime}$) for different ratios of
         $|a_{3g}|$ to $|a_{\gamma}|$: line 1 to 5 for $a_{3g}=0$,
         $|a_{3g}|=|a_{\gamma}|$, $|a_{3g}|=3.4|a_{\gamma}|$,
         $|a_{3g}|=5|a_{\gamma}|$ and $|a_{3g}|=10 |a_{\gamma}|$,
         respectively.}\label{fig:kratio}
\end{figure}

We display the effect from the continuum amplitude and
corresponding phase for $J/\psi$, $\psi^\prime$ and
$\psi^{\prime\prime}$ respectively. To do this, we calculate the
ratio 
\begin{equation} 
k(s) \equiv \frac{ \sigma^{\prime}_{exp}(s) - \sigma_{exp}(s) }
              { \sigma^{\prime}_{exp}(s) }
\label{eq:ratiok} 
\end{equation}
as a function of $\phi$ and $\phi^{\prime}$, as shown in
\Figure~\ref{fig:kratio}a for $\psi^\prime$ at
$\sqrt{s}=m_{\psi^\prime}+0.14$~MeV for $\Delta=1.3$~MeV. It can be
seen that for certain values of the two phases, $k$ deviates from 0,
or equivalently the ratio $\sigma^{\prime}_{exp}/ \sigma_{exp}$
deviates from 1, which demonstrates that the continuum amplitude is
non-negligible. By assuming there is no extra phase between
$a_{\gamma}$ and $a_c$ (\ie set $\phi=\phi^{\prime}$), we also work
out the $k$ values for different ratios of $|a_{3g}|$ to
$|a_{\gamma}|$, as shown in \Figure~\ref{fig:kratio}b: line 3 corresponds to
the numbers listed in \Table~\ref{tab:estimation}, line 1 is for pure
electromagnetic decay channels, and others are chosen to cover the
other possibilities of the ratio $|a_{3g}|$ to $|a_{\gamma}|$.

\subsubsection[Dependence on experimental conditions]
              {Dependence on experimental conditions}
\label{sec:dexp}

Here we emphasize the dependence of the observed cross-section in
$e^+e^-$ collision on the experimental conditions. The most
crucial ones are the accelerator energy spread and the beam energy
setting for the narrow resonances like $J/\psi$ and $\psi^\prime$.

\begin{figure}[hbt]
\begin{center}
\includegraphics[width=7.5cm,height=7cm]{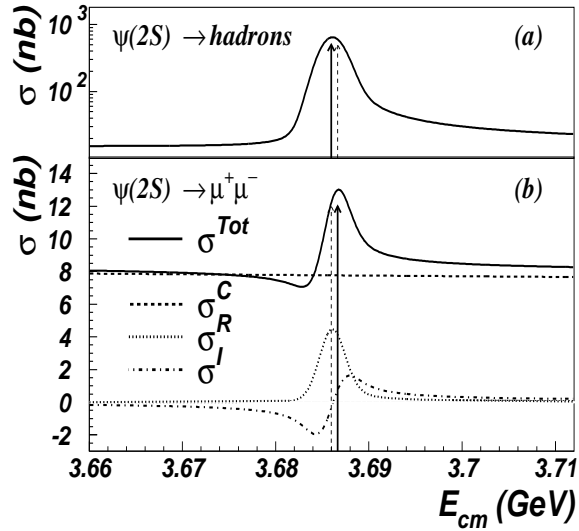}
\end{center}
\caption[Cross-sections in the vicinity of $\psi^\prime$ for inclusive
         hadrons and $\mu^+\mu^-$ final states] 
        {Cross-sections in the
         vicinity of $\psi^\prime$ for inclusive hadrons (a) and
         $\mu^+\mu^-$ (b) final states. The solid line with arrow
         indicates the peak position and the dashed line with arrow
         the position of the other peak. In (b), dashed line for QED
         continuum ($\sigma^C$), dotted line for resonance
         ($\sigma^R$), dash dotted line for interference($\sigma^I$),
         and solid line for total cross-section($\sigma^{Tot}$).}
\label{fig:cmphup}
\end{figure}

\Figure[b]~\ref{fig:cmphup} depicts the expected cross-sections of
inclusive hadrons and $\mu^+\mu^-$ pairs at $\psi^\prime$ in an
experimental setting under BEPC/BES condition. Two arrows in the
figure denote the different positions of the maximum heights of the
cross-sections. The height is reduced and the position of the peak is
shifted due to the radiative correction and the energy spread of the
collider. However, the energy smear hardly affects the continuum part
of the cross-section. The $\mu^+\mu^-$ channel is further affected by
the interference between resonance and continuum amplitude. As a
consequence, the relative contribution of the resonance and the
continuum varies as the energy changes.  In actual experiments, data
are naturally taken at the energy which yields the maximum inclusive
hadronic cross-section. This energy does not coincide with the maximum
cross-section of each exclusive mode. So it is important to know the
beam spread and beam energy precisely, which are needed in the
delicate task to subtract the contribution from $a_c$.

It is worth noting that in principle if $a_c$ is not considered
correctly, different experiments will give different results for
the same quantity, like the exclusive branching ratio of the
resonance, due to the dependence on beam energy spread and beam
energy setting.  The results will also be different for different
kinds of experiments, such as production of $J/\psi$ and
$\psi^\prime$ in $p \bar{p}$ annihilation, or in $B$ meson decays.
This is especially important since the beam spreads of different
accelerators are much different~\cite{phyreport} and charmonium
results are expected from $B$-factories.

\subsubsection[Implications to charmonium physics]
              {Implications to charmonium physics}
\label{sec:cccd}

With the non-resonance virtual photon amplitude taking into
account in the analysis of the data from $e^+e^-$ experiments,
some important conclusions in the charmonium physics could be
changed. In this section, we discuss the $\psi^\prime$,
$\psi^{\prime\prime}$ and $J/\psi$ decays.

In the pure electromagnetic decays of $\psi^\prime$, like
$\pi^+\pi^-$ or $\omega\pi^0$, depending on the energy resolution
of the $e^+e^-$ collider, a large fraction (\eg about 60\% for
$\Delta=1.3$~MeV) of the observed cross-section is due to
non-resonance continuum contribution. With the subtraction of this
contribution, the electromagnetic form factors (\eg $\pi^+\pi^-$
and $\omega\pi^0$) are changed substantially~\cite{formfactor}.

It has been known from experimental data that in two-body $J/\psi$
decays, the relative phase between the strong amplitude $a_{3g}$ and
electromagnetic (EM) amplitude $a_\gamma$ is orthogonal for the decay
modes $1^+0^-$ ($90^\circ$)~\cite{suzuki}, $1^-0^-$ ($(106 \pm
10)^\circ$)~\cite{phyreport,jousset,castro}, $0^-0^-$ ($(89.6 \pm
9.9)^\circ$)~\cite{phyreport,castro,suzuki2}, $1^-1^-$ ($(138 \pm
37)^\circ$)~\cite{phyreport} and $N\overline{N}$ ($(89 \pm
15)^\circ$)~\cite{castro,baldini}. It was argued that this large phase
follows from the orthogonality of three-gluon and one-photon virtual
processes~\cite{gerard}. But at first glance, the $\psi^\prime
\rightarrow 1^-0^-$ data does not seem to support the extension of
such orthogonality to $\psi^\prime$ decays. Here very small branching
fractions are reported for $\rho\pi$ and $K^{*+}K^-$ modes (at ${\cal
O} (10^{-5})$ ) while much larger branching fraction for
$K^{*0}\overline{K^0}$ mode (at ${\cal
O}(10^{-4})$)~\cite{besrop,cleocvp}. Since the amplitudes of these
three decay modes are expressed as~\cite{haber} 
\begin{equation}
\begin{array}{ccl}
A_{\rho\pi} &=& a_{3g}+a_{\gamma}, \\
A_{K^{*+}K^-} &=& a_{3g}+\epsilon+a_{\gamma}, \\
A_{K^{*0}\overline{K^0}} &=& a_{3g}+\epsilon-2a_{\gamma},
\end{array}\label{eq:arp} 
\end{equation} 
with $\epsilon$ a SU(3) breaking parameter, it suggests cancellation
between $a_{3g}$ and $a_{\gamma}$ in $A_{\rho\pi}$ and
$A_{K^{*+}K^-}$. This means the phase between $a_{3g}$ and
$a_{\gamma}$ is around $180^\circ$. But since the available data are
from $e^+e^-$ experiments, the amplitude $a_c$ must be included. To
explain the data, \Eq~(\ref{eq:arp}) should be replaced by:
\begin{equation}
\begin{array}{ccl}
A_{\rho\pi} &=& a_{3g}+a_{\gamma}+a_c , \\
A_{K^{*+}K^-} &=& a_{3g}+\epsilon+a_{\gamma}+a_c , \\
A_{K^{*0}\overline{K^0}} &=& a_{3g}+\epsilon-2(a_{\gamma}+a_c) ,
\end{array}\label{eq:arp1} 
\end{equation}
Instead of cancellation between $a_{3g}$ and $a_{\gamma}$ in
$A_{\rho\pi}$ and $A_{K^{*+}K^-}$, the observed cross-sections could
be due to the destructive interference between $a_{3g}$ and $a_c$ for
these two modes. On the other hand, the interference between these two
amplitudes is constructive for $K^{*0}\overline{K^0}$. Such
interference pattern happens if the phase between $a_{3g}$ and
$a_{\gamma}$ is $-90^\circ$, because on top of the resonance, the
phase between $a_{\gamma}$ and $a_c$ is $-90^\circ$. This means that
the orthogonality between $a_{3g}$ and $a_{\gamma}$ observed in
$J/\psi$ decays holds true in $\psi^\prime \rightarrow 1^-0^-$ decays,
and it has a negative sign~\cite{possiblephase}.  Similarly, with the
amplitude $a_c$ included, from the measured $\psi^\prime \rightarrow
\pi^+\pi^-$, $K^+K^-$ and $K^0_SK^0_L$~\cite{beskskl}, we know that in
$\psi^\prime \rightarrow 0^-0^-$ decays, the phase between $a_{3g}$
and $a_{\gamma}$ is either $(-82 \pm 29)^\circ$ or $(121 \pm
27)^\circ$~\cite{pspp}.

In the OZI suppressed $\psi^{\prime\prime}$ decays, MARK~III set an
upper limit of $\rho\pi$ production cross-section by $e^+e^-$
collision at this resonance to be less than 6.3~pb~\cite{zhuyn}.  On
the other hand, CLEO measured $e^+e^- \rightarrow \rho\pi$
cross-section at 3.67~GeV to be $(8.3^{+1.7}_{-1.4}\pm1.2)$~pb.
Scaled down to 3.770~GeV according to $1/s^2$, we expect the
non-resonance cross-section of $e^+e^- \rightarrow \gamma^*
\rightarrow \rho\pi$ to be $(7.5 \pm 1.8)$~pb, which is already
greater than the upper limit at the $\psi^{\prime\prime}$ peak. We
reach the conclusions~\cite{sipp2ropi}: (i) there must be destructive
interference between the $\psi^{\prime\prime}$ resonance and the
non-resonance virtual photon amplitudes, \ie the phase between the
strong and EM amplitude is around $-90^\circ$; (ii) the ${\cal
B}(\psi^{\prime\prime} \rightarrow \rho\pi)$ is roughly at $(6 \sim 7)
\times 10^{-4}$. This branching fraction coincides with the prediction
by $2S-1D$ mixing scenario which was proposed by Rosner to explain the
small $\rho \pi$ branching fraction in $\psi^\prime$
decays~\cite{rosner}. (In the original work of Ref.~\cite{rosner},
this branching fraction is $4.1 \times 10^{-4}$. But with the new
measurement of $J/\psi \rightarrow \rho\pi$ by BES~\cite{besjpsi}, it
becomes larger.) So with the amplitude $a_c$ being taken into account,
we find that this scenario is supported by experimental data. One
important prediction of this scenario is that the
$\psi^{\prime\prime}$ could have a large charmless decay branching
fraction (more than 10\%)~\cite{largebr}. In the search of the
exclusive charmless decays, the interference effect is important,
although there are some modes which do not couple with virtual photon,
like $K^0_SK^0_L$ which is purely from $\psi^{\prime\prime}$ decays
and is clean in such search~\cite{pspp2kskl}.

In this way, the correct treatment of the amplitude $a_c$ enables us
to reach two important conclusions in charmonium physics: (i) the
orthogonality between $a_{3g}$ and $a_{\gamma}$ can be extended from
$J/\psi$ decays to $\psi^\prime$ and OZI suppressed
$\psi^{\prime\prime}$ decays and the sign of the phase must be
negative; (ii) ${\cal B}(\psi^{\prime\prime} \rightarrow \rho \pi)$ is
consistent with the $2S-1D$ mixing scenario which is proposed to solve
the $\rho\pi$ puzzle in $J/\psi$ and $\psi^\prime$ decays.

As for $J/\psi$, the interference between the amplitude $a_c$ and the
resonance is at the order of a few percent at most. It is smaller than
the statistical and systematic uncertainties of current
measurements. Nevertheless, for future high precision experiments such
as CLEO-c~\cite{cleocycz} and BES~III~\cite{bes3}, when the accuracy
reaches a few per mille or even smaller level, it should be taken into
account.

\subsubsection[Summary and perspective]{Summary and perspective}
\label{sec:sum}

In summary, the continuum amplitude $a_c$, by itself or through
interference with the resonance, could contribute significantly to the
observed cross-sections in $e^+e^-$ experiments on charmonium
physics. Its treatment depends sensitively on the experimental
details, which has not been fully addressed in both $e^+e^-$
experiments and theoretical analyses. In principle, any experimental
measurement should subtract the contribution of the continuum
amplitude to get the physical quantity related to the
resonance. Unfortunately, up to now, most of the experiments just
neglect this contribution and the measured quantities are assumed to
be purely from resonance decays for almost all the channels studied,
or just subtract the continuum contribution incoherently without
considering the interference effect, at least at $J/\psi$ and
$\psi^\prime$. This potentially leaves a huge gap between theory and
experiments: the quantities which the experiments provide are not
exactly what the theory wants to understand.

The effect of the continuum amplitude in the physics analyses are
extensively examined in a series of papers published
recently~\cite{formfactor,possiblephase,pspp,sipp2ropi,wym,interf}: it
modifies the measurements of the $\pi^+\pi^-$ and $\omega \pi^0$ form
factors at $\psi^\prime$ significantly; it changes the fitting of the
relative phase between the strong and electromagnetic decay amplitudes
of $\psi^\prime$, it sheds light on the understanding of the
``$\rho\pi$ puzzle", and it decreases the observed $\rho\pi$
cross-section near the $\psi^{\prime\prime}$ resonance peak to a much
smaller level than the expectation from either pure continuum
contribution or estimation of the $\psi^{\prime\prime}$
non-$D\overline{D}$ decays. The recent large $J/\psi$ and
$\psi^\prime$ samples~\cite{bes2data} make these studies important due
to the improved statistical precision.

The effect of this continuum amplitude will become more significant in
the coming high luminosity experiments, such as CLEOc~\cite{cleocycz}
and BES~III~\cite{bes3}, in this energy region. To achieve high
precision to match the high statistics, the cross-section of each mode
in the vicinity of the resonance should be measured. This implies an
energy scan near the resonance peak at a few energy points with
considerably large statistics to allow a reasonable subtraction of the
continuum contribution via a fit to the line shape of the resonance.

The above argument also applies to the bottomonium states in the study
of their exclusive hadronic decays, where the maximum cross-sections
of the resonances are even smaller than those of the charmonium
states.

%\end{document}

\BLKP
%10.12.04

%\documentclass[11pt,twoside,amsmath,amssymb]{cernrep}
%\usepackage{graphicx,epsfig}
%\usepackage{here,cite,bm,amsmath,amssymb,amsfonts}
%\input{newcommand.tex}

%\begin{document}
\chapter{SPECTROSCOPY}
\label{chapter:spectroscopy}

%\centerline{\bf SPECTROSCOPY}

{\it Conveners:} G.~Bali, N.~Brambilla, R.~Mussa, J.~Soto \par\noindent
{{\it Authors:} G.~Bali, D.~Besson, A.~B\"ohrer, N.~Brambilla, P.~Cooper, 
 C.~Davies, E.~Eichten, S.~Eidelman, R.~Faustov,  T.~Ferguson,  R.~Galik, 
 S.~Godfrey, A.~Kronfeld, P.~Mackenzie,  C.~Morningstar, R.~Mussa, 
 V.~Papadimitriou, A.~Pineda, S.~Ricciardi, J.~-M.~Richard, E.~Robutti,  
 J.~Simone, T.~Skwarnicki, J.~Soto, G.~Stancari, Yu.~Sumino, 
 J.~Tseng, B.~Yabsley, Z.~Zhao}\par\noindent

\section[Theory introduction]
        {Theory introduction
         $\!$\footnote{Author: J.~Soto}} 
\label{sec:spthintrp}

Most theorists agree that QCD alone should describe the spectroscopy 
of heavy quarkonium. Nevertheless, there are important difficulties 
to do so in practise. One can roughly distinguish between two
approaches:
the phenomenological and the theoretical one.

The phenomenological approach attempts to model what are believed to be the
features of QCD relevant to heavy quarkonium with the aim to produce concrete
results which can be directly confirmed or falsified by experiment and may
guide experimental searches.
The theoretical approach tries to describe heavy quarkonium with 
QCD based calculations and/or approximations.

The basic tools of the phenomenological approach are potential models,
both non-relativistic and relativistic. The use of non-relativistic
potential models is justified by the fact that the bottom and, to a
lesser extent, the charm masses are large in comparison to $\lQ$, the
typical hadronic scale. Hence a quantum mechanical description of the
system based on two heavy quarks interacting through a suitable
potential appears reasonable. The potential is usually chosen in a way
that at short distances coincides with the weak coupling QCD one-gluon
exchange Coulomb potential and in the long range it incorporates
confinement, for instance, by including a linearly rising
potential. Since relativistic effects appear to be sizable for some
states, mostly in charmonium, models incorporating some relativistic
kinematics are also being used. Different models of quark confinement
may result in different classes of relativistic corrections.  For
states close to and beyond the two heavy-light meson threshold, the
potential models have to be complemented with these extra degrees of
freedom in order to account for possible mixing effects. Hybrid states
which are expected from QCD should also be incorporated by hand. The
phenomenological approaches will be described in \Section~\ref{sec:spphen}.
\shortpage

The theoretical approach aims at obtaining the spectrum of heavy
quarkonium from QCD.  This is in principle more complicated than
obtaining masses of light mesonic states since an additional large
scale $m$, the mass of the heavy quark, enters the calculation.  If we
assume that $m$ is much larger than any other scale in the system, in
particular $\lQ$, the heavy quark and antiquark are expected to move
slowly about each other at a relative velocity $v \ll 1$.  The system
becomes non-relativistic and hence splittings between states with the
same quantum numbers are expected to be of size $\sim mv^2$ whereas
hyperfine splittings are of order $\sim mv^4$, if one proceeds by
analogy to QED bound states (where $v\sim \alpha$).  If $v^2 \sim
0.1$, as expected in ground state bottomonium, a direct (lattice) QCD
calculation requires a precision significantly better than 10~\% to
detect spin-averaged masses and of more than $1~\%$ to resolve fine
structure splittings. Moreover, all these scales have to be resolved
on one and the same lattice, necessitating many lattice points.  This
is to be compared with light quarkonium where the splittings are a
leading order effect. Consequently, calculating the heavy quarkonium
spectrum from lattice QCD requires a tremendous computational effort,
which in some cases can be somewhat ameliorated with the introduction
of anisotropic lattices, as discussed in \Section~\ref{sec:spdqlc}.

Alternatively, it may be advisable to exploit the fact that $m$ is
large and $v$ small before attempting the computation.  This is most
efficiently done using non relativistic effective field theories. The
effective theory which takes into account that $m$ is much larger than
the remaining scales in the system is NRQCD
\cite{Caswell:1985ui,Thacker:1990bm,Bodwin:1994jh}.  Since $m \gg
\Lambda_{QCD}$, NRQCD can be made equivalent to QCD at any desired
order in $1/m$ and $\als (m) \ll 1$ by enforcing suitable matrix
elements to be equal at that order in both theories.  One may then
attempt a lattice calculation from NRQCD. What one gains now is that
the spin independent splittings are a leading order effect rather than
a $v^2$ one and the hyperfine splittings a $v^2$ correction (rather
than $v^4$). See \Section~\ref{sec:spnrqcdlc} for a detailed
discussion of these calculations.

NRQCD, however, does not fully exploit the fact that $v$ is small. In
particular, gluons of energy $\sim mv$, the typical relative
three-momentum of the heavy quarks, are still explicit degrees of
freedom in NRQCD whereas they can never be produced at energies $\sim
mv^2$. For lower lying states the scale $mv$ corresponds both to the
typical momentum transfer $k$ (inverse size of the system) and to the
typical relative three-momentum $p$. It is then convenient to
introduce a further effective theory where degrees of freedom of
energy $\sim k$ are integrated out. This EFT is called pNRQCD
\cite{Pineda:1997bj,Brambilla:1999xf}, see
\Section~\ref{sec:spnrqcd}. The degrees of freedom of pNRQCD depend on
the interplay of the scales $k$, $E\sim mv^2$ and $\lQ$. The weak and
strong coupling regimes are discussed respectively in
\Section~\ref{sec:spnrqcdwc} and \ref{sec:spnrqcdscc}. A related EFT for the
weak coupling regime, called vNRQCD \cite{Luke:1999kz}, will be
discussed in \Chapter~\ref{chapter:precisiondeterminations} (Standard
Model). Sum rules are also discussed in the same chapter in relation
to the calculation of the lowest energy levels in the spectrum.

The distribution of the theory contributions is as follows. We begin
with the theoretical approach and use the EFT philosophy as an
organizing principle. We shall arrange the contributions according to
the number of hypothesis that are done in order to obtain them from
QCD.  Hence, we shall start by contributions which rely on QCD only.
Next we will discuss contributions which may be embraced by NRQCD, and
finally contributions which may be embraced by pNRQCD. We would like
to emphasize that, if the relevant hypothesis are fulfilled, (i) NRQCD
and pNRQCD are equivalent to QCD, and (ii) each of these EFTs allows
to factorize a relevant scale, which further simplifies calculations.
All the states can in principle be studied from QCD, the main tool
being lattice techniques.  In practise, however, a number of
limitations exists, which are described in \Section~\ref{sec:spdqlc}.
Except for very high excitations (particularly in charmonium) for
which relativistic effects become important, these states can also be
studied from NRQCD, the main tool being again lattice techniques, see
\Section~\ref{sec:nrqcd}. States below and not too close to open flavour
threshold can also be studied using pNRQCD. A few of these, including
the $\Upsilon (1S)$ and $\eta_b (1S)$, can be studied by means of
analytical weak coupling techniques
(\Section~\ref{sec:spnrqcdwc}). The remaining ones can be studied
using pNRQCD in the strong coupling regime
(\Section~\ref{sec:spnrqcdscc}), which needs as an input
nonperturbative potentials to be calculated on the lattice.  We
continue next with the phenomenological approach, which mainly consist
of a description of potential models (\Section~\ref{sec:sppot}) and of
approaches to open flavour thresholds (\Section~\ref{sec:spopcha}). The
former provide good phenomenological descriptions for the states below
open flavour threshold whereas the latter are important for a good
description of excitations close or above the open flavour threshold,
in particular of the recently discovered $X(3872)$ charmonium
state. An effort has been made to link potential models to the
theoretical approach. Double (and triple) heavy baryons are also
discussed both in the theoretical (\Sections~\ref{sec:spqqql},
\ref{sec:spnrqcd3q}) and phenomenological approach
(\Section~\ref{sec:dbch-QQqq}).

\section[Theoretical approach]{Theoretical approach}
\label{sec:spta}

\subsection[Direct lattice QCD calculation]
           {Direct lattice QCD calculation
            $\!$\footnote{Author: G.~Bali}}
\label{sec:spdqlc}

\subsubsection{Methods}

(For an introduction to general QCD lattice methods cf.\
\Chapter~\ref{chapter:commontheoreticaltools}.)  When simulating
quarks with a mass $m$ on a lattice with lattice spacing $a$, one will
inevitably encounter $ma$ [or $(ma)^2$] corrections, which are of
order one, unless $m\ll a^{-1}$.  The Fermilab
group~\cite{El-Khadra:1996mp} have argued in favour of a
re-interpretation of the clover action, suggesting that physical
results can be obtained even for masses as large as $ma\approx 1$, see
also \Section~\ref{sec:spnrqcdlc} below.  However, still one would
either want to extrapolate such results to the continuum limit or at
least put them into the context of an effective field theory with two
large scales, in this case $m$ and $a^{-1}$. If interpreted as an EFT,
higher order terms have to be added and the matching coefficients to
QCD have to be determined to sufficiently high order in perturbation
theory, to reduce and estimate remaining systematic uncertainties.

In the quenched approximation, the condition $ma\ll 1$ can
be realized for charm quarks; however, at present
bottom quarks are still somewhat
at the borderline of what is possible.
One approach to tackle this problem is to introduce an anisotropy,
with a temporal lattice spacing $a_{\tau}$ smaller than
the spatial lattice spacing $a_{\sigma}=\xi a_{\tau}$, with
parameter $\xi>1$. The spatial lattice extent $L_{\sigma}a_{\sigma}$
has to be large enough
to accommodate the quarkonium state (whose size is of order
$r\simeq (mv)^{-1}$). With a sufficiently large $a_{\sigma}$ this is possible,
keeping the number of points $L_{\sigma}$ limited, while the temporal
lattice spacing can be chosen to be smaller than the quarkonium
mass in question, $a_{\tau}<M^{-1}$, at relative ease.
This means that anisotropic simulations
are naively cheaper by a factor $\xi^3$, compared to the isotropic analogue
with a lattice spacing $a=a_{\tau}$.

While at tree level the lattice spacing
errors are indeed of ${\cal O}[(ma_{\tau})^{n}]$, one loop corrections
mean that there will still be ${\cal O}[\alpha_{\rm s}(ma_{\sigma})^{n}]$ terms present:
only to the extent to which $\alpha_{\rm s}\xi^n$ is small,
the leading order lattice effects can be regarded as
${\cal O}[(ma_{\tau})^n]$. Furthermore, the anisotropy parameter $\xi$ has to
be determined consistently for the quark and gluon contributions to
the QCD action. Within the quenched approximation this problem factorizes:
one can first ``measure'' the gauge anisotropy by determining
the decay of purely gluonic spatial and temporal correlation functions.
Subsequently,
one can adjust the Fermionic anisotropy accordingly. This fine-tuning
does not come for free, in particular if the number of adjustable parameters
is larger than two. Consequently, no consistent nonperturbative ${\cal O}(a)$
improvement programme has been carried through so far, for non-trivial
anisotropies. While there might be a net gain from using
anisotropy techniques
in the quenched approximation, the parameter tuning becomes much
more delicate and costly once light sea quarks are included. In this case
the numerical matching of the anisotropy for light Fermions cannot be
disentangled from the gluonic one anymore.

\subsubsection{Results with relativistic heavy quarks}

We will first review results on the quenched bottomonium spectrum,
before discussing charmonia in the quenched approximation, on
anisotropic as well as on isotropic lattices and with sea quarks.

Only one bottomonium study with relativistic action has been performed
so far~\cite{Liao:2001yh}, employing lattices with anisotropies
$\xi=4$ and $\xi=5$, in the quenched approximation.  In this case, the
inverse lattice spacing, $a_{\tau}^{-1}$ was varied from 4.5~GeV up to
about 10.5~GeV. The lattice extents were typically of size
$L_{\sigma}a_{\sigma}\approx 1$~fm, however, they were not kept
constant when varying $a_{\tau}$ such that finite size effects are
hard to disentangle.  The spatial lattice sizes are also dangerously
close to the inverse confinement--deconfinement phase transition
temperature (cf. \Chapter~\ref{chapter:charm-beauty-in-media}).  After
using the $1^1P_1-1^3S_1$ splitting (identifying the $1^1P_1$ mass
with the spin averaged experimental $1\overline{{}^3P}$ states) to set
the lattice spacing and the $1^3S_1$ to adjust the $b$ quark mass,
qualitative agreement with the spin-averaged experimental spectrum is
observed.

For the $1^3S_1-1^1S_1$ splitting, where one might hope finite size effects
to largely cancel, the authors obtain the continuum extrapolated value
of $59\pm 20$~MeV. To leading order in pQCD,
this splitting is expected to be proportional to the wave function
density at the origin, multiplied by $\alpha_{\rm s}(\mu)$.
Adjusting the lattice spacing from spin-averaged splittings
amounts to matching the quenched lattice coupling
to the phenomenological one at a low energy scale $\ll\mu$. In the quenched
approximation $\alpha_{\rm s}(\mu)$ approaches zero faster as $\mu$ is increased
and hence $\alpha_{\rm s}(\mu)$ will be underestimated:
the quoted fine structure splitting represents a lower limit on the
phenomenological one.
Indeed, the analogous result for the charmonium case underestimates
the known experimental number by a factor 1.25--1.5, when setting
the scale in a similar way~\cite{Okamoto:2001jb,Choe:2003wx}.

Both, the Columbia group~\cite{Chen:2000ej,Liao:2002rj} as well as the CP--PACS
Collaboration~\cite{Okamoto:2001jb}
have studied the charmonium spectrum on anisotropic lattices.
The same anisotropic clover quark action was used as for the bottomonium
study discussed above, where the leading order lattice artefacts are
expected to be of ${\cal O}(\alpha_{\rm s}a_{\tau})$ and ${\cal O}(a_{\tau}^2)$.
The CP--PACS Collaboration studied the anisotropy, $\xi=3$,
on a set of four inverse lattice spacings $a_{\sigma}^{-1}$,
ranging from about 1 up to 2.8~GeV, on spatial volumes $(1.6\,\mbox{fm})^3$.
The Columbia group simulated four lattice spacings ranging from
about 0.8 up to 2~GeV at anisotropy $\xi=2$. They were able to vary
their volume from 1.5 up to 3.3~fm and found finite volume effects to
be below their statistical resolution.

\begin{table} 
\caption[Charmonium results in the quenched approximation]
        {Charmonium results in the quenched
          approximation~\cite{Okamoto:2001jb,Choe:2003wx,%
                              Chen:2000ej,Liao:2002rj},
          where the scale is such that $r_0^{-1}=394$~MeV. The purely
          statistical errors do not reflect the uncertainty in $r_0$,
          or due to quenching. All values are in units of
          MeV. Glueball
          masses~\cite{Morningstar:1999rf,Lucini:2001ej,Bali:1993fb}
          are included for comparison. The last three lines refer to
          spin-exotic (non-quark model) quantum numbers.}
\label{tab:chquench}
\renewcommand{\arraystretch}{0.9} 
\begin{center} 
\vspace*{0.1cm} 
\begin{tabular}{|c|c||c|c|c|c|c|} 
\hline 
$J^{PC}$&state&CP--PACS&Columbia&QCD-TARO&experiment&glueballs\\\hline
$0^{-+}$&$\eta_c$&    3013 (1)&3014 (4)&3010 (4) &2980(1)&2500(40)\\ 
&       $\eta_c'$&    3739(46)&3707(20)&         &3654(10)&3500(60)\\
$1^{--}$&$J/\psi$&    3085 (1)&3084 (4)&3087 (4)&3097&\\
&       $\psi(2S)$&   3777(40)&3780(43)&         &3686&3700(50)\\
$1^{+-}$&$h_c$     &  3474(10)&3474(20)&3528(25)&$m(1^3\overline{P})$=3525
&2830(30)\\
&       $h_c'$    &   4053(95)&3886(92)&         &---&\\
$0^{++}$&$\chi_{c_0}$&3408 &3413(10)&3474(15)&3415(1)&1720(30)\\
&       $\chi_{c_0}'$&4008(122)&4080(75)&       &---&2540(120)\\
$1^{++}$&$\chi_{c_1}$&3472 (9)&3462(15)&3524(16)&3511&\\
&        $\chi_{c_1}'$&4067(105)&4010(70)&       &---&\\
$2^{++}$&$\chi_{c_2}$&3503(24)&3488(11)&         &3556&2300(25)\\
&        $\chi_{c_2}'$&4030(180)&       &         &---&\\
$2^{-+}$&$1^1D_2$  &           &3763(22)&        &---&2975(30)\\
&                 &           &        &         &---&3740(40)\\
$2^{--}$&$1^3D_2$  &          &3704(33)&         &$X(3872)$ ???&3780(40)\\
$3^{--}$&$1^3D_3$  &          &3822(25)&         &---&3960(90)\\
$3^{+-}$&$1^1F_3$  &          &4224(74)&         &---&3410(40)\\
$3^{++}$&$1^3F_3$  &          &4222(140)&        &---&3540(40)\\
{\bf $0^{+-}$}&$H_0$&         &4714(260)&        &---&4560(70)\\
{\bf $1^{-+}$}&$H_1$&          &4366(64)&        &---&\\
{\bf $2^{+-}$}&$H_2$&          &4845(220)&       &---&3980(50)\\\hline
\end{tabular} 
\end{center} 
\renewcommand{\arraystretch}{1} 
\end{table}

We display the respective continuum-limit extrapolated results in
\Table~\ref{tab:chquench}. We also include results from the QCD--TARO
collaboration~\cite{Choe:2003wx}, with $\xi=1$.  The quark mass is set
such that the spin averaged $1\overline{S}$ state corresponds to
3067.6~MeV.  (Note that the present phenomenological value is slightly
higher than this.)  For comparison we convert the Columbia results
into units of $r_0^{-1}=394$~MeV. This scale is implicitly defined
through the static potential~\cite{Sommer:1993ce},
$\left.dV(r)/dr\right|_{r=r_0}=1.65$. It cannot directly be obtained
in experiment.  However, $r_0/a$ is easily and very precisely
calculable in lattice simulations. In the quenched approximation we
have to assume a scale error on spin averaged splittings of at least
10~\%, on top of the errors displayed in the Table.  We also include
glueball masses~\cite{Morningstar:1999rf,Lucini:2001ej,Bali:1993fb}
into the table. The last three lines incorporate spin-exotic $J^{PC}$
assignments ($c\bar{c}g$ hybrid mesons).

\begin{figure}
\begin{center}
\includegraphics[angle=270,width=0.8\textwidth]{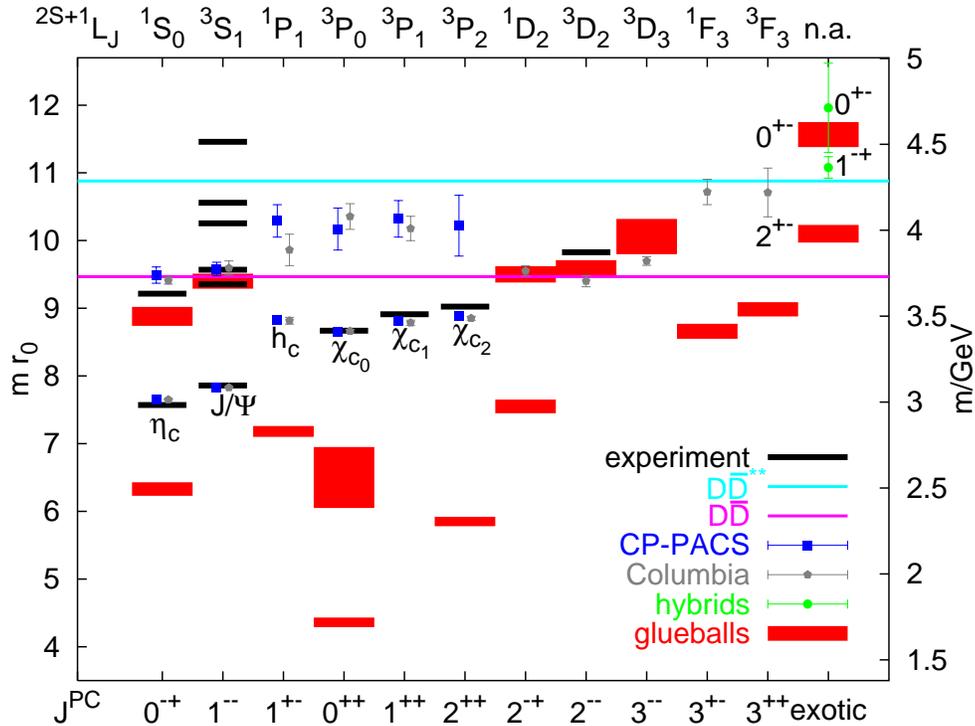}
\end{center}
\caption[The quenched charmonium spectrum]
        {The quenched charmonium spectrum
         (CP--PACS~\cite{Okamoto:2001jb},
         Columbia~\cite{Chen:2000ej,Liao:2002rj}),
         glueballs~\cite{Morningstar:1999rf,Lucini:2001ej,Bali:1993fb}
         and spin-exotic $c\bar{c}$-glue hybrids~\cite{Liao:2002rj},
         overlayed with the experimental spectrum.}
\label{fig:chquench}
\end{figure}

The anisotropic results are also displayed in
\Figure~\ref{fig:chquench}, borrowed from Ref.~\cite{Bali:2003tp},
where we plot the new $X(3872)$ state at $J^{PC}=2^{--}$, however,
this assignment is somewhat arbitrary.  As can be seen, where overlap
exists, the results from the three collaborations employing three
different anisotropies are consistent with each other. All S- and
P-wave fine structure splittings are underestimated, which is
expected in the quenched approximation.  The Columbia
group~\cite{Liao:2002rj} reported that the state created by the $J=1$
D-wave operator rapidly converged towards the mass of the vector
S-wave ground state. The same was observed in the case of the
$2^{++}$ F-wave with respect to the $\chi_{c_2}$ ground state: this
indicates that the charm quark mass is too light for $L$ to be a good
quantum number.

That the charm mass is not particularly heavy, in comparison
to typical scales of gluonic excitations, can also be seen
from the overlap between the glueball and charmonium spectra.
Once sea quarks are switched on, these glueballs will become unstable.
However, the presence of a background of such excitations
might very well affect spectrum and decays in some channels.
For instance the dominant decay of a vector charmonium is into
gluons, and it is quite conceivable that such a channel should
also couple to would-be glueballs.

When performing the Wick contractions of propagators of flavour
singlet states like charmonia, two contributions arise: a connected
one, with quark and antiquark propagating alongside each other, and a
disconnected (OZI suppressed) one, with annihilation and creation
diagrams of $c\bar{c}$.  In all charmonium simulations that have been
performed so far, with two notable
exceptions~\cite{McNeile:2004wu,deForcrand:2004ia}, the disconnected
diagram has been neglected. It is well known that OZI processes play a
role within the light pseudoscalar and scalar sectors.  This has also
been extensively studied on the
lattice~\cite{McNeile:2001cr,Struckmann:2000bt}.  In the case of
charmonia, in particular for S and D~waves, substantial
corrections due to mixing with intermediate gluonic states are a
possibility, even within the quenched approximation.  For states that
are close to threshold, in addition mixing with two-meson states will
occur, once sea quarks are included.

Charmonia have also been studied on isotropic lattices, within the
quenched
approximation~\cite{Boyle:1999dx,Bernard:1997ib,Choe:2003wx,McNeile:2004wu,deForcrand:2004ia},
and with sea quarks~\cite{diPierro:2003bu}.  The QCD--TARO
collaboration~\cite{Choe:2003wx} worked at tiny lattice spacings,
ranging from about 2~GeV down to 5~GeV. The results are consistent
with those obtained by the Columbia group~\cite{Liao:2002rj} and by
CP--PACS~\cite{Okamoto:2001jb}, but the use of an ${\cal O}(a)$
improved action allowed for a very well controlled continuum limit
extrapolation.  The quenched value, within the OZI approximation and
using $r_0^{-1}=394$~MeV to set the scale, is 77(2)(6)~MeV, with all
remaining systematic errors quoted.  This value would increase by
15~\% if the scale was set from the $1\overline{{}^3P}-1\overline{S}$,
still short of the experimental 117~MeV.

In an exploratory study, in which for the
first time the diagram that contains disconnected quark loops has been
included, McNeile and Michael~\cite{McNeile:2004wu} find evidence that
while the position of the ground state vector state appears to be
largely unaffected, the pseudoscalar mass is reduced by
an amount of the order of 20~MeV with respect to the
non-flavour singlet reference value. One explanation might be the
background of glueballs, c.f.\ \Figure~\ref{fig:chquench}.
A more recent study by
QCD--TARO~\cite{deForcrand:2004ia} confirms that the vector state remains
largely unaffected. They rule out an increase of the pseudoscalar
mass, however, a decrease by an amount of up to 20~MeV would not contradict
their data.

First studies~\cite{diPierro:2003bu} utilizing the AsqTad staggered
light quark action and approximating
$2+1$ flavours of sea quarks by taking roots of the Fermionic
determinant have been performed.
The light quark mass was varied down to about $m_s/6$.
The ${\cal O}(\alpha_{\rm s}a)$ clover action, in the Fermilab
heavy quark interpretation~\cite{El-Khadra:1996mp} was used.
Extrapolating to physical sea quark mass, a hyperfine structure
splitting of 97(2)~MeV is obtained, see also
\Section~\ref{sec:spnrqcdlc} below. This is an increase
of almost 40~\%, over their quenched reference value.  At least the
latter would have been somewhat smaller if normalized with respect to
$r_0$ rather than to the $\Upsilon'-\Upsilon$ splitting.  However,
OZI diagrams have been neglected and neither is the lattice spacing
dependence resolved as yet.  Clearly, a precision study of the
charmonium spectrum requires not only sea quarks but also flavour
singlet diagrams to be included.

\subsection[NRQCD]{NRQCD}
\label{sec:nrqcd}

NRQCD takes advantage that the masses of the charm and bottom quarks
are much larger than $\lQ$ in order to build an EFT which is
equivalent to QCD at any desired order in $1/m$ and $\als
(m)$. Starting from NRQCD two approaches may be followed for spectrum
computations: direct lattice calculations
(\Section~\ref{sec:spnrqcdlc}) or further integration of the soft
scale (the scale of the momentum transfer) to arrive at an EFT in
which only the ultrasoft degrees of freedom remain dynamical, pNRQCD
(\Section~\ref{sec:spnrqcd}).  An introduction to NRQCD is given in
\Chapter~\ref{chapter:commontheoreticaltools}, 
see also Refs.\ \cite{reveft,reveftgen,reveftii} for some
introduction to the nonrelativistic EFT formulation. An introduction
to lattice methods (quenched and unquenched) has been given in 
\Chapter~\ref{chapter:commontheoreticaltools}.

\subsubsection[Lattice NRQCD calculations with light sea quarks:
               Lattice NRQCD; $\Upsilon$ results with NRQCD; $\psi$ 
               results with the Fermilab method; The $B_c$ ground state]
              {Lattice NRQCD calculations with light sea quarks
               $\!$\footnote{Authors: C.~Davies, A.~Kronfeld, 
                             P.~Mackenzie, J.~Simone}}
\label{sec:spnrqcdlc}

The use of non-relativistic effective field theories permits the 
computer to handle only scales appropriate to the physics of the 
non-relativistic bound states without having to spend a lot
of computer power on the large scale associated 
with the heavy quark mass which is irrelevant to the bound state dynamics.
This makes the calculations more tractable so that many more hadron 
correlators can be calculated for better statistical precision. 
We will focus our discussion on the most recent calculations obtained
within this approach, which include light sea quarks.

On the lattice, heavy quark effects and discretisation effects are
intertwined.  One can treat them together by introducing an effective
Lagrangian~\cite{Kronfeld:2000ck,Kronfeld:2003sd}
\begin{eqnarray}
        {\cal L} = & - & \psi^\dagger\left[ \delta m + D_4 -
                \frac{\mbox{\boldmath$D$}^2}{2m} -
                \frac{c^{\rm lat}_4}{8m^3}\left(\mbox{\boldmath$D$}^2\right)^2 -
                \frac{w^{\rm lat}_1a^2}{6m}\sum_iD_i^4 -
                \frac{c^{\rm lat}_D}{8m^2}\left(
                        \mbox{\boldmath$D$}\cdot g\mbox{\boldmath$E$} -
                        g\mbox{\boldmath$E$}\cdot \mbox{\boldmath$D$} \right) 
        \right. \nonumber \\ & & \left. \hspace*{5em} - \,
                \frac{c^{\rm lat}_S}{8m^2}i\mbox{\boldmath$\sigma$}\cdot \left(
                        \mbox{\boldmath$D$}\times g\mbox{\boldmath$E$} +
                        g\mbox{\boldmath$E$}\times \mbox{\boldmath$D$} \right) -
                \frac{c^{\rm lat}_F}{2m}
                        \mbox{\boldmath$\sigma$}\cdot g\mbox{\boldmath$B$}
        \right] \psi + \cdots ,
        \label{eq:lat-spect-NRQCD}
\end{eqnarray}
similar to the standard (continuum) NRQCD Lagrangian, but note that
the derivative operators are `improved' on the lattice to remove
leading errors arising from the lattice spacing.  See also the
\Section~\ref{sec:HQactions} ``Heavy Quark Actions'' in
\Chapter~\ref{chapter:commontheoreticaltools}.  We have omitted the
term $\psi^{\dagger}m\psi$.

Compared to the NRQCD description of continuum QCD, an unimportant difference 
is the Euclidean metric ($D_4$ instead of $-iD_0$).
Also, unlike in dimensional regularization,
in lattice regularization
the mass shift $\delta m$ will in general be non-zero. However,
this cancels from mass differences and decay
amplitudes. Moreover, it can be determined nonperturbatively
from the $\Upsilon$ dispersion relation. Obviously, terms accompanied by
$w_i$ are lattice specific.
The essential difference is that the matching scale is provided by the
lattice spacing: the short-distance 
coefficients~$c^{\rm lat}_i$, $w^{\rm lat}_i$ and $\delta m$ depend on
$am$ and on the details of the chosen discretisation.
The matching of $c_i^{\rm lat}$ and $w_i^{\rm lat}$ is carried out
to some accuracy in $\alpha_{\rm s}$.
From \Eq~(\ref{eq:lat-spect-NRQCD}) one sees that the most important 
matching condition is to identify the kinetic mass~$m$ 
with the heavy quark mass in the lattice scheme,
and then tune the higher-dimension 
interactions.

One area of lattice QCD which has remained problematic is the handling of 
light quarks on the lattice. This is now
being addressed successfully and is critical to obtaining precision 
results of use to experiment.
In particular the problem is how to include 
the dynamical (sea) $u/d/s$ quark pairs that appear as a result of 
energy fluctuations in the vacuum. We can often
safely ignore $c/b/t$ quarks in the 
vacuum because they are so heavy, 
but we know that light quark pairs have significant effects, for example
in screening the running of the gauge coupling and in generating Zweig-allowed
decay modes for unstable mesons. 

Many calculations in the past have used the ``quenched approximation,''
attempting to compensate sea quark effects by {\em ad hoc} shifts 
in the bare coupling and (valence) quark masses.
The results then suffer from errors as large as 10--30\%. 
The error of the quenched approximation is not really quantifiable and
this is reflected by a lack of internal consistency when different kinds of
hadrons are used to fix the bare parameters.
This ambiguity plagues the lattice QCD
literature. 

\begin{figure}
\begin{center}
\includegraphics[height=.3\textheight]{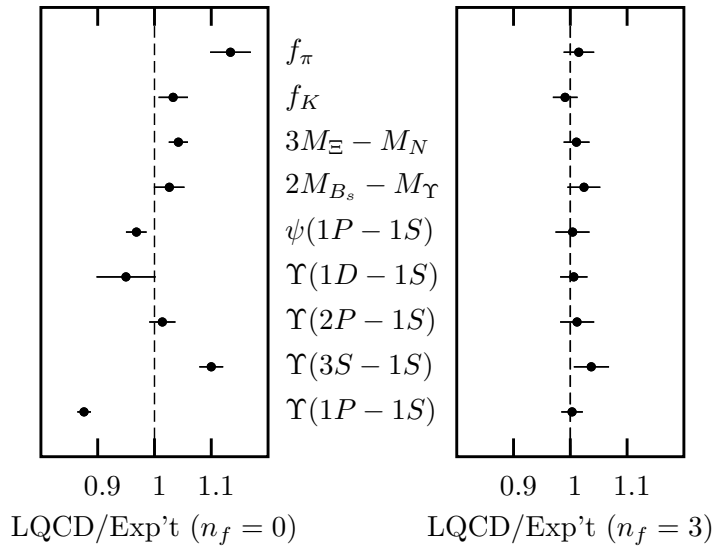}
\end{center}
\caption[Lattice QCD results divided by experiment]
        {Lattice QCD results divided by experiment for a range of
         ``gold-plated'' quantities which cover the full range of
         hadronic physics~\cite{Davies:2003ik}. The unquenched
         calculations on the right show agreement with experiment
         across the board, whereas the quenched approximation on the
         left yields systematic errors of $\cal{O}$(10\%).}
\label{fig:ratio}
\end{figure}
The MILC Collaboration recently have produced ensembles of 
gluon field configurations which include 2 degenerate
light sea quarks ($u,d$) and a heavier 
one ($s$)~\cite{Bernard:2001av}.
They rely on fast supercomputers and a new discretisation 
of the quark action: the improved staggered formalism~\cite{Lepage:1998vj}.
At quark masses small enough for reliable chiral extrapolations, 
staggered Fermions appear much faster than any other 
formulation of lattice Fermions. However, each flavour of staggered 
quarks is included in the sea by taking the fourth root of the staggered 
determinant and there are still theoretical issues to be resolved 
about this. 
Taking the $u$ and $d$ masses the 
same makes the lattice calculation much faster and 
leads to negligible errors in isospin-averaged 
quantities. The sea $s$ quark mass is chosen 
to be approximately correct based on earlier studies 
(in fact the subsequent analysis shows that it was slightly 
high and further ensembles are now being made with a lower value). 
The sea 
$u$ and $d$ quarks take a range of masses down 
as low as a sixth of the (real) $m_s$.
Ensembles are available at two different 
values of the lattice spacing, 0.12~fm and 0.09~fm, 
and the spatial lattice volume is $(2.5~{\rm fm})^3$,
reasonably large.
Analysis of hadronic quantities on 
these ensembles has been done by the 
MILC and HPQCD collaborations~\cite{Davies:2003ik}. 

There are 5 bare parameters of QCD relevant to this 
analysis: $\alpha_{\rm s}$, $m_{u/d}$, $m_s$, $m_c$ and $m_b$. 
Changing the bare $\alpha_{\rm s}$ changes the 
lattice spacing.
It is important to fix these 
parameters with the masses of ``gold-plated'' hadrons,
\ie hadrons which are well below their strong decay thresholds.
Such hadrons are well-defined experimentally and theoretically 
and should be accurately calculable in lattice QCD. 
Using them to fix parameters will then not introduce 
unnecessary additional 
systematic errors into lattice results for other quantities. 
This has not always been done in past lattice 
calculations, particularly in the quenched approximation.
It becomes an important issue when lattice QCD is 
to be used as a precision calculational tool. 
We use the radial excitation energy in the $\Upsilon$ 
system (\ie the mass splitting between the 
$\Upsilon^{\prime}$ and the $\Upsilon$) to fix the 
lattice spacing. This is a good quantity to use because 
it is very insensitive to all quark masses, including 
the $b$ quark mass (experimental values for this 
splitting are very similar for charmonium and bottomonium)
and so it can be determined without a complicated iterative 
tuning process. $m_{\pi}$, $m_K$, 
$m_{D_s}$ and $m_{\Upsilon}$ are used to fix the quark masses. 
Thus, quarkonium turns out to be a central part in this study.

Once the Lagrangian parameters are set, we can focus on the
calculation of other gold-plated masses and decay constants.  If QCD
is correct and lattice QCD is to work it must reproduce the
experimental results for these quantities precisely.
\Figure[b]~\ref{fig:ratio} shows that this indeed works for the
unquenched calculations with $u, d$ and $s$ quarks in the vacuum.  A
range of gold-plated hadrons are chosen which range from decay
constants for light hadrons through heavy-light masses to heavy
quarkonium. This tests QCD in different regimes in which the sources
of systematic error are very different and stresses the point that QCD
predicts a huge range of physics with a small set of parameters.

Refs.~\cite{Gottlieb:2003bt,Gray:2002vk,diPierro:2003bu,Aubin:2003ne}
give more details on the quantities shown in
\Figure~\ref{fig:ratio}. Here we concentrate on the spectrum of
bottomonium and charmonium states, using, respectively, lattice
NRQCD~\cite{Lepage:1992tx} and the Fermilab method for heavy
quarks~\cite{El-Khadra:1996mp}.  We include a brief discussion of the
$B_c$~mass, including the status of an ongoing unquenched calculation
using the MILC ensembles.

\vspace{2mm}
{\it
$\Upsilon$ results with NRQCD}\vspace{2mm}

\begin{figure}
\begin{center}
\makebox[0mm]{(a)}
\includegraphics[height=0.29\textheight]{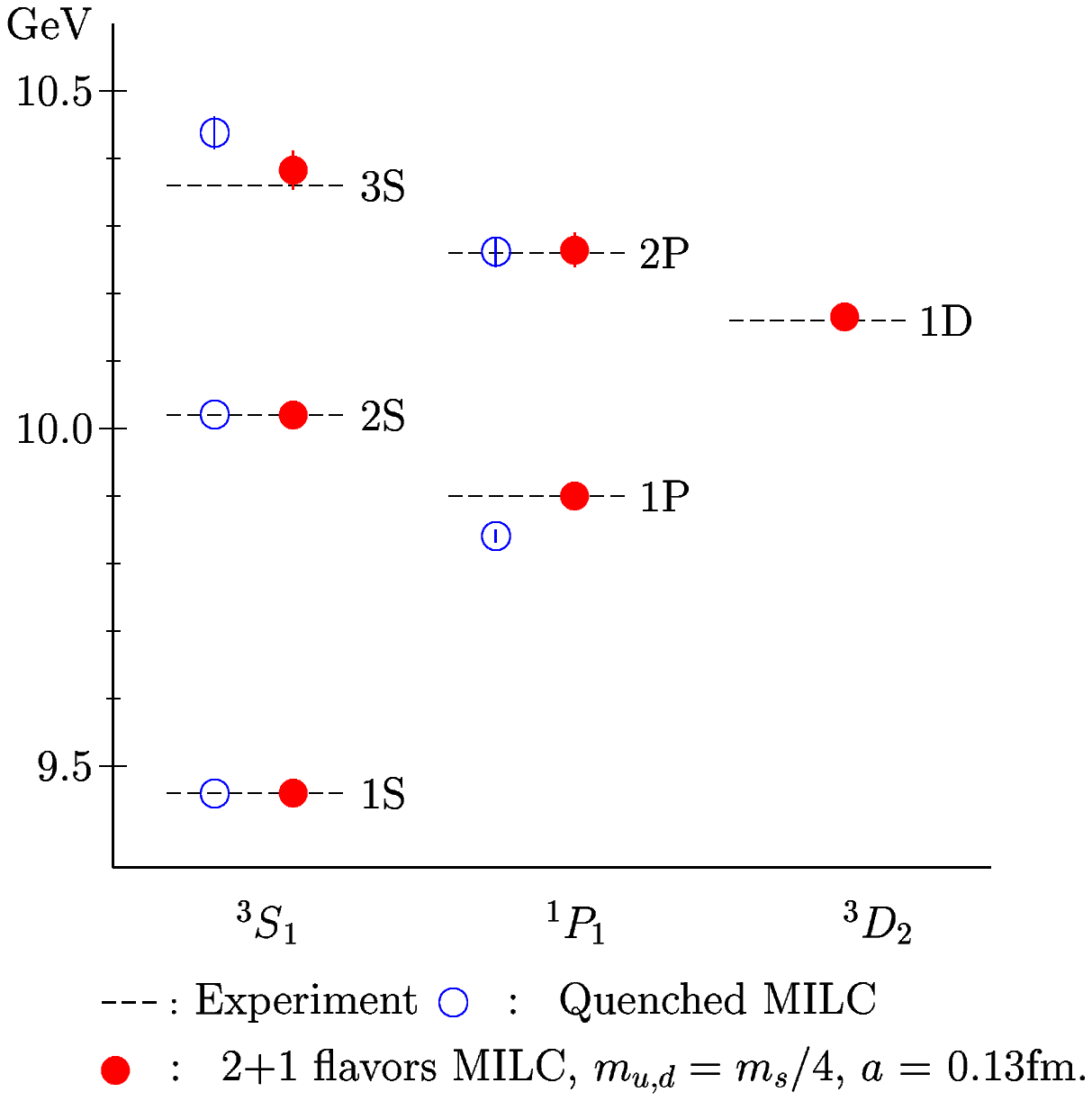}
\hfill
\makebox[0mm]{(b)}
\includegraphics[height=0.29\textheight]{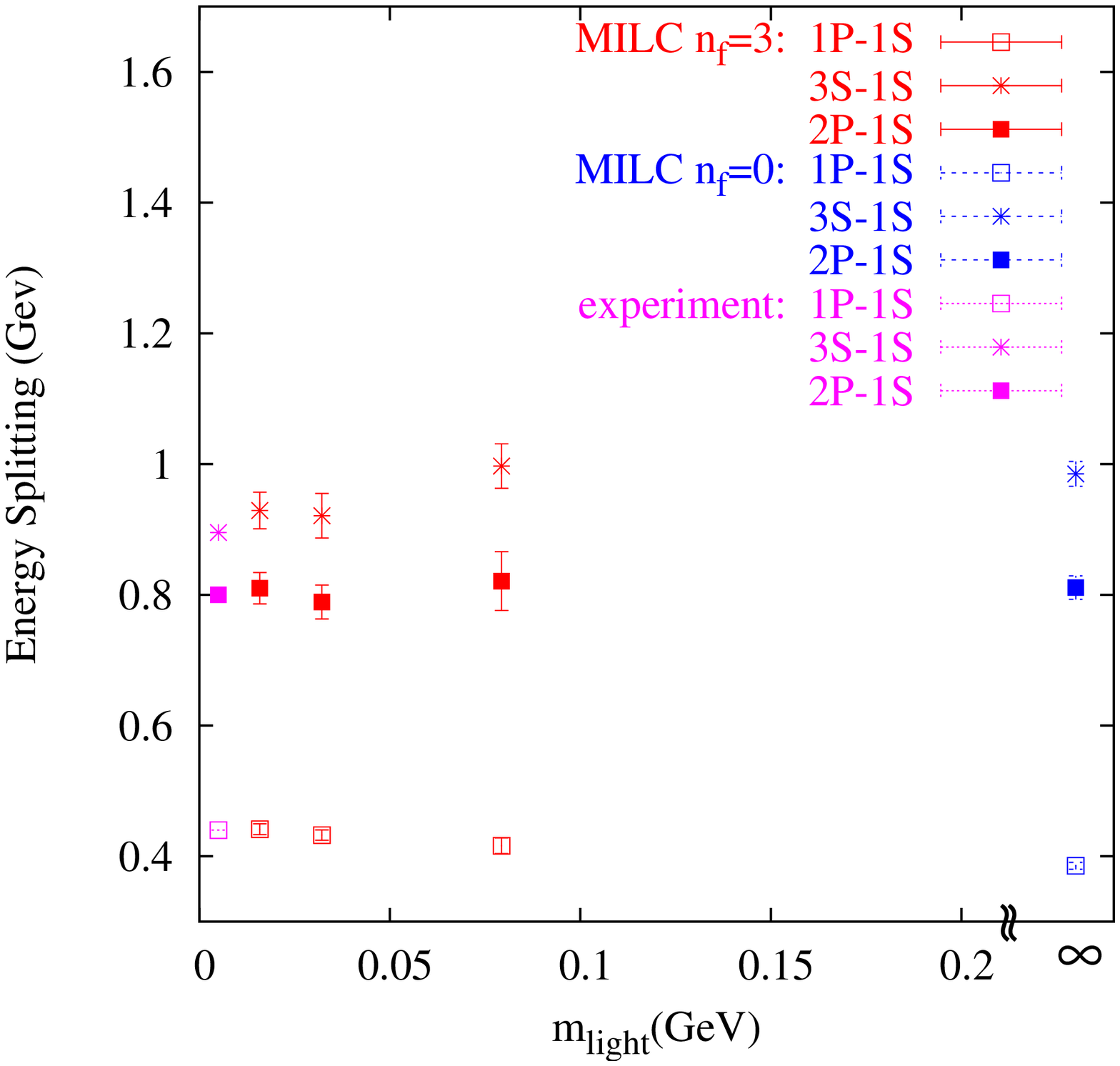}
\end{center}
\caption[Radial and orbital splittings in the $\Upsilon$ system 
         from lattice QCD]
        {Radial and orbital splittings in the $\Upsilon$ system from
         lattice QCD, using the $\Upsilon^{\prime}-\Upsilon$
         splitting and the $\Upsilon$ mass to fix the lattice spacing
         and the $b$-quark mass~\cite{Gray:2002vk}.  (a) Comparison of
         the quenched approximation (open circles) and QCD with $u, d$
         and $s$ sea quarks (filled) circles. Note that the 1S and 2S
         levels are used to fix the $b$ quark mass and lattice spacing
         respectively so are not predictions.  (b) Dependence of the
         splittings as a function of the of the bare sea $u/d$ quark
         mass.}
\label{fig:ups}
\end{figure}

\noindent
\Figure[b]~\ref{fig:ups}(a) shows the radial and orbital splittings~\cite{Gray:2002vk}
in the $b\bar{b}$ ($\Upsilon$) system for the quenched approximation
($n_f$ = 0) and with the dynamical MILC configurations with 3 flavours
of sea quarks.
We use the standard lattice NRQCD effective theory for the valence $b$
quarks~\cite{Lepage:1992tx}, which takes advantage of the
non-relativistic nature of the bound states. The lattice NRQCD action
used here is accurate through $v^4$ where $v$ is the velocity of the
$b$ quark in its bound state.  It also includes corrections to remove
discretisation errors at ${\cal O}(p^2a^2v^2)\sim {\cal O}(v^4)$, but
does not include ${\cal O}(\alpha_{\rm s} v^4)$ corrections to the
coefficients $c_i$ and $w_i$ in \Eq~(\ref{eq:lat-spect-NRQCD}),
which are subleading.  This means that spin-independent splittings,
such as radial and orbital excitations, are simulated through
next-to-leading-order in the velocity expansion and should be accurate
to around~1\%.  Thus, these splittings provide a very accurate test
not only of lattice QCD, but also of the effective-field theory
framework.  At present, the fine structure in the spectrum is only
correct through leading-order [which is ${\cal O}(v^4)$ in this case]
and more work must be done to bring this to the same level and allow
tests against, for example, the splittings between the different
$\chi_b$ states~\cite{Gray:2002vk}.  This is in progress.  Systematic
uncertainties due to such truncations have for instance been estimated
in Ref.~\cite{Bali:1998pi}, based on lattice potentials.

The $\Upsilon$ system is a good one for looking at the effects 
of sea quarks because we expect it to be 
relatively insensitive to sea quark masses. The momentum transfer 
inside an $\Upsilon$ is larger than any of the $u,d$ or $s$ masses 
and so we expect the radial and orbital splittings to simply 
count the number of sea quarks once they are reasonably light. 
\Figure[b]~\ref{fig:ups}(b) shows this to 
be true\,---\,the splittings are independent of the sea 
$u/d$ quark mass in the region we are working in.
Chiral extrapolation in the $u/d$ quark mass is immaterial in this case. 
Therefore, the left-most lattice points in \Figure~\ref{fig:ups}(b) are the
ones used in \Figures~\ref{fig:ratio} and~\ref{fig:ups}(a).

\vspace{2mm}
{\it
$\psi$ results with the Fermilab method}\vspace{2mm}

\begin{figure}
\begin{center}
\raisebox{56mm}{\makebox[2mm]{(a)}}
\raisebox{10mm}{\includegraphics[width=0.46\textwidth]{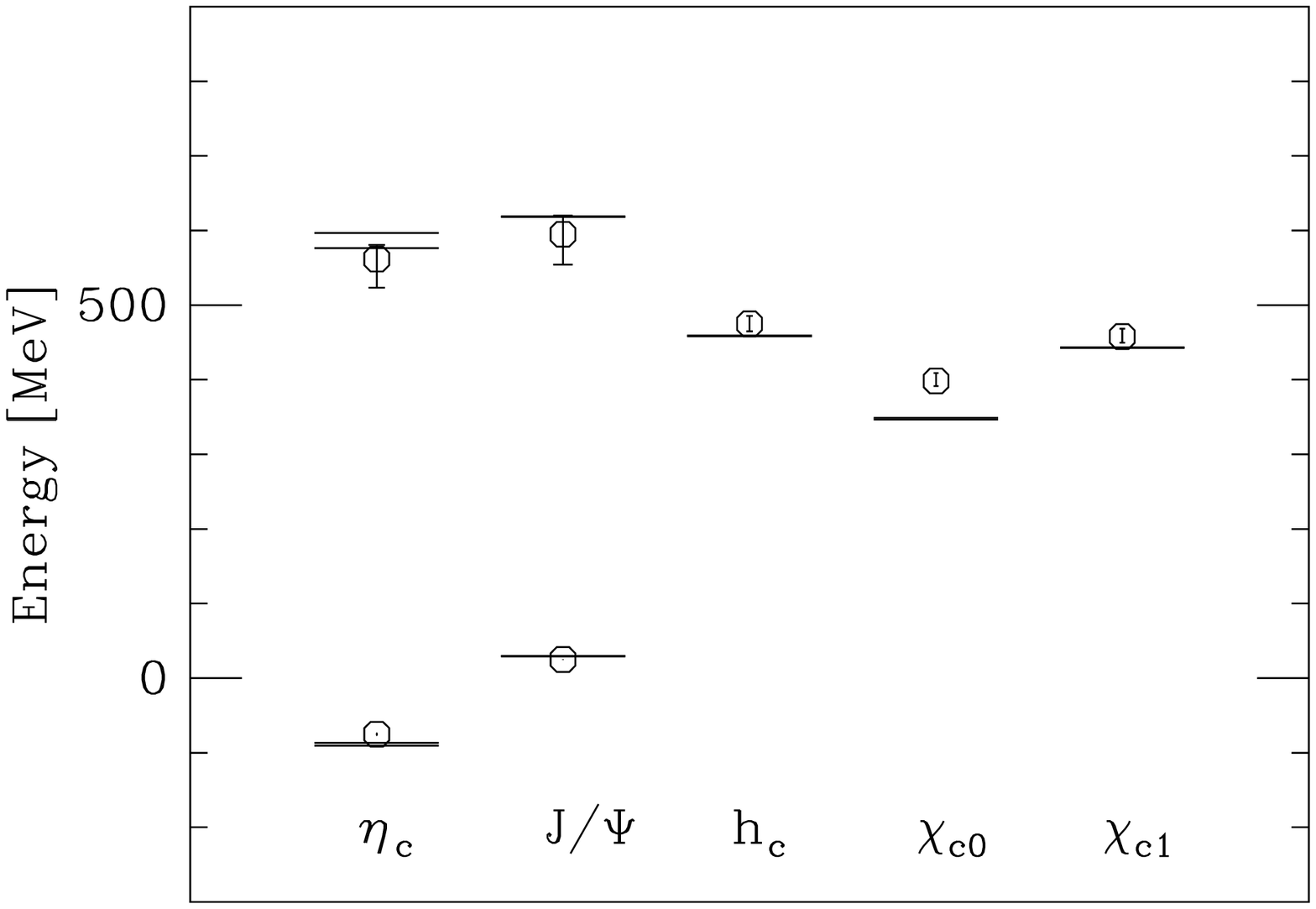}}
\hfill
\raisebox{56mm}{\makebox[2mm]{(b)}}
\includegraphics[width=0.46\textwidth]{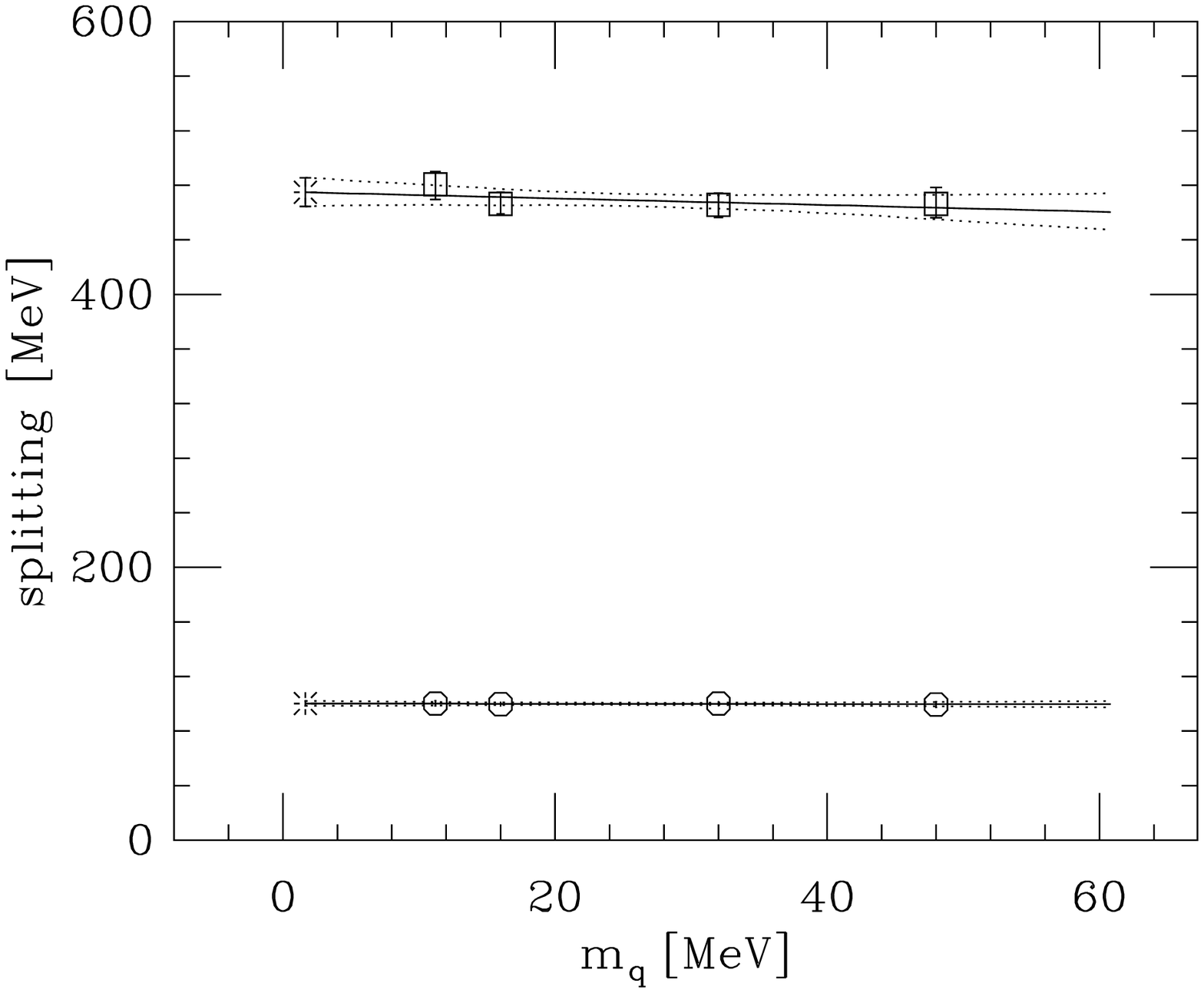}
\end{center}
\caption[Radial and orbital splittings in the charmonium system 
         from lattice QCD]
        {Radial and orbital splittings in the charmonium system from
         lattice QCD with 3 light sea quarks, fixing the lattice
         spacing from the $\Upsilon^{\prime}-\Upsilon$ splitting (as
         above), and the $c$ quark mass from the $D_s$
         mass~\cite{diPierro:2003bu}.  (a) Spectrum; (b) dependence on
         the sea quark mass.}
\label{fig:charmonium-spectrum}
\end{figure}

\noindent
\Figure[b]~\ref{fig:charmonium-spectrum} shows the spectrum of charmonium 
states below the $D\overline{D}$ threshold~\cite{diPierro:2003bu}.
In this plot the lattice spacing was fixed from the
$\Upsilon^{\prime}-\Upsilon$ splitting (as above), and the $c$ quark
mass was tuned to get the $D_s$ mass correct.  Therefore, these
results are obtained directly from QCD without adjusting any free
parameters.  For \Figure~\ref{fig:charmonium-spectrum}(a), the zero of
energy has been moved to the spin-averaged mass
$\bar{m}_\psi=\frac{1}{4}m_{\eta_c}+\frac{3}{4}m_{J/\psi}$.

These results are obtained using the Fermilab
method~\cite{El-Khadra:1996mp} for the charmed quark.  In this method
one starts with Wilson Fermions, but the discretisation effects are
controlled and understood using non-relativistic field theories, as in
\Eq~(\ref{eq:lat-spect-NRQCD}).  The non-relativistic interpretation
also has implications for how the action is improved.  In the notation
of \Eq~(\ref{eq:lat-spect-NRQCD}) the chromomagnetic interaction is
adjusted so that $c_F^{\rm lat}$ is correct at tree level.  However,
at higher order, there are $O[(m_ca)^2]\sim 10$\% and ${\cal
O}(\alpha_{\rm s})$ errors and some sign of these is seen in the
mismatch with experiment of the hyperfine splitting in
\Figure~\ref{fig:charmonium-spectrum}(a).  In the past such
discrepancies were masked by quenching errors, whereas now they can be
resolved.  Note that OZI violating
contributions~\cite{McNeile:2004wu,deForcrand:2004ia} are also
neglected currently.  They are expected to be small but a decrease of
up to 20~MeV in $m_{\eta_c}$ is not ruled out.

The Fermilab action can be systematically improved, and the theoretical 
work needed is in progress.
The most important new features are a one-loop calculation of the 
chromomagnetic coupling~\cite{Nobes:2003nc}, and a systematic 
enumeration of all operators needed for improvement 
through~$v^6$~\cite{Oktay:2002mj}.

\vspace{2mm}
{\it $B_c$ ground state}\vspace{2mm}

\noindent
In 1998 the lowest-lying bound state of $\bar{b}c$ quarkonium was
observed in semi-leptonic decays~\cite{Abe:1998wi}, yielding a mass of
$m_{B_c}=6.4\pm0.4$~GeV.  A more precise measurement with hadronic
decays is expected to come soon from Run~II of the Tevatron,
cf.~\Section~\ref{sec:spexBc}.
For lattice QCD, the $B_c$ is a `gold-plated' hadron and we have the
opportunity to predict its mass ahead of experiment.  Here we report
on a preliminary lattice calculation, building on the progress
detailed above.  In previous quenched calculations accurate result
could not be provided, due to the inconsistency of this approach
described above.

The method used in the present study was developed in a quenched
calculation~\cite{Shanahan:1999mv}, and follows almost immediately
from \Eq~(\ref{eq:lat-spect-NRQCD}).  As long as one may use the
effective Lagrangian to describe the charmed and bottom quarks on the
lattice, the meson mass satisfies~\cite{Kronfeld:2000ck},
\begin{equation}
        {M_1}_{B_c} = {m}_{\bar{b}} + {m}_{c} + B_{B_c},
\end{equation}
where $B_{B_c}$ is the binding energy of the $B_c$~meson.
The accuracy of the binding energy depends on how well the
coefficients $c_i^{\rm lat}$ have been adjusted.
The scheme- and scale-dependent quark masses
cancel from the relation~\cite{Shanahan:1999mv},
\begin{equation}
        {M_1}_{B_c} - {\textstyle\frac{1}{2}}
                \left[{M_1}_{\psi} + {M_1}_{\Upsilon}\right] =
        B_{B_c} - {\textstyle\frac{1}{2}}\left[B_{\psi} + B_{\Upsilon}\right].
\end{equation}
Note that within potential
models flavour independence implies that this combination is  small
and positive~\cite{Bertlmann:1979zs,Nussinov:1999sx}.
One can now predict the $B_c$ mass by adding back the 
experimental $\frac{1}{2}\left[M_{\psi} + M_{\Upsilon}\right]$.
A variant of this technique is to use the $D_s$ and $B_s$ masses instead 
of (half the) quarkonium masses.

An unquenched lattice calculation has recently been carried
out~\cite{Allison:2004be,Allison:2004mv}, using the MILC ensembles
discussed above.  Analyses at two light sea quark masses and two
values of the lattice spacing show a consistent picture, as expected.
Using the quarkonium baseline, Allison \emph{et al.}
find~\cite{Allison:2004be}
\begin{equation}
        M_{B_c} = 6304 \pm 4 \pm 11^{+18}_{-\;\,0}~{\rm MeV},
        \label{eq:mBcLattice}
\end{equation}
where the uncertainties are, respectively, from
statistics (after chiral extrapolation),
tuning of the heavy-quark masses, and
heavy-quark discretization effects.
The last is estimated from the mismatch of operators of order $v^4$ in
the effective Lagrangian and are dominated by the relativistic 
correction
$(\mbox{\boldmath$D$}^2)^2$.
The estimate is guided by potential models (and is the only change from
earlier conference reports~\cite{Allison:2004mv}).
The overall errors are so small because the lattice calculation has been
set up to focus on the binding-energy difference, and raw uncertainties
of several percent have been leveraged to the sub-percent level for the
mass itself.

This result can be checked with the heavy-light baseline,
$M_{B_c}=M_{D_s}+M_{B_s}+[B_{B_c}-(B_{D_s}+B_{B_s})]$,
with somewhat larger uncertainties.
Allison \emph{et al.} find~\cite{Allison:2004be}
\begin{equation}
        M_{B_c} = 6243 \pm 30 \pm 11^{+37}_{-\;\,0}~{\rm MeV}.
\end{equation}
The systematic uncertainties are larger with the heavy-light baseline
because there is less cancellation between the $B_c$ quarkonium and the
heavy-light $D_s$ and $B_s$.

The dominant uncertainties can be reduced by choosing more 
highly-improved
actions in lattice gauge theory, or by reducing the lattice spacing,
as discussed in Ref.~\cite{Allison:2004be}.

\subsubsection[Heavy hybrids on the lattice]
              {Heavy hybrids on the lattice
               $\!$\footnote{Author: C.~Morningstar}}
\label{sec:sphhl}

QCD suggests the existence of mesonic states in which the valence
quark-antiquark pair is bound by an \emph{excited} gluon field.  A
natural starting point in the quest to understand such states is the
heavy quark sector. The vastly different characteristics of the slow
massive heavy quarks and the fast massless gluons suggest that such
systems may be amenable to a Born--Oppenheimer treatment, similar to
diatomic molecules.  The slowly moving heavy quarks correspond to the
nuclei in diatomic molecules, whereas the fast gluon and light-quark
fields correspond to the electrons.  At leading order, the gluons and
light quarks provide adiabatic potentials $V_{Q\bar{Q}}(r)$, where $r$
is the quark--antiquark separation, and the behavior of the heavy
quarks is described by solving the Schr\"odinger equation separately
for each $V_{Q\bar{Q}}(r)$.  The Born--Oppenheimer approximation
provides a clear and unambiguous picture of conventional and hybrid
mesons: conventional mesons arise from the lowest-lying adiabatic
potential, whereas hybrid mesons arise from the excited-state
potentials.

The first step in a Born--Oppenheimer treatment of heavy quark mesons
is determining the gluonic terms $V_{Q\bar{Q}}(r)$.  Since familiar
Feynman diagram techniques fail and the Schwinger--Dyson equations
are intractable, the path integrals needed to determine $V_{Q\bar{Q}}(r)$
are estimated using Markov-chain Monte Carlo methods (Lattice QCD simulations).
The spectrum of gluonic excitations in the presence of a static
quark--antiquark pair has been accurately determined in lattice 
simulations \cite{JKM99,JKM03} which make use of anisotropic lattices,
improved actions, and large sets of operators with correlation matrix
techniques. These gluonic $V_{Q\bar{Q}}(r)$ levels may be classified
by the magnitude $\Lambda$ of the projection of the total angular momentum
${\bf J}_g$ of the gluon field onto the molecular axis, and by $\eta=\pm 1$,
the symmetry under charge conjugation combined with spatial inversion
about the midpoint between the quark and the antiquark.  States with
$\Lambda=0,1,2,\dots$ are denoted by $\Sigma, \Pi, \Delta, \dots$,
respectively.  States which are even (odd) under the above-mentioned
$CP$ operation are denoted by the subscripts $g$ ($u$).  An additional
$\pm$ superscript for the $\Sigma$ states refers to even or odd symmetry
under a reflection in a plane containing the molecular axis. 

In the leading Born--Oppenheimer approximation, one replaces
the covariant Laplacian $\bm{D}^2$ by an ordinary Laplacian $\bm{\nabla}^2$.
The error that one makes is equivalent to $1/M_Q$ and $1/M_Q^2$
corrections~\cite{m12}
to $V_{Q\bar{Q}}$
that go beyond the LBO and are suppressed by a factor $v^2$,
using perturbative NRQCD power counting rules.
The spin interactions of the 
heavy quarks are also neglected, and one solves the radial Schr\"odinger
equation:
\begin{equation}
 -\frac{1}{2\mu} \frac{d^2u(r)}{dr^2}
 + \left\{ \frac{\langle \bm{L}^2_{Q\bar{Q}}\rangle}{2\mu r^2}
   + V_{Q\bar{Q}}(r)\right\}u(r) = E\ u(r),
\end{equation}
where $u(r)$ is the radial wavefunction of the quark--antiquark pair
and $\mu$ denotes the reduced mass.
The expectation value in the centrifugal term is given in the 
adiabatic approximation by
\begin{equation}
  \langle \bm{L}^2_{Q\bar{Q}}\rangle=L(L+1)-2\Lambda^2
  +\langle\bm{J}_g^2\rangle,
\end{equation}
where
$\langle\bm{J}_g^2\rangle=0$ for the $\Sigma_g^+$ level and
$\langle\bm{J}_g^2\rangle=2$ for the $\Pi_u$ and $\Sigma_u^-$ levels.

\begin{figure}
\begin{center}
\includegraphics[width=.9\textwidth, bb=120 470 570 700,clip]{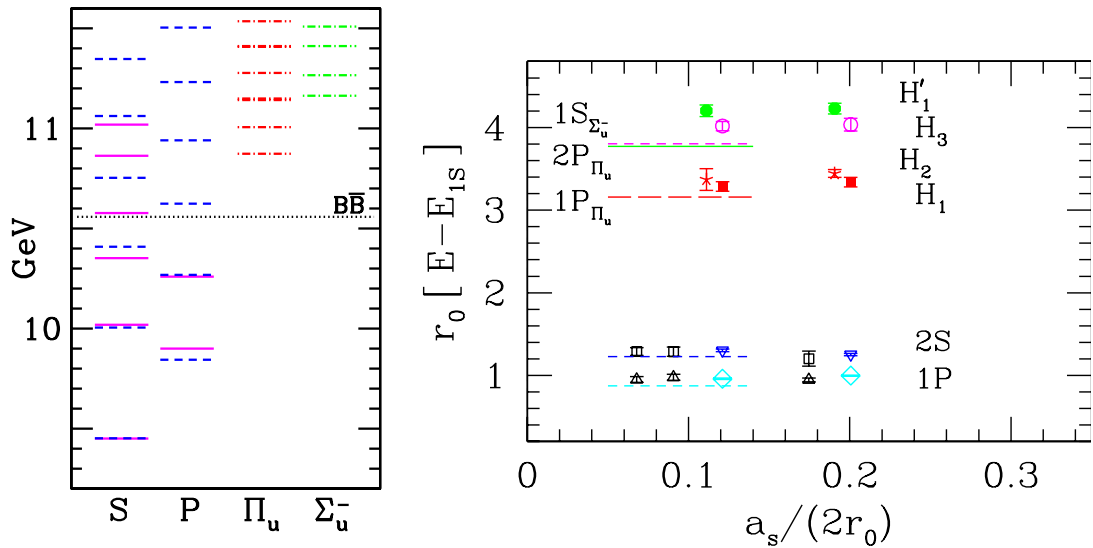}
\caption[cap1]
        {(Left) The spectrum of conventional and hybrid heavy-quark
         mesons in the leading Born--Oppenheimer approximation and
         neglecting light quarks (from Ref.\ \cite{JKM99}).
         Conventional $S$ and $P$ states are shown, as well as hybrids
         based on the $\Pi_u$ and $\Sigma_u^-$ adiabatic surfaces.
         Solid lines indicate spin-averaged experimental measurements.
         (Right) Simulation results from Ref.~\cite{JKM99} for two
         conventional and four hybrid bottomonium level splittings (in
         terms of $r_0^{-1}=450$~MeV and with respect to the $1S$
         state) against the lattice spacing $a_s$.  Predictions from
         the leading Born--Oppenheimer calculation, shown as horizontal
         lines, reproduce all of the simulation results to within
         10~\%, strongly supporting the validity of a Born--Oppenheimer
         picture for such systems at leading order.  Results from
         Ref.~\protect\cite{NRQCDWilson} using an NRQCD action with
         higher-order relativistic corrections are shown as hollow
         boxes and hollow upright triangles.
\label{fig:HQhybrids}}
\end{center}
\end{figure}

The leading-order Born--Oppenheimer spectrum of conventional $\bar{b}b$
and hybrid $\bar{b}gb$ states (in the absence of light quarks)
obtained from the above procedure is shown in
\Figure~\ref{fig:HQhybrids}.  Below the $\overline{B}B$ threshold, the
Born--Oppenheimer results agree well with the spin-averaged
experimental measurements of bottomonium states (any small
discrepancies essentially disappear once light quark loops are
included).  Above the threshold, agreement with experiment is lost,
suggesting significant corrections either from mixing and other
higher-order effects or (more likely) from light sea quark effects.

The validity of the Born--Oppenheimer picture relies on the smallness
of mixing between states based on different $V_{Q\bar{Q}}(r)$. In addition,
relativistic (including spin) corrections and
radiation of colour neutral objects such as glueballs and mesons are neglected.
In Ref.~\cite{JKM99} the LBO level splittings have been compared
with those determined from meson simulations using a 
non-relativistic (NRQCD) heavy-quark action.  The NRQCD
action included only a covariant temporal derivative and the leading
covariant kinetic energy operator; quark spin and $\bm{D}^4$ terms were
neglected. Differences between the two results originate from both
different ${\cal O}(1/M_Q)$ terms~\cite{m12} and from the automatic
inclusion of mixing effects between different adiabatic surfaces within
the NRQCD simulations. Naively one might expect the former effect
to be of ${\cal O}(v^2)\approx 10$~\%.
The level splittings (in terms of the hadronic scale $r_0$ and with respect
to the $1S$ state) of the conventional $2S$ and $1P$
states and four hybrid states were compared (see \Figure~\ref{fig:HQhybrids})
and indeed found to agree within $10 \%$, strongly supporting the validity
of the leading Born--Oppenheimer picture, at least in the absence of
light sea quarks and spin-effects.   

A very recent study \cite{burch} has demonstrated that the $\Upsilon$
ground state carries little admixture from hybrids, supporting the
LBO, at least in the sector that is governed by the ground state potential.
Using lowest-order lattice NRQCD to create heavy-quark propagators, a basis
of unperturbed S-wave and $\vert 1H\rangle$ hybrid states was formed.  
The $c_F \bm{\sigma\!\cdot\! B}/2M_Q$ spin interaction was then applied
at an intermediate time slice to compute the mixings between such
states due to this interaction in the quenched approximation.
Diagonalizing the resulting two-state Hamiltonian then yielded the
admixtures of hybrid configuration in the $\Upsilon$ and $\eta_b$.
For a reasonable range of $c_F$ values, the following results were
obtained:
$\langle 1H\vert\Upsilon\rangle \approx 0.076-0.11$ and
$\langle 1H\vert\eta_b\rangle \approx 0.13-0.19. $
Hence, hybrid mixings due to quark spin effects in bottomonium are very
small.  Even in charmonium, the mixings were found not to be large:
$\langle 1H\vert J/\Psi\rangle \approx 0.18-0.25 $ and
$\langle 1H\vert\eta_c\rangle \approx 0.29-0.4.$
Investigations of the mixing of hybrid states with radially excited
standard quarkonium states
which are energetically closer and spatially more extended are certainly an
exciting avenue of future research.

In the absence of light quark loops, one obtains a very dense spectrum
of mesonic states since the $V_{Q\bar{Q}}(r)$ potentials increase
indefinitely with $r$.  However, the inclusion of light quark loops
changes the $V_{Q\bar{Q}}(r)$ potentials.  First, there are slight
corrections at small $r$, and these corrections remove the small
discrepancies of the leading Born--Oppenheimer predictions with
experiment below the $B\overline{B}$ threshold seen in
\Figure~\ref{fig:HQhybrids}.  For large $r$, the inclusion of light
quark loops drastically changes the behavior of the $V_{Q\bar{Q}}(r)$
potentials: instead of increasing indefinitely, these potentials
eventually level off at a separation above 1~fm when the static
quark--antiquark pair, joined by gluonic flux, can undergo fission into
$(Q\bar{q})(\bar{Q}q)$, where $q$ is a light quark and $Q$ is a heavy
quark.  Clearly, such potentials cannot support the populous set of
states shown in \Figure~\ref{fig:HQhybrids}; the formation of bound
states and resonances substantially extending over 1~fm in diameter
seems unlikely.  A complete open-channel calculation taking the
effects of including the light quarks correctly into account has not
yet been done, but unquenched lattice simulations \cite{Bali:2000vr}
show that the $\Sigma_g^+$ and $\Pi_u$ potentials change very little
for separations below 1~fm when sea quarks are included. This makes it
conceivable that a handful of low-lying states whose wavefunctions do
not extend appreciably beyond 1~fm in diameter may exist as
well-defined resonances in nature.

In addition to such direct threshold effects there is the possibility of
transitions between different adiabatic surfaces, mediated by radiation
of pions and other light mesons or pairs of light mesons. A first
lattice study of such effects has been performed by McNeile and
Collaborators \cite{McNeile:2002az}.

A recent quenched calculation \cite{Liao:2001yh} of bottomonium hybrids using a
relativistic heavy-quark action on anisotropic lattices confirms the
predictions of the Born--Oppenheimer approximation, but admittedly,
the uncertainties in the simulation results are large.  These calculations
used a Symanzik-improved anisotropic gauge action and an improved
Fermion clover action. Quenched results on Charmonium hybrids obtained
by employing a relativistic quark actions \cite{Liao:2002rj} 
can be found in \Figure~\ref{fig:chquench} and \Table~\ref{tab:chquench} in
\Section~\ref{sec:spdqlc}.
The dominant decay channel for the lightest ($1^{-+}$) hybrid would be
into a $D$ and a $\overline{D}^{**}$ should it be heavier than
the respective threshold, and radiation of a light pseudoscalar or scalar
state if lighter.

A determination of the spectrum properly taking into account effects from
light quarks is still needed.  Taking the Born--Oppenheimer approximation
beyond leading order is also a project for future work.  Monte Carlo
computations of relevant matrix elements involving the gauge field
can not only facilitate the evaluation of higher-order terms in 
the Born--Oppenheimer expansion, but also provide valuable information on
the production and decays of these novel states.

\subsubsection[$QQq$ baryons on the lattice]
              {$QQq$ baryons on the lattice
               $\!$\footnote{Author: G.~Bali}}
\label{sec:spqqql}

While recent lattice results from several groups on three quark static
potentials exist~\cite{Bali:2000gf,Alexandrou:2001ip,Takahashi:2002bw,%
                       Okiharu:2003vt,Bornyakov:2004uv}, 
no such potentials have been calculated for the situation containing
two static sources at distance $r$, accompanied by a light quark, as
yet.  However, two groups have directly studied the situation for
$Q=c$, within the quenched approximation, one employing the so-called
D234 improved Wilson type action~\cite{Lewis:2001iz} as well as
NRQCD~\cite{Mathur:2002ce} on anisotropic lattices and the UKQCD
Collaboration employing the relativistic clover charm quark
action~\cite{Flynn:2003vz}.

In the NRQCD study~\cite{Mathur:2002ce} two lattice spacings,
$a\approx 0.15\,\mbox{fm},0.22\,\mbox{fm}$ and four light quark masses
have been realized and $bbq$, $ccq$ as well as $bqq$ and $cqq$ baryons
studied. No finite volume checks were performed and radiative
corrections to the NRQCD matching coefficients ignored. In the UKQCD
study~\cite{Flynn:2003vz} only one lattice spacing $a\approx 0.08$~fm
and one volume, $La\approx 2$~fm were realized. The light quark masses
scattered around the strange quark mass and both, singly
and doubly charmed baryons
were studied. All studies yield consistent results. The values quoted by
UKQCD are~\cite{Flynn:2003vz},
\begin{eqnarray}
\Xi_{cc}=3549(13)(19)(92)\,\mbox{MeV}\quad,\quad
\Omega_{cc}=3663(12)(17)(95)\,\mbox{MeV}\nonumber\\
\Xi^*_{cc}=3641(18)(08)(95)\,\mbox{MeV}\quad,\quad
\Omega^*_{cc}=3734(14)(08)(97)\,\mbox{MeV}.
\end{eqnarray}
The first errors are statistical, the second encapsulate uncertainties
in the chiral extrapolations and fit ranges. The third error represents
the uncontrolled systematics: finite $a$ effects,
finite volume effects and quenching, estimated by comparing the lattice
$\Lambda_c$ mass to the experimental result.

\subsection[pNRQCD]{pNRQCD $\!$\footnote{Authors: N.~Brambilla, J.~Soto}}
\label{sec:spnrqcd}

From the various dynamical scales that play a role in the heavy quarkonium systems, 
namely $m$, $mv$, $mv^2$ and $\lQ$, only the hard scale $m$
has been factorized in NRQCD and becomes explicit in 
its Lagrangian. Only the fact that $m \gg mv , mv^2 , \lQ$ is exploited
but no use is made of the scale separation, $mv \gg mv^2$.
A higher  degree of simplification is achieved by
building another effective theory, where degrees of freedom of
order $\sim mv$ are integrated out as well, 
\ie an  
EFT where only the ultrasoft degrees of freedom (with energies $\sim mv^2$) remain dynamical.
In this way a big simplification is obtained and analytic calculations of
the spectrum become feasible, at least in some dynamical regimes, at variance 
with NRQCD where the spectrum can only be obtained in a model independent
way by Lattice calculation.
 pNRQCD \cite{Pineda:1997bj,Brambilla:1999xf}  takes advantadge of the fact that for many non-relativistic systems
the scale associated to the size of the system $k\sim mv$ is much larger than the binding 
energy $E\sim mv^2$. Therefore it is possible to integrate  
out the scale of the momentum transfer $k$ in a way such that  pNRQCD is equivalent to NRQCD at any 
desired order in $E/k$, $k/m$ and $\als (\mu)$.
Two dynamical situations may occur here:  
(1) $k$ is much larger than $\lQ$,  (2) $k$ is of the order
of $\lQ$. In the first case
       the matching from NRQCD to pNRQCD may be performed in
 perturbation theory, expanding in terms of $\als$.  In the second situation,
  the matching has to be nonperturbative,
 \ie no expansion in $\als$ is allowed. We will refer to these two
limits as the weak and strong coupling regimes. Recalling that $k \sim r^{-1}
  \sim mv$, these two situations correspond  to systems with inverse
typical radius smaller or  bigger than $\lQ$, or systems
respectively dominated by the short range or long range (with respect
     to the confinement radius) physics.  We will consider these two situations
in the following two subsections.

 \subsubsection[Weak coupling regime: the pNRQCD Lagrangian; the
                static QCD potential; the heavy quarkonium energy levels;
                renormalization group improvement; nonperturbative effects.] 
               {Weak coupling regime
                $\!$\footnote{Authors: N.~Brambilla, J.~Soto}}
\label{sec:spnrqcdwc}

When $k \gg E >\sim \lQ$, we are in the perturbative matching regime
($v\sim \als (m\als)$). The scale $r\sim 1/(mv)$ is integrated out and
the pNRQCD Lagrangian consists of a singlet and an octet wave function
field interacting with respective potentials and coupled to ultrasoft
gluons.  The effective degrees of freedom are: $Q\bar{Q}$ states
(decomposed into a singlet and an octet wave function under colour
transformations) with energy of order of the next relevant scale,
$\lQ, mv^2$ and momentum ${\bf p}$ of order $mv$, plus ultrasoft
gluons $A_\mu({\bf R},t)$ with energy and momentum of order $\lQ,
mv^2$. All the gluon fields are multipole expanded (\ie  expanded in
$r$). The Lagrangian is then an expansion in the small quantities $
{p/m}$, ${ 1/(r m)}$ and ${\cal O}(\lQ, m v^2)\times r$.

The pNRQCD Lagrangian is given at the next to leading order (NLO) in
the multipole expansion by \cite{Brambilla:1999xf} (in the
centre-of-mass system):
\begin{eqnarray}
& &\hspace{-6mm}
{{\cal L}}_{\rm pNRQCD}=
  {\rm Tr} \Biggl\{ {\rm S}^\dagger \left( i\partial_0 - \frac{\mathbf{p}^2}{m} 
- V_s(r) -\sum_{n\geq 1} \frac{V_s^{(n)}}{m^n} \right) {\rm S} 
+ {\rm O}^\dagger \left( iD_0 - \frac{\mathbf{p}^2}{m} 
- V_o(r) - \sum_{n\geq 1} \frac{V_o^{(n)}}{m^n}  \right) {\rm O} \Biggr\}
\nonumber\\
& & \!\!\!\!\!\!\!\!
 + g V_A ( r) {\rm Tr} \left\{  {\rm O}^\dagger {\bf r} \cdot {\bf E} \,{\rm S}
+ {\rm S}^\dagger {\bf r} \cdot {\bf E} \,{\rm O} \right\} 
\label{eq:pnrqcd0}
   + g \frac{V_B (r)}{2} {\rm Tr} \left\{  {\rm O}^\dagger {\bf r} \cdot {\bf E} \, {\rm O} 
+ {\rm O}^\dagger {\rm O} {\bf r} \cdot {\bf E}  \right\} -\frac{1}{4} F^a_{\mu\nu}
F^{\mu \nu a}.  \label{eq:wc}
\end{eqnarray}
The $V_{s,o}^{(n)} , V_A , V_B$ are potentials, which play the role of
matching coefficients and contain the non-analytical dependence in
$r$, to be calculated in the matching between NRQCD and pNRQCD.
Poincar\'e invariance imposes relations among these matching
coefficients \cite{Brambilla:2003nt}.  To leading order in the
multipole expansion, the singlet sector of the Lagrangian gives rise
to equations of motion of the Schr\"odinger type. The other terms in
\Eq~(\ref{eq:pnrqcd0}) contain (apart from the Yang--Mills Lagrangian)
retardation (or non-potential) effects that start at the NLO in the
multipole expansion. At this order the non-potential effects come from
the singlet-octet and octet-octet interactions mediated by an
ultrasoft chromoelectric field.

Recalling that ${ r} \sim 1/(mv)$ and that the operators count like
the next relevant scale, ${\cal O}(mv^2,$ $\lQ)$, to the power of the
dimension, it follows that each term in the pNRQCD Lagrangian has a
definite power counting.  As a consequence of this power counting the
interaction of quarks with ultrasoft gluons is suppressed in the
Lagrangian by a factor $v$ ( by $g v$ if $mv^2 \gg \lQ$) with respect
to the LO.

The various potentials in \Eq~(\ref{eq:wc}) have been calculated at
different orders in the perturbative matching. $V_s$ is known to two
loops [${\cal O}(\als^3)$] \cite{Schroder:1998vy,Peter:1996ig} as well
as the leading log of the three loop contribution
\cite{Brambilla:1999qa}. $V_o$ is known to two loops (see York
Schr\"oder, private communications in Ref.\ \cite{Bali:2003jq}).
$V_s^{(1)}$ is known to two loops \cite{Kniehl:2001ju} and $V_s^{(2)}$
to one loop \cite{Kniehl:2002br}. $V_A$ and $V_B$ are known at tree
level \cite{Brambilla:1999xf} (and are independent of $r$) and have no
logs at one loop \cite{Pineda:2000gz}.

Note that the static limit of pNRQCD ($m\rightarrow \infty$) results
in a nontrivial theory (unlike in pNRQED), since both singlet and
octet fields remain dynamical and interact through ultrasoft
gluons. The static energy of two infinitely heavy sources
$V_{QCD}(r)$, which will be discussed below, can be obtained for small
$r$.  In fact, the coefficient of the infrared logarithmic
contribution to $V_{QCD}(r)$ first pointed out in Ref.\
\cite{Appelquist:es} was calculated using the static pNRQCD Lagrangian
\cite{Brambilla:1999qa}.

Given the Lagrangian in \Eq~(\ref{eq:pnrqcd0}) it is possible to
calculate the quarkonium energy levels. Contributions to the spectrum
originate both in quantum mechanical perturbation theory and in the
dynamics of ultrasoft gluons. The latter contributions contain
nonperturbative effects and this will be discussed in the
corresponding section below.

\subsubsection*{The static QCD potential\footnote{Author: Yu.~Sumino}}

For decades, the static QCD potential $V_{\rm QCD}(r)$, formally
defined from an expectation value of the Wilson loop, has been widely
studied for the purpose of elucidating the nature of the interaction
between heavy quark and antiquark.  The potential at short distances
can be computed by perturbative QCD, whereas its long distance shape
can be computed by lattice simulations.  (See
\Sections~\ref{sec:spnrqcdscc} and \ref{sec:spstatconf} for lattice
computations.)

Computations of $V_{\rm QCD}(r)$ in perturbative QCD have a long
history. The 1-loop and 2-loop corrections were computed in Refs.\
\cite{Appelquist:tw,Fischler:1977yf,Billoire:1979ih} and
\cite{Peter:1996ig,Schroder:1998vy,Melles:2000dq,Melles:2000ey,
Hoang:2000fm,Recksiegel:2001xq}, respectively.  The logarithmic
correction at 3-loops originating from the ultrasoft scale was first
pointed out in Ref.\ \cite{Appelquist:es} and computed in Refs.\
\cite{Brambilla:1999qa,Kniehl:1999ud}.  A renormalization-group (RG)
improvement of $V_{\rm QCD}(r)$ at next-to-next-to-leading log (NNLL)
was performed in Ref.\ \cite{Pineda:2000gz}.\footnote{ There are
estimates of higher-order corrections to the perturbative QCD
potential in various methods
\cite{Chishtie:2001mf,Pineda:2001zq,Cvetic:2003wk}.  }

Since the discovery \cite{Pineda:id,Hoang:1998nz,Beneke:1998rk} of the 
cancellation of ${\cal O}(\Lambda_{\rm QCD})$
renormalons between $V_{\rm QCD}(r)$ and twice the
quark pole mass\footnote{For similar work inside HQET see \cite{renheft}.},
the convergence of the perturbative series improved drastically and
much more accurate perturbative predictions 
of the potential shape became available.
This feature indicates the validity of the renormalon dominance
picture for the QCD potential and pole mass.
According to this picture, a perturbative
uncertainty of $V_{\rm QCD}(r)$, 
after cancelling the ${\cal O}(\Lambda_{\rm QCD})$ renormalon, is 
estimated to be 
${\cal O}(\Lambda_{\rm QCD}^3 r^2)$
at $r \ll \Lambda_{\rm QCD}^{-1}$ \cite{Aglietti:1995tg}.

An OPE of $V_{\rm QCD}(r)$ was developed  within the pNRQCD
framework \cite{Brambilla:1999xf}.
In this framework, residual renormalons, starting from 
${\cal O}(\Lambda_{\rm QCD}^3 r^2)$,
are absorbed into
the matrix element of a non-local operator
(non-local gluon condensate).
Then, in the multipole expansion at $r \ll \Lambda_{\rm QCD}^{-1}$, 
the leading nonperturbative contribution to the potential 
becomes ${\cal O}(\Lambda_{\rm QCD}^3 r^2)$ \cite{Brambilla:1999xf}.

Several studies 
\cite{Sumino:2001eh,Recksiegel:2001xq,Necco:2001gh,
Pineda:2002se,Recksiegel:2002um}
showed that perturbative
predictions for  $V_{\rm QCD}(r)$ agree well
with phenomenological potentials (determined from heavy quarkonium
spectroscopy) and lattice calculations of $V_{\rm QCD}(r)$, 
once the ${\cal O}(\Lambda_{\rm QCD})$ renormalon is accounted for.
Ref.\ \cite{Lee:2002sn} showed that 
also a Borel 
resummation of the perturbative series yields a potential shape
in agreement with lattice results if the ${\cal O}(\Lambda_{\rm QCD})$ 
renormalon is properly treated.
In fact the agreement holds within
the expected ${\cal O}(\Lambda_{\rm QCD}^3 r^2)$ uncertainty.\footnote{
This is true only
in the range of $r$ where the respective perturbative predictions are stable.
All perturbative predictions become uncontrolled 
beyond certain distances, typically around $r \sim \Lambda_{\rm QCD}^{-1}$.
}
These observations further support the validity of renormalon dominance
and of the OPE for $V_{\rm QCD}(r)$.

Qualitatively, 
the perturbative QCD potential becomes steeper than
the Coulomb potential as $r$ increases
(once the ${\cal O}(\Lambda_{\rm QCD})$ renormalon is cancelled).
This feature can be understood, within perturbative QCD, 
as an effect of the {\it running} of the strong coupling constant 
\cite{Brambilla:2001fw,Sumino:2001eh,Necco:2001gh}.

Using a scale-fixing prescription based on the renormalon dominance picture,
it was shown analytically \cite{Sumino:2003yp} that the
perturbative QCD potential approaches a ``Coulomb+linear''
form at large orders, up to an ${\cal O}(\Lambda_{\rm QCD}^3 r^2)$ uncertainty.
The ``Coulomb+linear'' potential can be computed systematically 
as more terms of perturbative series are included
via RG; up to NNLL,
it shows a convergence towards lattice results.

\medskip

{\it  Heavy quarkonium spectra \footnote{Author: Yu.~Sumino}}\\
\noindent
In recent years, perturbative computations of the heavy quarkonium spectrum 
(an expansion in $\alpha_{\rm s}$ and $\ln\alpha_{\rm s}$) 
have enjoyed a significant development.
A full computation of the spectrum
up to ${\cal O}(\alpha_{\rm s}^4 m)$ was performed in 
Refs.\ \cite{Pineda:1997hz,Titard:1994id}.
The spectra up to the same order for the system
with unequal heavy quark masses and
with non-zero quark mass in internal loops were computed, respectively,
in Refs.\ \cite{Brambilla:2000db,Brambilla:2001qk} and 
\cite{Hoang:2000fm,Brambilla:2001qk}.
Perturbative computations at higher orders were made possible by
the advent of effective field theories such as pNRQCD 
\cite{Pineda:1997bj,Brambilla:1999xf}
or vNRQCD \cite{Luke:1999kz}
and by the threshold expansion technique \cite{Beneke:1997zp}.
The ${\cal O}(\alpha_{\rm s}^5 m \ln \alpha_{\rm s} )$ term originating from 
the ultrasoft scale was computed in
Refs.\ \cite{Brambilla:1999qa,Brambilla:1999xj,Kniehl:1999ud}.
Ref.\ \cite{Pineda:2001ra,Hoang:2002yy} resummed the 
$\alpha_{\rm s}^4 m (\alpha_{\rm s} \ln \alpha_{\rm s} )^n$ terms.
The full Hamiltonian at the next-to-next-to-next-to-leading order 
was computed in Ref.\ \cite{Kniehl:2002br}.
Except for the 3-loop non-logarithmic term of the perturbative QCD 
potential,\footnote{
Estimates of the 3-loop correction to the QCD potential have been given
in various methods \cite{Chishtie:2001mf,Pineda:2001zq,Cvetic:2003wk}.
}
the energy levels of the $1S$ states were
computed up to ${\cal O}(\alpha_{\rm s}^5 m)$ from this Hamiltonian
\cite{Penin:2002zv}. The fine splittings have been calculated at NLO order 
${\cal O}(\alpha_{\rm s}^5 m)$ in \cite{Brambilla:2004wu}.

In the meantime, the discovery of the renormalon cancellation in the
quarkonium spectrum \cite{Pineda:id,Hoang:1998nz,Beneke:1998rk} led to
a drastic improvement of the convergence of the perturbative expansion
of the energy levels.  (See
\Chapter~\ref{chapter:precisiondeterminations} for precise
determinations of the heavy quark masses, as important applications.)
In Refs.~\cite{Brambilla:2001fw,Brambilla:2001qk} the whole structure
of the bottomonium spectrum up to ${\cal O}(\alpha_{\rm s}^4 m)$ was
predicted taking into account the cancellation of the ${\cal
O}(\Lambda_{\rm QCD})$ renormalons, and a good agreement with the
experimental data was found for the gross structure of the spectrum.
(Only the states below the threshold for strong decays were
considered.)  The consistency of the perturbative predictions with the
experimental data seems to indicate that, for bottomonium, the
momentum scale of the system is larger than $\Lambda_{\rm QCD}$, \ie
$m v \gg \Lambda_{\rm QCD}$, up to some of the $n=3$ states.  This is,
however, in apparent conflict with the fact that the leading
nonperturbative effects scale as a power $\ge 4$ of the principal
quantum number (see {\it Nonperturbative effects} below) and, hence,
are expected to be very important for any excited state.

Subsequently, in Refs.\ \cite{Recksiegel:2002za,Recksiegel:2003fm} a
specific formalism based on perturbative QCD was developed: using the
static QCD potential computed in Ref.\ \cite{Recksiegel:2001xq} and
taking into account the cancellation of the ${\cal O}(\Lambda_{\rm
QCD})$ renormalons, the Schr\"odinger equation was solved numerically
to determine the zeroth-order quarkonium wave function; all the
corrections up to ${\cal O}(\alpha_{\rm s}^5 m)$ for the fine and
hyperfine splittings have been included.  Good agreements were found
between the computed and the observed fine and hyperfine splittings of
the bottomonium and charmonium spectra, in addition to the gross
structure of the bottomonium spectrum\footnote{For technical reasons a
linear extrapolation of the potential at $r>4.5~{\rm GeV}^{-1}$ was
introduced in Ref.\ \cite{Recksiegel:2002za}.  This artefact was
eliminated in Ref.\ \cite{Recksiegel:2003fm}, in which it was also
shown that effects caused by the linear extrapolation of the potential
were minor.}.

\begin{table}
\caption[Predicted masses of $b\bar{b}$, $c\bar{c}$ and $b\bar{c}$ 
         states in perturbative QCD-based, 
         renormalon-subtracted computations]
        {Predicted masses of $b\bar{b}$, $c\bar{c}$ and $b\bar{c}$
         states in perturbative QCD-based, renormalon-subtracted
         computations.  BSV01 (and BV00) is the full perturbative
         computation up to ${\cal O}(\alpha_{\rm s}^4 m)$ without
         non-zero charm-mass corrections; BSV02 is the full
         perturbative computation up to ${\cal O}(\alpha_{\rm s}^4 m)$
         including non-zero charm-mass corrections; RS03 is based on a
         specific scheme and specific reorganization of perturbative
         series, incorporates full corrections up to ${\cal
         O}(\alpha_{\rm s}^4 m)$ in the individual levels and full
         corrections up to ${\cal O}(\alpha_{\rm s}^5 m)$ in the fine
         splittings, includes non-zero charm-mass corrections.  Errors
         shown in brackets represent $\sqrt{\delta_{\alpha_{\rm
         s}}^2+\delta_{\rm h.o.}^2 }$ (BSV01,BV00) and
         $\sqrt{\delta_{\alpha_{\rm s}}^2+\delta_{\rm h.o.}^2 +
         \delta_{m_c}^2}$ (BSV02), respectively, where
         $\delta_{\alpha_{\rm s}}$ originates from the error of
         $\alpha_{\rm s}(M_Z)$, $\delta_{\rm h.o.}$ is the error due
         to higher-order corrections, and $\delta_{m_c}$ is the error
         in the finite charm mass corrections. The errors do not
         include nonperturbative contributions estimates.  Numbers
         without errors are those without explicit or reliable error
         estimates in the corresponding works.}
\label{tab:predictedmasses}
\setlength{\extrarowheight}{4pt}
\renewcommand{\arraystretch}{1.0}
\begin{center}
\begin{tabular}{|>{\strut}c|c|llrr|}
\hline
State & expt & BSV01\cite{Brambilla:2001fw} & 
BSV02\cite{Brambilla:2001qk} & RS03\cite{Recksiegel:2002za}
& BV00\cite{Brambilla:2000db}
 \\
\hline
\multicolumn{6}{|c|}{$b\bar{b}$ states}\\
\hline
$1^3S_1$ & 9460 & 9460 & 9460 & 9460 &  \\
\hline
$1^3P_2$ & 9913 & 9916(59) & 10012(89) & 9956 & \\
$1^3P_1$ & 9893 & 9904(67) & 10004(86) & 9938&  \\
$1^3P_0$ & 9860 & 9905(56) &  9995(83) & 9915&  \\
\hline
$2^3S_2$ & 10023 & 9966(68) & 10084(102) & 10032& \\
\hline
$2^3P_2$ & 10269 &  & 10578(258) & 10270& \\
$2^3P_1$ & 10255 &  & 10564(247) & 10260& \\
$2^3P_0$ & 10232 & 10268 & 10548(239) & 10246& \\
\hline
$3^3S_1$ & 10355 & 10327(208) & 10645(298) & 10315& \\
\hline
\multicolumn{6}{|c|}{$c\bar{c}$ states}\\
\hline
$1^3S_1$ & 3097 & 3097 & & &\\
$1^1S_0$ & $2980(2)$ & 3056 & && \\
\hline
\multicolumn{6}{|c|}{$b\bar{c}$ states}\\
\hline
$1^1S_0$ & $6400(400)$ & 6324(22) & 6307(17) & & 6326 (29) \\
\hline
\end{tabular}
\end{center}
\end{table}

In \Table~\ref{tab:predictedmasses} particularly impressing is the
result for the perturbative calculation of the $B_c$ mass, that, with
finite charm mass effects included, is equal to $6307 \pm 17 {\rm
GeV}$ and is in complete agreement, inside errors and with small
errors, with lattice NRQCD unquenched result given in \Eq~(4).

These analyses have shown that the perturbative predictions of the
spectra agree with the corresponding experimental data within the
estimated perturbative uncertainties, and that the size of
nonperturbative contributions is compatible with the size of
perturbative uncertainties.

Although uncertainties of the perturbative predictions for the
individual energy levels grow rapidly for higher excited states, level
spacings among them have smaller uncertainties, since the errors of
the individual levels are correlated.  In particular, uncertainties of
the fine and hyperfine splittings are suppressed due to further
cancellation of renormalons.  These features enabled sensible
comparisons of the level structures including the excited states.

In predicting the spectrum, pNRQCD is a useful tool not only for fully
perturbative computations but also for factorizing short-distance
contributions into matching coefficients (perturbatively computable)
and nonperturbative contributions into matrix elements of operators
\cite{Brambilla:1999xf,m12}. This will be discussed in {\it
Nonperturbative effects} below.

\medskip

{\it The Renormalization group in heavy quarkonium spectroscopy
\footnote{Author: A.~Pineda}}\\
\noindent
In recent years, there has been a growing interest to perform
renormalization group analysis in heavy quarkonium
\cite{Luke:1999kz,Manohar:1999xd,Manohar:2000kr,Pineda:2000gz,Hoang:2000ib,
Hoang:2001rr,Pineda:2001ra,Pineda:2001et,Hoang:2002yy,Hoang:2003ns,
Pineda:2003be,Kniehl:2003ap,Penin:2004xi,Penin:2004ay}.  In many cases
this interest has been driven by the lack of convergence and strong
scale dependence one finds in the fixed (NNLO) analysis performed for
sum rules and $t$--$\bar t$ production near threshold (see
\Chapter~\ref{chapter:precisiondeterminations}). This problem has
turned out to be highly non-trivial. We will focus here on
computations related with spectroscopy.

The heavy quarkonium spectrum is known with NNLL accuracy
\cite{Pineda:2001ra,Hoang:2002yy}. These expressions have not yet been used
for phenomenological analysis of single heavy quarkonium states either in
bottomonium and charmonium systems. It would be very interesting to see
their effects on the spectra.

\begin{table}
\caption[Predicted fine and hyperfine splittings (in MeV) of
         $b\bar{b}$ and $c\bar{c}$ states in perturbative QCD-based,
         renormalon-subtracted computations]
        {Predicted fine and hyperfine splittings (in MeV) of
         $b\bar{b}$ and $c\bar{c}$ states in perturbative QCD-based,
         renormalon-subtracted computations.  $^3P_{\rm cog}$ denotes
         the centre of gravity of the triplet P-wave states.  PT88
         extracts the matrix elements of ${\cal O}(1/m^2)$ operators
         from the experimental values for the fine splittings, instead
         of computing them from perturbative QCD.  BSV01 is the full
         perturbative computation up to ${\cal O}(\alpha_{\rm s}^4 m)$
         without non-zero charm-mass corrections.  BSV02 is the full
         perturbative computation up to ${\cal O}(\alpha_{\rm s}^4 m)$
         including non-zero charm-mass corrections; RS03 and RS04 are
         based on specific schemes and specific reorganization of
         perturbative series, incorporate full corrections up to
         ${\cal O}(\alpha_{\rm s}^5 m)$ in the splittings, and include
         non-zero charm-mass corrections.  KPPSS03 and PPSS04 are the
         full NNLL computation [up to order $\alpha_{\rm s}^5 m \times
         \ (\alpha_{\rm s} \ln \alpha_{\rm s})^n)$] without non-zero
         charm-mass corrections.  Errors are shown in brackets when
         explicit and reliable estimates are given in the respective
         works. The errors do not include nonperturbative
         contributions estimates except in KPPSS03 and PPSS04 where
         they were roughly estimated using the multipole expansion.}
\label{tab:predictedsplittings}
\begin{center}
\setlength{\tabcolsep}{3pt}
\setlength{\extrarowheight}{4pt}
\renewcommand{\arraystretch}{1.0}
\begin{tabular}{|>{\strut}c|c|rrrrrrr|}
\hline
{\small Level splitting} & {\small expt} & 
{{\tiny PT88}{\small \cite{Pantaleone:1987qh}}} & 
{{\tiny BSV01}{\small \cite{Brambilla:2001fw}}} & 
{{\tiny BSV02} {\small \cite{Brambilla:2001qk}}} & 
{{\tiny RS03}{\small \cite{Recksiegel:2002za}}} & 
{{\tiny RS04}{\small\cite{Recksiegel:2003fm}}} & 
{{\tiny KPPSS03}{\small\cite{Kniehl:2003ap}}}&
{{\tiny PPSS04} {\small\cite{Penin:2004xi}}} \\
\hline
\multicolumn{9}{|c|}{$b\bar{b}$ states}\\
\hline
$1^3P_2$ -- $1^3P_1$ & 20 & & 12   & 8 & 18(10) & & &\\
$1^3P_1$ -- $1^3P_0$ & 33 & & $-1$ & 9 & 23(10) & & &\\
$2^3P_2$ -- $2^3P_1$ & 13 & &      & 16 & 11(10) & & &\\
$2^3P_1$ -- $2^3P_0$ & 23 & &      & 14 & 14(10) & & &\\
\hline
$1^3S_1$ -- $1^1S_0$ & & & & & & 44(11) & $39(11)^{+9}_{-8}$ &\\
$2^3S_1$ -- $2^1S_0$ & & & & & & 21(8) &  &\\
$3^3S_1$ -- $3^1S_0$ & & & & & & 12(9) &  &\\
\hline
$1^3P_{\rm cog}$ -- $1^1P_1$ & & $-0.5$ & & & & $-0.4(0.2)$ & &\\
$2^3P_{\rm cog}$ -- $2^1P_1$ & & $-0.4$ & & & & $-0.2(0.1)$ & &\\
\hline
\multicolumn{9}{|c|}{$c\bar{c}$ states}\\
\hline
$1^3P_2$ -- $1^3P_1$ & 46 & & & & & 43(24) & &\\
$1^3P_1$ -- $1^3P_0$ & 95 & & & & & 56(34) & &\\
\hline
$1^3S_1$ -- $1^1S_0$ & 118(1) & & & & & 88(26) & $104$& \\
$2^3S_1$ -- $2^1S_0$ & 32(10) & & & & & 38(36) &  &\\
\hline
$1^3P_{\rm cog}$ -- $1^1P_1$ & $-0.9$ & $-1.4$ & & & & $-0.8(0.8)$ & &\\
\hline
\multicolumn{9}{|c|}{$b\bar{c}$ states}\\
\hline
$1^3S_1$ -- $1^1S_0$ & & & & & &  & & $65(24)^{+19}_{-16}$ \\
\hline
\end{tabular}
\end{center}
\end{table}

The hyperfine splitting of the heavy quarkonium spectrum is known with LL
\cite{Hoang:2001rr,Pineda:2001et} and NLL accuracy for the
bottomonium and charmonium spectrum \cite{Kniehl:2003ap} and also
for the $B_c$ spectrum \cite{Penin:2004xi}. For those observables
a phenomenological analysis has been performed. The predictions
can be found in \Table~\ref{tab:predictedsplittings}.  The general trend
is that the introduction of these effects improves the agreement with
experiment (when experimental data are available). In particular, the
resummation of logarithms brings the perturbative prediction of the 
hyperfine
splitting of charmonium significantly closer to the experimental figure if
compared with a NLO computation. It is then possible to give predictions for
the hyperfine splitting of the ground state of bottomonium, and in
particular for the $\eta_b(1S)$ mass, as well as for the hyperfine
splitting of the $B_c$ ground state.
In these
computations a threshold mass was used (equivalent to the pole mass at this
order).
In any case, it should also be mentioned that the use of the $\MS$
mass may give a NLO value for the charmonium hyperfine splitting in 
agreement with experiment \cite{vairo04}.

As a final remark, for the bottomonium, charmonium and $B_c$ spectrum,
one should be careful, since the ultrasoft scale may run up to very low scales. On the other 
hand the general dependence on the renormalization scale appears to be the same no
matter whether we talk of toponium, bottomonium or charmonium. This may 
point to the fact that the same physics holds for all of them.

\medskip

{\it Nonperturbative effects\footnote{Authors: N.~Brambilla, J.~Soto}}\\ 
\noindent
Given the Lagrangian in \Eq~(\ref{eq:pnrqcd0}) it is possible to calculate
the full quarkonium energy levels at order $m\als^5$
\cite{Brambilla:1999xj,Kniehl:1999ud,Kniehl:2002br}.  At this order
the energy $E_n$ of the level $n$ receives contributions both from
standard quantum mechanics perturbation theory and from the
singlet-octet interaction (retardation effect) through ultrasoft
gluons. The latter reads
\begin{equation}
\delta E_{n}\vert_{us}  =  
-i \frac{g^2}{3 N_c}  \!\! \int_0^\infty \!\!\! dt \,  
 \langle n|  {\bf r} e^{it( E_n^{(0)} - h_o)} {\bf r}  | n \rangle  
\;  \langle {\bf E} (t) \,   {\bf E}  (0)  \rangle(\mu) .
\label{eq:energyleveln}
\end{equation}
being $E_n^{(0)}$ and $h_o$ the binding energy and the octet
Hamiltonian respectively, at leading order.  When we assume that the
chromoelectric fields have a typical scale $\sim \lQ$, the expression
(\ref{eq:energyleveln}) allows to discuss the nature of the leading
nonperturbative contributions.  Thus the integral in
(\ref{eq:energyleveln}) is a convolution of two objects: the
exponential with a typical scale $mv^2$ and the chromoelectric
correlator with a typical scale $\lQ$.  Depending on the relative size
of the two scales three different situations occur:
\begin{itemize}
\item if $mv^2 \gg \lQ$, the correlator reduces to the local gluon
      condensate and one recovers the result of Refs.\
      \cite{Voloshin:hc,Leutwyler:1980tn}, which is proportional to
      the sixth power of the principal quantum number.  The NLO
      nonperturbative contribution has been evaluated in Ref.\
      \cite{Pineda:1996uk}.  Note, however, that in this case the
      dominant contribution to the nonlocal chromoelectric correlator
      corresponds to fluctuations of order $mv^2$, which can be
      calculated perturbatively \cite{Brambilla:1999xj,Kniehl:1999ud}.

\item if $mv^2 \ll \lQ$, the exponential can be expanded and one
      obtains a quadratic short range nonperturbative potential
      \cite{Balitsky:iw,Brambilla:1999xf}. This potential absorbs the
      residual renormalons contained in the fully perturbative
      computations \cite{Brambilla:1999xf}.  For a Coulombic system,
      its expectation value grows as the fourth power of the principal
      quantum number.

\item if $mv^2 \sim \lQ$, no expansion can be performed and the
      nonlocal condensate has to be kept.  Its expectation value grows
      as the fourth power of the principal quantum number
      \cite{Brambilla:1999xj}.
\end{itemize}
Hence, both nonperturbative potentials and (non-potential) local
condensates are obtained from pNRQCD in the weak coupling regime for
different kinematical limits, see also \cite{Brambilla:2003ut}.

\subsubsection[Strong coupling regime] 
              {Strong coupling regime
               $\!$\footnote{Authors: N.~Brambilla, J.~Soto}}
\label{sec:spnrqcdscc}

When $k >\sim \lQ \gg E$, the pNRQCD Lagrangian consist of a singlet
wave function field interacting with a potential and with
pseudo-Goldstone bosons \cite{Brambilla:1999xf}.  The dynamics of the
singlet field $\rm S$ is described by the following Lagrangian (here,
we do not specialize to the centre-of-mass system) \cite{m12,sw}
\begin{equation}
\quad  {\cal L}_{\rm pNRQCD}= {\rm Tr} \,\left\{ {\rm S}^\dagger
   \left(i\partial_0- \frac{\mathbf{p}_1^2}{2m_1} - 
\frac{\mathbf{p}_2^2}{2m_2}
-V({\bf x}_1,{\bf x}_2, {\bf p}_1, {\bf p}_2)\right){\rm S}  \right\}
\label{eq:scr}
\end{equation}
The dynamics of the pseudo-Goldstone boson is given by the Chiral
Lagrangian \cite{Gasser:1983yg}.  The coupling of pseudo-Goldston
bosons with the singlet field has not been worked out yet. If we
ignore this coupling, we recover in \Eq~(\ref{eq:scr}) the structure
of non-relativistic potential models \cite{m12,sw}.  If we assume that
$V$ is analytical in $1/m$, the structure of the potential up to order
$1/m^2$ is
\begin{eqnarray}
&& \hspace{-14mm}
V({\bf x}_1,{\bf x}_2, {\bf p}_1, {\bf p}_2)
 = V^{(0)}(r)
+\frac{V^{(1,0)}(r)}{m_1} + \frac{V^{(0,1)}(r)}{m_2} + \frac{V^{(2,0)}}{m_1^2}
+ \frac{V^{(0,2)}}{m_2^2} + \frac{V^{(1,1)}}{m_1m_2}, \label{eq:hs} \\
&& \hspace{-14mm}
V^{(2,0)} = \frac{1}{2}\left\{{\bf p}_1^2,V_{{\bf p}^2}^{(2,0)}(r)\right\}
+ \frac{V_{\mathbf{L}^2}^{(2,0)}(r)}{r^2}{\bf L}_1^2
+ V_r^{(2,0)}(r) + V^{(2,0)}_{LS}(r){\bf L}_1\cdot{\bf S}_1, \label{eq:v20}\\
&& \hspace{-14mm}
V^{(0,2)} = \frac{1}{2}\left\{{\bf p}_2^2,V_{{\bf p}^2}^{(0,2)}(r)\right\}
+ \frac{V_{\mathbf{L}^2}^{(0,2)}(r)}{r^2}{\bf L}_2^2
+ V_r^{(0,2)}(r) - V^{(0,2)}_{LS}(r){\bf L}_2\cdot{\bf S}_2, \label{eq:v02}\\
&& \hspace{-14mm}
V^{(1,1)} = - \frac{1}{2}\left\{{\bf p}_1\cdot {\bf p}_2,V_{{\bf p}^2}^{(1,1)}(r)\right\}
- \frac{V_{\mathbf{L}^2}^{(1,1)}(r)}{2r^2}({\bf L}_1\cdot{\bf L}_2
+ {\bf L}_2\cdot{\bf L}_1)+ V_r^{(1,1)}(r) \nn \\ 
&& \hspace{-14mm}
+ V_{L_1S_2}^{(1,1)}(r){\bf L}_1\cdot{\bf S}_2 - V_{L_2S_1}^{(1,1)}(r){\bf L}_2\cdot{\bf S}_1
+ V_{S^2}^{(1,1)}(r){\bf S}_1\cdot{\bf S}_2 + V_{S_{12}}^{(1,1)}(r){\bf S}_{12}({\hat {\bf r}})
\label{eq:v11}, 
\end{eqnarray}
where $r=\vert {\bf r}\vert$, ${\bf r} = {\bf x}_1-{\bf x}_2$, ${\bf
L}_j \equiv {\bf r} \times {\bf p}_j$ and ${\bf S}_{12}({\hat {\bf
r}}) \equiv 12 {\hat {\bf r}}\cdot {\bf S}_1 \,{\hat {\bf r}}\cdot
{\bf S}_2 - 4 {\bf S}_1\cdot {\bf S}_2$.  The requisite of Poincar\'e
invariance imposes well defined relations among the spin-dependent and
velocity dependent potentials above
\cite{poincare0,wilsonrel,gromes84}.  If one further assumes that the
matching to NRQCD can be done in the $1/m$ expansion, the explicit
form of the potentials can be obtained in terms of Wilson loop
operators
\cite{m12,wilsonrel,gromes84,wilson74,eichten80,Chen:1994dg,wilsonpot}. We
display here some of them for illustration (for the form of all the
potentials see \cite{m12}).  For the static potential we have
\begin{equation}
V^{(0)}(r) \!\!=\!\! \lim_{T\to\infty}\frac{i}{T} \ln \langle W \rangle, 
\label{eq:v0}
\end{equation}
for the potential at order $1/m$
\begin{equation}
 V_{s}^{(1,0)}(r)   = \lim_{T\to\infty}
- \frac{g^2}{4 T}\int_{-T/2}^{T/2} \!\! dt \int_{-T/2}^{T/2} \!\!dt^\prime
\vert t -t^\prime \vert 
 \langle\!\langle {\bf E}(t) \cdot {\bf E}(t^{\prime})\rangle \! \rangle_c . 
\label{eq:v1E}
\end{equation}

At the order $1/m^2$ we display a potential contributing to the
spin-dependent (precisely the spin--orbit) relativistic corrections
\begin{eqnarray}
V_{LS}^{(2,0)}(r) \!\!&=&\!\!  \frac{c_F^{(1)}}{2r^2}i {\bf r}\cdot \lim_{T\rightarrow \infty}
\frac{1}{T} \int_{-T/2}^{T/2}\!\! dt \int_{-T/2}^{T/2}\!\!dt'' \, (t-t'') \,  
\lla g{\bf B}({\bf x}_1,t'') \times g{\bf E}({\bf x}_1,t) \rra \nn\\
& & + \frac{c_S^{(1)}}{2 r^2}{\bf r}\cdot (\bfnabla_r V^{(0)}),
\label{eq:vls20}
\end{eqnarray}
and a potential contributing to the spin-independent velocity dependent 
relativistic corrections
\begin{eqnarray}
V_{{\bf p}^2}^{(2,0)}(r) \!\!&=&\!\! \frac{i}{4}{\hat {\bf r}}^i{\hat {\bf r}}^j
\lim_{T\rightarrow \infty}\frac{1}{T} \int_{-T/2}^{T/2}\!\!dt \int_{-T/2}^{T/2}\!\!dt'' \, (t-t'')^2 \,  
\lla g{\bf E}^i({\bf x}_1,t'') \, g{\bf E}^j({\bf x}_1,t) \rra_c .
\end{eqnarray} 
The angular brackets $\langle \dots \rangle$ stand for the average value over the
Yang--Mills action, $W$ for the rectangular static Wilson loop of extension 
$r\times T$ (the time runs from $-T/2$ to $T/2$, the space coordinate from ${\bf x}_1$ 
to ${\bf x}_2$):
\be
W \equiv {\rm P} \exp\left\{{\displaystyle - i g \oint_{r\times T} \!\!dz^\mu A_{\mu}(z)}\right\},
\qquad 
dz^\mu A_{\mu} \equiv dz^0 A_0 - d{\bf z} \cdot {\bf A},
\ee
and $\langle\!\langle \dots \rangle\!\rangle 
\equiv \langle \dots W\rangle / \langle  W\rangle$; P is the path-ordering operator.
Moreover, we define the connected Wilson loop with $O_1(t_1)$, $O_2(t_2)$ and $O_3(t_3)$ operator 
insertions by: 
\begin{equation}
\hspace{-5mm} \lla O_1(t_1)O_2(t_2)\rra_c = 
\lla O_1(t_1)O_2(t_2)\rra - \lla O_1(t_1)\rra \lla O_2(t_2)\rra.
\end{equation} 
The operators ${\bf E}^i= F_{0i}$ and ${\bf B}^i=\epsilon^{ijk}F^{jk}/2$ 
($F_{\mu\nu} = \partial_\mu A_\nu - \partial_\nu A_\mu + ig[A_\mu,A_\nu]$)
are the chromoelectric and chromomagnetic field respectively.

Notice that the final result for the potentials (static and
relativistic corrections) appears factorized in a part containing the
high energy dynamics (and calculable in perturbation theory) which is
inherited from the NRQCD matching coefficients (the $c_j, d_j$,
cf.~\Section~\ref{sec:unonrqcd} on NRQCD in
\Chapter~\ref{chapter:commontheoreticaltools}),
and a part containing the low energy dynamics given in terms of Wilson
loops and chromo-electric and chromo-magnetic insertions in the Wilson
loop \cite{m12}. The inclusion of NRQCD matching coefficients solved
the inconsistency between perturbative one-loop calculations and the
Wilson loop approach which arose in the past \cite{rev,Chen:1994dg}.
The low energy contributions can be calculated on the lattice
\cite{bal,Bali:1997aj} or estimated in QCD vacuum models
\cite{rev,vac}.

Almost all the potentials given in \Eq~(\ref{eq:v11}) were evaluated
on the lattice in Refs.\ \cite{bal,Bali:1997aj}, but this is not so
for the potentials of order $1/m$, $V^{(1,0)}, V^{(0,1)}$.  It would
be very interesting to have such an evaluation (the perturbative one
exist at two loops \cite{Kniehl:2001ju}) since, phenomenologically,
they have not been considered up to now. In general, it would be very
interesting to have updated and more precise lattice calculations of
all the potentials.  We recall that these lattice calculations have
also a definite impact on the study of the properties of the QCD
vacuum in presence of heavy sources.  So far the lattice data for the
spin-dependent and spin-independent potentials are consistent with a
flux-tube picture, while it is only for the spin-dependent terms that
the so called scalar confinement is consistent with the lattice data
\cite{m12,rev,flux}.

It has recently been shown \cite{Brambilla:2003mu} that the assumption
that $V$ is analytic in $1/m$ is not correct. New non-analytic terms
arise due to the three-momentum scale $\sqrt{m\lQ}$.  These terms can
be incorporated into local potentials ($\delta^3 ({\bf r})$ and
derivatives of it) and scale as half-integer powers of
$1/m$. Moreover, it is possible to factorize these effects in a model
independent way and compute them within a systematic expansion in some
small parameters.  In any case, the corrections to the spectrum coming
from these non-analytical terms are subleading with respect to the
terms given in \Eq~(\ref{eq:hs}).

We emphasize that, in this regime, non-relativistic potential models,
as the ones discussed in \Section~\ref{sec:spphen} are demonstrated to
be EFTs of QCD, provided that the potentials used there are compatible
with the ones extracted from QCD (and the interaction with
pseudo-Goldstone bosons neglected).  It is a matter of debate,
however, which states in bottomonium and charmonium should be
considered as belonging to this regime. On one hand the mass should be
sufficiently lower than the heavy-light meson pair threshold to
justify the omission of higher Fock state effects. On the other hand
if the states are too low in mass then the perturbative matching
regime of \Section~\ref{sec:spnrqcdwc} will apply and the problem can be
further simplified.

Since the potentials are defined in an effective field theory framework
they are not plagued by the  inconsistency typically emerging in higher order 
calculations in  potential models. 
It is well known that at second order in quantum mechanical
perturbation theory the spin dependent terms result in a contribution
which is as large as the leading order one. This is due to the fact that
the resulting expression becomes ill-defined. Regulating it requires
to introduce a
cut-off (or dimensional regularization).
A large cut-off gives rise to a linear and 
to a logarithmic divergence. These divergences can be renormalized by 
redefining the coupling constant of a delta potential \cite{Lepage:1997cs}.
This is a mere reflection of the fact that when one matches QCD to NRQCD,
one expands the energy and three momentum. This induces infrared
divergences in the matching coefficients.
For quarkonium this happens in the calculation of a matching coefficient
of a four Fermion operator at two loops. If one uses a consistent
regularization scheme 
both for the QCD-NRQCD matching calculation and the quantum mechanical 
calculation in pNRQCD, the divergences exactly cancel and, at the  end  of
the  day, a  totally consistent  scale independent  result is  obtained
(for a QED example see Refs.\ \cite{Czarnecki:1999mw,Czarnecki:1998zv}).
 Notice that an EFT framework is crucial to understand this second 
order calculation and to render the result meaningful.

\subsubsection[The QCD static quark--antiquark spectrum and 
               the mechanism of confinement]
              {The QCD static spectrum and mechanism of confinement
               $\!$\footnote{Authors: N.~Brambilla, 
                             C.~Morningstar,  A.~Pineda}}
\label{sec:spstatconf}

The spectrum of gluons in the presence of a static quark--antiquark pair
has been extensively studied with high precision using lattice simulations.
Such studies involve the calculation of large sets of Wilson loops with a
variety of different spatial paths. Projections onto states of definite
symmetries are done, and the resulting energies are related to the static
quark--antiquark potential and the static hybrids potentials.  With accurate
results, such calculations provide an ideal testing ground for models 
of the QCD confinement mechanism. 

\medskip

\noindent
{\it The singlet  static energy } \\
The singlet static energy is the singlet static potential $\vs$.

\begin{figure}[t]
\begin{center}
\includegraphics[width=2.8in]{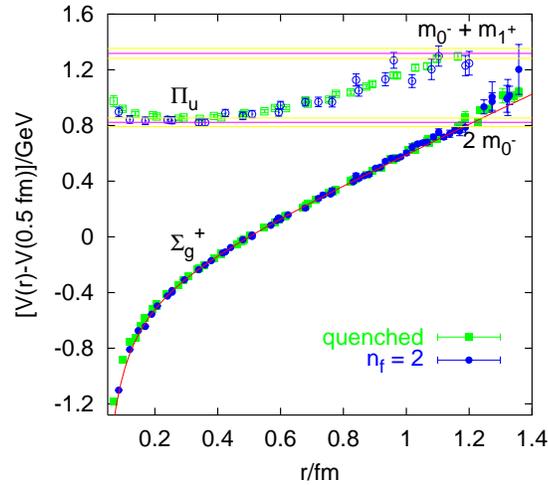}
\end{center}
\caption[The singlet static energy]
        {The singlet static energy (quenched and unquenched data)
         from Ref.\ \cite{Bali:2000vr}, see also \cite{Bolder:2000un}}
\label{fig:stringt0}
\end{figure}

In the plot\ref{fig:stringt0}, we report simulation results both with and
without light quark--antiquark pair creation. Such pair creation only
slightly modifies the energies for separations below 1~fm, but
dramatically affects the results around 1.2~fm, at a distance which is
too large with respect to the typical heavy quarkonium radius to be
relevant for heavy quarkonium spectroscopy.  At finite temperature,
the so-called string breaking occurs at a smaller distance
(cf. corresponding Section in
\Chapter~\ref{chapter:charm-beauty-in-media}, \emph{Media}).

\begin{figure}[t]
\begin{center}\setlength{\unitlength}{1mm}
   \begin{picture}(75,45)
   \put(0,0){\includegraphics[width=75mm]{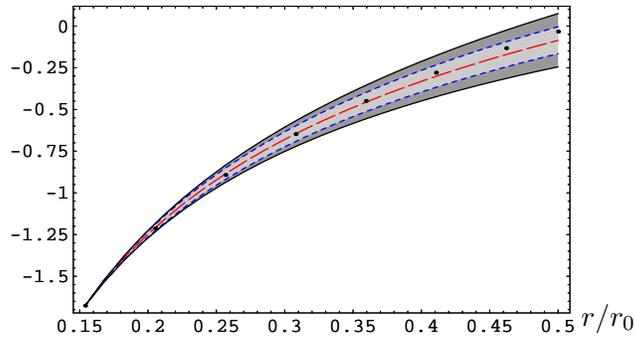}}
   \put(27,113){$r_0(V_{\RRS}(r)-V_{\RRS}(r')+E_{latt.}(r'))$}
   \put(76,1){$r/r_0$}
   \end{picture}
\end{center}
\caption[Plot of \ensuremath{r_0(V_{\RRS}(r)-V_{\RRS}(r')+E_{latt.}(r'))}
         versus \ensuremath{r}]
        {Plot of $r_0(V_{\RRS}(r)-V_{\RRS}(r')+E_{latt.}(r'))$ versus
         $r$ at three loops (estimate) plus the leading single
         ultrasoft log (dashed line) compared with the lattice
         simulations \cite{Necco:2001xg} $E_{latt.}(r)$. For the scale
         of $\als(\nu)$, we set
         $\nu=1/0.15399\,r_0^{-1}$. $\nu_{us}=2.5\,r_0^{-1}$ and
         $r'=0.15399\,r_0$. The inner and outer band are meant to
         estimate the errors in $\Lambda_{\MS}$ and perturbative. For
         further details see the main text.}
\label{fig:errors}
\end{figure}

One can study possible nonperturbative effects in the static potential
at short distances. As it has already been mentioned in the "static
QCD potential" subsection, the proper treatment of the renormalon
effects has made possible the agreement of perturbation theory with
lattice simulations (and potential models)
\cite{Sumino:2001eh,Recksiegel:2001xq,Necco:2001gh,
Pineda:2002se,Recksiegel:2002um,Lee:2002sn}. Here we would like to
quantify this agreement assigning errors to this comparison. In
particular, we would like to discern whether a linear potential with
the usual slope could be added to perturbation theory. In order to do
so we follow here the analysis of Ref.\
\cite{Pineda:2002se,Pineda:2003jv}, where the potential is computed
within perturbation theory in the Renormalon Subtracted scheme defined
in Ref.\ \cite{Pineda:2001zq}.  The comparison with lattice
simulations \cite{Necco:2001xg} in \Figure~\ref{fig:errors} shows that
nonperturbative effects should be small and compatible with zero,
since perturbation theory is able to explain lattice data within
errors.  The systematic and statistical errors of the lattice points
are very small (smaller than the size of the points). Therefore, the
main sources of uncertainty of our (perturbative) evaluation come from
the uncertainty in the value of $\Lambda_{\MS}$ ($\pm 0.48\,r_0^{-1}$)
obtained from the lattice \cite{Capitani:1998mq} and from the
uncertainty in higher orders in perturbation theory.  We show our
results in \Figure~\ref{fig:errors}.  The inner band reflects the
uncertainty in $\Lambda_{\MS}$ whereas the outer band is meant to
estimate the uncertainty due to higher orders in perturbation
theory. We estimate the error due to perturbation theory by the
difference between the NNLO and NNNLO evaluation. The usual confining
potential, $\delta V =\sigma r$, goes with a slope $\sigma=0.21 {\rm
GeV}^2$. In lattice units we take: $\sigma=1.35\, r_0^{-2}$. The
introduction of a linear potential at short distances with such slope
is not consistent with lattice simulations. This is even so after the
errors considered in \Figure~\ref{fig:errors} have been included.

At larger distances, $r\gg \lQ$, $\vs$ grows linearly, with the string
tension $\sigma =0.21 {\rm GeV}^2$.  Such a linear growth of the
energy is often taken as evidence that the gluon field forms a flux
tube whose dynamics can be described by an effective string theory.
However, it should be pointed out that a linearly growing potential
does not necessarily imply string formation; for example, the
spherical bag model also predicts a linearly rising potential for
moderate $r$.  It has been shown \cite{Luscher:1980fr} that the
formation of a string-like flux tube implies a characteristic and
universal $-\frac{\pi}{12 r}$ correction to the ground-state energy,
deriving from the zero-point energy of the transverse string
vibration.  Recent high precision simulations \cite{Luscher:2002qv}
(cf. also \cite{Caselle:2004er}) show that the coefficient of the
$1/r$ correction differs from $-\pi/12$ by $12 \%$.  The authors of
Ref.~\cite{Luscher:2002qv} introduce an {\em ad hoc} end-effect term
with a fit parameter $b$ to the effective string action to explain
this significant difference.  However, in a more recent paper
\cite{Luscher:2004}, these authors show that an open-closed string
duality relation requires $b=0$.  Furthermore, a simple resonance
model was used in Refs.~\cite{Kuti:2004a,Kuti:2004b} to show that the
Casimir energy expected from a string description could be reproduced
in a model in which string formation was not a good description,
concluding that no firm theoretical foundation for discovering string
formation from high precision ground state properties below the 1~fm
scale currently exists.

\medskip

\noindent
{\it Excitations of the static energy} 

The spectrum of gluons in the presence of a static
quark--antiquark pair provides valuable clues about 
the nonperturbative dynamics of QCD.
Adopting the viewpoint that the nature of the confining gluon field
is best revealed in its excitation spectrum, in
Ref.~\cite{JKM03}, recent advances in lattice simulation
technology, including anisotropic lattices, improved gauge actions,
and large sets of creation operators,were employed to investigate the static energies
of gluonic excitations between static quarks (hybrid static energies).

\begin{figure}[t]
\centering\includegraphics[width=70mm]{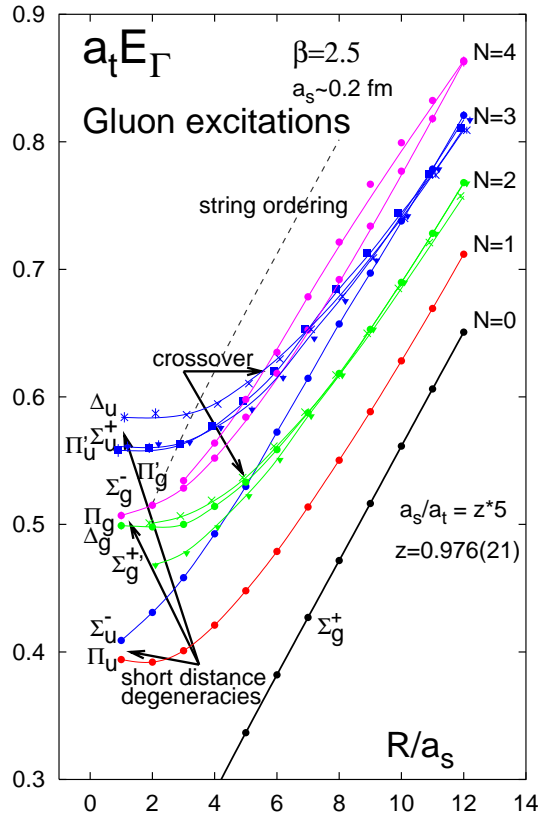}
\caption[The spectrum of gluonic excitations] 
        {The spectrum of gluonic excitations in the presence of a
        static quark--antiquark pair separated by a distance $R$ in
        4-dimensional $SU(3)$ gauge theory (from
        Ref.~\protect\cite{JKM03}).  Results are from one simulation
        for lattice spacing $a_s\sim 0.2$~fm using an improved action
        on a $(10^2\times 30)\times 60$ anisotropic lattice with
        coupling $\beta=2.5$ and bare aspect ratio $\xi=5$.  At large
        distances, {\em all} levels without exception are consistent
        with the expectations from an effective string theory
        description.  A dramatic level rearrangement is observed in
        the crossover region between $0.5-2.0$~fm. The dashed line
        marks a lower bound for the onset of mixing effects with
        glueball states.}
\label{fig:su3_spec}
\end{figure}

In NRQCD (as in QCD) the gluonic excitations between static quarks
have the same symmetries of a diatomic molecule plus charge
conjugation.  In the centre-of-mass system these correspond to the
symmetry group $D_{\infty h}$ (substituting the parity generator by
CP).  The mass eigenstates are classified in terms of the angular
momentum along the quark--antiquark axes ($|L_z| = 0,1,2, \dots$ which
traditionally are labelled as $\Sigma, \Pi, \Delta, \dots$), CP (even,
$g$, or odd, $u$), and the reflection properties with respect to a
plane passing through the quark--antiquark axes (even, $+$, or odd,
$-$).  Only the $\Sigma$ states are not degenerate with respect to the
reflection symmetry, see also \Section~\ref{sec:sphhl}.  In
\Figure~\ref{fig:su3_spec} we display lattice results of the hybrid
static energies $V_H$ obtained from Wilson loops with operators of the
appropriate symmetry inserted at the end points. \par

$D_{\infty h}$ is a subgroup of the rotational symmetry group $O(3)$ times
charge conjugation.
In  the short-range limit, $r \ll \lQ$, 
the hybrid energies approach so-called gluelump levels that can be
classified according to the usual $O(3)$ $J^{PC}$. The corresponding
operators can be 
explicitly constructed using pNRQCD in the static limit \cite{Brambilla:1999xf}. 
In  the case of pure gluodynamics,
the spectrum then consists of static energies which 
depend on {\bf r}. The energy units are provided by the only other scale in 
the problem, $\lQ$. 
The  gluelumps operators are of the type
${\rm Tr}\{{\rm O}H\}$, where ${\rm O} = O^aT^a$ corresponds 
to a quark--antiquark state in the adjoint representation (the octet) and 
$H = H^a T^a$ is a gluonic operator.
 By matching the QCD static hybrid 
operators into pNRQCD, we get the static energies (also called hybrids
 static potentials) $V_H$ 
of the gluelumps. At leading order in the multipole expansion, they read
\cite{Brambilla:1999xf}
\begin{equation}
V_H(r) = V_o(r) + \frac{1}{T_g^H},
\label{eq:potglue}
\end{equation}
being $T_g^H$ the correlation time of the corresponding gluelump correlator
$\langle H^a(t) \phi(t,0)^{\rm adj}_{ab}H^b(0)\rangle^{\rm non-pert.}$  
$\simeq$  $h \, e^{- i t/T_g^H}$. 
The lattice data confirm that (in the region in which decay into glueball
channels is not yet possible) all the $V_H$ behave like
$\vo=\frac{\als}{6 r}$ for $r\to 0$ cf.\ \Figure~\ref{fig:su3_spec} and
Ref.~\cite{Bali:2003jq}. 
The constant $T_g^H$ depends on the gluelump
operator  $H$, its inverse corresponds to the mass of the gluelump $H$.
Note that $T_g^H$ are scheme and scale dependent.
pNRQCD, in which $r$ is integrated out, predicts the short-range degeneracies,
\begin{equation}
\Sigma_g^{+\, \prime} \sim \Pi_g\;; \qquad
\Sigma_g^{-} \sim \Pi_g^{\prime} \sim \Delta_g\,; \quad
\Sigma_u^{-} \sim \Pi_u\;; \qquad
\Sigma_u^{+} \sim \Pi_u^{\prime} \sim \Delta_u \,.
\label{eq:dege}
\end{equation}  
This is confirmed by the lattice data, cf. \Figure~\ref{fig:su3_spec}.
Similar observations have also been previously 
made in the lattice theory in Ref.\ \cite{Foster:1998wu}.
It is interesting to notice that the hierarchy of the states, 
as displayed in \Figure~\ref{fig:su3_spec}, is reflected in the
dimensionality of the 
operators of pNRQCD~\cite{Brambilla:1999xf,Bali:2003jq}. 

By using only ${\bf E}$ and
${\bf B}$ fields and keeping only the lowest-dimensional representation 
we may identify the operator $H$ for the short-range hybrids called $\Sigma_g^{+\,'}$ 
(and $\Pi_g$) with ${\bf r}\cdot{\bf E}$ (and ${\bf r}\times{\bf E}$) 
and the operator $H$ for the short-range hybrids called $\Sigma_u^{-}$ 
(and $\Pi_u$) with ${\bf r}\cdot{\bf B}$ (and ${\bf r}\times{\bf B}$). 
Hence, the corresponding static energies for small $r$ are
$$
V_{\Sigma_g^{+\,'},\Pi_g}(r) = V_o(r) + \frac{1}{T_g^E}, \qquad\qquad
V_{\Sigma_u^{-},\Pi_u}(r) = V_o(r) + \frac{1}{T_g^B}.
$$
The lattice results of Ref.\ \cite{JKM03} show that, in the short range, 
$\!V_{\Sigma_g^{+\,'},\Pi_g}(r)\! >\! V_{\Sigma_u^{-},\Pi_u}(r)$. 
This supports the sum-rule prediction \cite{Dosch:1998th} that
the pseudovector 
hybrid lies lower than the vector one, \ie $T_g^E < T_g^B$
and the lattice evaluations of Refs.~\cite{Foster:1998wu,Bali:2003jq}.
In this way, in the short-distance limit, we can  relate the behavior of the 
energies for the gluonic excitations between static quarks with the large
time behavior 
of gluonic correlators.
We can extract results for 
gauge invariant two-point gluon field strength correlators
(which are also the relevant nonperturbative objects in the stochastic
vacuum model \cite{vac})
$
\langle 0|F^a_{\mu\nu}(t)\phi(t,0)^{\rm adj}_{ab}F^b_{\mu\nu}(0)|0 \rangle  
$, $\phi$ being the adjoint string.
One can parameterize these correlators in terms of two scalar functions:
$\langle 0|{\bf E}^a(t)\phi(t,0)^{\rm adj}_{ab}{\bf E}^b(0)|0 \rangle $
 and $ \langle 0|{\bf B}^a(t)\phi(t,0)^{\rm adj}_{ab}{\bf B}^b(0)|0 \rangle$
with correlations lengths: $T^E=1/\Lambda_{E}$ and $T^B=1/\Lambda_{B}$,
respectively. Note that while differences between
gluelump masses $\Lambda_H$ are universal the absolute normalization
is scheme- and scale-dependent~\cite{Bali:2003jq}.

The matching of pNRQCD to ($n_f=0$) QCD has been performed in the static limit
to ${\cal O}(\alpha_{\rm s}^3)$ in the lattice scheme
and the (scheme- and scale-dependent) gluelump masses
$\Lambda_H=1/T^H_g$ have been determined both, in the continuum limit
from short distance energy levels
and at finite lattice spacing from the gluelump spectrum~\cite{Bali:2003jq}.
Perfect agreement between these two determinations was found.
It would be highly desirable to have lattice determinations
at even shorter distances to further increase the precision of such
determinations, however, such calculations are rather challenging due to the
need to properly treat lower-lying glueball scattering states.

The behaviour of the hybrid static energies for large $r$ provides
further valuable information on the mechanism of confinement.  The
linearly rising ground-state energy is {\em not} conclusive evidence
of string formation \cite{flux}.  Computations of the gluon action
density surrounding a static quark--antiquark pair in $SU(2)$ gauge
theory also hint at flux tube formation
\cite{Bali:1994de}. Complementary information come from the study of
the static energies of the gluonic excitations between static quarks.
A treatment of the gluon field in terms of the collective degrees of
freedom associated with the position of the long flux might then be
sufficient for reproducing the long-wavelength physics.  If true, one
then hopes that the oscillating flux can be well described in terms of
an effective string theory \cite{Baker:2000ci}. In such a case, the
lowest-lying excitations are expected to be the Goldstone modes
associated with the spontaneously broken transverse translational
symmetry.  These modes are a universal feature of any low-energy
description of the effective QCD string and have energy separations
above the ground state given by multiples of $\pi/R$. A well-defined
pattern of degeneracies and level orderings among the different
symmetry channels form a very distinctive signature of the onset of
the Goldstone modes for the effective QCD string.

The spectrum of more than a dozen levels shown in \Figure~\ref{fig:su3_spec}
provides strong evidence that the gluon field can be well approximated by an
effective string theory for large separations $R$.  
For separations above 2~fm, the levels agree {\em without exception} with
the ordering and degeneracies expected from an effective string theory.
The gaps agree well with $N\pi/R$, but a fine structure remains,
offering the possibility to obtain details of the effective QCD
string action in future higher precision simulations.  For small $R<2$~fm,
the level orderings and degeneracies are not consistent with the expectations
from an effective string description, and the gaps differ appreciably from $N\pi/R$
with $N=1,2,3,\dots$.  Such deviations, as large as $50\%
$ or more, cannot be considered mere corrections, making the applicability
of an effective string description problematical.  Between 0.5 to 2~fm, a 
dramatic level rearrangement occurs.

Non-universal details of the underlying string description for large
separations, such as higher order interactions
and their couplings, are encoded in the fine structure of the spectrum
at large separations.  It is hoped that near future simulations will
have sufficient precision to be able to differentiate between
such corrections.  In the meantime, the excitation spectrum in
other space--time dimensions and other gauge theories, such as $SU(2)$
and $Z(2)$, are being explored \cite{Kuti:2003,Caselle:2004er}.

\subsubsection[pNRQCD for QQQ and QQq baryons]
              {pNRQCD for QQQ and QQq baryons
               $\!$\footnote{Author: N. Brambilla}}
\label{sec:spnrqcd3q}

In the case of a bound state formed by three heavy quarks, still a
hierarchy of physical scales similar to the quarkonium case
exists. Consequently, starting from a NRQCD description for each heavy
quark, it is possible to integrate out the scale of the momentum
transfer $\simeq mv$ and write the pNRQCD Lagrangian for heavy baryons
\cite{roesch,Soto:2003ft}.  Similarly to before two different
dynamical situations may occur: the momentum transfer is much larger
than $\lQ$, or it is of order $\lQ$. In the first case the matching is
perturbative and the Lagrangian is similar to \Eq~(\ref{eq:wc}) with
more degrees of freedom for the quark part: two octets, one singlet
and one decuplet (as it comes from the colour decomposition of $3
\times 3 \times 3$) \cite{roesch}.  In the second case the matching is
nonperturbative and the Lagrangian is similar to \Eq(\ref{eq:scr})
with only the three quark singlet as degree of freedom.  The
(matching) potentials are nonperturbative objects and their precise
expression in terms of static Wilson loop and (chromo)electric and
(chromo)magnetic insertions in static Wilson loops can be calculated
\cite{roesch}.  Experimental data for baryons composed by three quarks
are not existing at the moment, however lattice calculation of the
three quark potential exist
\cite{Bali:2000gf,Alexandrou:2001ip,Takahashi:2002bw}.

Baryons made by two heavy quarks and a light quark $QQq$ combine the
slow motion of the heavy quark with the fast motion of the light
quark. Thus a treatment combining in two steps an effective field
theory for the $QQ$ interaction and an effective field theory for the
$QQ$ degrees of freedom with the light quark is the most appropriate
one.  The interest of these states is also related to the fact that
the SELEX experiment recently announced the discovery of four doubly
charmed baryon states. This will be discussed in more detail in
\Section~\ref{sec:dbch-QQqq}.  The non relativistic motion of the two
heavy quarks is similar to quarkonium while the light quark is moving
relativistically around the slowly moving $QQ$. Since the $QQ$ is in a
colour antitriplet state, in the heavy quark limit the system is
similar to a $\bar{Q}q$ system.  However, the situation is much more
interesting because if one constructs first the EFT for the two heavy
quarks more degrees of freedom enter and depending on the dynamical
situation of the physical system, these degrees of freedom may or may
not have a role. In particular if we work under the condition that the
momentum transfer between the two heavy quarks is smaller than $\lQ$,
then we can construct a pNRQCD Lagrangian of the type
\Eq~(\ref{eq:wc}) with a triplet and a sextet as $QQ$ degrees of
freedom \cite{roesch}.  Such degrees of freedom, would also be
relevant for the study of double charmonia production
\cite{Ma:2003zk}.

\subsection[Threshold effects (EFT)]
           {Thresholds effects (EFT)
            $\!$\footnote{Authors: N.~Brambilla, J.~Soto}}
\label{sec:spnrqcdte}

For states for which $k\sim E \sim \lQ$, namely close or beyond threshold,
one has to stay at the NRQCD level. It is still an open question whether one 
can build a suitable EFT to study mixing and threshold effects.

For a confining potential (\eg harmonic oscillator), however, the
typical momentum transfer $k$ decreases with the principal quantum
number whereas both the typical relative three-momentum $p$ and the
binding energies increase. For some principal quantum number $n$, the
binding energy will become comparable to the momentum transfer and
hence $k \ge E$ will not hold anymore.  For these states pNRQCD is not
a good effective theory anymore (it may still remain a successful
model). This is expected to happen for states close to or higher than
the heavy-light meson pair threshold. There is no EFT beyond NRQCD
available for this regime at the moment.  Notice also that for some
$n$ the typical three momentum will become comparable to $m$ and hence
relativistic effects will not be small and NRQCD will not be a good
EFT anymore. This is expected to happen for states much higher than
the heavy-light meson pair threshold. Relativistic quark models like
the ones discussed in \Section~\ref{sec:spphen} are probably
unavoidable for this situation although it is not known at the moment
how to link them to QCD.

\section[Phenomenological approach]
        {Phenomenological approach $\!$\footnote{Author: S.~Godfrey}}
\label{sec:spphen}

From the discovery of charmonium states 
\cite{Aubert74,Augustin74,Abrams74}, QCD motivated
potential models have played an important role in understanding 
quarkonium spectroscopy 
\cite{Appelquist75,DeRujula75,Appelquist75b,Eichten75}.  The initial 
models describing charmonium spectroscopy, using a QCD motivated 
Coulomb plus linear confining potential with colour magnetic spin 
dependent interactions, have held up quite well.  This approach also 
provides a useful framework for refining our understanding of QCD and 
guidance towards progress in quarkonium physics.  The discovery of the 
$\Upsilon$ family of meson \cite{Herb77} was quickly recognized as 
a $b\bar{b}$ bound state whose spectroscopy was well described by the 
potential model picture used to describe the charmonium system.

In this section we give an overview of potential models of quarkonium
spectroscopy \cite{reviews}.  Most models
\cite{Eichten:ag,eichten78,godfrey85,stanley80,ebert03,gershtein94,fulcher91,fulcher99,fulcher94,gupta94,gupta96,zeng95}
have common ingredients.  Almost all such models are based on some
variant of the Coulomb plus linear potential confining potential
expected from QCD.  Quark potential models typically include one-gluon
exchange and most models also include the running constant of QCD,
$\alpha_{\rm s}(Q^2)$.  Finally, relativistic effects are often
included at some level
\cite{Eichten:ag,eichten78,godfrey85,stanley80,ebert03,gershtein94,fulcher91,fulcher99,fulcher94,
gupta94,gupta96,zeng95,Llanes-Estrada:2004wr,Baldicchi:2002qm,Allen:2002vd}.
At the minimum, all models we consider include the spin-dependent
effects that one would expect from one-gluon-exchange, analogous to
the Breit--Fermi interaction in QED, plus a relativistic spin--orbit
Thomas precession term expected of an object with spin (the quark or
antiquark) moving in a central potential.  Potential models have been
reasonably successful in describing most known mesons.  Although
cracks have recently appeared \cite{Barnes:2003vb,Eichten:2004uh}
these point to the need for including physics effects that have
hitherto been neglected such as coupled channel effects
\cite{Eichten:2004uh}.

In the next section we will give a brief introduction to quark 
potential models and attempt to describe the differences between 
models.  The subject is roughly thirty years old and a large 
literature on the subject exists.  It is impossible to cover 
all variants and we will almost totally neglect the considerable work 
that brought us to where we are today.  We apologise to all those 
whose work we do not properly cite and hope they understand.  In the 
next sections we compare the predictions of some 
models with experiment for the $c\bar{c}$, $b\bar{b}$ and $c\bar{b}$ 
mesons and point out variations in predictions 
and how they arise from the underlying model.  
\shortpage

\subsection[Potential models]
           {Potential models
            $\!$\footnote{Author: S. Godfrey}}
\label{sec:sppot}

Quarkonium potential models typically take the form of a Schr\"odinger 
like equation:
\begin{equation}
\label{eq:ham}
[T+V] \Psi = E\Psi
\end{equation}
where $T$ represents the kinetic energy term and $V$ the potential energy 
term.  We lump into these approaches the 
Bethe--Salpeter equation (\eg 
Ref.\ \cite{Roberts:2003kd,Baldicchi:2002qm})
 and quasi-potential approaches 
(\eg Ref.\ \cite{ebert03}).  

Different approaches have been used for the kinetic energy term 
ranging from the non-relativistic Schr\"odinger equation to relativistic kinetic energy 
\cite{godfrey85,gupta96,Jacobs:1986gv}
\begin{equation}
\label{eq:relkin}
T=\sqrt{p^2 + m_Q^2} + \sqrt{p^2 + m_{\bar{Q}}^2 }
\end{equation}
in the spinless Salpeter equation.

\subsubsection{The  potential}
The quark--antiquark potential is typically motivated by the properties 
expected from QCD \cite{wilson74,eichten80,Chen:1994dg,wilsonpot,wilsonrel,gromes84,rev,m12}
and while there are differences, most recent 
potentials show strong similarities.  It is worth pointing out that in 
the early days of quarkonium spectroscopy this was not obvious and 
much effort was expended in fitting different functional forms of the 
potential to the observed quarkonium masses.  In the end, the shape of 
the potentials converged to a form that one might expect from the
asymptotic limits of QCD and which has been qualitatively verified by 
Lattice QCD calculations \cite{bal} of the expression of the potentials
 obtained in the Wilson loop \cite{wilson74,eichten80,Chen:1994dg,wilsonpot,wilsonrel,gromes84,rev}
and in the EFT \cite{m12} approach. This is a great success of 
quarkonium phenomenology.

To derive the quarkonium potential we start with QCD where the 
gluons couple to quarks and to each other.  
The quark--gluon interaction is similar to the electron--photon 
interaction in quantum electrodynamics with the Born term for the 
$qq$ or $q\bar{q}$ interaction at short distance being the familiar $1/r$ 
form.  In contrast with QED the gluon self-coupling results in a slow 
decrease of the effective coupling strength at short distance.  In 
terms of the Fourier conjugate momentum the lowest order QCD 
corrections to $\alpha_{\rm s} = g_s^2/4\pi$ can be parametrized as 
\begin{equation}
\label{eq:alphas}
\alpha_{\rm s}(Q^2) =\frac{12\pi}{(33-2n_f) \ln(Q^2/\Lambda^2)}
\end{equation}
where $n_f$ is the number of Fermion flavours with mass below $Q$, and 
$\Lambda\sim \lQ$ is the characteristic scale of QCD measured to be $\sim 
200$~MeV.   At short distances one-gluon-exchange leads to the Coulomb 
like potential
\begin{equation}
\label{eq:coulomb}
V(r) = - \frac{4}{3} \frac{\alpha_{\rm s}(r)}{r}
\end{equation} 
for a $q\bar{q}$ pair bound in a colour singlet where the factor of 
$4/3$ arises from the SU(3) colour factors.  At short distances 
one-gluon-exchange becomes weaker than a simple Coulomb interaction.  

At momentum scales smaller than $\lQ$  which corresponds to a distance 
of roughly 1~fm, $\alpha_{\rm s}$ blows up and one-gluon-exchange is no 
longer a good representation of the $q\bar{q}$ potential.  The qualitative 
picture is that the chromoelectric lines of force bunch together into 
a {\it flux tube} which leads to a distance-independent force or a 
potential
\begin{equation}
\label{eq:linear}
V(r)=\sigma r.
\end{equation} 
This has been validated by Lattice QCD calculations.  
Phenomenologically,  every recent model which we will consider has 
found $\sigma \sim 0.18$~GeV$^2$.  

Numerous variations of the resulting Coulomb plus linear potential 
exist in the literature.  Some of the better known ones are the 
Cornell potential \cite{eichten78}, 
Richardson's potential \cite{richardson}, and
the Buchm\"uller Tye potential \cite{Buchmuller:1980su}.
Overall, the spin-independent features of quarkonium spectroscopy are well 
described by the potentials just described.  

Let us also mention that heavy quark mass corrections to the (static)
central (spin and velocity independent) potential exist, although they
have not yet been taken into account in potential models applications
so far. They correspond to $V^{(1,0)}_r$ , $V^{(2,0)}_r$ and
$V^{(1,1)}_r$ in \Section~\ref{sec:spnrqcdscc}.  Their expressions in
perturbation theory are known \cite{m12,Kniehl:2002br}.  Part of
$V^{(2,0)}_r$ and $V^{(1,1)}_r$ was included in the phenomenological
application to the spectrum in Refs.\
\cite{wilsonrel,gromes84,bal,Brambilla:1989ur}.

\subsubsection{Spin-dependent potentials}

Spin dependent multiplet splittings are an important test of the 
details of quarkonium models.  In particular, the nature 
of spin dependent potentials are decided by the Lorentz nature of the 
confining potentials \cite{eichten80,gromes84,flux,Brambilla:1992fx}.  
While there is general consensus that the short 
distance one-gluon-exchange piece is Lorentz vector and the linear 
confining piece is Lorentz scalar this is by no means universal and 
other possibilities are vigorously advocated.   Gromes described how 
to obtain the spin-dependent potentials given the Lorentz structure of 
the interaction \cite{gromes84}
and one can also use the prescription given in 
Berestetskij, Lifschitz and Pitaevskij \cite{berestetskij}.  
Simply put, one can obtain the form of the 
spin dependent interaction by Fourier transforming the on-shell 
$q\bar{q}$ scattering amplitude:
\begin{equation}
\label{eq:scatter}
M = [ \bar{u}(p_f') \Gamma u(p_i')] \; V(Q^2) \; [\bar{u}(p_f) \Gamma u(p_i)]
\end{equation}
where the $\Gamma$ matrices give the Lorentz structure of the 
interaction and $V(Q^2)$ is the Fourier transform of the 
spin-independent potential.  For example, for a Lorentz-vector 
interaction $\Gamma =\gamma^\mu$ and for a Lorentz-scalar interaction 
$\Gamma= {\rm I}$.  In principle other forms are possible with each giving 
rise to characteristic spin-dependent interactions. These can be found 
by expanding the scattering amplitude to order $(v/c)^2$ which 
corresponds to an expansion in inverse powers of quark masses. In the early 
years of quarkonium phenomenology they were all tried and it was found 
that the Lorentz-vector one-gluon-exchange plus Lorentz scalar linear 
confining potential gave the best agreement with experiment 
\footnote{Although other forms are still advocated.  See Ebert {\it et
al.} \cite{ebert00,ebert03}.}. Note that the form of the full QCD
potential at order $1/m^2$
\cite{m12,wilson74,eichten80,Chen:1994dg,wilsonpot,wilsonrel} has now
been obtained in the EFT (cf.\ \Section~\ref{sec:spnrqcdscc}), and
while the spin-dependent nonperturbative potential may correspond to a
scalar interaction in the language used above, the velocity-dependent
potentials do not fit such a picture.  The effective kernel is thus
not a simple scalar, precisely the dependence both on the momentum and
on the Lorentz structure is more involved than a pure convolution
(\ie only depending on the momentum transfer) scalar structure
\cite{flux,rev,m12,Brambilla:1992fx}. However, the spin dependency is
well approximated by a scalar interaction for phenomenological
applications. The QCD spin-dependent potentials are explicitly given
in \Section~\ref{sec:spnrqcdscc}.  A complete calculation of the spin
structure of the spectrum using the full expression given in 
\Section~\ref{sec:spnrqcdscc} does not yet exist.

To lowest order in $(v/c)^2$ the Lorentz-vector one-gluon-exchange
gives rise to terms familiar from one-photon exchange in atomic
physics.  The colour contact interaction, which in the language of
\Section~\ref{sec:spnrqcdscc} corresponds to taking $V^{(1,1)}_{S^2} (r)$ at
leading order in perturbation theory,
\begin{equation}
\label{eq:contact}
H^{cont}_{q\bar{q}} =  \frac{32\pi}{9} 
\frac{\alpha_{\rm s}(r)}{m_{q}m_{\bar{q}}} {\bf S}_q \cdot {\bf S}_{\bar{q}}
\, \delta^3 ({\bf r}) 
\end{equation}
gives rise to, for example the $J/\psi-\eta_c$ splitting.  The colour
tensor interaction, which in the language of
\Section~\ref{sec:spnrqcdscc} corresponds to taking $V^{(1,1)}_{{\bf
S}_{12}} (r)$ at leading order in perturbation theory,
\begin{equation}
\label{eq:tensor}
H^{ten}_{q\bar{q}} =  \frac{4}{3}
\frac{\alpha_{\rm s}(r)}{m_q m_{\bar{q}}} \frac{1}{r^3}
\left[{ 
\frac{3\,\mathbf{S}_q\cdot\mathbf{r}\;\mathbf{S}_{\bar{q}}\cdot\mathbf{r}}{r^2}
- \mathbf{S}_q \cdot \mathbf{S}_{\bar{q}} 
}\right] 
\end{equation}
contributes to splitting of $L\neq 0 $ spin triplet multiplets like 
the $\chi_{cJ}$ and $\chi_{bJ}$ multiplets.  The final spin dependent 
term is the spin orbit interaction which has two contributions. The 
first piece arises from the colour-magnetic one-gluon-exchange while 
the second piece is the Thomas precession term which is a 
relativistic effect for an object with spin moving in a central 
potential
\begin{equation}
\label{eq:so}
H^{s.o.}_{q\bar{q}} = H^{s.o.(cm)}_{q\bar{q}} + H^{s.o.(tp)}_{q\bar{q}}.
\end{equation}
The colour magnetic piece arising from one-gluon exchange is given by:
\begin{equation}
\label{eq:cmso}
H^{s.o.(cm)}_{q\bar{q}} = \frac{4}{3} \frac{\alpha_{\rm s}(r)}{r^3}
\left({ 
  \frac{\mathbf{S}_q}{m_q m_{\bar{q}}}
+ \frac{\mathbf{S}_{\bar{q}}}{m_q m_{\bar{q}}}
+ \frac{\mathbf{S}_q}{m_q^2}
+ \frac{\mathbf{S}_{\bar{q}}}{m_{\bar{q}}^2}
}\right) 
\cdot \mathbf{L} 
\end{equation}
and the Thomas precession term is given by
\begin{equation}
\label{eq:tpso}
H^{s.o.(tp)}_{q\bar{q}} = - \frac{1}{2r}
\frac{\partial H_{q\bar{q}}^{conf}}{\partial r}
\left( \frac{\mathbf{S}_q}{m_q^2}
 + \frac{\mathbf{S}_{\bar{q}}}{m_{\bar{q}}^2} \right) \cdot \mathbf{L}
\end{equation}
which includes a contribution from both the short distance $1/r$ piece
and the linear Lorentz-scalar confining potential.  In the language of
\Section~\ref{sec:spnrqcdscc}, both terms in (\ref{eq:so}) are
obtained by taking $V_{LS}^{(1,1)}$ at leading order in perturbation
theory and using the Gromes relation for $V_{LS}^{(2,0)}$.  In these
formulae $\alpha_{\rm s}(r)$ is the running coupling constant of QCD.

For mesons consisting of quarks with different flavours such as the
$B_c$ meson, charge conjugation is no longer a good quantum number so
states with different total spins but with the same total angular
momentum, like the $^3P_1 -^1P_1$ and $^3D_2 -^1D_2$ pairs
(i. e. $J=L$ for $L\geq 1$) can mix via the spin--orbit interaction or
some other mechanism.  \Eqs[b] (\ref{eq:cmso}) and (\ref{eq:tpso}) can
be rewritten to explicitly give the antisymmetric spin--orbit mixing
term:
\begin{equation}
\label{eq:mixso}
H_{s.o.}^- = + \frac{1}{4} 
 \left( \frac{4}{3} \frac{\alpha_{\rm s}}{r^3} - \frac{k}{r} \right)
 \left( \frac{1}{m_Q^2} - \frac{1}{m_{\bar{Q}}^2} \right) 
\; \mathbf{S}_-\cdot \mathbf{L}
\end{equation}
where $\mathbf{S}_- = \mathbf{S}_Q -\mathbf{S}_{\bar{Q}}$.
Consequently, the physical the physical $J=L$ ($J\geq 1$) states are
linear combinations of ${}^3L_J$ and ${}^1L_J$ states 
which we describe by the following mixing:
\begin{eqnarray}
\label{eq:mix}
L'  & = {}^1L_J\cos\theta_{nL} + {}^3L_J\sin\theta_{nL} \nonumber\\   
L & =- {}^1L_J\sin\theta_{nL} + {}^3L_J\cos \theta_{nL} 
\end{eqnarray}
where $L$ designates the relative angular momentum of the $Q\bar{Q}$ 
pair and the subscript is the total angular momentum of the $Q\bar{Q}$ 
which is equal to $L$.  Our notation implicitly implies $L-S$ 
coupling between the quark spins and the relative angular momentum.  
In the limit in which only one quark mass
is heavy, $m_Q\to \infty$, and the other one is light 
the states can be described by the total angular momentum of the 
light quark  which is subsequently coupled to the spin of the heavy quark.  
This limit gives rise to two doublets, one with $j=1/2$ and the 
other $j=3/2$ and corresponds to two physically independent mixing angles 
$\theta=-\tan^{-1}(\sqrt{2})\simeq -54.7^\circ$ and 
$\theta=\tan^{-1}(1/\sqrt{2})\simeq 35.3^\circ$ \cite{godfrey,barnes}. Some 
authors prefer to use the $j-j$ basis \cite{eichten94}
but we will follow the  $L-S$ eigenstates convention implied in 
the spin--orbit terms given above and include the 
$LS$ mixing as a perturbation. It is straightforward to
transform between the $L-S$ basis and the $j-j$ basis.  We note that
radiative transitions are particularly sensitive to the 
$^3L_L-^1L_L$ mixing angle with the predictions from the different 
models giving radically different results.  We also note that the 
definition of the mixing angles are fraught with ambiguities.  For 
example, charge conjugating $c\bar{b}$ into $b\bar{c}$ flips the 
sign of the angle and the phase convention depends on the
order of coupling $\mathbf{L}$, $\mathbf{S}_Q$ and $\mathbf{S}_{\bar{Q}}$ 
\cite{barnes}.

\subsubsection{Relativistic corrections}

The Hamiltonian with the spin-dependent terms as written above is
actually inconsistent as it stands as the terms more singular than
$r^{-2}$ are illegal operators in the Schr\"odinger equation.  This is
resolved by returning to the full scattering amplitude which has the
effect of smearing the coordinate $\mathbf{r}$ out over distances of the
order of the inverse quark mass and the strengths of the various
potentials become dependent on the momentum of the interacting quarks.
The smearing of the potentials has the consequence of taming the
singularities. Alternatively, if one regards this Hamiltonian in the
spirit of effective field theories, these singular operators are
subleading in any reasonable power counting, and hence they must be
treated as a perturbation. They may need regularization (smearing) at
higher orders of perturbation theory, which introduces a scale
dependence.  This scale dependence cancels against the one of higher
order NRQCD matching coefficients, see \Section~\ref{sec:spnrqcdscc},

From this starting point different authors 
\cite{Eichten:ag,eichten78,godfrey85,stanley80,ebert03,gershtein94,fulcher91,fulcher99,fulcher94,gupta94,gupta96,zeng95}
diverge in how they 
incorporate further relativistic corrections.  For example, 
Godfrey and Isgur (GI) \cite{godfrey85}
use the full relativistic scattering amplitude as 
the starting point but do not take it literally and instead 
parameterize the various relativistic effects.  
The relativistic smearing is described by a quark form factor and 
momentum dependent corrections are parametrized in a form that is in 
keeping with the generalities, if not the details, of the 
$q\bar{q}$ scattering 
amplitude.  The reasoning is that the scattering 
amplitudes are for free Dirac Fermions while quarks inside a hadron 
are strongly interacting and will have off-mass-shell behavior. 
In addition, in field theory the Schr\"odinger equation arises in the 
$q\bar{q}$ sector of Fock space by integrating over more complex 
components of Fock space such as $|q\bar{q}g\rangle$.  This 
integration will introduce additional momentum dependence in the 
$q\bar{q}$ potential not reflected 
in eq. (\ref{eq:scatter}).  
There are other deficiencies that arise from taking 
eq. (\ref{eq:scatter}) 
literally.  Thus,  GI use the full scattering amplitude as a 
framework on which to build a semiquantitative model of relativistic 
effects.   While they acknowledge 
that this procedure is not entirely satisfactory they argue that it 
enables them to successfully describe all mesons, from the lightest to 
the heaviest, in a unified framework.  

In contrast, the more recent work by Ebert, Faustov and Galkin 
performs an expansion in powers of velocity, including all 
relativistic corrections of order $v^2/c^2$, including retardation 
effects and one-loop radiative corrections \cite{ebert03,ebert00}.  
Ebert {\it et al} use a quasipotential approach in which the 
quasipotential operator of the quark--antiquark interaction is 
constructed with the help of the off-mass-shell scattering amplitude.
The expression they derived to describe the spin-independent and 
spin-dependent corrections are rather lengthy and we refer the reader 
to their papers \cite{ebert03,ebert00,gupta}.
They found that relativistic effects are important,
particularly in radiative transitions (which are outside the scope of 
this section).  

While the GI calculation \cite{godfrey85}
assumed a short distance Lorentz-vector 
interaction and a Lorentz-scalar confining potential Ebert {\it et al} 
 \cite{ebert03,ebert00}
employ a mixture of long-range vector and scalar linear confining 
potentials.  The effective long-range vector vertex includes an 
anomalous chromomagnetic moment of the quark, $\kappa$.  The fitted 
value for $\kappa$ results in the vanishing of the 
long-range magnetic contribution to 
the potential so that the long range confining potential is 
effectively Lorentz scalar. 

In both cases taking the non-relativistic limit recovers eqns.
(\ref{eq:contact}--\ref{eq:tpso}).
Despite differences in the details of the various approaches
most recent calculations are in fairly good agreement.

\subsubsection[Charm mass corrections to the bottomonium mass spectrum]
              {Charm mass corrections to the bottomonium mass spectrum
               $\!$\footnote{Author: R.~Faustov }}
\label{sec:cfinite}

For the calculation of the bottomonium mass spectrum it is
necessary to take into account 
additional corrections due to the non-zero mass of the charm quark
\cite{hm,eiras,Melles:2000dq,Brambilla:2001qk}. The one-loop correction to the one-gluon exchange
part of the static $Q\bar Q$ potential in QCD  
due to the finite $c$ quark mass is given by \cite{Melles:2000dq,efg2002}
\begin{equation}
  \label{eq:deltav}
  \Delta V_{m_c}(r)= -\frac49\frac{\alpha_{\rm s}^2(\mu)}{\pi
  r}\left[\ln(\sqrt{a_0}m_c r) +\gamma_E+E_1(\sqrt{a_0}m_c
  r)\right],\qquad E_1(x)=\int_x^\infty e^{-t}\ \frac{dt}t
\end{equation}  
where $\gamma_E\cong0.5772$ is the Euler constant  and $a_0\cong5.2$. 
Averaging of $\Delta V_{m_c}(r)$ over solutions of the relativistic
wave equation with the Cornell and Coulomb potentials
yields the bottomonium mass shifts presented in Table~\ref{tab:fmc}.
\begin{table}[htbp]
\caption{Charm mass corrections to the bottomonium masses (in MeV).}  
\label{tab:fmc}
\begin{center}
\begin{tabular}{|c|c|c|c|c|c|c|}
\hline
State& $1S$ & $1P$& $2S$ & $1D$ &
$2P$ & $3S$\\ 
\hline
$\left<\Delta V_{m_c}\right>_{\rm Cornell}^{\alpha_{\rm s}=0.22}$ \cite{efg2002}
& $-12$&$-9.3$ &$-8.7$& 
$-7.6$&$-7.5$&$-7.2$\\
\hline
$\left<\Delta V_{m_c}\right>_{\rm Coul}^{\alpha_{\rm s}=0.22}$&$-9.5$ &$-4.2$
&$-3.8$ &$-2.3$ &$-2.2$ &$-2.1$ \\
$\left<\Delta V_{m_c}\right>_{\rm Coul}^{\alpha_{\rm s}=0.3}$&$-20.7$ &$-9.7$
&$-8.8$ &$-5.5$ &$-5.2$ &$-4.9$ \\
\hline
$\left<\Delta V_{m_c}\right>_{\rm Coul}$ \cite{Brambilla:2001qk} &$-14.3$ &$-22.1$
&$-21.9$ &  & $-49$ & $-40.5$\\
$\alpha_{\rm s}(\mu)$ & 0.277 & 0.437 & 0.452 & & 0.733 & 0.698\\
\hline 
\end{tabular}
\end{center}
\end{table}

The \Table~\ref{tab:fmc} shows that for a fixed value of $\alpha_{\rm s}$ the
averaging with and without confining potential substantially differ
especially for the excited states. For growing $n=n_r+L+1$ the values
of $\left<\Delta V_{m_c}\right>$ slowly decrease for Cornell potential whereas
for the Coulomb potential with a fixed value of $\alpha_{\rm s}$ they fall
rapidly. The bottomonium mass spectrum with the account of the finite
charm mass corrections was obtained in Refs.~\cite{Brambilla:2001qk,ebert03}

\subsubsection[Coupled-channel effects]{Coupled-channel effects}
\label{sec:coupled}

An important ingredient that has not received the attention it
deserves but which has been brought to the forefront by some
spectacular recent failures of quark models are coupled channel
effects.  As the mass of a quarkonium state approaches the threshold
for decay to pairs of flavoured mesons, contributions from virtual
loops of the flavoured meson channels are expected to make important
contributions to masses and other meson properties
\cite{Eichten:ag,eichten78,Isgur:1999cd}.  These coupled channel
effects are expected to shift masses from naive quark model
predictions and to alter decay and production properties due to higher
order Fock-space components present in the wavefunctions.  These may
account for the discrepancies between quark model predictions and
those of the recently discovered and $X(3872)$ properties
\cite{Barnes:2003vb,Eichten:2004uh}.  There has been very little work
on this important subject since the original Cornell model
\cite{Eichten:ag,eichten78} and it is an important topic that needs to
be addressed \cite{Eichten:2004uh}.  For the charmonium example the
present situation is discussed in
\Section~\ref{sec:spopcha}.

\subsection[Comparison of models with experiment]
           {Comparison of models with experiment
            $\!$\footnote{Authors:S. Godfrey}}
\label{sec:cme}

\subsubsection[Bottomonium]{Bottomonium}

We start with the $b\bar{b}$ system as it has the most states observed
of any of the heavy quarkonium systems (see \Table~\ref{tab:bb}).  This
is due to the fact that threshold for the Zweig allowed decay to
$B\overline{B}$ lies above the $3S$ state.  The $J^{PC}=1^{--}$
$n^3S_1$ states are copiously produced in $e^+e^-$ annihilation and
can decay via $E1$ transitions to the $1^3P_J$ and $2^3P_J$
multiplets.  The masses of the $\chi_b$ states provide valuable tests
of the spin-dependence of the various models.  In particular, the
splittings of the $^3P_J$ masses are determined by the spin--orbit and
tensor terms which are sensitive to the presence of vector and scalar
interactions.  The Lorentz vector one-gluon-exchange plus Lorentz
scalar linear confinement gives a good description of the data (as
long as no velocity dependent corrections are included
\cite{Brambilla:1989ur,Barnes:1982eg}.

A test of potential models is their ability to predict as yet unseen 
properties correctly.  Most potential models predict that the lowest 
D-wave centre of gravity is around 10.16~GeV.  Although details of 
the multiplet splittings differ most models predict that the splittings 
are smaller than in the P-wave states.  Thus,  the observation of 
these states represents an important test of potential models. 
 
 Recently the CLEO collaboration 
has observed the first D-wave $b\bar{b}$ state in the cascade 
$\Upsilon(3S) \to$ $\chi_b' \gamma  \to  ^3D_J \gamma\gamma \to 
\chi_b \gamma \gamma \gamma \to \Upsilon(1S) \gamma \gamma \gamma 
\gamma $ \cite{Csorna:2002jg}.  
Due to expected transition probabilities (essentially 
reliable Clebsch factors)  it is believed that the observed state is 
the $J=2$, $1^3D_2$ state.  This is an important observation as it is 
able to distinguish among the various models \cite{Godfrey:2001vc}.  
Unfortunately this 
programme at CLEO  is completed and it is not clear when there will be 
another opportunity to search for more of the missing states.

So far no spin singlet $b\bar{b}$ state has been observed.  
The mass splittings between the singlet and triplet states is a key 
test of the applicability of perturbative quantum chromodynamics to 
the $b\bar{b}$ system and is a useful check of lattice QCD results.
The $\eta_b$ ($n^1S_0$) states can be produced via M1 radiative 
transitions from the $\Upsilon$ ($n^3S_1$) states, either unhindered 
or hindered, and via E1 radiative transitions from the $n^1P_1$ 
states \cite{Godfrey:2001eb}.   In the latter case, the decay chain 
would be $\Upsilon(3S) \to h_b(^1P_1) \pi\pi$ followed by $h_b\to 
\eta_b \gamma$.  The decay chain 
$\Upsilon(3S)\to h_b +\pi^0\to \eta_b + \pi^0 +\gamma$ is also 
possible \cite{Godfrey:2002rp}.  We note that there does not appear to 
be a consensus in the literature on the relative importance of the two 
$\Upsilon \to h_b$ hadronic transitions.
The decay chains proceeding via an intermediate $h_b$ 
would also be a means of observing the $h_b$ state.  
A recent run by CLEO did not lead to reports of the 
observation of the $\eta_b$ state although the limits straddles the 
range of predictions.  There is also the possibility that the $\eta_b$ 
can be observed by the Tevatron and LHC experiments.

\begin{table}[t]
\caption{Predicted and observed masses of $b\bar{b}$ states.
\label{tab:bb}}
\begin{center}
\begin{tabular}{|l|c|cccccc|} \hline
State           & expt  & GI85
                        & FU91
                        & EQ94
                        & GJ96
                        & EFG03
                        & ZVR95 \\
                &       & \cite{godfrey85}
                        & \cite{fulcher91}
                        & \cite{eichten94}
                        & \cite{gupta96}
                        & \cite{ebert03} 
                        & \cite{zeng95}\\
\hline
 $1^3{\rm S}_1 $ & 9460 & 9465  & 9459  & 9464 & 9460 & 9460 & 9460 \\
 $1^1{\rm S}_0 $ &      & 9402  & 9413  & 9377 & 9408 & 9400 & 9410 \\
\hline
 $1^3{\rm P}_2 $ & 9913 & 9897  & 9911  & 9886 & 9914 & 9913 & 9890  \\
 $1^3{\rm P}_1 $ & 9893 & 9876  & 9893  & 9864 & 9893 & 9892 & 9870  \\
 $1^3{\rm P}_0 $ & 9860 & 9847  & 9865  & 9834 & 9862 & 9863 & 9850 \\
 $1^1{\rm P}_1 $ &      & 9882  & 9900  & 9873 & 9901 & 9901 & 9880 \\
\hline
 $2^3{\rm S}_1 $ & 10023 & 10003& 10015 & 10007 & 10016 & 10023 & 10020  \\
 $2^1{\rm S}_0 $ &       & 9976 &  9992 & 9963 & 9991   & 9993 & 10000 \\
\hline
 $1^3{\rm D}_3 $ &      & 10155 & 10172 & 10130 &       & 10162 & 10150 \\
 $1^3{\rm D}_2 $ & 10162 & 10147& 10166 & 10126 &       & 10158 & 10150 \\
 $1^3{\rm D}_1 $ &       & 10138& 10158 & 10120 &       & 10153 & 10140 \\
 $1^1{\rm D}_2 $ &      & 10148 & 10167 & 10127 &       & 10158 & 10150 \\ 
\hline
 $2^3{\rm P}_2 $ & 10269 & 10261& 10269 & 10242 & 10270 & 10268 & 10280 \\
 $2^3{\rm P}_1 $ & 10255 & 10246& 10256 & 10224 & 10254 & 10255 & 10260 \\
 $2^3{\rm P}_0 $ & 10232 & 10226& 10234 & 10199 & 10229 & 10234 & 10240 \\
 $2^1{\rm P}_1 $ &      & 10250 & 10261 & 10231 & 10259 & 10261 & 10270 \\
\hline
 $3^3{\rm S}_1 $ & 10355 & 10354& 10356 & 10339 & 10358 & 10355 & 10390 \\
 $3^1{\rm S}_0 $ &      & 10336 & 10338 & 10298 & 10338 & 10328 & 10370 \\
\hline
\end{tabular}
\end{center}
\end{table}

\subsubsection[Charmonium]{Charmonium}

The discovery of the $J/\psi$ and $\psi'$ states revolutionized our 
understanding of hadron spectrocopy by demonstrating that they could 
be well described by potential models with the qualitative features 
expected from QCD (see \Table~\ref{tab:cc}).  
\shortpage[2]

The spin triplet $^3S_1$ states are produced copiously in $e^+e^-$ 
annihilation and the $^3P_J$ states are produced via $E1$ radiative 
transitions.
The $\chi_0 \; (^3P_0)$, $\chi_1 \; (^3P_1)$ and $\chi_2\; (^3P_2)$ 
$c\bar{c}$ states were first discovered in radiative 
decays from the $2^3S_1$ level (the $\psi(3685)$).  The $\chi$ states 
themselves undergo radiative transitions to the $J/\psi$ with measured 
partial widths in reasonable agreement with theoretical predictions once 
relativistic effects are taken into account.

The singlet states have been far more elusive.  The $1^1S_0$ 
state has been known for some time, seen in magnetic dipole ($M1$) 
transitions from both the $J/\psi$ and $\psi'$. 
In contrast, a strong claim for observation of
the $2^1S_0$ state has only occurred recently, first with its 
observation in the decay $B\to K \eta_c'$, $\eta_c'\to K_sK^+\pi^- $ 
by the Belle Collaboration \cite{Choi:2002na}
and its subsequent observation by Belle in the mass spectrum recoiling 
against $J/\psi$ in $e^+e^-$ annihilation \cite{Abe:2002rb}
and by CLEO  \cite{ernst}
and Babar \cite{Wagner:2003qb} in $\gamma \gamma $ collisions. 
While the mass 
measurement by Belle was higher than expected by most quark potential 
models, the current world average \cite{Eidelman:pdg2004}
is in reasonable agreement with theory. 

One place the models disagree is in the mass of the $1^1P_1$ 
state relative to the $1^3P_J$ cog \cite{Godfrey:2002rp}.  
However, the $^1P_1$ state has yet to be confirmed. 
The $1^3P_{cog}-1^1P_1$ splitting is 
dependent on the Lorentz structure of the interquark potentials and
relativistic corrections so that the $h_c$ 
mass measurement is an important test of 
perturbative QCD and more phenomenological quark potential models
which have a large variation of predictions.  
The decay chain 
$\psi'\to h_c +\pi^0\to \eta_c + \pi^0 +\gamma$ has been 
discussed as a possible mode of discovery of the 
$h_c$ \cite{Godfrey:2002rp}.  
Optimistically, one might hope 
that the current CLEO run will see evidence for the $h_c$ in this cascade. 

The charmonium D-wave states are predicted to lie above
$D\overline{D}$ threshold.  The $\psi(3770)$ is associated with the
$1^3D_1$ state.  It's leptonic width is larger than expected for a
pure $D$ state which is probably due to mixing with the $2^3S_1$ state
induced by tensor mixing or coupled channel effects.  The $1^3D_3$,
$1^3D_2$, and $1^1D_2$ are predicted to lie close in mass to the
$\psi(3770)$.  A $J^P=2^-$ state cannot decay to two $0^-$ particles
so the $1^3D_2$ and $^1D_2$ cannot decay to $D\overline{D}$ and are
expected to lie below the $D^*\overline{D}$ threshold.  They are
therefore expected to be narrow with prominent transitions to lower
$c\bar{c}$ states.  While there is no such conservation law for the
$1^3D_3$ state, recent calculations indicate that it should also be
relatively narrow, ${\cal O}$(MeV), due to the angular momentum
barrier \cite{Barnes:2003vb,Eichten:2004uh}.  It is therefore possible
that all $c\bar{c}$ D-wave states will be observed.  A $c\bar{c}$
state has recently been observed in $B$ decay, the $X(3872)$
\cite{Choi:2003ue}.  It's mass is higher than expected by quark models
which has led to considerable speculation about whether it is a
conventional $c\bar{c}$ state or a $D\bar{D}^*$ molecule
\cite{Swanson:2004pp}.  A number of tests have been proposed to sort
this out \cite{Barnes:2003vb,Eichten:2004uh} and experimental analysis
is in progress. Observation of the $\eta_{c2}$ and $\psi_{(2,3)}$
states would constrain spin-dependent interactions and provide
insights into the importance of coupled channel effects in the charm
threshold region.
\shortpage

\begin{table}[th]
\caption{Predicted and observed masses of $c\bar{c}$ states (in MeV).
\label{tab:cc}}
\begin{center}
\begin{tabular}{|l|c|cccccc|} \hline
State &  Expt
        & GI85
        & EQ94
        & FU91
        & GJ96
        & EFG03
        & ZVR95  \\ 
        &       & \cite{godfrey85} 
        & \cite{eichten94}
        & \cite{fulcher91}
        & \cite{gupta96}
        & \cite{ebert03} 
        & \cite{zeng95}\\ 
\hline
 $1^3{\rm S}_1 $ & $3096.87 \pm 0.04$ 
                        & 3098  & 3097  & 3104 & 3097 & 3096 & 3100 \\
 $1^1{\rm S}_0 $ & $2979.8 \pm 1.8$ 
                        & 2975  & 2980  & 2987 & 2979 & 2979 & 3000 \\
\hline
 $1^3{\rm P}_2 $ & $3556.18\pm 0.13$ 
                        & 3550  & 3507  & 3557 & 3557   & 3556 & 3540 \\
 $1^3{\rm P}_1 $ & $3510.51\pm 0.12$ 
                        & 3510  & 3486  & 3513 & 3511   & 3510 & 3500  \\
 $1^3{\rm P}_0 $ & $3415.0\pm 0.8$  
                        & 3445  & 3436  & 3404 & 3415   & 3424 & 3440  \\
 $1^1{\rm P}_1 $ &              
                        & 3517  & 3493  & 3529 & 3526 & 3526 & 3510 \\
\hline
 $2^3{\rm S}_1 $ & $3685.96\pm 0.09$ 
                        & 3676  & 3686  & 3670 & 3686 & 3686 & 3730 \\
 $2^1{\rm S}_0 $ & $3654\pm 10$ 
                        & 3623  & 3608  & 3584 & 3618 & 3588 & 3670 \\
\hline
 $1^3{\rm D}_3 $ &      & 3849  &       & 3884 &      & 3815 & 3830 \\
 $1^3{\rm D}_2 $ &      & 3838  &       & 3871 &      & 3813 & 3820 \\
 $1^3{\rm D}_1 $ & $3769.9\pm 2.5$ 
                        & 3819  &       & 3840 &      & 3798 & 3800 \\
 $1^1{\rm D}_2 $ &      & 3837  &       & 3872 &      & 3811 & 3820 \\ \hline
 $2^3{\rm P}_2 $ &      & 3979  &       &      &      & 3972 & 4020 \\
 $2^3{\rm P}_1 $ &      & 3953  &       &      &        & 3929 & 3990 \\
 $2^3{\rm P}_0 $ &      & 3916  &       &      &        & 3854 & 3940 \\
 $2^1{\rm P}_1 $ &      & 3956  &       &      &        & 3945 & 3990 \\
\hline
 $3^3{\rm S}_1 $ &      & 4100  &       &  &    & 4088 & 4180 \\
 $3^1{\rm S}_0 $ &      & 4064  &       &  &    & 3991 & 4130 \\
\hline
\end{tabular}
\end{center}
\end{table}

\subsubsection{$B_c$ mesons}

\begin{figure}
\begin{center}
\includegraphics[width=.8\textwidth]{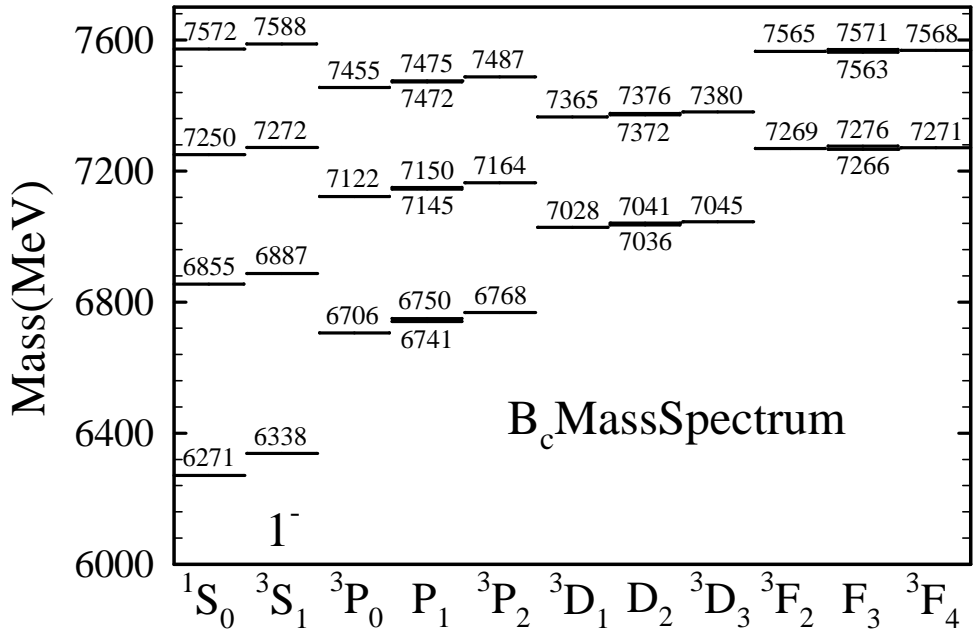}
\end{center}
\caption{$B_c$ spectrum.}
\label{fig:bc}
\end{figure}

The $B_c$ mesons provide a unique window into heavy quark dynamics.
Although they are intermediate to the charmonium and bottomonium
systems the properties of $B_c$ mesons are a special case in
quarkonium spectroscopy as they are the only quarkonia consisting of
heavy quarks with different flavours.  Because they carry flavour they
cannot annihilate into gluons so are more stable and excited $B_c$
states lying below $BD$ (and $BD^*$ or $B^*D$) threshold can only
undergo radiative or hadronic transitions to the ground state
pseudoscalar which then decays weakly.  This results in a rich
spectroscopy of narrow radial and orbital excitations
(\Figure~\ref{fig:bc} and \Table~\ref{tab:bc})
\cite{godfrey85,ebert03,gershtein94,fulcher99,gupta96,zeng95,eichten94,bcspec,gershtein95,gouz02,godfrey04,Baldicchi:2002qm}.
which are more stable than their charmonium and bottomonium analogues.
The hadronic transitions emitting two charged pions should offer a
good opportunity to reconstruct the excited $B_c$ state.

The discovery of the $B_c$ meson by the 
Collider Detector at Fermilab (CDF) Collaboration \cite{abe1998}
in $p\bar{p}$ collisions at $\sqrt{s}=1.8$~TeV has demonstrated the 
possibility of the experimental study of this system and has 
stimulated considerable interest in $B_c$ spectroscopy.
Calculations of $B_c$ cross-sections at hadron colliders
predict that 
large samples of $B_c$ states should be produced at the Tevatron 
and at the LHC opening up this new spectroscopy.  It should therefore 
be possible to start exploring $c\bar{b}$ spectroscopy at the 
Tevatron, producing $1P$ and $2S$ states and possibly even the 
D-wave states in sufficient numbers to be observed. At the LHC, with 
its higher luminosity, the D-wave $c\bar{b}$ states should be produced 
in a sizable number so that the LHC should allow the study of the 
spectroscopy and decay of $B_c$ mesons.  
  
\begin{table}[ht!]
\caption{Predicted $B_c$ masses and spin--orbit mixing angles (in MeV).}
\label{tab:bc}
\begin{center}
\begin{tabular}{|l|cccccccc|} \hline
State 
        & GI85
        & EFG03 
        & FU99
        & GKLT94
        & EQ94
        & GJ96
        & ZVR95 
        & Lattice \\ 
        & \cite{godfrey85} 
        &  \cite{ebert03} 
        & \cite{fulcher99}
        & \cite{gershtein94}
        & \cite{eichten94} 
        &  \cite{gupta96} 
        &  \cite{zeng95} 
        &  \\ 
\hline
$1^3S_1 $ & 6338 & 6332 & 6341 & 6317 & 6337 & 6308 & 6340 & $6321\pm 30$\\
$1^1S_0 $ & 6271 & 6270 & 6286 & 6253 & 6264 & 6247 & 6260 & 
                                                 $6280\pm30\pm190$\\ \hline
$1^3P_2 $ & 6768 & 6762 & 6772 & 6743 & 6747 & 6773 & 6760 & $6783\pm 30$ \\
$1 P_1' $ & 6750 & 6749 & 6760 & 6729 & 6730 & 6757 & 6740 & $6765\pm 30$\\
$1 P_1 $  & 6741 & 6734 & 6737 & 6717 & 6736 & 6738 & 6730 & $6743\pm 30$\\
$1^3P_0 $ & 6706 & 6699 & 6701 & 6683 & 6700 & 6689 & 6680 & $6727\pm 30$\\
$\theta_{1P}$ & 22.4$^\circ$ & 20.4$^\circ$ & 28.5$^\circ$  
                & 17.1$^\circ$  & $\sim 2^\circ$ & 25.6$^\circ$ 
                                        & & $33.4\pm 1.5^\circ$ \\ \hline
$2^3S_1 $ & 6887 & 6881 & 6914 & 6902 & 6899 & 6886 & 6900 & $6990\pm 80$ \\
$2^1S_0 $ & 6855 & 6835 & 6882 & 6867 & 6856 & 6853 & 6850 & $6960\pm 80$\\ \hline
$2^3P_2 $ & 7164 & 7156 &       & 7134 & 7153 &      & 7160 & \\
$2 P_1' $ & 7150 & 7145 &       & 7124 & 7135 &      & 7150 & \\
$2 P_1 $  & 7145 & 7126 &       & 7113 & 7142 &      & 7140 & \\
$2^3P_0 $ & 7122 & 7091 &       & 7088 & 7108 &      & 7100 & \\ 
$\theta_{2P}$ & 18.9$^\circ$ & 23.0$^\circ$ &   
                & 21.8$^\circ$  & $17^\circ$  &         &   & \\ \hline
$3^3S_1 $ & 7272 & 7235 &       &       & 7280 &     & 7280 & \\
$3^1S_0 $ & 7250 & 7193 &       &       & 7244 &     & 7240 & \\ \hline
$1^3D_3 $ & 7045 & 7081 & 7032  & 7007  & 7005 &     & 7040 & \\
$1 D_2' $ & 7036 & 7079 & 7028  & 7016  & 7012 &     & 7030 & \\
$1 D_2 $  & 7041 & 7077 & 7028  & 7001  & 7009 &     & 7020 & \\
$1^3D_1 $ & 7028 & 7072 & 7019  & 7008  & 7012 &     & 7010 & \\
$\theta_{1D}$ & 44.5$^\circ$ & -35.9$^\circ$    &       &  
                & 34.4$^\circ$   &      &   &  \\   \hline
$1^3F_4 $ & 7271 &      &       &       &       &    & 7250 & \\
$1 F_3' $ & 7266 &      &       &       &       &    & 7250 & \\
$1 F_3 $  & 7276 &      &       &       &       &    & 7240 & \\
$1^3F_2 $ & 7269 &      &       &       &       &    & 7240 & \\
$\theta_{1F}$ & 41.4$^\circ$ &  &       &  &  &   & & \\
\hline
\end{tabular}
\end{center}
\end{table}

\subsection[Coupling to open-charm channels]
           {Coupling to open-charm channels 
            $\!$\footnote{Authors: E. Eichten}}
\label{sec:spopcha}

\subsubsection{Theoretical models}

Near the threshold for open heavy flavour pair production, there are
significant nonperturbative contributions from light quark pairs to
the masses, wavefunctions and decay properties of physical $Q\bar Q$
states.  QCD sum rules \cite{Novikov:tn,Voloshin:1979uv} have been
used to obtain some results \cite{Azimov:1982ef, Voloshin:1987rp,
Kiselev:1994zy} and lattice QCD calculations extended into the
flavour-threshold region \cite{Baliinprep} should eventually give a
firm basis for predictions.  However, at present a more
phenomenological approach is required to provide a detailed
description of these effects.

The effects of light quark pairs near open heavy flavour threshold can
be described by coupling the potential model $Q\bar Q$ states to
nearby physical multibody states.  In this threshold picture, the
strong interactions are broken into sectors defined by the number of
valence quarks.  This separation is reminiscent of the Tamm--Dancoff
approximation \cite{TDA}.  The dynamics of the $Q\bar Q$ states (with
no valence light quarks, $q$) is described by the interaction ${\cal
H}_0$. Nonrelativistic potential models are normally used to determine
the properties of the resulting bound states in this sector. In this
framework excitations of the gluonic degrees of freedom would also be
contained the spectrum of ${\cal H}_0$.

The two meson sector $Q\bar q + q\bar Q$ are described by the
Hamiltonian ${\cal H}_2$.  In the simplest picture, ${\cal H}_2$ is
assumed to be described the low-lying spectrum of two free heavy-light
mesons.  The physical situation is more complex.  At large separation
between two mesons the interactions are dominated $t$-channel pion
exchanges.  For states very near threshold such as the X(3872)
charmonium state such pion exchange in attractive channels might have
significant effects on properties of the physical states
\cite{Braaten:2003he}.  At somewhat shorter distances, more
complicated interactions exist and new bound states might arise,
\eg molecular states \cite{DeRujula:1976qd,DeRujula:1977fs}.

Our command of quantum chromodynamics is inadequate to derive a
realistic description of the interactions, ${\cal H}_I$, that
communicate between the $Q\bar Q$ and $Q\bar q + q\bar Q$ sectors.
Two simple phenomenological models have been used to describe this
coupling: the Cornell coupled-channel model (CCC) and the vacuum quark
pair creation model (QPC).

The Cornell coupled-channel model for light quark pair creation
\cite{Eichten:ag} generalizes the Cornell $Q\bar{Q}$ model
\cite{eichten78} without introducing new parameters, writing the
interaction Hamiltonian as
\begin{equation}
     {\cal H}_{I} = \frac{3}{8} \sum_{a} 
     \int:\rho_{a}(\mathbf{r}) V(\mathbf{r} - 
     \mathbf{r}^{\prime})\rho_{a}(\mathbf{r}^{\prime}): 
     d^{3}{r}\,d^{3}{r}^{\prime}\; ,
     \label{eq:CCC}
\end{equation}
where $V$ is the quarkonium potential and $\rho_{a}(\mathbf{r}) =
\displaystyle\frac{1}{2}\psi^{\dagger}(\mathbf{r})\lambda_{a}\psi(\mathbf{r})$
is the colour current density, with $\psi$ the quark field operator and
$\lambda_{a}$ the octet of SU(3) matrices.  To generate the relevant
interactions, $\psi$ is expanded in creation and annihilation
operators (for up, down, strange and heavy quarks), but transitions
from two mesons to three mesons and all transitions that violate the
Zweig rule are omitted. It is a good approximation to neglect all
effects of the Coulomb piece of the potential in \Eq~(\ref{eq:CCC}).
It was shown that this simple model coupling charmonium to
charmed-meson decay channels gives a qualitative understanding of the
structures observed above threshold while maintaining the successes of
the single-channel $c\bar{c}$ analysis below threshold
\cite{eichten78}.

The characteristic  of the CCC model is the use of 
the time component of a long-range vector interaction between the 
heavy quarks colour densities rather than the Lorentz scalar confining interaction.  

The vacuum quark pair creation model (QPC).
This model was developed by Le Yaouanc {\it et. al.}
\cite{LeYaouanc:1972ae,LeYaouanc:1973xz,LeYaouanc:1975sa}
based on an earlier idea of Micu \cite{Micu:1968mk} that the light quark pair is 
produced from the vacuum with vacuum quantum numbers ${\rm J}^{\rm PC} = 0^{++}$.  
The model is also referred to as the \slj{3}{2}{0} model.
The form of the interaction Hamiltonian is
\begin{equation}
  {\cal H}_{I} = \gamma  \int \bar{\psi}\psi (\mathbf{r}) d^{3}{r}\,\; 
\label{eq:QPC}
\end{equation}
The constant $\gamma$ is a free parameter of the model.  This model
has been applied to the light meson states
\cite{Barnes:1996ff,Godfrey:1998pd}.  It was first applied above charm
threshold by the Orsay group \cite{LeYaouanc:1977ux}.

The main theoretical weakness of the QPC model is its failure to
reproduce the vanishing of the pair production amplitudes for a static
$Q\bar Q$ source at zero spatial separation.  The flux tube breaking
model \cite{Isgur:1984bm,Kokoski:1985is} somewhat addresses this
weakness. It has the same basic interaction as the QPC model
(\Eq~\ref{eq:QPC}) but the integration is only over a region near a
"string" between the $Q$ and $\bar Q$ positions. This region is
defined by a upper bound on the shortest distance between the pair
creation point and the string.  Detailed applications of QPC models to
the quarkonium systems are presently under investigation
\cite{Barnes:2004cz}.

There have been attempts to compare the various models for quark pair 
creation \cite{Ackleh:1996yt,Simonov:2002gx,Brambilla:1989ur}.  
At present the most studied system is the open charm 
threshold region and we will focus on that system below.
However, the same threshold effects 
are present in the $b\bar b$ states near $B\overline{B}$ threshold and 
$c\bar b$ states near $D\overline{B}$ threshold.  
A detailed comparison of the scaling behaviour between different 
heavy quark systems would provide valuable insight into the correct form for the 
coupling to light-quark pairs. 
 
\subsubsection{Mass shifts}

\begin{table}[th]
\caption[Charmonium spectrum]
        {Charmonium spectrum, including the influence of open-charm
         channels. All masses are in MeV. The penultimate column holds
         an estimate of the spin splitting due to tensor and
         spin--orbit forces in a single-channel potential model. The
         last column gives the spin splitting induced by communication
         with open-charm states, for an initially unsplit multiplet.
         From \cite{Eichten:2004uh}.
\label{tab:delM}}
\begin{center}
\begin{tabular}{|ccccc|}
\hline 
State & Mass & Centroid & $\begin{array}{c} {\rm Splitting} 
\\ {\rm (Potential)} \end{array}$ &  $\begin{array}{c} {\rm Splitting} 
\\ {(\rm Induced)} \end{array}$ \\
\hline
& & & & \\[-9pt]
$\begin{array}{c}
    1\slj{1}{1}{0} \\
    1\slj{3}{1}{1} 
\end{array}$ &
$\begin{array}{c} 2\,979.9 \\ 3\,096.9 
\end{array}$ & $3\,067.6$ & $\begin{array}{c} -90.5 
\\ +30.2 \end{array}$
   &  $\begin{array}{c} +2.8  \\ -0.9 \end{array} $\\ & & & & \\[-6pt]
$\begin{array}{c} 
1\slj{3}{2}{0} \\
1\slj{3}{2}{1} \\
1\slj{1}{2}{1} \\
1\slj{3}{2}{2}  \end{array}$ &  $\begin{array}{c}
3\,415.3\\
3\,510.5\\
3\,525.3\\
3\,556.2 \end{array}$
& $3\,525.3$ &  $\begin{array}{c} 
-114.9 \\ -11.6
\\ +1.5 \\ +31.9 \end{array}$ & 
$\begin{array}{c} +5.9 \\ -2.0 \\ +0.5 \\ -0.3
\end{array}$ \\
& & & & \\[-6pt]
$\begin{array}{c}
    2\slj{1}{1}{0} \\
    2\slj{3}{1}{1} 
\end{array}$ &
$\begin{array}{c} 3\,637.7 \\ 3\,686.0
\end{array}$ & $3\,673.9$ & $\begin{array}{c} -50.4 
\\ +16.8 \end{array}$
   &  $\begin{array}{c} +15.7  \\ -5.2 \end{array} $\\
   & & & & \\[-6pt]
$\begin{array}{c} 
 1\slj{3}{3}{1} \\
 1\slj{3}{3}{2} \\
 1\slj{1}{3}{2} \\
 1\slj{3}{3}{3}  \end{array}$ &  $\begin{array}{c}
 3\,769.9\\
 3\,830.6\\
 3\,838.0\\
 3\,868.3 \end{array}$
 & 
 (3\,815) & 
 $\begin{array}{c} -40 \\ 0 \\ 
 0 \\ +20 \end{array} $ & 
 $\begin{array}{c} -39.9 \\ -2.7\\ +4.2 \\ +19.0
 \end{array}$ \\ & & & & \\[-6pt]
 $\begin{array}{c} 
 2\slj{3}{2}{0} \\
 2\slj{3}{2}{1} \\
 2\slj{1}{2}{1} \\
 2\slj{3}{2}{2}  \end{array}$ &  $\begin{array}{c}
 3\,931.9\\
 4\,007.5\\
 3\,968.0\\
 3\,966.5 \end{array}$
 & 3\,968 & 
 $\begin{array}{c} -90 \\ -8 \\ 
 0 \\ +25 \end{array} $ & 
 $\begin{array}{c} +10 \\ +28.4 \\ -11.9 \\ -33.1
 \end{array}$ \\[3pt]
\hline
\end{tabular}
\end{center}
\end{table}

The mass $\omega$  of the quarkonium state $\psi$ in the presence of 
coupling to decay channels is given by:
\begin{equation} 
    [{\cal H}_0 + {\cal H}_2 + {\cal H}_I ] \psi = \omega \psi .
\label{eq:decayM}
\end{equation}
Above threshold $\omega$ has both a real (mass) and imaginary part (width). 
 
The basic coupled-channel interaction ${\cal H}_I$ (\Eq~(\ref{eq:CCC})
or \Eq~(\ref{eq:QPC})) appearing in \Eq~(\ref{eq:decayM}) is
independent of the heavy quarks spin, but the hyperfine splittings of
$D$ and $D^{*}$, $D_{s}$ and $D_{s}^{*}$, induce spin-dependent forces
that affect the charmonium states.  These spin-dependent forces give
rise to S--D mixing that contributes to the $\psi(3770)$ electronic
width, for example, and are a source of additional spin splitting.

The masses resulting from a full coupled channel analysis
\cite{Eichten:2004uh} in the CCC model are shown in the second column
of \Table~\ref{tab:delM}.  The parameters of the potential model
sector, ${\cal H}_0$, must be readjusted to fit the physical masses,
$\omega$, to the observed experimental values.  To compute the induced
splittings, the bare centroid of the spin-triplet states is adjusted
so that the physical centroid, after inclusion of coupled-channel
effects, matches the value in the middle column of
\Table~\ref{tab:delM}.  The centroid for the 1D masses is determined
by pegging the observed mass of the 1\slj{3}{3}{1} $\psi(3770)$.  For
the 2P levels, the bare centroid is adjusted so that the
2\slj{1}{2}{1} level lies at the centroid of a potential-model
calculation.  The assumed spin splittings in the single-channel
potential model are shown in the penultimate column and the induced
coupled channel spin splittings for initially unsplit multiplets are
presented in the rightmost column of \Table~\ref{tab:delM}.  The
shifts induced in the low-lying 1S and 1P levels are small.  For the
other known states in the 2S and 1D families, coupled-channel effects
are noticeable and interesting.

In a simple potential picture, the $\eta_{c}(2S)$ level lies below the
$\psi(2S)$ by the hyperfine splitting given by
\begin{equation} 
M(\psi{(2S)}) - M(\etac(2S)) 
    = \frac{|\psi(2S)(0)|^2}{|\psi(0)|^2} [M(\psi) - M(\etac)]\;. 
\end{equation} 
Using the observed 1S hyperfine splitting, $M(\psi) - M(\eta_{c}) =
117\mev$, one would find $M(\psi(2S)) - M(\eta_{c}(2S)) = 67\mev$,
which is larger than the observed $48.3 \pm 4.4\mev$, as is typical
for potential-model calculations.

One important result of coupling the open-charm threshold is that the
$\psip$ receives a downward shift of the nearby $D\overline{D}$, that
the $\etac^{\prime}$ does not get, as this state does not couple to
$D\overline{D}$.  This is implicitly present in the early Cornell
papers\cite{eichten78}, but the shift of spin singlets states was not
explicitly calculated.  The effect was first mentioned by Martin and
Richard\cite{Martin:1982nw,Martin:2003rr}, who calculated the size of
the effect.  Recent papers using the CCC model
interaction\cite{Eichten:2002qv, Eichten:2004uh} have confirmed this
behaviour.  In fact, the 2S induced shifts in \Table~\ref{tab:delM}
draw $\psi^{\prime}$ and $\eta_{c}^{\prime}$ closer by $20.9\mev$,
substantially improving the agreement between theory and experiment.
This suggests that the $\psi^{\prime}$--$\eta_{c}^{\prime}$ splitting
reflects the influence of virtual decay channels.

If the observed $X(3872)$ is a charmonium state, it is most naturally
interpreted as the 1\slj{3}{3}{2} or 1\slj{3}{3}{3} level
\cite{Barnes:2003vb,Eichten:2004uh}; if not, both these states remain
to be observed and the dynamics of ${\cal H}_2$ is significantly
richer.  As shown in \Table~\ref{tab:delM}, the coupling to open-charm
channels increases the 1\slj{3}{3}{2}--1\slj{3}{3}{1} splitting by
about $20\mev$, but does not fully account for the observed $102\mev$
separation between $X(3872)$ and $\psi(3770)$.  However the position
of the $3^{--}$ 1\slj{3}{3}{3} level turns out to be very close to
$3872\mev$.

\subsubsection{Mixing and physical state properties}

The physical states are not pure potential-model eigenstates but
include components with two virtual (real above threshold) open flavour meson
states.  Separating the physical state ($\psi$) into $Q\bar Q$ ($\psi_0$) and
two meson components ($\psi_2$), the resulting separation $\cal H$ by sector
leads to an effective Hamiltonian for the $\psi_0$ sector given by:
\begin{equation} 
\left[ {\cal H}_0 + {\cal H}_I^{\dagger} \frac{1}{\omega - {\cal H}_2 + i\epsilon}
{\cal H}_I \right] \psi_0 = \omega \psi_0 \; 
\label{eq:redH} 
\end{equation}

\begin{table}[th] 
\caption[Charmonium content of states near flavour threshold]
        {Charmonium content of states near flavour threshold.  The wave
         function $\psi$ takes account of mixing induced through open
         charm--anticharm channels. Unmixed potential-model eigenstates
         are denoted by $|n\sLj{2s+1}{L}{J}\!\rangle$. The coefficient
         of the dominant eigenstate is chosen real and positive. The
         1S, 1P, 2S, and 1\slj{3}{3}{1} states are evaluated at their
         physical masses. The remaining 1D states are considered at
         the masses in \Table~\ref{tab:delM}.  ${\cal Z}_{cc}$
         represents the $(c\bar{c})$ probability fraction of each
         state.}
\label{tab:wf}
\begin{center} 
\begin{tabular}{|l|l|c|} 
\hline
\hline
State & Major Components & ${\cal Z}_{c\bar c}$ \\
\hline
$\psi(1\slj{1}{1}{0})$ & $0.986\kst{1\slj{1}{1}{0}} - 0.042\kst{2\slj{1}{1}{0}} 
- 0.008\kst{3\slj{1}{1}{0}}$ 
&  $0.974$ \\
$\psi(1\slj{3}{1}{1})$ & $0.983\kst{1\slj{3}{1}{1}} - 0.050\kst{2\slj{3}{1}{1}}
- 0.009\kst{3\slj{3}{1}{1}}$ 
& $0.968$ \\[3pt]
$\psi(1\slj{3}{2}{0})$ & $0.919\kst{1\slj{3}{2}{0}} - 0.067\kst{2\slj{3}{2}{0}}
- 0.014\kst{3\slj{3}{2}{0}}$
& $0.850$ \\
$\psi(1\slj{3}{2}{1})$ & $0.914\kst{1\slj{3}{2}{1}} - 0.075\kst{2\slj{3}{2}{1}}
- 0.015\kst{3\slj{3}{2}{1}}$
& $0.841$ \\
$\psi(1\slj{1}{2}{1})$ & $0.918\kst{1\slj{1}{2}{1}} - 0.077\kst{2\slj{1}{2}{1}}
- 0.015\kst{3\slj{1}{2}{1}}$ 
& $0.845$ \\ 
$\psi(1\slj{3}{2}{2})$ & $0.920\kst{1\slj{3}{2}{2}} - 0.080\kst{2\slj{3}{2}{2}}
- 0.015\kst{3\slj{3}{2}{2}} - 0.002\kst{1\slj{3}{4}{2}}$ 
& $0.854$ \\[3pt] 
$\psi(2\slj{1}{1}{0})$ & $0.087\kst{1\slj{1}{1}{0}} +  0.883\kst{2\slj{1}{1}{0}}
 - 0.060\kst{3\slj{1}{1}{0}} - 0.016\kst{4\slj{1}{1}{0}}$ 
& $0.791$ \\ 
$\psi(2\slj{3}{1}{1})$ & $0.103\kst{1\slj{3}{1}{1}} + 0.838\kst{2\slj{3}{1}{1}} 
- 0.085\kst{3\slj{3}{1}{1}} - 0.017\kst{4\slj{3}{1}{1}}$ 
& $0.723$ \\
& $+ 0.040\kst{1\slj{3}{3}{1}} - 0.008\kst{2\slj{3}{3}{1}}$ & \\[3pt] 
$\psi(1\slj{3}{3}{1})$ & $0.694\kst{1\slj{3}{3}{1}} 
+ 0.097\,e^{0.935i\pi}\kst{2\slj{3}{3}{1}} 
+ 0.008\,e^{-0.668i\pi}\kst{3\slj{3}{3}{1}}$ 
& $0.520$ \\
& $+ 0.013\,e^{0.742i\pi}\kst{1\slj{3}{1}{1}}
+ 0.168\,e^{0.805i\pi}\kst{2\slj{3}{1}{1}} +
0.014\,e^{0.866i\pi}\kst{3\slj{3}{1}{1}}$ & \\
& $+ 0.012\,e^{-0.229i\pi}\kst{4\slj{3}{1}{1}}$ & \\
$\psi(1\slj{3}{3}{2})$ & $0.754\kst{1\slj{3}{3}{2}} - 0.084\kst{2\slj{3}{3}{2}}
- 0.011\kst{3\slj{3}{3}{2}}$ 
& $0.576$ \\ 
$\psi(1\slj{1}{3}{2})$ & $0.770\kst{1\slj{1}{3}{2}}
- 0.083\kst{2\slj{1}{3}{2}} - 0.012\kst{3\slj{1}{3}{2}}$
& $0.600$ \\ 
$\psi(1\slj{3}{3}{3})$ & $0.812\kst{1\slj{3}{3}{3}} 
+0.086\,e^{0.990i\pi}\kst{2\slj{3}{3}{3}}
+0.013\,e^{-0.969i\pi}\kst{3\slj{3}{3}{3}}$ 
& $0.667$ \\
 & $+ 0.007\,e^{0.980i\pi}\kst{4\slj{3}{3}{3}}
+ 0.016\,e^{0.848i\pi}\kst{1\slj{3}{5}{3}}$ & \\
\hline
\hline
\end{tabular} 
\end{center} 
\end{table}
 
Solving \Eq~(\ref{eq:redH}) in the $Q\bar Q$ sector 
determines the mixing between the potential
model states and coupling to decay channels. 
This approach has been
described in detail \cite{eichten78}  for the CCC model
with ${\cal H}_I$ (\Eq~\ref{eq:CCC}).  
An effective Hamiltonian approach has also been considered 
in the QPC model\cite{Isgur:1999cd}.  

The results for the low-lying $c\bar c$ states is shown in
\Table~\ref{tab:wf} for the CCC model. The overall probability for the
physical state to be in the $c\bar c$ sector, denoted ${\cal Z}_{c\bar
c}$, decreases as open charm threshold is approached.  For states
above threshold the mixing coefficients become complex.  These mixing
effects contribute to observed S--D mixing as well as modifying
radiative transition rates\cite{Rosner:2001nm,Rosner:2004mi}.  A more
detailed discussion of these effects appear in the Decay section.

\subsubsection[Zweig-allowed strong decays]
{Zweig-allowed strong decays}

Once the mass of a resonance is given, the coupled-channel 
formalism yields reasonable predictions for the other resonance 
properties.  
Eichten, Lane and Quigg\cite{Eichten:2004uh} have 
estimated the strong decay rates within the CCC model for 
all the charmonium levels that populate the threshold region
between $2M_D$ and $2M_{D^*}$. 
For 1\slj{3}{3}{1} state $\psi^{\prime\prime}(3770)$, 
which lies some $40 {\rm Mev}$ above charm threshold,  
they obtain $\Gamma( \psi^{\prime\prime}(3770) \rightarrow 
D\overline{D}) = 20.1\mev$, to be compared with the PDG's 
fitted value of $23.6 \pm 2.7\mev$\cite{Eidelman:pdg2004}.
The natural-parity 1\slj{3}{3}{3} state can decay into 
$D\overline{D}$, but its F-wave decay is suppressed
by the centrifugal barrier factor. 
The partial width is only $0.77\mev$ at a mass 
of $3868\mev$ and the 1\slj{3}{3}{3} may be discovered as 
a narrow $D\overline{D}$ resonance up to a mass of about $4000\mev$.

Barnes and Godfrey\cite{Barnes:2003vb} have estimated the decays of
several of the charmonium states into open charm, using the
\slj{3}{2}{0} model.  Their estimates of open-charm partial decay widths into 
$D\overline{D}$ are $42.8\mev$ for the 1\slj{3}{3}{1} state
and $3.6\mev$ for a 1\slj{3}{3}{3} state at a mass of $3868\mev$.
They did not carry out a coupled-channel analysis
which makes a direct comparison of models more difficult. 
Detailed comparisons (\eg Ackleh, Barnes and Swanson\cite{Ackleh:1996yt})
between various light quark pair creation models are highly desirable. 

Estimates for decay widths of the $1^{--}$ charmonium states above
open-charm threshold in the \slj{3}{2}{0} model have recently been
reported by Barnes\cite{Barnes:2004cz}.  The comparison with
experimentally extracted values is shown in \Table~\ref{tab:decaySD}.
Along with the current PDG values for the total widths of $c\bar c$
resonances, a reanalysis by Seth\cite{Seth:2004py} of the existing
experimental data is also shown in \Table~\ref{tab:decaySD}.
\longpage 

The resonance decay widths are determined from fitting measurements
of $\Delta R$ in $e^+e^-$ annihilation to a model for each resonance
including radiative corrections. 
This whole procedure is complicated by 
its dependence on the resonance shape, \ie the expected non  
Breit--Wigner nature of the partial widths for radially excited resonances.
It may be more useful for theorists to produce a model of
$\Delta R$ for direct comparison with data. Greater resolving power
between models is possible if the contribution from each individual 
open heavy flavour final state is separately reported. 

For the CCC model, the structure of $\Delta R(b\bar b)$ in the threshold
region was studied in the original Cornell group works
\cite{Eichten:ag,eichten78} and later extended to the
$\Delta R(b\bar b)$ in the threshold region \cite{Eichten:ce}.
The structure of $\Delta R(c\bar c)$ and $\Delta R(b\bar b)$ 
has also been studied in QPC models\cite{Heikkila:1983wd}.
There are also some attempts to compare the different 
models\cite{Byers:1989zn,Byers:1994dx}.

\begin{table}[th]
\caption[Open-charm strong decay modes of the $1^{--}$ states]
        {Open-charm strong decay modes of the $1^{--}$ states.
        Experimental widths from the PDG\cite{Eidelman:pdg2004} and a
        recent analysis of Seth\cite{Seth:2004py}.  The theoretical
        widths using the QPC model\cite{Barnes:2004cz} and the CCC
        model\cite{Eichten:2004uh} are shown. For the $\psi(4159)$
        some S wave plus P wave charmed meson two body channels are
        also open.}
\label{tab:decaySD}
\begin{center}
\begin{tabular}{|l|l|l|l|l|l|}
\hline
State & Mode & \multicolumn{2}{|c|}{$\Gamma_{\rm EXP}$ (MeV)} 
             & \multicolumn{2}{|c|}{$\Gamma_{\rm THEORY}$ (MeV)} \\ 
      &      & PDG & Seth & QPC Model & CCC model \\ 
\hline
$\psi(3770)$ (${}^3$D$_1$) 
         &  DD             &                 &              & 42.8  & 20.1 \\   
\hline
         &  total          & $23.6 \pm 2.7 $ &              & 42.8  & 20.1 \\
\hline
\hline
$\psi(4040)$  ($3\, {}^3$S$_1$) 
         &  DD             &                 &              & 0.1  & \\
         &  DD$^*$         &                 &              & 33.  & \\
         &  D$_s$D$_s$     &                 &              & 8.   & \\
         &  D$^*$D$^*$     &                 &              & 33.  & \\
\hline
         &  total          & $52 \pm 10 $    & $88 \pm 5$   & 74.  & \\
\hline
\hline
$\psi(4159)$   ($2\, {}^3$D$_1$) 
         &  DD             &                 &              & 16.  & \\
         &  DD$^*$         &                 &              & 0.4  & \\
         &  D$^*$D$^*$     &                 &              & 35.  & \\
         &  D$_s$D$_s$     &                 &              & 8.   & \\
\hline
         &  total          & $78 \pm 20 $    & $107 \pm 8$  & 73.  & \\
\hline
\hline
$\psi(4415)$   ($4\, {}^3$S$_1$) 
         &  DD             &                 &              & 0.4  & \\
         &  DD$^*$         &                 &              & 2.3  & \\
         &  D$^*$D$^*$     &                 &              & 16.  & \\
         &  D$_s$D$_s$     &                 &              & 1.3  & \\
         &  D$_s$D$_s^*$   &                 &              & 2.6  & \\
         &  D$_s^*$D$_s^*$ &                 &              & 0.7  & \\
\hline
         &  total          & $43 \pm 15 $    & $119 \pm 15$ &      & \\
\hline
\hline
\end{tabular}
\end{center}
\end{table} 

Experiments can also search for additional 
narrow charmonium states in neutral combinations of charmed 
mesons and anticharmed mesons.  The most likely candidates correspond to 
the $1\slj{3}{3}{3}$, $2\slj{3}{2}{2}$, and 
$1\slj{3}{4}{4}$ levels\cite{Eichten:2002qv,Eichten:2004uh,Quigg:2004nv}. 
These detailed analyses of the 
$c\bar c$ system can be extended to the $b\bar{b}$ system, 
where it may be possible to see discrete threshold-region states 
in direct hadronic production.  

\subsection[QQq states and molecules]
           {QQq states and molecules $\!$\footnote{Author: J. M. Richard}}
\label{sec:spphenqqq}

\subsubsection{Doubly charmed baryons}\label{dbch:QQq}
The earliest studies on $QQq$ baryons were based on the flavour group
SU(4)$_\mathrm{F}$, as an extension of SU(3))$_\mathrm{F}$.  After the
discovery of hidden and naked charm, some classic papers were written
on hadrons with charm, including a section on $(ccq)$ states
\cite{Gaillard:1974mw,DeRujula:ge}.

Now, our ideas on flavour symmetry have evolved. The conventional
SU($n$)$_{\rm F}$ approach, with elegant mass formulae, is replaced by
\emph{flavour independence}. The potential between two quarks is
generated by their colour, and flavour enters only in recoil
corrections through the quark mass, mainly for describing the fine and
hyperfine structure.

Flavour independence was the main guide line of the detailed studies
of $(QQq)$ baryons made in the 80's and later
\cite{Fleck:mb,Fleck:ma,% Savage:di,Savage:pr,%
Bagan:1994dy,Roncaglia:1995az,Silvestre-Brac:bg,%
Gershtein:nx,Kiselev:2000jb,Tong:1999qs}: the dynamics tuned for
mesons, light baryons and single-charm baryons was tentatively
extrapolated to the $(QQq)$ sector.  More papers came after the recent
findings at SELEX (cf. the experimental part of this chapter), for
instance Ref.~\cite{Gelman:2002wf}, where a link is made with
double-charm exotics, to be discussed shortly.

To study confinement, $(QQq)$ baryons are perhaps the most interesting of ordinary hadrons, as they
combine    two extreme regimes in a single bag:
\begin{enumerate}
\item the  slow relative motion of two heavy quarks, as in charmonium,
\item the fast motion of a light quark. Remember that the electron moves faster
in hydrogen than in positronium. Similarly, a light quark is likely more
relativistic in heavy-light hadrons than in light mesons.
\end{enumerate}

In the $(QQq)$ wave function, the average $QQ$ separation is smaller than the
$Qq$ one. This leads to envisage approximations. One of them consists of 
replacing the full three-body calculation by a two-step procedure where one
first  calculates the $QQ$ mass, by solving a two-body problem, and then 
estimates  the $QQ-q$ mass by solving another two-body problem.
 The second step is rather safe. The finite-size corrections are
small. For instance, they cancel out exactly for the harmonic oscillator. As for
the first step, one should be aware that the $QQ$ potential is {\em effective},
since it contains both the direct $QQ$ interaction and a contribution from the
light quark. For instance, in the harmonic oscillator model,  1/3 of the $QQ$
interaction comes from the light quark, and neglecting this term results into an
underestimation of energies and spacings by a factor $\sqrt{3/2}$. 
Another limitation to the quark--diquark picture, is that the diquark is not
frozen. The first excitations of $QQq$ occur inside the diquark. So one should
recalculate the properties of the diquark for each level.

Another way to take advantage of the large mass ratio $M/m$ is to use
the Born--Oppenheimer approximation, as done, \eg by Fleck and
Richard \cite{Fleck:mb}. For a given $QQ$ separation $r_{12}$, the
two-centre problem is solved for the light quark, with proper reduced
mass. The ground-state energy $E_0(r_{12})$, supplemented by the
direct $QQ$ interaction, provides the adiabatic potential
$V_{QQ}$. Solving the 2-body problem with this potential gives the
first levels. The adiabatic potential built out of the second
``electronic'' energy $E_1(r_{12})$ leads to a second series of
levels. This is very similar to the spectroscopy of H$_2^+$ in atomic
physics.

Within explicit potential models, the Born--Oppenheimer approximation
can be checked against an accurate solution of the 3-body problem,
using for instance a systematic hyperspherical expansion. The
approximation is excellent for $(bbq)$ and $(ccq)$, with $q=u$, $d$ or
$s$, or even for $(ssu)$ or $(ssd)$ \cite{Fleck:mb,Richard:uk}.

In Ref.~\cite{Fleck:mb}, $(ccq)$ masses were estimated from a specific
variant of the bag model, already used for charmed mesons. The results
turn out to be rather sensitive to details such as centre-of-mass
corrections, value of the bag constant, \etc Other bag-model
calculations have been performed \cite{Ponce:gk}.

Potential models, on the other hand, tend to give very stable results,
when the parameters are varied while maintaining a reasonable fit of
lighter hadrons.  One typically obtains:
\begin{itemize}
\item a ground-state near or slightly above $3.6\,$GeV for the $(ccu)$
      or $(ccd)$ ground state,
\item a hyperfine splitting of about $80\,$MeV between the spin 3/2
      and spin 1/2 states,
\item the first orbital excitation about $300\,$MeV above the ground-state,
\item the first $(ccs)$ state near $3.7\,$GeV
\end{itemize} 

Note that models tuned to $(cqq)$ or lighter baryons might
underestimate the short-range $QQ$ attraction. If models are adjusted
to $(c\bar{c})$ spectroscopy, there is an ambiguity on how to
translate it to $cc$.  The usual recipe stating that
\begin{equation}
V_{QQ}=\frac{1}{2}V_{Q\bar{Q}}~,
\end{equation}
implies pairwise forces mediated by colour-octet exchanges. Small,
non-confining, colour-singlet exchanges, as well as three-body forces might
complicate the issue.
  
Most existing calculations are of rather exploratory nature, since
made when double charm was considered as science fiction, or far
future. Meanwhile, the art of QCD has made significant progress. One
could retain from simple potential models that the Born--Oppenheimer
approximation provides an adequate framework. The effective $QQ$
potential could be estimated from relativistic models or from lattice
calculations, similar to those of the $Q\bar{Q}$ potential or the
effective $QQ$ potential in exotic $(QQ\bar{q}\bar{q})$ mesons, to be
discussed shortly. It is hoped that the new experimental results will
stimulate such calculations.

The literature already contains approaches somewhat more ambitious
than simple bag or non-relativistic potential models: relativistic
models \cite{Ebert:2002ig}, QCD sum rules \cite{Bagan:1994dy}, string
picture \cite{Gershtein:nx}, \etc  The lattice QCD approach is
presented in \Section~\ref{sec:spqqql} and the EFT one is presented 
in \Section~\ref{sec:spnrqcdte}

The appearance of the $D_{s,J}^*$ state not very far above the ground
state $D_s$ of meson with flavour content $(c\bar{s})$ has stimulated
several studies on the dynamics of light quarks in a static colour
field.  In Ref.~\cite{Bardeen:2003kt}, it is suggested that the same
phenomenon will occur for double-charm baryons. On this respect the
doubling of states in the preliminary data by SELEX is of particular
interest.

\subsubsection[Exotic mesons with double charm]
{Exotic mesons with double charm}\label{sec:dbch-QQqq}

The physics of multiquarks, though it benefits from a dramatic revival
since the tentative discovery of a light pentaquark, remains penalized
by the confusion about baryonium states in the late 70's and early
80's.  This is actually a difficult field, where speculations about
confinement mechanisms should be combined with delicate few-body
calculations.

The $H$ dibaryon  \cite{Jaffe:1976yi}, and the heavy  pentaquark $P$ proposed independently 
by Lipkin \cite{Lipkin:1987sk} and the Grenoble group  \cite{Gignoux:1987cn}, owe their tentative stability to
chromomagnetic forces, schematically \cite{DeRujula:ge}
\begin{equation}
\label{eq:chromo} H_{\mathrm cm}= -C
\sum_{i<j} \frac{\bfsigma_i \cdot \bfsigma_j\,\tilde{\lambda}_i 
                 \cdot \tilde{\lambda}_j}{m_i m_j}
           \delta^{(3)}(\mathbf{r}_{ij})~,
\end{equation}                   
or its bag model analogue \cite{Chodos:1974je}, that describes the
observed hyperfine splittings such as $\Delta-N$ or
$J/\Psi-\eta_c$. The astute observation by Jaffe \cite{Jaffe:1976yi}
is that this operator provides a binding
$(ssuudd)-2(sud)\sim-150\,\mathrm{MeV}$ to the $H=(ssuudd)$ dibaryon
with spin and isospin $J=I=0$. This estimate, however, relies on
SU(3)$_\mathrm{F}$ flavour symmetry and $\langle \delta^{(3)}({\bf
r}_{ij}) \rangle$ being independent of $(i,j)$ pair and borrowed from
the wave function of ordinary baryons. Relaxing these hypotheses, and
introducing kinetic energy and spin-independent forces in the 6-body
Hamiltonian, and a realistic estimate of short-range correlations,
usually spoils the stability of $H$
\cite{Rosner:1985yh,Karl:cg,Fleck:ff}. The existence of $H$ is
nowadays controversial. It has been searched in many experiments,
without success so far. For instance, the doubly-strange hypernucleus
${}_\Lambda^{\phantom{6}}{\!}_\Lambda^6\!\mathrm{He}$ is not observed
to decay into $H+\alpha$ \cite{Gal:2002pn}.

 If the calculation made for the $H$ is repeated in the limit where
$m(Q)\to\infty$, the same binding $
(\bar{Q}qqqq)-(\bar{Q}q)-(qqq)\sim-150\,\mathrm{MeV}$ is obtained for
the pentaquark $(\bar{Q}qqqq)$, $qqqq$ being in a SU(3)$_\mathrm{F}$
triplet \cite{Lipkin:1987sk,Gignoux:1987cn}. All corrections, again,
tend to weaken this binding \cite{Fleck:ff,Karl:uf} so it is not
completely sure that the actual pentaquark is stable. See, also,
\cite{Leandri:su}.

After the tentative discovery of a light pentaquark state at about
1.53~GeV, with flavour content $(uudd\bar{s})$, and possible partners
with strangeness $S=-2$, many authors have revisited the possibility
of stable or metastable pentaquarks with heavy antiflavour. See, for
instance Refs.\
\cite{Jaffe:2003sg,Stewart:2004pd,Liu:2004qx,Zhu:2004xa,Oka:2004xh,Pochodzalla:2004up}.
In the light pentaquark, the binding is achieved by the chiral
dynamics of light quarks. A forerunner in this field was Stancu
\cite{Stancu:1998sm}, who proposed positive-parity pentaquarks with a
heavy antiquark in a simple potential model where the chromomagnetic
interaction is replaced by a short-range spin-flavour interaction
which looks like the exchange of Goldstone bosons between quarks.

In short, there are still many open issues for the $H$ dibaryon, the
pentaquarks, as well as for possible light scalar mesons made out of
two quarks and two antiquarks.  This is, however, more of the domain
of light-quark spectroscopy.

More than twenty years ago, another mechanism for multiquark binding
was proposed. It was pointed out that current confining potentials
applied to a $(QQ\bar{q}\bar{q})$ system put its mass below the
dissociation threshold into $(Q\bar{q})+(Q\bar{q})$, provided the mass
ratio $m(Q)/m(q)$ is large enough \cite{Ader:1981db}. This {\it
chromoelectric} binding was studied by several authors, in the context
of flavour-independent potentials
\cite{Heller:cb,Heller:bt,Carlson:hh,%
Zouzou:qh,Lipkin:1986dw,Brink:ic,Brink:as,%
Janc:2000vq,Janc:2003cf,Gelman:2002wf} % or with lattice QCD
\cite{Mihaly:1996ue,Michael:1999nq} (see, also,
\cite{Green:1997mv,Green:1998nt}), with a remarkable convergence
towards the same conclusion. This somewhat contrasts with the
confusion in other sectors of multiquark spectroscopy.

Let us consider, indeed, the limit of a purely flavour-independent
potential $V$ for $(QQ\bar{q}\bar{q})$. The situation becomes similar
to that of exotic four-body molecules $(M^+,M^+,m^-,m^-)$, all of them
using the very same Coulomb potential when $M$ and $m$ are varied.
The hydrogen molecule with $M\gg m$ is much more stable than the
positronium molecule Ps$_2$ with $M=m$. If one decomposes the 4-body
Hamiltonian as
\begin{equation}
{\cal H}_4=\left[\frac{M^{-1}+m^{-1}}{4}
\left(\mathbf{p}_1^2 + \mathbf{p}_2^2 + \mathbf{p}_3^2 + 
      \mathbf{p}_4^2\right) + V\right] +
\frac{M^{-1}-m^{-1}}{4}
\left(\mathbf{p}_1^2 + \mathbf{p}_2^2 - \mathbf{p}_3^2 - \mathbf{p}_4^2
\right)~,
\end{equation}
the first term, even under charge conjugation, corresponds to a
rescaled equal-mass system with {\it the same threshold} as ${\cal
H}_4$. The second term, which breaks charge conjugation, improves the
energy of ${\cal H}_4$ (one can applies the variational principle to
${\cal H}_4$ using the symmetric ground state of the first term as a
trial wave function). In the molecular case, the second term changes
the marginally bound Ps$_2$ (or rescaled copy) into the deeply bound
H$_2$. In quark models, an unbound $(qq\bar{q}\bar{q})$ becomes a
stable $(QQ\bar{q}\bar{q})$.

The effective $QQ$ potential has been estimated by Rosina et al.\
\cite{Janc:2000vq} in the framework of empirical potential models, and
by Mihaly et al.\ \cite{Mihaly:1996ue} and Michael et al.\ (UKQCD)
\cite{Michael:1999nq}, who used lattice simulations of QCD. The
question is obviously: is the $c$ quark heavy enough to make
$(cc\bar{q}\bar{q})$ bound when $q=u$ or $d$?  At this point, the
answer is usually negative, most authors stating that $b$ is required
to bind $(QQ\bar{q}\bar{q})$ below its $(Q\bar{q})+(Q\bar{q})$
threshold.
 
There is, however, another mechanism: pion-exchange or, more
generally, nuclear-like forces between hadrons containing light quarks
or antiquarks. This effect was studied by several authors, in
particular T{\"o}rnqvist \cite{Tornqvist:1991ks,Tornqvist:ng}, Manohar
and Wise \cite{Manohar:1992nd}, and Ericson and Karl
\cite{Ericson:1993wy}. In particular a $D$ and $D^*$ can exchange a
pion, this inducing an attractive potential. It is weaker than in the
nucleon--nucleon case, but what matters for a potential $g V(r)$ to
bind, is the product $gm$ of the strength $g$ and reduced mass $m$.
It is found that $(DD^*)$ is close to be bound, while binding is
better established for $(BB^*)$. The result depends on how sharply the
long-range potential is empirically regularised at short distances.

A lattice calculation such as those of
Refs.~\cite{Mihaly:1996ue,Michael:1999nq} contains in principle all
effects. In practice, the pion is unphysically heavy such that
long-range forces are perhaps not entirely included. Explicit quark
models such as \cite{Janc:2000vq} make specific assumptions about
interquark forces, but do not account for pion exchange.  In our
opinion, a proper combination of long- and short-range forces should
lead to bind $(DD^*)$, since each component is almost sufficient by
itself. This is presently under active study.
 
There is a further possibility to build exotic, multicharmed
systems. If the interaction between two charmed mesons is slightly too
weak to lead to a bound state (this is presumably the case for $(DD)$,
since pion exchange does not contribute here), it is likely that the
very same meson--meson interaction binds three or more mesons. This is
known as the phenomenon of ``Boromean'' binding.

For instance, in atomic physics, neither two ${}^3$He atoms nor a
${}^3$He atom and a ${}^4$He atom can form a binary molecule, even at
vanishing temperature, but it is found that ${}^3\mathrm{He}
{}^3\mathrm{He} {}^4\mathrm{He}$ is bound
\cite{Bressanini2002}. Similarly, in nuclear physics, the isotope
${}^6$He is stable against evaporating two neutrons, or any other
dissociation process, while ${}^5$He is unstable. In a 3-body picture,
this means that $(\alpha,n,n)$ is stable, while neither $(\alpha,n)$
nor $(n,n)$ have a stable bound state. In short, binding three
constituents is easier than two.

\subsection[Quarkonium hybrids]
           {Quarkonium hybrids $\!$\footnote{Author: S. Godfrey}}

The existence of gluonic excitations in the hadron spectrum is one of
the most important unanswered questions in hadron physics.  Hybrid
mesons form one such class which consists of a $q\bar{q}$ with an
excited gluonic degree of freedom.  Their spectroscopy are discussed
extensively in this Chapter.  Recent observations of charmonium states
in exclusive $B$-meson decays
\cite{cleo95,belle02a,babar02,Choi:2002na,babar02b,babar02c} suggest
that charmonium hybrid mesons ($\psi_g$) \cite{giles77} with mass
$\sim$4~GeV may be produced in $B$-decay via $c\bar{c}$ colour octet
operators \cite{close98,chiladze98}.  Some of these states are likely
to be narrow with clean signatures to $J/\psi \pi^+\pi^-$ and $J/\psi
\gamma$ final states.  The unambiguous discovery of such a state would
herald an important breakthrough in hadronic physics, and indeed, in
our understanding of Quantum Chromodynamics, the theory of the strong
interactions.  In this section we give a brief overview of charmonium
hybrid properties and and suggest search strategies for charmonium
hybrids at existing B-factories \cite{close03b}.

\subsubsection{Spectroscopy}

Lattice gauge theory and hadron models predict a rich 
spectroscopy of charmonium hybrid mesons 
\cite{giles77,Isgur:1984bm,isgur85,barnes95,hybrids,Bernard:1997ib,perantonis90,juge97,Liao:2002rj,con-glue}.  
For example,
the flux tube model predicts 8 low lying hybrid states
in the 4 to 4.2~GeV mass region with
 $J^{PC}=0^{\pm\mp}$, $1^{\pm\mp}$, $2^{\pm\mp}$, and $1^{\pm\pm}$.
Of these states the 
$0^{+-}$, $1^{-+}$, and $2^{+-}$ have exotic quantum numbers;
quantum numbers not consistent with the constituent quark model.
The flux-tube model predicts $M(\psi_g) \simeq 4 - 4.2$~GeV \cite{isgur85,barnes95};  
lattice QCD predictions for the $J^{PC}=1^{-+}$ state 
range from 4.04~GeV to 4.4~GeV \cite{Bernard:1997ib,perantonis90} with
 a recent quenched
lattice QCD calculation \cite{Liao:2002rj} finding 
$M(1^{-+})=4.428\pm 0.041$~GeV.   
These results have 
the $ 1^{-+}$ lying in the vicinity of the $D^{**}D$ threshold of 4.287~GeV.
There is the tantalising possibility that
the $ 1^{-+}$ could lie 
below $D^{**}D$ threshold and therefore be relatively narrow.  
% We refer the reader to the spectroscopy chapter for further details.

\subsubsection{Decays}

There are three important decay modes for charmonium hybrids:
  (i) the Zweig allowed fall-apart mode $\psi_g \to 
D^{(*,**)}\bar{D}^{(*,**)}$ \cite{ikp85,page97,page98};  
(ii) the cascade 
to conventional $c\bar{c}$ states, of the type 
$\psi_g \to (c\bar{c})(gg) \to (c\bar{c})$ $+(\hbox{light hadrons})$
and $\psi_g \to (c\bar{c})+\gamma$
\cite{close95}; (iii) decays to light hadrons via intermediate gluons,
$\psi_g \to (ng)$ $\to$ $\hbox{light hadrons}$, analogous to 
$J/\psi\to  \hbox{light hadrons}$ and $\eta_c\to  \hbox{light hadrons}$.
Each mode plays a unique role.  
$\psi_g$ hybrids with exotic $J^{PC}$ quantum numbers offer the most 
unambiguous signal 
since they do not mix with conventional quarkonia.

\paragraph{(i) Decays to $D^{(*)}D^{(*)}$:}
In addition to $J^{PC}$ selection rules (for example, 
$2^{-+}$ and $2^{--}$ decay to $D\bar{D}$ are forbidden by parity
and the exotic hybrid $\psi_g(0^{+-}) $ decays to
$D^{(*)}D^{(*)}$ final states are forbidden by $P$ and/or $C$ conservation)
a general feature 
of most models of hybrid meson decay is that decays to two mesons with 
the same spatial wave function are suppressed \cite{page97b}.  
The dominant coupling of charmonium hybrids is to excited states, in 
particular $D^{(*)}(L=0)+D^{**}(L=1)$ states
for which the threshold is $\sim 4.3$~GeV.  This 
is at the kinematic limit for most mass predictions so 
that decays into the preferred $D^{(*)}D^{**}$ states 
are expected to be significantly suppressed if not outright 
kinematically forbidden.
A refined version of the Isgur Kokoski Paton flux
model \cite{ikp85}  predicts partial 
widths 
of 0.3--1.5~MeV  depending on the $J^{PC}$ of the hybrid \cite{page98}.  
These widths are
quite narrow for charmonia of such high mass.
If the hybrid masses are 
above $D^{**}$ threshold then the total widths increase to 4--40~MeV for 
4.4~GeV charmonium hybrids which are still relatively narrow for 
hadron states of such high mass.  
The challenge is 
to identify decay modes that can be reconstructed by experiment.

\paragraph{(ii) Decays to $(c\bar{c})+(\hbox{light hadrons})$:}
The 
$\psi_g \to (c\bar{c})+(\hbox{light hadrons})$ 
mode offers the cleanest signature for $\psi_g$ observation if its
branching ratio is large enough.
In addition, a small total width also offers the possibility that the 
radiative branching ratios into  $J/\psi$, $\eta_c$, $\chi_{cJ}$, and 
$h_c$
could be significant and offer a clean signal for the detection of 
these states.

For masses below $D D^{**}$ threshold the 
cascade decays
$\psi_g \to (\psi,\; \eta_c, \ldots) +(gg)$ and annihilation decays
$\psi_g (C=+) \to (gg) \to \hbox{light hadrons}$
will dominate.  If the masses of exotic $J^{PC}$ states are 
above  $D D^{**}$ threshold their widths are 
also expected to be relatively narrow for states of such high mass, 
in which case
cascades to conventional $c\bar{c}$ states transitions of the type 
$\psi_g \to (\psi,\;\psi')+(\hbox{light hadrons})$
should have significant branching ratios \cite{close95}
making them important signals to look for in $\psi_g$ searches.
In the Kuang--Yan formalism \cite{kuang81} the matrix elements for 
hadronic  transitions between conventional quarkonia are related to 
hybrid-conventional quarkonium hadronic transitions. 
A not unreasonable assumption is that the
partial widths for the decays 
$\psi_g(1^{-+}) \to \eta_c +(\pi\pi, \eta, \eta') $ and
$\psi_g(0^{+-},2^{+-})\to J/\psi +(\pi\pi,\eta, \eta')$
will be similar 
in magnitude to $(c\bar{c})\to \pi\pi J/\psi$ and 
$(c\bar{c})\to \eta J/\psi$, of ${\cal O}(10-100)$~keV.

Estimates of radiative 
transitions involving hybrids with light quarks 
\cite{page95,close03} found that the $E1$ transitions 
between hybrid and conventional states to be comparable in magnitude 
to transitions between conventional mesons. 
While neither calculation can be applied directly to $c\bar{c}$ 
one might take this to suggest that 
the partial widths for 
$\psi_g(1^{-+}) \to \gamma + (J/\psi, h_c)$ 
and $\psi_g (0^{+-},2^{+-}) \to \gamma + (\eta_c, \chi_{cJ})$
are the same order of magnitude %in size
as transitions between conventional 
charmonium states.  However, a recent flux-tube model calculations by
Close and Dudek 
\cite{close03} found that the $\Delta S=0$ E1 transitions to hybrids 
only occur for charged particles, and hence would vanish for 
$c\bar{c}$.
The $\Delta S=1$ M1 transitions can occur, but 
are non-leading and less well defined.
Estimates \cite{close03} for their widths are ${\cal O}(1-100)$~keV. 
Clearly, given our general lack of understanding of radiative transitions 
involving hybrids, the measurement of these transitions, 
$\psi_g \to (c\bar{c}) \gamma$, has important implications for model 
builders.

\paragraph{(iii) Decays to light hadrons:}
Decays of the type $\psi_g  \to \hbox{light hadrons}$ offer the 
interesting possibility of producing light exotic mesons.
Estimates of annihilation widths to light hadrons will be order of 
magnitude guesses at best due to uncertainties in wavefunction effects 
and QCD corrections.  We estimate the annihilation widths 
$\Gamma[\psi_g (C=-) \to \hbox{ light hadrons}]$
and $\Gamma [c\bar{c}(C=+) \to \hbox{ light hadrons}]$ by 
comparing them to $\Gamma(\psi' \to  \hbox{ light hadrons})$ and 
$\Gamma (\eta_c' \to \hbox{ light hadrons})$.
The light hadron production rate from $\psi_g(C=-)$ decays 
is suppressed by one power of $\alpha_{\rm s}$ with respect to $\psi_g(C=+)$ 
decays.  This very naive assumption gives
$\Gamma[\psi_g (C=-) \to \hbox{ light hadrons}]\sim {\cal O}(100)$~keV
and $\Gamma [c\bar{c}(C=+) \to \hbox{ light hadrons}]\sim {\cal 
O}(10)$~MeV \cite{gr02}.  These widths could be 
smaller 
because the 
$q\bar{q}$ pair in hybrids 
is expected to be separated by a distance of order 
$1/\Lambda_{QCD}$ resulting in a smaller annihilation rate than the 
S-wave $\psi'$ and $\eta_c'$ states.

\subsubsection{Hybrid production}

Recent developments in both theory and experiment lead us to
expect that charmonium hybrids will be produced in $B$ decays.  
The partial widths for $B\to c\bar{c} +X$, with $c\bar{c}$ representing 
specific final states such as $J/\psi$, $\psi'$, $\chi_{c0}$,
$\chi_{c1}$, $\chi_{c2}$, $^3D_2$, $^1D_2$ \etc, 
have been calculated in the NRQCD formalism 
\cite{bodwin92,Bodwin:1994jh,beneke98,ko96,yuan97,ko97} which
factorizes the decay mechanism into short (hard) and nonperturbative (soft)  
contributions.  The hard contributions are fairly well understood but 
the soft contributions, included as colour singlet and colour 
octet matrix elements, have model dependent uncertainties.
Insofar as hybrid 
$c\bar{c}$ wavefunctions have a non-trivial colour representation they 
can be produced via a colour octet intermediate state.  
Chiladze {\it et al.} \cite{chiladze98} estimated the branching ratio 
${\cal B}[B\to \psi_g(0^{+-}) + X ] \sim 10^{-3}$ for $M \sim$ 4~GeV
(though recent quenched lattice calculations suggest $M(0^{+-})=4.70\pm 0.17$~GeV,
and hence will be inaccessible).  
Close {\it et al.}  \cite{close98} estimate a 
similar branching ratio to $1^{-+}$ and argued that if $M_g < 4.7$~GeV,
the total branching ratio
to $\psi_g$ for all $J^{PC}$ could be % ${\cal O}(1\%)$
${\cal B} [\psi_g (\hbox{all }J^{PC}) +X ]\sim {\cal O}(1\%)$.  
Thus, using two different approaches for estimating 
${\cal B}[B\to \psi_g +X]$ both 
Chiladze {\it et al.} \cite{chiladze98} and 
Close {\it et al.}  \cite{close98} obtain similar results.
Both calculations estimate ${\cal B}$'s of ${\cal O}(0.1-1\%)$ which
are comparable to the ${\cal B}$'s for conventional $c\bar{c}$ states.  

\subsubsection{Experimental signatures}

The decays discussed above lead to a
number of possible signals:   $\psi_g \to 
D^{(*)}D^{(*,**)}$, $\psi_g(0^{+-},2^{+-})$ $\to$  $J/\psi$ $+$  
$(\pi^+\pi^-,\eta,\eta')$, 
$\psi_g(1^{-+}) \to \eta_c  +  (\pi^+\pi^-,\eta,\eta')$,  
$\psi_g \to (c\bar{c}) \gamma$, 
and $\psi_g \to \hbox{light hadrons}$.
Of the possible decay modes,  $\psi_g \to J/\psi \pi^+\pi^-$,
$\psi_g \to J/\psi \eta$, and $\psi_g\to (c\bar{c})\gamma$ 
give distinctive and easily reconstructed signals.  In the former case,
the subsequent decay, $J/\psi\to e^+e^-$ and $\mu^+\mu^-$ offers 
a clean tag for the event so that searches for peaks 
in the invariant mass distributions $M(e^+e^- \pi^-\pi^+)-M(e^+e^-)$ 
is a promising search strategy for hybrids.
Both the $0^{+-}$ and $2^{+-}$ should decay via the $\psi_g\to J/\psi 
\pi\pi$ cascade.  
For the $\psi_g$ lying below $DD^{**}$ threshold 
combining estimates of ${\cal B}(B\to \psi_g +X) \simeq 10^{-3}$ 
and 
${\cal B}[\psi_g(2^{+-}) \to J/\psi \pi^+\pi^-] \simeq 0.2$
with the PDG value of
${\cal B}(\psi\to \ell^+ \ell^-) = 11.81\% $  and the Babar detection 
efficiency we estimate that for 100~fb$^{-1}$ of integrated luminosity
each experiment should 
observe roughly 50 events.  If the $2^{+-}$ lies above the $DD^{**}$ 
threshold the ${\cal B}$ for $2^{+-}\to J/\psi \pi\pi $ decreases 
significantly to $2.6\times10^{-2}$
lowering the expected number to about 6 events.
Similarly,  for the $0^{+-}$ hybrid we estimate roughly 1200 events if 
it lies below threshold but only 5 events once the $DD^{**}$ decay 
modes open up.  

The $1^{-+}$ state is expected to be the lightest exotic $c\bar{c}$ 
hybrid \cite{Bernard:1997ib,Liao:2002rj}
and therefore the most likely to lie below $DD^{**}$ threshold. 
However, in this case the cascade goes to $\eta_c \pi\pi$, a more 
difficult final state to reconstruct.
Estimates of the relevant partial widths are 
${\cal B}(B\to \psi_g +X) \simeq 10^{-3}$ and 
${\cal B}(\psi_g(1^{-+} \to \eta_c \pi^+\pi^-) \simeq 9\times10^{-3}$.
The Babar collaboration studied the decay $B\to\eta_c K$ by observing 
the $\eta_c$ in $KK\pi$ and $KKKK$ final states.  Combining the PDG 
values for the ${\cal B}$'s to these final states with the Babar detection 
efficiencies of roughly 15\% and 11\% respectively 
we estimate that for 100~fb$^{-1}$ each experiment should 
observe roughly 10 events. If the $1^{-+}$ lies above the $DD^{**}$
threshold, the ${\cal B}$ for $1^{-+}\to \pi\pi \eta_c$ decreases to 
$3\times10^{-3}$ lowering the expected number to about 3 events.

The radiative transition, $\psi_g (1^{-+})\to \gamma J/\psi$, also has a 
distinct signal if it has a significant branching ratio.
The conservative value of 
$\Gamma(\psi_g(1^{-+}) \to \gamma J/\psi)\simeq 1$~keV,  
yields a rather small ${\cal B}$ for this transition.  On the other hand, a 
monochramatic photon offers a clean tag with a high efficiency.  One 
could  look for peaks in $M(\mu^+\mu^- 
\gamma) - M(\mu^+\mu^-)$.   Babar observed  $\chi_{c1}$ 
and $\chi_{c2}$ this way \cite{babar02} obtaining
$\simeq 394$ $\chi_{c1}$'s and $\simeq 1100$ $\chi_{c2}$'s
with a 20.3 fb$^{-1}$ data sample and an
efficiency of about 20 \% for the $J/\psi \gamma$ 
final state \cite{babar02}.  So although the rate may be too 
small to observe, given the potential payoff, it is probably worth 
the effort to perform this search.  

Experiments might also look for charmonium hybrids in invariant mass 
distributions of light hadrons.  For example,
Belle observed the $\chi_{c0}$ by looking at the invariant mass 
distributions from the decays $\chi_{c0}\to \pi^+\pi^-$
and $\chi_{c0}\to K^+K^-$ \cite{belle02a}.  They found 
efficiencies of 21\% for  $\chi_{c0}\to \pi^+\pi^-$
and 12.9\% for $\chi_{c0}\to K^+K^-$, obtaining $\sim 16$ events in 
the former case and $\sim 9$ in the  latter.  

The decay to charmed mesons also needs to be studied.  
Because there are more particles in the final state 
it will be more difficult to reconstruct the charmonium hybrid. 
On the other hand, with sufficient 
statistics these channels will be important for measuring the $\psi_g$ 
quantum numbers and distinguishing their properties from conventional 
$c\bar{c}$ states.

\subsubsection{Summary and future opportunities}

The fundamental problem with all the estimates given above is
that they 
are based on models that have not been tested against experiment.  
Observing a charmonium hybrid and measuring its properties is 
necessary to test these calculations.  It may be that the models are 
correct but it is also possible that they have totally missed the mark. 

Establishing the existence of mesons with explicit gluonic degrees of 
freedom is one of the most important challenges in strong 
interaction physics.  As demonstrated by the discovery of the 
$\eta_c(2S)$ in $B$ decay, $B$ decays offer a promising approach to 
discovering charmonium hybrid mesons.  
We have focused on how to search for these states in 
$B$-decay.  Other possibilities are $1^{--}$ hybrids produced in 
$e^+e^-$ annihilation.  These would likely mix with conventional 
vector quarkonium states so that it would be very difficult to 
distinguish them from conventional states. And recently the Belle 
collaboration
observed the $\eta_c'$ in double charm production in $e^+e^-$ 
collisions.   Part of the GSI upgrade is 
to study and search for charmonium states in $p\bar{p}$ annihilation.  
It is quite possible that hybrids can be studied once the PANDA 
project comes to fruition.  
While there is no question that the estimates for the various partial 
widths are crude, the essential point is that these states are expected 
to be relatively narrow and that distinctive final states are likely 
to have observable branching ratios.  Given how much we can learn by 
finding these states we strongly advocate that some effort be devoted 
to their searches.  In the long term, with the various facilities 
mentioned above,  we 
should be able to open up and study an exciting new spectroscopy.

\section[Introduction to experimental spectroscopy]
        {Introduction to experimental spectroscopy
         $\!$\footnote{Author: R.~Mussa}}

The experimental spectroscopy review is made of four Sections on
charmonia and bottomonia, followed by a Section on $B_c$, and one on
the $ccq$ systems.  The paragraphs follow a hyerarchical structure,
based on the precision reached in the knowledge of the parameters of
these states.  Therefore we start from the vector states ($\psi$'s and
$\Upsilon$'s), which were first discovered, have the narrowest widths,
and are easiest to produce and detect. At present, with the resonant
depolarization technique, it is possible to know these masses with
absolute precision between 10 and 100~keV, and these states are widely
used as calibration tools for HEP detectors.

\Section~\ref{sec:spexfine} scans through triplet P-wave states 
(known as $\chi_c$'s and $\chi_b$'s), which were discovered from 
radiative transitions of upper vector excitations.  
$\chi_c$'s  could not be precisely studied before 
the 90's, when direct access to the formation of these states in $\bar{p}p$ 
annihilations allowed to reach 100--200~keV precisions on their masses, and 
$\approx 10\%$ resolution of their total widths.
The first two Sections allow to realize that the S and P wave states of both 
ortho-charmonium and -bottomonium constitute a very solid, well established
system of resonant states. These narrow resonances can be detected with 
very small  or negligible experimental background and have reached the 
mature stage, from a barely spectroscopical point of view. 

In contrast, all S=0 states are a very active field of research 
for spectroscopy. The best known among those, $\eta_c(1S)$ (described in 
\Section~\ref{sec:spexetac12})
despite being produced with a wide variety of techniques, has still an
uncertainty above 1~MeV on the measured mass, and a rapid progress is
expected to happen in the next few years. Same can be said of the
recently re-discovered $\eta_c(2S)$, described in
\Section~\ref{sec:spexetac12} which greatly benefits from the advent
of the new generation of B-factories.  The hyperfine splitting on
charmonium S states is then approaching maturity.  On the other side,
the large amount of data taken by CLEO at $\Upsilon(1,2,3S)$ energies
did not yield so far to the discovery of $\eta_b$ states. A
comprehensive review of these searches, also performed at LEP
experiments and CDF, is then given in \Section~\ref{sec:spexetab}.
The elusive singlet P state of charmonium, named $h_c$, has been
extensively searched by the $p\bar{p}$ experiments, resulting in
inconclusive evidences; its saga is described in
\Section~\ref{sec:spexhc}. With the advent of B-factories, its search
has regained interest.

Being right across the first open charm threshold, charmonium D-wave 
multiplets still lack a complete understanding, while the first evidence 
of bottomonium D state comes from the recent CLEO~III run at 
$\Upsilon(3S)$, described in \Section~\ref{sec:spexfineDbot}.
The phenomenology of all the other vector orbital excitations is still quite 
unclear as the different thresholds open up: R scans between 3.7 and 4.7~GeV
 are reviewed in \Section~\ref{sec:spexopencha}.
  Further studies on these states have regained priority after the discovery 
of the narrow state X(3872), seen by Belle, and confirmed by BaBar, 
CDF and D0. An overview on the 
experimental evidences of this resonance, as well as the current experimental 
attempts to clarify its nature and its quantum numbers, 
is given separately in \Section~\ref{sec:spexX3872}. 
Despite its most likely interpretation as one of the two above mentioned D 
states, other possible assignments of this resonance, extensively described 
in the theory chapter, span from orbital excitations of P wave states to 
molecular charmonia, opening a wide number of possible searches in this energy
region. 

Another field of research which can bloom in the next years, mostly thanks to 
large samples of B states taken at the Tevatron as well as HERA-B, is the 
study of the $B_c$. Despite the weak decay of its ground state may accomunate
this object to the heavy light mesons, the mass of its two components suggests
that the spectrum of its excited states can be quite similar to the one of 
charmonium and bottomonium. The experimental evidence of the ground state of 
such system and  the searches for its excitations are described in 
\Section~\ref{sec:spexBc}.

The last Section is devoted to another class of bound states which
share a set of similarities with the heavy quarkonia.  The evidence of
the doubly charmed baryons claimed by Fermilab experiment E781 is
still rather weak and is described in \Section~\ref{sec:spexdcb}; further
searches, possibly by the B-factories, are needed before speculating
on their phenomenology.

\section{High precision measurements of vector state masses and widths}

\subsection[Charmonia]{Charmonia $\!$\footnote{Author:S.~Eidelman}}

\begin{figure}
\centering\includegraphics[width=\linewidth]{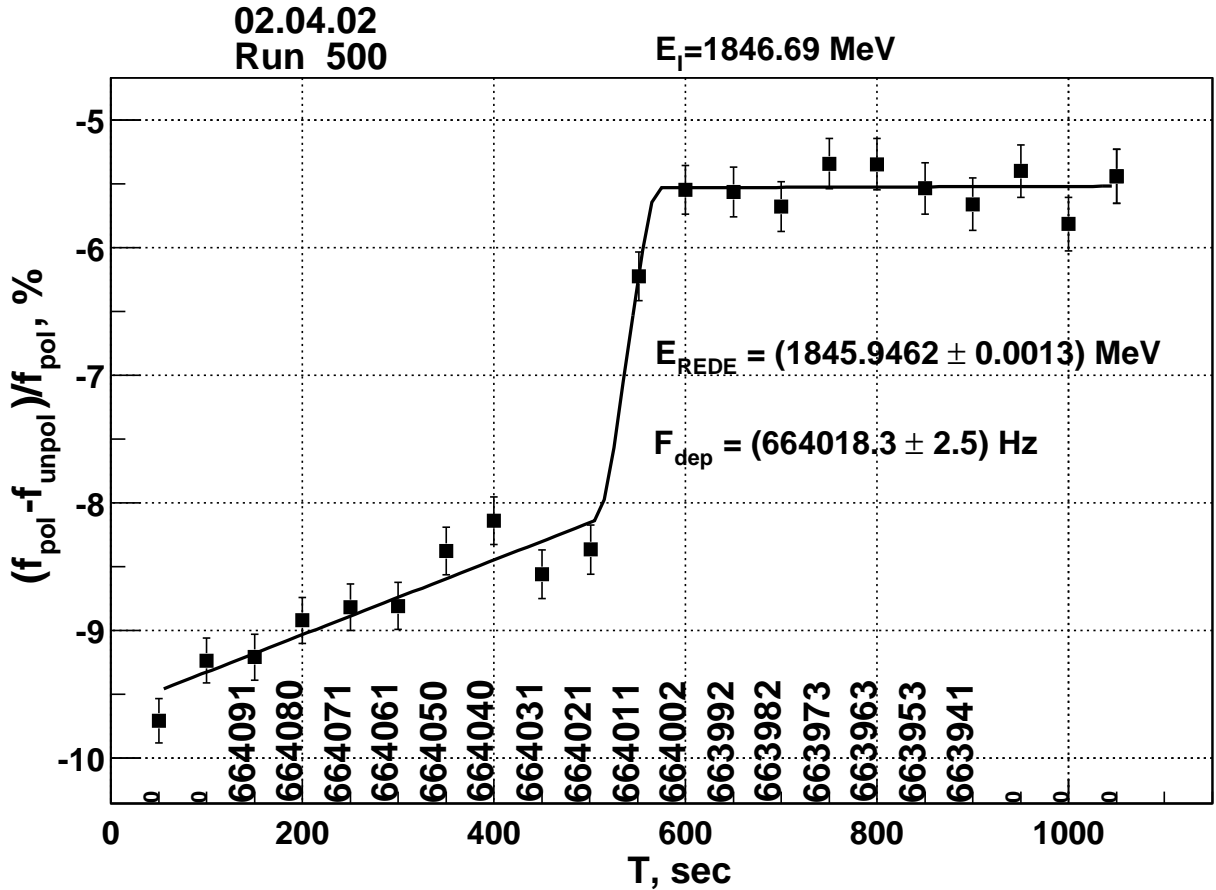}
\caption{The variation of the coincidence rate ratio for the polarized
         and unpolarized beams.}
\label{fig:Jump}
\end{figure}

The first precise measurement of the $J/\psi(1S)$ and $\psi(2S)$ meson
masses \cite{Zholents:1980qu} set the mass scale in the range around 3
GeV which provided a base for the accurate determination of the
charmonium state location.  The method of resonant depolarization,
described in Appendix~\ref{sec:redepol} of
\Chapter~\ref{chapter:commonexperimenttools}, has been developed in
Novosibirsk and first applied to the $\phi$ meson mass measurement at
the VEPP-2M storage ring~\cite{BUK}.  Later it was successfully used
to measure masses of the $\psi$-~\cite{Zholents:1980qu} and
$\Upsilon$-meson
family~\cite{Artamonov:1982mb,Artamonov:1983vz,Baru:1993tb}, see also
Ref.~\cite{Artamonov:2000cz}, in which the values of the masses were
rescaled to take into account the change of the electron mass value.
The accuracy of the $J/\psi(1S)$ meson mass measurement was later
improved in the Fermilab $p\bar{p}$-experiment
E760~\cite{Armstrong:1992wu} to $1.2\cdot 10^{-5}$ using the
$\psi(2S)$ mass value from Ref.~\cite{Zholents:1980qu}.  The new high
precision measurement \cite{Aulchenko:2003qq} of the $J/\psi$ and
$\psi'$ meson masses has been performed at the collider VEPP-4M using
the KEDR detector\cite{Anashin:2002uj}.  The polarimeter unit was
installed in the technical straight section of VEPP-4M and consisted
of the polarimeter\,---\,two scintillation counters detecting electron
pairs of the intrabeam scattering whose rate is spin-dependent
(Touschek effect~\cite{TOUSCHEK}) and the TEM wave-based
depolarizer~\cite{Blinov:2002ib}.  The characteristic jump in the
relative rate of scattered electrons at the moment of resonant
depolarization is $3\div3.5$\% with the statistical error of
0.3--0.4\% for the beam polarization degree higher than $50$\%.
Typical behavior of the rate ratio is shown in \Figure~\ref{fig:Jump}.

The characteristic uncertainty of the beam energy calibration due to the
depolarization procedure is 1.5~keV. 

The first part of the experiment consisted of three scans of the
$J/\psi(1S)$ region (the integrated luminosity  
$\approx 40$~nb$^{-1}$, 
the beam energy spread $\sigma_E \approx 0.6$~MeV) and
three scans of the $\psi(2S)$ region 
(the integrated luminosity $\approx 76$~nb$^{-1}$, 
$\sigma_E \approx 0.9$~MeV).
Then the betatron and synchrotron dumping decrements of VEPP-4M
were rearranged to reduce the energy spread down to 0.45~MeV and
the fourth scan of $J/\psi(1S)$  was performed (the integrated 
luminosity is  $\approx 10$~nb$^{-1}$).
The goal of this was the verification of systematic errors connected
with the collider operating mode and the beam energy spread.

The beam polarization time in the VEPP-4M ring is about 100 hours at
the $J/\psi(1S)$-energy. For the energy calibration runs, the beam spent
the time sufficient for the polarization in the booster ring VEPP-3
(2.5 hours at $J/\psi(1S)$    and about 1 hour at $\psi(2S)$ ) and was 
injected to VEPP-4M
 without essential loss of the polarization degree.

During the scan the data were acquired at seven energies around the
resonance peak.  Before data acquisition,
the beam energy calibration was made at point 1 to fix the 
current energy scale.
At points 2--6 the calibrations before and after data taking were
performed with the opposite direction of the depolarizer frequency
scan. The point 7 requires no energy calibration.

On completion of the scan the VEPP-4M magnetization cycle was
performed and the whole procedure was repeated. 
The energy dependence of the resonance cross-section was fitted 
taking into account the interference with continuum and 
radiative corrections.
The results obtained can be presented in the form
$$
 M_{J/\psi(1S)} - M_{J/\psi(1S)}^{PDG} = \,\ 47\pm 10\pm 7\
 {\rm keV},
$$
$$
 M_{\psi(2S)}\ -\ M_{\psi(2S)}^{PDG} = 151\pm 25\pm 9\  {\rm keV},
$$
demonstrating the agreement with the world average values
taking into account their uncertainties of $\pm 40$~keV and
$\pm 90$~keV, respectively~\cite{Eidelman:pdg2004}.
The following mass values have been obtained: 
$$
    M_{J/\psi(1S)} = 3096.917 \pm 0.010 \pm 0.007\ {\rm MeV},
$$
$$
    M_{\psi(2S)} = 3686.111 \pm 0.025 \pm 0.009\ {\rm MeV}.
$$
The relative measurement accuracy reached $4\cdot 10^{-6}$ for 
the $J/\psi(1S)$,
$7\cdot 10^{-6}$ for the $\psi(2S)$ and is approximately 3 times
better than that of the previous precise experiments in~\cite{Zholents:1980qu} and
\cite{Armstrong:1992wu}. 

The new result for the mass difference is
$$
M_{\psi(2S)} - M_{J/\psi(1S)} =  589.194 \pm 0.027 \pm 0.011\
{\rm MeV.}
$$

Substantial improvement in the beam energy accuracy obtained
by the presented experiment sets a
new standard of the mass scale in the charmonium range.   

\subsection[Bottomonia]{Bottomonia $\!$\footnote{Author: S.~Eidelman}}

Development of the resonant depolarization method suggested and first
realized in Novosibirsk~\cite{BUK,Derbenev:1980gp} also allowed high
precision measurements of the resonance masses in the $\Upsilon$
family.  The MD-1 group in Novosibirsk carried out three independent
measurements of the $\Upsilon(1S)$
mass~\cite{Artamonov:1982mb,Artamonov:1983vz,Baru:1986mi,Baru:1993tb}. The
$\Upsilon(1S)$ mass was also measured by the CUSB collaboration in
Cornell~\cite{MacKay:1984kv}.  Their result was by $0.63 \pm 0.17$~MeV
or $3.8\sigma$ lower than that of MD-1. The reasons of this
discrepancy are not clear, however, when the MD-1 group performed a
fit of the CUSB results using the Novosibirsk procedure (in
particular, it included a new method or calculating radiative
corrections according to~\cite{Kuraev:1985hb} instead of the older
approach of Ref.~\cite{Jackson:1975vf}), the difference between the
two results decreased to $0.32 \pm 0.17$~MeV or $1.9\sigma$ only.
    
The mass of the $\Upsilon(2S)$ meson was measured by the MD-1 group in
Novosibirsk~\cite{Artamonov:1983vz,Baru:1986mi} and two groups in
DESY\,---\,ARGUS and Crystal Ball~\cite{Barber:1983im}. Both groups in
DESY obtained the mass value consistent with that in Novosibirsk, the
average being $0.5 \pm 0.8$~MeV lower than that of MD-1.

The mass of the $\Upsilon(3S)$ meson was measured by the MD-1 group
only~\cite{Artamonov:1983vz,Baru:1986mi}. As in the case of the
$\Upsilon(2S)$ meson, a systematic error of the measurement was less
than 0.2~MeV, much smaller than the statistical one.
   
Finally, in 2000 all the results on the mass of the
$\psi$~\cite{Zholents:1980qu,Zholents:1981vv} and
$\Upsilon$~\cite{Artamonov:1982mb,Artamonov:1983vz,Baru:1986mi,Baru:1993tb,Baru:1986mi}
family resonances were updated~\cite{Artamonov:2000cz} to take into
account a more precise value of the electron
mass~\cite{Cohen:1973bc,Cohen:1987fr} (for the $\psi$ family an
additional correction has been made to take into account the new way
of calculating radiative corrections~\cite{Kuraev:1985hb}).  In
\Table~\ref{tab:ups} we summarize the information on these experiments
presenting for each detector the number of energy points and the
energy range studied, the integrated luminosity and the final value of
the mass.  The results after the update mentioned above are shown in
parentheses.
\begin{table}[th]
\caption{Mass Measurements in the $\Upsilon$ Meson Family}
\label{tab:ups}
\begin{center}
\begin{tabular}{|c|c|c|c|c|c|}
\hline
Resonance & Collider & N of Points &  Detector & $\int{\rm Ldt}$,  
& Mass, MeV \\
 & & $\sqrt{s}$, MeV & Reference & pb$^{-1}$   & \\
\hline
$\Upsilon(1S)$ & VEPP-4 & 43 & MD-1~\cite{Baru:1993tb} & 2.0 
& $9460.59 \pm 0.09 \pm 0.05$    \\
& & 9420--9490 & (\cite{Artamonov:2000cz}) & &  ($9460.51 \pm 0.09 \pm 0.05$)    \\
\cline{2-6}
   & CESR & 13 & CUSB~\cite{MacKay:1984kv} & 0.285 & $9459.97 \pm 0.11 \pm 0.07$ \\
   &      & 9446--9472 & & &  \\
\hline
$\Upsilon(2S)$ & VEPP-4  & 37 & MD-1~\cite{Baru:1986mi} & 0.6  
& $10023.6 \pm 0.5$  \\
& & 9980--10075 & (\cite{Artamonov:2000cz}) & & ($10023.5 \pm 0.5$)   \\
\cline{2-6}
   & DORIS & 13 & ARGUS~\cite{Barber:1983im} & 2.0  & $10023.43 \pm 0.45$ \\
   &  & 9960--10040 & Cr. Ball~\cite{Barber:1983im} & 2.0 & $10022.8 \pm 0.5$ \\
   &     &  & Average~\cite{Barber:1983im} &  & $10023.1 \pm 0.4 \pm 0.5$ \\
\hline
$\Upsilon(3S)$ & VEPP-4  & 35 & MD-1~\cite{Baru:1986mi} & 1.25   
& $10355.3 \pm 0.5$  \\
& & 10310--10410 & (\cite{Artamonov:2000cz}) & &  ($10355.2 \pm 0.5$)    \\
\hline
\end{tabular}
\end{center}
\end{table}

\section[Spin averaged and fine splittings]
        {Spin averaged and fine splittings}
\label{sec:spexfine}

\subsection[Charmonium P states: COG and fine splittings]
           {Charmonium P states: COG and fine splittings
            $\!$\footnote{Authors: R.~Mussa, G.~Stancari}
}
\def\logtwo{\rm ln(2)}
\def\grb{\frac{\Gamma_{res}}{\Gamma_{beam}}}
\def\grbq{\frac{\Gamma_{res}^2}{\Gamma_{beam}^2}}
\def\Gio{\Gamma_{io}}
\def\thcm{\theta_{CM}}

The most precise determinations of mass and width come from the
 study of charmonium spectroscopy by direct formation of $\bar{c} c$ states in 
$\bar{p}p$ annihilation at the Fermilab Antiproton Source 
(experiments E760 and E835). 
The E760 collaboration measured the resonance parameters of the 
$\chicj{1}$ and $\chicj{2}$\cite{Armstrong:1991yk}.

For both E760 and E835-I, the transition energy of the Antiproton
Accumulator was close enough to the $\chicj{0}$ mass to prevent stable
running with large stacks in this energy region. Nevertheless, a few
stacks were decelerated to the $\chicj{0}$ region at the end of Run I,
yielding an unexpectedly high rate of $\jpsi\gamma$ events. The
Accumulator underwent a major upgrade between 1997 and 2000, shifting
the transition energy \cite{mcginnis:2003} and allowing a smooth
running at the $\chicj{0}$, with substantial increase in statistics
\cite{Bagnasco:2002si}, and a better control of systematics.
 
A new measurement of the $\chicj{1}$ parameters was made in year 2000, with 
roughly 15 times more statistics than the predecessor experiment E760.
The $\chicj{2}$ parameters were also remeasured with statistics comparable to
those of experiment E760. This report includes the new results, in 
publication, not yet included in the PDG.

\begin{figure}
\begin{center}  
\includegraphics[width=9.0cm]{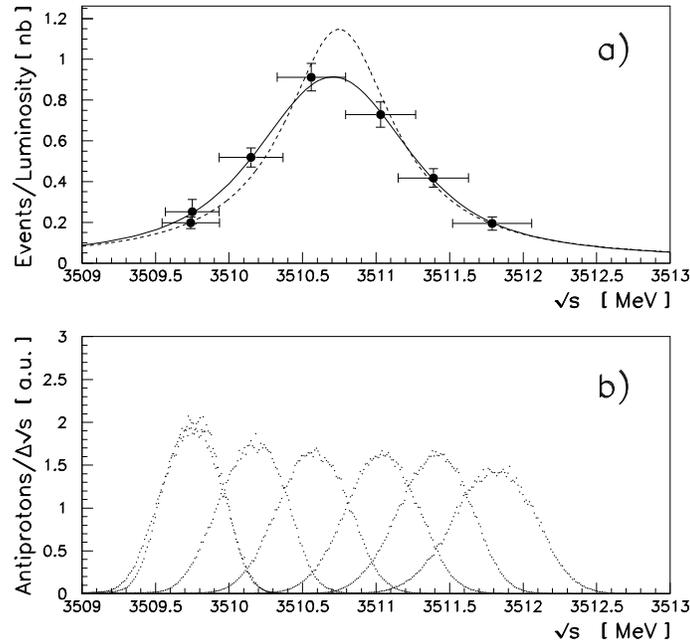} 
\end{center}
\caption{Measured cross-section at each data point,  
excitation curve (full line) and deconvoluted resonance curve (dotted line)
for one scan at the $\chicj{1} $ ; 
plotted in the lower part of the figure 
are the beam energy profiles corresponding to each data point}
\label{fig:giulio}
\end{figure}

The effect of scanning a narrow resonance with a beam of
comparable width is show in  \Figure~\ref{fig:giulio}, where the
excitation curve for one scan at the $\chicj{1}$ is compared with 
the deconvoluted Breit Wigner shape and the measured beam energy profiles 
for each point. 

In mass and width measurements, the systematic error comes from uncertainties 
on auxiliary variables measured concurrently to data taking 
(changes in beam orbit length, efficiency and luminosity at each energy 
point), as well as the absolute calibration of the beam energy.
The absolute calibration of the beam energy is deduced from the absolute 
calibration of the orbit length,  done using $\psi(2S)$
scans, and assuming 3686.000 for the mass of this state.
The more precise determination recently done at VEPP-4, documented in the 
previous section, implies a systematic shift (up) of 70, 83, 89~keV of the
$\chicj{0,1,2}$ measurements respectively.
The systematic error on $\chi_c$ masses from $\psi(2S)$ mass 
determination reduces  then to 16,19,20~keV respectively, and is 
now negligible if compared to the other sources, which are uncorrelated 
when we merge different scans.
The  impact of radiative corrections to account for proton bremsstrahlung
is still well below other systematic errors; it was estimated  
 using the expression:
\[
 \sigma_{BW}^{rad}(\beta,s)= \beta\int_0^{\frac{\sqrt{s}}{2}}
\frac{dk}{k} \left(\frac{2k}{\sqrt{s}}\right)^\beta \sigma_{BW}(s-2k\sqrt{s})
\]
with 
\bea
 \beta &=& \frac{2\alpha}{\pi} \times \left({\frac{s-2m_p^2}{\sqrt{s(s-4m_p^2)}}}
\times \ln \frac{s +  \sqrt{s(s-4m_p^2)}}
{s - \sqrt{s(s-4m_p^2)}} - 1 \right) 
\nn\\
&=& 
6.7 \times 10^{-3}(\chicj{0}), \quad
7.0 \times 10^{-3}(\chicj{1}), \quad
7.2 \times 10^{-3}(\chicj{2}). 
\nn
\eea

Systematic shifts on masses are $\Delta
m(\chicj{0,1,2})=-0.06,-0.01,-0.02 \, {\rm MeV}/c^2$; the shifts on
total widths are $\Delta\Gamma / \Gamma \approx -1\%$ for all $\chi_c$
states.

\begin{table}[th]
\caption{Parameters of $\chi_c$ states from E760, E835, and BES}
\label{tab:chisumm}
\begin{center}
\renewcommand{\tabcolsep}{1pc} 
\renewcommand{\arraystretch}{1.2} 
\begin{tabular}{@{}cllrrr}
\hline 
 R & Expt.& Mass(MeV/c$^2$) & $\Gamma$(MeV) & 
Ref. \\
\hline
 $\chicj{0}$ & & & &  \\
 & BES     & 3414.1$\pm$0.6$\pm$0.8 & 14.3$\pm$2.0$\pm$3.0 &\cite{Bai:1998cw}\\
 & E835 & 3415.4$\pm$0.4$\pm$0.2 & 9.9$\pm$1.0$\pm$0.1 &\cite{Bagnasco:2002si}\\
 & E835 & 3414.7$\pm$0.7$\pm$0.2 & 8.6$\pm$1.7$\pm$0.1 &\cite{Andreotti:2003sk}\\
 & PDG 2004 & 3415.19$\pm$0.34 & 10.1$\pm$0.8  & \cite{Eidelman:pdg2004}\\
\hline
 $\chicj{1}$ & & & &  \\
 & E760 & 3510.61$\pm$0.10$\pm$0.02 & 0.88$\pm$0.11$\pm$0.08 & \cite{Armstrong:1991yk}\\
 & PDG 2004 & 3510.59$\pm$0.10 &  0.91$\pm$0.13 & \cite{Eidelman:pdg2004} \\
 & E835 & 3510.725$\pm$0.065$\pm$0.018 & 0.88$\pm$0.06$\pm$0.09 & \cite{chinew}\\
\hline
 $\chicj{2}$ & & & &  \\
 & E760 & 3556.24$\pm$0.07$\pm$0.12 & 1.98$\pm$0.17$\pm$0.07 & \cite{Armstrong:1991yk} \\
 & PDG 2004 & 3556.26$\pm$0.11 & 2.11$\pm$0.16 & \cite{Eidelman:pdg2004} \\
 & E835 & 3556.10$\pm$0.09$\pm$0.17 & 1.93$\pm$0.19$\pm$0.09 & \cite{chinew}\\
\hline
\end{tabular}
\end{center}
\end{table}

E835 could also measure the $\chicj{0}$ excitation curve in the  
$\ppbar\to\pi^0\pi^0$ channel, exploting the amplification due to 
interference with continuum. The measurement is compatible with 
result obtained in $\psi\gamma$ and of course has correlated systematic 
errors.

A measurement of mass  \cite{Bai:1998cw} and width  \cite{Bai:1998gh} 
with accuracy almost comparable to the one obtained in $\ppbar$ annihilations 
was made by BES on the $\chicj{0}$, exploiting the sample of 3.8M 
$\psip$ decays to various decay channels. There are not yet mass and width 
measurements of $\chi_c$ states from the 14M $\psip$ sample.
\Table~\ref{tab:chisumm} summarizes the most accurate results on masses 
and widths at present. Statistical errors on $\chicj{1,2}$ masses 
are obtained from gaussian sums of errors from event statistics and 
errors from orbit length measurements; the latter are dominant, therefore
future improvements will require to push fractional errors on orbit lengths
below $10^{-6}$. 
 In the case of $\chicj{0}$ there is still room for improvement:
ten times more statistics at the $\chicj{0}$ in a $\ppbar$ annihilation 
experiment could take errors on masses down to 200~keV, and on widths down 
to 3\%. To reach a comparable level on narrow $\chi_b$ states is very 
challenging, and will require new ideas.

It is finally possible to present the results on P states by
calculating the spin independent ($M_{COG}$), spin--orbit ($h_{LS}$)
and tensor ($h_T$) terms of the $\ccbar$ Hamiltonian. All values are
summarized in \Table~\ref{tab:chicsum}.

\begin{table}[th]
\caption{Fine splittings between $\chi_c$  states}
\label{tab:chicsum}
\begin{center}
\begin{tabular}{lc}
\hline
 & $\ccbar(n=1)$ \\
\hline
$M_{COG}$ (in MeV)  \\
$\Delta M_{21}=M(\chicj{2})-M(\chicj{1})$ (in MeV) 
& 45.6$\pm$0.2 \\
$\Delta M_{10}=M(\chicj{1})-M(\chicj{0})$ (in MeV) 
& 95.3$\pm$0.4 \\
$\rho(\chi)=\Delta M_{21}/\Delta M_{10}$ & 
0.470$\pm$0.003 \\
$h_{T}$ (in MeV)              & 10.06$\pm$0.06  \\
$h_{LS}$ (in MeV)              & 34.80$\pm$0.09  \\
\hline
\end{tabular}
\end{center}
\end{table}

\subsection[Bottomonium P states: COG and Fine splittings]
           {Bottomonium P states: COG and Fine splittings 
            $\!$\footnote{Author: T.~Skwarnicki}}
\shortpage

After discovery of the $\Y(1S)$, $\Y(2S)$ and $\Y(3S)$ resonances
at the fixed target $pN$ experiment 
at Fermilab in 1997 \cite{UpsilonDiscovery}
the first two were observed a year later at the $e^+e^-$
storage ring DORIS at DESY \cite{UpsilonsDORIS}. 
Since DORIS energy reach was stretched well beyond its design,
the $\Y(3S)$ could not be reached. The limited statistics 
and limited photon detection capabilities of the detectors prevented
observation of the $\chi_{bJ}(1P)$ states via E1 photon transitions
from $\Y(2S)$ at that time.
Energy range of another $e^+e^-$ storage ring, CESR at Cornell University,
was extended high enough to reach the $\Y(3S)$ in 1982.
The CUSB detector at CESR had sufficient photon detection resolution
in NaI(Tl)/Lead-glass calorimeter
to discover the three $\chi_{bJ}(2P)$ states in inclusive photon
spectrum in $\Y(3S)$ decays \cite{chib2pdisci}. 
The $J=1$ and $J=2$ states were also
observed in two-photon cascade, $\Y(3S)$ $\to$ $\gamma\chi_{bJ}(2P)$,
$\chi_{bJ}(2P)\to\gamma\Y(nS)$ ($n=1,2$), followed by
$\Y(nS)\to l^+l^-$, where $l^+l^-$ stands for $e^+e^-$ or $\mu^+\mu^-$
\cite{chib2pdisce}.
The latter ``exclusive'' approach eliminates all photon backgrounds
from $\pi^0$s copiously produced in hadronic decays of $b\bar b$
states, but results in low signal statistics. 
In fact, the $J=0$ is very difficult to observe this way 
since it has larger gluonic annihilation width, which suppresses
branching ratios for radiative transitions.
\begin{figure}[p]
\begin{center}
\includegraphics[width=86mm]{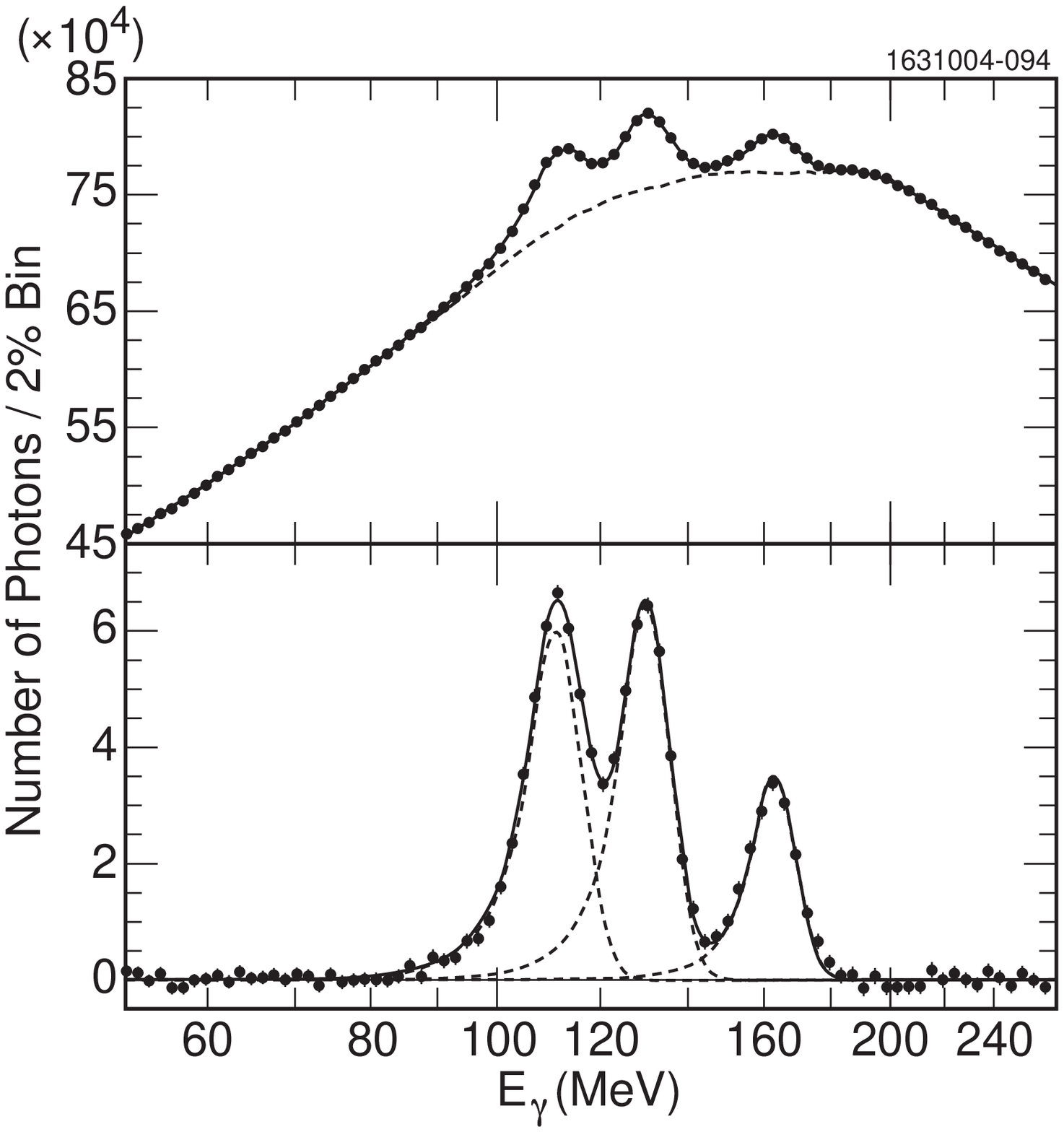}
\end{center}
\caption[Fit to the $\Y(2S)\to\gamma\chi_{bJ}(1P)$ ($J=2,1,0$) photon
         lines in the CLEO~III data]
        {Fit to the $\Y(2S)\to\gamma\chi_{bJ}(1P)$ ($J=2,1,0$) photon
         lines in the CLEO~III data.  The points represent the data
         (top plot).  Statistical errors on the data are smaller than
         the point size.  The solid line represents the fit.  The
         dashed line represents total fitted background.  The
         background subtracted data (points with error bars) are shown
         at the bottom.  The solid line represents the fitted photon
         lines together. The dashed lines show individual photon
         lines.}
\label{fig:2sg1p}
\medskip
\begin{center}
\includegraphics[width=86mm]{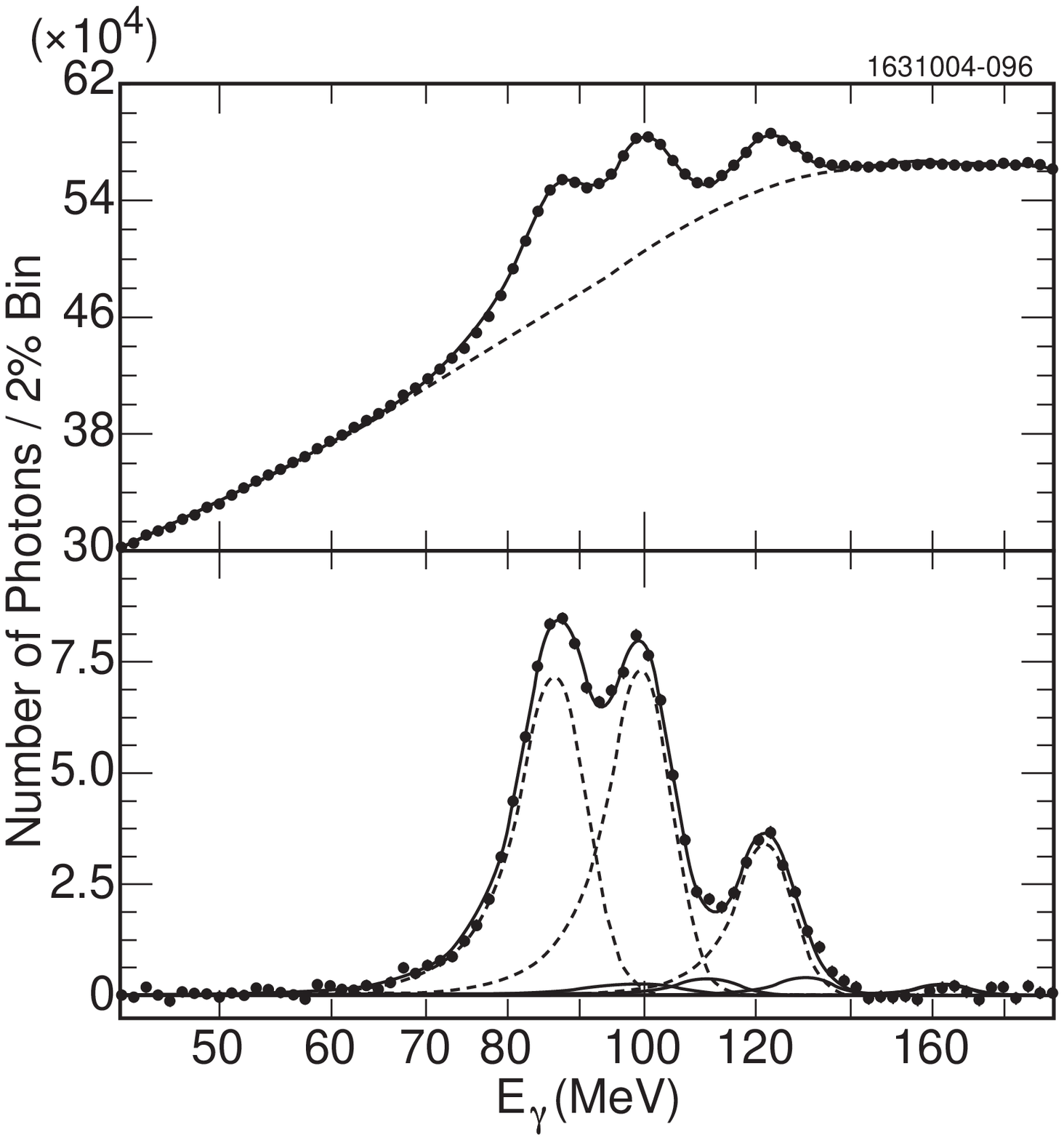}
\end{center}
\caption[Fit to the $\Y(3S)\to\gamma\chi_{bJ}(2P)$ ($J=2,1,0$) photon
         lines in the CLEO~III data]
        {Fit to the $\Y(3S)\to\gamma\chi_{bJ}(2P)$ ($J=2,1,0$) photon
         lines in the CLEO~III data.  See caption of
         \Figure~\ref{fig:2sg1p} for the description.  Small solid-line
         peaks in the bottom plot show the
         $\chi_{bJ}(2P)\to\gamma\Y(1D)$ and
         $\Y(2S)\to\gamma\chi_{bJ}(1P)$ contributions.  }
\label{fig:3sg2p}
\end{figure}
A year later the CUSB experiment produced similar evidence for
$\chi_{bJ}(1P)$ states in the $\Y(2S)$ data \cite{chib1pdiscovery}.
The $J=2$ and $J=1$ states were also observed by the CLEO 
experiment in inclusive photon spectrum, with photons 
reconstructed in the tracking system after conversion
to $e^+e^-$ pairs at the beam-pipe \cite{CLEOIchib}.

Meanwhile DORIS accumulated more data at the $\Y(2S)$ resonance
with two new detectors: magnetic spectrometer ARGUS, and 
NaI(Tl)-calorimeter Crystal Ball, which previously explored
photon spectroscopy in charmonium at SPEAR.
The Crystal Ball confirmed the CUSB results on the $\chi_{bJ}(1P)$
states \cite{CBchib}, 
though the $J=0$ photon line was observed at a different energy, 
soon confirmed by ARGUS via photon conversion technique \cite{ARGUSchib}.
Analysis of angular correlation in $\gamma\gamma l^+l^-$
by Crystal Ball established spin assignment 
to the observed $\chi_{b2}(1P)$ and $\chi_{b1}(1P)$ states \cite{CBspin}.
\shortpage
Next round of improvements in experimental results came about
a decade later from the CESR upgraded to higher luminosity 
and upgraded CUSB and CLEO experiments. 
The CUSB-II detector was equipped with compact
BGO calorimeter. The CLEO~II collaboration built large CsI(Tl)
calorimeter which was put inside the superconductive magnet.
Both experiments improved the results on $\chi_{bJ}(2P)$ 
states, with the increased $\Y(3S)$ data size \cite{cleocusb}.

A few years later the CLEO~II experiments took a short $\Y(2S)$ run.
Even though the number of $\Y(2S)$ resonance decays was not
much larger than in the previous measurements, the results on
$\chi_{bJ}(1P)$ states were substantially improved \cite{CLEOIIchib}
thanks to much larger photon detection efficiency of 
well-segmented CLEO~II calorimeter. 
\begin{figure}[p]
\begin{center}
\vskip-0.7cm
Average: $(110.4\pm0.3)$~MeV \hfill
Average: $(129.6\pm0.3)$~MeV \\[-1.2truecm]
\hbox{
\includegraphics[width=7cm,height=7cm]{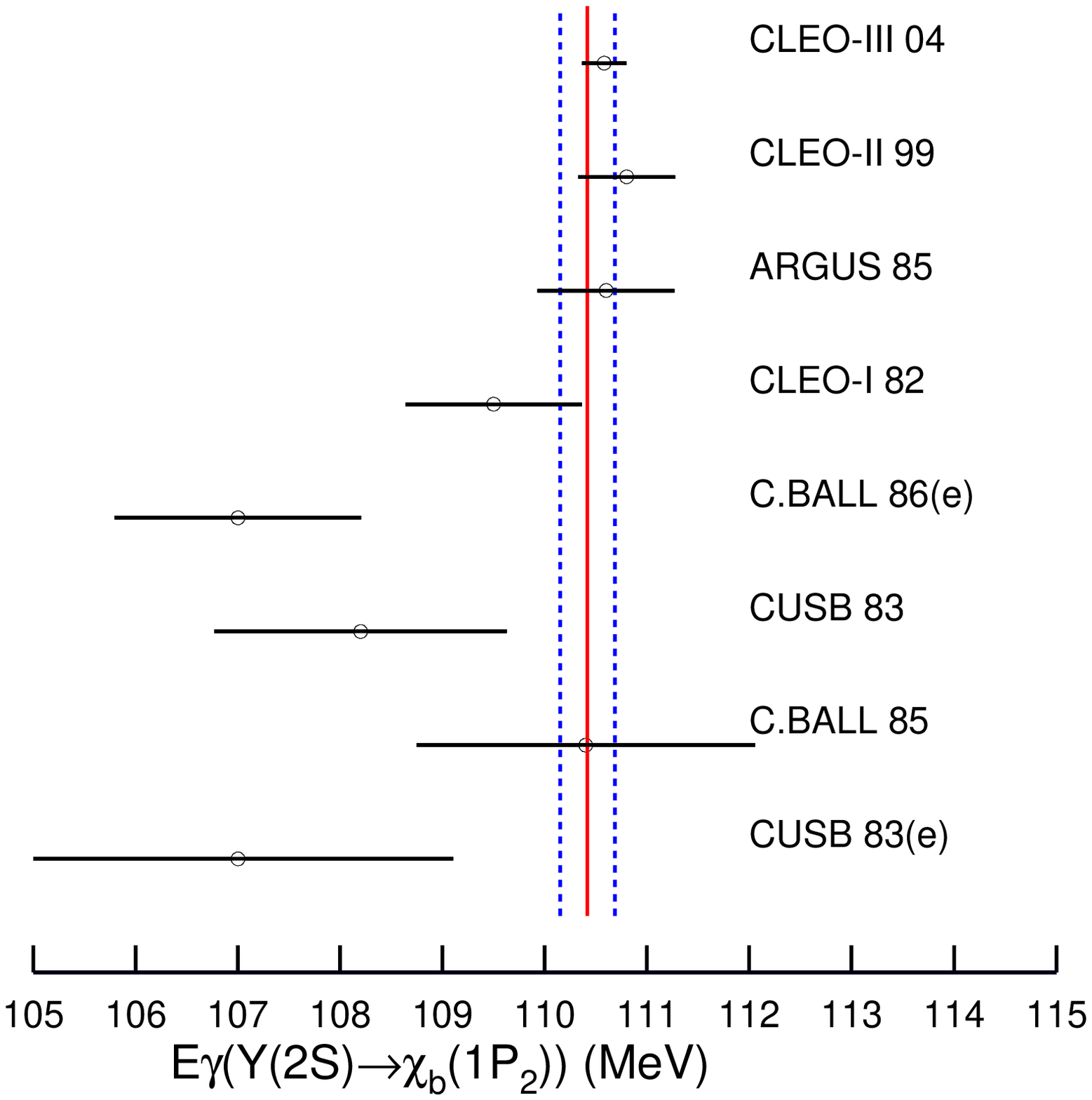}
\hskip-0.5truecm
\includegraphics[width=2cm,height=7.14cm]{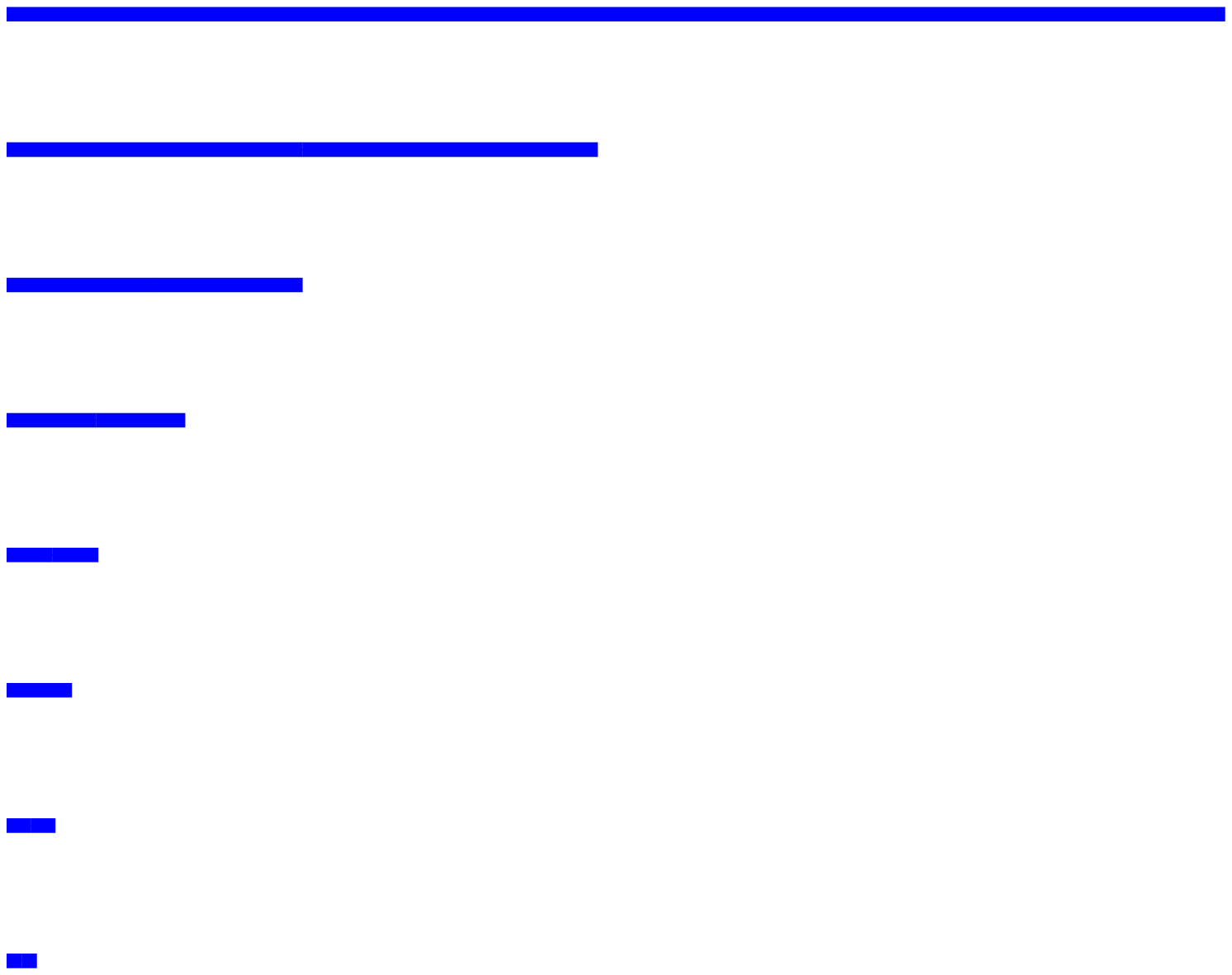}
\hskip-1.0truecm
\includegraphics[width=7cm,height=7cm]{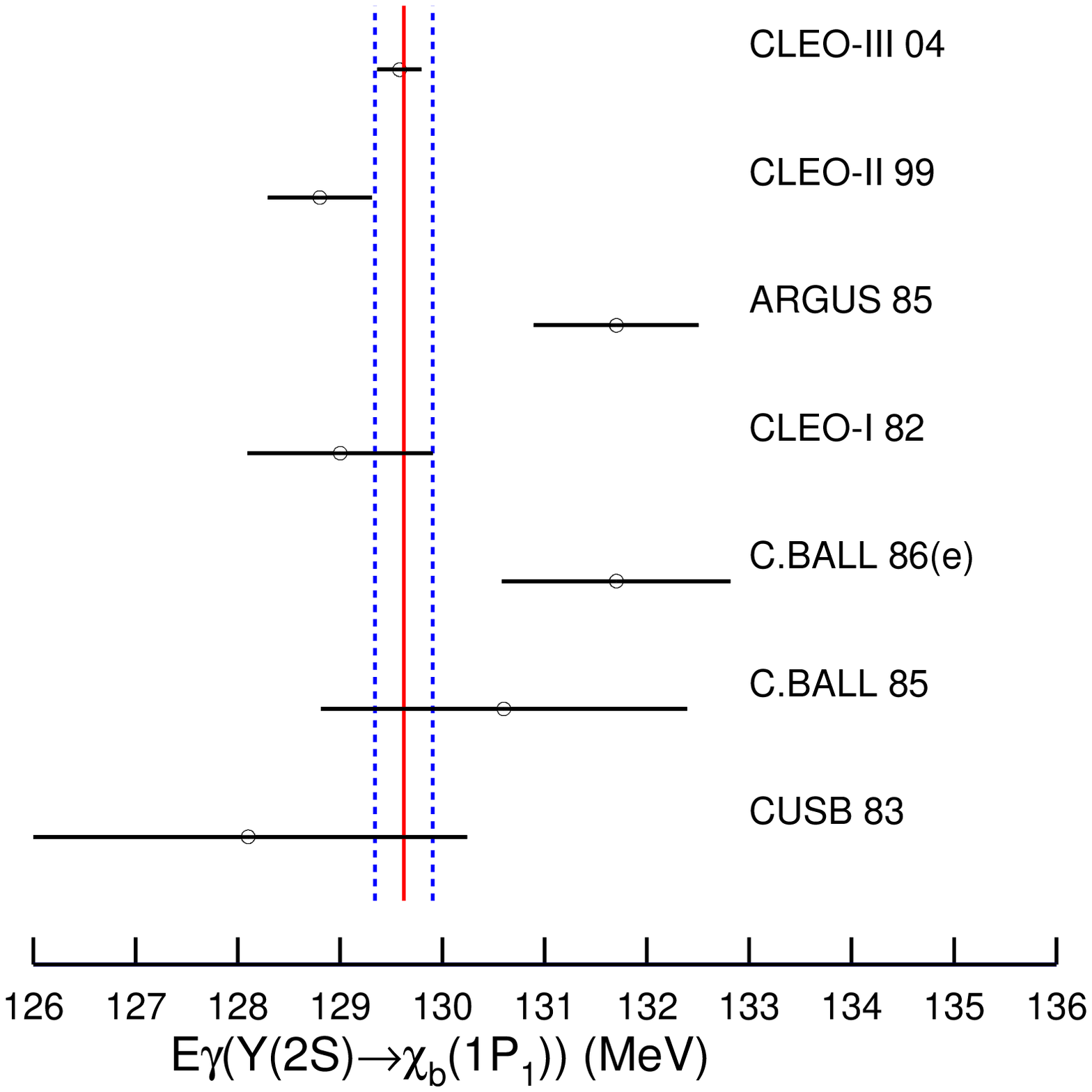}
\hskip-0.5truecm
\includegraphics[width=2cm,height=7.14cm]{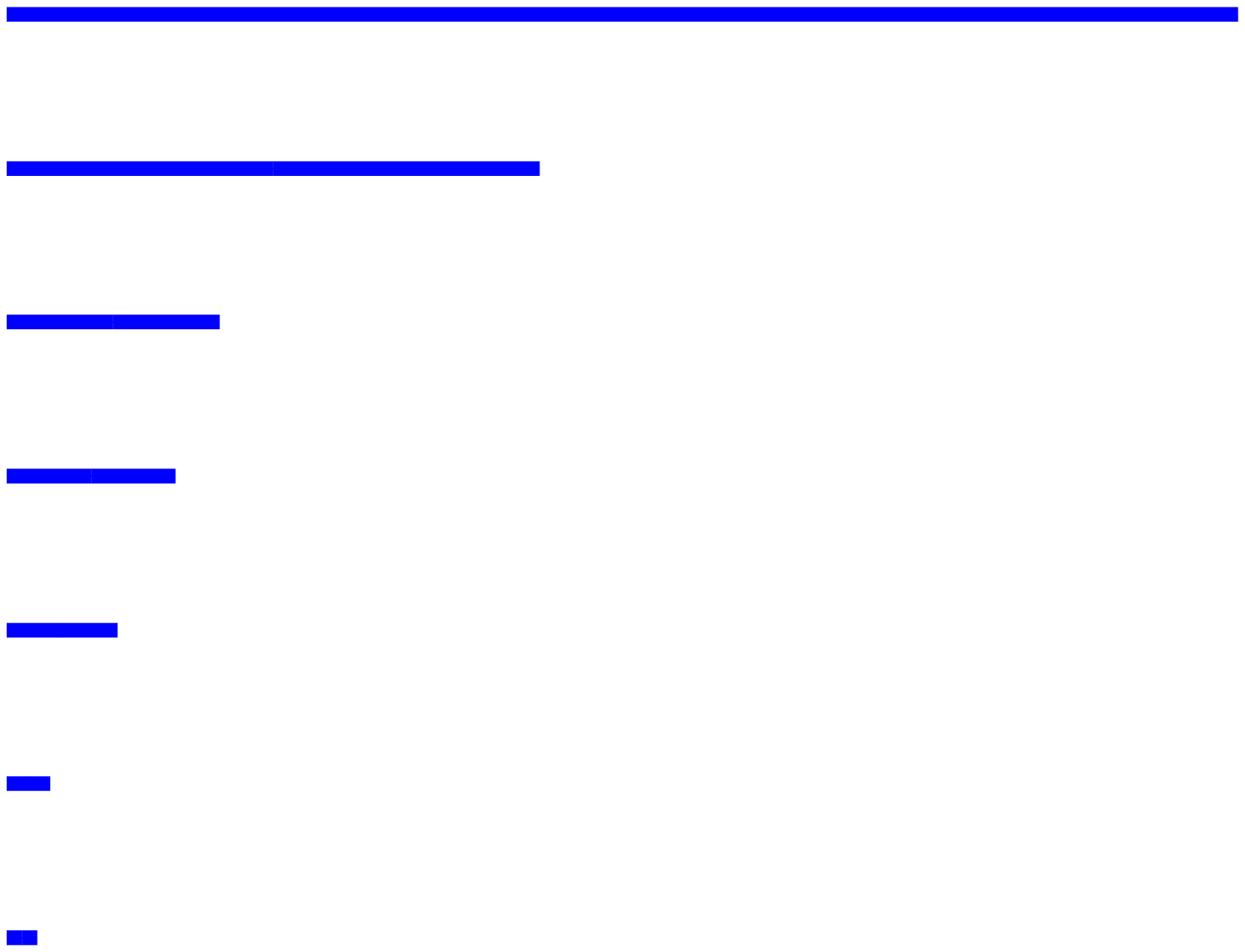}
}
\vskip0.1cm
Average: $(162.4\pm0.4)$~MeV \hfill
\phantom{Average: $(111.1\pm1.1)$~MeV} \\[-1.2truecm]
\hbox{
\includegraphics[width=7cm,height=7cm]{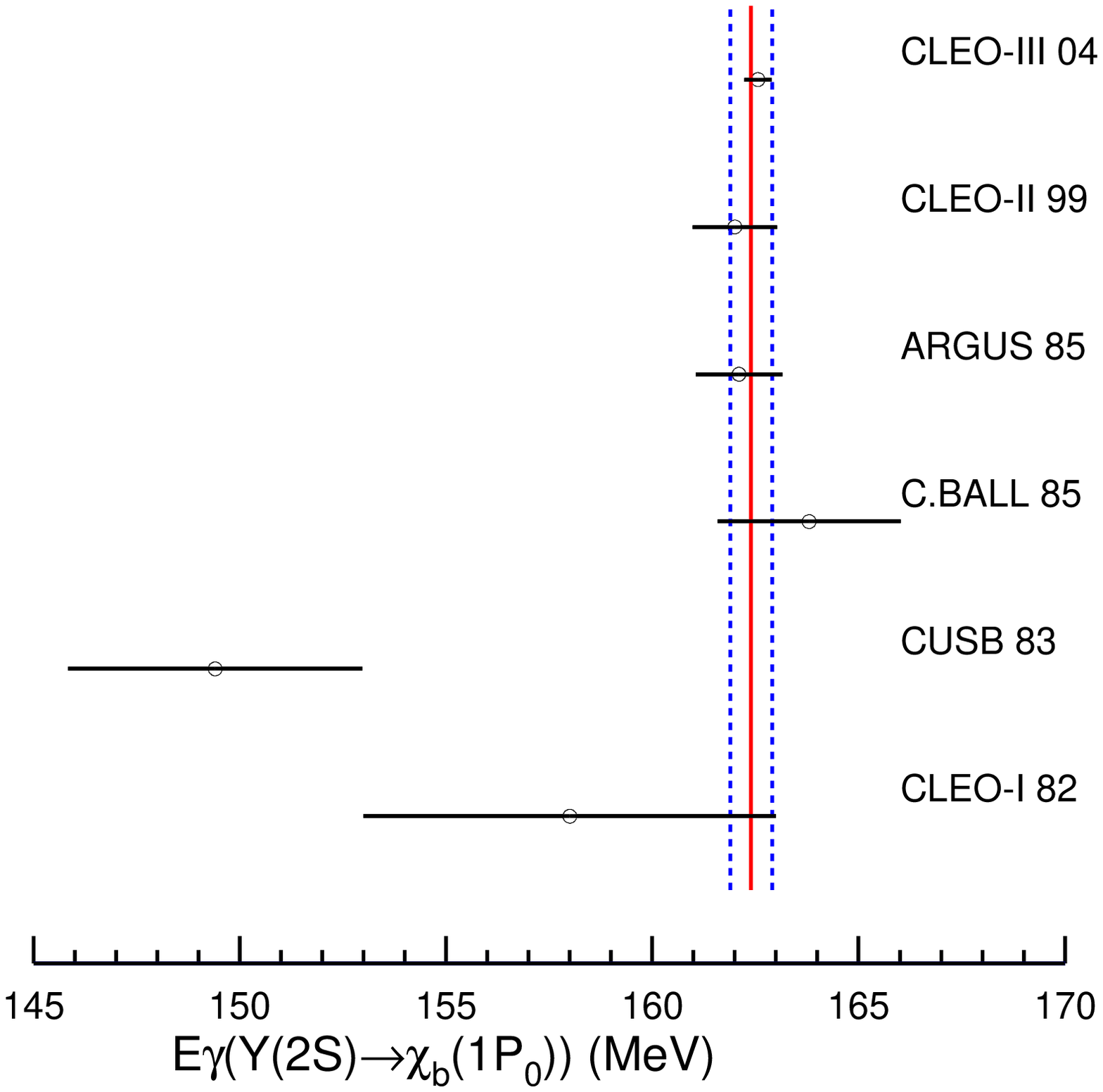}
\hskip-0.5truecm
\includegraphics[width=2cm,height=7.14cm]{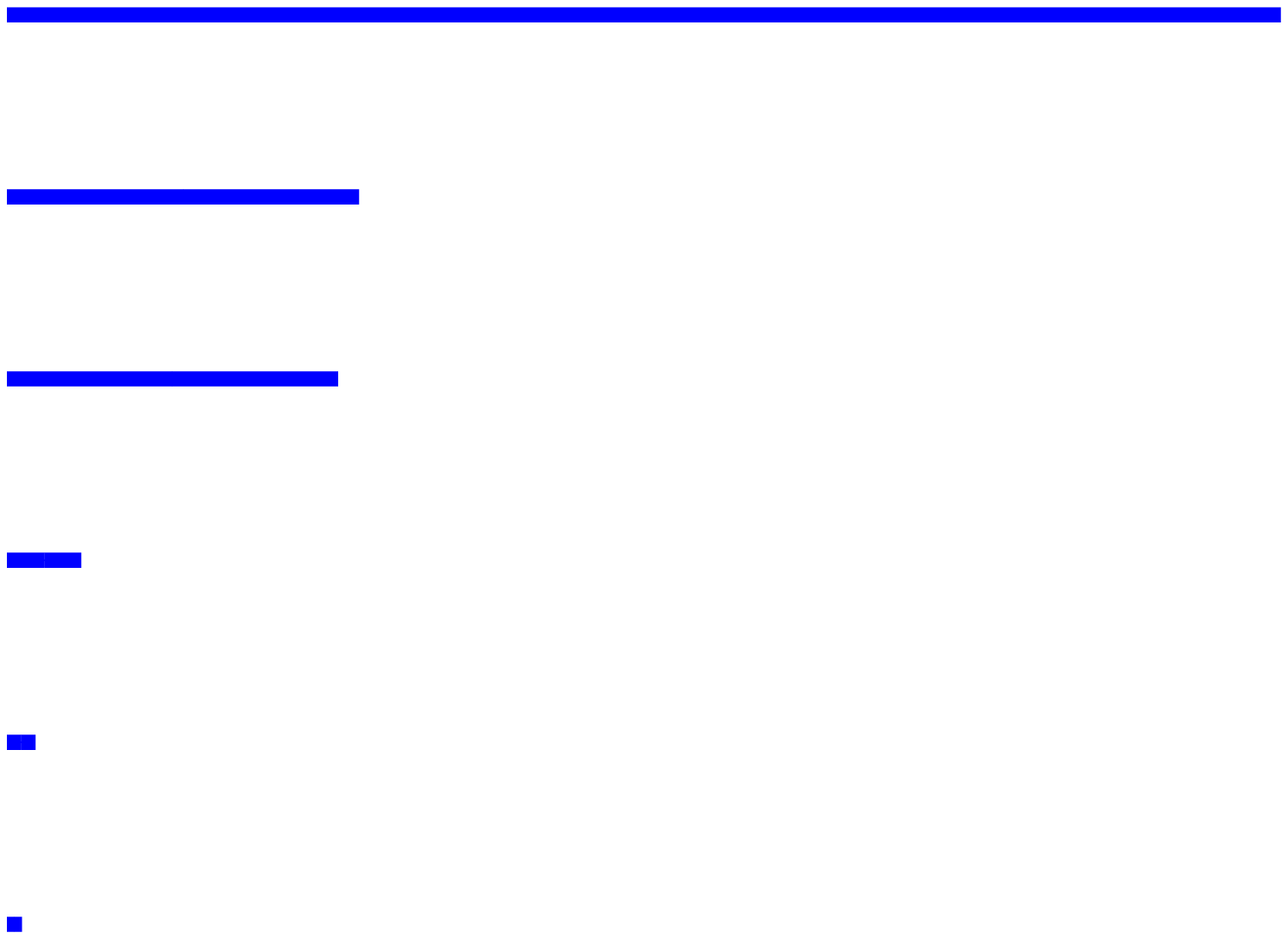}
}
\vskip-0.2cm
\caption{Measurements of the photon energies 
         in $\Y(2S)\to\gamma\chi_{bJ}(1P)$.
         The vertical bars indicate the 
         world average value (solid) and its error
         (dashed). These are also listed on top.
         The thick horizontal bars to the right of the name of
         the experiment
         give the relative weight of each experiment
         into the average value.
         Photon energy measurements from analyses of
         exclusive $\gamma\gamma l^+l^-$ events are
         indicated with an ``(e)'' after the date of the
         publication.}
\vskip-0.6cm
\label{fig:e2s1pj}
\end{center}
\end{figure}

\begin{figure}[p]
\vskip-0.7cm
Average: $(86.1\pm0.2)$~MeV \hfill
Average: $(99.3\pm0.2)$~MeV \\[-1.2truecm]
\hbox{
\includegraphics[width=7cm,height=7cm]{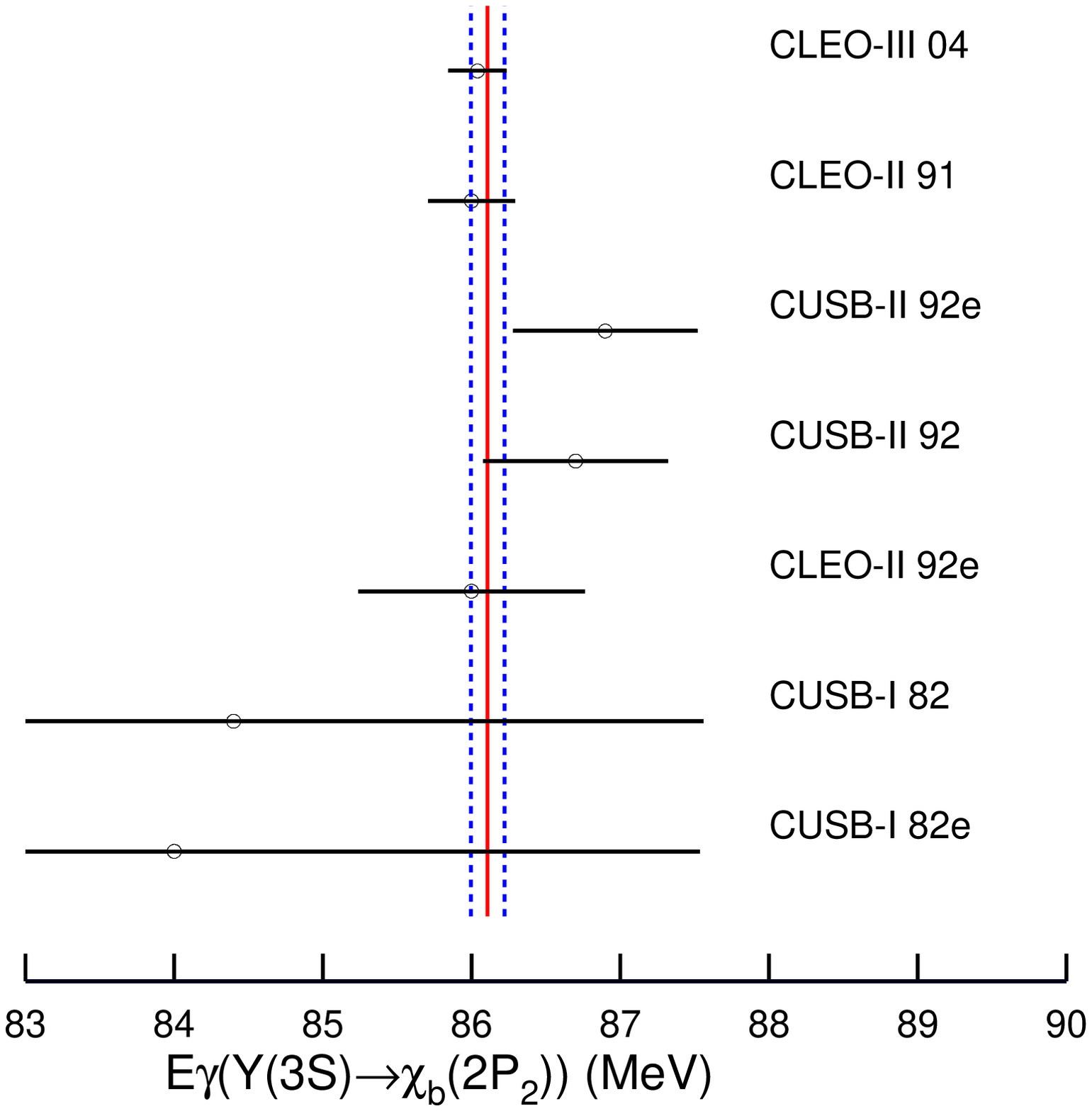}
\hskip-0.5truecm
\includegraphics[width=2cm,height=7.14cm]{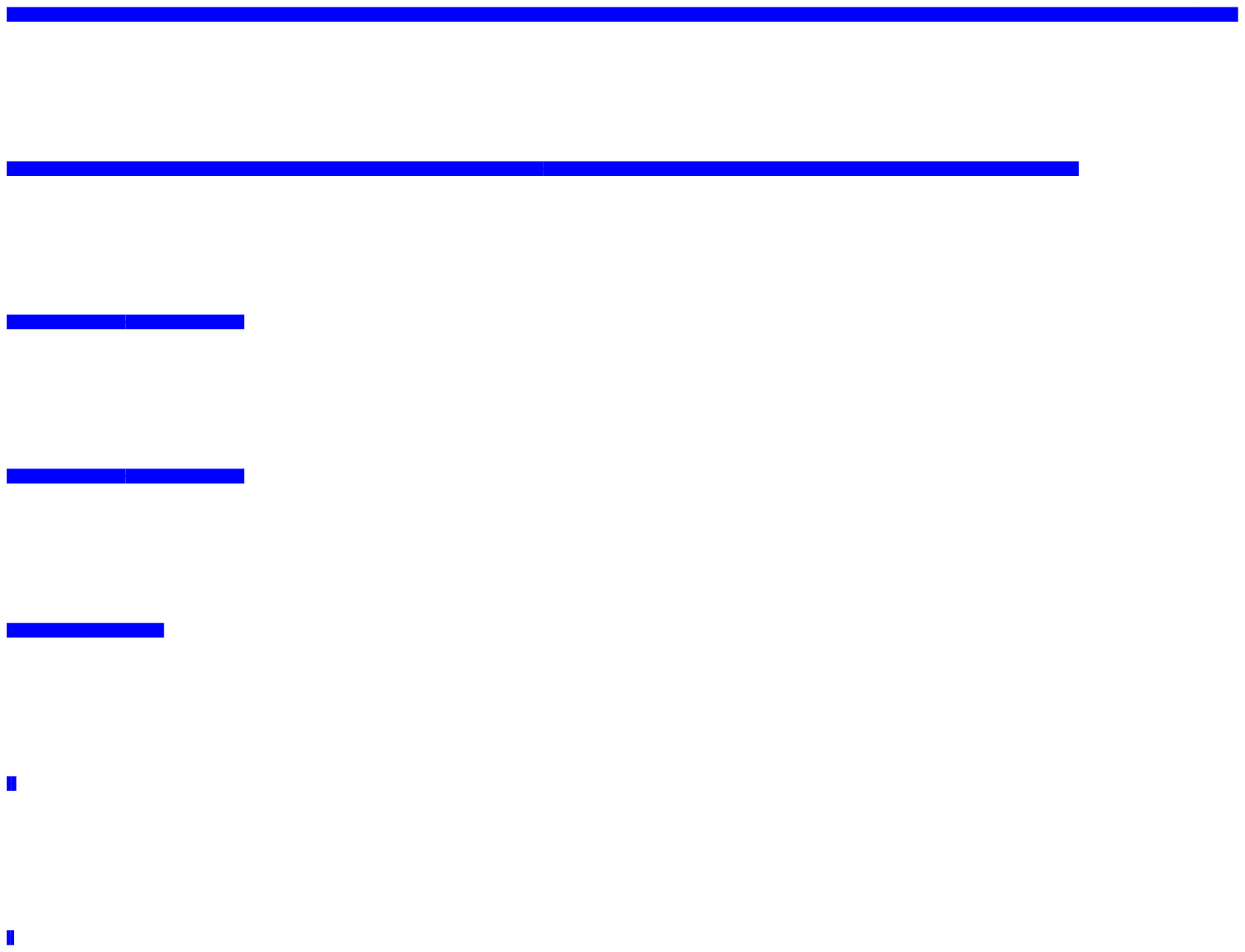}
\hskip-1.0truecm
\includegraphics[width=7cm,height=7cm]{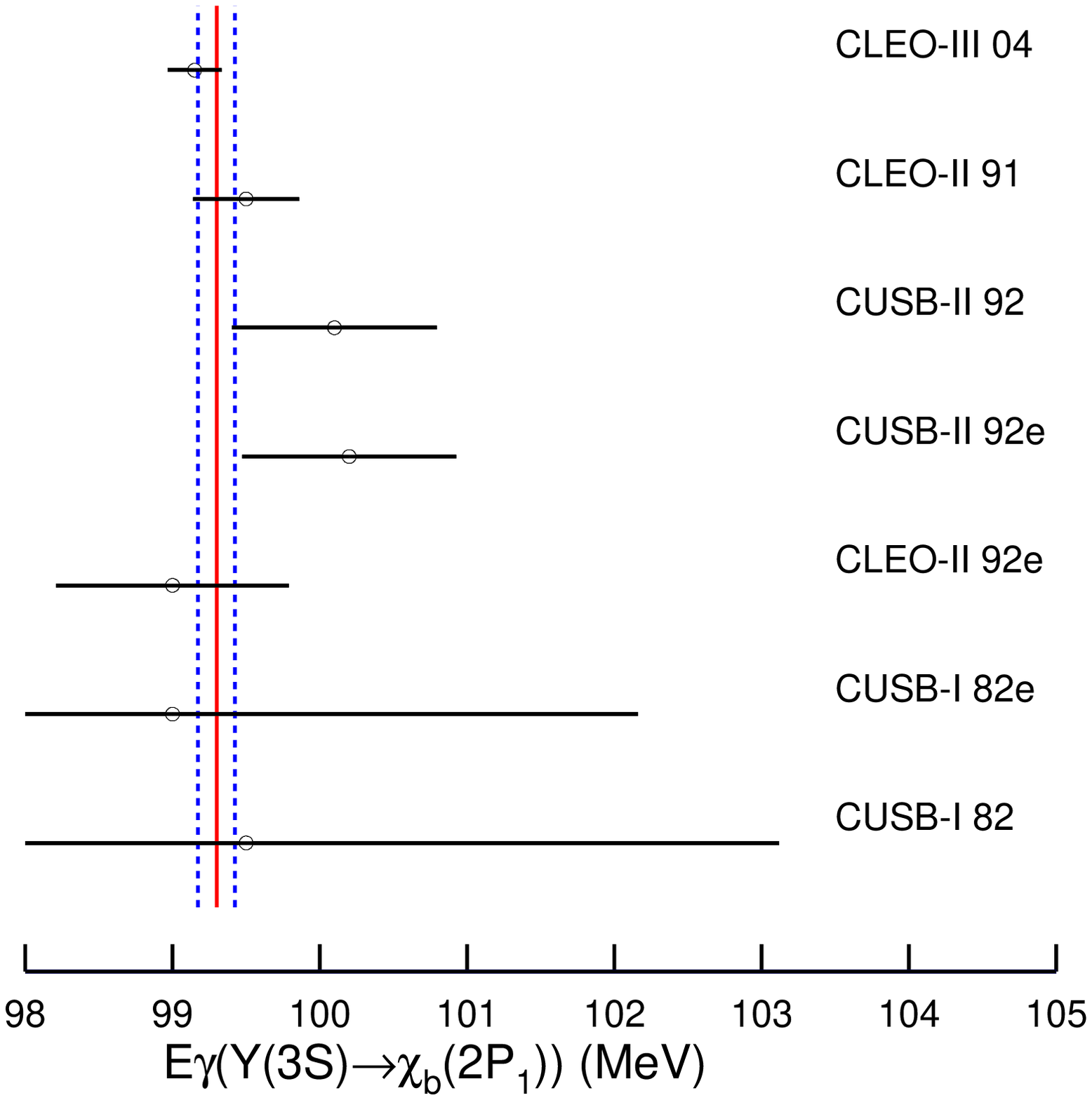}
\hskip-0.5truecm
\includegraphics[width=2cm,height=7.14cm]{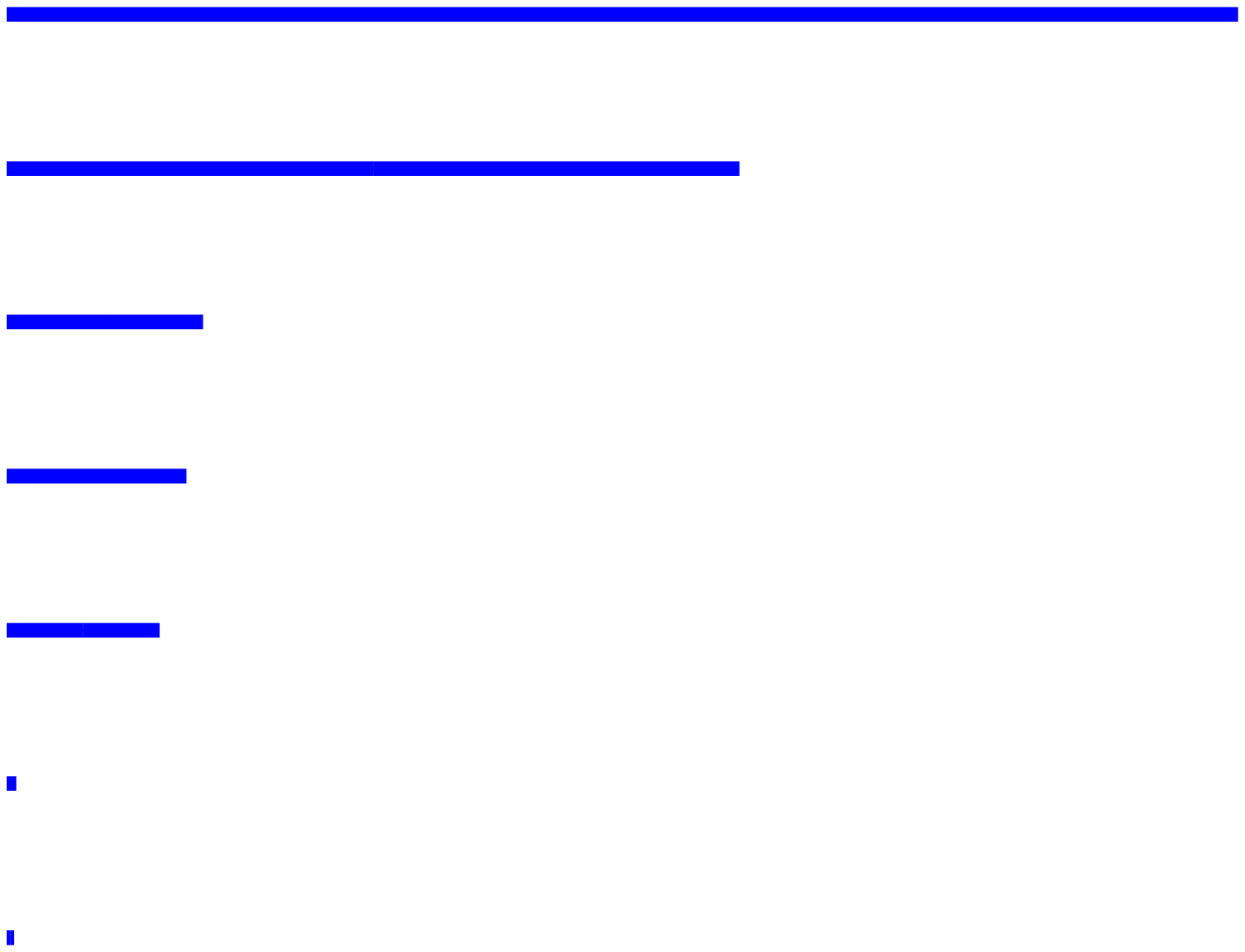}
}
\vskip0.1cm
Average: $(121.9\pm0.4)$~MeV \hfill
\phantom{Average: $(111.1\pm1.1)$~MeV} \\[-1.2truecm]
\hbox{
\includegraphics[width=7cm,height=7cm]{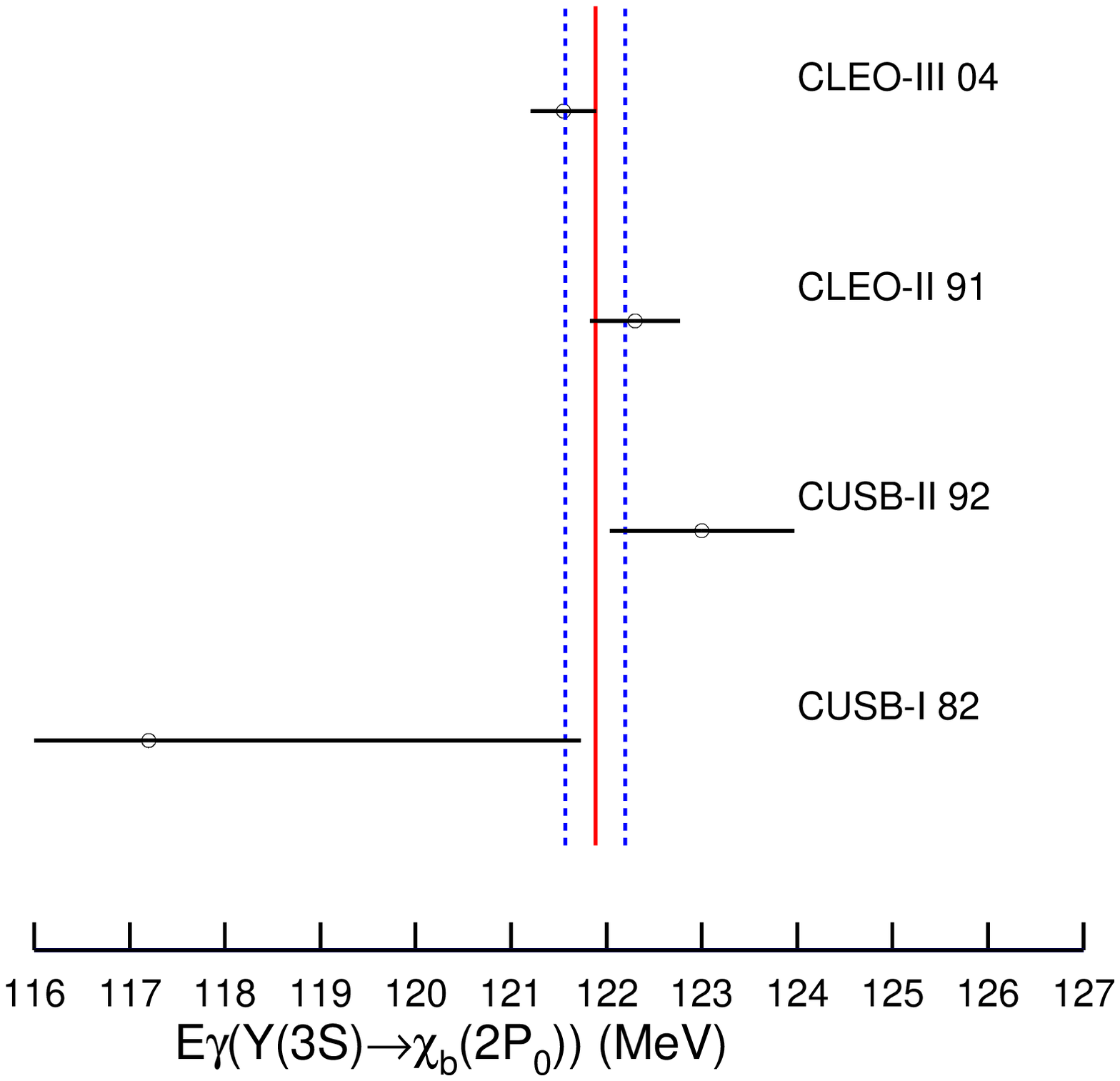}
\hskip-0.5truecm
\includegraphics[width=2cm,height=7.14cm]{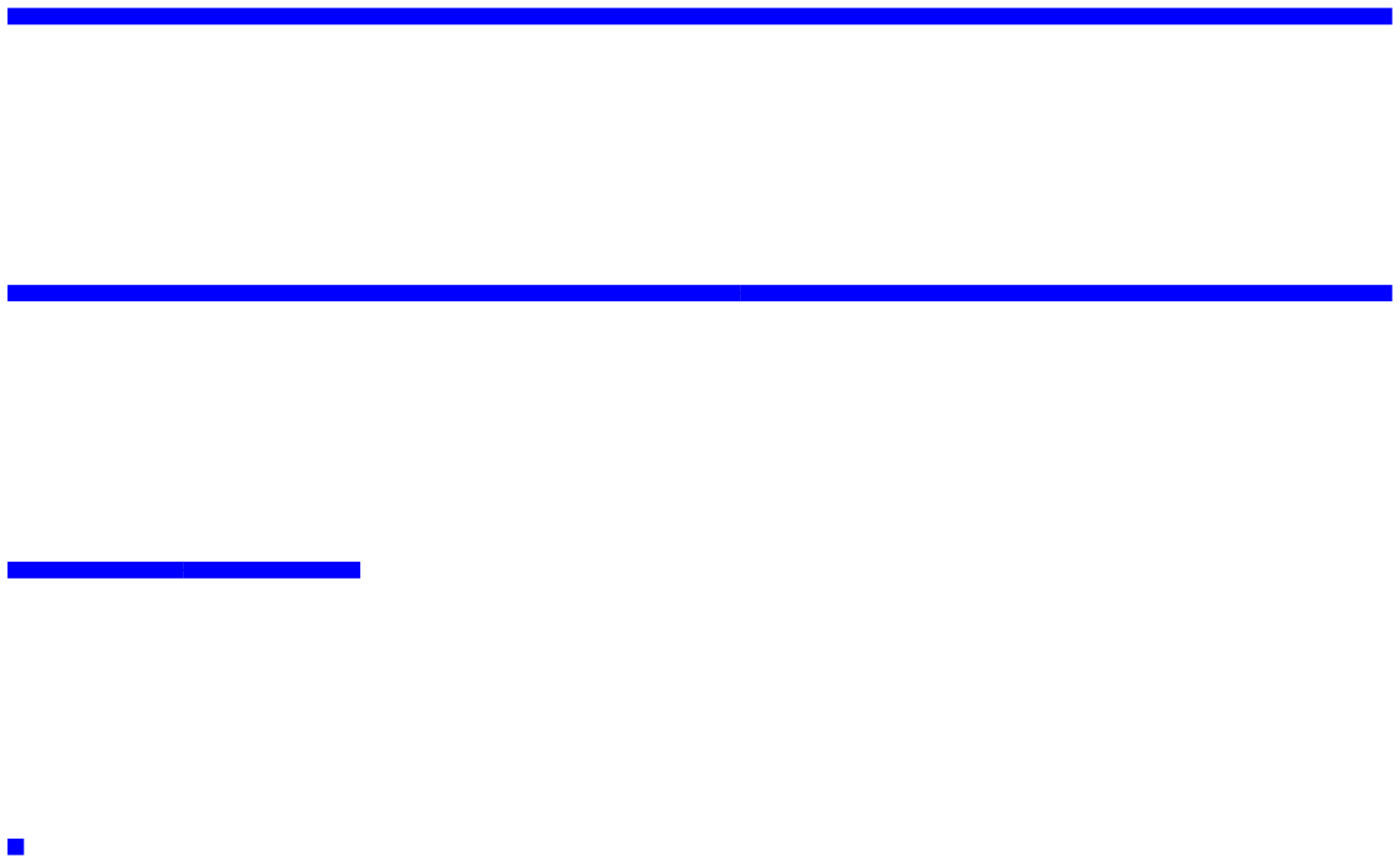}
}
\vskip-0.2cm
\caption{Measurements of the photon energies 
         in $\Y(3S)\to\gamma\chi_{bJ}(2P)$.
         The vertical bars indicate the 
         world average value (solid) and its error
         (dashed). These are also listed on top.
         The thick horizontal bars to the right of the name of
         the experiment
         give the relative weight of each experiment
         into the average value.
         Photon energy measurements from analyses of
         exclusive $\gamma\gamma l^+l^-$ events are
         indicated with an ``(e)'' after the date of the
         publication.}
\vskip-0.6cm
\label{fig:e3s2pj}
\end{figure}

CESR continued to improve its luminosity via the storage ring upgrades.
Its running time was exclusively devoted to $B$-meson physics
with data taken at the $\Y(4S)$ resonance.
The CLEO tracking and particle identification systems were
replaced, while the CsI(Tl) calorimeter was preserved.
After the $B$ physics program at CESR had ended, the CLEO~III
detector accumulated large samples at the narrow $\Y(nS)$
resonances. Number of collected $\Y(2S)$ and $\Y(3S)$ resonant
decays was increased by an order of magnitude.
Analysis of inclusive photon spectra has been recently
completed \cite{CLEOIIIchib}. 
Photon lines due to $\Y(2S)\to\gamma\chi_{bJ}(1P)$ and
$\Y(3S)\to\gamma\chi_{bJ}(2P)$ observed in inclusive
photon spectrum are shown in 
\Figure~\ref{fig:2sg1p} and  
\Figure~\ref{fig:3sg2p} respectively.
Determination of energies of these photon lines
is limited by the systematic error in calibration of the
calorimeter. The latter was improved in CLEO~III by
analysis of the $\psi(2S)$ photon spectrum obtained with the
same detector \cite{CLEOpsip}.
Since the photon energies in 
$\psi(2S)\to\gamma\chi_{cJ}(1P)$ transitions are precisely know
from the scans of the resonant cross-sections in $e^+e^-$
($\psi(2S)$) or $\bar pp$ ($\chi_{cJ}$) collisions, 
the $\psi(2S)$ photon lines were turned into the calibration
points. 
\shortpage

Comparisons of the photon energies for $\Y(2S)\to\gamma\chi_{bJ}(1P)$
and $\Y(3S)\to\gamma\chi_{bJ}(2P)$ determined in various experiments,
together with the world average values, are shown in
\Figure~\ref{fig:e2s1pj} and \Figure~\ref{fig:e3s2pj} respectively.
The masses of the $\chi_{bJ}(1P)$ ($\chi_{bJ}(2P)$) states can be
calculated from these photon energies and the masses of $\Y(2S)$
($\Y(3S)$).  The errors on the latter are significant, thus the errors
on the masses of the $\chi_{bJ}(nP)$ states are strongly correlated
between different values of $J$.  These need to be properly taken into
account when calculating the centre-of-gravity mass and fine-splitting
parameters.  The results are tabulated in \Table~\ref{tab:chibsum}.
\begin{table}[!h]
\caption[Masses and fine splittings for the $\chi_b(nP)$  states]
        {Masses and fine splittings for the $\chi_b(nP)$ states
         obtained from the world average values.  The values of $\rho$
         given in brackets come from the CLEO~III measurements
         \cite{CLEOIIIchib} and have smaller errors than the world
         average values since cancellations in the systematic errors
         of photon energies for different $J$ values are properly
         considered.}
\label{tab:chibsum}
\begin{center}
\begin{tabular}{lc}
\hline
 & $\bbbar(n=1)$ \\
\hline
$M(\chi_{b2})$ &  9912.2$\pm$0.4 (in MeV)  \\ 
$M(\chi_{b1})$ &  9892.8$\pm$0.4 (in MeV)  \\ 
$M(\chi_{b0})$ &  9859.5$\pm$0.5 (in MeV)  \\ 
$M_{COG}$      &  9899.9$\pm$0.4 (in MeV)  \\
$\Delta M_{21}=M(\chi_{b2})-M(\chi_{b1})$ (in MeV) 
& 19.4$\pm$0.4 \\
$\Delta M_{10}=M(\chi_{b1})-M(\chi_{b0})$ (in MeV) 
& 33.3$\pm$0.5 \\
$\rho(\chi)=\Delta M_{21}/\Delta M_{10}$ & 
0.584$\pm$0.016 (0.574$\pm$0.012) \\
$h_{T}$ (in MeV)               &  3.27$\pm$0.08  \\
$h_{LS}$ (in MeV)              & 13.64$\pm$0.14  \\
\hline
 & $\bbbar(n=2)$ \\
\hline
$M(\chi_{b2})$ &  10268.7$\pm$0.5 (in MeV)  \\ 
$M(\chi_{b1})$ &  10255.4$\pm$0.5 (in MeV)  \\ 
$M(\chi_{b0})$ &  10232.6$\pm$0.6 (in MeV)  \\ 
$M_{COG}$      &  10260.3$\pm$0.5 (in MeV)  \\
$\Delta M_{21}=M(\chi_{b2})-M(\chi_{b1})$ (in MeV) 
& 13.3$\pm$0.3 \\
$\Delta M_{10}=M(\chi_{b1})-M(\chi_{b0})$ (in MeV) 
& 22.8$\pm$0.4 \\
$\rho(\chi)=\Delta M_{21}/\Delta M_{10}$ & 
0.583$\pm$0.020 (0.584$\pm$0.014) \\
$h_{T}$ (in MeV)               & 2.25$\pm$0.07  \\
$h_{LS}$ (in MeV)              & 9.35$\pm$0.12  \\
\hline
\end{tabular}
\end{center}
\end{table}

\newpage
\subsection[Bottomonium D states]{Bottomonium D states $\!$\footnote{Author:T. ~Skwarnicki}
}
\label{sec:spexfineDbot}
The lowest radial excitations of the D states in charmonium have
masses above the the $D\bar D$ meson threshold.  The lightest member
of the spin-triplet is a vector state. It is identified with the
$\psi(3770)$ state, which is a third $c\bar c$ resonance observed in
the $e^+e^-$ cross-section.  Unlike the $J/\psi(1S)$ and the
$\psi(2S)$ resonances, the $\psi(3770)$ is broad because it decays to
$D\bar D$ meson pairs.  Since, the coupling of the $D$ state to
$e^+e^-$ is expected to be small, its large $e^+e^-$ cross-section is
attributed to a significant mixing between the $2S$ and $1D$
$J^{PC}=1^{--}$ states.  Whether the narrow $X(3872)$ state is one of
the other members of the $1D$ family is a subject of intense
disputes. The $J\!=\!2$ states (the spin triplet and the spin singlet)
are narrow below the $D\bar D^*$ threshold, since they can't decay to
$D\bar D$. The $J\!=\!3$ state can decay to $D\bar D$ but, perhaps,
its width is sufficiently suppressed by the angular momentum barrier
\cite{Barnes:2003vb}.  In all scenarios, masses of all $1D$ states
must be strongly affected by the proximity of open-flavour thresholds
via coupled channel effects.

\begin{figure}[t]
\begin{center}
\includegraphics[width=.7\linewidth]{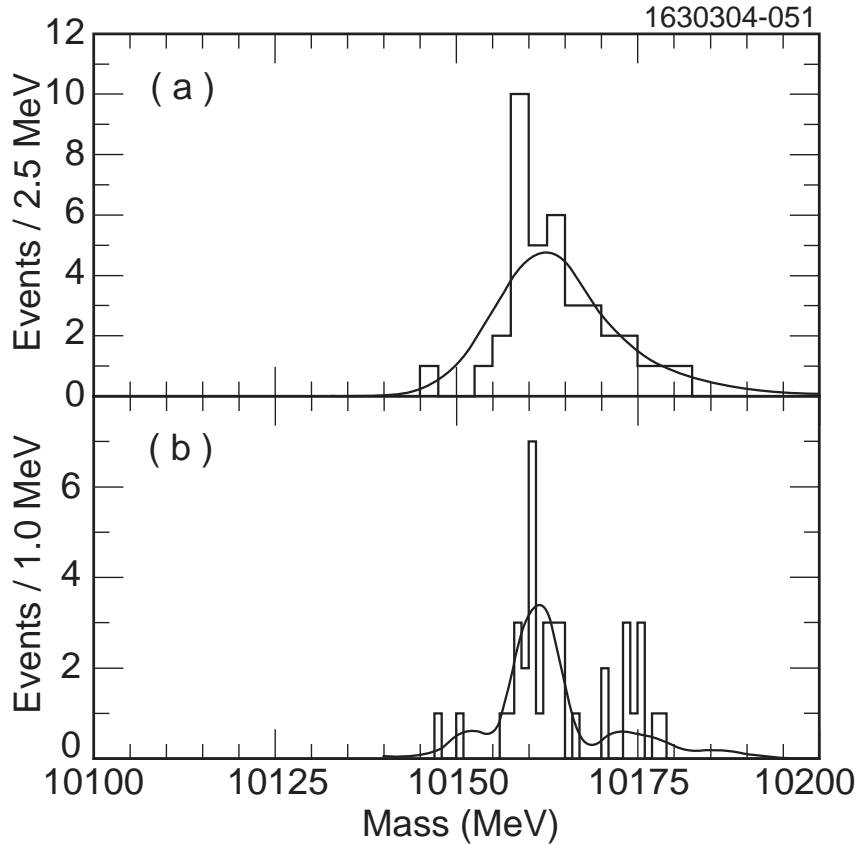}
\caption{Distributions of the measured $\Y(1D)$ mass in the CLEO~III 
data \cite{CLEOIII1D} using 
(a) the recoil mass against the two lowest energy photons,
(b) the fit implementing the $\chi_{bJ'}(2P)$, $\chi_{bJ}(1P)$, $\Y(1S)$ mass
constraints.
The results of fits for a single $\Y(1D)$ state are superimposed.
The mass-constraint method produces satellite peaks because of ambiguities in 
$J'$ and $J$ values.}
\label{fig:1dmass}
\end{center}
\end{figure}

In contrast, the $1D$ states of bottomonium are well below the
open-flavour threshold, thus their masses are easier to predict
theoretically.  Unfortunately, the mixing of the $2S$ and $1D$
$J^{PC}=1^{--}$ states is expected to be small for bottomonium.  Not
surprisingly, the $J\!=\!1$ $1D$ $b\bar b$ state has not been observed
in $e^+e^-$ collisions.  The spin-triplet states are accessible from
the $\Y(3S)$ resonance by two subsequent E1 photon transitions via
intermediate $\chi_{bJ}(2P)$ states.  Energies of photons in the
$\chi_{bJ}(2P)\to\gamma \Y(1D)$ transitions fall in the same range as
the dominant $\Y(3S)\to\gamma \chi_{bJ}(2P)$ photon lines.  Therefore,
they cannot be resolved in the inclusive photon spectrum. Two-photon
coincidence is of not much help, since the photon background from
$\pi^0$ decays is very large in $\Y(3S)$ decays.  Nevertheless, the
$\Y(1D)$ states have been discovered by CLEO~III in the $\Y(3S)$
decays \cite{CLEOIII1D}.  The photon backgrounds are removed by using
the ``exclusive'' approach (see the previous section), in which the
three additional decays are required, $\Y(1D)\to\gamma \chi_{bJ}(1P)$,
$\chi_{bJ}(1P)\to\gamma\Y(1S)$, $\Y(1S)\to l^+l^-$.  Since the product
branching ratio for these five subsequent decays is rather small
\cite{Godfrey:2001vc,kwros}, the large CLEO~III sample of the $\Y(3S)$
resonances was essential for this measurement. After suppression of
the $\Y(3S)\to\pi^0\pi^0\Y(1S)$ and 4-photon cascades via the
$\chi_{bJ}(2P)$, $\Y(2S)$, $\chi_{bJ}(1P)$ states 38 $1D$ candidates
are observed in the CLEO~III data.  The mass of the $1D$ state is
estimated by two different techniques, as shown in
\Figure~\ref{fig:1dmass}.  In both cases, the mass distribution
appears to be dominated by production of just one state.  The
theoretical and experimental clues point to the $J\!=\!2$ assignment.
The mass of the $\Y_2(1D)$ state is measured by CLEO~III to be:
$(10161.1\pm0.6\pm1.6)$~MeV.

Masses of the other bottomonium $1D$ states remain unknown. However,
the fine structure of the $1D$ spin-triplet is predicted to be small.
All potential model calculations predict the $\Upsilon_2(1D)$ mass to
be between $0.5$ and $1.0$~MeV lower than the centre-of-gravity
(c.o.g.)  mass for this triplet \cite{Godfrey:2001vc}.  Adding this
theoretical input, CLEO obtains $(10162\pm2)$~MeV for the c.o.g.\
mass, where they assigned an additional uncertainty of 1~MeV to the
correction for the $1^3D_2-$c.o.g.\ mass difference.

The CLEO~III also looked for $\Y(1D)\to\pi^+\pi^-\Y(1S)$ and
$\Y(1D)\to\eta\Y(1S)$ transitions. No evidence for such decays
was found and upper limited were set \cite{CLEOIII1D}.
The upper limit on $\Y(1D)$ $\to$ $\pi^+$ $\pi^-$ $\Y(1S)$ 
rules out rather large width for this transition predicted
by the Kuang--Yan model \cite{kuang81,1dpipi}.
 
\clearpage
\section{Hyperfine splittings}

\subsection[$\eta_c$(1,2S): comparison of all measurements]
           {$\eta_c(1,2S)$:  comparison of all measurements
            $\!$\footnote{Authors: R.~Galik, R.~Mussa, S.~Ricciardi}}
\label{sec:spexetac12}

Despite the large variety of available data on the $\eta_c(1S)$, 
the precise  determination of its mass and width is still an 
open problem. It is likely that unexpected systematic errors 
be present in some of these measurements. It is worth to compare 
the subsets of measurements of masses and widths of 
the $\eta_c$ done with the same reaction, before comparing the 
large variety of techniques which allowed to measure this state, each one 
with its own dominant systematic error.
The two states share most of the decay channels, therefore 
 the same analysis is usually applied to extract their signal.

\subsubsection[$\eta_c$(1S) in \jpsi{} and \psip{} decays]
{ $\eta_c$(1S) in \jpsi\ and \psip{} decays}

\begin{figure}
\begin{minipage}{.48\linewidth}
   \includegraphics[width=\linewidth]{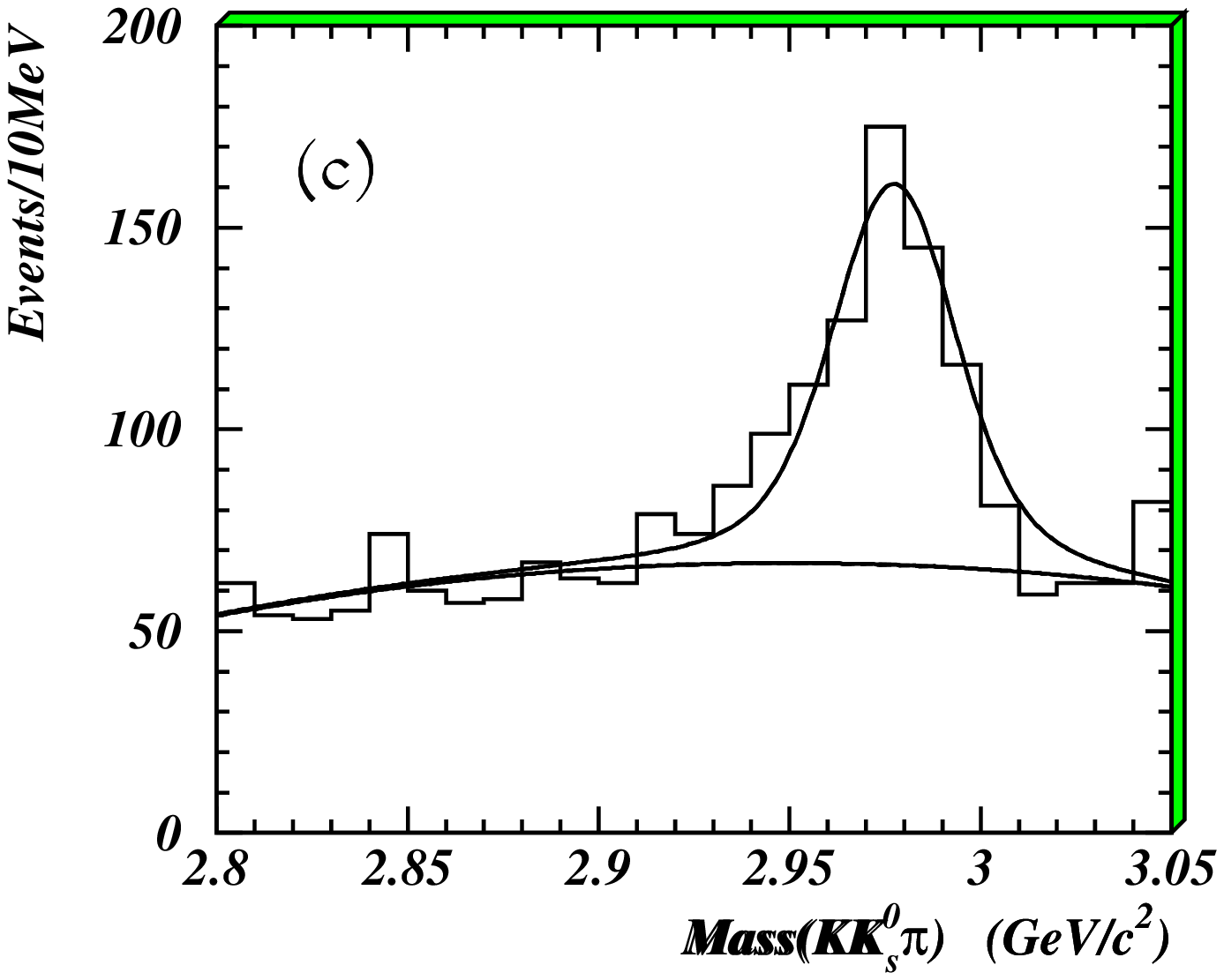}
\end{minipage}
\hfill
\begin{minipage}{.48\linewidth}
   \includegraphics[width=\linewidth]{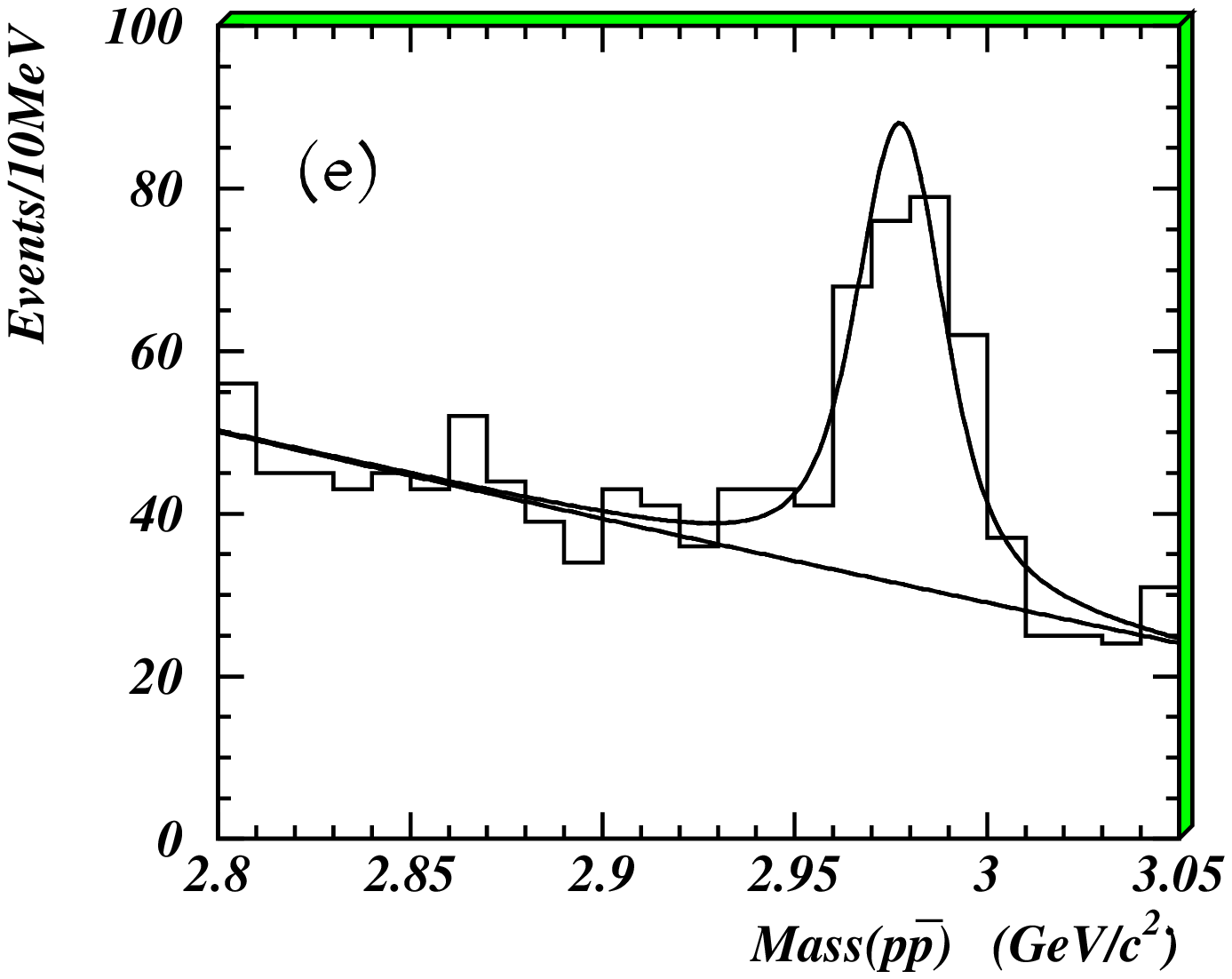}
\end{minipage}
\caption{Invariant mass distributions for $K_{S}^{0}K^{\pm}\pi^{\mp}$(left)
 and $\ppbar$(right)  events from BES}
\label{fig:etac-bes}
\end{figure}

\shortpage
The \etac parameters have been extracted from the radiative
transitions of \Jpsi\ and \psip\ by a large number of experiments:
while Crystal Ball (and more recently CLEO-c) studied the inclusive
photon spectrum, Mark~II and III, DM2, BES studied the invariant mass
distributions of decay products in reactions with 2 or 4 charged
tracks and 0 to 2 neutral pions.  The samples taken in the 80's and
early 90's were recently overwhelmed by the 58~M BES sample.
\Table~\ref{tab:etacmasses} summarizes the mass and width measurements
done in the past 20 years.  The $\etac$ peak is observed in the
invariant mass of the following decay modes: $\KS K^\pm\pi^\mp$,
$\pi^+\pi^-\pi^+\pi^-$, $\pi^+\pi^-K^+K^-$, $K^+K^-K^+K^-$, $\ppbar$.
\Figure[b]~\ref{fig:etac-bes} shows two of these distributions.

\begin{table}[!h]
\caption[The world largest samples of $\jpsi$ and $\psip$]
        {The world largest samples of $\jpsi$ and $\psip$
         used for the determination of the $\etac$ mass and width.}
\label{tab:etacmasses}
\renewcommand{\arraystretch}{1.3} 
\begin{center}
\begin{tabular}{lcccc}
\hline
Expt. &        MarkIII    &     DM2    &   BES~I   &    BES~II\\
\hline 
 year        &  1986      &    1991    &   2000   &   2003\\
Mass(\mevcc) &  2980.2$\pm$1.6 &   2974.4$\pm$1.9  
            &  2976.3$\pm$2.3$\pm$1.2 &  2977.5$\pm$1.0$\pm$1.2 \\
Width(MeV)  &  10.1$^{+33.0}_{-8.2}$ & -- & 11.0$\pm$8.1$\pm$4.1 & 17.0$\pm$3.7$\pm$7.4 \\ 
Sample & 2.7M\jpsi & 8.6M\jpsi & 3.8M\psip+7.8M\jpsi &   58M\jpsi \\
\hline
\end{tabular}
\end{center}
\renewcommand{\arraystretch}{1} 
\end{table}

A cut on the kinematic fit to the exclusive  hypothesis (referred as 
\jpsi\  veto) is applied, to reject direct \jpsi\  decays to the 
same channels, or feed-down from other decay channels, such as
$(\omega,\phi)\pi\pi$, $\omega K^+K^-$, $\gamma K^0_SK^0_S$ .
The systematic errors on mass determination come mostly from the mass scale 
calibration (0.8 \mevcc, calculated by comparing $K^0_S$, $\phi$ and even 
$\chi_c$ masses with PDG values) and from the $\jpsi$ veto.
The $\jpsi$ veto is also the dominant source of systematics on the 
total width determination: 5.6 out of 7.4\mevcc. 

\subsubsection[$\eta_c$(1S) in $\ppbar$ annihilations]
              {$\eta_c$(1S) in $\ppbar$ annihilations}
\shortpage

The $\eta_c$ was investigated in $p\bar{p}$ annihilation only in the
$\gamma\gamma$ channel, which is affected by a substantial feeddown
from the continuum reactions $\pi^0\pi^0$ and $\pi^0\gamma$: both
reactions are sharply forward-backward peaked.  The number of 'signal'
events is 12 in R704, 45 in E760 and 190 in E835, which respectively
took 0.7,3.6,17.7 pb$^{-1}$ of data in the $\eta_c$ mass region.  It
is worth to stress the fact that an increasing amount of integrated
luminosity was taken away from the peak , in order to better
understand the size and nature of the non resonant background.  The
experiment E835 can discriminate a $\pi^0$ from a single photon with
96.8\% efficiency: this reduces the feed-down to
0.1\%$\sigma_{\pi^0\pi^0}$+ 3.2\% $\sigma_{\pi^0\gamma}$ at
$\sqrt{s}=2984 \; {\rm MeV}/c^2$.

\begin{figure}
\begin{center}
\begin{minipage}{.48\linewidth}
   \includegraphics[width=\linewidth]{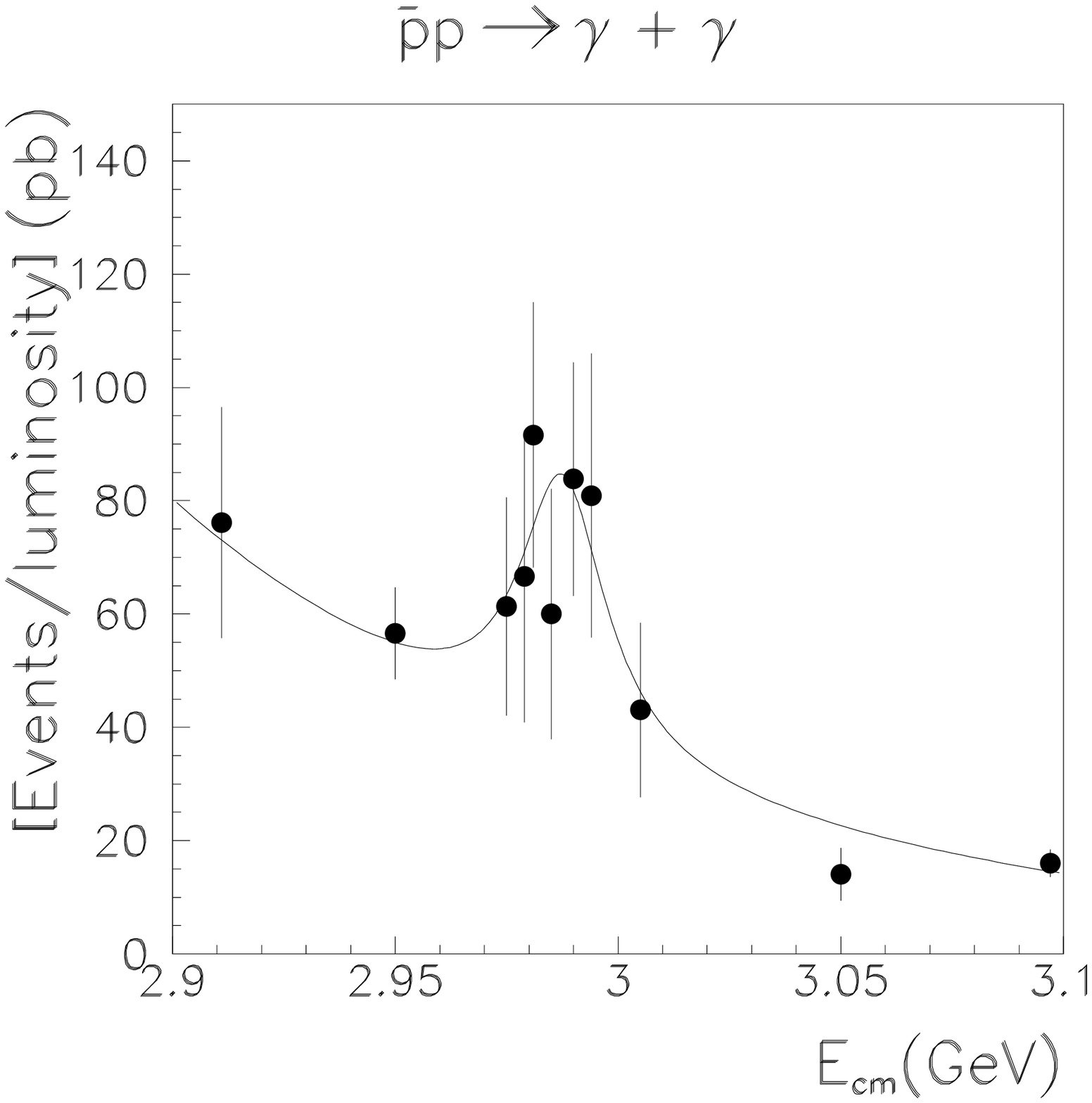}
\end{minipage}
\hfill
\begin{minipage}{.48\linewidth}
   \includegraphics[width=\linewidth]{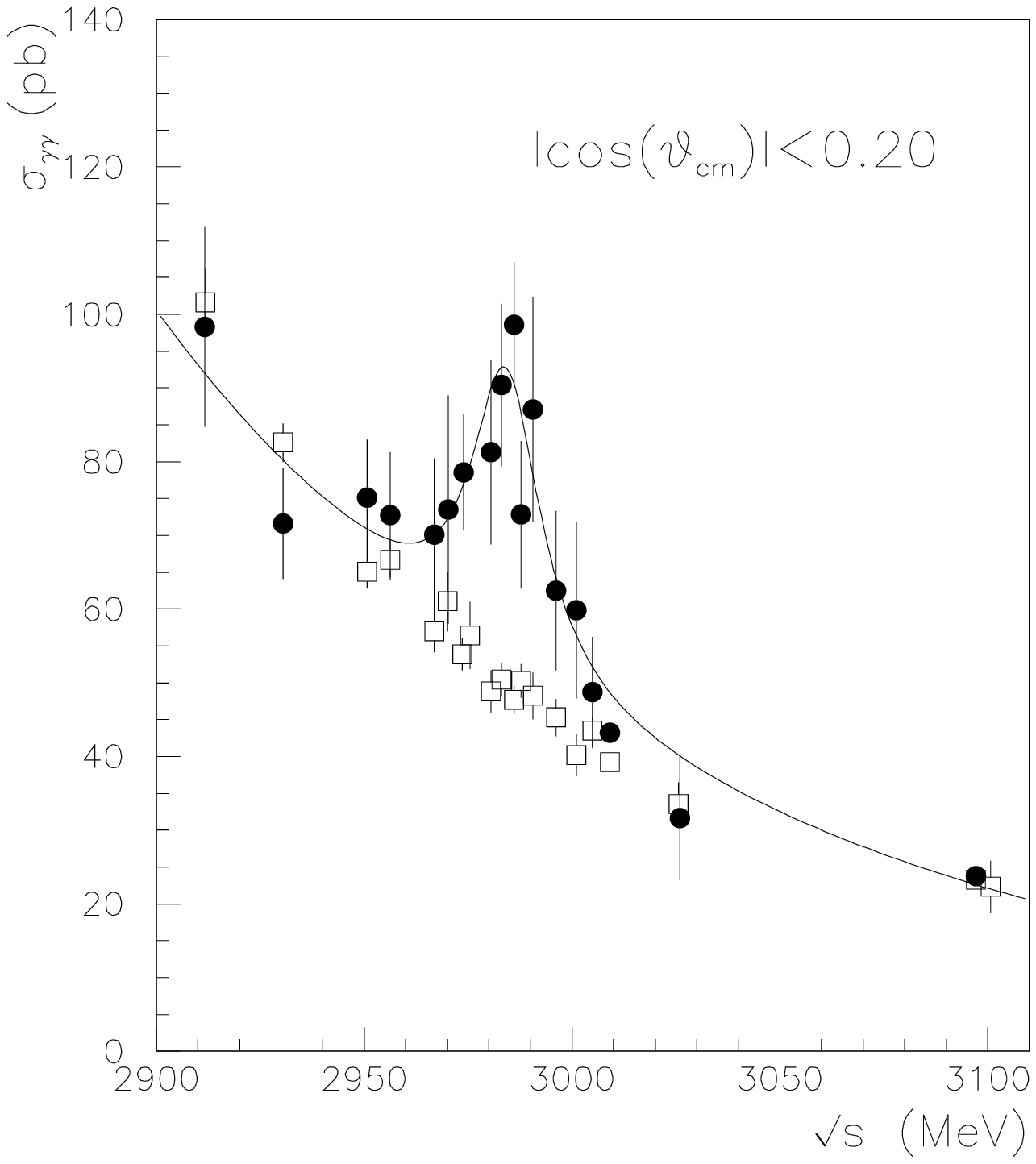}
\end{minipage}
\end{center}
\caption[Cross-section observed by E760 and E835 for the reaction 
         $\ppbar\to\gamma\gamma$]
        {Cross-section (black dots) observed by E760 (left) and
         E835(right) for the reaction $\ppbar\to\gamma\gamma$ in the
         region with cos$\theta_{CM}<$0.25(E760), 0.2(E835). The blank
         squares show the expected feed-down from
         $\pi^0\pi^0,\pi^0\gamma$.}
\label{fig:etac-e835}
\end{figure}
The very small sample taken by R704 in the resonant region ends up with 
a remarkably small result on the $\eta_c$ width: all this is based on the 
{\it ansatz} to have a small background. Such hypothesis was strongly 
disconfirmed by E760, therefore the R704 result  is affected by a very 
large hidden systematic error. The statement is even stronger, if we take
into account that the R704 fiducial region was extended up to 
$cos(\theta_{cm,\pi^0})=0.35$, where the feeddown dominates, 
and the detector did not have full azimuthal coverage  
(thus introducing an even larger feeddown).

E835 precisely measured the $\pi^0\gamma$ and $\pi^0\pi^0$
cross-section: the feeddown from these reactions can account for most
of the background. E835 could not exclude the existence of a residual
tiny $\gamma\gamma$ continuum, which can in principle interfere with
the resonant reaction, but is not large enough to shift the mass peak
beyond the statistical error.  \Figure[b]~\ref{fig:etac-e835}, on the
right, shows both signal and feed-down cross-section observed in
E835. A power law dependence on energy was assumed for the background,
in the fits. The choice of background parametrization and of the
fiducial region for the signal are the dominant sources of systematic
error, which amounts to 1 $\mevcc$ on the mass and 2~MeV on the width.
A comparative summary of $\ppbar$ measurements on $\eta_c(1S)$
parameters can be found in \Table~\ref{tab:etac-ppb}.

\begin{table}[!h]
\caption{Comparison of E760 and E835  results.}
\label{tab:etac-ppb}
\begin{center}
\begin{tabular}{lrr} \hline 
Expt. & E760 & E835    
 \\ \hline 
{\cal L}dt (pb$^{-1}$) & 3.6 & 17.7 \\
$m(\etac)$(\mevcc)     &   2988.3 \PM 3.3&   2984.1 \PM 2.1 \PM 1.0 
\\ $\Gamma(\etac)$ (\mevcc)  & 23.9$^{+12.6}_{-7.1}$  
& 20.4$^{+7.7}_{-6.7}\pm2.0$
\\
\hline \hline
\end{tabular}
\end{center}
\end{table}

E760 and E835 also searched for the $\eta_c(2S)$ state in the energy
range $3575\mevcc < \sqrt{s} < 3660\mevcc$, putting a 90\% CL upper
limit at $\simeq$ 0.4 eV on ${\cal B}(\etac(2S)\to\ppbar) \times
\Gamma(\etac(2S)\to\gamma\gamma)$.

\subsubsection[$\eta_c$(1S,2S) in B decays]{$\eta_c$(1,2S) in B decays}
\shortpage

\begin{center}
\begin{figure}
\begin{minipage}{7.5cm}
   \includegraphics[width=7.5cm]{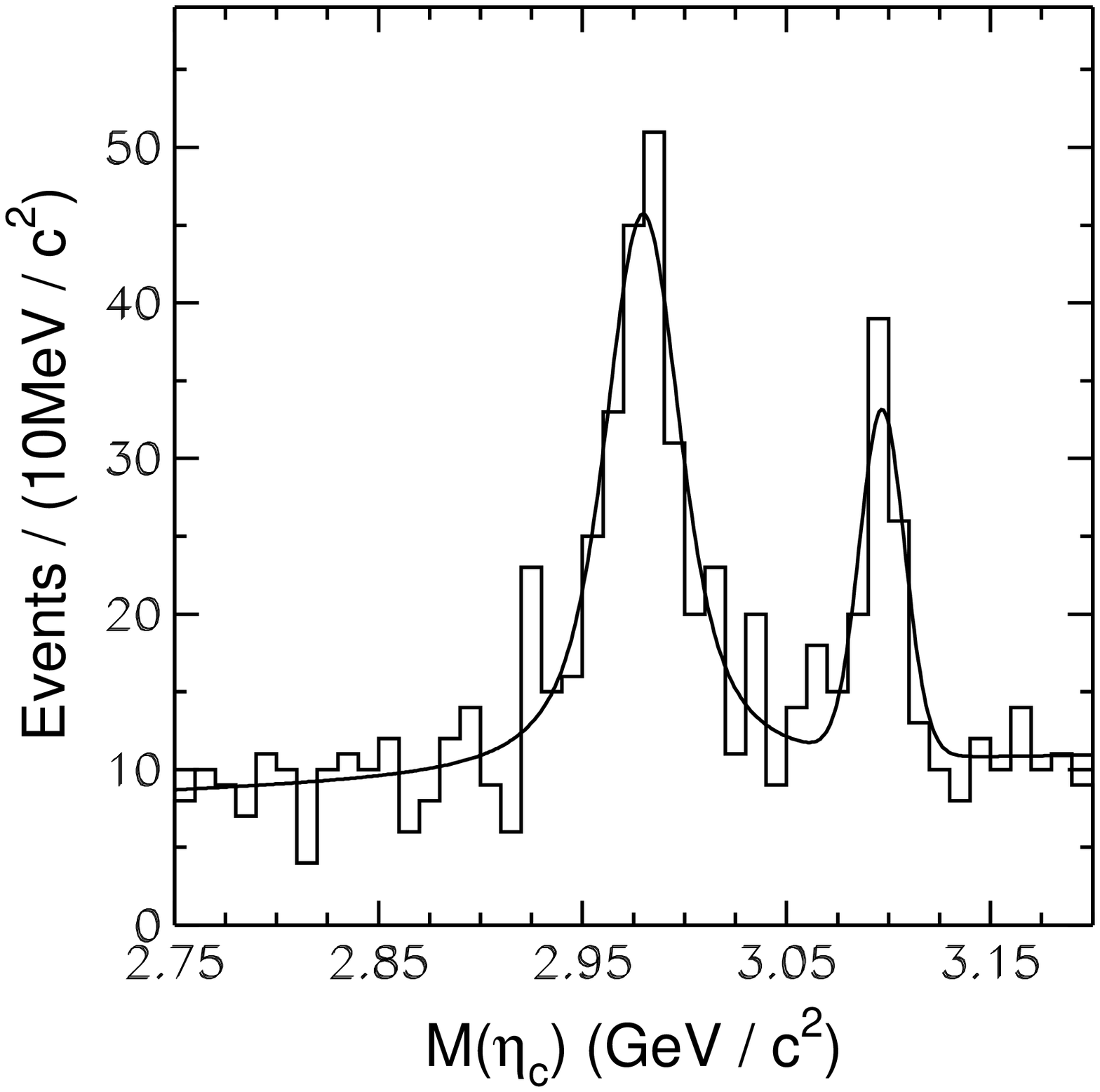}
\end{minipage}
\hfill
\begin{minipage}{7.5cm}
\includegraphics[width=7.5cm,scale=0.55]{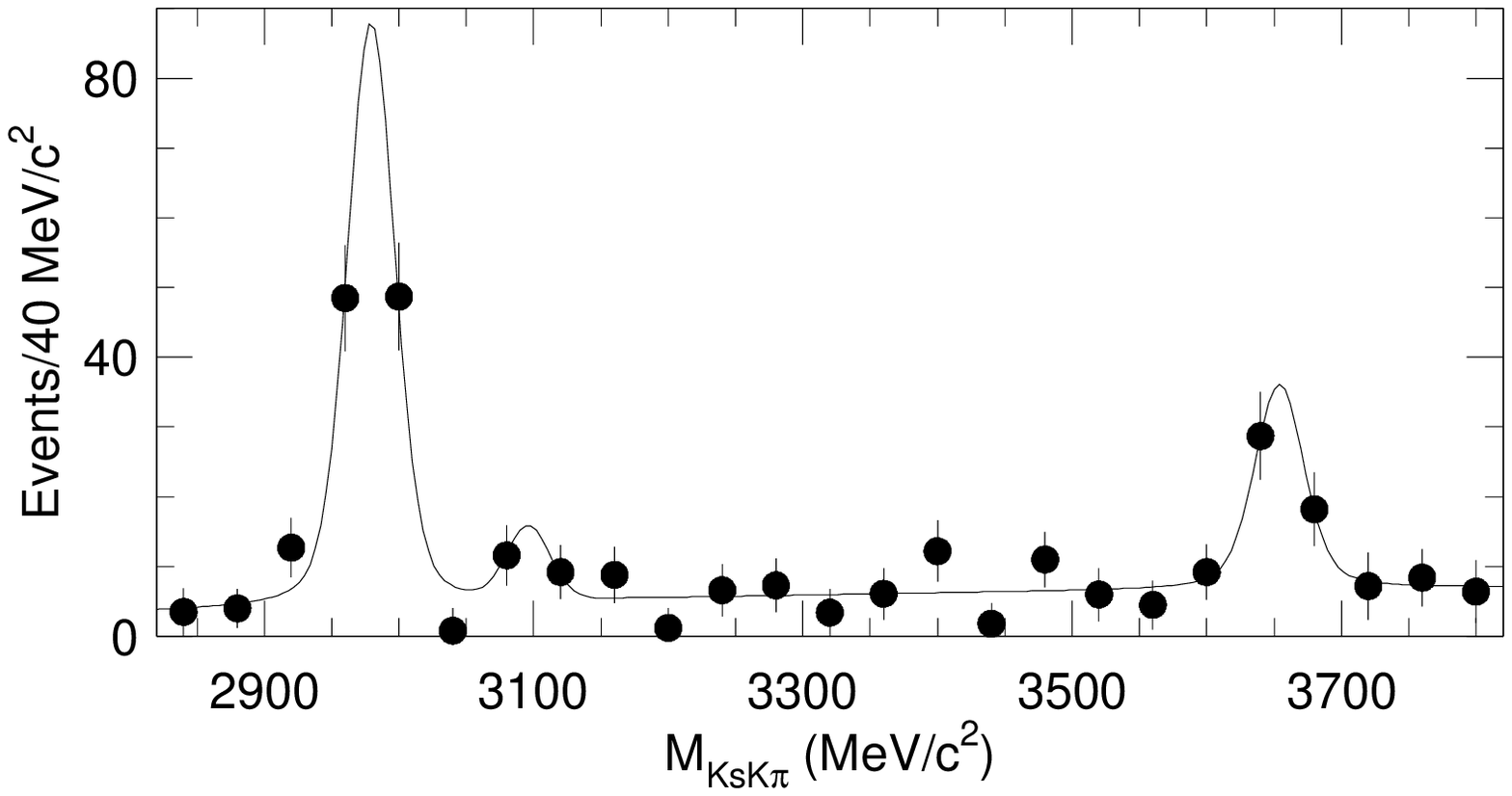}   
\includegraphics[width=7.5cm,scale=0.55]{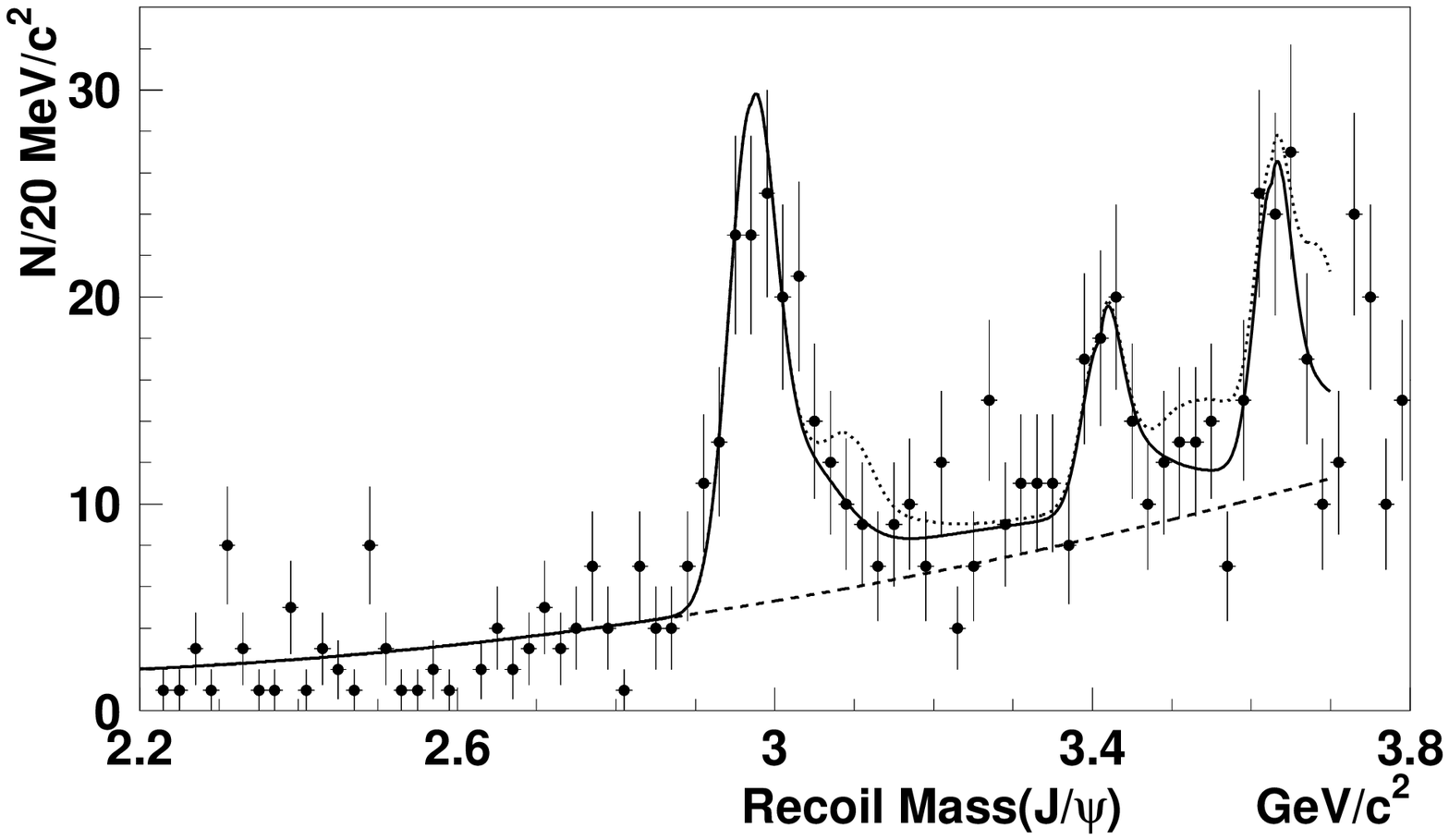}
\end{minipage}
\caption[Distribution of reconstructed B decays]
        {On the left: distribution of reconstructed B decays to
         $\eta_c(1S)$ and $\jpsi$, in the common final state
         $K^0_SK^\mp\pi^\pm$, from refs.  \cite{Fang:2002gi}. On the
         right: Belle observed the $\eta_c(2S)$ both in B decays (top,
         from ref.\cite{Choi:2002na}) and in double $\ccbar$ (bottom,
         from ref.\cite{Abe:2003ja}).}
\label{fig:etac-belle}
\end{figure}
\end{center}

In the last years, the B-factories have exploited the B meson decays
to charmonium as a new powerful tool for the measurement of the
$\eta_c$ mass \cite{Fang:2002gi}, as well as for the discovery of
$\eta_c(2S)$ and the measurement of its mass.  Exclusive decays of
both B$^0$ and B$^+$ mesons were detected with the $\eta_c$
reconstructed in the $K^0_SK^\mp\pi^\pm$, $K^+K^-\pi^0$,
$K^{*0}K^\mp\pi^\pm$, $\bar{p}p$ decay channels.  Exploiting common
decay modes, it was possible to measure the mass difference between
$J/\psi$ and $\eta_c$, \Figure~\ref{fig:etac-belle} (left) shows the
invariant mass distribution of decay products from $B\rightarrow K+X$
in the 2.75--3.2~GeV/c$^2$ region: $J/\psi$ and $\eta_c$ peaks are
clearly visible.  Fitting the distribution with a Breit--Wigner
convoluted with a MonteCarlo generated resolution function, it was
possible to extract a value of 2979.6$\pm$2.3$\pm$1.6~MeV/c$^2$ for
the mass , and a total width of 29$\pm$8$\pm$6~MeV (from a sample of
182$\pm$25 events, out of 31.3~M $B\bar{B}$ pairs).  The systematic
errors include the effect of varying the bin size as well as the shape
of background, and the difference between data and MC generated
detector resolutions.

The  $K^0_SK^\mp\pi^\pm$ final state is an ideal place to look for the 
 $\eta_c(2S)$, a state which was awaiting confirmation since its first 
and only observation by Crystal Ball in the inclusive photon spectrum from 
$\psip$ decays. In 2002, the Belle collaboration reported the evidence 
of $\etacp$ production via the exclusive processes $B^+\to K^+ \etacp$ and
 $B^0\to K^0_S \etacp$. Given the suppression of the 
$\psip\to K_S K^\pm\pi^\mp$ decay, contamination from the process 
$B\to K\psip$ is estimated to be negligible. The first evidence 
\cite{Choi:2002na}  of the 
$\etacp$ came from a sample of 44.8M$B\bar{B}$ pairs, using the exclusive 
channel $B\to K(K^0_SK^-\pi^+)$. A likelihood function based on the angle
between the B candidate and the $e^+e^-$ axis, and on the transverse momenta
of the other tracks with the respect to the B candidate thrust axis, was used 
to suppress any background from continuum processes. Given a good B candidate,
the feeddown from $B\to D(D_s)+X$ was reduced by cutting at 
$|M_{K\pi}-M_D|>10 \mevcc$  and $|M_{K_SK^+}-M_{D_s}|>10 \mevcc$; the
 feeddown from $B\to K^*+X$ was reduced by cutting at 
$|M_{K\pi}-M_{K^*}|>50 \mevcc$, as the $\etac(nS)\to KK^*$ component
 is expected to be suppressed by the angular momentum barrier.
The mass for the $\etacp$ was measured to be 3654\PM6\PM8 \mevcc,
with systematic error coming mostly from the choice of binning.
A 90\%C.L. upper limit on the width at 55~MeV was given.

\subsubsection[$\eta_c$(1S,2S) in $\gaga$ fusion]
              {$\eta_c$(1S) in $\gaga$ fusion}

The $e^{+}e^{-}$ collider detectors collecting data in the
$\Upsilon(4S)$ region (CLEO, BaBar, BELLE) have good ``reach'' to
produce $C=+1$ charmonium states through two-photon fusion.  These are
states such as the $\eta_c$ and $\chi_c$ which are not produced
directly in the $e^{+}e^{-}$ annihilation process.  Such
$\gamma\gamma$ interactions strongly peak at low $q^{2}$ so that the
scattered lepton are not detected (``untagged'' events) and the
photons are approximately real.  For instance, in CLEO the active
detector elements go to within $22^{\circ}$ of the beam axis, or
$|$cos$\theta$$|$$<$$0.93$; this means that untagged events all have
photons with $Q^{2}$ less than roughly 1~GeV$^{2}$, and usually {\it
much} less.\footnote{The one published {\it tagged} CLEO analysis
started at $Q^{2} = 1.5 $GeV$^{2}$.}  Both CLEO and \babar\ have thus
recently studied the reactions:
\[
\gamma\gamma\to (\eta_c/\eta_c^{\prime})\to K_{S}^{0} K^{\pm} \pi^{\mp}~.
\]
The \etac\ is known to be coupled to two photons (${\cal B}(\etac \to
\gamma \gamma) \sim 5 \cdot 10^{-4}$).  An estimate of the two-photon
production rate of \etacp suggests that also the radial excitation
could be identified in the current $e^+e^-$
$B$-factory~\cite{Barnes:1996my}.  The regions of the detector
acceptance occupied by such $\gamma\gamma$ fusion reactions and the
competing initial state radiation (ISR for short, also called
``radiative return'') processes are quite dissimilar for a symmetric
collider experiment such as CLEO and the asymmetric $B$-factories.
Given this and the differing sources of systematic uncertainties, the
\babar\ and CLEO results are rather independent.

The CLEO analysis used $\approx 14$ fb$^{-1}$ and $\approx 13$
fb$^{-1}$ of data taken with the CLEO II and CLEO III detectors,
respectively, mostly near the $\Upsilon$(4S) resonance.  The particle
identification systems and tracking chambers in these two
configurations are quite different, so these can be considered truly
independent experiments.  The preliminary results were first shown at
the April 2003 APS meeting and submitted \cite{ernst:2003bk} to the
EPS meeting of that summer; final results have recently been submitted
for publication ~\cite{Asner:2003wv}.  The \babar\ collaboration has
both preliminary~\cite{Wagner:2003qb} and final
results~\cite{Aubert:2003pt}, based on a sample of data corresponding
to an integrated luminosity of about $90 \invfb$.  In the CLEO
analysis, these events are characterized by lots of missing energy and
momentum, but very little transverse momentum ($p_{T}$) of the
hadronic system and very little excess energy in the detectors. The
selection criteria included that $p_{\mathrm T} < 0.6 \gevc$, that
there were no additional charged tracks, and that the unassociated
energy in the electromagnetic calorimeter was less than 200~MeV
(300~MeV) for CLEO II (CLEO III).  The CLEO mass spectra are shown in
\Figure~\ref{fig:CLEO-Babar}(a,b) , clearly indicating evidence for
both the $\eta_c$ and $\eta_c^{\prime}$. Fits to these spectra
(polynomial backgrounds, Breit--Wigner line shapes, double-Gaussian
detector resolution functions) yielded the results shown in
Table~\ref{tab:CLEOetacpr}.

In the \babar\ analysis, events are selected by requiring four charged
particles with total transverse momentum $p_{\mathrm T} < 0.5 \gevc$
and total energy in the laboratory frame $E_{\mathrm tot} < 9 \gev$,
in order to suppress $e^+e^- \to q \bar{q}$ events. One track is
required to be identified as a kaon and pairs of oppositely charged
tracks are used to reconstruct $\KS\to\pip\pim$ decays.  The $\KS \Kp
\pim$ vertex is fitted, with the \KS mass constrained to the world
average value.

\Figure[b]~\ref{fig:CLEO-Babar}~(c) shows the resulting $\KS \Kp \pim$
invariant mass spectrum.  The presence of a peak at the \jpsi\ mass is
due to ISR events, where a photon is emitted in the initial state, and
a backward-going \jpsi\ is produced, its decay products falling into
the detector acceptance because of the Lorentz boost of the centre of
mass.  A fit to this distribution with a sum of a smooth background
shape, a Gaussian function for the \jpsi\ peak and the convolution of
a non-relativistic Breit--Wigner shape with a Gaussian resolution
function for the \etac peak, gives: $m(\jpsi) - m(\etac)$ = (114.4
$\pm$ 1.1) \mevcc, $m(\jpsi)$ = (3093.6 $\pm$ 0.8) \mevcc, ${\mathrm
\Gamma}(\etac)$ = (34.3 $\pm$ 2.3 \mevcc), $\sigma(\jpsi)$ = (7.6
$\pm$ 0.8) \mevcc.  The numbers of \etac and \jpsi\ events are
respectively 2547 $\pm$ 90 and 358 $\pm$ 33.

The results from B-factories can be compared in Table~\ref{tab:comparison}.

For CLEO, the three major sources of systematic uncertainty in the
masses of these singlets are (i) comparisons of masses of 
the $K_{S}^{0}$ (in $\pi^{+}\pi^{-}$), 
the $D^{0}$ (in $K_{S}^{0}\pi^{+}\pi^{-}$), and
the $D^{+}$ (in $K^{+}\pi^{+}\pi^{-}$) between CLEO data
and the Particle Data Group compilations, (ii) dependences
on fitting shapes used for background and for signal, and
(iii) the observed shifts between mass values used as input
to the Monte Carlo simulations and the  mass values
reconstructed.  In obtaining the widths of these mesons, the
dominant source of possible bias is the shape assumed for the
background.

In \babar, the \etac mass resolution $\sigma(\etac)$ is constrained by
the close \jpsi~peak; the small difference (0.8 \mevcc) observed
between $\sigma(\jpsi)$ and $\sigma(\etac)$ in the simulation is taken
into account in the fit to data.  The simulation is also used to check
for possible bias in the fitted masses.  The \etac and \jpsi mass
peaks are shifted by the same amount (1.1 \mevcc) in the simulation,
therefore the bias does not affect the mass difference.  The
systematic error on the mass accounts for an uncertainty on $m(\jpsi)
- m(\etac)$ due to the background subtraction, and for an uncertainty
associated to the different angular distributions of the \jpsi and the
\etac.  The systematic error on the width is dominated by the
uncertainty in the background-subtraction and in the mass resolution.

\subsubsection{Overview on all results}

\begin{figure}[p]
\begin{center}
\begin{minipage}{7.5cm}
   \includegraphics[width=7.5cm]{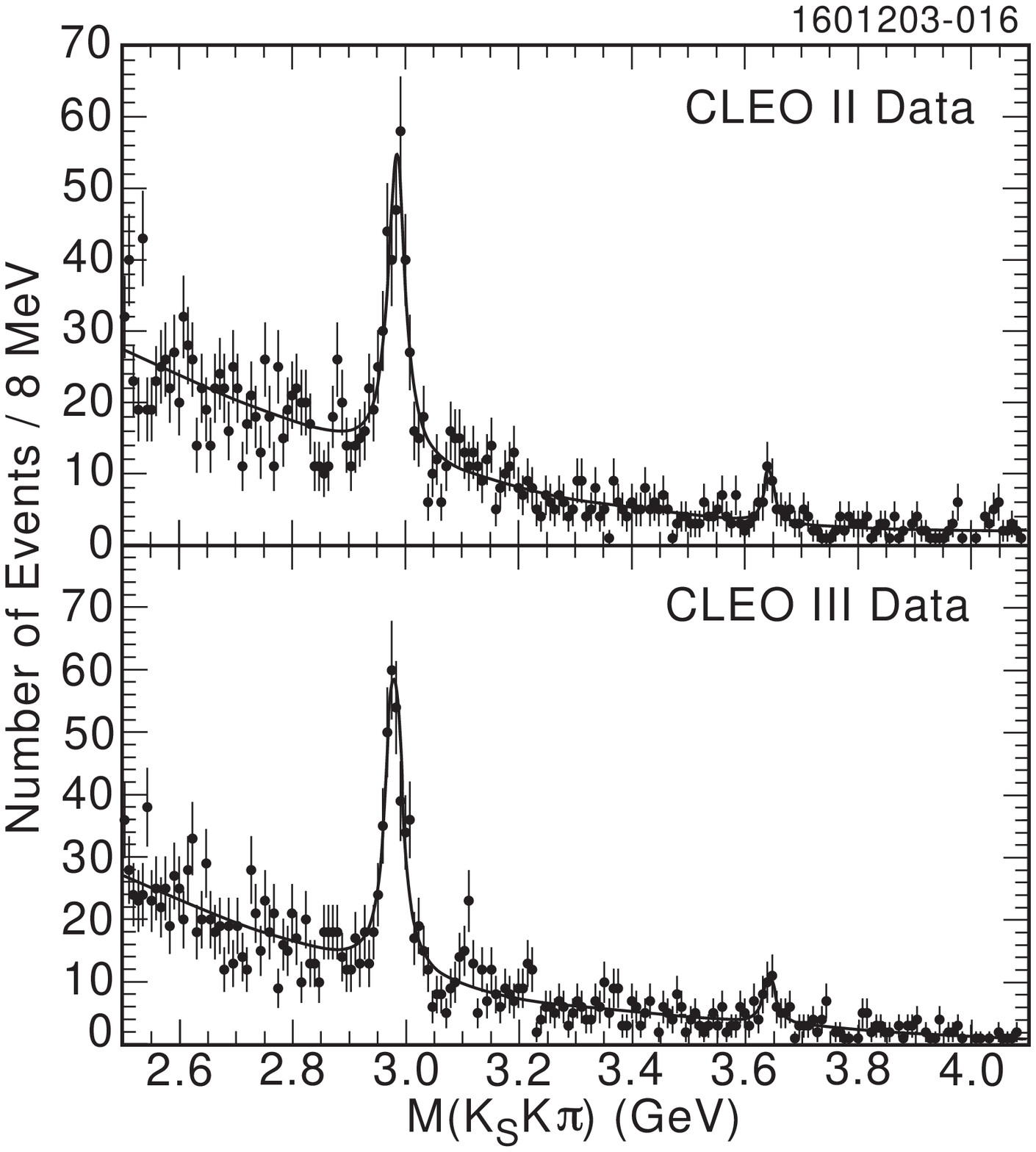}
\end{minipage}
\begin{minipage}{7.5cm}
    \includegraphics[width=7.5cm,scale=0.55]{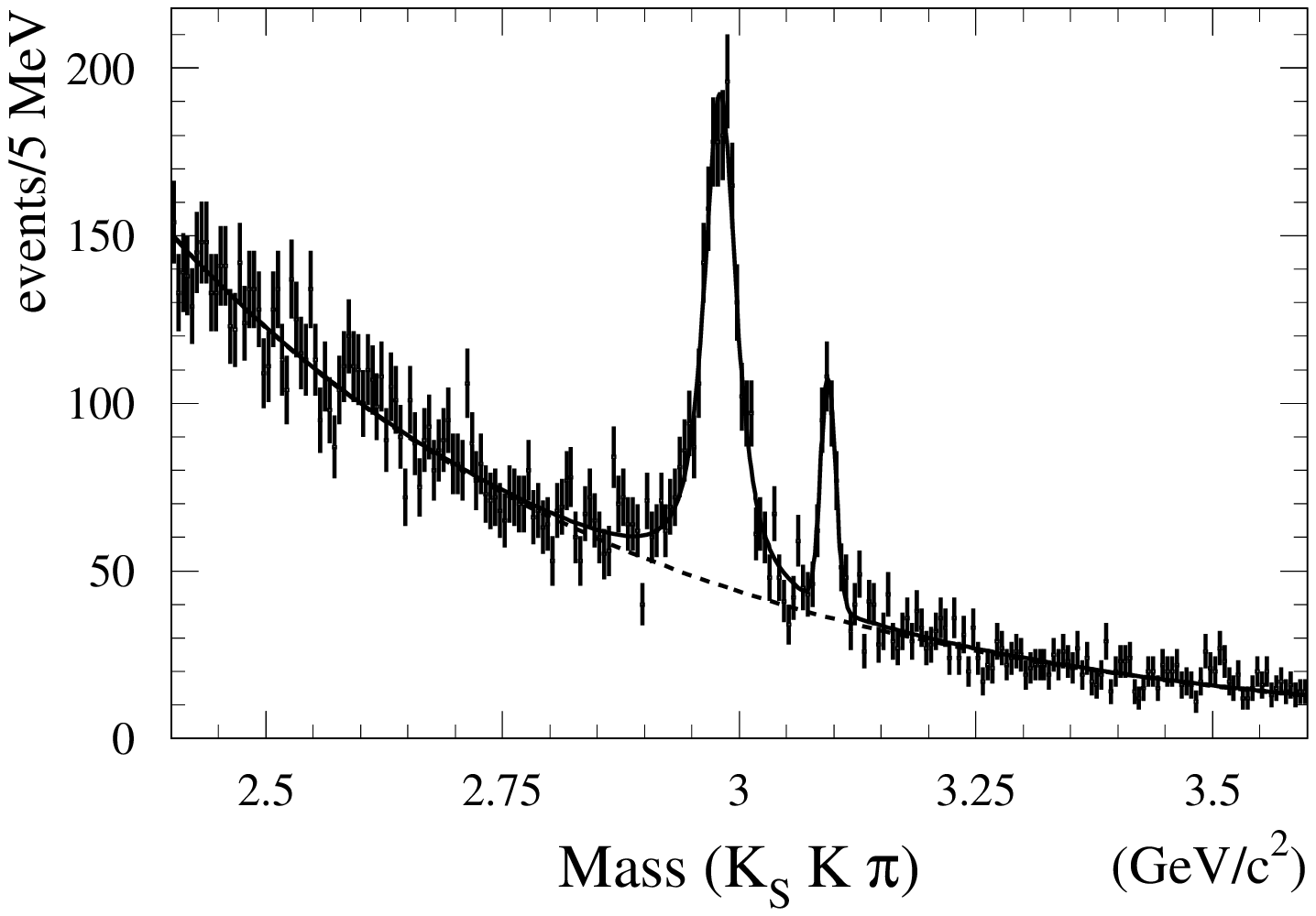}
    \includegraphics[width=7.5cm,scale=0.55]{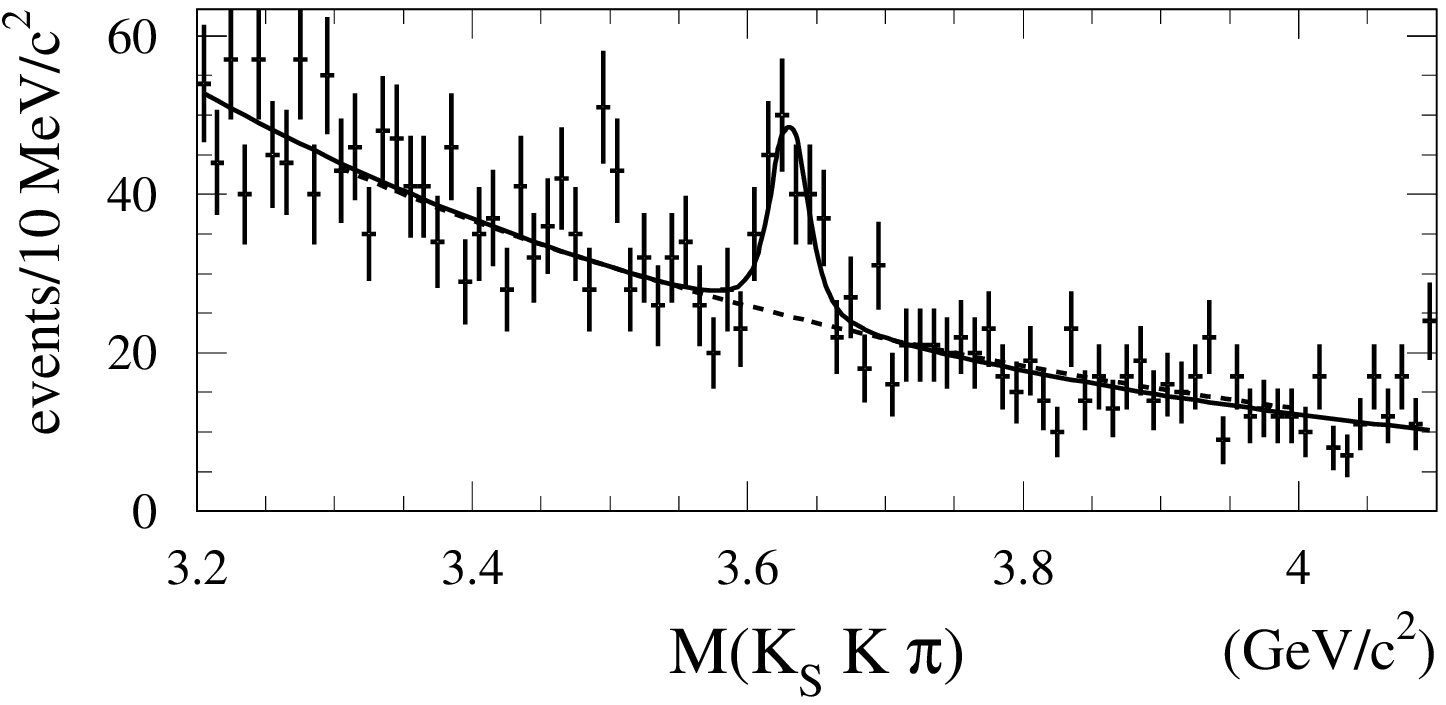}
\end{minipage}
\end{center}
\caption[Invariant mass distributions for $K_{S}^{0}K^{\pm}\pi^{\mp}$ events]
        {Invariant mass distributions for $K_{S}^{0}K^{\pm}\pi^{\mp}$
         events from (a) CLEO II, (b) CLEO III; from \babar in the (c)
         \etac (and \jpsi) region and (d) \etacp region.  The results
         from the fit are superimposed.}
\label{fig:CLEO-Babar}

\begin{center}
\begin{minipage}{7cm}
    \includegraphics[width=\linewidth]{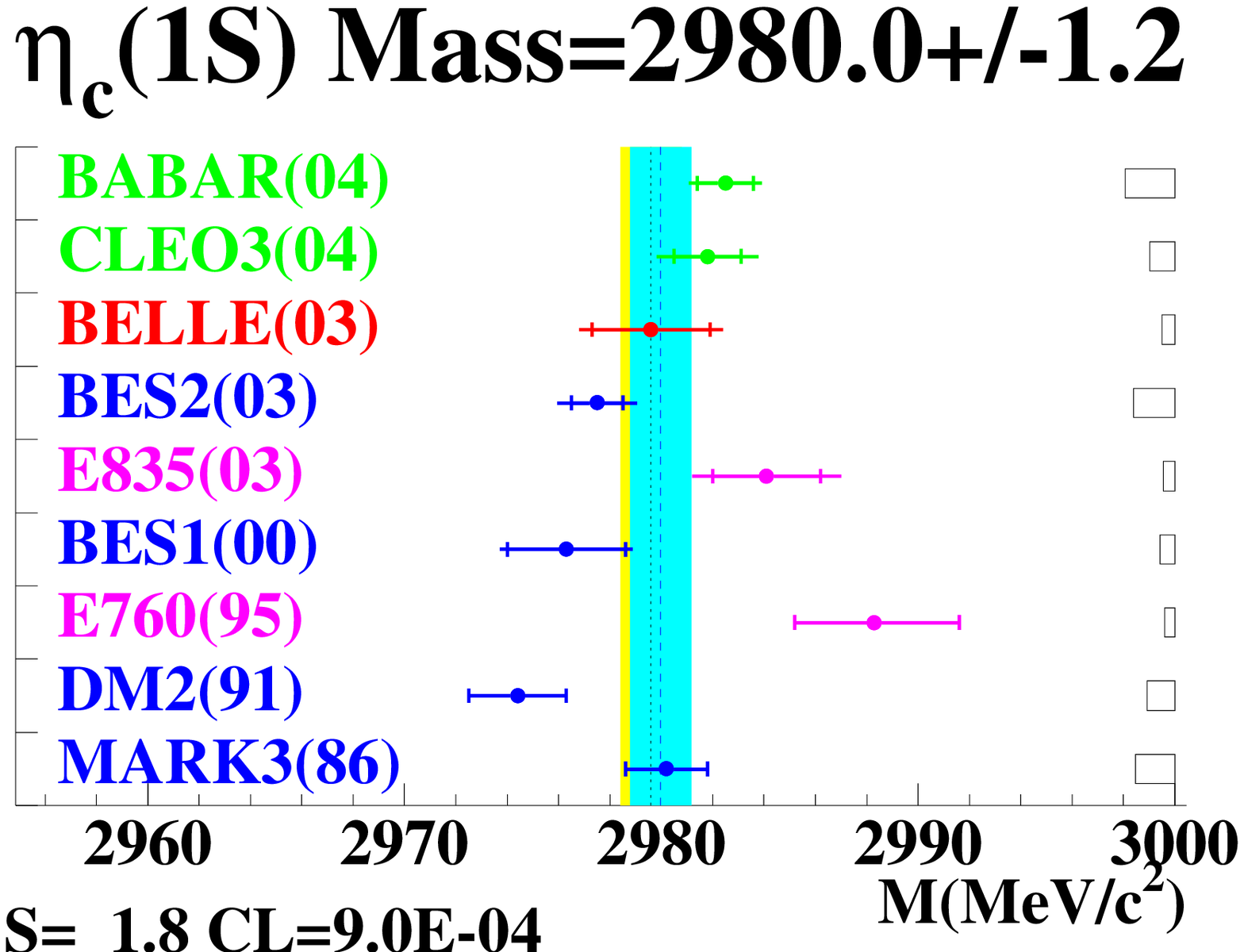}
\end{minipage}
\hfill
\begin{minipage}{8cm}
    \includegraphics[width=\linewidth]{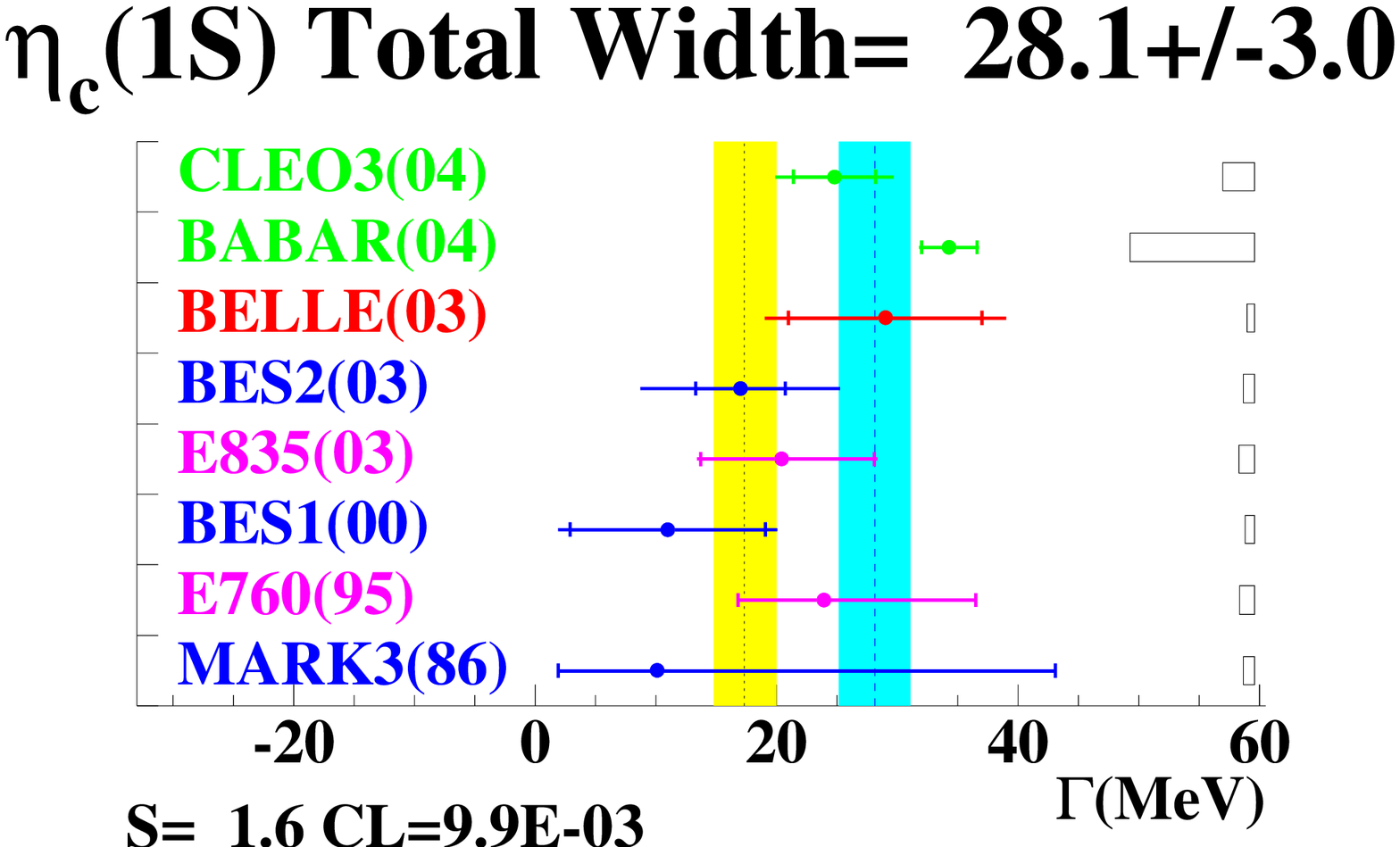}
\end{minipage}
\end{center}
\caption{Mass and width fits for $\etac(1S)$}
\label{fig:sum-etac1}

\begin{center}
\begin{minipage}{7cm}
    \includegraphics[width=\linewidth]{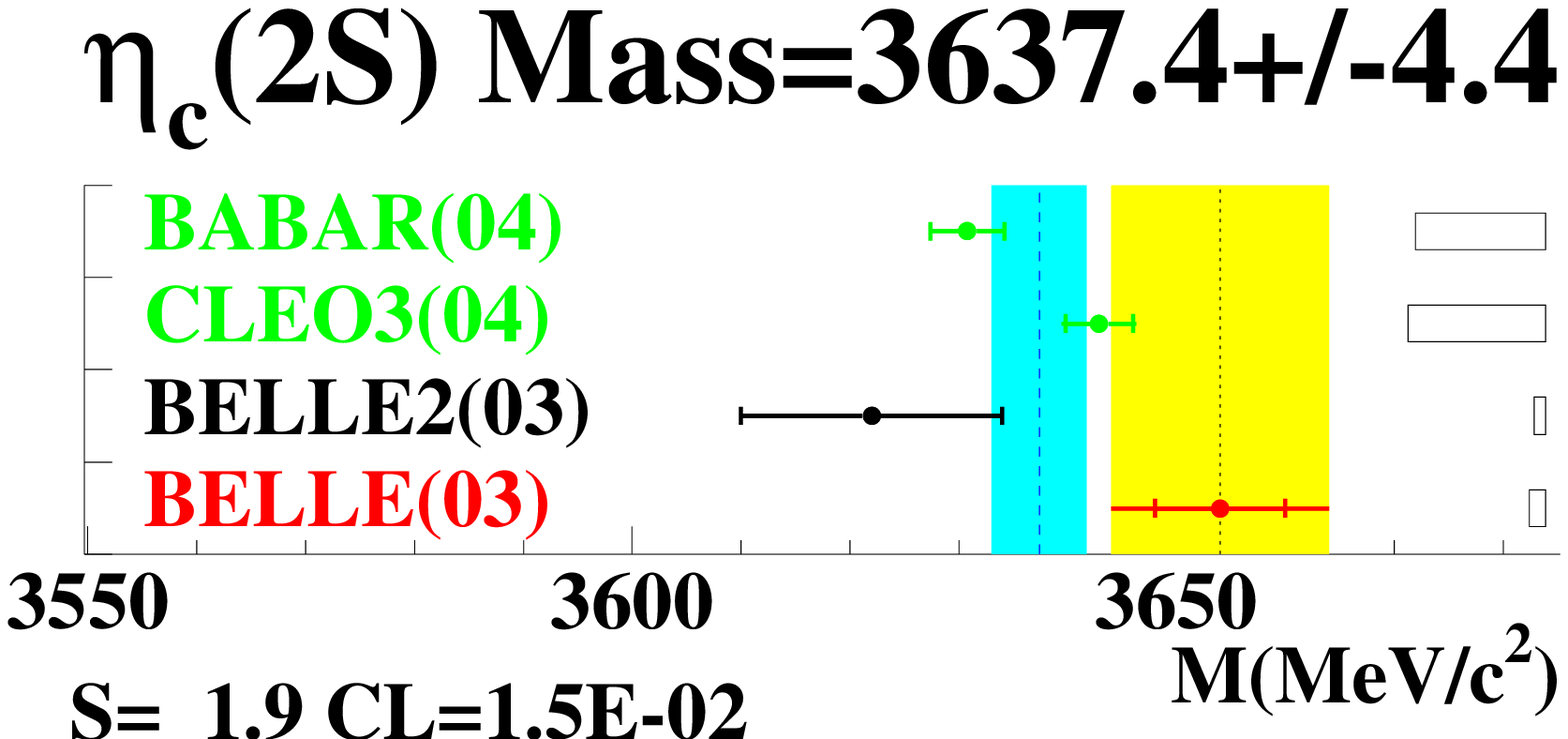}
\end{minipage}
\hfill
\begin{minipage}{8cm}
    \includegraphics[width=\linewidth]{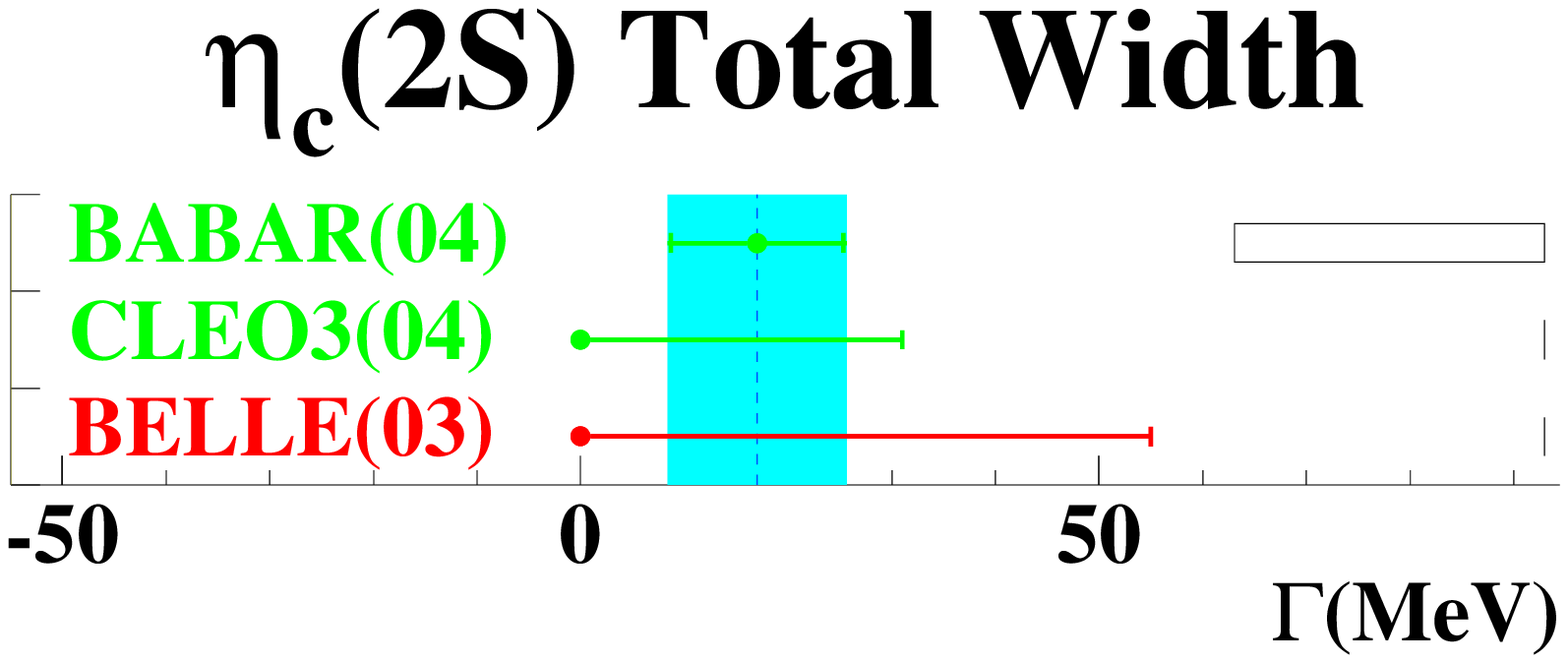}
\end{minipage}
\end{center}
\caption{Mass and width fits for $\etac(2S)$}
\label{fig:sum-etac2}
\end{figure}

\Table~\ref{tab:etac-fits} and \Figures~\ref{fig:sum-etac1} and
\ref{fig:sum-etac2} summarize the results of an attempt to fit the
mass of the $\etac(1S)$ by using (a) all measurements quoted in this
review, (b) only measurements published in the last 5 years, and
results from (c) $\psi(1,2S)$ decays, (d) $\ppbar$ annihilation, (e)
B-factories.  The only {\it rationale} for dataset (b) is to exclude
samples that were superseded by new data taken by the same experiment.
A scale factor S was applied on the $\sigma$'s whenever the confidence
level of the $\chi^2$ obtained from the fits was below 10\%.  The
results are then compared with the values found in PDG 2004.  The
B-factories have been arbitrarily grouped together, despite they use
different techniques.

Despite the substantial improvement in statistics, and the new ways to 
explore the $\eta_c(nS)$ states which came from the B-factories, a
discrepancy between results obtained by different techniques remains. 
The increase in statistics has been surely beneficial in understanding 
systematic effects. Nonetheless , crosschecks between all different 
measurement techniques will be even more vital in the future, 
when statistic errors will be further reduced. 
Hopefully both asymmetric B-factories will be able 
to do internal crosschecks of the results from $\gamma\gamma$ fusion and
from B-decays. CLEO-c will be able to crosscheck the $\gamma\gamma$ 
measurement by CLEO~III with one from $\psi(1,2S)$ decays.

\begin{table}[p]
\caption{Various theoretical estimates for the mass splitting
  $\Delta m = m(\Upsilon) - m(\eta_b)$.}
\label{tab:mass}
\begin{center}
\begin{tabular}{|l|rcl|c|}
\hline
 & \multicolumn{3}{c|}{$\Delta m\;[{\rm MeV}/c^2]$} & Ref \\
\hline
\hline
lattice NRQCD & 19 & -- & 100 & \cite{latnrqcd,Bali:2000gf,El-Khadra:1995ei,
Marcantonio:2000fc,Spitz:1999tu,Manke:2000dg} \\
lattice potential & 60 & -- & 110 & \cite{Bali:1997am} \\
pQCD & 36 & -- & 55 & \cite{Brambilla:2001fw,pqcd} \\
$1/m$ expansion & 34 & -- & 114 & \cite{Narison:1995tw} \\
potential model & 57 & -- & 141 & \cite{potrosner}\cite{pot,Barnes:2001rt,Eichten:1994gt,Ebert:2000bi} \\
\hline
\end{tabular}
\end{center}

\medskip

\caption[Summary of results for $\eta_{c}$ and $\eta_{c}^{\prime}$]
        {Summary of the results for $\eta_{c}$ and $\eta_{c}^{\prime}$
         for both CLEO~II and CLEO~III data sets. The errors shown are
         statistical only.}
\label{tab:CLEOetacpr}
\renewcommand{\arraystretch}{1.3} 
\begin{center}
\begin{tabular}{|c|c|c||c|c|} \hline
& \multicolumn{2}{c||}{CLEO II} &  \multicolumn{2}{c|}{CLEO III} \\
\hline
   &  $\eta_{c}$ &  $\eta_{c}^{\prime}$ &
$\eta_{c}$ &  $\eta_{c}^{\prime}$ \\ \hline
Yield (events) & 282$\pm$30 &28$^{+13}_{-10}$ & 310 $\pm$29&
33$^{+14}_{-11}$\\
Mass (MeV) &  2984.2$\pm$2.0 &3642.4$\pm$4.4 & 2980.0$\pm$1.7&
3643.4$\pm$4.3\\
Width (MeV) &  24.7$\pm$5.1 &3.9$\pm$18.0 & 24.8$\pm$4.5 & 8.4 $\pm$17.1
\\
significance & 15.1$\sigma$ & 4.4$\sigma$ & 17.0$\sigma$& 4.8$\sigma$\\
\hline \hline
$R(\eta_{c}^{\prime}/\eta_{c})$ & \multicolumn{2}{c||} 
{0.17$\pm$0.07}  & \multicolumn{2}{c|}{0.19$\pm$0.08} \\ \hline 
\end{tabular}
\end{center}

\medskip

\renewcommand{\arraystretch}{1} 

\caption{Comparison of CLEO, \babar\ and Belle results.}
\label{tab:comparison}
\begin{center}
\begin{tabular}{lrrr} \hline \hline
Expt. & CLEO & \babar\   & Belle  \\
{\cal L}dt($fb^{-1}$) & 13+14 & 90 & 29.1~\cite{Fang:2002gi},
31.3\cite{Choi:2002na}
 \\ \hline $m(\etac)$(\mevcc)                   
&   2981.8 \PM 1.3 \PM 1.5
&   2982.5 \PM 1.1 \PM 0.9 
&   2979.6 \PM 2.3 \PM 1.6~\cite{Fang:2002gi}
 \\ $\Gamma(\etac)$ (MeV)             
& 24.8 \PM 3.4 \PM 3.5
& 34.3 \PM 2.3 \PM 0.9   
&   29 \PM 8 \PM 6~\cite{Fang:2002gi}
 \\ \hline $m(\etacp)$(\mevcc)                  
& 3642.9 \PM 3.1 \PM 1.5
& 3630.8 \PM 3.4 \PM 1.0 
&    3654 \PM 6 \PM 8~\cite{Choi:2002na}
 \\ $\Gamma(\etacp)$ (MeV)
&  $<$31 (90\%CL)               
& 17.0 \PM 8.3 \PM 2.5   
&      $<$55 (90\%CL)~\cite{Choi:2002na}
\\
\hline \hline
\end{tabular}
\end{center}

\medskip

\caption{Fits of all $\etac$ mass measurements}
\label{tab:etac-fits}
\begin{center}
\begin{tabular}{lrrr} \hline
Dataset & Mass(\mevcc) & S & C.L. \\
\hline
(a) ALL                  & 2980.0 \PM 1.2 & 1.82 & 0.09\% \\
(b) ALL after 1999       & 2980.4 \PM 1.2 & 1.44 & 6.6\% \\
(c) $\psi(1,2S)$ decays  & 2977.5 \PM 0.9 & 1(1.38) & 13\% \\
(d) $\ppbar$             & 2984.5 \PM 1.6 & 1(1.05) & 33\% \\
(e) B-factories          & 2981.9 \PM 1.1 & 1(0.65) & 65\% \\
PDG 2004                 & 2979.6 \PM 1.2 & 1.7   & 0.1\%  \\
\hline
\end{tabular}
\end{center}
\end{table}

\clearpage

\subsection[$\eta_b$(nS) and $h_b$(nP): searches]
           {$\eta_b$(nS) and $h_b$(nP): searches 
            $\!$\footnote{Authors: A.~B\"ohrer, T.~Ferguson, J.~Tseng}}
\label{sec:spexetab}

Over twenty-five years after the discovery of the $\Upsilon(1S)$, no
pseudoscalar $b\overline{b}$ states have been conclusively uncovered.  In
recent years, the search has been conducted at CLEO, LEP, and CDF,
using both inclusive and exclusive methods.

The inclusive CLEO search~\cite{Athar:2004dn} identifies distinctive
single photons with its high-resolution CsI electromagnetic
calorimeter.  These photons are signatures of $\Upsilon$ radiative
decays, in this case $\Upsilon(3S)\rightarrow\eta_b\gamma$,
$\Upsilon(2S)\rightarrow\eta_b\gamma$,
$\Upsilon(3S)\rightarrow\eta_b'\gamma$, and $\Upsilon(3S)\rightarrow
h_b\pi^0$ or $h_b\pi^+\pi^-$ followed by $h_b\rightarrow\eta_b\gamma$.
Godfrey and Rosner have pointed out that these hindered M1 transitions
could have observable branching ratios, in spite of their small
associated matrix elements, because of their large phase
space~\cite{Godfrey:2001vc}.

\begin{figure}[p]
\begin{center}
\includegraphics[width=.49\linewidth]{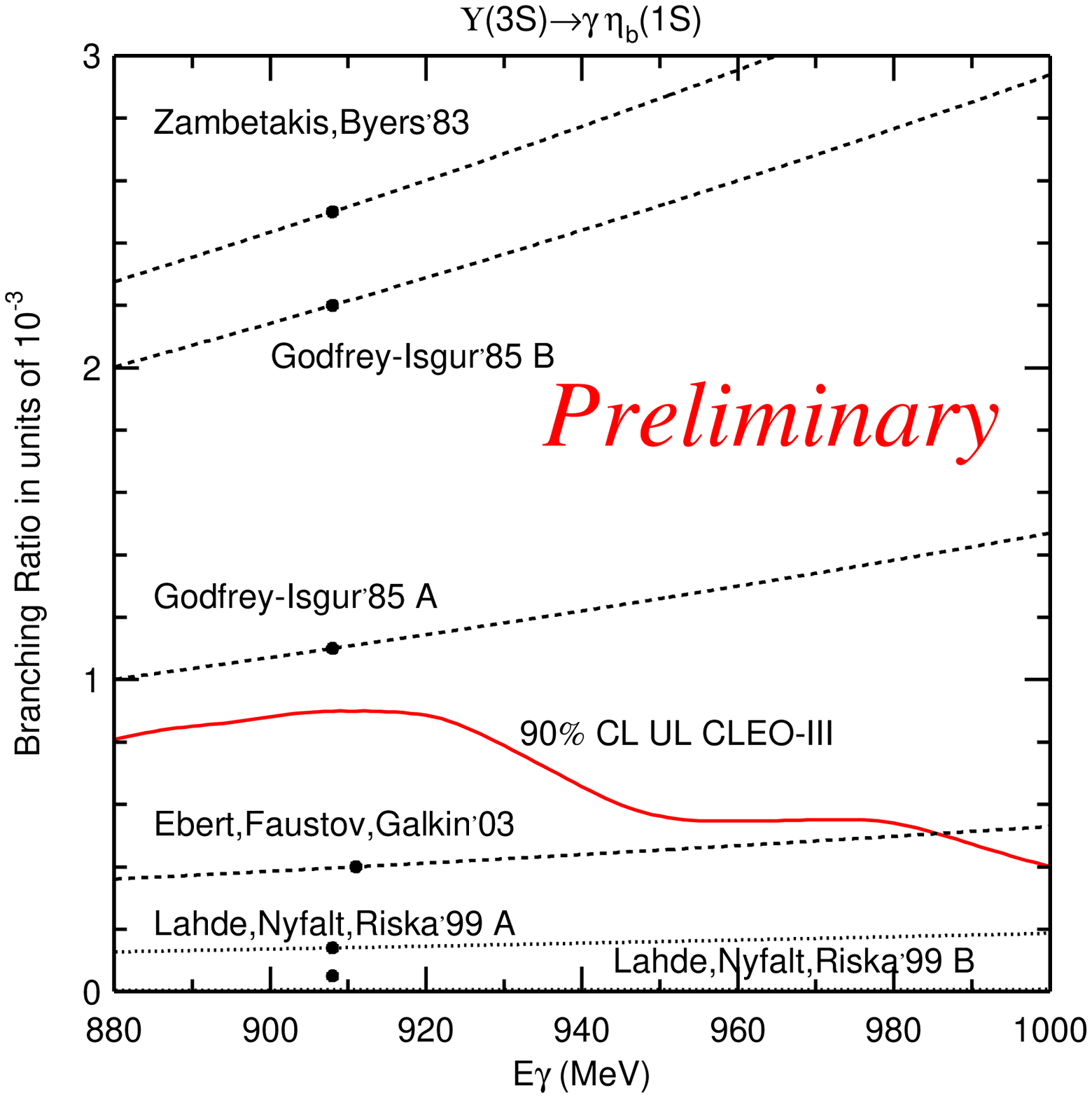}
\hfill
\includegraphics[width=.49\linewidth]{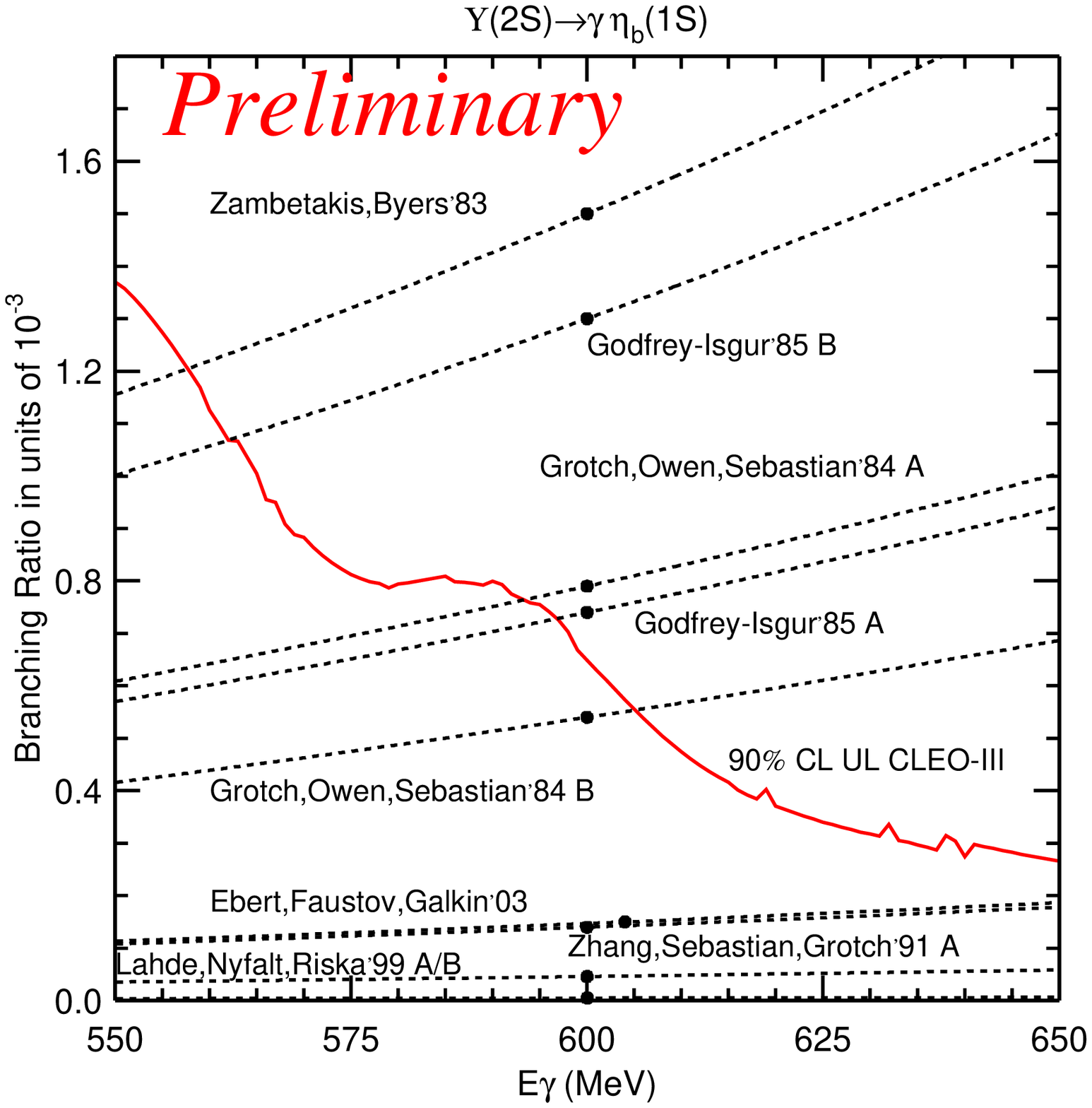}\\[5mm]
\includegraphics[width=.49\linewidth]{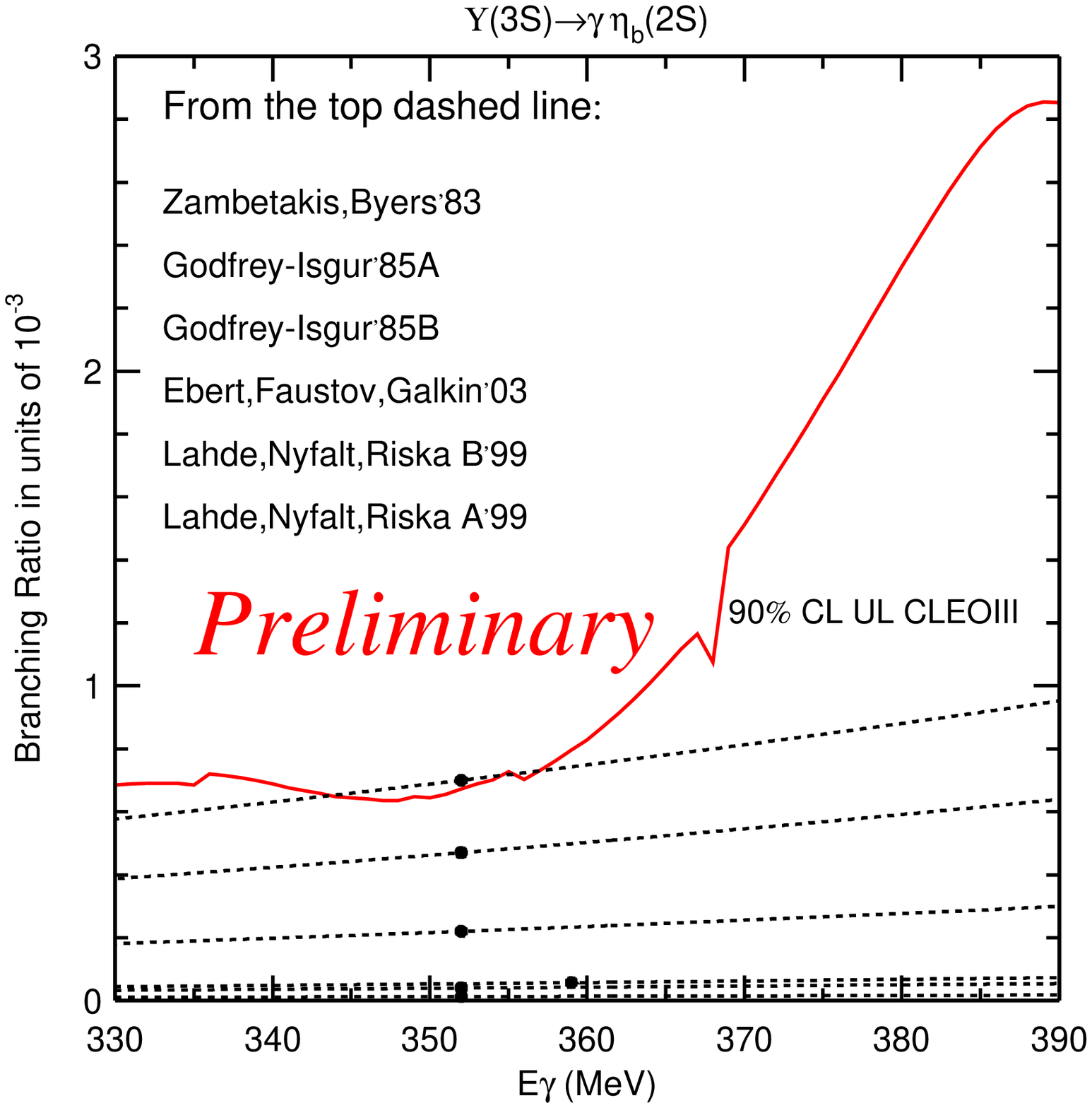}
\hfill
\includegraphics[width=.49\linewidth]{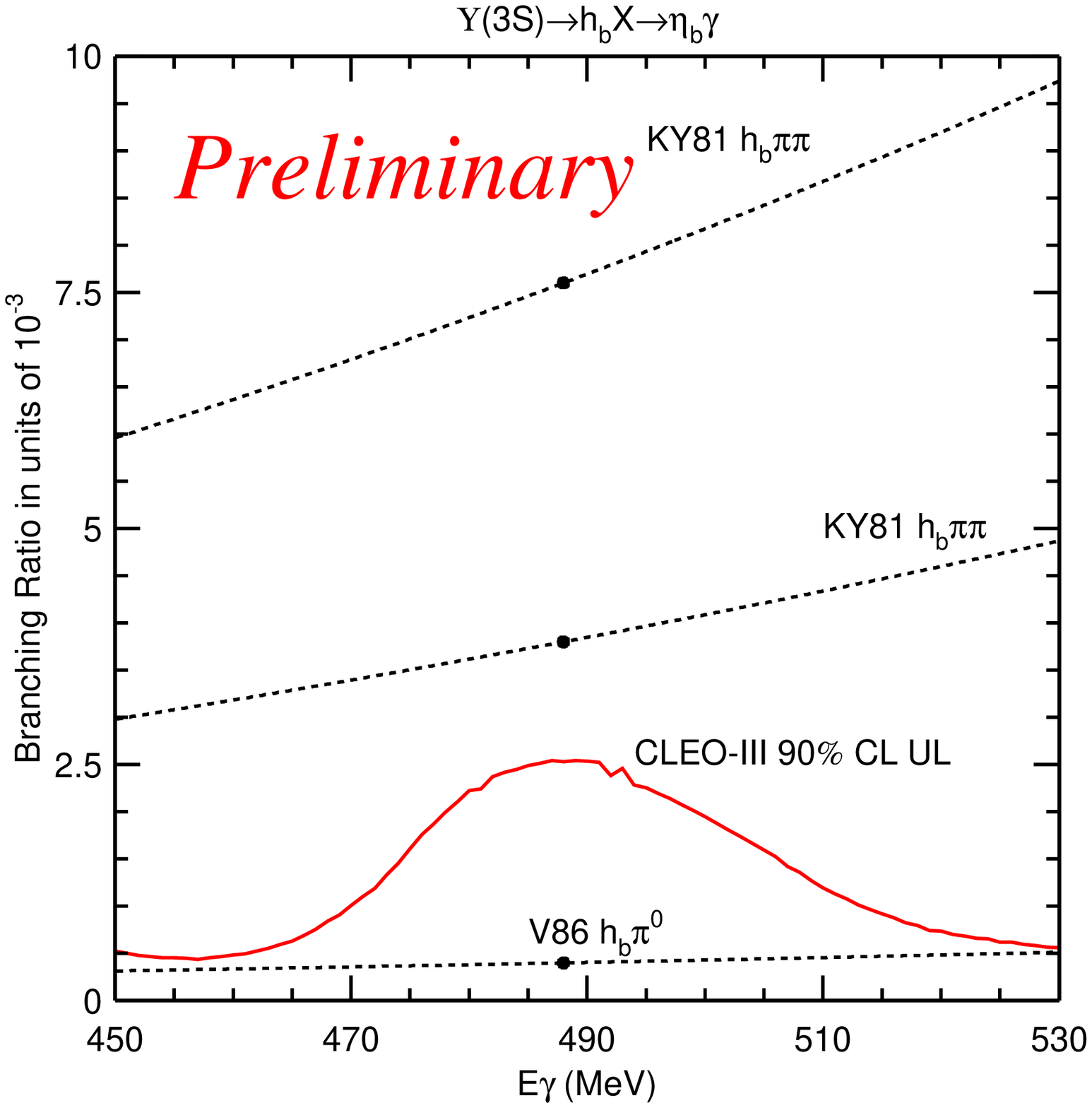}
\end{center}
\caption[CLEO 90\% C.L. upper limits on ${\cal
         B}(\Upsilon(3S)\rightarrow\eta_b\gamma)$ ${\cal
         B}(\Upsilon(2S)\rightarrow\eta_b\gamma)$ ${\cal
         B}(\Upsilon(3S)\rightarrow\eta_b'\gamma)$ and ${\cal
         B}(\Upsilon(3S)\rightarrow h_b\pi^0, h_b\pi^+\pi^-)\times
         {\cal B}(h_b\rightarrow\eta_b\gamma)$]
        {CLEO 90\% C.L. upper limits on ${\cal
         B}(\Upsilon(3S)\rightarrow\eta_b\gamma)$ (top left), ${\cal
         B}(\Upsilon(2S)\rightarrow\eta_b\gamma)$ (top right), ${\cal
         B}(\Upsilon(3S)\rightarrow\eta_b'\gamma)$ (bottom left), and
         ${\cal B}(\Upsilon(3S)\rightarrow h_b\pi^0,
         h_b\pi^+\pi^-)\times {\cal B}(h_b\rightarrow\eta_b\gamma)$
         (bottom right) as a function of the photon energy $E_\gamma$,
         along with various theoretical
         predictions~\cite{Zambetakis:1983te,Godfrey:1985xj,
         Lahde:1998ee,Ebert:2002pp,Godfrey:2002rp}.}
\label{fig:cleo}
\end{figure}

No evidence of a signal for any of the above modes has been seen in
the total $2.4\;{\rm fb}^{-1}$ of data taken at
the $\Upsilon(2S)$ and $\Upsilon(3S)$ resonances between 2001 and 2002,
corresponding to roughly six million decays of each resonance.
\Figure[b]~\ref{fig:cleo} shows the resulting 90\% C.L. upper
limits on the branching
fractions.  Several of the theoretical predictions shown can be ruled out.

It has been shown that with the full data samples of LEP 2, the
$\eta_b(1S)$ might be detected in two-photon
events~\cite{Boehrer:2001ef, Boehrer:2003yb}.  The $\eta_b$ is fully
reconstructed with four, six, or eight charged decay products and
possibly a $\pi^0$.  In the expected mass range, for which estimates
are listed in Table~\ref{tab:mass}, the corresponding invariant mass
distribution is rapidly decreasing, and the background from $\tau$
pairs can be kept small.
\shortpage[2]

\begin{figure}[p]
\centering\includegraphics[width=\linewidth]{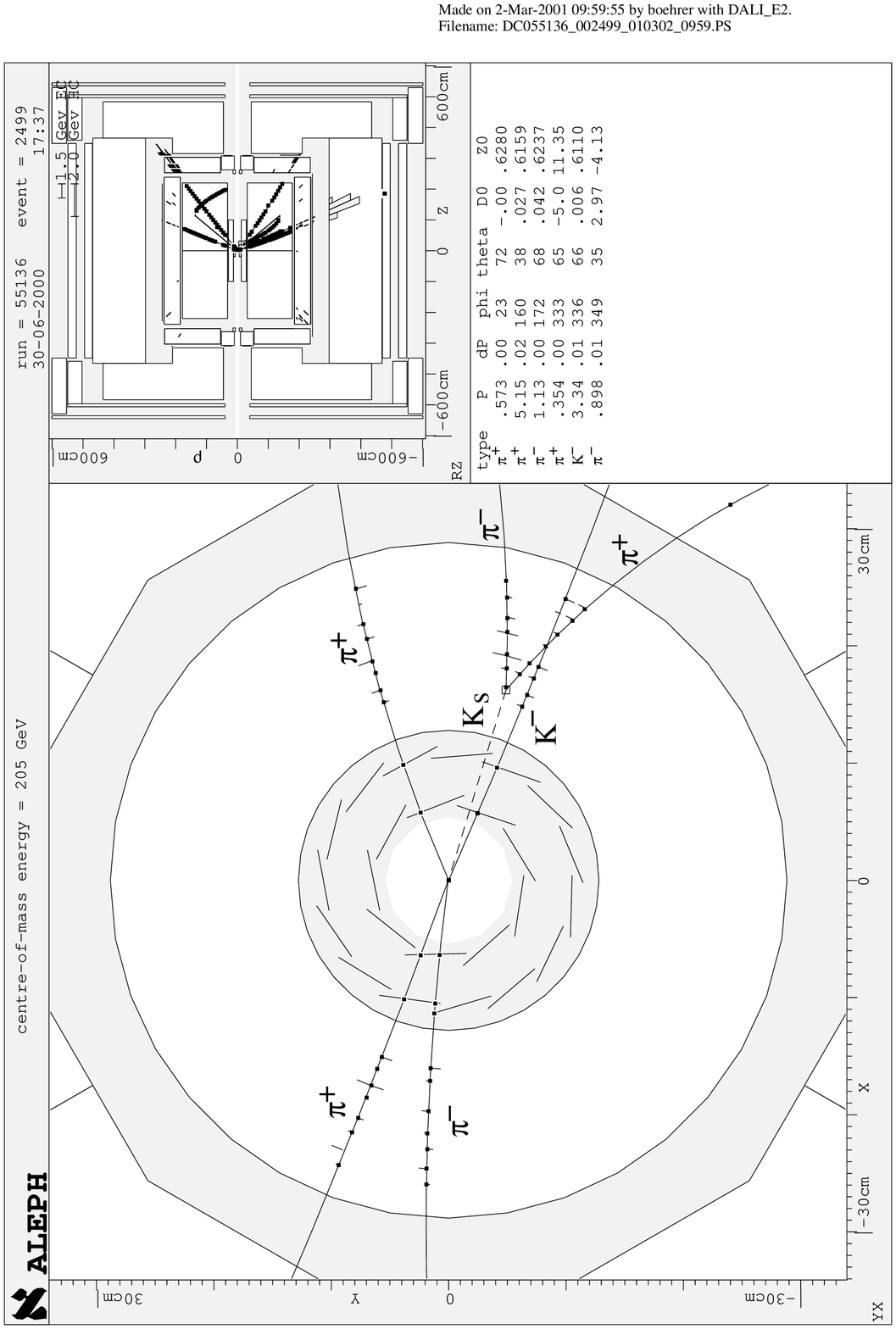}
%\epsffile[40 36 571 792]{fig2-etabbest-bw}
\caption{$\eta_b\rightarrow K^0_S K^-\pi^+\pi^+\pi^-$
  candidate at ALEPH, with a reconstructed
  mass of $9.30\pm 0.03\;{\rm GeV}/c^2$.}
\label{fig:aleph}
\end{figure}

Table~\ref{tab:lep} summarizes the results for ALEPH, L3, and DELPHI.
The search by ALEPH~\cite{Heister:2002if} in an $800\;{\rm MeV}/c^2$
window turned up one candidate, shown in \Figure~\ref{fig:aleph}, with
an excellent mass resolution of $30\;{\rm MeV}/c^2$ at a mass of
$9.30\pm 0.03\;{\rm GeV}/c^2$.  The signal expectation is about 1.6
events over one background event.

\begin{table}[!h]
\caption{95\% C.L. upper limits on the $\eta_b$ two-photon partial
width times branching ratio into various hadronic states from searches
at LEP.}
\label{tab:lep}
\begin{center}
\begin{tabular}{|l|l|l|c|}
\hline
Expt & final state & $\Gamma_{\gamma\gamma}\times{\cal B}$ (keV) & Ref \\
\hline
\hline
ALEPH & 4 charged & $< 0.048$ & \cite{Heister:2002if} \\
      & 6 charged & $< 0.132$ & \cite{Heister:2002if} \\
\hline
L3 & $K^+K^-\pi^0$ & $<2.83$ & \cite{Levtchenko:2004ku} \\
   & 4 charged & $< 0.21$ & \cite{Levtchenko:2004ku} \\
   & 4 charged $\pi^0$ & $<0.50$ & \cite{Levtchenko:2004ku} \\
   & 6 charged & $< 0.33$ & \cite{Levtchenko:2004ku} \\
   & 6 charged $\pi^0$ & $<5.50$ & \cite{Levtchenko:2004ku} \\
   & $\pi^+\pi^-\eta'$ & $<3.00$ & \cite{Levtchenko:2004ku} \\
\hline
DELPHI & 4 charged & $<0.093$ & \cite{Sokolov:2004kv} \\
       & 6 charged & $<0.270$ & \cite{Sokolov:2004kv} \\
       & 8 charged & $<0.780$ & \cite{Sokolov:2004kv} \\
\hline
\end{tabular}
\end{center}
\end{table}

L3 has reported an analysis, considered close to final, in six decay
modes~\cite{Levtchenko:2004ku}.  Six candidates are found, compatible
with an expected background of 2.5 events.  The mass measurement is
dominated by the detector resolution of about $300\;{\rm MeV}/c^2$.

Recently, DELPHI has also reported preliminary
results~\cite{Sokolov:2004kv}.  A total of seven candidates are found
in a search window of $400\;{\rm MeV}/c^2$.  The expected background
level is 5.5 events, and the mass resolution roughly $120\;{\rm
MeV}/c^2$.

\begin{figure}[t]
\begin{center}
\begin{minipage}{5in}
\epsfxsize=5in
\epsffile[1 256 522 512]{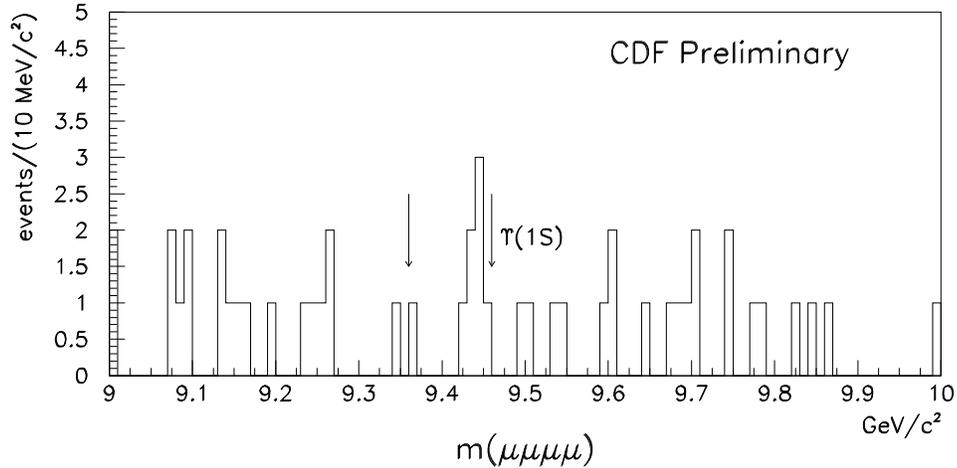}
\end{minipage}
\caption{The 4-muon invariant mass distribution from $J/\psi J/\psi$
events in CDF Run~1 data.  The search window, the upper side of which
is the world-averaged $\Upsilon(1S)$ mass~\cite{Eidelman:pdg2004}, is marked by arrows.}
\label{fig:cdf}
\end{center}
\end{figure}

CDF has searched for the exclusive decay $\eta_b\rightarrow J/\psi
J/\psi$, where both $J/\psi$'s decay to muon pairs, in the full
1992--96 ``Run~1'' data sample of about $100\;{\rm
pb}^{-1}$~\cite{Tseng:2003md}.  The mass spectrum is shown in
\Figure~\ref{fig:cdf}; in this region, the mass resolution is about
$10\;{\rm MeV}/c^2$.  A small cluster of seven events can be seen,
where 1.8 events are expected from background.  The statistical
significance of the cluster is estimated to be $2.2\sigma$.  A simple
fit to the mass distribution gives $9445\pm 6(stat)\;{\rm MeV}/c^2$ as
the mass of the cluster, where the error is only statistical.  The
mass difference relative to $\Upsilon(1S)$ is well to the low side of
the theoretical expectation.  If this cluster is due to $\eta_b$
decay, then the product of its production cross-section and decay
branching fractions is near the upper end of
expectations~\cite{Braaten:2000cm}.

The existence of the $\eta_b$ is a solid prediction of the
quark model, and its mass one of the most tractable to calculate.
Both its existence and mass remain, for the present time,
open questions.  Some data at completed experiments remain to be
published, however, while Run~2 is well underway at the Fermilab Tevatron.

\subsection[$h_c$: searches]
           {$h_c$: searches $\!$\footnote{Authors: R.~Mussa, D.~Besson}}
\label{sec:spexhc}

\def\logtwo{\rm ln(2)}
\def\grb{\frac{\Gamma_{res}}{\Gamma_{beam}}}
\def\grbq{\frac{\Gamma_{res}^2}{\Gamma_{beam}^2}}
\def\Gio{\Gamma_{io}}
\def\thcm{\theta_{CM}}

The search of the singlet state of P wave charmonium (dubbed $\hc$) 
poses a unique experimental challenge for a variety of reasons:

\begin{itemize}
\item it cannot be resonantly produced in \ep annihilation;

\item it cannot be reached via E1 radiative transitions from $\psip$;
      C-parity conservation forbids the transition from a $1^{--}$ to
      a $1^+-$ state.

\item its production in $\psip$ hadronic decays to $\hc\pi^0$ is
      isospin violating and has a small phase space available (if
      $M_\hc=M_{COG}$, $p_{\pi^0} = 86 \mev/c$; the two Doppler
      broadened photons will have and energy between 30 and 100~MeV in
      the $\psip$ rest frame.  In \ep machines, the sensitivity on
      slow pions is not just affected by the physical backgrounds from
      other $\psip$ decays, but also by the large combinatorial
      background with low energy uncorrelated photons from the beam.

\item its production in B decays via the intermediate state $\etacp$,
      which can decay radiatively (E1) to $\hc$, is suppressed by the
      large hadronic width of the $\etacp$.

\item its detection in the $\jpsi\pi^0$ decay mode, from $\psip$ and B
      decays, as well as in hadroproduction, is shadowed by the more
      copious decay $\chicj{1,2}\to\gamma\jpsi$, with an extra photon
      accidentally matching the $\pi^0$ mass; this is also the most
      likely explanation of the signal seen in $\Jpsi\pi^0$ by
      experiment E705, in 300~GeV/c $\pi^\pm$ and proton interactions
      on a lithium target \cite{Antoniazzi:1993jz}.

\item its formation in $\ppbar$ annihilation {\it may} be suppressed
      by helicity selection rule, but the same rule would forbid
      $\chicj{0}$ and $\etac$ formation , against the experimental
      evidence.

\item its production in exclusive B decays {\it may} be suppressed as
      ${\cal B}(B\to\chicj{0}K$); if such selection mechanism does not
      apply, a search of $\hc$ via its E1 decay to $\eta_c$ may soon
      give positive results.
\end{itemize}
Such elusive state was extensively searched for in formation from
$\ppbar$ annihilations: searching for a resonance which has a width
expected to be between the $\psi$ and $\chicj{1}$ but with an expected
${\cal B}$ to detectable EM decay channels of interests which is 100
to 1000 times weaker than the radiative decay of $\chicj{1}$, \ie
expected cross-sections between 1 and 10 picobarns.  Experiment R704
at CERN \cite{Baglin:1986yd} observed the signal:
\[
\Gamma(\hc\to\ppbar)\times{\cal B}(\hc\to\jpsi + X)\times{\cal B}(\jpsi\to\ep) = 
0.14^{+0.15}_{-0.06} \; {\rm eV}
\] 
at a nominal mass of 3525.4$\pm$0.8$\pm$0.5 , which should be shifted
down 0.8 \mevcc after comparing the $\chi_c$ measurements done by the
two experiments.

Experiment E760 at Fermilab \cite{Armstrong:1992ae} observed the signal:
\[
\Gamma(\hc\to\ppbar)\times{\cal B}(\hc\to\jpsi + \pi^0)\times{\cal B}(\jpsi\to\ep) 
= 0.010\pm 0.003 \; {\rm eV}
\] 
at a nominal mass of 3526.2 $\pm$ 0.15,  
and did not see events in the channels $\jpsi\pi^+\pi^-,\jpsi\pi^0\pi^0$
E760 also determined a level of continuum for the inclusive reaction 
which was consistent with the one observed by R704.

In channels with such low statistics, a large amount of integrated
luminosity taken to precisely quantify the background level is
crucial. Such an issue was taken very seriously in E760, and even more
in E835.  To complicate the experimental situation, the signal
observed by E760 is expected to be comparable to the $\jpsi\pi^0$
continuum, as predicted in reference \cite{Gaillard:1982zm}, from soft
pion radiation.  It is hard to predict how interference between the
resonant and continuum amplitude can distort the lineshape.

\begin{center}
\begin{figure}[t]
\begin{minipage}{.48\linewidth}
   \includegraphics[width=\linewidth]{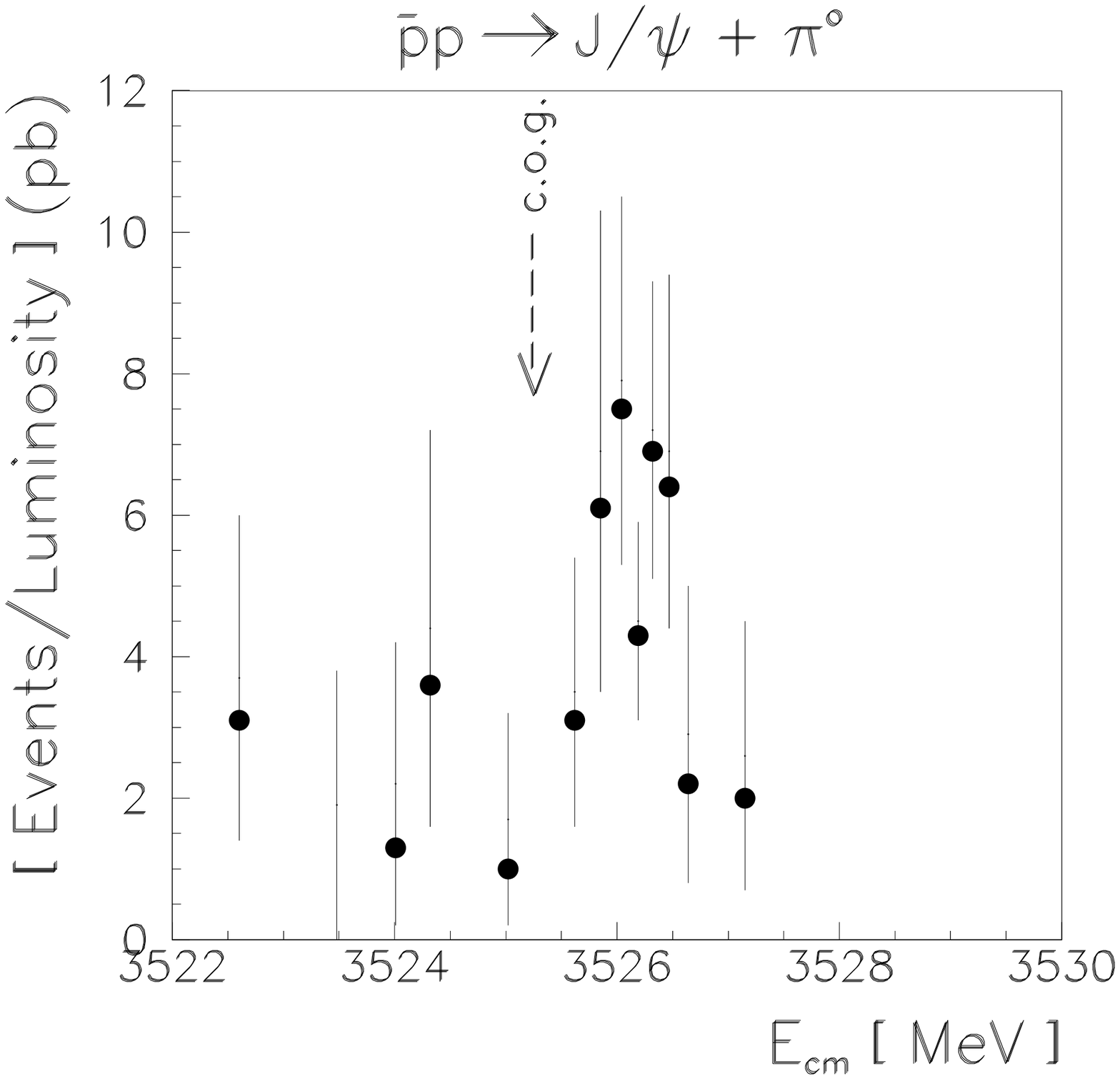}
\end{minipage}
\hfill
\begin{minipage}{.48\linewidth}
   \includegraphics[width=\linewidth]{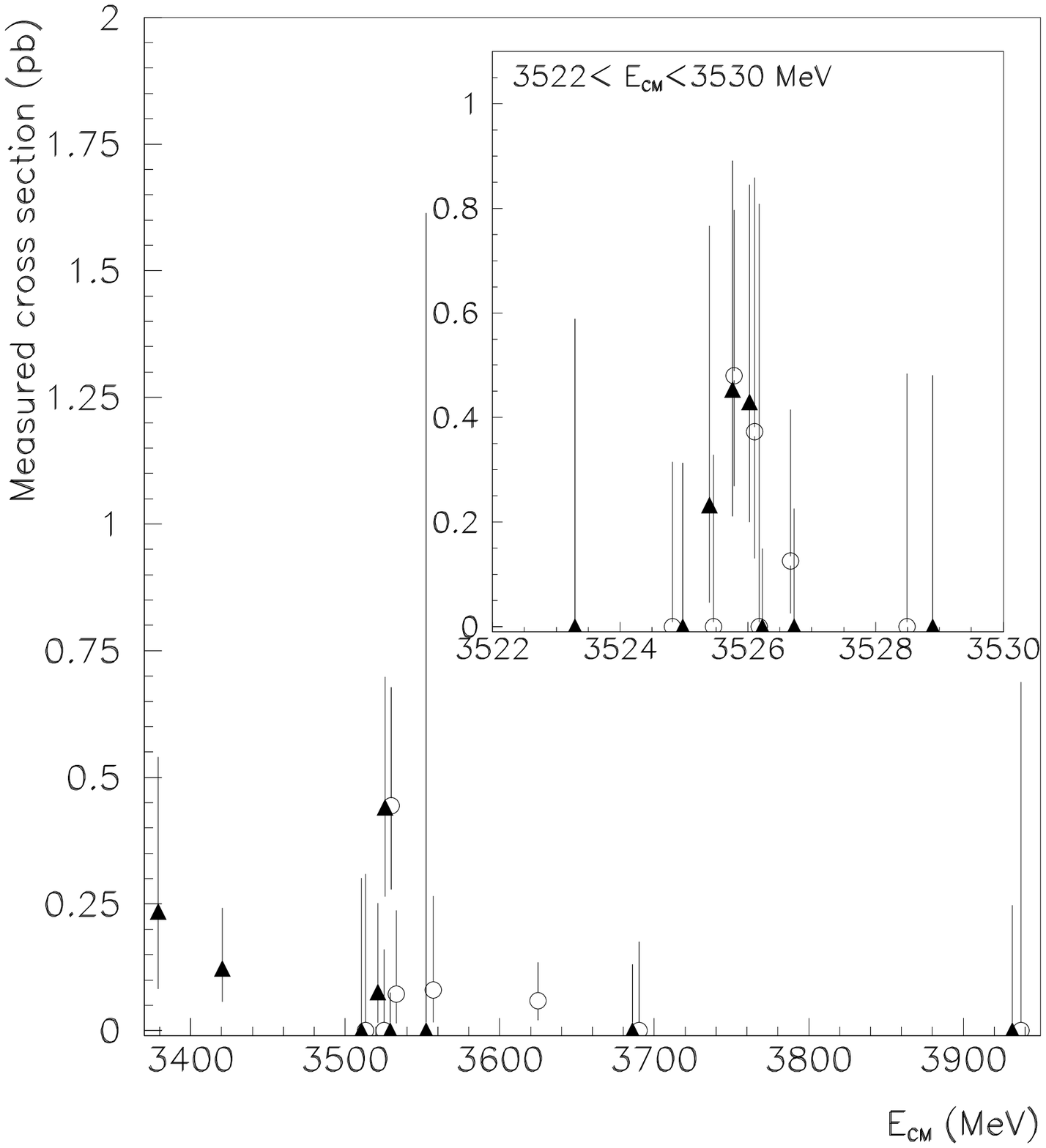}
\end{minipage}
\caption[Cross-section observed by E760 for the reaction
         $\ppbar\to\jpsi\pi^0$ in the COG region] 
        {Cross-section (black dots) observed by E760(left) for the
         reaction $\ppbar\to\jpsi\pi^0$ in the COG region; E835 could
         not confirm this evidence and observed the hint on the right
         in the $\gamma\etac\to 3 \gamma$ channel.}
\label{fig:hc-e835}
\end{figure}
\end{center}

E835 took 6 times more data with respect to E760, to confirm the observation of
$\hc$ and possibly measure the width as well as its decay ratios to other 
channels:the probably dominant decay mode to $\etac\gamma$ was studied, 
relying upon the rare $\etac$ decays to $\gaga$. 
The first data set, 50 pb$^{-1}$ taken in 1996, proved lately to be affected by
an anomaly in the beam positioning system, which prevented to determine the 
absolute energy calibration of the machine better than 200~KeV.
A second data taking period in year 2000 allowed to accumulate a comparable 
sample of data, but with 150~KeV resolution on the CM energy determination.

The E835 experiment, despite the 6 times larger statistics, could not
confirm the $\jpsi\pi^0$ evidence observed by E760. On the other side,
a hint of a signal is observed in the $3\gamma$ channel \cite{e835hc}
Very tight cuts were applied in order to reject hadronic backgrounds
from reactions with two neutral mesons in the final state. In the
$3\gamma$ Dalitz plot, invariant masses of all pairs were requested to
be above 1~GeV$/c^2$, to reject backgrounds from
$\pi^0,\eta,\eta^\prime,\omega$.  As the recoil photon angular
distribution is expected to behave as $\sin\theta_{CM}^2$ on the
resonance, a cut at $\cos\theta_{CM}<0.5 $ was imposed. This allowed
to suppress most of the two meson background, which is prevalently
forward-backward peaked.  13 events out of 29~pb$^{-1}$ are observed
in a $\delta M=0.5 \mevcc$ wide bin between 3.5257 and 3.5262~\mevcc,
while 3 events are observed in the remaining data between the
$\chicj{1}$ and the $\chicj{2}$ (87~pb$^{-1}$).  The statistical
significance of the excess is between 1 and 3$\times 10^{-3}$, with
different hypotheses on the resonance width.  If the excess is not a
statistical fluctuation, assuming a total width of 0.5~MeV, it is
possible to measure $\Gamma(h_c\to\ppbar){\cal B}(h_c\to\etac\gamma) =
10.4\pm3.7\pm3.4 \, {\rm eV} $, where the systematic error comes from
the statistical error on ${\cal B}(\eta_c\to\gamma\gamma)$), at a mass
$M(h_c)=3525.8\pm0.2\pm0.2 \mevcc$.  The CLEO Collaboration has
preliminary evidence \cite{cleohc} for the spin singlet $h_{c}$
($1^{1}P_{1}$) in looking at $\sim 3 \times 10^{6}$ decays of the
$\psi^{\prime}$(3686).  This state is seen in two independent
analyses, both of which use the decay chain $\psi^{\prime}\to\pi^{0}
h_{c}$ followed by $h_{c}\to\gamma\eta_{c}$: one analysis is {\it
in}clusive and the other uses six dominant {\it ex}clusive decays of
the $\eta_{c}$.

\begin{figure}
  \centerline{\hbox{ \hspace{0.2cm}
    \includegraphics[width=6.0cm]{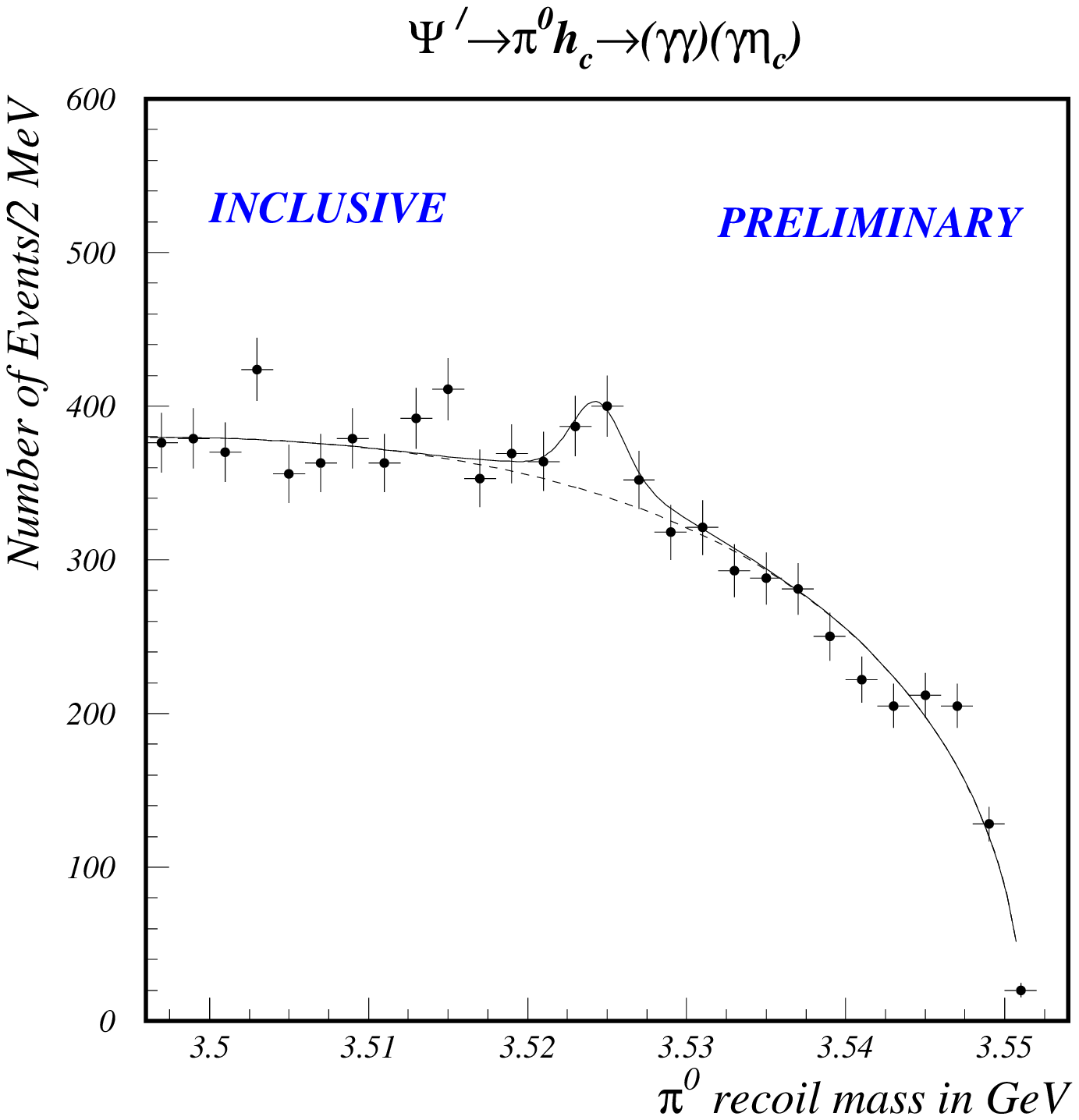}
    \hspace{0.3cm}
    \includegraphics[width=6.0cm,angle=270,origin=br]{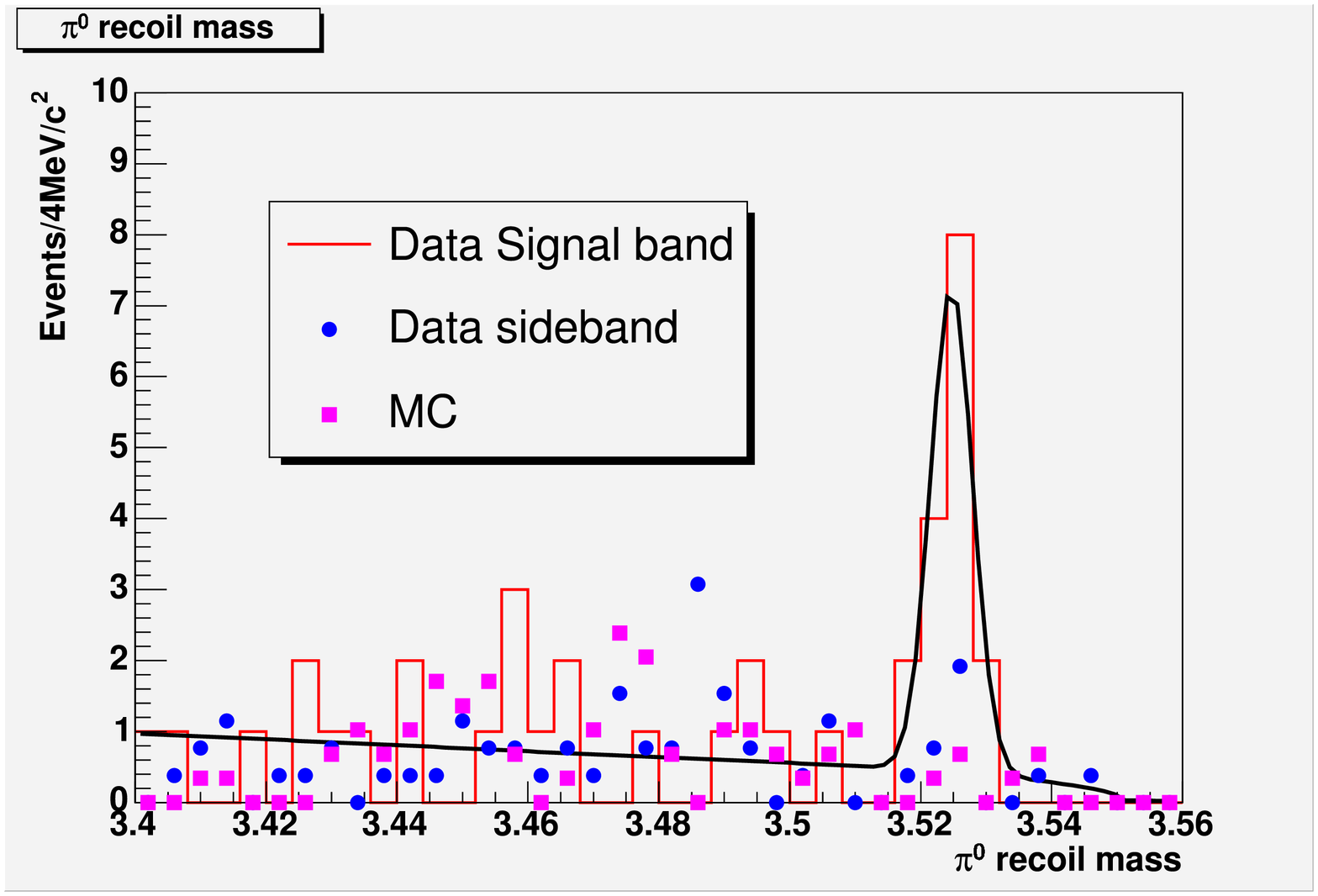}
   }}  
\caption{The preliminary evidence from CLEO for the $h_c$ singlet.
On the left is the recoil mass spectrum against the $\pi^{0}$ in the
{\it in}clusive analysis that only uses that pion and the photon
in the $E1$  decay $h_{c}\to\gamma\eta_{c}$.  To the right is the
same variable for the {\it ex}clusive analysis in which six of the
dominant $\eta_{c}$ decay modes are used. In both cases the fits are 
to an ARGUS background and a resolution function from Monte Carlo studies.
The exclusive plot also shows data events from the $\eta_{c}$ sideband
region and Monte Carlo simulation events of other $\psi^{\prime}$ decays.}
 \label{fig:hc} 
\end{figure}

The inclusive analysis shows an enhancement at over $3\sigma$ 
significance at a mass of $3524.4 \pm 0.7_{stat}$~MeV.  The systematic
uncertainty is $\sim 1$~MeV.  The left plot in \Figure~\ref{fig:hc} 
shows the fit of the data to the resolution function from Monte Carlo
simulation and an ``ARGUS'' background shape.

Shown in the right panel of that figure is the exclusive analysis,
with a 
statistical significance of $\sim 5\sigma$.  The figure shows 
the data with, again, a fit to an ARGUS background and detector
resolution function.  Also shown are the events from the sideband
of the invariant mass spectrum of the $\eta_{c}$ reconstruction
and the spectrum from a $\psi^{\prime}$ Monte Carlo simulation that
does not include the $h_{c}$ decay chain.  Further
checks on backgrounds peaking in the
signal region are under way.  The mass from the exclusive analysis is
$3524.4 \pm 0.9_{stat}$ Mev, with systematic studies ongoing.  
All of these CLEO results on the $h_{c}$ are considered
preliminary.
As a final remark, we can comment that the 20 years old search for
this state is not over yet, and its evidence is still weak.  It is
therefore necessary to (a) consolidate the evidence for such a state
from either B or $\psi(2S)$ decays , (b) to measure its mass at better
than 1--2~MeV, (c) to prove its coupling to $\ppbar$, before planning
to precisely measure its mass, total width and partial widths in
formation from $\ppbar$ annihilations.

\section{States close to open flavour thresholds}

\subsection[$R$ values between 3.7 and 5~GeV]
           {$R$ values between  3.7 and 5~GeV
            $\!$\footnote{Author: Z.~Zhao}}
\label{sec:spexopencha}

The $R$ value to be discussed in this section is one of the most 
fundamental quantities in particle physics that is defined as,
\begin{equation}                                                               
R=\frac{\sigma (e^+e^-\rightarrow hadrons)}{\sigma
(e^+e^-\rightarrow\mu^+\mu^-)}
\end{equation}     

$R$ value is expected to be constant so long as the centre-of-mass
(c.m.)  energy $E_{cm}$ does not overlap with resonances or the
threshold of production of a new quark flavour.  A thorough review of R
measurements on the full energy range can be found in
\Chapter~\ref{chapter:precisiondeterminations}, while this subsection
focuses on its complex structure in the energy region between 3.7~GeV
and 5~GeV.

The most striking feature of the $R$ values below 5~GeV is the complex
structure in the energy region between 3.7~GeV and 4.5~GeV. Besides
the resonance of $\psi(3770)$, broad resonance like structures peaking
at around 4.04, 4.1 and 4.41~GeV have not been well understood in
terms of their components and decay channels. These resonances near
the charm threshold were observed more than 20 years ago
~\cite{Burmester:1976mn,Brandelik:1978ei,Siegrist:1976br,Rapidis:1977cv,%
       Peruzzi:1977ms,Siegrist:1981zp,Abrams:1979cx,Bacino:1977uh}.
\Table~\ref{tab:respara} lists the resonance parameters reported by
these experiments.

\begin{table}[!h]
\caption{Resonance parameters measured for the broad 
         structures between 3.7 and 4.5~GeV}
\label{tab:respara} 
\begin{center}
\begin{tabular}{|c|c|l|l|l|} \hline\hline
Resonance & Experiment&Mass(MeV) &$\Gamma_{tot}$(MeV)&$\Gamma_{ee}$(eV)\\
\hline
             &MARK~I & $3772  \pm~6  $ & $28\pm~8    $ & $345 \pm~85$ \\
$\psi(3770)$ &DELCO  & $3770  \pm~6  $ & $24\pm~5    $ & $180 \pm~60$ \\
             &MARK~II& $3764  \pm~5  $ & $24\pm~5    $ & $276 \pm~50$ \\ 
             &BES(~\cite{Eidemuller:2002ru})   & $3772.7\pm~1.6$ & $24.4\pm~4.3$ & $190 \pm~25$ \\\hline
$\psi(4040)$ &DASP   & $4040  \pm~10 $ & $52\pm~10   $ & $750 \pm~150$\\ 
             &BES(~\cite{Eidemuller:2002ru})  & $4050.4\pm~4.3$ & $98.5\pm~12.8$ &$1030\pm~110$\\ 
             &BES(~\cite{Seth:2004py})  & $4040\pm~1$ & $89\pm~6$ &$911\pm~130$\\ 
             &CB(~\cite{Seth:2004py})      & $4037\pm~2$ & $85\pm~10$ & $ 880 \pm 110$ \\ \hline
$\psi(4160)$ &DASP   & $4159  \pm~20 $ & $78\pm~20   $ & $770 \pm~230$\\ 
             &BES(~\cite{Eidemuller:2002ru})    & $4166.5 \pm~6.1 $ & $55.9\pm~12.3$ & $370\pm~81 $ \\ 
             &BES(~\cite{Seth:2004py})    & $4155\pm~5 $ & $107\pm~16$ & $840\pm~130 $ \\ 
             &CB(~\cite{Seth:2004py})      & $4151\pm~4$ &  $   107\pm~10$ & $ 830 \pm 80$ \\ \hline
$\psi(4415)$ &DASP   & $4417  \pm10  $ & $66\pm~15   $ & $490 \pm~130$ \\
             &MARK~I & $4414  \pm~7  $ & $33\pm~10   $ & $440 \pm~140$ \\
             &BES(~\cite{Eidemuller:2002ru})    & $4429.4 \pm~8.5 $ & $86.0\pm~20.9  $ & $390 \pm~74$ \\ 
             &BES(~\cite{Seth:2004py})    & $4429\pm~9 $ & $118\pm~35$ & $640\pm~230 $ \\ 
             &CB(~\cite{Seth:2004py})      & $4425\pm~6$ &  $   119\pm~16$ & $ 720 \pm 110$ \\\hline \hline
\end{tabular}
\end{center}
\end{table}

\subsubsection{PLUTO measurement between 3.1 and 4.8~GeV}
The PLUTO Collaboration measured $R$ values with the magnetic detector
PLUTO at the $e^+e^-$ storage ring DORIS between 3.1 and 4.8~GeV
c.m. energy.  A superconducting coil procedures a 2T magnetic field
parallel to the beam axis.  Inside coil there are 14 cylindrical
proportional wire chambers and two lead converter, a 2~mm converter at
radius 37.5~cm and a 9~mm converter at radius 59.4~cm.  Two or more
charged tracks are triggered and selected as hadronic event
candidates.  The background from beam--gas interaction and cosmic ray
events is subtracted using the distribution of reconstructed event
vertices alone the beam direction. Monte Carlo events are generated
according to isotropic phase space to determine the detection
efficiency for the hadronic events. An external luminosity monitor
system is employed to observe the beam luminosity. The uncertainty of
the luminosity measurement is about $\pm 5\%$. The systematic error in
$R$ values is estimated to be about 12\%.  PLUTO results agree with
those of the SLAC--LBL group within systematic errors, but is about
10--15\% lower than those of SLAC--LBL on the narrow $J/\psi$ resonance
and higher energies. However, the agreement on the energy dependence
and the structure of the $R$ values is quite good. The accuracy of
PLUTO's measurement is limited by systematic error, which amount to
almost one unit in $R$ in the broad resonance region. The resonance
parameters of the broad resonances cannot be determined with such a
limited accuracy and energy points.

\subsubsection{DASP measurement between 3.6 and 5.2~GeV}
DASP Collaboration measured $R$ values at c.m. energy between 3.6 and
5.2~GeV with a non magnetic inner detector of the double arm
spectrometer DASP, which has similar trigger and detection
efficiencies for photon and charged particles. The inner detector of
DASP is mounted between the two magnet arms of DASP. It is azimuthally
divided into eight sectors, six of which consist of scintillation
counters, proportional chambers, lead scintillator sandwiches and tube
chambers, and the remaining two facing the magnet aperture, have only
scintillation counter and proportional chambers. Tracks are recorded
over solid angle of 62\% for photon and 76\% of $4\pi$ for charged
particles. DASP collected a total integrated luminosity of 7500
$nb^{-1}$, which was determined by small angle Bhabha scattering
measured by four identical hodoscopes with an uncertainty of 5\%. The
additional normalization uncertainty is estimated to be 15\%. The
uncertainties of the detection efficiencies for the hadronic events is
about 12\%. Three peaks centred around 4.04, 4.16 and 4.42 are
observed. The data are insufficient to resolve structures between 3.7
and 4.5~GeV. By making a simplifying assumption that the cross-section
can be described by an incoherent sum of Breit--Wigner resonances and a
non resonant background, DASP reported resonance parameters as listed
in \Table~\ref{tab:respara}.

\subsubsection{SLAC--LBL measurement between 2.6 and 7.8~GeV}

SLAC--LBL group did a $R$ scan with MARK~I at SPEAR which operated at
c.m. energy between 2.6 and 7.8~GeV with peak luminosity between
$10^{29}$ and $10^{31}$~cm$^{-2}$ sec$^{-1}$.  MARK~I was a general
purpose collider detector of the first generation. Its solenoidal
magnet provide a near uniform magnetic field of $3891\pm~1$ G over a
volume 3.6~m long and 3.3~m in diameter. A pipe counter consisting of
four hemicylindrical plastic counters surrounding the vacuum pipe were
used to reduce the trigger rate of cosmic ray.  Two sets of
proportional wire chambers on the outside of the pipe counters had
spacial resolution of 700 $\mu$m. Four modules of concertric
cylindrical wire spark chambers were the main tracking elements of the
detector, which gave a spacial resolution in the azimuthal direction
of 340 $\mu$m, 1.0 and 0.5~cm for the $2^0$ and $4^0$ stereo gaps,
respectively. Outside the spark chamber was an array of 48 plastic
scintillation counters with a width of 20~cm each. The time-of-flight
for this system was about 480 psec. An array of 24 shower counters
made of five layers, each consisting of 0.64~cm of pilot F
scintillator and 0.64~cm of lead. The energy resolution measured with
Bhabha events was $\Delta E/E=35\%/\sqrt E$. The muon-identification
spark chamber, the end-cap spark chamber, and the photon-detection
capabilities of the shower counters were not used in this analysis.
The $R$ values and the corresponding resonance parameters in the
energy region between 3.4 and 5.5~GeV is plotted together with those
from PLUTO and DASP in \Figure~\ref{fig:resall} (right).

MARK~I studied exclusive decay channels on the resonance at 4040~MeV
and reported~\cite{Goldhaber:1977qn} $PsPs:PsV:VV = 0.05 \pm 0.03 : 1
: 32 \pm 12$, where $Ps$ represents $D$ meson and V stands for $D^*$
meson.  These early results stimulated a variety of theoretical
interpretations.

\subsubsection{BES measurement between 2 to 5~GeV}

BES Collaboration has done a $R$ scan with updated Beijing
Spectrometer (BES~II) at Beijing Electron--Positron Collider(BEPC).

The trigger efficiencies, measured by comparing the responses to different 
trigger requirements in $R$ scan data and special runs taken at the $J/\psi$ 
resonance, are determined to be 99.96\%, 99.33\% and 99.76\% for Bhabha, 
dimuon and hadronic events, respectively. 

BES's measurement first selects charged tracks, then hadronic events
with charged tracks equal and greater than two. The number of hadronic
events and the beam-associated background level are determined by
fitting the distribution of event vertices along the beam direction
with a Gaussian for real hadronic events and a polynomial of degree
two for the background.

The subtraction of the beam-associated backgrounds is cross checked by
applying the same hadronic event selection criteria to separated-beam
data.

A new Monte Carlo event generator called LUARLW is developed together
with LUND group for the determination of detection efficiencies of the
hadronic events~\cite{Andersson:1999ui}. LUARLW removes the
extreme-high-energy approximations used in JETSET's string
fragmentation algorithm. The final states simulated in LUARLW are
exclusive in contrast to JETSET, where they are inclusive. In
addition, LUARLW uses fewer free parameters in the fragmentation
function than JETSET. Above 3.77~GeV, the production of charmed mesons
is included in the generator according to the Eichten
Model~\cite{Eichten:1979ms,Chen:2000tv}.

\begin{figure}[t]
  \begin{minipage}{.46\linewidth}
    \includegraphics[width=\linewidth]{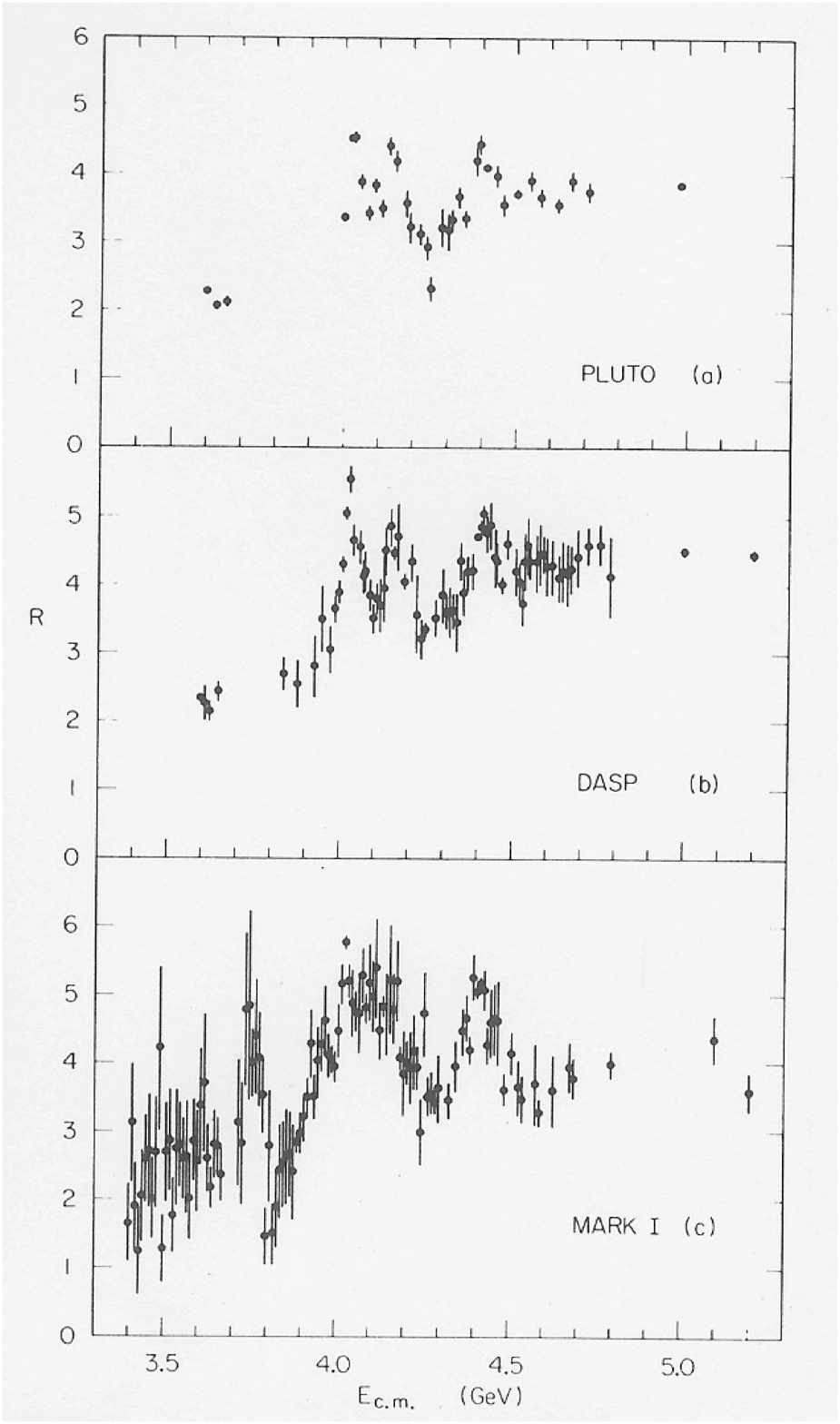}
  \end{minipage}
\hfill
  \begin{minipage}{.52\linewidth}
    \includegraphics[width=\linewidth]{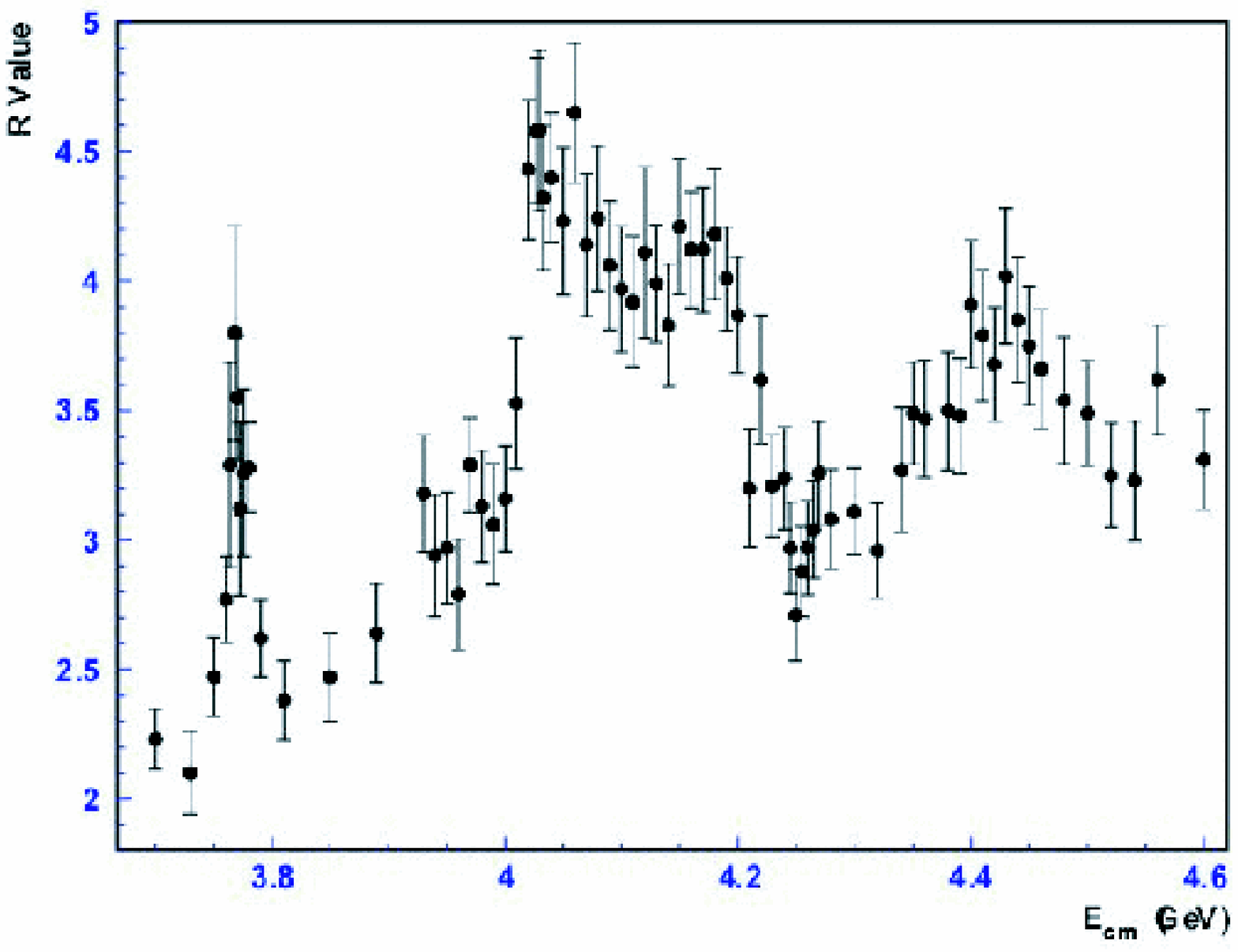}
  \end{minipage}
    \caption[]{$R$ Values between 3.7 and 5~GeV from PLUTO, DASP and MARKI (left), and BES (right)
 experiments.}
    \label{fig:resall}
\end{figure}

Different schemes for the radiative corrections were compared
~\cite{Berends:1980jk,Bonneau:1971mk,Kuraev:1985hb,Edwards:1990pc}.
Below charm threshold the four different schemes agree with each other
to within 1\%.  Above charm threshold, where resonances are important,
the agreement is within 1 to 3\%.  The formalism of
Ref.~\cite{Edwards:1990pc} is used in our calculation, and differences
between it and the schemes described in Ref.~\cite{Kuraev:1985hb} are
included in the systematic errors. In the calculation of the radiative
correction above charm threshold, where the resonances are broad and
where the total width of the resonance is related to the energy, we
take the interference between resonances into account.  The integrated
luminosity is determined to a precision of 2--3\% from the number of
large-angle Bhabha events selectedusing only the BSC energy
deposition.  \Figure[b]~\ref{fig:resall} (right) shows the BES $R$ scan
results between 3.6 and 4.6~GeV.

Previously, BES Collaboration measured cross-section for charm meson
production, using 22.3 pb$^{-1}$ of $e^+e^-$ data collected with BES~I
at $\sqrt{S}$=4.03 and 15 pb$^{-1}$ at 4.14~GeV~\cite{Bai:1999tv}. The
charmed mesons used in this measurement are $D^0$ and $D^+$, of which
the number of signal events are selected by fitting the inclusive
$K^-\pi^+$ and $K^-\pi+\pi^+$ invariant mass distribution with
Gaussian as signal plus a third order of polynomial background.
Taking into account the detection efficiency, the correction of
initial state radiation, and quote the corresponding braching ratio
from PDG1998, BES reported their results as shown in
\Table~\ref{tab:besCharmCross}, together with that predicted by the
coupled channel model.  
\shortpage

\begin{table}[!h]
\caption[Comparison of tree level cross-section measurement of BES]
        {Comparison of tree level cross-section measurement of BES
         with predictions of the coupled channel model. experimental
         $D_s$ cross-section is taken from early work.}
\label{tab:besCharmCross} 
\begin{center}
\begin{tabular}{|c|c|c|} \hline
$\sqrt{s}$=4.03~GeV                   &  Experiment                 &  Coupled channel model \\ \hline       
$\sigma_{D^0}$+$\sigma_{\bar{D^0}}$     &  19.9$\pm$0.6$\pm$2.3 nb    &  18.2 nb               \\
$\sigma_{D^+}$+$\sigma_{D^-}$           &   6.5$\pm$0.2$\pm$0.8 nb    &   6.0 nb               \\
$\sigma_{D_s^+}$+$\sigma_{D_s^-}$       &   0.81$\pm$0.16$\pm$0.27 nb &  11.6 nb               \\
$\sigma_{charm}$                      &  13.6$\pm$0.3 $\pm$1.5 nb     &  12.9 nb               \\ \hline

$\sqrt{s}$=4.14~GeV                   &  Experiment                 &  Coupled channel model \\  \hline      
$\sigma_{D^0}+\sigma_{\bar{D}^0}$     &   9.3$\pm$2.1$\pm$1.1 nb    &  15.1 nb               \\
$\sigma_{D^+}+\sigma_{D^-}$           &   1.9$\pm$0.9$\pm$0.2 nb    &   4.5 nb               \\
$\sigma_{D_s^+}+\sigma_{D_s^-}$       &   1.64$\pm$ 0.39$\pm$ 0.42 nb   &   1.85 nb               \\
$\sigma_{charm}$                      &   6.4$\pm$1.2$\pm$0.7 nb    &  10.7 nb               \\ \hline 
\end{tabular}
\end{center}
\end{table}

\subsubsection{Remarks and prospects}
DASP data agree with those of PLUTO resonabl well in shape but exceed
their cross-sections by about half a unit in R above 4~GeV.  In
magnitude DASP's data are in closer agreement with those of SLAC--LBL
but show some difference in the finer details of the energy
dependence. For example, SLAC--LBL data didn't resolve the structure at
4.16~GeV. The total width measured by SLAC--LBL is smaller than that of
DASP measurement. Despite of these discrepances, the difference
observed among the three experiments are within the systmatic errors
quoted.
\shortpage

BES's $R$ scan is done with a newer generation detector and $e^+e^-$
collider as compared with the previous measurements, and has about 80
points in the energy region from 3.7 to 5~GeV.  Because of this fine
scan in this energy region that contributes most to the precision
evaluation of $\alpha_{QED}(M_Z)$.

BES also fitted resonances as a Breit--Wigner shape with different
continue background and takes into account the energy-dependence of
resonance width and the coherence of the
resonance~\cite{huanggs}. BES's preliminary results are consistent
with those of previous measurements for the peak positions at 3.77,
4.04, 4.16 and 4.42~GeV, and show larger $\Gamma_{tot}$ of the
resonances at 4.04 and 4.42~GeV and smaller $\Gamma_{ee}$ of the
resonaces at 3.77 and 4.16~GeV.

Fitting BES's $R$ data between 3.7 and 4.6~GeV (75 data points) with
Breit--Wigner resonances and none resonant background based on
perturbative QCD~\cite{Eidemuller:2002ru}, one obtain resonance
parameters as listed in \Table~\ref{tab:respara}. The results from
this fit has similar conclusion as the one from BES's, except that
$\Gamma_{tot}$ is no longer larger than the other measurements of the
resonance at 4.42~GeV.

\begin{figure}[t]
\centering\includegraphics[width=.58\linewidth]{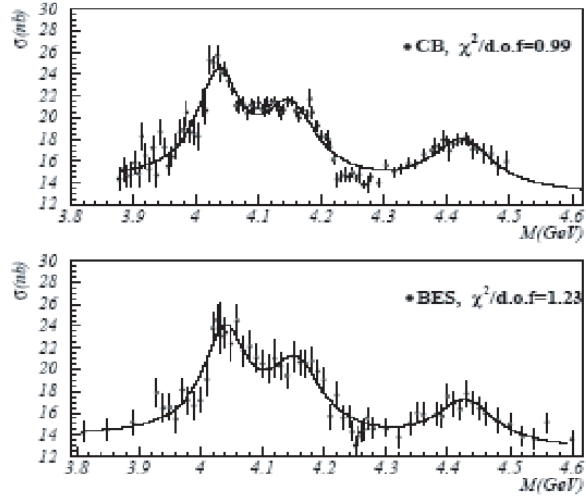}
\caption[]{Refit the R data of CB and BES~II.} 
\label{fig:CB-BES}
\end{figure}

Recently, Kamal K. Seth refitted resonance parameters of the higher
vector states of charmonium with existing $R$
data~\cite{Seth:2004py}. Three Breit--Wigner resonances plus background
that is parametrized with a linear function. He shows that the Crystal
Ball (CB) and BES measurements are in excellent agreement. The
analysis of the CB and BES data leads to consistent resonance
parameters for the three vector resonances above the $D\bar{D}$
threshold.  The masses of the three resonances determined by him in
general agree with PDG, but have much smaller errors. However, the
total widths of these three resonances determined by this work are
about 67\%, 37\% and 179\% larger than those adopted by PDG. The
corresponding electron widths determined by this work are 23\%, 8\%
and 51\% larger with about a factor of 2 less errors.
\Figure[b]~\ref{fig:CB-BES} shows the fits to CB and BES data.

A factor of 2 to 3 reduction in uncertainty in the energy region of
3--5~GeV significantly improved the experimental situation, providing
an opportunity to directly test QCD sum role where the notion of
quark--hadron duality (QHD) plays a dominant
role~\cite{Eidemuller:2002ru}, and evaluate charm quark mass via
experimental data to a precision below 10\%.  However, BES's data is
still not enough, in terms of both statistics and systematic error
restriction, to provide a clear picture of the broad resonance
structures. To fully understand the complicated structures at the
energies between 3.7 and 4.5~GeV, one needs to:
\begin{itemize}
\item[--] perform the $R$ scan with smaller energy steps and higher
          statistic in the entire energy region to a precision around 2--3\%.
\item[--] collect data at the peak positions with high enough
          statistics to study the exclusive decay channels of the resonances.
\end{itemize}
These could be the important physics topics for CLEO-c at CESR-c and
BES~III at BEPC~II~\cite{cleoc,besiii}.  Both CLEO-c and BES~III may
have the ability to clarify the ambiguity that has been bothering
physicists for over 20 years.

\subsection[X(3872): discovery and interpretations]
           {X(3872): discovery and interpretations 
            $\!$\footnote{Author: B.~Yabsley}}
\label{sec:spexX3872}

The $X(3872)$ is a narrow state decaying into $\pi^+ \pi^- J/\psi$,
with a mass $M_X \sim 3872\,\mathrm{MeV}$.  Given the observed final
state and the observed mass, in the charmonium region, it is natural
to assume that the $X(3872)$ is itself a charmonium state.  It has
however proved difficult to identify the $X(3872)$ with any of the
expected narrow $c\bar{c}$ mesons, leading to suggestions that it may
be a more exotic particle. In this section, we briefly review the
discovery and known properties of the $X(3872)$, and the difficulties
they create for its interpretation.

\subsubsection{Discovery, confirmation, and properties}

\begin{figure}
  \begin{center}
    \includegraphics[angle=-90,width=0.9\columnwidth]{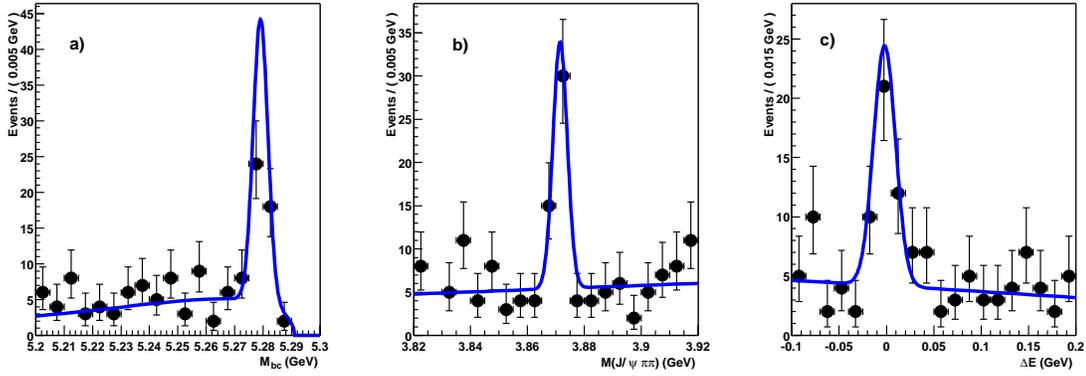}
  \end{center}
  \caption{From the Belle discovery paper~\cite{Choi:2003ue}:
        projections of the data (points with error bars) and the results
        of an unbinned maximum likelihood fit (solid curve) for the
        $X(3872) \to \pi^+ \pi^- J/\psi$ signal region. 
        The variables
        (a) beam-constrained mass $M_{bc} = \sqrt{E_{{\rm beam}}^2 - p_B^2}$,
        (b) invariant mass $M_{\pi^+ \pi^- J/\psi}$, and
        (c) energy difference $\Delta E = E_B - E_{{\rm beam}}$,
        are those used in the fit; $E_B$ and $p_B$ are the energy and momentum
        of the $B^\pm \to K^\pm \pi^+ \pi^- J/\psi$ candidate,
        and $E_{{\rm beam}}$ the energy of either $e^\pm$ beam, 
        in the $e^+ e^-$ centre-of-mass system.}
  \label{fig:x3872-belle}
\end{figure}

\begin{figure}
  \begin{center}
    \begin{tabular}{c@{\hspace*{0.05\columnwidth}}c}
      \includegraphics[width=0.4\columnwidth]{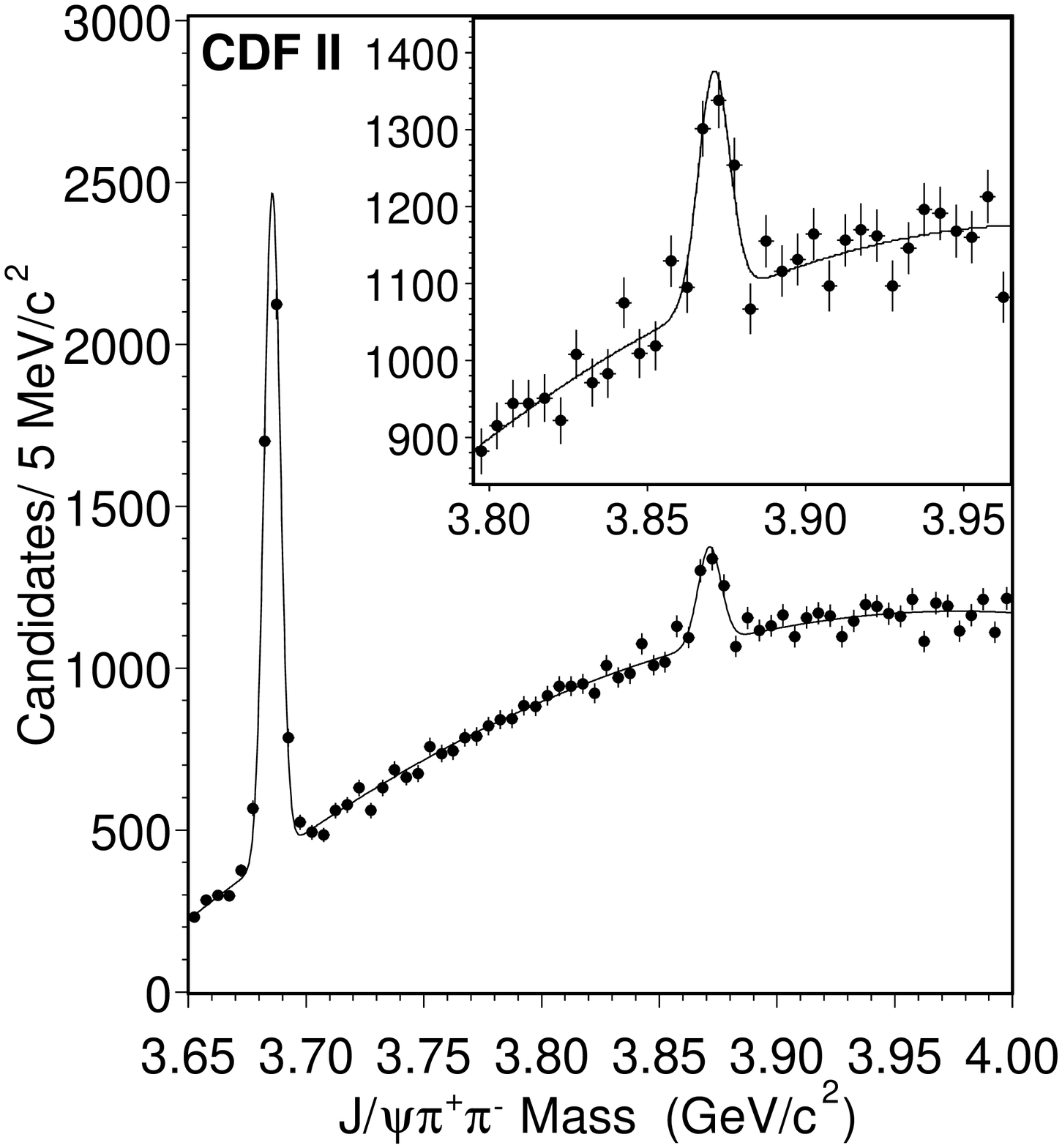}  &
          \includegraphics[width=0.47\columnwidth]{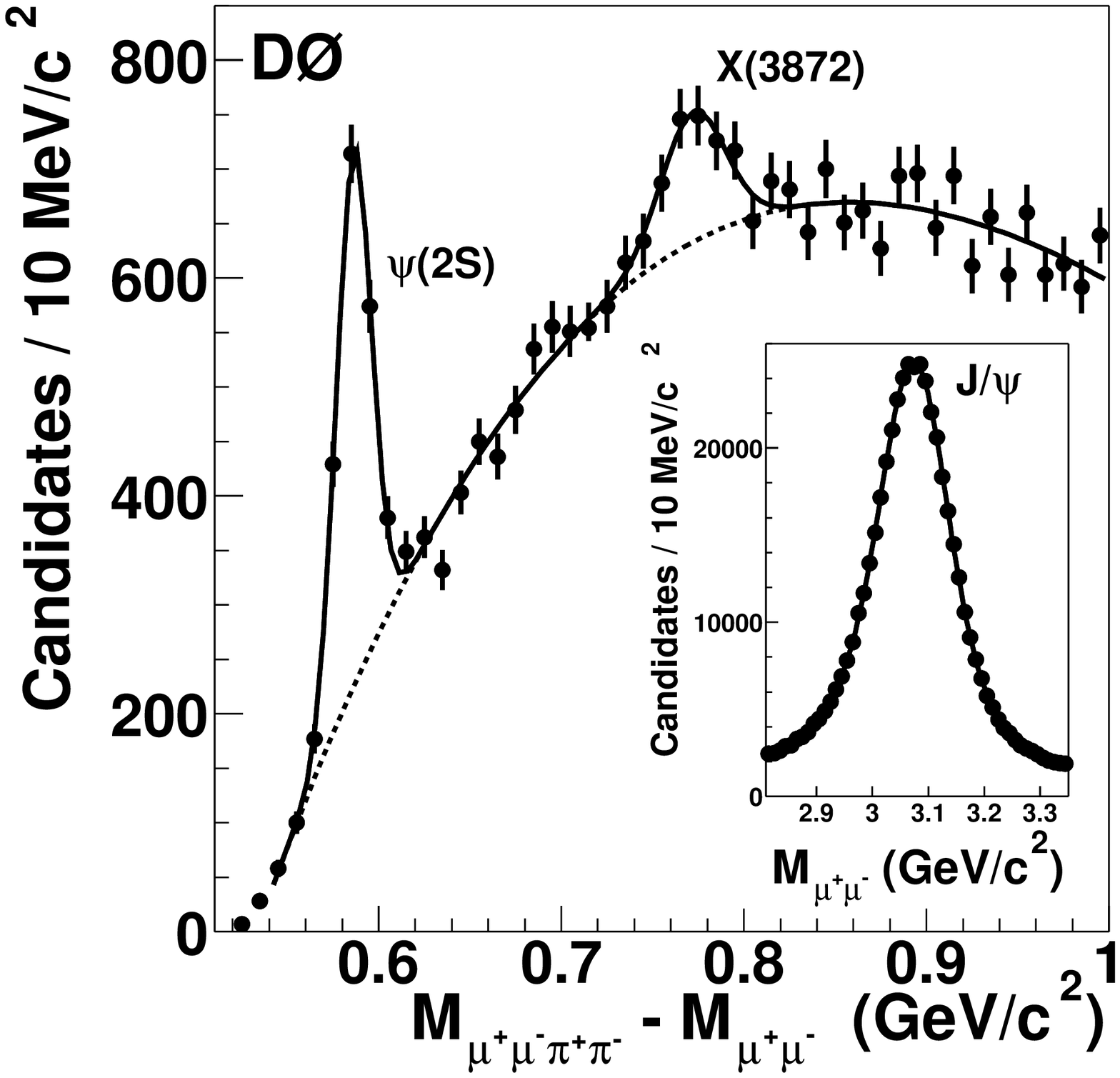}
    \end{tabular}
  \end{center}
  \caption{Confirmation of the $X(3872)$ in inclusive $p\overline{p}$
        collisions by CDF~\cite{Acosta:2003zx} (left) and D0~\cite{Abazov:2004kp} (right).
        In each case peaks due to
        $\psi'$ and $X(3872) \to \pi^+ \pi^- J/\psi$ can be clearly seen;
        the insets show (left) an enlargement of the $X(3872)$ region and
        (right) the mass distribution for the $J/\psi \to \mu^+ \mu^-$
        candidates used in the analysis.}
  \label{fig:x3872-ppbar}
\end{figure}

\begin{figure}[p]
  \begin{center}
    \includegraphics[width=0.6\columnwidth,bb=0 35 425 425]{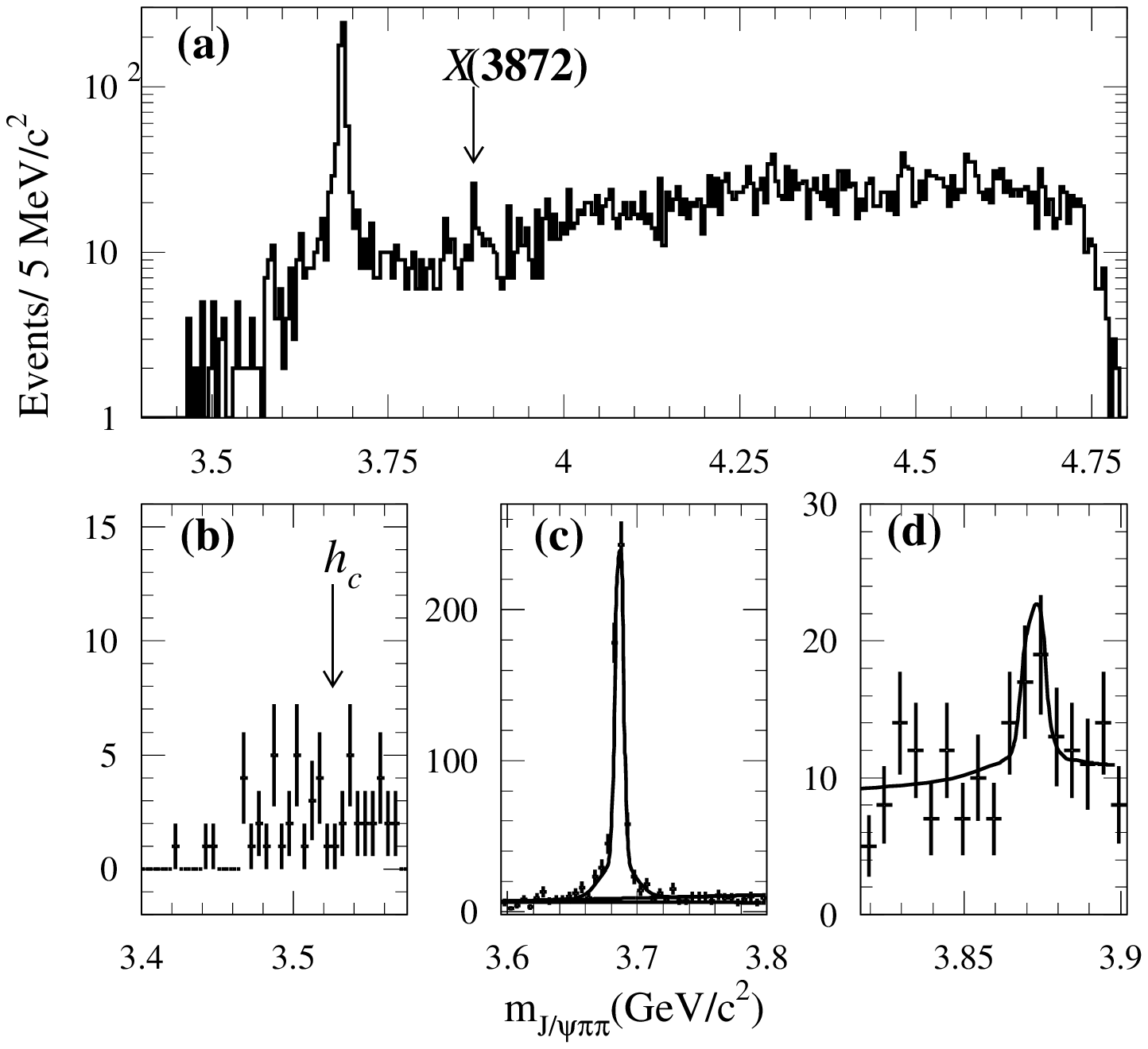}
  \end{center}
  \caption{Confirmation of the $X(3872)$ in $B^\pm \to K^\pm \pi^+ \pi^- J/\psi$
        decay from BaBar~\cite{Aubert:2004ns}.
        Distributions of the $\pi^+ \pi^- J/\psi$ invariant mass are shown for
        $B$ candidates in (a) the $B$-signal region, together with expanded
        views of the (b) $h_c$, (c) $\psi(2S)$, and (d) $X(3872)$ mass regions.
        In (c) and (d), the results of an unbinned maximum likelihood fit
        to the data are superimposed as a solid curve.}
  \label{fig:x3872-babar}
 \medskip
  \begin{center}
    \includegraphics[width=0.6\columnwidth,bb=0 0 552 490]{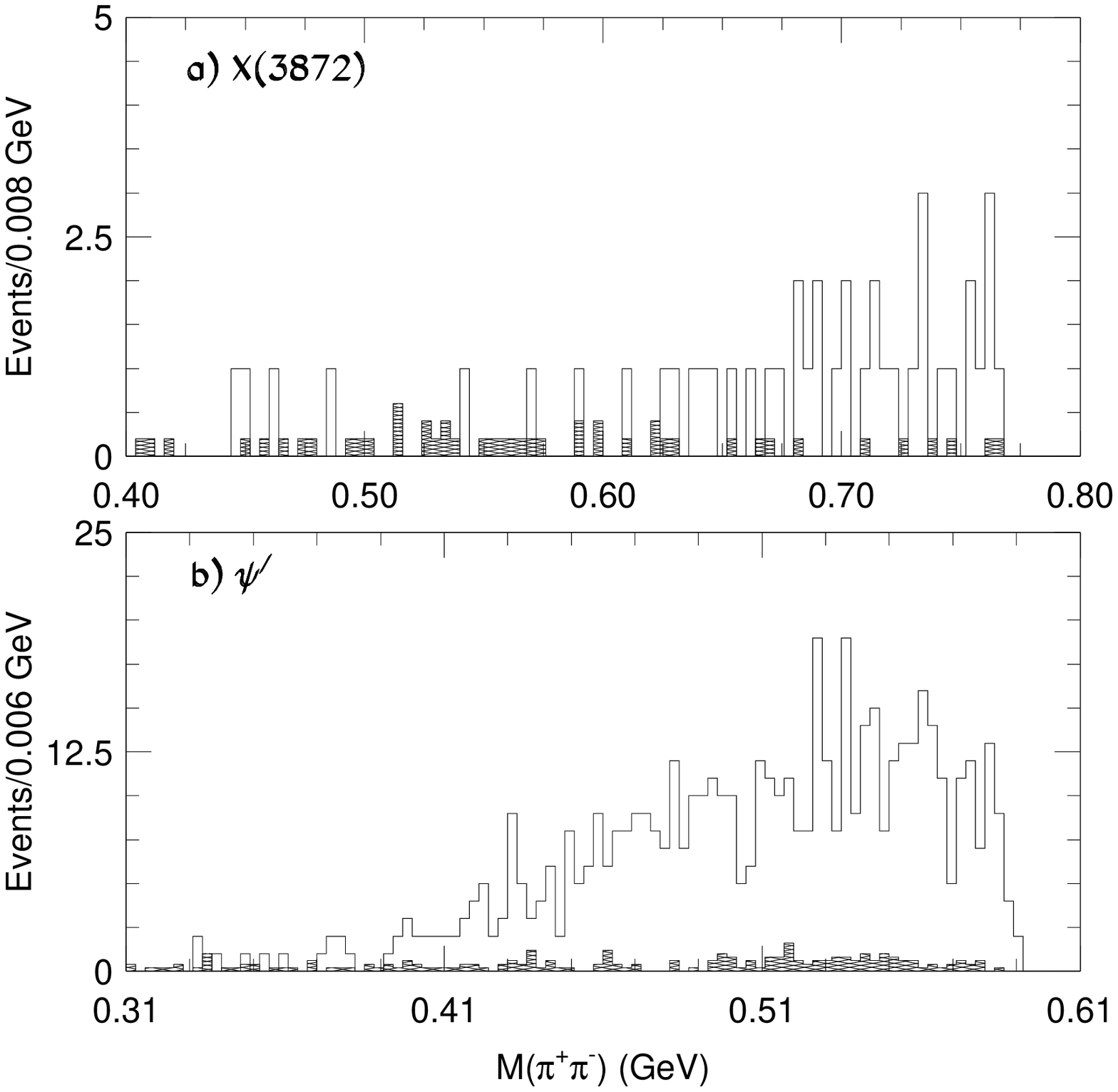}
  \end{center}
  \caption{From~\cite{Choi:2003ue}:
        $M(\pi^+\pi^-)$  distribution for events in the
        (a) $M(\pi^+\pi^-J/\psi)=3872\,\mathrm{MeV}$ signal region and
        (b) the $\psi'$ region in Belle data.
        The shaded histograms are sideband data
        normalized to the signal-box area.
        Note the different horizontal scales.}
  \label{fig:x3872-belle-Mpipi}
\end{figure}

The $X(3872)$ was discovered by the Belle collaboration in a study of 
$B^\pm \to K^\pm \pi^+ \pi^- J/\psi$ decays~\cite{Choi:2003ue}.
In addition to the well-known $\psi'$, a second peak was seen in the
$M(\pi^+ \pi^- J/\psi)$ distribution;
the results of an unbinned maximum likelihood fit to the $X(3872)$
signal region in $M$, and two other variables which peak in the case of
$B^\pm \to K^\pm \pi^+ \pi^- J/\psi$ decay, are shown in
\Figure~\ref{fig:x3872-belle}. A yield of $35.7 \pm 6.8$ events was observed,
with high significance ($10.3\sigma$), and the width of the mass peak was found
to be consistent with the detector resolution.
As the measured mass is well above the $D\overline{D}$ open charm threshold,
the narrow width implies that decays to $D\overline{D}$ are forbidden;
Belle~\cite{Abe:2003zv} reports
$\Gamma(X(3872)$ $\to$ $ D\overline{D})/\Gamma(X(3872)$ $\to\pi^+\pi^-
J/\psi)$ $ < 7$
(90\% CL), to be compared with a corresponding value $> 160$ for the
$\psi(3770)$~\cite{Bai:2003hv}.
Comparing decays via the $X(3872)$ to those via the $\psi'$,
Belle finds a considerable production rate in $B$ decays,
with product branching ratio
\be
  \frac {{\cal B}(B^+ \to K^+ X(3872)) \times
         {\cal B}(X(3872) \to \pi^+\pi^- J/\psi)}
        {{\cal B}(B^+ \to K^+ \psi') \times
         {\cal B}(\psi' \to \pi^+ \pi^- J/\psi)}        
        = 0.063 \pm 0.012\,{\rm (stat)} \pm 0.007\,{\rm (syst)}.
\\
  \label{eq:x3872-belle-branching}
\ee
The observation has been confirmed in inclusive $p\overline{p}$ collisions by
CDF~\cite{Acosta:2003zx} and D0~\cite{Abazov:2004kp},
as shown in \Figure~\ref{fig:x3872-ppbar},
and in exclusive $B$ meson decays by BaBar~\cite{Aubert:2004ns},
shown in \Figure~\ref{fig:x3872-babar}.
The observed masses are consistent, with a weighted average value
\begin{equation}
  M_X = (3871.9 \pm 0.5)\,\mathrm{MeV}
  \label{eq:x3872-av-mass}
\end{equation}
across the four experiments~\cite{Choi:2003ue,Acosta:2003zx,Abazov:2004kp,Aubert:2004ns}.
Each of CDF, D0, and BaBar likewise find a width consistent with the detector
resolution; the only limit is that inferred by Belle~\cite{Choi:2003ue},
\begin{equation}
  \Gamma_X < 2.3\,\mathrm{MeV}\; (90\%\;{\rm CL}).
  \label{eq:x3872-belle-width}
\end{equation}

Belle finds a $M(\pi^+ \pi^-)$ distribution concentrated at the kinematic
boundary~\cite{Choi:2003ue},
coinciding with the $\rho$ mass (\Figure~\ref{fig:x3872-belle-Mpipi}).
This is confirmed by CDF~\cite{Acosta:2003zx}, who find little signal with
$M(\pi^+ \pi^-) < 500\,\mathrm{MeV}$; BaBar report that their statistics
are too small to allow a clear conclusion, but do not exclude a concentration
at the boundary~\cite{Aubert:2004ns}.

\subsubsection{Decay modes and interpretation of the $X(3872)$}

The Belle collaboration has performed searches for various 
decay modes~\cite{Choi:2003ue,Olsen:2004fp} and an angular distribution
study~\cite{Olsen:2004fp}, to compare $X(3872)$ properties with those 
of predicted, but so far unseen, charmonium states. They restrict their
attention to states with
\begin{enumerate}
  \item expected masses~\cite{Godfrey:1985xj} within 200~MeV of
        $M_X \simeq 3872\,\mathrm{MeV}$;
  \item unnatural quantum numbers $J^P = 0^-,\, 1^+,\, 2^-,\ \ldots$
        since decays to $D\overline{D}$ are not seen; and
  \item spin angular momentum $J < 3$, since the state is seen in exclusive
        $B \to K X(3872)$ production with a significant rate, making high
        $J$ unlikely (\emph{cf.}\ $B^+ \to K^+ \chi_{c2}$, still
        not observed, and $B^+ \to K^+ \chi_{c0}$ and $K^+ \chi_{c1}$ with
        branching fractions $(6\sim7) \times 10^{-4}$).
\end{enumerate}
The $1^3D_3$ state, $\psi_3$, is also studied
following suggestions~\cite{Barnes:2003vb,Eichten:2004uh} that the rate
for $\psi_3 \to D\overline{D}$, suppressed by an $L = 3$ angular momentum
barrier, may be low.

The search therefore includes
the $C=-1$ states $\psi_2,\, h_c^\prime$, and $\psi_3$, and
the $C=+1$ states $\eta_{c2},\, \chi_{c1}^\prime$, and $\eta_c^{\prime\prime}$.
The observation of decays to $\pi^+ \pi^- J/\psi$ favors $C=-1$,
for which this mode is isospin-conserving; this would imply
$\Gamma(X \to \pi^0 \pi^0 J/\psi) \simeq \frac{1}{2}
 \Gamma(X \to \pi^+ \pi^- J/\psi)$.
On the other hand, the observed concentration of events at
$M(\pi^+ \pi^-) \approx m_\rho$ suggests that the decay may proceed
via $X(3872) \to \rho J/\psi$, an isospin-violating process;
this requires $C=+1$ and forbids the decay to $\pi^0 \pi^0 J/\psi$. 
A study of the $\pi^0 \pi^0 J/\psi$ final state is therefore important.

\subsubsection{Searches for radiative decays}

\begin{figure}
  \begin{center}
    \includegraphics[angle=-90,width=0.7\columnwidth,bb=410 38 572 731,clip]
        {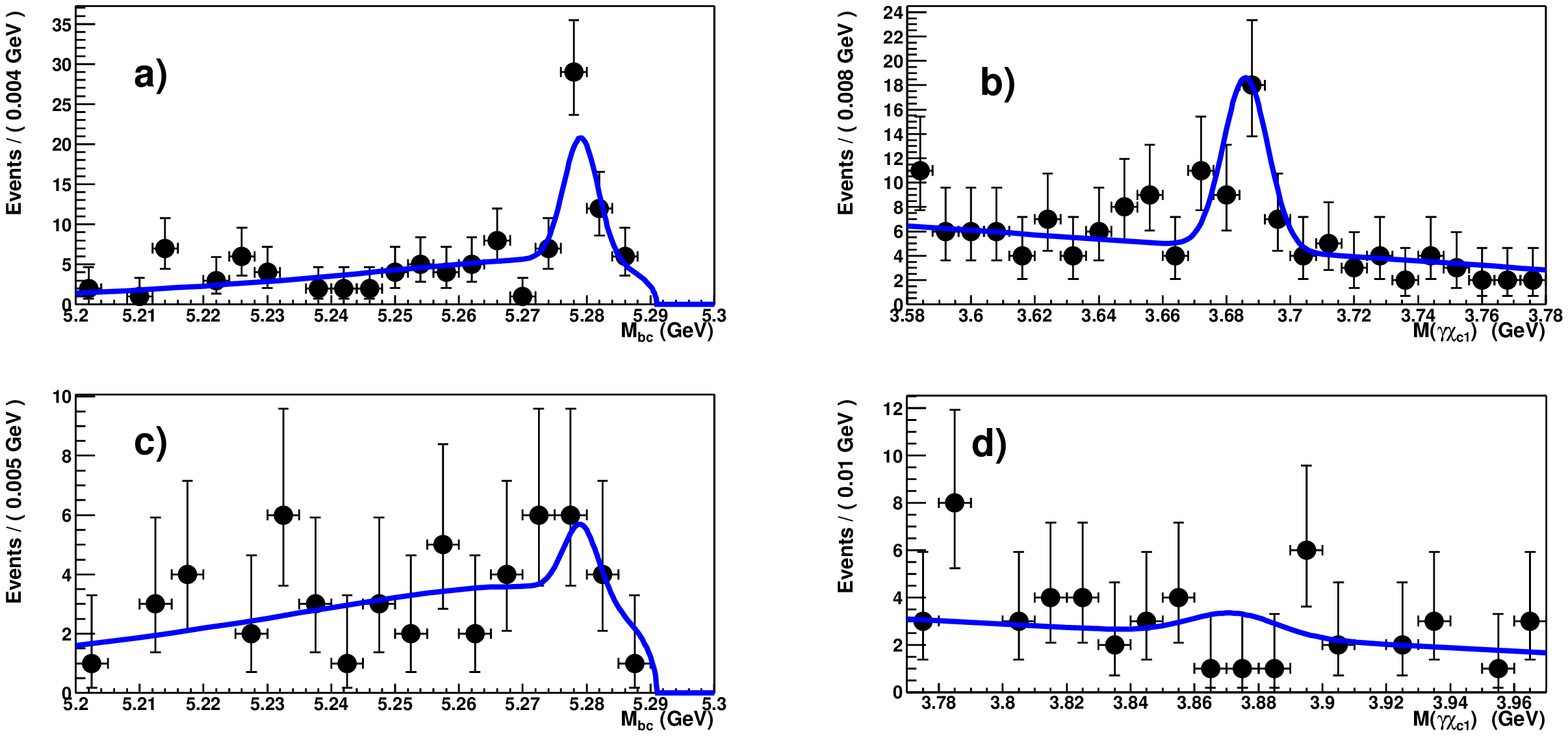}    \\
    \includegraphics[width=0.7\columnwidth]{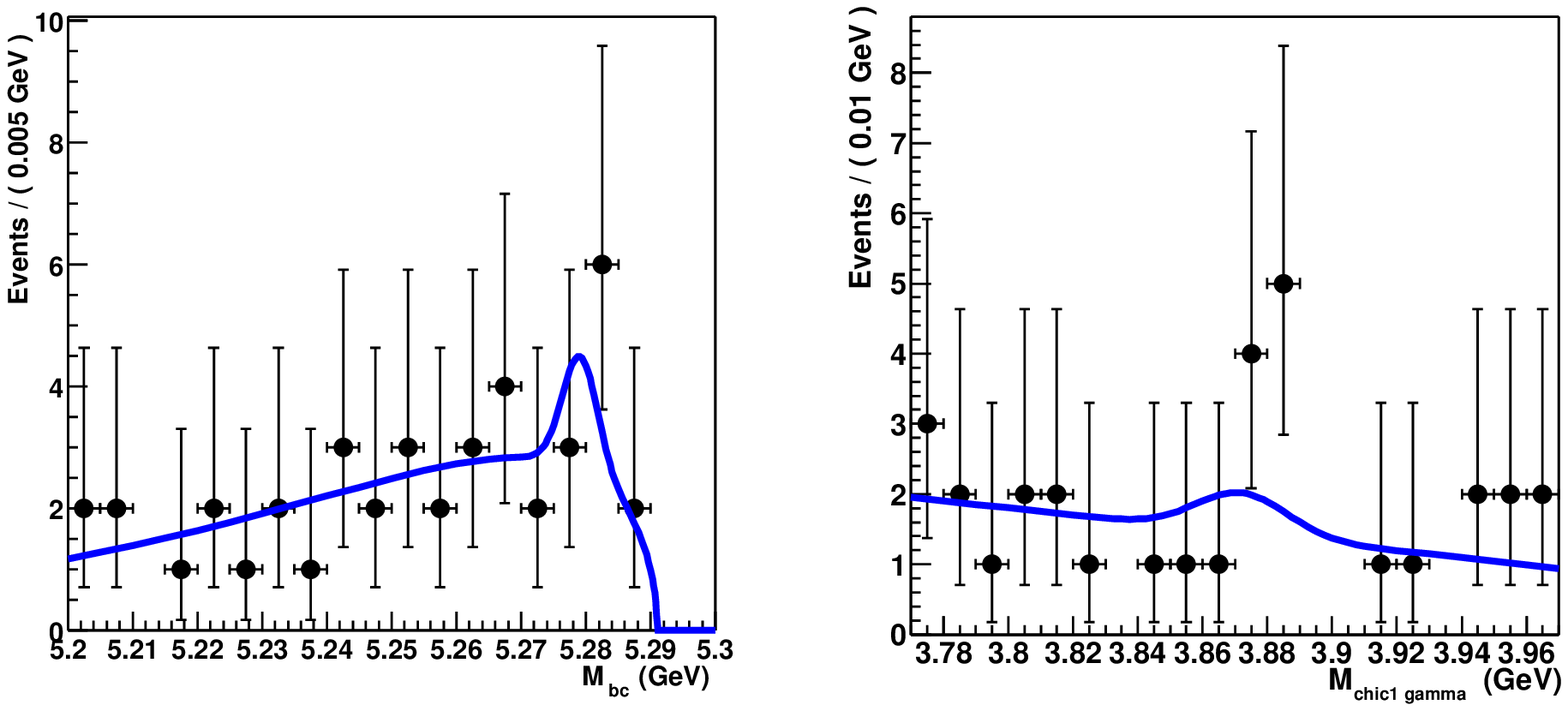}
  \end{center}
  \caption{(Upper plots) Signal band projections for the beam-constrained ($B$-candidate) mass $M_{bc}$,
        and charmonium candidate mass $M_{\gamma\chi_{c1}}$, in the Belle search for decays
        $B^+ \to K^+ X(3872)$, $X\to \gamma\chi_{c1}$~\cite{Choi:2003ue}; 
        the results of an unbinned maximum likelihood fit are superimposed.
        The signal yield, $3.7 \pm 3.7 \pm 2.2$, is consistent with zero.
        (Lower plots) Similar distributions in the search for decays to $\gamma\chi_{c2}$~\cite{Olsen:2004fp}; 
        the fitted yield is $2.9 \pm 3.0 \pm 1.5$ events.}
  \label{fig:x3872-belle-gamma-chi}
\end{figure}

If the $X(3872)$ is identified with the $1 {}^3D_2$ ($\psi_2$) state,
the decay to $\gamma \chi_{c1}$ is an allowed E1 transition with a
large partial width, calculated to be $\Gamma(X \to \gamma \chi_{c1})
\simeq 210\,\mathrm{keV}$ in potential models, taking coupled channel
effects into account~\cite{Barnes:2003vb,Eichten:2004uh}.  Similarly,
the partial width for $1 {}^3D_3\; (\psi_3) \to \gamma \chi_{c2}$ is
calculated to be $\sim 300\,\mathrm{keV}$.  This is to be compared to
the partial width for $\psi_{2,3} \to \pi^+ \pi^- J/\psi$, expected to
be equal to the $\psi(3770)$ partial width for both states.
Ref.~\cite{Olsen:2004fp} conservatively assumes $\Gamma(\psi(3770) \to
\pi^+ \pi^-J/\psi) < 130\,\mathrm{keV}$, leading to predictions
$\Gamma(X \to \gamma \chi_{cJ})/\Gamma(X \to \pi^+ \pi^-J/\psi) > 1.6$
for $\psi_2 \to \gamma\chi_{c1}$ and $> 2.3$ for $\psi_3 \to \gamma
\chi_{c2}$ respectively.  Belle has performed searches for $X(3872)$
decays to these final states (see
\Figure~\ref{fig:x3872-belle-gamma-chi}), setting upper limits on the
branching ratios (at 90\% CL) of $0.89$ for
$\gamma\chi_{c1}$~\cite{Choi:2003ue}, and $1.1$ for
$\gamma\chi_{c2}$~\cite{Olsen:2004fp}, below these expectations.
Other considerations disfavor these states.
If the $X$ is the $\psi_2$, its separation from the $\psi(3770)$, $\Delta M = 102\,\mathrm{MeV}$,
is larger than present theory can accomodate~\cite{Eichten:2004uh}.
The $\psi_3$ mass is expected to be similar. 
Production $B^+ \to K^+ \psi_3$ is also expected to be suppressed relative to other
$K^+ (c\bar{c})$ decays, due to the high spin $J=3$, whereas the data implies a comparable rate
if $X(3872) = \psi_3$~\cite{Olsen:2004fp}.

Another radiative decay search, for $X(3872) \to \gamma J/\psi$,
tests the $X(3872) = 2 {}^3P_1\; (\chi^\prime_{c1})$ assignment~\cite{Olsen:2004fp}.
The partial width $\Gamma(\chi^\prime_{c1} \to \gamma J/\psi)$,
for $M_{\chi^\prime_{c1}} = 3872\,\mathrm{MeV}$,
is expected to be 11~KeV in the potential model~\cite{Barnes:2003vb},
possibly reduced by coupled channel effects~\cite{Eichten:2004uh}.
To estimate the partial width for the isospin-violating process
$\chi^\prime_{c1} \to \pi^+ \pi^- J/\psi$, we take
the isospin-violating hadronic charmonium transition $\psi' \to \pi^0 J/\psi$,
with $\Gamma \simeq 0.3\,\mathrm{keV}$: the ratio
$\Gamma(X \to \gamma J/\psi)/\Gamma(X \to \pi^+ \pi^- J/\psi)$ should then be ${\cal O}(10)$.
The Belle search places an upper limit of $0.40\;(90\%\;{\rm CL})$ on this ratio,
inconsistent with the expected value.
The $\chi^\prime_{c1}$ mass is predicted~\cite{Barnes:2003vb,Eichten:2004uh}
to be $3930\sim 3990\,\mathrm{MeV}$ or greater, 
likewise inconsistent with the $X(3872)$.

\subsubsection{Studies of angular distributions} 
\begin{figure}
  \begin{center}
    \includegraphics[width=0.7\columnwidth]{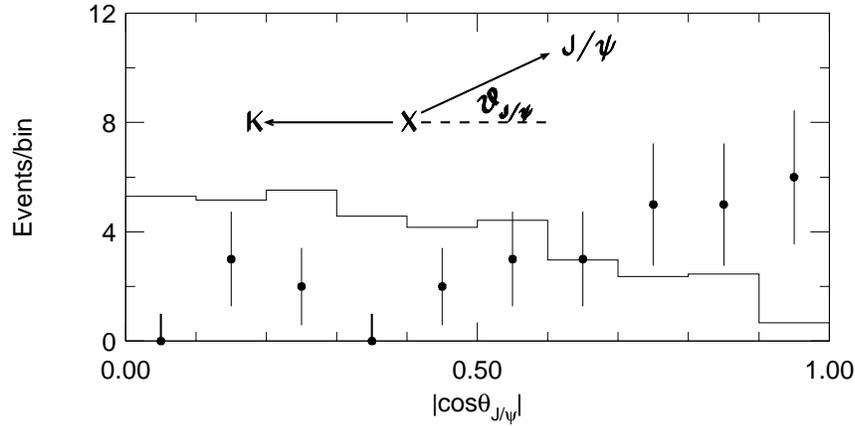}
  \end{center}
  \caption{From~\cite{Olsen:2004fp}: 
        $|\cos^2 \theta_{J/\psi}|$ distribution for $X(3872) \to \pi^+ \pi^- J/\psi$ events
        in the data (points with error bars),
        and assuming $J^{PC} = 1^{+-}$ in the Monte Carlo (histogram);
        background events, determined using sidebands, are included in the Monte Carlo.}
  \label{fig:x3872-belle-angular}
\end{figure}

The $X(3872)$ will be produced polarized in the reaction $B^\pm \to
K^\pm X(3872)$ for any spin $J_X > 0$, as both the initial state $B$
and the accompanying $K$ mesons are spin zero. Angular distributions
of the final state particles can therefore distinguish between
different quantum number assignments $J^{PC}$ for the $X(3872)$.  If
the $X$ is the $h_c^\prime$, with $J^{PC} = 1^{+-}$, decays should be
distributed as $(1-\cos^2 \theta_{J/\psi}) \mathrm{d}\cos
\theta_{J/\psi}$~\cite{Pakvasa:2003ea}, where $\theta_{J/\psi}$ is the
angle between the $J/\psi$ and the negative of the $K$ momentum
vectors in the $X(3872)$ rest frame.  In the Belle
study~\cite{Olsen:2004fp}, the data tend to peak near $\cos
\theta_{J/\psi} = 1$, where the $1^{+-}$ expectation is zero; both the
data and expectation are shown in
\Figure~\ref{fig:x3872-belle-angular}. The poor fit to the data
($\chi^2/dof = 75/9$) rules out any $1^{+-}$ assignment for the
$X(3872)$, including the $h_c^\prime$; this state in any case has an
expected mass well above 3872~MeV.  Further angular studies of other
modes are foreseen.
\subsubsection{Other searches} If $X(3872) = 1 {}^1D_2\; (\eta_{c2})$,
the isospin conserving transition $\eta_{c2} \to \pi^+ \pi^- \eta_c$ should be
much more common than $\eta_{c2} \to \pi^+ \pi^- J/\psi$,
which is isospin violating;
the branching fraction ${\cal B}(X \to \pi^+ \pi^- J/\psi)$
would be ${\cal O}(1\%)$ or less, implying a large $B \to K X(3872)$ rate.
This seems unlikely but can be tested by searching for the 
$X(3872) \to \pi^+ \pi^- \eta_c$ decay.

Similar considerations apply if $X(3872) = \eta_c^{\prime\prime}$:
the branching fraction to $\pi^+ \pi^- J/\psi$ should be small,
although in this case (with the $\eta_c^{\prime\prime}$ below 
open charm threshold) the dominant decay would be into two gluons,
less convenient for a search.
Assuming that such a state would have a similar width to the 
$\eta_c$ ($17 \pm 3\,\mathrm{MeV}$)~\cite{Eidelman:pdg2004},
which also predominantly decay via two gluons, it is already disfavored
by the $2.3\,\mathrm{MeV}$ upper limit on the $X(3872)$ width.
Given $M_{\psi(3S)} = (4040 \pm 10)\,\mathrm{MeV}$,
$M_{\eta_c^{\prime\prime}} = 3872\,\mathrm{MeV}$ also implies a large
$\psi(3S)-\eta_c^{\prime\prime}$ mass splitting, $\sim 168\,\mathrm{MeV}$,
contrary to the expectation that the splitting will decrease with increasing
radial quantum number (\emph{cf.}\ 
$M_{\psi'}-M_{\eta_c^\prime} = 48\,\mathrm{MeV}$~\cite{Skwarnicki:2003wn} and
$M_{\psi}-M_{\eta_c} = 117\,\mathrm{MeV}$)~\cite{Eidelman:pdg2004}.

\subsubsection{Summary}
The status of the six candidates considered by Belle~\cite{Olsen:2004fp}
is summarized in Table~\ref{table:x3872-belle-candidates}: some are already
excluded by the data, and none is a natural candidate. Significant further
information is expected once searches for other decays become available;
the search for $X(3872) \to \pi^0 \pi^0 J/\psi$ is particularly important.
Already however the lack of a natural charmonium candidate that fits the data
suggests two possibilities:
(1) that the theory used to predict charmonium properties is flawed, or
(2) that the $X(3872)$ is not a conventional $(c\bar{c})$ state.
As the $X(3872)$ mass is very close to the $M_{D^0} + M_{D^{*0}}$ threshold,
a $D\overline{D}^{*}$ bound state is a natural
candidate~\cite{Voloshin:2003nt,Pakvasa:2003ea,Tornqvist:2004qy,Close:2003sg,Wong:2003xk,
Swanson:2004pp,Braaten:2004rn}.
\begin{table}[!t]
\caption[Some properties of candidate charmonium states for 
         $X(3872)$ and comparison with data]
        {From~\cite{Olsen:2004fp}: Some properties of candidate
         charmonium states for the $X(3872)$, and a summary of the
         comparison with data.  Mass predictions are taken
         from~\cite{Godfrey:1985xj}, and width predictions computed
         from~\cite{Barnes:2003vb}, using a 3872~MeV mass value; the
         predicted width for the $\eta_c^{\prime\prime}$ is taken to
         be the same as the $\eta_c$ width.  Masses and widths are
         shown in~MeV.}
  \label{table:x3872-belle-candidates}
  \renewcommand{\arraystretch}{1.2}
  \small
\begin{center}
  \begin{tabular}{lclccl}
    \\ \hline
    state       &  alias        &$J^{PC}$& $M_{{\rm pred}}$
                                                        &  $\Gamma_{{\rm pred}}$
                                                                &  comment                      \\ \hline
    $1 {}^3D_2$ & $\psi_2$      &$2^{--}$&  $3838$      & $0.7$ &  Mass wrong; $\Gamma_{\gamma \chi_{c1}}$ too small    \\
    $2 {}^1P_1$ & $h_c^\prime$  &$1^{+-}$&  $3953$      & $1.6$ &  Ruled out by $|\cos\theta_{J/\psi}|$ distribution    \\
    $1 {}^3D_3$ & $\psi_3$      &$3^{--}$&  $3849$      & $4.8$ &  $M,\,\Gamma$ wrong;
                                                                        $\Gamma_{\gamma \chi_{c2}}$ too small;
                                                                        $J$ too high                                    \\
    $1 {}^1D_2$ & $\eta_{c2}$   &$2^{-+}$&  $3837$      & $0.9$ &  ${\cal B}(\pi^+ \pi^- J/\psi)$ expected
                                                                        to be very small                                \\
    $2 {}^3P_1$ & $\chi_{c1}^\prime$
                                &$1^{++}$&  $3956$      & $1.7$ &  $\Gamma_{\gamma J/\psi}$ too small                   \\
    $3 {}^1S_0$ & $\eta_c^{\prime\prime}$
                                &$0^{-+}$&  $4060$    &$\sim 20$&  Mass and width are wrong                     \\ \hline
  \end{tabular}
\end{center}
\end{table}

\section[The observation of the $B_c$ meson at CDF and D0]
        {The observation of the $B_c$ meson at CDF and D0 
         $\!$\footnote{Author: V.~Papadimitriou}}
\label{sec:spexBc}

\begin{figure}[t]
\begin{center}
   \includegraphics[width=7.5cm]{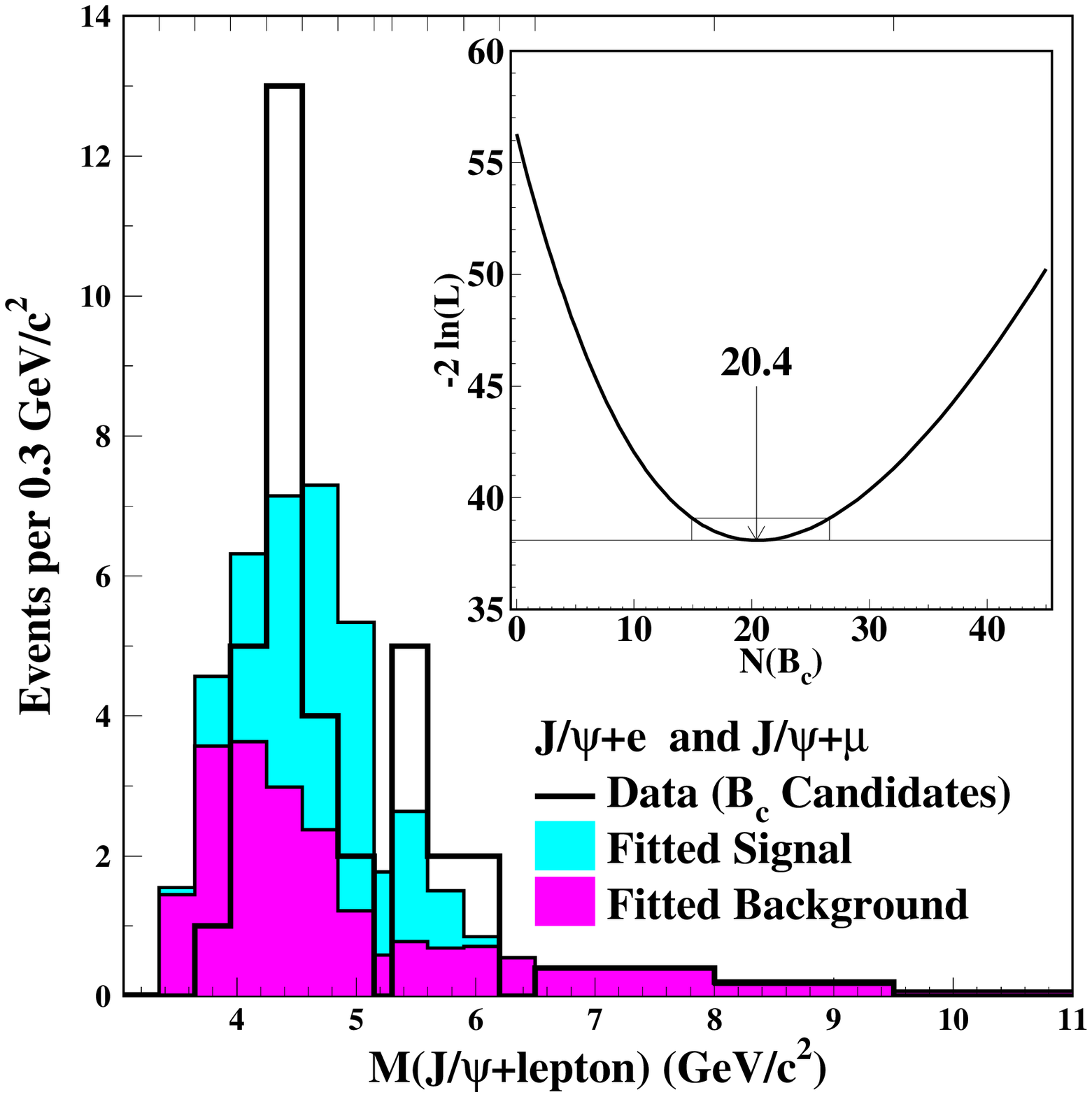}
\hfill
   \includegraphics[width=7.5cm]{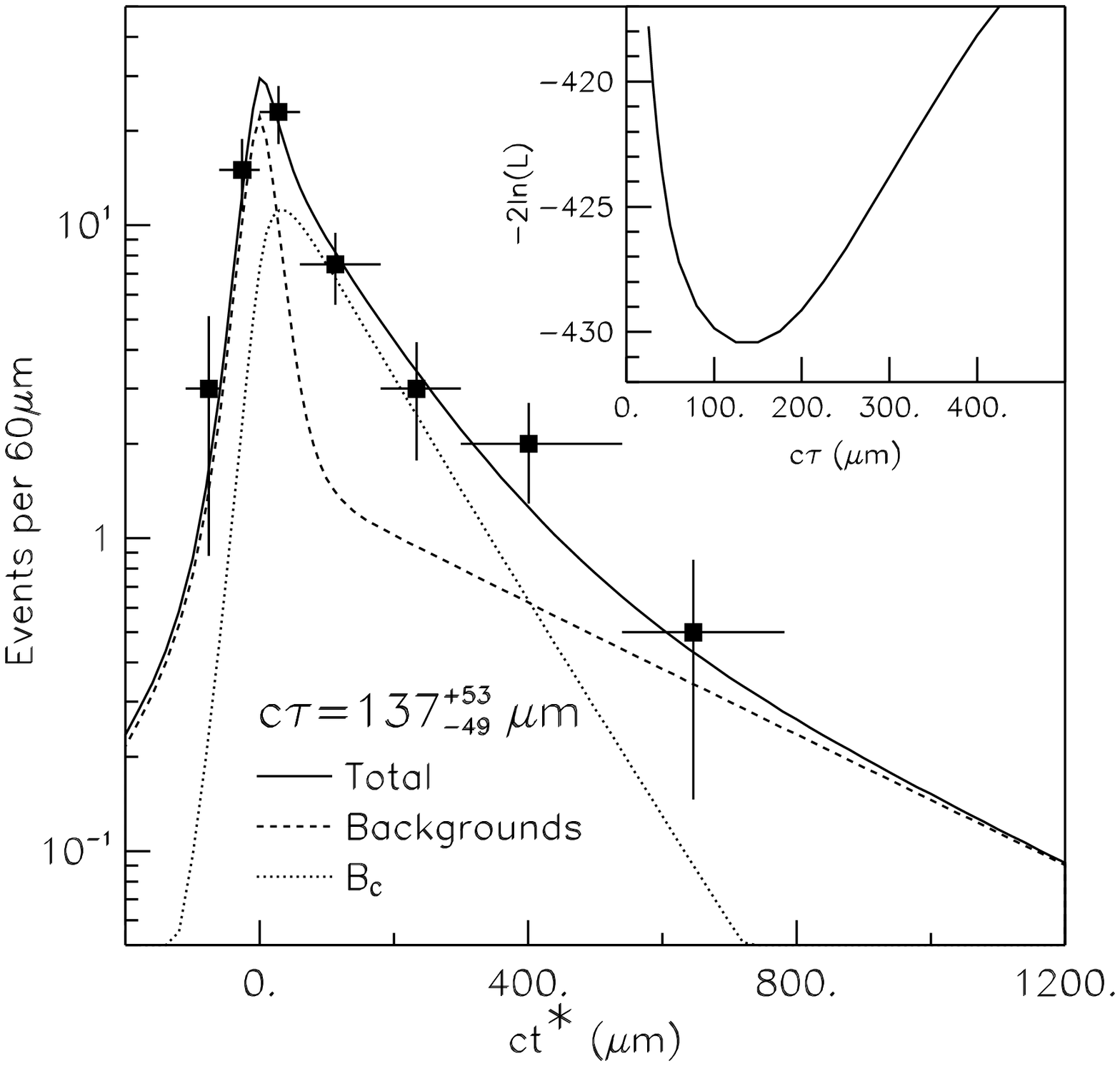}
\end{center}
\caption[$J/\psi l$ mass]
        {On the left, (a) the histogram of the $J/\psi l$ mass that
         compares the signal and background contributions determined
         in the likelihood fit to the combined data for $J/\psi e$ and
         $J/\psi \mu$. The mass bins, indicated by tick marks at the
         top, vary in width.  The total $B_c^{\pm}$ contribution is
         20.4$^{+6.2}_{-5.5}$ events. The inset shows the behavior of
         the log-likelihood function $-$2Ln(L) vs the number of $B_c$
         mesons.  On the right, (b) the distribution in $ct^*$ for the
         combined $J/\psi e$ and $J/\psi \mu$ data along with the
         fitted curve and contributions to it from signal and
         background. The inset shows the log-likelihood function vs
         $c\tau$ for the $B_c$ meson.  }
\label{fig:bc_cdf}
\end{figure}

The CDF Collaboration has observed the ground state of the
bottom-charm meson $B_c^{\pm}$ via the decay mode $B_c^{\pm}
\rightarrow J/\psi l^{\pm}\nu$ and measured its mass, lifetime and
production cross-section~\cite{Abe:1998wi,Abe:1998fb}.  The
measurement was done at the Tevatron, in Run I, at $\sqrt(s) = 1.8$
TeV.  \Figure[b]~\ref{fig:bc_cdf}a shows the mass spectra for the
combined $J/\psi e$ and $J/\psi \mu$ candidate samples, the combined
backgrounds, and the fitted contribution from the $B_c^{\pm}
\rightarrow J/\psi l^{\pm}\nu$ decay. The fitted number of $B_c$
events is 20.4$^{+6.2}_{-5.5}$, out of which 12.0$^{+3.8}_{-3.2}$ come
for the electron sample and 8.4$^{+2.7}_{-2.4}$ from the muon sample.A
fit to the same distribution with background alone was rejected at the
level of 4.8 standard deviations. The $B_c^{\pm}$ mass was measured to
be equal to 6.40$\pm$0.39(stat.)$\pm$0.13(syst.)~GeV/c$^2$.

A measure of the time between production and decay of a $B_c^{\pm}$ meson is
\begin{equation}
ct^* \equiv \frac{M(J/\psi l) \cdot L_{xy}(J/\psi l)}
{|p_T(J/\psi l)|}
\end{equation}
where $ L_{xy}$ is the distance between the beam centroid and the
decay point of the $B_c^{\pm}$ candidate in the transverse plane and
projected along the $J/\psi l$ direction, and $p_T(J/\psi l)$ is the
tri-lepton transverse momentum. \Figure[b]~\ref{fig:bc_cdf}b shows the
$ct^*$ distribution for the data, the signal and the background
distributions. The mean proper decay length $c\tau$ and hence the
lifetime $\tau$ of the $B_c^{\pm}$ meson was obtained from the above
distribution.  It was determined that $c\tau = 137^{+53}_{-49} (\rm
stat.) \pm 9 (syst.)~\mu$m or $\tau = 0.46^{+0.18}_{-0.16} (\rm stat.)
\pm 0.03 (syst.)$ ps.

\begin{figure}[t]
\begin{center}
   \includegraphics[width=7.5cm]{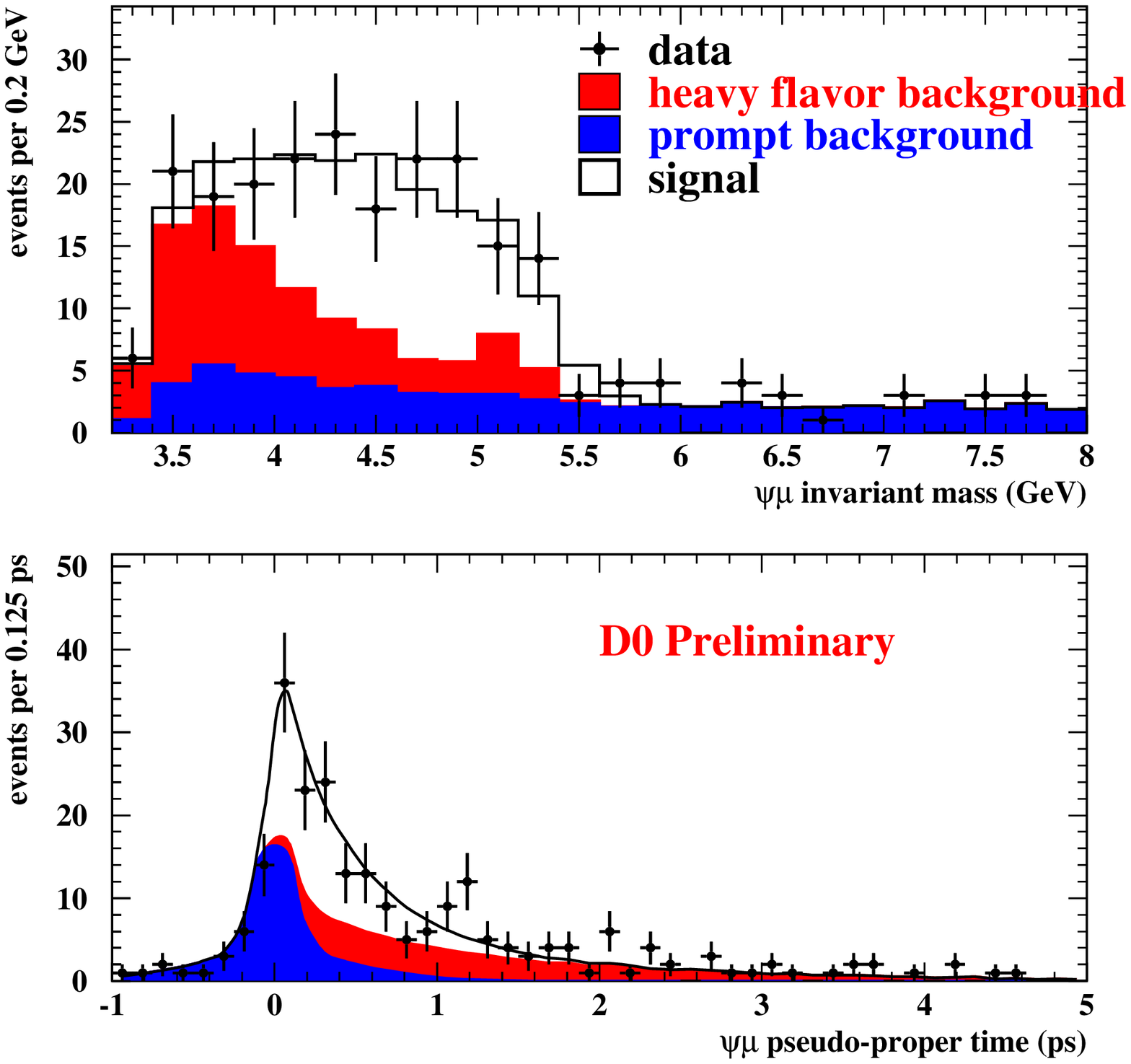}
\hfill
   \includegraphics[width=7.5cm]{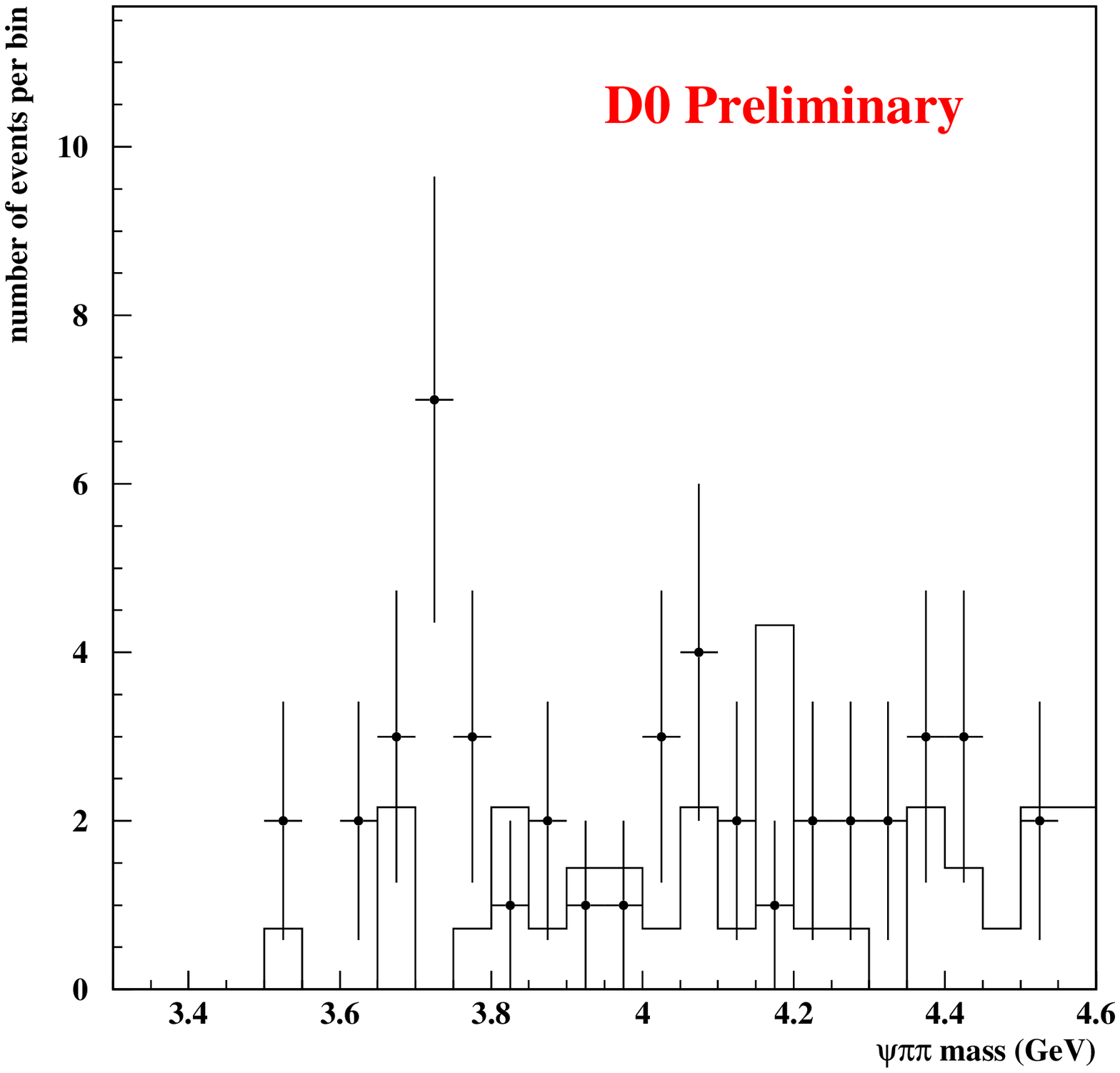}
\end{center}
\caption[The $\jpsi\mu$ invariant mass and pseudo-proper time distributions]
        {The $\jpsi\mu$ invariant mass (top left) and pseudo-proper
         time distributions (bottom left) of the $B_c\to\jpsi\mu X$
         candidates (points with error bars) from D0.  The signal
         MonteCarlo events, generated with a mass of 5.95~GeV$/c^2$,
         as well as the most likely background sources are shown as
         solid histograms.  The right plot shows the $\jpsi\pi^+\pi^-$
         invariant mass of $\jpsi\pi^+\pi^-\mu X$ events that have
         M($\jpsi\pi^+\pi^-\mu$) between 4 and 6~GeV$/c^2$. The
         background (solid histogram) is estimated from events outside
         this mass range.}
\label{fig:bc_d0}
\end{figure}

Recently \cite{D0Bc} the D0 collaboration has reported the observation
of a $B_c$ signal in the decay mode $B_c^{\pm} \rightarrow J/\psi
\mu^{\pm}\nu$, from a sample of 210 pb$^{-1}$ of data taken during Run
II, at $\sqrt{s}=1.96$~TeV.  The dimuon coming from $\jpsi$ was
required to be within 0.25 from the $\jpsi$ mass, and a third muon
track was required to come from the same vertex.  The analysis yielded
95$\pm$12$\pm$11 events, at a mass $ M(B_c^\pm) =
5.95^{+0.14}_{-0.13}$(stat.)$\pm0.34$(syst.)~GeV/c$^2$.  The lifetime
was calculated to be $ \tau(B_c^\pm)$ $ =$ $0.448^{+0.123}_{-0.096}$
(stat.) $\pm 0.121$(syst.) ps.  Fitted mass and lifetime are found to
be uncorrelated.  \Figure[b]~\ref{fig:bc_d0}(left) shows the invariant
mass and pseudo-proper time distributions of the events.  The analysis
accounts for the possible contribution from
$B_c\to\psi(2S)\mu^{\pm}\nu$ on the inclusive sample.  As shown in
\Figure~\ref{fig:bc_d0}(right), it is estimated that about 15 events
are due to this component, and the systematic errors are obtained by
varying this contribution from 0 to 30 events.  In the near future,
the mass uncertainty can be improved to better than 5(50)~MeV/c$^2$ by
CDF(D0) by using hadronic exclusive decay channels.

\section[Evidence for doubly charmed baryons at SELEX]
        {Evidence for doubly charmed baryons at SELEX 
         $\!$\footnote{Author:P.~Cooper}}
\label{sec:spexdcb}

The addition of the charmed quark to the ($uds$) triplet extends the
flavour symmetry of the baryon octet and decuplet from SU(3) to SU(4).
There is strong experimental evidence for all the predicted baryon
states which contain zero or one valence charmed
quark~\cite{Eidelman:pdg2004}.  We review here the first experimental
evidence for one of the six predicted baryon states which contain two
valence charmed quarks, the doubly charmed baryons.  There have been
many predictions of the masses and other properties of these
states~\cite{HQEH,SW,Korner,BPhys}.  The properties of doubly charmed
baryons provide a new window into the structure of baryonic matter.

\subsection{The SELEX experiment}
The SELEX experiment uses the Fermilab 600~$\GeV /c$ charged hyperon beam  to
produce charm particles in a set of thin foil targets of Cu or diamond.  The
three-stage magnetic spectrometer is shown elsewhere~\cite{Thesis,SELEX}.  The
most important features are: (a) the high-precision, highly redundant, silicon 
vertex detector that provides an average proper time resolution of 20~$\fsec$ 
for single-charm particle decays, (b) a 10~m long Ring-Imaging Cherenkov 
(RICH) detector that separates $\pi$ from K up to 165~$\GeV /c$~\cite{RICH}, 
and (c) a high-resolution tracking system that has momentum resolution of 
${\sigma}_{P}/P<1\%$ for a \mbox{200\,$\GeV /c$} reconstructed 
$\Lambda_{c}^{+}$.  

The experiment selected charm candidate events using an online secondary 
vertex algorithm which required full track reconstruction for measured
fast tracks. An event was written to tape if all the fast tracks in the
event were inconsistent with having come from a single primary
vertex.  This filter passed 1/8 of all interaction triggers and had
about $50\%$ efficiency for otherwise accepted charm decays.  The
experiment recorded data from $15.2 \times 10^{9}$ inelastic
interactions and wrote $1 \times 10^{9}$ events to tape using both
positive and negative beams. $67\%$ of events were induced by $\Sigma^{-}$,
$13\%$ by $\pi^{-}$, and $18\%$ by protons.

\subsection{Search strategy}
A Cabibbo-allowed decay of a doubly charmed baryon must have a 
net positive charge and contain a charmed quark, a strange quark and a baryon. 
We chose to search for decay modes like \ccdlcki \  with an intermediate 
\kpi secondary vertex between the primary vertex and the \lc vertex and \\
\dpk with an intermediate $pK^-$ secondary vertex between the primary 
vertex and the $D^+$ .  

Events were analyzed for evidence of a secondary vertex composed of an
opposite-signed pair between the primary and the single charm decay point.  
We used all tracks not assigned to the single charm candidate in the search.  
The new secondary vertex had to have an acceptable fit $\chi^2$ and a 
separation of at least 1$\sigma$ from the new primary.
These cuts were developed and fixed in previous searches for short-lived 
single-charm baryon states.  We have applied them here without change.  
As a background check we also kept wrong-sign combinations in which the mass 
assignments are reversed.  

\subsection{$\lcki$ Search results and significance}

\begin{figure}[t]
\begin{center}
\includegraphics[width=.52\linewidth]{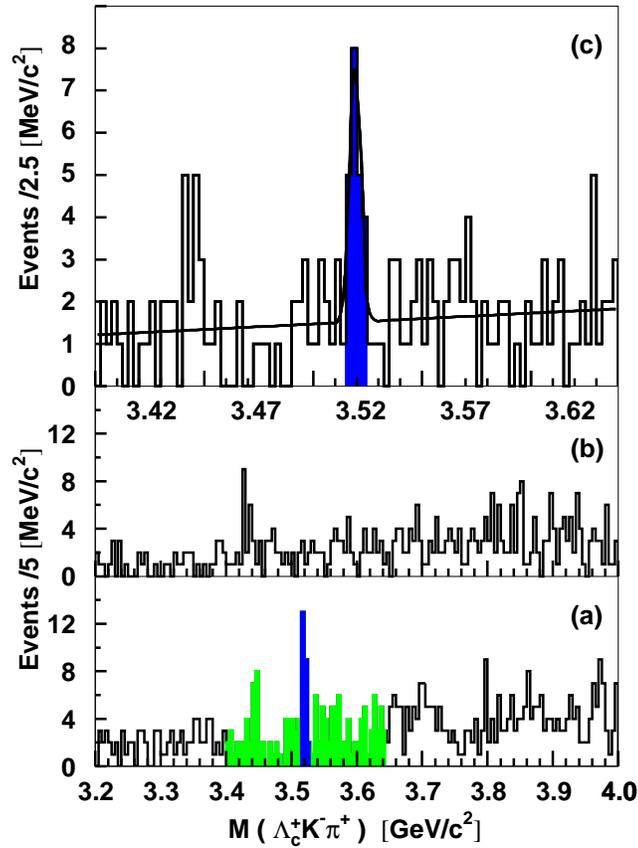}
\end{center}
\caption{(a) The $\lc K^- \pi^+$ mass distribution in 5~$\MeV /c^2$ bins.
             The shaded region 3.400--3.640~$\GeV /c^2$ contains the signal 
             peak and is shown in more detail in (c).
         (b) The wrong-sign combination  $\lc K^+ \pi^-$ mass distribution in 
             5 $\MeV /c^2$ bins.
         (c) The signal (shaded) region ($\Nevt$ events) and sideband mass 
             regions with $\Ntot$ total events in 2.5 $\MeV /c^2$ bins.  
             The fit is a Gaussian plus linear background.}
\label{fig:mass}
\end{figure}

The signal and wrong-sign backgrounds are shown in \Figure~\ref{fig:mass}.
There is a obvious peak at a mass of \Mass $\pm$ \dMass \mvc.
The number of events in the signal region shown is $\Nevt$ events.
We estimate the number of expected background events in the signal region 
from the sidebands as $\Nback \pm \dBack$ events.  This determination has a 
(Gaussian) statistical uncertainty, solely from counting statistics.
The single-bin significance of this signal is the excess in the signal region 
divided by the total uncertainty in the background estimate: 
$\Nsig / \sqrt{6.1 + 0.5^2} = 6.3\sigma$.  The Poisson probability 
of observing at least this excess, including the Gaussian uncertainty in the 
background, is $1.0 \times 10^{-6}$.  The overall probability of  
observing an excess at least as large as the one we see anywhere in the 
search interval is $1.1 \times 10^{-4}$.  This result is published in 
reference~\cite{SELEX}.

\subsection{$\dpk$ search}

After the discovery and publication of the $\lcki$ signal we sought to confirm
the discovery in another decay mode which was likely to have a significant 
branching ratio.  Obvious choices were $\Xi_{c}^{+}\pi^{+}\pi^{-}$ and 
\ $\dpk$.  Since the SELEX $D^+$ signal is large and well studied we began 
with it.

\begin{figure}[t]
\begin{center}
\includegraphics[width=75mm]{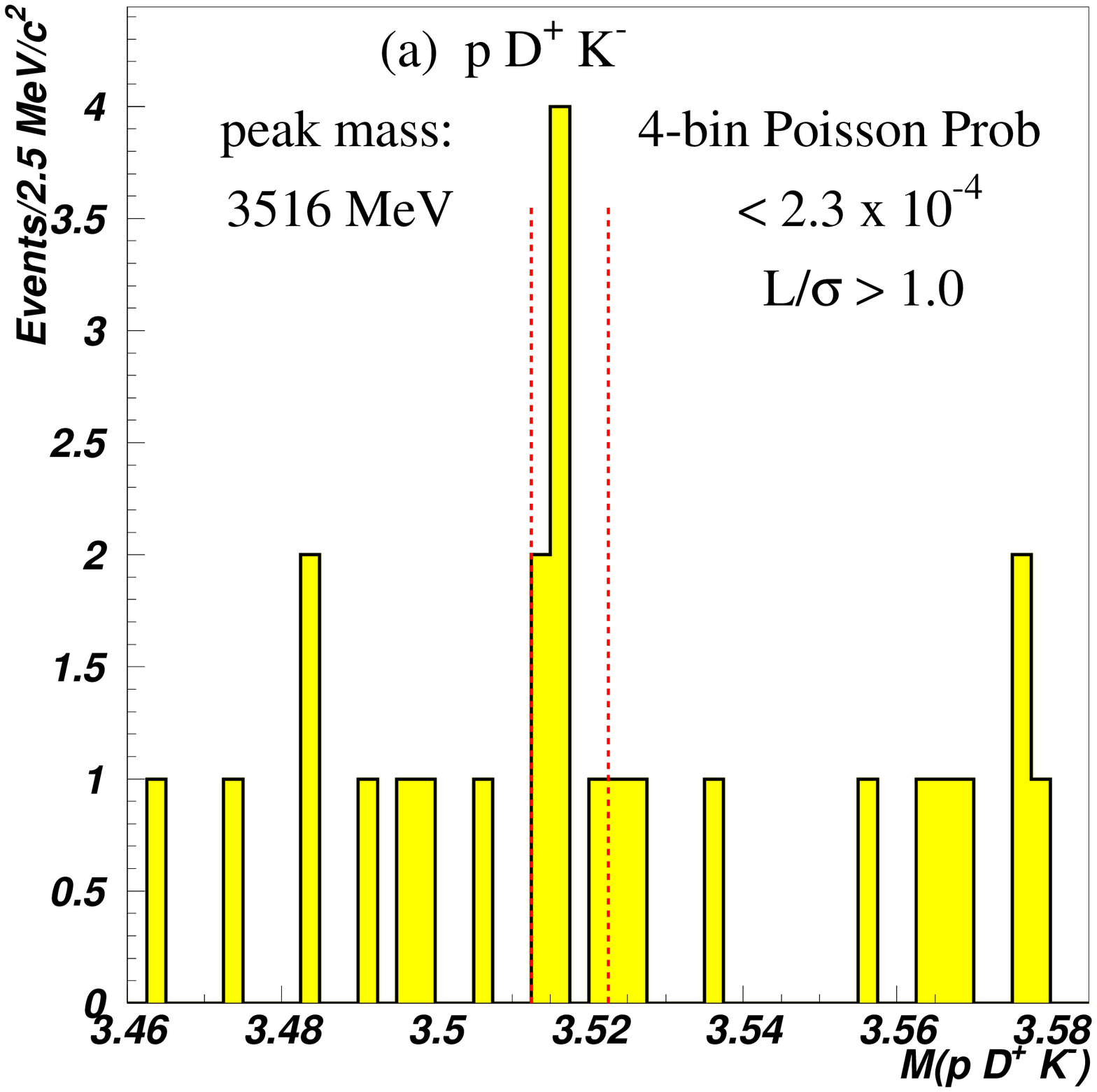}
\hfill
\includegraphics[width=75mm]{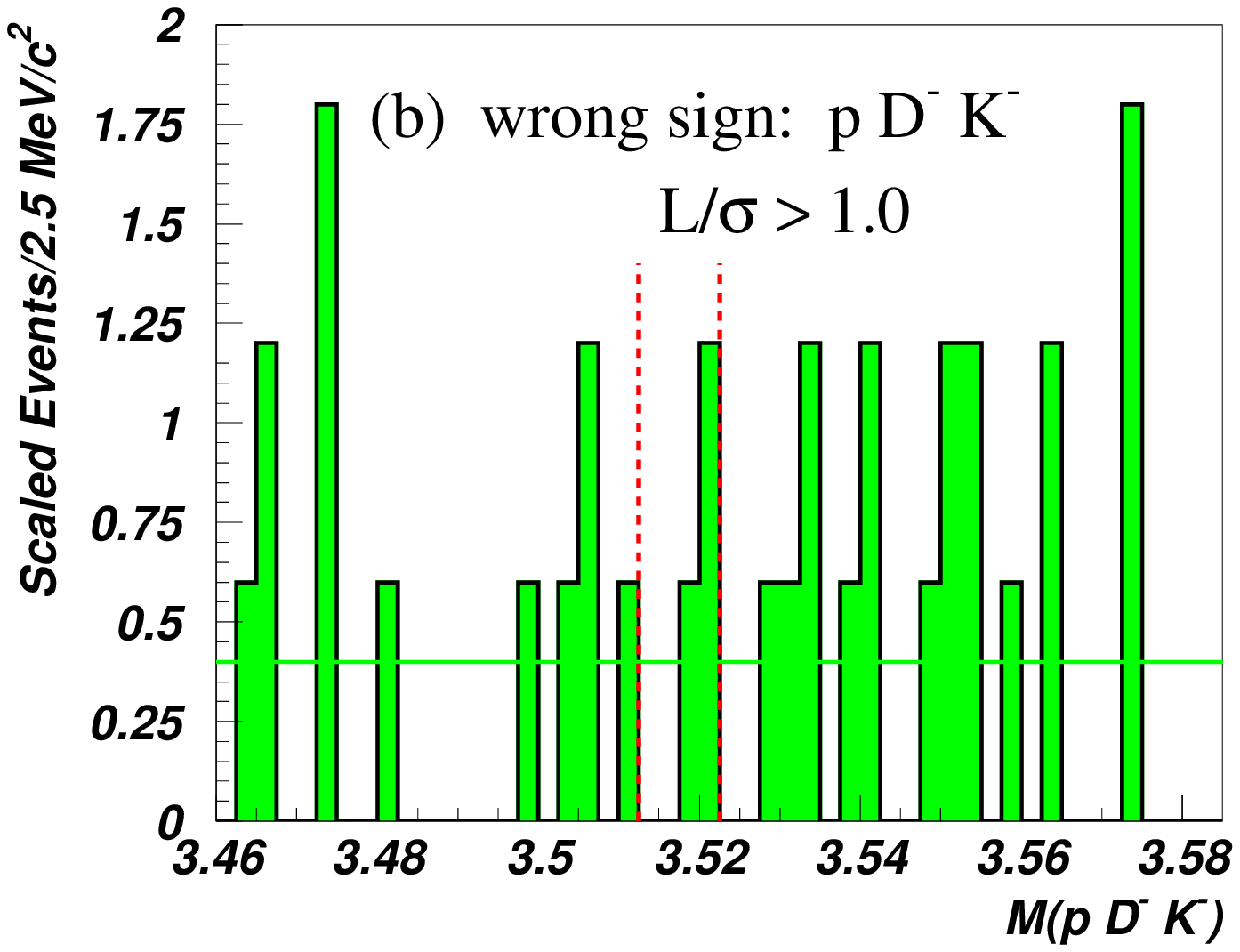}
\end{center}
\caption[\dpk mass distribution]
        {(a) \dpk mass distribution for right-sign mass combinations.  
         (b) Wrong-sign events with a ${D^- K^+}$, scaled by 0.6. 
         The line shows a maximum likelihood fit to occupancy.}
\label{fig:dpkmass}
\end{figure}

A similar analysis technique~\cite{SELEX2} resulted in the signal and 
wrong-sign background shown in \Figure~\ref{fig:dpkmass}.
In this new decay mode we observe an excess of $\NsigD$ events over an 
expected background of $\BackD \pm \dBackD$ events.  The Poisson probability 
that a background fluctuation can produce the apparent signal is less than 
$\Pois$.  The observed mass of this state is \MassD $\pm$ \dMassD~\mvc, 
consistent with the published result.  Averaging the two results gives a mass 
of \amass $\pm$ \damass~\mvc.  The observation of this new weak decay mode 
confirms the previous suggestion that this state is a double charm baryon.  
The relative branching ratio 
$\Gamma$(\dpk)/$\Gamma$(\lcki) = 0.078 $\pm$ 0.045.

The lifetime of this state in both decay modes
is very short; less than 33~$\fsec$ at 90\% confidence.  
The properties of these two signals are consistent with each other.
SELEX reports an independent confirmation of the double charm baryon \ccd,
previously seen in the \lcki decay mode, via the observation of its decay
\dpkn.  

\subsection{Conclusions}

The \ccd({\it ccd}) \ doubly charmed baryon has now been observed by SELEX 
in two decay modes at a mass of \amass $\pm$ \damass~\mvc with a lifetime less 
than 33~$\fsec$ at 90\% confidence.  Analysis continues with SELEX data to 
searchfor additional decays modes for this state and to search for the two 
other doubly-charmed baryons ground states expected: 
\ccu({\it ccu}) and \ccs({\it ccs}).

\section[Summary and outlook]
        {Summary and outlook
         $\!$\footnote{Authors: G.~Bali, N.~Brambilla, R.~Mussa, J.~Soto}}

It took few years , after the discovery of the $\jpsi$, to sketch 
the  spectroscopical pattern of the narrow ortocharmonium and ortobottomonium
states: the experimental determination of such energy levels is extremely 
good, all states are know with precisions better than 1~MeV. 
On the other side, the experimental history of spin singlet states has 
started to clear up only in the recent years, but open puzzles remain:
\begin{itemize}
\item the total width of the $\eta_c(1S)$ (the ground state of charmonium)
  is as large as the one of the  $\psi(3770)$, which can decay to open charm:
\item after 16 years, the real $\eta_c(2S)$ has been observed, disconfirming 
an evidence by Crystal Ball that misled theory calculations on hyperfine 
splittings for more than a decade.
\item two compatible evidences of the $h_c$ state have been found in the
last year, and may bring to an end the saga of the spin singlet P state; 
a concrete strategy to consolidate this observation is now needed.
\item none of the 5 spin singlet states in the bottomonium system 
has been found yet; given the absence of scheduled running time on 
narrow $\Upsilon$ states in the near future, it is necessary to 
elaborate smarter search strategies to spot these states at asymmetric 
B-factories or hadron colliders . 
\end{itemize}
\vskip 0.2truecm
The quest to complete the experimental spectra  is now extending to the 
higher excitations:
\begin{itemize}
\item the search for narrow D-states resulted in the discovery of the 
$\Upsilon(1D)$ states in CLEO~III,  
and to the observation of the intriguing X(3872) meson by Belle; 
while the bottomonium state falls well in the expected pattern, there 
is a wide variety of speculations on the nature of the X(3872).
\item the need to achieve a deeper understanding of the region just 
above open charm threshold, together with the improvement of the experimental
tools, will allow to disentangle each single contribution to the R ratio, 
hopefully clarifying the puzzles opened by the pioneering studies of Mark~II. 
\end{itemize}
\smallskip
As we have seen in this chapter the application of EFTs 
of QCD to heavy quarkonium has considerably increased our 
understanding of these systems from a fundamental point of view.
NRQCD has allowed, on the one hand, for efficient lattice calculations of the
masses of the bottomonium and charmonium states below open heavy flavour threshold. 
On the other hand, it has paved the way to pNRQCD, which provides, in the 
strong coupling regime, a rigorous link from QCD to potential models 
for states below open heavy flavour threshold.
In the weak coupling regime, pNRQCD has allowed to carry out 
higher-order calculations and to implement renormalization group resummations 
and renormalon subtractions in a systematic way. 
This regime appears to be applicable at least for the $\Upsilon (1S)$ and
$\eta_b (1S)$. Interestingly, as discussed in \Section~\ref{sec:spnrqcdwc}
 (\Tables~\ref{tab:predictedmasses} and \ref{tab:predictedsplittings})
even some excited states can be reproduced  in perturbation theory (inside
the errors of the perturbative series). 

The  most challenging  theoretical problem  at present  is the  description of
states above  open heavy flavour  threshold. The recent discovery  of $X(3872)$
has   made  clear  that   potential  models   suffer  from   large  systematic
uncertainties in this region and  that the inclusion of, at least, heavy-light
meson degrees  of freedom is necessary.  Although NRQCD holds  in this region,
extracting information  from it  on the lattice  is not simple,  since besides
heavy quarkonium,  heavy-light meson pairs  and hybrid states populate  it. It
would  be important to  develop theoretical  tools in  order to  bring current
phenomenological approaches into QCD based ones.

In order to stimulate progress in heavy quarkonium spectroscopy, 
we shall try to pose a number of questions, and try to provide 
what we believe to be reasonable answers to them, from the theory 
and experimental point of view.

\medskip

$\bullet$ {\bf Q.} What does theory need from experiment?

\medskip

{\bf A1}{(\it TH:)} Discovery and good mass measurements of the missing states below open heavy flavour threshold.

{\bf A1}{(\it EXP:)} Concerning triplet S and P states of neutral heavy quarkonia,
experimental measurements are mature and ahead of theory.  
Concerning the singlet S and P states, charmonia are under very active 
investigation at present, 
and probably will be nailed down to less than .5~MeV/c$^2$ in the near future,
with an active cooperation amid experimental groups. 
In bottomonium, the situation looks less promising: only Tevatron experiments
have currently some chance to detect the missing (narrow?) states, while 
CLEO~III searches turned out no plausible candidates, and showed that more 
luminosity is needed at $\Upsilon(1,2,3S)$, but none of the active B-factories
is planning to shift out of $\Upsilon(4S)$. 

The experimental study of the spectrum of the charged heavy
quarkonium, the $B_c$, has not started yet. The ground state has been
seen by CDF and confirmed by D0, but via its semileptonic decay, which
yield still very wide uncertainties on the mass (0.4~GeV/c$^2$). The
experimental search for an exclusive, non leptonic mode is under way
and will allow to know this state with accuracies better than
5~MeV/c$^2$ in the near future.  Beyond this, most experimental
efforts will be focused on finding the dominant decay modes of the
ground state. The search for the $B_c^*$, which decays dominantly to
$B_c$ via M1 radiation of a soft photon , will be extremely
challenging for current Tevatron experiments, due to the high
combinatorial background and to the low efficiency on low energy
photons. Same can be said for the P states , which are expected to
decay to $B_c^{(*)}$ via dipion emission. It is hard to predict
whether the hadronic B-factories, BTEV and LHCB, will be able to
contribute to these spectroscopical studies.  The issue should be
discussed in
\Chapter~\ref{chapter:futureexpfacilities}.

\medskip

{\bf A2}{(\it TH:)} Thorough analysis of the region above open heavy flavour 
threshold in search for quarkonium states, hybrid states, molecules and other exotica.

{\bf A2}{(\it EXP:)} The BES~II R scan and the surprises from the
asymmetric B-factories (X(3872) and double $\ccbar$ production) have
ignited new experimental and theoretical interest in this physics
region.  The CLEO-c running at $\psi(3770)$ and above $D_s\bar{D_s}$
threshold has a very large physics potential for heavy quarkonium
studies.  At the same time, B factories can benefit from a large
variety of techniques to identify new charmonium states: (a) inclusive
ones , such as $\jpsi$ and $\psip$ recoil in double $\ccbar$, or K
recoil in tagged B decays; (b) exclusive ones , such as $B\to
(\psi,\eta_c)X K^{(*)} $ (for narrower states), $B\to D^{(*)} \bar{
D^{(*)} } K^{(*)} $ (for wider states).

Some discovery potential is to be expected also from hadron colliders, where
the large, very clean samples of D mesons can be used as starting point 
to search for peaks in  $D\bar D$ invariant mass combinations. 

\medskip

$\bullet$ {\bf Q.} What does experiments need from theory?

\medskip
{\bf A1}{(\it EXP:)} In spectroscopy,  two are the crucial issues in the search of 
missing states: (a) a good understanding of the production/formation  
mechanisms; (b) a comprehensive set of decay channels, with solid predictions 
on the partial widths. The two issues are related between each other, and 
to the hot topics of the next chapter.

There is NOT an infinite number of ways to produce charmonium, less 
for bottomonium, much less for $B_c$:
these couplings deserve a higher level of understanding, both 
 theoretically and experimentally. This is much more important, when we do
want to understand whether we can get some deeper insight from the non 
observation of a missing state. It must be noticed that most production
mechanisms are not fully understood, and/or lead to wrong predictions.

A limited set of processes, then, deserve  deeper theoretical understanding: 
\begin{itemize}
\item 
  M1 hindered radiative transitions: relativistic corrections are dominant
in these processes that are the main gateways to $\eta_b$'s.
   
\item 
  isospin violating hadronic transitions: it is now very important to 
establish a physical relation between $\psi(2S)\to h_c\pi^0$ and 
$h_c\to\jpsi\pi^0$. This can help clarifying the consistency between 
the two (still weak) evidences.

\item 
  factorization in  B decays: exclusive B decays to K+$\ccbar$ were expected 
to yield $0^{-+},1^{--},1^{++}$ charmonia, and, in smaller quantities,
$0^{++},2^{++}$. The prediction holds for the second, but $\chicj{0}$'s
are produced as copiously as $\chicj{1}$'s . The understanding of the
effective selection rules can help to set limits on the $h_c$ production,
and to find the possible quantum numbers of the X(3872) meson.

\item 
  coupling to exclusive $\ppbar$: helicity selection rule in perturbative 
QCD forbids the formation of $\eta_c$, $\chi_{c0}$, $h_c$   from $\ppbar$ 
annihilations; no suppression is observed in the first two cases, and the 
third is under active investigation. It is auspicable that recent developments 
in NRQCD can help to explain the $\ppbar $ coupling and make testable 
predictions on the coupling to $\eta_c(2S)$ and X(3872).

\item 
  the double $\ccbar$ selection rules are not yet clear: so far only scalars
and pseudoscalars were observed  recoiling against the $\jpsi$ . This process
has already allowed an indipendent confirmation of the 
$\eta_c(2S)$ observation. By understanding the dynamics, one can converge
more rapidly on the determination of the quantum numbers of any bump that 
shows up in the $\jpsi$ recoil spectrum.
 
\end{itemize}

\medskip

Within theory one may ask the following questions:

\medskip

$\bullet$ {\bf Q.}
What does the phenomenological approach need from the theoretical one?

\medskip

{\bf A} $\ast$ That the theory clearly points out the most relevant feautures that 
should be implemented in phenomenological models.

\medskip

$\bullet$ {\bf Q.}
 What does the theoretical approach need from the phenomenological one?

\medskip

{\bf A} $\ast$ To point out shortcomings in models which are relevant to experimental 
observations.

\medskip

Within the theoretical approach:

\medskip

$\bullet$ {\bf Q.}  What does EFTs need from lattice?

\medskip

$\ast$ Calculation of the correlators which parameterize nonperturbative 
effects in the weak coupling\par \indent
 regime of pNRQCD.

\medskip

$\ast$ Calculation of the various potentials which enter pNRQCD in the strong coupling regime.

\medskip

$\bullet$ {\bf Q.} What does lattice need from EFT?

\medskip

$\ast$ Calculation of the NRQCD matching coefficients in lattice regularizations.

\medskip

$\ast$ Chiral extrapolations.

\medskip

Let us next describe the future development which are desirable within each particular approach.

\medskip
\noindent From the side of the EFT the priority ``to-do'' list is:
\begin{itemize}

\item Develop a suitable EFT for the region above open heavy flavour
      threshold.

\item Include the effects of virtual pions.  Pions should be included
      in the strong coupling regime of pNRQCD as ultrasoft degrees of
      freedom and their effect on the spectrum should be investigated.

\item A systematic investigation of the structure of the renormalon
      subtractions in NRQCD matching coefficients and in the
      perturbative potentials.
\end{itemize}

\medskip\noindent
For what concerns lattice calculations the priority
practical lattice "to-do list" is:

\begin{itemize}
\item Further investigations of sea quark effects, in particular on
      charmonia and also in bottomonia, including charm mass effects.

\item Calculation of threshold effects in charmonia and bottomonia,
      first using lattice potentials, then a multichannel analysis in
      lattice NRQCD/QCD.

\item Further investigations of OZI suppressed contributions, in
      particular in the PS charm-sector.

\item Mixing of charmonia and would-be glueballs.

\item Doubly charmed baryons.

\item $QQq$ potentials.

\end{itemize}
\medskip
From the side of phenomenological models the wish list includes:
\begin{itemize}
\item The major deficiency of these models is that they only include
      the $q\bar{q}$ components of the Fock space expansion and
      totally neglect higher Fock space components which can be
      included as coupled channel effects.  These are expected to be
      most prominent for states close to threshold.
\end{itemize}
\medskip
From the side of experiments we  need:
\begin{itemize}
\item to clarify the nature of the X(3872) state, fully exploiting the
      four running experiments that see this state.

\item to strengthen the $h_c$ evidence , by asking for an active
      collaboration between experiments , in order to intensify the
      checks which certify the compatibility between the two recent
      evidences.

\item to support further cross checks on the systematic errors on the
      masses of pseudoscalar charmonia: both BaBar and Belle should
      already have a large sample of $\gamma\gamma\to\eta_c(1,2S)$, to
      measure with high precision both states.

\item to search for doubly charmed baryons in asymmetric B-factories,
      as well as at the Tevatron.

\item to measure, at CLEO-c, the coupling of the $\psipp$ and the
      $\Upsilon(1,2,3S)$ states to $\ppbar$, to quantify the
      perspectives to study charmonium at open charm threshold and
      bottomonium with antiproton beams.

\item to support further $\eta_b$ searches at the Tevatron, and to
      strengthen the physics case for further running at narrow
      bottomonium energies.

\end{itemize}

%\end{document}

\BLKP
%10/12/2004

%%%%%%%%%%%%%%
%                            CHAPTER IV: Decay 
%%%%%%%%%%%%%%

%\documentclass[11pt,twoside]{cernrep} 
%\usepackage{epsfig} 
%\usepackage{graphicx} 
%\usepackage{here,cite,amssymb}
%\usepackage{bm}
%\include{rlFig}

%\input{newcommand.tex}

%\begin{document}

%Counter commands
%\setcounter{page}{1}
%\setcounter{chapter}{3}
%\setcounter{secnumdepth}{3}
%\setcounter{tocdepth}{3}
%\setcounter{section}{4}
%\setcounter{subsection}{3}

\chapter{DECAY}
{{\it Conveners:}   E.~Eichten, C.~Patrignani, A.~Vairo}\par\noindent
{{\it Authors:} D.~Z.~Besson, E.~Braaten, A.~Deandrea,
E.~Eichten,  T.~Ferguson, F.~A.~Harris,
V.~V.~Kiselev, P.~Kroll, Y.-~P.~Kuang,  A.~Leibovich, S.~L.~Olsen,
C.~Patrignani, A.~Vairo}
\label{chapter:decay}

%%%%%%%%%%%%%%  
%Introduction
%%%%%%%%%%%%%%  
\section[Introduction]{Introduction $\!$\footnote{Author: A.~Vairo}}

The study of decay observables has witnessed in the last years a
remarkable progress. New experimental measurements, mainly coming from
Belle, BES, CLEO and E835 have improved existing data on inclusive
(\Section~\ref{sec:ID} and \ref{sec:IDr}), electromagnetic
(\Section~\ref{sec:ID}) and several exclusive (\Section~\ref{sec:exdec}) decay
channels as well as on several electromagnetic
(\Section~\ref{emsec:intro}) and hadronic (\Section~\ref{sec:HT}) transition
amplitudes.  In some cases the new data have not only led to a
reduction of the uncertainties but also to significant shifts in the
central values. Also the error analysis of several correlated
measurements has evolved and improved our determination of quarkonium
branching fractions (\Section~\ref{sec:brmeas}). New data have also
led to the discovery of new states.  These have been mainly discussed
in
\Chapter~\ref{chapter:spectroscopy}.

From a theoretical point of view several heavy quarkonium decay
observables may be studied nowadays in the framework of effective
field theories of QCD. These have been introduced in
\Chapters~\ref{chapter:commontheoreticaltools} and
\ref{chapter:spectroscopy}.  In some cases, like inclusive and
electromagnetic decay widths, factorization of high and low energy
contributions has been achieved rigorously.  In some others, where
more degrees of freedom, apart from the heavy-quarkonium state, are
entangled and the problem becomes quite complicated, models are still
used to some extent and factorization formulas, if there are, are on a
less solid ground.  There is room there for new theoretical
developments.  High energy contributions can be calculated in
perturbation theory.  Low energy matrix elements, which may include,
among others, heavy quarkonium wave functions, colour-octet matrix
elements, correlators, overlap integrals in radiative transitions,
multipole gluon emission factors, can be determined either by suitable
fitting of the data or on the lattice or by means of potential
models. They typically set the precision of the theoretical
determinations.

In each of the following sections we will have a first part where the
theoretical framework is reviewed and the basic formalism set up and a
second part that summarizes the phenomenological applications and
presents the experimental status. In the last section of the chapter,
\Section~\ref{sec:BC}, we will discuss decay modes of the $B_c$. There are
no data available yet (apart from the lifetime), but $B_c$ will be
copiously produced at future hadron colliders. This system,
differently from bottomonium and charmonium, decays only weakly.
Therefore, it opens in quarkonium physics a window to some of the
electroweak parameters of the Standard Model.

The outline of the chapter is the following.  We will start in
\Section~\ref{sec:brmeas} by making some general remarks on the
determination of quarkonium branching ratios from experiments.  In
\Section~\ref{sec:ID} we will discuss inclusive and electromagnetic
decay widths, in \Section~\ref{sec:IDr} $\Upsilon$ inclusive radiative
decays, in \Section~\ref{sec:exdec} exclusive decays, in
\Section~\ref{emsec:intro} radiative and in \Section~\ref{sec:HT} 
hadronic transitions.  Finally, \Section~\ref{sec:BC} will be devoted
to the decays of the $B_c$.

%%%%%%%%%%%%%%  
%BF
%%%%%%%%%%%%%%  
\section[Branching ratio measurements]
        {Branching ratio measurements 
         $\!$\footnote{Author: C.~Patrignani}}
\label{sec:brmeas}

The measurement of branching ratios (or partial widths) ${\cal B}$ is
deceptively simple: the total number of events observed in a given
final state $N^{\rm obs}_{Q\bar Q\to f}$ is proportional to the total
number of events produced $N^{\rm prod}_{Q\bar Q}$ for that particular
resonance:
\begin{equation}
N^{\rm obs}_{Q\bar Q\to {\rm f}} ={\rm eff}\times N^{\rm prod}_{Q\bar Q}\times
{\cal B}(Q\bar Q\to {\rm f}),
\end{equation}
$N^{\rm prod}_{Q\bar Q}$ in turn needs to be measured by counting some
specific events.  In most cases, depending on the process under study
and the analysis strategy, $N^{\rm prod}_{Q\bar Q}$ is calculated from
the number of events observed in a given ``reference'' final state
$N^{\rm obs}_{Q\bar Q\to {\rm Ref}}$:
\[
N^{\rm prod}_{Q\bar Q} =\frac{N^{\rm obs}_{Q\bar Q\to {\rm Ref}}}{{\rm eff'}\;{{\cal
      B}_{\rm Ref}}}.
\] 
The reported value of ${\cal B}(Q\bar Q\to {\rm f})$ will therefore use  
${\cal B}_{\rm Ref}$ as reported by some previous experiment:
\begin{equation}
{\cal B}(Q\bar Q\to {\rm f}) =
\frac{N^{\rm obs}_{Q\bar Q\to {\rm f}}}{N^{\rm obs}_{Q\bar Q\to {\rm Ref}}}\;
\frac{{\rm eff'}}{{\rm eff }}\;{{\cal B}_{\rm Ref}}.
\label{eq:brmeas-wrong}
\end{equation}

As discussed in \cite{Patrignani:2001as}, there are a number of
potentially dangerous consequences in this procedure.  First of all
different experiments might use the same reference mode, so their
values of ${\cal B}$ are not independent.  Even worse, the ${\cal
B}(Q\bar Q\to {\rm f})$ reported in \Eq~(\ref{eq:brmeas-wrong}) will
also be (hiddenly) correlated to the normalization ${\rm Ref'}$ chosen
by the previous experiment(s) where ${\cal B}_{\rm Ref}$ had been
measured, and ultimately may depend on some other branching ratio
${\cal B}'_{\rm Ref'}$.  Such hidden correlations are hard to identify
and can have pernicious consequences on the evaluation of ${\cal B}'$
based on independent measurements from different experiments.

For precision determination of branching ratios or partial widths, it is 
important to know the normalization used in each measurement and to quote explicitly
the quantity that is indeed directly measured by each experiment
\begin{equation}
\frac{{\cal B}(Q\bar Q\to {\rm f})}{{\cal B}_{\rm Ref}}=
\frac{N^{\rm obs}_{Q\bar Q\to {\rm f}}}{N^{\rm obs}_{Q\bar Q\to {\rm Ref}}}
\frac{{\rm eff'}}{{\rm eff }},
\label{eq:brmeas-correct}
\end{equation}
\ie the ratio or product of branching ratios (even of different particles), 
which is most directly related to the event yield. 
Many experiments could also provide measurements of ratios of branching ratios
\begin{equation}
R_{\cal B}(f/f')=\frac{{\cal B}(Q\bar Q\to {\rm f})}{{\cal B}(Q\bar Q\to {\rm f'})},
\label{eq:brmeas-Rdef}
\end{equation}
which do not depend on the normalization, and where usually also a number of 
other systematics cancel. 

With the increased statistical precision that is to be expected
in the next few years, it will become increasingly important for an appropriate
branching ratio and partial width evaluation that individual measurements 
are reported according to \Eq~(\ref{eq:brmeas-correct})
and whenever possible also as in \Eq~(\ref{eq:brmeas-Rdef}).
In order to perform the best estimate based on a set of measurements from different
experiments, it might also become important to take into account the systematic errors
that are common to all measurements performed by the same experiment. 
An appropriate choice of a set of independent measurements of (\ref{eq:brmeas-correct}) and 
(\ref{eq:brmeas-Rdef}) from each experiment is likely the best option for a global
fit to quarkonium branching ratios. 
A comparison of $R_{\cal B}(f/f')$ that could  be directly measured
by virtually all experiments, could also help understand possible
systematic effects, which are going to be the limiting factor on branching ratio 
determinations.

Here, we briefly outline the experimental techniques and analysis strategies
adopted to determine these branching ratios with emphasis on the corresponding possible 
normalization choices, as a necessary ingredient to understand possible mutual
dependencies and constraints.

\subsection[Branching ratios measured in $e^+e^-$ formation
  experiments]{Branching ratios measured in $e^+e^-$ formation experiments}
\label{sec:brmeas-ee}
$e^+e^-$ formation experiments are undoubtedly the most important tool to 
investigate charmonium and bottomonium branching ratios by a variety
of techniques.
In these experiments the $n^3S_1$ quarkonium states can be directly formed and
the ${\cal B}(n^3S_1\to {\rm f})$ 
are determined either normalizing to a specific decay mode,
\ie providing a direct measurement of 
$\displaystyle\frac{{\cal B}(n^3S_1\to {\rm f})}{{\cal B}(n^3S_1\to {\rm Norm})}$,
or measuring the number of $n^3S_1$ by performing
a scan of the resonance.

The usual choice for the normalization channel is the inclusive
hadronic decay mode, which is close to 100\% for all resonances, \ie
it provides to a good approximation an absolute normalization.
However, it requires subtraction of the non resonant hadronic
cross-section whose yield (at the given running condition) must be
calculated taking into account the interference with the resonance.
When the total number of events is determined by a scan of the
resonance (which also provides measurements of $\Gamma_{\rm tot}$,
${\cal B}_{\ell\ell}$ and ${\cal B}_{\rm hadr}$), there is in
principle a possible correlation of the branching ratio to the values
for these quantities that is likely small if the scan has many points,
but should not be overlooked.  As stressed in
\Chapter~\ref{chapter:commonexperimenttools},
\Section~\ref{sec:interf}, interference with the continuum for any
specific final state might introduce sizeable corrections.  A
measurement of the ratio $R_{\cal B}(f/{\rm Norm})$ across the
formation energy of the resonance is needed to understand the
interference and its impact on branching ratios.

All other states are studied in hadronic or radiative decays, and the number of events produced
for each state must be determined using the appropriate $n^3S_1$ branching ratio:
\begin{eqnarray}
N^{\rm prod}_{n'3P_J} & = & N^{\rm prod}_{n^3S_1}\times
{\cal B}(n^3S_1\to \gamma\, n'^3P_J ),
\label{eq:brmeas-eechi}\\
N^{\rm prod}_{n'1S_0} & = & N^{\rm prod}_{n^3S_1}\times
{\cal B}(n^3S_1\to \gamma\, n'^1S_0).
\label{eq:brmeas-eeeta}
\end{eqnarray}
Thus, for $^3P_J$ and $^1S_0$ states these experiments can only directly measure
the ratios $R_{\cal B}(f/f')$ and the following combinations of branching ratios:
\begin{eqnarray} & &
\frac{{\cal B}(n^3S_1\to \gamma\, n'^3P_J)}{{\cal B}(n^3S_1\to {\rm Norm})}
\,{\cal B}(n^3P_J\to {\rm f}),\\
&& \frac{{\cal B}(n^3S_1\to \gamma\, n'^1S_0)}{{\cal B}(n^3S_1\to {\rm Norm})}
\,{\cal B}(n'^1S_0\to {\rm f}).
\end{eqnarray}

On the other hand, since the ${\cal B}(\psi(2S)\to J/\psi \pi^+\pi^-)$
is reasonably large, and the events can be easily selected by just
reconstructing the $\pi^+\pi^-$ recoiling against the $J/\psi$, 
absolute measurements of $J/\psi$ 
branching ratios have been obtained based on ``tagged'' $J/\psi$ samples:
\begin{equation}
{\cal B}(J/\psi\to {\rm f}) =\frac{eff_{\pi^+\pi^-\,X}}{eff_{\pi^+\pi^-\,{\rm f}}}
\frac{N^{\rm obs}(\psi'\to (\pi^+\pi^-)_{\rm recoil} {\rm f})}
{N^{\rm obs}(\psi'\to (\pi^+\pi^-)_{\rm recoil}\,X)}. 
\end{equation}
From the experimental point of view this is a particularly clean measurement, 
since the efficiency ratio can be determined with high precision. 
With the increased CLEO~III samples, it would be interesting to fully exploit
the possibility of using ``tagged''
$\Upsilon(2S)$ and $\Upsilon(3S)$ samples to perform absolute 
$\Upsilon(1S)$ and $\Upsilon(2S)$ branching ratios determinations.  

Radiative decay branching ratios (\eg direct $1^{--}\to \gamma\, X$ and 
$1^{--}\to \gamma\, X\to \gamma\gamma X'$) have also been directly measured.

In all cases, photon candidates that are likely to originate from $\pi^0$
are not considered ($\pi^0$ veto), and the efficiency correction relies
on Monte Carlo, and ultimately on the event generator used to
model the particle multiplicities, and the angular and momentum distributions.

Despite efforts to tune JETSET \cite{Sjostrand:1995iq} fragmentation
parameters to reproduce specific classes of inclusive events (\eg
hadronic events in the continuum \cite{Hu:2001wp} below $D\bar D$
threshold or $J/\psi$, $\psi(2S)$ decays \cite{Chen:2000tv}), there
are simply not enough experimentally measured $\chi_c$, $\chi_b$,
$\eta_c$, $\eta_b$ decays to light hadrons ($l.h.$) to compare these
models with.  That could eventually become a limiting systematic to
these measurements.

\subsection[Branching ratios and partial widths measured in $p\bar p$
  formation experiments]{Branching ratios and partial widths measured in $p\bar p$ formation experiments}
\label{sec:brmeas-ppbar}
In these experiments \cite{Garzoglio:2004kw}
a scan of the resonance allows direct measurements of mass,
total width and ${\cal B}(p\bar p){\cal B}_{\rm f}$ for all
charmonium resonances.\footnote{The $p\bar p$ branching ratios of bottomonium states
are likely 3 orders of magnitudes smaller than for charmonium, and only
when a measurement will be available, it will be possible to judge on the
feasibility of such experiments.} 
For resonances whose natural width is comparable or smaller than the beam width 
(${\cal O}(700\,{\rm MeV})$ for E760 and E835), the product  
${\cal B} (p\bar p){\cal B}_{\rm f}$ is highly correlated to the
total width, and the quantity $\Gamma(p\bar p){\cal B}_{\rm f}$ is more
precisely determined. 
By detecting the resonance formation in more than one final state,
the ratio of branching ratios $R_{\cal B}(f/f')$
can be determined independently from the
total width and ${\cal B}(p\bar p)$, in general with small systematic errors 
since the final state is fully reconstructed, and the angular distribution only 
depends on a limited number of decay and formation amplitudes. 
Interference effects with the continuum could
affect the measurement of ${\cal B} (p\bar p){\cal B}_{\rm f}$ and $R_{\cal B}(f/f')$,
but as in $e^+e^-$ experiments, their relevance could be estimated by a measurement of 
$R_{\cal B}(f/f')$ across the formation energy of the resonance. 
Unfortunately, only a few highly characteristic final states of charmonium ($e^+e^-$, 
$J/\psi\,X$, $\gamma\gamma$) can be detected by these experiments, because of the
large hadronic non-resonant cross-section. 

Recently, a pioneering study of $p\bar p\to\pi^0\pi^0$
\cite{Andreotti:2003sk} and $\eta\eta$ differential cross-sections at
the $\chi_{c0}$ formation energy has shown that also selected
exclusive two-body hadronic decays can be successfully measured.  The
interference with the continuum could be successfully exploited by the
next generation of $p \bar p $ annihilation experiments to extend the
knowledge of $\chi_c$ and $\eta_c$ branching ratios to baryons or
light hadrons.

\subsection{Branching ratios and partial widths measured 
            in two-photon reactions}
\label{sec:brmeas-gamgam}

The number of events observed for a specific final state is
proportional to $\Gamma_{ \gamma \gamma}{\cal B}_{\rm f}\times {\cal
L}_{\gamma\gamma}$, where the effective two-photon luminosity function
${\cal L}_{\gamma\gamma}$ (see
\Chapter~\ref{chapter:commonexperimenttools},
\Section~\ref{sec:gagalum})
is calculated by all experiments using the same formalism (even if not
all using the same generator). The only directly measurable quantity
is
\begin{equation}
\Gamma_{ \gamma \gamma}{\cal B}_{\rm f},
\label{eq:brmeas-gamgam}
\end{equation}
or (if more than one final state is detected) $R_{\cal B}(f/f')$.  The
theoretical uncertainties in ${\cal L}_{\gamma\gamma}$ are largely
common to all experiments and that should be taken into account for
future high statistics measurements.  It might be worth mentioning
here that the values reported in the past by different experiments for
the $\Gamma_{\gamma\gamma}$, derived from their measurement of
(\ref{eq:brmeas-gamgam}), are not independent and they are not always
easily comparable since some of them are obtained by a weighted
average of many decay modes, which are individually poorly known.

\subsection[Branching ratios and partial widths measured by radiative return
  (ISR)]{Branching ratios and partial widths measured by radiative return (ISR)}
\label{sec:brmeas-ISR}
Because of initial state radiation (ISR, also referred to as hard photon emission or
radiative return), $e^+e^-$ colliders 
are effectively at the same time (asymmetric) colliders for all
$\sqrt{s}$ energies below nominal collision energy.
The effective luminosity (and therefore event yields) can be sizeable
\cite{Benayoun:1999hm} and
can be determined quite accurately by counting $\mu\mu\gamma$
events, for which precise expressions (and event generators
based on them) are commonly available.
The major advantage of this technique is that $e^+e^-\to X$
can be measured simultaneously and under uniform detector conditions
over a broad range of $\sqrt{s}$. And they ``come for free'' at any
of the $e^+e^-$ factories, which are expected to collect large data samples.

The main interest is the measurement of $R$, but for any exclusive final state
those experiments could obtain a direct measurement of 
$\Gamma_{ e^+ e^-}{\cal B}_{\rm f}$ for any resonance whose mass is lower 
than the collision energy, and, again by detecting more than one final state,
$R_{\cal B}(f/f')$.
To date only BES \cite{Bai:1998qe} and BaBar \cite{Aubert:2003sv} have
used this technique to measure 
$\Gamma(\psi'\to e^+ e^-){\cal B}(\psi'\to J/\psi\pi\pi)$ and
$\Gamma(J/\psi\to e^+ e^-){\cal B}(J/\psi\to \mu^+\mu^-)$ respectively.
Measurements of $\Gamma_{e^+ e^-}{\cal B}_{l^+l^-}$ would provide
important constraints on both the total width and $\Gamma_{e^+e^-}$ for all
$1^{--}$ states, providing at the same time an important cross check for 
possible systematic errors.

\subsection[Branching ratios measured in $B$ decays]{Branching ratios measured in $B$ decays}
\label{sec:brmeas-Bdec}
Asymmetric $B$ factories focused originally on exclusive $B$ decays to final 
states involving a $c\bar c$ as the cleanest modes to study CP violation.   

With the impressive amount of data collected so far (more than 500~fb$^{-1}$ as of summer
2004 adding Belle and BaBar) and ${\cal B}(B\to c\bar c\, X)$ of order  $10^{-3}$,
both experiments are collecting larger and larger samples of exclusive $B$ decays
to charmonia, and they are obviously interested in reconstructing them into
as many different final states as possible.
The same is true for D0 and CDF, since the preliminary reconstruction of highly 
characteristic exclusive charmonium (and bottomonium) final states is needed for 
other analyses. 

For charmonium the quantity directly measured by these experiments is
\begin{equation}
{\cal B}(B\to c\bar c \,X)\times {\cal B} (c\bar c\to {\rm f}),
\label{brmeas:Bfact}
\end{equation}
and again from the number of fully reconstructed events
into different final states these experiments can directly measure $R_{\cal B}(f/f')$
for a variety of final states and for virtually all quarkonium states.
Even if the precision might not always compete with other techniques, the wide range of
possible $R_{\cal B}(f/f')$ measurements, with likely different sources of systematic 
errors, would certainly be important in evaluating quarkonium branching ratios, in 
particular for those states ($\chi_Q$ and $\eta_Q$) whose branching ratios are largely 
unknown.

\subsection{Indirect determinations as a tool to 
            investigate systematic effects}
\label{sec:brmeas-sys}

The possibilities offered by the mutual constraints posed by
measurements of different products or ratios of branching ratios have
so far been only partially exploited.

The first advantage is that branching ratios measured by different
techniques have different sources of systematic errors, and the
comparison can provide insight on how to nail them down.  The current
best estimate for ${\cal B}(\chi_{c2}\to\gamma J/\psi)$
\cite{Eidelman:2004wy} is largely determined by measurements of
$\Gamma(\chi_{c2}\to p\bar p){\cal B}(\chi_{c2}\to \gamma J/\psi)$,
$\Gamma(\chi_{c2}\to \gamma \gamma){\cal B}(\chi_{c2}\to \gamma
J/\psi)$ and ${\cal B}(\chi_{c2}\to \gamma\gamma)/ {\cal
B}(\chi_{c2}\to \gamma J/\psi)$, to the point that these measurements
indirectly constrain the estimate of ${\cal B}(\psi'\to \gamma
\chi_{c2})$ to a value significantly lower than the world average of
direct measurements, since the product ${\cal B}(\psi'\to \gamma
\chi_{c2}){\cal B}(\chi_{c2}\to\gamma J/\psi)$ has been measured with
high precision.\footnote{New more precise measurements of ${\cal
B}(\psi'\to \gamma \chi_{c2})$ might in turn provide constraints for
${\cal B}(\psi'\to \gamma \chi_{c2}){\cal B}(\chi_{c2}\to\gamma
J/\psi)$}

The other advantage is that measurements of different product and
ratios of branching ratios pose constraints on their values: for
$\chi_{c0}$ at present the partial widths $\Gamma_{\gamma\gamma}$ and
$\Gamma_{\gamma J/\psi}$ are known to $\approx$10\%
\cite{Eidelman:2004wy}, even if none of the many measurements more or
less directly related to these quantities ($\Gamma$,
$\Gamma_{\gamma\gamma}{\cal B}_{4\pi}$,
$\Gamma_{\gamma\gamma}/\Gamma_{\gamma J/\psi}$, $\Gamma_{\gamma
J/\psi}{\cal B}_{p\bar p}$, ${\cal B}(\psi'\to\gamma \chi_{c0})$,
${\cal B}(\psi'\to\gamma \chi_{c0}){\cal B}_{p\bar p}$, ${\cal
B}(\psi'\to\gamma \chi_{c0}){\cal B}_{\gamma J/\psi}$ and others) is
individually known much better than that.

The proposed next generation of $p\bar p$ experiments with extended
PID ability could provide invaluable information by measuring $p\bar
p\to p\bar p$ differential cross-section at the $\eta_c$ (and possibly
at the $\chi_{c0}$). This would provide a direct measurement of ${\cal
B}(c\bar c\to p\bar p)$, indirectly constraining the radiative
$J/\psi$ (and $\psi'$) M1 transitions from the well measured ${\cal
B}(J/\psi\to \gamma\eta_c\to \gamma p\bar p)$.  Since at present the
$\approx$30\% uncertainty in ${\cal B}(J/\psi\to \gamma\eta_c)$ is the
major source of uncertainty in all $\eta_c$ branching ratios, this
will also directly affect all $\eta_c$ branching ratios.

With the increased statistics available at $B$ factories it might soon become
possible to determine at least some of the ${\cal B}(B\to c\bar c\, l.h.)$ 
branching ratios without explicitly reconstructing the charmonium. 
In this case, simultaneous measurements of the same $B$ decay mode in 
exclusive final states ${\cal B}(B\to c\bar c\, l.h.) {\cal B}(c\bar c\, \to {\rm f})$
would allow $B$ factories to directly measure ${\cal B}(c\bar c\, \to {\rm f})$ from
\Eq~(\ref{brmeas:Bfact}). Considering that the photon in $\psi(2S)\to\gamma\eta_c(2S)$ is 
very soft and that this inclusive radiative transition will likely be 
difficult to measure for both CLEO-c and BES~III, this might well be the best 
way of determining the $\eta_c(2S)$ branching ratios, 
and indirectly determining the partial width for the M1 
$\psi(2S)\to\gamma\eta_c(2S)$ transition itself.

%%%%%%%%%%%%%%  
%ID
%%%%%%%%%%%%%%  
\section[Electromagnetic and inclusive decays into light particles]
        {Electromagnetic and inclusive decays into light particles 
         $\!$\footnote{Authors: T.~Ferguson, C.~Patrignani, A.~Vairo}}
\label{sec:ID}

\subsection{Theoretical framework}
\label{sec:secnrqcd}
\shortpage

The main dynamical mechanism of heavy-quarkonium decay into light
particles is quark--antiquark annihilation. Since this happens at a
scale $2m$ ($m$ is the heavy quark mass), which is perturbative, the
heavy quarks annihilate into the minimal number of gluons allowed by
colour conservation and charge conjugation. The gluons subsequently
create light quark--antiquark pairs that form the final state hadrons:
$Q\bar{Q} \to n g^* \to m(\qqb)$. Values of $n$ are given for various
quarkonia in \Table~\ref{tab:exdec-tab1}; for comparison the minimal number
of photons into which a $Q\bar{Q}$ pair can annihilate is also listed.
Experimentally this fact is reflected by the narrow width of the heavy
quarkonia decays into hadronic channels in a mass region where strong
decays typically have widths of hundreds of MeV.  As an example let us
consider the $\jpsi$ decay into light hadrons.  Following
\cite{exdec:appel75}, this process is regarded as the decay into three
real gluons.  The calculation of this width leads to the result 
\begin{eqnarray}
\Gamma(\jpsi\to l.h.) & = & \frac{10}{81}\, \frac{\pi^2-9}{\pi
e_c^2}\, \frac{\as^3}{\aem^2}\, \Gamma(\jpsi\to e^+e^-) = 205\; {\rm
keV}\, \left(\frac{\as}{0.3}\right)^3\,.
\label{eq:exdec-tot}
\ea 
Although this value is somewhat larger than the experimental one it
explains the narrowness of the hadronic decays of the quarkonia.
Corrections like relativistic, $\as$ or colour-octet ones, may lead to
a better agreement with experiment. A systematic way to include these
corrections is provided by nonrelativistic effective field theories of
QCD.

\begin{table}[ht]
\caption[Quantum numbers of quarkonium states]
        {Quantum numbers of quarkonium states and the minimal number
         of virtual gluons and photons into which they can
         annihilate. The subscript $d$ refers to a gluonic
         colour-singlet state that is totally symmetric under
         permutations of the gluons.}
\label{tab:exdec-tab1}
\renewcommand{\arraystretch}{1.2}
 \begin{center} 
  \begin{tabular}{|c||c|c||c|c|} \hline
             & $^{2S+1}L_J$ & $I^G$($J^{PC}$)& gluons & photons  \\ \hline \hline
     $\eta_{c}$, $\eta_b$  & $^1S_0$ & $0^+$($0^{-+}$)&   2g   & 2$\gamma$ \\ \hline
     $\jpsi$, $\Upsilon(1S)$        & $^3S_1$ & $0^-$($1^{--}$)&(3g)$_d$&  $\gamma$ \\ \hline
     $h_{c}$,  $h_b$    & $^1P_1$ & $0^-$($1^{+-}$)&(3g)$_d$&  $3\gamma$ \\ \hline
    $\chi_{c 0}$, $\chi_{b 0}$ & $^3P_0$ & $0^+$($0^{++}$)&   2g   & 2$\gamma$ \\ \hline
    $\chi_{c 1}$, $\chi_{b 1}$  & $^3P_1$ & $0^+$($1^{++}$)&   2g   & 2$\gamma$ \\ \hline
    $\chi_{c 2}$, $\chi_{b 2}$  & $^3P_2$ & $0^+$($2^{++}$)&   2g   & 2$\gamma$ \\ \hline
   \end{tabular}
 \end{center} 
\renewcommand{\arraystretch}{1.0}
\end{table}

In an effective field theory language\footnote{We refer to
\Chapter~\ref{chapter:commontheoreticaltools} for a basic introduction
to effective field theories and NRQCD.}, at scales lower than $m$
heavy-quarkonium annihilation is resolved as a contact interaction.
This is described at the Lagrangian level by four-fermion operators
whose matching coefficients develop an imaginary part.  Consequently,
the annihilation width of a heavy quarkonium state $|H\rangle$ into
light particles may be written as
\begin{equation}
\Gamma(H \rightarrow light \; particles) = 2\, {\rm Im}\,  \langle H  | {\cal L}_{\psi\chi} | H \rangle,
\label{eq:decaywidth}
\end{equation}
where ${\cal L}_{\psi\chi}$ is given by \Eq~(\ref{eq:4fermiondim6}) of
\Chapter~\ref{chapter:commontheoreticaltools} up to four-fermion
operators of dimension 6.  The low-energy dynamics is encoded in the
matrix elements of the four-fermion operators evaluated on the
heavy-quarkonium state.  If one assumes that only heavy-quarkonium
states with quark-antiquark in a colour-singlet configuration can
exist, then only colour-singlet four-fermion operators contribute and
the matrix elements reduce to heavy-quarkonium wave functions (or
derivatives of them) calculated at the origin. This assumption is
known as the ``colour-singlet model''.  Explicit calculations show that
at higher order the colour-singlet matching coefficients develop
infrared divergences (for P~waves this happens at NLO
\cite{Barbieri:1976fp}).  In the colour-singlet model, these do not
cancel in the expression of the decay widths.  It has been the first
success of NRQCD \cite{Caswell:1985ui,Bodwin:1994jh} to show that the
Fock space of a heavy-quarkonium state may contain a small component
of quark--antiquark in a colour-octet configuration, bound with some
gluonic degrees of freedom (the component is small because operators
coupling transverse gluons with quarks are suppressed by powers of $v
\ll 1$, $v$ being the heavy-quark velocity in the centre-of-mass
frame), that due to this component, matrix elements of colour-octet
four-fermion operators contribute and that exactly these contributions
absorb the infrared divergences of the colour-singlet matching
coefficients in the decay widths, giving rise to finite results
\cite{Bodwin:1992ye,Bodwin:1994jh}.  NRQCD is now the standard
framework to study heavy-quarkonium inclusive decays.
\shortpage

The NRQCD factorization formulas are obtained by separating
contributions coming from degrees of freedom of energy $m$ from those
coming from degrees of freedom of lower energy. In the case of
heavy-quarkonium decay widths, they have been rigorously proved
\cite{Bodwin:1994jh}.  High-energy contributions are encoded into the
imaginary parts of the four-fermion matching coefficients,
$f,g_{1,8,ee,\gamma\gamma, ...}(^{2S+1}L_J)$ and are ordered in powers
of $\als$ (coefficients labeled with $ee$, $\gamma\gamma$, ... refer
to pure electromagnetic decays into $e^+e^-$, $\gamma \gamma$, ...).
Low-energy contributions are encoded into the matrix elements of the
four-fermion operators on the heavy-quarkonium states $|H\rangle$
($\langle \dots \rangle_{H} \equiv \langle H | \dots | H \rangle$).
These are, in general, nonperturbative objects, which can scale as
powers of $\lQ$, $mv$, $mv^2$, ... (\ie of the low-energy dynamical
scales of NRQCD).  Therefore, matrix elements of higher dimensionality
are suppressed by powers of $v$ or $\lQ/m$.  Including up to
four-fermion operators of dimension 8, the NRQCD factorization
formulas for inclusive decay widths of heavy quarkonia into light
hadrons, which follow from \Eq~(\ref{eq:decaywidth}), read
\cite{Bodwin:1994jh,Bodwin:1992ye}:
\begingroup
\allowdisplaybreaks
\begin{eqnarray}
&&\hspace{-8mm}
\Gamma(V_Q (nS) \rightarrow l.h.) = \frac{2}{m^2}\Bigg( 
{\rm Im\,}f_1(^3 S_1) \,  \langle O_1(^3S_1)\rangle_{V_Q(nS)}
\nn
\\
&&\hspace{-8mm}
\qquad\qquad\qquad\qquad
+ {\rm Im\,}f_8(^3 S_1)\, \langle O_8(^3S_1)\rangle_{V_Q(nS)}
+ {\rm Im\,}f_8(^1 S_0)\, \langle O_8(^1S_0)\rangle_{V_Q(nS)} 
\nn
\\
&&\hspace{-8mm}
\qquad\qquad\qquad\qquad
+ {\rm Im\,}g_1(^3 S_1)\,
\frac{\langle {\mathcal P}_1(^3S_1)\rangle_{V_Q(nS)}}{m^2}
+ {\rm Im\,}f_8(^3 P_0)\,
\frac{\langle O_8(^3P_0)\rangle_{V_Q(nS)}}{m^2}
\nn
\\
&&\hspace{-8mm}
\qquad\qquad\qquad\qquad
+ {\rm Im\,}f_8(^3 P_1)\,
\frac{\langle O_8(^3P_1)\rangle_{V_Q(nS)}}{m^2}
+ {\rm Im\,}f_8(^3 P_2)\,
\frac{\langle O_8(^3P_2)\rangle_{V_Q(nS)}}{m^2}\Bigg),
\label{eq:gamma1}
\\
&&
\nn 
\\
&&\hspace{-8mm}
\Gamma(P_Q (nS) \rightarrow l.h.) = \frac{2}{m^2}\Bigg( 
{\rm Im\,}f_1(^1 S_0)\,   \langle O_1(^1S_0)\rangle_{P_Q(nS)}
\nn
\\
&&\hspace{-8mm}
\qquad\qquad\qquad\qquad
+ {\rm Im\,}f_8(^1 S_0)\, \langle O_8(^1S_0)\rangle_{P_Q(nS)}
+ {\rm Im\,}f_8(^3 S_1)\, \langle O_8(^3S_1)\rangle_{P_Q(nS)} 
\nn
\\
&&\hspace{-8mm}
\qquad\qquad\qquad\qquad
+ {\rm Im\,}g_1(^1 S_0)\,
\frac{\langle {\mathcal P}_1(^1S_0)\rangle_{P_Q(nS)}}{m^2}
+ {\rm Im\,}f_8(^1 P_1)\,
\frac{\langle O_8(^1P_1)\rangle_{P_Q(nS)}}{m^2} \Bigg),
\\
&&
\nn
\\
&&\hspace{-8mm}
\Gamma(\chi_Q(nJS)  \rightarrow l.h.)= 
\frac{2}{m^2}\Bigg( {\rm Im \,}  f_1(^{2S+1}P_J)\, 
\frac{\langle O_1(^{2S+1}P_J ) \rangle_{\chi_Q(nJS)}}{m^2}
\nn
\\
&&\hspace{-8mm}
\qquad\qquad\qquad\qquad
+ {\rm Im \,} f_8(^{2S+1}S_S) \,\langle O_8(^1S_0 ) \rangle_{\chi_Q(nJS)} \Bigg).
\label{eq:gammachiLH}
\end{eqnarray}
At the same order the electromagnetic decay widths are given by:
\begin{eqnarray} 
&&\hspace{-8mm}
\Gamma(V_Q (nS) \rightarrow e^+e^-)= \frac{2}{m^2}\Bigg( 
{\rm Im\,}f_{ee}(^3 S_1)\,   \langle O_{\rm EM}(^3S_1)\rangle_{V_Q(nS)}
\nn
\\
&&\hspace{-8mm}
\qquad\qquad\qquad\qquad\qquad
+ {\rm Im\,}g_{ee}(^3 S_1)\,
\frac{\langle {\mathcal P}_{\rm EM}(^3S_1)\rangle_{V_Q(nS)}}{m^2}\Bigg),
\\
&&
\nn
\\
&&\hspace{-8mm}
\Gamma(P_Q (nS) \rightarrow \gamma\gamma)= \frac{2}{m^2}\Bigg( 
{\rm Im\,}f_{\gamma\gamma}(^1 S_0)\,   
\langle O_{\rm EM}(^1S_0)\rangle_{P_Q(nS)}
\nn
\\
&&\hspace{-8mm}
\qquad\qquad\qquad\qquad\qquad
+ {\rm Im\,}g_{\gamma\gamma}(^1 S_0)\,
\frac{\langle {\mathcal P}_{\rm EM}(^1S_0)\rangle_{P_Q(nS)}}{m^2} \Bigg),
\\
&&
\nn
\\
&&\hspace{-8mm}
\Gamma(\chi_Q(nJ1)  \rightarrow \gamma\gamma)= 
2 \, {\rm Im \,}  f_{\gamma\gamma}(^3P_J)\, 
\frac{\langle O_{\rm EM}(^3P_J )\rangle_{\chi_Q(nJ1)}}{m^4}, 
\quad J=0,2\,.
\label{eq:gammachi}
\end{eqnarray}
\endgroup 
The symbols $V_Q$ and $P_Q$ indicate respectively the vector
and pseudoscalar S-wave heavy quarkonium and the symbol $\chi_Q$ the
generic P-wave quarkonium (the states $\chi_Q(n10)$ and
$\chi_Q(nJ1)$ are usually called $h_Q((n-1)P)$ and
$\chi_{QJ}((n-1)P)$, respectively).

The operators $O,{\mathcal P}_{1,8,{\rm EM}}(^{2S+1}L_J)$ are the
dimension $6$ and $8$ four-fer\-mion o\-pe\-ra\-tors of the NR\-QCD
Lagrangian. They are classified by their transformation properties
under colour as singlets ($1$) and octets ($8$), and under spin ($S$),
orbital ($L$) and total angular momentum ($J$).  The operators with
the subscript EM are the colour-singlet operators projected on the QCD
vacuum. The explicit expressions of the operators can be found in
\cite{Bodwin:1994jh} (or listed in Appendix A of
\cite{Brambilla:2002nu}). The dimension 6 operators are also given in
\Eq~(\ref{eq:4fermiondim6}) of
\Chapter~\ref{chapter:commontheoreticaltools}.

In general different power countings are possible at the level of
NRQCD, due to the fact that different scales ($mv$, $\lQ$, $mv^2$,
$\sqrt{m\lQ}$, ...) are still dynamically entangled
\cite{powercounting,Pineda:2000sz}. Likely different power countings
will apply to different physical systems. Therefore, the relative
importance of the different matrix elements that appear in
\Eqs~(\ref{eq:gamma1})--(\ref{eq:gammachi}) may change in going from
lower to higher quarkonium states and from bottomonium to charmonium.
Whatever the power counting is, the pseudoscalar and vector state
decay widths are dominated by the colour-singlet matrix elements, which
contribute at order $mv^3$.  The hadronic P-state decay widths have
two contributions (the colour-singlet and colour-octet matrix elements),
which contribute at the same order $mv^5$, if we assume that a
fraction $v$ of the P-state wave function projects onto the
colour-octet operator.

Since NRQCD is an expansion in two small parameters ($\als$ and $v$),  
progress comes typically from (1) improving the perturbative 
series of the matching coefficients either by fixed order calculations 
or by resumming large contributions (large logs or large contributions 
associated to renormalon singularities); 
(2) improving the knowledge of the NRQCD matrix elements either by direct 
evaluation, which may be obtained by fitting the experimental data, by lattice 
calculations, and by models, or by exploiting the hierarchy of scales still entangled 
in NRQCD and constructing EFTs of lower energy.

\subsubsection[The perturbative expansion]{The perturbative expansion}
\label{sec:perturexpan}

The imaginary parts of the four-fermion matching coefficients have
been calculated over the last twenty years to different levels of
precision.  Up to order $\als^3$ the imaginary parts of $f_8(^1S_0)$,
$f_1(^{3}P_1)$, and $f_8(^3P_J)$ can be found in
\cite{Petrelli:1998ge}, the imaginary parts of $f_8(^3S_1)$,
$f_8(^1P_1)$ in \cite{Maltoniphd} and the imaginary part of
$f_1(^1S_0)$ in \cite{Petrelli:1998ge,Huang:1996fa}.  Two different
determinations of $f_1(^3P_0)$ and $f_1(^3P_2)$ exist at NLO in
\cite{Petrelli:1998ge} and \cite{BCGRpwave}.  The imaginary part of
$f_1(^3S_1)$ has been calculated (numerically) up to order $\als^4$ in
\cite{Mackenzie:1981sf}. The imaginary part of $g_1(^3S_1)$ at order
$\als^3$ can be found in \cite{Gremm:1997dq}, the imaginary part of
$g_1(^1S_0)$ at order $\als^2$ in \cite{Bodwin:1994jh}. Where the
electromagnetic coefficients are concerned, the imaginary part of
$f_{ee}(^3 S_1)$ has been calculated up to order $\al^2\als^2$ in
\cite{Czarnecki:1997vz,Beneke:1997jm}, the imaginary parts of
$f_{\gamma\gamma}(^1 S_0)$ and $f_{\gamma\gamma}(^3 P_{0,2})$ up to
order $\al^2\als$ can be found in \cite{Petrelli:1998ge,BCGRswave} and
$g_{ee}(^3 S_1)$ and $g_{\gamma\gamma}(^1 S_0)$ up to order $\al^2$ in
\cite{Bodwin:1994jh}. A complete list of the above matching
coefficients at our present level of knowledge can be found in
Appendix A of \cite{Vairo:2003gh}.  The LL running for the imaginary
parts of the matching coefficients of the four-fermion NRQCD operators
of dimension 6 and 8 have been obtained in\cite{Brambilla:2002nu} and
can be read there in Appendix C.  The tree-level matching of dimension
9 and 10 S-wave operators can be found in \cite{Bodwin:2002hg}.  The
tree-level matching of dimension 9 and 10 electromagnetic P-wave
operators can be found in \cite{Ma:2002ev}.

The convergence of the perturbative series of the four-fermion
matching coefficients is often poor. While the large two-loop
contribution of ${\rm Im\,}f_{ee}(^3 S_1)$ seems to be related, at
least in the bottomonium case, to the factorization scale and,
therefore, may be put presumably under control via renormalization
group improvement techniques \cite{Beneke:1997jm,Penin:2004ay}, large
corrections appearing in other S-wave decay channels have been
ascribed to renormalon-type contributions \cite{chen}.  There is no
such study so far for P-wave decays.

\subsubsection[The relativistic expansion]{The relativistic expansion}
\label{sec:rel}
The NRQCD matrix elements may be fitted to the experimental decay data
\cite{exdec:man95,Maltoni:2000km,mussa} or calculated on the lattice
\cite{Bodwin:2001mk,Bodwin:1996tg}. The matrix elements of
colour-singlet operators can be linked at leading order to the
Schr\"odinger wave functions at the origin
\cite{Bodwin:1994jh}\footnote{ This statement acquires a precise
meaning only in the context of pNRQCD, see \Section~\ref{sec:pnrqcd}.}
and, hence, may be evaluated by means of potential models
\cite{Eichten:1995ch} or potentials calculated on the lattice
\cite{Bali:2000gf}.  In \cite{Maltoni:2000km} by fitting to the
charmonium P-wave decay data it was obtained that $\langle
O_1(^1P_1)\rangle_{h_c(1P)} \approx 8.1\times 10^{-2}~\hbox{GeV}^5$
and $\langle O_8(^1S_0)\rangle_{h_c(1P)} \approx 5.3\times
10^{-3}~\hbox{GeV}^3$ in the $\overline{\rm MS}$ scheme and at the
factorization scale of 1.5~Gev.  In the quenched lattice simulation of
\cite{Bodwin:1996tg} it was obtained that $\langle
O_1(^1S_0)\rangle_{\eta_c(1S)} \approx 0.33~\hbox{GeV}^3$, $\langle
O_1(^1P_1)\rangle_{h_c(1P)} \approx 8.0\times 10^{-2}~\hbox{GeV}^5$
and $\langle O_8(^1S_0)\rangle_{h_c(1P)} \approx 4.7\times
10^{-3}~\hbox{GeV}^3$ in the $\overline{\rm MS}$ scheme and at the
factorization scale of 1.3~Gev.  In the lattice simulation of
\cite{Bodwin:2001mk} and in the three light-quark flavours
extrapolation limit it was obtained that $\langle
O_1(^1S_0)\rangle_{\eta_b(1S)} \approx 4.1~\hbox{GeV}^3$, $\langle
O_1(^1P_1)\rangle_{h_b(1P)} \approx 3.3~\hbox{GeV}^5$ and $\langle
O_8(^1S_0)\rangle_{h_b(1P)} \approx 5.9\times 10^{-3}~\hbox{GeV}^3$ in
the $\overline{\rm MS}$ scheme and at the factorization scale of 4.3
GeV.

It has been discussed in \cite{Ma:2002ev} and \cite{Bodwin:2002hg},
that higher-order operators, not included in the formulas
(\ref{eq:gamma1})--(\ref{eq:gammachi}), even if parametrically
suppressed, may turn out to give sizable contributions to the decay
widths.  This may be the case, in particular, for charmonium, where
$v^2 \sim 0.3$, so that relativistic corrections are large, and for
P-wave decays where the above formulas provide, indeed, only the
leading-order contribution in the velocity expansion.  In fact it was
pointed out in \cite{Ma:2002ev} (see also \cite{Vairo:2002iw}) that if
no special cancellations among the matrix elements occur, then the
order $v^2$ relativistic corrections to the electromagnetic decays
$\chi_{c0} \to \gamma\gamma$ and $\chi_{c2}\to \gamma\gamma$ may be as
large as the leading terms.

In \cite{Gremm:1997dq,Maltoni:2000km} it was also noted that the
numerical relevance of higher-order matrix elements may be enhanced by
their multiplying matching coefficients.  This is, indeed, the case
for the decay width of S-wave vector states, where the matching
coefficients multiplying the colour-octet matrix elements (with the
only exception of ${\rm Im} f_8(^3P_1)$) are enhanced by $\als$ with
respect to the coefficient ${\rm Im} f_1(^3S_1)$ of the leading
colour-singlet matrix element.

In the bottomonium system, 14 S- and P-wave states lie below the open
flavour threshold ($\Upsilon(nS)$ and $\eta_b(nS)$ with $n=1,2,3$;
$h_b(nP)$ and $\chi_{bJ}(nP)$ with $n=1,2$ and $J=0,1,2$) and in the
charmonium system 8 ($\psi(nS)$ and $\eta_c(nS)$ with $n=1,2$;
$h_c(1P)$ and $\chi_{cJ}(1P)$ with $J=0,1,2$).  For these states
Eqs. (\ref{eq:gamma1})--(\ref{eq:gammachi}) describe the decay widths
into light hadrons and into photons or $e^+e^-$ in terms of 46 NRQCD
matrix elements (40 for the S-wave decays and $6$ for the P-wave
decays), assuming the most conservative power counting.  More matrix
elements are needed if higher-order operators are included.

\subsubsection[pNRQCD]{pNRQCD}
\label{sec:pnrqcd}

The number of nonperturbative parameters may be reduced by integrating
out from NRQCD degrees of freedom with energy lower than $m$, since
each degree of freedom that is integrated out leads to a new
factorization. Eventually, one ends up with pNRQCD
\cite{Pineda:1997bj,Brambilla:1999xf}, where only degrees of freedom
of energy $mv^2$ are left dynamical. In the context of pNRQCD, the
NRQCD four-fermion matrix elements can be written either as
convolutions of Coulomb amplitudes with non-local correlators (in the
dynamical situation $mv^2 \simg \lQ$) or as products of wave functions
at the origin by non-local correlators (in the dynamical situation
$mv^2 \ll \lQ$).

The first situation may be the relevant one at least for the
bottomonium ground state
\cite{Brambilla:1999xf,Brambilla:1999xj,Brambilla:2000am}.  In the
limiting case $mv^2 \gg \lQ$, the correlators reduce to local
condensates and explicit formulas have been worked out in
\cite{TY,Pineda:1996uk}.  Concerning the perturbative calculation of
the electromagnetic decay widths, the NLL renormalization group
improved expression can be found in \cite{Pineda:2001et} and has been
used in a phenomenological analysis in \cite{Pineda:2003be}. The
perturbative wave functions at the origin at NNLO order can be found
in \cite{mel}. Recently, a full NNLL analysis has been carried out in
\cite{Penin:2004ay}; the authors predict
$\Gamma(\eta_b\to\gamma\gamma) /\Gamma(\Upsilon(1S)\to e^+e^-) = 0.502
\pm 0.068 \pm 0.014$, where the first error is an estimate of the
theoretical uncertainty and the second reflects the uncertainty in
$\als$.  We also mention that there exists a determination of
$\Gamma(\Upsilon(2S)$ $\to$ $e^+e^-)$ $/$ $\Gamma(\Upsilon(1S)\to
e^+e^-)$ in lattice NRQCD with 2+1 flavours of dynamical quarks
\cite{Gray:2002vk}.  The calculated ratio is still far from the
experimental result, although the unqueching has considerably reduced
the discrepancy.

The last situation is expected to be the relevant one for most of the
existing excited heavy-quarkonium states (with the possible exception
of the lowest bottomonium states) and has been studied in
\cite{Brambilla:2001xy,Brambilla:2002nu,Brambilla:2003mu}.  However, a
general consensus on the above assignments of heavy-quarkonium states
to dynamical regions has not been reached yet (see also
\Chapter~\ref{chapter:spectroscopy}).

At leading order in the $v$ and $\lQ/m$ expansion, the colour-singlet
matrix elements can be expressed in terms of the wave functions at the
origin only \cite{Bodwin:1994jh,Brambilla:2002nu}:
\begin{eqnarray}
\label{eq:O1S}
&&\langle O_1(^3S_1)\rangle_{V_Q(nS)}=
\langle O_1(^1S_0)\rangle_{P_Q(nS)}=
\langle O_{\rm EM}(^3S_1)\rangle_{V_Q(nS)}
\nn\\
&&\qquad\qquad\qquad~~~~
=\langle O_{\rm EM}(^1S_0)\rangle_{P_Q(nS)}= 
C_A \frac{|R^{(0)}_{n0}({0})|^2}{2\pi}, 
\label{eq:O1P}
\\
&&
\langle O_1(^{2S+1}P_J ) \rangle_{\chi_Q(nJS)} = 
\langle O_{\rm EM}(^{2S+1}P_J ) \rangle_{\chi_Q(nJS)}  
= \frac{3}{2}\frac{C_A}{\pi} |R^{(0)\,\prime}_{n1}({0})|^2,
\end{eqnarray}
where $R_{n\ell}^{(0)}$ is the zeroth-order radial part of the
heavy-quarkonium wave function, obtained from the pNRQCD Hamiltonian
\cite{Brambilla:2000gk,Pineda:2000sz} and $C_A = N_c = 3$.

In the situation $mv^2 \ll \lQ$ there are no dynamical gluons at energies 
of order $mv^2$. Under the conditions that: (a) all
higher gluonic excitations between the two heavy quarks develop a mass
gap of order $\lQ$, (b) threshold effects are small, and (c)
contributions coming from virtual pairs of quark--antiquark with
three-momentum of order $\sqrt{m\lQ}$ are
subleading,\footnote{Condition (b) may be problematic for the
$\psi(2S)$, whose mass is very close to the $D\bar{D}$ production
threshold.}  the NRQCD colour-octet matrix elements relevant for
\Eqs~(\ref{eq:gamma1})--(\ref{eq:gammachi}) can be written at leading order
in the $v$ and $\lQ/m$ expansion as\cite{Brambilla:2001xy,Brambilla:2002nu}:
\begin{eqnarray}
&&\hspace{-5mm}
\langle O_8(^3S_1)\rangle_{V_Q(nS)}=
\langle O_8(^1S_0)\rangle_{P_Q(nS)}
=C_A \frac{|R^{(0)}_{n0}({0})|^2}{2\pi}
\left(- \frac{2 (C_A/2-C_F) {\mathcal E}^{(2)}_3}{3 m^2 }\right),
\\
&&\hspace{-5mm}
\langle O_8(^1S_0)\rangle_{V_Q(nS)}=
\frac{\langle O_8(^3S_1)\rangle_{P_Q(nS)}}{3}
=C_A \frac{|R^{(0)}_{n0}({0})|^2}{2\pi}
\left(-\frac{(C_A/2-C_F) c_F^2{\mathcal B}_1}{3 m^2 }\right),
\\
&&\hspace{-5mm}
\frac{\langle O_8(^3P_J) \rangle_{V_Q(nS)}}{2J+1}=
\frac{\langle O_8(^1P_1)\rangle_{P_Q(nS)}}{9}
= C_A \frac{|R^{(0)}_{n0}({0})|^2}{2\pi}
\left(-\frac{(C_A/2-C_F) {\mathcal E}_1}{9}\right),
\\
&&\hspace{-5mm}
\langle O_8(^1S_0)\rangle_{\chi_Q(nJS)} 
= \frac{T_F}{3}
\frac{\vert R^{(0)\,\prime}_{n1}({0})\vert^2}{\pi m^2} {\mathcal E}_3, 
\label{eq:matoct}
\end{eqnarray}
where $c_F$ stands for the chromomagnetic matching coefficient, which
is known at NLL \cite{Amoros:1997rx}, $C_F = (N_c^2-1)/(2N_c) = 4/3$
and $T_F = 1/2$.  Therefore, at the considered order, the colour-octet
matrix elements factorize into the product of the heavy-quarkonium
wave function with some chromoelectric and chromomagnetic correlator
(Wilson lines connecting the fields are not explicitly shown, but
understood):
\begin{eqnarray} 
&& \hspace{-5mm}
{\cal E}_n = 
\frac{1}{N_c}\int_0^\infty \!\! dt \, t^n 
\langle {\rm Tr} (g{\bf E}(t)\cdot g{\bf E}(0))\rangle,
\qquad\quad
{\cal B}_n = 
\frac{1}{N_c}\int_0^\infty  \!\! dt \, t^n 
\langle {\rm Tr} (g{\bf B}(t)\cdot g{\bf B}(0))\rangle, \\
&& \hspace{-5mm}
{\cal E}^{(2)}_3 = \frac{1}{4 N_c}
\int_0^\infty  \!\! dt_1\int_0^{t_1} \!\!  dt_2\int_0^{t_2}  \!\! dt_3 \,(t_2-t_3)^3 
\bigg\{
\langle {\rm Tr} (\{g{\bf E}(t_1)\cdot, g{\bf E}(t_2)\}\, \{g{\bf E}(t_3)\cdot, g{\bf 
E}(0)\})\rangle_c
\nn
\\
&&  
~~~~~~~~~~~~~~~~~~~~~~~~~~~~~~~~~~~~~~~~~~~~~~~~~~~~~~
- \frac{4}{N_c}\langle {\rm Tr}(g{\bf E}(t_1)\cdot g{\bf E}(t_2))\, {\rm
Tr}(g{\bf E}(t_3)\cdot g{\bf E}(0))\rangle_c
\bigg\} ,
\end{eqnarray}
where
\begin{eqnarray}
&& \hspace{-5mm}
\langle {\rm Tr} \left( g{\bf E}(t_1)\cdot g{\bf E}(t_2) \; g{\bf E}(t_3)\cdot g{\bf
E}(0)\right) \rangle_c =
\langle {\rm Tr} \left( g{\bf E}(t_1)\cdot g{\bf E}(t_2) \; g{\bf E}(t_3)\cdot g{\bf
E}(0)\right)\rangle
\nn
\\
&&
~~~~~~~~~~~~~~~~~~~~~~~~~~~~~~~~~~~~~~~~~~~~~~~~~~~~~~
- \frac{1}{N_c} \langle {\rm Tr} (g{\bf E}(t_1)\cdot g{\bf E}(t_2))\rangle
\langle {\rm Tr} (g{\bf E}(t_3)\cdot g{\bf E}(0))\rangle. 
\end{eqnarray}
These correlators are universal in the sense that they do not depend 
on the heavy-quarkonium state and, hence, may be calculated once and for 
all, either by means of lattice simulations \cite{correlatorlat},
or specific models of the QCD vacuum \cite{correlator}, or extracted from 
some set of experimental data \cite{Brambilla:2001xy}. 

Finally, at leading order the matrix elements of the ${\mathcal P}_1$ operators 
can be written as:
\begin{eqnarray}
&&
\langle {\mathcal P}_1(^3S_1)\rangle_{V_Q(nS)}=
\langle {\mathcal P}_1(^1S_0)\rangle_{P_Q(nS)}=\langle {\mathcal P}_{\rm EM}(^3S_1)\rangle_{V_Q(nS)}
\nn\\
&& \qquad\qquad\qquad
=\langle {\mathcal P}_{\rm EM}(^1S_0)\rangle_{P_Q(nS)}
=C_A \frac{|R^{(0)}_{n0}({0})|^2}{2\pi}
\left(m E_{n0}^{(0)} -{\mathcal E}_1 \right),
\label{eq:P13S1}
\end{eqnarray}
where $E_{n0}^{(0)} \simeq M - 2m \sim m v^2$ is the leading-order
binding energy.  \Eq[b]~(\ref{eq:P13S1}) reduces to the formula obtained
in \cite{Gremm:1997dq} if the heavy-quarkonium state satisfies also
the condition $mv \gg \lQ$.

The leading corrections to the above formulas come from 
quark--antiquark pairs of three momentum of order $\sqrt{m\lQ}$.
The existence of this degree of freedom in the heavy-quarkonium system 
has been pointed out in \cite{Brambilla:2003mu}, where the 
leading correction to \Eq~(\ref{eq:O1S}) has been calculated.  

The pNRQCD factorization formulas reduce, when applicable, the number
of nonperturbative parameters needed to describe heavy-quarkonium
decay widths \cite{Brambilla:2002nu}.  In particular, using charmonium
data to extract ${\cal E}_3$, in Ref.~\cite{Brambilla:2001xy} it was found 
${\cal E}_3(1\hbox{GeV}) = 5.3^{+3.5}_{-2.2}$, where the errors 
account for the experimental uncertainties only. This value has been used to predict P-wave
bottomonium inclusive decay widths in \cite{Brambilla:2001xy,Vairo:2002nh}.
We will come back to this in \Section~\ref{sec:idchib}.

\subsection[Experimental status]{Experimental status}
\label{sec:exp}
This section is a snapshot of the current status of various
experimental results on the electromagnetic and inclusive hadronic
decays of heavy-quarkonium states. The results come from the CLEO
experiment at CESR, the BES experiment at BEPC and E835 at Fermilab.

\subsubsection[$\Upsilon$ widths]{$\Upsilon$ widths}
Crucial parameters for any heavy-quarkonium state are its total width
and its hadronic and three leptonic partial widths.  For the three
$\Upsilon$ bound states, since their total widths, $\Gamma_{\rm tot}$, are
much less than the energy spread of the CESR machine ($\approx$ 4~MeV)
where they are studied, the procedure is to scan over each resonance
measuring the hadronic and $\mu^+\mu^-$ rates.  Then we use:
\begin{equation}
 \int \sigma_{\rm had} \; dE_{\rm cm} \; \propto \; \left ( \frac{\Gamma_{ee} \; 
\Gamma_{\rm had}}{\Gamma_{\rm tot}} \right ) \;\;\; \mbox{and} \;\;\;
{\cal B}_{\mu\mu} \; = \; \frac{\Gamma_{\mu\mu}}{\Gamma_{\rm tot}}.
\end{equation}
Assuming lepton universality, we have: $\Gamma_{\rm tot} \; = \;
\Gamma_{\rm had} \; + \; 3\;\Gamma_{\ell\ell}$.  This allows us to solve
for the total width and the partial widths into electrons and hadrons:
\begin{equation}
\Gamma_{ee} \;=\; \frac{(\Gamma_{ee}\Gamma_{\rm had}/\Gamma_{\rm tot})}{1
  \;-\; 3\;{\cal B}_{\mu\mu}}, \;\;\;\;\;\;\;\;\;
\Gamma_{\rm tot} = \frac{\Gamma_{ee}}{{\cal B}_{\mu\mu}}, \;\;\;\;\;\;\;\;\;
\Gamma_{\rm had} \;=\; \Gamma_{\rm tot}(1 - 3{\cal B}_{\mu\mu})\:. 
\end{equation}
Once the total width is known, the partial width into $\tau^+\tau^-$
can then be determined from its respective branching ratio. The
current experimental status from the 2004 PDG \cite{Eidelman:2004wy} is
shown in \Table~\ref{tab:Ups1}.

\begin{table}[htbp]
\caption[Present PDG values for the parameters of the $\Upsilon$ states]
        {Present PDG values \cite{Eidelman:2004wy} for the parameters of
         the $\Upsilon$ states.} 
\label{tab:Ups1} 
\renewcommand{\arraystretch}{1.3} 
\setlength{\tabcolsep}{4pt} 
\begin{center} 
\begin{tabular}{|c||c|c|c|c|} 
\hline 
Resonance & $\Gamma_{\rm tot}$ ({\rm keV})(\% error) & $\Gamma_{ee}$ ({\rm keV})(\%
error)  & ${\cal B}_{\mu\mu}$(\%)(\% error) & ${\cal B}_{\tau\tau}$(\%)(\% error) \\ \hline
$\Upsilon$(1S) & 53.0 $\pm$ 1.5 (2.8\%)& 1.314 $\pm$ 0.029 (2.2\%)
& 2.48 $\pm$ 0.06 (2.4\%) & 2.67 $\pm$ 0.15 (5.6\%) \\ \hline
$\Upsilon$(2S) & 43 $\pm$ 6 (14\%) & 0.576 $\pm$ 0.024 (4.2\%) &  
1.31 $\pm$ 0.21 (16\%) & 1.7 $\pm$ 1.6 (94\%)  \\ \hline
 $\Upsilon$(3S) & 26.3 $\pm$ 3.4 (13\%) & $-$  & 1.81
  $\pm$ 0.17 (9.4\%) & $-$ \\ \hline
\end{tabular} 
\end{center} 
\renewcommand{\arraystretch}{1} 
\end{table} 

The PDG does not use the 1984 CLEO measurement of $\Gamma_{ee}(3S) =
0.42 \pm 0.05$~keV  because new radiative corrections have
now been accepted which were not used in that analysis, thus
invalidating the measurement.  From the large percentage errors on
many of the quantities in the table, it is obvious that there is much room for
improvement.  To this end, the CLEO~III detector devoted a large amount of running at
each of the three $\Upsilon$ resonances, as shown in \Table~\ref{tab:Ups2}. 

\begin{table}[htbp]
\caption[Summary of CLEO~III running at the three $\Upsilon$ bound states]
        {Summary of the CLEO~III running at the three $\Upsilon$ bound states.}
\label{tab:Ups2} 
\renewcommand{\arraystretch}{1.3} 
\begin{center} 
\begin{tabular}{|c||c|c|c|} 
\hline 
Resonance & $\int$ L dt (fb$^{-1}$) & Number of Decays (M) & Factor
Increase Over CLEO!II \\ \hline
$\Upsilon$(1S) & 1.2 & 29 & 15 \\ \hline
$\Upsilon$(2S) & 0.9 & 6.0 & 12 \\ \hline
$\Upsilon$(3S) & 1.5 & 6.5 & 14 \\ \hline
\end{tabular} 
\end{center} 
\renewcommand{\arraystretch}{1} 
\end{table} 

\noindent All the results from this running have not yet been
finalized, but new measurements of the muonic branching ratios for the
3 bound $\Upsilon$ states have been published \cite{CLEO_muon_04}. These new
measurements are shown in \Table~\ref{tab:newmuon}, along with the
corresponding new values for the total widths.  The new $\Upsilon(2S)$
and $\Upsilon(3S)$ muonic branching ratio measurements are substantially
higher than previous results, giving correspondingly
smaller total widths for these resonances.

\begin{table}[htbp]
\caption[New CLEO measurements of muonic branching ratios for the 
         3 $\Upsilon$ states]
        {New CLEO measurements \cite{CLEO_muon_04} of the muonic
         branching ratios for the 3 $\Upsilon$ states, along with
         their statistical and systematic errors and the corresponding
         new values for the total widths.}
\label{tab:newmuon} 
\renewcommand{\arraystretch}{1.3} 
\begin{center} 
\begin{tabular}{|c||c|c|} 
\hline 
Resonance & ${\cal B}_{\mu\mu}$(\%)(\% error) & $\Gamma_{\rm tot}$ ({\rm keV})(\% error) \\ \hline
$\Upsilon$(1S) & 2.49 $\pm$ 0.02 $\pm$ 0.07 (2.8\%) & 52.8 $\pm$ 1.8 (3.4\%)  \\ \hline
$\Upsilon$(2S) & 2.03 $\pm$ 0.03 $\pm$ 0.08 (4.0\%) & 29.0 $\pm$ 1.6 (5.5\%) \\ \hline
$\Upsilon$(3S) & 2.39 $\pm$ 0.07 $\pm$ 0.10 (5.1\%) & 20.3 $\pm$ 2.1 (10.3\%)  \\ \hline
\end{tabular} 
\end{center} 
\renewcommand{\arraystretch}{1} 
\end{table} 

From the number of detected hadronic and leptonic events and a
knowledge of the CLEO detector performance, estimates of the final
statistical and systematic errors for the other resonance parameters
can be made.  These are shown in \Table~\ref{tab:errUps}.  Thus, once the
analyses are complete, there will be a tremendous improvement in our
knowledge of the basic parameters of the $\Upsilon$ bound-state
resonances.

\begin{table}[htbp]
\caption[Expected fractional errors for various quantities from the
         eventual CLEO~III measurements]
        {Expected fractional errors for various quantities from the
         eventual CLEO~III measurements.} 
\label{tab:errUps} 
\renewcommand{\arraystretch}{1.3} 
\begin{center} 
\begin{tabular}{|c||c|c|c|} 
\hline 
Parameter & Statistical Error & Systematic Error & Total Error \\ \hline
$\Gamma_{ee} \Gamma_{\rm had}/\Gamma_{\rm tot}$ & 1\% & 2.5\% & 3\% \\ \hline
$\Gamma_{ee}$ & 2\% & 2\% & 3\% \\ \hline
${\cal B}_{\tau\tau}$ & 2\% & 3\% & 4\% \\ \hline
$\Gamma_{\rm tot}$ & 2\% & 3\% & 4\% \\ \hline
\end{tabular} 
\end{center} 
\renewcommand{\arraystretch}{1} 
\end{table}

\subsubsection{$J/\psi$ and $\psi(2S)$ widths}

In the last two years the knowledge of both $J/\psi$ and $\psi(2S)$
parameters has improved.  In 2002, the BES collaboration reported
results \cite{Bai:2002zn} from a new scan of the $\psi(2S)$ resonance,
corresponding to an integrated luminosity of 1.15~pb$^{-1}$ and 114k
$\psi(2S)$ hadronic decays.  In 2004 BaBar has presented the first
measurement of $\Gamma_{ee}{\cal B}_{\mu\mu}$ \cite{Aubert:2003sv}
from ISR production of $J/\psi$ in 88.4~fb$^{-1}$ taken at the
$\Upsilon(4S)$ resonance.  \Table \ref{tab:psip} lists the values of
the widths and leptonic branching ratios for $J/\psi$ and $\psi(2S)$
from PDG\cite{Eidelman:2004wy}.

\begin{table}[htbp]
\caption[Present PDG values for the parameters of the $J/\psi$ and
         $\psi(2S)$ states]
        {Present PDG values \cite{Eidelman:2004wy} for the parameters
         of the $J/\psi$ and $\psi(2S)$ states.}
\label{tab:psip} 
\renewcommand{\arraystretch}{1.3} 
\setlength{\tabcolsep}{1mm}
\begin{center} 
\begin{tabular}{|c||c|c|c|c|} 
\hline 
Resonance & $\Gamma_{\rm tot}$ ({\rm keV})(\% error) & $\Gamma_{ee}$ ({\rm keV})(\%
error)  & ${\cal B}_{\mu\mu}$(\%)(\% error) & ${\cal B}_{\tau\tau}$(\%)(\% error) \\ \hline
$J/\psi$ & 91.0 $\pm$ 3.2 (3.5\%)& 5.40 $\pm$ 0.15$\pm$0.07 (3.1\%)
& 5.88 $\pm$ 0.10 (1.7\%) & --- \\ \hline
$\psi$(2S) & 281 $\pm$ 17 (6\%) & 2.12 $\pm$ 0.12 (9\%) &  
0.73 $\pm$ 0.08 (11\%) & 0.28$\pm$ 0.07 (25\%)  \\ \hline
\end{tabular} 
\end{center} 
\renewcommand{\arraystretch}{1} 
\end{table}

\subsubsection{Two-photon partial widths measurements}
\label{sec:ggmeas}

Experimental determinations of two-photon partial widths of quarkonia
depend on measurements of products and ratios of branching ratios 
performed by more than one experiment, and the best
estimate is obtained from a global fit to directly measured quantities
as it is done by the PDG \cite{Eidelman:2004wy}.  
When more measurements are available, subsets of measurements
may allow a direct extraction of the value for $\Gamma_{\gamma\gamma}$,
in general with a larger error than a global fit.  But this can
be useful both as a cross check for the global fit and to  
identify which measurements could yield improvements.    

The simplest case is the $\chi_{c2}$, where direct measurements of three 
independent quantities allows one to extract
$\Gamma_{\gamma\gamma}$ and $\Gamma_{J/\psi\gamma}$:
\begin{equation}
\Gamma=2.00\pm0.18\,{\rm MeV}, 
\label{eq:ggmeas:chi2W}
\end{equation}
\begin{equation}
\Gamma_{\gamma\gamma}{\cal B}_{J/\psi\gamma}=121\pm13\, {\rm eV}\;, 
\label{eq:ggmeas:chi2pr}
\end{equation}
and 
\begin{equation}
\frac{{\cal B}_{\gamma\gamma}}{{\cal B}_{J/\psi\gamma}}=(1.02\pm0.15)\cdot10^{-3},
\label{eq:ggmeas:chi2rat}
\end{equation}
where experimental values are world averages\cite{Eidelman:2004wy} except 
in \Eq~(\ref{eq:ggmeas:chi2rat}) where we averaged the E835 result with the ratio of
${\cal B}_{p\bar p}{\cal B}_{\gamma\gamma}$ and ${\cal B}_{p\bar p}{\cal B}_{J/\psi\gamma}$ measured by 
E760\cite{Armstrong:1991yk,Armstrong:1992qe}.
The product of \Eq~(\ref{eq:ggmeas:chi2W}), \Eq~(\ref{eq:ggmeas:chi2pr}), and 
\Eq~(\ref{eq:ggmeas:chi2rat}), yields 
$\Gamma_{\gamma\gamma}=0.50\pm0.05\;$keV, while taking 
\Eq~(\ref{eq:ggmeas:chi2pr}) multiplied by \Eq~(\ref{eq:ggmeas:chi2W}) and divided by 
\Eq~(\ref{eq:ggmeas:chi2rat}), we would obtain 
$\Gamma_{J/\psi\gamma}=490\pm50\,$keV, or 
${\cal B}_{J/\psi\gamma}=0.244\pm0.024$. 
The global fit to all measurements\cite{Eidelman:2004wy}
(including all other measurements related to ${\cal B}_{J/\psi\gamma}$)
improves on $\Gamma_{J/\psi\gamma}=430\pm40\,$keV and
${\cal B}_{J/\psi\gamma}=0.202\pm0.017$, but has almost no effect on
$\Gamma_{\gamma\gamma}=0.52\pm0.05\;$keV, indicating that the measurements
considered above are the only ones relevant to $\Gamma_{\gamma\gamma}$.

The case for $\chi_{c0}$ is similar to that of the $\chi_{c2}$,
even if apparently more complicated.
The world average of total width measurements is\cite{Eidelman:2004wy}
\begin{equation}
\Gamma=10.2\pm0.9\,{\rm MeV}.
\label{eq:ggmeas:chi0W}
\end{equation}
There is a measurement of 
\begin{equation}
\Gamma_{\gamma\gamma}{\cal B}_{2\pi^+2\pi^-}=75\pm13\pm8\;{\rm eV}~
\hbox{\cite{Eisenstein:2001xe}},
\label{eq:ggmeas:chi0fourpi}
\end{equation} 
and measurements (from a single experiment) of 
${\cal B}_{p\bar p}{\cal B}_{\gamma\gamma}$\cite{Andreotti:2004ru} and
${\cal B}_{p\bar p}{\cal B}_{\pi^0\pi^0}$\cite{Andreotti:2003sk}, from which
we can calculate (assuming isospin symmetry) the ratio
\begin{equation}
\frac{{\cal B}_{\gamma\gamma}}{{\cal B}_{\pi\pi}}=0.043\pm0.011\;.
\label{eq:ggmeas:chi0rat1}
\end{equation}
Even if ${{\cal B}_{\pi\pi}}$ and ${{\cal B}_{2\pi^+2\pi^-}}$ are not
directly measured, their ratio can be determined from 
quantities measured by a single experiment (in this case 
BES\cite{Bai:1998gh,Bai:1998cw,Bai:2001ha}):
\begin{equation}
\frac{{\cal B}_{\pi\pi}}{{\cal B}_{2\pi^+2\pi^-}}=0.47\pm0.10\,.
\label{eq:ggmeas:chi0rat2}
\end{equation}
This means that we can extract $\Gamma_{\gamma\gamma}=3.9\pm0.8\;$keV
from the product of  the four quantities in
Eqs. (\ref{eq:ggmeas:chi0W}), (\ref{eq:ggmeas:chi0fourpi}), 
(\ref{eq:ggmeas:chi0rat1}), and (\ref{eq:ggmeas:chi0rat2}). 
Notice that including MARK~II measurements in the evaluation of
\Eq~(\ref{eq:ggmeas:chi0rat2}) would give $\Gamma_{\gamma\gamma}=3.1\pm0.8$.
The global fit 
(which does not include the new measurement of
${\cal B}_{p\bar p}{\cal B}_{\gamma\gamma}$\cite{Andreotti:2004ru})
yields a significantly more precise value 
$\Gamma_{\gamma\gamma}=2.6\pm0.5\;$keV,
indicating
that in this case there are other measurements that are relevant,
such as ${\cal B}(\psi(2S)\to\gamma \chi_{c0}\to 3\gamma)$.

The case for $\eta_c(1S)$ and $\eta_c(2S)$ is different.
To date these states have been observed in two-photon
reactions with direct measurement of 
\begin{eqnarray}
\eta_c(1S): ~~& \Gamma_{\gamma\gamma}{\cal B}_{K\bar K\pi}& =0.48\pm0.06\;{\rm keV}\,,
\label{eq:ggmeas:etacgg}\\
\eta_c(2S): ~~& \Gamma_{\gamma\gamma}{\cal B}_{K\bar K\pi}& =73\pm 23\;{\rm eV} \; 
\hbox{\cite{Asner:2003wv}} \,.
\label{eq:ggmeas:etac2gg}
\end{eqnarray}
The $\eta_c(1S)$ has also been observed in $\bar p p $ annihilations with direct measurement of
\begin{equation}
{\cal B}_{\gamma\gamma}{\cal B}_{p\bar p}=(0.26\pm0.05)\times10^{-6} .
\label{eq:ggmeas:etacpp}
\end{equation}
In this case there are no measurements of the ratio of branching ratios
between the $\gamma\gamma$ and any other decay mode, so it is necessary
to use the values of ${\cal B}_{K\bar K\pi}$ or ${\cal B}_{p\bar p}$ 
that (for $\eta_c(1S)$ only) are determined by
\begin{equation}
{\cal B}_{X}=\frac{{\cal B}(J/\psi\to\gamma \eta_c\to\gamma\,{X})}{
                    {\cal B}(J/\psi\to\gamma \eta_c)}, 
\label{eq:ggmeas:brrad}
\end{equation}
with precision limited by the $\approx 30\%$ uncertainty in 
${\cal B}(J/\psi\to\gamma \eta_c)$ that is to date a common 
systematic to all two-photon partial widths of $\eta_c(1S)$.
Since no measurement is yet available for the $\eta_c(2S)$ branching ratio to 
${K\bar K\pi}$, its $\Gamma_{\gamma\gamma}$ cannot be determined.

The most obvious strategy to increase the precision on $\Gamma_{\gamma\gamma}$
is to improve the measurements for quantities used in its determination.
But based on the case of $\chi_{c2}$ discussed above,
a major improvement could be obtained by 
measuring the pair of quantities $\Gamma_{\gamma\gamma}{\cal B}_X$ 
and ${{\cal B}_{\gamma\gamma}}/{{\cal B}_{X}}$  for more than one
final state $X$. $B$ factories can reasonably measure to $<10\%$ precision  
$\Gamma_{\gamma\gamma}{\cal B}_X$ for more than one final state.
It is also reasonable that total widths will be more precisely measured in 
$p\bar p$ experiments, thus the question is whether it is possible to
measure to better than $10\%$ the ratios ${{\cal B}_{\gamma\gamma}}/{{\cal B}_{X}}$.
How well can BES and CLEO measure $\psi(2S)$ or $J/\psi$ to 3$\gamma$?
How well can $\bar p p\to \gamma\gamma$ be measured and  
what are the channels that could be measured in these experiments
simultaneously to $\bar p p\to \gamma\gamma$? 
With a magnetic detector, $p\bar p\to \phi\phi$ is the obvious
choice, but interference with two-body non-resonant reactions
may offer other opportunities (\eg $p\bar p\to p\bar p$).
The goal of $<5\%$ precision on two-photon widths is not unreasonable.

\subsubsection[$\chi_b$ widths]{$\chi_b$ widths}
\label{sec:idchib}
Since the $\chi_b(2P_J)$ states are not produced directly in
$e^+ e^-$ annihilations, their hadronic widths cannot be measured
using the same technique as for the $S$ states.  However, we can use the
fact that the partial width for their photonic E1 transitions to the
$\Upsilon (2S)$ state are proportional to a common matrix element
squared times a phase space factor of $E_\gamma^3$ (see Secs.~\ref{sec:emssec-et}
and \ref{sec:emssec-ettf}, $E_\gamma=k$).  
Thus, from measuring the individual photon energies and branching ratios for the
decays $\chi_b$($2P_J$) $\rightarrow$ $\Upsilon(2S) + \gamma$, along
with the branching ratios for $\chi_b(2P_J) \rightarrow 
\Upsilon(1S) + \gamma$, we can measure the ratio of the
$\chi_b(2P_J)$ hadronic partial widths, $\Gamma({\rm had})$.  We first
use:
\begin{equation}
{\cal B}(2S) \;=\; \frac{\Gamma(2S)}{\Gamma(1S)\;+\;\Gamma(2S)
  \;+\:\Gamma({\rm had})},
\end{equation}
where ${\cal B}(2S) = {\cal B}(\chi_b(2P_J) \rightarrow 
\Upsilon(2S) + \gamma)$ and ${\cal B}(1S) = {\cal B}(\chi_b(2P_J)
\rightarrow \Upsilon(1S) + \gamma)$ are the two E1 branching
ratios, and $\Gamma(2S)$ and $\Gamma(1S)$ are the corresponding partial
widths.  Then, since $\Gamma(2S)/\Gamma(1S) = {\cal B}(2S)/{\cal B}(1S)$, we can
solve for the hadronic partial width, obtaining:
\begin{equation}
\Gamma({\rm had}) \;=\; \Gamma(2S) \left [ \frac{1 - {\cal B}(1S)}{{\cal B}(2S)} \;-\; 1
  \right ]. 
\end{equation}
\noindent Making the assumption mentioned above that the partial
widths for E1 transitions of different $J$ states to the same $\Upsilon$
state should be proportional to a common matrix element squared times $E_\gamma^3$,  
we obtain an expression for the ratio of hadronic partial
widths for two different $\chi_b(2P_J)$ states.  For example, for $J$
= 0 and $J$ = 2, we get:
\begin{equation}
\frac{\Gamma_{\rm had}(2P_0)}{\Gamma_{\rm had}(2P_2)} \;=\; \left (
\frac{E_\gamma(2P_0 \rightarrow 2S + \gamma)}{E_\gamma(2P_2
  \rightarrow 2S + \gamma)} \right )^3 
\; \left ( \frac{ \frac{1 - {\cal B}(1S)_0}{{\cal B}(2S)_0} - 1}{ \frac{1 - {\cal B}(1S)_2}{{\cal B}(2S)_2} - 1} \right ),
\end{equation}
where ${\cal B}(2S)_0 = {\cal B}(\chi_b(2P_0) \rightarrow \Upsilon(2S)
+ \gamma)$, etc.  Using this technique and the E1 branching ratios
given in \Section~\ref{sec:emssec-ettf}, CLEO~III finds the ratio of the
$J$ = 0 to the $J$ = 2 hadronic widths to be:
\begin{equation}
 \frac{\Gamma_{\rm had}(2P_0)}{\Gamma_{\rm had}(2P_2)} \;=\; 6.1 \; \pm \; 2.8. 
\label{eq:cl02}
\end{equation}
For the $J$ = 1 and $J$ = 2 states, CLEO~III measures:
\begin{equation}
\frac{\Gamma_{\rm had}(2P_1)}{\Gamma_{\rm had}(2P_2)} \;=\; 0.25 \; \pm \; 0.09.
\label{eq:cl12}
\end{equation}
Since the $J$ = 1 state cannot annihilate into two massless
gluons, to first order its hadronic width is expected to be suppressed
by one order of $\alpha_{\rm s}$ compared to the $J$ = 2 state.  The
measurement confirms this suppression.

As discussed in \Section~\ref{sec:secnrqcd}, at leading order in the
heavy-quark velocity expansion, the above ratios depend on a
colour-octet matrix element.  One can consider the combination
\begin{equation}
\frac{\Gamma_{\rm had}(2P_0) - \Gamma_{\rm had}(2P_1)}{\Gamma_{\rm had}(2P_2)
  - \Gamma_{\rm had}(2P_1)},
\label{eq:fr0121}
\end{equation}
which is completely determined by perturbative QCD \cite{Bodwin:1992ye}. Using
(\ref{eq:cl02}) and (\ref{eq:cl12}), this ratio is measured by CLEO~III to be:
\begin{equation}
\frac{\Gamma_{\rm had}(2P_0) - \Gamma_{\rm had}(2P_1)}{\Gamma_{\rm had}(2P_2)
  - \Gamma_{\rm had}(2P_1)} \;=\;
7.8 \;\pm\; 3.8.
\label{eq:cl0121}
\end{equation}
LO QCD predicts 15/4 = 3.75 for this ratio, and NLO QCD about 7, 
which is quite consistent with (\ref{eq:cl0121}). However, the
combination (\ref{eq:fr0121}) distinguishes between bottomonium and
charmonium only at NNLO, while the ratios (\ref{eq:cl02}) and
(\ref{eq:cl12}) do so at NLO.  A direct determination of these ratios has
been done in the framework of \mbox{pNRQCD}, as discussed in
\Section~\ref{sec:pnrqcd}, using the factorization formula (\ref{eq:matoct})
and fixing the nonperturbative constant to the value found from
charmonium data. The result at NLO is ${\Gamma_{\rm had}(2{\rm
P}_0)}/{\Gamma_{\rm had}(2{\rm P}_2)} \;\simeq\; 4.0$, consistent with
(\ref{eq:cl02}), and ${\Gamma_{\rm had}(2{\rm P}_1)}/{\Gamma_{\rm
had}(2{\rm P}_2)} \;\simeq\; 0.50$, which is somewhat larger than
(\ref{eq:cl12}) \cite{Brambilla:2001xy,Vairo:2002nh}.

CLEO cannot resolve the individual photon lines for the similar decays
from the $\Upsilon(3S)$ to the $\chi_b(1P_J)$ states (see
Sec.~\ref{sec:emssec-ettf}).  However,
we can use the quite old $\chi_b(1P_J) \rightarrow \Upsilon(1S) +
\gamma$ branching ratios from the PDG \cite{Eidelman:2004wy} for $J = 1$
and 2 (the $J = 0$ branching ratio is very small, given the large
hadronic width of that state). In this case, the ratio of the
hadronic widths for the two states can be found from:
\begin{equation}
\frac{\Gamma_{\rm had}(1P_1)}{\Gamma_{\rm had}(1P_2)} \;=\; \left (
\frac{E_\gamma(1P_1 \rightarrow 1S + \gamma)}{E_\gamma(1P_2
\rightarrow 1S + \gamma)} \right )^3 \; \left ( \frac{ \frac{1}{{\cal
B}(1S)_1} - 1} { \frac{1}{{\cal B}(1S)_2} - 1} \right ).
\end{equation}
This leads to the result:
\begin{equation}
\frac{\Gamma_{\rm had}(1P_1)}{\Gamma_{\rm had}(1P_2)} \;=\; 0.46 \; \pm \; 0.20,
\end{equation}
showing again the suppression of the $J = 1$ state's hadronic width
compared to the $J = 2$, albeit with larger errors in this case.

\subsubsection[$\chi_c$ widths]{$\chi_c$ widths}

The $\chi_c$ states are also not directly produced in $e^+e^-$
annihilations.  However, in this case an extremely powerful
alternative method has been used to measure their masses and total
widths.  In an experimental technique first pioneered by experiment
R704 at CERN, and continued by experiments E760 and E835 at the
Fermilab Antiproton Accumulator, a stochastically cooled
$\overline{p}$ beam collides with a hydrogen gas jet target.  In the
subsequent $p\overline{p}$ annihilations, all $J^{PC}$ states can be
formed via 2 or 3 gluons.  Thus, the P-wave charmonium states are
directly accessible.  By scanning the proton beam energy over each
resonance, the mass and total width of each $P$ state can be measured
with extremely high accuracy.
  
As mentioned in \Section~\ref{sec:ggmeas}, these experiments have also
measured products or ratios of branching ratios that help constrain
the radiative and $\gamma\gamma$ widths of those states.
\Table~\ref{tab:ChiW} shows the current best estimates of the
$\chi_{c}$ widths, using data from PDG\cite{Eidelman:2004wy}. E835 is
finalizing the analysis of the scans of the $\chi_{c1}$ and
$\chi_{c2}$ resonances\cite{E835id}, with an anticipated precision of
$\approx 7\%$ on $\chi_{c1}$ and $\chi_{c2}$ total widths.

\begin{table}[t]
\caption[Widths of $\chi_c$ states from PDG]
        {Widths of $\chi_c$ states from PDG\cite{Eidelman:2004wy}.} 
\label{tab:ChiW}
\renewcommand{\arraystretch}{1.3} 
\begin{center} 
\begin{tabular}{|c||c|c|c|} 
\hline 
Resonance & $\Gamma_{\rm tot}$ ({\rm MeV})(\% error) & $\Gamma(\gamma J/\psi)$ ({\rm keV})(\%
error)  & $\Gamma(\gamma\gamma)$(\%)(\% error) \\ \hline
$\chi_{c0}$ & 10.1 $\pm$ 0.8 (8\%)   & 119 $\pm$ 16 (13\%) & 2.6 $\pm$ 0.5 (19\%) \\ \hline
$\chi_{c1}$ & 0.91 $\pm$ 0.13 (14\%) & 290 $\pm$ 50 (17\%) &  ---  \\ \hline
$\chi_{c2}$ & 2.11 $\pm$ 0.16 (8\%)  & 430 $\pm$ 40 (9\%) &  0.52 $\pm$0.05 (10\%) \\ \hline
\end{tabular} 
\end{center} 
\renewcommand{\arraystretch}{1} 
\end{table} 

In order to show the impact of the new measurements of the $\chi_c$
widths, in \Table~\ref{tab:comp} we compare the PDG 2000
\cite{Groom:2000in} with the PDG 2004 \cite{Eidelman:2004wy}
determinations of different ratios of hadronic and electromagnetic
widths (similar ratios have been considered in the previous section
for the $\chi_b$ case). There have been sizable shifts in some central
values and considerable reductions in the errors. In particular, the
error on the ratio of the electromagnetic $\chi_{c0}$ and $\chi_{c2}$
widths has been reduced by about a factor 10, while in all other
ratios the errors have been reduced by a factor 2 or 3.  The
considered ratios of hadronic and electromagnetic widths do not depend
at leading order in the velocity expansion (see
\Eqs~(\ref{eq:gammachiLH}) and (\ref{eq:gammachi})) on any
nonperturbative parameter. Therefore, they can be calculated in
perturbation theory.  The last two columns of \Table~\ref{tab:comp}
show the result of a leading and next-to-leading order calculation
respectively. Despite the fact that the convergence is not always very
good and that, therefore, the NLO calculation should be taken with
some care (see also \Section~\ref{sec:perturexpan}), all data now
clearly prefer (and are consistent with) NLO results.

\begin{table}[t]
\caption[Comparison of ratios of $\chi_{cJ}$ partial widths]
        {Comparison of ratios of $\chi_{cJ}$ partial widths.  The
         experimental values PDG 2004 are obtained from the world
         averages of \cite{Eidelman:2004wy}, with the assumption
         $\Gamma(\chi_{c0}\to
         l.h.)\approx\Gamma(\chi_{c0})=10.1\pm0.8$~MeV,
         $\Gamma(\chi_{c1}\to l.h.)\approx\Gamma(\chi_{c1})\left[1-
         {\cal B}(\chi_{c1}\to\gamma J/\psi)\right]$
         =0.62$\pm$0.10~MeV, $\Gamma(\chi_{c2}\to
         l.h.)\approx\Gamma(\chi_{c2})\left[1- {\cal
         B}(\chi_{c2}\to\gamma J/\psi)\right]$ =1.68$\pm$0.15~MeV.
         Similarly the experimental values PDG 2000 have been obtained
         from \cite{Groom:2000in}.  The chosen ratios do not depend at
         leading order in the velocity expansion on octet or singlet
         matrix elements. The LO and NLO columns refer to a leading
         and next-to-leading order calculation done at the
         renormalization scale $2m_c$ with the following choice of
         parameters: $m_c=1.5$~Gev and $\als(2m_c)=0.245$.}
\label{tab:comp} 
\begin{center}
\begin{tabular}{|c||c|c||c|c|}
\hline
Ratio &{ PDG 2004} &{ PDG 2000} &LO & NLO 
 \\
\hline
 & & & & \\
\(\displaystyle \frac{\Gamma(\chi_{c0}\to\gamma\gamma)}
                     {\Gamma(\chi_{c2}\to\gamma\gamma)}\)
 & 5.1$\pm$1.1 &  13$\pm$10  &  3.75  & $\approx$ 5.43   \\
 & & & & \\
\hline
 & & & & \\
\(\displaystyle \frac{\Gamma(\chi_{c2}\to l.h.) - \Gamma(\chi_{c1}\to l.h.)}
                     {\Gamma(\chi_{c0}\to\gamma\gamma)}\)
& 410$\pm$100 &  270$\pm$200  &  $\approx$ 347  & $\approx$ 383   
\\
 & & & & \\
\hline
 & & & & \\
\(\displaystyle \frac{\Gamma(\chi_{c0}\to l.h.) - \Gamma(\chi_{c1}\to l.h.)}
                     {\Gamma(\chi_{c0}\to\gamma\gamma)}\)
& 3600$\pm$700 &  3500$\pm$2500  &  $\approx$ 1300  & $\approx$ 
2781   \\
 & & & & \\
\hline
 & & & & \\
\(\displaystyle \frac{\Gamma(\chi_{c0}\to l.h.) - \Gamma(\chi_{c2}\to l.h.)}
                     {\Gamma(\chi_{c2}\to l.h.) - \Gamma(\chi_{c1}\to l.h.)}\)
& 7.9$\pm$1.5 &  12.1$\pm$3.2  &  2.75  & $\approx$ 6.63   \\
 & & & & \\
\hline
 & & & & \\
\(\displaystyle \frac{\Gamma(\chi_{c0}\to l.h.) - \Gamma(\chi_{c1}\to l.h.)}
                     {\Gamma(\chi_{c2}\to l.h.) - \Gamma(\chi_{c1}\to l.h.)}\)
& 8.9$\pm$1.1 &  13.1$\pm$3.3  &  3.75  & $\approx$ 7.63   \\
 & & & & \\
\hline
\end{tabular}
\end{center}
\end{table}

\subsubsection[$\Upsilon$(1S) $\rightarrow$ $\gamma$ + X and $\Upsilon$(1S)
  $\rightarrow$ X]{$\Upsilon$(1S) $\rightarrow$ $\gamma$ + X and $\Upsilon$(1S) $\rightarrow$ X}

There has been much theoretical interest lately in trying to predict
the direct photon energy distribution for $\Upsilon$(1S) $\rightarrow$
$\gamma$ + X inclusive decays \cite{Fleming:2002rv}.  See the following section.
The last reported measurement was from the CLEO~II experiment in 1997
\cite{Nemati:1996xy}, based on 1.4 million $\Upsilon$(1S) decays.
Besides the photon energy spectrum, they measured the ratio:
\begin{equation}
\frac{\Gamma(\gamma gg)}{\Gamma(ggg)} \;=\; (2.75 \;\pm\; 0.04
\;\pm\; 0.15)\:\%\:,
\end{equation}
which allowed a fairly accurate determination of
$\Lambda_{\overline{\rm MS}}$ and $\alpha_{\rm s}$.  Given the small
statistical errors in these measurements, it is doubtful that the CLEO
III experiment will repeat them using their 29 million $\Upsilon$(1S)
decays.  Rather, the emphasis will be on detailed studies of exclusive
$\gamma$ + X decays of the $\Upsilon$(1S), especially the search for
possible glueball candidates.

For measurements of the inclusive production of various hadronic
particle types from the $\Upsilon$(1S), one must go back to a 1985
paper by the CLEO~I experiment \cite{Behrends:1984be}, based on only 50k
$\Upsilon$(1S) decays.  They measured the average multiplicities and
momentum distributions of $\pi$, $K$, $\rho$, $K^*$, $\phi$, $p$,
$\Lambda$ and $\Xi$ in $\Upsilon$(1S) decays and compared them to
those from the nearby continuum.  The only addition to these results
was a 2003 CLEO~II measurement \cite{Artuso:2002px} of the inclusive
$\eta^\prime$ production from the $\Upsilon$(1S), based on 1.9 million
decays and motivated by the large observed $B \rightarrow \eta^\prime + X$ 
branching ratio.

\section[Inclusive radiative decays]
        {Inclusive radiative decays 
         $\!$\footnote{Author: A.~Leibovich}}
\label{sec:IDr}
The radiative inclusive decay of heavy quarkonium has been
investigated for about a quarter century.  Here we will study
$\Upsilon\to X\gamma$ decays in particular.  The direct radiative
decay is calculated by using the operator product expansion, where the
operators are the same nonperturbative matrix elements that appear in
the inclusive decay to hadrons (see \Section~\ref{sec:secnrqcd}). Thus
we obtain an expansion in the velocity, $v$, of the heavy quarks. The
rate is written as
\begin{equation}
\frac1{\Gamma_0}\frac{d\Gamma^{\rm dir}}{dz} = \sum_n C_n(M,z) 
   \langle \Upsilon \vert O_n \vert \Upsilon \rangle,
\end{equation}
where $M = 2m_b$, $z = 2E_\gamma/M$, the $C_n(z,M)$ are short 
distance Wilson coefficients,
calculable in perturbation theory, and the NRQCD matrix elements scale
with a certain power in $v$.  The lowest order contribution is the 
colour-singlet $^3S_1$ operator, where the quark--antiquark pair annihilate 
into a photon and two gluons.  Therefore, in the $v\to 0$ limit, we obtain 
the colour-singlet model calculation of Ref. \cite{firstRad}.  
At higher order in the velocity expansion, there are direct
contributions from the colour-octet matrix elements
\cite{Maltoni:1999nh}.  The decay through a colour-octet matrix element
can occur at one lower order in $\alpha_{\rm s}$, with the $b\bar{b}$
decaying to a photon and a single gluon.  

However, this calculation is only valid in the intermediate range of photon 
energies ($0.3 \lsim z \lsim 0.7$).
For low photon energies, $z\lsim 0.3$, the major photon production mechanism
is fragmentation~\cite{Catani:1994iz,Maltoni:1999nh}.  At large photon energies, 
$z\gsim0.7$, the perturbative~\cite{Maltoni:1999nh} and nonperturbative 
expansions~\cite{Rothstein:1997ac} both break down.

\subsection[Photon fragmentation]{Photon fragmentation}
The inclusive photon spectrum can be written as a sum of a direct and
a fragmentation contribution~\cite{Catani:1994iz},
\begin{equation}
\frac{d\Gamma}{dz}= \frac{d\Gamma^{\rm dir}}{dz}
  +\frac{d\Gamma^{\rm frag}}{dz},
\end{equation}
where in the direct term the photon is produced in the hard
scattering, and in the fragmentation term the photon fragments from a
parton produced in the initial hard scattering.  The fragmentation
contribution has been well studied in Ref. \cite{Maltoni:1999nh}.

Catani and Hautmann pointed out the importance of fragmentation for
the photon spectrum in quarkonium decays \cite{Catani:1994iz}.  The
fragmentation rate can be written as
\begin{equation}
\frac{d\Gamma^{\rm frag}}{dz} = 
  \sum_{a = q,\bar q, g} \int_z^1\frac{dx}{x} \frac{d\Gamma_a}{dx}
    D_{a\gamma}\left(\frac{z}{x},M\right),
\end{equation}
where the rate to produce parton $a$, $d\Gamma_a/dx$, is convoluted
with the probability that the parton fragments to a photon,
$D_{a\gamma}$, with energy fraction $z/x$.  The rate to produce parton
$a$ can again be expanded in powers of $v$ \cite{Maltoni:1999nh}, with
the leading term being the colour-singlet rate for an $\Upsilon$ to
decay to three gluons,
\begin{equation}
\frac{d\Gamma^{\rm frag}_{\rm LO}}{dz} = 
\int_z^1\frac{dx}{x} \frac{d\Gamma_{ggg}}{dx} 
D_{g\gamma}\left(\frac{z}{x},M\right).
\end{equation}
At higher orders in $v$, there are three colour-octet fragmentation
contributions, where the photon can fragment off either a quark or a gluon.

The partonic rates must be convoluted with the fragmentation functions,
$D_{a\gamma}(z,M)$.  The $M$-dependence of the fragmentation functions
can be predicted using perturbative QCD via Altarelli--Parisi evolution
equations.  However, the solution depends on nonperturbative
fragmentation function at some input scale $\Lambda$, which must be
measured from experiment.  This has been done by the ALEPH
collaboration for the $D_{q\gamma}$ fragmentation function
\cite{Buskulic:1995au}, but the $D_{g\gamma}$ fragmentation function 
is unknown, so at this point it must be modeled.

\subsection[Resumming the large $z$ contribution]
           {Resumming the large $z$ contribution}
\label{sec:resumminglargez}

The colour-octet contributions to the rate are the first subleading
terms in the operator product expansion.  Diagrammatically, these
contributions occur for the quark--antiquark pair annihilating into
a photon back-to-back with a gluon.  Thus the ${}^1S_0$ and 
${}^3P_0$ colour-octet contributions 
begin as a delta function of $(1-z)$~\cite{Maltoni:1999nh}.  If we
look at the integrated rate near the endpoint, the colour-octet contributions 
are as important as the ``leading" colour-singlet piece, in the region
$1-v^2 \lsim z \leq 1$.   Perturbative
corrections to the colour-octet contributions have large kinematic logarithms, which
destroy the perturbative expansion.  
The $\alpha_{\rm s}$ correction to the leading colour-singlet rate was
calculated numerically in Ref. \cite{Kramer:1999bf}. It leads to small
corrections over most of phase space; however, in the endpoint region
the corrections are of order the leading contribution.  Thus both
higher orders in $v$ and in $\alpha_{\rm s}$ are not suppressed in the 
endpoint region.  Both the nonperturbative and perturbative series 
break down.  

This breakdown at large $z$ is due to NRQCD not including collinear
degrees of freedom.  
In the endpoint region, the outgoing gluons are moving back-to-back to
the photon, with large energy and small invariant mass (\ie a
collinear jet).
The correct effective field theory is a
combination of NRQCD for the heavy degrees of freedom and the
soft-collinear effective theory
(SCET)~\cite{Bauer:2001ew,Bauer:2001ct} for the light
degrees of freedom.  

\begin{figure} 
\begin{center} 
\includegraphics[width=0.4\textwidth,clip]{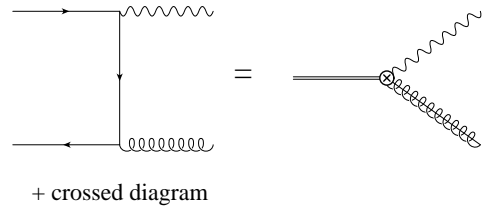} 
\end{center} 
\caption{Matching QCD onto NRQCD and SCET.  The double lines
represents the $\Upsilon$, while the spring with the line through it
represent a collinear gluon.
\label{fig:comatching}} 
\end{figure} 

SCET is an effective field theory describing collinear fields interacting
with soft degrees of freedom.  It is thus the appropriate effective field
theory to use when there are energetic particles moving with small 
invariant mass, such as $\Upsilon\to X\gamma$ in the endpoint region.
We therefore use NRQCD to describe the quarkonium, and SCET to 
describe the jet of collinear particles.  The invariant mass of the jet
of particles is $p^2\sim M_\Upsilon^2(1-z)$, which is small as $z\to 1$. 
In SCET there are three mass scales: the hard scale, which for this
process is $\mu_h \sim M_\Upsilon$, the collinear scale, which is
$\mu_c \sim M_\Upsilon\sqrt{1-z}$, and the ultrasoft scale, 
$\mu_u \sim M_\Upsilon(1-z)$.  These scales are widely separated in 
the endpoint region.  SCET allows us to separate the physics
coming from the disparate scales.

To calculate, the QCD process is matched onto operators in SCET and
NRQCD.  For example, the matching for the colour-octet channel is
pictured in \Figure~\ref{fig:comatching}.  Then to resum the kinematic
logarithms, we use the renormalization group equations in SCET, by
evolving from $\mu_h$ to $\mu_u$.  So we first renormalize the
operators in SCET, and calculate the anomalous dimensions in the usual
way.  Then by running the SCET operators to the ultrasoft scale, the
logarithms of $1-z$ are summed.

\begin{figure} 
\begin{center} 
\includegraphics[width=0.6\textwidth,clip]{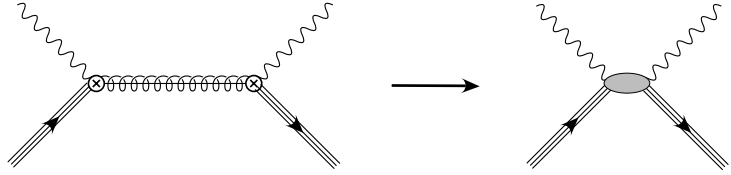} 
\end{center} 
\caption{The leading OPE: tree level matching of the time ordered
product in the collinear-soft theory to a nonlocal operator in the
soft theory.
\label{fig:leadingope}} 
\end{figure} 

The colour-singlet process does not run below the collinear scale.
This is because the ultrasoft gluons cannot couple to the
colour-singlet jet or the incoming colour-singlet quarkonium.  This fact
was first pointed out by Hautmann~\cite{Hautmann:2001yz}.  However,
there are still logarithms that are generated between the hard and
collinear scales~\cite{Photiadis:1985hn,Fleming:2002rv}.  For the
colour-octet processes~\cite{Bauer:2001rh}, at the collinear scale
$\mu_c$ we integrate out collinear modes.  Since there are collinear
particles in the final state, we first perform an OPE for the
inclusive $\Upsilon$ radiative decay rate in the endpoint region, and
match onto the large energy effective theory~\cite{Dugan:1990de}. The
result is a nonlocal OPE in which the two currents are separated along
a light-like direction. Diagrammatically this is illustrated in
\Figure~\ref{fig:leadingope}.  This is run to the ultrasoft scale, at
which point we are left with a nonperturbative shape function, which
describes the movement of the heavy quark--antiquark pair within the
meson.  This function is precisely what was shown to occur in
Ref.~\cite{Rothstein:1997ac}.  Unfortunately, this nonperturbative
function is unknown, and must be modeled.

\begin{figure} 
\begin{center} 
\includegraphics[width=0.6\textwidth,clip]{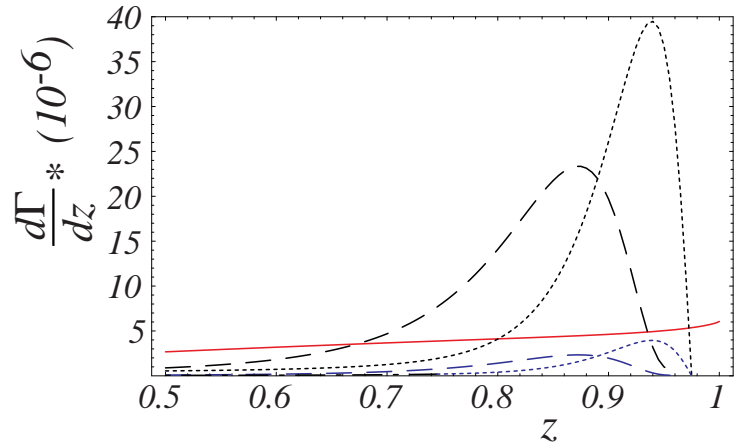} 
\end{center} 
\caption{The differential decay spectra in the region $0.5<z$. The
dashed curves are the fully resummed colour-octet result 
convoluted with a model for the shape
function for two choices of the colour-octet matrix elements. The larger
curves have the colour-octet matrix elements suppressed by $v^4/10$, 
while the lower curves have $v^4/100$.  In addition we
have interpolated the fully resummed result with the next-to-leading
order result in the region away from the endpoint. The dotted curves are the next-to-leading
result convoluted with the structure function for two choices of the matrix
elements. The solid curve is the tree-level colour-singlet contribution.
\label{fig:spec}} 
\end{figure} 

Before we proceed we need the NRQCD matrix elements.  We can extract
the colour-singlet matrix elements from the $\Upsilon$ leptonic
width. The colour-octet matrix elements are more difficult to
determine.  NRQCD predicts that the colour-octet matrix elements scale
as $v^4$ compared to the singlet matrix elements.  In
Ref.~\cite{Petrelli:1998ge} it was argued that an extra factor of $1/2
N_c$ should be included.  By looking at the shape of the resummed
colour-octet rates, it appears that these channels would give a
contribution an order of magnitude too large compared with the data in
the endpoint region if they were even as small as $v^4/2 N_c$ times
the colour-singlet, as shown in \Figure~\ref{fig:spec}, so we will set
them to zero.  This eliminates two of the three possible colour-octet
matrix elements, leaving the $^3S_1$.  It also eliminates the
dependence at this order on the unknown shape functions and the
largest dependence on the unknown fragmentation function,
$D_{g\gamma}$.  We set the colour-octet $^3S_1$ matrix element to be
$v^4$ suppressed compared to the colour-singlet matrix element
extracted from the leptonic width, where we use $v^2=0.08$.  This
colour colour-octet matrix element does not give a large contribution in
the large $z$ region, but is important at low $z$, due to the
fragmentation function $D_{q\gamma}$.

\begin{figure}[t] 
\begin{center} 
\includegraphics[width=0.57\textwidth]{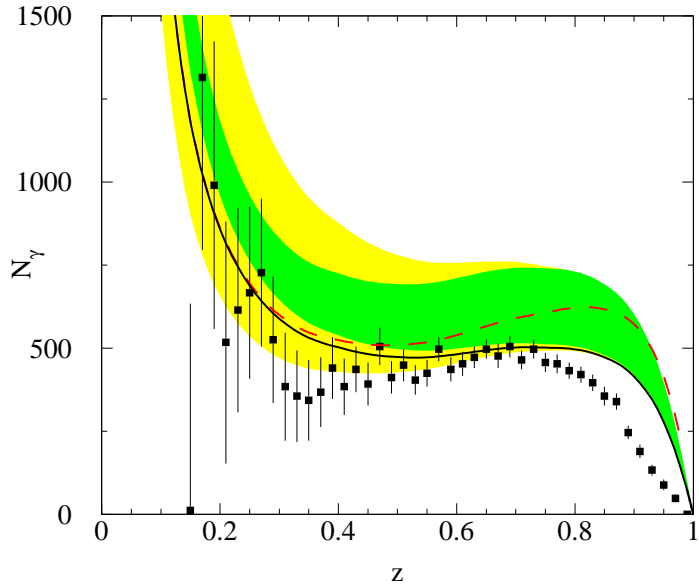} 
\end{center} 
\caption{The inclusive photon spectrum, compared with data
\cite{Nemati:1996xy}.  The theory predictions are described in the
text.  The plot is from Ref.~\cite{Fleming:2002rv}.}
\label{fig:comparedata}
\end{figure} 

The CLEO collaboration measured the number of photons in inclusive
$\Upsilon(1S)$ radiative decays~\cite{Nemati:1996xy}.  The data does
not remove the efficiency or energy resolution and is the number of
photons in the fiducial region, $|\cos\theta|<0.7$.  In order to
compare our theoretical prediction to the data, we integrate over the
barrel region and convolute with the efficiency that was modeled in
the CLEO paper.  We do not do a bin-to-bin smearing of our prediction.

In \Figure~\ref{fig:comparedata} we compare the prediction to the
data.  The error bars on the data are statistical only.  The dashed
line is the direct tree-level plus fragmentation result, while the
solid curve includes the resummation of the kinematic logarithms.  For
these two curves we use the $\alpha_{\rm s}$ extracted from these
data, $\alpha_{\rm s}(M_\Upsilon) = 0.163$, which corresponds to
$\alpha_{\rm s}(M_Z) = 0.110$ \cite{Nemati:1996xy}.  The shape of the
resummed result is much closer to the data than the tree-level curve,
though it is not a perfect fit.  We also show the resummed plus
fragmentation result, using the PDG value of $\alpha_{\rm s}(M_Z)$,
including theoretical uncertainties, denoted by the shaded region.  To
obtain the darker band, we first varied the choice of $m_b$ between
$4.7 {\rm\ GeV} < m_b < 4.9 {\rm\ GeV}$ and the value of $\alpha_{\rm
s}$ within the errors given in the PDG, $\alpha_{\rm s}(M_Z) =
0.1172(20)$ \cite{Hagiwara:2002fs}.  We also varied the collinear
scale, $\mu_c$ from $M\sqrt{(1-z)/2} < \mu_c < M\sqrt{2(1-z)}$.
Finally, the lighter band also includes the variation, within the
errors, of the parameters for the quark to photon fragmentation
function extracted by ALEPH \cite{Buskulic:1995au}.  The low $z$
prediction is dominated by the quark to photon fragmentation coming
from the colour-octet $^3S_1$ channel.  We did not assign any error to
the colour-octet $^3S_1$ matrix elements.  Since it is unknown, there
is a very large uncertainty in the lower part of the prediction that
we decided not to show.  Recently, colour-octet ${}^1S_0$ and ${}^3P_0$
contributions, calculated in the weak-coupling regime, have been
included in the analysis \cite{GarciaiTormo:2004jw}.  They appear to
improve the agreement with the data in the end-point region.  Also
recently operator mixing between the gluon jet, considered here, and
the quark--antiquark jet has been considered in
\cite{Fleming:2004rk}.

%%%%%%%%%%%%%%  
%ED
%%%%%%%%%%%%%% 
\section[Exclusive decays]
        {Exclusive decays 
         $\!$\footnote{Author: P.~Kroll (with contributions 
         from C.~Patrignani)}}
\label{sec:exdec}
Exclusive charmonium decays have been investigated within QCD by many
authors, \eg \cite{exdec:dun80,exdec:bro81,exdec:che82,exdec:bai85}.
As already argued at the beginning of \Section~\ref{sec:secnrqcd} the
dominant dynamical mechanism is $\ccbar$ annihilation into the minimal
number of gluons allowed by symmetries and subsequent creation of
light quark--antiquark pairs forming the final state hadrons.

In hard exclusive reactions higher Fock-state contributions are
usually suppressed by inverse powers of the hard scale, $Q$, appearing
in the process ($Q\sim\mc$ for exclusive charmonium decays), as
compared to the valence Fock-state contributions. Hence, higher
Fock-state contributions are expected to be negligible in most
cases. It has turned out, however, that higher Fock states of the
charmonium play an important role in understanding the production (see
\Chapter~\ref{chapter:production}) and the inclusive decays of charmonium (see
\Section~\ref{sec:secnrqcd}).  As shown in \cite{Bodwin:1994jh} the
long-distance matrix elements can there be organized into a hierarchy
according to their scaling with $v$, the typical velocity of the $c$
quark in the charmonium. The velocity expansion can also be applied to
exclusive charmonium decays \cite{exdec:BKS}.  The Fock expansions of
the charmonium states start (in the power counting of
\cite{Bodwin:1994jh}) as \ba |\jpsi\rangle & = &
\underbrace{|\ccbar_{1}({}^3S_1)\rangle} +
\underbrace{|\ccbar_{8}({}^3P_J)\, g \rangle} +
\underbrace{|\ccbar_{8}({}^3S_1)\, gg \rangle} + \dots , \nn\\ &&
\hspace{5mm}{{\cal O}(1)}\hspace{15mm}{{\cal O}({\it v})}\hspace{15mm}{{\cal O}({\it v}^2)}
\nn\\[0.2em]
|\;\;\eta_{c}\;\rangle & = & 
\underbrace{|\ccbar_{1}({}^1S_0)\rangle}
+ \underbrace{|\ccbar_{8}({}^1P_1)\, g\rangle}
+ \underbrace{|\ccbar_{8}({}^1S_0)\, gg \rangle}
+ \dots,
\nn\\
&&  
\hspace{5mm}{{\cal O}(1)}\hspace{15mm}{{\cal O}({\it v})}\hspace{15mm}{{\cal O}({\it v}^2)}
\nn \\[0.2em]
  |\;\chi_{c J}\rangle & = & 
\underbrace{|\ccbar_{1}({}^3P_J)\rangle}
+ \underbrace{|\ccbar_{8}({}^3S_1)\, g \rangle}
+ \dots, 
\label{eq:exdec-Fockexpansion}
\\
&&  
\hspace{5mm}{{\cal O}(1)}\hspace{15mm}{{\cal O}({\it v})}
\nn
\end{eqnarray}
where the subscripts at the $\ccbar$ pair specify whether it is in a
colour-singlet ($1$) or colour-octet ($8$) state; ${\cal O}(1)$, ${\cal O}(v)$
and ${\cal O}(v^2)$ are the orders to which the corresponding Fock
states contribute, once evaluated in a matrix element.
The amplitude for a two-body decay of a charmonium state satisfies a factorization formula, 
which separates the scale $\mc$ from the lower momentum scales. The
decay amplitude is then expressed as a convolution of a partonic
subprocess amplitude that involves the scale $\mc$, the charmonium
wave function for the initial state that involves scales of order
$\mc\, v$ and lower, and a factor that takes into account the light
hadron wave functions for the final state. This factor involves only
the scale $\Lambda_{\rm QCD}$. In the formal limit of $\mc\to\infty$
the dominant terms in the factorization formula involve the minimal
number of partons in the hard scattering, which is given by the valence
quarks of the hadrons participating in the considered process. Terms 
involving additional partons in the initial state are suppressed by 
powers of $v$ while terms involving additional partons in the final
state are suppressed by  powers of $\Lambda_{\rm QCD}/\mc$. Moreover,
in this limit of an asymptotically large charm quark mass, the valence
quarks of a light hadron move collinear with it, their transverse quark
momenta can be neglected. In this situation the soft parton--hadron 
transition is described by a leading-twist distribution amplitude,
$\Phi(x,\mu_F)$, for finding valence quarks in the hadron, each
carrying some fraction $x_i$ of the hadron's momentum and for which
the quark helicities sum up to the hadronic one. The distribution
amplitudes,  which represent light-cone wave functions integrated over
transverse momenta up to a factorization scale $\mu_F$ of order $\mc$
\cite{exdec:bro81,exdec:che82}, are the only  
nonperturbative input required in the calculation of decay amplitudes 
along these lines. The convolution formula in such a leading-twist
calculation of a charmonium decay into a pair of hadrons $h_1, h_2$ reads
\begin{equation}
M \= \int [dx]_N [dy]_N [d^3 k]_{N^\prime}\,\Phi_1(x,\mu_F)\, \Phi_2(y,\mu_F)\, 
          T_H(x,y,\mc,\mu_F) \, \Psi_c(k)\,,
\label{eq:exdec-fac-formula}
\end{equation}
where $x(y)$ represents the set of independent momentum fractions for an
$N$-particle Fock state of a light hadron and $\Psi_c$ is the
charmonium wave function for an $N^\prime$-particle Fock state. 
$k$ denotes the set of momenta of the particles in that Fock state. 
Soft and hard physics is separated at the factorization scale $\mu_F$.
    
\begin{figure}[t]
\begin{center}
\includegraphics[width=12cm,bbllx=91pt,bblly=520pt,bburx=555pt,%
bbury=753pt,clip=true]{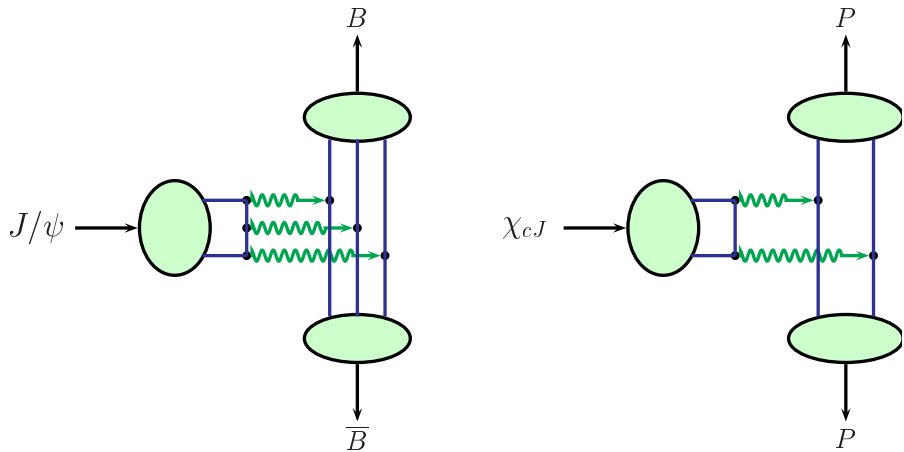}    
\caption[Typical lowest-order Feynman graphs for $\jpsi$ and
         $\chi_{cJ}$ decays]
        {Typical lowest-order Feynman graphs for $\jpsi$ decays into a
         baryon-antibaryon pair (left) and $\chi_{cJ}$ decays into a
         pair of pseudoscalar mesons (right). The wavy lines represent
         gluons.}
\label{fig:exdec-graphs} 
\end{center}
\end{figure}

The relative strength of various contributions to specific decay
processes can be easily estimated. Typical lowest-order Feynman graphs
are shown in \Figure~\ref{fig:exdec-graphs}.  A P-wave $\ccbar$ pair
requires a power of the $c$-quark's relative momentum ${\bf k}$ ($k
\sim \mc v$) from the hard scattering amplitude, which is to be
combined with a ${\bf k}$ from the P-wave charmonium spin wave
function in a $k^2$. In contrast to ${\bf k}$ itself, a term
proportional to $k^2$ does not lead to a vanishing contribution after
the ${\bf k}$ integration. Since, for dimensional reasons, ${\bf k}$
is to be replaced by ${\bf k}/\mc$ the subprocess amplitude involving
a P-wave $\ccbar$ pair, is of order $v$. Combining this fact with
the Fock-state expansion (\ref{eq:exdec-Fockexpansion}), one finds for
the amplitude of $\chi_{cJ}$ decays into, say, a pairs of pseudoscalar
mesons ($P$) the behaviour
\begin{equation}
M(\chi_{cJ}\to PP) \= a_1\,\as^2 v + a_8\, \as^2
                 \big (v\sqrt{\as}\big )
                     +{\cal O}(v^2)\,,
\label{eq:exdec-ampPP} 
\end{equation}
where the $a_i$ are process-typical constants.
For the reaction $\jpsi\to B\overline{B}$ ($B$ stands for baryon), 
on the other hand, one has
\begin{equation}
M(\jpsi\to B\overline{B}) \= \tilde{a}_1\, \as^3 + \tilde{a}_8\, \as^3
                     v\big (v\sqrt{\as}\big )
        +\tilde{b}_8\, \as^3\; v^2 \as + {\cal O}(v^3)\,.   
\label{eq:exdec-ampBB}
\end{equation}
Or, for the $\eta_c$ decaying for instance into a scalar ($S$) and a
pseudoscalar meson
\begin{equation}
{\cal M}(\eta_{c}\to S P) \=  \hat{a}_1\, \as^2+ \hat{a}_8\, \as^2 v\big (v\sqrt{\as}\big )
          +  \hat{b}_8 \,\as^2 (v\sqrt{\as})^2 + {\cal O}(v^3)\,.
\label{eq:exdec-ampSP}
\end{equation}
Thus, one sees that in the case of the $\chi_{cJ}$ the
colour-octet contributions $\propto a_8$ are not suppressed by powers of either 
$v$ or $1/\mc$ as compared to the contributions from the valence Fock 
states \cite{exdec:BKS}. For charmonium decays $\sqrt{\as}$ is large and
does not suppress the colour-octet contribution considerably. Hence, 
the colour-octet contribution, \ie the next higher Fock state of the 
charmonium state, has to be included for a consistent analysis of 
P-wave charmonium decays. The situation is different for $\jpsi$
decays into baryon--antibaryon pairs or $\eta_c\to SP$: higher Fock 
state contributions first start at ${\cal O}(v^2)$. Moreover, there is no 
obvious enhancement of the corresponding subprocess amplitudes, 
they appear with at least the same power of $\as$ as the valence Fock 
state contributions. Thus, despite of the fact that $\mc$ is not very 
large and $v$ not small ($v^2\simeq 0.3$), it seems 
reasonable to expect small higher Fock-state contributions to the 
baryonic decays of the $\jpsi$.

\begin{figure}[t]
\begin{center}
\includegraphics[width=14cm,bbllx=140pt,bblly=612pt,bburx=470pt,%
bbury=682pt,clip=]{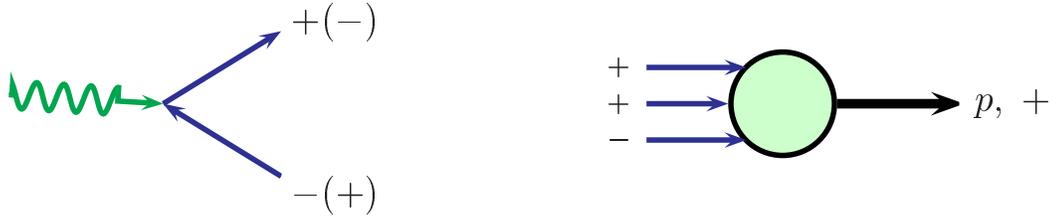}    
\caption[Helicity configurations in the creation of a light $\qqb$
         pair and for a leading-twist parton--proton transition]
        {Helicity configurations in the creation of a light $\qqb$
         pair (left) and for a leading-twist parton--proton transition
         (right).}
\label{fig:exdec-hel}
\end{center}
\end{figure}

The leading-twist formation of the light hadrons in the final state
has implications for their helicity configurations. As a consequence of the vector
nature of QCD (and QED) time-like virtual gluons (or photons) 
create light, (almost) massless quarks and antiquarks in opposite
helicity states, see \Figure~\ref{fig:exdec-hel}. To leading-twist
accuracy such partons form the valence quarks of the light hadrons
and transfer their helicities to them (see \Figure~\ref{fig:exdec-hel}). 
Hence, the total hadronic helicity is zero
\begin{equation}
\lambda_{1} + \lambda_{2} \= 0\,.
\label{eq:exdec-hsr}
\end{equation}
The conservation of hadronic helicities is a dynamical consequence of
QCD (and QED) which holds to leading-twist order.
The violation of helicity conservation in a decay process signals the
presence of higher-twist, higher Fock state and/or soft, non-factorizable
contributions. Such processes (\eg $\jpsi\to \rho\pi$, $\eta_c\to
\rho\rho$) have indeed been observed experimentally
with often sizeable branching ratios. For the two-meson channels
involving pseudoscalar ($P$) and vector mesons ($V$)
they are characterized by 
\begin{equation}
    (-1)^{J_c}\, P_c \neq (-1)^{J_1+J_2}\, P_1 P_2\,,
\end{equation}
where $J_i$ and $P_i$ are the spin and parity of the meson $i$.
The amplitudes for processes of this kind are proportional to the 
Levi-Civita tensor, $\varepsilon$, which is to be contracted in all
possible ways with the available Lorentz vectors, namely the two independent
light hadron momenta, $p_1$ and $p_2$, and the polarization 
vectors (or tensors) of the light vector mesons and the charmonium
state. As an example let us consider the process $\jpsi\to VP$, for
which the amplitude reads
\begin{equation}
{\cal M}_{\lambda_V,\lambda_{\jpsi}}(\jpsi\to VP) \= \frac{A}{M^2_{\jpsi}}\,
\varepsilon(p_1,p_2,\epsilon^*(\lambda_V),\epsilon(\lambda_{\jpsi}))\,, 
\end{equation}
where $A$ is a constant. Now, in the rest frame of the decaying meson,
the polarization vector of a helicity zero vector meson can be
expressed by a linear combination of the two final state momenta. The
number of independent Lorentz vectors is, therefore, insufficient to
contract the Levi-Civita tensor with the consequence of a vanishing
amplitude for processes involving longitudinally polarized vector
mesons. Thus, hadronic helicity conservation (\ref{eq:exdec-hsr}) is
violated in $\jpsi\to VP$ decays. By the same argument longitudinally
polarized vector mesons are forbidden in the decay $\eta_c\to
VV$. Since angular momentum conservation requires the same helicity
for both vector mesons, hadronic helicity is not conserved in the case
of transversally polarized vector mesons, too.  With similar arguments
the processes $\chi_{c1}, h_c\to VV$ and $\chi_{c2}\to VP$ are also
forbidden to leading twist order.  We note that hadronic helicity
conservation does also not hold in $\eta_c$ and $\chi_{c0}$ decays
into baryon--antibaryon pairs where, in the charmonium rest frame,
angular momentum conservation requires
$\lambda_{B\phantom{\overline{B}}}\hspace*{-0.4cm} \=
\lambda_{\overline{B}}$. A systematic investigation of higher-twist
contributions to these processes is still lacking despite some
attempts of estimating them, for a review see \cite{exdec:CZ}.  Recent
progress in classifying higher-twist distribution amplitudes and
understanding their properties \cite{exdec:braun90,exdec:braun} now
permits such analyses. The most important question to be answered is
whether or not factorization holds for these decays to higher-twist
order. It goes without saying that besides higher-twist effects, the
leading-twist forbidden channels might be under control of other
dynamical mechanisms such as higher Fock state contributions or soft
power corrections. In \Section~\ref{sec:exdec-rule} a variety of such
mechanisms will be discussed.

\begin{figure}[h]
\centerline{\includegraphics[width=60mm]{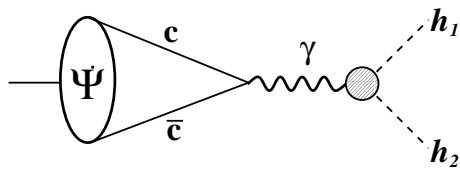}}
\caption[Electromagnetic $\psi(nS)$ decays into pairs of hadrons]
        {Electromagnetic $\psi(nS)$ decays into pairs of hadrons. The
         shaded blob indicates a time-like electromagnetic transition
         form factor.}
\label{fig:exdec-elm} 
\end{figure}

Next, let us consider $G$-parity and isospin. $G$-parity or 
isospin-violating decays are not strictly forbidden since they can 
proceed through electromagnetic $\ccbar$ annihilation and may receive 
contributions {}from the isospin-violating part of QCD. The latter 
contributions, being related to the $u-d$ quark mass difference, seem 
to be small \cite{exdec:che82}. $G$-parity or isospin-violating decays 
of $C$-even charmonia (\eg $\eta_{c},\chi_{c 1}, \chi_{c 2} \to PV$  
for non-strange final state mesons) have not been observed
experimentally as yet \cite{Eidelman:2004wy}. Proceeding on the assumption that
these decays are dominantly mediated by $\ccbar\to 2\gamma^*\to PV$, 
this is understandable. They are suppressed by a factor $(\aem/\as)^4$ 
as compared to the $G$-parity and isospin allowed decays of the 
$C$-even charmonia and their decay widths are therefore extremely small. 
Channels involving strange mesons (\eg $K K^*$), are also expected 
to be strongly suppressed by virtue of $U$-spin invariance. For $\jpsi$
decays the situation is different. Many $G$-parity violating (\eg
$\pi^+\pi^-$) or isospin-violating (\eg $\omega\pi^0$) decays have
been observed, the experimental branching ratios being of the order 
of $10^{-4}$--$10^{-3}$ \cite{Eidelman:2004wy}. As compared to $G$-parity and 
isospin allowed $\jpsi$ decays they are typically suppressed by factors 
of about $10^{-2}$--$10^{-1}$ in accord with what is expected for an 
electromagnetic decay mechanism, see \Figure~\ref{fig:exdec-elm}.
An overview over the allowed and forbidden charmonium decays into
pseudoscalar and vector mesons is given in \Table~\ref{tab:exdec-dec}. 

\begin{table}
\caption[Charmonium decays into $PP$, $PV$ and $VV$ meson pairs]
        {Charmonium decays into $PP$, $PV$ and $VV$ meson pairs.  The
         symbols $0$, $\epsilon$, $\surd$ denote channels forbidden by
         angular momentum and parity conservation, forbidden to
         leading-twist accuracy, and allowed, respectively. The
         brackets indicate that these channels violate either
         $G$-parity or isospin invariance for non-strange mesons.}
\label{tab:exdec-dec}
\renewcommand{\arraystretch}{1.2}
 \begin{center}
  \begin{tabular}{|c||c|c|c|} \hline 
              &  $\;PP\;$&  $\;PV\;$    &$\;VV\;$    
                \\ \hline \hline 
   $\eta_{c}$&   $0$    &$(\surd)$  &$\epsilon$  \\ \hline
     $\jpsi$    &($\surd$)&$\epsilon$&($\surd$)    \\ \hline
 $h_{c}$     &   $0$    & $\surd$  & $\epsilon$ \\ \hline
 $\chi_{c 0}$&$\surd$&   $0$    &$\surd$    \\ \hline
 $\chi_{c 1}$&  $0$   &($\surd)$  &$\epsilon$  \\ \hline
 $\chi_{c 2}$&$\surd$&$(\epsilon)$ &$\surd$    \\ \hline
   \end{tabular}
\end{center}
\renewcommand{\arraystretch}{1.0}
\end{table}

All what we have discussed so far holds for exclusive bottomonium decays
as well. The situation is even better in this case. Due to the larger
mass of the $b$ quark, corrections to the leading-twist QCD results 
for bottomonium decays are probably reasonably small. Thus, the data on
branching ratios can be expected to exhibit the pattern of
leading-twist predictions. 
Exclusive quarkonium decays constitute an interesting laboratory 
for investigating corrections to the leading-twist lowest-order
approach from various sources such as power and higher-twist
corrections as well as higher Fock-state contributions. 
A systematic study of such is still lacking.

\subsection[Decays of \jpsi ~and \psip into two meson]{Decays of
  $\jpsi$ and $\psi(2S)$ into two mesons $\!$\footnote{Author: E.~Braaten}}
\label{sec:exdec-rule}
The most dramatic unsolved problem in 
quarkonium physics is probably the $\rho$--$\pi$ puzzle.
In  analyzing the 2-body decays of the  $\jpsi$ and $\psi(2S)$
into two light hadrons $h_1$ and $h_2$,
it is convenient to consider the following quantity:
\begin{equation}
\kappa [h_1 h_2] = 
        \frac{{\cal B}(\psi(2S)\to h_1h_2)} {{\cal B}(\jpsi\to h_1h_2)}\,
        \frac{{\cal B}(\jpsi\to e^+e^-)} {{\cal B}(\psi(2S)\to e^+e^-)}\,
        \frac{\varrho[\jpsi h_1 h_2]}{\varrho[\psi(2S) h_1 h_2]} ,
\label{eq:exdec-kappa12}
\end{equation}
where 
\begin{equation}
\varrho [H h_1 h_2] = \sqrt{1 - 2 (M_{h_1}^2 + M_{h_2}^2)/M_H^2
                        + (M_{h_1}^2 - M_{h_2}^2)^2/M_H^4}\,.
\label{eq:exdec-phase}
\end{equation}
is a phase space factor that depends on the masses of 
the hadrons $H$, $h_1$, and $h_2$.
As will be explained shortly, very simple theoretical 
considerations lead to the expectation that this quantity 
should be close to 1 for all light hadrons $h_1$ and $h_2$:
\begin{equation}
\kappa [h_1 h_2] = 1.
\label{eq:exdec-kappapred}
\end{equation}
This prediction was once referred to as the 12\% rule because the 
experimental value of the ratio of the electronic branching fractions
of the $\psi(2S)$ and $\jpsi$ was at one time near 12\%.
That experimental value is now $15 \pm 2$\%.
The last factor in (\ref{eq:exdec-kappa12})  is a phase space factor 
that is close to 1 for hadrons whose masses are much smaller than 
that of the $\jpsi$.  Thus the prediction (\ref{eq:exdec-kappapred})
implies that the ratio of the branching fractions of the 
$\psi(2S)$ and $\jpsi$ into $h_1 h_2$ should be near 15\%.
All the baryon--antibaryon decay modes that have been measured are compatible 
with the prediction (\ref{eq:exdec-kappapred}), see Sec.~\ref{sec:exdec-bbbar}.
Some two-meson decay modes are compatible with this prediction,
but there are others for which it 
is badly violated.  The most severe violation that has been observed is in 
the $\rho \pi$ decay mode.  The first hint of this problem was seen by the 
Mark II collaboration in 1983 \cite{exdec:Franklin:1983ve}.  The decay $J/\psi 
\to \rho \pi$, with a branching fraction of about 1.3\%, is the largest 
2-body hadronic decay mode of the $J/\psi$.  
In contrast, the partial width for $\psi (2S) \to \rho \pi$ is so small that
this decay was not observed until very recently
by the CLEO and BES collaborations \cite{Adam:2004pr,Ablikim:2004bd}.
The branching fraction is measured to be $0.46\pm0.09$,
and the ratio defined in (\ref{eq:exdec-kappa12}) is
$\kappa[\rho \pi] = 0.028 \pm 0.006$.
The dramatic discrepancy between this result and the prediction
in \Eq~(\ref{eq:exdec-kappapred}) is the $\rho -\pi$ puzzle.

We proceed to explain the assumptions underlying the prediction 
(\ref{eq:exdec-kappapred}).  Because there is a nonzero amplitude for the 
$J/\psi$ to be a pure $c \bar c$ state, the matrix element for its decay 
into two light hadrons $h_1$ and $h_2$ can be expressed in the form
\begin{equation}
{\cal M} (J/\psi \to h_1 h_2) = \int \frac {d^3 p} {(2 \pi)^3} \psi_{J/\psi} 
({\bf p}) {\cal A} (c ({\bf p}) \bar c (- {\bf p}) \to h_1 h_2) ,
\label{eq:exdec-MJpsi12}
\end{equation}
where $\psi_{J/\psi} ({\bf p})$ is the momentum-space wave function 
for the pure $c\bar c$ component of the $J/\psi$.  
This can be regarded as an exact formula that defines the amplitude 
${\cal A} ( c \bar c \to h_1 h_2)$.  It relies on the fact that 
wave functions satisfy integral equations, so even if there are other 
components of the $J/ \psi$ wave function besides $c \bar c$,
the iteration of the integral equation will eventually 
produce a pure $c \bar c$ state.
The annihilation of the $c \bar c$ pair produces an intermediate state
consisting of partons with momenta of order $m_c$,  
which is much larger than either the momentum scale $p \sim m_c v$ 
for the $c \bar c$ wave function of the $J/ \psi$ 
or the scale $\Lambda_{\rm QCD}$ associated with the wave functions 
of the light hadrons $h_1$ and $h_2$.
If the factored expression in (\ref{eq:exdec-MJpsi12}) also corresponds 
to a separation of small 
momenta associated with the wave function of $J / \psi$ from small momenta 
associated with the wave functions of $h_1$ and $h_2$, 
then the amplitude ${\cal A}$ in (\ref{eq:exdec-MJpsi12}) should be insensitive 
to the value of ${\bf p}$.  It 
can be approximated by its value at ${\bf p} = 0$ up to corrections 
suppressed by powers of $v$ and $\Lambda_{\rm QCD}/m_c$:
\begin{equation}
{\cal A} (c ({\bf p}) \bar c (- {\bf p}) \to h_1 h_2) \approx 
{\cal A} (c ({\bf 0}) \bar c ({\bf 0}) \to h_1 h_2) .
\label{eq:exdec-Ap0}
\end{equation}
With this approximation, the matrix element (\ref{eq:exdec-MJpsi12}) reduces to
\begin{equation}
{\cal M}(J/\psi \to h_1 h_2) \approx \psi_{J/\psi} ({\bf r} = 0) {\cal A} 
(c ({\bf 0}) \bar c ({\bf 0}) \to h_1 h_2),
\label{eq:exdec-MJpsiapprox}
\end{equation}
where $\psi_{J/\psi} ({\bf r})$ is the coordinate-space wave function 
for the pure $c\bar c$ component of 
$J/\psi$.  The decay rate then has the factored form
\begin{equation}
\Gamma (J/\psi \to h_1 h_2) \approx 
\left| \psi_{J/\psi} ({\bf r} = 0) \right|^2 
\left| {\cal A} (c ({\bf 0}) \bar c ({\bf 0}) \to h_1 h_2) \right|^2 
\frac{\varrho[\jpsi h_1 h_2]} {16 \pi M_{J/\psi}}  .
\label{eq:exdec-GamJpsiAB}
\end{equation}
The corresponding expression for the decay $\psi(2S) \to h_1 h_2$ differs 
only in the mass and the wave function factor.  These factored expressions 
apply equally well to decays into $e^+ e^-$.  Taking the ratio of decay 
rates in (\ref{eq:exdec-kappa12}), we obtain the prediction $\kappa [h_1 h_2] 
= 1$ for any light hadrons $h_1$ and $h_2$.  Any significant deviation of 
$\kappa [h_1 h_2]$ from 1 indicates a breakdown of the approximation 
(\ref{eq:exdec-Ap0}).

An important reference point for the prediction (\ref{eq:exdec-kappapred}) is 
provided by the (leading twist) asymptotic predictions of perturbative QCD 
\cite{exdec:bro81,exdec:che82}.  These predictions are most easily 
described using a ratio $R$ defined by
\begin{equation}
R_\jpsi[h_1h_2] = 
\frac{\Gamma(J/\psi \to h_1 h_2)} {\Gamma (J/\psi \to e^+e^-)} .
\label{eq:exdec-R12def}
\end{equation}
The asymptotic predictions for this ratio depend on the helicities 
$\lambda_1$ and $\lambda_2$ of the two hadrons $h_1$ and $h_2$.  
If the hadrons are mesons and the decay proceeds via 
the annihilation process $c \bar c \to ggg$, the prediction for the 
scaling behavior of the ratio is
\begin{equation}
R_\jpsi [h_1 (\lambda_1) h_2 (\lambda_2)] \sim \frac{\alpha_s^6 (m_c)} {\aem^2} 
\left ( \frac {\Lambda_{\rm QCD}} {m_c} \right)^{4 + 2| \lambda_1 + 
\lambda_2 |} \label{eq:exdec-R12lambda} .
\end{equation}
If the decay proceeds via the annihilation process $c \bar c \to \gamma^*$, 
the prefactor $\als^6 / \aem^2$ is replaced by $\als^2$.  The 
scaling behavior (\ref{eq:exdec-R12lambda}) illustrates one of the basic 
qualitative features of the asymptotic QCD predictions:  light hadron 
helicity conservation.  The dominant decay modes are predicted to 
satisfy the helicity selection rule (\ref{eq:exdec-hsr}).  
In the case of 
the decay $J/\psi \to \rho \pi$, the helicity of the pion is $\lambda_\pi = 
0$ and the helicity of the $\rho$ is constrained by Lorentz invariance to 
be $\lambda_\rho = \pm 1$.  Thus this decay necessarily violates the 
helicity selection rule, and its rate is predicted to be suppressed by 
$\Lambda^2_{\rm QCD} / m_c^2$ relative to modes that are compatible with 
the helicity selection rule.  
But $\rho \pi$ is observed to be the largest 
2-body decay mode of the $J/\psi$.  This appears to be a clear violation of 
the asymptotic PQCD predictions.  An understanding of the 
$\rho$--$\pi$ puzzle may have important implications for the relevance of 
asymptotic PQCD to charmonium decays.

The dramatic failure of the prediction (\ref{eq:exdec-kappapred}) 
in some channels indicates a breakdown of the approximation 
(\ref{eq:exdec-Ap0}) for either the $J/ \psi$ 
decay or the $\psi (2S)$ decay or both.  The contribution to the amplitude 
${\cal A}$ from the annihilation of $c \bar c$ into 3 hard gluons or a 
virtual photon should be insensitive to the relative momentum ${\bf p}$ of 
the $c \bar c$ pair.  The failure of the prediction (\ref{eq:exdec-kappapred}) 
indicates that at least one other dynamical mechanism must be involved.
The sensitivity of the amplitude to ${\bf p}$ could arise from a 
fluctuation of the charmonium state into some component of the wave function 
other than $c \bar c$.  
In a hadronic basis, this fluctuation can be expressed in terms of 
mixing of the charmonium state with other hadrons.  
In a parton basis, it can be expressed in terms of $c \bar c$ 
annihilation from a higher Fock state that includes soft gluons. 

Many explanations for the $\rho \pi$ puzzle have been proposed.  The small 
upper bound on $\kappa [\rho \pi]$ can be explained either by an 
enhancement of the rate for $J/\psi \to \rho \pi$ or by a suppression of 
the rate for $\psi (2S) \to \rho \pi$.  The enhancement of $J/ \psi \to 
\rho \pi$ relative to $\psi (2S) \to \rho \pi$ could occur through mixing 
of $J / \psi$ with another narrow state that has a much larger branching 
fraction into $\rho \pi$.  One such possibility is
\begin{enumerate}

\item mixing of $J /\psi$ with a narrow glueball 
        \cite{exdec:Hou:1982kh,exdec:Brodsky:1987bb}.

\end{enumerate}
Direct searches have failed to reveal any evidence for such a glueball.  
The suppression of $\psi (2S) \to \rho \pi$ relative to $J/\psi \to \rho 
\pi$ could be explained if the decay is dominated by a particular component 
of the wave function that is suppressed for $\psi (2S)$ relative to 
$J/\psi$.  The possibilities include

\begin{enumerate}
\setcounter{enumi}{1}

\item  suppression of the $c \bar c$ wave function at the origin for a 
        component of $\psi (2S)$ in which the $c \bar c$ 
        is in a colour-octet $^3S_1$ state \cite{exdec:chen98},

\item  suppression of the $\omega \phi$ component of 
        $\psi(2S)$ \cite{exdec:impli}.

\end{enumerate}
The suppression of $\psi (2S) \to \rho \pi$ relative to $J / 
\psi \to \rho \pi$ could be explained if the amplitude is dominated by two 
components of the wave function that nearly cancel in the case of $\psi 
(2S)$ but not for $J/\psi$.  The possibilities include
\begin{enumerate}
\setcounter{enumi}{3}

\item  cancellation between $c \bar c$ and $D \bar D$ components of 
        $\psi (2S)$ \cite{exdec:Suzuki:fs},

\item  cancellation between $c \bar c$ and glueball components of 
        $\psi (2S)$ \cite{exdec:Suzuki:fs},

\item  cancellation between S-wave $c \bar c$ and D-wave $c \bar c$ 
        components of $\psi (2S)$ \cite{Rosner:2001nm}.
\end{enumerate}
This last proposal leads to the very simple and unambiguous 
prediction that the D-wave charmonium state $\psi (3770)$ should have a 
branching fraction into $\rho \pi$ of about $4 \times 10^{-4}$
\cite{Rosner:2001nm}.
A recently proposed explanation for the $\rho$--$\pi$ puzzle is a 
\begin{enumerate}
\setcounter{enumi}{6}

\item  cancellation between the amplitudes for the resonant process $e^+e^-
  \to \psi(2S) \to \rho\pi$ and the direct process $e^+e^- \to \rho\pi$. See
  Sec.~2.8.5.
\end{enumerate}
This proposal predicts that the observed suppression of $\psi(2S)\to\rho\pi$
relative to $J/\psi \to \rho\pi$ is specific to $e^+e^-$ annihilation and
should not occur for other charmonium production processes, such as $B$-meson decay.

\begin{table}
\caption[Comparison of $J/\psi$ and $\psi'$ branching ratios to VP,
         PP, PA, VS, VV and VT mesons]
        {Comparison of $J/\psi$ and $\psi'$ branching ratios to VP,
         PP, PA, VS, VV and VT mesons. Unless specified data are from
         PDG \cite{Eidelman:2004wy}.  Where specified we have included
         in the averages recent data on $\psi(2S)$ decays from
         BES\cite{Ablikim:2004ky,Ablikim:2004bd,Ablikim:2004sf,Ablikim:2004kv}
         and CLEO\cite{Adam:2004pr}, the latter derived from reported
         ratios of branching ratios using values in
         PDG\cite{Eidelman:2004wy}.}
\label{tab:exdec-rhopi}
\renewcommand{\arraystretch}{1.2}   
\begin{center}
\begin{tabular}{|l  | c |c | c |}
\hline
Decay mode $h_1 h_2$ & ${\cal B}(J/\psi \to h_1 h_2)$ & ${\cal B}(\psi'\to h_1 h_2)$ &
$\kappa[h_1 h_2]$ \\
&($\times 10^{4}$)&($\times 10^{4}$) & (\Eq~\ref{eq:exdec-kappa12})  \\
\hline
\hline %PV
$\varrho\pi$               & $127\pm 9$    & $0.46\pm0.09$ \cite{Adam:2004pr}\cite{Ablikim:2004bd}
                                           &  $0.028\pm0.006$ \\ 
$\omega\pi^0$              & $4.2\pm0.6$   & $0.22\pm0.09$ \cite{Adam:2004pr}\cite{Ablikim:2004kv}
                                           &  $0.40\pm0.17$   \\ 
$\varrho\eta$              & $1.93\pm0.23$ & $0.23\pm0.12$\cite{Adam:2004pr}\cite{Ablikim:2004kv}
                                           & $0.9\pm0.5$     \\ 
$\omega\eta$               & $15.8\pm 1.6$ & $<0.11$\cite{Ablikim:2004sf}
                                           & $<0.06$ \\ 
$\phi\eta$                 & $6.5\pm0.7$   & $0.35\pm0.11$\cite{Adam:2004pr}\cite{Ablikim:2004sf}
                                           & $0.40\pm0.13$ \\
$\varrho\eta'(958)$        & $1.05\pm0.18$ & $0.19^{0.16}_{-0.11}\pm0.03$\cite{Ablikim:2004kv}
                                           & $2.5\pm0.9$ \\
$\omega\eta'(958)$         & $1.67\pm0.25$ & $<0.81$\cite{Ablikim:2004sf}
                                           & $<4.3$ \\   
$\phi\eta'(958)$           & $3.3\pm0.4$   & $0.33\pm0.13\pm0.07$\cite{Ablikim:2004sf}
                                           & $0.71\pm0.33$      \\ 
$K^{*}(892)^\mp K^\pm$     & $50\pm 4$     & $0.26\pm0.11$\cite{Adam:2004pr}\cite{Ablikim:2004ky}
                                           & $0.039\pm0.017$  \\
$\bar K^{*}(892)^0 K^0$+c.c.& $42\pm 4$    & $1.55\pm0.25$\cite{Adam:2004pr}\cite{Ablikim:2004ky}
                                           & $0.28\pm0.05$         \\
\hline %PP
$\pi^+\pi^-$              & $1.47\pm0.23$ & $0.8\pm0.5$   & $4.3\pm2.7$  
                                                                    \\ %2003 no update
$K^+K^-$                  & $2.37\pm0.31$ & $1.0\pm0.7$    & $3.2\pm2.3$  
                                                                    \\ %2003 no update
$K^0_SK^0_L$              & $1.46\pm0.26$  & $0.52\pm0.07$ & 
                                                      $2.7\pm0.6$   \\ %%% BES UPDATE
\hline %AP
$\pi^\pm b_1(1235)^\mp$   & $30\pm 5$  & $3.9\pm1.6$ (incl.\cite{Adam:2004pr})
                                       & $1.0\pm0.4$\\
$\pi^0 b_1(1235)^0$       & $23\pm 6$  & $4.0^{+0.9}_{-0.8}\pm0.6$ \cite{Adam:2004pr}
                                       & $1.3\pm0.5$\\
$K^\pm K_1(1270)^\mp$     & $<30$      & $10.0\pm2.8$ & $>1.7 $      \\%2003 no update
$K^\pm K_1(1400)^\mp$     & $38\pm14$  & $<3.1$       & $<0.8 $     \\%2003 no update
\hline %VS
$\omega f_0(980)\to\omega \pi\pi$   & $1.1\pm0.4$ &   &    
                                              \\ %2003 no update (*.78 as in listing)
$\phi f_0(980)\to\phi \pi\pi$       & $2.5\pm0.7$ & $0.60\pm0.22$ & $1.7\pm0.8$   
                                                         \\ %*.78 NB:  Psi(2S) Errata
$\omega f_0(1710)\to\omega K\bar K$ & $4.8\pm1.1$ &   &    \\ %2003 no update
$\phi f_0(1710)\to\phi K\bar K$     & $3.6\pm0.6$ &               &  \\%2003 no update
\hline % VV
$\omega f_1(1420)$     & $6.8\pm2.4$ &   &    \\ %2003 no update
$\phi f_1(1285)$       & $2.6\pm0.5$ &   &    \\ %2003 no update
\hline %VT All of these have been upated 
$\omega f_2(1270)$          & $43\pm6$    & $2.1\pm0.6$\cite{exdec:BES-VT}  
                                                           & $0.34\pm0.11 $  \\
$\varrho a_2(1320)$         & $109\pm22$  & $2.6\pm0.9$\cite{exdec:BES-VT}  
                                                           & $0.17\pm0.07 $  \\
$K^{*}(892)^0 \bar 
K_2^{*}(1430)^0$ + c.c.     & $67\pm26$   & $1.9\pm0.5$\cite{exdec:BES-VT}  
                                                           & $0.19\pm0.09 $  \\
$\phi f_2'(1525)$           & $12.3\pm2.1$& $0.44\pm0.16$\cite{exdec:BES-VT} 
                                                           & $0.22\pm0.09 $  \\
\hline
\end{tabular}
\end{center}
\renewcommand{\arraystretch}{1.0}   
\end{table}

It is reasonable to expect that a definitive solution to the
$\rho$--$\pi$ puzzle should also explain the deviations of $\kappa[h_1
h_2]$ from the prediction 1 for other hadrons $h_1$ and $h_2$.  The
existing measurements of the branching fractions into two mesons for
$J/\psi$ and $\psi(2S)$ are shown in \Table~\ref{tab:exdec-rhopi}.
While many of the values of $\kappa[h_1 h_2]$ are compatible with 1,
there are modes other than $\rho \pi$ for which $\kappa$ is
significantly smaller than 1, such as $\rho a_2$, and and there are
modes for which $\kappa$ is significantly greater than 1, such as
$K_S^0 K_L^0$.

One clue to the mechanism is how $\kappa[h_1 h_2]$ depends on the
$J^{PC}$ quantum numbers for hadrons $h_1 h_2$ with the same flavour
quantum numbers as $\rho \pi$.  As can be seen in
\Table~\ref{tab:exdec-rhopi}, there also seems to be suppression in
the vector-tensor (VT) channel $\rho a_2$, but there seems to be no
significant suppression in the axial vector-pseudoscalar (AP) channel
$b_1 \pi$ or in the pseudoscalar-pseudoscalar (PP) channel $\pi^+
\pi^-$.  The absence of any suppression in the channel $\pi^+ \pi^-$
is to be expected, because it proceeds predominantly through $c \bar
c$ annihilation into a single photon, and therefore the approximation
(\ref{eq:exdec-MJpsiapprox}) should hold.
 
Another clue to the suppression mechanism is the pattern of
$\kappa[h_1 h_2]$ for different radial excitations of mesons with the
same $J^{PC}$ quantum numbers.  An example is the AP decay modes
$K^\pm K_1^\mp$ for different $K_1$ resonances.  The mode $K^\pm
K_1(1400)^\mp$ has been observed in $J/\psi$ decays but not in
$\psi(2S)$ decays.  The mode $K^\pm K_1(1270)^\mp$ has been observed
in $\psi(2S)$ decays but not in $J/\psi$ decays.  The lower bound on
$\kappa$ for $K^\pm K_1(1270)^\mp$ is significantly greater than the
upper bound on $\kappa$ for $K^\pm K_1(1400)^\mp$.  This demonstrates
that whether $\kappa$ is suppressed or enhanced relative to the
prediction (\ref{eq:exdec-kappapred}) is not determined solely by the
$J^{PC}$ quantum numbers of the mesons.

The suppression pattern in a given channel as a function of the flavour 
quantum numbers should also provide important clues to the 
suppression mechanism.  The channel for which the most measurements 
are available is the VP channel. The decay amplitude for 
$J\ /\psi \to VP$ can be resolved 
into 3 terms with distinct flavour structures:
\begin{itemize}
\item{a flavour-connected amplitude $g$ with quark structure $(q_i \bar q_j) 
( q_j \bar q_i)$,}
\item{a flavour-disconnected amplitude $h$ with quark structure 
$(q_i \bar q_i) ( q_j \bar q_j)$,}
\item{an electromagnetic amplitude $e$ with quark structure $Q_{ik}(q_i 
\bar q_j) ( q_j \bar q_k)$ where $Q$ is the light quark charge matrix.}
\end{itemize}
For example, the amplitude for $J\ /\psi \to \rho \pi$ is proportional to $g + e$.
A quantitative analysis should also take into account SU(3) symmetry 
breaking from the strange quark mass and U$_{\rm A}$(1) symmetry breaking from 
the triangle anomaly.
In the case of $J/\psi$, there are enough precise measurements of VP decays 
to completely determine the flavour decomposition of the amplitude 
\cite{exdec:Seiden:1988rr,exdec:Bramon:1997mf}. 
The conclusion is that $|e|$ and $|h|$ are comparable in magnitude and 
about an order of magnitude smaller than $|g|$.

The analogous flavour decomposition for $\psi (2S) \to VP$ expresses the 
decay amplitudes as a linear combination of amplitudes $g^\prime, 
h^\prime$, and $e^\prime$ with distinct flavour structures.  The same 
reasoning that led to the prediction $\kappa [h_1 h_2] = 0$ implies that 
these amplitudes $g^\prime, h^\prime$ and $e^\prime$ should differ from the 
corresponding amplitudes $g_1, h_1$ and $e$ for $J/\psi$ by the factor
\begin{equation}
\left( \frac {M_{\psi (2S)} \Gamma ( \psi (2S) \to e^+ e^-)} 
           {M_{J/\psi}    \Gamma ( J/\psi \to e^+ e^-)} \right)^{1/2} 
\approx 0.70.
\end{equation}
However, the measurement $\kappa [ \rho \pi] \approx 0.028$ implies
$| g^\prime + e^\prime| \approx 0.12 | g + e|$.
Since $|g| \gg |e|$, this requires $| g^\prime |$ to be suppressed
relative to $0.70 | g |$.
A mechanism for such a suppression was proposed in Ref.~\cite{exdec:chen98}.
If $g^\prime$ was so strongly suppressed that it was
small compared to $|e^\prime |$, it would make the rate for
$\psi (2S) \to \rho \pi$ comparable to electromagnetic processes
such as $\psi (2S) \to \omega \pi^0$.
The stronger suppression of $\psi (2S) \to \rho \pi$ that is observed
requires that $g^\prime$ and $e^\prime$ be comparable
in magnitude and to have phases such that there is a
further cancellation in the sum $g^\prime + e^\prime$.

The CLEO collaboration has recently presented the first evidence 
for two-body decays of the $\Upsilon(1S)$ \cite{exdec:Dytman:2003qx}.
They observed signals with a statistical significance of greater than
$5 \sigma$ for decays into $\phi f_2'(1525)$ and $\bar K K_1(1400)$.
The decay of $\Upsilon(1S)$ into $\bar K K_1(1270)$ is observed to be 
suppressed relative to $\bar K K_1(1400)$, which is the same 
pattern observed in $J/\psi$ decays.
The CLEO collaboration  also set upper limits on other decay modes,
the strongest of which is ${\cal B}(\Upsilon(1S) \to \rho \pi) < 4 \times 10^{-6}$.

\subsection[Decays of \jpsi ~and \psip into baryon--antibaryon]{Decays
of $\jpsi$ and $\psi(2S) $ into baryon--antibaryon}
\label{sec:exdec-bbbar}
As we already discussed these decays seem to be dominated by hard physics where the charm and anticharm quark
annihilate into gluons at short distances. In a leading-order
calculation of decay widths for the $B\overline{B}$ channels
contributions from higher charmonium Fock states can be neglected
since they only produce ${\cal O}(v^2)$ corrections, see \Eq~(\ref{eq:exdec-ampBB}); 
contributions from higher baryon Fock states are
suppressed by powers of $1/\mc$.  For consistency, the masses of the
$\jpsi$ and $\psi(2S)$ are to be replaced by $2\mc$ (except in phase
space factors) since the energy for the binding of a $\ccbar$ pair in a
charmonium state is an ${\cal O}(v^2)$ effect. The only soft physics
information on the charmonium state needed in a calculation to lowest 
order in $v$ is its decay constant. The corresponding electronic decay 
widths 
\begin{equation}
 \Gamma(\jpsi\to e^+e^-) \= \frac{4\pi}{3}\, \frac{e_c^2\, \aem^2\,
   f_\jpsi^2}{M_\jpsi}\,,
\label{eq:exdec-electronic}
\end{equation}
provide their values: $f_{\jpsi}=409\, {\rm MeV}$, 
$f_{\psi(2S)}=282\, {\rm MeV}$. The other soft physics information
required is the leading-twist baryon distribution amplitude. As can 
be shown \cite{exdec:dzi88} the proton is described by one independent
distribution amplitude, $\Phi^{\;p}_{123}(x)$, to leading-twist
accuracy. The set of subscripts $1,2,3$ refers to the quark configuration
$u_+\, u_-\, d_+$ of a proton with positive helicity. The
distribution amplitudes for other valence quark configurations in the
proton  are obtained by permutations of the subscripts. Since flavour
SU(3) is a good symmetry, only mildly broken by quark mass effects,
it is reasonable to assume that the other members of the lowest-lying 
baryon  octet are also described by only one independent distribution
amplitude, which, up to flavour symmetry breaking effects, is the same
as the proton one.

To start with and for orientation, we present the leading-twist result
for the width of the decays
of transversely polarized $\jpsi$s, as for instance are produced in
$e^+e^-$ annihilations, into proton--antiproton pairs. The width,  
evaluated from the asymptotic form of
the baryon wave function $\Phi_{\rm AS}^B=120\,x_1\,x_2\,x_3$, reads 
\begin{equation}
\Gamma(\jpsi\to \ppbar) \= \frac{5^6\, 2^{10}}{3^5}\, \pi^5\, \as(\mc)^6\,
        \frac{\varrho[\jpsi \, \ppbar]}{M_\jpsi}\, 
        \left(\frac{f_\jpsi\, f_p^2}{\mc^4}\right)^2\, I^2_{\rm AS}\,,
\label{eq:exdec-ppasy}
\end{equation}
where
\begin{equation}
I_{\rm AS} \= 6\int [dx]_3\, [dy]_3 \, \frac{x_1 y_3}
    {[x_1(1-y_1)+(1-x_1)y_1][x_3(1-y_3)+(1-x_3)y_3]}\,.
\label{eq:exdec-inte}
\end{equation}
The normalization parameter $f_p$ represents the proton's light-cone
wave function for zero spatial separation of the quarks. Strictly
speaking, it is defined by \cite{exdec:bro80}
\begin{equation}
\frac{f_p(\mu_F)}{8\sqrt{n_c!}}\, \Phi^{\;p}_{123}(x,\mu_F) \=
                 \int^{\mu_F} [d^2 k_\perp]_3 \Psi^{\;p}_{123}(x,k_\perp)\,,
\end{equation}
with
\begin{equation}
    \int [dx]_3 \Phi^{\;p}_{123}(x,\mu_F) \= 1\,.
\end{equation}
Both the distribution amplitude and $f_p$ are subject to evolution
\cite{exdec:bro80}. A typical value for $f_p$ is 
$\simeq 6\times 10^{-3}\, {\rm GeV}^2$ \cite{exdec:che82,exdec:bolz95}. 
Evaluating the branching ratio from (\ref{eq:exdec-ppasy}),
(\ref{eq:exdec-inte}), one obtains 
\begin{equation}
{\cal B}(\jpsi\to\ppbar) \= 1.5 \times 10^{-3}\;
                   \left(\frac{\as}{0.4}\right)^3\,
                   \left(\frac{1.5\, {\rm GeV}}{\mc}\right)^7\,
             \left(\frac{f_p}{6\times 10^{-3}\, {\rm GeV}^2}\right)^4\,,
\end{equation}
which is in quite good agreement with experiment, see
\Table~\ref{tab:exdec-jpsi}.  The predictions for the branching ratio
are more robust than that from the $\jpsi\to \ppbar$ decay widths
since the total $\jpsi$ width is dominated by the decays into light
hadrons. Hence, according to (\ref{eq:exdec-tot}) and
(\ref{eq:exdec-electronic}), the branching ratios approximately scale as
$1/m_c^7$ and $\as^3$.
 
In previous calculations \cite{exdec:che82,exdec:ste94}   
of the $\jpsi\to \ppbar$ decay width, distribution amplitudes
have been employed that are strongly concentrated in the end-point
regions where one of the momentum fractions is small. The use of such
distribution amplitudes has been heavily criticized
\cite{exdec:isg89}. Due to their properties the bulk of the
amplitude for the subprocess $\ccbar\to 3g^*\to 3(\qqb)$ is
accumulated in the soft end-point regions where the use of perturbative QCD
is inconsistent. Moreover, such distribution amplitudes lead to extremely
strong contributions to the decay amplitude and require compensation
by small values of $\as$, typically in the range of $0.2$--$0.3$. Such
values are unrealistically small with regard to the characteristic
scales available in charmonium decays. For an average gluon virtuality
of about $1\,{\rm GeV}^2$ one would expect $\as$ to be rather $0.4$. 

Recent theoretical \cite{exdec:braun,exdec:pet03} and phenomenological
\cite{exdec:bolz95} studies provide evidence that the proton
distribution amplitude is close to the asymptotic form for baryons:
the end-point concentrated forms seem to be obsolete. 
In a recent analysis of the $\jpsi$ and $\psi(2S)$
decays into baryon--antibaryon pairs \cite{exdec:bolz97} use is made of
the phenomenological proton distribution amplitude proposed in \cite{exdec:bolz95}
\begin{equation}
\Phi^{\;p}_{123}(x,\mu_0) \= \Phi_{\rm AS}^B\; \frac12\, (1+3x_1)\,,
\label{eq:exdec-pda}
\end{equation}
which is valid at the factorization scale $\mu_0=1\, {\rm GeV}$. This
distribution amplitude goes along with the normalization parameter
$f_p(\mu_0)=6.64 \times 10^{-3}\, {\rm GeV}^2$. 
In \cite{exdec:bolz97} the distribution amplitude (\ref{eq:exdec-pda})
has been suitably generalized to the cases of hyperons and decuplet baryons
by allowing for flavour symmetry breaking due to the effect of the strange
quark mass. Instead of the collinear approximation as used in
\cite{exdec:che82,exdec:ste94} or in (\ref{eq:exdec-ppasy}), the modified
perturbative approach \cite{exdec:BLS} is applied in \cite{exdec:bolz97}. 
In this approach quark transverse momenta are retained and Sudakov
suppressions, comprising those gluonic radiative corrections not
included in the evolution of the distribution amplitude, are taken
into account. The advantage of the modified perturbative approach is
the strong suppression of the soft end-point regions where perturbative 
QCD cannot be applied. If distribution amplitudes close to the asymptotic
form are employed the difference between a calculation on the basis of
the collinear approximation and one within the modified perturbative
approach is, however, not substantial given that the $\jpsi\to B\overline{B}$ 
amplitude is anyhow not very sensitive to the end-point regions. 
This is in marked contrast to the case of the proton's electromagnetic 
form factor \cite{exdec:berg95}. On the other hand, a disadvantage
of the modified perturbative approach is that the full baryon light-cone 
wave function is needed and not just the distribution amplitude. In 
\cite{exdec:bolz97} the transverse momentum dependence of the baryon
wave functions has been parameterized by a simple Gaussian  
\begin{equation}
        \propto  \exp{\left[ - a_B^2\, \sum k_{\perp
              i}^2/x_i\right]}\,,
\label{eq:exdec-gaussian}
\end{equation}
where a value of $0.75\,{\rm GeV}^{-1}$ has been adopted for the
transverse size parameter $a_B$. For the decuplet baryons a somewhat
larger value has been used ($0.85\,{\rm GeV}^{-1}$).  Calculating the
subprocess amplitude from the Feynman graphs shown in
\Figure~\ref{fig:exdec-graphs} and working out the convolution of
subprocess amplitude and baryon wave functions, one obtains the widths
$\Gamma_{3g}$ for the $\jpsi$ decays into $B\overline{B}$ pairs
mediated by the hard annihilation process $\ccbar\to 3g^*\to 3(\qqb
)$.  The results are listed and compared to experimental data in
\Table~\ref{tab:exdec-jpsi}.  In addition to the three-gluon
contribution there is also an isospin symmtetry violating
electromagnetic one generated by the subprocess $\ccbar\to\gamma^*\to
3(q\bar{q})$, see \Figure~\ref{fig:exdec-elm}. According to
\cite{exdec:bolz97} this contribution is probably small, of the order
of a few percent only. An important ingredient in this estimate of the
size of the electromagnetic contribution is the agreement of the
experimental widths for $\jpsi$ decays into $n\bar{n}$ and $\ppbar$
within the errors \cite{Eidelman:2004wy}. The contributions from the
$\ccbar\to g^* g^* \gamma^* \to 3(q\bar{q})$ to the baryon--antibaryon
channels amount to less than $1\%$ of the three-gluon contribution and
can be neglected.

\begin{table}
\caption[Results for $\jpsi$ and $\psi(2S)$ branching ratios for
         $B\overline{B}$ channels]
        {Results for $\jpsi$ and $\psi(2S)$ branching ratios for
         $B\overline{B}$ channels in units of $10^{-3}$ and $10^{-4}$,
         respectively. The three-gluon contributions, taken from
         \cite{exdec:bolz97}, are evaluated from $\mc=1.5$~Gev, and
         the one-loop $\as$ with $\Lambda_{\rm QCD}=210$~MeV. Unless
         specified data are taken from
         Ref.\,\cite{Eidelman:2004wy}. For the $J/\psi\to p\bar p$ we
         have included the recent BES measurement
         \cite{exdec:BES-Jppbar} in the average.  The theoretical
         branching ratios are evaluated using
         $\Gamma(\jpsi)=91.0\pm3.2\,{\rm keV}$\cite{Eidelman:2004wy}.}
\label{tab:exdec-jpsi}
\renewcommand{\arraystretch}{1.2}
  \begin{center}
  \begin{tabular}{|c||c|c|c|c|c|c|} \hline
    \rule{0cm}{8mm} channel & $p\overline p\;\;$ &
    $\Sigma^0\overline{\Sigma}{}^0\;\;$ & $\Lambda\overline \Lambda\;\;$ &
    $\Xi^-\overline{\Xi}{}^+\;\;$ & $\Delta^{++}\overline{\Delta}{}^{--}\;\;$ &
    $\Sigma^{*-}\overline{\Sigma}{}^{*+}$ \\ \hline\hline
    ${\cal B}_{3g}(\jpsi)$  & 1.91 & 1.24 & 1.29 & 0.69 & 0.72 & 0.45\\ \hline
    ${\cal B}_{\rm exp}$
\cite{Eidelman:2004wy}
& $2.16\pm 0.08$
            & $1.27\pm 0.17$ &  $1.30\pm 0.12$ & $0.90\pm 0.20$ &
                              $1.10\pm 0.29$ & $1.03\pm 0.13$ \\ \hline
 ${\cal B}_{3g}(\psi(2S))$  & 2.50 & 1.79 & 1.79 & 1.11 & 1.07 & 0.80\\ \hline
    ${\cal B}_{\rm exp}$
\cite{Eidelman:2004wy}
 & $2.07\pm 0.31$
            & $1.2\pm 0.6$ &  $1.81\pm 0.34$ & $0.94\pm 0.31$ &
                                   $1.28\pm 0.35$ & $1.10\pm 0.40$ \\ \hline
  \end{tabular}
  \end{center}
\renewcommand{\arraystretch}{1.0}
\end{table}

The widths for the corresponding decays of the $\psi(2S)$ are easily
obtained within the perturbative approach by rescaling the $\jpsi$
ones by the ratio of the electronic $\psi(2S)$ and  $\jpsi$ decay
widths, the $15\%$ rule, \ie \Eq~(\ref{eq:exdec-kappa12}) with 
$\kappa[B\overline{B}]=1$, holds strictly in the approach put forward 
in \cite{exdec:bolz97}. The results obtained that way are also quoted 
in \Table~\ref{tab:exdec-jpsi}. Good agreement between theory and
experiment \cite{Eidelman:2004wy} is observed. Predictions of the
absolute value of a decay width are subject to many uncertainties, see 
(\ref{eq:exdec-ppasy}) while ratios of any two $B\overline{B}$ decay widths 
are robust since most of the uncertainties cancel to a large extent.  
It is to be emphasized that the  $\psi(2S)$ and  $\jpsi$ decay widths
do not scale as $(M_{\jpsi}/M_{\psi(2S)})^8 \simeq 1/4$ as suggested
in \cite{exdec:bro81} since the subprocess amplitude in a calculation 
to lowest order in the charm quark velocity (see (\ref{eq:exdec-ampBB})) 
has to be calculated with $2\mc$ and not with the bound state mass. 

Bottomonium decays into $B\overline{B}$ pairs can be calculated along
the same lines. The hard scale is now provided by the $b$-quark mass.
Hence, relativistic and higher-twist corrections are expected to be
smaller than in the charmonium case. But, as it turns out, the
predicted decay widths for the baryonic channels are very small.
Approximately, \ie ignoring the fact that the $k_\perp$-dependent
suppression of the three-gluon contribution is perhaps a bit different
in the two cases, one finds the following rescaling formula
\begin{eqnarray}
\Gamma(\Upsilon\to B\overline{B}) & \= & 
\frac{\varrho[\Upsilon B \overline{B}]}{\varrho[\jpsi B \overline{B}]}\, 
\frac{\Gamma(\Upsilon\to e^+e^-)}{\Gamma(\jpsi\to e^+e^-)} \nn\\
         &\times& \left(\frac{e_c}{e_b}\right)^2\,
         \left(\frac{\as(\mb)}{\as(\mc)}\right)^6\; 
         \left(\frac{\mb}{\mc}\right)^8\; \Gamma(\jpsi\to
         B\overline{B})\,.
\end{eqnarray}
Using $\mb=4.5\,{\rm GeV}$ one obtains, for instance, a value of
$0.02\,{\rm eV}$ for the $\Upsilon\to \ppbar$ decay width, which value
corresponds to a branching ratio of $0.3 \times 10^{-7}$ well below the
present experimental upper bound \cite{Eidelman:2004wy}. 

It goes without saying that the hard contributions, $\Gamma_{3g}$, to
the $\jpsi$ and $\psi(2S)$ decays into $B\overline{B}$ pairs respect
the helicity sum rule (\ref{eq:exdec-hsr}), \ie the amplitude for the
production of baryon and antibaryon in equal helicities states
vanishes. Measurements of the angular distribution in $e^+ e^-\to
\jpsi,\,\psi(2S)\,\to B_8\overline{B}_8$
\begin{equation}
\frac{d \Gamma}{d\cos{\vartheta}} \propto 1+ \alpha_{B_8}\, \cos^2{\vartheta}\,,
\label{eq:exdec-angle}
\end{equation}
where $B_8$ is any member of the lowest-lying baryon octet and
$\vartheta$ the c.m.s.\ production angle, allow for a test of this
prediction. In the formal limit of an infinitely heavy charm quark
$\alpha_{B_8}=1$ as a consequence of hadronic helicity conservation
\cite{exdec:bro81}. The available data
\cite{exdec:BES-Jppbar,exdec:DM2,exdec:MRK2-bar,%
      exdec:E760-psiw,exdec:E835-Conf}, listed in
\Table~\ref{tab:exdec-alphas}, tell us that only a fraction of about
$10\%$ of the total number of $B_8\overline{B}_8$ pairs are produced
with the same helicity of baryon and antibaryon. This observation is
in fair agreement with hadronic helicity conservation. The production
of $B_8\overline{B}_8$ pairs with equal helicities has been modeled as
a constituent quark \cite{exdec:car87,exdec:mur95} and/or hadron mass
effect \cite{exdec:clau82}, both the effects are part of the ${\cal
O}(v^2)$ and higher-twist/power corrections. Also electromagnetic
effects in $\alpha_B$ have been investigated. For results we refer to
\Table~\ref{tab:exdec-alphas}.
\begin{table}
\caption[Experimental and theoretical results for the parameter
         $\alpha_{B_8}$ in $\jpsi, \psi(2S)\to B_8\overline{B}_8$]
        {Experimental and theoretical results for the parameter
         $\alpha_{B_8}$ in $\jpsi, \psi(2S)\to B_8\overline{B}_8$ as
         defined in \Eq~(\ref{eq:exdec-angle}). Experimental values
         obtained averaging data from BES \cite{exdec:BES-Jppbar}, DM2
         \cite{exdec:DM2}, MARK~II \cite{exdec:MRK2-bar}, E760
         \cite{exdec:E760-psiw} and E835 \cite{exdec:E835-Conf}.}
\label{tab:exdec-alphas}
\renewcommand{\arraystretch}{1.2}
  \begin{center}
  \begin{tabular}{|l||c|c|c|} \hline
    \rule{0cm}{8mm} $\alpha_{B_8}(J/\psi)$ & $p\overline p\;\;$ &
 $\Lambda\overline \Lambda\;\;$ & 
    $\Sigma^0\overline{\Sigma}{}^0\;\;$ 
 \\ \hline\hline
   Predicted:   \cite{exdec:clau82}& 0.46 & 0.32 & 0.31 \\   
   \phantom{Predicted:}   \cite{exdec:car87} (no e.m. corr)& 0.69 & 0.51 & 0.43 \\   
   \phantom{Predicted:}   \cite{exdec:car87} (incl. e.m. corr)& 0.70 &  &  \\\hline  
   Experiment: $J/\psi$  & $0.66 \pm 0.05$ & $0.65 \pm 0.19$ & $0.26 \pm 0.30$\\\hline   
   \phantom{Experiment:} $\psi(2S)$  & $0.68 \pm 0.14$ &  &  \\ \hline   
  \end{tabular}
  \end{center}
\renewcommand{\arraystretch}{1.0}   
\end{table}

\subsection[Hadronic two-body decays of the $\eta_c$]{Hadronic two-body decays of the $\eta_c$}
Such decays of the $\eta_c$ have been observed in experiment only for
the $B\overline{B}$ and $VV$ channels, upper bounds exist for a few
others like $a_0(980)\pi$. Decays into $PP$ and $PV$ have not been
observed, they are either strictly forbidden or strongly suppressed,
see \Table~\ref{tab:exdec-dec}. As noted at the beginning of this
section the $B\overline{B}$ and $VV$ channels are forbidden to
leading-twist accuracy since hadronic helicity conservation
(\ref{eq:exdec-hsr}) is in conflict with angular momentum conservation
for these processes. In contrast to the expectation from the
leading-twist approximation the measured branching ratios are rather
large ($10^{-3}$--$10^{-2}$). We repeat, it is worthwhile to explore
the role of higher-twist baryon and vector meson wave functions in the
decays of the $\eta_c$ \cite{exdec:braun90,exdec:braun}.

\begin{figure}[t,h,b]
\begin{center}
\includegraphics[width=6.5cm,bb=125 575 383 694,clip=]{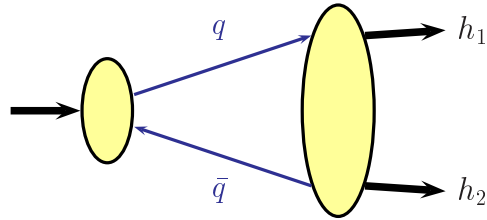}    
\caption{The mixing mechanism for charmonium decays into light hadrons.}  
\label{fig:exdec-mix} 
\end{center}
\end{figure}

In \cite{exdec:impli} a mixing approach for the explanation of these 
$\eta_c$ decays has been advocated. As is well-known the
$U_{\rm A}(1)$ anomaly leads to mixing among the pseudoscalar mesons
$\eta - \eta^\prime -\eta_c$ \cite{exdec:fri77,Chao:1989yp}. This mixing can 
adequately be treated in the quark-flavour mixing scheme \cite{exdec:fel98}
where one starts from the quark-flavour basis and assumes that
the basis states and their decay constants follow the same pattern of 
mixing with common mixing angles. This assumption is supported by an 
analysis of the $\gamma - \eta$ and $\gamma-\eta^\prime$ transition form 
factors at large momentum transfer \cite{exdec:fel97}. The quark-flavour 
basis states are defined by the flavour content of their valence Fock
states
\begin{equation}
\eta_q\, \to\, (u\bar{u} + d\bar{d})/\sqrt{2}\,, \qquad
\eta_s\, \to\, s\bar{s}\,, \qquad  \eta_{c0}\, \to\, \ccbar\,.
\end{equation}
The admixture of the light quarks to the $\eta_c$, which we need here
in this work,
is controlled by a mixing angle $\theta_c$ \cite{exdec:fel98}
\begin{equation}
|\eta_c\rangle \= |\eta_{c0}\rangle\, - \,\frac{\theta_c}{\sqrt{1+y^2/2}}\; \left[\,
  |\eta_q\rangle + \frac{y}{\sqrt{2}}\, |\eta_s\rangle\, \right]\,.
\label{eq:exdec-eta-etac}
\end{equation}
The ratio of the basis decay constants $f_q$ and $f_s$ is denoted by $y$ 
\begin{equation}
 y \= f_q/f_s\,.
\label{eq:exdec-ymix}
\end{equation} 
According to \cite{exdec:fel98}, its value amounts to 0.81 while
$\theta_c=-1^\circ \pm 0.1^\circ$. The light-quark admixture to the
$\eta_c$ (\ref{eq:exdec-eta-etac}) is somewhat smaller than estimates
given in \cite{exdec:fri77} but slightly larger than quoted in
\cite{exdec:chao97}. In combination with the strong vertex $\qqb\to
VV$ this small light-quark component of the $\eta_c$ suffices to
account for the $VV$ decays. In the spirit of this dynamical mechanism
(see \Figure~\ref{fig:exdec-mix}) the invariant amplitude, $A$, for
the $\eta_c\to VV$ decays can be parameterized as
\begin{equation}
A(\eta_c \to VV) \= C^{\rm mix}_{VV}\; \sigma_{VV}\; F_{\rm mix}(s=M^2_{\eta_c})\,.
\end{equation}
It is related to the decay width by
\begin{equation}
\Gamma(\eta_c \to VV) \= \frac1{32\pi S_{VV}}\,
\frac{\varrho[\eta_c V V]^3}{M_{\eta_c}}\, \left|A(\eta_c \to VV)\right|^2\,.
\label{eq:exdec-VVwidth}
\end{equation} 
The statistical factor for the decay into a pair of identical
particles is denoted by $S_{VV}$. The mixing factor
$C^{\rm mix}_{VV}$ embodies the mixing of the $\eta_c$ with the basis 
states $\eta_q$ and $\eta_s$ (\ref{eq:exdec-eta-etac}). These factors are
quoted in \Table~\ref{tab:exdec-VV}. Flavour symmetry breaking effects
in the transitions $\eta_{\,i}\to VV$ ($i=q,s$) are absorbed in the
factor $\sigma_{VV}$. As a simple model for it one may take the square 
of the vector meson's decay constants as a representative of SU(3)
violations in these transitions ($f_\rho=216\, {\rm MeV}$,
$f_\omega=195\, {\rm MeV}$, $f_\phi=237\, {\rm MeV}$, 
$f_{K^*}=214\, {\rm MeV}$). In order to have a dimensionless
quantity, $f_V^2$ is scaled by the squared vector meson mass 
\begin{equation}
    \sigma_{VV} \= \left(\frac{f_V}{M_V}\right)^2\,.
\end{equation}
Ratios of decay widths are free of the unknown transition form
factor $F_{\rm mix}$. With respect to the $\rho^0\rho^0$ channel one finds
for the other uncharged vector mesons channels
\begin{equation}
\frac{\Gamma(\eta_c\to V^0V^0)}{\Gamma(\eta_c\to\rho^0\rho^0)} \= 
      \frac{2}{S_{V^0V^0}}\; \left(C^{\rm mix}_{V^0V^0}\right)^2 \; 
      \left(\frac{\sigma_{V^0V^0}}{\sigma_{\rho^0\rho^0}}\right)^2\;
      \left(\frac{\varrho[\eta_c V^0V^0]} 
      {\varrho[\eta_c \rho^0 \rho^0]} \right)^3\,.
\end{equation} 
The theoretical and experimental results on the ratios are listed in
\Table~\ref{tab:exdec-VV}. Reasonable agreement between theory and
experiment can be seen although the errors are large.  Assuming a
monopole behaviour for the transition form factor $F_{\rm mix}$ and
fitting its strength to the $\rho\rho$ data, one obtains a value that
is in accord with the concept of mixing.

\begin{table}
\caption[Mixing factors and experimental and theoretical ratios 
         of decay widths for ${\eta_{c}} \to V^0V^0$]
        {Mixing factors as well as experimental and theoretical ratios
         of decay widths for ${\eta_{c}} \to V^0V^0$. The ratios are
         quoted with respect to the $\rho^0\rho^0$ channel ($C^{\rm
         mix}_{\rho^0\rho^0}=1$).  Experimental ratios are calculated
         taking into account the common systematics.}
\label{tab:exdec-VV}
\renewcommand{\arraystretch}{1.2}   
\begin{center}
\begin{tabular}{|c  || c || c |  c|}
\hline
$VV$ & $ C_{VV}^{\rm mix}$ & 
         $R_{\rm th}$ & $R_{\rm exp}$ \\
\hline
\hline
$\omega\omega $ & $1$ & $0.63$  & $< 0.37$ \cite{exdec:MRK3a} \\
& & & $<0.75$ \cite{exdec:DM2-91}\\
\hline
$K^{*0} {\overline {K}}^{*0}$  
  & $ (1+y^2)/2$ & $0.61$ & $0.47\pm 0.09$ \cite{exdec:DM2-91}\\
 & & &  $0.55 \pm 0.27$ \cite{exdec:MRK3a}\\
$\phi\phi $ 
  & $ y^2$ & $0.13$ & $0.93 \pm 0.33$ \cite{exdec:MRK3a} \\
 & & &  $0.35 \pm 0.10$ \cite{exdec:DM2-91}\\
 & & &  $0.30\pm 0.10$ \cite{exdec:BES-03}\\
 & & &  $0.21 \pm 0.14$ \cite{exdec:BELLE-03}
\\ \hline 
\end{tabular}
\end{center}
\renewcommand{\arraystretch}{1.0}   
\end{table}

The mixing approach can also be applied to the $\eta_c$ decays into
baryon--antibaryon pairs. It seems that at least the
$p\bar{p}$ channel for which the decay width has been measured, is
also controlled by the mixing mechanism \cite{exdec:impli}.

\subsection[The decays of the $\chi_{cJ}$ and the role of the colour-octet contribution]{The 
decays of the $\chi_{cJ}$ and the role of the colour-octet contribution} 
\label{sec:exdec-octet}
The colour-singlet contribution to $\chi_{cJ}$ decays into pairs of
pseudoscalar or vector mesons is well-known, it has been calculated
several times \cite{exdec:dun80,exdec:che82,exdec:CZ}. The
convolution of wave functions and hard subprocess amplitudes, which are 
to be calculated from Feynman graphs as shown in \Figure~\ref{fig:exdec-graphs}, 
leads to a decay width for the $\pi^+\pi^-$ channel as ($J=0,2$)
\begin{eqnarray}
\Gamma(\chi_{cJ}\to \pi^+\pi^-)& \= & 2\, \frac{\pi^2}{3^5}\, 
    \frac{\varrho[\chi_{cJ} \, \pi\pi]}{M_{\chi_{cJ}}}\, \frac{f_\pi^4}{m_c^7}\,
                |R^{\,\prime}_{\chi_{cJ}}(0)|^2\, \as^4(\mc) \nn\\
&\times& \left|a_J + b_J\, B_2^\pi(\mc) + c_J\, B_2^\pi(\mc)^2\right|^2\,,
\label{eq:exdec-PPwidth-singlet}
\end{eqnarray}
where the parameters $a_J$, $b_J$ and $c_J$ are analytically
calculable real numbers in the leading-twist approximation; they
represent the convolution of distribution amplitudes an subprocess
amplitude. The parameter $a_0$, for instance, reads
\begin{equation}
   a_0 \= 27 \pi^2/2 - 36\,.
\end{equation}
The representation (\ref{eq:exdec-PPwidth-singlet}) also holds in the
modified perturbative approach but the parameters are then complex valued.  
The constant $B_2^\pi(\mu_0)$ is the first coefficient of the
expansion of the leading-twist pion distribution amplitude upon
Gegenbauer polynomials $C_n^{3/2}$ \cite{exdec:bro80}
\begin{equation}
\Phi_\pi \= \Phi^M_{\rm AS} \left[\,1\,+ \, \sum_{n=2,4,\cdots} \,B_n^\pi(\mu_F)\,
        C^{3/2}_n(2x-1) \,\right]\,,
\label{eq:exdec-Pda}
\end{equation}
where $ \Phi^M_{\rm AS}$ is the asymptotic form of a meson
distribution amplitude
\begin{equation}
\Phi^M_{\rm AS} \= 6 x (1-x)\,,
\end{equation}
and 
\begin{equation}
B_n(\mu_F) \= \left(\frac{\ln(\mu_F^2/\Lambda_{\rm QCD}^2)} 
{\ln(\mu_0^2/\Lambda_{\rm QCD}^2)}\right)^{\gamma_n} \, B_n(\mu_0)\,.
\end{equation}
In \Eq~(\ref{eq:exdec-PPwidth-singlet}) terms of order higher than 2 in the
expansion are neglected and the factorization scale dependence of the
Gegenbauer coefficient $B_2$ is controlled by $\gamma_2=-50/81$. As the
starting scale of the evolution, $\mu_0$, a value of $1\, {\rm GeV}$ is taken. 
Finally, $f_\pi$ ($=132\, {\rm MeV}$) is the pion decay constant and 
$R^{\,\prime}_{\chi_{cJ}}(0)$ ($=0.22\, {\rm GeV}^{5/2}$ \cite{exdec:man95, exdec:buch81}) 
denotes the derivative of the nonrelativistic radial $\ccbar$ wave
functions at the origin (in coordinate space). As usual a
normalization factor $f_\pi/(2\sqrt{6})$ is pulled out from the
distribution amplitude.   

The distribution amplitude of the pion is fairly well-known by now 
from analyses of the $\pi^0-\gamma$ transition form factor. It is
close to the asymptotic form of a meson distribution amplitude 
\cite{exdec:rau}. Deviations from that form are difficult to estimate
since they 
strongly depend on details of the analysis such as whether or not NLO,  
higher-twist corrections or tranverse degrees of freedom are taken
into account \cite{exdec:rau,exdec:DKV1}. But in any case the Gegenbauer
coefficient $B_2^{\pi}$ seems to be small in magnitude. Combining the
results from different analyses of the $\pi^0-\gamma$ transition form
factor, one may conclude that $|B_2^{\pi}| \lsim 0.1$ at $\mu_0 \= 1\, {\rm GeV}$. 
Taking first $B_2^{\pi}=0$ in (\ref{eq:exdec-PPwidth-singlet}), 
one evaluates from (\ref{eq:exdec-PPwidth-singlet}) the branching ratio
\begin{equation}
{\cal B}(\chi_{c0(2)}\to\pi^+\pi^-) \simeq 0.31\, (0.10) \times 10^{-3}\,
       \left(\frac{\as}{0.4}\right)^2\; \left(\frac{1.5\, {\rm GeV}}{\mc}\right)^3\,.
\label{eq:exdec-cpp}
\end{equation}
The majority of the widths of the $\chi_{c0}$ and $\chi_{c2}$ come
from decays into light hadrons. The contribution coming from the decay 
of a colour-singlet $\ccbar$ into real gluons is given by \cite{exdec:man95} 
\begin{equation}
\Gamma(\chi_{cJ}\to l.h.)\; \propto \; |R^\prime_{\chi_{cJ}}(0)|^2 \,
  \frac{\as^2}{\mc^4}\,.
\end{equation}
Therefore, the branching ratios approximately scale as given in
(\ref{eq:exdec-cpp}) and not as in (\ref{eq:exdec-PPwidth-singlet}). 
The $\ccbar$ wave function $R^\prime_{\chi_{cJ}}(0)$ 
almost cancel in the ratio. Otherwise its well-known scaling properties
\cite{Quigg:1979vr} would have to be taken into account as well.  

\begin{figure}[ht]
\centerline{\includegraphics[width=90mm,bbllx=35pt,bblly=513pt,bburx=400pt,
    bbury=798pt,clip=]{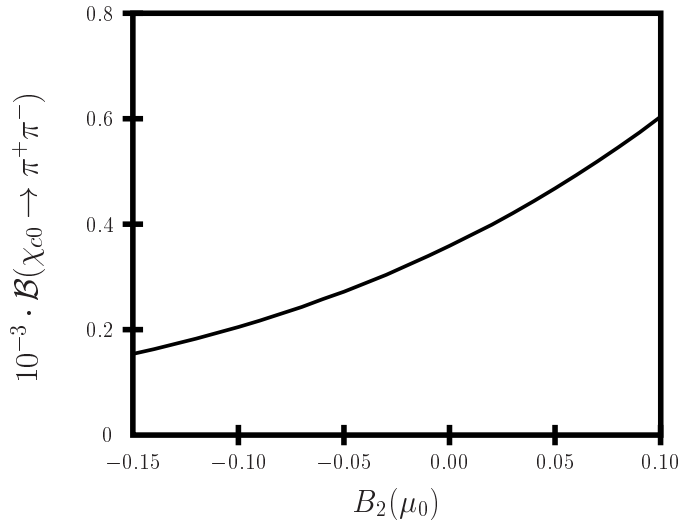}}
\caption[Dependence of the leading-twist
         colour-singlet contribution]
        {Dependence of the leading-twist colour-singlet contribution to
         the $\chi_{c0}\to \pi^+\pi^-$ branching ratio on the
         expansion parameter $B_2^\pi$ of the pion distribution
         amplitude at the scale $\mu_0=1\, {\rm GeV}$. The evolution
         of $B_2^\pi$ is evaluated from $\Lambda_{\rm QCD}=200\, {\rm
         MeV}$.}
\label{fig:exdec-col8}
\end{figure}

The variation of the branching ratio with the
Gegenbauer coefficient $B_2^{\pi}$ is displayed in \Figure~\ref{fig:exdec-col8}. 
One can conclude that, stretching all parameters
($B_2^\pi$, $\as$, $\mc$) to the extreme, the predictions for 
${\cal B}(\chi_{c0(2)}$ $\to$ $\pi^+\pi^-)$ from the colour-singlet
contribution to leading-twist accuracy stay a factor 3--6 below the
data. Results of similar magnitude are found within the modified
perturbative approach.

Thus, there is obviously room for the colour-octet contributions (see
(\ref{eq:exdec-ampPP})), \ie from the subprocess $\ccbar g \to 2
(\qqb)$. A first attempt to include the colour-octet contribution has
been undertaken in \cite{exdec:BKS}. This calculation, performed
within the modified perturbative approach \cite{exdec:BLS}, is based
on a very rough model for the colour-octet $\chi_{cJ}$ wave function,
the new ingredient of this calculation. Despite this the authors of
Ref.~\cite{exdec:BKS} were able to show that the combined
colour-singlet and -octet contributions are likely large enough to
account for the data \cite{Eidelman:2004wy,Bai:1998cw}, see
\Table~\ref{tab:exdec-chi}. The calculation of the $\chi_{cJ}\to
\pi^+\pi^-$ decay width can be generalized to other pseudoscalar meson
channels with results of similar quality as for the $\pi\pi$
channels. For the $\eta^\prime\eta^\prime$ channel an additional
two-gluon Fock component of the $\eta^\prime$ is to be taken into
account whose leading-twist distribution amplitude has recently been
extracted from a NLO analysis of the $\eta -\gamma$ and $\eta^\prime
-\gamma$ transition form factor \cite{exdec:passek,exdec:ali}. For the
$\eta\eta$ channel the two-gluon contribution is probably negligible.

\begin{table}[tbh]
\caption[Comparison of theoretical and experimental branching ratios
         for various $\chi_{cJ}$ decays into pairs of light hadrons]
        {Comparison of theoretical and experimental branching ratios
         for various $\chi_{cJ}$ decays into pairs of light
         hadrons. The theoretical values have been computed within the
         modified perturbative approach, colour-singlet and -octet
         contributions are taken into account
         ($B_2^\pi=B_2^\eta=B_1^K=0$, $B_2^K=-0.1$, baryon wave
         functions (\ref{eq:exdec-pda}), (\ref{eq:exdec-gaussian})). The
         branching ratios are quoted in units of $10^{-3}$ for the
         mesonic channels and $10^{-5}$ for the baryonic ones. Data
         taken from \cite{Eidelman:2004wy}. The values listed for
         $\ppbar$ branching rates do not include the most recent
         values $\left(27.4^{+4.2}_{-4.0}\pm 4.5\right)\, \cdot \,
         10^{-5}$, $\left(5.7^{+1.7}_{-1.5}\pm 0.9\right)\, \cdot \,
         10^{-5}$ and $\left(6.9^{+2.5}_{-2.2}\pm 1.1\right)\, \cdot
         \, 10^{-5}$ measured by BES \cite{exdec:BES-ppbar} for
         $\chi_c0$, $\chi_{c1}$ and $\chi_{c2}$ respectively.}
\label{tab:exdec-chi}
\renewcommand{\arraystretch}{1.2}
\begin{center}
\begin{tabular}{|c  || c |c |}
\hline
process & theory & experiment \\
\hline
\hline
${\cal B}(\chi_{c0}\to\,\pi^+\,\pi^-\,)$ & $3.0\,$ \cite{exdec:BKS} &
                                                     $4.9\pm0.6$    \\ \hline
${\cal B}(\chi_{c2}\to\,\pi^+\,\pi^-\,)$ & $1.8\,$ \cite{exdec:BKS} &
                                                  $1.77\pm0.27$     \\ \hline
${\cal B}(\chi_{c0}\to K^+ K^-)$   & $2.4\,$ \cite{exdec:BKS} &
                                                        $6.0\pm0.9$ \\ \hline
${\cal B}(\chi_{c2}\to K^+ K^-)$   & $1.4\,$ \cite{exdec:BKS} &
                                                       $0.94\pm0.21$\\ \hline
${\cal B}(\chi_{c0}\to\,\eta\;\eta\,)$   & $2.0\,$ \cite{exdec:BKS} &
                                                         $2.1\pm1.1$\\ \hline
${\cal B}(\chi_{c2}\to\,\eta\;\eta\,)$   & $1.3\,$ \cite{exdec:BKS} &
                                                       $<1.5$ \\\hline \hline
${\cal B}(\chi_{c0}\to\, p\; \bar{p}\,)$ & $-$  & $22.4\pm 2.7$  \\\hline
${\cal B}(\chi_{c1}\to\, p\; \bar{p}\,)$ & $6.4\,$ \cite{exdec:wong} &
                                                         $7.2\pm1.3$\\ \hline
${\cal B}(\chi_{c2}\to\, p\; \bar{p}\,)$ & $7.7\,$ \cite{exdec:wong} &
                                                         $6.8\pm0.7$\\ \hline
${\cal B}(\chi_{c0}\to\, \Lambda\, \overline{\Lambda}\,)$ & $-$ &
                                             $47\pm 16$  \\ \hline
${\cal B}(\chi_{c1}\to\, \Lambda\, \overline{\Lambda}\,)$ & $3.8\,$ \cite{exdec:wong}
                                          & $26\pm12$ \\ \hline
${\cal B}(\chi_{c2}\to\, \Lambda\, \overline{\Lambda}\,)$ & $3.5\,$ \cite{exdec:wong}
                                          & $34\pm17$ \\ \hline
\end{tabular}
\end{center}
\renewcommand{\arraystretch}{1.0}
\end{table}

The colour-singlet contribution to the decays $\chi_{cJ}\to p\bar{p}$
($J=1,2$) has been investigated by many authors
\cite{exdec:che82,exdec:CZ,exdec:ste94,exdec:dam85}. Employing the
proton distribution amplitude (\ref{eq:exdec-pda}) or a similar one,
one again finds results that are clearly below experiment, which again
signals the lack of the colour-octet contributions. An analysis of the
$\chi_{c1(2)}$ decays into the octet and decuplet baryons along the
same lines as for the pseudoscalar meson channels \cite{exdec:BKS} has
been carried through by Wong \cite{exdec:wong}. The branching ratios
have been evaluated from the baryon wave functions
(\ref{eq:exdec-pda}), (\ref{eq:exdec-gaussian}) and the same
colour-octet $\chi_{cJ}$ wave function as in \cite{exdec:BKS}. Some of
the results obtained in \cite{exdec:wong} are shown and compared to
experiment in \Table~\ref{tab:exdec-chi}. As can be seen from the
table the results for the $p\bar{p}$ channels are in excellent
agreement with experiment while the branching ratios for
$\Lambda\overline{\Lambda}$ channels are much smaller than experiment
\cite{exdec:BES-LLbar} although the errors are large. A peculiar fact
has to be noted: the experimental $\Lambda\overline{\Lambda}$
branching ratios are larger than the proton--antiproton ones although
there is agreement within two standard deviations.

The present analyses of the $\chi_{cJ}$ decays suffer from the rough
treatment of the colour-octet charmonium wave function. As we mentioned
before a reanalysis of the decays into the $PP$ and $B\overline{B}$
channels as well as an extension to the $VV$ ones is required. Our
knowledge of the colour-octet wave function has been improved recently
due to the intense analyses of inclusive processes involving
charmonia, \eg \cite{exdec:inc}.  This new information may be used
to ameliorate the analysis of the $\chi_{cJ}\to PP, B\overline{B}$
decays and, perhaps, to reach a satisfactory quantitative
understanding of these processes.  We finally want to remark that the
colour-octet contribution does not only play an important role in the
$\chi_{cJ}$ decays into $PP$ and $B\overline{B}$ pairs but potentially
also in their two-photon decays
\cite{exdec:man95,exdec:ma02,Ma:2002ev} (see also
\Section~\ref{sec:ID}).

The leading-twist forbidden $\chi_{c0}\to B\overline{B}$ decays have
sizeable experimental branching ratios, see
\Table~\ref{tab:exdec-chi}.  There is no reliable theoretical
interpretation of these decays as yet. The only proposition
\cite{exdec:anselmino} is the use of a diquark model, a variant of the
leading-twist approach in which baryons are viewed as being composed
of quarks and quasi-elementary diquarks. With vector diquarks as
constituents one may overcome the helicity sum rule
(\ref{eq:exdec-hsr}). The diquark model in its present form, however,
contends with difficulties. Large momentum transfer data on the Pauli
form factor of the proton as well as a helicity correlation parameter
for Compton scattering off protons are in severe conflict with
predictions from the diquark model.

\subsection[Radiative decays of charmonia into light hadrons]{Radiative 
decays of charmonia into light hadrons}
\label{sec:exdec-radiative}

First let us consider the process $\jpsi\to\gamma\pi^0$. The apparently
leading contribution to it is generated by the subprocess $\ccbar\to\gamma\,
g^*g^*\to \gamma q\bar{q}$, which, in principle, leads to a decay width
of order $\as^4$. However, due to the pion's flavour content
$\propto u\bar{u} - d\bar{d}$ this contribution exactly
cancels to zero in the limit of massless quarks. A VDM contribution
$\jpsi\to \rho\pi$ followed by a $\rho - \gamma$ conversion
\cite{exdec:CZ} seems to dominate this process. Indeed, an estimate of
the VDM contribution leads to a branching ratio of $3.3 \times 10^{-5}$, 
which compares favorably with the experimental result of $(3.9 \pm 1.3) \times 10^{-5}$ 
\cite{Eidelman:2004wy}. Analogue estimates of
the $\gamma\eta$ and $\gamma\eta^\prime$ branching ratios lead to similar
values, about $1 \times 10^{-5}$, which fall short of the experimental
results by two orders of magnitude. The solution of this discrepancy
is a gluonic contribution, which occurs as a consequence of the U$_{\rm
  A}$(1) anomaly; it formally presents a power correction. According
to Novikov et al.\ \cite{exdec:nov80}, the photon is emitted by the
$c$ quark with a subsequent annihilation of the $\ccbar$ pair into
lighter quarks through the effect of the anomaly. The creation of the 
corresponding light quarks is controlled by the gluonic matrix element $\langle
0|\as G\widetilde{G}\,|\eta^{(\prime)}\rangle$ where $G$ is the gluon field
strength tensor and $\widetilde{G}$ its dual. Photon emission from the
light quarks is negligible as can be seen from the smallness of the
$\gamma\pi$ width. This mechanism leads to the following width for the radiative
$\jpsi$ decay into $\eta^{(\prime)}$ \cite{exdec:nov80} 
\begin{equation}
\Gamma(\jpsi\to\gamma\eta^{(\prime)}) \= \frac{2^5}{5^2 3^{8}}\; \pi
               e_c^2 \aem^3 \,
           \varrho[\jpsi \gamma \eta^{(\prime)}]\,      
           \left(\frac{M_\jpsi}{\mc^2}\right)^4\,
        \frac{|\langle \,0\,|\frac{\as}{4\pi}\, G\widetilde{G}\,|\, 
              \eta^{(\prime)}\,\rangle|^2}{\Gamma(\jpsi\to e^+e^-)}\,.
\label{eq:exdec-G-rad}
\end{equation}
In the quark-flavour mixing scheme the gluonic matrix element for the
$\eta$ is given by \cite{exdec:uppsala}
\begin{equation}
\langle \,0\,|\frac{\as}{4\pi}\, G\widetilde{G}|\,\eta\,\rangle \= 
                   -\sin{\theta_8}\, \sqrt{2+y^2}\, f_q\, a^2\,.
\label{eq:exdec-gluonic}
\end{equation}
For the $\eta^\prime$ matrix element $\sin{\theta_8}$ is to be replaced
by $\cos{\theta_8}$. The angle $\theta_8$ controls the mixing
of the octet decay constants. In \cite{exdec:fel98} the various mixing
parameters have been determined; their values amount to: 
\begin{equation}
\theta_8 \= -21.2^\circ\,; \quad f_q \= 1.07 f_\pi\,; \quad a^2 \=
0.265\, {\rm GeV}^2\,; \quad \phi \= 39.3^\circ\,. 
\end{equation}
The latter angle is the mixing angle in the quark-flavour basis. 
The parameter $y$ has been defined in \Eq~(\ref{eq:exdec-ymix}).
Evaluating the decay width or rather the branching ratio from these
parameter values, one obtains
\begin{equation}
{\cal B}(\jpsi\to\gamma\eta) \= 3.7 \times 10^{-4}\, \left(\frac{1.5\,
  {\rm GeV}}{\mc}\right)^7 \,.
\label{eq:exdec-ge}
\end{equation}
The comparison with the experimental value of $(8.6\pm 0.8) \cdot 10^{-4}$ 
\cite{Eidelman:2004wy} reveals that the order of magnitude is
correctly predicted. As happens frequently in exclusive charmonium
decays the charm-quark mass appears to a high power in the theoretical
estimates of branching ratios with the consequence of large
uncertainties in the predicted values. With regard to the fact that
the total $\jpsi$ decay width is dominated by the decays into light
hadrons (\ref{eq:exdec-tot}), the power of $\mc$ in (\ref{eq:exdec-ge}) is
approximately seven. The mass of the $\jpsi$ appears 
in (\ref{eq:exdec-G-rad}) through a pole saturation of a 
QCD sum rule \cite{exdec:nov80}; it should not be replaced by $2\mc$. 

While the calculation of the individual decay widths is not easy, 
ratios of the $\eta$ and $\eta^\prime$ widths can be reliably 
predicted from $\eta-\eta^\prime$ mixing. 
Using the quark-flavour mixing scheme again, one finds from
(\ref{eq:exdec-G-rad}) and (\ref{eq:exdec-gluonic}) the following
ratios for radiative $\jpsi$ decays \cite{exdec:fel98}
\begin{equation}
\frac{{\cal B}(\jpsi\to\gamma\eta^\prime)}{{\cal B}(\jpsi\to\gamma\eta)} \=
                  \cot^2{\theta_8}\;
                \left (\frac{\varrho[\jpsi \gamma\eta^\prime]}
                          {\varrho[\jpsi \gamma\eta]}\right)^3\,. 
\label{eq:exdec-eepratio}
\end{equation}
The extension to the $\eta_c$ is also
possible. With (\ref{eq:exdec-eta-etac}) one obtains
\begin{equation}
\frac{{\cal B}(\jpsi\to\gamma\eta^\prime)}{{\cal B}(\jpsi\to\gamma\eta_{c})}
            \= \theta_{c}^2\cos^2{\theta_8}\;
                \left(\frac{\varrho[\jpsi \gamma \eta^\prime]}
                  {\varrho[\jpsi \gamma\eta_c]} \right)^3\,.
\label{eq:exdec-rad}
\end{equation}
This approach leads to the following numerical results:
\begin{eqnarray}
\frac{{\cal B}(\jpsi\to\gamma\eta^\prime)}{{\cal B}(\jpsi\to\gamma\eta)} 
          & \= & 5.39\,, \quad\quad  {\rm Exp:}\;\; 5.0\;\pm 0.6\;
                                               \hbox{
\cite{Eidelman:2004wy}
}\,,\nn\\ 
\frac{{\cal B}(\jpsi\to\gamma\eta^\prime)}{{\cal B}(\jpsi\to\gamma\eta_{c})}
         & \= & 0.48\,, \quad \quad {\rm Exp:} \;\; 0.33 \pm 0.1\, \hbox{
\cite{Eidelman:2004wy}
}\,.
\end{eqnarray}
Due to the large uncertainties in the angle $\theta_c$ the prediction
for the second ratio has an error of about $20\%$.

It is tempting to extend the anomaly dominance to the case of the
radiative $\Upsilon$ decays. One obtains
\begin{equation}
\frac{{\cal B}(\Upsilon\to\gamma\eta^\prime)}{{\cal
    B}(\Upsilon\to\gamma\eta)} \= 6.51\,, \qquad 
\frac{{\cal B}(\Upsilon\to\gamma\eta^\prime)}{{\cal
    B}(\Upsilon\to\gamma\eta_c)} \= 3.5\times 10^{-4}\,.
\end{equation}
Comparison with experiment is not yet possible, only upper bounds
exist for the individual branching ratios. Doubts have, however, been
raised by Ma \cite{exdec:ma0202} on the validity of this approach for
the $\Upsilon$ decays. Generalizing the result for the $\jpsi$ case
(\ref{eq:exdec-G-rad}) appropriately, one finds a too large branching 
ratio, namely $\simeq 8.3 \times 10^{-5}$, as compared to the 
experimental upper limit of $\leq 1.6 \times 10^{-5}$ \cite{Eidelman:2004wy}. 
The estimate advocated for by Ma, is based on the assumption of
scale independence of the gluonic matrix element. With regard to the 
well separated scales $\mc$ and $\mb$ this assumption is suspicious. 
Nonetheless, the investigation of the $\Upsilon\to \gamma\eta^{(\prime)}$ 
decays is to be addressed further. Of interest would also be an
investigation of the radiative $h_c$ decays into pseudoscalar mesons. 
It is likely that these decays are under control of the same dynamical 
mechanism as the corresponding $\jpsi$ decays. Results analogue to
(\ref{eq:exdec-eepratio}), (\ref{eq:exdec-rad}) would then hold.  
Instead of the decays into pseudoscalar mesons one may also explore 
radiative quarkonium decays into scalar mesons. As is well-known
scalar mesons may have sizeable glue--glue Fock components
\cite{exdec:chase}, they may even be glueballs although they likely have
sizeable admixtures of light quarks \cite{exdec:ochs,exdec:close}. It 
would be interesting to unravel the dynamics mediating these decays. 
For first attempts see for instance \cite{exdec:close,exdec:ma02a}. 

The decays $\jpsi\to\rho\eta^{(\prime)}$ can be treated analogously to
the radiative decays. Since in these processes $G$-parity is not
conserved, they proceed through $\ccbar\to\gamma^*$. On account of the
flavour content of the $\rho$ meson, the
$\gamma^*\to\rho\eta^{(\prime)}$ transition only probes the $\eta_q$
component of the $\eta^{(\prime)}$ if OZI-suppressed contributions are
neglected. Hence,
\begin{equation}
\frac{{\cal B}(\jpsi\to\rho\eta^\prime)}{{\cal B}(\jpsi\to\rho\eta)} \=
            \tan^2{\phi}\,
\left(\frac{\varrho[\jpsi \rho\eta^\prime]}{\varrho[\jpsi
\rho\eta]}\right)^3\,,
\label{eq:exdec-rho}
\end{equation}
the $\rho-\eta_q$ form factor cancels in the ratio. \Eq[b]~(\ref{eq:exdec-rho}) 
leads to $0.52$ for the ratio of the decay widths
while the experimental value is $0.54\pm 0.21$ \cite{Eidelman:2004wy}.  

Finally, we want to mention the radiative $\jpsi$ decay into a
proton--antiproton pair. Recently, an enhancement near $2
M_p$ in the invariant mass spectrum of $\ppbar$ pairs has been observed
while  $\jpsi\to \pi^0 \ppbar$ behaves regular near the $\ppbar$
threshold \cite{exdec:BES03}. The combination of both the results
hints at a peculiar behaviour of the $\ppbar$ pair in an isospin-zero state.
The enhancement observed in  $\jpsi\to \gamma\ppbar$ parallels similar
anomalies near the $\ppbar$ threshold. They have been reported by
Belle \cite{exdec:BELLE02} for the decays $B^+\to K^+\ppbar$ and 
$\overline{B}^0\to D^0 \ppbar$. An anomalous threshold behaviour is
also seen in the proton's time-like form factor 
\cite{exdec:fenice}, in the charged pion spectrum from 
$\bar{p} d\to\pi^-\pi^0p$ and $\pi^+\pi^-n$ reactions \cite{exdec:bridges} 
and in the real part of the elastic proton--antiproton forward amplitude 
\cite{exdec:LEAR}. 

Frequently these anomalies have been associated with narrow $\ppbar$
bound states. Indeed, an analysis of the BES
provides evidence for an S-wave bound state with a mass of $1859
{\textstyle{{+3\phantom{0}}\atop{-10}}} ({\rm stat})
{\textstyle{{+5\phantom{0}}\atop{-25}}} ({\rm syst}) \; {\rm MeV}$
and a total width less than $30$~MeV \cite{exdec:BES03}.
A P-wave bound state instead of an S-wave one cannot be excluded
from the BES data. This BES result is very close to findings from an
analysis of $\bar{p} d$ reactions \cite{exdec:dal97} (a bound
state mass of $1870\, {\rm MeV}$ and a width of $10\,{\rm MeV}$) and
from a proton--antiproton forward dispersion relation \cite{exdec:schw89} 
(mass: $1852\, {\rm MeV}$, width: $35\, {\rm MeV}$). In the CERN WA56 
experiment \cite{exdec:ferrer}, on the other hand, a narrow peak 
(mass $2.02\, {\rm GeV}$) has been observed in the $\ppbar$
invariant mass spectrum of the reaction $\pi^- p\to p_f \pi^-
[\ppbar]$ where $p_f$ is a fast forward going proton. 
Puzzling is, however, the fact that this peak is not seen in $\jpsi\to
\gamma\ppbar$ \cite{exdec:BES03} while there is no indication of a 
threshold enhancement in the WA56 measurement. 
Several authors \cite{exdec:zou03} have pointed out that the dynamics of
the low-energy $\ppbar$ system such as pion exchange or the physics 
inherent in the effective range expansion, provides an important
contribution to the threshold enhancement. 
An appealing mechanism has been suggested by Rosner
\cite{exdec:rosner}. He assumes that the partonic subprocess in the
process $\jpsi\to\gamma\ppbar$ is
$\ccbar\to \gamma gg$ followed by a nonperturbative $gg\to (\ppbar)_S$
transition where the subscript indicates a $\ppbar$ pair in a resonant
S-state. Rosner further assumes that the corresponding $B$ decays,
for instance $B^+\to K^+\ppbar$, receives a substantial contribution
associated with the subprocess $\bar{b}\to \bar{s} gg$ and the same
nonperturbative $gg\to (\ppbar)_S$ transition as for
$\jpsi\to\gamma\ppbar$. Producing an $\eta^\prime$ through this
mechanism instead of the proton--antiproton pair leads to similar
contributions except that now a different gluonic matrix element 
occurs, see (\ref{eq:exdec-G-rad}). In ratios of these processes most
details cancel and, according to Rosner, one arrives at
\begin{equation}
\frac{{\cal B}(B^+\to K^+(\ppbar)_S)|_{gg}}
                  {{\cal B}(B^+\to K^+\eta^\prime)|_{gg}} \= 
\frac{\varrho[B^+ K^+ (\ppbar)_S]}{\varrho[B^+ K^+ \eta^\prime]} 
\left(\frac{\varrho[\jpsi \gamma\eta^\prime]}
                     {\varrho[\jpsi \gamma (\ppbar)_S]}\right)^3
     \frac{{\cal B}(\jpsi\to\gamma (\ppbar)_S)}
                   {{\cal B}(\jpsi\to\gamma\eta^\prime)} \,.
\end{equation}
The $gg$ subscript at the $B$-meson matrix elements is meant as a hint that
there might be other non-negligible contributions to the $B$ decays
than those from the subprocess $\bar{b}\to \bar{s} gg$. 
This mechanism relates the threshold enhancement in  $B^+\to K^+\ppbar$ 
to that in $\jpsi\to\gamma\ppbar$. Using the experimental information 
on the latter process, Rosner found that this mechanism provides a 
substantial fraction of the first one. It is to be stressed that the
ratio of $B^{+(0)}$ decays into $K^{+(0)} \eta^\prime$ and  $K^{+(0)} \eta$
are not in conflict with this interpretation.

%%%%%%%%%%%%%%  
%RT
%%%%%%%%%%%%%% 
\section[Electromagnetic transitions]
        {Electromagnetic transitions $\!$\footnote{Author: E.~Eichten}}
\label{emsec:intro}

For quarkonium states, $Q_1\bar{Q_2}$, above the ground state but
below threshold for strong decay into a pair of heavy flavoured mesons,
electromagnetic transitions are often significant decay modes.  In
fact, the first charmonium states not directly produced in $e^+e^-$
collisions, the $\chi_c^J$ states, were discovered in photonic
transitions of the $\psi^{\prime}$ resonance.  Even today, such
transitions continue to be used to observe new quarkonium states
\cite{Bonvicini:2004yj}.

\subsection[Theoretical framework]{Theoretical framework}

\subsubsection{Effective Lagrangian}
\label{sec:emssec-theory}

The theory of electromagnetic transitions between these quarkonium
states is straightforward.  Much of the terminology and techniques are
familiar from the study of EM transitions in atomic and nuclear
systems. The photon field ${\bf A}_{\rm em}^{\mu}$ couples to charged
quarks through the electromagnetic current:
\begin{equation} 
  j_{\mu} \equiv \sum_{i = {\rm u, d, s}} j^i_{\mu} + \sum_{i = {\rm c, b, t}} j^i_{\mu} \,.
\end{equation} 
The heavy valence quarks ($c,b,t$) can be described by the usual
effective action:
\begin{equation}
\label{eq:nrqcd}
{\cal L}_{\rm NRQCD} = \psi^\dagger \Biggl\{ i D_0 + \,\frac{\mathbf{D}^2}{2 m}
+ c_F\, g \frac{\mathbf{\bfsigma \cdot B}}{2 m}
+ c_D \, g \frac{\left[\mathbf{D} \cdot, \mathbf{E} \right]}{8 m^2}
+ i c_S \, g \frac{\mathbf{\bfsigma \cdot \left[D \times, E \right]}}{8 m^2}
+ \dots 
\Biggr\} \psi \,,
\end{equation}
where the ${\bf E}$ and ${\bf B}$ fields are the chromoelectric and
chromomagnetic fields.  Corrections to the leading NR behaviour are
determined by the expansion in the quark and antiquark velocities.
For photon momentum small compared to the heavy quark masses, the form
of the EM interaction (in Coulomb gauge) is determined in the same way
as the NRQCD action itself
\cite{Caswell:1985ui,Bodwin:1994jh,Thacker:1990bm,Brambilla:1999xf},
the leading order terms are:
\begin{equation} 
\label{eq:emcur}
  {\bf j}\cdot{\bf A}_{\rm em}  = e_Q \psi^\dagger \Biggl\{
     \frac{\{\mathbf{D} \cdot , \mathbf{A}_{\mathrm{em}} \}}{2 m} + 
       (1+\kappa_Q)\, \frac{\mathbf{\bfsigma \cdot B}_{\mathrm{em}}}{2 m} + 
   \dots \Biggr\} \psi \,.
\end{equation}

The first term of \Eq~(\ref{eq:emcur}) produces electric and the
second magnetic transitions.  The coefficient $\kappa_Q$ is a possible
anomalous magnetic moment for the heavy quark.  It is a perturbative
quantity at the level of NRQCD, but may get nonperturbative
contributions in going to lower energy effective field theories, once
the scale $\lQ$ has been integrated out. Since we may assume that
potential models are an attempt to mimic such theories, we will
interpret in this last way the quantity $\kappa_Q$ that appears there
and will be used in the following.

For quarkonium systems, light quarks ($u,d,s$) only contribute to
internal quark loops, described perturbatively at short distance and
as virtual pairs of heavy flavour mesons at large distance. In the
SU(3) limit the total contribution from light quarks vanishes since
its EM current has no SU(3) singlet part.  Hence, to leading order in
SU(3) breaking these contributions can be ignored. We return to these
corrections in Sec.~\ref{sec:em-vloops}.

\subsubsection{Transition amplitudes}
\label{sec:emssec-amps}

Within a $\bar{Q_2}Q_1$ quarkonium system, the electromagnetic
transition amplitude is determined by the matrix element of the EM
current, $\langle f|j^{\mu}_{\rm em}|i \rangle$, between an initial
quarkonium state, $i$, and a final state $f$.  Including the emission
of a photon of momentum $k$ and polarization $\epsilon_{\gamma}$, the
general form of the transition amplitude is the sum of two terms
\begin{equation}
\label{eq:tramp}
 {\cal M}(i \rightarrow  f) = 
[{\bf M}^{(1)}(i \rightarrow f) + {\bf M}^{(2)}(i \rightarrow f)] \cdot {\epsilon_{\gamma}(k)}, 
\end{equation}
where in the term ${\bf M}^{(1)}$ the photon is emitted off the 
quark $Q_1$ with mass  $m_1$ and charge $e_1$, 
\begin{equation}
\label{eq:amp1}
 {\bf M}^{(1)}(i \rightarrow f)  =  \frac{e_1}{2 m_1} 
      \int d^3x \langle i | Q_1^{\dagger}(x)({{\bf D}, \exp{(i{\bf x\cdot k})}} 
        + (1+\kappa_{Q_1}){\bf \bfsigma \times k} \exp{(i{\bf x\cdot k})} ) Q_1(x) | f  \rangle,   
\end{equation}
and in the corresponding term ${\bf M}^{(2)}$ the photon is emitted off 
the antiquark $\bar{Q_2}$ with mass $m_2$ and charge $-e_2$.

Electromagnetic transition amplitudes can be computed from first principles in
Lattice QCD\cite{LQCD}.  Preliminary studies\cite{Duncan:1996xy} have even 
included electromagnetic interactions directly into Lattice QCD simulations.
However, these transitions for quarkonium systems have not yet been computed. 
Various relations between transitions also arise from 
QCD sum rules\cite{Colangelo:2000dp}. 

Although other calculational models, \eg using the MIT bag model
\cite{Baacke:1981zx}, have been explored, only potential model
approaches provide the detailed predictions for the strength of
individual transition amplitudes needed to compare with experiments.
The remainder of this section will focus on the issues within
potential model approaches.

Within nonrelativistic (NR) potential models, a quarkonium state is
characterized by a radial quantum number, $n$, orbital angular
momentum, $l$, total spin, $s$, and total angular momentum, $J$. In
the NR limit the spin dependence decouples from the spatial
dependence. The spatial wave function for a NR state, $\psi(x)$, can
be expressed in terms of a radial wave function, $u_{nl}(r)$ and an
orbital angular momentum dependence by:
\begin{equation}
  \psi (x) = Y_{lm}(\theta, \phi) \frac{u_{nl}(r)}{r } .
\end{equation}
The spatial dependence of EM transition amplitudes reduces to
expectation values of various functions of quark position and momentum
between the initial and final state wave functions.  Expanding
\Eq~(\ref{eq:amp1}) in powers of photon momentum generates the
electric and magnetic multipole moments.  This is also an expansion in
powers of velocity.  The leading order transition amplitudes are
electric dipole (E1) or magnetic dipole (M1).

\subsubsection{Electric transitions}
\label{sec:emssec-et}

Electric transitions do not change quark spin.  The lowest NR order
transition is the electric dipole (E1) transition. These transitions
have $\Delta l = \pm 1$ and $\Delta s = 0$. To compute the E1
transition amplitudes $\exp{(i{\bf x\cdot k})}$ can be replaced by $1$
in electric transition term in \Eq~(\ref{eq:amp1}).  Separating out
the overall centre of mass motion of the system, the quark momentum
operator, $i{\bf D}/m_Q$, can be replaced by the commutator, $[h, {\bf
x}]$, of the bound state Hamiltonian, $h$, with the quark position
operator, ${\bf x}$.  Finally, the Hamiltonian acting on the initial
or final state is simply the mass of that state.  To leading NR order,
this is equal to the momentum of the final photon $k = (M_i^2 -
M_f^2)/(2 M_i)$.  The E1 radiative transition amplitude between
initial state ($n\slj{2s+1}{\ell}{J}$), $i$, and final state
($n^{\prime}\slj{2s^{\prime}+1}{\ell^{\prime}}{J^{\prime}}$), $f$, is
\cite{Eichten:1978tg}:
\begin{eqnarray}
{\bf M}^e({i \rightarrow f})_{\mu} &=& \delta_{s,s'} (-1)^{s + J + J'+ 1 + M'} k
  \sqrt{(2 J+1)(2J'+1)(2l+1)(2l'+1)}  
\nn\\
&&     \left ( \begin{tabular}{ccc}
                $J'$  & $1$  & $J$ \\
                $-M'$ &$\mu$ & $M$\\
                \end{tabular} \right )   
       \left ( \begin{tabular}{ccc}
                $l'$  & $1$  & $l$ \\
                $0$ &$0$ & $0$ \\
                \end{tabular} \right )   
       \left\{ \begin{tabular}{ccc}
                $l$ & $s_a$ & $J$ \\
                $J'$ & 1 & $l'$\\
                \end{tabular} \right\}  
   \,\langle e_Q \rangle \,{\cal E}_{if},   
\label{eq:ME1}
\end{eqnarray}
where $\langle e_Q \rangle = (e_1m_2-e_2m_1)/(m_1+m_2)$ and the
overlap integral ${\cal E}_{if}$ is
\begin{equation}
\label{eq:Eif0}
{\cal E}_{if} = \int_0^\infty \!\!\!\!dr \, u_{n\ell}(r) r 
u_{n^{\prime}\ell^{\prime}}(r).
\end{equation}
If the full photon momentum dependence in \Eq~(\ref{eq:amp1}) is
retained (even through this is formally a higher order relativistic
corrections); the overlap integral ${\cal E}$ for $m_1=m_2$ and
$e_1=-e_2=e_Q$ is given by
\begin{equation}
\label{eq:Eifsize}
{\cal E}_{if} =
\frac{3}{k}\int_0^\infty \!\!\!\!dr \, u_{n\ell}(r)
u_{n^{\prime}\ell^{\prime}}(r) \left[\frac{kr}{2} j_0\left(\frac{kr}{2}\right) -
j_1\left(\frac{kr}{2}\right)\right] + {\cal O}(k/m).
\end{equation}
The spin averaged decay rate is given by
\begin{equation}
   \Gamma(i \stackrel{\mathrm{E1}}{\longrightarrow} f+\gamma) =
        \frac{4\alpha e_{Q}^{2}}{3}(2J^{\prime}+1){\rm S}^{\rm E}_{if} k^{3}|\mathcal{E}_{if}|^{2}, 
\label{RD:E1trans}
\end{equation}
where the statistical factor ${\rm S}^{\rm E}_{if}={\rm S}^{\rm E}_{fi}$ is
\begin{equation}
{\rm S}^{\rm E}_{if} = \max{(\ell,\ell^{\prime})} \left\{
          \begin{array}{ccc}
            J & 1 & J^{\prime}  \\
            \ell^{\prime} & s & \ell
            \end{array}\right\}^{2}\; .
\end{equation}

\subsubsection{Magnetic transitions}
\label{sec:emssec-mt}

Magnetic transitions flip the quark spin. The 
M1 transitions have $\Delta l = 0$ and the amplitude is given by: 
\begin{eqnarray}
\lefteqn{{\bf M}^m({i \rightarrow f})_{\mu} = \delta_{\ell, \ell^{\prime}} 
 (-1)^{l+J^{\prime}+J+l+\mu+ M^{\prime}} 
3 \sqrt{(2J+1)(2J^{\prime}+1)(2s+1)(2s^{\prime}+1)}} 
\nonumber \\
 & & \sum_{\nu,\sigma} k_{\sigma} 
\left ( \begin{tabular}{ccc}
$1$ & $1$ & $1$ \\
$-\mu$ & $\sigma$ & $\nu$ \\
\end{tabular} \right )  
\left ( \begin{tabular}{ccc}
$J^{\prime}$ & $J$ & $1$ \\
$-M^{\prime}$ & $M$ & $\nu$ \\
\end{tabular} \right )  
\left\{ \begin{tabular}{ccc}
$s$ & $l$ & $J$ \\
$J^{\prime}$ & 1 & $s^{\prime}$ \\
\end{tabular} \right\}  
\left\{
\begin{tabular}{ccc}
 1 & 1/2 & 1/2 \\
1/2 & $s$ & $s^{\prime}$ \\
\end{tabular} \right\}  \nonumber \\
 & & ~~~~~~~~~~~\left[\frac{e_1}{m_1}+ (-1)^{s+s'} \frac{e_2}{m_2}\right]~ {\mathcal M}_{if } \;, 
\end{eqnarray}
where for equal mass quarks the overlap integral ${\cal M}$ is given by
\begin{equation}
\label{eq:mtxm1}
{\cal M}_{if} = (1+\kappa_Q) \int_0^\infty dr \, u_{n\ell}(r) u_{n^{\prime}\ell}^{\prime}(r) 
                 \, j_0\left(\frac{kr}{2}\right)   + {\cal O}(k/m) \;. 
\end{equation}
The spin-flip radiative transition rate between an initial state   
($n\sLj{2s+1}{\ell}{J}$), $i$, and a final state 
($n^{\prime}\sLj{2s^{\ell}+1}{S}{J^{\prime}}$), $f$, is:
\begin{equation}
    \Gamma(i \stackrel{\mathrm{M1}}{\longrightarrow} f + \gamma) =
        \frac{4\alpha e_{Q}^{2}}{3m_{Q}^{2}}(2J^{\prime}+1) k^{3} 
         {\rm S}^{\rm M}_{if}|\mathcal{M}_{if}|^{2}, 
\end{equation}
where the statistical factor ${\rm S}^{\rm M}_{if}={\rm S}^{\rm M}_{fi}$ is
\begin{equation}
{\rm S}^{\rm M}_{if} = 6 (2s+1)(2s^{\prime}+1)\left\{ \begin{array}{ccc}
                               J & 1 & J^{\prime}  \\
                               s^{\prime} & \ell & s
                                    \end{array}\right\}^{2} 
                        \left\{ \begin{array}{ccc}
                               1 & \frac{1}{2} & \frac{1}{2} \\
                               \frac{1}{2} & s^{\prime} & s 
                                    \end{array}\right\}^{2}\; . 
\end{equation}
For $l=0$ transitions, $S^{\rm M}_{if} = 1$.

\subsubsection[Relativistic corrections]{Relativistic corrections}
The leading relativistic corrections for electric transitions 
have been considered by a number of authors
\cite{Feinberg:hk, Sucher:wq, Kang:yw, McClary:1983xw, Zambetakis:1983te,
Grotch:bi, Grotch:1984gf, Lahde:2002wj, Ebert:2002pp}.
A general form was derived by Grotch, Owen and Sebastian \cite{Grotch:1984gf}.
For example, for the equal mass quark--antiquark $\cc$ and $\bb$ systems 
the E1 transition amplitude is $\langle f|{\bf X}_0 + {\bf X}_1|i \rangle$, 
\begin{eqnarray}
 {\bf X}_0 &=& e_Q {\bf r},   \nonumber \\
 {\bf X}_1 &=& -i \frac{k e_Q}{2 m_Q} \left( \frac{1}{10}\left(\{r^2,{\bf p}\}
   -\frac{1}{2}[{\bf r},[{\bf r}\cdot,{\bf p}]]\right) - \frac{\kappa_Q}{2}({\bf r}
   \times {\bf S}) \right )  \;, 
\end{eqnarray}
where $\kappa_Q$ is the quark anomalous magnetic moment and 
${\bf p}$ is the relative momentum. The decay rate then has the general form:
\begin{equation}
\Gamma^{\rm E1} = \Gamma^{\rm E1}_{\rm NR}(1 + R1 + R2 + R3), 
\label{eq:EMrel}
\end{equation}
where $R1$ are corrections due to the modification of the
nonrelativistic wave functions, $R2$ originates from the 
relativistic modification of the transition operator and $R3$ 
are the finite size corrections (arising from the plane wave expansion
for the emitted photon).
For the $1^3P_J \rightarrow 1^3S_1$ E1 transition:
\begin{eqnarray}  
R1 &=& 2 E_1^J + (E_1^J)^2, \nonumber \\
R2 &=& \frac{k \kappa_Q}{2 m_Q} \left[\frac{J (J+1)}{2} - 2\right], \\
R3 &=& -\frac{1}{10}(M_i-M_f)^2 E_2 + \frac{k}{8 m_Q} E_3, \nonumber \; 
\end{eqnarray}  
where 
\begin{eqnarray}  
\label{eq:E1vcorr}
E_1 &=& \frac{\displaystyle \int_0^{\infty} dr\,r\, \left[u_{10}^{(0)}(r) 
              u_{11}^{(1)J}(r) + u_{10}^{(1)}(r) u_{11}^{(0)}(r)\right]}
             {{\cal E}_{if}},  \nonumber \\
E_2 &=& \frac{\displaystyle \int_0^{\infty} dr\,r^3\, u_{10}^{(0)}(r) 
              u_{11}^{(0)}(r)}
             {{\cal E}_{if}}, \\ 
E_3 &=& \frac{\displaystyle \int_0^{\infty} dr\,r\, 
              \left[u_{10}^{(0)}(r)\left(2r\frac{d}{dr}u_{11}^{(0)}(r) -
                    u_{11}^{(0)}(r)\right)  
                    - \left(2r\frac{d}{dr}u_{10}^{(0)}(r) - 
                    u_{10}^{(0)}(r)\right)u_{11}^{(0)}(r)\right]}
             {{\cal E}_{if}}, \nonumber 
\end{eqnarray} 
and $u^{(1)}(r)$ is the first order relativistic correction to the NR
(reduced) radial wave function, $u^{(0)}(r)$.

The corrections for M1 transitions are more complicated and depend
explicitly on the structure of the nonrelativistic potential.
Assuming that the potential can be decomposed into three terms $V(r) =
V_p(r) + (1-\eta) V_v(r) + \eta V_s(r)$, \ie  a perturbative part
$V_p(r)$ and a (nonperturbative) confining part, which is a linear
combination of a Lorentz vector $V_v(r)$ and a scalar $V_s(r)$ term,
the expression $|{\cal M}_{if}|^2$ in \Eq~(\ref{eq:mtxm1}) is
replaced by \cite{Grotch:bi} $|I_1+I_2+I_3+I_4|^2$, where for $S$ wave
transitions in $\bar QQ$ systems:
\begin{eqnarray}
\label{eq:M1vcorr}
 I_1  &=& \int_0^{\infty} dr\, u_{n'0}^{(0)}(r) u_{n0}^{(0)}(r) \left[(1+\kappa_Q)j_0\left(\frac{kr}{2}\right) + 
             \frac{k(1+2\kappa_Q)}{4m_Q}\right]\,, \nonumber \\ 
 I_2  &=& \int_0^{\infty} dr\, u_{n'0}^{(0)}(r) u_{n0}^{(0)}(r) \left[-(1+\kappa_Q)\frac{{\bf p}^2}{2m_Q^2} 
            - \frac{{\bf p}^2}{3m_Q^2}\right] \,,\\
 I_3  &=& \int_0^{\infty} dr\, u_{n'0}^{(0)}(r) u_{n0}^{(0)}(r) \left[\frac{\kappa_Q r}{6m_Q}
    \frac{\partial (V_p + (1-\eta) V_v)}{\partial r}\right] \,,\nonumber \\ 
 I_4  &=& \int_0^{\infty} dr\, u_{n'0}^{(0)}(r) u_{n0}^{(0)}(r) \left[-\frac{\eta V_s}{m_Q}
       j_0\left(\frac{kr}{2}\right)\right] \,. \nonumber 
\end{eqnarray} 
Further details of these relativistic corrections can been found at the original 
papers of Feinberg and Sucher\cite{Feinberg:hk, Sucher:wq, Kang:yw}, 
Zambetakis and Byers\cite{Zambetakis:1983te} 
and Grotch and Sebastian \cite{Grotch:bi, Grotch:1984gf}. 
General treatments of relativistic corrections for all quarkonium states 
can be found in recent works \cite{Lahde:2002wj, Ebert:2002pp}.

\subsection[E1 transitions]{E1 transitions}
\label{sec:E1}
Since the discovery of the $\jpsi$  and $\psi^{\prime}$ resonances 
in November 1974, E1 transitions have played an important
theoretical and experimental role in quarkonium physics.
Initial theoretical papers on charmonium\cite{Eichten:1974af,Appelquist:yr}
predicted the $1P$ states in the $\cc$ system and suggested that the
triplet $1P$ states could be observed through the E1 transitions from 
the $\psi^{\prime}$ resonance. 
In fact, explicit calculations of the $2S\rightarrow 1P$
and $1P \rightarrow 1S$ E1 transition amplitudes ${\cal E}_{if}$ by 
the Cornell group\cite{Eichten:1974af}
agree within 25\% with present experimental values\cite{Gottfried:1997uc}.

\begin{figure}
\begin{center}
\includegraphics[width=.85\textwidth]{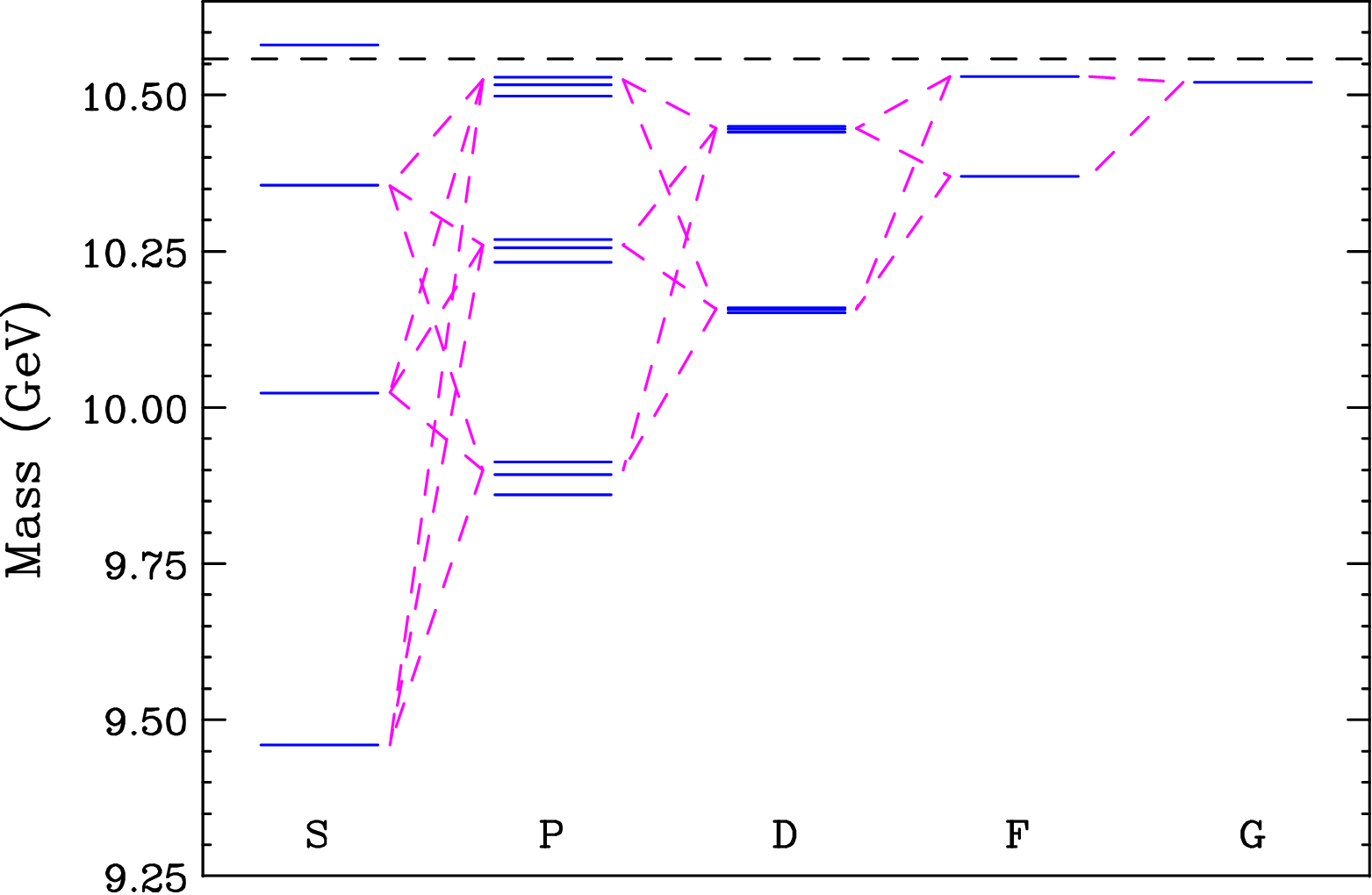}
\caption[E1 transitions in the narrow spin triplet $\bb$ states]
        {E1 transitions in the narrow spin triplet $\bb$ states.  For
         each S--P transition indicated there are three individual
         transitions (one for each $P_J$ state); while for transitions
         involving any other pair of orbital angular momenta
         (P--D, D--F, F--G, ...)  there are six individual
         transitions ($\Delta J = 0, \pm 1$).
\label{fig:level}}
\end{center}
\end{figure}

Today there is a wealth of theoretical predictions 
and experimental data on E1 transitions.
Many E1 transitions have been observed in the  
$\cc$, $\bb$ and more are expected. 
For example, \Figure~\ref{fig:level} shows the E1 
transitions from narrow spin triplet states in the 
$\bb$ system. Transitions occur between two states differing 
in L by one and J by zero or one; thus for the $\bb$ system 
there are a total of 99 E1 transitions, 30 of which are theoretically accessible 
in $e^+e^-$ collisions from the $\Y$(2S) and $\Y$(3S) resonances.

\subsubsection[Model predictions]{Model predictions}

The theoretical models used to calculate the E1 transitions can be
classified by the following two considerations: (1) What
nonrelativistic potential was used?  and (2) Which relativistic
corrections (as shown in \Eq~(\ref{eq:EMrel})) were included in the
calculations?

An early choice for the potential was the Cornell model
\cite{Eichten:1974af,Eichten:1976jk,Eichten:1978tg,Eichten:1979ms,Eichten:1994gt}.  
Here the exchange interaction was the time component of a vector 
with a Coulomb short range part $-K/r$ plus 
a linear $r/a^2$ long range confining part.
The Coulomb part was modified to agree with perturbative QCD at
short distance by Buchm\"uller and Tye\cite{Buchmuller:1980bm,exdec:buch81}. 
Other simple forms for the potential, 
logarithmic\cite{Quigg:1979vr,Quigg:1981bj}
and power law \cite{Martin:1980,Grant:1993uf}, were also proposed.     

In the NRQCD limit the quark--antiquark interaction is spin independent,
but including relativistic corrections introduces dependencies on the 
Lorentz structure of the potential.  Of particular importance is the vector 
versus scalar nature of the long-range confining interaction.
Many modern theoretical calculations assume a long range scalar 
confining potential\cite{Buchmuller:1981fr}
or a linear combination of the form $\eta V_S(r) + (1-\eta) V_V(r)$
\cite{Pignon:1977cf,McClary:1983xw,Ebert:2002pp}. 
Moxhay and Rosner\cite{Moxhay:1983vu} assumed an additional long 
range tensor force.  

The second consideration is the extent of the inclusion of the
relativistic corrections.  Some calculations are essentially
nonrelativistic. These calculations often include some finite size
effects ($R3$ of \Eq~(\ref{eq:EMrel})) by retaining the form for
${\cal E}_{if}$ given in \Eq~(\ref{eq:Eifsize})
\cite{Eichten:1974af,Eichten:1976jk,Eichten:1978tg,Eichten:1979ms,
Kwong:1988ae, Eichten:1994gt}.  Other models also include relativistic
corrections to the wave functions ($R1$ of \Eq~(\ref{eq:EMrel}))
either perturbatively or nonperturbatively.  The relativistic quark
model of Godfrey and Isgur\cite{Godfrey:xj} is an example in this
class.  Gupta, Radford and Repko computed the relativistic corrections
using only the gluon exchange interactions of QCD perturbation
theory\cite{Gupta:1982kp,Gupta:1984jb,Gupta:1986xt}.  Many models
include the full relativistic
corrections\cite{McClary:1983xw,Grotch:1984gf,Bander:1984ew,Moxhay:1983vu,
Daghighian:1987ru,Lahde:2002wj}.

\begin{table}
\caption[E1 transition rates for low-lying $\cc $ states]
        {E1 transition rates for low-lying $\cc $ states. The measured
         masses are used for observed states. The mass values used for
         the $^3D_2$, $^1D_2$ and $^3D_3$ states are suggested by the
         coupled channel calculations of Ref. \cite{Eichten:2004uh}.
         The E1 rates are shown for the (NR) model described in the
         text.  The variation of results for ${\cal E}_{if}$ with
         inclusion of relativistic corrections is shown for two models
         of Ref. \cite{Ebert:2002pp} with scalar confinement (RA) and
         a mixture of vector and scalar confining terms (RB).}
\label{tab:ccE1}
\renewcommand{\arraystretch}{1.3}
\begin{center}
\begin{tabular}{@{}llcccrrr@{}}
\hline
\hline
\multicolumn{2}{c}{Transition} & $k$ & ${\rm S}^{\rm E}_{if}$ 
& $\Gamma (i \rightarrow f)$(NR) &${\cal E}_{if}$(NR)
&${\cal E}_{if}$(RA)&${\cal E}_{if}$(RB)\\
\multicolumn{2}{c}{$i \stackrel{{\rm E1}}{\longrightarrow} f$} & $(\mev)$ & & $(\kev)$ 
& $(\gev^{-1})$ & $(\gev^{-1})$ & $(\gev^{-1})$ \\ 
\hline
 $1^3P_0( 3.415)$ & $1^3S_1( 3.097)$ & $304$ & $\frac{1}{9}$ & $120$ &$1.724$&$2.121$&$1.720$ \\ 
 $1^3P_1( 3.511)$ & $1^3S_1( 3.097)$ & $389$ & $\frac{1}{9}$ & $241$ &$1.684$&$1.896$&$1.767$ \\ 
 $1^1P_1( 3.526)$ & $1^1S_0( 2.979)$ & $504$ & $\frac{1}{3}$ & $482$ &$1.615$&$1.742$&$1.742$ \\ 
 $1^3P_2( 3.556)$ & $1^3S_1( 3.097)$ & $430$ & $\frac{1}{9}$ & $315$ &$1.661$&$1.596$&$1.689$ \\ 
 $2^3S_1( 3.686)$ & $1^3P_0( 3.415)$ & $261$ & $\frac{1}{9}$ & $47.0$ &$-2.350$&$-2.296$&$-1.775$ \\ 
 $2^1S_0( 3.638)$ & $1^1P_1( 3.526)$ & $110$ & $\frac{1}{3}$ & $35.1$ &$-2.469$&$-2.126$&$-2.126$ \\ 
 $2^3S_1( 3.686)$ & $1^3P_1( 3.511)$ & $171$ & $\frac{1}{9}$ & $42.8$ &$-2.432$&$-2.305$&$-1.782$ \\ 
 $2^3S_1( 3.686)$ & $1^3P_2( 3.556)$ & $127$ & $\frac{1}{9}$ & $30.1$ &$-2.460$&$-2.362$&$-1.901$ \\ 
 $1^3D_1( 3.770)$ & $1^3P_0( 3.415)$ & $338$ & $\frac{2}{9}$ & $299$ &$2.841$&$2.718$&$2.802$ \\ 
 $1^3D_1( 3.770)$ & $1^3P_1( 3.511)$ & $250$ & $\frac{1}{18}$ & $99.0$ &$2.957$&$2.799$&$2.969$ \\ 
 $1^3D_1( 3.770)$ & $1^3P_2( 3.556)$ & $208$ & $\frac{1}{450}$ & $3.88$ &$3.002$&$3.016$&$3.348$ \\ 
 $1^3D_2( 3.831)$ & $1^3P_1( 3.511)$ & $307$ & $\frac{1}{10}$ & $313$ &$2.886$&$2.593$&$2.593$ \\ 
 $1^3D_2( 3.831)$ & $1^3P_2( 3.556)$ & $265$ & $\frac{1}{50}$ & $69.5$ &$2.940$&$2.781$&$2.991$ \\ 
 $1^1D_2( 3.838)$ & $1^1P_1( 3.526)$ & $299$ & $\frac{2}{15}$ & $389$ &$2.896$&$2.610$&$2.610$ \\ 
 $1^3D_3( 3.872)$ & $1^3P_2( 3.556)$ & $303$ & $\frac{2}{25}$ & $402$ &$2.892$&$2.508$&$2.402$ \\ 
\hline
\hline
\end{tabular}
\end{center}
\renewcommand{\arraystretch}{1}
\end{table}

\begin{table}[p]
\caption[E1 transition rates for low-lying spin triplet $ \bb $ states]
        {E1 transition rates for low-lying spin triplet $ \bb $ states.}
\label{tab:bb1tE1}
\renewcommand{\arraystretch}{1.3}
\begin{center}
\begin{tabular}{@{}llcccrrr@{}}
\hline
\hline
\multicolumn{2}{c}{Transition} & $k$ & ${\rm S}^{\rm E}_{if}$ & $\Gamma (i \rightarrow f)$(NR) 
& ${\cal E}_{if}$(NR) &${\cal E}_{if}$(RA) & ${\cal E}_{if}$(RB)\\
\multicolumn{2}{c}{$i \stackrel{{\rm E1}}{\longrightarrow} f$} & $(\mev)$ & &
$(\kev)$ & $(\gev^{-1})$ & $(\gev^{-1})$ & $(\gev^{-1})$ \\ 
\hline
 $1^3P_0( 9.860)$ & $1^3S_1( 9.460)$ & $392$ & $\frac{1}{9}$ & $22.2$ &$1.013$&$1.205$&$1.178$ \\ 
 $1^3P_1( 9.893)$ & $1^3S_1( 9.460)$ & $423$ & $\frac{1}{9}$ & $27.8$ &$1.010$&$1.175$&$1.163$ \\ 
 $1^3P_2( 9.913)$ & $1^3S_1( 9.460)$ & $442$ & $\frac{1}{9}$ & $31.6$ &$1.007$&$1.124$&$1.137$ \\ 
 $2^3S_1(10.023)$ & $1^3P_2( 9.913)$ & $110$ & $\frac{1}{9}$ & $2.04$ &$-1.597$&$-1.800$&$-1.778$ \\ 
 $2^3S_1(10.023)$ & $1^3P_1( 9.893)$ & $130$ & $\frac{1}{9}$ & $2.00$ &$-1.595$&$-1.781$&$-1.759$ \\ 
 $2^3S_1(10.023)$ & $1^3P_0( 9.860)$ & $162$ & $\frac{1}{9}$ & $1.29$ &$-1.590$&$-1.803$&$-1.781$ \\ 
 $1^3D_1(10.151)$ & $1^3P_2( 9.913)$ & $236$ & $\frac{1}{450}$ & $0.564$ &$1.896$&$2.104$&$2.104$ \\ 
 $1^3D_1(10.151)$ & $1^3P_1( 9.893)$ & $255$ & $\frac{1}{18}$ & $10.7$ &$1.890$&$2.050$&$2.050$ \\ 
 $1^3D_1(10.151)$ & $1^3P_0( 9.860)$ & $287$ & $\frac{2}{9}$ & $20.1$ &$1.880$&$2.106$&$2.106$ \\ 
 $1^3D_2(10.157)$ & $1^3P_2( 9.913)$ & $241$ & $\frac{1}{50}$ & $5.46$ &$1.894$&$2.048$&$2.048$ \\ 
 $1^3D_2(10.157)$ & $1^3P_1( 9.893)$ & $261$ & $\frac{1}{10}$ & $20.5$ &$1.888$&$1.999$&$1.999$ \\ 
 $1^3D_3(10.160)$ & $1^3P_2( 9.913)$ & $244$ & $\frac{2}{25}$ & $22.6$ &$1.893$&$1.979$&$1.979$ \\ 
 $2^3P_0(10.232)$ & $1^3D_1(10.151)$ & $ 81$ & $\frac{2}{9}$ & $1.13$ &$-1.723$&$-1.740$&$-1.740$ \\ 
 $2^3P_0(10.232)$ & $2^3S_1(10.023)$ & $207$ & $\frac{1}{9}$ & $9.17$ &$1.697$&$1.872$&$1.855$ \\ 
 $2^3P_0(10.232)$ & $1^3S_1( 9.460)$ & $743$ & $\frac{1}{9}$ & $10.9$ &$0.272$&$0.214$&$0.239$ \\ 
 $2^3P_1(10.255)$ & $1^3D_2(10.157)$ & $ 98$ & $\frac{1}{10}$ & $1.49$ &$-1.720$&$-1.751$&$-1.751$ \\ 
 $2^3P_1(10.255)$ & $1^3D_1(10.151)$ & $104$ & $\frac{1}{18}$ & $0.593$ &$-1.718$&$-1.721$&$-1.721$ \\ 
 $2^3P_1(10.255)$ & $2^3S_1(10.023)$ & $229$ & $\frac{1}{9}$ & $12.4$ &$1.688$&$1.837$&$1.831$ \\ 
 $2^3P_1(10.255)$ & $1^3S_1( 9.460)$ & $764$ & $\frac{1}{9}$ & $12.0$ &$0.274$&$0.228$&$0.216$ \\ 
 $2^3P_2(10.268)$ & $1^3D_3(10.160)$ & $108$ & $\frac{2}{25}$ & $2.25$ &$-1.717$&$-1.763$&$-1.763$ \\ 
 $2^3P_2(10.268)$ & $1^3D_2(10.157)$ & $111$ & $\frac{1}{50}$ & $0.434$ &$-1.717$&$-1.737$&$-1.737$ \\ 
 $2^3P_2(10.268)$ & $1^3D_1(10.151)$ & $117$ & $\frac{1}{450}$ & $0.034$ &$-1.715$&$-1.766$&$-1.766$ \\ 
 $2^3P_2(10.268)$ & $2^3S_1(10.023)$ & $242$ & $\frac{1}{9}$ & $14.5$ &$1.682$&$1.792$&$1.797$ \\ 
 $2^3P_2(10.268)$ & $1^3S_1( 9.460)$ & $776$ & $\frac{1}{9}$ & $12.7$ &$0.274$&$0.207$&$0.218$ \\ 
 $3^3S_1(10.355)$ & $2^3P_2(10.268)$ & $ 86$ & $\frac{1}{9}$ & $2.40$ &$-2.493$&$-2.663$&$-2.644$ \\ 
 $3^3S_1(10.355)$ & $2^3P_1(10.255)$ & $100$ & $\frac{1}{9}$ & $2.20$ &$-2.489$&$-2.607$&$-2.586$ \\ 
 $3^3S_1(10.355)$ & $2^3P_0(10.232)$ & $122$ & $\frac{1}{9}$ & $1.35$ &$-2.479$&$-2.608$&$-2.582$ \\ 
 $3^3S_1(10.355)$ & $1^3P_2( 9.913)$ & $433$ & $\frac{1}{9}$ & $0.015$ &$0.016$&$0.063$&$0.045$ \\ 
 $3^3S_1(10.355)$ & $1^3P_1( 9.893)$ & $452$ & $\frac{1}{9}$ & $0.008$ &$0.011$&$0.063$&$0.045$ \\ 
 $3^3S_1(10.355)$ & $1^3P_0( 9.860)$ & $483$ & $\frac{1}{9}$ & $0.001$ &$0.004$&$0.063$&$0.045$ \\ 
\hline
\hline
\end{tabular}
\end{center}
\renewcommand{\arraystretch}{1}
\end{table}

Differences in theoretical assumptions and experimental input for the
various potential model calculations of E1 transitions make it
difficult to draw sharp conclusions from the level of agreement of a
particular model and experimental data.  However, it is known that
there is usually very little model variation in the NR predictions
(lowest order) if the models are fit to the same
states\cite{Kwong:1988ae}.  The only exceptions are transitions where
the overlap integral ${\cal E}_{if}$ exhibits large dynamical
cancellations.  Therefore, to compare the variations in results due to
the inclusion of relativistic corrections from a common base, three
models for E1 radiative transitions are presented, which are fit with
the same input masses.  First a reference Cornell
model\cite{Eichten:1979ms} (NR), with parameters ($a$ and $K$)
adjusted to fit the COG positions of the 1S, 1P and 2S states in each
of the $\cc$ and $\bb$ systems\cite{Eichten:2002qv}.  Here E1
transitions are computed with ${\cal E}_{if}$ given in
\Eq~(\ref{eq:Eifsize}), \ie  with only finite size relativistic
corrections included.  Second, a recent model by Ebert, Faustov and
Galkin\cite{Ebert:2002pp} with full relativistic corrections in two
cases: (RA) $\eta = 1$ (scalar confinement) and (RB) $\eta = -1$ (a
fitted mixture of scalar and vector confinement).

The results for ${\cal E}_{if}$ are shown for the $\cc$ narrow states
in \Table~\ref{tab:ccE1}. The size of the relativistic corrections to
${\cal E}_{if}$ shown in \Table~\ref{tab:ccE1} vary as much as $\pm
25$\%. This variation is perfectly consistent with naive expectations
for $v^2/c^2$ corrections.  McClary and Byers\cite{McClary:1983xw}
first emphasized that because of the node in the radial wave function
of the $2S$ state the overlap ${\cal E}_{\rm 2S,1P}$ is particularly
sensitive to relativistic corrections in the $\cc$ system.  The
significant leptonic width for the $\Psi(3770)$ resonance implies that
there is a sizeable S-D mixing between the $2^3S_1$ and $1^3D_1$
states. This mixing arises both from the usual relativistic correction
terms and coupling to strong decay channels and will affect the
$\Psi(3686) \rightarrow 1^3P_J$ and $\Psi(3770) \rightarrow 1^3P_J$ E1
transitions rates (See \Section~\ref{sec:em-Dwave}).  For the $1D$
states there may be additional large effects on rates associated with
this coupling to nearby strong decay channels. (See
\Section~\ref{sec:em-vloops}.)

Results for narrow $\bb$ states accessible from the $\Y$(3S) or
$\Y$(2S) resonances are shown for spin-triplets in
\Table~\ref{tab:bb1tE1} and for the spin-singlets in
\Table~\ref{tab:bb1sE1}.  The typical size of the relativistic
corrections for ${\cal E}_{if}$ are approximately half as large as in
the corresponding $\cc$ transition. This is again as expected, since
$\langle v^2/c^2 \rangle$ is smaller in the $\bb$ system. There is a
notable exception for the overlap ${\cal E}_{\rm 3S,1P}$.  In the NR
limit this overlap is less than 5\% of any other S--P overlap.
This dynamical accident makes these transition rates very sensitive to
the details of wave functions and relativistic corrections, which are
{\it not} well under control theoretically.

Finally, for completeness, radiative transitions involving $\bb$
states not accessible from the $3S$ states are shown in
\Table~\ref{tab:bb2tE1}.  Only the NR rates are shown. One observes
large dynamical cancellations for the overlap ${\cal E}_{\rm 3P,1D}$
and to a lesser extent in the overlaps ${\cal E}_{\rm 3P,1S}$, ${\cal
E}_{\rm 2D,1P}$ and ${\cal E}_{\rm 3P,2S}$.

\begin{table}
\caption{E1 transition rates for low-lying spin singlet $ \bb $ states.}
\label{tab:bb1sE1}
\renewcommand{\arraystretch}{1.3}
\begin{center}
\begin{tabular}{@{}llcccrrr@{}}
\hline
\hline
\multicolumn{2}{c}{Transition} & $k$ & ${\rm S}^{\rm E}_{if}$ 
& $\Gamma (i \rightarrow f)$(NR) &${\cal E}_{if}$(NR)
&${\cal E}_{if}$(RA)&${\cal E}_{if}$(RB)\\
\multicolumn{2}{c}{$i \stackrel{{\rm E1}}{\longrightarrow} f$} & $(\mev )$ & 
& $(\kev )$ & $(\gev^{-1} )$ & $(\gev^{-1} )$ & $(\gev^{-1} )$ \\ 
\hline
 $1^1P_1( 9.900)$ & $1^1S_0( 9.400)$ & $487$ & $\frac{1}{3}$ & $41.8$ &$1.001$&$1.149$&$1.149$ \\ 
 $2^1S_0( 9.990)$ & $1^1P_1( 9.900)$ & $ 90$ & $\frac{1}{3}$ & $1.99$ &$-1.600$&$-1.743$&$-1.743$ \\ 
 $1^1D_2(10.157)$ & $1^1P_1( 9.900)$ & $254$ & $\frac{2}{15}$ & $25.3$ &$1.891$&$2.002$&$2.002$ \\ 
 $2^1P_1(10.260)$ & $2^1S_0( 9.990)$ & $266$ & $\frac{1}{3}$ & $19.0$ &$1.671$&$1.817$&$1.817$ \\ 
 $2^1P_1(10.260)$ & $1^1D_2(10.157)$ & $102$ & $\frac{2}{15}$ & $2.29$ &$-1.719$&$-1.782$&$-1.782$ \\ 
 $3^1S_0(10.328)$ & $2^1P_1(10.260)$ & $ 68$ & $\frac{1}{3}$ & $2.10$ &$-2.498$&$-2.571$&$-2.571$ \\ 
 $3^1S_0(10.328)$ & $1^1P_1( 9.900)$ & $419$ & $\frac{1}{3}$ & $0.007$ &$0.010$&$0.064$&$0.064$ \\ 
\hline
\hline
\end{tabular}
\end{center}
\renewcommand{\arraystretch}{1}
\end{table}

\begin{table}[p]
\caption[E1 transition rates for the remaining spin triplet $\bb$ states]
        {E1 transition rates for the remaining spin triplet $\bb$
         states.  For each $(n^{\prime}$ and $\ell^{\prime})$ only the
         final state $J^{\prime}$ with the largest rate is shown. The
         transition rates for spin-singlet $\bb$ states differ from
         the corresponding spin triplet rates by the ratio of
         statistical factors ${\rm S}^{\rm E}(s=0)/{\rm S}^{\rm
         E}(s=1)$: 3, 2/3, 9/8 and 16/15 for S--P, P--D, D--F
         and F--G transitions respectively.}
\label{tab:bb2tE1}
\renewcommand{\arraystretch}{1.23}
\begin{center}
\begin{tabular}{@{}llccrc@{}}
\hline
\hline
\multicolumn{2}{c}{Transition} & $k$ & ${\rm S}^{\rm E}_{if}$ & ${\cal E}_{if}$ & $\Gamma (i \rightarrow f)$ \\
\multicolumn{2}{c}{$i \stackrel{{\rm E1}}{\longrightarrow} f$} & $(\mev )$ & & $(\gev^{-1} )$ & $(\kev )$ \\ 
\hline
 $1^3F_2(10.370)$ & $1^3D_1(10.151)$ & $217$ & $\frac{3}{25}$ & $2.681$ &$28.5$ \\ 
 $1^3F_3(10.370)$ & $1^3D_2(10.157)$ & $211$ & $\frac{8}{105}$ & $2.684$ &$27.8$ \\ 
 $1^3F_4(10.370)$ & $1^3D_3(10.160)$ & $208$ & $\frac{3}{49}$ & $2.686$ &$30.0$ \\ 
 $2^3D_1(10.441)$ & $1^3F_2(10.370)$ & $ 71$ & $\frac{3}{25}$ & $-1.904$ &$0.833$ \\ 
 $2^3D_1(10.441)$ & $2^3P_0(10.232)$ & $207$ & $\frac{2}{9}$ & $2.487$ &$13.1$ \\ 
 $2^3D_1(10.441)$ & $1^3P_0( 9.860)$ & $565$ & $\frac{2}{9}$ & $0.288$ &$3.60$ \\ 
 $2^3D_2(10.446)$ & $1^3F_3(10.370)$ & $ 76$ & $\frac{8}{105}$ & $-1.903$ &$0.907$ \\ 
 $2^3D_3(10.450)$ & $1^3F_4(10.370)$ & $ 80$ & $\frac{3}{49}$ & $-1.902$ &$1.09$ \\ 
 $2^3D_3(10.450)$ & $2^3P_2(10.268)$ & $180$ & $\frac{2}{25}$ & $2.506$ &$15.8$ \\ 
 $2^3D_3(10.450)$ & $1^3P_2( 9.913)$ & $524$ & $\frac{2}{25}$ & $0.278$ &$4.80$ \\ 
 $3^3P_0(10.498)$ & $2^3D_1(10.441)$ & $ 57$ & $\frac{2}{9}$ & $-2.584$ &$0.884$ \\ 
 $3^3P_0(10.498)$ & $3^3S_1(10.355)$ & $142$ & $\frac{1}{9}$ & $2.308$ &$5.47$ \\ 
 $3^3P_0(10.498)$ & $1^3D_1(10.151)$ & $341$ & $\frac{2}{9}$ & $-0.047$ &$0.063$ \\ 
 $3^3P_0(10.498)$ & $2^3S_1(10.023)$ & $464$ & $\frac{1}{9}$ & $0.351$ &$4.44$ \\ 
 $3^3P_0(10.498)$ & $1^3S_1( 9.460)$ & $986$ & $\frac{1}{9}$ & $0.137$ &$6.46$ \\ 
 $3^3P_1(10.516)$ & $2^3D_2(10.446)$ & $ 70$ & $\frac{1}{10}$ & $-2.579$ &$1.22$ \\ 
 $3^3P_1(10.516)$ & $3^3S_1(10.355)$ & $160$ & $\frac{1}{9}$ & $2.295$ &$7.71$ \\ 
 $3^3P_1(10.516)$ & $1^3D_2(10.157)$ & $353$ & $\frac{1}{10}$ & $-0.050$ &$0.060$ \\ 
 $3^3P_1(10.516)$ & $2^3S_1(10.023)$ & $481$ & $\frac{1}{9}$ & $0.355$ &$5.06$ \\ 
 $3^3P_1(10.516)$ & $1^3S_1( 9.460)$ & $1003$ & $\frac{1}{9}$ & $0.137$ &$6.86$ \\ 
 $1^3G_3(10.520)$ & $1^3F_2(10.498)$ & $ 22$ & $\frac{4}{49}$ & $3.812$ &$0.068$ \\ 
 $1^3G_4(10.520)$ & $1^3F_3(10.498)$ & $ 22$ & $\frac{5}{84}$ & $3.812$ &$0.069$ \\ 
 $1^3G_5(10.520)$ & $1^3F_4(10.498)$ & $ 22$ & $\frac{4}{81}$ & $3.812$ &$0.074$ \\ 
 $3^3P_2(10.529)$ & $2^3D_3(10.450)$ & $ 79$ & $\frac{2}{25}$ & $-2.576$ &$1.96$ \\ 
 $3^3P_2(10.529)$ & $3^3S_1(10.355)$ & $172$ & $\frac{1}{9}$ & $2.284$ &$9.63$ \\ 
 $3^3P_2(10.529)$ & $1^3D_3(10.160)$ & $363$ & $\frac{2}{25}$ & $-0.053$ &$0.082$ \\ 
 $3^3P_2(10.529)$ & $2^3S_1(10.023)$ & $494$ & $\frac{1}{9}$ & $0.358$ &$5.54$ \\ 
 $3^3P_2(10.529)$ & $1^3S_1( 9.460)$ & $1014$ & $\frac{1}{9}$ & $0.138$ &$7.16$ \\ 
 $2^3F_2(10.530)$ & $2^3D_1(10.441)$ & $ 89$ & $\frac{3}{25}$ & $3.337$ &$3.02$ \\ 
 $2^3F_3(10.530)$ & $2^3D_2(10.446)$ & $ 84$ & $\frac{8}{105}$ & $3.340$ &$2.69$ \\ 
 $2^3F_4(10.530)$ & $2^3D_3(10.450)$ & $ 80$ & $\frac{3}{49}$ & $3.342$ &$2.62$ \\ 
 $2^3F_2(10.530)$ & $1^3G_3(10.520)$ & $ 10$ & $\frac{4}{49}$ & $-2.262$ &$0.003$ \\ 
 $2^3F_3(10.530)$ & $1^3G_4(10.520)$ & $ 10$ & $\frac{5}{84}$ & $-2.262$ &$0.003$ \\ 
 $2^3F_4(10.530)$ & $1^3G_5(10.520)$ & $ 10$ & $\frac{4}{81}$ & $-2.262$ &$0.003$ \\ 
\hline
\hline
\end{tabular}
\end{center}
\renewcommand{\arraystretch}{1}
\end{table}

\clearpage

\subsubsection[Comparison with experiment: $S$ and $P$ states]
              {Comparison with experiment: $S$ and $P$ states 
               $\!$\footnote{Authors: E.~Eichten, T.~Ferguson}} 
\label{sec:emssec-ettf}

\begin{figure}[p]
\begin{center}
  \includegraphics[width=105mm]{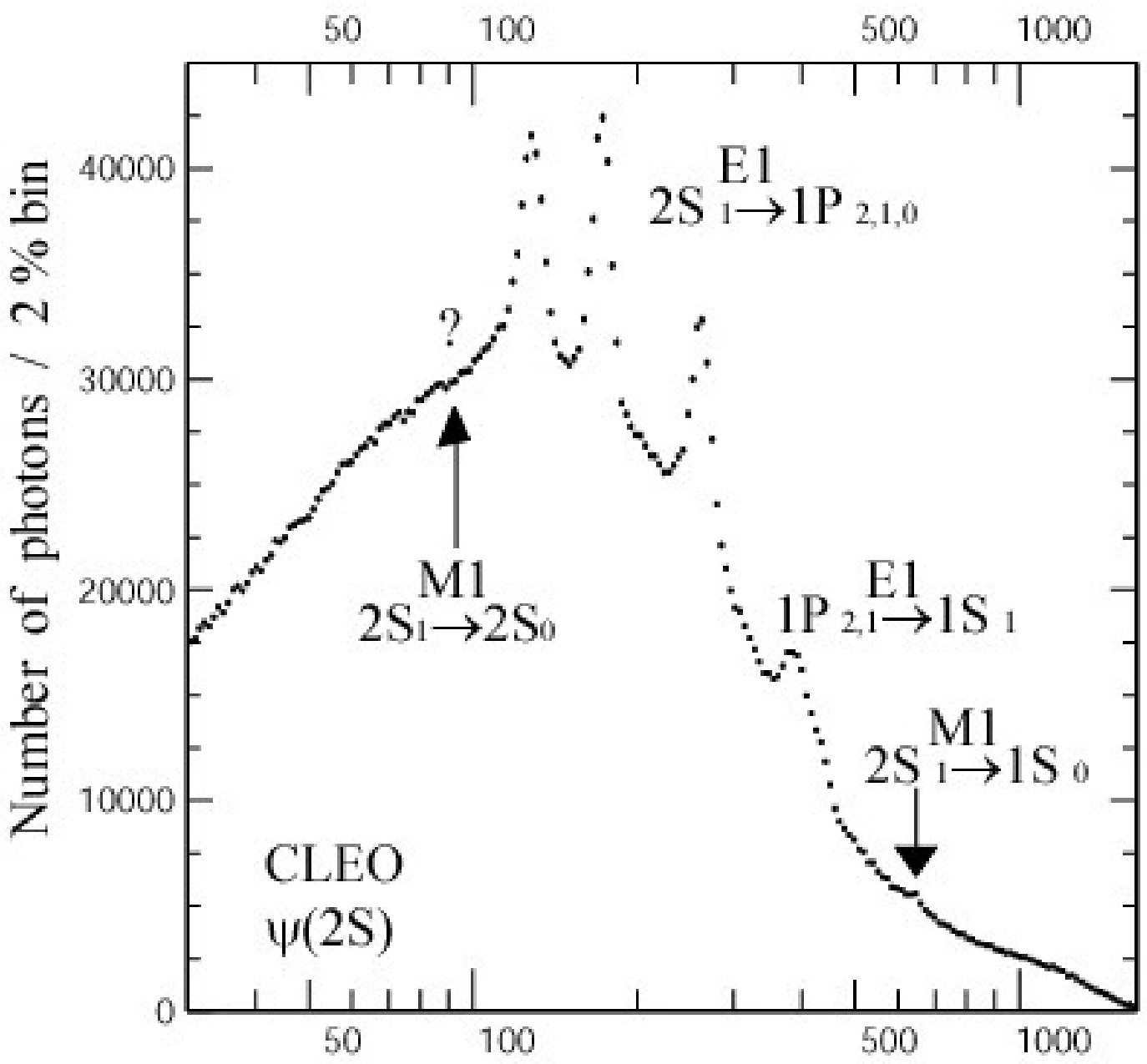}\\[2mm]
  \includegraphics[width=105mm]{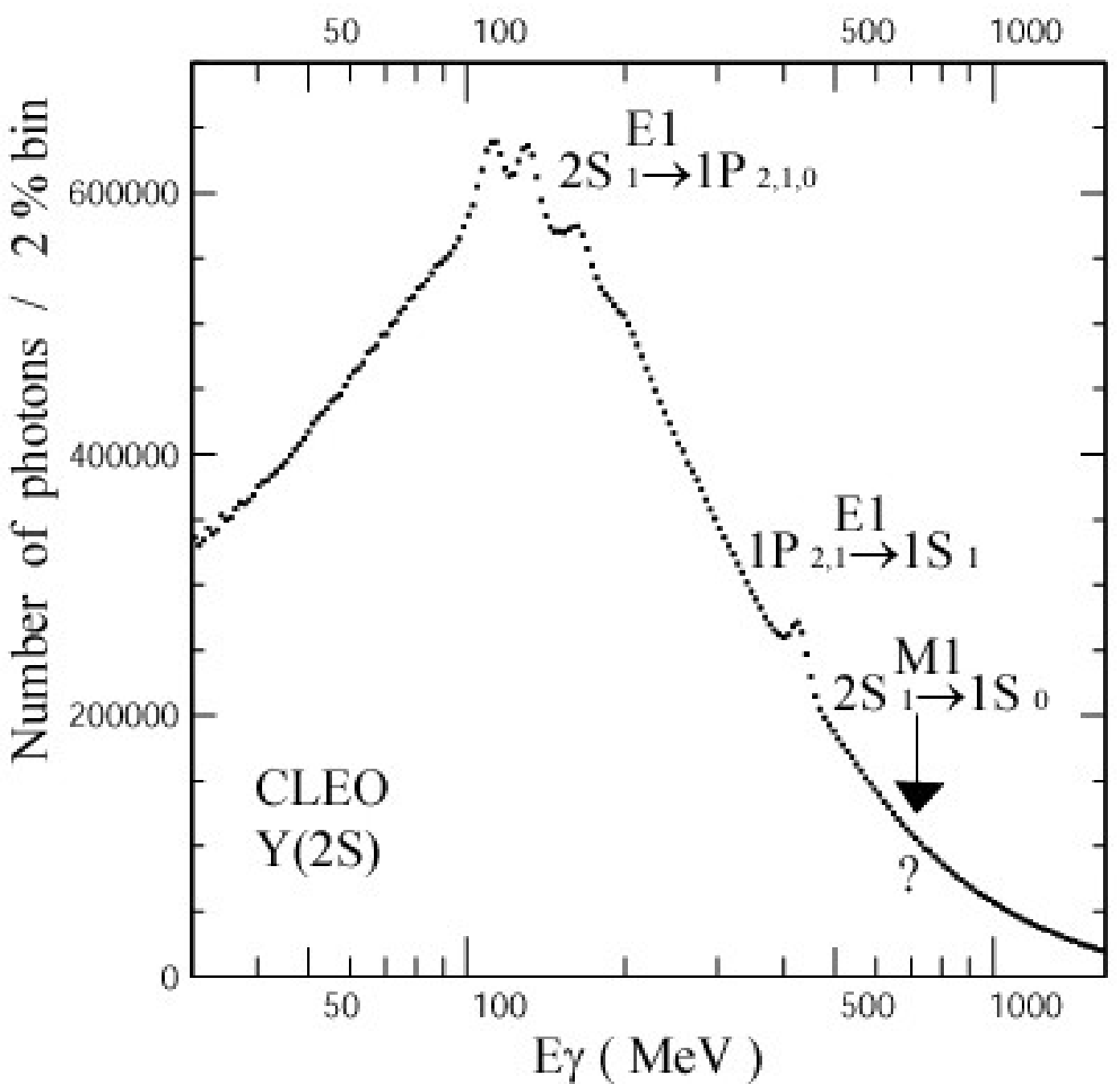}
\end{center}
\caption[Inclusive photon spectrum from $2^3S_1$ decays
         in the $\cc$ (top) and $\bb$ (bottom) systems]
        {Inclusive photon spectrum from $2^3S_1$ decays in the $\cc$
         (top) and $\bb$ (bottom) systems measured with the CLEO
         detector.  The data correspond to about 1.5M $\psi(2S)$ and
         9M $\Y$(2S) decays. From Skwarnicki\cite{Skwarnicki:2003wn}.}
\label{fig:gspec}
\end{figure}

There is now extensive data on electromagnetic transitions among heavy
quarkonium states.  \Figure[b]~\ref{fig:gspec} shows the inclusive photon
spectra from the $\cc$ and $\bb$ $2\slj{3}{1}{1}$ decays measured with
the CLEO detector\cite{Skwarnicki:2003wn}.  This section provides a
snapshot of the current status of various S--P transitions.  New
data come mainly from the CLEO experiment at CESR.

In the NR limit the overlap $\dipmtx{nS}{n'P_J} = \left | \langle
n'P_J|r|nS \rangle \right |$ is independent of $J$.  Experimentally,
it is useful to define averages over $J$ by
\begin{eqnarray}
 {\dipmtx{nS}{n^{\prime}P}} ({\rm avg})  &=&
   \sqrt{\frac{\B(nS\to\gamma n'P_J)\,\Gamma_{\rm tot}(nS) }{{ D\, \sum_J (2J+1) E_\gamma(nS\to n'P_J)^3 }}} 
\\   
 {\dipmtx{nP}{n^{\prime}S}} ({\rm avg})  &=&
   \sqrt{\frac{ \B(nP_J\to\gamma n'S)\,\Gamma_{\rm tot}(nP_J) }{{ D\, \sum_J E_\gamma(nP_J\to n'S)^3 }}}    
\nonumber 
\end{eqnarray}
where $D=4/3\,\alpha\,{e_b}^2{\rm S}^{\rm E}_{^3P_J,^3S_1}$. These
quantities reduce to the usual overlaps in the NR limit. In order to
see the relativistic corrections (which vary with $J$) it is also
useful to define ratios, $\dipmtx{nS}{n'P_J}/\dipmtx{nS}{n'P}({\rm
avg})$.  Given the total width of the initial state these overlaps can
be determined directly from experimental branching ratios.  The
experimental results for the $\cc$ and $\bb$ states are shown in
\Table~\ref{tab:e1exp}.  These results are extracted from the world
average results for $\B(\chic(1P_J)\to\gamma\jpsi)$ and
$\B(\psi(2S)\to\gamma\chic(1P_J))$. Also shown are recent results from
CLEO-c for $\B(\psi(2S)\to\gamma\chic(1P_J))$
transitions\cite{Athar:2004dn}. Results for
$\B(\Y(3S)\to\gamma\chib(2P_J))$ and $\B(\Y(2S)\to\gamma\chib(1P_J)$
are taken from Ref.~\cite{Eidelman:2004wy}.  The E1 transitions show
clear evidence of $J$ dependence and, hence, relativisitic corrections
in $S$ state transitions. The largest relativistic effects are in the
$2^3S_1$ to $1^3P_J$ $\cc$ transitions.

With their large $\Y$(3S) data sample and excellent CsI
electromagnetic calorimeter, the CLEO~III experiment has been able to
measure the E1 photon transitions from the $\Y$(3S) to the
$\chi_b$(2P$_J$) states, and the subsequent photon decays of those
states to the $\Y$(2S) and $\Y$(1S).  They identify exclusive
$\gamma\gamma\ell^+\ell^-$ events, which are consistent with photon
transitions through the $\chi_b$(2P$_J$) states to the $\Y$(2S) or
$\Y$(1S), followed by the leptonic decay of the $\Y$.  This provides a
very clean signal with little background.  \Tables~\ref{tab:2p2s} and
\ref{tab:2p1s} give a summary of their preliminary results
\cite{CLEOphoton} on the product branching ratios, along with
comparisons with the previous CLEO~II \cite{Crawford:1992vs} and CUSB
\cite{Heintz:1992cv} measurements. Then, by using the world average
values for the $\Y$(3S) $\rightarrow$ $\chi_b$(2P$_J$) + $\gamma$ and
$\Y$ leptonic branching ratios, the $\chi_b$(2P$_J$) $\rightarrow$
$\Y$ + $\gamma$ branching ratios can be obtained.

\begin{table}[p]
\caption[Measured E1 overlap integrals for S--P transitions]
        {Measured E1 overlap integrals for S--P
         transitions. Transition rates use branching ratios and total
         widths from PDG04 world averages \cite{Eidelman:2004wy}
         except for second set of values for the $\cc$ transition
         $2^3S_1 \rightarrow 1^3P_J$, which uses branching ratios from
         recent results of CLEO-c\cite{Athar:2004dn}.}
\label{tab:e1exp}
\renewcommand{\arraystretch}{1.3}
\begin{center}
\begin{tabular}{|ll|c|ccc|}
\hline
\multicolumn{2}{|c|}{Transition} & 
  $|{\cal E}_{\rm avg}|$ 
   &\multicolumn{3}{c|}{${\cal E}_{J}/{\cal E}_{\rm avg}$}  \\
\multicolumn{2}{|c|}{$i \stackrel{\rm{E1}}{\longrightarrow} f$} & 
   $ (\gev^{-1})$ & $J=0$ & $J=1$ & $J=2$ \\ 
\hline
\multicolumn{6}{|c|}{$\cc$} \\ 
\hline
 $1^3P_J$ & $1^3S_1$ & $1.87\pm 0.07$ 
    & $0.92\pm 0.05$ & $0.99\pm 0.06$ & $1.04\pm 0.03$ \\
 $2^3S_1$ & $1^3P_J$ & $1.78\pm 0.07$ 
    & $0.94\pm 0.04$ & $1.01\pm 0.05$ & $1.07\pm 0.05$ \\ 
          &          & $1.94\pm 0.07$  
    & $0.90\pm 0.02$ & $0.97\pm 0.03$ & $1.19\pm 0.04$ \\
\hline
\multicolumn{6}{|c|}{$\bb$} \\ 
\hline
 $3^3S_1$ & $2^3P_J$ & $2.75\pm 0.21$ 
    & $0.92\pm 0.06$ & $1.06\pm 0.05$ & $1.02\pm 0.06$ \\ 
 $2^3S_1$ & $1^3P_J$ & $1.94\pm 0.18$ 
    & $0.92\pm 0.06$ & $1.09\pm 0.05$ & $0.98\pm 0.06$ \\ 
\hline
\end{tabular}
\end{center}
\renewcommand{\arraystretch}{1}

\caption[CLEO~III preliminary results for $\Y$(3S) $\rightarrow$
         $\gamma$ $\chi_b$(2P$_J$) $\rightarrow$ $\gamma\gamma$
         $\Y$(2S) $\rightarrow$ $\gamma\gamma\ell^+\ell^-$]
        {CLEO~III preliminary results \cite{CLEOphoton} for $\Y$(3S)
         $\rightarrow$ $\gamma$ $\chi_b$(2P$_J$) $\rightarrow$
         $\gamma\gamma$ $\Y$(2S) $\rightarrow$
         $\gamma\gamma\ell^+\ell^-$, along with comparisons with CLEO
         II \cite{Crawford:1992vs} and CUSB \cite{Heintz:1992cv}.}
\label{tab:2p2s}
\renewcommand{\arraystretch}{1.0} 
\begin{center} 
\begin{tabular}{|c|c|c|c|c|} 
\hline 
Parameter (units) & Ref. &J = 2 & J = 1 & J = 0 \\ \hline 
$\cal B$($\gamma\gamma\ell^+\ell^-$) (10$^{-4}$) 
  & \cite{CLEOphoton} & $2.73 \pm 0.15 \pm 0.24$ & $5.84 \pm 0.17 \pm 0.41$ & $0.17 \pm 0.06 \pm 0.02$ \\
\cline{2-5}
  & \cite{Crawford:1992vs}  & $2.49 \pm 0.47 \pm 0.31$ & $5.11 \pm 0.60 \pm 0.63$ & $< 0.60$ \\ 
\cline{2-5}
  & \cite{Heintz:1992cv}   & $2.74 \pm 0.33 \pm 0.18$ & $3.30 \pm 0.33 \pm 0.19$ & $0.40 \pm 0.17 \pm 0.03$ \\ 
\hline
$\cal B$($\Y$(3S) $\to$ $\gamma\gamma\Y$(2S)) (\%) 
  & \cite{CLEOphoton} & $2.20 \pm 0.12 \pm 0.31$ & $4.69 \pm 0.14 \pm 0.62$ & $0.14 \pm 0.05 \pm 0.02$ \\ 
\hline
$\cal B$($\chi_b$(2P$_J$) $\to$ $\gamma\Y$(2S)) (\%) 
  & \cite{CLEOphoton} & $19.3 \pm 1.1 \pm 3.1$   & $41.5 \pm 1.2 \pm 5.9$ & $2.59 \pm 0.92 \pm 0.51$ \\ 
\hline
\end{tabular} 
\end{center} 

\medskip

\caption[CLEO~III preliminary results for $\Y$(3S) $\rightarrow$
         $\gamma$ $\chi_b$(2P$_J$) $\rightarrow$ $\gamma\gamma$
         $\Y$(1S) $\rightarrow$ $\gamma\gamma\ell^+\ell^-$]
        {CLEO~III preliminary results \cite{CLEOphoton} for $\Y$(3S)
         $\rightarrow$ $\gamma$ $\chi_b$(2P$_J$) $\rightarrow$
         $\gamma\gamma$ $\Y$(1S) $\rightarrow$
         $\gamma\gamma\ell^+\ell^-$, along with comparisons with CLEO
         II \cite{Crawford:1992vs} and CUSB
         \cite{Heintz:1992cv}.}
\label{tab:2p1s}
\renewcommand{\arraystretch}{1.0}
\begin{center} \setlength{\tabcolsep}{3pt}
\begin{tabular}{|c|c|c|c|c|} 
\hline 
Parameter (units)& Ref. & J = 2 & J = 1 & J = 0 \\ \hline 
$\cal B$($\gamma\gamma\ell^+\ell^-$) (10$^{-4}$) 
 & \cite{CLEOphoton} & $1.93 \pm 0.12 \pm 0.17$ & $3.19 \pm 0.13 \pm 0.18$ & $< 0.16$ \\ 
\cline{2-5}
 & \cite{Crawford:1992vs} & $2.51 \pm 0.47 \pm 0.32$ & $3.24 \pm 0.56 \pm 0.41$ & $< 0.32$ \\ 
\cline{2-5}
 & \cite{Heintz:1992cv}  & $1.98 \pm 0.28 \pm 0.12$ & $2.34 \pm 0.28 \pm 0.14$ & $0.13 \pm 0.10 \pm 0.03$ \\ 
\hline
$\cal B$($\Y$(3S) $\rightarrow$ $\gamma\gamma\Y$(1S)) (\%) 
 & \cite{CLEOphoton} & $0.79 \pm 0.05 \pm 0.07$ & $1.31 \pm 0.05 \pm 0.08$ & $< 0.08$ \\ 
\hline
$\cal B$($\chi_b$(2P$_J$) $\rightarrow$ $\gamma\Y$(1S)) (\%) 
 &  \cite{CLEOphoton} & $7.0 \pm 0.4 \pm 0.8$ & $11.6 \pm 0.4 \pm 0.9$ & $< 1.44$ \\ 
\hline
\end{tabular} 
\end{center} 
\renewcommand{\arraystretch}{1} 
\caption[CLEO~III preliminary results for $\Y$(3S) $\rightarrow$
         $\gamma$ $\chi_b$(1P$_J$) $\rightarrow$ $\gamma\gamma$
         $\Y$(1S) $\rightarrow$ $\gamma\gamma\ell^+\ell^-$, along with
         comparisons with the CUSB]
        {CLEO~III preliminary results \cite{CLEOphoton} for $\Y$(3S)
         $\rightarrow$ $\gamma$ $\chi_b$(1P$_J$) $\rightarrow$
         $\gamma\gamma$ $\Y$(1S) $\rightarrow$
         $\gamma\gamma\ell^+\ell^-$, along with comparisons with the
         CUSB experiment \cite{Heintz:1992cv}. The values are summed
         over the J = 1 and J = 2 transitions.}
\label{tab:1p1s}
\renewcommand{\arraystretch}{1.0}
\begin{center} \setlength{\tabcolsep}{3pt}
\begin{tabular}{|c|c|c|} 
\hline 
Parameter & Ref. & J = 1 and 2 Combined \\ \hline 
$\cal B$($\gamma\gamma\ell^+\ell^-$) (10$^{-4}$) 
  &  \cite{CLEOphoton} & $0.520 \pm 0.054 \pm 0.052$  \\ 
\hline
$\cal B$($\Y$(3S) $\rightarrow$ $\gamma\gamma\Y$(1S)) (\%) 
  &  \cite{CLEOphoton} & $0.241 \pm 0.022 \pm 0.021$ \\ 
\cline{2-3}
  & \cite{Heintz:1992cv}  & $0.12 \pm 0.04 \pm 0.01$ \\ 
\hline
\end{tabular} 
\end{center} 
\renewcommand{\arraystretch}{1} 
\end{table} 

For the similar transitions through the $\chi_b$(1P$_J$) states:
$\Y$(3S) $\rightarrow$ $\gamma\chi_b$(1P$_J$), $\chi_b$(1P$_J$)
$\rightarrow$ $\gamma\Y$(1S), the photon lines for the different
J states cannot be resolved, due to the finite crystal energy
resolution. The J = 0 branching ratio is expected to be small, given
the large hadronic width of this state.  So CLEO~III gives a combined
product branching ratio, summed over the J = 1 and J = 2 states.  The
results are shown in \Table~\ref{tab:1p1s}.

We can extract the $ \ecalif{1P}{3S} $ matrix element from
the photon transitions via the $\chib(1P)$ states:
\begin{equation}
\label{eq:e13S1P}
 \dipmtx{1P}{3S}({\rm avg}) = 
\sqrt{
      \frac{ \B(3S\to\gamma 1P, 1P\to\gamma 1S)\,\Gamma_{\rm tot}(3S) }{
           { D\, \sum_J (2J+1) E_\gamma(1P_J\to 1S)^3 \B(1P_J\to\gamma 1S) }}
}.       
\end{equation}
This formula assumes that the matrix element is spin independent.
Taking $\B(3S\to\gamma 1P, 1P\to\gamma 1S)$ from \Table~\ref{tab:1p1s}
and the world average values for the other quantities from
PDG04\cite{Eidelman:2004wy}, we obtain:
\[
 \dipmtx{1P}{3S}({\rm avg})  = (0.050\pm0.006)\, {\rm GeV}^{-1}\,.
\]
The error here includes the statistical and systematic uncertainties
on all quantities added in quadrature.  The averaging is only over $J
= 1$ and $J = 2$.

Results for the values of ${\cal E}$(avg) in the $\bb$ $P$ system are
compared to various potential model predictions in
\Table~\ref{tab:radmatrix}.  We also include results for $
\dipmtx{3S}{2P} $ and $ \dipmtx{2S}{1P} $ from \Table~\ref{tab:e1exp}
extracted from the world average results for ${\cal
B}(\Y(3S)\to\gamma\chib(2P_J))$ and ${\cal
B}(\Y(2S)\to\gamma\chib(1P_J)$\cite{Eidelman:2004wy}.  While most of
the potential models have no trouble reproducing the large matrix
elements, $ \dipmtx{3S}{2P} $, $ \dipmtx{2S}{1P}$, which show also
little model dependence, only a few models predict $ \dipmtx{3S}{1P} $
in agreement with measurements.  Clearly, the latter transition is
highly sensitive to the underlying description of $\bb$ states as
discussed above.

The branching ratios given in the
\Tables~\ref{tab:2p2s}--\ref{tab:1p1s} can also be used to measure the
ratios of various E1 matrix elements, which can then be compared to
different potential model predictions. First, the ratio of the matrix
elements for the decay of the same $\chi_b(2P_J)$ state to different
$\Y$ states can be found using:
\begin{equation}
\frac{{\cal E}_{2{{\rm P}_{\rm J}},{\rm 1S}}}{{\cal E}_{2{{\rm P}_{\rm J}},{\rm 2S}}}
\;=\; \sqrt{\frac{{\cal B}(3S \;\rightarrow\; \gamma 2P_J,
    2P_J\;\rightarrow\;\gamma 1S)}{{\cal B}(3S \;\rightarrow\; \gamma 2P_J,
    2P_J\;\rightarrow\;\gamma 2S)} \; \left ( \frac{E_\gamma(2P_J
    \;\rightarrow\;2S)}{E_\gamma(2P_J
    \;\rightarrow\;1S)} \right )^3}
\end{equation}
With this method, the following values are obtained:
\begin{equation}
\frac{{\cal E}_{2{{\rm P}_2},{\rm 1S}}}{{\cal E}_{2{{\rm P}_2},{\rm 2S}}}
\;=\; 0.105 \;\pm\; 0.004 \;\pm\;0.006, \;\;\;\;\;\;
\frac{{\cal E}_{2{{\rm P}_1},{\rm 1S}}}{{\cal E}_{2{{\rm P}_1},{\rm 2S}}}
\;=\; 0.087 \;\pm\; 0.002 \;\pm\;0.005,
\end{equation}
\begin{equation}
\frac{{\cal E}_{2{{\rm P}_2},{\rm 1S}}}{{\cal E}_{2{{\rm P}_2},{\rm 2S}}}
\;/\; 
\frac{{\cal E}_{2{{\rm P}_1},{\rm 1S}}}{{\cal E}_{2{{\rm P}_1},{\rm 2S}}}
\;=\; 1.21 \;\pm\;0.06,
\;\;\;\;\;\; 
\frac{{\cal E}_{2{{\rm P}_{1,2}},{\rm 1S}}}{{\cal E}_{2{{\rm P}_{1,2}},{\rm 2S}}}
\;=\; 0.096 \;\pm\; 0.002 \;\pm\;0.005.
\end{equation}

\begin{table}
\caption[Comparison of average E1 matrix elements and their ratios
         predicted by different potential models
         with measurements from $\bb$ data]
        {Comparison of average E1 matrix elements and their ratios
         predicted by different potential models with measurements
         from $\bb$ data. ``NR'' denotes nonrelativistic calculations
         and ``rel'' refers to models with relativistic corrections.
         The first set of model entries are the reference models
         considered here. The second set is a selection of other
         models taken from Ref. \cite{CLEOphoton}.}
\label{tab:radmatrix}
\def\1#1{\multicolumn{1}{c}{#1}}
\def\2#1#2{\multicolumn{#1}{c|}{#2}}
\def\3#1#2{\multicolumn{#1}{c|}{#2}}
\def\etal{et al.}
\begin{center}
\begin{tabular}{|l|cc|cc|cc|cc|}
\hline
      & \multicolumn{2}{c|}{$|{\cal E}_{\rm 3S,2P}|$} & \multicolumn{2}{c|}{$|{\cal E}_{\rm 2S,1P}|$} &
     \multicolumn{2}{c|}{$|{\cal E}_{\rm 3S,1P}|$} & \multicolumn{2}{c|}{$|{\cal E}_{\rm 2P,1S}|$} \\
      & & & & & &  &\multicolumn{2}{c|}{$\overline{|{\cal E}_{\rm 2P,2S}|}$}  \\
\cline{2-9} 
      & \multicolumn{2}{c|}{${\rm GeV}^{-1}$} & \multicolumn{2}{c|}{${\rm GeV}^{-1}$} 
        & \multicolumn{2}{c|}{${\rm GeV}^{-1}$} & \multicolumn{2}{c|}{}\\
\hline\hline
DATA &  \multicolumn{2}{c|}{$2.7\pm 0.2$} & \multicolumn{2}{c|}{$1.9\pm 0.2$} 
        & \multicolumn{2}{c|}{$0.050\pm 0.006$} & \multicolumn{2}{c|}{$0.096\pm 0.005$}  \\
\cline{2-9}
     & \multicolumn{4}{c|}{World Average} & \multicolumn{4}{c|}{CLEO~III\cite{CLEOphoton}} \\
\hline
\hline
Model & NR & rel & NR & rel & NR & rel & NR & rel \\
\hline
NR    & 2.5 &     & 1.6 &     & 0.014 &       & 0.16 &      \\
RA    &     & 2.6 &     & 1.8 &       & 0.063 &      & 0.12 \\
RB    &     & 2.6 &     & 1.8 &       & 0.045 &      & 0.12 \\
\hline
Kwong, Rosner \cite{Kwong:1988ae}     
& 2.7 &     & 1.6 &     & 0.023 &       & 0.13 &      \\
Fulcher \cite{Fulcher:kx}         
& 2.6 &     & 1.6 &     & 0.023 &       & 0.13 &      \\
B\"uchmuller \etal \cite{Buchmuller:1980bm,exdec:buch81} 
& 2.7 &     & 1.6 &     & 0.010 &       & 0.12 &      \\
Moxhay, Rosner \cite{Moxhay:1983vu}    
& 2.7 & 2.7 & 1.6 & 1.6 & 0.024 & 0.044 & 0.13 & 0.15 \\
Gupta \etal \cite{Gupta:1986xt}      
& 2.6 &     & 1.6 &     & 0.040 &       & 0.11 &      \\
Gupta \etal \cite{Gupta:1982kp,Gupta:1984jb}      
& 2.6 &     & 1.6 &     & 0.010 &       & 0.12 &      \\
Fulcher \cite{Fulcher:is}         
& 2.6 &     & 1.6 &     & 0.018 &       & 0.11 &      \\
Danghighian \etal \cite{Daghighian:1987ru}
& 2.8 & 2.5 & 1.7 & 1.3 & 0.024 & 0.037 & 0.13 & 0.10 \\
McClary, Byers \cite{McClary:1983xw}    
& 2.6 & 2.5 & 1.7 & 1.6 &       &       & 0.15 & 0.13 \\
Eichten \etal  \cite{Eichten:1979ms}  
& 2.6 &     & 1.7 &     & 0.110 &       & 0.15 &      \\
Grotch \etal \cite{Grotch:1984gf}     
& 2.7 & 2.5 & 1.7 & 1.5 & 0.011 & 0.061 & 0.13 & 0.19 \\
\hline
\end{tabular}
\end{center}
\end{table}

To compare to potential model predictions, the last number above is an
average over the J = 1 and J = 2 values.  In the nonrelativistic
limit, the E1 matrix elements should not depend on J.  Since the
values for the J = 1 and J = 2 matrix elements differ by 3.5 standard
deviations, there appears to be evidence for relativistic effects in
the $\bb$ system in both the $S$ and $P$ states transitions.  Again
these results are compared to various potential model predictions in
\Table~\ref{tab:radmatrix}.  Predictions for the ratio
${\ecalif{2P}{1S}}/{\ecalif{2P}{2S}}$ are very model dependent, but
somewhat higher than the experimental values.

Overall, the comparison of the measured matrix elements and the predictions of 
various potential models shows that the recent theoretical
calculations that incorporate relativistic effects are better at reproducing
the data \cite{CLEOphoton,Skwarnicki:2003wn}.

\subsubsection[$D$ states]{$D$ states}
\label{sec:em-Dwave}

In the $\cc$ system, the $1^3D_1$ and $1^3D_3$ states are above $D\bar
D$ threshold and have open flavour strong decays. The $J=2$ states
$1^3D_2$ and $1^1D_2$ are below (or at) the $D^*\bar D+D\bar D^*$
threshold and are expected to be narrow.  In all cases, the coupling
to real and virtual strong decay channels is likely to significantly
alter the potential model radiative transition rates shown in
\Table~\ref{tab:ccE1}. (We will discuss this further in
\Section~\ref{sec:em-vloops} below.)  One effect of these couplings is
that the $\psi(3770)$ state will not be a pure $1^3D_1$ state but will
have a sizeable mixing component with the $2^3S_1$ state:
\begin{equation}
\label{eq:sdmix}
\psi(3770) = \cos(\phi) |1^3D_1 \rangle + \sin(\phi) |2^3S_1 \rangle \;.
\end{equation}
Using the measured leptonic width of the $\psi(3770)$ and resolving a
two-fold ambiguity in favor of the value of the mixing angle favored
by Cornell coupled channel calculations \cite{Eichten:1979ms}, Rosner
finds \cite{Rosner:2001nm} $\phi = (12\pm 2)^{\circ}$.  Employing the
NR results of \Table~\ref{tab:ccE1}, the ratios of E1 transitions to
various $\chic$ states are:
\begin{eqnarray}
 \frac{\Gamma(\psi(3770) \rightarrow \gamma {\chic}_1)}{\Gamma(\psi(3770) 
                 \rightarrow \gamma {\chic}_0)} 
&=& 1.32 \left (\frac{-\frac{\sqrt{3}}{2} + \tan(\phi)}
                  {\sqrt{3} + \tan(\phi)}\right )^2, \nonumber \\
 \frac{\Gamma(\psi(3770) \rightarrow \gamma {\chic}_2)}{\Gamma(\psi(3770) 
                \rightarrow \gamma {\chic}_0)} 
&=& 1.30 \left (\frac{\frac{\sqrt{3}}{10} + \tan(\phi)}
                 {\sqrt{3} + \tan(\phi)}\right )^2. 
\end{eqnarray}
Measuring these branching ratios is experimentally challenging. [The
only existing data is contained in an unpublished Ph. D. thesis based on MARK III 
data\cite{Zhu:1988}.]  CLEO-c may be able to determine some of these 
branching ratios in the near future.   
 
In the $\bb$ system CLEO~III\cite{Bonvicini:2004yj} has presented
evidence for the production of $\Y(1D)$ sta\-tes in the four-photon
cascade (see \Figure~\ref{fig:level}), $\Y(3S)\to\gamma\chi_b(2P)$,
$\chi_b(2P)\to\gamma\Y(1D)$, $\Y(1D)\to\gamma\chi_b(1P)$,
$\chi_b(1P)\to\gamma\Y(1S)$, followed by the $\Y(1S)$ annihilation
into $e^+e^-$ or $\mu^+\mu^-$.

In addition to the four-photon cascade via the $\Y(1D)$ states, they observe
events with the four-photon cascade via the $\Y(2S)$ state:
$\Y(3S)$ $\to$ $\gamma\chi_b(2P)$, 
$\chi_b(2P)$ $\to$ $\gamma\Y(2S)$, 
$\Y(2S)$ $\to$ $\gamma\chi_b(1P)$, 
$\chi_b(1P)$ $\to$ $\gamma\Y(1S)$,
$\Y(1S)\to l^+l^-$
The product branching ratio for this entire decay sequence
(including $\Y(1S)\to l^+l^-$)
is predicted
to be $3.84\cdot10^{-5}$\cite{Godfrey:2001eb}, thus
comparable to the predicted $\Y(1D)$ production rate.
In the four-photon cascade via the $\Y(1D)$ the second highest energy photon is
due to the third transition, while in these cascades  
the second highest energy photon
is due to the second photon transition (see \Figure~\ref{fig:level}).  
This allows the discrimination of the $\Y(1D)$ signal from
the $\Y(2S)$ background events.

CLEO~III\cite{Bonvicini:2004yj} finds their data are dominated by the
production of one $\Y(1D)$ state consistent with the $J=2$ assignment
and a mass $(10161.1\pm0.6\pm1.6)$~MeV, which is consistent with the
predictions from potential models and lattice QCD calculations.

The signal product branching ratio obtained is
$\B(\gamma\gamma\gamma\gamma\LL)_{\Y(1D)}
=(2.5\pm0.5\pm0.5)\cdot10^{-5}$.  The first error is statistical,
while the second one is systematic.  The significance of the signal is
10.2 standard deviations.  This branching ratio is consistent with the
theoretically estimated rates.  Godfrey and Rosner
\cite{Godfrey:2001eb}, summing over $\Y(1D_{1,2,3})$ contributions,
obtained $3.76\times 10^{-5}$; while the predicted rate
\cite{Godfrey:2001eb,Kwong:1988ae} for the $\Y(1D_2)$ state alone is
$2.6\times 10^{-5}$.

Forming the ratio of $\Y(1D)$ to $\Y(2S)$ four-photon cascades would
allow the measurement in a fairly model independent way of the
estimate of the total width of the $\Y(1D)$ state, if the individual
$\Y(2P_J)$ and $\Y(1P_{J'})$ transitions could be resolved.

\subsection[M1 transitions]{M1 transitions}
\label{sec:M1}

For M1 transitions, the leading order NRQCD prediction 
for the overlap ${\cal M}_{if}$ is independent of the potential model.
The spin independence and orthogonality of states guarantee 
that the spatial overlap matrix is one for states within the same multiplet 
and zero for allowed transitions between multiplets, which have different radial quantum numbers. 

Including relativistic corrections, \eg finite size corrections, will spoil
these exact results and induce a small overlap between states 
with different radial quantum numbers. Such $n \ne n^{\prime}$ transitions are denoted hindered.

\subsubsection[Model predictions]{Model predictions}

Within the (NR) model used for the E1 transitions (\ie a
nonrelativistic treatment except for finite size corrections and
$\kappa_Q = 0$) the M1 transition rates and overlap matrix elements
${\cal M}$ for $\cc$ and $\bb$ $S$ states are shown in
\Table~\ref{tab:m1NR}.

Numerous papers have considered these M1 transitions including full
relativistic
corrections\cite{Zambetakis:1983te,Grotch:1984gf,Godfrey:xj,
Zhang:et,Godfrey:2001eb,Ebert:2002pp,Lahde:2002wj}.  There are several
sources of uncertainty that contribute making M1 transitions
particularly complicated to calculate.  In addition to the usual
issues associated with the form of the long range potential there is
the unknown value for the anomalous magnetic moment for the quark
($\kappa_Q$).  Furthermore, the results depend explicitly on the quark
masses and on other details of the potential (see
\Eqs~\ref{eq:M1vcorr}). For the models (RA) and (RB) used for the E1
transitions, $\kappa_Q = -1$.  The theoretical uncertainty in the
value of $\kappa_Q$ will eventually be reduced by lattice calculations
in quarkonium systems.

\subsubsection[Comparison with experiment]{Comparison with experiment}

\begin{figure}
\centering\includegraphics[width=.9\textwidth]{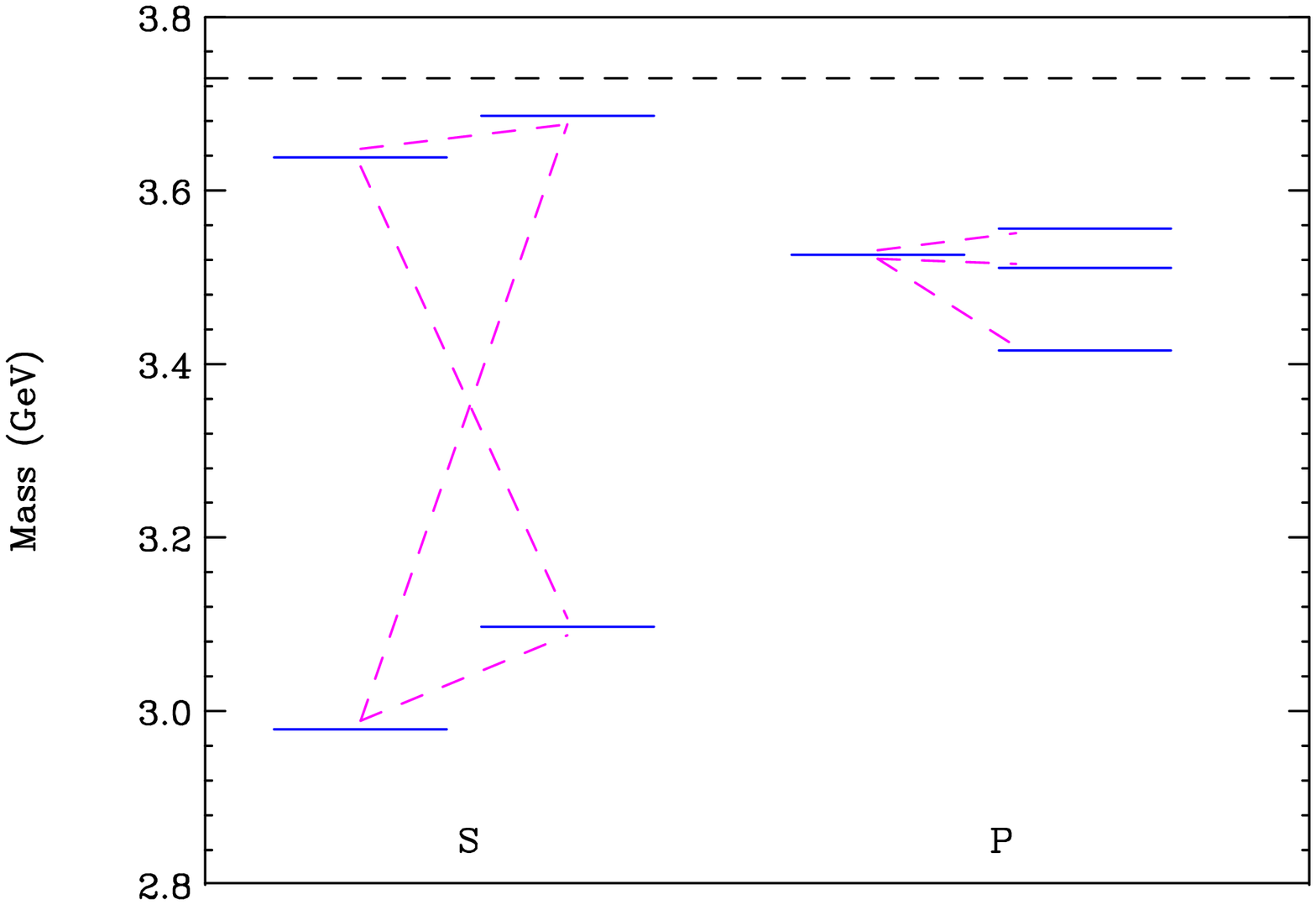}
\caption{Allowed M1 transitions in the narrow $\cc$. The $1P$ transition rates are 
unobservably small ($\approx 1 eV$).
\label{fig:ccM1}}
\end{figure}

M1 transitions have only been observed in the $\cc$ system.  The
allowed transitions in the $\cc$ system below threshold are shown in
\Figure~\ref{fig:ccM1}.  The transitions within the $1P$ system are
tiny ($\approx 1$ eV).  Only the $\jpsi \rightarrow \etac$ and $\psip
\rightarrow \etac$ are observed experimentally \cite{Eidelman:2004wy}.
For the $\bb$ system CLEO \cite{Mahmood:2002jd} sees no evidence for
the hindered M1 transition $\Upsilon({\rm 3S}) \rightarrow
\eta_b$(1S).  The $90\%$ cl upper bound on the branching ratio varies
from $4-6 \times 10^{-4}$ depending on the mass splitting. For the
expected splitting $\approx 910 \mev$ the bound is $5.3 \times
10^{-4}$\cite{Mahmood:2002jd}. This rules out a number of older
models\cite{Zambetakis:1983te, Godfrey:xj}. A comparison of the
experimental results with a variety of more modern models is shown in
\Table~\ref{tab:M1comp}.  For each model the assumptions for the
mixture of scalar and vector confinement and the value of $\kappa_Q$
is exhibited explicitly. For the model of Lahde\cite{Lahde:2002wj} the
results are also shown without including the exchange term (NEX).
This (NEX) piece neglects the time ordering of photon emission and
potential interaction, which vanishes in the NR limit.  Generally
models with a scalar confining interaction and/or a sizable negative
anomalous quark magnetic moment are favored.

\begin{table}[p]
\caption[M1 transition rates for S-wave quarkonium states using the NR model]
        {M1 transition rates for S-wave quarkonium states using the
         NR model described in text.  Finite size corrections are
         included in the calculation of ${\cal M}$ (see 
         \Eq~(\ref{eq:mtxm1})) and $\kappa_Q = 0$.}
\label{tab:m1NR}
\renewcommand{\arraystretch}{1.2}
\begin{center}
\begin{tabular}{llccr}
\hline
\hline
\multicolumn{2}{c}{Transition} & $k$ & $\Gamma (i \rightarrow f)$(NR) & ${\cal M}_{if}$(NR)\\
\multicolumn{2}{c}{$i \stackrel{{\rm M1}}{\longrightarrow} f$} & $(\mev )$ & $(eV )$ &  \\ 
\hline
\multicolumn{5}{c}{$\cc$} \\
\hline
 $1^3S_1( 3.097)$ & $1^1S_0( 2.979)$ & $115$ & $1,960$  &$0.998$ \\ 
 $2^3S_1( 3.686)$ & $2^1S_0( 3.638)$ & $ 48$ & $140$ &$0.999$ \\ 
 $2^1S_0( 3.638)$ & $1^3S_1( 3.097)$ & $501$ & $538$ &$0.033$ \\ 
 $2^3S_1( 3.686)$ & $1^1S_0( 2.979)$ & $639$ & $926$ &$0.053$ \\ 
\hline
\hline
\multicolumn{5}{c}{$\bb$} \\
\hline
 $1^3S_1( 9.460)$ & $1^1S_0( 9.400)$ & $ 60$ & $8.953$ &$1.000$ \\ 
 $2^1S_0( 9.990)$ & $1^3S_1( 9.460)$ & $516$ & $2.832$ &$0.013$ \\ 
 $2^3S_1(10.023)$ & $2^1S_0( 9.990)$ & $ 33$ & $1.509$ &$1.000$ \\ 
 $2^3S_1(10.023)$ & $1^1S_0( 9.400)$ & $604$ & $2.809$ &$0.018$ \\ 
 $3^1S_0(10.328)$ & $2^3S_1(10.023)$ & $300$ & $0.620$ &$0.014$ \\ 
 $3^1S_0(10.328)$ & $1^3S_1( 9.460)$ & $831$ & $3.757$ &$0.007$ \\ 
 $3^3S_1(10.355)$ & $3^1S_0(10.328)$ & $ 27$ & $0.826$ &$1.000$ \\ 
 $3^3S_1(10.355)$ & $2^1S_0( 9.990)$ & $359$ & $0.707$ &$0.019$ \\ 
 $3^3S_1(10.355)$ & $1^1S_0( 9.400)$ & $911$ & $2.435$ &$0.009$ \\ 
\hline
\hline
\end{tabular}
\end{center}

\bigskip

\caption[Comparison of M1 transition overlaps with experiment for
         various models]
        {Comparison of M1 transition overlaps with experiment for
         various models.  The transition overlap $I \equiv
         \frac{M(1^3S_1)}{2m_Q}{\cal M}_{if}$ is from nS spin triplet
         to the n'S spin singlet $S$ states in the $\cc $ and $\bb $
         systems. The experimental upper bounds are $90$\% cl.}
\label{tab:M1comp}
\renewcommand{\arraystretch}{1.2}
\setlength{\tabcolsep}{4pt}
\begin{center}
\begin{tabular}{|l|rr|rr|rrrr|}
\hline
\multicolumn{3}{|c|}{Type} 
   & \multicolumn{6}{c|}{Transition $I_{n,n'}$} \\
\hline
Model & \multicolumn{2}{c|}{parameters} 
   & \multicolumn{2}{c|}{$(n,n')$ [$\cc $]} & \multicolumn{4}{c|}{$(n,n')$ [$\bb $]} \\
\hline
   & $\eta$ & $\kappa_Q$ & $(1,1)$ & $(2,1)$ & $(1,1)$ & $(2,1)$ & $(3,1)$ & $(3,2)$ \\
\hline
Cornell\cite{Eichten:1979ms} & NR & $0$ & $0.84$ & $0.028$ 
        & $0.92$ & $0.017$ & $0.007$ & $0.018$ \\
GOS84\cite{Grotch:1984gf} & $0$ & $0$ & $0.86$ & $0.075$ 
        & $0.88$ & $0.058$ & -- & -- \\
                            & $0$ & $-1$ & $0.58$ & $0.054$ 
        & $0.081$ & $0.007$ & -- & -- \\
                            & $1$ & $0$ & $0.65$ & $0.127$ 
        & $0.91$ & $0.048$ & -- & -- \\
                            & $1$ & $-1$ & $0.39$ & $0.029$ 
        & $0.049$ & $0.021$ & -- & -- \\
EFG02\cite{Ebert:2002pp} & $0$ & $0$ & $0.84$ & $0.036$ 
        & $0.91$ & $0.018$ & $0.013$ & $0.016$ \\
                           & $1$ & $0$ & $1.06$ & $0.027$ 
        & $1.08$ & $0.011$ & $0.009$ & $0.007$ \\
                           & $-1$ & $-1$ & $0.62$ & $0.045$ 
        & $0.75$ & $0.025$ & $0.026$ & $0.017$ \\
Lahde02\cite{Lahde:2002wj} & NEX & $0$ & $0.87$ & $0.011$ 
        & $0.92$ & $0.020$ & $0.009$ & $0.016$ \\
                             & $1$ & $0$ & $0.67$ & $0.049$ 
        & $0.88$ & $0.032$ & $0.014$ & $0.037$ \\
\hline
EXP & & & $0.66\pm 0.10$ & $0.042\pm 0.004$ 
        & & $< 0.045$ & $< 0.020$ & $< 0.080$ \\
Ref & & & \cite{Eidelman:2004wy} & \cite{Eidelman:2004wy}\cite{Athar:2004dn}
    &  & \cite{Skwarnicki:2003wn} & \cite{Mahmood:2002jd} 
       & \cite{Skwarnicki:2003wn} \\
\hline
\end{tabular}
\end{center}
\renewcommand{\arraystretch}{1}
\end{table}

\subsection[Higher order corrections]{Higher order corrections}
\label{sec:corr}

\subsubsection[Higher multipole contributions]{Higher multipole contributions}

In lowest order, only the E1 amplitude contributes to $\chi_c$ states
radiative transitions.  In higher order in $v^2/c^2$ a M2 amplitude
contributes for $J = 1,2$ and an E3 amplitudes is also possible for
the $J=2$ state.  To order $v^2/c^2$ these M2 and E3 corrections to
the dominant E1 term can contribute to angular distributions but
cannot contribute to total decay rates.  This comes from the
orthogonality of terms in the multipole expansion.

It was originally suggested by Karl, Meshkov and Rosner\cite{Karl:1980wm} that
these corrections can be studied by measuring the angular correlations between
the two photons in the transition 
$\psip \rightarrow \chic +\gamma \rightarrow \jpsi + 2\gamma$.
These effects can also be studied for $\chic$ states directly produced in 
hadron collisions, $B$ decays or $\bar pp$ annihilation 
by measuring the combined angular distributions of the photon
and the $l^+ l^-$ pair produced in the subsequent $\jpsi$ decay.
The details of these correlations have been calculated by Sebastian, 
Grotch and Ridener\cite{Sebastian:1992xq}. 

For the photon transition from a ${\chic}_J$ state there are $J+1$ 
normalized helicity amplitudes, $A_{\nu}$. 
Defining $|a| = \sqrt{E1^2+M2^2+E3^2}$, 
$a_1 = E1/|a|$, $a_2 = M2/|a|$ and 
$a_3 = E3/|a|$ the relation between helicity and multipole
amplitudes is:
\begin{equation}
  A_{\nu} = \sum_\ell a_\ell \left (\frac{2\ell+1}{2J+1} \right )^{\frac{1}{2}} \langle \ell,1;1,\nu-1|J,\nu \rangle \,.
\end{equation} 

Allowing for an anomalous magnetic moment $\kappa_c$ and
mixing between the $2S$ and $1D$ states the theoretical predictions for  
\[ 
\psip \rightarrow {\chic}_J + \gamma ~~~{\rm and}~~~ {\chic}_J
\rightarrow \jpsi + \gamma 
\]
are shown in \Table~\ref{tab:M2E3} along with a comparison with
present experimental results.  The S--D mixing parameter is ${\cal
E}_{\rm 2S,1P} X = -\tan{\phi}~{\cal E}_{\rm 1D,1P}$ where $\phi$ is
defined in \Eq~(\ref{eq:sdmix}). In the notation of
\Eq~(\ref{eq:E1vcorr}) the other model dependent parameter is defined
by ${\cal E}_{\rm 1D,1P} Y = \displaystyle \int dr\,r\,
\left(r\frac{du_{12}^{(0)}}{dr}-u_{12}^{(0)}\right)\,u_{11}^{(0)}$.

As can be seen from \Table~\ref{tab:M2E3} a nonzero E3 amplitude in
the $\psip \rightarrow {\chic}_2 + \gamma$ decay is evidence for
S--D mixing in the $\psip$.  Also note that the M2 term is
sensitive to a possible anomalous magnetic moment, $\kappa_c$, for the
charm quark.  The recent BES results\cite{Ablikim:2004qn} for the M2
and E3 contributions do not differ significantly from zero.
Additional high statistics studies of these angular distributions will
be necessary to determine the size of S--D mixing and shed light on the
magnitude of the charm quark magnetic moment.

\subsection{Coupling to virtual decay channels}
\label{sec:em-vloops}

When light quark loops are included in the description of quarkonium
systems, the physical quarkonium states are not pure potential-model
eigenstates and the effects of coupling to real and virtual
heavy-light meson pairs must be included.  Our command of quantum
chromodynamics is inadequate to derive a realistic description of the
interactions that communicate between the $\bar QQ$ and $\bar Qq+\bar
qQ$ sectors.  However, the physical picture is that wave functions
corresponding to physical states are now linear combinations of
potential-model $\bar QQ$ eigenstates plus admixtures of open
heavy-flavour-meson pairs.  The open heavy-flavour pieces have the
spatial structure of bound states of heavy-flavour mesons: they are
virtual contributions for states below threshold (see
\Section~\ref{sec:spopcha} in \Chapter~\ref{chapter:spectroscopy} 
for more details).

\begin{table}[p]
\caption[M2 and E3 multipole amplitudes for radiative transitions
         involving $\chic$ states]
        {M2 and E3 multipole amplitudes for radiative transitions
         involving $\chic$ states. The values of X and Y are model
         dependent and are defined in the text. Note $X=0$ if no
         S--D mixing.}
\label{tab:M2E3}
\renewcommand{\arraystretch}{1.2} 
\begin{center} 
\begin{tabular}{|c|l|c|c|} 
\hline 
 & \multicolumn{3}{|c|}{${\chic}_J \rightarrow \jpsi + \gamma$} \\
\hline 
 $J$ & theory\cite{Sebastian:1992xq} & E835\cite{Ambrogiani:2001jw} & PDG04
\cite{Eidelman:2004wy} \\
\hline 
 $2$ & $a_2 \approx -\frac{\sqrt{5}}{3} \frac{k}{4m_c}(1+\kappa_c)$ 
    & $-0.093 ^{+0.039}_{-0.041} \pm 0.006$ & $-0.13 \pm 0.05$ \\
 $2$ & $a_3 \approx 0$ 
    & $0.020 ^{+0.055}_{-0.044} \pm 0.009$ & $0.011 ^{+0.041}_{-0.033}$ \\
 $1$ & $a_2 \approx - \frac{k}{4m_c} (1 + \kappa_c)$ 
    & $0.002 \pm 0.032 \pm 0.004$ & $-0.002 ^{+0.008}_{-0.017}$ \\
\hline 
\hline 
 & \multicolumn{3}{|c|}{$\psip \rightarrow {\chic}_J + \gamma$} \\
\hline 
$J$ & \multicolumn{2}{|l|}{theory\cite{Sebastian:1992xq}} & BES\cite{Ablikim:2004qn} \\
\hline 
 $2$ & \multicolumn{2}{|l|}{$a_2 \approx -\frac{\sqrt{3}}{2\sqrt{10}} \frac{k}{m_c}
      [(1 + \kappa_c)(1+\frac{\sqrt{2}}{5}X) - i\frac{1}{5}X]/[1-\frac{1}{5\sqrt{2}}X]$}  
  & $-0.051^{+0.054}_{-0.036}$ \\
 $2$ & \multicolumn{2}{|l|}{$a_3 \approx -\frac{12\sqrt{2}}{175} \frac{k}{m_c}X
        [1+\frac{3}{8}Y]/[1-\frac{1}{5\sqrt{2}}X]$}  
  & $-0.027^{+0.043}_{-0.029}$ \\
 $1$ & \multicolumn{2}{|l|}{$a_2 \approx - \frac{k}{4m_c} 
   [(1 + \kappa_c)(1+\frac{2\sqrt{2}}{5}X) + i\frac{3}{10}X]/[1+\frac{1}{\sqrt{2}}X]$}  
  &  \\
\hline
\end{tabular} 
\end{center} 
\renewcommand{\arraystretch}{1} 

\bigskip

\caption[Calculated and observed rates for E1 radiative transitions 
         among charmonium levels]
        {Calculated and observed rates for E1 radiative transitions
         among charmonium levels from Ref.~\cite{Eichten:2004uh}.
         Values in {\it italics} result if the influence of open-charm
         channels is not included.  Measured rates are shown for
         comparison. Experimental values are calculated from world
         averages \cite{Eidelman:2004wy}, except for ${\cal
         B}(\psi'\to\gamma^3P_J)$ whose values are taken from the
         recent CLEO-c measurement\cite{Athar:2004dn}.}
\label{tab:radtranstw}
\renewcommand{\arraystretch}{1.25} 
\begin{center} 
\begin{tabular}{|c|clclcl|} 
\hline
Transition  & $k_\gamma$ & width & $k_\gamma$ & width & $k_\gamma$ & width \\
  & $(\mev )$  & (keV) & $(\mev )$  & (keV) & $(\mev )$  & (keV) \\
\hline
\hline
  & \multicolumn{6}{|c|}{P state} \\ 
\hline
S state  & \multicolumn{2}{c}{$\chi_{c2}$} 
  & \multicolumn{2}{c}{$\chi_{c1}$} & \multicolumn{2}{c|}{$\chi_{c0}$} \\
\hline
$\jpsi$ & $429$ & $\mathit{300} \to 287$  
      & $390$ & $\mathit{228} \to 216$  & $303$ & $\mathit{113} \to 107$ \\[3pt] 
[exp] & & $430\pm 40$
  & & $290\pm 50$ & & $119 \pm 16$ \\[6pt]
$\psi^{\prime}$ & $129$ &  $\mathit{23} \to 23$ 
     & $172$ & $\mathit{33} \to 32$ & $261$ &  $\mathit{36} \to 38$  \\[3pt]
[exp] & & $25.9\pm 2.1$
 & & $25.5\pm 2.2$ & & $26.2\pm 2.6$ \\[6pt]
\hline
\hline
  & \multicolumn{6}{|c|}{P state} \\ 
\hline
D state & \multicolumn{2}{c}{$\chi_{c2}$} 
  & \multicolumn{2}{c}{$\chi_{c1}$} & \multicolumn{2}{c|}{$\chi_{c0}$} \\
\hline
$1\slj{3}{3}{1}(3770)$ & $208$ & $\mathit{3.2} \to 3.9$ 
   & $251$ & $\mathit{183} \to 59$ & $338$ & $\mathit{254} \to 225$ \\[3pt]
$1\slj{3}{3}{1}(3815)$ & $250$ & $\mathit{5.5} \to 6.8$ 
   & $293$ & $\mathit{128} \to 120$ & $379$ & $\mathit{344} \to 371$  \\[6pt]
$1\slj{3}{3}{2}(3815)$ & $251$ & $\mathit{50} \to 40$ 
   & $293$ & $\mathit{230} \to 191$ & & \\[3pt] 
$1\slj{3}{3}{2}(3831)$ & $266$ & $\mathit{59} \to 45$ 
   & $308$ & $\mathit{264} \to 212$ & & \\[3pt]
$1\slj{3}{3}{2}(3872)$ & $303$ & $\mathit{85} \to 45$ 
   & $344$ & $\mathit{362} \to 207$ & & \\[6pt]
$1\slj{3}{3}{3}(3815)$ & $251$ & $\mathit{199} \to 179$ & & & & \\[3pt]
$1\slj{3}{3}{3}(3868)$ & $303$ & $\mathit{329} \to 286$ & & & & \\[3pt]
$1\slj{3}{3}{3}(3872)$ & $304$ & $\mathit{341} \to 299$ & & & & \\[3pt]
\hline
\end{tabular}
\end{center}
\renewcommand{\arraystretch}{1.0} 
\end{table}

Far below heavy flavour threshold, the nonrelativistic potential model
is a good approximation to the dynamics of the $\bar QQ$ system.  For
excited states above the first few levels, the coupling of $\bar QQ$
to heavy-flavour-meson pairs modifies wave functions, masses, and
transition rates. In particular, this modifies electromagnetic
transition rates considered in the previous subsections.  In addition
to these contributions, which involved photon coupling to a heavy
(anti)quark, the contributions of light quark currents can no longer
be neglected.  The mass differences among the $\bar Qu$, $\bar Qd$ and
$\bar Qs$ mesons, induce large SU(3) symmetry breaking effects. This
destroys the cancellation among the light quark EM current
contributions.

To compute the E1 radiative transition rates, we must  
take into account both the standard $(\bar QQ)$ $\to$ $(\bar QQ)\gamma$ 
transitions and the transitions between (virtual) decay channels in 
the initial and final states.  This second set of transitions contains 
light quark contributions for states near threshold.  
Recently, the effects of configuration mixing on
radiative decay rates in the $\cc$ system were reexamined\cite{Eichten:2004uh} 
within the Cornell coupled-channel formalism. 
A full outline of the calculational procedure appears in
Refs.~\cite{Eichten:1978tg,Eichten:1979ms}.  
(In particular, see \Section~IV.B of Ref.~\cite{Eichten:1979ms}.) 

Expectations for E1 transition rates among spin-triplet levels are
shown in \Table~\ref{tab:radtranstw}.  Both the rates
calculated between single-channel potential-model eigenstates (in
italics) and the rates that result from the Cornell coupled-channel
model are shown, to indicate the influence of the open-charm channels.

The 1\slj{3}{3}{1} transition rates at the mass of $\psi(3770)$
and at the predicted 1\slj{3}{3}{1} centroid, $3815\mev$, are shown.  
For the $\psi(3770)$, with its total width of about $24\mev$, the
$1\slj{3}{3}{1}(3770) \to \chi_{c0}\,\gamma(338)$ transition might
someday be observable with a branching fraction of 1\%.
For the 1\slj{3}{3}{2} and 1\slj{3}{3}{3} levels, the
radiative decay rates were calculated at the predicted 1\slj{3}{3}{1} centroid,
$3815\mev$, at the mass calculated for the states ($3831\mev$ and
$3868\mev$, respectively), and at the mass of $X(3872)$.  
The model reproduces the trends of transitions to and from the $\chi_{c}$ 
states in broad outline.  For these low-lying states, the mixing through 
open-charm channels results in a mild reduction of the rates. 

This study was done in the Cornell coupled channel model. It would
be useful to do a similarly detailed study of these effects in other models.

\subsection[$B_c$ states]{$B_c$ states}

Quarkonium systems with unequal quark and antiquark masses, \ie $B_c$
mesons, are theoretical interesting, but are not easily accessible in
$e^+e^-$ collisions.  They can be produced in significant numbers in
hadron collisions (see \Chapter~\ref{chapter:production},
\Section~\ref{sec:prodsec-Bc}).  CDF has reported the discovery of the
ground state $B_c$ meson via its semileptonic weak decay
\cite{Abe:1998wi}.  Theoretical calculations for E1 and M1 radiative
transitions have been presented by a number of
authors\cite{Eichten:1994gt,Gershtein:1994jw,Fulcher:1998ka,Ebert:2002pp}
even though the whole excitation spectrum remains to be observed
experimentally.

%%%%%%%%%%%%%%  
%HT
%%%%%%%%%%%%%% 
\section[Hadronic transitions]{Hadronic transitions $\!$\footnote{Authors: 
D.~Z.~Besson, A.~Deandrea, F.~A.~Harris, Y.-P.~Kuang, S.~L.~Olsen}} 
\label{sec:HT}

\subsection[Theoretical approaches]{Theoretical approaches} 
Hadronic transitions (HTs)
\begin{equation}
\Phi_i\to\Phi_f+h
\end{equation}
are important decay modes of heavy quarkonia [$\Phi_i$, $\Phi_f$
and $h$ stand for the initial-, final-state quarkonia and the
emitted light hadron(s)]. For instance, the branching ratio of
$\psi^\prime\to J/\psi+\pi+\pi$ is approximately $50\%$. In the
$c\bar{c}$ and $b\bar{b}$ systems, the typical mass difference
$M_{\Phi_i}-M_{\Phi_f}$ is around a few hundred MeV, so that the
typical momentum of $h$ is low. In the single-channel picture,
the light hadron(s) $h$ are converted from the
gluons emitted by the heavy quark $Q$ or antiquark $\bar{Q}$ in
the transition. The typical momentum of the emitted gluons is
too low for perturbative QCD to be reliable.
Certain nonperturbative approaches are thus needed for studying
HTs. In the following, we briefly review two feasible approaches:
namely, the {\it QCD multipole expansion} (QCDME) and the {\it
Chiral Lagrangian for Heavy Mesons}.

\null\noindent
{\it A.~~ QCD Multipole expansion}

Heavy $Q\bar{Q}$ bound states can be calculated by solving the
Schr\"odinger equation with a given potential model.
For $c\bar{c}$ and $b\bar{b}$ quarkonia, the typical radius is
$a=\sqrt{\langle r^2\rangle}\sim {\cal O}(10^{-1})$~fm.
With such a small radius, the idea of QCDME can be applied to the
soft gluon emissions in HTs. QCDME is an expansion in powers of
$~\bm x\cdot\bm\nabla~$ operating on the gluon field, where $\bm
x$ is the separation between $Q$ and $\bar{Q}$ in the quarkonium.
For a gluon with a typical momentum $k\sim{\rm few~hundred~MeV}$,
the expansion parameter is actually $~ak\sim {\cal O}(10^{-1})$, ensuring
convergence\footnote{We know from classical
electrodynamics that the coefficient of the $(ak)^l$ term in the
multipole expansion contains a factor
$\displaystyle\frac{1}{(2l+1)!!}$. Hence the expansion actually
works better than what might be expected by simply estimating the size
of $(ak)^l$.}. Note that the convergence of QCDME does not depend
on the value of the QCD coupling constant. 

QCDME has been studied by many authors \cite{Gottfried,ME,VZNS,Yan,KYF}. The gauge invariant
formulation is given in Ref. \cite{Yan}. Let $\psi(x)$ and $A^a_\mu(x)$ be the quark and gluon fields.
Following Refs. \cite{Yan,KYF}, we introduce
\begin{equation}
\Psi({\bm x},t)\equiv U^{-1}({\bm x},t)\psi(x),~~~~
\frac{\lambda_a}{2}A^{a\prime}_\mu({\bm x},t)\equiv
U^{-1}({\bm x},t)\frac{\lambda_a}{2}A^a_\mu(x)U({\bm x},t)
-\frac{i}{g_s}U^{-1}({\bm x},t)\partial_\mu U({\bm x},t),
\end{equation}
with
\begin{equation}
U({\bm x},t)=P\exp\bigg[ig_s\int^{\bm x}_{\bm X}
\frac{\lambda_a}{2}{\bm A}^a({\bm x}^\prime,t)\cdot
d{\bm x}^\prime\bigg],
\end{equation}
in which $P$ is the path-ordering operation, the path is
along the straight-line connecting the two ends, and ${\bm
X}\equiv {(\bm x_1+\bm x_2)}/2$ is the c.o.m. coordinate of
$Q$ and $\bar{Q}$. It is shown in Ref. \cite{Yan} that, in the Lagrangian,
$\Psi({\bm x},t)$ serves as the {\it dressed} ({\it constituent}) quark field.
Now we make the multipole expansion \cite{Yan}
\begin{equation}
A^{a\prime}_0({\bm x},t)=A^{a\prime}_0({\bm X},t)-({\bm x}-{\bm X})\cdot
{\bm E}^a({\bm f X},t)+\cdots,~~
{\bm A}^{a\prime}({\bm X},t)=-\frac{1}{2}({\bm x}-{\bm X})
\times {\bm B}^a({\bm X},t)+\cdots,
\end{equation}
where ${\bm E}^a$ and ${\bm B}^a$ are colour-electric and
colour-magnetic fields, respectively. The Hamiltonian is then \cite{Yan}
\begin{equation}
H^{\rm eff}_{\rm QCD}=H^{(0)}_{\rm QCD}+H^{(1)}_{\rm QCD},
\label{eq:4.1.4}
\end{equation}
with $H^{(0)}_{\rm QCD}$
the sum of the kinetic and potential energies of the heavy quarks, and
\begin{equation}
H^{(1)}_{\rm QCD}=H_1+H_2,~~~~
H_1\equiv Q_a A^a_0(\bm X,t),~~~~
H_2\equiv -{\bm d}_a\cdot{\bm E}^a(\bf X,t)-
{\bm m}_a\cdot{\bf B}^a(\bm X,t)+\cdots,\hspace{0.5
cm}
\end{equation}
in which $Q_a$, ${\bm d}_a$, and ${\bm m}_a$ are the colour charge,
colour-electric dipole moment, and colour-magnetic dipole moment of the
$Q\bar{Q}$ system, respectively.  \Eq[b]~(\ref{eq:4.1.4}) is regarded as
an effective Hamiltonian \cite{Yan}.  Considering that the heavy quark
may have an anomalous magnetic moment, we take $g_E$ and $g_M$ to
denote the effective coupling constants for the electric and magnetic
multipole gluon emissions (MGE), respectively.

We shall take $H^{(0)}_{\rm QCD}$ as the zeroth order
Hamiltonian, and take $H^{(1)}_{\rm QCD}$ as a perturbation. This is
different from the ordinary perturbation theory since
$H^{(0)}_{\rm QCD}$ is not a free field Hamiltonian.
The general formula for the $S$ matrix element
in this expansion has been given in Ref. \cite{KYF}, which is
\begin{equation}
\langle f|S|i\rangle=-i2\pi\delta(E_f+\omega_f-E_i)
\bigg\langle  f\bigg|H_2\frac{1}{E_i-H^{(0)}_{\rm QCD}
+i\partial_0-H_1}\cdots \frac{1}{E_i-H^{(0)}_{\rm QCD}+i\partial_0-H_1}H_2\bigg|i\bigg \rangle,
\label{eq:4.1.6}
\end{equation}
where $\omega_f$ is the energy of the emitted gluons. Explicit
evaluations of the $S$ matrix elements
in various cases will be presented in \Section~\ref{sec:4.2}.

\null\noindent
{\it B.~~Chiral Lagrangian for heavy mesons}

In the effective Lagrangian approach one can
construct a heavy meson multiplet field analogous to the one introduced for
heavy-light mesons. Symmetry-breaking terms can be easily
added to the formalism as we shall see in the following.
As in the single heavy quark case, an effective
Lagrangian describing the low-momentum interactions of heavy quarkonia with
light mesons can be written down. The heavy quarkonium multiplets are
described by a simple trace formalism \cite{hq5}.
Parity $P$ and charge conjugation $C$, which determine selection rules for
electromagnetic and hadronic transitions
are exactly conserved quantum numbers for quarkonium, together with $J$.
If spin-dependent interactions are neglected, it is natural to describe the
spin singlet $m\,^{1}l_{J}$ and the spin triplet $m\,^{3}l_{J}$ by means of a
single multiplet $J(m,l)$. For the case $l=0$, when the triplet $s=1$
collapses into a single state with total angular momentum $J=1$,
this is readily realized:
\begin{equation}
J= \frac{(1+\HTslash v)}{2}[H_{\mu}\gamma^\mu-\eta\gamma_5]
\frac{(1-\HTslash v)}{2}~~.
\end{equation}
Here $v^{\mu}$ denotes the four velocity associated to the multiplet $J$;
$H_{\mu}$ and $\eta$ are the spin 1 and spin 0 components respectively; the
radial quantum number has been omitted.
The expressions for the general wave $J^{\mu_1 \ldots \mu_l}$
can be found in Appendix C of Ref. \cite{report}.

For illustrative purpouses let us start by considering
radiative transitions, whose analysis can be easily carried out in terms of the multiplet field introduced above. 
The Lagrangian for radiative decays is:
\begin{equation}
{\cal L}=\sum_{m,n} \delta (m,n) \langle{\overline{J}}(m)\; J_{\mu} (n)\rangle v_{\nu}
F^{\mu\nu} + h.c.,
\label{eq:q3.1}
\end{equation}
where a sum over velocities is understood, $F^{\mu\nu}$ is the electromagnetic
tensor, the indices $m$ and $n$ represent the radial quantum numbers, $J(m)$
stands for the multiplet with radial number $m$ and $\delta (m,n)$ is a
dimensional parameter (the inverse of a mass), to be fixed from
experimental data and which also depends on the heavy flavour.
The Lagrangian (\ref{eq:q3.1}) conserves parity and charge conjugation
and is invariant under the spin transformation. It reproduces
the electric dipole selection rules $\Delta \ell=\pm 1$ and $\Delta s=0$.
It is straightforward to obtain the corresponding radiative widths:
\begin{equation}
\Gamma (^3P_J \to \;^3S_1 \gamma)= {\frac {\delta^2} {3\pi}} k^3 {\frac
{M_{S_1}} {M_{P_J}}},
\label{eq:q3.2}
\end{equation}
\begin{equation}
\Gamma (^3S_1 \to \;^3P_J \gamma)= {\frac {(2J+1)} {9\pi}} \delta^2 k^3
{\frac {M_{S_1}} {M_{P_J}}},
\label{eq:q3.3}
\end{equation}
\begin{equation}
\Gamma (^1P_1 \to \;^1S_0 \gamma)= {\frac {\delta^2} {3\pi}} k^3 {\frac
{M_S} {M_P}},
\label{eq:q3.4}
\end{equation}
where $k$ is the photon momentum. Once the radial numbers $n$ and $m$
have been fixed, the Lagrangian (\ref{eq:q3.1}) describes four no
spin-flip transitions with a single parameter; this allows three
independent predictions.  The previous decay widths can be compared
with those of the electric transitions of Sec.~\ref{sec:emssec-et} and
in particular with formula (\ref{RD:E1trans}). The ratio of the masses
in the previous widths should be put to one in the nonrelativistic
limit and the free parameter $\delta$ of the effective Lagrangian
encodes the information of the overlap integral of equation
(\ref{eq:Eif0}).

The effective heavy-meson description of quarkonium does not seem
to present special advantages to describe heavy quarkonium annihilation.
In the following we shall concentrate on quarkonium hadronic transitions.

The heavy quark spin symmetry leads to general relations for the differential
decay rates in hadronic transitions among quarkonium states that
essentially reproduce the results of a QCD double multipole expansion
for gluonic emission. Further use of chiral symmetry leads to differential
pion decay distributions valid in the soft regime \cite{noi0,noi1}.
At the lowest order in the
chiral expansion for the emitted pseudoscalars we find a selection
rule allowing only for even (odd) number of emitted pseudoscalars for
transitions between quarkonium states of orbital angular momenta different by
even (odd) units. Such a rule can be violated by higher chiral terms, by
chiral breaking, and by terms breaking the heavy quark spin symmetry. Specialization
to a number of hadronic transitions reproduces (by elementary tensor
construction) the known results from the expansion in gluon multipoles,
giving a simple explanation for the vanishing of certain coefficients, 
which would otherwise be allowed in the chiral expansion. In certain cases,
such as for instance $~^3P_0 \to ~^3P_2 \pi\pi$, $~^3P_1 \to ~^3P_2 \pi\pi$,
or D--S transitions via $2\pi$, the final angular and mass distributions are
uniquely predicted from heavy quark spin and the
lowest-order chiral expansion.

An important class of hadronic transitions between heavy-quarkonium states
is provided by the decays with emission of two pions, for example:
\begin{equation}
\psi' \rightarrow J/\psi \, \pi \pi \,.
\label{eq:4.1.8}
\end{equation}
To describe these processes one can use the chiral symmetry for the
pions and the heavy-quark spin symmetry for the heavy states.
The first of these is expected to
hold when the pions have small energies. 
We notice that the velocity superselection rule applies
at $q^2=q^2_{\rm max}$, when the energy transfer to the pion
is maximal. Therefore, we expect these approximations to be valid in the whole
energy range only if $q^2_{\rm max}$ is small.

Nonetheless a number of interesting properties of these transitions can
be derived  on the basis of the heavy quark symmetry alone. Therefore,
before deriving the pion couplings by means of chiral symmetry,
we discuss the implications of the heavy quark spin symmetry in
hadronic transitions.

As an example, we consider transitions of the type $~^3S_1 \to ~^3S_1+h$
and $~^1S_0 \to ~^1S_0+h$, where $h$ can be light hadrons, photons, etc.
By imposing the heavy quark spin symmetry, one is led to describe
these processes by an interaction Lagrangian:
\begin{equation}
{\cal L}_{SS'}=\langle J'\bar{J} \rangle\Pi_{SS'}+h.c.~~,
\end{equation}
where the dependence upon the pion field is contained in the 
as-yet-unspecified operator $\Pi_{SS'}$. It is immediate to derive from
${\cal L}_{SS'}$ the averaged modulus square matrix elements for
the transitions $^3S_1\to^3S_1+h$ and $^1S_0\to^1S_0+h$ with an
arbitrary fixed number of pions in the light final state $h$.
We obtain:
\begin{equation}
|{\cal M} (^3S_1\to^3S_1+h)|^2_{\rm av.}=
|{\cal M} (^1S_0\to^1S_0+h)|^2_{\rm av.}=4 M_S M_{S'}|\Pi_{SS',h}|^2,
\end{equation}
where $M_S$ and $M_S'$ are the average masses of the two
S-wave multiplets; $\Pi_{SS',h}$ is the appropriate
tensor for the emission of the light particles $h$, to be calculated from the
operator $\Pi_{SS'}$. By denoting with $d\Gamma$ the generic differential
decay rate, we have:
\begin{equation}
d\Gamma(^3S_1\to^3S_1+h)=
d\Gamma(^1S_0\to^1S_0+h)~~.
\end{equation}

This is the prototype of a series of relations, which can be derived
for hadronic transitions as a consequence of the spin independence of the
interaction terms. In all the known cases they coincide with those calculated
in the context of a QCD double multipole expansion. We note, however, that
we do not even need to specify the nature of the operator $\Pi$,
which may depend on light fields different from the pseudoscalar
mesons (\eg the photon,
or a light hadron, etc), provided that the interaction term we are building
is invariant under parity, charge conjugation, and the other
symmetries relevant to the transition considered.

A useful symmetry that can be used in processes involving light quarks
is the chiral symmetry. It is possible to build up an
effective Lagrangian, which allows to study transitions among quarkonium states
with emissions of soft light pseudoscalars,
considered as the Goldstone bosons of the spontaneously broken chiral
symmetry.

The light mesons are described as pseudo-Goldstone bosons, included in the
matrix $\Sigma=\xi^2$, where we use the standard notation of chiral
perturbation theory. Frequently occurring quantities are the
functions of $\xi$ and its derivatives ${\cal A}_\mu$ and ${\cal V}_\mu$
given by:
\begin{equation}
{\cal V}_\mu=\frac{1}{2}\left(\xi^\dagger\partial_\mu \xi+ \xi
\partial_\mu \xi^\dagger\right)~~~~~~{\rm and}~~~~~~
{\cal A}_\mu=\frac{1}{2}\left(\xi^\dagger\partial_\mu \xi - \xi
\partial_\mu \xi^\dagger\right).
\end{equation}
The octet of vector resonances ($\rho$, etc.) can be introduced as the gauge
multiplet associated with the hidden group SU(3)$_H$ (see Ref. \cite{bando}),
designated as $\rho_\mu$ in the following.

By imposing the heavy quark spin symmetry, parity and charge
conjugation invariance, and by assuming that the pseudoscalar meson
coupling are described by the lowest order (at most two derivatives)
chiral invariant operators, we can establish the following selection
rules for hadronic transitions:
\begin{eqnarray}
&& {\rm even~ number~ of~ emitted~ pseudoscalars} \leftrightarrow\Delta
l=0,2,4,... 
\nn\\
&& {\rm odd~ number~ of~ emitted~ pseudoscalars} 
\leftrightarrow
\Delta l=1,3,5,...
\end{eqnarray}

In fact the spin independent operator describing $\Delta l=0,2,4,...$
transitions has charge conjugation $C=+1$.
On the other hand, the lowest order, chiral invariant terms with
positive charge conjugation are:
\begin{equation}
\langle {\cal A}_\mu {\cal A}_\nu \rangle,~~~~~~~~
\langle ({\cal V}_\mu-\rho_\mu) ({\cal V}_\nu-\rho_\nu) \rangle,
\end{equation}
whose expansion contains an even number of pseudoscalar mesons.
Spin independence of the interaction, on the other hand, requires that the
$\Delta l=1,3,5,...$ transitions are described by $C=-1$ operators.
At the lowest order we can form just one chiral invariant term with $C=-1$:
\begin{equation}
\langle {\cal A}_\mu ({\cal V}_\nu-\rho_\nu) \rangle,
\end{equation}
whose expansion contains an odd number $(\ge 3)$of pseudoscalar mesons.

This selection rule is violated at higher orders of the chiral expansion
or by allowing for terms that explicitly break the heavy quark or
the chiral symmetries.

To further characterize the hadronic transitions respecting chiral
symmetry, we consider below explicit expressions for the most general
operators $\Pi_{ll'}$. For simplicity, we limit ourselves to those
contributing
to two or three pion emissions:
\begin{eqnarray}
\Pi_{SS'} &=&
A_{SS'} \langle {\cal A}_\rho {\cal A}^\rho \rangle
+B_{SS'} \langle (v\cdot {\cal A})^2 \rangle,
\nn\\
\Pi_{PS}^\mu &=&
D_{PS}~\epsilon^{\mu\nu\rho\sigma} v_\nu \langle {\cal A}_\rho
({\cal V}_\sigma-\rho_\sigma) \rangle,
\nn\\
\Pi_{PP'}^{\mu\nu} &=&
A_{PP'} \langle {\cal A}_\rho {\cal A}^\rho \rangle g^{\mu\nu}
+B_{PP'} \langle (v\cdot {\cal A})^2 \rangle g^{\mu\nu}
+C_{PP'}\langle {\cal A}^\mu {\cal A}^\nu \rangle,
\nn\\
\Pi_{DS}^{\mu\nu} &=&
C_{DS}\langle {\cal A}^\mu {\cal A}^\nu \rangle.
\end{eqnarray}
The constants $A_{ll'}$, $B_{ll'}$, $C_{ll'}$ and $D_{ll'}$ are
arbitrary parameters of dimension $({\rm mass})^{-1}$, to be fixed
from experiment. One can easily derive amplitudes, decay rates and
distributions for the corresponding hadronic transitions.

For instance, the amplitude for the decay (\ref{eq:4.1.8}) is given by:
\begin{equation}
{\cal M}(^3S_1\to^3S_1+\pi\pi)=\frac{4i\sqrt{M_S M_{S'}}}{f_\pi^2}
\epsilon' \cdot \epsilon^* \left( A_{SS'} p_1 \cdot
p_2 +B_{SS'} v\cdot p_1 v\cdot p_2 \right)
\label{eq:4.1.16}
\end{equation}
where $\epsilon$ and $\epsilon '$ are the polarisation vectors of
quarkonium states; $p_1$, $p_2$ are the momenta of the two pions.
It is well known that the use of chiral symmetry arguments
leads to a general amplitude for the process in question, which
contains a third independent term given by:
\begin{equation}
C_{SS'} \frac{4 i \sqrt{M_S M_{S'}}}{f_\pi^2}
\left(\epsilon'\cdot p_1 \epsilon^*\cdot p_2+
\epsilon'\cdot p_2 \epsilon^*\cdot p_1 \right)~~.
\end{equation}
In the nonrelativistic limit in QCDME,
Yan \cite{Yan} finds $C_{SS'}=0$. It is interesting
to note that, within the present formalism, this result is an
immediate consequence of the chiral and heavy quark spin symmetries.
However, these symmetries are not exact 
and corrections to the symmetry limit are expected.

In the chiral Lagrangian (CL) approach, the $\pi^0 - \eta -
\eta^\prime$ mixings can be derived, which should be taken into
account in predicting single pseudoscalar meson transitions of heavy
quarkonia (cf. \Section~\ref{sec:4.2}). Let us define
\begin{equation}
{\hat m}\equiv
\left (\begin{array}{ccc}
m_u & 0 & 0  \\
0 & m_d & 0  \\
0 & 0 & m_s
\end{array}\right ).
\end{equation}
The Lagrangian that gives mass to the pseudoscalar octet (massless in the chiral limit) and causes
$\pi^0-\eta$ mixing is
\begin{equation}
{\cal L}_{m}=\lambda_0 \langle\hat m (\Sigma + \Sigma^{\dagger})\rangle,
\end{equation}
and that giving rise to the mixing of $\eta^\prime$ with $\pi^0$ and $\eta$ is
\begin{equation}
{\cal L}_{\eta\eta'}=\frac{i f_\pi}{4} \tilde\lambda 
\langle\hat m(\Sigma - \Sigma^{\dagger})\rangle
\eta',
\end{equation}
where $\hat\lambda$ is a parameter with the dimension of a mass. At first order in 
the mixing angles the physical states $\tilde\pi^0$, $\tilde\eta$, and $\tilde\eta^\prime$ 
determined from the above Lagrangians are:
\begin{equation}
{\tilde \pi}^0 = \pi^0 +\epsilon \eta +\epsilon' \eta',~~~~~~
{\tilde \eta}=\eta -\epsilon \pi^0 +\theta \eta',~~~~~~ \\
{\tilde \eta}'=\eta' -\theta \eta -\epsilon' \pi^0,
\label{eq:4.1.21}
\end{equation}
in which the mixing angles are
\begin{equation}
\epsilon=\frac{(m_d -m_u) \sqrt{3}}{4 (m_s-\dd\frac{m_u+m_d}{2})},~~~~~~
\epsilon'=\frac{\tilde\lambda (m_d -m_u)}{\sqrt{2} (m^2_{\eta'}- 
m^2_{\pi^0})},~~~~\theta =\sqrt{\frac{2}{3}}~~~\frac{\tilde\lambda \dd\left(m_s-\frac{m_u+m_d}{2}
\right)}{m^2_{\eta'}- m^2_{\eta}}.
\label{eq:4.1.22}
\end{equation}

\subsection[Predictions for hadronic transitions in the single-channel
  approach]{Predictions for hadronic transitions in the single-channel
  approach}
\label{sec:4.2}

\begin{figure}[!h]
\begin{center}
\includegraphics[width=0.35\textwidth]{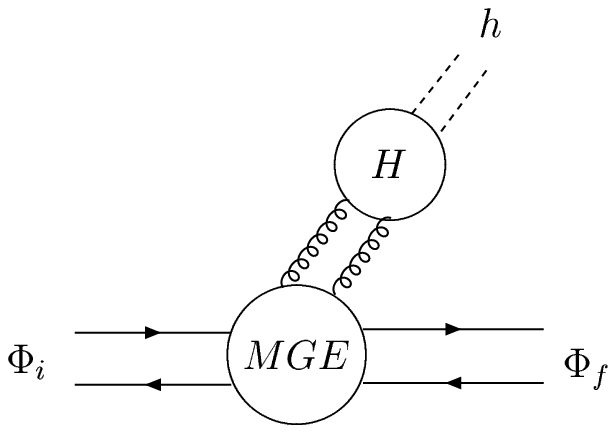}
\end{center}
\caption{Diagram for a typical hadronic transition in the single-channel QCDME approach.
\label{fig:htFig}}
\end{figure}

In this section, we give the predictions for HTs in the single-channel
approach.  In this approach, the amplitude of HT is diagrammatically
shown in \Figure~\ref{fig:htFig} in which there are two complicated
vertices: namely, the MGE vertex of the heavy quarks and the vertex of
hadronization (H) describing the conversion of the emitted gluons into
light hadrons. In the following, we shall treat them separately.

Let us first consider the HT processes $n_i^3S_1\to
n_f^3S_1+\pi+\pi$. To lowest order, these are double electric-dipole
transitions (E1E1). The transition amplitude can be obtained from the
$S$ matrix element (\ref{eq:4.1.6}). After some algebra, we obtain
\cite{KYF,Yan,KY81}
\begin{equation}
{\cal M}_{E1E1}=i\frac{g_E^2}{6}\sum_{KLK^\prime L^\prime}
\langle\Phi_f h|{\bm x}\cdot{\bm E}
|K L\rangle
\bigg
\langle  K L\bigg|\frac{1}{E_i-H^{(0)}_{\rm QCD}-iD_0}\bigg|K^\prime L^\prime
\bigg \rangle
\langle K^\prime L^\prime
|{\bm x}\cdot{\bm E}
|\Phi_i\rangle,
\label{eq:4.2.1}
\end{equation}
where $(D_0)_{bc}\equiv \delta_{bc}\partial_0-g_s f_{abc}A^a_0$, and
$|KL\rangle$ is the intermediate state with principal quantum number
$K$ and orbital angular momentum $L$. According to the angular
momentum selection rule, $L=L^\prime=1$. The intermediate states in
the HT are the states after the emission of the first gluon and before
the emission of the second gluon (cf. \Figure~\ref{fig:htFig}),
\ie they are states with a gluon and a colour-octet $Q\bar{Q}$.  These
are the so-called hybrid states. It is difficult to calculate these
states from the first principles of QCD. So we shall take a reasonable
model for it. The model should {\it reasonably reflect the main
properties of the hybrid states} and should {\it contain as few free
parameters as possible} in order not to affect the predictive power of
the theory. The quark confining string (QCS) model \cite{QCS}
satisfies these requirements \footnote{Another possible model
satisfying the requirements is the MIT bag model for the hybrid
states, which can also lead to reasonable predictions \cite{LK87}.}
Explicit calculations with the QCS are given in Ref. \cite{KY81}; the
transition amplitude (\ref{eq:4.2.1}) then becomes
\begin{equation}
{\cal M}_{E1E1}=i\frac{g_E^2}{6}\sum_{KL}\frac{\langle\Phi_f|x_k|KL\rangle
\langle KL| x_l|\Phi_i\rangle}
{E_i-E_{KL}}\langle \pi\pi|E^a_k E^a_l|0\rangle,
\label{eq:4.2.2}
\end{equation}
We see that, in this approach, the transition amplitude contains two
factors: namely, the heavy quark MGE factor (the summation) and the H
factor $\langle\pi\pi|E^a_k E^a_l|0\rangle$. The first factor can be
calculated for a given potential model. Let us now consider the second
factor. Its scale is the light hadron mass scale, which is very low
(highly nonperturbative), and there is, therefore, no currently
reliable way of calculating it from the first principles of QCD. Thus
we take a phenomenological approach based on PCAC and the soft pion
technique in Ref. \cite{BrownCahn}. From the standard tensor
reduction, this H factor can be written as \cite{KY81}
\begin{equation}
\frac{g_E^2}{6}\langle\pi_\alpha(q_1)\pi_\beta(q_2)|E^a_k E^a_l|0\rangle
=\frac{\delta_{\alpha\beta}}{\sqrt{(2\omega_1)(2\omega_2)}}
\bigg[C_1\delta_{kl}q^\mu_1 q_{2\mu}
+C_2\bigg(q_{1k}q_{2l}+q_{1l}q_{2k}-\frac{2}{3}\delta_{kl}{\bm q}_1
\cdot{\bm q}_2\bigg)\bigg],
\label{eq:4.2.3}
\end{equation}
where $C_1$ and $C_2$ are two unknown constants. For a given $\pi\pi$
invariant mass $M_{\pi\pi}$, the $C_1$ term is isotropic (S-wave),
while the $C_2$ term is angular dependent (D-wave).  In the
nonrelativistic single-channel (NRSC) approach, orbital angular
momentum conservation leads to the conclusion that the MGE factor is
proportional to $\delta_{kl}$. Thus only the $C_1$ term contributes to
the S-state to S-state transitions\footnote{This is consistent
with the CL approach in the nonrelativistic limit ($v=0$)
[cf. (\ref{eq:4.1.16})].}. In this case, the $n_i^3S_1\to
n_f^3S_1+\pi+\pi$ transition rate can be expressed as \cite{KY81}
\begin{equation}
\Gamma(n_i^3S_1\to n_f^3S_1~\pi~\pi)=|C_1|^2G|f^{111}_{2010}|^2,
\label{eq:4.2.4}
\end{equation}
where $G$ is a phase-space factor given in Ref. \cite{KY81} and
\begin{equation}
f^{LP_iP_f}_{n_il_in_fl_f}\equiv \sum_{K}\frac{\int R_f(r)r^{P_f}R^*_{KL}(r)r^2dr
\int R^*_{KL}(r^\prime)
r^{\prime P_i}R_i(r^\prime)r^{\prime 2}dr^\prime}{M_i-E_{KL}},
\label{eq:4.2.5}
\end{equation}
with $R_i$, $R_f$, and $R_{KL}$ the radial wave functions of
the initial, final, and intermediate states, respectively.

There is only one overall unknown constant $C_1$ left in this
transition amplitude, which can be determined by taking the
well-measured HT rate $\Gamma(\psi^\prime\to J/\psi\pi\pi)$.
The updated experimental values are \cite{Eidelman:2004wy}
\begin{equation}
\Gamma_{\rm tot}(\psi^\prime)=281\pm 17~{\rm keV},~~
{\cal B}(\psi^\prime\to J/\psi\pi^+\pi^-)=(31.7\pm 1.1)\%,~~
{\cal B}(\psi^\prime\to J/\psi\pi^0\pi^0)=(18.8\pm 1.2)\%.
\label{eq:4.2.6}
\end{equation}

\begin{table}[t]
\caption[The values of $|C_1|^2$ and the predicted $\pi\pi$ transition
         rates]
        {The values of $|C_1|^2$ and the predicted $\pi\pi$ transition
         rates (in keV) determined for the $\Upsilon$ system using the
         Cornell model and the BGT model. The corresponding updated
         experimental values of the transition rates
         \cite{Eidelman:2004wy} are also listed for comparison.
         $\pi\pi$ stands for the sum over all the $\pi^+\pi^-$ and
         $\pi^0\pi^0$ channels.}
\label{tab:c1ht}
\begin{center}
\tabcolsep 0.6cm
\begin{tabular}{cccc}
\hline\hline
 &Cornell&BGT&Expt.\\
 \hline
$~~~~~~|C_1|^2\hspace{2.8cm}$&$83.4\times 10^{-6}$&$67.8\times 10^{-6}$&\\
$\Gamma(\Upsilon^\prime\to\Upsilon\pi\pi)$~(keV)~~&8.6~~~&7.8~~~&$~~~12.0~\pm 1.8~~~~~~$\\
$\Gamma(\Upsilon^{\prime\prime}\to\Upsilon\pi\pi)$~(keV)&0.44&1.2~~~&$1.72\pm 0.35$\\
$\Gamma(\Upsilon^{\prime\prime}\to\Upsilon^\prime\pi\pi)$~(keV)&0.78&0.53&$1.26\pm 0.40$\\
\hline\hline
\end{tabular}
\end{center}
\end{table}
Given these, we can then predict all the S-state to S-state
$\pi\pi$ transitions rates in the $\Upsilon$ system.  Let us take the
Cornell \cite{Eichten:1978tg,Eichten:1979ms} and the
Buchm\"uller--Grunberg--Tye (BGT)
\cite{exdec:buch81,Buchmuller:1980bm} potential models as examples to
show the extracted $|C_1|$ values and the predicted rates in the
$\Upsilon$ system.  The results are listed in
\Table~\ref{tab:c1ht}\footnote{The updated results listed in
    \Table~ref{tab:c1ht} are roughly larger than those in Ref.~\cite{KY81}
    by a factor of 1.3 since the updated input data $\Gamma(\psi^\prime\to
    J/\psi\pi\pi)$ is larger than the old experimental value used in
    Ref. \cite{KY81} by the same factor of 1.3.} in which the experimental
errors are dominated by the uncertainty of the total width. We see
that the BGT model predicted ratios
$\Gamma(\Upsilon^{\prime\prime}\to\Upsilon\pi\pi)/
\Gamma(\Upsilon^\prime\to \Upsilon\pi\pi)\approx 1.2/7.8=0.15$ and
$\Gamma(\Upsilon^{\prime\prime}\to\Upsilon^\prime\pi\pi)/\Gamma(\Upsilon^\prime
\to\Upsilon\pi\pi)$ $\approx 0.53/7.8=0.07$ are close to the
corresponding experimental values 1.72/12.0=0.14 and
1.26/12.0=0.11. However, the predicted absolute partial widths are
smaller than the experimental values by roughly a factor of
50--75\%. Moreover, when the $M_{\pi\pi}$ distributions are
considered, the situation will be more complicated.  We shall deal
with these issues in \Section~\ref{sec:4.3}.

Note that the phase space factor $G$ in
$\Upsilon^{\prime\prime}\to\Upsilon\pi\pi$ is much larger than that in
$\Upsilon^\prime\to \Upsilon\pi\pi$, $G(\Upsilon^{\prime\prime}\to$
$\Upsilon\pi\pi)/$ ${G(\Upsilon^\prime\to \Upsilon\pi\pi)}$ $=33$
\cite{KY81}. One may naively expect that
$\Gamma(\Upsilon^{\prime\prime}\to\Upsilon\pi\pi)>\Gamma(\Upsilon^\prime\to
\Upsilon\pi\pi)$. However, we see from \Table~\ref{tab:c1ht} that the
measured $\Gamma(\Upsilon^{\prime\prime}\to\Upsilon\pi\pi)
/\Gamma(\Upsilon^\prime\to \Upsilon\pi\pi)\approx 0.14$. The reason
why the predicted ratio is close to the experimental value is that the
contributions from various intermediate states to the overlap
integrals in the summation in $f^{111}_{3010}$ [cf.  (\ref{eq:4.2.5})]
{\it drastically cancel} each other due to the fact that the
$\Upsilon^{\prime\prime}$ wave function contains two nodes.  This is
{\it characteristic} of such intermediate state models (QCS or bag
model).

The decays $~n_i^3S_1\to n_f^3S_1+\eta~$ are
dominated by E1M2 transitions. We can predict the
ratios $R^\prime\equiv\Gamma(\Upsilon^\prime\to\Upsilon\eta)/\Gamma(\psi^\prime\to J/\psi\eta)$ and 
$R^{\prime\prime}\equiv\Gamma(\Upsilon^{\prime\prime}\to\Upsilon\eta)/
\Gamma(\psi^\prime\to J/\psi\eta)$:
\begin{equation}
R^\prime=\displaystyle \frac{\bigg(\bigg|\displaystyle
\frac{f^{111}_{2010}(b\bar{b})}{m_b}\bigg|^2|{\bf q}(b\bar{b)}|^3\bigg)}{\bigg(
\bigg|\displaystyle\frac{f^{111}_{2010}(c\bar{c})}{m_c}\bigg|^2
|{\bf q}(c\bar{c})|^3\bigg)},~~~~~~~~~~
R^{\prime\prime}=\displaystyle\frac{\bigg(\bigg|\displaystyle
\frac{f^{111}_{3010}(b\bar{b})}{m_b}\bigg|^2|{\bf q}(b\bar{b)}|^3\bigg)}{\bigg(
\bigg|\displaystyle\frac{f^{111}_{2010}(c\bar{c})}{m_c}\bigg|^2
|{\bf q}(c\bar{c})|^3\bigg)},
\end{equation}
where $\bm q$ is the momentum of $\eta$. The BGT model predicts~
$R^\prime=0.0025,~R^{\prime\prime}=0.0013.$
Recently BES has obtained an accurate measurement of $\Gamma(\psi^\prime\to J/\psi\eta)$ and 
$\Gamma(\psi^\prime\to J/\psi~\pi^0)$ 
\cite{besggJ} (see \Section~\ref{sec:4.6}A). With the new BES data and the bounds on 
$\Gamma(\Upsilon^\prime\to\Upsilon\eta)$ and 
$\Gamma(\Upsilon^{\prime\prime}\to\Upsilon\eta)$ \cite{Eidelman:2004wy}, the experimental bounds are 
$R^\prime|_{\rm exp}<0.0098,~R^{\prime\prime}|_{\rm exp}<0.0065$ \cite{besggJ}.
The predictions are consistent with these bounds.

An interesting prediction in the CL approach is the prediction for the ratio
\begin{equation}
R=\frac {\Gamma (\psi' \to J/\psi~ \pi^0)}{\Gamma(\psi' \to J/\psi~ \eta )},
\label{eq:4.2.8}
\end{equation}
which provides a measure of the light-quark mass ratio
$r=(m_d-m_u)/(m_s-(m_u+m_d)/2)$.
This belongs to the class of hadronic transitions,
which violate heavy quark spin symmetry (HQSS)\cite{noi1}.
For heavy mesons, there are only two types of operators that break HQSS.
In the parent's rest frame, the most general
spin symmetry breaking term is of the form ${\bf a}\cdot\bfsigma$,
where $\bfsigma$ are the Pauli matrices.
In an arbitrary frame one observes that any $\Gamma$-matrix
sandwiched between two projectors $(1+\HTslash v)/2$, or $(1-\HTslash v)/2$
can be expressed in terms of $\sigma_{\mu\nu}$ sandwiched between the same projectors:
$$
\frac{1+\HTslash v}{2} 1\frac{1+\HTslash v}{2}=\frac{1+\HTslash
  v}{2},~~~~~~~~~~~~~~~~
\frac{1+\HTslash v}{2}\gamma_5\frac{1+\HTslash v}{2}=0,~~~~~~~~~~~~~~~~
\frac{1+\HTslash v}{2}\gamma_\mu\frac{1+\HTslash v}{2}=v_\mu\frac{1+\HTslash
v}{2},
$$
$$
\hspace{-0.14cm}\frac{1+\HTslash v}{2}\gamma_\mu\gamma_5\frac{1+\HTslash v}{2}= 
\frac{1}{2}\epsilon_{\mu\nu\alpha\beta}v^\nu\frac{1+\HTslash v}{2}
\sigma^{\alpha\beta}\frac{1+\HTslash v}{2},
\frac{1+\HTslash v}{2}\gamma_5\sigma_{\mu\nu}\frac{1+\HTslash v}{2}= 
-\frac{i}{2}\epsilon_{\mu\nu\alpha\beta}\frac{1+\HTslash v}{2}
\sigma^{\alpha\beta}\frac{1+\HTslash v}{2};
$$
there are analogous relations with $(1+\HTslash v)/2\to(1-\HTslash v)/2$.
We use here $\epsilon_{0123}=+1$. Let us define
\begin{equation}
\sigma_{\mu\nu}^{(\pm)}=\frac{1 \pm \HTslash v}{2}\sigma_{\mu\nu}
\frac{1 \pm \HTslash v}{2}~~.
\end{equation}
In the parent's rest frame, $\sigma_{\mu\nu}^{(\pm)}$ reduce to Pauli matrices. From 
the previous identities it follows that the most general spin symmetry
breaking terms are of the form
$G_1^{\mu\nu}\sigma_{\mu\nu}^{(+)}$, or $G_2^{\mu\nu}\sigma_{\mu\nu}^{(-)}$,
with $G_i^{\mu\nu}$ two arbitrary antisymmetric tensors. 
One expects that any insertion of the operator
$\sigma_{\mu\nu}^{(\pm)}$ gives a suppression factor $1/m_Q$.

Using partial conservation of axial-vector current, Ioffe and Shifman
\cite{Ioffe} give the prediction
\begin{equation}
R=\frac{27}{16} \left[\frac{{\bf p}_{\pi}}{{\bf p}_{\eta}}\right]^3 
\bigg[\frac{m_d-m_u}{m_s-(m_u+m_d)/2}\bigg]^2.
\label{eq:4.2.10}
\end{equation}
The new BES experiment (see \Section~\ref{sec:4.6}A) \cite{besggJ}
provides a new precision value of $R$.  With the conventional values
of the current quark masses, the prediction of (\ref{eq:4.2.10}) is
smaller than the BES value by about a factor of 3 \cite{besggJ}.  So
(\ref{eq:4.2.10}) should be regarded as an order of magnitude estimate.

The calculation of $R$ in the CL approach is straightforward. 
The most general spin breaking Lagrangian for the processes $\psi' \to J/\psi 
\pi^0 , \eta$ is
\begin{equation}
{\cal L}=
i\epsilon_{\mu\nu\rho\lambda} \left[ \langle J'\sigma^{\mu\nu}\bar{J}\rangle -
\langle\bar{J}\sigma^{\mu\nu}J'\rangle\right] v^{\rho}
~\partial^{\lambda} \left[\dd\frac{i A}{4} \langle\hat m (\Sigma -\Sigma^{\dagger})\rangle
+B\eta ' \right] +h.c. \,.
\label{eq:4.2.11}
\end{equation}
The couplings $A$ and $B$ have dimension $({\rm mass})^{-1}$; the $B$
term contributes to the ratio (\ref{eq:4.2.8}) via the mixing $\pi^0
-\eta '$ and $\eta -\eta '$.  There are no terms with the insertion of
two $\sigma$ terms; the two P and C conserving candidates $
\epsilon_{\mu\nu\rho\lambda} \left[ \langle
J'\sigma^{\mu\tau}\bar{J}\sigma_{\tau}^{~\nu}\rangle +
\langle\bar{J}\sigma^{\mu\tau}J'\sigma_{\tau}^{~\nu}\rangle\right]
v^{\rho} \partial^{\lambda} \langle\hat m (\Sigma
-\Sigma^{\dagger})\rangle$ and $\epsilon_{\mu\nu\rho\lambda}$ $\left[
\langle J'\sigma^{\mu\nu}\bar{J}\sigma^{\rho\lambda}\rangle\right.$
$+$
$\left. \langle\bar{J}\sigma^{\mu\nu}J'\sigma^{\rho\lambda}\rangle\right]$
$\langle\hat m (\Sigma -\Sigma^{\dagger})\rangle $ both vanish.

Using the Lagrangian (\ref{eq:4.2.11}) and taking into account the
mixings (\ref{eq:4.1.21}) and (\ref{eq:4.1.22}), we can calculate the
ratio (\ref{eq:4.2.8})
\begin{equation}
R=\frac{27}{16} \left[\frac{{\bf p}_\pi}{{\bf p}_\eta}\right]^3 
\bigg[\frac{m_d-m_u}{m_s-(m_u+m_d)/2}\bigg]^2
\left[\frac{
1+\dd\frac{2 B}{3 A}\frac{\hat\lambda f_{\pi}}{m^2_{\eta '}-m^2_{\pi^0}}}
{1+\dd\frac{B}{A}\frac{\hat\lambda f_{\pi}}{m^2_{\eta 
'}-m^2_{\eta}}}\right]^2~~.
\label{eq:4.2.12}
\end{equation}
If we neglect the $\pi^0-\eta'$ and $\eta-\eta'$ mixings,
(\ref{eq:4.2.12}) reduces to the simple result (\ref{eq:4.2.10}).  So far
$B/A$ in (\ref{eq:4.2.12}) is not determined yet. Taking the
$\eta-\eta^\prime$ mixing angle $\theta_P\approx -20^\circ$
\cite{Eidelman:2004wy} and using the new BES data on $R$
\cite{besggJ}, one can determine $B/A$ from (\ref{eq:4.2.12}):
$B/A=-1.42\pm0.12$ or $-3.11\pm 0.15$ \cite{besggJ}.

The $\pi\pi$ transitions between P-wave quarkonia, $2^3 P_{J_i}\to
1^3 P_{J_f}+\pi+\pi$, have been studied in Ref. \cite{KY81}.  The
obtained transition rates $\Gamma(2^3P_{J_i}\to 1^3P_{J_f}\pi\pi)$ are
of the order of $10^{-1}$--$10^{-2}$~keV~\cite{KY81}. The relations
between different $\Gamma(2^3P_{J_i}\to 1^3P_{J_f}\pi\pi)$ reflect the
symmetry in the E1E1 multipole expansion \cite{Yan}, so that
experimental tests of these relations are of special interest.

In the CL approach, the single pseudoscalar meson transitions between
heavy quarkonia states such as
\begin{equation}
^3P_{J'} \to \;^3P_J \pi^0 \;~{\rm and}~ \;^3P_J \eta 
\label{eq:4.2.13}
\end{equation}
are chiral-breaking but spin conserving \cite{noi1}, which are
important for transitions forbidden in the SU(3) $\times$ SU(3)
symmetry limit.

To first order in the chiral breaking mass matrix we consider the quantities:
\begin{equation}
\langle\hat m(\Sigma +\Sigma^{\dagger})\rangle \;~{\rm and}~ \;
\langle\hat m(\Sigma - \Sigma^{\dagger})\rangle.
\end{equation}
The first quantity is parity even, while the second is parity odd;
both have $C=+1$.

The only term spin-conserving and of leading order in the current
quark masses contributing to the transition (\ref{eq:4.2.13}) is
\begin{equation}
\langle J_\mu {\bar J}_\nu \rangle v_\rho \epsilon^{\mu\nu\rho\sigma} \partial_\sigma
\left[ \alpha \frac{i f_\pi}{4} \langle\hat m(\Sigma - \Sigma^{\dagger})\rangle + \beta
f_\pi\eta'
\right],
\label{eq:4.2.15}
\end{equation}
where $\alpha$ and $\beta$ are coupling constants of dimensions $({\rm
mass})^{-2}$.  The direct coupling to $\eta'$ contributes through the
mixing (\ref{eq:4.1.21}).  The spin symmetry of the heavy sector gives
relations among the modulus square matrix elements of the transitions
between the two P-wave states. In particular we find that
\begin{equation}
|{\cal M}|^2 (^3P_0 \to ^3P_0\pi)
=|{\cal M}|^2 (^3P_2 \to ^3P_0\pi)=0,
\label{eq:4.2.16}
\end{equation}
and that all non-vanishing matrix elements can be expressed in terms of
$^3P_0 \to ^3P_1\pi$:
\begin{eqnarray}
&&
|{\cal M}|^2 (^3P_1 \to ^3P_1\pi)=\frac{1}{4}
|{\cal M}|^2 (^3P_0 \to ^3P_1\pi),~~~~~~
|{\cal M}|^2 (^3P_1 \to ^3P_2\pi)=\frac{5}{12}
|{\cal M}|^2 (^3P_0 \to ^3P_1\pi),\hspace{0.9cm}
\nn\\
&&
|{\cal M}|^2 (^3P_2 \to ^3P_2\pi)=\frac{3}{4}
|{\cal M}|^2 (^3P_0 \to ^3P_1\pi),~~~~~~
|{\cal M}|^2 (^1P_1 \to ^1P_1\pi)=|{\cal M}|^2 (^3P_0 \to ^3P_1\pi),
\label{eq:4.2.17}
\end{eqnarray}
where $\pi$ stays for $\pi^0$ or $\eta$. The relations (\ref{eq:4.2.17})  
can be generalized for any spin-conserving transition between $l=1$ multiplets,
leading to the same results as the QCD double multipole expansion \cite{Yan}.
Predictions for widths can be easily obtained from (\ref{eq:4.2.15}).

Now we consider the $\pi\pi$ transitions of D-wave quarkonia.
Theoretical studies of HTs of D-wave quarkonia have been carried out
by several authors in different approaches leading to quite different
predictions \cite{BLMN,KY81,Moxhay,Ko,Rosner:2003ty,KY90,Kuang02}. We
briefly review the approach in Refs. \cite{KY90,Kuang02}, and compare
the predictions with recent experimental results.

Since the $\psi(3770)$ (or $\psi^{\prime\prime}$) lies above the
$D\bar{D}$ threshold, it is believed that it
decays mainly into $D\bar{D}$ \cite{Eidelman:2004wy}.
Experimental observations show that the directly measured
$e^+e^-\to\psi(3770)$ cross-section and the
$e^+e^-\to\psi(3770)\to D\bar{D}$ cross-section are different \cite{different},
suggesting considerable
non-$D\bar{D}$ decay modes of $\psi(3770)$.
$\psi(3770)\to J/\psi~\pi\pi$ is one possibility.

If $\psi(3770)$ is regarded as a pure $1D$
state, the predicted leptonic width will be smaller
than the experimental value by an order of magnitude. The
$\psi(3770)$ is often regarded as a mixture of the $1D$ and $2S$ states
\cite{KY90,Kuang02,Godfrey}:
$~~\psi^\prime = |2S\rangle \cos\theta + |1D \rangle \sin\theta,~~
\psi(3770) = -|2S\rangle \sin\theta + |1D\rangle \cos\theta$.
$\theta$ can be determined by
fitting the ratio of the leptonic widths of $\psi^\prime$ and
$\psi(3770)$.
The determination of $\theta$ in the Cornell potential model \cite{Eichten:1978tg,Eichten:1979ms}
and the improved QCD motivated potential model by Chen and Kuang (CK) \cite{CK92}
(which leads to more successful phenomenological results) are:
$~\theta=-10^\circ$ (Cornell) and $~\theta=-12^\circ$ (CK).

The rate of this E1E1 transition is \cite{KY90}
\begin{equation}
\Gamma(\psi(3770)\to J/\psi\pi\pi)=|C_1|^2\bigg[\sin^2\theta ~G(\psi^\prime)
~|f^{111}_{2010}(\psi^\prime)|^2
+\frac{4}{15}\bigg|\frac{C_2}{C_1}\bigg|^2\cos^2\theta ~H(\psi^{\prime\prime})
~|f^{111}_{1210}(\psi^{\prime\prime})|^2\bigg].\hspace{0.4cm}
\label{eq:4.2.18}
\end{equation}
Since there is no available data to determine $C_2$, we take an
approximation to estimate $C_2$.  In Ref. \cite{KY81}, it is assumed
that $\langle\pi\pi|E^a_kE^a_l|0\rangle\propto \langle
gg|E^a_kE^a_l|0\rangle$, \ie  that the factor describing the
conversion of the two gluons into $\pi\pi$ is approximately
independent of the pion momenta in the HTs under consideration. In
this approximation, we obtain \cite{KY81}
\begin{equation}
C_2\approx 3C_1.
\end{equation}
So it is possible that $C_2/C_1\sim {\cal O}(1)$. 
\begin{table}[ht]
\caption[The predicted transition rate $\Gamma(\psi(3770)\to
         J/\psi~\pi^+\pi^-)$]
        {The predicted transition rate $\Gamma(\psi(3770)\to
         J/\psi~\pi^+\pi^-)$ (in keV) in the Cornell model and the CK
         model with the updated input data (\ref{eq:4.2.6}). The
         corresponding branching ratios are listed in the brackets
         using the total width of $\psi(3770)$ given in
         Ref.~\cite{Eidelman:2004wy}.}
\label{tab:HT2}
\tabcolsep 1.5cm
\vspace{0.4cm}
\begin{tabular}{cc}
\hline\hline
&$\Gamma(\psi(3770)\to J/\psi\pi^+\pi^-)$~(keV)\\
 &Cornell\hspace{3.4cm}CK\\
 \hline
$C_2\approx 3C_1$&\hspace{0.4cm}139~[$(0.59\pm0.07)\%$]\hspace{1.2cm}147~[$(0.62\pm0.07)\%$]\\
$C_2\approx C_1$&\hspace{0.6cm}26~[$(0.11\pm0.01)\%$]\hspace{1.4cm}32~[$(0.14\pm0.02)\%$]\\
\hline\hline
\end{tabular}
\end{table}

For comparison, we list the predicted rate $\Gamma(\psi(3770)\to
J/\psi~\pi^+\pi^-)$ with $C_2/C_1=3$ and $C_2/C_1$ $=$ $1$ in
\Table~\ref{tab:HT2}.\footnote{The values listed in
\Table~\ref{tab:HT2} are larger than those given in
Refs.~\cite{KY90,Kuang02} since the updated input data values are
larger.}  Note that S--D mixing only affects a few percent of the
rate, so that the rate is essentially $\Gamma(\psi(2D)\to
J/\psi~\pi^+\pi^-)$.

Recently, BES has measured the rate $\Gamma(\psi(3770)\to
J/\psi+\pi^++\pi^-)$ based on 27.7 pb$^{-1}$ data of $\psi(3770)$. The
result is $\Gamma(\psi(3770)\to J/\psi+\pi^++\pi^-)=80\pm32\pm21$~keV
\cite{Bai:2003hv} (see \Eq~(\ref{eq:4.6.5}) in
\Section~\ref{sec:4.6}C). \Eq[b]~(\ref{eq:4.2.18}) is in agreement with
the central value of the BES result with $C_2/C_1\approx
2$. Considering the large error in the BES experiment, $C_2/C_1$ can
still be in the range $0.8\le C_2/C_1\le 2.8$. We expect more precise
future measurements to give a better determination of $C_2/C_1$.

For the $\Upsilon$ system, the state mixings are much smaller \cite{UQM}. 
Neglecting such mixings, the predicted 
rate of $\Upsilon(1^3D_1)\to \Upsilon\pi\pi$ in the Cornell model
with $C_2/C_1=3$ was 
$\Gamma(\Upsilon(1^3D_1)\to \Upsilon\pi\pi)$ $\approx$ $24~{\rm keV}$ \cite{KY81}.
Taking the central value $C_2/C_1\approx 2$ determined from
BES data, the prediction is
$\Gamma(\Upsilon(1^3D_1)\to \Upsilon\pi\pi)\approx 11~{\rm keV}$.
Considering the above range of $C_2/C_1$, we predict
1.8~keV $\le$ $\Gamma(\Upsilon(1^3D_1)$ $\to$  $\Upsilon\pi\pi)$ $\le$  21~keV.

HTs are useful processes to investigate the $h_c$ [or $\psi(1^1P_1)$]
and $h_b$ [or $\Upsilon(1^1P_1)$] states. $h_c$ and $h_b$ are of
special interest since the difference between the mass of the $1^1P_1$
state and the centre-of-gravity of the $1^3P_J$ states gives useful
information about the spin-dependent interactions between $Q$ and
$\bar{Q}$.  The possibilities to detect $h_c$ and $h_b$ at $e^+e^-$
colliders, in $^3S_1\to\pi^0\, ^1P_1$, $^1P_1\to \pi\pi\, ^3S_1$, and
$^1P_1\to \pi^0\, ^3S_1$ transitions have been studied in
Refs.~\cite{CrystalBall83,KY81,KTY88,Kuang02,VZNS}; $h_c$ could also
be detected at the $B$ factories \cite{Suzuki:2002sq}, depending on
the value for the $B\to h_c\,K$ branching ratio.  So far, the $h_b$
has not been experimentally found, while the $h_c$ has probably been
observed, based on recent preliminary results presented by
CLEO\cite{Tomaradze-hc} and E835\cite{Patrignani:2004nf}.  CLEO has
observed significant excess of events in $\psi(2S)\to \pi^0
h_c\to\pi^0\gamma\eta_c$, in both exclusive and inclusive $\eta_c$
decays. E835 has a significant excess of events in $\bar p p\to
h_c\to\eta_c\gamma\to3\gamma$.  The mass of the CLEO and E835
candidates are compatible, and very close to the
centre-of-gravity. For more details we refer to
\Chapter~\ref{chapter:spectroscopy}.

\subsection[Nonrelativistic coupled-channel approach to  hadronic transitions]
{Nonrelativistic coupled-channel approach to  hadronic transitions}
\label{sec:4.3}

\begin{figure}[ht]
\begin{center}
\includegraphics[width=0.28\textwidth,clip]{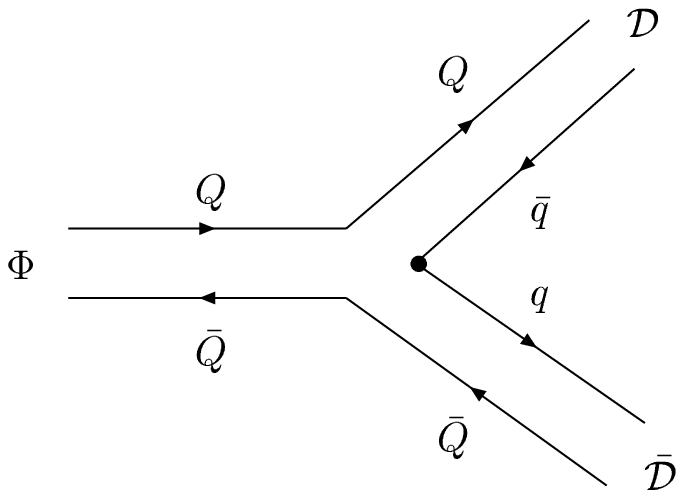}
\end{center}
\caption{Coupling of the heavy quarkonium $\Phi$ to its decay 
         channel ${\cal D}\bar{\cal D}$.}
\label{fig:phi-d-d}
\end{figure}

Since a heavy quarkonium $\Phi$ lying above the threshold can decay
into a pair of heavy flavour mesons ${\cal D}\bar{\cal D}$ [${\cal D}$
stands for $D$ mesons (for $c\bar{c}$) and $B$ mesons (for
$b\bar{b}$)], there must exist $\Phi$--${\cal D}$--$\bar{\cal D}$
couplings as shown in \Figure~\ref{fig:phi-d-d}.

A complete theory should include not only the part describing $\Phi$,
but also the part corresponding to the ${\cal D}\bar{\cal D}$ sector
as well. Such a theory is the so-called coupled-channel (CC) theory.

\begin{figure}[t]
\begin{center}
\includegraphics[width=0.72\textwidth,clip]{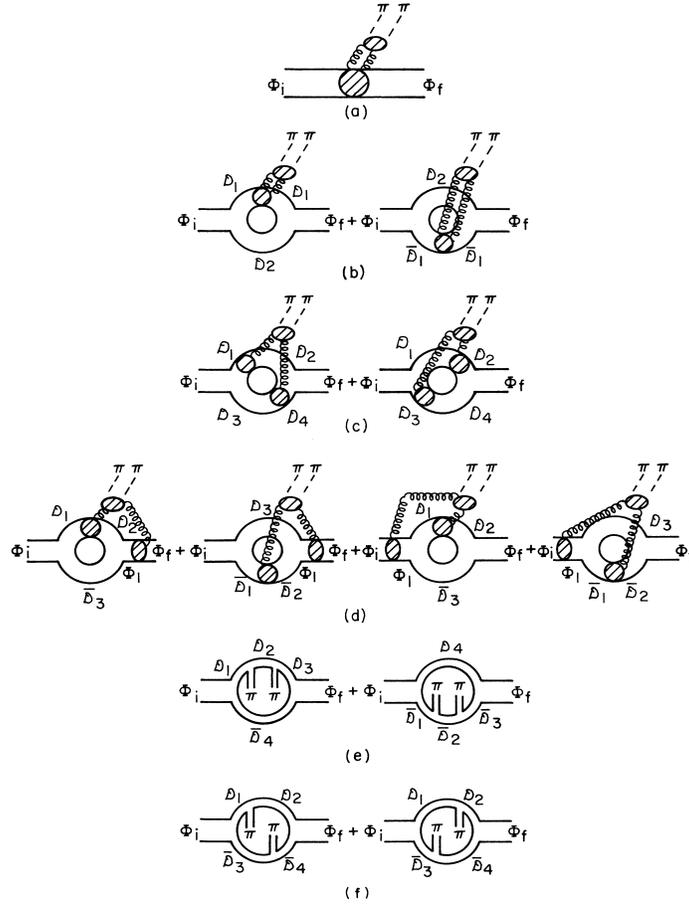}
\end{center}
\vskip -1.5cm
\caption{Diagrams for hadronic transitions in the CC
approach. Quoted from Ref. \cite{ZK91}.}
\label{fig:ZK91Fig1}
\end{figure}

It is hard to study the $\Phi$--${\cal D}$--$\bar{\cal D}$ vertex from
the first principles of QCD, since it is the vertex of three bound
states. There are various models describing CC effects; the two
well-accepted models are the Cornell CC model (CCCM)
\cite{Eichten:1978tg,Eichten:1979ms,CCCM} and the unitary quark model
(UQM) \cite{UQM}. The $\Phi$--${\cal D}$--$\bar{\cal D}$ vertex in the
UQM is taken to be the $^3P_0$ quark-pair-creation (QPC) mechanism
\cite{QPC}.  The parameters in the UQM are carefully adjusted so that
the model gives a better fit to the $c\bar{c}$ and $b\bar{b}$ spectra,
leptonic widths, etc. It is shown that the QPC model gives acceptable
results even for OZI-allowed productions of light mesons
\cite{QPC,QPCappl}, which is relevant in the calculation of the HT
amplitudes in the CC theory.

The formulation of the theory of HTs in the framework of the UQM was
given in Ref. \cite{ZK91}.  The Feynman diagrams for $n_i^3S_1\to
n_f^3S_1\pi\pi$ are shown in \Figure~\ref{fig:ZK91Fig1}.  We see that there
are more channels of $\pi\pi$ transitions in this theory than in the
single-channel theory.
\Figures[b]~\ref{fig:ZK91Fig1}(a)--\ref{fig:ZK91Fig1}(d) are based on
the QCDME mechanism; we designate this the MGE part.
\Figures[b]~\ref{fig:ZK91Fig1}(e) and \ref{fig:ZK91Fig1}(f) are based on
a new $\pi\pi$ transition mechanism via QPC; we designate this the QPC
part.  \Figure[b]~\ref{fig:ZK91Fig1}(a) is similar to \Figure~\ref{fig:htFig}
but with state mixings, so the single-channel amplitude mentioned in
\Section~\ref{sec:4.2} is only a part of \Figure~\ref{fig:ZK91Fig1}(a).

\begin{table}[ht]
\caption[$\Gamma(\Upsilon^\prime\to \Upsilon\pi\pi)$,
         $\Gamma(\Upsilon^{\prime\prime}\to \Upsilon\pi\pi)$, and
         $\Gamma(\Upsilon^{\prime\prime}\to\Upsilon^\prime\pi\pi)$
         predicted in CC theory] {$\Gamma(\Upsilon^\prime\to
         \Upsilon\pi\pi)$, $\Gamma(\Upsilon^{\prime\prime}\to
         \Upsilon\pi\pi)$, and
         $\Gamma(\Upsilon^{\prime\prime}\to\Upsilon^\prime\pi\pi)$
         predicted in CC theory, with $\cos\vartheta=-1$ and $-0.676$,
         together with the updated experimental values
         \cite{Eidelman:2004wy}.  $\pi\pi$ stands for the sum over all
         the $\pi^+\pi^-$ and $\pi^0\pi^0$ channels.}
\label{tab:HT3}
\tabcolsep 0.45cm
\begin{center}
\begin{tabular}{ccccc}
\hline\hline
 &&Theory&&Expt.\\
 &$\cos\vartheta=-1$&&$\cos\vartheta=-0.676$&\\
 \hline
$\Gamma(\Upsilon^\prime\to\Upsilon\pi\pi)$~(keV)~~&14~~~&&13~~~&$~~~12.0~\pm 1.8~~~~~~$\\
$\Gamma(\Upsilon^{\prime\prime}\to\Upsilon\pi\pi)$~(keV)&~1.1&&~1.0&$1.72\pm 0.35$\\
$\Gamma(\Upsilon^{\prime\prime}\to\Upsilon^\prime\pi\pi)$~(keV)&~0.1&&~~0.3&$1.26\pm 0.40$\\
\hline\hline
\end{tabular}
\end{center}
\end{table}

Since state mixings and the QPC vertices are all different in the
$c\bar{c}$ and the $b\bar{b}$ systems, the predictions for
the$\Upsilon$ HT rates by taking the input (\ref{eq:4.2.6}) will be
different from those in the single-channel theory. Such predictions
were studied in Ref. \cite{ZK91}. Note that for a given QPC model, the
QPC part is fixed, while the MGE part still contains an unknown
parameter $C_1$ after taking the approximation $C_2\approx
3C_1$. Since there is interference between the MGE and the QPC parts,
the phase of $C_1$ is important; explicitly, $C_1=|C_1|~\displaystyle
e^{i\vartheta}$.  The data of the HT rate and $M_{\pi\pi}$
distribution in $\psi^\prime\to J/\psi~\pi\pi$ can be taken as inputs
to determine $C_1$ and $\vartheta$\cite{ZK91}. Considering the error
bars in the $M_{\pi\pi}$ distribution, $\vartheta$ is restricted in
the range ~$-1\le \cos\vartheta\le-0.676$ \cite{ZK91}. The predicted
transition rates in the $\Upsilon$ system are listed in
\Table~\ref{tab:HT3} together with the experimental results for
comparison. We see that the obtained $\Gamma(\Upsilon^\prime\to
\Upsilon\pi\pi)$ is in good agreement with the experiment, and the
results of $\Gamma(\Upsilon^{\prime\prime}\to \Upsilon\pi\pi)$ and
$\Gamma(\Upsilon^{\prime\prime}\to \Upsilon^\prime\pi\pi)$ are in
agreement with the experiment at the level of $2\sigma$ and
$2.4\sigma$, respectively.

Next we look at the predicted $M_{\pi\pi}$ distributions. It is
pointed out in Ref. \cite{Albrecht:1986gb} that there is a tiny
difference between the measured $M_{\pi\pi}$ distributions in
$\psi^\prime\to J/\psi\pi\pi$ and $\Upsilon^\prime\to
\Upsilon\pi\pi$. In the single-channel theory, the formulas for these
two $M_{\pi\pi}$ distributions are the same.  In the CC theory, once
$C_1$ and $\vartheta$ are determined, the $M_{\pi\pi}$ distribution of
$\Upsilon^\prime\to\Upsilon\pi\pi$ is uniquely determined. It is shown
in Ref. \cite{ZK91} that the prediction fits the experiment
\cite{Albrecht:1986gb} very well

However, the situation of the $M_{\pi\pi}$ distributions of
$\Upsilon^{\prime\prime} \to\Upsilon\pi^+\pi^-$ and
$\Upsilon^{\prime\prime}\to\Upsilon^\prime\pi^+\pi^-$ are more
complicated. Comparison of the CC predictions with the CLEO experiment
will be shown in \Section~\ref{sec:4.5}E.

\subsection[Application of the QCD multipole expansion to radiative decays of the $J/\psi$]{
            Application of the QCD multipole expansion to radiative decays of the $J/\psi$}
\label{sec:4.4}
In the above sections, QCDME is applied to
various HTs in which $\Phi_i$ and $\Phi_f$ are composed of the
same heavy quarks. In this case, the dressed (constituent) quark
field $\Psi({\bf x},t)$ needs not actually to be quantized. Now
we generalize QCDME theory to processes
including changes of
heavy quark flavour and heavy quark pair
annihilation or creation, for which the quantization of $\Psi({\bf
x},t)$ is needed. This has been studied in Ref. \cite{KYF} with the
electroweak interactions included as well.

\begin{figure}[ht]
\begin{center}
\includegraphics[width=0.75\textwidth,clip]{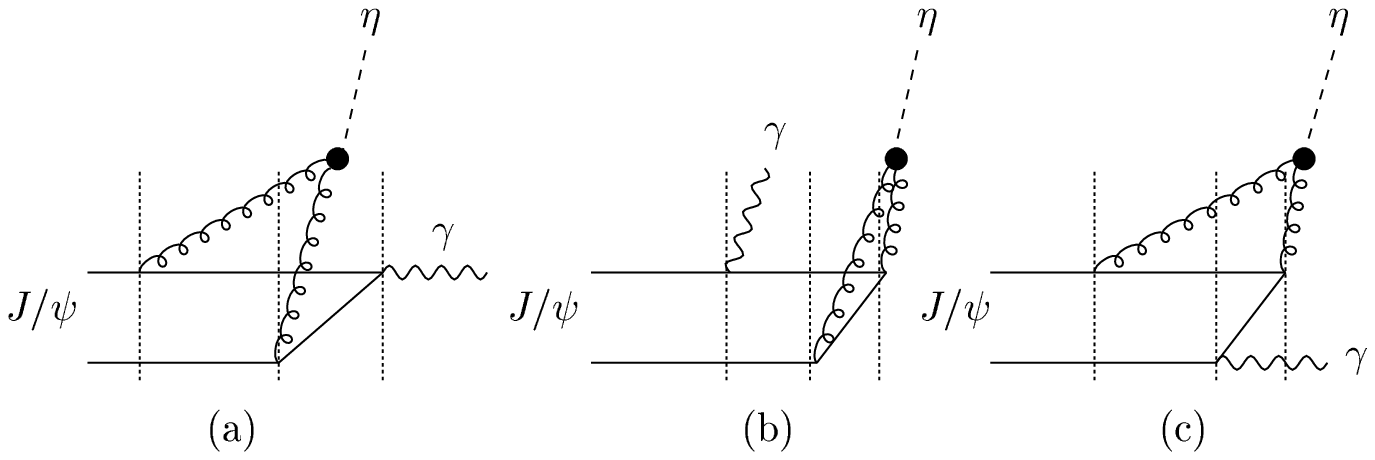}
\end{center}
\caption{Feynman diagrams for the radiative decay process
         $J/\psi\to\gamma+\eta$.}
\label{fig:psi-ge}
\end{figure}

An example of application of such a theory is $J/\psi\to\gamma \eta$
(see \Section~\ref{sec:exdec-radiative} for a discussion in the
framework of Ref. \cite{exdec:nov80}).  This process has been studied
in the framework of perturbative QCD and the nonrelativistic quark
model in Ref. \cite{KKKS}, but the predicted rate is significantly
smaller than the experimental value. The $\eta$ momentum in this
process is $q_\eta=1.5$~Gev. If $\eta$ is converted from two emitted
gluons from the heavy quark, the typical gluon momentum is then $k\sim
q_\eta/2\sim 750$~MeV. At this momentum scale perturbative QCD does
not work well but QCDME works \cite{KYF}. The Feynman diagrams for
this process in the QCDME approach are shown in \Figure~\ref{fig:psi-ge}, in
which the intermediate states marked between two vertical dotted lines
are all treated as bound states. In this sense this approach is
nonperturbative.

Since this process is dominated by E1M2 transition; 
the transition rate depends on the pseudoscalar nonet mixing angle $\theta_P$. 
Taking the value $\theta_P\approx -20^\circ$ determined from
the $\eta\to\gamma\gamma$ and $\eta^\prime\to\gamma\gamma$ rates
\cite{Eidelman:2004wy}, we obtain \cite{KYF}
\begin{equation}
\Gamma(J/\psi\to\gamma\eta)=0.041\bigg(\frac{\alpha_M}{\alpha_E}\bigg)~{\rm keV},~~~~~~
{\cal B}(J/\psi\to\gamma\eta)=(4.7\pm 0.2)\times 10^{-4}\bigg(\frac{\alpha_M}{\alpha_E}\bigg).
\end{equation}
With the reasonable value $\alpha_M/\alpha_E=1.8$, 
the predicted branching ratio can agree with the experimental value 
$B_{\rm exp}(J/\psi\to\gamma\eta)=(8.6\pm 0.8)\times 10^{-4}$ \cite{Eidelman:2004wy}. 
To avoid the uncertainties from $\alpha_M/\alpha_E$ and $\theta_P$, 
we take the ratio of $\Gamma(J/\psi\to\gamma\eta)$ to another E1M2 
transition rate $\Gamma(\psi^\prime\to J/\psi\eta)$. The theoretical prediction
is \cite{KYF}
\begin{equation}
R_\eta\equiv \frac{\Gamma(J/\psi\to\gamma\eta)}{\Gamma(\psi^\prime\to J/\psi\eta)}=0.012.
\label{eq:4.4.2}
\end{equation}
In $R_\eta$, uncertainties in the H factors cancel, 
so $R_\eta$ offers a direct test of the MGE mechanism. 
(\ref{eq:4.4.2}) is in agreement with the experimental value
$R_\eta|_{\rm exp}=0.009\pm 0.003$ \cite{Eidelman:2004wy} at the $1\sigma$ level.

This approach can also be applied to $J/\psi\to\gamma\eta^\prime$. With $\theta_P\approx -20^\circ$,
we obtain
\begin{equation}
R_{\eta^\prime}\equiv 
\frac{\Gamma(J/\psi\to\gamma\eta^\prime)}{\Gamma(\psi^\prime\to J/\psi \eta)}
=\bigg|\frac{{\bf q}(J/\psi\to\gamma\eta^\prime)}{{\bf q}(J/\psi\to \gamma\eta))}\bigg|^3
\bigg|\frac{m_{\eta^\prime}^2(\sqrt{2}\cos\theta_P+\sin\theta_P)}
{m_\eta^2(\cos\theta_P-\sqrt{2}\sin\theta_P)}\bigg|^2 R_\eta=0.044.
\end{equation}
This is also in agreement with the experimental value
$R_{\eta^\prime}|_{\rm exp}=0.044\pm 0.010$ \cite{Eidelman:2004wy}.

We would like to mention that this approach is not suitable for
$\Upsilon\to\gamma\eta$ since the typical gluon momentum in this
process is $k\sim q_\eta/2\sim 2.4$~Gev, appropriate for perturbative QCD,
but not for QCDME.

\subsection[Hadronic transition experiments in the $b\bar{b}$ system]
           {Hadronic transition experiments in the $b\bar{b}$ system}
\label{sec:4.5}

\null\noindent
{\it A. Experimental analysis of hadronic transitions --- bottomonium}

\begin{figure}[t]
\centerline{\includegraphics[width=8cm]{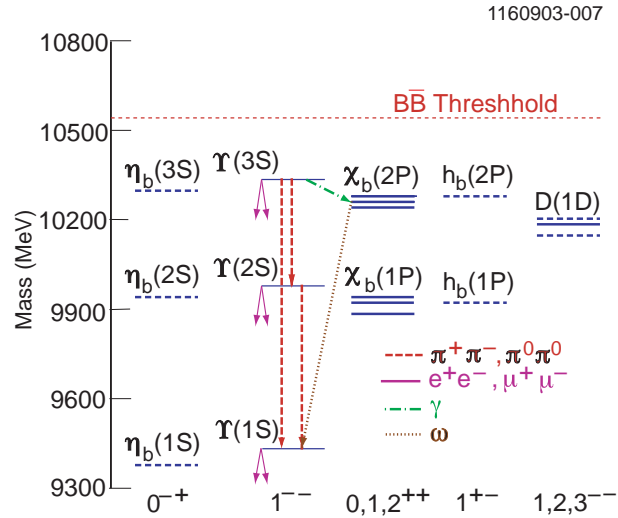}}
\caption{Allowed photon, dipion, and omega transitions allowed
within the $b\bar{b}$ system.}
\label{fig:bbbar-spectroscopy}
\end{figure}

We see from \Eq~(\ref{eq:4.2.2}) that, in the framework of QCDME, the
transition amplitude contains an MGE factor and a H factor. Selection
rules, as well as the limited phase space, restrict the possible
transitions. A summary of the rich spectroscopy afforded by bottomonia
is shown in \Figure~\ref{fig:bbbar-spectroscopy}

The principal experimental observables here are the partial widths for
the transitions between bottomonia and the Dalitz plot variables: the
$\pi\pi$ and $\Upsilon\pi$\ invariant mass spectra, and the angular
distributions between final-state particles.  To measure the
transition $\Upsilon^{\prime\prime}\to\Upsilon\pi\pi$, for example, in
electron--positron annihilation data (where $\Upsilon^{\prime\prime}$
is produced at rest, and polarized along the beam axis), one can use
the constraint that the $\Upsilon$ energy can be inferred directly
from the measurement of the pion four-momenta to calculate the mass
recoiling against the dipion system.  As with the $\gamma\gamma$
cascades, one differentiates the ``exclusive'' case in which the
$\Upsilon$ decays to a clean, background-free topology, such as
$\mu^+\mu^-$ or $e^+e^-$, from the ``inclusive'' case in which all
events are accepted, and one calculates the mass recoiling against all
oppositely-signed dipion pairs.  In the former case, one, therefore,
selects events consistent with the cascade:
$\Upsilon^{\prime\prime}\to\Upsilon\pi\pi$, $\Upsilon\to l^+l^-$,
allowing one to isolate a very clean sample, but at the expense of
lower overall efficiency owing to the small ($\sim 2\%$) dileptonic
BR's of the final state $\Upsilon$'s.

\null\noindent{\it B. Branching ratios and partial widths}

\begin{figure}[t]
\centerline{\includegraphics[width=7.6cm]{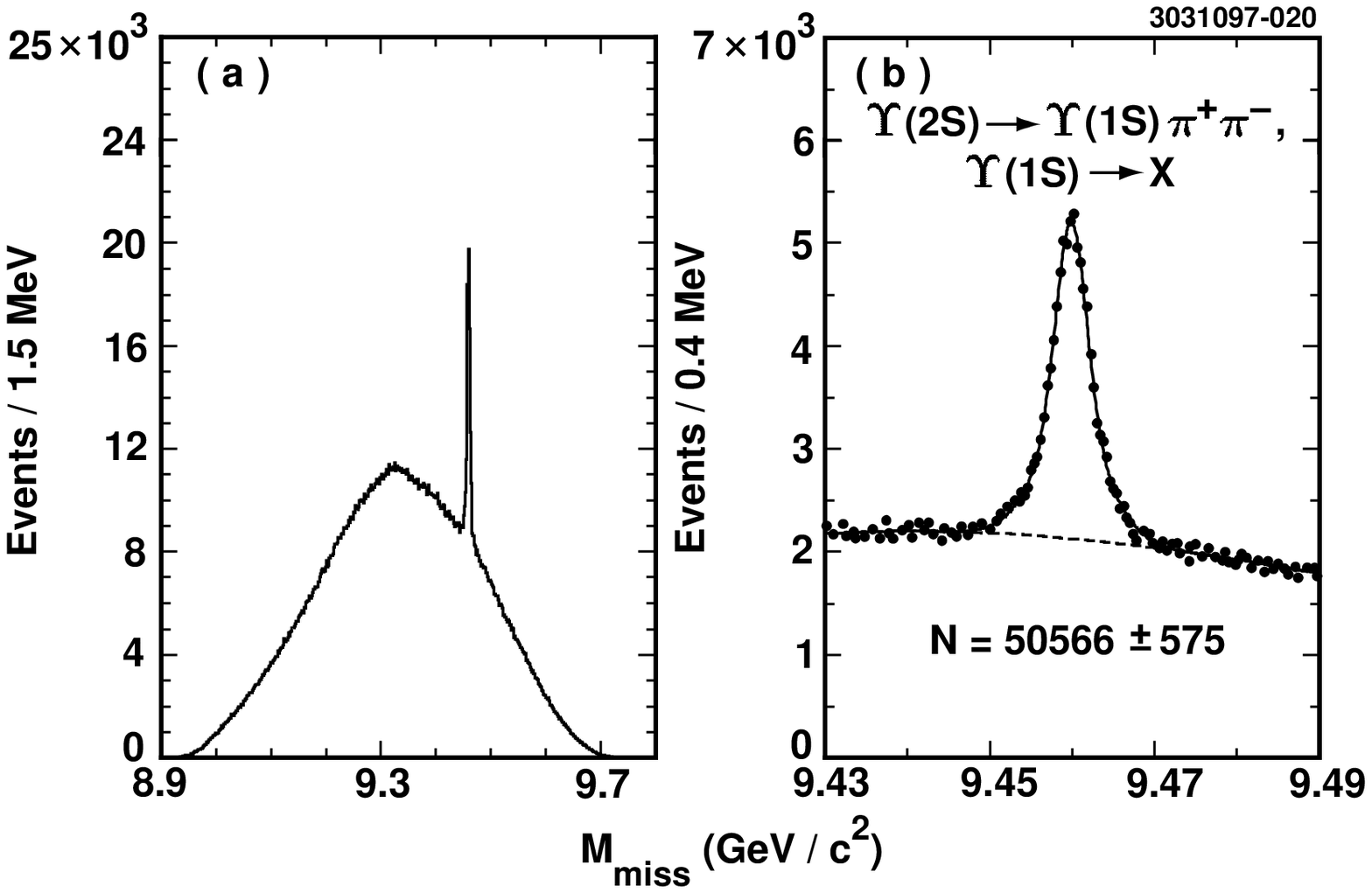}}
\caption{Mass recoiling against two oppositely charged tracks,
assumed to be pions, for data taken at the $\Upsilon^\prime$ resonance.}
\label{fig:2spp1sll}

\medskip

\centerline{\includegraphics[width=7.6cm]{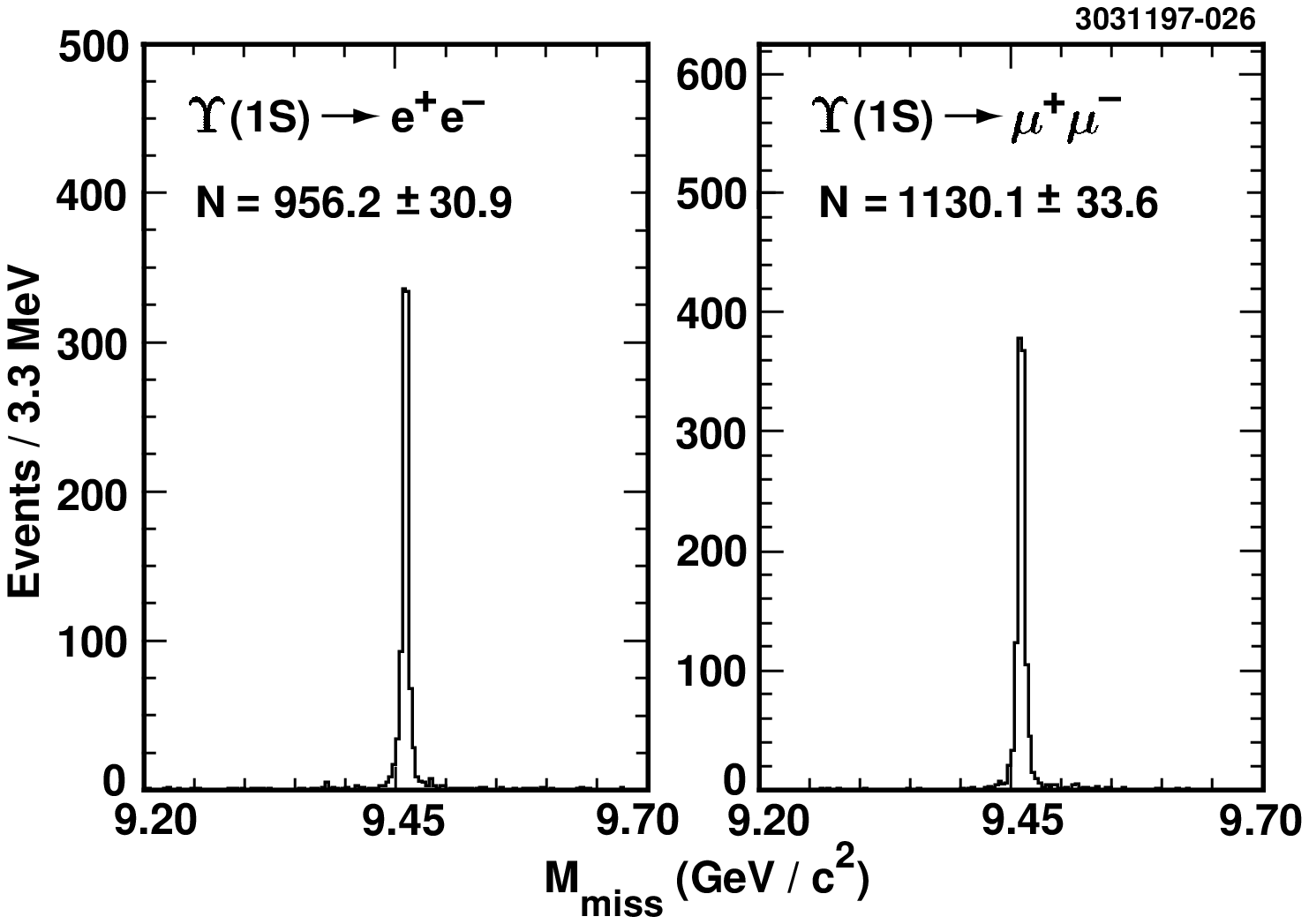}}
\caption{Mass recoiling against two oppositely charged tracks,
assumed to be pions, for data taken at the $\Upsilon^\prime$ resonance, with
the additional restriction that there be exactly four charged tracks in
the event, and that the two most energetic charged tracks be consistent
with $e^+e^-$ or $\mu^+\mu^-$.}
\label{fig:2spp1sinc}
\end{figure}

The CLEO~II mass spectra recoiling against charged dipions, for data
taken at the $\Upsilon^\prime$ \cite{CLEO9491}, are shown in
\Figures~\ref{fig:2spp1sll} and \ref{fig:2spp1sinc}, and illustrate the
trade-off between the higher statistical power of the inclusive data
sample vs. the better signal-to-noise of the exclusive data
sample.\footnote{Because of the poor signal-to-noise ratio, the
$\Upsilon^{\prime\prime}\to\Upsilon\pi^0\pi^0$ transitions cannot be
studied inclusively.}

Branching ratios are calculated based directly on the number of events
found in each peak. Predictions for the partial widths in the
nonrelativistic single-channel and coupled-channel theories are shown
in \Tables~\ref{tab:c1ht} and \ref{tab:HT3}.  In addition to CLEO,
the tabulated branching ratios for $\Upsilon^\prime\to\Upsilon\pi\pi$
also include measurements made by the ARGUS \cite{Albrecht:1986gb},
CLEO~I \cite{CLEO_2s1s_84}, CUSB-I \cite{CUSB_2s1s_84}, and Crystal
Ball \cite{XBAL_2s1s_85} collaborations.  The CLEO~II collaboration
are also able to derive estimates for the transition rates for
$\Upsilon^{\prime\prime}\to\Upsilon^\prime+X$ by performing a hand
scan of the events it reconstructs in
$\Upsilon^{\prime\prime}\to\Upsilon^\prime+X$,
$\Upsilon^\prime\to\Upsilon\pi^+\pi^-$, $\Upsilon\to l^+l^-$, and
using the unitarity constraint that the sum of the dipion transitions
plus the radiative transitions must saturate the overall
$\Upsilon^{\prime\prime}\to\Upsilon^\prime+X$ decay rate to determine
X.  These values have been compiled along with the direct observation
of the $\Upsilon^{\prime\prime}\to\Upsilon^\prime\pi^0\pi^0$ and
$\Upsilon^{\prime\prime}\to\Upsilon^\prime\pi^+\pi^-$ transitions.
According to isospin symmetry, the $\pi^+\pi^-$ transition rate is
expected to be twice that of the $\pi^0\pi^0$ transition, modulo the
ratios of available phase space ($\pi^0\pi^0$/$\pi^+\pi^-$) (1.36 for
$\Upsilon^{\prime\prime}\to\Upsilon^\prime\pi^0\pi^0$ and 1.02 for
$\Upsilon^{\prime\prime}\to\Upsilon\pi^0\pi^0$).  The measurements to
date are generally consistent with this expectation, with the
exception of $\Upsilon^{\prime\prime}\to\Upsilon^\prime\pi^+\pi^-$.
Curiously, despite an inability to match the dipion mass distributions
for the $\Upsilon^{\prime\prime}\to\Upsilon\pi\pi$ transitions
(Secs. \ref{sec:4.2} and \ref{sec:4.3}), the QCDME approach gives a
better match for this partial width than for
$\Upsilon^{\prime\prime}\to\Upsilon^\prime\pi\pi$.

\null\noindent
{\it C. Angular distributions}

In the nonrelativistic limit, orbital angular momentum and spin are
separately conserved.  The spin of a bottomonium resonance produced at
$e^+e^-$ colliders lies along the beam axis. In
$\Upsilon(nS)\to\Upsilon\pi^+\pi^-$, the orbital angular momentum
between the pions, or the orbital angular momentum between the dipion
system and $\Upsilon$ is a useful observable in addition to the
polarization of $\Upsilon$.  Predictions for the populations of the
allowed angular momentum states have been made for both the $\psi$
system as well as the $\Upsilon$ system \cite{BrownCahn,Cahn75}.  All
measurements to date (\eg by verifying in exclusive events that the
angular distribution of the leptons relative to the beam axis follows
$dN/d(cos\theta)\sim1+cos^2\theta$) from ARGUS, CLEO, and CUSB give
strong evidence that the daughter $\Upsilon$ is indeed polarized along
the beam axis in the dipion transitions, and are consistent with an
S-wave decay. The other allowed amplitude is a possible D-wave
contribution in the dipion system [cf. \Eq~(\ref{eq:4.2.3})].
Convincing evidence for a large D-wave component of the dipion system
has not yet been presented, although it has received some theoretical
attention \cite{Chakko92,Belanger89,MorganPen75}, and suggestions for
non-S-wave anisotropy are found in both the
$\Upsilon^{\prime\prime}\to\Upsilon\pi^+\pi^-$\cite{CLEO9491} and
$\Upsilon^\prime\to\Upsilon\pi^+\pi^-$ data
\cite{CLEO_2Spp1S98,Albrecht:1986gb}, both of which show $\sim
2\sigma$ indications of a D-wave contribution at the few percent level
\cite{CLEO9491}. Mapping out the ratio of D-wave to S-wave
amplitudes as a function of dipion mass in the
$\Upsilon^{\prime\prime}$ system is a project requiring substantially
more statistics than have been accumulated to date; expectations are
that a D-wave amplitude would be more observable at low values of
invariant mass, corresponding to higher energy release in the
$\Upsilon^{\prime\prime}$ decay. Such an analysis is currently
underway at CLEO and should mature within the next year.

\null\noindent
{\it D. Single pion transitions}

For dipion transitions Yan \cite{Yan}, collaborating with Kuang \cite{KY81}, and their work later
extended by Zhou and Kuang \cite{ZK91}, estimated the magnitude of
the second piece of the product matrix element,
the hadronization term of the transition amplitude.
An immediate consequence of the multipole approach is the 
expected suppression of the case $X=\eta$ relative to $X=\pi\pi$. The former system 
has the wrong quantum numbers for two $E1$ gluons, and proceeds in
lowest order as either $E1\cdot M2$ or $M1\cdot M1$ in QCDME. Since the 
mass dependence of the chromomagnetic
transitions goes as $m^{-4}$ ($m$ = quark mass),
QCDME, therefore predicts that the ratio for 
${\cal B}(\Upsilon^\prime\to\Upsilon\eta)/{\cal B}(\Upsilon^\prime\to\Upsilon\pi\pi)$
should be substantially smaller than the ratio 
${\cal B}(\psi'\to\psi\eta)/{\cal B}(\psi'\to\psi\pi\pi$). 
By contrast, if the ratio of $\pi^+\pi^-$ to $\eta$ 
transitions were governed by phase space 
alone, the $\eta$ transition would be about 15\%
of the $\pi^+\pi^-$ transition for
$\Upsilon^\prime\to\Upsilon$. 
The most recent CLEO analysis yielded an upper limit:
${\cal B}(\Upsilon^\prime\to\Upsilon\eta)<0.0028$, in qualitative agreement
with the rule given above.

The isospin-violating decay $\psi$(2S)$\to\pi^0\psi$(1S) and the 
M1 transition $\psi$(2S)$\to\eta\psi$(1S) have been observed in the
charmonium sector; searches for the corresponding transitions in the
bottomonium sector have resulted only in the upper limit:
$\Upsilon^\prime\to\Upsilon\pi^0<$0.11\%. 
The typically poorer energy resolution in neutral particle measurements, 
coupled with small predicted branching fractions,
makes observation of such decays difficult.

\null\noindent
{\it E. Dipion mass spectra}

\begin{figure}[t]
\centerline{\includegraphics[width=7cm]{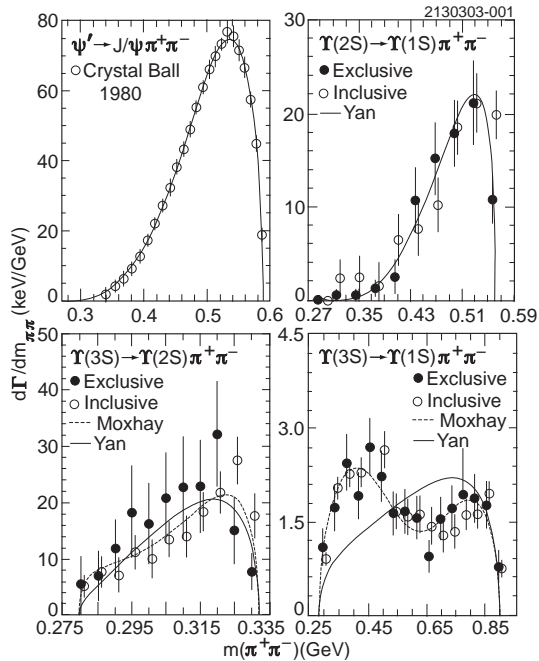}}
\caption[]{Left: $\pi^+\pi^-$ recoil mass spectra from CLEO~II
using data taken at the peak of the $\Upsilon^{\prime\prime}$ for
a) exclusive transitions, and b) inclusive transitions, as indicated 
\cite{CLEO9491}.
Right: $\pi^0\pi^0$ recoil mass spectrum for exclusives with the 
indicated cuts on dilepton mass \cite{CLEO9491}.}
\label{fig:pipirecmasCLEO93} 
\end{figure}

The dipion mass spectra are calculated directly from the pion
four-momenta.  As stated before, the invariant mass spectra are
expected to peak at high mass values. This is, in fact, what is
observed for the transition $\Upsilon^\prime\to\Upsilon\pi^+\pi^-$, as
shown in \Figure~\ref{fig:pipirecmasCLEO93}, and entirely consistent
with an exhaustive study of this process by the ARGUS
collaboration\cite{Albrecht:1986gb}.  Also shown in
\Figure~\ref{fig:pipirecmasCLEO93} are the $\pi^0\pi^0$ mass
spectra for $\Upsilon^{\prime\prime}\to\Upsilon\pi^0\pi^0$ and
$\Upsilon^{\prime\prime}\to\Upsilon^\prime\pi^0\pi^0$.

The current data 
show peaking at high mass for the 
$\Upsilon^{\prime\prime}\to\Upsilon^\prime\pi^+\pi^-$ and
$\Upsilon^\prime\to\Upsilon\pi\pi$ transitions, 
consistent with the
expectation for S$\to$S transitions (and also consistent
with charmonium results). This is the process for 
which the multipole expansion model, owing to the 
smallness of the expansion parameter, claims to have 
the greatest predictive power. However, the 
$\pi^0\pi^0$ and $\pi^+\pi^-$ invariant mass distributions in the 
$\Upsilon^{\prime\prime}\to\Upsilon\pi^+\pi^-$ transition show a
``double bump'' structure that disagrees with the gluon field multipole
expansion model as well as with the expectation
that the matrix element for a transition with these quantum numbers
should approach zero at threshold. 
This is perhaps an indication that the average value of $Q^2$ is too
large to make predictions reliably using the multipole model. It may
also be an indication that a low-mass 0++ scalar (\eg the $\sigma$) may
be contributing to the intermediate state.

There have been various attempts to explain the double-peaked
shape. Ref. \cite{Voloshin-Truong,Belanger89,ABSZ} assumed the
existence of a four-quark state $\Upsilon_1$, which enhances the
low-$M_{\pi\pi}$ region.  So far such a resonance is not found
experimentally. Ref.  \cite{Lipkin-Tuan-Moxhay} assumed a large QPC
part in the $\Upsilon^{\prime\prime}\to \Upsilon\pi\pi$ amplitude
whose interference with the MGE part may form a double-peaked shape.
However, the systematic calculation shown in \Section~\ref{sec:4.3}
does not support this assumption.  Recently, another attempt
considering certain models for a $\sigma$ meson resonance around 500
MeV in the final state $\pi\pi$ interactions \cite{Ishida01,Uehara}
have been proposed. By adjusting the free parameters in the models,
the CLEO data on the $M_{\pi\pi}$ distributions can be fitted.
However, the model need to be tested in other processes.  Therefore,
the HT $\Upsilon^{\prime\prime}\to \Upsilon\pi\pi$ is still an
interesting process needing further investigation.

\null\noindent
{\it F. Three-pion transitions}

\begin{figure}[t]
\begin{center}
\includegraphics[width=7cm]{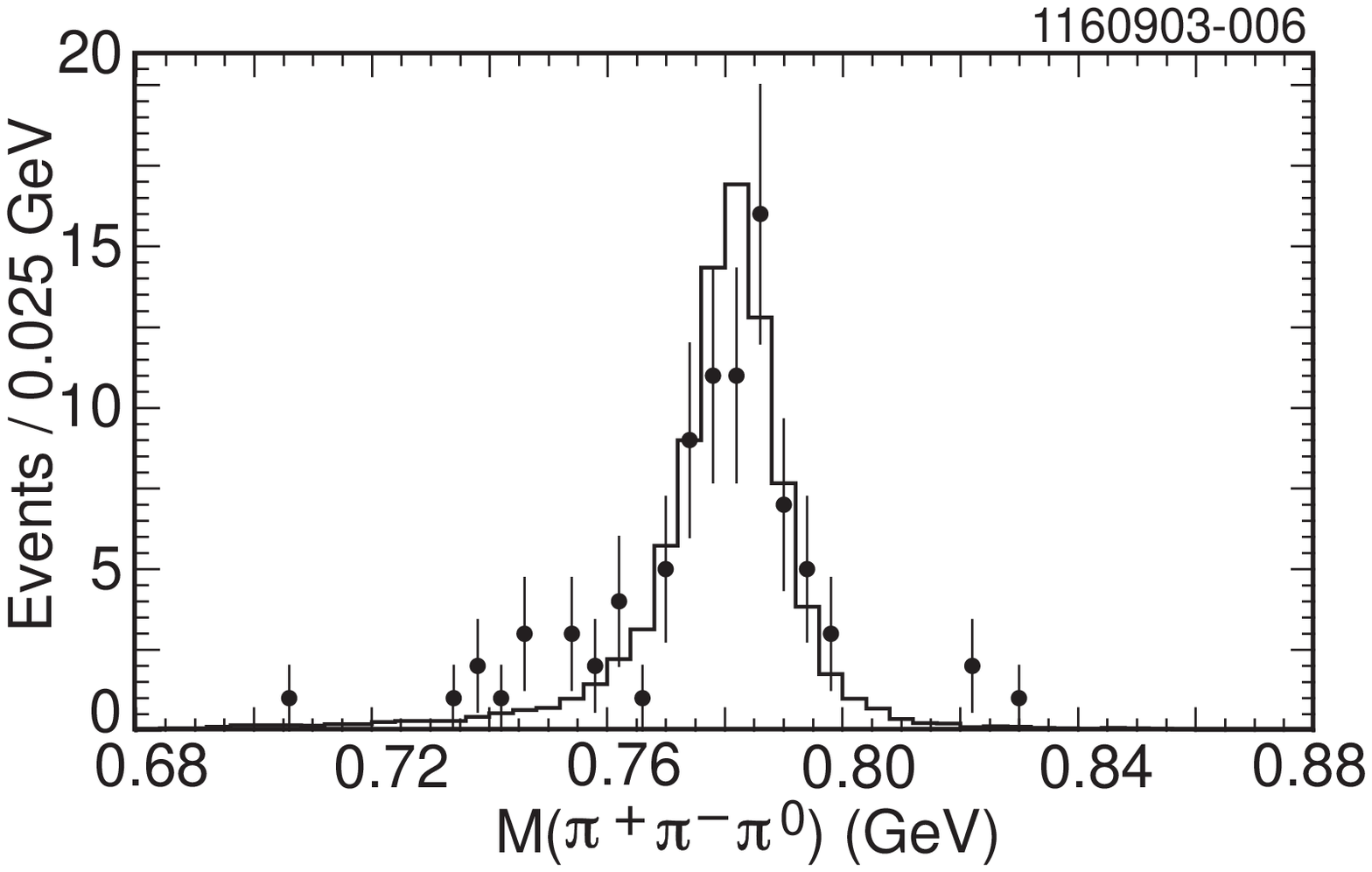}
\end{center}
\caption{Invariant mass of three pions in events consistent with
         $\Upsilon^{\prime\prime}\to\chi_b'\gamma$;
         $\chi_b'\to\Upsilon\pi^+\pi^-\pi^0$.}
\label{fig:chib-omega-1S-omega}

\begin{center}
\includegraphics[width=7cm]{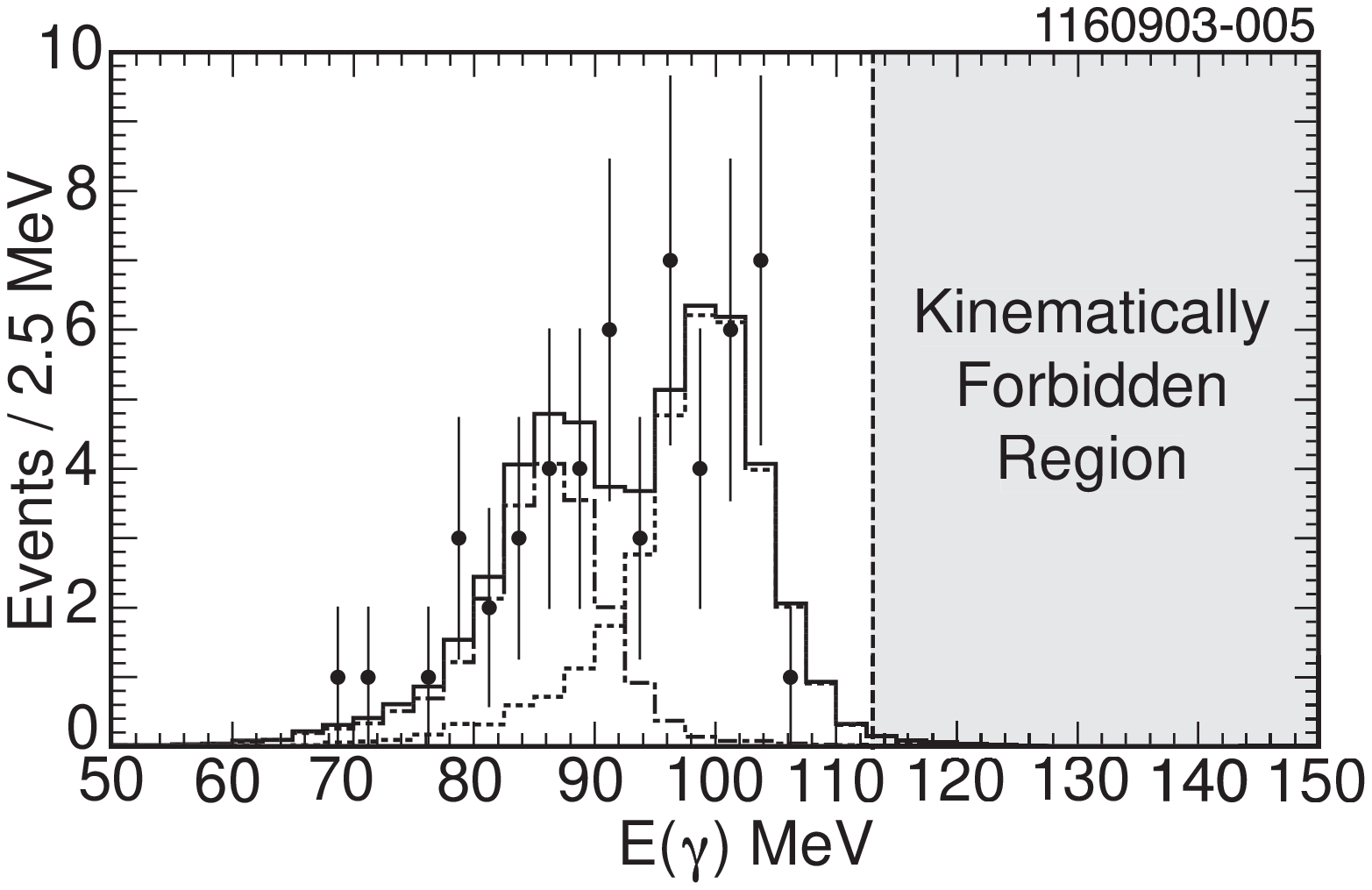}
\end{center}
\caption{Photon energy distribution for the previous figure's
         data sample, showing fits to the J=2 and J=1 states. Transitions from
         the J=0 state are kinematically forbidden (as indicated).}
\label{fig:chib-omega-1S-Egam}
\end{figure}

With their large $\Upsilon^{\prime\prime}$ data sample in hand, the
CLEO collaboration is able to probe beyond the now-familiar dipion
transitions. Of particular interest are $\omega$-mediated transitions,
which have been long-suggested as a possible path to the $\eta_b$,
via: $\Upsilon^{\prime\prime}\to\eta_b\omega$.  In QCDME, by colour
conservation, this must correspond to three E1 gluon emissions.
Although direct decays $\Upsilon^{\prime\prime}\to\eta_b\omega$ were
not found, CLEO has observed significant production of $\Upsilon$ via
$\Upsilon^{\prime\prime}\to\chi_b'(2P)\gamma$,
$\chi_b'(2P)\to\Upsilon\omega$, as shown in
\Figures~\ref{fig:chib-omega-1S-omega} and \ref{fig:chib-omega-1S-Egam}.

What is actually observed are two recoil
mass peaks, corresponding to decays from the $\chi_b'$(2P) (J=2) and (J=1)
states. In fact, large partial widths for such decays had been predicted
(albeit indirectly) in the original QCDME formulation of Gottfried.
As pointed out by Voloshin, 
since the $\omega$ is spin 1, the matrix element should be
largely independent of the spin of the parent 2P, consistent with observation.
The measured branching fractions (${\cal B}(\chi_b'(J=2)\to\Upsilon\omega)
=(1.0\pm0.3\pm0.1)$$\%$ 
and ${\cal B}(\chi_b'(J=1)\to\Upsilon\omega) = (1.6\pm0.3\pm0.24)$$\%$ 
are unexpectedly large, given the limited phase space for these decays.

\null\noindent
{\it G. Hadronic transitions from the $\Upsilon(4S)$}

Observation of hadronic transitions from the $\Upsilon(4S)$, interesting
on its own merits, would provide essential information on the $\Upsilon(4S)$
wave function. Since the $\Upsilon(4S)$ resonance is above the threshold
for $B\bar{B}$ production, measurement of
the dipion transitions, with partial widths a factor
$10^{-4}$ smaller than the dominant strong decays to open bottom, require
data samples of order
$10^{8}$ $\Upsilon(4S)$ events. The BaBar and Belle
experiments now have accumulated samples of 100M $\Upsilon(4S)$ events and
may produce the first signals for such dipion transitions soon. 
CLEO have produced the most recent results on these transitions,
resulting only in upper limits: 
$\Upsilon(4S)\to\Upsilon^\prime\pi\pi<0.039$$\%$;
$\Upsilon(4S)\to\Upsilon\pi\pi<0.012$$\%$. Interest in such decays has recently
been promoted by the BES claim of the corresponding decay in the charmonium
sector: $\psi(3770)\to J\psi\pi^+\pi^-$.

\null\noindent
{\it H. Unanswered questions}

Aside from a first-principles explanation of the dipion mass spectrum
in the $\Upsilon^{\prime\prime}\to\Upsilon\pi^+\pi^-$ spectrum (such a
three-body decay does not, unfortunately, easily lend itself to
lattice gauge techniques), much experimental work remains. Among the
dipion transitions one would like to observe are the $\eta$
transitions between the $S$ states, or one of the two dipion
transitions involving the singlet $1^1P_1$ state: the
isospin-violating decay $\Upsilon^{\prime\prime}\to h_b(1^1P_1)\pi^0$,
or $\Upsilon^{\prime\prime}\to h_b\pi^+\pi^-$~\footnote{For this
$S\rightarrow P$ transition, Kuang \& Yan predict a dipion mass
distribution that peaks at $low$ values of invariant mass. This is
understood by the following argument: such a transition
$1^-\rightarrow 0^+1^+$ can only proceed in P~wave, which suppresses
the high mass region.}, as well as the dipion transitions between the
$\chi_b$ states: $\chi_b'\rightarrow\chi_b\pi\pi$. Owing to the larger
total widths of the $\chi_b'$ (J=2 and J=0) states relative to the J=1
state, the first observation of this decay might be expected in the
transition between the J=1 states. Transitions at higher order in
QCDME, \eg $\Upsilon^{\prime\prime}\to\eta_b\omega$ (E1E1M1
transition), and also HT to the $\eta_b$, which is accessible through
two routes, each of which involves a radiative and a hadronic
transition: either $\Upsilon^{\prime\prime}\to h_b(1^1P_1)\pi\pi$;
followed by $h_b(1^1P_1)\to\eta_b\gamma$, or
$\Upsilon^{\prime\prime}\to\chi_b'\gamma$;
$\chi_b'\rightarrow\eta_b\pi^+\pi^-$, would both help complete our
picture of heavy quark spectroscopy (see
\Chapter~\ref{chapter:spectroscopy}).  Also extremely interesting
would be the observation of HTs from the recently discovered triplet
$D$-bottomonia states ($\Upsilon(1^3D_J)$),
\eg $\Upsilon(1^3D_2)\to\Upsilon\pi^+\pi^-$, or
$\Upsilon(1^3D_2)\to\Upsilon(1S)\eta$. Currently, only an upper limit
exists for the product branching fraction:
$\Upsilon^{\prime\prime}\to\chi'_{b,J=2}\gamma$, $\chi'_{b,J=2}\to
1^3D$, $1^3D\to\Upsilon\pi\pi$ of $1.1\times 10^{-4}$ for the J=2
D-state, and $2.7\times 10^{-4}$, including all the D-states. A
90\% c.l. upper limit is also set for the same decay chain, but with
an $\eta$ rather than dipion transition, of $2.3\times 10^{-4}$.

\subsection[Hadronic transition experiments in the $c\bar{c}$ system]{Hadronic
  transition experiments in the $c\bar{c}$ system}
\label{sec:4.6}

Hadron transitions in the charmonium system where there is
experimental information include $\pp \rt J/\psi~\pi^0$, $\pp \rt
J/\psi~\eta$, and $\pp \rt J/\psi~\pi \pi$. Recently evidence has been
presented on $\psi(3770) \rt J/\psi~\pi^+ \pi^- $ decays, and very
recently, Belle announced the discovery of the $X(3872)$
\cite{belle3870}, which is detected via $X(3872) \rt J/\psi~\pi^+
\pi^-$, making it another means to study hadronic transitions.  Here
recent experimental results on $\pp \rt J/\psi~\pi^0$, $\pp \rt
J/\psi~\eta$, $\pp \rt J/\psi~\pi \pi$ and $\psi(3770) \rt
J/\psi~\pi^+ \pi^-$ will be summarized. We will shortly mention the
$X(3872) \rt J/\psi~\pi^+ \pi^- $ transition, which has been discussed
in detail in \Chapter~\ref{chapter:spectroscopy},
\Section~\ref{sec:spexX3872}.

\null\noindent
{\it A.~~~~{$\pp \rt J/\psi~\pi^0$, $\pp \rt J/\psi~\pi^0~\pi^0$ and $\pp \rt J/\psi~\eta$ }}

Experimental results for the processes $\psip\ra \J~\pi^0$ and $\J~\eta$
are few and were mainly taken in the 1970s and
80s \cite{MK1,CNTR2,DASP,MK2,CB}. 
Recently, however, BES, using a sample of $(14.0\pm 0.6)\times10^6$ $\psip$
events collected with the BES~II detector~\cite{besii}, studied $\psip$
decaying into $\J(\pi^0,\eta)$, with $\pi^0$ and $\eta$ decaying to
two photons, and $\J$ to lepton pairs~\cite{besggJ}.
Events with two charged tracks identified as an electron pair or muon
pair and two or three photon candidates are selected.  A five
constraint (5C) kinematic fit to the hypothesis $\psip\ra\ggll$ with
the invariant mass of the lepton pair constrained to $\J$ mass is
performed, and the fit probability is required to be greater than
0.01.
 
\begin{figure}[t]
\begin{center}
\includegraphics[height=5.5cm,width=8cm]{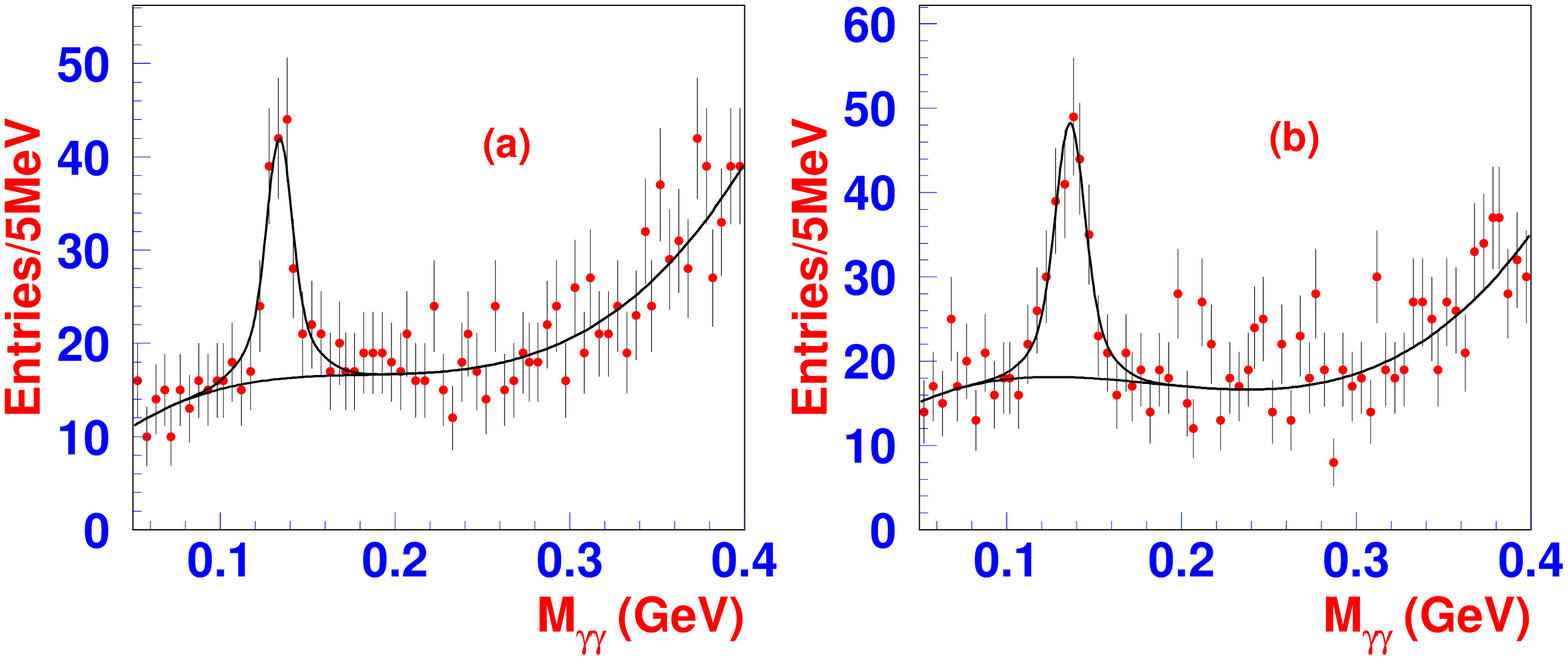}
\end{center}
\caption[Two-photon invariant mass distribution for candidate
         $\psip\ra\Jpi$ events for $\ggee$ and $\gguu$]
        {Two-photon invariant mass distribution for candidate
         $\psip\ra\Jpi$ events for (a) $\ggee$ and (b) $\gguu$.}
\label{fig:MJpiee}

\begin{center}
\includegraphics[height=4.cm,width=7.5cm]{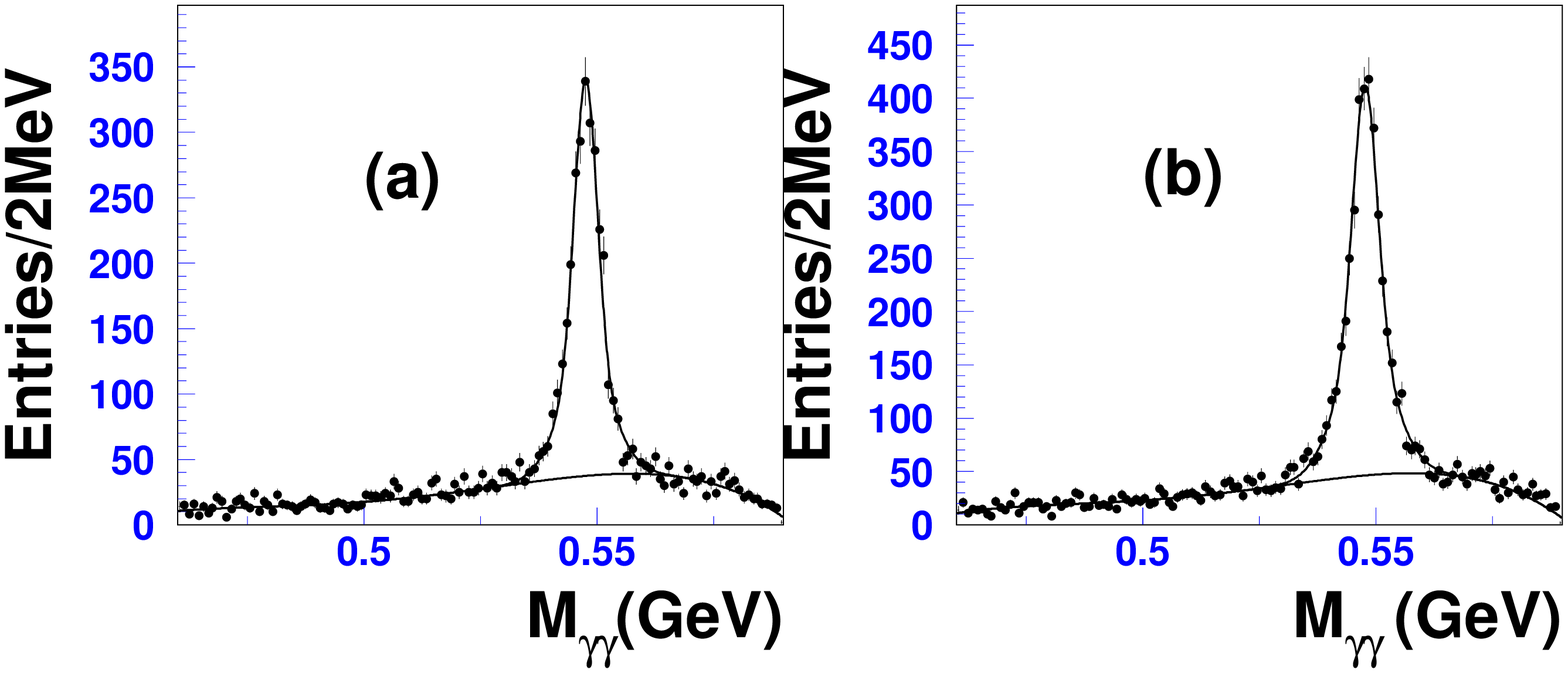}
\end{center}
\caption[Two-photon invariant mass distribution for candidate
         $\psip\ra\Jeta$ events for $\ggee$ and $\gguu$]
        {Two-photon invariant mass distribution for candidate
         $\psip\ra\Jeta$ events for (a) $\ggee$ and (b) $\gguu$.}
\label{fig:MJetaee}
\end{figure}

To remove the huge background from $\psip\ra\gx$ under the
$\psip\ra\J~\pi^0$ signal, the invariant mass of the highest energy
gamma and the $J/\psi$, $\MgJ$ is required to be less than 3.49 or
greater than 3.58~Gev/c$^2$.  \Figure[b]~\ref{fig:MJpiee} shows, after
this requirement, the distribution of invariant mass, $\Mgg$, where
the smooth background is due to $\ptochic$ and $J/\psi~\pi^0 \pi^0
$. A Breit Wigner with a double Gaussian mass resolution function to
describe the $\pi^0$ resonance plus a third-order background
polynomial is fitted to the data.

In the $\psip\ra\J~\eta$ channel, the main backgrounds are from
$\psip\ra\J~\pi^0\pi^0$ and $\gx$. By requiring $\MgJ<3.49$~Gev/c$^2$,
most background from $\psip\ra \gx$ is removed.  The resultant plot
shown in \Figure~\ref{fig:MJetaee} shows a clear $\eta$ signal
superimposed on background, mainly from $\psip\ra\Jpipi$.  A fit is
made using a Breit--Wigner resonance convoluted with a mass resolution
function for the $\eta$ signal plus a polynomial background, where the
width of the $\eta$ is fixed to its Particle Data Group (PDG) value
\cite{Eidelman:2004wy} and the background function is determined from
$\psip\ra\J~\pi^0\pi^0$ Monte Carlo simulated events that satisfy the
same criteria as the data.

\begin{table}[htbp]
\caption{Recent BES results on $\psip\ra\J~\pi^0$ and $\psip\ra\J~\eta$.}
\label{tab:BESresults} 
\doublerulesep 0.5pt
\begin{center}
{\footnotesize{
\begin{tabular}{c|cc|cc}              \hline        \hline
Channel&\multicolumn{2}{c}{$\jpsi~\pi^0$}&\multicolumn{2}{c}{$\jpsi~\eta$}\\\hline
Final state&$\ggee$&$\gguu$&$\ggee$&$\gguu$\\\hline
Number of events&$123\pm 18$&$155\pm 20$&$2465\pm 101$&$3290\pm 148$\\
Efficiency ($\%$)&11.21&13.34&26.94&34.07\\
Sys. error (\%)&9.68&8.77&8.54&8.40\\
Correction factor&0.962&0.974&0.962&0.974\\\hline
BR (\%)&$0.139\pm 0.020\pm 0.013$&$0.147\pm 0.019\pm
0.013$&$ 2.91\pm 0.12\pm 0.21$&$3.06\pm 0.14\pm 0.25 $\\\hline
Combine BR (\%)&\multicolumn{2}{|c|}{$0.143\pm 0.014\pm 0.013$}
&\multicolumn{2}{|c}{$2.98\pm 0.09\pm 0.23$}\\
PDG (\%)\cite{Eidelman:2004wy}&\multicolumn{2}{|c|}{$0.096\pm 0.021 $}&\multicolumn{2}{|c}
{$3.16\pm 0.22$}\\\hline\hline\hline
\end{tabular}
}}
\end{center}
\end{table}

Using the fitting results and the efficiencies and correction factors
for each channel, the branching fractions listed in
\Table~\ref{tab:BESresults} are determined. The BES ${\cal
B}(\psip\ra\J~\pi^0)$ measurement has improved precision by more than
a factor of two compared with other experiments, and the
$\psip\ra\J~\eta$ branching fraction is the most accurate single
measurement. The BES ${\cal B}(\psip\ra\J~\pi^0)$ agrees better with
the Mark~II result \cite{MK2} than with the Crystal Ball result
\cite{CB}. For the comparison of the BES result with related
theoretical predictions, see \Section~\ref{sec:4.2}.

In another recent BES analysis \cite{Xjpsi}, based on a sample of
approximately $4 \times 10^6$ $\pp$ events obtained with the BES~I
detector~\cite{besI}, a different technique is used for measuring
branching fractions for the inclusive decay $\pp \rt J/\psi \: {\rm
anything} $, and the exclusive processes for the cases where $X =
\eta$ and $X = \pi\pi$.  Inclusive $\mu^+ \mu^-$ pairs are
reconstructed, and the number of $\pp \rt J/\psi X$ events is
determined from the $J/\psi \rt \mu^+ \mu^-$ peak in the $\mu^+ \mu^-$
invariant mass distribution.  The exclusive branching fractions are
determined from fits to the distribution of masses recoiling from the
$J/\psi$ with Monte Carlo determined distributions for each individual
channel.

\begin{figure}
\begin{center}
\includegraphics[height=52mm]{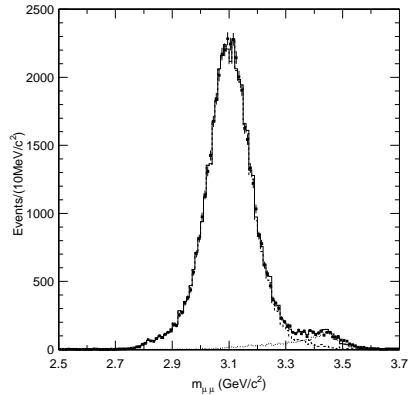}
\end{center}
\caption[Distribution of dimuon invariant mass $m_{\mu \mu}$]
        {Distribution of dimuon invariant mass, $m_{\mu \mu}$, for
         events that pass the $J/\psi \rt \mu^+ \mu^-$ kinematic
         fit. Dots with error bars are data.  Also shown is the fit
         (solid histogram) to the distribution with signal (long
         dashed histogram) and background (short dashed histogram)
         shapes.}
\label{fig:fit_mmumu}
\end{figure}

Selected events are required to have more than one and less than six
charged tracks and must have two identified muon tracks with zero net
charge.  The two muon tracks must
satisfy a one constraint kinematic fit to the $J/\psi$ mass.  Shown in
\Figure~\ref{fig:fit_mmumu} is the dimuon invariant mass distribution,
$m_{\mu \mu}$, for these events.  A clear peak at the
$J/\psi$ mass is evident above background.

\begin{figure}[t]
\begin{center}
\begin{minipage}[t]{.54\textwidth}
\begin{center}
\includegraphics[width=70mm]{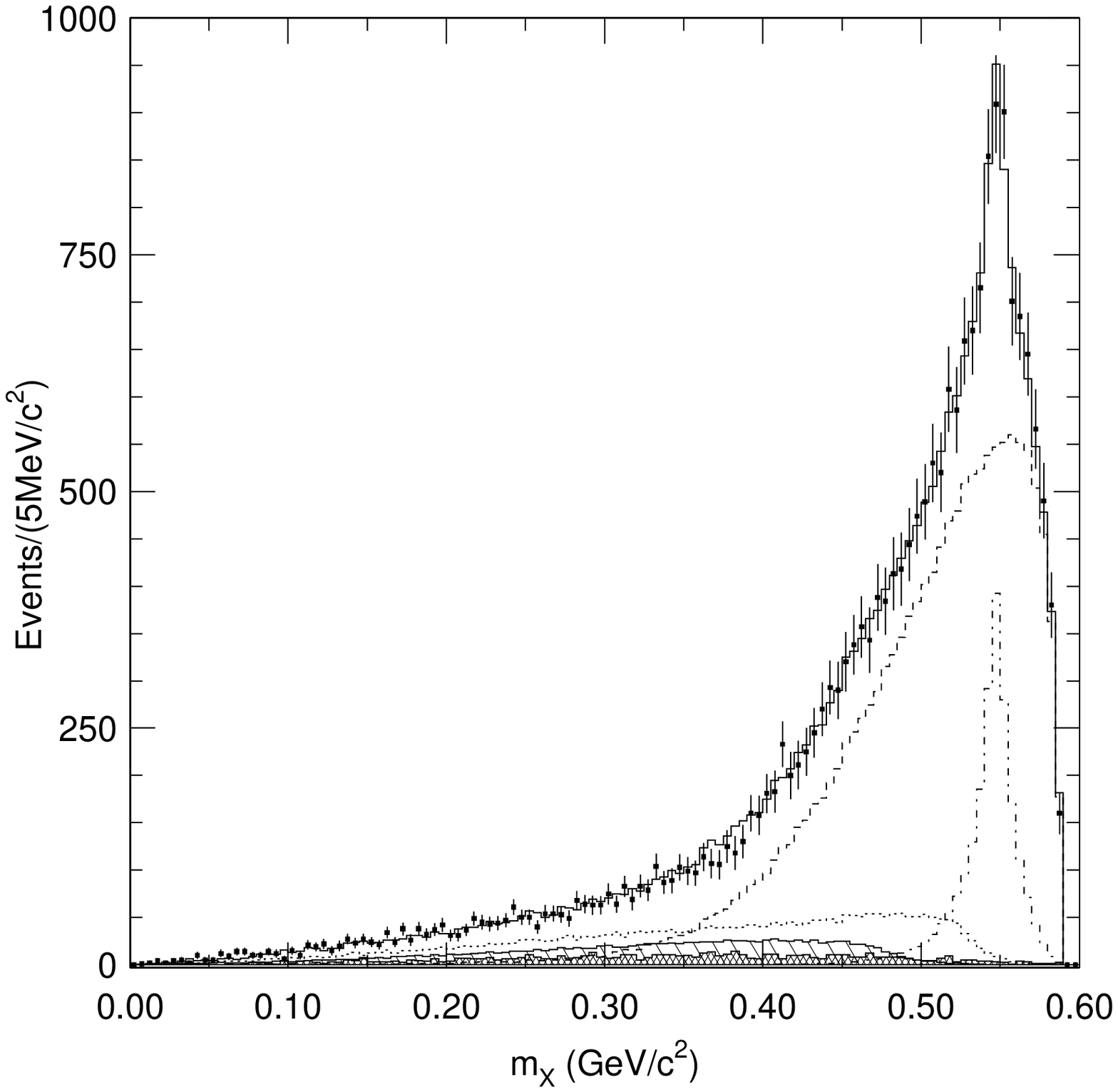}
\end{center}
\caption[Fit of the $m_X$ distribution events with
         no additional charged tracks]
        {Fit of the $m_X$ distribution events with no additional
         charged tracks.  Shown are the data (points with error bars),
         the component histograms, and the final fit.  For the
         components, the large, long-dash histogram is $\psi(2S) \rt
         J/\psi \pi \pi$, the narrow, dash--dot histogram is $\psi(2S)
         \rt J/\psi \eta $, the broad, short-dashed histogram is $\pp
         \rt \gamma \chi_{c1}, \chi_{c1} \rt \gamma J/\psi$, the
         broad, hatched histogram is $\pp \rt \gamma \chi_{c2},
         \chi_{c2} \rt \gamma J/\psi$, and the lowest cross-hatched
         histogram is the combined $e^+ e^- \rt \gamma \mu^+ \mu^-$
         and $e^+ e^- \rt \psi(2S), \psi(2S) \rt (\gamma)\mu^+ \mu^-$
         background. The final fit is the solid histogram.}
\label{fig:none}
\end{minipage} \hfill
\begin{minipage}[t]{.44\textwidth}
\begin{center}
\includegraphics[width=70mm]{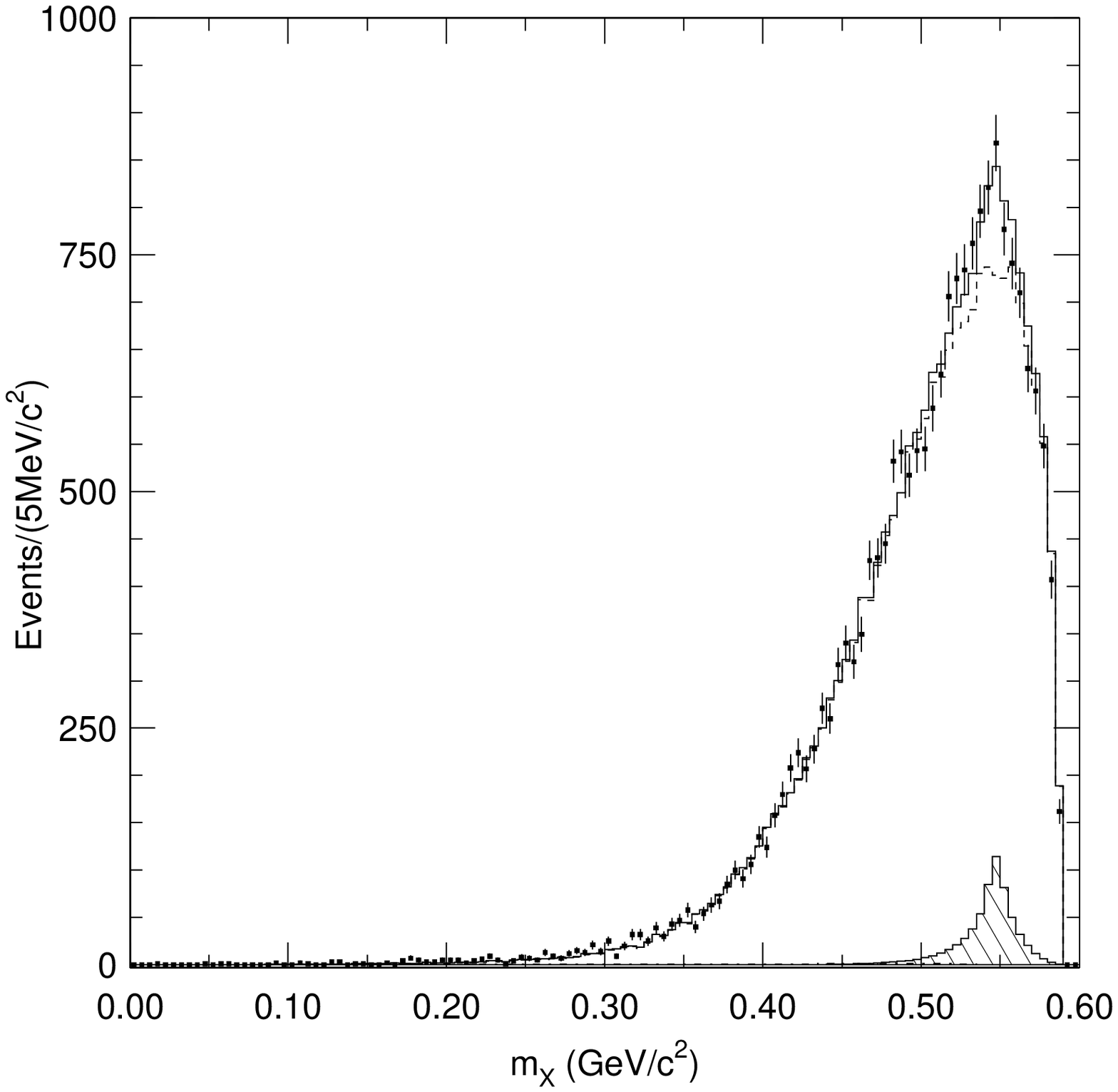}
\end{center}
\caption[Fit of the $m_X$ distribution for events with any number of
         additional charged tracks]
        {Fit of the $m_X$ distribution for events with any number of
         additional charged tracks.  Shown are the data (points with
         error bars), the component histograms, and the final fit
         (solid histogram). The dashed histogram is $\psi(2S) \rt
         J/\psi \pi^+ \pi^-$, and the hatched histogram is $\psi(2S)
         \rt J/\psi \eta$.  There is very little evidence for
         $\psi(2S) \rt \gamma \chi_{c1/2}, \chi_{c1/2} \rt \gamma
         J/\psi$.  This distribution is composed predominantly of
         $\psi(2S) \rt J/\psi \pi^+ \pi^-$.}
\label{fig:one}
\end{minipage}
\end{center}
\end{figure}

The mass recoiling against the $J/\psi$ candidates, $m_X$ is
determined from energy and momentum conservation.  In order to
distinguish $\psi(2S) \rt J/\psi \pi^+ \pi^-$ and $\psi(2S) \rt J/\psi
\pi^0 \pi^0$ events, separate $m_X$ histograms are made for events
with no additional charged tracks and those with additional charged
tracks.  To reduce background and improve the quality of the track
momentum measurements, events used for this part of the analysis are
required to have a kinematic fit $\chi^2 < 7$.  The $m_X$ histograms
for events with and without additional charged tracks, selected
according to the above requirements, are shown in
\Figures~\ref{fig:none} and \ref{fig:one}.

To determine the number of exclusive decays and separate $\psi(2S) \rt
J/\psi \pi^0 \pi^0$ and $\psi(2S) \rt J/\psi \pi^+ \pi^- $ events,
$m_X$ histograms for events with and without additional charged
tracks, shown in \Figures~\ref{fig:none} and \ref{fig:one}, are fit
simultaneously.  Contributions from the $\psi(2S) \rt \gamma
\chi_{c0}, \chi_{c0} \rt \gamma J/\psi$ are expected to be very small
\cite{Eidelman:2004wy} and are not included in the fit.  The influence
of $\pp \rt J/\psi \pi^0 $ is also small, indeed there is no
indication of such a component in \Figure~\ref{fig:none}, and this
channel is also not included.  The $m_X$ distributions for $\psi(2S)
\rt \gamma \chi_{c1}, \chi_{c1} \rt \gamma J/\psi$, $\psi(2S) \rt
\gamma \chi_{c2}, \chi_{c2} \rt \gamma J/\psi$ , and the background
are broad and rather similar in shape, as can be seen in
\Figure~\ref{fig:none}.  Since these are difficult to distinguish, the
$\chi_{c2}$ to $\chi_{c1}$ ratio is constrained using calculated
efficiencies and the PDG world average branching fractions for the two
processes.

To avoid a number of systematic errors, the channels of interest are
normalized to the observed number of $J/\psi \pi^+ \pi^- $ events;
ratios of the studied branching fractions to that for ${\cal B}(\pp
\rt J/\psi \pi^+ \pi^- )$ are reported.  The advantage of normalizing
in this way is that many of the muon selection systematic errors
largely cancel, as well as the systematic error due to the $\chi^2$
requirement.

\begin{table}[ht]
\caption[Final branching ratios and branching fractions]
        {Final branching ratios and branching fractions.  PDG04-exp
         results are single measurements or averages of measurements,
         while PDG04-fit are results of their global fit to many
         experimental measurements. For the value marked with an
         asterisk, the PDG gives the reciprocal.  The BES results in
         the second half of the table are calculated using the PDG
         value of ${\cal B}_{\pi \pi} = {\cal B}(\pp \rt J/\psi \pi^+
         \pi^- ) = (31.7 \pm 1.1)\%$.}
\label{tab:results} 
\begin{center}
\begin{tabular} {|l|c|c|c|} \hline
Case                                      &   This result &
PDG04-exp & PDG04-fit  \\ \hline
${\cal B}(J/\psi \; {\rm anything})/{\cal B}_{\pi \pi}$   & $1.867  \pm 0.026  \pm 0.055$ & $2.016
\pm 0.150$ \cite{armstrong} &$1.821 \pm 0.036^*$   \\
${\cal B}(J/\psi\pi^0 \pi^0)/{\cal B}_{\pi \pi}$ & $0.570 \pm 0.009 \pm
0.026$ & -- & $0.59 \pm 0.05$ \\
${\cal B}(J/\psi \eta)/{\cal B}_{\pi \pi}$ & $0.098 \pm  0.005 \pm
0.010$ & $0.091 \pm 0.021$ \cite{MK2}& $0.100 \pm 0.008$\\ \hline
${\cal B}(J/\psi \; {\rm anything})$ (\%)  & $59.2  \pm 0.8 \pm 2.7  $ & $55 \pm 7$ & $57.6 \pm 2.0$ \\ 
 ${\cal B}(J/\psi \pi^0 \pi^0)$ (\%)  &  $18.1 \pm 0.3 \pm 1.0 $ & -- & $18.8 \pm 1.2$\\
 ${\cal B}(J/\psi \eta)$  (\%)    &  $3.11 \pm 0.17 \pm 0.31 $ & $2.9 \pm
0.5$ & $3.16 \pm 0.22$\\ \hline
\end{tabular}
\end{center}
\end{table}

The final branching fraction ratios and branching fractions are shown
in \Table~\ref{tab:results}, along with the PDG results, including
their experimental averages and global fit results.  For the ratio of
${\cal B}(\psi(2S)$ $\rt$ $J/\psi \pi^0 \pi^0)$ to ${\cal B}(\psi(2S) \rt J/\psi \pi^+
\pi^-)$, the PDG does not use the previous experimental results and
gives no average value.  For the other branching fraction ratios, only
one measurement exists for each, and \Table~\ref{tab:results} lists the
single measurements quoted by the PDG. The results for ${\cal B}(J/\psi
\;{\rm anything})/{\cal B}(\psi(2S)$ $\rt$ $J/\psi \pi^+ \pi^-)$ and ${\cal B}( J/\psi
\eta)/{\cal B}(\psi(2S) \rt J/\psi \pi^+ \pi^- )$ have smaller errors than
the previous results.

To determine the branching fractions, the ratios are multiplied by the
PDG value for ${\cal B}(\pppp) = (31.7 \pm 1.1)$\%.  The agreement for
both the ratios of branching fractions and the calculated branching
fractions using the PDG result for ${\cal B}(\pp \rt J/\psi \pi^+ \pi^-)$
with the PDG fit results is good, and the determination of ${\cal B}(
J/\psi \eta)$ agrees well with the determination from $\psi(2S)  \rt \gamma
\gamma J/\psi$ decays above.
\shortpage

\null\noindent
{\it B.~~~~{$\pp \rt J/\psi~\pi^+ \pi^-$}}

The process $\pp\rt \J~\pi^+\pi^-$, is the largest decay mode of the $\psi(2S)$
\cite{Eidelman:2004wy}.  Early investigation of this decay by Mark I
\cite{abrams} found that the $\pi^+ \pi^-$ mass distribution was
strongly peaked towards higher mass values, in contrast to what was
expected from phase space.  Further, angular distributions strongly
favored S-wave production of $J/\psi~\pi \pi$, as well as an S-wave
decay of the dipion system.
The challenge of describing the mass spectrum attracted considerable
theoretical interest~\cite{BrownCahn,voloshin2,Gottfried,Yan,KY81,KTY88,VZNS}.

The $\pp \rt J/\psi~\pi \pi$ decay was studied by BES \cite{dist_pub},
using 22,800 almost background free exclusive $\ppll$ events, where
$l$ signifies either $e$ or $\mu$, from a data sample of $3.8 \times
10^6$ $\psi(2S)$ decays.

The angular distributions were fit using the general decay amplitude
analysis of Cahn \cite{Cahn75}.  The decay can be described in terms
of partial wave amplitudes, $M_{l,L,S}$, where ${\bf l}$ is the $\pi
\pi$ angular momentum, ${\bf L}$ is $J/\psi$ X ($X \rt \pi^+ \pi^-$)
angular momentum, ${\bf S}$ is the channel spin (${\bf S} = {\bf s} +
{\bf l}$), and ${\bf s}$ is the spin of the $J/\psi$.  Parity
conservation and charge conjugation invariance require both $L$ and
$l$ to be even.  The partial waves can be truncated after a few
terms. Considering only $M_{001}$, $M_{201}$, and $M_{021}$
\cite{cahnerror}:
\begin{equation}
\frac{d \Gamma}{d \Omega_{J/\psi}} \propto [ |M_{001}|^2 + |M_{201}|^2
+ \frac{1}{4}|M_{021}|^2(5 - 3 \cos^2 \theta_{J/\psi}^*)
+ \frac{1}{\sqrt 2}\Re \{M_{021}M_{001}^*\}(3\cos^2
\theta_{J/\psi}^*-1)], 
\label{eq:4.6.1}
\end{equation}
\begin{equation}
\frac{d \Gamma}{d \Omega_{\pi}} \propto [ |M_{001}|^2
 + \frac{1}{4}|M_{201}|^2(5 - 3\cos^2 \theta_{\pi}^*)
 + |M_{021}|^2
+ \frac{1}{\sqrt 2}\Re \{M_{201}M_{001}^*\}(3\cos^2 \theta_{\pi}^*-1)],
\end{equation}
\begin{equation}
\frac{d \Gamma}{d \Omega_{\mu}} \propto [ |M_{001}|^2(1 
+ \cos^2 \theta_{\mu}^*) +\frac{1}{10}(|M_{201}|^2 + |M_{021}|^2)(13
+ \cos^2 \theta_{\mu}^*)], 
\label{eq:4.6.3}
\end{equation}
where $\theta_{J/\psi}^*$ is the polar angle of the ${J/\psi}$
relative to the beam direction in the lab, $\theta_{\pi}^*$ is the
angle between the momenta of $J/\psi$ and $\pi^+$ in the rest frame of
the $\pi \pi$ system, and $\theta_{\mu}^*$ is the angle between the
beam direction and $\mu^+$ in the rest frame of the $J/\psi$.  The $d
\Omega$'s are measured in their respective rest frames, and the
$M_{l,L,S}$ are functions of $m_{\pi \pi}$.
\shortpage

\begin{figure}[t]
\begin{center}
\includegraphics[width=.57\linewidth]{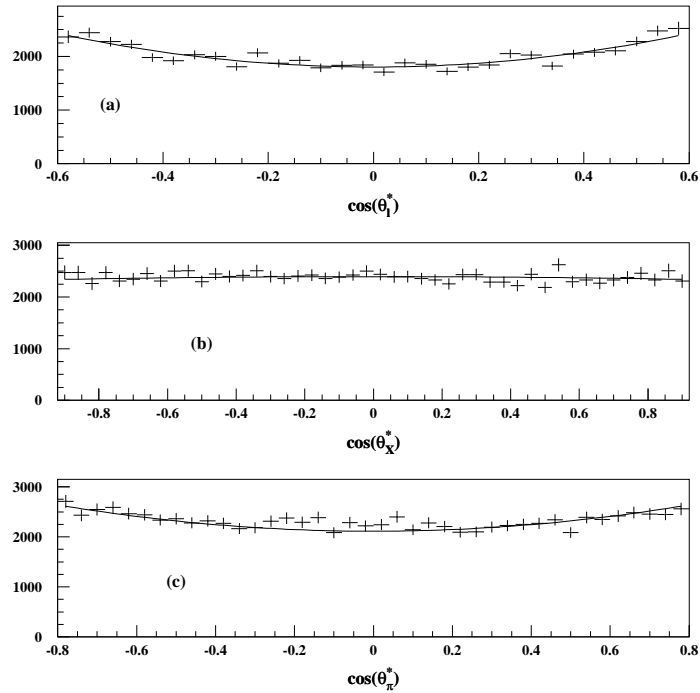}
\end{center}
\caption[Angular distributions of $\cos \theta_{\mu}^*$, 
         $\cos \theta_X^*$, and $\cos \theta_{\pi}^*$]
        {Angular distributions of {\bf (a)} $\cos \theta_{\mu}^*$,
         {\bf (b)} $\cos \theta_X^*$, and {\bf (c)} $\cos
         \theta_{\pi}^*$.  The fit shown uses the partial wave
         analysis description of Cahn~\cite{cahnerror}.}
\label{fig:simul1da}
\end{figure}

There are three complex numbers to be obtained.  According to Cahn, if
the $\psi(2S)$ and $J/\psi$ are regarded as inert, then $M_{l,L,S} =
e^{i\delta^0_l(m_{\pi \pi})} |M_{l,L,S}|$, where $\delta^0_l(m_{\pi
\pi})$ is the isoscalar phase shift for quantum number $l$.  The phase
angles are functions of $m_{\pi \pi}$.  Interpolating the S-wave,
isoscalar phase shift data found in Ref.~\cite{Belanger89,Ishida}, BES
took $\delta^0_0$ to be $\approx 45^{\circ}$ and $\delta^0_2 \approx
0$.  Using these values as input, BES obtained the combined fit to
\Eqs~(\ref{eq:4.6.1})--(\ref{eq:4.6.3}), shown in
\Figure~\ref{fig:simul1da}. The fit yields a nonzero result for
$|M_{201}|$, indicating that the dipion system contains some D-wave,
which is shown by the non-flat angular distribution for $\cos
\theta_{\pi}^*$ seen in \Figure~\ref{fig:simul1da}.  On the other hand
$|M_{021}|/|M_{001}|$, which measures the amount of D-wave of the
$J/\psi$--$X$ system relative to the S-wave, is consistent with
zero, which is indicated by the flat angular distribution for $\cos
\theta_X^*$ shown in \Figure~\ref{fig:simul1da} \null\vspace{0.2cm}

Observation of a small D-wave contribution is interesting
theoretically since, as we have seen in \Eqs~(\ref{eq:4.2.16}) and
(\ref{eq:4.2.4}), there is only S-wave contribution in the NRSC
approach, \ie {\it the existence of a small D-wave contribution
implies that the present NRSC theory should be improved to contain
systematic relativistic and coupled-channel contributions}.

\begin{figure}[t]
\begin{center}
\begin{minipage}[b]{.51\textwidth}
\centering\includegraphics[width=\linewidth]{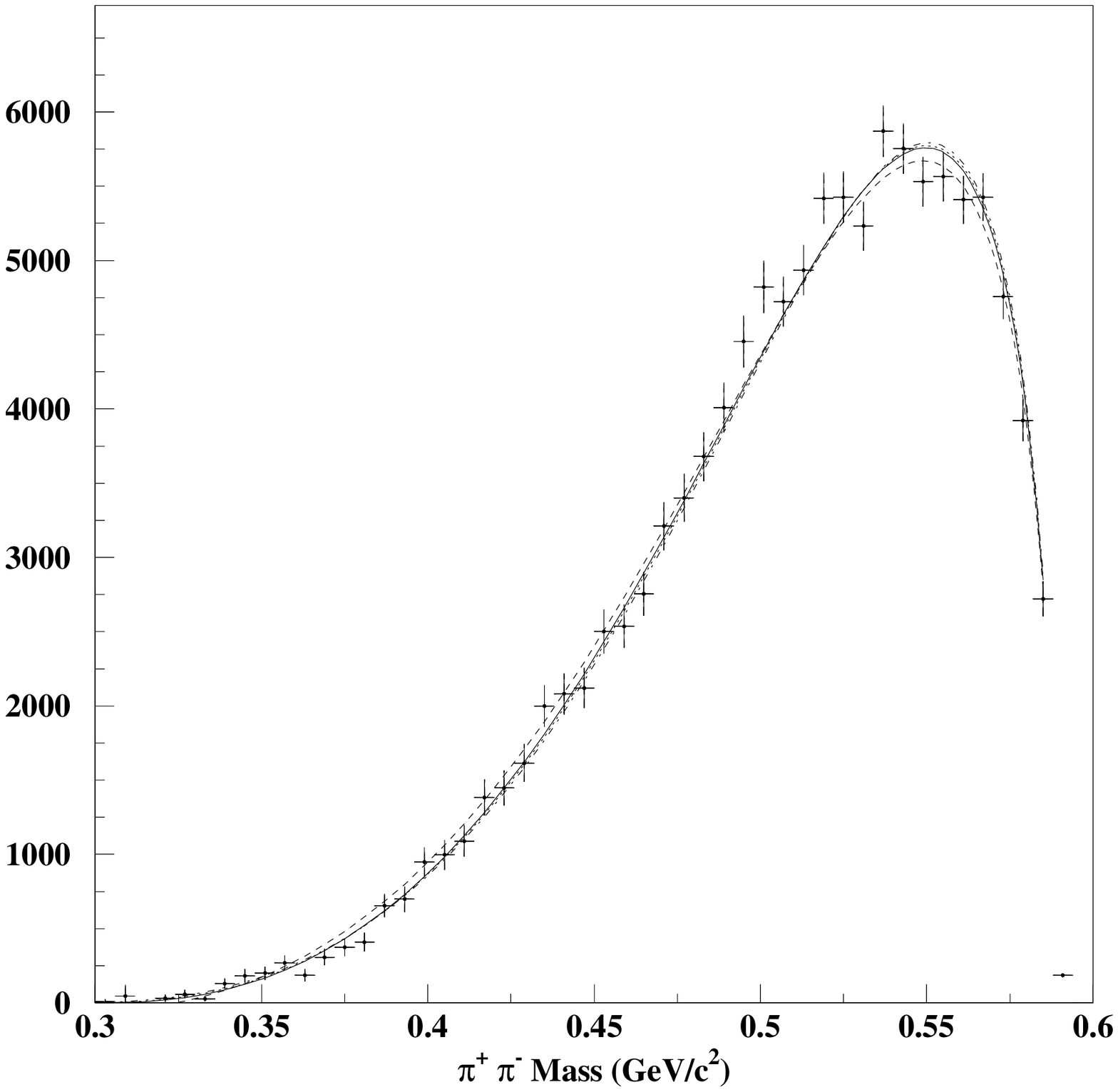}
\end{minipage}\hfill
\begin{minipage}[b]{.46\textwidth}
\setlength{\unitlength}{1mm}%
\centering\includegraphics[height=8cm]{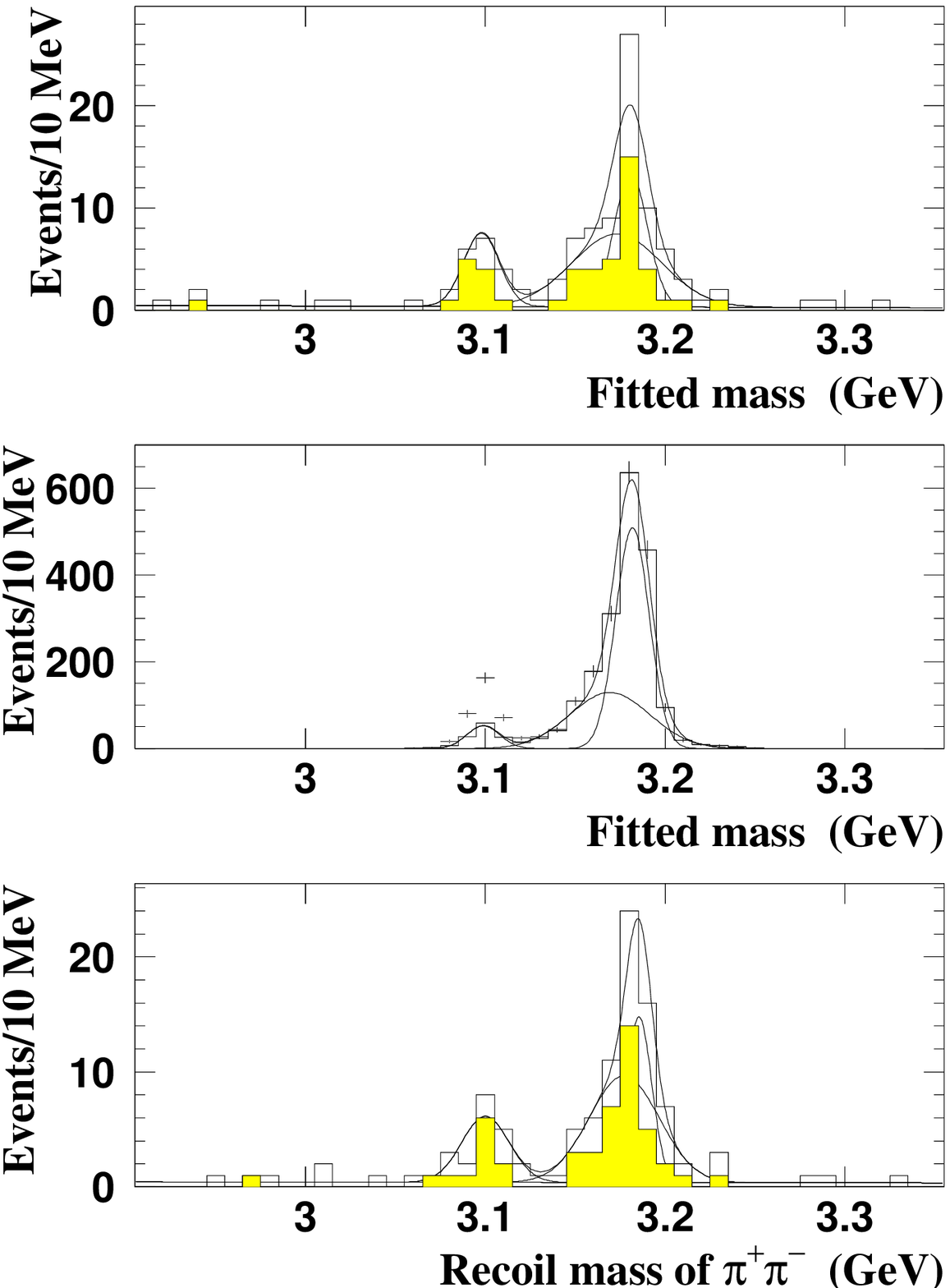}
\begin{picture}(0.0001,0.0001)
\put(-45,74){\bf{ (a)}}
\put(-45,46){\bf{ (b)}}
\put(-45,20){\bf{ (c)}}
\end{picture}\end{minipage}\\
\begin{minipage}[t]{.51\textwidth}
\caption[Fits to the $m_{\pi \pi}$ distribution from 
         $\pp$ $\rt$ $\pi^+ \pi^-$ $J/\psi$]
        {Fits to the $m_{\pi \pi}$ distribution from $\pp$ $\rt$
         $\pi^+ \pi^-$ $J/\psi$.  The points are the data corrected
         for efficiency, and the curves are the fit results.  The
         smooth curve is the Novikov--Shifman model.  The other models,
         which are described in Ref.~\cite{dist_pub}, give similar
         results.}
\label{fig:jingyun_mpipi}
\end{minipage}~
\begin{minipage}[t]{.46\textwidth}
\caption[Distributions of dilepton masses for data and Monte
         Carlo]
        {Distributions of dilepton masses for (a) data and (b) Monte
         Carlo sample for events passing the selection for $\psi(3770)
         \rt \pi^+ \pi^- J/\psi$. The hatched histogram in (a) is for
         $J/\psi \rightarrow \mu^+\mu^-$, while the open one is for
         $J/\psi \rightarrow e^+e^-$. The histogram in (b) is for
         $\psi(2S)\rightarrow J/\psi \pi^+\pi^-$, while the points
         with error bars are the sum of $\psi(3770)\rightarrow J/\psi
         \pi^+\pi^-$ and $\psi(2S)\rightarrow J/\psi \pi^+\pi^-$.  (c)
         Distribution of mass recoiling against the $\pi^+\pi^-$
         system calculated using measured momenta for events that pass
         the kinematic fit requirement, where the hatched histogram is
         for $J/\psi \rightarrow \mu^+\mu^-$ and the open one is for
         $J/\psi \rightarrow e^+e^-$.}
\label{fig:3770}
\end{minipage}
\end{center}
\end{figure}

The $m_{\pi \pi}$ invariant mass spectrum has been fit with the
Novikov--Shifman model and other models, as shown in
\Figure~\ref{fig:jingyun_mpipi}. As can be seen, they give nearly
identical fits.
%%%%%$$$$$$$$$$$$

Mannel and Urech have constructed an effective Lagrangian using chiral
symmetry arguments to describe the decay of heavy excited S-wave
spin-1 quarkonium into a lower S-wave spin-1 state
\cite{Mannel:1995jt}.  Using total rates, as well as the invariant
mass spectrum from Mark II via ARGUS \cite{Albrecht:1986gb}, the
parameters of this theory have been obtained.  More recently,
M. L. Yan \etal ~\cite{mlyan} have pointed out that this model allows
D-wave contributions.  BES fit the joint $\cos
\theta_{\pi}^*$--$m_{\pi \pi}$ distribution using the amplitude of
Mannel and Urech.  The results are given in Ref.~\cite{dist_pub},
along with the results from Ref.~\cite{Mannel:1995jt} which are based
on ARGUS--Mark II \cite{Albrecht:1986gb}.
\shortpage

\null\noindent
{\it C.~~~~ $\psi(3770) \rt J/\psi~\pi^+ \pi^-$}

BES has reported evidence for $\psi(3770) \rightarrow
J/\psi~\pi^+\pi^-$ based on 
$27.7$ pb$^{-1}$ of data taken in the centre-of-mass (c.m.) 
energy region around 3.773~Gev 
using the BES~II detector \cite{Bai:2003hv}.

To search for the decay of $\psi(3770) \rightarrow J/\psi~\pi^+\pi^-$,
$J/\psi \rightarrow e^+e^-$ or $\mu^+\mu^-$, $\mu^+\mu^- \pi^+ \pi^-$ and
$e^+e^- \pi^+ \pi^-$ candidate events are selected.
They are required to have four charged tracks with 
zero total charge.
Each track is required to have a good helix fit, to be consistent
with originating from the primary event vertex, and to satisfy $|\cos
\theta|<0.85$, where $\theta$ is the polar angle.
Pions and leptons must satisfy particle identification requirements.

In order to reduce background and improve momentum resolution, events
are subjected to four-constraint kinematic fits to either the $e^+e^-
\rightarrow \mu^+\mu^- \pi^+ \pi^-$ or the $e^+e^- \rightarrow e^+e^-
\pi^+ \pi^-$ hypothesis.  Events with a confidence level greater than
1\% are accepted.  \Figure[b]~\ref{fig:3770}(a) shows the dilepton
masses determined from the fitted lepton momenta of the accepted
events.  There are clearly two peaks. The lower mass peak is mostly
due to $\psi(3770) \rightarrow J/\psi~\pi^+\pi^-$, while the higher
one is produced via radiative return to the peak of the $\psi(2S)$.

A maximum likelihood fit to the mass distribution in
\Figure~\ref{fig:3770}(a), using a Gaussian function to describe the
peak near the $J/\psi$ mass, two Gaussian functions to represent the
second peak from radiative return to the $\psi(2S)$ peak, and a
polynomial to represent the broad background, yields a signal of
$17.8\pm 4.8$ events with a significance of $6.2~\sigma~$.
\shortpage

Backgrounds from QED radiative processes with $\gamma$ conversion,
two-photon backgrounds, such as $e^+e^- \rightarrow e^+e^-\mu^+\mu^-$
(where the slow muons are misidentified as pions) and $e^+e^-
\rightarrow e^+e^-\pi^+\pi^-$, and $e^+e^- \rightarrow \tau^+\tau^-$,
are negligibly small, as are $J/\psi~\pi^+\pi^-$ events produced in
the continuum process, $e^+ e^- \rt l^+l^-\pi^+\pi^-$.  However, there
is a contribution from $\psi(2S) \rightarrow J/\psi~\pi^+\pi^-$ that
can pass the event selection criteria and yield fitted dilepton masses
around 3.097~Gev.  This is the main background to $\psi(3770)
\rightarrow J/\psi~\pi^+\pi^-$, as shown in
\Figure~\ref{fig:3770}(b). Here the histogram shows the dilepton mass
distribution for $\psi(2S) \rightarrow J/\psi~ \pi^+\pi^-$ from a
Monte Carlo simulation.  The higher peak is due to the radiative
return to the $\psi(2S)$ peak, and the lower peak is from the tail of
the $\psi(2S)$.  The points with error bars show the total
contribution from $\psi(2S)$ and $\psi(3770)$ production and decay.
From the simulation, it is estimated that $6.0 \pm 0.5 \pm 0.6$ out of
$17.8 \pm 4.8$ events in the peak near 3.1~Gev in
\Figure~\ref{fig:3770}(a) are due to $\psi(2S) \rightarrow J/\psi~
\pi^+\pi^-$, where the first error is statistical and the second one
is the systematic error arising from the uncertainty in the $\psi(2S)$
resonance parameters.

With the calculated cross-sections for $\psi(3770)$ production at each
energy point around 3.773~Gev and the corresponding luminosities,
the total number of $\psi(3770)$ events
in the data sample is determined to be
$ N_{\psi(3770)}^{\rm prod} =  (1.85 \pm 0.37) \times 10^5$,
where the error is mainly due to the uncertainty in the
observed cross-section for $\psi(3770)$ production.
The detection efficiency for the decay channel is determined
to be $\epsilon_{\psi(3770) \rightarrow  J/\psi~\pi^+\pi^-,
J/\psi \rightarrow l^+l^-}=0.160 \pm 0.002$, where the error is statistical.
Using these numbers and the known branching fractions for $J/\psi
\rightarrow e^+e^-$ and  $\mu^+\mu^-$ \cite{Eidelman:2004wy}, the branching fraction
for the non-$D \bar D$ decay
$\psi(3770) \rightarrow J/\psi~\pi^+\pi^-$ is measured to be
\begin{equation}
{\cal B}(\psi(3770) \rightarrow J/\psi~\pi^+\pi^-) =
 (0.34 \pm 0.14 \pm 0.08)\%, 
\end{equation}
where the first error is statistical and the second systematic.
Using $\Gamma_{\rm tot}$ from the PDG \cite{Eidelman:2004wy},
this branching fraction corresponds to a partial width of
\begin{equation}
\Gamma(\psi(3770) \rightarrow J/\psi~\pi^+\pi^-) =
(80 \pm 32 \pm 21)~~{\rm {keV}}. 
\label{eq:4.6.5}
\end{equation}
The dominant systematic uncertainty is due to the uncertainty in the total
number of $\psi(3770)$ produced ($\pm 24\%$ ). 
Other systematic uncertainties are due to the efficiency
($\pm 10\%$), the background 
shape ($\pm 6\%$), and $\psi(2S) \rt J/\psi~\pi^+ \pi^- $ background
subtraction ($\pm 7\%$).

CLEOc has analyzed a sample of $\psi(3770)$ decays ($4.5 \times
10^4$) \cite{Skwarnicki:2003wn}.  Although the sample is smaller, they have a
larger detection efficiency (37\%).  They find two events in the
signal region, consistent with the estimated background, and set a
preliminary upper limit ${\cal B}(\psi(3770) \rt J/\psi~\pi^+ \pi^- ) <
0.26\%$ (90\% CL).  The result does not confirm the BES result, but is
not inconsistent with it either. 
CLEOc is now analyzing a sample of about 50 pb$^{-1}$, and the situation
should be better understood when this is completed. See \Section~\ref{sec:4.2} for the 
comparison of the BES result with the related theoretical prediction.

\null\noindent
{\it D.~~~~$X(3872) \rt J/\psi~\pi^+ \pi^- $}

The Belle group has recently reported the observation of the
$X(3872)$, a charmonium-like particle with mass $3872.0\pm0.8$~MeV
that decays to $J/\psi~\pipi$~\cite{belle3870}.  For a review on the
charmonium assignments (and their problems) for the $X(3872)$ we refer
to \Chapter~\ref{chapter:spectroscopy}, \Section~\ref{sec:spexX3872},
and \cite{Olsen:2004fp}.

The $\pipi$ invariant mass distribution for this process, shown in
\Figure~\ref{fig:x3872-belle-Mpipi}(a) in
\Chapter~\ref{chapter:spectroscopy}, has a stronger concentration at
high mass values than QCDME~\cite{Yan} expectations for D-wave to
S-wave transitions, and is also more pronounced than that seen in the
S-wave to S-wave $\psip\rt\jp~\pipi$ process, which is shown in
\Figure~\ref{fig:x3872-belle-Mpipi}(b).  This concentration at high
$\pipi$ masses in $X(3872)\rt\jp~\pipi$ has been experimentally
confirmed by the CDF experiment~\cite{Acosta:2003zx}.

%%%%%%%%%%%%%%
%Bc
%%%%%%%%%%%%%%
\section[Decays of the $\boldsymbol B_{\boldsymbol c}$]{Decays of the
  $\boldsymbol B_{\boldsymbol c}$ $\!$\footnote{Author: V.~V.~Kiselev}}
\label{sec:BC} 
Besides new spectroscopy, production and decay observables, 
the investigation of the long-lived heavy quarkonium
$B_c$, the pseudoscalar ground state of the $\bar b c$ system,
provides the possibility to get model-independent information on
some electroweak parameters, like the CKM matrix elements, in the
heavy quark sector \cite{CPBc,CPBcKis}. The first experimental
observation of the $B_c$ meson by the CDF collaboration
\cite{Abe:1998wi,cdf} confirmed the theoretical predictions (and
postdictions) on its mass, production rate and lifetime
\cite{revbc,Brambilla:2000db,Capstick:1989ra,Fulcher:1998ka,
Ebert:2002pp,Eichten:1994gt,Gershtein:1994jw,OPEBc,KT,KLO,KKL,exBc}.
Tevatron \cite{BRunII} and LHC \cite{LHCB} will provide in the
near future new data with increased statistics, opening the field
to full experimental investigation and systematic test of the theory.

Decays of the $B_c$ meson were considered in the pioneering paper by
Bjorken of 1986 \cite{Bj}. A lot of work has been done after that in
order to understand long-lived doubly heavy hadrons.\footnote{Reviews
on the physics of $B_c$ meson and doubly heavy baryons can be found in
Refs.
\cite{revbc,Fulcher:1998ka,Ebert:2002pp,Eichten:1994gt,Gershtein:1994jw}
and \cite{QQq}, respectively. For the doubly heavy baryons see also
\Chapter~\ref{chapter:spectroscopy}.} Surprisingly, the Bjorken's
estimates of total widths and various branching fractions are close to
what is evaluated now in a more rigorous way. The $B_c$ properties
determined by the strong interactions can be investigated in the
framework of effective field theories for heavy quarkonia, \ie  NRQCD
\cite{Bodwin:1994jh,NRQCDbc}, potential NRQCD
\cite{Pineda:1997bj,Brambilla:1999xf} or vNRQCD \cite{vNRQCD} (see
also \Chapters~\ref{chapter:commontheoreticaltools},
\ref{chapter:spectroscopy} and
\ref{chapter:precisiondeterminations}). In contrast to the Wilson
coefficients, the hadronic matrix elements of operators composed by
the effective fields of the nonrelativistic heavy quarks cannot be
evaluated in a perturbative manner. So, one has to use nonperturbative
methods such as QCD sum rules (SR) \cite{QCDSR}, operator product
expansion (OPE) for inclusive estimates and potential models (PM).

The measured $B_c$ lifetime is equal to 
\begin{equation}
\tau[B_c] =
0.46^{+0.18}_{-0.16}\pm 0.03\; {\rm ps,} 
\label{eq:tauBcmeas} 
\end{equation}
which is close to the value expected by Bjorken. The $B_c$ decays
were, at first, calculated in PM
\cite{PMBc,PMK,PML,PML2,vary,chch,ivanov,ISGW2,narod,CdF}.  We do not
distinguish here among relativistic and nonrelativistic PM,
light-front, Bethe--Salpeter or quasi-potential approaches,
calculations with or without confined quark-propagators and so on,
because (1) relativistic corrections to the initial and final state
heavy quarkonium form factors are suppressed by powers of the heavy
quark velocity (at least, by a factor 10); (2) light mesons in the
final states are usually factorized, and corrections to the
factorization are small; (3) heavy-light mesons in the final states
are quite accurately described by potential models adjusted to the
decays of such mesons. As a consequence the different models agree on
most of the decay channels.

The results of PM for the $B_c$ lifetime agree with each others
after having been adjusted on the semileptonic decays of the $B$
mesons. The OPE evaluations of inclusive decays give lifetime and
widths \cite{OPEBc} in agreement with PM, if one sums up the
dominant exclusive modes. On the other hand, SR of QCD gave at
first semileptonic $B_c$ widths, which were one order of magnitude
smaller than those of PM and OPE \cite{QCDSRBc}. The reason was
identified in the missing Coulomb resummation
\cite{PML,KT,KLO,KKL}. At present, all mentioned approaches give
close results for the lifetime and decay modes of the $B_c$ if
similar sets of parameters are used. Nevertheless, various
questions remain open:

\begin{itemize}
\item What is the appropriate normalization point of the
      non-leptonic weak Lagrangian in the $B_c$ decays? 
\item What are the values of the masses for the $c$ and $b$ quarks
      that have to be used (see in this respect
      \Chapter~\ref{chapter:precisiondeterminations})?
\item What are the implications of the NRQCD symmetries for the $B_c$
      form factors?
\item How consistent is our understanding of hadronic matrix elements
      characterizing the $B_c$ decays with the data from other heavy
      hadrons?
\end{itemize}

In the following of this section we shortly review the $B_c$
decays by summarizing the theoretical predictions in the different
frameworks and discussing how direct experimental measurements can
help to answer the above questions.

\subsection[$\boldsymbol B_{\boldsymbol c} $ lifetime and inclusive decay
  rates]{$\boldsymbol B_{\boldsymbol c} $ lifetime and inclusive decay rates}

The $B_c$ decay processes can be divided into three classes
\cite{OPEBc}:

1) the $\bar b$-quark decay with the spectator $c$ quark,

2) the $c$-quark decay with the spectator $\bar b$ quark and

3) the annihilation channel $B_c^+\rightarrow l^+\nu_l (c\bar s,
u\bar s)$, where $l=e,\; \mu,\; \tau$.

\noindent In the $\bar b \to \bar c c\bar s$ decays one separates
also the Pauli interference with the $c$ quark from
the initial state. In accordance with the given classification,
the total width is the sum over the partial widths
\begin{equation}
\Gamma (B_c\rightarrow X)=\Gamma (b\rightarrow X) +\Gamma
(c\rightarrow X)+\Gamma \mbox{(ann.)}+\Gamma\mbox{(PI)}.
\end{equation}
We will see that the dominant contribution to the $B_c$ lifetime
is expected to be given by the charmed quark decays ($\approx 70\%$),
the $b$-quark decays and the weak annihilation are expected
to add about 20\% and 10\%, respectively, while the Pauli
interference term gives a valuable contribution in the $\bar b\to
c\bar c s$ decays at the level of $-1.5\%$, which we have included
in the $b$-quark decay fraction. 
The above percentages were obtained in \cite{exBc}. Somewhat different figures may be obtained 
in different approaches, \eg C.~H.~Chang {\it et al.} obtain in \cite{OPEBc} 
about $70\%$ for the fraction of $c$-quark decays, 
about $22\%$ for the fraction of $b$-quark decays without Pauli interference, 
about $17\%$ for the fraction of weak annihilation and 
about $-9\%$ for the fraction of the Pauli interference.

The annihilation width, $\Gamma\mbox{(ann.)}$, can
be reliably estimated in the framework of inclusive approaches.
Let us consider, for instance, the effective weak interaction Hamiltonian 
in the quark transition $b\to c \bar u d$:
\begin{equation}
H_{\rm eff}=\frac{G_F}{2
\sqrt{2}}V_{cb}V_{ud}^*\{C_+(\mu)O_{+}+C_-(\mu)O_-\},
\end{equation}
with 
\begin{equation}
O_{\pm}=\bar u_i\gamma_{\nu}(1-\gamma_5)d_i\;
\bar b_j\gamma^{\nu}(1-\gamma_5)c_j \pm 
\bar u_i\gamma_{\nu}(1-\gamma_5)d_j \; 
\bar b_i\gamma^{\nu}(1-\gamma_5)c_j,
\end{equation}
where $i,j$ are colour indices. The factors $C_{\pm}(\mu)$
account for the corrections induced by hard gluons to the corresponding
four-fermion operators. A review on the
evaluation of $C_{\pm}(\mu)$ can be found in \cite{Buchalla:1995vs}. 
The normalization condition is given by $C_\pm (m_b)=1$. A natural choice for 
$\mu$ in decays with given initial and final hadronic states should correspond 
to the scale at which the hadronic matrix elements are evaluated. We also define 
\begin{equation}
\begin{array}{l}
a_1(\mu)= \displaystyle\frac{1}{2 N_{c}}
\big[C_+(\mu)(N_c+1)+C_-(\mu)(N_c-1)\big],\\[3mm]
a_2(\mu)= \displaystyle\frac{1}{2 N_{c}}
\big[C_+(\mu)(N_c+1)-C_-(\mu)(N_c-1)\big].
\end{array}
\end{equation}
Then, we obtain 
\begin{equation}
\Gamma \mbox{(ann.)} =\sum_{i=\tau,c}\frac{G^2_F}{8\pi} |V_{bc}|^2f^2_{B_c}M m^2_i
(1-m^2_i/m^2_{Bc})^2\cdot C_i\;, 
\end{equation}
where $f_{B_c}\approx 400$~MeV (see below), $C_\tau = 1$ for the
$\tau^+\nu_\tau$-channel, $C_c =3|V_{cs}|^2a_1^2 $ for the $c\bar
s$-channel, and the gluon corrections for the annihilation into
hadrons go in the factor $a_1=1.22\pm 0.04$ (see
\cite{Buchalla:1995vs}).  This estimate of the quark contribution does
not depend on a hadronization model, since a large energy release, of
the order of the meson mass, takes place. Moreover, one can see that
the contributions from light leptons and quarks can be neglected.

As for the non-annihilation decays, in the approach of the
OPE for the quark currents of weak decays
\cite{OPEBc}, one takes into account $\als$ corrections to the
free quark decays and uses the quark--hadron duality for the final
states. Then one considers the matrix element for the transition
operator over the  meson state. The latter allows one also to take
into account the effects caused by the motion and virtuality of
the decaying quark inside the meson because of the interaction
with the spectator. In this way the $\bar b\to \bar c c\bar s$
decay mode turns out to be suppressed almost completely due to the
Pauli interference with the charm quark from the initial state.
Besides, the $c$-quark decays with the spectator $\bar b$ quark
are essentially suppressed in comparison with the free quark
decays because of the large binding energy in the initial state.

\begin{figure}[t]
\setlength{\unitlength}{0.8mm}
\begin{center}
\begin{picture}(120,89)
\put(0,5){\epsfxsize=120\unitlength \epsfbox{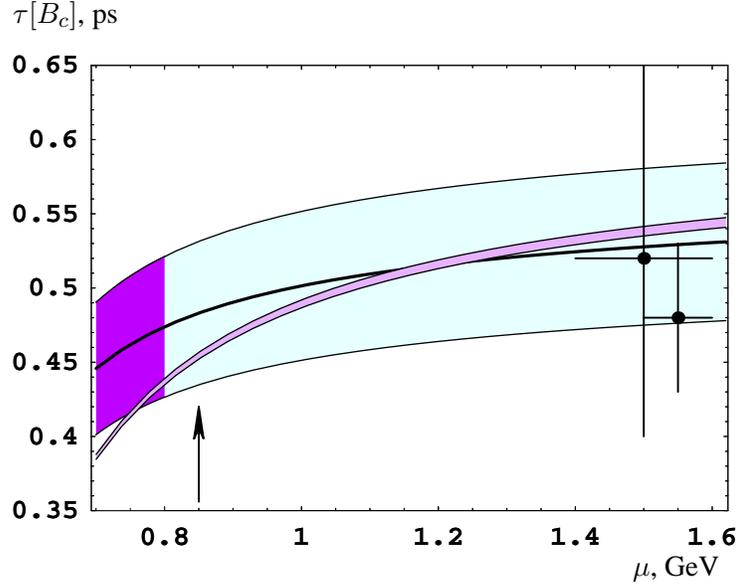}}
\put(0,92){$\tau[{B_c}] $, ps} \put(103,0){$\mu$, GeV}
\end{picture}
\end{center}
\caption[The $B_c$ lifetime calculated in QCD sum rules versus the
         scale of the hadronic weak Lagrangian in the decay 
         of the charm quark]
        {The $B_c$ lifetime calculated in QCD sum rules versus the
         scale of the hadronic weak Lagrangian in the decay of the
         charm quark.  The wide shaded region, taken from
         Ref.~\cite{KKL}, shows the uncertainty of the semi-inclusive
         estimates, the dark shaded region is the preferable choice as
         given by the lifetimes of charmed mesons. The dots represent
         the values in the OPE approach by M.~Beneke and G.~Buchalla
         (left point) and A.~Onishchenko (right point) taken from
         Refs.~\cite{OPEBc}. The narrow shaded region represents the
         result of \cite{exBc} obtained by summing up the exclusive
         channels with a variation of the hadronic scale in the decays
         of the $\bar b$ in the range of $1 <\mu_b < 5$~Gev. The arrow
         points to the preferable prescription of $\mu =0.85$~Gev as
         discussed in \cite{KKL}.} 
\label{fig:life}
\end{figure}

In an exclusive approach it is necessary to sum up widths of
different decay modes calculated in potential models. Considering
the semileptonic decays due to the $\bar b \to \bar c l^+\nu_l$
and $c\to s l^+\nu_l$ transitions, one finds that the hadronic
final states are practically saturated by the lightest $1S$ state
in the $(\bar c c)$ system, \ie $\eta_c$ and $J/\psi$ and the
$1S$ states in the $(\bar b s)$ system, \ie $B_s$ and $B_s^*$.
Further, the $\bar b\to \bar c u\bar d$ channel, for example, can
be calculated through the decay width of $\bar b \to \bar c
l^+\nu_l$, taking into account colour factors and hard gluon
corrections to the four-quark interaction. It can be also obtained
as a sum over the widths of decays to $(u\bar d)$ bound states.

The results of the calculation of the $B_c$ total width in the
inclusive OPE and exclusive PM approaches give values that are
consistent with each other, if one takes into account the most
significant uncertainty, which is related to the choice of the
quark masses (especially of the charm quark):
\begin{equation}
\left.\tau[B_c^+]\right._{\mbox{\small\sc ope,\,pm}}= 0.55\pm
0.15\; \mbox{ps}.
\end{equation}
So, for instance, M.~Beneke and G.~Buchalla using OPE \cite{OPEBc}
give the estimate $0.4$--$0.7$ ps (see \Figure~\ref{fig:life}), which
slightly corrects a result by I.~Bigi \cite{OPEBc}: $0.4$ ps. As for
the potential approach, despite huge differences in details of
exclusive estimates, models usually give a lifetime close to
$0.4$--$0.6$ ps, although the estimates strongly depend on the choice
of the charm quark mass. We refer to the pioneering paper by
M.~Lusignoli and M.~Masetti \cite{PMBc}.  The obtained value agrees
with the measured one (\ref{eq:tauBcmeas}).  In \Table~\ref{tab:inc}
the reader may find summarized several theoretical results for
inclusive decay channels.

\begin{table}[t]
\caption[Summary of theoretical predictions in various approaches for
         the branching ratios of the $B_c$ decay modes]
        {Summary of theoretical predictions in various approaches for
         the branching ratios of the $B_c$ decay modes calculated in
         the framework of the inclusive OPE approach (see M.~Beneke
         and G.~Buchalla in \cite{OPEBc}), by summing up the exclusive
         modes in potential models (for instance, in the model of
         \cite{PMK,PML} used in \cite{KKL}) and according to the
         semi-inclusive estimates in the sum rules of QCD and NRQCD
         \cite{KLO,KKL,exBc}.}
\label{tab:inc}
\begin{center}
\begin{tabular}{|l|c|c|c|}
\hline
$B_c$ decay mode & OPE, \%  & PM, \% & SR, \%\\
\hline
$\bar b\to \bar c l^+\nu_l$ & $3.9\pm 1.0$  & $3.7\pm 0.9$  & $2.9\pm 0.3$\\
$\bar b\to \bar c u\bar d$  & $16.2\pm 4.1$ & $16.7\pm 4.2$ & $13.1\pm 1.3$\\
$\sum \bar b\to \bar c$     & $25.0\pm 6.2$ & $25.0\pm 6.2$ & $19.6\pm 1.9$\\
$c\to s l^+\nu_l$           & $8.5\pm 2.1$  & $10.1\pm 2.5$ & $9.0\pm 0.9$\\
$c\to s u\bar d$            & $47.3\pm 11.8$& $45.4\pm 11.4$& $54.0\pm 5.4$\\
$\sum c\to s$               & $64.3\pm 16.1$& $65.6\pm 16.4$& $72.0\pm 7.2$\\
$B_c^+\to \tau^+\nu_\tau$   & $2.9\pm 0.7$  & $2.0\pm 0.5$  & $1.8\pm 0.2$\\
$B_c^+\to c\bar s$          & $7.2\pm 1.8$  & $7.2\pm 1.8$  & $6.6\pm 0.7$\\
\hline
\end{tabular}
\end{center}
\end{table}

The OPE estimates of inclusive decay rates agree with recent
semi-inclusive calculations in the sum rules of QCD and NRQCD
\cite{KLO,KKL}, where one assumes the saturation of hadronic final
states by the ground levels in the $c\bar c$ and $\bar b s$ systems as
well as the factorization that allows to relate the semileptonic and
hadronic decay modes. The Coulomb resummation plays an essential role
in the $B_c$ decays and removes the disagreement between the estimates
in sum rules and OPE. In contrast to OPE, where the basic uncertainty
is given by the heavy quark masses, these parameters are fixed by the
two-point sum rules for bottomonia and charmonia, so that the accuracy
of SR calculations for the total width of the $B_c$ is determined by
the choice of the scale $\mu$ for the hadronic weak Lagrangian in
decays of charmed quarks. We show this dependence in
\Figure~\ref{fig:life}, where ${m_c}/{2} < \mu < m_c$. The dark shaded
region corresponds to the scales preferred by the data on the charmed
meson lifetimes.  Choosing the scale in the $c\to s$ decays of $B_c$
to be equal to $\mu^2_{B_c} \approx (0.85\; {\rm GeV})^2$, putting
$a_1(\mu_{B_c}) =1.20$ and neglecting the contributions of a nonzero
$a_2$ in the charmed quark decays, in the framework of semi-inclusive
sum rule calculations one obtains \cite{KKL}
\begin{equation}
\left.\tau[B_c]\right._{\mbox{\small\sc sr}} = 0.48\pm 0.05\;{\rm
ps},
\end{equation}
which agrees with the direct sum of exclusive channels presented in
the next sections. In \Figure~\ref{fig:life} we also show the
exclusive estimate of the lifetime from Ref.~\cite{exBc}.

\subsection[Exclusive decays]{Exclusive decays}

Typical values for the exclusive decay branching ratios of the $B_c$,
as obtained in QCD SR \cite{KKL,exBc}, are shown in
\Table~\ref{tab:common} at given values of the factors $a_{1,2}$ and
lifetime. The uncertainty of such predictions is about 15\%, and the
numbers essentially agree with most of the potential models within the
theoretical uncertainties of the QCD SR estimates.  In square bracket
we show the marginal deviations from the central values obtained in
some potential models.

\begin{table}[p]
\caption[QCD SR predictions for the branching ratios of 
         exclusive $B_c^+$ decays]
        {QCD SR predictions \cite{KKL,exBc} for the branching ratios
         of exclusive $B_c^+$ decays with the choice of factors:
         $a_1^c =1.20$ and $a_2^c=-0.317$ in the non-leptonic decays
         of the $c$ quark, and $a_1^b =1.14$ and $a_2^b=-0.20$ in the
         non-leptonic decays of the $\bar b$ quark. The lifetime of
         the $B_c$ is taken $\tau[B_c] \approx 0.45$ ps.  The
         uncertainty of the widths is estimated to be about 15\%.  The
         numbers in square brackets show the marginal values obtained
         in some potential models \cite{narod,CdF,verma,Faust'}.  The
         maximal difference is of one order of magnitude.}
\label{tab:common}
\begin{center}
\begin{tabular}{|l|rr|}
\hline
~~~~~Mode & \multicolumn{2}{c|}{BR, \%}\\
\hline
 $B_c^+ \rightarrow \eta_c e^+ \nu$
 & 0.75 & [0.5]\\
 $B_c^+ \rightarrow \eta_c \tau^+ \nu$
 & 0.23 & [0.2]\\
 $B_c^+ \rightarrow \eta_c^\prime e^+ \nu$
 & 0.020 & [0.05]\\
 $B_c^+ \rightarrow \eta_c^\prime \tau^+ \nu$
 & 0.0016 & [--]\\
 $B_c^+ \rightarrow J/\psi e^+ \nu $
 & 1.9 & [1]\\
 $B_c^+ \rightarrow J/\psi \tau^+ \nu $
 & 0.48 & [0.35]\\
 $B_c^+ \rightarrow \psi^\prime e^+ \nu $
 & 0.094 & [0.2]\\
 $B_c^+ \rightarrow \psi^\prime \tau^+ \nu $
 & 0.008 & [--]\\
 $B_c^+ \rightarrow  D^0 e^+ \nu $
 & 0.004 & [0.02]\\
 $B_c^+ \rightarrow  D^0 \tau^+ \nu $
 & 0.002 & [0.08]\\
 $B_c^+ \rightarrow  D^{*0} e^+ \nu  $
 & 0.018  & [0.004]\\
 $B_c^+ \rightarrow  D^{*0} \tau^+ \nu  $
 & 0.008 & [0.016]\\
 $B_c^+ \rightarrow  B^0_s e^+ \nu  $
 & 4.03  & [1]\\
 $B_c^+ \rightarrow B_s^{*0} e^+ \nu  $
 & 5.06 & [1.2]\\
  $B_c^+ \rightarrow B^0 e^+ \nu  $
 & 0.34 & [0.08]\\
 $B_c^+ \rightarrow B^{*0} e^+ \nu  $
 & 0.58 & [0.15]\\[1mm]
\hline \hline
 $B_c^+ \rightarrow \eta_c \pi^+$
 & 0.20 & [0.12]\\
 $B_c^+ \rightarrow \eta_c \rho^+$
 & 0.42 & [0.3]\\
 $B_c^+ \rightarrow J/\psi \pi^+$
 & 0.13 & [0.08]\\
 $B_c^+ \rightarrow J/\psi \rho^+$
 & 0.40 & [0.2]\\
 $B_c^+ \rightarrow \eta_c K^+ $
 & 0.013 & [0.008]\\
 $B_c^+ \rightarrow \eta_c K^{*+}$
 & 0.020 & [0.018]\\
  $B_c^+ \rightarrow J/\psi K^+$
 & 0.011 & [0.007]\\
 $B_c \rightarrow J/\psi K^{*+}$
 & 0.022 & [0.016]\\
 $B_c^+ \rightarrow D^+
\overline D^{\hspace{1pt}\raisebox{-1pt}{$\scriptscriptstyle 0$}}$
 & 0.0053 & [0.0018]\\
 $B_c^+ \rightarrow D^+
\overline D^{\hspace{1pt}\raisebox{-1pt}{$\scriptscriptstyle
*0$}}$
 & 0.0075 & [0.002]\\
 $B_c^+ \rightarrow  D^{\scriptscriptstyle *+}
\overline D^{\hspace{1pt}\raisebox{-1pt}{$\scriptscriptstyle 0$}}$
 & 0.0049 & [0.0009]\\
 $B_c^+ \rightarrow  D^{\scriptscriptstyle *+}
\overline D^{\hspace{1pt}\raisebox{-1pt}{$\scriptscriptstyle
*0$}}$
 & 0.033 & [0.003]\\
 $B_c^+ \rightarrow D_s^+ \overline
D^{\hspace{1pt}\raisebox{-1pt}{$\scriptscriptstyle 0$}}$
 & 0.00048 & [0.0001]\\
 $B_c^+ \rightarrow D_s^+
\overline D^{\hspace{1pt}\raisebox{-1pt}{$\scriptscriptstyle
*0$}}$
 & 0.00071 & [0.00012]\\
 $B_c^+ \rightarrow  D_s^{\scriptscriptstyle *+} \overline
D^{\hspace{1pt}\raisebox{-1pt}{$\scriptscriptstyle 0$}}$
 & 0.00045 & [0.00005]\\
 $B_c^+ \rightarrow  D_s^{\scriptscriptstyle *+}
\overline D^{\hspace{1pt}\raisebox{-1pt}{$\scriptscriptstyle
*0$}}$
 & 0.0026 & [0.0002]\\
 $B_c^+ \rightarrow \eta_c D_s^+$
 & 0.28 & [0.07]\\
\hline
\end{tabular}
\begin{tabular}{|l|rr|}
\hline
~~~~~Mode & \multicolumn{2}{c|}{BR, \%}\\
\hline
  $B_c^+ \rightarrow \eta_c D_s^{*+}$
 & 0.27 & [0.07]\\
 $B_c^+ \rightarrow J/\psi D_s^+$
 & 0.17 & [0.05]\\
 $B_c^+ \rightarrow J/\psi D_s^{*+}$
 & 0.67 & [0.5]\\
 $B_c^+ \rightarrow \eta_c D^+$
 & 0.015 & [0.04]\\
 $B_c^+ \rightarrow \eta_c D^{*+}$
 & 0.010 & [0.002]\\
 $B_c^+ \rightarrow J/\psi D^+$
 & 0.009 & [0.002]\\
 $B_c^+ \rightarrow J/\psi D^{*+}$
 & 0.028 & [0.014]\\
 $B_c^+ \rightarrow B_s^0 \pi^+$
 & 16.4 & [1.6]\\
 $B_c^+ \rightarrow B_s^0 \rho^+$
 & 7.2 & [2.4]\\
 $B_c^+ \rightarrow B_s^{*0} \pi^+$
 & 6.5 & [1.3]\\
 $B_c^+ \rightarrow B_s^{*0} \rho^+$
 & 20.2 & [11]\\
 $B_c^+ \rightarrow B_s^0 K^+$
 & 1.06 & [0.2]\\
 $B_c^+ \rightarrow B_s^{*0} K^+$
 & 0.37 & [0.13]\\
 $B_c^+ \rightarrow B_s^0 K^{*+}$
 & -- & \\
 $B_c^+ \rightarrow B_s^{*0} K^{*+}$
 & -- & \\
 $B_c^+ \rightarrow B^0 \pi^+$
 & 1.06 & [0.1]\\
 $B_c^+ \rightarrow B^0 \rho^+$
 & 0.96 & [0.2]\\
 $B_c^+ \rightarrow B^{*0} \pi^+$
 & 0.95 & [0.08]\\
 $B_c^+ \rightarrow B^{*0} \rho^+$
 & 2.57 & [0.6]\\
 $B_c^+ \rightarrow B^0 K^+$
 & 0.07 & [0.01]\\
 $B_c^+ \rightarrow B^0 K^{*+}$
 & 0.015 & [0.012]]\\
 $B_c^+ \rightarrow B^{*0} K^+$
 & 0.055 & [0.006]\\
 $B_c^+ \rightarrow B^{*0} K^{*+}$
 & 0.058 & [0.04]\\
 $B_c^+ \rightarrow B^+ \overline{K^0}$
 & 1.98 & [0.18]\\
 $B_c^+ \rightarrow B^+ \overline{K^{*0}}$
 & 0.43 & [0.09]\\
 $B_c^+ \rightarrow B^{*+} \overline{K^0}$
 & 1.60 & [0.06]\\
 $B_c^+ \rightarrow B^{*+} \overline{K^{*0}}$
 & 1.67 & [0.6]\\
 $B_c^+ \rightarrow B^+ \pi^0$
 & 0.037 & [0.004]\\
 $B_c^+ \rightarrow B^+ \rho^0$
 & 0.034 & [0.01]\\
 $B_c^+ \rightarrow B^{*+} \pi^0$
 & 0.033 & [0.003]\\
 $B_c^+ \rightarrow B^{*+} \rho^0$
 & 0.09 & [0.03]\\[0mm]
 \hline\hline
 $B_c^+ \rightarrow \tau^+ \nu_\tau$
 & 1.6 & \\
 $B_c^+ \rightarrow c \bar s$
 & 4.9 & \\
\hline
\end{tabular}
\end{center}
\end{table}

In addition to the decay channels with a $J/\psi$ well detectable
through its leptonic mode, one could expect significant information on
the dynamics of $B_c$ decays coming from channels with a single heavy
mesons, if the experimental efficiency is good enough to extract a
signal from the cascade decays. Since decays to excited charmonia in
the final state (like P~waves) \cite{Chang:2001pm,Pakh}, radiative
leptonic modes \cite{radlep} and some rare decays \cite{rare} are out
of reach for the experimental facilities of the nearest future, we do
not display them in \Table~\ref{tab:common}.

In \cite{exBc} the $\bar b$ decay to the doubly charmed states is
predicted to give
\begin{equation}
{\cal B}(B_c^+\to \bar c c\,c\bar s)
\approx 1.39\%. 
\end{equation}
Comparing the width with the estimate from the spectator decay \cite{OPEBc},
\begin{eqnarray}
\left.\Gamma(B_c^+\to \bar c c\,c\bar s)\right|_{\mbox{\sc sr}}
&\approx & 20\cdot
10^{-15}\,\mbox{GeV},\\[2mm]
\left.\Gamma(B_c^+\to \bar c c\,c\bar s)\right|_{\rm spect.}
&\approx & 90\times 10^{-15}\,\mbox{GeV},
\end{eqnarray}
we see that they differ by a factor of about $1/4.5$. The SR
result is in agreement with an estimate in OPE by M.~Beneke and
G.~Buchalla of \cite{OPEBc}, though the uncertainty is quite large
($\approx$ 60\%) due to the mentioned uncertainty in the renormalization
point as well as in the charm quark mass.

At present we can say that an accurate direct measurement of the
$B_c$ lifetime can provide information on the masses
of the $c$ and $b$ quarks and the normalization point of the
non-leptonic weak Lagrangian in the $B_c$ decays (the $a_1$ and
$a_2$ factors). The experimental study of semileptonic decays and
the extraction of ratios of form factors can test the spin
symmetry of NRQCD and HQET and decrease the uncertainties in the
corresponding theoretical evaluation of the quark parameters as well
as the hadronic matrix elements. The measurement of branching
fractions for semileptonic and non-leptonic modes and their
ratios can give information on the values of the factorization
parameters, which depend again on the normalization of the
non-leptonic weak Lagrangian. The charm quark counting in the
$B_c$ decays is related to the overall contribution of $b$ quark
decays as well as to the suppression of 
$\bar b\to c\bar c \bar s$ transitions because of the destructive interference, 
whose value depends on the nonperturbative parameters (roughly said, the
leptonic constant) and on the non-leptonic weak Lagrangian.

\subsubsection[Semileptonic decays]{Semileptonic decays}
The semileptonic decay rates estimated in the QCD sum rules for
3-point correlators \cite{SR3pt} are underestimated in
\cite{QCDSRBc}, because large Coulomb-like corrections were not
taken into account. The recent analysis of SR in \cite{KT,KLO,KKL}
decreased the uncertainty, so that the estimates agree with the
calculations in the potential models.

{\it (A) Coulomb resummation}

For the heavy quarkonium $\bar b c$, where the relative quark
velocity is small, Coulomb-like $\als/v$ corrections are important
and have to be resummed.  It is well known that taking into
account these corrections in two-point sum rules numerically
enhances the Born value of the spectral density by a factor two or
three \cite{QCDSR}.
\shortpage

{\it (B) Primary modes}

In practice, the most important information comes from the $\psi$ mode, 
since this charmonium is clearly detected in experiments \cite{Abe:1998wi,cdf}.
In addition to the investigation of various form factors and their
dependence on the momentum transfer squared, 
the measurement of the decay to $\psi^\prime$, could answer the question of the
reliability of QCD predictions for the decays to excited states.
At present, finite energy sum rules predict the width of the $B_c^+\to
\psi^\prime l^+ \nu$ decay in reasonable agreement with 
potential models if one takes into account an uncertainty of about
50\%.

{\it (C) Relations between the form factors}

In the limit of infinitely heavy quark masses, the NRQCD and HQET
Lagrangians possess spin symmetry. The most familiar implication of
such symmetry is the common Isgur--Wise function determining the form
factors in the semileptonic decays of single heavy hadrons. In
contrast to weak decays with a light spectator quark, the $B_c$ decays
to $\psi$, $\eta_c$ and $B_s^{(*)}$ involve the heavy spectator, so
that the spin symmetry works only at recoil momenta close to zero,
where the spectator enters the heavy hadron in the final state with no
hard gluon rescattering. Hence, we expect relations between the form
factors in the vicinity of zero recoil. The normalization of the form
factor is not fixed, as it is in decays of hadrons with a single heavy
quark, since the heavy quarkonia wave functions are flavour
dependent. In practice, the ratios of form factors, which are fixed at
zero recoil, are expected to exhibit a dependence on the momentum
transfer squared, which is not significant in actual numerical
estimates in the restricted region of the physical phase space. The SR
estimates of the form factors show a good agreement with the
expectations, whereas the deviations can be traced back to the
difference in the $q^2$ evolution of the form factors from the zero
recoil point. This can be neglected within the accuracy of the SR
method for the transitions of $B_c\to \bar c c$, as shown in
\cite{KLO}. The $1/m_Q$ deviations from the symmetry relations in the
decays of $B_c^+\to B_s^{(*)} e^+\nu$ are about 10--15\%, as found in
the QCD sum rules considered in \cite{KKL}. Form factors for specific
decay channels have been considered also in \cite{chch,Chang:2001pm}.

The combinations of relations derived in \cite{KLO,KKL} reproduce
the only equality in \cite{Jenk}, which was found for each mode in
the strict limit of $v_1=v_2$ also considered by Sanchis--Lozano in
\cite{Sanchis-Lozano:1994vh}.

\subsubsection[Leptonic decays]{Leptonic decays}

The dominant leptonic decay of the $B_c$ is given by the $\tau
\nu_\tau$ mode (see \Table~\ref{tab:inc}). However, it has a low
experimental efficiency of detection because of the hadronic
background in the $\tau$ decays. Recently, in
Refs.~\cite{radlep} the enhancement of muon and electron channels
in the radiative modes has been studied. The additional photon 
removes the helicity suppression for the leptonic decay of
pseudoscalar particles, leading to an increase of the muonic
mode by about a factor two.

{\it (A) Leptonic constant of $B_{c}$}

In NRQCD the calculation of the leptonic constant for the heavy
quarkonium with two-loop accuracy requires the two-loop matching of 
the NRQCD currents with the currents in full QCD,
\[
J_\nu^{\Rsub QCD}= \bar Q_1 \gamma_5\gamma_\nu Q_2, \;\;\; {\cal
J}_\nu^{\Rsub NRQCD} = -\chi^\dagger  \psi \; v_\nu,\qquad
J_\nu^{\Rsub QCD} = {\cal K}(\mu_{\rm hard}; \mu_{\rm fact})\;
{\cal J}_\nu^{\Rsub NRQCD}(\mu_{\rm fact}), 
\]
where the scale $\mu_{\rm hard}$ gives the normalization point for
the matching of NRQCD with full QCD, while $\mu_{\rm fact}$
denotes the normalization point for the calculations in
perturbation theory in NRQCD.

For the pseudoscalar heavy quarkonium composed of heavy quarks
with different flavours, the Wilson coefficient ${\cal K}$ has been 
calculated with two-loop accuracy in Refs.~\cite{braflem} and
\cite{ov}. In NRQCD the current ${\cal J}_\nu^{\Rsub NRQCD}$ has
nonzero anomalous dimension, so that we find
\begin{equation}
\langle 0| {\cal J}_\nu^{\Rsub NRQCD}(\mu) |\bar Q Q \rangle =
{\cal A}(\mu)\; v_\nu f_{\bar Q Q }^{\Rsub NRQCD} M_{\bar Q Q },
\end{equation}
where, in terms of nonrelativistic quarks, the leptonic constant
for the heavy quarkonium is given by the well-known relation with
the wave function at the origin.

\begin{figure}
\setlength{\unitlength}{0.6mm}
\begin{center}
\begin{picture}(100,110)
\put(-5,5){\epsfxsize=100\unitlength \epsfbox{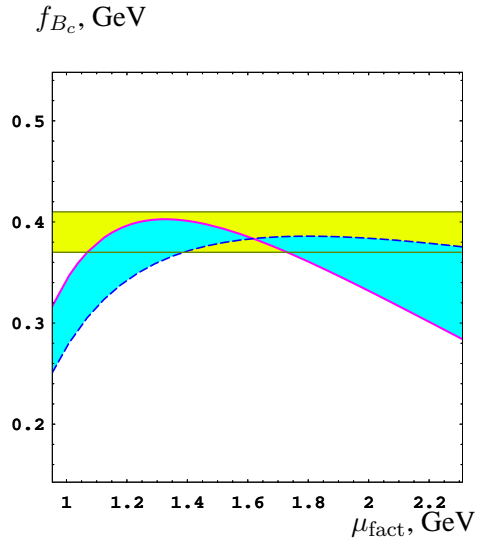}}
\put(70,0){$\mu_{\rm fact}$, GeV} \put(0,112){$f_{B_c}$, GeV}
\end{picture}
\end{center}
\caption[The leptonic constant of the $B_c$ is presented versus
         the soft scale of normalization]
        {The leptonic constant of the $B_c$ is presented versus the
         soft scale of normalization. The shaded region restricted by
         curves corresponds to the change of the hard scale from
         $\mu_{\rm hard} = 3$~Gev (the dashed curve) to $\mu_{\rm
         hard} = 2$~Gev (the solid curve) with the initial condition
         for the evolution of the normalization factor ${\cal
         A}(\mu_{\rm fact})$, ${\cal A}(1.2\;{\rm GeV})=1$ and ${\cal
         A}(1\;{\rm GeV})=1$ respectively, in the nonrelativistic
         current matrix element. The horizontal band is the region
         expected from the QCD sum rules \cite{sr,KT,QCDSRBc} and
         scaling relations for the leptonic constants of heavy
         quarkonia \cite{scale}. In the overlap region, the leptonic
         constant of $B_c$ depends weakly on the parameters.}
\label{fig:fB_c}
\end{figure}

Following the method described in \cite{KKO,KLPS}, one can estimate
the wave function of the $\bar b c$ quarkonium using the static
potential given in \cite{KKO}. Details of the calculations 
can be found in \cite{fBc-V}. The result of the calculation of the
$B_c$ leptonic constant is shown in \Figure~\ref{fig:fB_c}. 
The final result of the two-loop calculation is
\begin{equation}
f_{B_c} = 395\pm 15\; {\rm MeV},
\end{equation}
which is close to an early estimate by S.~Capstick and S.~Godfrey
in \cite{Capstick:1989ra}.

The result on $f_{B_c}$ is in agreement with the scaling relation
derived from the quasi-local QCD sum rules \cite{scale}, which use
the regularity in the heavy quarkonium mass spectra, \ie the fact
that the splitting between the quarkonium levels after averaging
over the spins of the heavy quarks depends weakly on the quark
flavours. So, the scaling law for S-wave quarkonia has the form
\begin{equation}
\frac{f_n^2}{M_n}\;\left(\frac{M_n}{M_1}\right)^2\;
\left(\frac{m_1+m_2}{4\mu_{12}}\right)^2
 = \frac{c}{n},
\end{equation}
where $n$ is the radial quantum number, $m_{1,2}$ the masses of
the heavy quarks composing the quarkonium, $\mu_{12}$ the reduced
mass and $c$ a dimensional constant independent on both the quark
flavours and $n$. The accuracy depends on the heavy quark masses,
and is discussed in detail in \cite{scale}. The parameter $c$
can be extracted from the known leptonic constants of $\psi$ and
$\Upsilon$.

\subsubsection{Non-leptonic modes}

With respect to the inclusive non-leptonic widths, which can be
estimated in the framework of quark--hadron duality (see \Table~\ref{tab:inc}), the calculation of exclusive modes usually involves
factorization \cite{Dugan:1990de,fact}, which, as expected, can be
quite accurate for the $B_c$, since the quark--gluon sea is
suppressed in the heavy quarkonium. Thus, the important parameters
are the factors $a_1$ and $a_2$ in the non-leptonic weak
Lagrangian, which depend on the normalization point.

The agreement of QCD SR estimates for the non-leptonic decays of the
charm quark in the $B_c$ with the values predicted by potential models
is rather good for the direct transitions with no permutation of colour
lines, \ie processes involving the factor $a_1$ in the non-leptonic
amplitude. In contrast, the sum rule predictions are significantly
enhanced in comparison with the values calculated in potential models
for transitions with colour permutation, \ie for processes involving
the factor $a_2$ (see \Table~\ref{tab:common}). Further, for
transitions $\bar b\to c\bar c s$ where the interference is
significantly involved the size of the interference is about 35--50\%
of the width evaluated by neglecting interference terms. These
estimates are in agreement with the potential models of
Refs.~\cite{chch,narod}.

At large recoils as in $B_c^+\to \psi \pi^+(\rho^+)$, the spectator
picture of transition can be broken by hard gluon exchanges
\cite{Gers}. The spin effects in such decays were studied in
\cite{Pakh}. Typically recoil effects are taken into account to some
extent in any relativistic approach like \cite{chch}.

For the widths of non-leptonic $c$-quark decays the sum rule estimates
are typically greater than those of potential models\footnote{See also
the recent discussions on the $B_c$ decays in
\cite{verma,giri,Nobes,Mannel,Ma,Ivanov2,Faust'}.}. In this respect we
note that the QCD SR calculations are consistent with the inclusive
ones. Summing up the calculated exclusive widths, the total width of
the $B_c$ meson is shown in \Figure~\ref{fig:life}, which points to a
good agreement of the exclusive calculations with those of OPE and
semi-inclusive estimates.

Another interesting point is the possibility to extract the
factorization parameters $a_1$ and $a_2$ in the $c$-quark decays
by measuring the ratios of widths 
 \begin{equation}
 \frac{\Gamma(B_c^+\to B^{(*)+}\bar K^{(*)0})}{\Gamma(B_c^+\to B^{(*)0} K^{(*)+})} =
\left|\frac{V_{cs}}{V_{cd}^2}\right|^2\frac{a^2_2}{a^2_1}, 
\end{equation}
where one should take identical sets of pseudoscalar and vector
states in both decays. This procedure can give a test for the
factorization approach itself.

The suppressed decays caused by the flavour changing neutral
currents were studied in \cite{rare}.

\medskip

{\it (A) CP violation in $B_c$ decays}

CP violation in $B_c$ decays can be investigated in the same
way as in $B$ decays. The expected CP asymmetry of ${\cal
A}(B_c^\pm \to J/\psi D^\pm)$ is about $4\times 10^{-3}$, when the
corresponding branching ratio is suppressed as $10^{-4}$
\cite{CPBc}. Therefore, the direct study of CP violation in $B_c$
decays is practically difficult because of the low relative yield
of $B_c$ with respect to ordinary $B$ mesons:
$\sigma(B_c)/\sigma(B) \sim 10^{-3}$.

As mentioned at the beginning, the $B_c$ meson is expected to be copiously produced 
in future colliders. In such circumstances a possible challenge is whether one could
get an opportunity to extract some information about the CKM
unitarity triangle from the $B_c$ in a model independent
way. Indeed, there is such an opportunity for the angle $\gamma$ using the strategy of
the reference triangles \cite{Gronau} in the decays of doubly heavy
hadrons. This strategy for the study of CP violation in $B_c$
decays was originally proposed by M.~Masetti \cite{CPBc},
independently investigated by R.~Fleischer and D.~Wyler
\cite{CPBc} and extended to the case of doubly heavy baryons in
\cite{CPQQq}. Other possibilities include the lepton tagging of $B_s$
in the $B_c^\pm\to B_s^{(*)} l^\pm \nu$ decays for the study of
mixing and CP violation in the $B_s$ sector \cite{Quigg:1993xg}, 
and a possible transverse polarization of the $\tau$ lepton 
in $B_c \to \tau \bar{\nu}_\tau \gamma$ \cite{giri}.

\begin{figure}[t]
  \begin{center}
\setlength{\unitlength}{1.3mm} \hspace*{2cm}
\begin{picture}(90,25)
\put(-10,0){\epsfxsize=55\unitlength \epsfbox{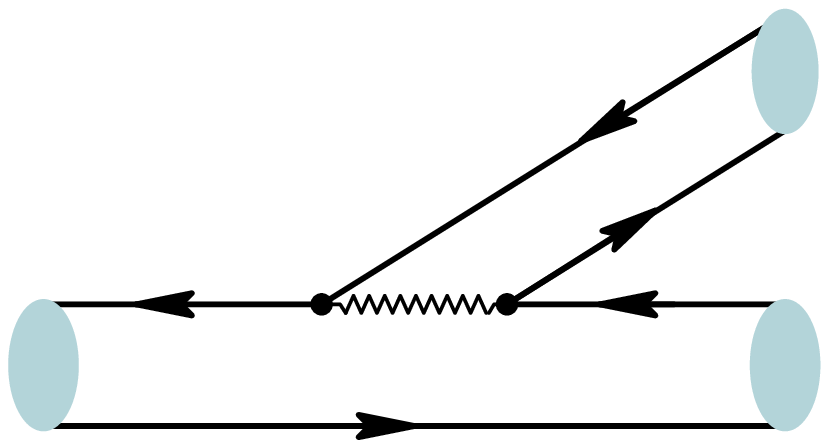}}
\put(33,0){\epsfxsize=55\unitlength \epsfbox{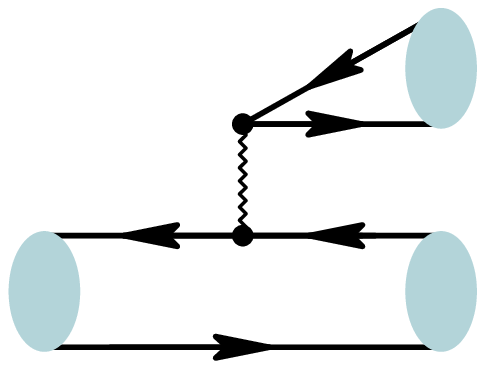}}
\put(-2.3,6.5){$B^+_{c}$} \put(34.,6.5){$D^+_{s}$}
\put(34.2,21){$D^0$} \put(47.2,6.55){$B^+_{c}$}
\put(67.5,6.6){$D^0$} \put(67.5,18.1){$D^+_{s}$}
\put(5.,11.5){$\scriptstyle b$} \put(15.,2.){$\scriptstyle c$}
\put(26.,20.8){$\scriptstyle u$} \put(27,15.){$\scriptstyle c$}
\put(29,11.5){$\scriptstyle s$} \put(53.5,12){$\scriptstyle b$}
\put(63.5,11.5){$\scriptstyle u$} \put(57.5,2){$\scriptstyle c$}
\put(63.2,20.2){$\scriptstyle s$} \put(63,14.2){$\scriptstyle c$}
\end{picture}
  \end{center}
    \caption{The diagrams of $\bar b$-quark decay contributing to the weak
     transition $B^+_c\to D^0 D_s^+$.}
    \label{fig:1}

\medskip

  \begin{center}
\setlength{\unitlength}{1.3mm} \hspace*{0cm}
\begin{picture}(90,25)
\put(-10,2){\epsfxsize=55\unitlength \epsfbox{1g}}
\put(42,1){\epsfxsize=55\unitlength \epsfbox{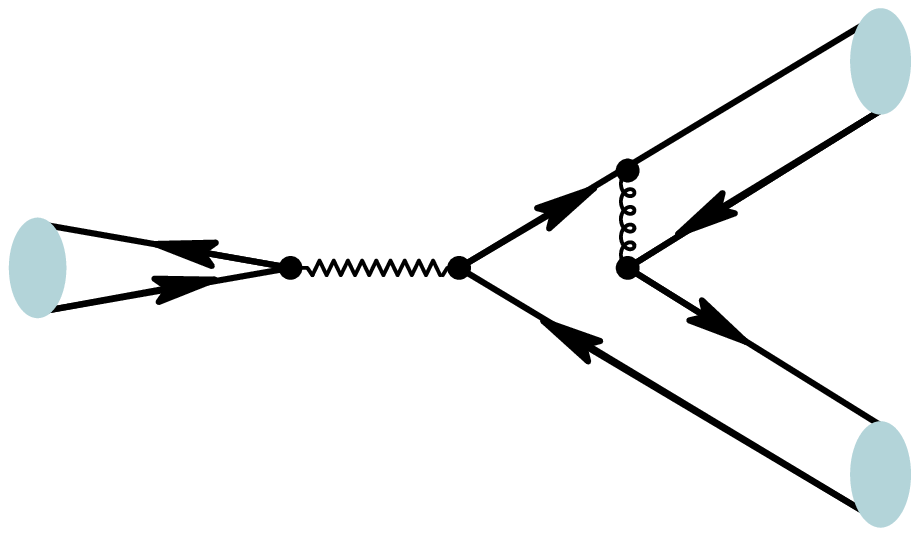}}
\put(-2.3,8.5){$B^+_{c}$} \put(34.,8.5){$D^+_{s}$}
\put(34.2,23){$\bar D^0$} \put(48.2,13.9){$B^+_{c}$}
\put(87.5,4.2){$D^+_{s}$} \put(87.5,23.6){$\bar D^0$}
\put(5.,13.5){$\scriptstyle b$} \put(15.,4.){$\scriptstyle c$}
\put(26.,22.8){$\scriptstyle c$} \put(27,17.){$\scriptstyle u$}
\put(29,13.5){$\scriptstyle s$} \put(56.5,16.7){$\scriptstyle b$}
\put(73.5,9.5){$\scriptstyle s$} \put(56.5,12){$\scriptstyle c$}
\put(73.2,18.7){$\scriptstyle u$} \put(83,16.8){$\scriptstyle c$}
\put(83,12.){$\scriptstyle c$}
\end{picture}
  \end{center}
    \caption{The diagrams of $\bar b$-quark decay contributing to the weak
     transition $B^+_c\to \bar D^0 D_s^+$.}
    \label{fig:1a}
\end{figure}

The triangle strategy is based on the direct determination of
absolute values for the set of four decays, at least: the decays
of the hadron into the tagged $D^0$ meson, the tagged $\bar D^0$ meson,
the tagged CP-even state of $D^0$, and the decay of the anti-hadron into the tagged CP-even
state of $D^0$. To illustrate the point, let us consider the
decays
$$
B^+_c\to D^0 D_s^+ \quad\mbox{and}\quad B^+_c\to \bar D^0 D_s^+.
$$
The corresponding diagrams with the decay of $\bar b$-quark are
shown in \Figures~\ref{fig:1} and \ref{fig:1a}.

The exclusive modes do not have penguin terms at the leading
order in the Fermi constant $G_F$ that we consider here.
However, the diagram with the weak annihilation of two
constituents, \ie the charmed quark and beauty antiquark in the
$B_c^+$ meson, can contribute at the next order in $\als$ as shown
in \Figure~\ref{fig:1a} for the given final state. Nevertheless, we
see that such diagrams have the same weak-interaction structure as
at tree level. The magnitude of the $\als$ corrections to the
absolute values of the corresponding decay widths is discussed in
\cite{CPBcKis}. We expect the sides of the reference-triangles to
be of the same order of magnitude, which makes the method an
attractive way to extract the \mbox{angle $\gamma$}.

The predictions of QCD sum rules for the exclusive decays of $B_c$
are summarized in \Table~\ref{tab:common2} at fixed values of $a_{1,2}$
and lifetime. For the sake of completeness and comparison we show
the estimates for the channels with the neutral $D$ meson and
charged $D^+$ as well as for the vector states in addition to the
pseudoscalar ones.

First, we see that the similar decay modes without the strange
quark in the final state can also be used, in principle, for the
extraction of $\gamma$, however, these channels are more
problematic since the sides of the reference-triangles
significantly differ from each other\footnote{The ratio of widths
is basically determined by $|V_{cb} V_{ud} a_2|^2/|V_{ub} V_{cd}
a_1|^2\sim 110$, if we ignore the interference effects.}, so that
the measurements have to be extremely accurate to get useful
information on the angle, which means that one has to accumulate a
huge statistics for the dominant mode.

Second, the decay modes with the vector neutral $D$ meson in the
final state are useless for the purpose of the CKM measurement
under the discussed approach. However, the modes with the vector
charged $D^*$ and $D_s^*$ mesons can be important for the
extraction of $\gamma$. For instance, one could consider the modes
$D^{*+}\to D^0 \pi^+$ and $D^0\to K^-\pi^+$.
The neutral charmed meson should be carefully treated in order to
avoid misidentification with the primary one. Otherwise, one could
use the mode with the neutral pion, $D^{*+}\to D^+ \pi^0$, whose
detection in an experimental facility could be problematic. The
same considerations apply to the vector meson $D_s^{*+}$, whose
radiative electromagnetic decay is also problematic for the
detection, since the photon could be easily lost. On the other
hand, the loss of the photon does not disturb the analysis in the
case of fully reconstructed $D_s^+$ and $B_c^+$.

\begin{table}[t]
\caption[QCD SR predictions for the branching ratios of exclusive
         $B_c^+$ decays]
        {QCD SR predictions \cite{CPBcKis} for the branching ratios of
         exclusive $B_c^+$ decays with the choice of factors:
         $a_1^b=1.14$ and $a_2^b=-0.20$ in the non-leptonic decays of
         $\bar b$ quark. The lifetime of the $B_c$ is taken $\tau[B_c]
         \approx 0.45$ ps. For comparison we show in square brackets
         minimal values estimated in the potential models of
         \cite{CPBc}.}
\label{tab:common2}
\begin{center}
\begin{tabular}{|l|rr|}
\hline
~~~~~Mode & \multicolumn{2}{c|}{BR, $10^{-6}$}\\
\hline
 $B_c^+ \rightarrow D^+
\overline D^{\hspace{1pt}\raisebox{-1pt}{$\scriptscriptstyle 0$}}$
 & 53 & [18]\\
 $B_c^+ \rightarrow D^+
\overline D^{\hspace{1pt}\raisebox{-1pt}{$\scriptscriptstyle
*0$}}$
 & 75 & [20]\\
 $B_c^+ \rightarrow  D^{\scriptscriptstyle *+}
\overline D^{\hspace{1pt}\raisebox{-1pt}{$\scriptscriptstyle 0$}}$
 & 49 & [9]\\
 $B_c^+ \rightarrow  D^{\scriptscriptstyle *+}
\overline D^{\hspace{1pt}\raisebox{-1pt}{$\scriptscriptstyle
*0$}}$
 & 330 & [30]\\
 $B_c^+ \rightarrow D_s^+ \overline
D^{\hspace{1pt}\raisebox{-1pt}{$\scriptscriptstyle 0$}}$
 & 4.8 & [1]\\
 $B_c^+ \rightarrow D_s^+
\overline D^{\hspace{1pt}\raisebox{-1pt}{$\scriptscriptstyle
*0$}}$
 & 7.1 & [1.2]\\
 $B_c^+ \rightarrow  D_s^{\scriptscriptstyle *+} \overline
D^{\hspace{1pt}\raisebox{-1pt}{$\scriptscriptstyle 0$}}$
 & 4.5 & [0.5]\\
 $B_c^+ \rightarrow  D_s^{\scriptscriptstyle *+}
\overline D^{\hspace{1pt}\raisebox{-1pt}{$\scriptscriptstyle
*0$}}$
 & 26 & [2]\\
\hline
\end{tabular}
\begin{tabular}{|l|rr|}
\hline
~~~~~Mode & \multicolumn{2}{c|}{BR, $10^{-6}$}\\
\hline
 $B_c^+ \rightarrow D^+
D^{\hspace{1pt}\raisebox{-1pt}{$\scriptscriptstyle 0$}}$
 & 0.32 & [0.1]\\
 $B_c^+ \rightarrow D^+
D^{\hspace{1pt}\raisebox{-1pt}{$\scriptscriptstyle *0$}}$
 & 0.28 & [0.07]\\
 $B_c^+ \rightarrow  D^{\scriptscriptstyle *+}
D^{\hspace{1pt}\raisebox{-1pt}{$\scriptscriptstyle 0$}}$
 & 0.40 & [0.4]\\
 $B_c^+ \rightarrow  D^{\scriptscriptstyle *+}
D^{\hspace{1pt}\raisebox{-1pt}{$\scriptscriptstyle *0$}}$
 & 1.59 & [0.4]\\
 $B_c^+ \rightarrow D_s^+
D^{\hspace{1pt}\raisebox{-1pt}{$\scriptscriptstyle 0$}}$
 & 6.6 & [1.7]\\
 $B_c^+ \rightarrow D_s^+
D^{\hspace{1pt}\raisebox{-1pt}{$\scriptscriptstyle *0$}}$
 & 6.3 & [1.3]\\
 $B_c^+ \rightarrow  D_s^{\scriptscriptstyle *+}
D^{\hspace{1pt}\raisebox{-1pt}{$\scriptscriptstyle 0$}}$
 & 8.5 & [8.1]\\
 $B_c^+ \rightarrow  D_s^{\scriptscriptstyle *+}
D^{\hspace{1pt}\raisebox{-1pt}{$\scriptscriptstyle *0$}}$
 & 40.4 & [6.2]\\
\hline
\end{tabular}
\end{center}
\end{table}

In the BTeV \cite{BRunII} and LHCb \cite{LHCB} experiments one
expects the $B_c$ production at the level of several billion
events. Therefore, one expects $10^4$--$10^5$ decays of $B_c$ in the
gold-plated modes under interest. The experimental challenge is
the efficiency of detection. One usually gets a 10\% efficiency
for the observation of distinct secondary vertices outstanding
from the primary vertex of beam interaction. Next, one has to take
into account the branching ratios of $D_s$ and $D^0$ mesons. This
efficiency crucially depends on whether one can detect the neutral
kaons and pions or not. So, for the $D_s$ meson the corresponding
branching ratios grow from 4\% (no neutral $K$ and $\pi$) to 25\%.
The same interval for the neutral $D^0$ is from 11\% to 31\%. The
detection of neutral kaon is necessary for the measurement of
decay modes into the CP-odd state $D_2$ of the neutral $D^0$
meson, however, one can omit this cross-check channel from the
analysis in dealing with the CP-even state $D_1$. The
corresponding intervals of branching ratios reachable by the
experiments are from 0.5\% to 1.3\% for the CP-even state and from
1.5\% to 3.8\% for the CP-odd state of $D^0$. A pessimistic
estimate for the product of branching ratios is about $2\times
10^{-4}$, which results in $2$--$20$ reconstructed events.

%\end{document}

\BLKP
%9/12/04
% Last mod 29 May 2005 mg
\chapter {PRODUCTION}
\label{chapter:production}
{{\it Conveners:} G.T.~Bodwin, E.~Braaten, M.~Kr\"amer, A.B.~Meyer,
V.~Papadimitriou}\par\noindent 
{{\it Authors:} G.T.~Bodwin, E.~Braaten, C.-H.~Chang, M.~Kr\"amer,
J.~Lee, A.B.~Meyer, V.~Papadimitriou, R.~Vogt}
\par\noindent

%%%%%%%%%%%%%%%%%%%%%%%%%%%%%%%%%%%%%%%%%%%%%%%%%%%%%%%%%%%%%%
%\input{secnrqcdintro}
\section{Formalism for inclusive quarkonium production}
\label{sec:prodsec-nrqcd}

\subsection{NRQCD factorization method}
\label{sec:prodsec-nrqcdfact}

In both heavy-quarkonium annihilation decays and hard-scattering
production, large energy--momentum scales appear. The heavy-quark mass
$m$ is much larger than $\Lambda_{\rm QCD}$, and, in the case of
production, the transverse momentum $p_T$ can be much larger than
$\Lambda_{\rm QCD}$ as well. This implies that the associated values of
the QCD running coupling constant are much less than one.
($\alpha_s(m_c)\approx 0.25$ and $\alpha_s(m_b)\approx 0.18$.)
Therefore, one might hope that it would be possible to calculate the
rates for heavy quarkonium decay and production accurately in
perturbation theory. However, there are clearly low-momentum,
nonperturbative effects associated with the dynamics of the quarkonium
bound state that invalidate the direct application of perturbation
theory.

In order to make use of perturbative methods, one must first separate
the short-distance/high-momentum, perturbative effects from the
long-distance/low-momentum, nonperturbative effects\,---\,a process which is
known as ``factorization.'' One convenient way to carry out this
separation is through the use of the effective field theory
Nonrelativistic QCD (NRQCD) \cite{Caswell:1985ui,Thacker:1990bm,Bodwin:1994jh}.
NRQCD reproduces full QCD accurately at momentum scales of order $mv$
and smaller, where $v$ is the typical heavy-quark velocity in the bound
state in the CM frame. ($v^2\approx 0.3$ for charmonium, and 
$v^2\approx 0.1$ for bottomonium.) Virtual processes involving momentum
scales of order $m$ and larger can affect the lower-momentum processes,
and their effects are taken into account through the short-distance
coefficients of the operators that appear in the NRQCD action.

Because $Q\bar Q$ production occurs at momentum scales of order $m$ or
larger, it manifests itself in NRQCD through contact interactions. As a
result, the inclusive cross-section for the direct production of the
quarkonium $H$ at large transverse momentum ($p_T$ of order $m$ or
larger) in hadron or $ep$ colliders or at large momentum in the CM frame
($p^*$ of order $m$ or larger) in $e^+e^-$ colliders can be written as a
sum of products of NRQCD matrix elements and short-distance coefficients:
\begin{equation}
\sigma[H]=\sum_n \sigma_n(\Lambda) \langle {\cal O}_n^H(\Lambda) \rangle.
\label{eq:prod-fact}
\end{equation}
Here, $\Lambda$ is the ultraviolet cutoff of the effective theory, the
$\sigma_n$ are short-distance coefficients, and the $\langle {\cal
O}_n^H \rangle$ are vacuum-expectation values of four-fermion
operators in NRQCD. There is a formula analogous to
\Eq~(\ref{eq:prod-fact}) for inclusive quarkonium annihilation rates,
except that the vacuum-to-vacuum matrix elements are replaced by
quarkonium-to-quarkonium matrix elements \cite{Bodwin:1994jh}.

The short-distance coefficients $\sigma_n(\Lambda)$ in
(\ref{eq:prod-fact}) are essentially the process-dependent partonic
cross-sections to make a $Q\bar Q$ pair, convolved with parton
distributions if there are hadrons in the initial state. The $Q\bar Q$
pair can be produced in a colour-singlet state or in a colour-octet
state. Its spin state can be singlet or triplet, and it also can have
orbital angular momentum. The short-distance coefficients are
determined by matching the square of the production amplitude in NRQCD
to full QCD. Because the scale of the $Q\bar Q$ production is of order
$m$ or greater, this matching can be carried out in perturbation
theory.

The four-fermion operators in \Eq~(\ref{eq:prod-fact}) create a $Q\bar
Q $ pair in the NRQCD vacuum, project it onto a state that in the
asymptotic future consists of a heavy quarkonium plus anything, and
then annihilate the $Q\bar Q$ pair. The vacuum matrix element of such
an operator is the probability for a $Q\bar Q$ pair to form a
quarkonium plus anything.  These matrix elements are somewhat
analogous to parton fragmentation functions. They contain all of the
nonperturbative physics associated with evolution of the $Q\bar Q$
pair into a quarkonium state. An important property of the matrix
elements, which greatly increases the predictive power of NRQCD, is
the fact that they are universal, \ie process independent.

The colour-singlet and colour-octet four-fermion operators that appear
in \Eq~(\ref{eq:prod-fact}) correspond to the evolution into a
colour-singlet quarkonium of a $Q\bar Q$ pair created at short distance
in a colour-singlet state or a colour-octet state, respectively. In the
case of decay, the colour-octet matrix elements have the interpretation
of the probability to find the quarkonium in a Fock state consisting
of a $Q\bar Q$ pair plus some number of gluons. It is a common
misconception that colour-octet production proceeds through the
production of a higher Fock state of the quarkonium.  However, in the
leading colour-octet production mechanisms, the gluons that neutralize
the colour are not present at the time of the creation of the
colour-octet $Q \bar Q$ pair, but are emitted during the subsequent
hadronization process.  The production of the quarkonium through a
higher Fock state requires the production of gluons that are nearly
collinear to the $Q\bar Q$ pair, and it is suppressed by additional
powers of $v$.

NRQCD power-counting rules allow one to organize the sum over
operators in \Eq~(\ref{eq:prod-fact}) as an expansion in powers of
$v$. Through a given order in $v$, only a finite set of matrix
elements contributes. Furthermore, there are simplifying relations
between matrix elements, such as the heavy-quark spin symmetry and the
vacuum-saturation approximation, that reduce the number of independent
matrix elements \cite{Bodwin:1994jh}.  Some examples of relations
between colour-singlet matrix elements that follow from heavy-quark
spin symmetry are
\begin{eqnarray}
\langle{\cal O}^{J/\psi}_1({}^3S_1)\rangle &=&
3 \;\langle{\cal O}^{\eta_c}_1({}^1S_0)\rangle,
\label{eq:s1-symmetry}
\\ 
\langle{\cal O}^{\chi_{cJ}}_1({}^3P_J)\rangle &=& 
{\textstyle\frac{1}{3}} (2J+1) \langle{\cal O}^{h_c}_1({}^1P_1)\rangle. 
\label{eq:p1-symmetry}    
\end{eqnarray}                                                  
These relations hold up to corrections of order $v^2$.  The prefactors
on the right side of
\Eqs~(\ref{eq:s1-symmetry})--(\ref{eq:p1-symmetry}) are just ratios of
the numbers of spin states.  Since the operators in
\Eqs~(\ref{eq:s1-symmetry}) and (\ref{eq:p1-symmetry}) have the same
angular momentum quantum numbers as the $Q \bar Q$ pair in the
dominant Fock state of the quarkonium, the vacuum-saturation
approximation can be used to express the matrix elements in terms of
the squares of wave functions or their derivatives at the origin, up
to corrections of order $v^4$.  heavy-quark spin symmetry also gives
relations between colour-octet matrix elements, such as
\begin{eqnarray}
\langle{\cal O}^{J/\psi}_8({}^3S_1)\rangle &=&
3 \; \langle{\cal O}^{\eta_c}_8({}^1S_0)\rangle,
\label{eq:s8-symmetry}
\\
\langle{\cal O}^{J/\psi}_8({}^1S_0)\rangle &=&
\langle{\cal O}^{\eta_c}_8({}^3S_1)\rangle, 
\\
\langle{\cal O}^{J/\psi}_8({}^3P_J)\rangle &=&
{\textstyle\frac{1}{3}} (2J+1) \langle{\cal O}^{\eta_c}_8({}^1P_1)\rangle,
\\
\langle{\cal O}^{\chi_{cJ}}_8({}^3S_1)\rangle &=& 
{\textstyle\frac{1}{3}} (2J+1) \langle{\cal O}^{h_c}_8({}^1S_0)\rangle.
\label{eq:p8-symmetry}
\end{eqnarray}                                                  
These relations hold up to corrections of order $v^2$. The prefactors
on the right side of \Eqs~(\ref{eq:s8-symmetry})--(\ref{eq:p8-symmetry}) are
again just ratios of the numbers of spin states.  The
vacuum-saturation approximation is not applicable to colour-octet
matrix elements.

The relative importance of the terms in the factorization formula in
\Eq~(\ref{eq:prod-fact}) is determined not only by the sizes of the
matrix elements but also by the sizes of the coefficients $\sigma_n$
in \Eq~(\ref{eq:prod-fact}).  The size of the coefficient depends on
its order in $\alpha_s$, colour factors, and dimensionless kinematic
factors, such as $m^2/p_T^2$.

The NRQCD factorization approach is sometimes erroneously called the
``colour-octet model,'' because colour-octet terms are expected to
dominate in some situations, such as $J/\psi$ production at large
$p_T$ in hadron colliders.  However, there are also situations in
which colour-singlet terms are expected to dominate, such as $J/\psi$
production in continuum $e^+ e^-$ annihilation at the $B$ factories.
Moreover, NRQCD factorization is not a model, but a rigorous
consequence of QCD in the limit $\Lambda_{\rm QCD}/m\rightarrow 0$.

A specific truncation of the NRQCD expansion in
\Eq~(\ref{eq:prod-fact}) could be called a model, although, unlike
most models, it is in principle systematically improvable.  In
truncating at a given order in $v$, one can reduce the number of
independent matrix elements by making use of approximate relations,
such as \Eqs~(\ref{eq:s1-symmetry})--(\ref{eq:p1-symmetry}) and
\Eqs~(\ref{eq:s8-symmetry})--(\ref{eq:p8-symmetry}).  The simplest
truncation of the NRQCD expansion in \Eq~(\ref{eq:prod-fact}) that is
both phenomenologically viable and corresponds to a consistent
truncation in $v$ includes four independent NRQCD matrix elements for
each S-wave multiplet (one colour-singlet and three colour-octet) and
two independent NRQCD matrix elements for each P-wave multiplet (one
colour-singlet and one colour-octet).  We will refer to this truncation
as the standard truncation in $v$.  For the S-wave charmonium
multiplet consisting of $J/\psi$ and $\eta_c$, one can take the four
independent matrix elements to be $\langle {\cal
O}^{J/\psi}_1({}^3S_1) \rangle$, $\langle {\cal O}^{J/\psi}_8({}^1S_0)
\rangle$, $\langle {\cal O}^{J/\psi}_8({}^3S_1) \rangle$, and $\langle
{\cal O}^{J/\psi}_8({}^3P_0) \rangle$.  Their relative orders in $v$
are $v^0$, $v^3$, $v^4$, and $v^4$, respectively. It is convenient to
define the linear combination
\begin{equation}
M_k^H = 
\langle {\cal O}^H_8({}^1S_0) \rangle +
 \frac{k}{m_c^2} \langle {\cal O}^H_8({}^3P_0) \rangle \, ,
\label{eq:prod-lincomb}
\end{equation}
because many observables are sensitive only to the linear combination
of these two colour-octet matrix elements corresponding to a specific
value of $k$.  These four independent matrix elements can be used to
calculate the cross-sections for the $\eta_c$ and each of the 3 spin
states of the $J/\psi$. Thus, this truncation of NRQCD gives
unambiguous predictions for the polarization of the $J/\psi$.  For the
P-wave charmonium multiplet consisting of $\chi_{c0}$, $\chi_{c1}$,
$\chi_{c2}$, and $h_c$, we can take the two independent matrix
elements to be $\langle {\cal O}^{\chi_{c0}}_1({}^3P_0) \rangle$ and
$\langle {\cal O}^{\chi_{c0}}_8({}^3S_1) \rangle$.  Their orders in
$v$ relative to $\langle {\cal O}^{J/\psi}_1({}^3S_1) \rangle$ are
both $v^2$.  These two independent matrix elements can be used to
calculate the cross-sections for each of the 12 spin states in the
P-wave multiplet.  Thus, this truncation of NRQCD gives unambiguous
predictions for the polarizations of the $\chi_{c1}$, $\chi_{c2}$, and
$h_c$.
\shortpage

The NRQCD {\it decay matrix} elements can be calculated in lattice
simulations \cite{Bodwin:1993wf,Bodwin:1994js,Bodwin:1996tg,%
Bodwin:1996mf,Bodwin:2001mk} or determined from
phenomenology. However, it is not yet known how to formulate the
calculation of production matrix elements in lattice simulations, and,
so, the production matrix elements must be obtained
phenomenologically.  In general, the production matrix elements are
different from the decay matrix elements. The exceptions are the
colour-singlet production matrix elements in which the $Q \bar Q$ pair
has the same quantum numbers as the quarkonium state, such as those in
\Eqs~(\ref{eq:s1-symmetry}) and (\ref{eq:p1-symmetry}). They can be
related to the corresponding decay matrix elements through the
vacuum-saturation approximation, up to corrections of relative order
$v^4$ \cite{Bodwin:1994jh}.  Phenomenological determinations of the
production matrix elements for charmonium states are given in
\Section~\ref{sec:prodsec-tevatroncharm}.

The proof of the factorization formula in \Eq~(\ref{eq:prod-fact})
relies both on NRQCD and on the all-orders perturbative machinery for
proving hard-scattering factorization.  A detailed proof does not yet
exist, but work is in progress \cite{qiu-sterman}. At a large
transverse momentum ($p_T$ of order $m$ or larger), corrections to
hard-scattering factorization are thought to be of order
$(mv)^2/p_T^2$ (not $m^2/p_T^2$) in the unpolarized case and of order
$mv/p_T$ (not $m/p_T$) in the polarized case. At a small transverse
momentum, $p_T$ of order $mv$ or smaller, the presence of soft gluons
in the quarkonium binding process makes the application of the
standard factorization techniques problematic. It is not known if
there is a factorization formula for $d\sigma/dp_T^2$ at small $p_T$
or for $d\sigma/dp_T^2$ integrated over $p_T$.

In practical calculations of the rates of quarkonium decay and
production, a number of significant uncertainties arise. In many
instances, the series in $\alpha_s$ and $v$ in the factorization
formula in \Eq~(\ref{eq:prod-fact}) converge slowly, and the
uncertainties from their truncation are large\,---\,sometimes $100\%$ or
larger. In addition, the matrix elements are often poorly determined,
either from phenomenology or lattice measurements, and the important
linear combinations of matrix elements vary from process to process,
making tests of universality difficult. There are also large
uncertainties in the heavy-quark masses (approximately 8\% for $m_c$
and approximately 2.4\% for $m_b$) that can be very significant for
quarkonium rates that are proportional to a large power of the mass.

Many of the largest uncertainties in the theoretical predictions, as
well as some of the experimental uncertainties, cancel in the ratios
of cross-sections. Examples in charmonium production are the ratio
$R_\psi$ of the inclusive cross-sections for $\psi(2S)$ and $J/\psi$
production and the ratio $R_{\chi_c}$ of the inclusive cross-sections
for $\chi_{c1}$ and $\chi_{c2}$ production.  These ratios are defined
by
\begin{eqnarray}
R_\psi &=& \frac{\sigma[\psi(2S)]}{\sigma[J/\psi]} \, ,
\label{eq:prod-psirat}
\\
R_{\chi_c} &=& \frac{\sigma[\chi_{c1}]}{\sigma[\chi_{c2}]} \, .
\label{eq:prod-chirat}
\end{eqnarray}
Other useful ratios are the fractions $F_H$ of $J/\psi$'s that come
from decays of higher quarkonium states $H$.  The fractions that come
from decays of $\psi(2S)$ and from $\chi_c(1P)$ are defined by
\begin{eqnarray}
F_{\psi(2S)} &=& 
{\rm Br}[\psi(2S) \to J/\psi + X] \; \frac{\sigma[\psi(2S)]}{\sigma[J/\psi]},
\label{eq:FJpsipsi2S}
\\
F_{\chi_c} &=& \sum_{J=0}^2
{\rm Br}[\chi_{cJ}(1P) \to J/\psi + X] \;
        \frac{\sigma[\chi_{cJ}(1P)]}{\sigma[J/\psi]}.
\label{eq:prod-chifrac}
\end{eqnarray}
The $J=0$ term in (\ref{eq:prod-chifrac}) is usually negligible, because the 
branching fraction $\rm{Br}[\chi_{c0}\to J/\psi\,X]$ is so small.
The fraction of $J/\psi$'s that are produced directly
can be denoted by $F_{J/\psi}$.
\shortpage

Another set of observables in which many of the uncertainties cancel out
consists of polarization variables, which can be defined as ratios
of cross-sections for the production of different spin states of the same 
quarkonium. The angular distribution of
the decay products of the quarkonium depends on the spin state of the 
quarkonium. The polarization of a $1^{--}$
state, such as the $J/\psi$, can be measured from the
angular distribution of its decays into lepton pairs. Let $\theta$ be
the angle in the $J/\psi$ rest frame between the positive lepton
momentum and the chosen polarization axis.
The most convenient choice of polarization axis depends on the process.
The differential cross-section has the form
\begin{equation}
\frac{ d\sigma}{d(\cos \theta)} \propto 1+\alpha\cos^2\theta,
\label{eq:prod-alphadef}
\end{equation}
which defines a polarization variable $\alpha$
whose range is $-1 \le \alpha \le +1$. We can define longitudinally 
and transversely polarized $J/\psi$'s to be ones whose spin components 
along the polarization axis are 0 and $\pm 1$, respectively. 
The polarization variable $\alpha$ can 
then be expressed as $(1-3\xi)/(1+\xi)$, where $\xi$ is the 
fraction of the $J/\psi$'s that are longitudinally polarized.
The value $\alpha=1$ corresponds to $J/\psi$ with 100\% transverse
polarization, while $\alpha=-1$ corresponds to $J/\psi$ with 100\%
longitudinal polarization. 

One short-coming of the NRQCD factorization approach is that, at
leading order in $v$, some of the kinematics of production are treated
inaccurately. Specifically, the mass of the light hadronic state that
forms during the evolution of the $Q\bar Q$ pair into the quarkonium
state is neglected, and no distinction is made between $2m$ and the
quarkonium mass. While the corrections to these approximations are
formally of higher order in $v$, they can be important numerically in
the cases of rapidly varying quarkonium-production distributions, such
as $p_T$ distributions at the Tevatron and $z$ distributions at the $B$
factories and HERA near the kinematic limit $z=1$. These effects can be
taken into account through the resummation of certain operator matrix
elements of higher order in $v$ \cite{Beneke:1997qw}. The resummation
results in universal nonperturbative shape functions that give the
probability distributions for a $Q\bar Q$ pair with a given set of
quantum numbers to evolve into a quarkonium with a given fraction of the 
pair's momentum. The shape functions could, in principle, be extracted 
from the data for one process and applied to another
process. Effects from resummation of logarithms of $1-z$ and model shape
functions have been calculated for the process $e^+e^-\to J/\psi+X$
\cite{Fleming:2003gt}. For shape functions that satisfy the
velocity-scaling rules, these effects are comparable in size. It may be
possible to use this resummed theoretical prediction to extract the
dominant shape function from the Belle and BaBar data for $e^+e^-\to
J/\psi+X$ and then use it to make predictions for $J/\psi$
photoproduction near $z=1$~\cite{Beneke:1999gq}. 

\subsection{Colour-singlet model}
\label{sec:prodsec-nrqcdCSM}

The colour-singlet model (CSM) was first proposed shortly after the
discovery of the $J/\psi$. The initial applications were to $\eta_c$
and $\chi_c$ production through two-gluon fusion
\cite{Einhorn:1975ua,Ellis:1976fj,Carlson:1976cd,Kuhn:1979kb}. Somewhat
later, the CSM was applied to the production of $J/\psi$ and $\eta_c$
in $B$-meson decays \cite{DeGrand:wf,Kuhn:1979zb,Wise:1979tp} and to
the production of $J/\psi$ plus a gluon
\cite{Chang:1979nn,Baier:1981zz,Baier:1981uk,Baier:1983va,%
Berger:1980ni,Keung:1981gs} through two-gluon fusion and photon--gluon
fusion.  The CSM was taken seriously until around 1995, when
experiments at the Tevatron showed that it under-predicts the
cross-section for prompt charmonium production in $p \bar p$
collisions by more than an order of magnitude.  An extensive review of
the colour-singlet model can be found in Ref.~\cite{Schuler:1994hy}.

The colour-singlet model can be obtained from the NRQCD factorization
formula in \Eq~(\ref{eq:prod-fact}) by dropping all of the colour-octet
terms and all but one of the colour-singlet terms. The term that is
retained is the one in which the quantum numbers of the $Q \bar Q$ pair
are the same as those of the quarkonium. The CSM production matrix
elements are related to the corresponding decay matrix elements by the
vacuum-saturation approximation, and, so, they can be determined from
annihilation decay rates. Thus, the CSM gives absolutely
normalized predictions for production cross-sections. 
The heavy-quark spin symmetry relates the CSM
matrix elements of the $4(2L+1)$ states within an
orbital-angular-momentum multiplet with quantum number $L$. 
Thus, the CSM also gives nontrivial predictions for polarization.

In the case of an S-wave state, the CSM term in \Eq~(\ref{eq:prod-fact})
is the one whose matrix element is of leading order in $v$. However,
owing to kinematic factors or factors of $\alpha_s$ in the short-distance
coefficients, the CSM term may not be dominant. In the case of a
P-wave state or a state of higher orbital angular momentum, the CSM
term is only one of the terms whose matrix element is of leading
order in $v$. For these states, the CSM leads to infrared divergences
that cancel only when one includes colour-octet terms whose matrix
elements are also of leading order in $v$. Thus, the CSM is
theoretically inconsistent for quarkonium states with nonzero orbital
angular momentum.

\subsection{Colour-evaporation model}
\label{sec:prodsec-nrqcdCEM}

The colour evaporation model (CEM) was first proposed in 1977 
\cite{Fritzsch:1977ay,Halzen:1977rs,Gluck:1977zm,Barger:1979js}
and has enjoyed considerable phenomenological success. 
In the CEM, the cross-section for a quarkonium state $H$
is some fraction $F_H$ of the cross-section for producing
$Q \bar Q$ pairs with invariant mass below the $M \bar M$ threshold, 
where $M$ is the lowest mass meson containing the heavy quark $Q$. 
(The CEM parameter $F_H$ should not be confused with the fraction
of $J/\psi$'s that come from decay of $H$.)
This cross-section has an upper limit on the 
$Q \bar Q$ pair mass but no constraints on the 
colour or spin of the final state.  The $Q \bar Q$ pair 
is assumed to neutralize its colour by interaction with the
collision-induced colour field, that is, by ``colour evaporation.'' The
$Q$ and the $\bar Q$ either combine with light quarks to produce
heavy-flavoured hadrons or bind with each other to form quarkonium.  If
the $Q \bar Q$ invariant mass is less than the heavy-meson threshold
$2m_M$, then the additional energy that is needed to produce
heavy-flavoured hadrons can be obtained from the nonperturbative colour
field. Thus, the sum of the fractions $F_H$ over all quarkonium states
$H$ can be less than unity. The fractions $F_H$ are assumed to be
universal so that, once they are determined by data, they can be used to
predict the cross-sections in other processes and in other kinematic
regions.
\longpage

In the CEM at leading order in $\alpha_s$, the production cross-section 
for the quarkonium state $H$ in collisions of the light hadrons 
$h_A$ and $h_B$ is
\begin{eqnarray}
\lefteqn{\sigma_{\rm CEM}^{\rm (LO)}[h_A h_B \to H + X]  =} \nonumber\\
&& F_H \sum_{i,j} 
\int_{4m^2}^{4m_M^2} d\hat{s}
\int \! dx_1 dx_2~f_i^{h_A}(x_1,\mu)~f_j^{h_B}(x_2,\mu)~ 
\hat\sigma_{ij}(\hat{s})~\delta(\hat{s} - x_1x_2s)  \, , 
\label{eq:prod-sigtil}
\end{eqnarray} 
where $ij = q \bar q$ or $g g$, $\hat s$ is the square of the partonic
centre-of-mass energy, and $\hat\sigma_{ij}(\hat s)$ is the
$ij\rightarrow Q\bar Q$ subprocess cross-section. The leading-order
calculation cannot describe the quarkonium $p_T$ distribution, since
the $p_T$ of the $Q \bar Q$ pair is zero at LO. At NLO in $\alpha_s$,
the subprocesses $ij\rightarrow k Q \bar Q$, where $i$, $j$, and $k$
are light quarks, antiquarks, and gluons, produce $Q \bar Q$ pairs
with nonzero $p_T$. Complete NLO calculations of quarkonium production
in hadronic collisions using the CEM have been carried out in
Refs.~\cite{Gavai:1994in,Schuler:1996ku}, using the exclusive $Q \bar
Q$ production code of Ref.~\cite{Mangano:kq} to obtain the $Q \bar Q$
pair distributions. The resulting values of the 
parameters $F_H$ are given 
in \Section~\ref{sec:prodsec-fixed-targetCEM}. There are also
calculations in the CEM beyond LO that use only a subset of the NLO
diagrams~\cite{Amundson:1996qr} and calculations that describe the
soft colour interaction within the framework of a Monte Carlo event
generator~\cite{Edin:1997zb}. Calculations beyond LO in the CEM have
also been carried out for $\gamma p$, $\gamma\gamma$ and
neutrino--nucleon collisions and for $Z^0$ decays
\cite{Eboli:1998xx,Eboli:2003fr,Eboli:2003qg,Eboli:2001hc,Gregores:1996ek}.  
Apparently, the colour-evaporation model has not been applied to
quarkonium production in $e^+ e^-$ annihilation.
\longpage

The most basic prediction of the CEM is that the ratio of the
cross-sections for any two quarkonium states should be constant,
independent of the process and the kinematic region. Some variations
in these ratios have been observed. For example, the ratio of the
cross-sections for $\chi_c$ and $J/\psi$ are rather different in
photoproduction and hadroproduction. Such variations present a serious
challenge to the status of the CEM as a quantitative phenomenological
model for quarkonium production.

In some papers on the Colour Evaporation Model
\cite{Amundson:1996qr}, the collision-induced colour field that
neutralizes the colour of the $Q \bar Q$ pair is also assumed to
randomize its spin. This leads to the prediction that the quarkonium
production rate is independent of the quarkonium spin. This prediction
is contradicted by measurements of nonzero polarization of the
$J/\psi$, the $\psi(2S)$, and the $\Upsilon(nS)$ in several
experiments. The assumption of the randomization of the $Q \bar Q$ spin
also implies simple spin-counting ratios for the cross-sections 
for the direct production of
quarkonium states in the same orbital-angular-momentum multiplet.
For example, the CEM with spin randomization predicts that 
the direct-production cross-sections for charmonium satisfy 
$\sigma_{\rm dir}[\eta_c]:\sigma_{\rm dir}[J/\psi] = 1:3$ and  
$\sigma_{\rm dir}[\chi_{c0}]:\sigma_{\rm dir}[\chi_{c1}]:
        \sigma_{\rm dir}[\chi_{c2}] = 1:3:5$. 
The inclusive cross-sections need not satisfy these spin-counting relations
if there is significant feeddown from decay of higher quarkonium states, 
as is the case for $J/\psi$.
Deviations from the predicted spin-counting ratio for
$\chi_{c1}$ to $\chi_{c2}$ have been observed.
One might conclude that the CEM is ruled out by the
observations of nonzero polarization 
and of deviations from the spin-counting relations. 
On the other hand, the assumption of
the randomization of the $Q \bar Q$ spin is really independent of the
assumption of colour evaporation. 
Some proponents of the CEM omit the assumption of spin randomization.
Alternatively, since the CEM is just a model, one can simply declare 
it to apply only to spin-averaged cross-sections.
In the remainder of this chapter, when we mention the predictions of the
CEM for the relative production rates of quarkonium states that differ
only in their spin or total-angular-momentum quantum numbers, we are
referring to the version of the CEM that includes the assumption of spin
randomization.

There is a simple correspondence between
the CEM and the NRQCD factorization approach. The CEM amounts to the
assumption that an NRQCD production matrix element 
$\langle {\cal O}_n^H(\Lambda) \rangle$
is proportional to the expectation value of
the operator that is obtained by replacing the projector onto the
hadronic state $H$ with a projector onto the set of $Q\bar Q$ states
with invariant mass less
than $2m_M$. In addition to an integral over the $Q\bar Q$ phase space,
the projector contains sums over the $Q\bar Q$ spins and colours.
The only dependence on
the quarkonium $H$ is through a common factor $F_H$ in the
proportionality constant for each NRQCD matrix element. Since, in this
picture, the probability of forming a specific quarkonium state $H$ is
independent of the colour and spin state of the $Q \bar Q$ pair, NRQCD
matrix elements that differ only  by colour and spin quantum numbers are
equal up to simple group theory factors. This picture also implies a
hierarchy of NRQCD matrix elements according to their
orbital-angular-momentum quantum number $L$. In the integration over the
$Q\bar Q$ phase space of an NRQCD operator with orbital-angular-momentum
quantum number $L$, the leading term scales as $k^{2L+1}$, where $k$ is
the $Q$ or $\bar Q$ momentum in the $Q\bar Q$ rest frame. The difference
$s_{\rm max}- 4 m^2$ is proportional to $k^2$. Hence, there is an
orbital-angular-momentum suppression factor $[(s_{\rm max}- 4 m^2)/4
m^2]^L \sim v^{2L}$ in the matrix elements.\footnote{From the
perspective of NRQCD, the upper limit $s_{\rm max} = 4m_M^2$ on the $Q
\bar Q$ invariant mass that traditionally has been used in the CEM is
quite arbitrary.  Any choice that satisfies $s_{\rm max}- 4m_Q^2 \sim 4
m_Q^2 v^2$ leads to the same velocity-scaling rules.} That is, the CEM
implies that S-wave NRQCD matrix elements dominate and that those with
orbital-angular-momentum quantum number $L\ge 1$ are suppressed as
$v^{2L}$. One way to test the assumptions of the CEM is to extract the
NRQCD matrix elements from data and compare them with the predictions of
the CEM. 

The qualifier NLO in ``the CEM at NLO'' is somewhat misleading. As is
described in \Section~\ref{sec:prodsec-nrqcdmge}, the NLO cross-section for
$Q \bar Q$ production that is used in computing the CEM predictions is
accurate through order $\alpha_s^3$, which is next-to-leading order at
zero $p_T$, but leading order at nonzero $p_T$. This is the same
accuracy in $\alpha_s$ as the existing predictions in the NRQCD
factorization approach. The NLO $Q\bar Q$ $p_T$ distribution is singular
at $p_T=0$, but integrable.  The existing NLO calculations in the CEM
obtain a smooth $p_T$ distribution at small $p_T$ by using a smearing
prescription to mimic the effects of multiple gluon emission. The
smearing has a significant effect on the shape of the $p_T$
distribution, except at very large $p_T$.

\subsection{Multiple gluon emission}
\label{sec:prodsec-nrqcdmge}

Multiple gluon emission can be very important for transverse momentum
distributions, distributions near kinematic limits, and in situations in
which production near threshold is important.  For example, a
fixed-order perturbative calculation typically gives a transverse
momentum distribution $d \sigma/dp_T^2$ for quarkonium that includes
terms proportional to $\delta(p_T^2)$ and $1/p_T^2$ that are singular as
$p_T \rightarrow 0$. (However, the distribution has a well-behaved
integral over $p_T$.) This singular distribution becomes a smooth one
when the effects of multiple gluon emission are taken into account to
all orders in perturbation theory. Several methods, which we now
describe, have been developed to take into account some of these
effects.

{\it Resummation} methods sum, to all orders in $\alpha_s$,  certain
logarithmically enhanced terms that are associated with soft- and
collinear-gluon emission. The resummations can be carried out at various
levels of precision in the logarithmic enhancements, that is, in leading 
logarithmic (LL) order, in next-to-leading logarithmic (NLL) order,
\etc Resummation can, in principle, be extended to arbitrarily high
precision in the logarithmic enhancements. However, in practice, it is
seldom carried out beyond LL or NLL accuracy. Generally, logarithms of
$p_T^2/M^2$ have the largest effect on $p_T$ distributions
\cite{Collins:1984kg}, although logarithms of the available partonic
energy above threshold (threshold logarithms) and logarithms of
$s/p_T^2$ (small-$x$ logarithms) can also be important for particular
processes and kinematic regions\footnote{For a general discussion of
resummation techniques for logarithms of $p_T^2/M^2$ and threshold
logarithms, see Ref.~\cite{Contopanagos:1996nh}.}. Because arbitrarily
soft or collinear gluon emissions are resummed, the resummed expressions
depend on nonperturbative functions. This dependence lessens as the mass
and transverse momentum scales of the process increase, and it may be
insignificant at large masses and/or transverse momenta. Some
practical disadvantages of the resummation method are that it has to be
reformulated, to some extent, for every process and that it usually does
not yield results that are fully differential in all of the kinematic
variables. Since resummation calculations retain only soft and collinear
logarithmically enhanced terms, they generally do not describe
accurately processes in which hard gluons are emitted at large
angles\,---\,so called ``Mercedes events.'' This situation can be remedied
to some extent by combining resummation with exact next-to-leading order
(NLO) calculations, which retain all contributions associated with gluon
emission at NLO, not just logarithmically enhanced contributions
\cite{Cacciari:1998it}. 

{\it Parton-shower Monte Carlos} share with resummation methods the
approach of modeling multiple gluon emission by retaining certain
logarithmically enhanced terms in the cross-section. The Monte Carlos
take into account a finite, but arbitrarily large, number of gluon
emissions. The original implementations of shower Monte Carlo methods,
such as ISAJET \cite{Paige:fb,Paige:2003mg}, generally treat only the
leading collinear logarithmic enhancements correctly, while more recent
implementations, such as \textsc{Pythia}
\cite{Sjostrand:1985ys,Sjostrand:1986hx} and HERWIG
\cite{Marchesini:1987cf,Marchesini:1991ch} treat both the leading
collinear and soft logarithmic enhancements correctly. Generally, the
showering processes are cut off so that they do not become so soft or
collinear as to be nonperturbative in nature. The showering may then be
supplemented with nonperturbative models that describe the 
hadronization of the
partons. A practical advantage of the
shower Monte Carlo approach is that it is generally applied easily to
any Born-level production process. Furthermore, it produces results that
are differential in all of the kinematic variables that are associated
with the final-state particles. Hence, it lends itself to the
application of experimental cuts. As is the case with resummation
methods, the shower Monte Carlo approach does not yield an accurate
modeling of processes in which hard gluons are emitted at large angles.
A partial remedy for this problem is to use shower Monte Carlos in
conjunction with exact NLO calculations, rather than LO calculations.
Recently, important progress has been made in this direction
\cite{Frixione:2002ik,Frixione:2002bd,Frixione:2003ei,%
Frixione:2004wy,Kramer:2003jk,Soper:2003ya}.  In contrast with
resummation methods, some shower Monte Carlos do not take into account
virtual gluon emission. Such shower Monte Carlos do not yield reliable
estimates of the total cross-section.

The {\it $k_T$-factorization} method is an attempt to take into account
initial-state radiation through parton distributions that depend the
parton's transverse momentum $k_T$, as well as on the parton's
longitudinal momentum fraction $x$. It generally gives answers that are
very different from those of collinear factorization.  The
$k_T$-dependent parton distributions are not very well known
phenomenologically, and there are possibly unresolved theoretical
issues, such as the universality of the $k_T$-dependent parton
distributions.

The {\it $k_T$-smearing} method is a phenomenological model for multiple
initial-state radiation.  As in the $k_T$-factorization method, the
$k_T$ smearing method makes use of $k_T$-dependent parton distributions.
It is assumed that the distribution factors into the $x$-dependent PDF's
that are defined by collinear factorization and a Gaussian distribution
in the transverse momentum $k_T$.  The width $\langle k_T^2 \rangle$ of
the Gaussian can be treated as a process-dependent phenomenological
parameter.  One advantage of this model is that it is easy to implement.
On the other hand, while this model may capture some of the crude
features of soft- and collinear-gluon emission, it is probably incorrect
in detail: resummation methods and shower Monte Carlos
yield transverse-momentum distributions that have longer tails than
those of a Gaussian distribution. The impact of a parton shower on the
quarkonium transverse momentum distribution is, in general, larger than
for the Gaussian $k_T$ smearing, and it extends out to larger values of 
$p_T$.

\subsection{Production in nuclear matter}
\label{sec:prodsec-nrqcdnuc}

The existing factorization ``theorems'' for quarkonium production in
hadronic collisions are for cold hadronic matter. These theorems
predict that nuclear matter is ``transparent'' for $J/\psi$ production
at large $p_T$. That is, at large $p_T$, all of the nuclear effects
are contained in the nuclear parton distributions. The corrections to
this transparency are of order $(mv)^2/p_T^2$ for unpolarized
cross-sections and of order $mv/p_T$ for polarized cross-sections.

The effects of transverse-momentum kicks from multiple elastic
collisions between active partons and spectators in the nucleons are
among those effects that are suppressed by $(mv)^2/p_T^2$. Nevertheless,
these multiple-scattering effects can be important because the
production cross-section falls steeply with $p_T$ and because the number
of scatterings grows linearly with the length of the path through the
nuclear matter. Such elastic interactions can be expressed in terms of
eikonal interactions \cite{Bodwin:1988fs} or higher-twist matrix
elements \cite{Qiu:2001hj}.

Inelastic scattering of the quarkonium by the nuclear matter is also an
effect of higher order in $(mv)^2/p_T^2$. However, it can become
dominant when the amount of nuclear matter that is traversed by the
quarkonium is sufficiently large. Factorization breaks down when the
length $L$ of the quarkonium path in the nucleus satisfies
\begin{equation}
L \gsim \frac{\mathrm{Min}(z_Q,z_{\bar Q})P_{\mathrm{onium}}^2}
{M_A (k_T^{\rm tot})^2},
\end{equation}
where $M_A$ is
the mass of the nucleus, $z$ is the parton longitudinal momentum
fraction, $P_{\rm onium}$ is the momentum of the quarkonium in the
parton CM frame, and $k_T^{\rm tot}$ is the accumulated 
transverse-momentum ``kick'' from passage through the nuclear matter. 
This condition for the break-down of factorization is similar to
``target-length condition'' in Drell--Yan production
\cite{Bodwin:1981fv,Bodwin:1984hc}. Such a breakdown of
factorization is observed in the Cronin effect at low $p_T$ and in
Drell--Yan production at low $Q^2$, where the cross-section is
proportional to the nucleon number raised to a power less than unity.

It is possible that multiple-scattering effects may be
larger for colour-octet production than for colour-singlet production.
In the case of colour-octet production, the pre-quarkonium $Q\bar Q$
system carries a nonzero colour charge and, therefore, has a larger 
amplitude to exchange soft gluons with spectator partons.

At present, there is no complete, rigorous theory to account for all of
the effects of multiple scattering, and we must resort to
``QCD-inspired'' models. A reasonable requirement for models is that
they be constructed so that they are compatible with the factorization
result in the large-$p_T$ limit. Many models treat interactions of the
pre-quarkonium with the nucleus as on-shell scattering (Glauber
scattering). This assumption should be examined carefully, as on-shell
scattering is known, from the factorization proofs, not to be a valid
approximation in leading order in $(mv)^2/p_T^2$.

\section{Quarkonium production at the Tevatron}
\label{sec:prodsec-tevatron}
Charmonium and bottomonium are produced copiously in high energy hadron
colliders. The present and future hadron colliders include
\begin{itemize}

\item the Tevatron, 
a $p \bar p$ collider operating at Fermilab with centre-of-mass energy 
of 1.8~TeV in Run~I and 1.96~TeV in Run~II,
 
\item RHIC, 
a heavy-ion or $p p$ collider operating at Brookhaven with 
centre-of-mass energy of up to 200~GeV per nucleon--nucleon collision,

\item 
the LHC at CERN, a $p p$ collider under construction at CERN 
with centre-of-mass energy of 17~TeV.
\end{itemize}
In this section, we focus on the Tevatron, because it has produced the
most extensive and precise data on quarkonium production.  The
photoproduction of quarkonium at high-energy $p\bar{p}$, $pp$, and
heavy ion colliders is discussed in
\Chapter~\ref{chapter:charm-beauty-in-media} of this report.

\subsection{Charmonium cross-sections}
\label{sec:prodsec-tevatroncharm}

%%%%%%%%%%%%%%%%%%%%%%%%%%%%%%%%%%%%%%%%%%%%%%%%%%%%%%%%%%%%%%%%%%%%%%%%%%%%%%%%%%%%%%%%%%%%%
\begin{figure}[p]
\begin{center}
\vspace{-0.5cm}
\includegraphics[width=9cm]{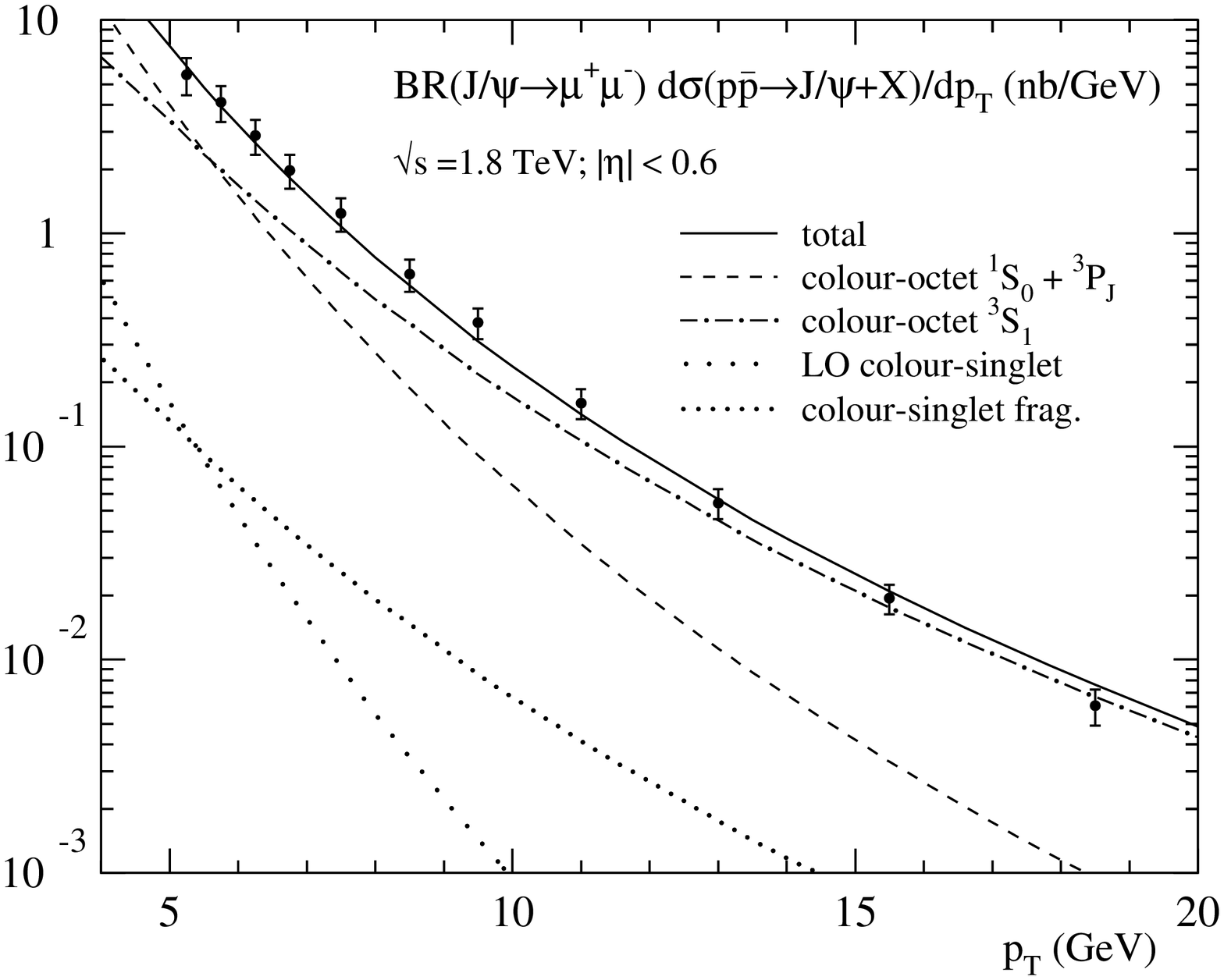}
\vspace{-0.5cm}
\includegraphics[width=9cm]{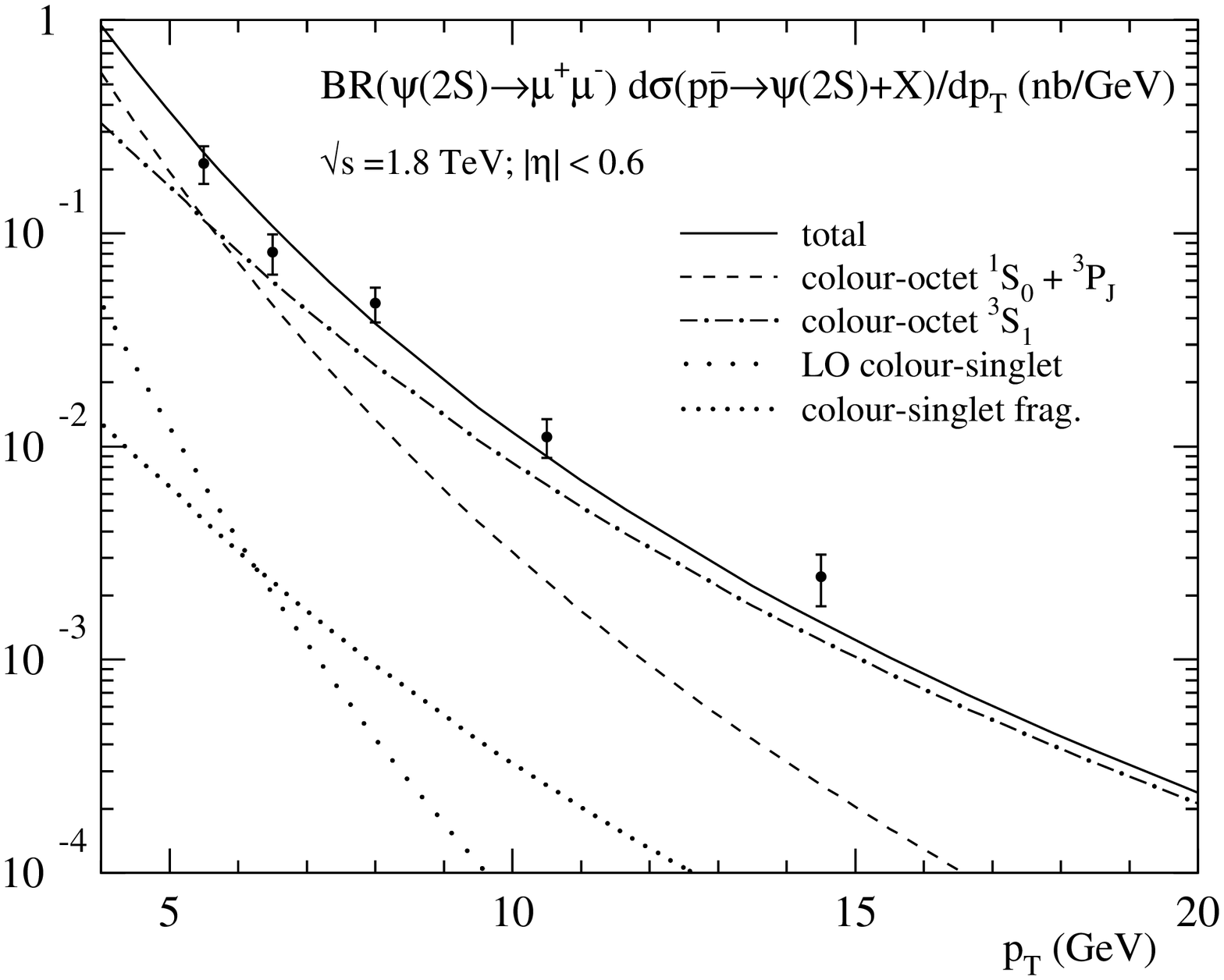}
\vspace{-0.5cm}
\includegraphics[width=9cm]{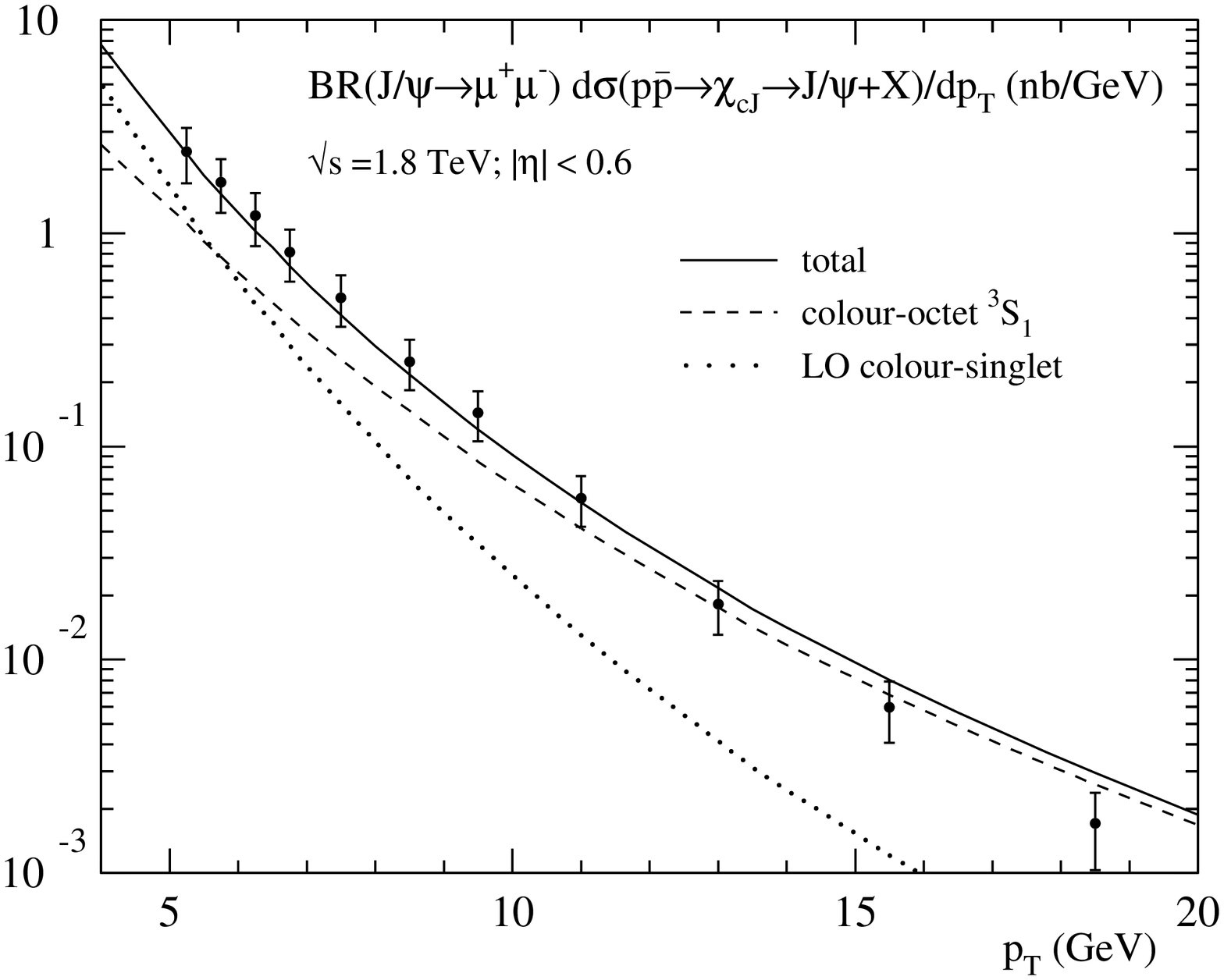}
\end{center}
\caption[Differential cross-sections for production of direct
         $J/\psi$, prompt $\psi(2S)$, and prompt $J/\psi$ from 
         decay of $\chi_c$]
        {Differential cross-sections for the production of direct
         $J/\psi$ (top), prompt $\psi(2S)$ (middle), and prompt
         $J/\psi$ from decay of $\chi_c$ (bottom) at the Tevatron as a
         function of $p_T$.  The data points are CDF measurements from
         Run I \cite{Abe:1997jz,Abe:1997yz}.  The dotted curves are
         the CSM contributions.  The solid curves are the NRQCD
         factorization fits, and the other curves are individual
         colour-octet contributions to the fits. From
         Ref.~\cite{Kramer:2001hh}.}
\label{fig:tevatron_psi}
\end{figure}
%%%%%%%%%%%%%%%%%%%%%%%%%%%%%%%%%%%%%%%%%%%%%%%%%%%%%%%%%%%%%%%%%%%%%%%%%%%%%%%%%%%%%%%%%%%%%

In high energy collisions, charmonium is produced both through direct
production mechanisms and through decays of other hadrons. In the case
of charmonium production through $B$-hadron decays, the charmonium is
produced at a secondary vertex, and a vertex detector can be used to
identify this contribution to the measured production rate.  We refer
to the inclusive cross-section for production of a charmonium state
with the contribution from $B$ decays removed as the {\it prompt}
cross-section. The prompt cross-section includes both the direct
production of the charmonium and its production through decays of
higher charmonium states.

In Run I of the Tevatron, the CDF collaboration measured the prompt
cross-sections for the production of several charmonium states
in $p \bar p$ collisions at
a centre-of-mass energy of 1.8~TeV \cite{Abe:1997jz,Abe:1997yz}. The
CDF data for production of direct $J/\psi$, prompt $\psi(2S)$, and
prompt $J/\psi$ from decay of $\chi_c$ are shown in
\Figure~\ref{fig:tevatron_psi}. In the CDF analysis, prompt $J/\psi$'s
that do not come from decays of $\psi(2S)$ or $\chi_c$ were assumed to
be produced directly. 

At non-vanishing transverse momentum, the leading parton processes for
producing charmonium ($ij \rightarrow c \bar{c} + k$, where $i$, $j$,
and $k$ are light quarks, antiquarks, and gluons) occur at order
$\alpha_s^3$. The colour-singlet-model (CSM) predictions
are shown as dotted lines in \Figure~\ref{fig:tevatron_psi}.  In the top
two panels of \Figure~\ref{fig:tevatron_psi}, the more steeply falling
dotted lines are the predictions of the CSM at leading order in
$\alpha_s$. The other dotted lines in the top two panels of
\Figure~\ref{fig:tevatron_psi} are contributions of higher order in
$\alpha_s$ involving gluon fragmentation.  As can be seen in the top
panel of \Figure~\ref{fig:tevatron_psi}, the gluon-fragmentation
contribution renders the shape of the CSM prediction for direct
$J/\psi$ production roughly compatible with the CDF data. However, the
normalization is too small by more than an order of magnitude. There
is a similar discrepancy in the normalization for prompt $\psi(2S)$
production, as can be seen in the middle panel of
\Figure~\ref{fig:tevatron_psi}. In the case of production of prompt
$J/\psi$ from decay of $\chi_c$, which is shown in the bottom panel of
\Figure~\ref{fig:tevatron_psi}, the discrepancy is less dramatic, but the
cross-section is still under-predicted by the CSM.  The large
discrepancies between the measurements and the CSM predictions for the
production cross-section for S-wave charmonium states rules out the
CSM as a credible model for quarkonium production.

%%%%%%%%%%%%%%%%%%%%%%%%%%%%%%%%%%%%%%%%%%%%%%%%%%%%%%%%%%%%%%%%%%%%%%%%%%%%%%%%%%%%%%%%%%%%%
\begin{table}
\caption[NRQCD production matrix elements for charmonium states]
        {NRQCD production matrix elements for charmonium states
         obtained from the transverse momentum distributions at the
         Tevatron~\cite{Kramer:2001hh}. The errors quoted are
         statistical only.}
\label{tab:me-1}
\begin{center}
\renewcommand{\arraystretch}{1.5}
\[
\begin{array}{|c|ccc|}
\hline\hline
 H & \langle {\cal{O}}_1^{H} \rangle  & \langle
 {\cal{O}}^{H}_8({}^3S_1) \rangle  &
 M_{3.5}^{H}\\ \hline
 J/\psi   & 1.16~{\rm GeV^3} & (1.19 \pm 0.14)\times 10^{-2}~{\rm GeV}^3 &  
 (4.54 \pm 1.11)\times 10^{-2}~{\rm GeV}^3 \\[-1mm] 
 \psi(2S) & 0.76~{\rm GeV^3} & (0.50 \pm 0.06)\times 10^{-2}~{\rm GeV}^3 & 
 (1.89 \pm 0.46)\times 10^{-2}~{\rm GeV}^3 
 \\[-1mm]
 \chi_{c0} & 0.11~{\rm GeV^5} & (0.31 \pm 0.04)\times 10^{-2}~{\rm GeV}^3 & 
 \\[1mm] \hline \hline
\end{array}
\]
\renewcommand{\arraystretch}{1.0}
\end{center}
\end{table}
%%%%%%%%%%%%%%%%%%%%%%%%%%%%%%%%%%%%%%%%%%%%%%%%%%%%%%%%%%%%%%%%%%%%%%%%%%%%%%%%%%%%%%%%%%%%%

According to the NRQCD factorization approach, the charmonium
production cross-section contains not only the CSM terms, which are
absolutely normalized, but also colour-octet terms, whose
normalizations are determined by colour-octet matrix elements.  In the
case of $J/\psi$ and $\psi(2S)$ production, the most important
colour-octet matrix elements are $\langle{\cal
O}^H_8({}^3S_1)\rangle$, $\langle{\cal O}^H_8({}^3P_0)\rangle$, and
$\langle{\cal O}^H_8({}^1S_0)\rangle$. At large $p_T$, the $J/\psi$
and $\psi(2S)$ cross-sections are dominated by gluon fragmentation
into colour-octet ${}^3S_1$ charm pairs~\cite{Braaten:1995vv}, which
falls as $d\hat{\sigma}/dp_T^2\sim1/p_T^4$. The colour-octet ${}^1S_0$
and ${}^3P_J$ channels are significant in the region $p_T\;
\rlap{\lower 3.5 pt \hbox{$\mathchar \sim$}} \raise 1pt \hbox {$<$}\;
10$~GeV, but fall as $d\hat{\sigma}/dp_T^2\sim1/p_T^6$ and become
negligible at large $p_t$.  Because the ${}^1S_0^{(8)}$ and
${}^3P_J^{(8)}$ short-distance cross-sections have a similar $p_t$
dependence, the transverse momentum distribution is sensitive only to
the linear combination $M^H_k$ defined in (\ref{eq:prod-lincomb}),
with $k\approx 3$.  As can be seen in the top panel of
\Figure~\ref{fig:tevatron_psi}, a good fit to the normalization and
shape of the direct $J/\psi$ cross-section can be obtained by
adjusting $\langle{\cal O}^{J/\psi}_8({}^3S_1)\rangle$ and
$M^{J/\psi}_{3.5}$. As is shown in the middle panel of
\Figure~\ref{fig:tevatron_psi}, a similarly good fit to the prompt
$\psi(2S)$ cross-section can be obtained by adjusting the
corresponding parameters for $\psi(2S)$.  In the case of production of
the $\chi_{cJ}$ states, the most important colour-octet matrix element
is $\langle{\cal O}^H_8({}^3S_1)\rangle$.  As can be seen in the
bottom panel of \Figure~\ref{fig:tevatron_psi}, the fit to the
cross-section for production of prompt $J/\psi$ from decay of $\chi_c$
can be improved by adjusting $\langle{\cal
O}^{\chi_{c0}}_8({}^3S_1)\rangle$.  \Table~\ref{tab:me-1} shows the
values of the quarkonium matrix elements that are obtained in the fit
of Ref.~\cite{Kramer:2001hh,Beneke:1996yw}.  The colour-singlet matrix
elements are taken from the potential-model calculation of
Refs.~\cite{Buchmuller:1980su,Eichten:1995ch}.  The colour-octet
matrix elements have been extracted from the CDF data
\cite{Abe:1997jz,Abe:1997yz}.  The CTEQ5L parton distribution
functions \cite{Lai:1999wy} were used, with renormalization and
factorization scales $\mu=(p_T^2+4 m_c^2)^{1/2}$ and
$m_c=1.5\,$GeV. The Altarelli--Parisi evolution has been included for
the $\langle{\cal O}^{\chi_{c0}}_8({}^3S_1)\rangle$ fragmentation
contribution. See Ref.~\cite{Beneke:1996yw} for further details. The
extraction of the various colour-octet matrix elements relies on the
differences in their $p_T$ dependences. Smaller experimental error
bars could help to resolve the different $p_T$ dependences with
greater precision.

%%%%%%%%%%%%%%%%%%%%%%%%%%%%%%%%%%%%%%%%%%%%%%%%%%%%%%%%%%%%%%%%%%%%%%%%%%%%%%%%%%%%%%%%%%%%%
\begin{figure}[t]
\begin{center}
\includegraphics[width=\textwidth]{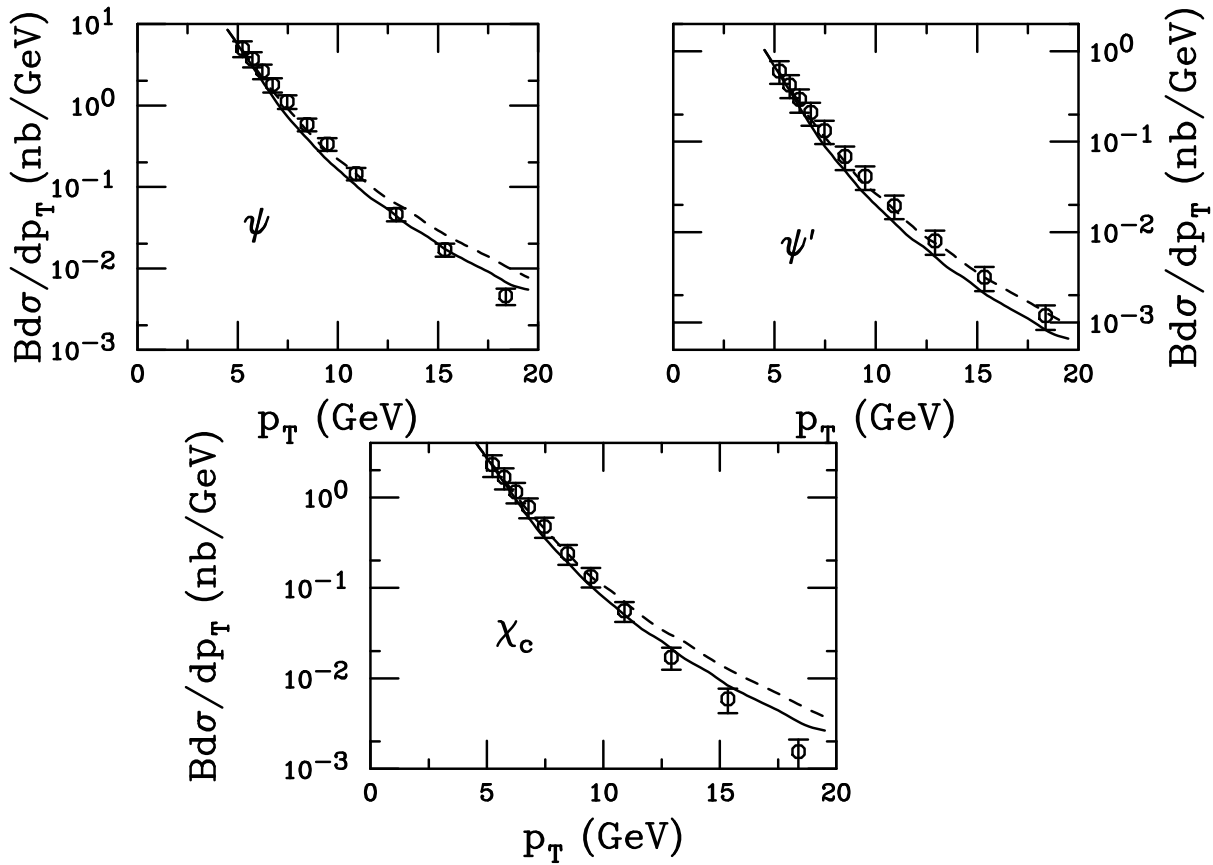}
\end{center}
\caption{Differential cross-sections for production of direct $J/\psi$
         (top left), prompt $J/\psi$ from decays of $\psi(2S)$ (top
         right), and prompt $J/\psi$ from decays of $\chi_c$ (bottom)
         at the Tevatron as a function of $p_T$.  The data points are
         the CDF measurements \cite{Abe:1997jz,Abe:1997yz}.  The
         dotted and solid curves are the CEM predictions at NLO with
         $\langle k_T^2\rangle = 2.5$~GeV$^2$, using the first and
         fourth charmonium parameter sets in
         \Table~\ref{tab:prodsec-qqbparams}.  }
\label{fig:psicdfptdep}
\end{figure}
%%%%%%%%%%%%%%%%%%%%%%%%%%%%%%%%%%%%%%%%%%%%%%%%%%%%%%%%%%%%%%%%%%%%%%%%%%%%%%%%%%%%%%%%%%%%%

The normalization and the shape of the prompt charmonium cross-section
at the Tevatron can also be described reasonably well by the
colour-evaporation model (CEM). The CEM parameters can be fixed by
fitting to the data from $pN$ collisions and by using the measured
branching fractions for charmonium decays. The predictions of the CEM
at next-to-leading order in $\alpha_s$ (NLO) can be calculated using
the NLO parameter sets that are described in
\Section~\ref{sec:prodsec-fixed-targetCEM}. The normalization of the predicted
cross-section for prompt $J/\psi$ production is in reasonable
agreement with the CDF data from Run~I. The shape can be brought into
good agreement by adding $k_T$ smearing, with $\langle k_T^2\rangle =
2.5$~GeV$^2$. In \Figure~\ref{fig:psicdfptdep}, the resulting CEM
predictions are compared with the CDF charmonium data for production
of direct $J/\psi$, prompt $J/\psi$ from decay of $\psi(2S)$, and
prompt $J/\psi$ from decay of $\chi_c$. The predictions are all in
good agreement with the CDF data.

In the case of the S-wave production matrix elements, the NRQCD
velocity-scaling rules predict that 
\begin{equation} 
\frac{ \langle{\cal O}_8\rangle}{\langle{\cal O}_1\rangle}
\sim \frac{v^4}{2N_c},
\end{equation} 
where this estimate includes colour factors that are associated with
the expectation values of the NRQCD operators, as advocated by
Petrelli {\it et al.} \cite{Petrelli:1997ge}. As can be seen from
\Table~\ref{tab:me-1}, the extracted colour-octet matrix elements are
roughly compatible with this estimate [$v^4/(2N_c)\approx 0.015$].
However, a much more stringent test of the theory is to check the
universality of the extracted matrix elements in other processes. In
the case of the P-wave production matrix elements, the velocity
scaling rules yield the estimate
\begin{equation}
\frac{\langle{\cal O}_8\rangle}{\langle{\cal O}_1\rangle/m_c^2}
\sim \frac{v^0}{2N_c}.
\end{equation}
The P-wave colour-octet matrix element in \Table~\ref{tab:me-1} is
somewhat smaller than this estimate would suggest. That is also the
case for the matrix elements that appear in P-wave quarkonium
decays, which have been determined phenomenologically
\cite{Maltoni:2000km} and in lattice calculations
\cite{Bodwin:1993wf,Bodwin:1994js,Bodwin:1996tg,Bodwin:1996mf,Bodwin:2001mk}.

% ----------------------------------------------------------------------

In \Table~\ref{tab:me-2}, we show matrix elements for $J/\psi$ production
that have been obtained from various other fits to the transverse
momentum distribution. We see that there is a large uncertainty that
arises from the dependence of the matrix elements on the factorization
and renormalization scales, as well as a large dependence on the choice
of parton distributions. The  extracted values of the colour-octet matrix
elements (especially $M_k$) are very sensitive to the small-$p_T$
behavior of the cross-section and this, in turn, leads to a sensitivity
to the behavior of the gluon distribution at small $x$. Furthermore, the
effects of multiple gluon emission are important, and their
omission in the fixed-order perturbative calculations leads to
overestimates of the sizes of the matrix elements. In
\Table~\ref{tab:me-2}, one can see the results of various attempts to
estimate the effects of multiple gluon emission. Sanchis--Lozano (S)
and Kniehl and Kramer (KK) made use of parton-shower Monte Carlos,
while Petrelli (P) and Sridhar, Martin, and Stirling (SMS) employed
models containing Gaussian $k_T$ smearing. In addition, Sanchis--Lozano
included a resummation of logarithms of $p_T^2/m^2$. H\"agler,
Kirschner, Sch\"afer, Szymanowski, and Teryaev (HKSST) used the
$k_T$-factorization formalism to resum large logarithms in the limit $s
\gg 4m_c^2$. (See also the calculations by Yuan and Chao \cite{Yuan:2000cp,Yuan:2000qe}.) 
Similar large dependences on the choices of factorization and
renormalization scales, parton distributions, and multiple gluon
emission can be seen in the matrix elements that have been extracted
from the $\psi(2S)$ and $\chi_c$ transverse momentum
distributions. See Ref.~\cite{Kramer:2001hh} for details.

Effects of corrections of higher order in $\alpha_s$ are a further
uncertainty in the fits to the data in \Table~\ref{tab:me-2}. Such
corrections are known to be large in the case of charmonium decays. In
the case of charmonium production, a new channel for colour-singlet
production, involving $t$-channel gluon exchange, first appears in order
$\alpha_s$ and could yield a large correction. Maltoni and Petrelli
\cite{Petrelli:1999rh} have found that real-gluon corrections to
colour-singlet ${}^3S_1$ production give a large contribution.
Next-to-leading order (NLO) corrections in $\alpha_s$ for 
$\chi_{c0}$ and $\chi_{c2}$
production have been calculated \cite{Petrelli:1997ge}, as have
NLO corrections for the fragmentation process
\cite{Beneke:1995yb,Ma:1995ci,Braaten:2000pc}. Large corrections from
the resummation of logarithms of $p_T^2/m^2$ in the fragmentation of
partons into quarkonium have also been calculated
\cite{Cacciari:1994dr,Braaten:1994xb,Roy:1994ie,Sanchis-Lozano:1999um}.

Similar theoretical uncertainties arise in the extraction of the NRQCD
production matrix elements for the $\psi(2S)$ and $\chi_c$ states. The
statistical uncertainties are larger for $\psi(2S)$ and $\chi_c$
production than for $J/\psi$ production. We refer the reader to
Ref.~\cite{Kramer:2001hh} for some examples of the NRQCD matrix elements
that have been extracted for these states.

%%%%%%%%%%%%%%%%%%%%%%%%%%%%%%%%%%%%%%%%%%%%%%%%%%%%%%%%%%%%%%%%%%%%%%%%%%%%%%%%%%
\begin{table}
\caption[$J/\psi$ production matrix elements]
        {$J/\psi$ production matrix elements in units of
         $10^{-2}$~GeV${}^3$~\cite{Kramer:2001hh}.  The first error
         bar is statistical; the second error bar (where present) is
         obtained by varying the factorization and renormalization
         scales.}
\label{tab:me-2}
\begin{center}
\renewcommand{\arraystretch}{1.3}
\begin{tabular}{|c|cc|ccc|}
\hline\hline
 \mbox{Reference} & \multicolumn{2}{c|}{\mbox{PDF}} & $\langle
 {\cal{O}}^{J/\psi}_8({}^3S_1) \rangle$ &
 $M_{k}^{J/\psi} $ & $ k $ \\ \hline\hline
 \multicolumn{6}{|c|}{\mbox{LO collinear factorization}} \\ \hline
 {\rm CL} \cite{Cho:1995ce} &
 \multicolumn{2}{c|}{\mbox{MRS(D0)~\cite{Martin:1992zi}}} & $ 0.66 \pm
 0.21 $ & $6.6 \pm 1.5$ & 3 \\ \hline &
 \multicolumn{2}{c|}{\mbox{CTEQ4L~\cite{Lai:1996mg}}} & $1.06 \pm
 0.14^{+1.05}_{-0.59}$ & $4.38 \pm 1.15^{+1.52}_{-0.74}$ & \\
 {\rm BK~\cite{Beneke:1996yw}} &
 \multicolumn{2}{c|}{\mbox{GRV-LO(94)~\cite{Gluck:1994uf}}} & $ 1.12 \pm
 0.14^{+0.99}_{-0.56} $ & $3.90 \pm 1.14^{+1.46}_{-1.07}$ & 3.5 \\ &
 \multicolumn{2}{c|}{\mbox{MRS(R2)~\cite{Martin:1996as}}} & $ 1.40 \pm
 0.22^{+1.35}_{-0.79} $ & $10.9 \pm 2.07^{+2.79}_{-1.26}$ & \\ \hline &
 \multicolumn{2}{c|}{\mbox{MRST-LO(98)~\cite{Martin:1998sq}}} & $ 0.44
 \pm 0.07 $ & $ 8.7 \pm 0.9$  & \\
 \raisebox{2ex}[-2ex]{BKL~\cite{Braaten:1999qk}} &
 \multicolumn{2}{c|}{\mbox{CTEQ5L~\cite{Lai:1999wy}}} & $0.39 \pm 0.07$
 & $6.6 \pm 0.7 $  & \raisebox{2ex}[-2ex]{3.4} \\[0.5mm] \hline\hline
 \multicolumn{6}{|c|}{\mbox{Parton shower radiation}} \\ \hline &
 \multicolumn{2}{c|}{\mbox{CTEQ2L~\cite{Tung:ua}}} & $0.96 \pm 0.15$
 & $1.32 \pm 0.21 $ & \\ {\rm S~\cite{Sanchis-Lozano:1999um}} &
 \multicolumn{2}{c|}{\mbox{MRS(D0)~\cite{Martin:1992zi}}} & $0.68 \pm
 0.16$ & $1.32 \pm 0.21$ & 3  \\ &
 \multicolumn{2}{c|}{\mbox{GRV-HO(94)~\cite{Gluck:1994uf}}} & $0.92 \pm
 0.11$ & $0.45 \pm 0.09$ & \\ \hline {\rm KK~\cite{Kniehl:1998qy}} &
 \multicolumn{2}{c|}{\mbox{CTEQ4M~\cite{Lai:1996mg}}} & $0.27 \pm 0.05$
 & $0.57 \pm 0.18 $ & 3.5 \\[0.5mm] \hline \hline
 \multicolumn{6}{|c|}{\mbox{$k_T$-smearing}} \\ \hline & & $ \langle k_T
 \rangle \mbox{[GeV]} $ & & & \\ & & 1 & $1.5\pm 0.22$ & $8.6\pm 2.1$ & \\
 \raisebox{2ex}[-2ex]{P~\cite{Petrelli:1999rh}} &
 \raisebox{2ex}[-2ex]{CTEQ4M~\cite{Lai:1996mg}} & 1.5 & $1.7 \pm 0.19$ & $
 4.5 \pm 1.5 $ & \raisebox{2ex}[-2ex]{3.5}\\ \hline & & 0.7 & $ 1.35 \pm
 0.30 $ & $ 8.46 \pm 1.41 $ & \\ \raisebox{2ex}[-2ex]{SMS~\cite{Sridhar:1998rt}}
 & \raisebox{2ex}[-2ex]{MRS(D$'_-$)~\cite{Martin:1992zi}} & 1 & $ 1.5
 \pm 0.29 $  & $ 7.05 \pm 1.17 $ & \raisebox{2ex}[-2ex]{3} \\[0.5mm]
 \hline\hline \multicolumn{6}{|c|}{\mbox{$k_T$-factorization}} \\
 \hline {\rm HKSST1~\cite{Hagler:2000eu}} &
 \multicolumn{2}{c|}{\mbox{KMS~\cite{Kwiecinski:1997ee}}} & $ \approx
 0.04 \pm 0.01 $  & $ \approx 6.5 \pm 0.5 $ & 5\\ \hline \hline
\end{tabular}
\renewcommand{\arraystretch}{1.0}
\end{center}

\medskip

\caption{The fractions $F_H$ of prompt $J/\psi$ mesons that are
         produced by the decay of higher charmonium states $H$ and the
         fraction $F_{J/\psi}$ that are produced directly.}
\label{tab:prodsec-Jpsifractions}
\begin{center}
\begin{tabular}{|c|c|} 
\hline \hline
$H$ & $F_H$ (in \%) \\ 
\hline
$J/\psi$        & $64 \pm 6$ \\ 
$\psi(2S)$      & $7 \pm 2$ to $15\pm 5$ \\ 
$\chi_c(1P)$    & $29.7 \pm 1.7(\hbox{stat.}) \pm 5.7(\hbox{sys.})$ \\
\hline \hline
\end{tabular}
\end{center}
\end{table}
%%%%%%%%%%%%%%%%%%%%%%%%%%%%%%%%%%%%%%%%%%%%%%%%%%%%%%%%%%%%%%%%%%%%%%%%%%%%%%%%%%

The CDF collaboration has measured the fraction of prompt $J/\psi$'s
that come from decays of $\psi(2S)$ and $\chi_c(1P)$ states and the
fractions that are produced directly \cite{Abe:1997yz}.  The CDF
measurements were made for $J/\psi$'s with transverse momentum $p_T >
4$~GeV and pseudo-rapidity $|\eta| < 0.6$.  The fractions, which are
defined in \Eqs~(\ref{eq:FJpsipsi2S}) and (\ref{eq:prod-chifrac}),
are given in \Table~\ref{tab:prodsec-Jpsifractions}.  The fraction of
$J/\psi$'s that are directly produced is approximately constant over
the range 5~GeV $< p_T <$ 15~GeV.  The fraction from decays of
$\psi(2S)$ increases from $(7 \pm 2)$\% at $p_T = 5$~GeV to $(15 \pm
5)$\% at $p_T = 15$~GeV.  The fraction from decays of $\chi_c(1P)$
seems to decrease slowly over this range of $p_T$.  Such variations
with $p_T$ are counter to the predictions of the colour-evaporation
model.

The CDF collaboration has also measured the ratio of the prompt $\chi_{c1}$
and $\chi_{c2}$ cross-sections at the Tevatron \cite{Affolder:2001ij}.
The measured value of the ratio $R_{\chi_c}$ defined in 
\Eq~(\ref{eq:prod-chirat}) is
\begin{equation}
R_{\chi_c} = 
1.04 \pm 0.29(\hbox{stat.}) \pm 0.12(\hbox{sys.}).
\label{eq:Rchi12Tev}
\end{equation}
The $\chi_{c2}$ and $\chi_{c1}$ were observed through their radiative decays
into a $J/\psi$ and a photon, which were required to have transverse
momenta exceeding 4~GeV and 1~GeV, respectively. The colour-evaporation
model predicts that this ratio should be 
close to the spin-counting ratio $3/5$, 
since the feeddown from the $\psi(2S)$ is small.
The NRQCD factorization fit to the prompt $\chi_c$ cross-section in the
region $p_T> 5$~GeV implies a ratio of $0.9\pm 0.2$
\cite{maltoni-chi-ratio}. The CDF result slightly favors the
NRQCD factorization prediction.

%%%%%%%%%%%%%%%%%%%%%%%%%%%%%%%%%%%%%%%%%%%%%%%%%%%%%%%%%%%%%%%%%%%%%%%%%%%%%%%%%%%%%%%%%%%%%
\begin{figure}[t]
\begin{center}
    \includegraphics[width=.48\textwidth]{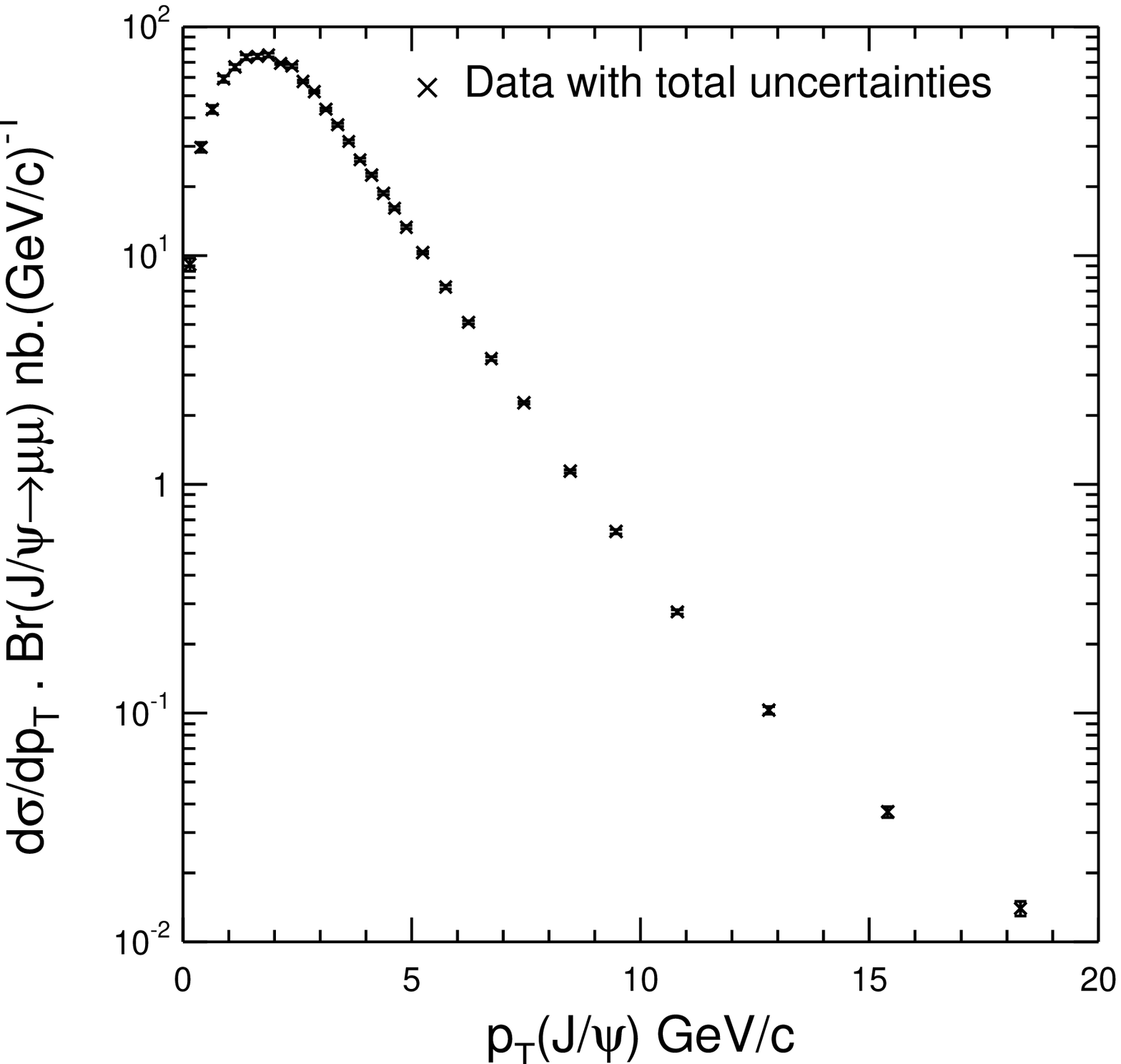}
    \hfill
    \includegraphics[width=.48\textwidth]{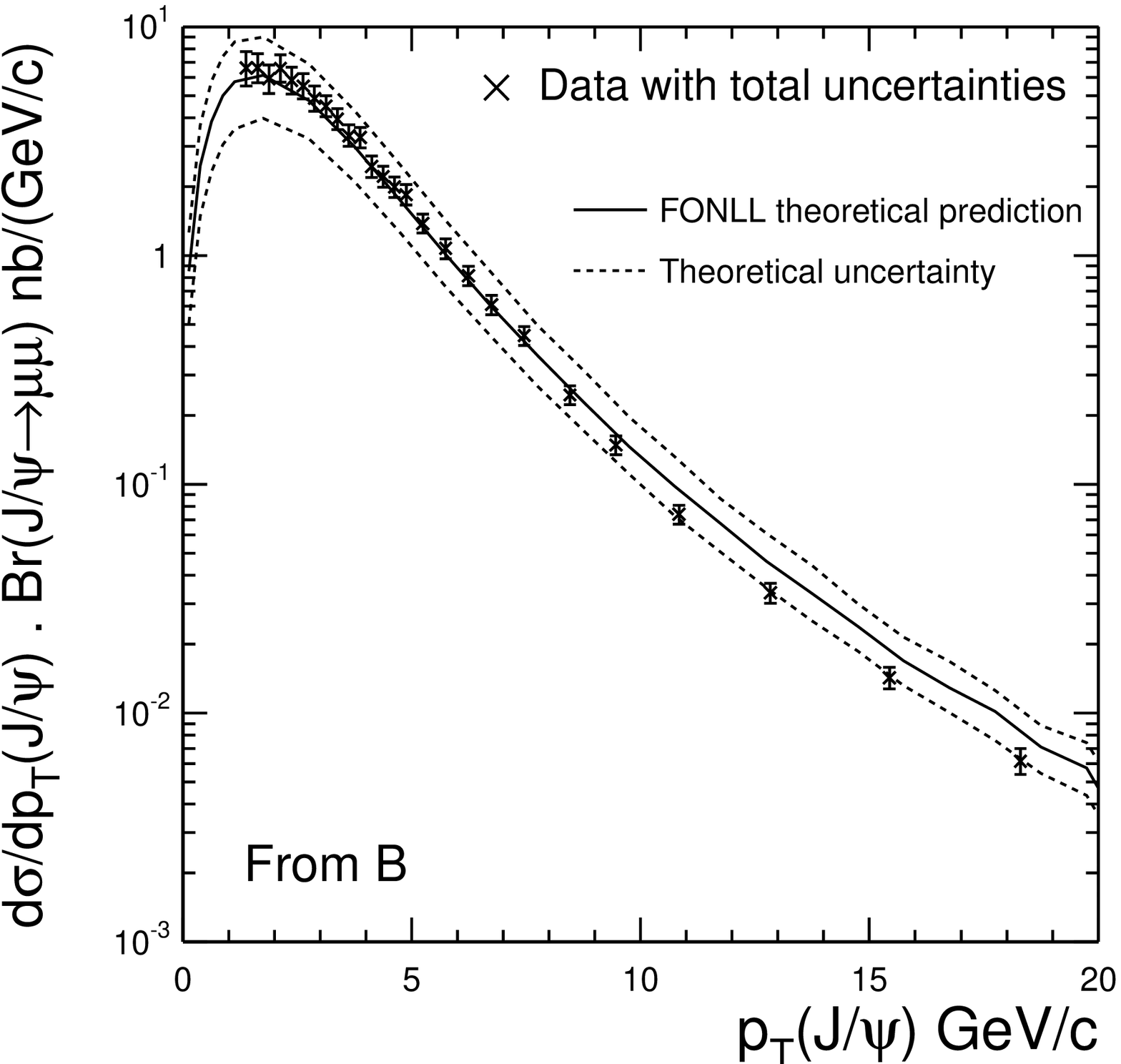}
\end{center}
 \caption[Differential inclusive cross-section for $p\bar{p}
          \rightarrow J/\psi X$]
         {Differential inclusive cross-section for $p\bar{ p}
          \rightarrow J/\psi X$ (left).  Differential cross-section
          distribution of $J/\psi$ events from $b$-hadron decay
          (right).  Both cross-sections are plotted as a function of
          the transverse momentum $p_T$ of the $J/\psi$ and are
          integrated over the rapidity range $|y(J/\psi)|<0.6$.
\label{fig:CDF_xsec} }
\end{figure}
%%%%%%%%%%%%%%%%%%%%%%%%%%%%%%%%%%%%%%%%%%%%%%%%%%%%%%%%%%%%%%%%%%%%%%%%%%%%%%%%%%%%%%%%%%%%%

Charmonium production data from Tevatron Run~II have recently become
available. Using a 39.7~pb$^{-1}$ data sample from Run~II, the CDF
Collaboration has measured the inclusive cross-section for $J/\psi$
production and subsequent decay into $\mu^+\mu^-$ \cite{CDF_LP03}.
The inclusive cross-section includes both prompt $J/\psi$'s and
$J/\psi$'s from decays of $b$-hadrons.  The inclusive differential
cross-section as a function of $p_T$ for rapidity $|y|<0.6$ has been
obtained down to zero transverse momentum and is shown in the left
panel of \Figure~\ref{fig:CDF_xsec}. The total integrated
cross-section for inclusive $J/\psi$ production in $p\bar{p}$
interactions at $\sqrt{s} = 1.96$~TeV is measured to be
\begin{equation}
\sigma[p\bar{p} \rightarrow J/\psi X, |y(J/\psi)|<0.6 ] \; 
= 4.08 \pm 0.02({\rm stat}) \pm 0.36(\rm {syst})~{\mu\rm b}.
\end{equation}
These new measurements await comparison
with updated theoretical calculations in the low $p_T$ region.

Using a sample of 4.7~pb$^{-1}$ of Run II data, the D0 collaboration
has verified that the $J/\psi$ cross-section is independent of the
rapidity of the $J/\psi$ for a rapidity range 0$~<|y|<~$2.  
This analysis has been performed for $p_T(J/\psi)
>5$~GeV and $p_T(J/\psi) > 8$~GeV~\cite{D0_LP03}.
The CDF and D0 collaborations have performed studies of forward 
 differential $J/\psi$ production cross-sections 
in the pseudo-rapidity regions
2.1$~<|\eta (J/\psi)| <~$2.6 and 2.5$~\leq |\eta (J/\psi)| \leq~$3.7, 
respectively, using their Run I data  \cite{CDF_Iforw, D0_Iforw}. 

  Using 39.7~pb$^{-1}$ of the Run II data, the CDF Collaboration has also
measured the differential cross-section as a function of $p_T$ and the
cross-section integrated over $p_T$ for the production of $b$-hadrons
that decay in the channel $H_b \rightarrow J/\psi X$ \cite{CDF_LP03}.
The differential cross-section multiplied by the branching fraction for
$J/\psi \to \mu^+ \mu^-$ is shown in the right panel of
\Figure~\ref{fig:CDF_xsec}. A recent QCD prediction that is based on a fixed
order (FO) calculation plus a resummation of next-to-leading order logs
(NLL) \cite{Cacciari:2004} is overlaid. The cross-section integrated
over $p_T$ was found to be
\begin{equation}
\sigma[p\bar{p} \rightarrow H_b X, p_T(J/\psi)> 1.25~ {\rm GeV}, 
|y(J/\psi)|<0.6] \;
= 28.4 \pm 0.4({\rm stat}) ^{+4.0}_{-3.8} (\rm {syst})~{\mu\rm b}.
\end{equation}
This measurement can be used to extract the total inclusive $b$-hadron
cross-section.

\subsection{Bottomonium cross-sections}
\label{sec:prodsec-tevatronbottom}

%%%%%%%%%%%%%%%%%%%%%%%%%%%%%%%%%%%%%%%%%%%%%%%%%%%%%%%%%%%%%%%%%%%%%%%
\begin{figure}[t]
\begin{center}
\includegraphics[width=.94\linewidth]{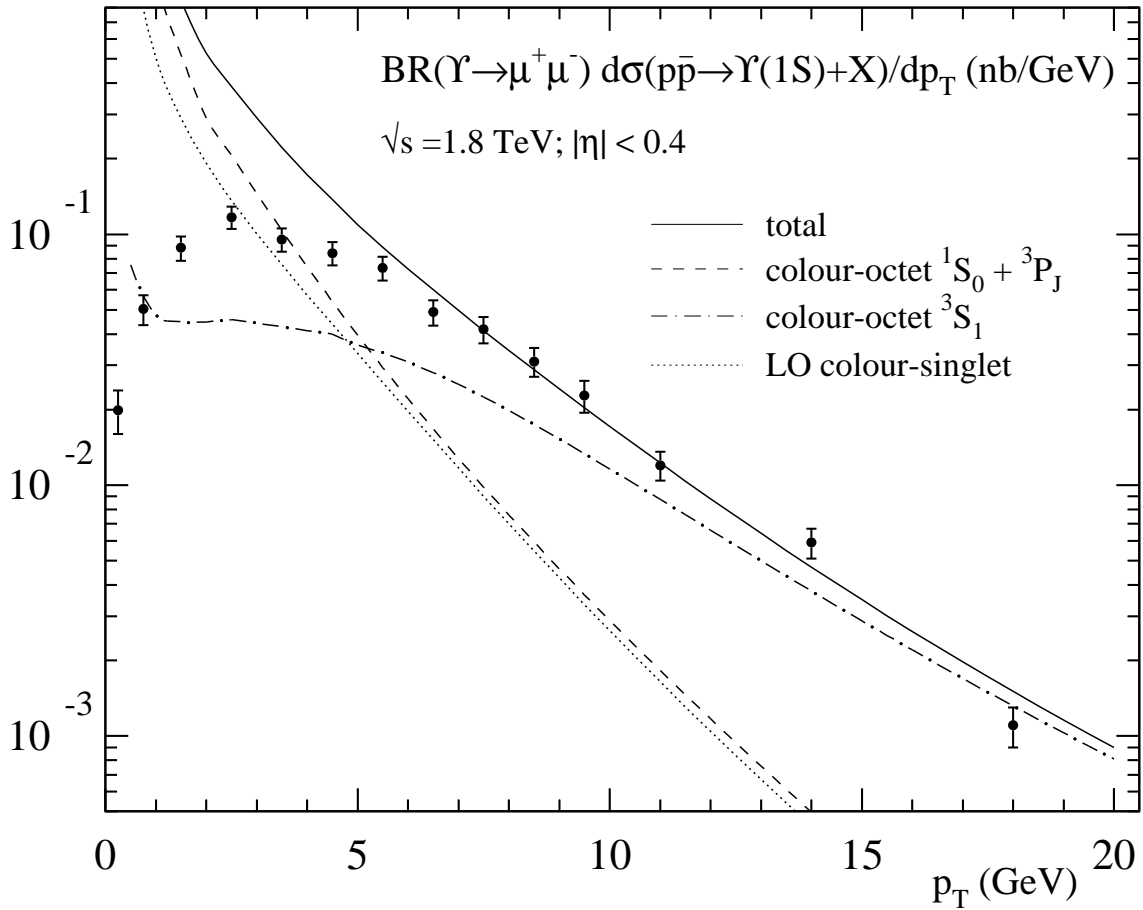}
\caption{Inclusive $\Upsilon(1S)$ cross-section at the Tevatron as a
function of $p_T$. The data points are the CDF measurements
\cite{Abe:1997jz}. The solid curve is the NRQCD factorization fit, and
the other curves are individual contributions to the
NRQCD factorization fit.  From
Ref.~\cite{Kramer:2001hh,LHC-workshop}.}
\label{fig:tevatron_ups1S}
\end{center}
\end{figure}
%%%%%%%%%%%%%%%%%%%%%%%%%%%%%%%%%%%%%%%%%%%%%%%%%%%%%%%%%%%%%%%%%%%%%%%

Using Run I data, the CDF Collaboration has reported inclusive
production cross-sections for the $\Upsilon(1S)$, $\Upsilon(2S)$ and
$\Upsilon(3S)$ states in the region 0 $< p_T < 20$~GeV
\cite{Acosta:2001gv}.  The rates of inclusive
production of the $\Upsilon(1S)$, $\Upsilon(2S)$ and $\Upsilon(3S)$
states for $p_T > 4$~GeV were found to be higher than the rates
predicted by CSM calculations by a factor of about five. 
Inclusion of colour-octet production mechanisms within the 
NRQCD framework can account for the observed cross-sections for 
$p_T > 8$~GeV~\cite{Cho:1995vh,Cho:1995ce,LHC-workshop,Braaten:2000cm}, 
as is shown for $\Upsilon(1S)$ production in \Figure~\ref{fig:tevatron_ups1S}. 
An accurate description of the $\Upsilon$ cross-section in the low-$p_T$
region requires NLO corrections and a resummation of multiple gluon 
radiation. A fit to the CDF data using a parton shower Monte Carlo 
to model the effects of multiple gluon emission has given much 
smaller values of the colour-octet matrix elements that are compatible 
with zero~\cite{Domenech:2000ri}.

%%%%%%%%%%%%%%%%%%%%%%%%%%%%%%%%%%%%%%%%%%%%%%%%%%%%%%%%%%%%%%%%%%%%%%%%%%%%%%%%%%%%%%%%%%%%%
\begin{figure}[t]
\begin{center}
\includegraphics[width=\textwidth]{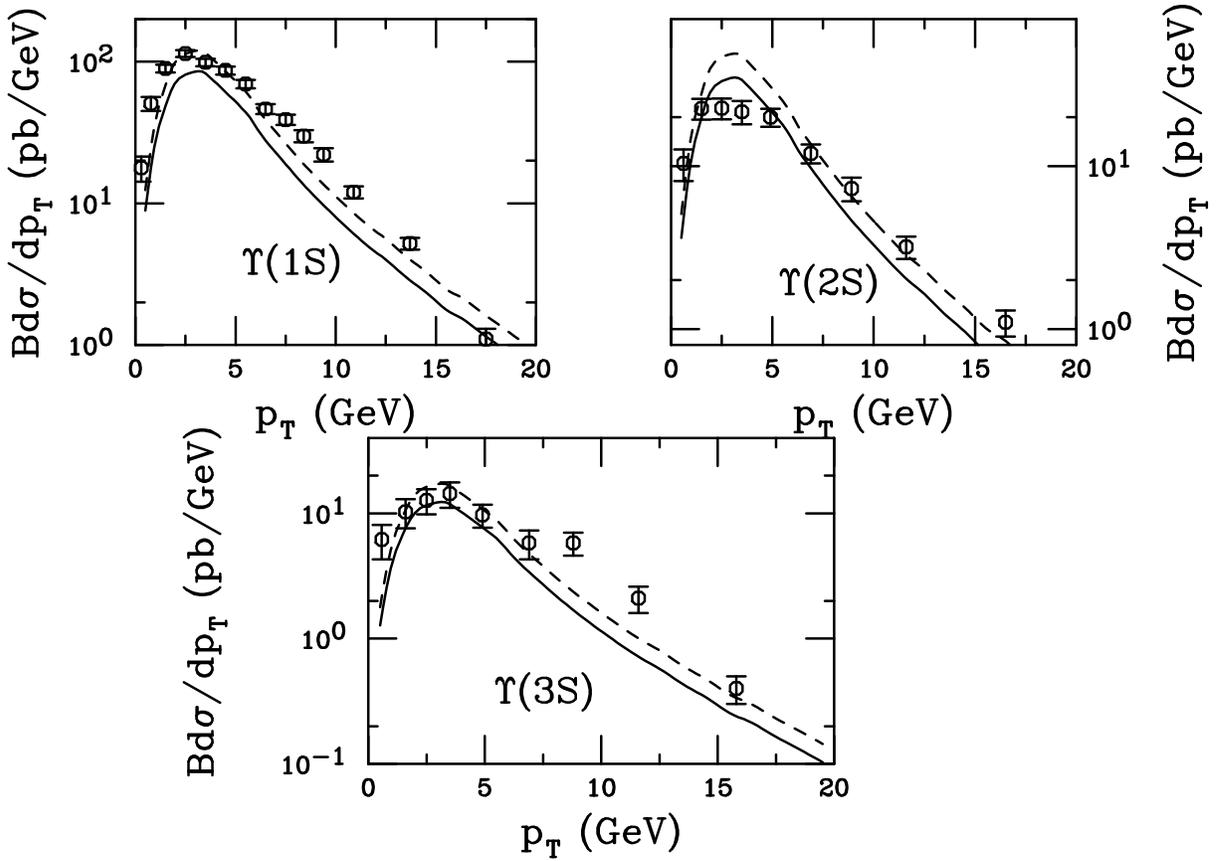}
\end{center}
\caption{Differential cross-sections for $\Upsilon(1S)$ (top left),
         $\Upsilon(2S)$ (top right), and $\Upsilon(3S)$ (bottom) at
         the Tevatron as a function of $p_T$. The data points are the
         CDF measurements \cite{Acosta:2001gv}. The solid curves are
         the CEM predictions at NLO with $\langle k_T^2\rangle = 3.0$~GeV$^2$, 
         using the first bottomonium parameter set in
         \Tables~\ref{tab:prodsec-qqbparams}. The dashed curves are
         multiplied by a $K$-factor of 1.4.}
\label{fig:upscdfptdep}
\end{figure}
%%%%%%%%%%%%%%%%%%%%%%%%%%%%%%%%%%%%%%%%%%%%%%%%%%%%%%%%%%%%%%%%%%%%%%%%%%%%%%%%%%%%%%%%%%%%%

The normalization and the shape of the bottomonium cross-sections at the
Tevatron can also be described reasonably well by the colour-evaporation
model (CEM).  The CEM predictions are compared
with the CDF data for $\Upsilon(1S)$, $\Upsilon(2S)$, and
$\Upsilon(3S)$ in \Figure~\ref{fig:upscdfptdep}. Most of the relevant
parameters can be fixed completely by fitting data from $pN$
collisions and by using measured branching fractions for bottomonium
decays.  The predictions of the CEM at NLO that are shown in
\Figure~\ref{fig:upscdfptdep} have been calculated using the NLO parameter
sets that are described in \Section~\ref{sec:prodsec-fixed-targetCEM}.  The
predicted cross-sections for $\Upsilon(1S)$ and $\Upsilon(3S)$
production are a little below the data; the normalizations can be
improved by multiplying the cross-sections by a K-factor of 1.4.  The
shapes have been brought into good agreement with the data by
including $k_T$ smearing, with $\langle k_T^2\rangle = 3.0$
GeV$^2$.  This value of $\langle k_T^2\rangle$ is a little larger
than the value $\langle k_T^2\rangle = 2.5$~GeV$^2$ that gives the
best fit to the charmonium cross-sections.  

%%%%%%%%%%%%%%%%%%%%%%%%%%%%%%%%%%%%%%%%%%%%%%%%%%%%%%%%%%%%%%%%%%%%%%%%%%%%%%%%%%%%%%%%%%%%%
\begin{figure}[t]
\begin{center}
\includegraphics[width=.32\linewidth]{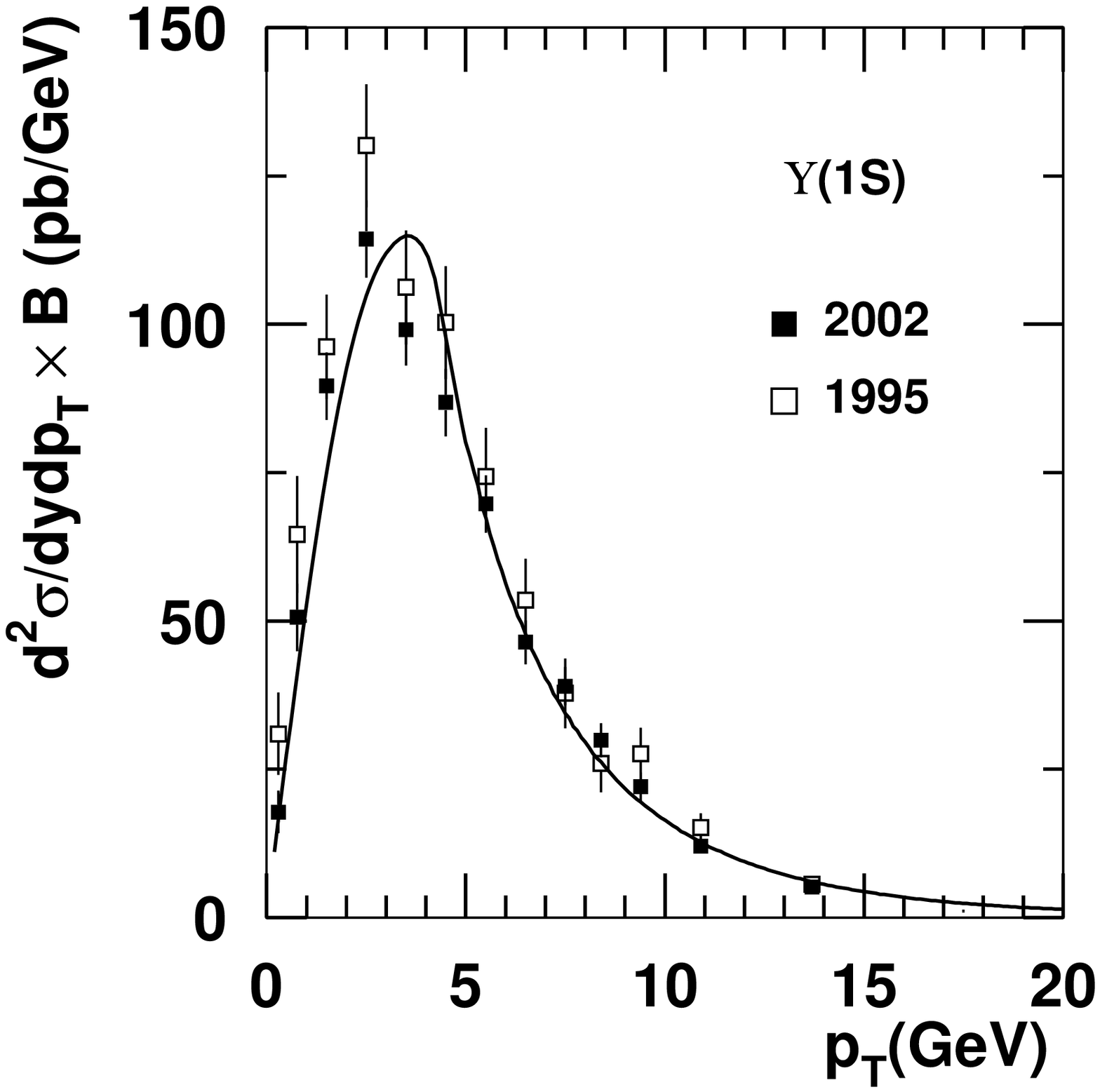}
\hfill
\includegraphics[width=.32\linewidth]{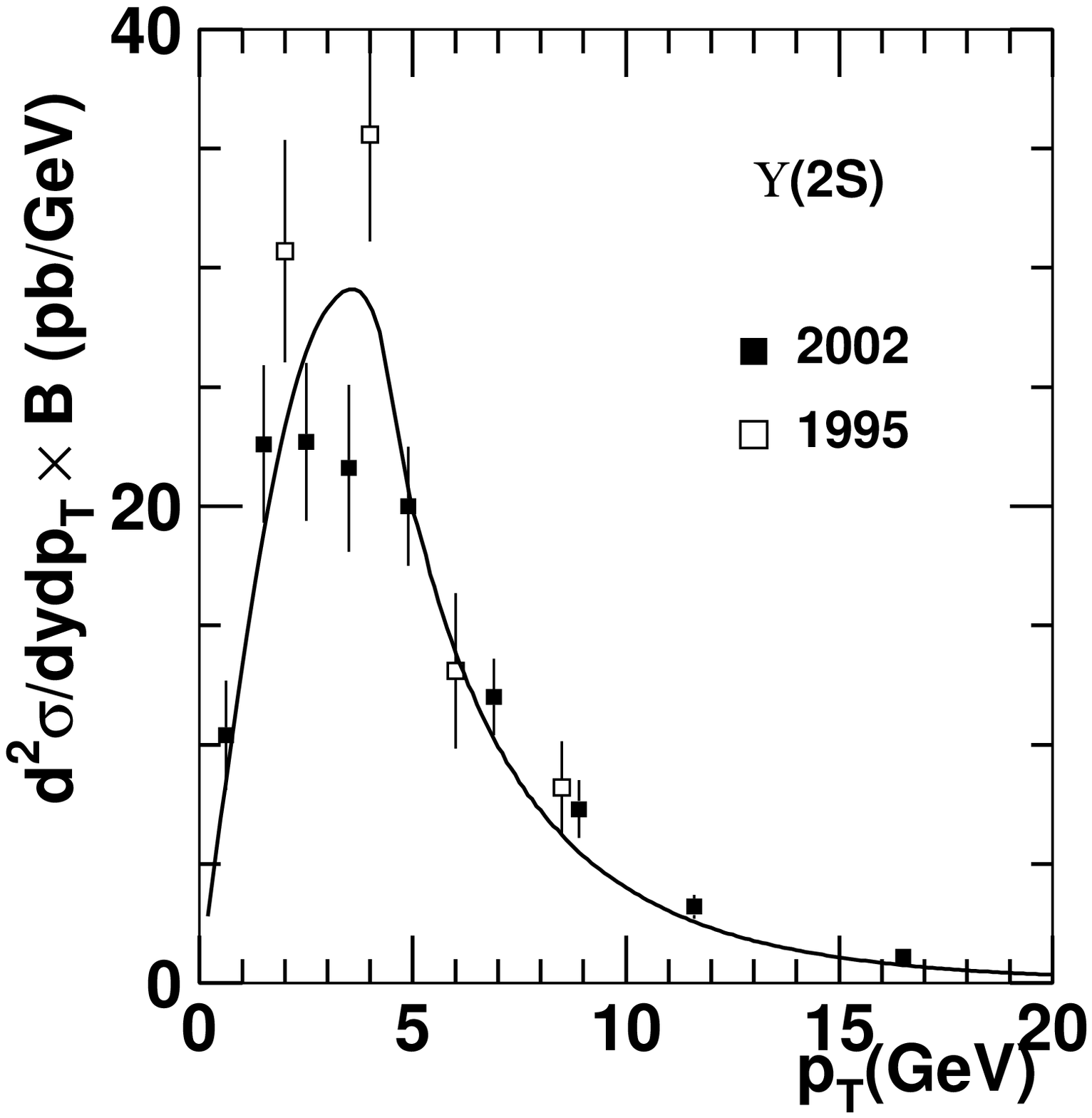}
\hfill
\includegraphics[width=.32\linewidth]{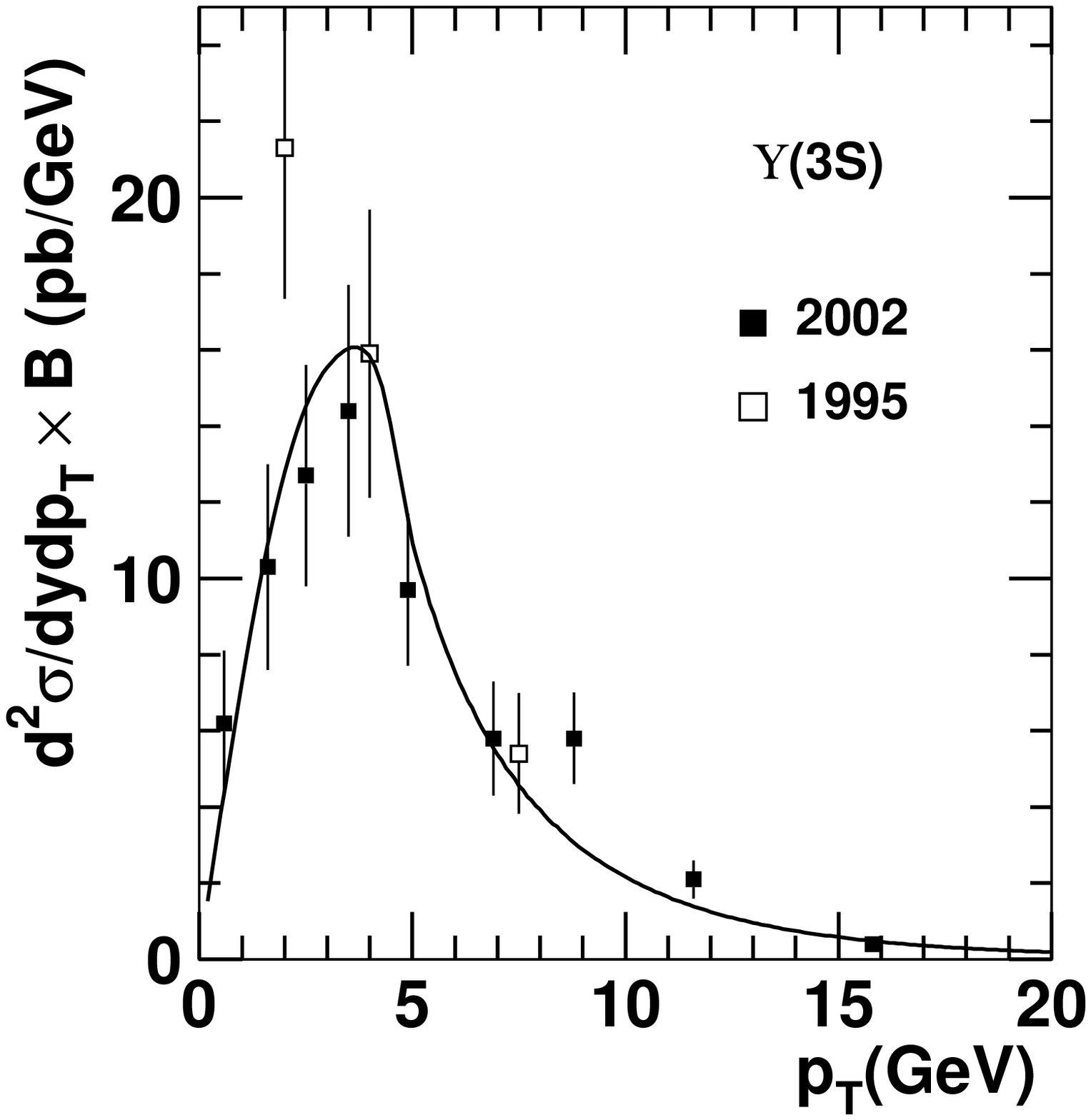}
\end{center}
\caption[Calculated differential cross-sections times leptonic branching
         fractions $B$]
        {Calculated differential cross-sections times leptonic
         branching fractions $B$, evaluated at $y = 0$, as functions
         of transverse momentum for hadronic production of (a)
         $\Upsilon(1S)$, (b) $\Upsilon(2S)$, and (c) $\Upsilon(3S)$
         \cite{Berger:2004cc}, along with CDF
         data~\cite{Abe:1995an,Acosta:2001gv} at $\sqrt S = 1.8$~TeV.
         The solid lines show the result of the full calculation. The
         1995 CDF cross-sections are multiplied a factor $0.88$.}
\label{fig:berger-qiu-wang}
\end{figure}
%%%%%%%%%%%%%%%%%%%%%%%%%%%%%%%%%%%%%%%%%%%%%%%%%%%%%%%%%%%%%%%%%%%%%%%%%%%%%%%%%%%%%%%%%%%%%

A recent calculation of the production cross-sections for the
$\Upsilon(1S)$, $\Upsilon(2S)$, and $\Upsilon(3S)$ at the Tevatron
combines a resummation of logarithms of $M_{\Upsilon}^2/p_T^2$ with a
calculation at leading order in $\alpha_s$ in what is, in essence, the
colour-evaporation model \cite{Berger:2004cc}.  The resummation of the
effects of multiple gluon emission in the CEM has some simplifications
that do not occur in the NRQCD factorization approach.  The results of
the calculation of Ref.~\cite{Berger:2004cc} are shown, along with CDF
data, in \Figure~\ref{fig:berger-qiu-wang}. The resummation of
logarithms of $M_{\Upsilon}^2/p_T^2$ allows the calculation to
reproduce the shape of the data at small $p_T$. The normalizations
have been adjusted to obtain the best fit to the data.  the best fit
to the data.

%%%%%%%%%%%%%%%%%%%%%%%%%%%%%%%%%%%%%%%%%%%%%%%%%%%%%%%%%%%%%%%%%%%%%%%%%%%%%%%%%%
\begin{table}[ht]
\caption{The fractions $F_H$ of $\Upsilon(1S)$ mesons that are
         produced by the decay of a higher bottomonium state $H$ and
         the fraction $F_{\Upsilon(1S)}$ that are produced directly.}
\label{tab:prodsec-Upsfractions}
\begin{center}
\begin{tabular}{|c|c|} 
\hline \hline
$H$ & $F_H$ (in \%) \\ 
\hline
$\Upsilon(1S)$  & $50.9 \pm 8.2(\hbox{stat.}) \pm 9.0(\hbox{sys.})$ \\ 
$\Upsilon(2S)$  & $10.7^{+7.7}_{-4.8}$ \\ 
$\Upsilon(3S)$  & $ 0.8^{+0.6}_{-0.4}$ \\ 
$\chi_b(1P)$    & $27.1 \pm 6.9(\hbox{stat.}) \pm 4.4(\hbox{sys.})$ \\
$\chi_b(2P)$    & $10.5 \pm 4.4(\hbox{stat.}) \pm 1.4(\hbox{sys.})$ \\
$\chi_b(3P)$    & $< 6$ \\
\hline \hline
\end{tabular}
\end{center}
\end{table}
%%%%%%%%%%%%%%%%%%%%%%%%%%%%%%%%%%%%%%%%%%%%%%%%%%%%%%%%%%%%%%%%%%%%%%%%%%%%%%%%%%

The CDF Collaboration has also reported the fractions of $\Upsilon(1S)$
mesons, for $p_T> 8$~GeV, that come from decays of $\chi_b(1P)$, 
$\chi_b(2P)$, $\chi_b(3P)$, $\Upsilon(2S)$, and $\Upsilon(3S)$ and the 
fraction that originate from direct production \cite{Affolder:2000nn}. 
The fractions from decays of $\Upsilon(nS)$ and for $\chi_b(nP)$ 
are defined by
\begin{eqnarray}
F_{\Upsilon(nS)} &=& 
{\rm Br}[\Upsilon(nS) \to \Upsilon(1S) + X] \; 
\frac{\sigma[\Upsilon(nS)]}{\sigma[\Upsilon(1S)]},
\label{eq:FUpsUps}
\\
F_{\chi_b(nP)} &=& \sum_{J=0}^3
{\rm Br}[\chi_{bJ}(nP) \to \Upsilon(1S) + X] \;
\frac{\sigma[\chi_{bJ}(nP)]}{\sigma[\Upsilon(1S)]}.
\label{eq:FUpschi}
\end{eqnarray}
The fraction of $\Upsilon(1S)$'s that are produced directly can be
denoted by $F_{\Upsilon(1S)}$.  The fractions are given in
\Table~\ref{tab:prodsec-Upsfractions}.

\subsection{Polarization} 
\label{sec:prodsec-tevatronpol}

%%%%%%%%%%%%%%%%%%%%%%%%%%%%%%%%%%%%%%%%%%%%%%%%%%%%%%%%%%%%%%%%%%%%%%%%%%%%%%%%%%%%%%%%%%%%%
\begin{figure}[t]
\begin{center}
\includegraphics[width=.46\textwidth]{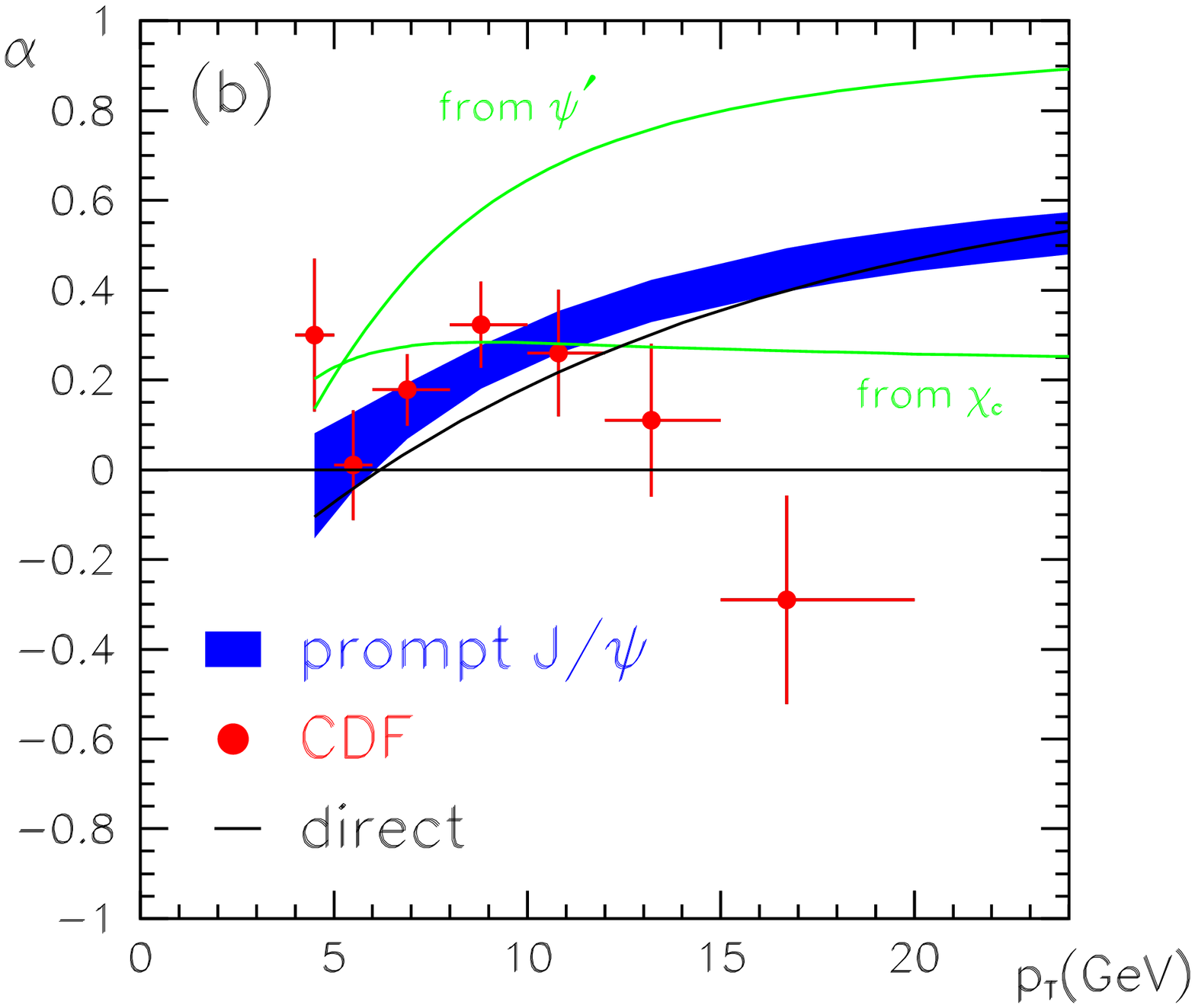}
\quad
\includegraphics[width=.46\textwidth]{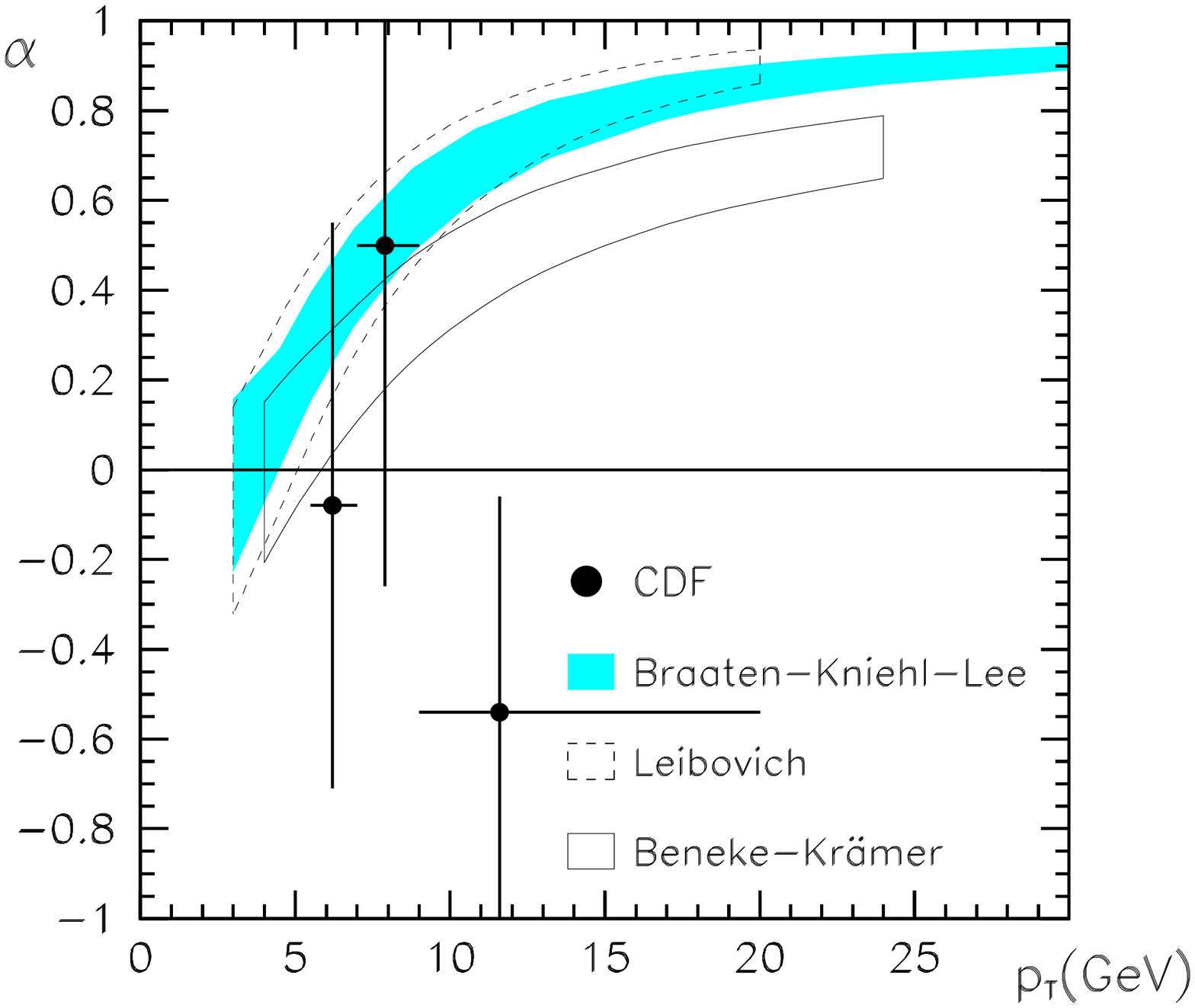} \\
\end{center}
\caption{Polarization variable $\alpha$ for prompt $J/\psi$ (left) and
         for prompt $\psi(2S)$ (right) at the Tevatron as a function
         of $p_T$.  The data points are the CDF measurements from Run
         I \cite{Affolder:2000nn}. In the left panel (prompt
         $J/\psi$), the band is the NRQCD factorization prediction of
         Ref.~\cite{Braaten:1999qk}, and the other curves are the
         values of $\alpha$ for individual components of the prompt
         $J/\psi$ signal. In the right panel (prompt $\psi(2S)$), the
         bands are various NRQCD factorization
         predictions~\cite{Beneke:1996yw,Leibovich:1996pa,Braaten:1999qk}.}
\label{fig:pol}
\end{figure}
%%%%%%%%%%%%%%%%%%%%%%%%%%%%%%%%%%%%%%%%%%%%%%%%%%%%%%%%%%%%%%%%%%%%%%%%%%%%%%%%%%%%%%%%%%%%%

The polarization of the quarkonium contains important information 
about the production mechanism. The polarization variable $\alpha$ 
for a $1^{--}$ state, such as $J/\psi$, $\psi(2S)$, or $\Upsilon(1S)$, 
is defined by \Eq~(\ref{eq:prod-alphadef}), where the angle $\theta$
is measured with respect to some polarization axis. 
At a hadron collider, a convenient choice of the polarization axis
is the direction of the boost vector from the quarkonium rest frame 
to the centre-of-momentum frame of the colliding hadrons.  

The NRQCD factorization approach gives a simple prediction for the 
polarization variable $\alpha$ at very large transverse momentum.
The production of a quarkonium with $p_T$ that
is much larger than the quarkonium mass is dominated by gluon
fragmentation\,---\,a process in which the quarkonium is formed in the
hadronization of a gluon that is created with even larger transverse
momentum. The NRQCD factorization approach predicts that the dominant
gluon-fragmentation process is gluon fragmentation into a $Q\bar Q$ pair
in a colour-octet ${}^3S_1$ state.  The fragmentation probability for
this process is of order $\alpha_s$, while the fragmentation
probabilities for all other processes are of order $\alpha_s^2$ or
higher. The NRQCD matrix element for this fragmentation process is
$\langle{\cal O}^H_8({}^3S_1)\rangle$. At large $p_T$, the fragmenting
gluon is nearly on its mass shell, and, so, is transversely polarized.
Furthermore, the velocity-scaling rules predict that the colour-octet
$Q\bar Q$ state retains the transverse polarization as it evolves into
an S-wave quarkonium state \cite{Cho:1994ih}, up to corrections of
relative order $v^2$. Radiative corrections and colour-singlet production
dilute the quarkonium polarization somewhat
\cite{Beneke:1995yb,Beneke:1996yw}. In the case of $J/\psi$ production,
feeddown from higher quarkonium states is also important
\cite{Braaten:1999qk}. Feeddown from $\chi_c$ states is about 30\% of
the $J/\psi$ sample and dilutes the polarization. Feeddown from the
$\psi(2S)$ is about 10\% of the $J/\psi$ sample and is largely
transversely polarized. Despite these various diluting effects, a
substantial polarization is expected at large $p_T$, and its detection
would be a ``smoking gun'' for the presence of the colour-octet production
mechanism. In contrast, the colour-evaporation model predicts zero
quarkonium polarization.

The CDF measurement of the $J/\psi$ polarization as a function of
$p_T$ \cite{Affolder:2000nn} is shown in the left panel of
\Figure~\ref{fig:pol}, along with the NRQCD factorization prediction
\cite{Braaten:1999qk}.  The observed $J/\psi$ polarization is in
agreement with the prediction, except for the highest $p_T$
bin. However, the prediction of increasing polarization with
increasing $p_T$ is not in evidence. The CDF data
\cite{Affolder:2000nn} and the NRQCD factorization prediction 
\cite{Beneke:1996yw,Leibovich:1996pa,Braaten:1999qk} for $\psi(2S)$
polarization are shown in the right panel of \Figure~\ref{fig:pol}.  The
theoretical analysis of $\psi(2S)$ polarization is simpler than for
the $J/\psi$, since feeddown does not play a r\^ole. However, the
experimental statistics are not as good for the $\psi(2S)$ as for
$J/\psi$. Again, the expectation of increasing polarization with
increasing $p_T$ is not confirmed.

Because the polarization depends on ratios of matrix elements, some of
the theoretical uncertainties are reduced compared with those in the
production cross-section. The polarization is probably not strongly
affected by multiple gluon emission or $K$-factors. Uncertainties
from contributions of higher-order in $\alpha_s$ could conceivably
change the rates for the various spin states by a factor of two. 
Therefore, it is important to carry out the NLO calculation, but that
calculation is very difficult technically and is computing intensive. 
order-$v^2$ corrections to parton fragmentation to quarkonium can be
quite large. Bodwin and Lee \cite{Bodwin:2003wh} have found that the $v^2$
corrections to gluon fragmentation to $J/\psi$ are about $+70\%$ for
the colour-singlet channel and $-50\%$ for the colour-octet channel.
The colour-singlet correction shifts $\alpha$ down by about 10\% at the
largest $p_T$. Since the colour-octet matrix element is fit to Tevatron
data, the $v^2$ correction merely changes the size of the matrix element
and has no immediate effect on the theoretical prediction. An additional
theoretical uncertainty comes from the presence of order-$v^2$ spin-flip
processes in the evolution of the $Q\bar Q$ pair into the quarkonium.
It could turn out that spin-flip contributions are large, either
because their velocity-scaling power laws happen to have large
coefficients or because, as has been suggested in
Refs.~\cite{Beneke:1997av,Brambilla:1999xf,Fleming:2000ib,
Sanchis-Lozano:2001rr,Brambilla:2002nu}, the
velocity scaling rules themselves need to be modified. Then spin-flip
contributions could significantly dilute the $J/\psi$ polarization.
Nevertheless, it is is difficult to see how there could not be
substantial polarization in $J/\psi$ or $\psi(2S)$ production for
$p_T>4m_c$.\footnote{It has been argued that re-scattering interactions
between the intermediate charm-quark pair and a co-moving colour field
could yield unpolarized quarkonium~\cite{Marchal:2000wd,Maul:2001fw}.
The theoretical analysis of these effects, however, relies on several
simplifying assumptions, and further work is needed to establish the
existence of re-scattering corrections in charmonium hadroproduction at
large $p_T$.}

%%%%%%%%%%%%%%%%%%%%%%%%%%%%%%%%%%%%%%%%%%%%%%%%%%%%%%%%%%%%%%%%%%%%%%%%%%%%%%%%%%%%%%%%%%%%%
\begin{figure}[t]
\begin{center}
\includegraphics[width=.85\linewidth]{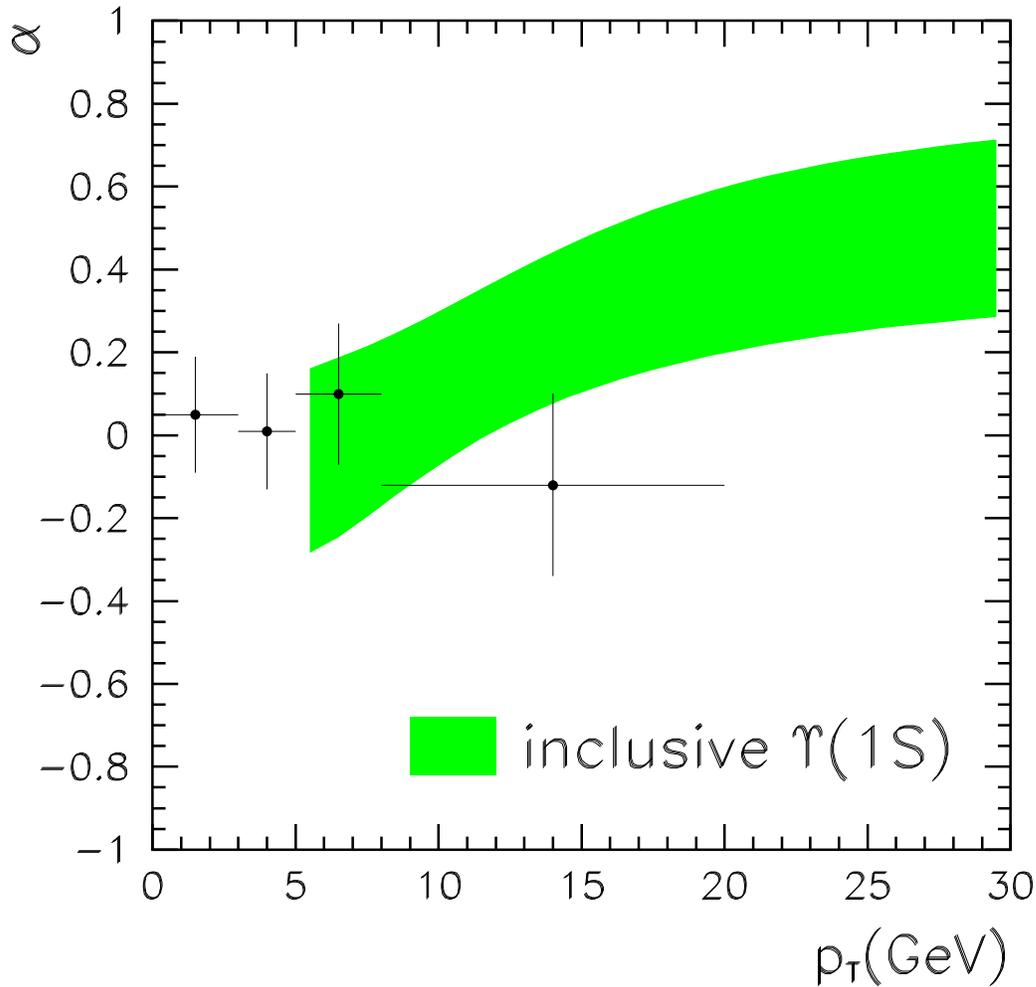}
\caption{ Polarization variable $\alpha$ for inclusive $\Upsilon(1S)$
production at the Tevatron as a function of $p_T$. The data points
are the CDF measurements from Run I \cite{Acosta:2001gv}.
The theoretical band represents 
the NRQCD factorization prediction \cite{Braaten:2000gw}.}
\label{fig:upsilon-pol}
\end{center}
\end{figure}
%%%%%%%%%%%%%%%%%%%%%%%%%%%%%%%%%%%%%%%%%%%%%%%%%%%%%%%%%%%%%%%%%%%%%%%%%%%%%%%%%%%%%%%%%%%%%

The CDF data for $\Upsilon$ polarization is shown in 
\Figure~\ref{fig:upsilon-pol}, along with the NRQCD factorization prediction.
Averaging over a range of $p_T$, the CDF
Collaboration finds $\alpha=-0.06\pm 0.20$ for $1\hbox{
GeV}<p_T<20\hbox{~GeV}$ \cite{Cropp:1999ub,Papadimitriou:2001bb}, which
is consistent with the NRQCD factorization prediction
\cite{Braaten:2000gw}.
In comparison with the prediction for $J/\psi$ polarization, the
prediction for $\Upsilon$ polarization has smaller $v$-expansion
uncertainties. However, in the case of $\Upsilon$ production, the
fragmentation mechanism does not dominate until relatively large values
of $p_T$ are reached, and, hence, the transverse
polarization is predicted to be small for $p_T$ below about 10~GeV. 
Unfortunately, the current Tevatron data sets run out of statistics 
in the high-$p_T$ region.

\subsection{Prospects for the Tevatron Run II}

Run~II at the Tevatron will provide a substantial increase in luminosity
and will allow the collider experiments to determine the $J/\psi$,
$\psi(2S)$ and $\chi_c$ cross-sections more precisely and at larger
values of $p_t$. An accurate measurement of the $J/\psi$ and $\psi(2S)$
polarization at large transverse momentum will be the most crucial test
of NRQCD factorization. In addition, improved data on the $J/\psi$ and
$\psi(2S)$ cross-sections will help to reduce some of the ambiguities in
extracting the colour-octet matrix elements.
 
With increased statistics it might be possible to access other
charmonium states such as the $\eta_c(nS)$ or the $h_c(nP)$.  heavy-quark
spin symmetry provides approximate relations between the
nonperturbative matrix elements that describe spin-singlet and
spin-triplet states. The matrix elements for $\eta_c(nS)$ are related to
those for $\psi(nS)$, while the leading matrix elements for $h_c(nP)$
can be obtained from those for $\chi_c(nP)$. [See
\Eqs~(\ref{eq:s1-symmetry}--\ref{eq:p8-symmetry}).] Within NRQCD, the rates for
$\eta(nS)$ and $h(nP)$ production can thus be predicted unambiguously in
terms of the nonperturbative matrix elements that describe the $J/\psi$,
$\psi(2S)$ and $\chi_c$ production cross-sections. A comparison of the
various charmonium production rates would therefore provide a stringent
test of NRQCD factorization and the heavy-quark spin symmetry.  The
cross-sections for producing the $\eta_c$ and the $h_c$ at Run~II of the
Tevatron are large~\cite{Mathews:1998nk,Sridhar:1996vd}, but the
acceptances and efficiencies for observing the decay modes on which one
can trigger are, in general, small, and detailed experimental studies
are needed to quantify the prospects.  Other charmonium processes that
have been studied in the literature include the production of D-wave
states~\cite{Qiao:1997wb}, $J/\psi$ production in association with
photons~\cite{Kim:1997bb,Mathews:1999ye}, and double gluon fragmentation
to $J/\psi$ pairs~\cite{Barger:1996vx}.
 
The larger statistics expected at Run~II of the Tevatron will also allow
the collider experiments to improve the measurements of the bottomonium
cross-sections. As yet undiscovered states, such as the $\eta_b(1S)$,
could be detected, for example, in the decay $\eta_b \to J/\psi +
J/\psi$~\cite{Braaten:2000cm} or in the decay $\eta_b \to D^* +
D^{(*)}$~\cite{Maltoni:2004hv}, and the associated production of
$\Upsilon$ and electroweak bosons might be
accessible~\cite{Braaten:1999th}. If sufficient statistics can be
accumulated, the onset of transverse $\Upsilon(nS)$ polarization may be
visible at $p_{T,\Upsilon}\gsim 15$~GeV.

\section{Quarkonium production in fixed-target experiments}
\label{sec:prodsec-fixed-target}

\subsection{Cross-sections}
\label{sec:prodsec-fixed-targetxsec}

\begin{figure}[t]
\begin{center}
$ \begin{array}{cc}
\includegraphics[width=7.5cm]{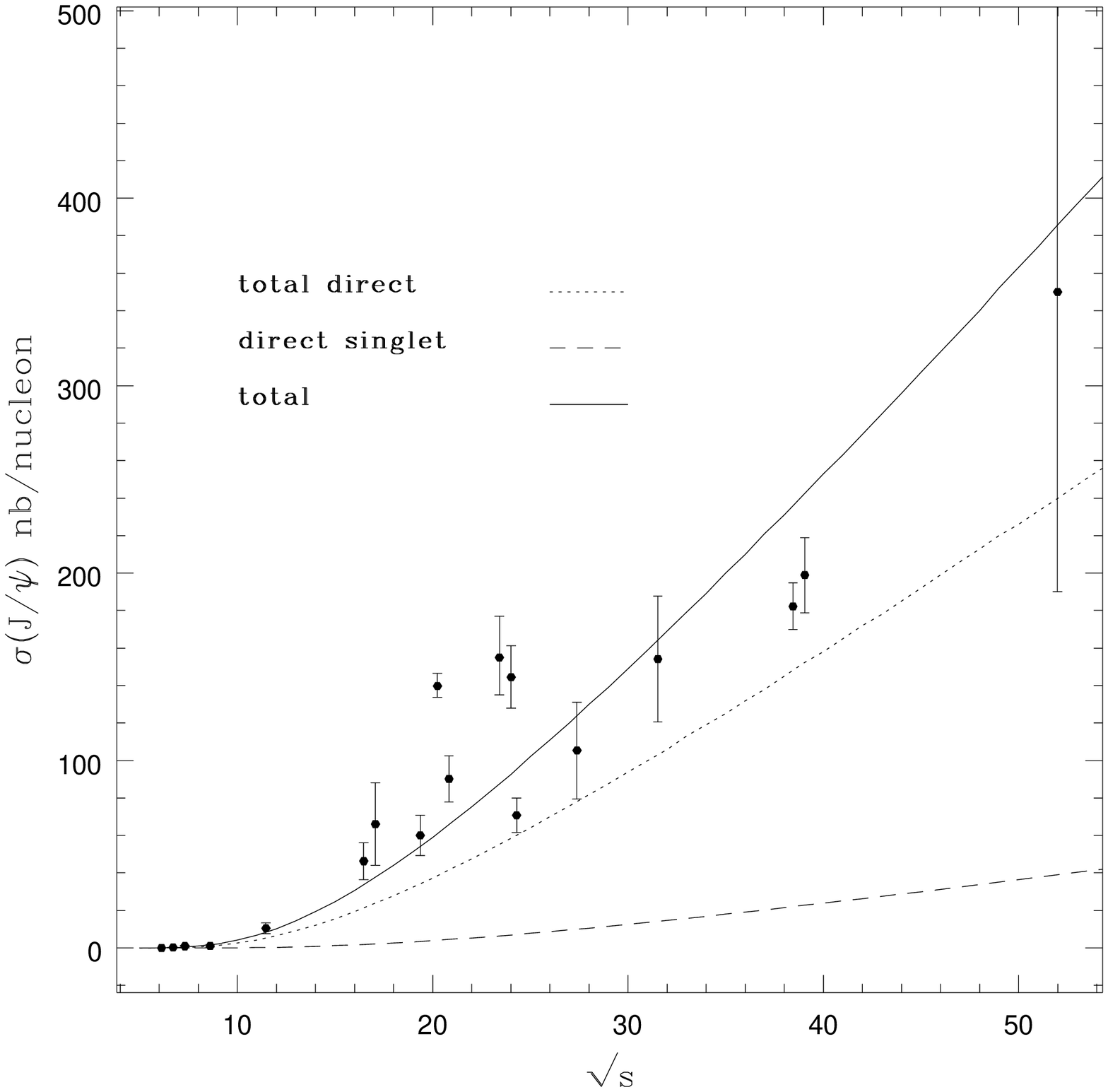}
&
\includegraphics[width=7.5cm]{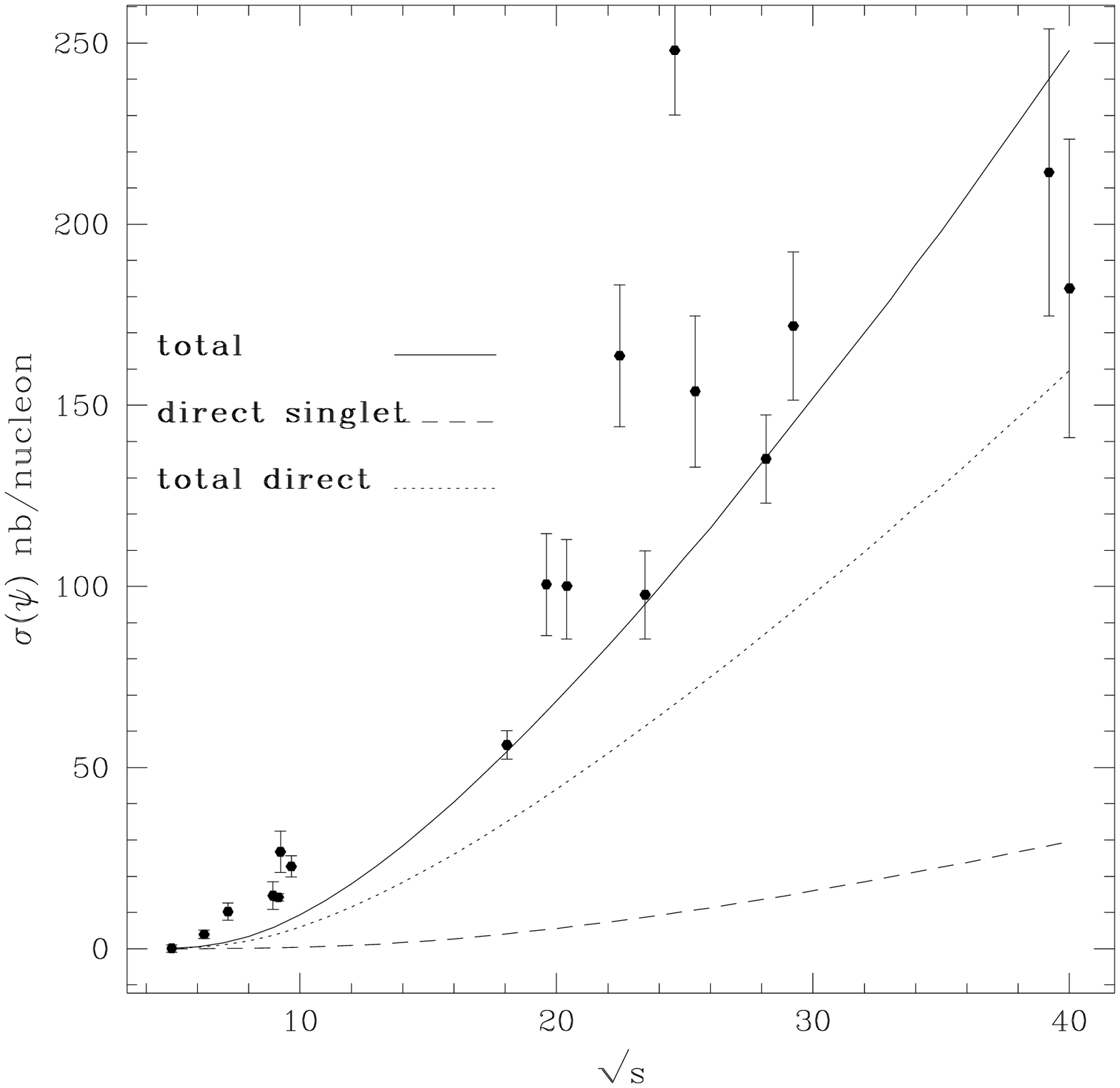}
\end{array} $
\caption{
Forward cross-section ($x_F>0$) for $J/\psi$ production 
in $pN$ collisions (left) and $\pi N$ collisions (right).  
The curves are the CSM predictions for direct $J/\psi$ (dashed lines), 
the NRQCD factorization predictions for direct $J/\psi$ with 
$M_7^{J/\psi}=3.0\times 10^{-2}$~GeV$^3$ (dotted lines), 
and the inclusive cross-sections for $J/\psi$ including 
radiative feed-down from $\chi_{cJ}$ and $\psi(2S)$ 
(solid lines).  From Ref.~\cite{Beneke:1996tk}.
}
\label{fig:fixed-target-psi}
\end{center}
\end{figure}

\begin{figure}[t]
\begin{center}
$ \begin{array}{cc}
\includegraphics[width=7.5cm]{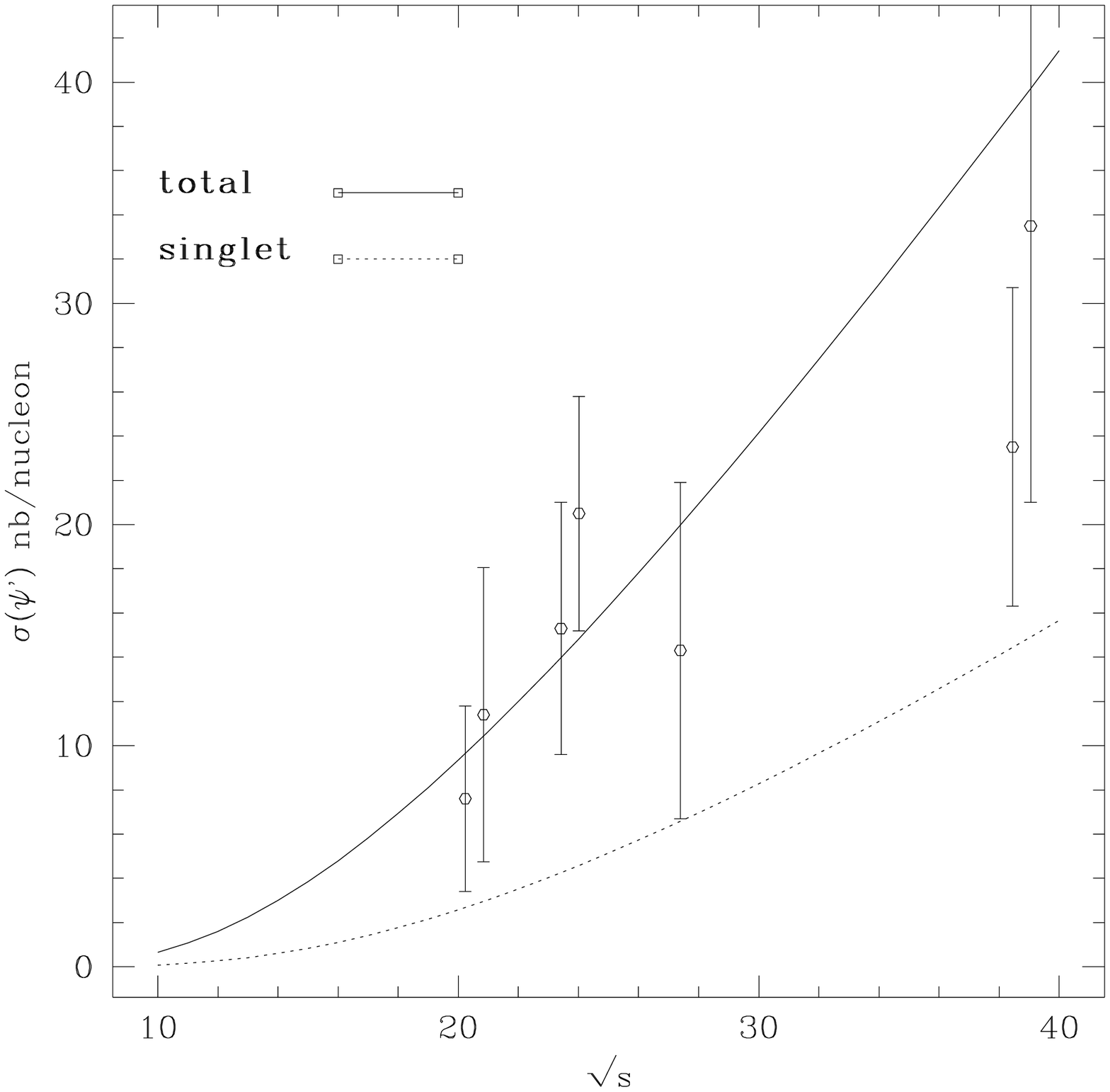} 
&
\includegraphics[width=7.5cm]{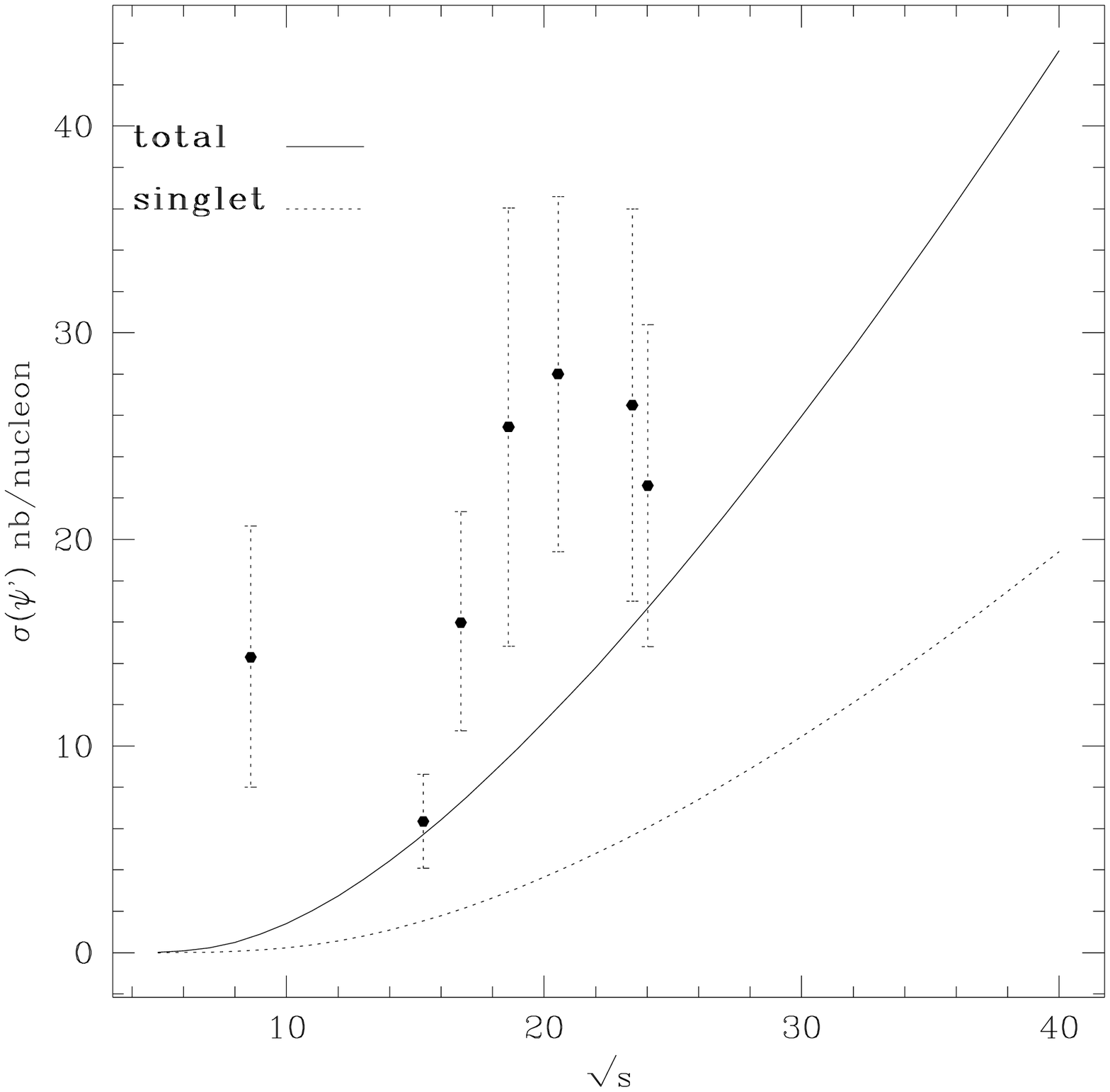}
\end{array} $
\caption{
Forward cross-section ($x_F>0$) for $\psi(2S)$ production 
in $pN$ collisions (left) and $\pi N$ collisions (right).  
The curves are the CSM predictions (dotted lines) 
and the NRQCD factorization predictions with 
$M_7^{\psi(2S)}=5.2\times 10^{-3}$~GeV$^3$ (solid lines).  
From Ref.~\cite{Beneke:1996tk}.
}
\label{fig:fixed-target-pri}
\end{center}
\end{figure}

Several collaborations have made predictions for fixed-target
quarkonium production within the NRQCD factorization formalism
\cite{Beneke:1996tk,Tang:1996rm,Gupta:1996ut}. The predictions of
Ref.~\cite{Beneke:1996tk} for $J/\psi$ and $\psi(2S)$ production in
$pN$ collisions are shown, along with the experimental data, in the
left panels of \Figures~\ref{fig:fixed-target-psi} and
\ref{fig:fixed-target-pri}. The calculation is at leading-order in
$\alpha_s$ and uses the standard truncation in $v$ that is described
in \Section~\ref{sec:prodsec-nrqcdfact}.  The data are from the
compilation in Ref.~\cite{Schuler:1994hy}, with additional results
from Refs.~\cite{Tzamarias:1990ij,Schub:1995pu,Alexopoulos:1995dt}. In
the case of $pN$ production of $J/\psi$, the data clearly require a
colour-octet contribution, in addition to a colour-singlet
contribution.  In the case of $\psi(2S)$ production, it is less clear
that a colour-octet contribution is essential. One should keep in mind
that the colour-singlet contribution is quite uncertain, owing to
uncertainties in the values of $m_c$ and the renormalization scale
\cite{Beneke:1997av}.  One can reduce these uncertainties by
considering the ratio of the cross-sections for direct and inclusive
$J/\psi$ production, which is predicted to be approximately 0.6 in the
NRQCD factorization approach and approximately 0.2 in the
colour-singlet model \cite{Beneke:1997av}.  Clearly, experiment favors
the NRQCD factorization prediction. However, the prediction for the
ratio depends on our knowledge of feed-down from $\chi_c$ states, and,
as we shall see, NRQCD factorization predictions for $\chi_c$
production in fixed-target experiments are not in good agreement with
the data.

%%%%%%%%%%%%%%%%%%%%%%%%%%%%%%%%%%%%%%%%%%%%%%%%%%%%%%%%%%%%%%%%%%%%%%%%%%%%%%%%%%%%%%%%%%%
%%%%%%%%%%%%%%%%%%%%%%%%%%%%%%%%%%%%%%%%%%%%%%%%%%%%%%%%%%%%%%%%%%%%%%%%%%%%%%%%%%%%%%%%%%%

%%%%%%%%%%%%%%%%%%%%%%%%%%%%%%%%%%%%%%%%%%%%%%%%%%%%%%%%%%%%%%%%%%%%%%%%%%%%%%%%%%%%%%%%%%%
%%%%%%%%%%%%%%%%%%%%%%%%%%%%%%%%%%%%%%%%%%%%%%%%%%%%%%%%%%%%%%%%%%%%%%%%%%%%%%%%%%%%%%%%%%%

In fixed-target production of $J/\psi$ and $\psi(2S)$ at leading order
in $\alpha_s$ (LO), the relevant production matrix elements are
$\langle{\cal O}^H_8({}^3S_1)\rangle$, $\langle{\cal
O}^H_8({}^1S_0)\rangle$, and $\langle{\cal O}^H_8({}^3P_0)\rangle$,
but the cross-section is sensitive only to the linear combination
$M^H_k$ defined in (\ref{eq:prod-lincomb}) with $k\approx 7$.  The
fits of the LO predictions for $J/\psi$ and $\psi(2S)$ production in
$pN$ collisions \cite{Beneke:1996tk} yield $M_7^{J/\psi}=3.0\times
10^{-2}\hbox{~GeV}^3$ and $M_7^{\psi(2S)}=5.2\times
10^{-3}\hbox{~GeV}^3$.  The corrections at next-to-leading order in
$\alpha_s$ (NLO) give a large $K$-factor in the colour-octet
contributions \cite{Petrelli:1997ge}. A fit to the data using the NLO
result for the colour-octet contributions gives
$M_{6.4}^{J/\psi}=1.8\times 10^{-2}\hbox{~GeV}^3$ and
$M_{6.4}^{\psi(2S)}=2.6\times 10^{-3}\hbox{~GeV}^3$
\cite{Maltoni:2000km}.  The NLO value of $M_{6.4}^{J/\psi}$ is about a
factor $2$ smaller than the LO value of $M_7^{J/\psi}$. Note that the
NLO fit uses CTEQ4M \cite{Lai:1996mg} parton distributions, while the
LO fit uses the CTEQ3L \cite{Lai:1994bb} parton distributions. The LO
result for $M^{J/\psi}$ is somewhat smaller than the LO result from
the Tevatron, and the NLO result for $M^{J/\psi}$ is somewhat larger
than the parton-shower result from the Tevatron.  However, given the
large uncertainties in these quantities, the agreement is
reasonable. It should also be remembered that the Tevatron
cross-sections are sensitive to $M_k^H$ with $k \approx 3$ rather than
$k \approx 7$, and, so, comparisons are somewhat uncertain. Attempts
to constrain this uncertainty are hampered by the fact that the
$\overline{\rm MS}$ matrix elements need not be positive. One can also
question whether hard-scattering factorization holds for the total
cross-section, which is dominated by small
$p_T$-contributions. Furthermore, kinematic corrections from the
difference between $2m$ and the quarkonium mass may be large.

The predictions of Ref.~\cite{Beneke:1996tk} for $J/\psi$ and
$\psi(2S)$ production in $\pi N$ collisions are shown, along with the
experimental data, in the right panels of 
\Figures~\ref{fig:fixed-target-psi} and \ref{fig:fixed-target-pri}. 
The calculation is at leading-order in $\alpha_s$ and uses the
standard truncation in $v$ that is described in
\Section~\ref{sec:prodsec-nrqcdfact}. Again, the data are from the compilation
in Ref.~\cite{Schuler:1994hy}, with additional results from
Refs.~\cite{Tzamarias:1990ij,Schub:1995pu,Alexopoulos:1995dt}. In the
NRQCD predictions in \Figures~\ref{fig:fixed-target-psi} and
\ref{fig:fixed-target-pri}, the values of $M_7$ that are used are the
ones that were obtained from the fits to the $pN$ production data. The
$\pi N$ production data clearly show an excess over these predictions
that cannot be accounted for by the colour-octet contributions. This
discrepancy has been discussed extensively in
Ref.~\cite{Schuler:1994hy}, and it may reflect our lack of knowledge
of the gluon distribution in the pion or the presence of different
higher-twist effects in the proton and the pion. Such higher-twist
effects are not accounted for in the standard NRQCD factorization
formulas, which are based on leading-twist hard-scattering
factorization.

Some of the largest uncertainties in the predictions cancel out
if we consider ratios of cross-sections.  
The uncertainties in the NRQCD factorization predictions 
can still be very large. They arise
from uncertainties in the colour-octet matrix elements, uncalculated
corrections of higher order in $v$ and $\alpha_s$, and uncertainties
from the choices of renormalization and factorization scales. In
addition, one can question whether hard-scattering factorization holds
for the cross-section integrated over $p_T$.

%%%%%%%%%%%%%%%%%%%%%%%%%%%%%%%%%%%%%%%%%%%%%%%%%%%%%%%%%%%%%%%%%%%%%%%%%%%%%%%%%%%%%%%%%%%
\begin{table}[ht]
\caption[Experimental results for the ratio $R_\psi$]
        {Experimental results for the ratio $R_\psi$ of the inclusive
         cross-sections for $\psi(2S)$ and $J/\psi$ production.}
\label{tab:psi} 
\addtolength{\arraycolsep}{0.2cm}
\renewcommand{\arraystretch}{1.25} 
\begin{center}
\begin{tabular}{|c|ccc|} 
\hline
\hline
\mbox{Experiment} & \mbox{beam/target} & $\sqrt{s}/\mbox{GeV}$ & 
$R_\psi$ \\
\hline 
\mbox{E537}~\cite{Tzamarias:1990ij}& $\bar p\mbox{W}$ & $15.3$ 
& $0.185\pm 0.0925$\\ 
\mbox{E705}~\cite{Arenton:pw}& $p\mbox{Li}$ & $23.7$ & $0.14\pm 0.02 \pm 
0.004\pm 0.02$\\ 
\mbox{E705}~\cite{Arenton:pw}& $\bar p\mbox{Li}$ & $23.7$ & $0.25\pm 
0.22 \pm 0.007\pm 0.04$\\ 
\mbox{E771}~\cite{Alexopoulos:1995dt}& $p\mbox{Si}$ & $38.8$
& $0.14\pm 0.02$ \\ 
\mbox{HERA-B}~\cite{Spengler:2004gr} & p\mbox{(C, W)} & $41.5$ 
&$0.13\pm 0.02 $\\ 
\hline
\mbox{E537}~\cite{Tzamarias:1990ij}& $\pi^-\mbox{W}$ & $15.3$ 
& $0.2405\pm 0.0650$\\ 
\mbox{E673}~\cite{Hahn:tz}    & $\pi\mbox{Be}$ &$20.6$&$0.20\pm 0.09$\\ 
\mbox{E705}~\cite{Arenton:pw} & $\pi^+\mbox{Li}$ &$23.7$ &
$0.14\pm 0.02\pm 0.004\pm 0.02$\\ 
\mbox{E705}~\cite{Arenton:pw} & $\pi^-\mbox{Li}$ &$23.7$ &
$0.12\pm 0.03\pm 0.03\pm 0.02$\\ 
\mbox{E672/706}~\cite{Gribushin:1995rt} & $\pi^-\mbox{Be}$ &$31.1$
&$0.15\pm 0.03\pm 0.02$\\ 
\hline
\hline 
\end{tabular}
\end{center}
\end{table} 
%%%%%%%%%%%%%%%%%%%%%%%%%%%%%%%%%%%%%%%%%%%%%%%%%%%%%%%%%%%%%%%%%%%%%%%%%%%%%%%%%%%%%%%%%%%

The $\psi(2S)$ to $J/\psi$ ratio $R_\psi$ is defined in
\Eq~(\ref{eq:FJpsipsi2S}).  The experimental results for $R_\psi$ from
fixed-target experiments are compiled in \Table~\ref{tab:psi}.  The
result from experiment E673 is obtained by dividing the observed
fraction of $J/\psi$'s from decays of $\psi(2S)$ by the branching
fraction for $\psi(2S) \to J/\psi X$ given by the Particle Data Group
\cite{Eidelman:2004wy}.  The result from experiment E771 is obtained
by dividing the observed ratio of the products of the cross-sections
and the branching fractions into $\mu^+ \mu^-$ by the ratio of the
branching fractions into $\mu^+ \mu^-$ given by the Particle Data
Group \cite{Eidelman:2004wy}.  The NRQCD factorization approach gives
the values $R_\psi=0.16$ for both $pN$ collisions and $\pi^-N$
collisions \cite{Beneke:1996tk}.  The colour-singlet model gives
$R_\psi=0.14$ for $pN$ collisions and $R_\psi=0.16$ for $\pi^-N$
collisions \cite{Beneke:1996tk}. In the colour-evaporation model, this
ratio is simply an input. Thus the ratio $R_\psi$ is not able to
discriminate between any of these approaches.

%%%%%%%%%%%%%%%%%%%%%%%%%%%%%%%%%%%%%%%%%%%%%%%%%%%%%%%%%%%%%%%%%%%%%%%%%%%%%%%%%%%%%%%%%%%
\begin{table}[ht]
\caption[Experimental results for the fraction of $J/\psi$'s]
        {Experimental results for the fraction of $J/\psi$'s from
         $\chi_c$ decay, $F_{\chi_c}$, and the
         $\chi_{c1}$ to $\chi_{c2}$ ratio, $R_{\chi_c}$. In view of
         the experimental uncertainties, no attempt has been made to
         rescale older measurements to account for the latest $\chi_c$
         branching fractions.  Modified version of a table from
         Ref.~\cite{Beneke:1997av}.}
\label{tab:chi}
\addtolength{\arraycolsep}{0.2cm}
\renewcommand{\arraystretch}{1.25}
\begin{center}
\begin{tabular}{|c|cccc|} 
\hline
\hline
\mbox{Experiment} & \mbox{beam/target} & $\sqrt{s}/\mbox{GeV}$ & 
$F_{\chi_c}$ &$R_{\chi_c}$ \\
\hline 
\mbox{E673}~\cite{Bauer:yf} & $p\mbox{Be}$ & $19.4/21.7$ &$0.47\pm 0.23$ 
&$0.24\pm 0.28$ \\ 
\mbox{E705}~\cite{Antoniazzi:1993yf}& $p\mbox{Li}$ & $23.7$ & ---
&$0.08_{-0.15}^{+0.25}$ \\ 
\mbox{E705}~\cite{Arenton:pw}& $p\mbox{Li}$ & $23.7$ & $0.30\pm 0.04$
&--- \\ 
\mbox{E771}~\cite{Alexopoulos:1999wp}& $p\mbox{Si}$ & $38.8$ &---
& $0.53\pm 0.20\pm 0.07$ \\ 
\mbox{HERA-B}~\cite{Spengler:2004gr} & p\mbox{(C, W)} & $41.5$ 
&$0.32\pm 0.06\pm 0.04 $&--- \\ 
\hline
\mbox{WA11}~\cite{Lemoigne:1982jc}& $\pi\mbox{Be}$ & $18.6$ 
&$0.305\pm 0.050$ & $0.68\pm 0.28$ \\ 
\mbox{E673}~\cite{Bauer:yf} & $\pi\mbox{Be}$ &$18.9$ &$0.31\pm 0.10$
& $0.96\pm 0.64 $\\ 
\mbox{E673}~\cite{Hahn:tz}    & $\pi\mbox{Be}$ &$20.6$&$0.37\pm 0.09$
& $0.9\pm 0.4$\\ 
\mbox{E705}~\cite{Antoniazzi:1993yf} & $\pi\mbox{Li}$ &$23.7$ &--- &
$0.52_{-0.27}^{+0.57}$ \\ 
\mbox{E705}~\cite{Arenton:pw} & $\pi^+\mbox{Li}$ &$23.7$ &
$0.40\pm 0.04$ &--- \\ 
\mbox{E705}~\cite{Arenton:pw} & $\pi^-\mbox{Li}$ &$23.7$ &
$0.37\pm 0.03$ &--- \\ 
\mbox{E672/706}~\cite{Koreshev:1996wd} & $\pi^-\mbox{Be}$ &$31.1$
&$0.443\pm 0.041\pm 0.035$ & $0.57\pm 0.18\pm 0.06$ \\ 
\hline
\hline 
\end{tabular}
\end{center}
\end{table} 
%%%%%%%%%%%%%%%%%%%%%%%%%%%%%%%%%%%%%%%%%%%%%%%%%%%%%%%%%%%%%%%%%%%%%%%%%%%%%%%%%%%%%%%%%%%

The fraction $F_{\chi_c}$ of $J/\psi$'s that come from $\chi_c$ decays
is defined in \Eq~(\ref{eq:prod-chifrac}).  The experimental results
for $F_{\chi_c}$ from fixed-target experiments are compiled in
\Table~\ref{tab:chi}.  The NRQCD factorization approach gives the
values $F_{\chi_c}=0.27$ for $pN$ collisions and $F_{\chi_c}=0.28$ for
$\pi^-N$ collisions \cite{Beneke:1996tk}. The colour-singlet model
gives $F_{\chi_c}=0.68$ for $pN$ collisions and $F_{\chi_c}=0.66$ for
$\pi^-N$ collisions \cite{Beneke:1996tk}. In the colour-evaporation
model, this ratio is simply an input.  Clearly, the experimental
results favor the NRQCD factorization approach over the colour-singlet
model.  The most precise results from $pN$ fixed target experiments
are compatible with the Tevatron result in
\Table~\ref{tab:prodsec-Jpsifractions}.  The most precise results from
$\pi N$ fixed target experiments are somewhat larger.

The $\chi_{c1}$ to $\chi_{c2}$ ratio $R_{\chi_c}$ is defined in
\Eq~(\ref{eq:prod-chirat}).  There are substantial variations among
the NRQCD factorization predictions for $R_{\chi_c}$ in fixed-target
experiments. Beneke and Rothstein \cite{Beneke:1996tk} give the values
$R_{\chi_c}=0.07$ for $pN$ collisions and $R_{\chi_c}=0.05$ for
$\pi^-N$ collisions. Their calculation is carried out at leading order
in $\alpha_s$ and uses the standard truncation in $v$ that is
described in \Section~\ref{sec:prodsec-nrqcdfact}. Beneke and
Rothstein \cite{Beneke:1996tk} suggest that corrections to
hard-scattering factorization may be large.  Beneke
\cite{Beneke:1997av} gives the estimate $R_{\chi_c}\approx 0.3$ for
both $pN$ and $\pi N$ collisions. This estimate is based on the
assumption that the ${}^3P_2$ and ${}^3P_0$ colour-octet matrix
elements dominate the $\chi_{c1}$ production. It is consistent with
the estimate in Ref.~\cite{Gupta:1997me}, once that estimate is
modified to take into account the dominant colour-singlet channel in
$\chi_{c2}$ production \cite{Beneke:1997av}. Maltoni
\cite{Maltoni:2000km} gives central values of $R_{\chi_c}$ for $pN$
collisions that range from $R_{\chi_c}=0.04$ to $R_{\chi_c}=0.1$ as
the beam energy ranges from 200~GeV to 800~GeV. Maltoni's calculation
takes into account matrix elements at leading order in $v$, but
contains corrections of next-to-leading order in $\alpha_s$. His
calculation displays a very large dependence on the renormalization
scale. In summary, the existing predictions for $R_\chi$ based on
NRQCD factorization are in the range 0.04--0.3 for both $pN$ and $\pi
N$ collisions. The colour-singlet model predicts that
$R_{\chi_c}\approx 0.05\hbox{--}0.07$ for both $pN$ and $\pi N$
collisions \cite{Beneke:1996tk,Beneke:1997av}. The colour-evaporation
model predicts that $R_{\chi_c} \simeq 3/5$
\cite{Amundson:1995em,Amundson:1996qr}.

The experimental results for $R_{\chi_c}$ are compiled in
\Table~\ref{tab:chi}. As can be seen, the data are somewhat
inconsistent with each other.  The results from the most precise
experiments are significantly smaller than the Tevatron result in
\Eq~(\ref{eq:Rchi12Tev}).  There seems to be a trend toward larger
values of $R_{\chi_c}$ in $\pi N$ experiments than in $pN$
experiments. Such a dependence on the beam type is contrary to the
predictions of the colour-evaporation model. It also would not be
expected in the NRQCD factorization approach, unless there is an
unusual enhancement in the $q\bar q$ production channel
\cite{Beneke:1997av}. Both the $pN$ and $\pi N$ data yield results
that are significantly larger than the predictions of the
colour-singlet model. The $pN$ experiments seem to favor the NRQCD
factorization predictions, while the $\pi N$ experiments seem to favor
the colour-evaporation prediction. However, in light of the large
theoretical and experimental uncertainties, no firm conclusions can be
drawn.

\subsection{Polarization}
\label{sec:prodsec-fixed-targetpol}

%%%%%%%%%%%%%%%%%%%%%%%%%%%%%%%%%%%%%%%%%%%%%%%%%%%%%%%%%%%%%%%%%%%%%%%%%%%%%%%%%%%%%%%%%%%
\begin{table}[ht]
\caption[Experimental results for polarization variable $\alpha$
         in $J/\psi$ production]
        {Experimental results for the polarization variable $\alpha$
         in $J/\psi$ production. Modified version of a table from
         Ref.~\cite{Spengler:2004gr}.}
\label{tab:fixed-target-pol}
\addtolength{\arraycolsep}{0.2cm}
\renewcommand{\arraystretch}{1.25} 
\begin{center}
\begin{tabular}{|c|ccc|} 
\hline
\hline
\mbox{Experiment} & \mbox{beam/target} & \mbox{Beam Energy}/\mbox{GeV}
&$\alpha$  \\ 
\hline \mbox{E537}~\cite{Tzamarias:1990ij} & $(\pi, p)$\mbox{(Be, Cu, W)} 
& $125$ &$0.024$--$0.032$\\
\mbox{E672/706}~\cite{Gribushin:1999ha} & p\mbox{Be} & $530$ 
& $0.01\pm 0.15$ \\
\mbox{E672/706}~\cite{Gribushin:1999ha}& p\mbox{Be} & $800$ 
& $-0.11\pm 0.15$ \\
\mbox{E771}~\cite{Introzzi:yi}& p\mbox{Si} & $800$ & $-0.09\pm 0.12$ \\
\mbox{E866}~\cite{Chang:2003rz}& p\mbox{Cu} & $800$ & $0.069\pm 0.08$\\
\mbox{HERA-B}~\cite{Spengler:2004gr} & p\mbox{(C, W)} & $920$ 
& $(-0.5,\, +0.1)\pm 0.1$ \\ 
\hline
\hline 
\end{tabular} 
\end{center}
\end{table}
%%%%%%%%%%%%%%%%%%%%%%%%%%%%%%%%%%%%%%%%%%%%%%%%%%%%%%%%%%%%%%%%%%%%%%%%%%%%%%%%%%%%%%%%%%%

The polarization variable $\alpha$ for $J/\psi$ production is defined
by the angular distribution in \Eq~(\ref{eq:prod-alphadef}).  In
fixed-target experiments, the most convenient choice of the
polarization axis is the direction of the boost vector from the
$J/\psi$ rest frame to the lab frame.  Experimental results for
$\alpha$ are shown in \Table~\ref{tab:fixed-target-pol}.  The
prediction of the NRQCD factorization approach is
$0.31<\alpha<0.63$~\cite{Beneke:1996tk}. Both the theoretical
prediction and the data include feeddown from $\chi_c$ states. The
prediction is largely independent of the target and beam types. It was
made specifically for the beam energy 117~GeV. However, the energy
dependence of the prediction is quite mild, and the prediction would
be expected to hold with little error even at a beam energy of
800~GeV. The colour-singlet model predicts a substantial transverse
polarization \cite{Vanttinen:1994sd}. The colour-evaporation model
predicts that $\alpha = 0$ for all processes. There are also specific
predictions for the HERA-B experiment in which the region of small
$p_T$ is excluded.  The predictions for the range $p_T=1.5$--$4$~GeV
are $\alpha=0$--$0.1$ in the NRQCD factorization approach and
$\alpha=0.2$--$0.4$ in the colour-singlet
model~\cite{lee-2000}. Experimental results for the polarization
variable $\alpha$ in $J/\psi$ production are shown in
\Table~\ref{tab:fixed-target-pol}. The data from the conventional
fixed-target experiments are consistent with $\alpha=0$ and favor the
prediction of the colour-evaporation model over the predictions of
NRQCD factorization or the colour-singlet model \cite{Beneke:1996tk}.
At the smaller values of $p_T$, one can question whether resummation
of the perturbation series is needed and whether hard-scattering
factorization would be expected to hold. The HERA-B data are also
consistent with $\alpha=0$ and favor the predictions of the NRQCD
factorization approach and the colour-evaporation model over the
prediction of the colour-singlet model.

There is also a measurement of the polarization of $\psi(2S)$ in a
fixed-target experiment. The E615 experiment measured $\alpha$ for
$\psi(2S)$ mesons produced in $\pi N$ collisions at 253~GeV
\cite{Heinrich:zm}. The data yield $-0.12<\alpha<0.16$, while the
prediction of the NRQCD factorization approach is
$0.15<\alpha<0.44$~\cite{Beneke:1996tk}.

%%%%%%%%%%%%%%%%%%%%%%%%%%%%%%%%%%%%%%%%%%%%%%%%%%%%%%%%%%%%%%%%%%%%%%%%%%%%%%%%%%
\begin{figure}[t]
\begin{center}
\includegraphics[width=.9\linewidth]{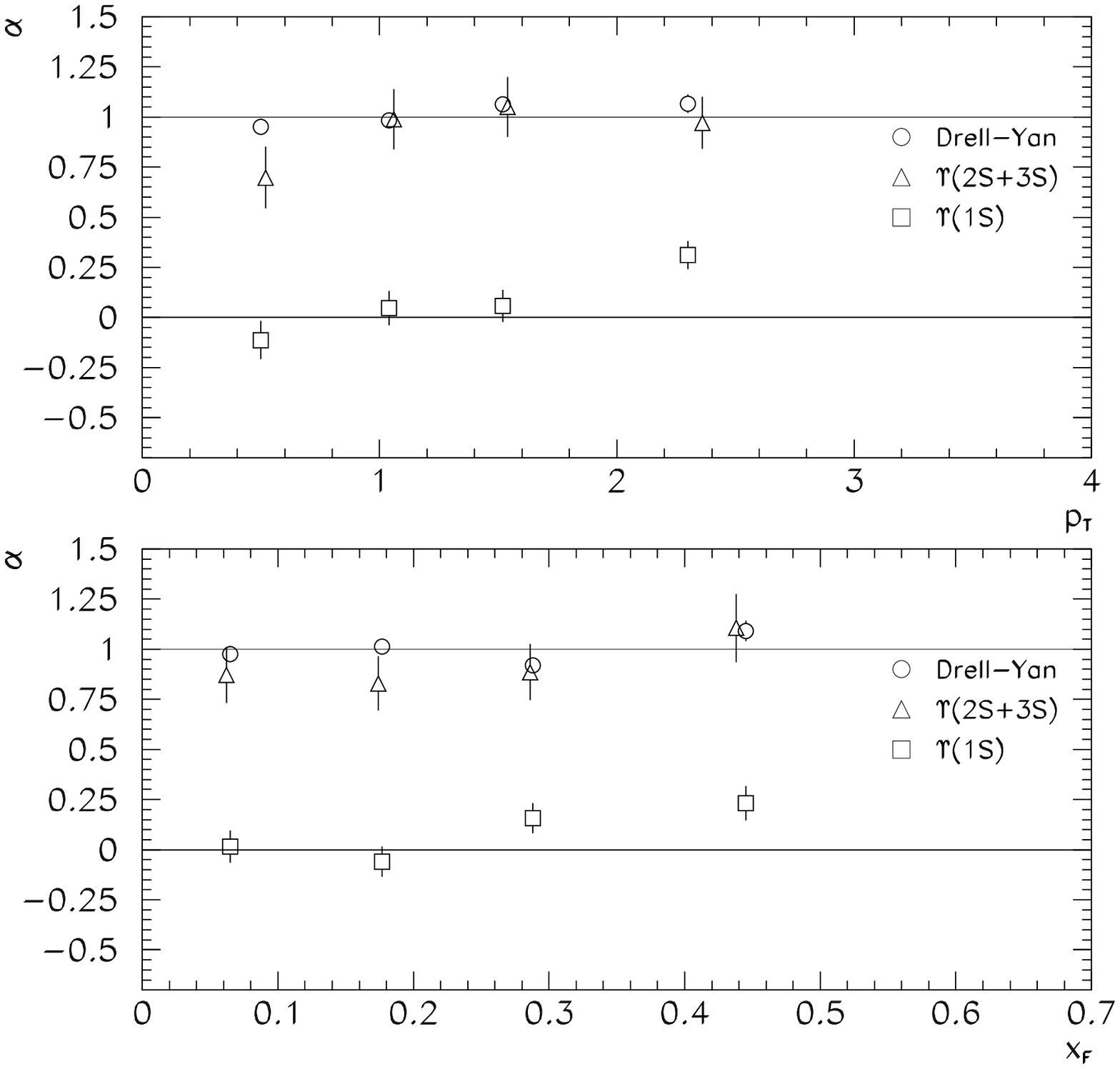}
\caption{Polarization of $\Upsilon$ mesons and Drell--Yan pairs as a function
of $p_T$ and $x_F$ in p--Cu collisions in the E866 experiment. From
Ref.~\cite{Brown:2000bz}.}
\label{fig:e866-upsilon-pol}
\end{center}
\end{figure}
%%%%%%%%%%%%%%%%%%%%%%%%%%%%%%%%%%%%%%%%%%%%%%%%%%%%%%%%%%%%%%%%%%%%%%%%%%%%%%%%%%

The E866/NuSea experiment has studied the production of dimuons in the
collision of 800~GeV protons with copper \cite{Brown:2000bz}. The
experiment used the angular distributions of dimuons in the mass range
8.1--15.0~GeV to measure the polarization variable $\alpha$ for
Drell--Yan pairs, for $\Upsilon(1S)$ mesons, and for a mixture of
$\Upsilon(2S)$ and $\Upsilon(3S)$ mesons. The data cover the kinematic
ranges 0.0 $< x_F <$ 0.6 and $p_T <$ 4.0~GeV. The results for the
polarization variable $\alpha$ as a function of $p_T$ and $x_F$ are
shown in \Figure~\ref{fig:e866-upsilon-pol}. The $\Upsilon(1S)$ data
show almost no polarization at small $x_F$ and $p_T$, but show nonzero
transverse polarization at either large $p_T$ or large $x_F$. A fit at
the $\Upsilon$(1S) mass for a polarization that is independent of
$x_F$ and $p_T$ gives $\alpha =$0.07 $\pm$ 0.04. This observation is
substantially smaller than a prediction that is based on the NRQCD
factorization approach, which gives $\alpha$ in the range
0.28--0.31~\cite{Kharchilava:1998wa,Tkabladze:1999mb}. However, it
also disagrees with the prediction of the colour-evaporation model that
the polarization should be zero \cite{Amundson:1996qr}. The most
remarkable result from this experiment is that the $\Upsilon(2S)$ and
$\Upsilon(3S)$ were found to be strongly transversely polarized, with
the polarization variable $\alpha$ close to its maximal value $\alpha
= +1$ for all $x_F$ and $p_T$, as in the case of Drell--Yan pairs. This
result provides strong motivation for measuring the polarizations of
the $\Upsilon(2S)$ and $\Upsilon(3S)$ at the Tevatron to see if these
states are also produced with substantial polarizations in $p \bar p$
collisions.

It has been proposed that $\chi_2$ production in hadron collisions at
zero $p_T$ may serve as a tool to measure the polarized gluon structure
function of the proton \cite{Cortes:1988ww}. This proposal relies on the
assumption that $\chi_2$ production at zero $p_T$ is dominated by gluon
fusion, and it requires that at least one of the colliding hadrons be
polarized, as is the case, for example, in the RHIC polarized program.
 
\subsection{Colour-evaporation-model parameters}
\label{sec:prodsec-fixed-targetCEM}

%%%%%%%%%%%%%%%%%%%%%%%%%%%%%%%%%%%%%%%%%%%%%%%%%%%%%%%%%%%%%%%%%%%%%%%%%%%%%%%%%%
%%%%%%%%%%%%%%%%%%%%%%%%%%%%%%%%%%%%%%%%%%%%%%%%%%%%%%%%%%%%%%%%%%%%%%%%%%%%%%%%%%

%%%%%%%%%%%%%%%%%%%%%%%%%%%%%%%%%%%%%%%%%%%%%%%%%%%%%%%%%%%%%%%%%%%%%%%%%%%%%%%%%%
\begin{table}[t]
\caption{Inclusive CEM parameters $F_{J/\psi}$ and $F_{\Upsilon(1S)}$
         from Ref.~\cite{Bedjidian:2003gd} for various choices of
         PDF's, quark masses (in~GeV), and scales.  }
\label{tab:prodsec-qqbparams}
\begin{center}
\begin{tabular}{|llcc||llcc|}
\hline \hline
PDF & $m_c$ & $\mu/m_{cT}$ & $F_{J/\psi}$ & 
PDF & $m_b$ & $\mu/m_{bT}$ & $F_{\Upsilon(1S)}$ \\ \hline
MRST HO & 1.2 & 2   & 0.0144  &
MRST HO & 4.75 & 1   & 0.0276 \\
MRST HO & 1.4 & 1   & 0.0248  &
MRST HO & 4.5  & 2   & 0.0201 \\
CTEQ 5M & 1.2 & 2   & 0.0155  &
MRST HO & 5.0  & 0.5 & 0.0508 \\
GRV 98 HO & 1.3 & 1 & 0.0229  &
GRV 98 HO & 4.75 & 1 & 0.0225 \\ 
\hline \hline
\end{tabular}
\end{center}

\caption[Ratios of the direct CEM parameters]
        {Ratios of the direct CEM parameters $F_H^{\rm dir}$ to the
         inclusive CEM parameter $F_{J/\psi}$ in the case of
         charmonium states and to the inclusive CEM parameter
         $F_{\Upsilon(1S)}$ in the case of bottomonium states. From
         Ref.~\protect \cite{Digal:2001ue}.}
\label{tab:prodsec-ratios}
\begin{center}
\begin{tabular}{|c|ccccc|} 
\hline \hline
$H$ & $J/\psi$ & $\psi(2S)$ & $\chi_{c1}$ & $\chi_{c2}$ & \\ \hline
$F_H^{\rm dir}/F_{J/\psi}$       & 0.62 & 0.14 & 0.60 & 0.99 & \\ 
\hline \hline
$H$ & $\Upsilon(1S)$ & $\Upsilon(2S)$ & $\Upsilon(3S)$ 
        & $\chi_b(1P)$ & $\chi_b(2P)$ \\ \hline
$F_H^{\rm dir}/F_{\Upsilon(1S)}$ & 0.52 & 0.33 & 0.20 & 1.08 & 0.84 \\ 
\hline \hline
\end{tabular}
\end{center}
\end{table}
%%%%%%%%%%%%%%%%%%%%%%%%%%%%%%%%%%%%%%%%%%%%%%%%%%%%%%%%%%%%%%%%%%%%%%%%%%%%%%%%%%

Data from $pp$ and $pA$ collisions have been used to extract the
parameters $F_H$ of the colour-evaporation model.  (The CEM parameter
$F_H$ should not be confused with the fraction of $J/\psi$'s that come
from decay of $H$.)  The results of these extractions are given in
\Tables~\ref{tab:prodsec-qqbparams} and~\ref{tab:prodsec-ratios}.  The
numerical values of the CEM parameters $F_H$ that are obtained by
fitting data depend on the choices of the parton densities (PDF's),
the heavy quark mass $m_Q$, the renormalization/factorization scale
$\mu$, and the order in $\alpha_s$ of the calculation. The CEM
parameters have been calculated using several sets of parton densities
\cite{Martin:1998sq,Lai:1999wy,Gluck:1998xa}, quark masses, and scales
\cite{Vogt:2002vx,Vogt:2002ve} that reproduce the measured $Q \bar Q$
cross-section. In these calculations, the scale $\mu$ was set to a
constant times $m_{QT} = (m_Q^2 + p_T^2)^{1/2}$, where $p_T$ is the
sum of the transverse momenta of the $Q$ and the $\bar Q$.

We first describe the extraction of the CEM parameters $F_H$ for
charmonium states. The inclusive cross-section for $J/\psi$ production
has been measured in $pp$ and $pA$ collisions up to $\sqrt{s} = 63$
GeV.  The data are of two types: the forward cross-section,
$\sigma(x_F > 0)$, and the cross-section at zero rapidity,
$d\sigma/dy|_{y=0}$.  These cross-sections include the feeddown from
decays of $\chi_{cJ}$ and $\psi(2S)$. The parameters $F_{J/\psi}$ that
were obtained by fitting the inclusive $J/\psi$ cross-sections
measured in $pp$ and $pA$ collisions are given in
\Table~\ref{tab:prodsec-qqbparams} for four sets of PDF's and
parameters. The ratio of the parameter $F_H^{\rm dir}$ for the direct
production of a charmonium state $H$ to the parameter $F_{J/\psi}$ for
the inclusive production of $J/\psi$ can be determined from the
measured ratios of the inclusive cross-sections for $H$ and $J/\psi$
using the known branching fractions for the feeddown decays. These
ratios are given in \Table~\ref{tab:prodsec-ratios} for various
charmonium states.

A similar procedure can be used to determine the CEM parameters $F_H$
for bottomonium states. In most data on $pp$ and $pA$ collisions below
$\sqrt{s} = 100$~GeV, only the sum of the $\Upsilon(1S)$,
$\Upsilon(2S)$, and $\Upsilon(3S)$ cross-sections weighted by their
branching fractions to decay into lepton pairs is reported. A fit to
the lepton-pair cross-section in the $\Upsilon$ region at zero
rapidity therefore gives a linear combination of the inclusive
parameters $F_{\Upsilon(nS)}$ weighted by the branching fractions
$B[\Upsilon(nS) \to \ell^+ \ell^-]$.  The inclusive parameters
$F_{\Upsilon(1S)}$ given in \Table~\ref{tab:prodsec-qqbparams} were
extracted by using the known branching fractions and the measured
ratios of the inclusive cross-sections for $\Upsilon(nS)$ in $p \bar
p$ collisions at the Tevatron \cite{Affolder:1999wm}. The ratios of
the parameters $F_H^{\rm dir}$ for the direct production of a
bottomonium state $H$ to the parameter $F_{\Upsilon(1S)}$ for the
inclusive production of $\Upsilon(1S)$ that were obtained in
Ref.~\cite{Gunion:1996qc} have been updated in
Ref.~\cite{Digal:2001ue} by using recent CDF data on $\chi_b$
production and are given in \Table~\ref{tab:prodsec-ratios}.

%%%%%%%%%%%%%%%%%%%%%%%%%%%%%%%%%%%%%%%%%%%%%%%%%%%%%%%%%%%%%%%%%%%%%%%%%%%%%%%%%%
\begin{figure}[t]
\begin{center}
\includegraphics[width=.9\linewidth]{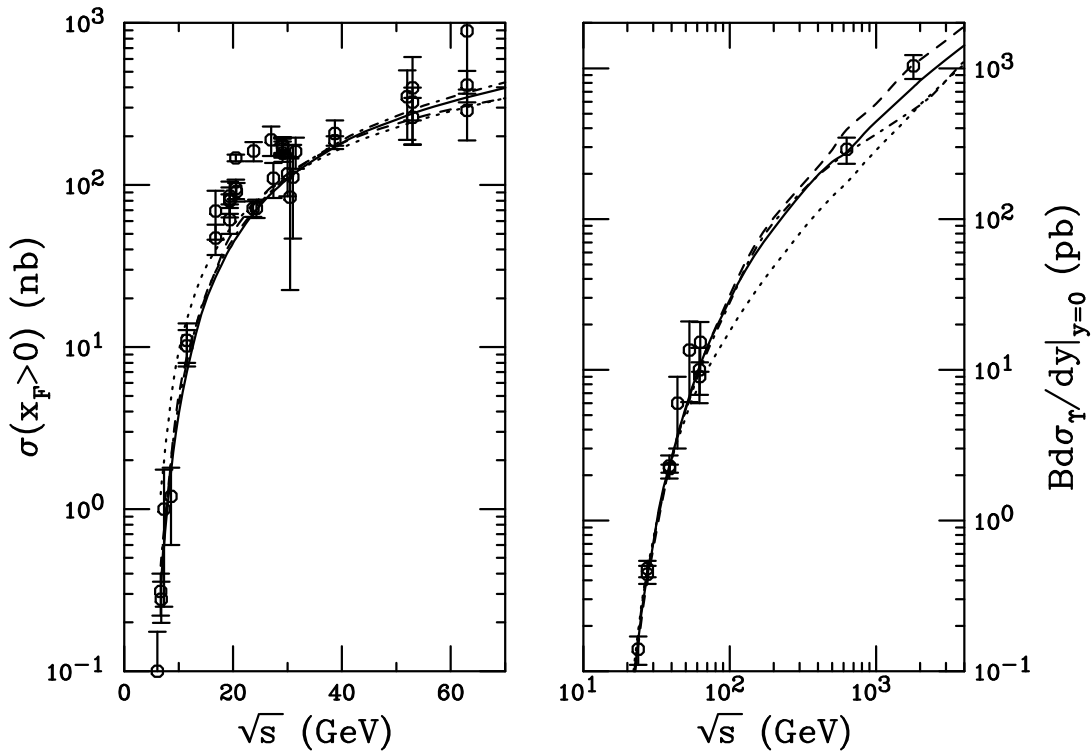}
\end{center}
\caption[Forward $J/\psi$ production cross-section]
        {Forward $J/\psi$ production cross-section (left) and weighted
         average of the $\Upsilon(nS)$ production cross-sections at
         zero rapidity (right) as a function of the centre-of-mass
         energy $\sqrt{s}$. The $J/\psi$ data are from $pp$
         experiments and from $pA$ experiments with light targets $A
         \leq 12$. It has been assumed that the cross-sections scale
         as $A^{0.9}$.  The low-energy $\Upsilon$ data are from $pp$
         and $pA$ experiments. It has been assumed that the cross-sections
         are linear in $A$. The high-energy $\Upsilon$ data are from
         $p \bar p$ experiments.  The curves are the cross-sections
         calculated to NLO in the CEM using the four
         charmonium parameter sets and the four bottomonium parameter
         sets in \Table~\ref{tab:prodsec-qqbparams}.}
\label{fig:psiupsfixt}
\end{figure}
%%%%%%%%%%%%%%%%%%%%%%%%%%%%%%%%%%%%%%%%%%%%%%%%%%%%%%%%%%%%%%%%%%%%%%%%%%%%%%%%%%

The forward cross-section for $J/\psi$ and the weighted cross-section
at zero rapidity for $\Upsilon(nS)$ are shown as a function of the
centre-of-mass energy in \Figure~\ref{fig:psiupsfixt}. The energy
dependence of both cross-sections is well reproduced by the CEM at
NLO. All of the CEM parameter sets give good fits to the data for
$\sqrt{s} \leq 63$~GeV, but their predictions for $\Upsilon(nS)$
differ by up to a factor of two when extrapolated to 2~TeV. The
extrapolation of the forward $J/\psi$ cross-section to 2~TeV cannot be
compared with data from Run I of the Tevatron because the lepton-$p_T$
cut excludes a measurement of the cross-section for $J/\psi$ in the
region $p_T < 5$~GeV that dominates the integrated cross-section.

%%%%%%%%%%%%%%%%%%%%%%%%%%%%%%%%%%%%%%%%%%%%%%%%%%%%%%%%%%%%%%%%%%%%%%%%%%%%%%%%%%

%\input{sechera}
\section{Quarkonium production at HERA} 
\label{sec:prodsec-hera}

\subsection{Inelastic photoproduction of charmonium}
\label{sec:prodsec-hera:gp}

\begin{figure}[h]
\begin{center}
\includegraphics[width=15cm]{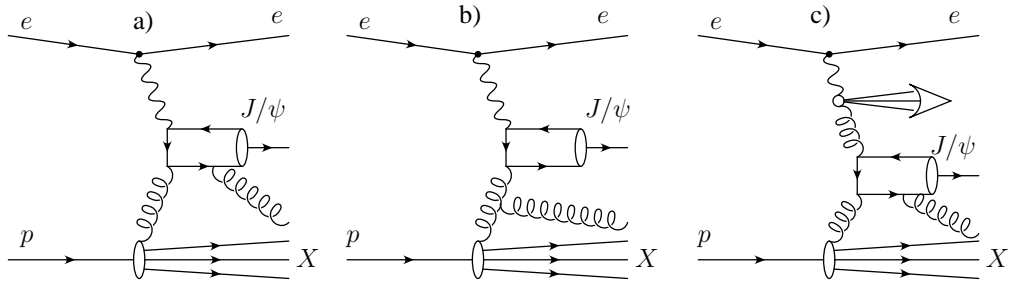}
\end{center}
\caption[Generic Feynman diagrams for inelastic \jpsi\ production]
        {Generic Feynman diagrams for inelastic \jpsi\ production.
         a,b) direct-photon processes; c) resolved-photon process. In
         diagrams a) and c), the \ccbar\ pair leading to the formation
         of the \jpsi\ can be in a colour-singlet or a colour-octet
         state while in b) it can only be in a colour-octet
         state. Additional soft gluons emitted during the
         hadronization process are not shown.}
\label{fig:herafeyn}
\end{figure}

At the $ep$ collider HERA, the inelastic charmonium production process
is dominantly virtual-photon-gluon fusion: a photon emitted from the
incoming electron or positron interacts with a gluon from the proton
to produce a \ccbar\ pair that subsequently forms a charmonium state.
In photoproduction, the photon virtuality $Q^2$ is small and the
photon is quasi-real.  In this case, the photon can either couple to
the $c$ quark directly (``direct'' processes,
\Figure~\ref{fig:herafeyn}a or b) or it can interact via its hadronic
component (``resolved'' processes, \Figure~\ref{fig:herafeyn}c).  Many
models have been suggested to describe inelastic charmonium production
in the framework of perturbative QCD, such as the colour-singlet model
(CSM) \cite{Berger:1980ni,Baier:1981uk,Baier:1981zz,Baier:1983va}
described in \Section~\ref{sec:prodsec-nrqcdCSM}, the colour-evaporation
model \cite{Halzen:1977rs,Eboli:1998xx} described in
\Section~\ref{sec:prodsec-nrqcdCEM}, and soft colour interactions
\cite{Edin:1997zb}.

%%%%%%%%%%%% Colour singlet and NRQCD

For $J/\psi$ and $\psi(2S)$ photoproduction, the CSM calculations are
available to next-to-leading order\cite{Kramer:1994zi,Kramer:1995nb}.
These are performed using standard hard-scattering factorization in
which the gluon density depends only on the momentum fraction $x$.
Alternatively, using the CSM, inelastic \jpsi\ production can be
modeled in the $k_T$-factorization approach (see
\Section~\ref{sec:prodsec-nrqcdmge}) using an unintegrated
($k_T$-dependent) gluon density in the proton.

Theoretical calculations based on the NRQCD factorization approach 
\cite{Caswell:1985ui,Thacker:1990bm,Bodwin:1994jh} are 
available in leading order. For \jpsi\ and $\psi(2S)$ photoproduction
at HERA, these have been performed by Cacciari and Kr\"amer
\cite{Cacciari:1996dg}, Beneke, Kr\"amer, and V\"anttinen
\cite{Beneke:1998re}, Amundson, Fleming, and Maksymyk
\cite{Amundson:1996ik}, Ko, Lee, and Song \cite{Ko:1996xw},
Godbole, Roy, and Sridhar \cite{Godbole:1995ie}, and Kniehl and
G.~Kramer \cite{Kniehl:1997fv,Kniehl:1997gh}. The theoretical
calculations use the standard truncation in $v$, in which 
the independent NRQCD matrix elements are 
$\langle {\cal O}_1^{J/\psi}({}^3S_1) \rangle$, $\langle {\cal
O}_8^{J/\psi}({}^1S_0)
\rangle$, $\langle {\cal O}_8^{J/\psi}({}^3S_1) \rangle$, and $\langle
{\cal O}_8^{J/\psi}({}^3P_0) \rangle$.  The relative strength of the
colour-octet contributions depends crucially on the size of the
corresponding NRQCD matrix elements. Unfortunately the values of the
matrix elements $\langle {\cal O}_8^{J/\psi}({}^1S_0) \rangle$ and
$\langle {\cal O}_8^{J/\psi}({}^3P_0) \rangle$, which are most
important in \jpsi\ and $\psi(2S)$ photoproduction at HERA, still show
large uncertainties. (See \Section~\ref{sec:prodsec-tevatroncharm} and
Ref.~\cite{Kramer:2001hh}.)

The theoretical predictions are sensitive to a number of input
parameters, \eg the parton distributions, the values of $\alpha_s$,
and the charm-quark mass $m_c$, as well as the choice of the
renormalization and factorization scales. In the NRQCD factorization
approach, the values of the colour-octet NRQCD matrix elements are
additional parameters.  The comparison with the data in the NRQCD
approach also suffers from the uncertainties associated with LO
calculations. Next-to-leading-order corrections might change the
results substantially. Although the NLO terms have not been calculated
in the NRQCD approach, effects that are similar to those in the CSM
may be expected, in which the NLO terms lead to an increase in the
cross-section of typically a factor two, with a strong
\ensuremath{p_{T,\psi}} dependence.

\begin{figure}[p]
\begin{center}
\includegraphics[width=9cm]{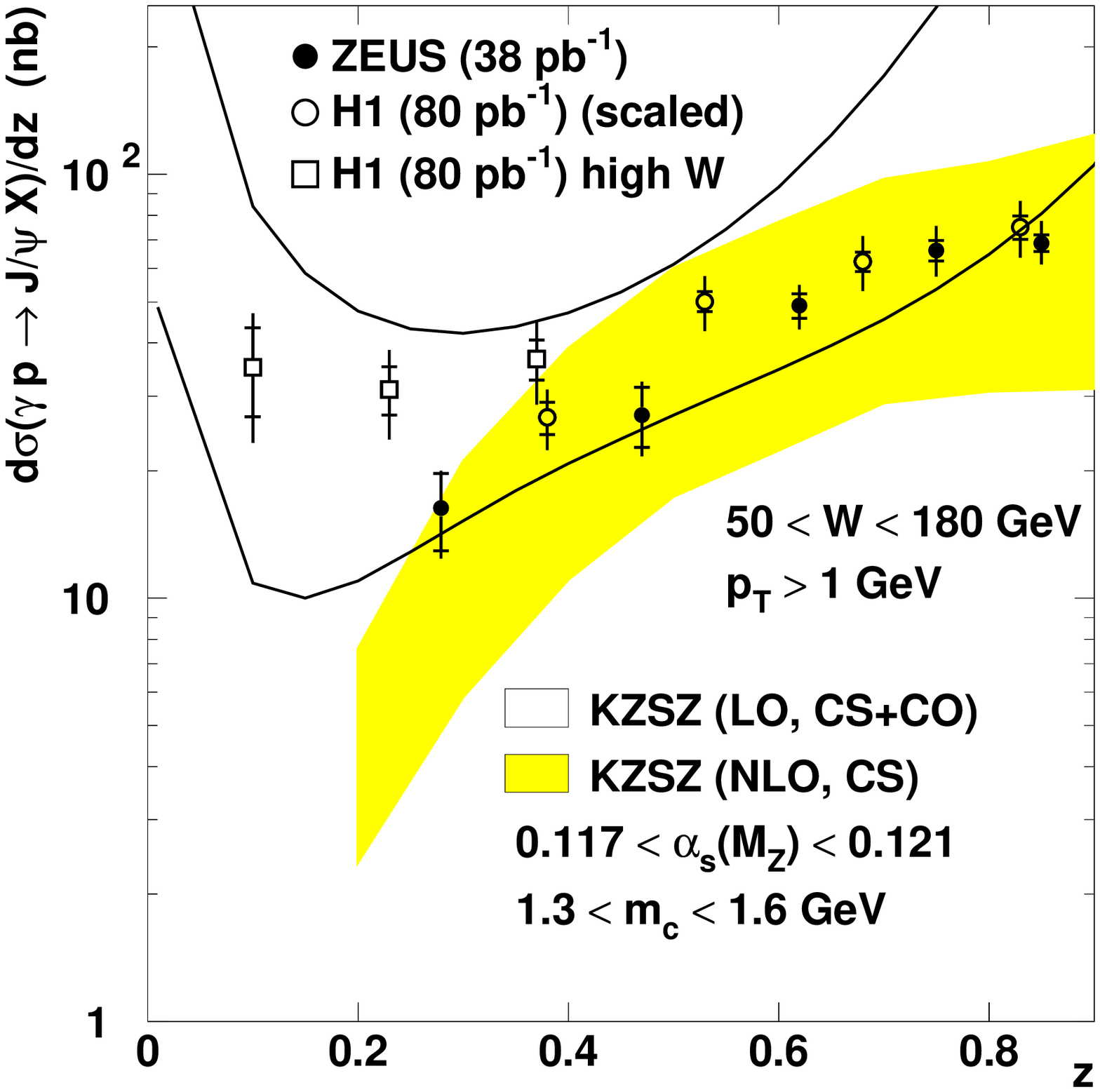}
\end{center}
\caption[The rate for inelastic $J/\psi$ photoproduction at HERA
         as a function of $z$]
        {The rate for inelastic $J/\psi$ photoproduction at HERA as a
         function of $z$. The open band represents the LO NRQCD
         factorization prediction~\cite{Kramer:2001hh}. The solid band
         represents the NLO colour-singlet
         contribution~\cite{Kramer:1995nb,Kramer:2001hh}. The data
         points are from the H1 \cite{Adloff:2002ex} and ZEUS
         \cite{Chekanov:2002at} measurements.}
\label{fig:photo-prod}

\begin{center}
\includegraphics[width=9cm]{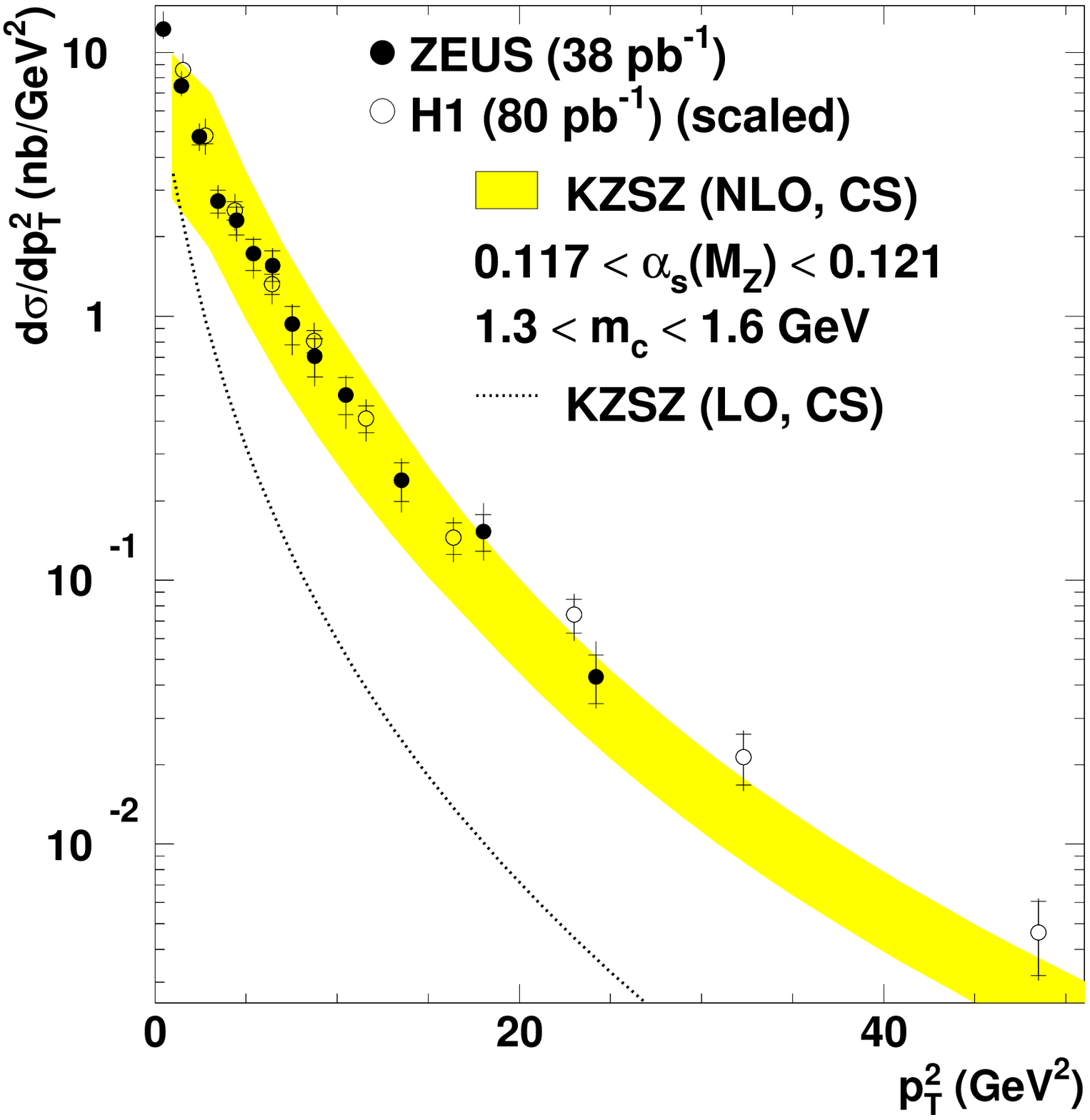}
\end{center}
\caption[The rate for inelastic $J/\psi$ photoproduction at HERA as a
         function of $p_{T,\psi}$]
        {The rate for inelastic $J/\psi$ photoproduction at HERA as a
         function of $p_{T,\psi}$. The solid band represents the NLO
         colour-singlet
         contribution~\cite{Kramer:1995nb,Kramer:2001hh}. The dotted
         line is the LO colour-singlet contribution. The data points
         are from the H1~\cite{Adloff:2002ex} and ZEUS
         \cite{Chekanov:2002at} measurements.}
\label{fig:photo-prod-csm}
\end{figure}

\Figure[b]~\ref{fig:photo-prod} shows the measurements of the prompt
\jpsi\ cross-section by the H1 collaboration \cite{Adloff:2002ex} 
and the ZEUS collaboration \cite{Chekanov:2002at}, compared with the
theoretical predictions given in Ref.~\cite{Kramer:2001hh}.
The variable $z$ denotes the fraction of the photon energy that is
transferred to the $J/\psi$ and is defined as
\begin{equation}
z=\frac{(E-p_z)_{J/\psi}}{(E-p_z)_{\rm hadrons}}, 
\label{eq:zdef}
\end{equation}
where $E$ and $p_z$ in the numerator are the
energy and $z$-component of the momentum of the $J/\psi$ and  
$E$ and $p_z$ in the denominator are the sums of the energies 
and $z$-components of the momenta of all the hadrons in the final state.

The $J/\psi$ data points shown in \Figure~\ref{fig:photo-prod} are not corrected
for feeddown processes, such as diffractive and inelastic production
of $\psi(2S)$ mesons ($\approx 15\% $), the production of $b$ hadrons
with subsequent decays to \jpsi\ mesons, or feeddown from the
production of $\chi_c$ states. The latter two contributions are
estimated to contribute between 5\% at medium $z$ and 30\% at the
lowest values of $z$.
The open band in \Figure~\ref{fig:photo-prod} represents the sum of the
colour-singlet and colour-octet contributions, calculated in leading
order in QCD perturbation theory.  The uncertainty is due to the
uncertainty in the colour-octet NRQCD matrix elements. The NRQCD
prediction deviates from the data near $z=1$, owing to the large
colour-octet contribution in that region.  The shaded band shows the
calculation of the colour-singlet contribution to next-to-leading order
in $\alpha_s$
\cite{Kramer:1994zi,Kramer:1995nb}. The NLO corrections increase 
the colour-singlet contribution by about a factor of two, so that it
accounts for the data quite well without the inclusion of a
colour-octet contribution.

Uncertainties in $m_c$ could lower the colour-singlet contribution by
about a factor of two, leaving more room for colour-octet
contributions.  In the experimental data, the cut $p_{T,\psi}>1$~GeV
is employed. One can question whether hard-scattering factorization is
valid at such small values of $p_{T,\psi}$. However, the data
differential in $p_{T,\psi}$ are compatible with colour-singlet
production alone at large $p_{T,\psi}$
(\Figure~\ref{fig:photo-prod-csm}).

The next-to-leading-order QCD corrections are crucial in describing the
shape of the transverse-momentum distribution of the $J/\psi$. The NLO
colour-singlet cross-section includes processes such as $\gamma + g \to
(c\bar{c}) + g  g $, which are dominated by $t$-channel gluon exchange
and scale as $\alpha_s^3 m_c^2 / p_{T,\psi}^6$. At
$p_{T,\psi}\;\rlap{\lower 3.5 pt \hbox{$\mathchar \sim$}} \raise 1pt
\hbox {$>$}\;m_c$ their contribution is enhanced with respect to the
leading-order cross-section, which scales as $\sim \alpha_s^2
m_c^4/p_{T,\psi}^8$. The comparison with the experimental data in
\Figure~\ref{fig:photo-prod-csm} confirms the importance of the
higher-order corrections.

\begin{figure}[t]
\begin{center}
\includegraphics[width=16cm]{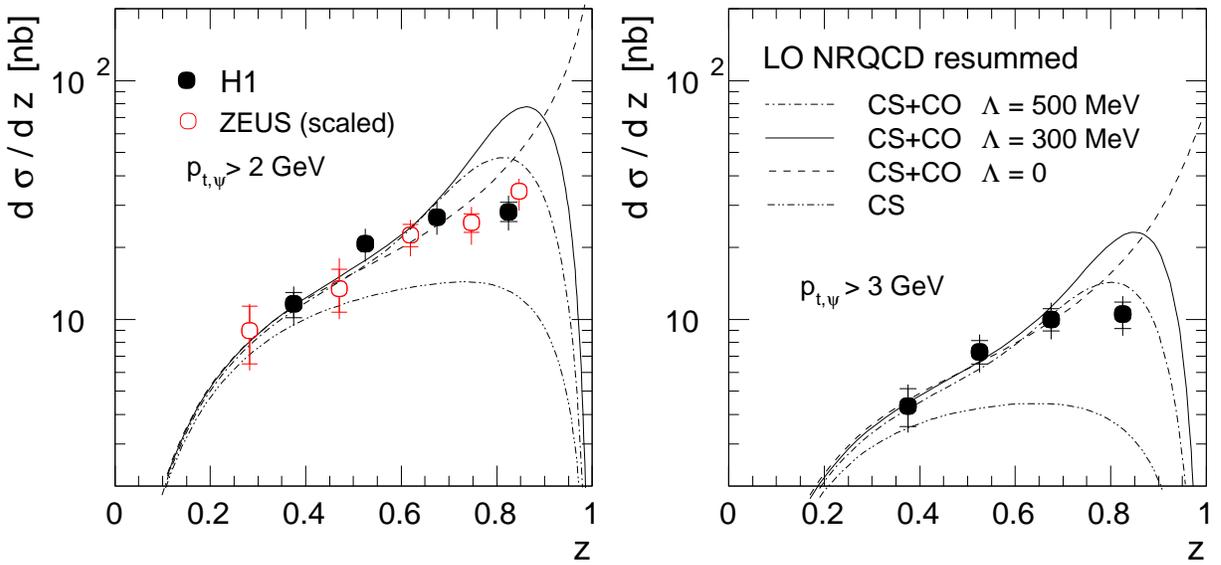}
\end{center}
\caption[Differential cross-sections ${\rm d}\sigma/{\rm d}z$
         ($60<W_{\gamma p}<240\,{\rm GeV}$) for $p_{T,\psi}>2\,{\rm
         GeV}$ and $p_{T,\psi}>3\,{\rm GeV}$]
        {Differential cross-sections ${\rm d}\sigma/{\rm d}z$
         ($60<W_{\gamma p}<240\,{\rm GeV}$) for $p_{T,\psi}>2\,{\rm
         GeV}$ (left panel) and $p_{T,\psi}>3\,{\rm GeV}$ (right
         panel) in comparison with NRQCD calculations that include
         colour-octet and colour-singlet contributions and resummations
         of soft contributions at high $z$ \cite{Beneke:1999gq}. The
         curves correspond to three values of the parameter $\Lambda$:
         $\Lambda=0$, \ie no resummation (dashed line),
         $\Lambda=300$~MeV (solid line), and $\Lambda= 500$~MeV
         (dash--dotted line). The theoretical curves have been scaled
         with a common factor 2 in the left panel and 3 in the right
         panel.}
\label{fig:gammap-beneke}
\end{figure}

At large $z$, the emission of soft gluons in the conversion of the
\ccbar\ pairs to \jpsi\ mesons is suppressed, owing to phase-space
limitations. Furthermore, the velocity expansion of the NRQCD
factorization approach is expected to break down \cite{Beneke:1997qw}.
These effects are not taken into account in the theoretical
calculation that is shown in \Figure~\ref{fig:photo-prod}.  In
Ref.~\cite{Beneke:1999gq}, a resummation of the nonrelativistic
expansion was carried out, leading to a decrease of the predicted
cross-section at high $z$.  The resummation involves a parameter
$\Lambda$ that describes the energy in the \ccbar\ rest frame that is
lost by the \ccbar\ system in its conversion into the \jpsi\ meson.
In \Figure~\ref{fig:gammap-beneke}, the measured cross-sections
$d\sigma/dz$ for ${p_{T,\psi}}>2$~GeV and for ${p_{T,\psi}}>3$~GeV are
compared with the results of these resummed calculations.  The
calculated curves have been roughly normalized to the data points at
low $z$. The resummed calculation for $\Lambda=500$~MeV gives an
acceptable description of the data at \ensuremath{p_{T,\psi}}$ > 3$
GeV, while the agreement between data and calculation is still poor
for \ensuremath{p_{T,\psi}}$ > 2$~GeV or for lower $\Lambda$ values.

Effects from resummation of logarithms of $1-z$ and model shape
functions have also been calculated for the process $e^+e^-\to
J/\psi+X$ \cite{Fleming:2003gt}. It may be possible to use this
resummed theoretical prediction to extract the dominant shape function
from the Belle and BaBar data for $e^+e^-\to J/\psi+X$ and then use it
to make predictions for $J/\psi$ photoproduction near $z=1$.

Measurements of the $J/\psi$ production cross-section at large $z$ are 
available from H1~\cite{Aktas:2003zi} and from ZEUS~\cite{Chekanov:2002at}. 
In this region, the contribution from diffractively produced $J/\psi$ 
mesons is expected to be large, as is discussed below 
in \Section~\ref{sec:subsechera:diffjpsi}.

\begin{figure}[t]
\begin{center}
\includegraphics[width=14cm]{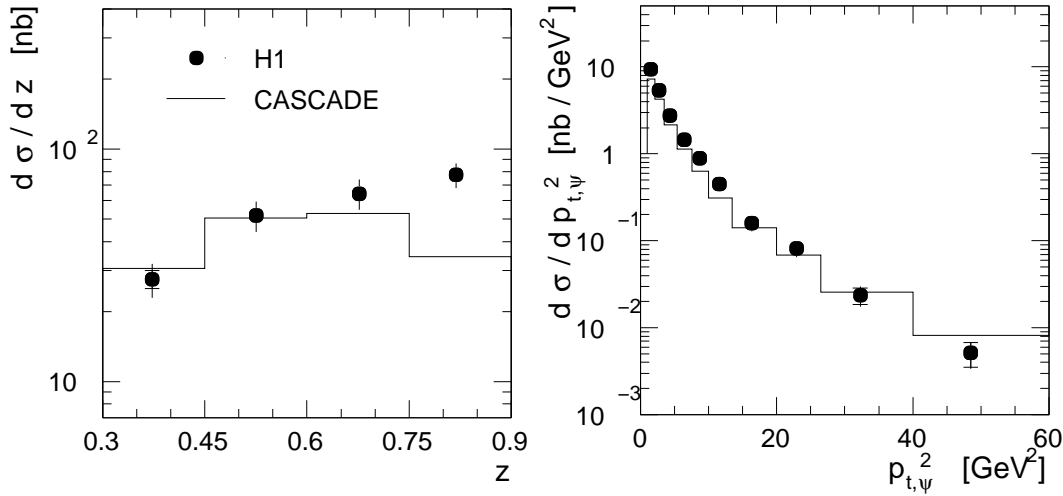}
\end{center}
\caption[Inelastic \jpsi\ production in the region $60<W_{\gamma
         p}<240$~GeV, $0.3<z<0.9$, and $p_{T,\psi}^2>1$~GeV$^2$]
        {Inelastic \jpsi\ production in the region $60<W_{\gamma
         p}<240$~GeV, $0.3<z<0.9$, and $p_{T,\psi}^2>1$~GeV$^2$, in
         comparison with a $k_T$-factorization model implemented in
         the Monte Carlo generator
         CASCADE\cite{Jung:2001hk,Jung:2001hx}. Left panel:
         differential cross-section ${\rm d}\sigma/{\rm d}z$; right
         panel: $d\sigma/d\ensuremath{p_{T,\psi}^2}$ in the range
         $0.3<z<0.9$.}
\label{fig:gammap-cascade}
\end{figure}

The ZEUS Collaboration has also measured the $\psi'$ to $J/\psi$
cross-section ratio \cite{Chekanov:2002at} in the range $0.55 < z <
0.9$ and $50 < W < 180$~GeV. It is found to be consistent with being
independent of the kinematic variables $z$, $p_{t,\psi}$ and $W$, as
is expected if the underlying production mechanisms for the $J/\psi$
and the $\psi'$ are the same.  An average value
$\sigma(\psi')/\sigma(J/\psi)= 0.33\pm0.10 ({\rm stat.})
^{+0.01}_{-0.02} ({\rm syst.})$ is found which compares well with the
prediction from the leading-order colour-singlet model
\cite{Kramer:1994zi}.

The $k_T$-factorization approach %~\cite{Catani:1990eg,Collins:1991ty}
(see \Section~\ref{sec:prodsec-nrqcdmge}) has recently been applied
successfully to the description of a variety of
processes\cite{Jung:2001hk,Jung:2001hx,Saleev:1994fg}. In this
approach, the \jpsi\ production process is factorized into a
$k_T$-dependent gluon density and a matrix element for off-shell
partons. A leading-order calculation within this approach is
implemented in the Monte Carlo generator
CASCADE~\cite{Jung:2001hk,Jung:2001hx}.
\Figure[b]~\ref{fig:gammap-cascade} shows a comparison of the data with
the predictions from the $k_T$-factorization approach. Good agreement
is observed between data and predictions for $z\lsim 0.8$. At high $z$
values, the CASCADE calculation underestimates the cross-section. This
may be due to missing higher-order effects, or missing relativistic
corrections, which are not available for the off-shell matrix element.
It could also indicate a possible missing colour-octet contribution.
The CASCADE predictions for the the \ensuremath{p_{T,\psi}^2}
dependence of the cross-section (\Figure~\ref{fig:gammap-cascade}c)
fit the data considerably better than the collinear LO
calculations. This improved fit is due to the transverse momentum of
the gluons from the proton, which contributes to the transverse
momentum of the \jpsi\ meson. Note, however, that the NLO
colour-singlet calculation in collinear factorization
\cite{Kramer:1995nb} also describes the \ensuremath{p_{T,\psi}^2}
distribution.

The polarization of the \jpsi\ meson is expected to differ in the
various theoretical approaches discussed here and could in principle
be used to distinguish between them, independently of normalization
uncertainties. The general decay angular distribution can be
parametrized as
\begin{equation}
  \frac{d\Gamma(J/\psi\to l^+l^-)}{d\Omega}
  \propto
  1 + \lambda \cos^2\theta + \mu \sin 2\theta \cos\phi
  + \frac{\nu}{2} \sin^2\theta \cos 2\phi,
\end{equation}
where $\theta$ and $\phi$ refer to the polar and azimuthal angle of the
$l^+$ three-momentum with respect to a coordinate system that is defined
in the $J/\psi$ rest frame. (See, for example, Ref.~\cite{Beneke:1998re}
for details.) The parameters $\lambda, \mu, \nu$ can be calculated
within NRQCD or the CSM as a function of the kinematic variables, 
such as $z$ and $p_{T,\psi}$.

\begin{figure}[p]
\begin{center}
\includegraphics[width=.48\textwidth]{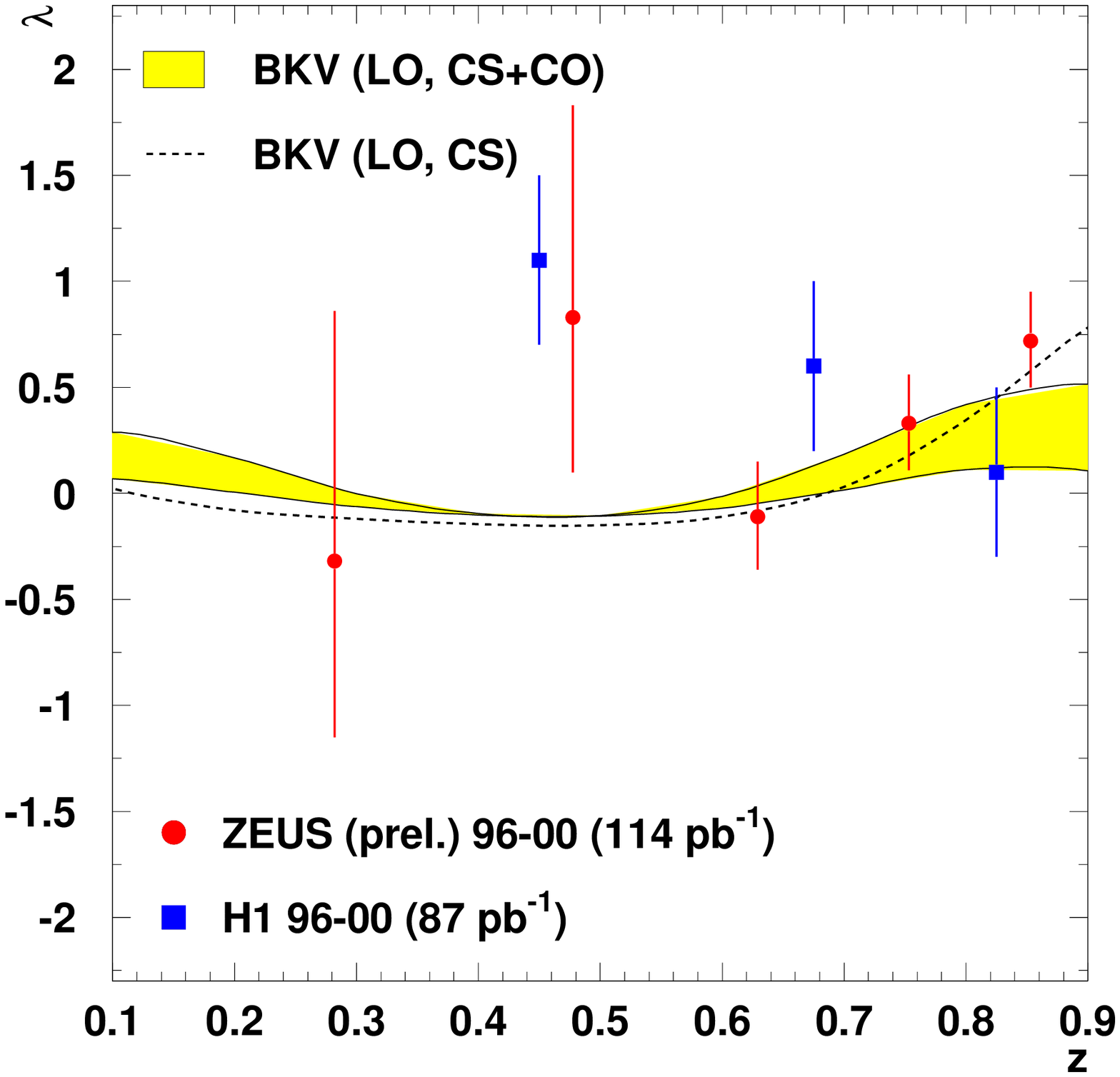}\hfill
\includegraphics[width=.48\textwidth]{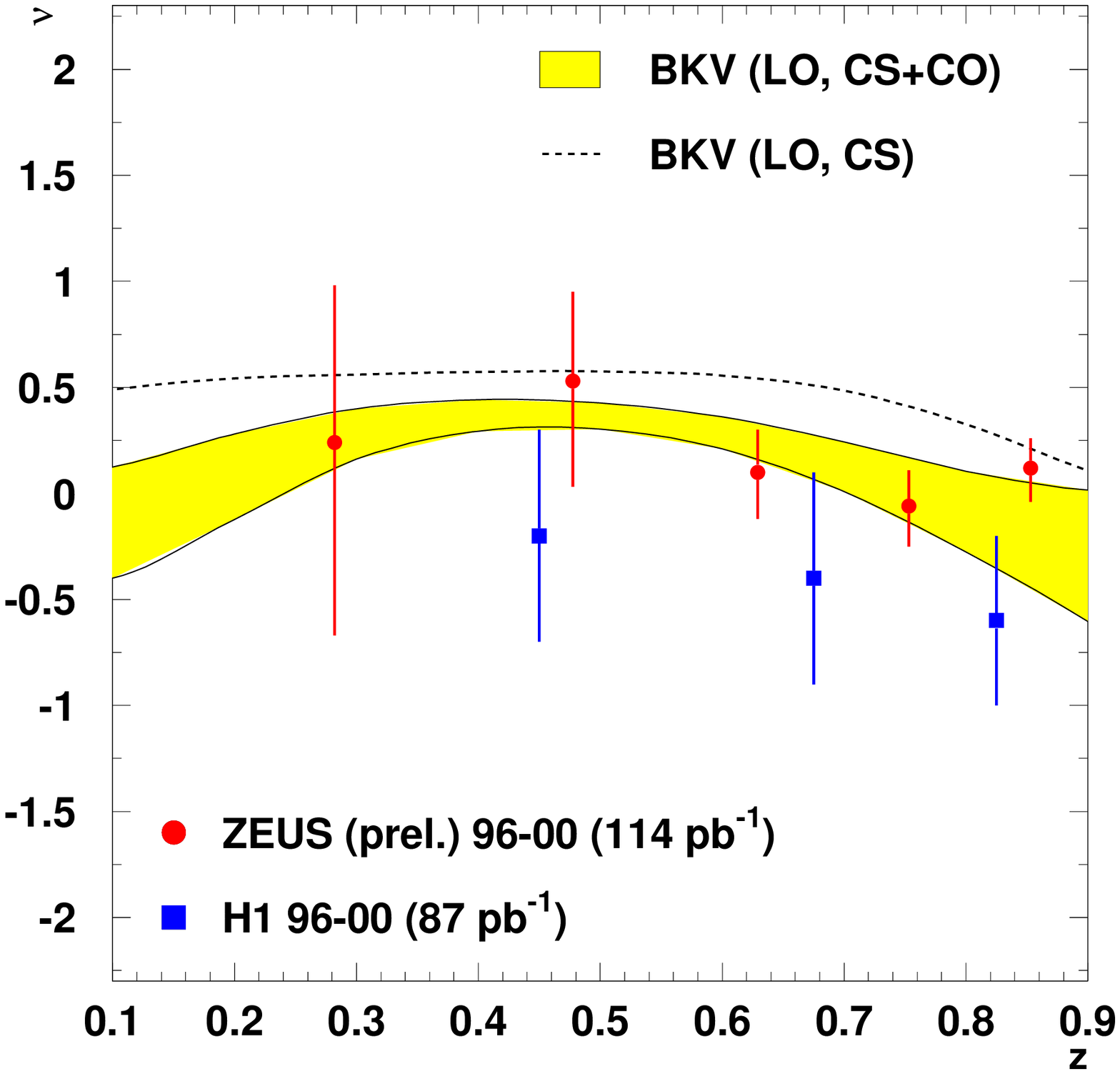}\\[4mm]
\includegraphics[width=.48\textwidth]{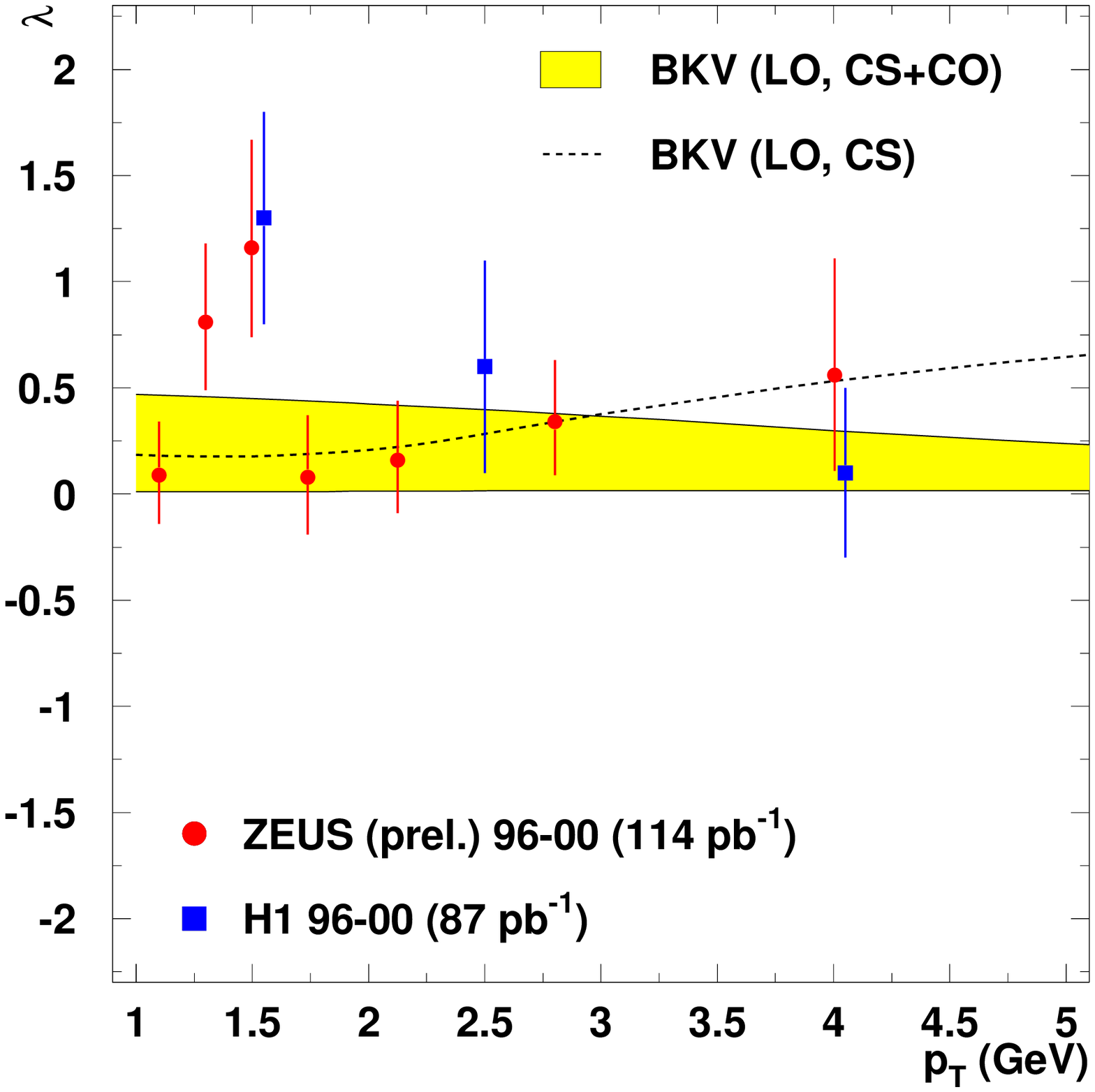}\hfill
\includegraphics[width=.48\textwidth]{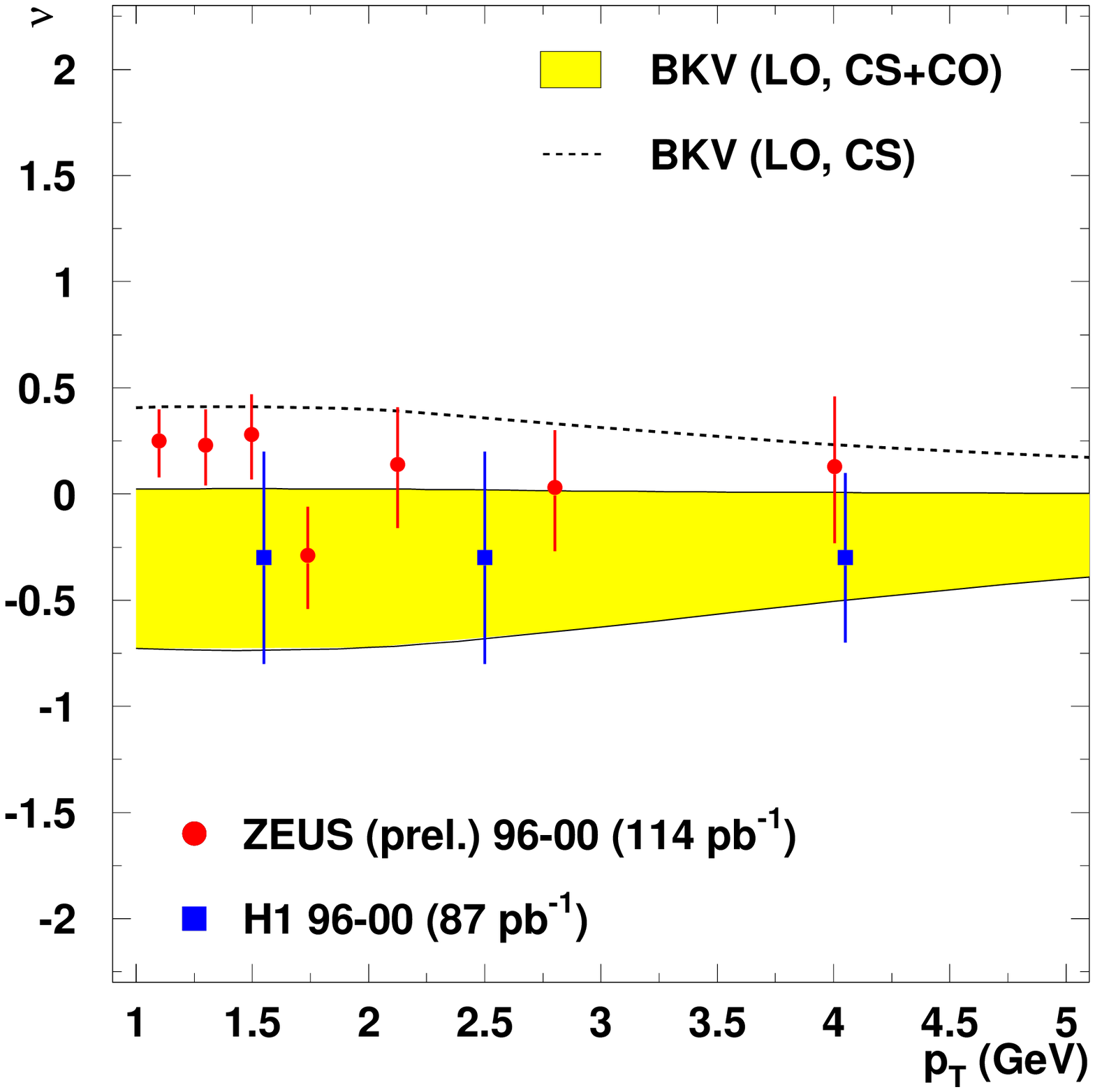}
\end{center}
\caption[Polarization parameters $\lambda$ and $\nu$]
        {Polarization parameters $\lambda$ (left panels) and $\nu$
         (right panels) in the target rest frame as functions of $z$
         (top panels) and \ensuremath{p_{T,\psi}} (bottom panels). The
         error bars on the data points correspond to the total
         experimental error. The theoretical calculations shown are
         from the NRQCD approach \cite{Beneke:1998re} (shaded bands)
         with colour-octet and colour-singlet contributions, while the
         curves show the result from the colour-singlet contribution
         separately.  }
\label{fig:gammap-pola}
\end{figure}

In \Figure~\ref{fig:gammap-pola}, the data are shown, together with
the results from two LO calculations: the NRQCD prediction, including
colour-octet and colour-singlet contributions \cite{Beneke:1998re}, and
the colour-singlet contribution alone.  A calculation that uses a
$k_T$-factorization approach and off-shell gluons is also
available~\cite{Baranov:1998af}.  In contrast to the predictions shown
in the \Figure~\ref{fig:gammap-pola}, in which $\lambda$ is zero or
positive, the prediction of the $k_T$-factorization approach is that
$\lambda$ should become increasingly negative toward larger values of
$p_{t,J/\psi}$, reaching $\lambda \sim -0.5$ at $p_{T,\psi}=6$~GeV.
However, at present, the errors in the data preclude any firm
conclusions. In this range of $p_{T,\psi}$ none of the calculations
predicts a decrease in $\lambda$ with increasing $z$.  In order to
distinguish between full NRQCD and the colour-singlet contribution
alone, measurements at larger \ensuremath{p_{T,\psi}} are
required. The measured values of $\nu$, for which no prediction is
available from the $k_T$-factorization approach, favor the full NRQCD
prediction.

\subsection{Inelastic electroproduction of charmonium}
\label{sec:prodsec-hera:dis}

As in photoproduction, inelastic leptoproduction of $J/\psi$ mesons at
HERA ($e+p\rightarrow e+J/\psi+X$) is dominated by virtual-photon-gluon
fusion ($\gamma^*g\rightarrow \ccbar$). In leptoproduction, or deep
inelastic $ep$-scattering (DIS), the exchanged photon has a nonzero
virtuality $Q^2=-q^2$, where $q$ is the four-momentum of the virtual
photon. For events with a photon virtuality of $Q^2\gtrsim 1$~GeV$^2$,
the electron scattering angle is large enough for the electron to be
detected in the central detectors.

The analysis of leptoproduction at finite $Q^2$ has experimental and
theoretical advantages in comparison with the analysis of
photoproduction. At high $Q^2$, theoretical uncertainties in the models
decrease and resolved-photon processes are expected to be negligible.
Furthermore, the background from diffractive production of charmonia is
expected to decrease faster with $Q^2$ than the inelastic process, and
the distinct signature of the scattered lepton makes the inelastic
process easier to detect.

\begin{figure}[p]
\centering\includegraphics[width=.65\linewidth]{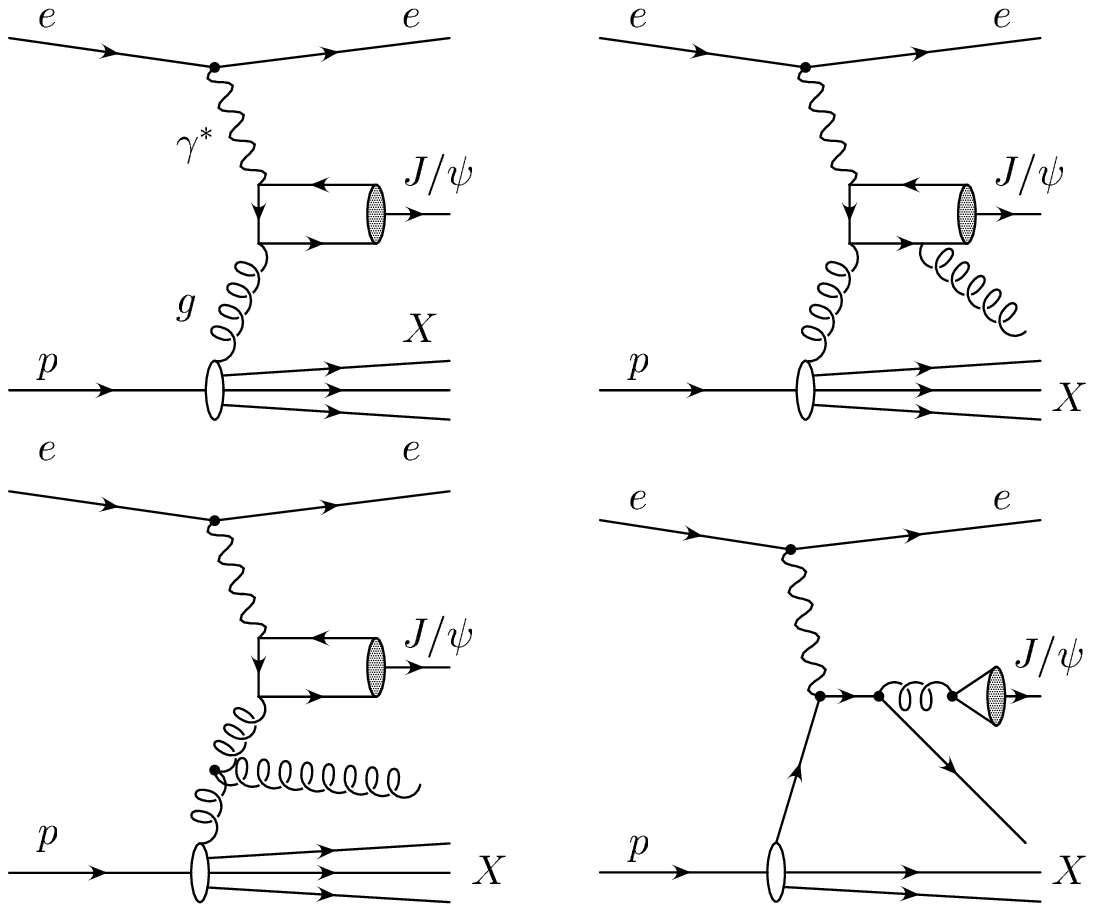}
\caption{Generic diagrams for charmonium production 
 mechanisms: photon--gluon fusion via a ``$2\rightarrow 1$''
 process (top left diagram) and ``$2\rightarrow 2$'' processes 
 (remaining diagrams). 
 All the diagrams contribute via colour-octet mechanisms, 
 while the top right diagram can also contribute
 via the colour-singlet mechanism.  Additional soft gluons emitted during
 the hadronization process are not shown.}
\label{fig:ep-kniehl}

\bigskip

\centering\includegraphics[width=.65\linewidth]{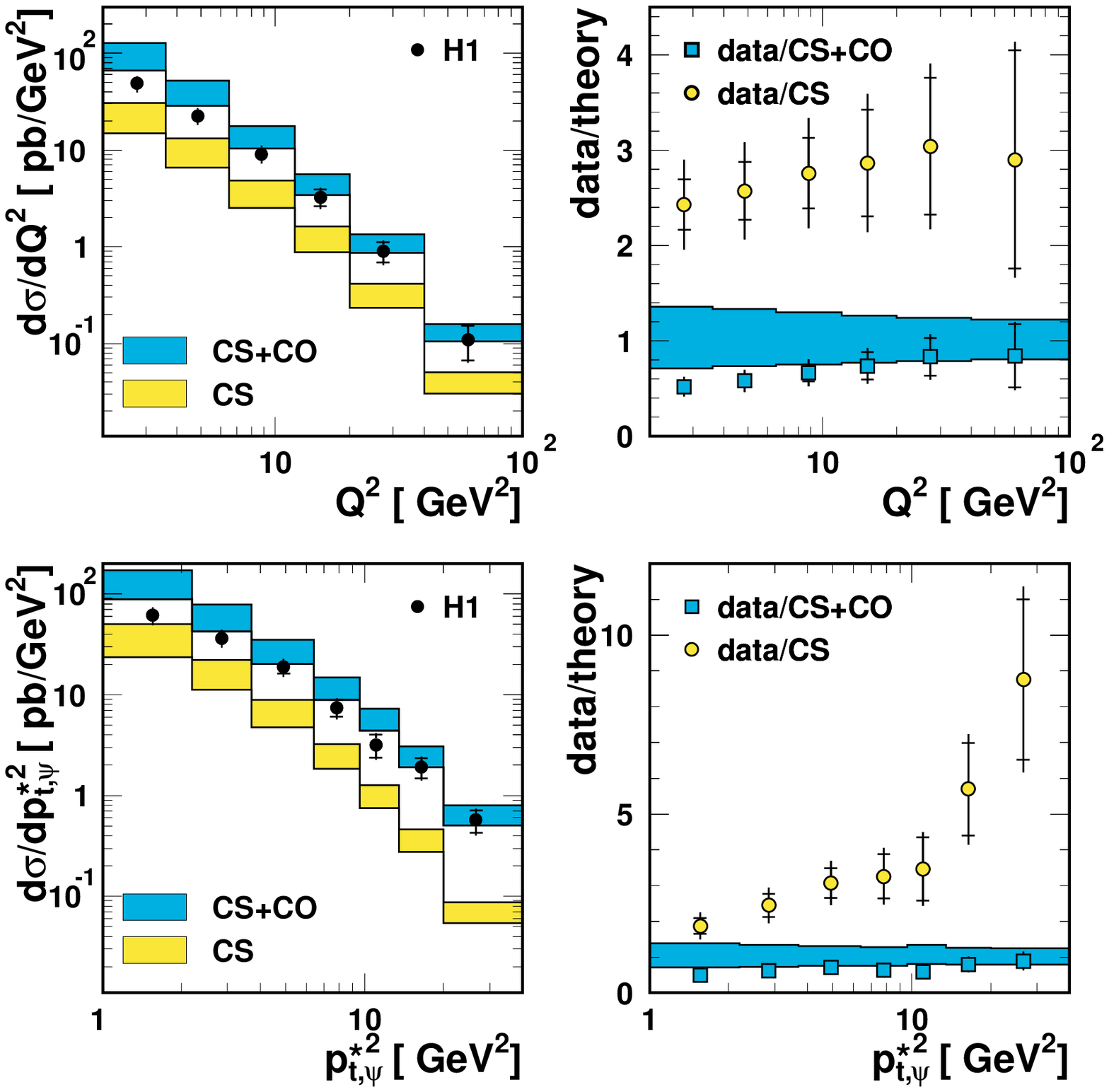}
\caption[Differential cross-sections $d\sigma / dQ^2$ 
         and $d\sigma/dp_{T,\psi}^{*2}$]
        {Differential cross-sections $d\sigma / dQ^2$ (top left panel)
         and $d\sigma/dp_{T,\psi}^{*2}$ (bottom left panel) and the
         corresponding ratios of data to theory (right panels).  The
         data from H1~\cite{Adloff:2002ey} are compared with the NRQCD
         calculation~\cite{Kniehl:2001tk} (CO+CS, dark band) and the
         colour-singlet contribution~\cite{Kniehl:2001tk} (CS, light
         band).}
\label{fig:ep-q2}
\end{figure}

A first comparison between data and NRQCD calculations was presented
in Ref.~\cite{Adloff:1999zs}.  The NRQCD calculations in
Ref.~\cite{Adloff:1999zs} were performed by taking into account only
``$2\rightarrow 1$'' diagrams (see the top left diagram of
\Figure\ref{fig:ep-kniehl}) \cite{Fleming:1997fq}, and disagreement
between data and theory was observed both in the absolute values of
the cross-sections and in their shapes as functions of the variables
that were studied.

More recently, the cross-section for $J/\psi$ production in
deep-inelastic $ep$ scattering at HERA was calculated in the NRQCD
factorization approach at leading order in $\alpha_s$ by Kniehl and
Zwirner \cite{Kniehl:2001tk}, taking into account diagrams of the type
``$2\rightarrow 2$'', as are shown in the top right and bottom diagrams
of \Figure~\ref{fig:ep-kniehl}. The calculation made use of the matrix
elements of Ref.~\cite{Braaten:1999qk} and MRST98LO \cite{Martin:1998sq}
and CTEQ5L \cite{Lai:1999wy} parton distributions.

\begin{figure}
\begin{center}
\includegraphics[width=65mm]{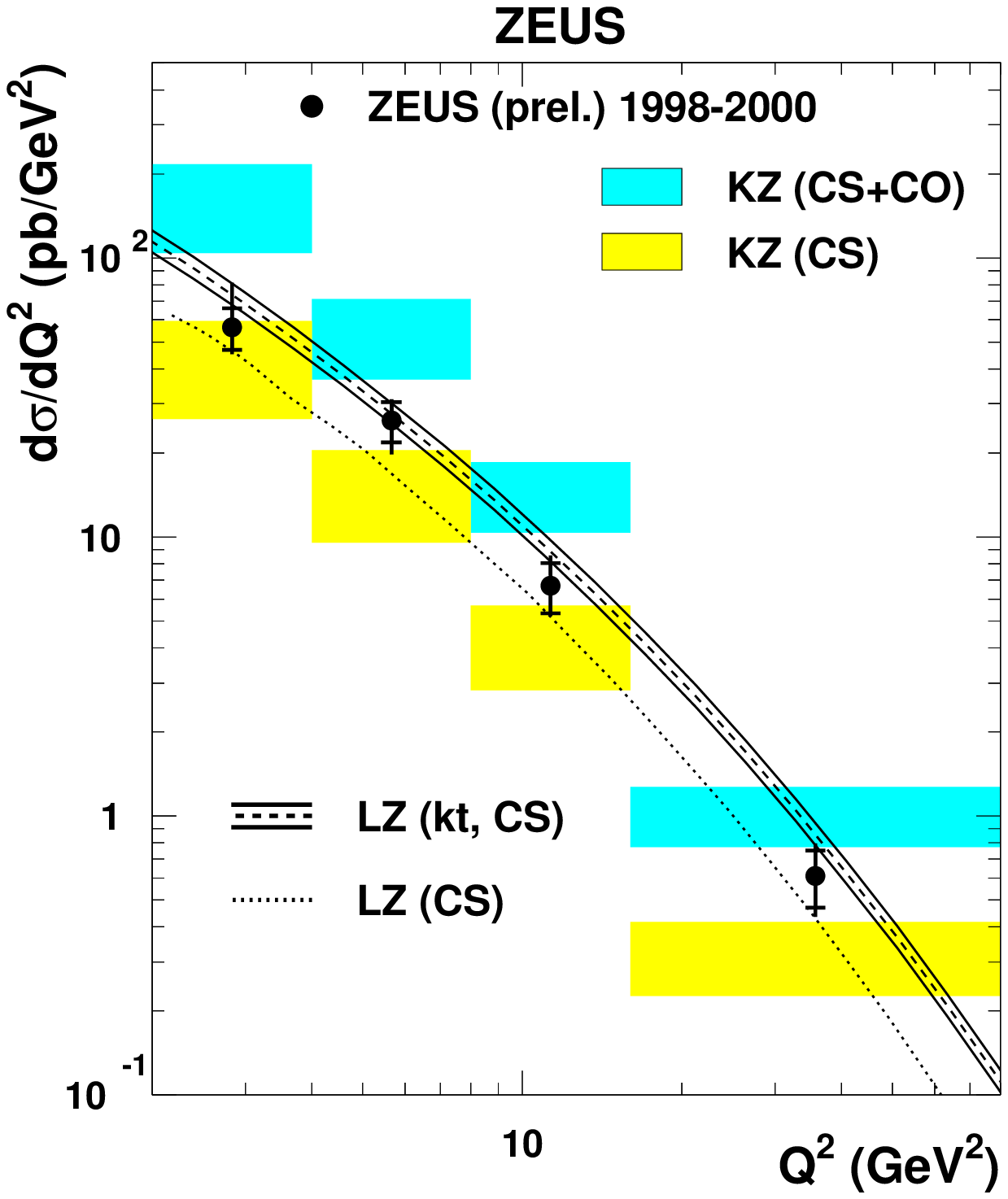}\qquad
\includegraphics[width=65mm]{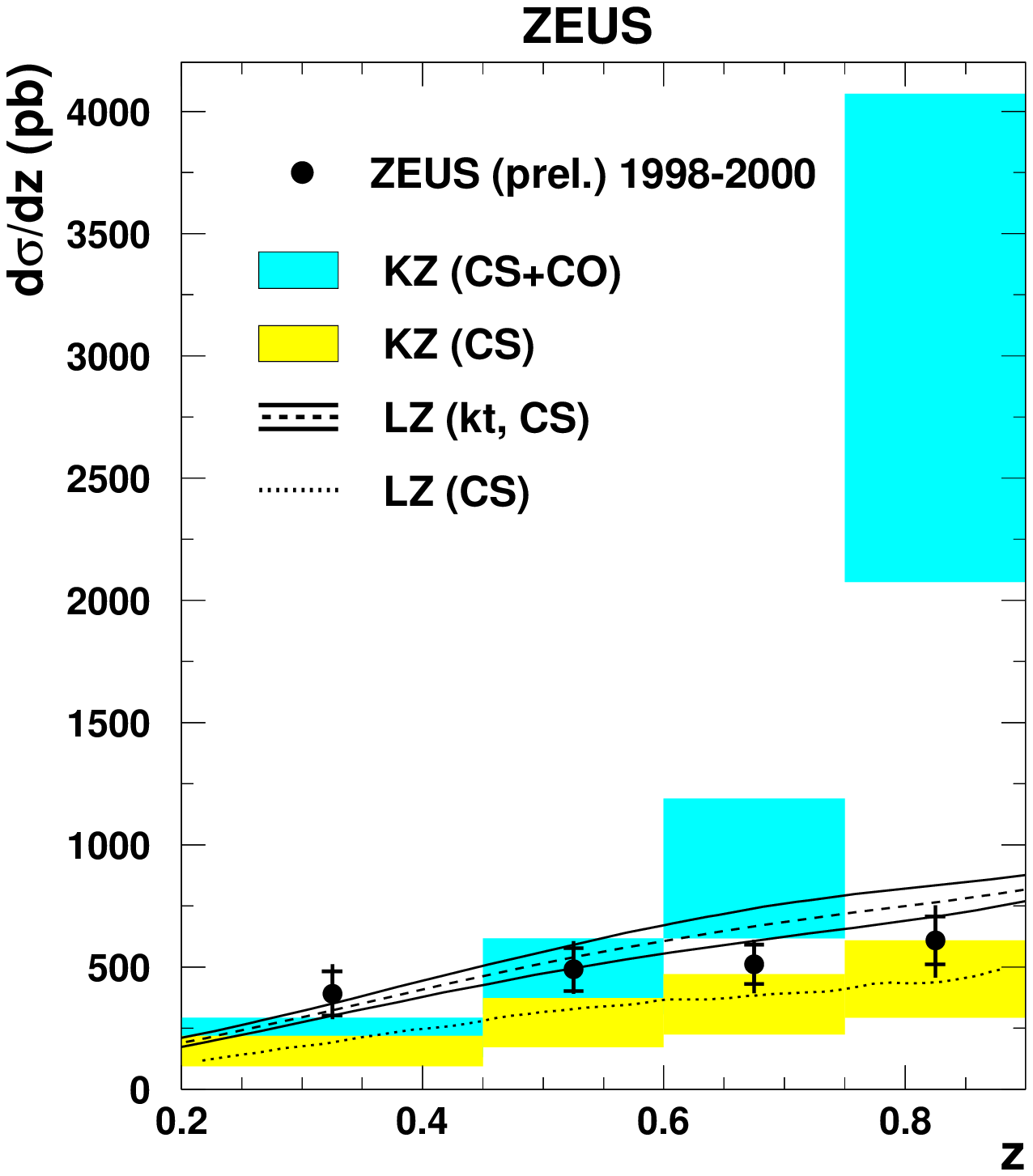}
\end{center}
\caption[Differential cross-sections $d\sigma / dQ^2$ 
         and $d\sigma/dz$ and theory predictions]
        {Differential cross-sections $d\sigma / dQ^2$ (left) and
         $d\sigma/dz$ (right) and theory predictions.  The data from
         ZEUS~\cite{zeus-eps03-565} are compared with the NRQCD
         calculation~\cite{Kniehl:2001tk} (CO+CS, dark band), the
         colour-singlet contribution (CS, light band), and with the
         prediction LZ(kt,CS) from the $k_T$-factorization approach
         within the CSM~\cite{Lipatov:2002tc}. The solid lines delimit
         the uncertainties, and the dashed line show the central
         values. The CSM prediction LZ (CS) in the
         collinear-factorization approach, as given by the authors of
         Ref.~\cite{Lipatov:2002tc}, is also shown (dotted line). }
\label{fig:zeus-q2z}
\end{figure}

\begin{figure}[p]
\begin{center}
\includegraphics[height=.87\textheight]{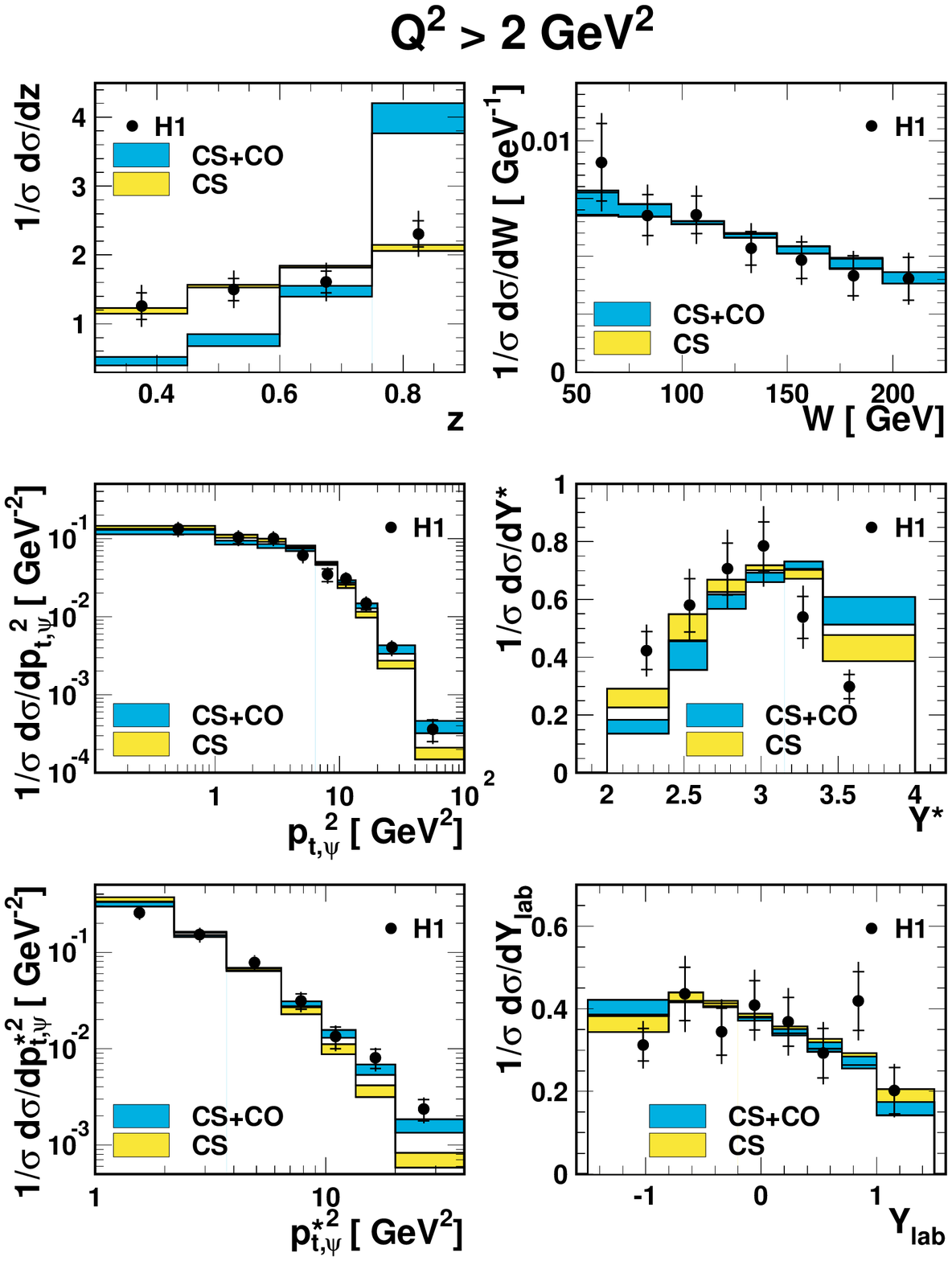}
\end{center}
\caption[Normalized differential cross-sections]
        {Normalized differential cross-sections. $1/\sigma\,d\sigma /
         dz$ (top left panel), $1/\sigma\,d\sigma / dW$ (top right
         panel), $1/\sigma\,d\sigma / dp_{T,\psi}^2$ (middle left
         panel), $1/\sigma\,d\sigma /dY^\ast$ (middle right panel),
         $1/\sigma\,d\sigma / dp_{T,\psi}^{*2}$ (bottom left panel),
         and $1/\sigma\,d\sigma /dY_{lab}$ (bottom right panel).  The
         histograms show calculations for inelastic $J/\psi$
         production within the NRQCD factorization
         approach\protect\cite{Kniehl:2001tk}, which have been
         normalized to the integrated cross-section. The dark band
         represents the sum of CS and CO contributions, and the light
         band represents the CSM contribution alone.  These
         contributions are normalized separately. The error bands
         reflect the theoretical uncertainties.}
\label{fig:ep-z}
\end{figure}

In \Figure~\ref{fig:ep-q2}, the results of this calculation are plotted
as a function of $Q^2$ and $p^2_{T,\psi}$, along with the H1 data
\cite{Adloff:2002ey}. The NRQCD results that are shown in
\Figure~\ref{fig:ep-q2} include the contributions from the colour-octet
channels $^{3}\!S_1$, $^{3}\!P_{J=0,1,2}$, $^{1}\!S_0$, as well as
from the colour-singlet channel $^{3}\!S_1$. The contribution of the
colour-singlet channel is also shown separately. The values of the
NRQCD matrix elements were determined from the distribution of
transverse momenta of \jpsi\ mesons produced in \ppbar\
collisions~\cite{Braaten:1999qk}.%
\footnote{\label{fn} The extracted values for the NRQCD matrix
          elements depend on the parton distributions.  For the set
          MRST98LO \cite{Martin:1998sq}, the values are $\langle {\cal
          O}_1^{J/\psi}(^3\!S_1)\rangle=1.3\pm 0.1 \,$GeV$^3$,
          $\langle {\cal O}_8^{J/\psi}(^3\!S_1)\rangle=(4.4\pm0.7)
          \times 10^{-3} \,$GeV$^3$ and $M_{3.4}^{J/\psi}=(8.7\pm 0.9)
          \times 10^{-2}\,$GeV$^3$, where $M_{3.4}^{J/\psi}$ is the
          linear combination of two NRQCD matrix elements that is
          defined in \Eq~(\ref{eq:prod-lincomb}).}
The bands include theoretical uncertainties, which originate from the
uncertainty in the charm-quark mass $m_c=1.5\pm0.1$~GeV, the variation
of renormalization and factorization scales by factors 1/2 and 2, and
the uncertainties in the NRQCD matrix elements, all of which result
mainly in normalization uncertainties that do not affect the shapes of
the distributions.

\Figure[b]~\ref{fig:zeus-q2z} shows the differential electroproduction
cross-sections for $J/\psi$ mesons as functions of $Q^2$ and $z$, as
measured by the ZEUS collaboration~\cite{zeus-eps03-565}.  The ZEUS
data, which are consistent with the H1 results shown in
\Figure~\ref{fig:ep-q2}, are compared with predictions in the framework of
NRQCD (CS+CO)~\cite{Kniehl:2001tk} and also with predictions in the
$k_T$-factorization approach in which only the colour-singlet
contribution (CS) is included~\cite{Lipatov:2002tc}.  As in
\Figure~\ref{fig:ep-q2}, the uncertainties in the NRQCD calculations are
indicated in \Figure~\ref{fig:zeus-q2z} as shaded bands.  For the
prediction within the $k_T$-factorization approach (LZ(kt,CS)), only
one of the sources of uncertainty is presented, namely the uncertainty
in the pomeron intercept $\Delta$, which controls the normalization of
the unintegrated gluon density.

In \Figure~\ref{fig:ep-z}, the normalization uncertainties of the
theory, which are dominant, are removed by normalizing the
differential cross-sections measured by H1~\cite{Adloff:2002ey} and
the theory predictions to the integrated cross-sections in the
measured range for each distribution.  The comparisons in
\Figures~\ref{fig:ep-q2}--\ref{fig:ep-z} indicate that the
colour-singlet contribution follows the shape of the data from H1 and
ZEUS quite well.  In general, the CSM predictions are below the H1 and
ZEUS data, but are consistent with the data, given the uncertainties,
both in shape and normalization.  However, the differential
cross-sections as a function of the transverse momentum squared of the
$J/\psi$ are too steep compared to the data (lower left plot in
\Figure~\ref{fig:ep-q2}).  A similar observation was made for
photoproduction (\Section~\ref{sec:prodsec-hera:gp},
\Figure~\ref{fig:photo-prod-csm}), in which the LO CSM calculation is
too steep and the NLO CSM calculation is found to describe the data
well.  The $z$ distribution (\Figures~\ref{fig:zeus-q2z} and
\ref{fig:ep-z}) is very poorly described by the full calculation that
includes colour-octet contributions, while the colour-singlet
contribution alone reproduces the shape of the data rather well.  The
failure of the colour-octet calculations could be due to the fact that
resummations of soft-gluons are not included here.  It is worth noting
that the calculation of Kniehl and Zwirner disagrees with a number of
previous results
\cite{Korner:1982fm,Guillet:1987xr,Merabet:sm,Krucker:1995uz,Yuan:2000cn},
which themselves are not fully consistent.

\subsection{Diffractive vector meson production}
\label{sec:subsechera:diffjpsi}

\begin{figure}
\begin{center}
\includegraphics[width=7cm]{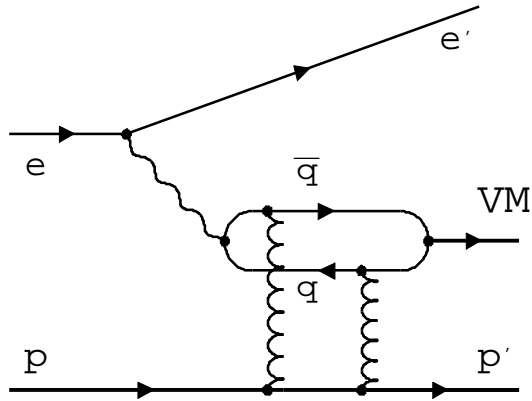}
\end{center}
\caption{Diagram for diffractive charmonium production via exchange of two gluons
 in a colour-singlet state.}
\label{fig:ep-feyn-diff}
\end{figure}

\begin{figure}
\includegraphics[width=16cm]{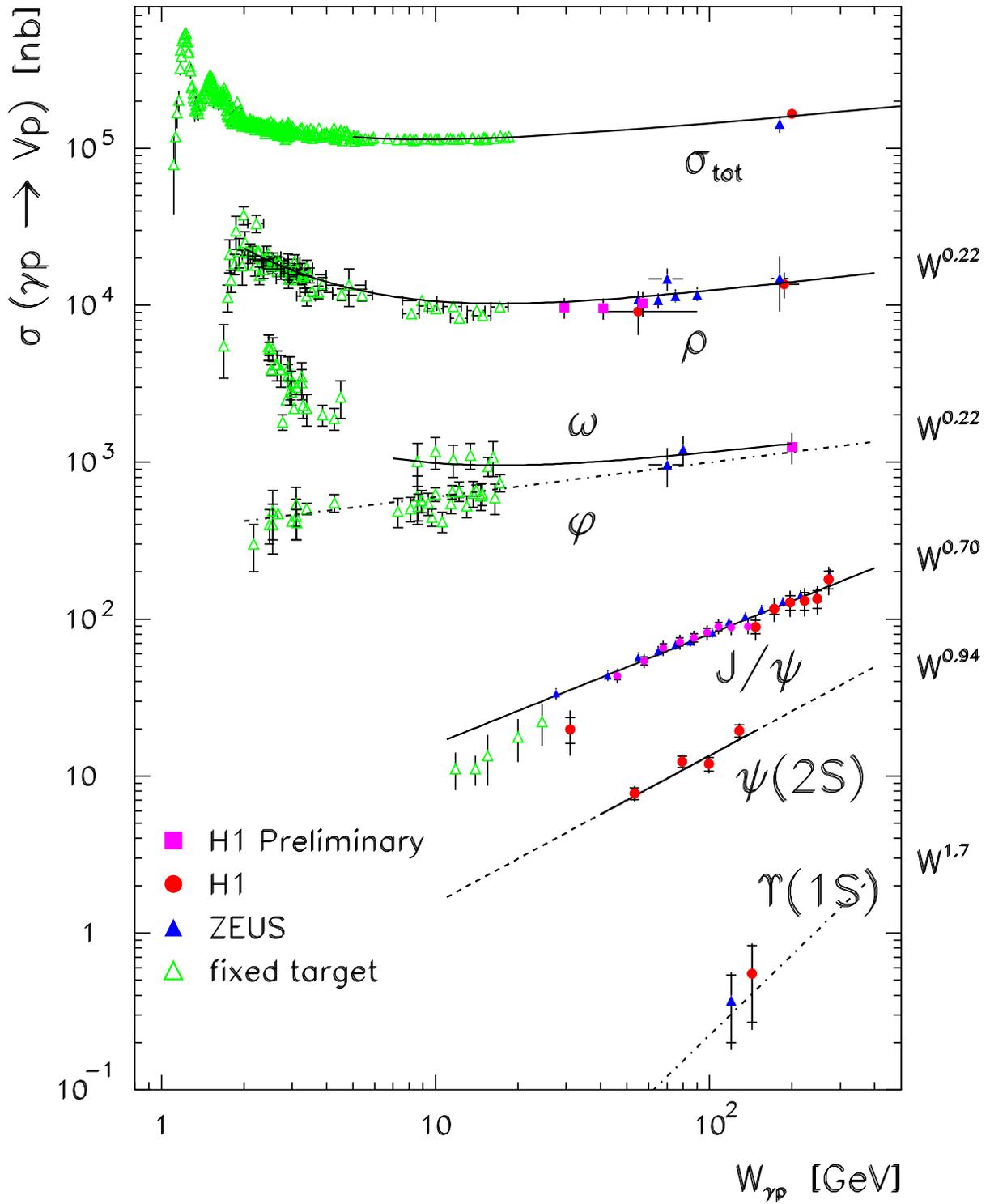}
\caption{Total cross-section and cross-sections for production of 
various vector mesons in $\gamma p$ collisions as a function of 
$W_{\gamma p}$, as measured at HERA and in fixed-target experiments.}
\label{fig:gp-vmdiffractive}
\end{figure}

At HERA, the dominant production channel for quarkonia with quantum
numbers of real photons (\ie $J^{PC}=1^{--}$) is through diffractive
processes.
In perturbative QCD, the diffractive production of vector mesons can
be modeled in the proton rest frame by a process in which the photon
fluctuates into a $q\bar{q}$ pair at a long distance from the proton
target.  The $q\bar{q}$ subsequently interacts with the proton via a
colour-singlet exchange, \ie in lowest order QCD via the exchange of a
pair of gluons with opposite colour (see
\Figure~\ref{fig:ep-feyn-diff})~\cite{Ryskin:1992ui,
Brodsky:1994kf, %
Collins:1996fb, %
Collins:1997sr, %
Teubner:1999pm, %
Bartels:2000ze, % confinement
Hayashigaki:2004iw}.
At small $|t|$, where $t$ is the
momentum-transfer-squared at the proton vertex, the elastic process in
which the proton stays intact dominates.  Toward larger values of
$|t|$, the dissociation of the proton into a small-invariant-mass
state becomes dominant. Measurements of diffractive vector-meson
production cross-sections and helicity structure from the
H1~\cite{Aid:1996bs,Adloff:1999kg, Adloff:1999zs,Adloff:2000vm,
Adloff:2000nx,Adloff:2002tb,
Adloff:2002re,Aktas:2003zi,h1-ichep04-6-0180} and
ZEUS~\cite{Breitweg:1998nh,Breitweg:1998ki,
Breitweg:1999jy,Breitweg:1999fm,Breitweg:2000mu,Chekanov:2002rm,
Chekanov:2002xi,Chekanov:2004mw} collaborations are available for
$\rho^0$, $\omega$, $\phi$, $J/\psi$, $\psi'$, and $\Upsilon$
production, spanning the ranges of $0 \simeq Q^2 < 100 $~GeV$^2$, $ 0
\simeq |t| < 20 $~GeV$^2$, and $ 20 < W_{\gamma p} < 290 $~GeV.
($W_{\gamma p}$ is the $\gamma p$ centre-of-mass energy.) In
\Figure~\ref{fig:gp-vmdiffractive}, the elastic photoproduction
cross-sections are shown.  Perturbative calculations in QCD are
available for the kinematic regions in which at least one of the
energy scales $\mu^2$ (\ie $Q^2$, $M_V^2$ or $|t|$) is large and the
strong-coupling constant $\alpha_s(\mu^2)$ is
small~\cite{Frankfurt:1997fj,%
McDermott:1999fa,%
Frankfurt:2000ez,% gamma p (high W)
Ryskin:1995hz,% gluon density
Martin:1997wy,% skewed parton densities
Martin:1999wb,% sigma_l/sigma_t
Ivanov:2004vd% dipole calculations
}.

In the presence of such a `hard' scale, QCD predicts a steep rise of the
cross-section with $W_{\gamma p}$.  At small $Q^2$, $|t|$ and meson
masses $M_V$, vector-meson production is known to show a
non-perturbative ``soft'' behavior that is described, for example, by
Regge-type models~\cite{Regge:mz, Regge:1960zc, Sakurai:1969ss,
% VDM 
Donnachie:1992ny,Donnachie:1998gm}.
Toward larger values of $|t|$, in the leading logarithmic approximation,
diffractive $J/\psi$ production can be described by the effective
exchange of a gluonic ladder. At sufficiently low values of Bjorken $x$
(\ie large values of $W_{\gamma p}$), the gluon ladder is expected to
include contributions from BFKL
evolution~\cite{FKL:1976, Kuraev:fs, BL:1978, Mueller:1994jq, % bfkl
Nikolaev:1993th % qcd pomeron 
}, as well as from DGLAP evolution~\cite{Gribov:ri}.

Experimentally, diffractive events are generally distinguishable from
inelastic events, since, aside from meson-decay products, only a few
final-state particles are produced in the central rapidity range in
proton dissociation and no particles are produced in the central
rapidity range in elastic diffraction.
The elasticity variable $z$ defined in \Eq~(\ref{eq:zdef}) 
is often used to demark
the boundary between the elastic and inelastic regions, with a typical
demarcation for $J/\psi$ production being $z>0.95$ for the diffractive
region and $z<0.95$ for the inelastic region. However, at large $z$,
there is actually no clear distinction between inelastic $J/\psi$
production and diffractive $J/\psi$ production 
in which the proton dissociates
into a final state with large invariant mass, owing to the fact that the
two processes can produce the same final-state particles.
In the region of large $z$, both inelastic and
diffractive processes are expected to contribute to the cross-section.
In calculations that are based on the NRQCD factorization approach, the
cross-section increases toward large $z$, owing to large contributions
from colour-octet $c \bar c$ pairs, as is explained in
\Section~\ref{sec:prodsec-hera:gp}. These contributions are, however,
substantially reduced when one takes into account multiple soft gluon
emission, \eg in resummation calculations~\cite{Beneke:1999gq}. 
At the same time, calculations in perturbative QCD that assume a
diffractive colour-singlet exchange are capable of describing the
production cross-sections at large
$z$~\cite{Aktas:2003zi,Chekanov:2002rm,zeus-eps03-549}. 
A unified description in QCD
of the large $z$ region, taking into account both inelastic and
diffractive contributions, has yet to be developed.

\subsection{Prospects for the upgraded HERA collider}

With the HERA luminosity upgrade, a wealth of new quarkonium data will
become available. The existing \jpsi\ and $\psi(2S)$ measurements can
be improved and extended into new kinematic regions, and other
quarkonium final states may become accessible.  The future analyses of
quarkonium production at HERA offer unique possibilities to test the
theoretical framework of NRQCD factorization. 
It should be noted here that calculations to next-to-leading order, which
are not yet available in the framework of NRQCD factorization, could 
be an essential ingredient in a 
full quantitative understanding of charmonium production at HERA, and
also at other experiments, such as those at the Tevatron.
Some of the most interesting reactions will be discussed briefly below. See
Refs.~\cite{Cacciari:1996dy,Kramer:2001hh} for more details.

The measurement of inelastic {\em $\chi_c$ photoproduction} is a
particularly powerful way to discriminate between NRQCD and the
colour-evaporation model. The assumption of a single, universal
long-distance factor in the colour-evaporation model implies a
universal $\sigma[\chi_c]/\sigma[J/\psi]$ ratio.  A large $\chi_c$
cross-section is predicted for photon--proton collisions.  The ratio
of $\chi_c$ production to $J/\psi$ production is expected to be
similar to that at hadron colliders, for which $\sigma[\chi_c] /
\sigma[J/\psi] \approx 0.5$~\cite{Abe:1997yz}. In NRQCD, on the other
hand, the $\sigma[\chi_c]/\sigma[J/\psi]$ ratio is process-dependent
and strongly suppressed in photoproduction. Up to corrections of
${\cal O}(\alpha_s,v^2)$ one finds that~\cite{Kramer:2001hh}
\begin{equation}
\frac{\sigma[\gamma p \to \chi_{cJ}\,X]}
     {\sigma[\gamma p \to J/\psi \,X]}
\approx \frac{15}{8}\, (2J+1)\, \frac{\langle {\cal O}^{\chi_{c0}}_8
   (^3\!S_1)\rangle}
   {\langle {\cal O}^{J/\psi}_1(^3\!S_1)\rangle}
 \approx (2J+1)\,0.005 ,
\end{equation}
where the last approximation makes use of the NRQCD matrix elements
that are listed in \Table~\ref{tab:me-1}. A search for $\chi_c$
production at HERA that results in a cross-section measurement or an
upper limit on the cross-section would probe directly the colour-octet
matrix element $\langle {\cal O}^{\chi_J}_8(^3\!S_1)\rangle$ and would
test the assumption of a single, universal long-distance factor that is
implicit in the colour-evaporation model.

The inclusion of colour-octet processes is crucial in describing the
photoproduction of the {\em spin-singlet states} $\eta_c(1S)$,
$\eta_c(2S)$, and $h_c(1P)$. With regard to the P-wave state $h_c$,
the colour-octet contribution is required to cancel the infrared
divergence that is present in the colour-singlet
cross-section~\cite{Fleming:1998md}. The production of the $\eta_c$,
on the other hand, is dominated by colour-octet processes, since the
colour-singlet cross-section vanishes at leading-order, owing to
charge-conjugation invariance~\cite{Hao:1999kq,Hao:2000ci}, as is the
case for $\chi_c$ photoproduction. The cross-sections for
$\eta_c(1S)$, $\eta_c(2S)$, and $h_c(1P)$ photoproduction are
sizable~\cite{Fleming:1998md,Hao:1999kq}, but it is not obvious that
these particles can be detected experimentally, even with
high-statistics data.

The energy spectrum of {\em $J/\psi$'s produced in association with a
photon} via the process $\gamma p \to J/\psi + \gamma\,X$ is a
distinctive probe of colour-octet
processes~\cite{Kim:1993at,Cacciari:1996dy,Mehen:1997vx,Cacciari:1997zu}.
In the colour-singlet channel and at leading-order in $\alpha_s$, $J/\psi
+ \gamma$ can be produced only through resolved-photon interactions. The
corresponding energy distribution is therefore peaked at low values of
$z$. The intermediate-$z$ and large-$z$ regions of the energy spectrum
are  expected to be dominated by the colour-octet process $\gamma g \to
c\bar c_8({}^3\!S_1)\,\gamma$. Observation of a substantial fraction of
$J/\psi + \gamma$ events at $z\;\rlap{\lower 3.5 pt \hbox{$\mathchar
\sim$}} \raise 1pt \hbox {$>$}\; 0.5$ would provide clear evidence for
the presence of colour-octet processes in quarkonium photoproduction.
Experimentally, this measurement is very difficult due to the
large background from photons from $\pi^0$ decays which are produced
in the final state.

With the significant increase in statistics at the upgraded HERA
collider, it might be possible to study {\em inelastic photoproduction
of bottomonium states} for the first time. The large value of the
$b$-quark mass makes the perturbative QCD predictions more reliable than
for charm production, and the application of NRQCD should be on safer
ground for the bottomonium system, in which $v^2 \approx 0.1$. However,
the production rates for $\Upsilon$ states are suppressed compared with
those for $J/\psi$ by more than two orders of magnitude at HERA\,---\,a
consequence of the smaller $b$-quark electric charge and the phase-space
reduction that follows from the larger $b$-quark mass.

\section{Quarkonium production at LEP}
\label{sec:prodsec-lep}

\subsection{$J/\psi$ production}

The LEP collider was used to study $e^+ e^-$ collisions at the $Z^0$
resonance. Charmonium was produced at LEP through direct production
in $Z^0$ decay, through the decays of $b$ hadrons from $Z^0$ decay,
and through $\gamma \gamma$ collisions.  
The contributions from the decays of $b$ hadrons can be separated from 
those from direct production by using a vertex detector.
Charmonium that is produced directly will be referred to as ``prompt.''

%%%%%%%%%%%%%%%%%%%%%%%%%%%%%%%%%%%%%%%%%%%%%%%%%%%%%%%%%%%%%%%%%%%%%%%%%%%%%%%%%%%%%%%%
\begin{figure}[t]
\begin{center}
\includegraphics[width=10cm]{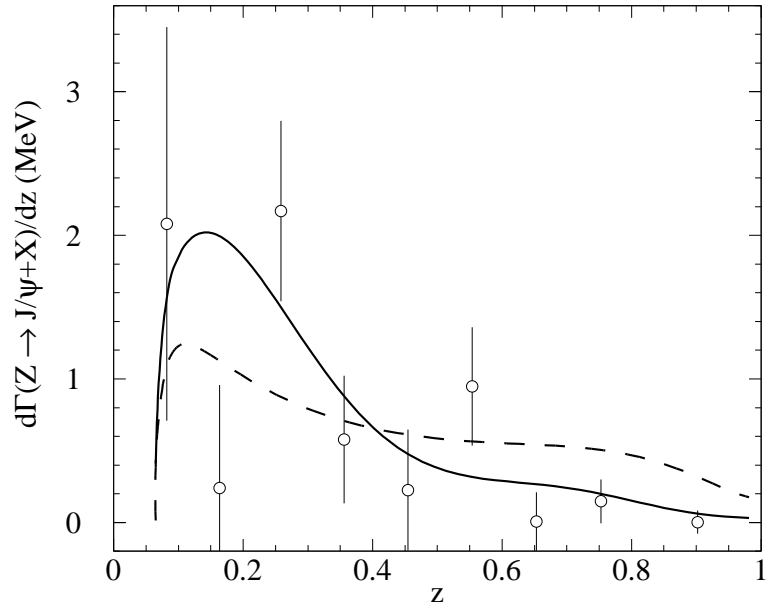}
\caption{
Differential rate $d \Gamma/ dz$ for inclusive decay of $Z^0$ 
into $J/\psi$. The data is from the ALEPH collaboration 
\cite{ALEPH:1997zj}.  The dashed line is the sum of the tree-level 
colour-singlet and colour-octet terms.  The solid line  is an 
interpolation between resummed calculations in the small-$z$ 
and large-$z$ regions. From Ref.~\cite{Boyd:1998km}. 
}
\label{fig:boyd-lep}
\end{center}
\end{figure}
%%%%%%%%%%%%%%%%%%%%%%%%%%%%%%%%%%%%%%%%%%%%%%%%%%%%%%%%%%%%%%%%%%%%%%%%%%%%%%%%%%%%%%%%

%%%%%%%%%%%%%%%%%%%%%%%%%%%%%%%%%%%%%%%%%%%%%%%%%%%%%%%%%%%%%%%%%%%%%%%%%%%%%%%%%%%%%%%%
\begin{figure}[t]
\begin{center}
\includegraphics[width=.75\linewidth]{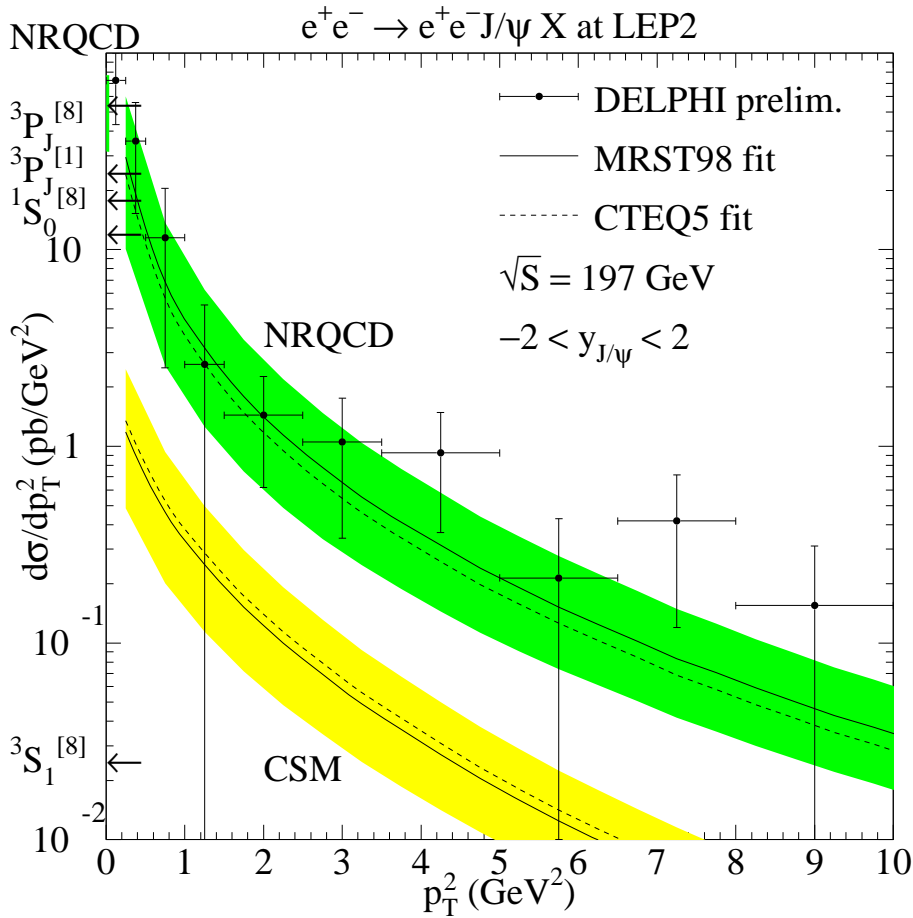}
\caption{
Differential cross-section for the process ${\gamma
\gamma\rightarrow J/\psi +X}$ as a function of $p_T^2$. 
The data points are from the DELPHI Collaboration
\cite{Todorova-Nova:2001pt,Abdallah:2003du}. The upper set of curves is the
NRQCD factorization predictions, and the lower set of curves is the
colour-singlet model prediction. The solid and dashed curves correspond
to the MRST98LO \cite{Martin:1998sq} and CTEQ5L \cite{Lai:1999wy} parton
distributions, respectively. The arrows indicate the relative
contributions at $p_T=0$ from parton processes $ij \to c \bar c$,
which were ignored in the analysis. 
Here $ij = \gamma \gamma$, $gg$, or $q \bar q$. 
From Ref.~\cite{Klasen:2001cu}.}
\label{fig:gamma-gamma-psi}
\end{center}
\end{figure}
%%%%%%%%%%%%%%%%%%%%%%%%%%%%%%%%%%%%%%%%%%%%%%%%%%%%%%%%%%%%%%%%%%%%%%%%%%%%%%%%%%%%%%%%

In $Z^0$ decay, the dominant mechanism for charmonium production is the 
decay of the $Z^0$ into $b \bar b$, followed by the fragmentation 
of the $b$ or $\bar b$ into a heavy hadron and the subsequent
decay of the heavy hadron into charmonium.
The inclusive branching fraction of the $Z^0$ into a charmonium state
$H$ is to a good approximation the product of the branching fraction 
for $Z^0 \to b \bar b$, a weighted average of the inclusive branching 
fractions of $b$ hadrons into $H$, and a factor of two to account for the 
$b$ and the $\bar b$:
\begin{equation}
{\rm Br}[ Z^0 \to H X] \approx 
2 \; {\rm Br}[ Z^0 \to b \bar b ] \sum_B D_{b \to B} \; {\rm Br}[B \to H X].
\end{equation}
The branching fraction for the $b$ hadron $B$ to decay into a state
that includes $H$ is weighted by the probability $D_{b \to B}$ for a
45~GeV $b$ quark to fragment into $B$.  The inclusive branching
fractions for $Z^0$ decay into several charmonium states have been
measured.  Since these measurements have more to do with $b$-hadron
decay than $Z^0$ decay, they are presented in
\Section~\ref{sec:prodsec-bdecays}.

The ALEPH, DELPHI, L3, and OPAL collaborations at LEP have measured
the inclusive branching fraction of $Z^0$ into prompt $J/\psi$
\cite{ALEPH:1997zj,Abreu:1994rk,Wadhwa:1998mt,Alexander:1996jp}.
In the NRQCD factorization approach, there are two
mechanisms that dominate direct $J/\psi$ production.  The first is
$Z^0$ decay into $c \bar c$, followed by the fragmentation of the $c$
or $\bar c$ into $J/\psi$ via the colour-singlet channel $c \bar
c_1({}^3S_1)$. The second is $Z^0$ decay into $q \bar g g$, followed by
the fragmentation of the gluon into $J/\psi$ via the colour-octet channel
$c \bar c_8({}^3S_1)$.  Boyd, Leibovich, and Rothstein
\cite{Boyd:1998km} have used the results from the four LEP
collaborations to extract the colour-octet matrix element: $\langle{\cal
O}^{J/\psi}_8({}^3S_1) \rangle =(1.9\pm 0.5_{stat}\pm1.0_{theory})
\times 10^{-2}\hbox{ GeV}^3$. This is about a factor of two larger than
the Tevatron value and has smaller theory errors, but feeddown from
$\chi_c$ and $\psi(2S)$ states was not taken into account in the
theoretical analysis. Boyd, Leibovich, and Rothstein \cite{Boyd:1998km}
also carried out a resummation of the logarithms of $M_Z^2/M_\psi^2$ and
$z^2$, where $z = 2E_{c \bar c}/m_Z$. Their result for the resummed $z$
distribution for prompt $J/\psi$ production is compared with
data from the ALEPH collaboration \cite{ALEPH:1997zj} in
\Figure~\ref{fig:boyd-lep}. Their analysis predicts an enhancement in the
production rate near $z=0.15$. The uncertainties in the data are too
large to make a definitive statement about the presence or absence of
this feature.

The inclusive cross-section for ${\gamma \gamma\rightarrow J/\psi +X}$
at LEP has been measured by the DELPHI Collaboration
\cite{Todorova-Nova:2001pt,Abdallah:2003du}. The cross-section at nonzero
$p_T$ has been computed at leading order in $\alpha_s$. 
The computation includes the direct-photon process 
$\gamma \gamma \to (c \bar c ) + g$, which is of order $\alpha^2 \alpha_s$, 
the single-resolved-photon process $i \gamma \to (c \bar c ) + i$, 
which is of order $\alpha \alpha_s^2$, 
and the double-resolved-photon process $i j \to (c \bar c ) + k$, 
which is of order $\alpha_s^3$
\cite{Ma:1997bi,Japaridze:1998ss,Godbole:2001pj,Klasen:2001mi,Klasen:2001cu}.
(Here, $ij = gg$, $g q$, $g \bar q$, or $q \bar q$.) Note that all
processes contribute formally at the same order in perturbation theory since
the leading behavior of the parton distributions in the photon is
$\propto \alpha/\alpha_s$. The contribution to the 
$\gamma\gamma\rightarrow J/\psi +X$ cross-section at LEP
that is by far dominant numerically is that
from single-resolved processes, \ie photon--gluon fusion.

The results of the LO computation~\cite{Klasen:2001cu} are
shown in \Figure~\ref{fig:gamma-gamma-psi}. The computation uses the
NRQCD matrix elements of Ref.~\cite{Braaten:1999qk}. Theoretical
uncertainties were estimated by varying the renormalization and
factorization scales by a factor two and by incorporating the effects
of uncertainties in the values of the colour-octet matrix elements. As
can be seen from \Figure~\ref{fig:gamma-gamma-psi}, the comparison with
the DELPHI data
\cite{Todorova-Nova:2001pt,Abdallah:2003du} clearly favors the
NRQCD factorization approach over the colour-singlet model. However,
the comparison of \Figure~\ref{fig:gamma-gamma-psi} is based on a
leading-order calculation. It is known from the related process of
$J/\psi$ photoproduction at HERA, which is also dominated by
photon--gluon fusion, that the LO colour-singlet cross-section fails to
describe the $J/\psi$ data at nonzero $p_T$. Inclusion of the NLO
correction, however, brings the colour-singlet prediction in line with
experiment. Similarly large NLO corrections can be expected for
${\gamma \gamma\rightarrow J/\psi +X}$ production at LEP, and a
complete NLO analysis is needed before firm conclusions on the
importance of colour-octet contributions can be drawn.  A first step in
this direction has been taken recently in Ref.~\cite{Klasen:2004tz},
where the NLO corrections to the direct process $\gamma \gamma \to (c
\bar c ) + g$ have been calculated. 

\subsection{$\Upsilon(1S)$ production}

The OPAL collaboration has measured the inclusive branching fraction for
the decay of $Z^0$ into $\Upsilon(1S)$ \cite{Alexander:1995vh}. The
NRQCD factorization prediction for ${\rm Br}[Z^0\rightarrow \Upsilon(1S)
+ X]$ is $5.9\times 10^{-5}$ \cite{Cho:1995vv}. The colour-singlet-model
prediction is $1.7\times 10^{-5}$
\cite{Keung:1980ev,Kuhn:1981jy,Abraham:1989ri,Hagiwara:1991mt,%
Braaten:1993mp,Cho:1995vv}. The experimental result from OPAL
\cite{Alexander:1995vh} is $[1.0\pm 0.4(\hbox{stat.})\pm
0.1(\hbox{sys.}) \pm 0.2(\hbox{prod.\ mech.})]\times 10^{-4}$.
This is compatible with the NRQCD factorization prediction, but not with
the colour-singlet-model prediction.

\section{Charmonium production in $e^+e^-$ annihilations at 10.6~GeV}
\label{sec:prodsec-eecontinuum}
\
The $B$ factories have proved to be a rich source of data on charmonium
production in $e^+ e^-$ annihilation. The $B$ factories operate near the
peak of the $\Upsilon(4S)$ in order to maximize the production rate for
$B$ mesons, but about 75\% of the events are continuum $e^+e^-$
annihilation events. The enormous data samples that have been accumulated
compensate for the relatively small cross-sections for 
$e^+ e^-$ annihilation into states that include charmonium.

\subsection{$J/\psi$ production}

%%%%%%%%%%%%%%%%%%%%%%%%%%%%%%%%%%%%%%%%%%%%%%%%%%%%%%%%%%%%%%%%%%%%%%%%%%%%%%%%%%%%%%
\begin{figure}
\begin{center}
\begin{tabular}{@{}c@{\quad}c@{}}
\includegraphics[width=77mm]{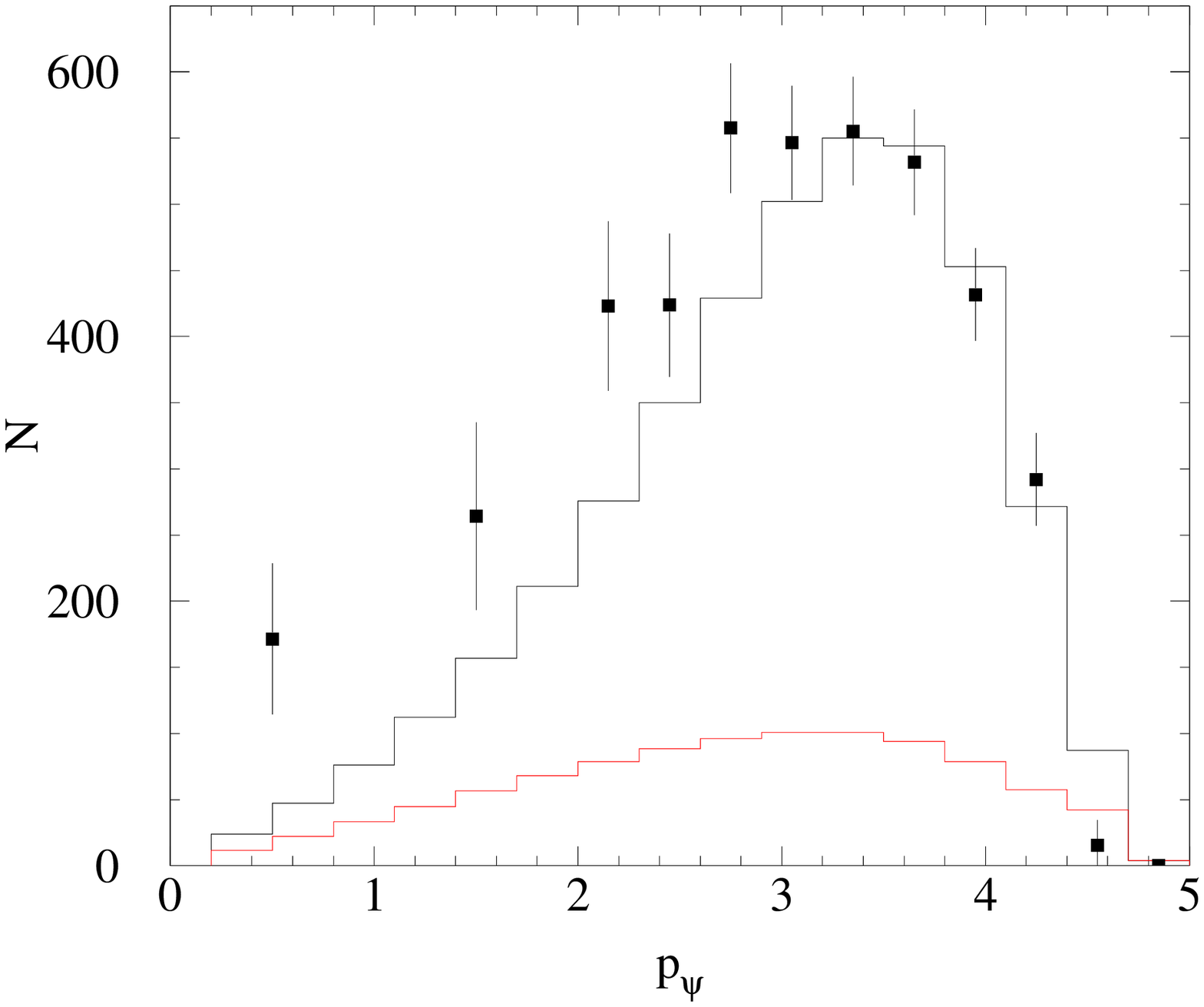}
&
\includegraphics[width=77mm]{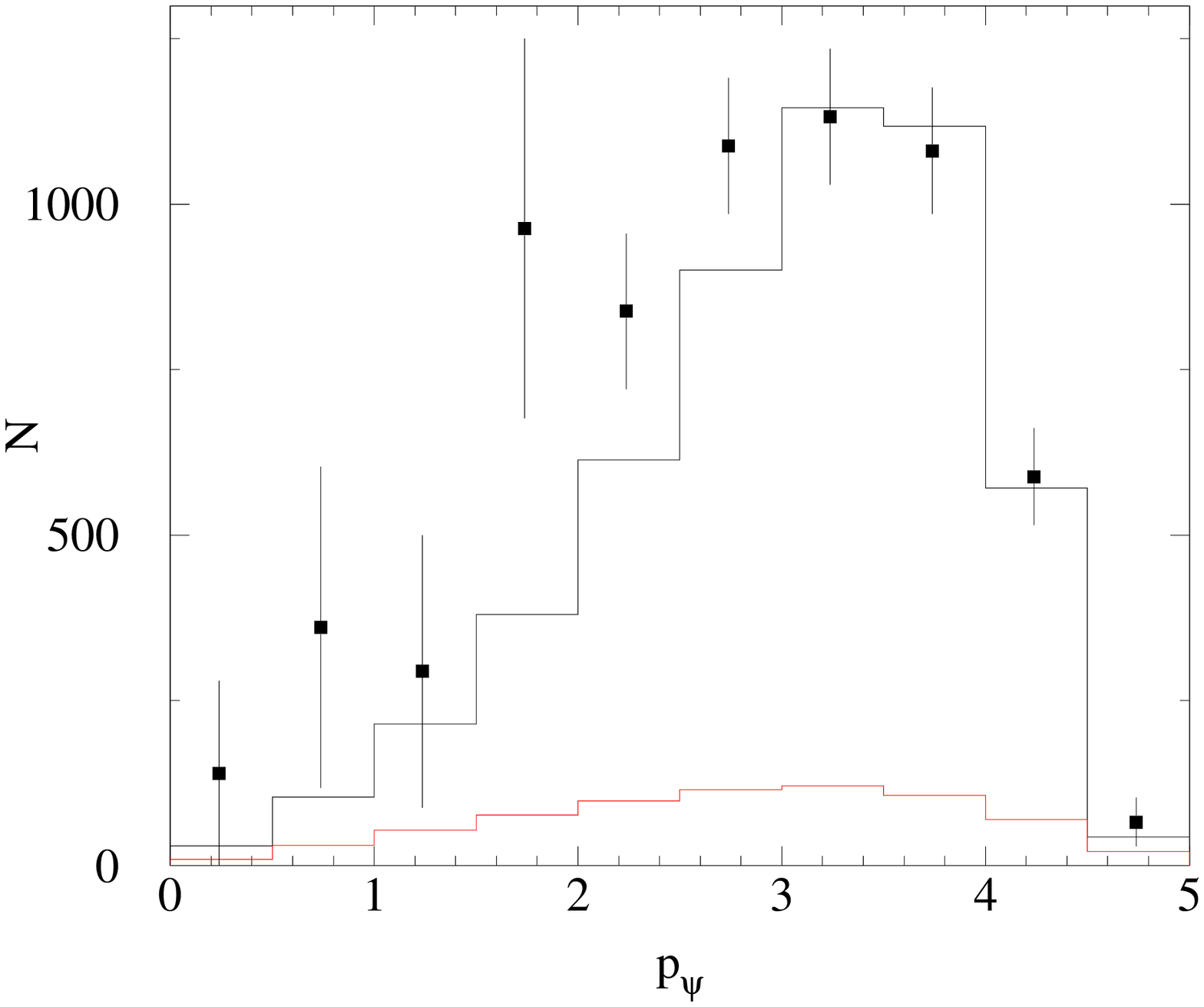} \\
(a) & (b)
\end{tabular}
\end{center}
\caption[$J/\psi$ production rate in $e^+e^-$ annihilation at $10.6$~GeV]
        {$J/\psi$ production rate in $e^+e^-$ annihilation at
         $10.6$~GeV as a function of $p^* = p_{\psi}$, the $J/\psi$
         momentum in the CM frame. The vertical axis is the number of
         $J/\psi$ events per 0.5~GeV$/c$. The data points are from (a)
         the Belle Collaboration \cite{Abe:2001za} and (b) the BaBar
         Collaboration \cite{Aubert:2001pd}. The upper lines are the
         sum of the leading-order colour-singlet contribution and the
         colour-octet contribution, which includes a resummation of
         logarithms of $1-z$ and a phenomenological shape
         function. The lower lines are the leading-order colour-singlet
         contribution alone. From Ref.~\cite{Fleming:2003gt}.}
\label{fig:belle-babar}
\end{figure}
%%%%%%%%%%%%%%%%%%%%%%%%%%%%%%%%%%%%%%%%%%%%%%%%%%%%%%%%%%%%%%%%%%%%%%%%%%%%%%%%%%%%%%

The Belle and BaBar Collaborations have measured the inclusive
cross-section $\sigma[e^+e^-\rightarrow J/\psi\, X]$. The Belle
Collaboration obtains $2.52\pm 0.21\pm 0.21\hbox{~pb}$
\cite{Abe:2001za}, while the BaBar Collaboration obtains $1.47\pm
0.10\pm 0.13\hbox{~pb}$ \cite{Aubert:2001pd}. The leading-order parton
process in the colour-singlet model is $e^+ e^- \to (c \bar c) + gg$,
which is of order $\alpha^2 \alpha_s^2$. The leading colour-octet
contributions in the NRQCD factorization approach come from $e^+ e^-$
annihilation into $(c \bar c) + g$, which is order $\alpha^2
\alpha_s$, and into $(c \bar c) + q \bar q$ and $(c \bar c) + g g$,
which are order $\alpha^2 \alpha_s^2$.  The prediction for the
cross-section $\sigma[e^+e^-\rightarrow J/\psi\,X]$ in the
colour-singlet model is $0.45-0.81\hbox{~pb}$
\cite{Cho:1996cg,Yuan:1996ep,Yuan:1997sn,Schuler:1998az}, while the
NRQCD factorization prediction is $1.1-1.6\hbox{~pb}$
\cite{Yuan:1996ep,Yuan:1997sn,Schuler:1998az}. There is a $3\sigma$
discrepancy between the experiments, but the NRQCD factorization
prediction seems to be favored.  The discrepancies between the two
experiments in this and other measurements may be due partly to
differences in cuts that were used to suppress contributions from
processes in which the charmonium is not produced by annihilation of
$e^+$ and $e^-$ with a centre-of-mass energy of 10.6~GeV.  These
include radiative-return processes, in which the $e^+$ or $e^-$ loses
a substantial fraction of its momentum by radiating a collinear photon
before the collision, virtual photon radiation, in which the $e^+$ or
$e^-$ radiates a virtual photon that becomes a $J/\psi$ or $\psi(2S)$,
and two-photon collisions, which produce $\eta_c$, $\chi_{c0}$, and
$\chi_{c2}$.

The momentum distribution of the $J/\psi$ provides information about the
production mechanism. The momentum of the $J/\psi$ in the CM frame
can be characterized in terms of its magnitude $p^*$ and its angle
$\theta^*$ with respect to the beam direction. The Belle
\cite{Abe:2001za} and BaBar \cite{Aubert:2001pd} measurements for the
differential cross-section for $J/\psi$ production as a function of
$p^*$ are shown in \Figure~\ref{fig:belle-babar}. The colour-singlet prediction,
which is shown in the lower curves in \Figure~\ref{fig:belle-babar}, is far too
small to describe the data. The measurements from Belle and BaBar do not
show any enhancement at the maximum value of $p^*$, as might be expected
from the colour-octet process $e^+ e^- \to (c \bar c) + g$ that is of
leading order in $\alpha_s$. However, there are two effects that are
expected to modify the leading-order result. The first effect is that
the $v$ expansion of NRQCD breaks down near the kinematic maximum value
of $p^*$. Resummation of the expansion is required
\cite{Beneke:1997qw,Beneke:1999gq}, and it leads to a nonperturbative
shape function \cite{Beneke:1997qw}, which smears out the peak in the
leading-order result. A second effect near the maximum value of $p^*$ is
that there are large logarithms of $1-z$, where $z=E_{c\bar c}/E_{c\bar
c}^{\rm max}$, that must also be resummed. The effect of that
resummation is again to smear out the peak in the leading-order result.
A resummation of logarithms of $1-z$ has been combined with a
phenomenological shape function in Ref.~\cite{Fleming:2003gt}. The
results of this calculation are shown in the upper curves in
\Figure~\ref{fig:belle-babar}. The shape function has been chosen to fit the
Belle and BaBar data. The normalization of the shape function is fixed
by the colour-octet NRQCD matrix elements, which were taken to be
$\langle{\cal O}^{J/\psi}_8({}^1S_0)\rangle= \langle{\cal
O}^{J/\psi}_8({}^3P_0)\rangle=6.6\times10^{-2}$~GeV.  These values of
the colour-octet matrix elements are consistent with data from
photoproduction and hadroproduction
\cite{Beneke:1996tk,Amundson:1996ik}. As can be seen, the resummations of
the $v$ expansion and the logarithms of $1-z$ produce reasonable fits to
the data. The resummation prediction is not expected to be valid at
small values of $p^*$. It should also be kept in
mind that hard-scattering factorization may not hold unless
$p^*\gg\Lambda_{\rm QCD}$. While the comparison of the resummed theory
with experiment indicates that it is plausible that the NRQCD
factorization approach can describe the experimental data, the
theoretical results rely heavily on the phenomenological shape function,
whose shape is tuned to fit the data. The resummed theory will receive a
much more stringent test when a phenomenological shape function that has
been extracted from the $e^+e^-$ data is used to predict the $J/\psi$
production cross-section in some other process, for example,
photoproduction at HERA.

%%%%%%%%%%%%%%%%%%%%%%%%%%%%%%%%%%%%%%%%%%%%%%%%%%%%%%%%%%%%%%%%%%%%%%%%%%%%%%%%%%%%%%
\renewcommand{\arraystretch}{1.1}
\begin{table}[ht]
\caption[Angular asymmetry variable $A$ and polarization variable $\alpha$]
        {Angular asymmetry variable $A$ and polarization variable
         $\alpha$ for various ranges of the CM momentum $p^*$ of the
         $J/\psi$ in $e^+ e^- \to J/\psi X$ at $\sqrt{s} = 10.6$~GeV,
         as measured by the Belle \cite{Abe:2001za} and BaBar
         \cite{Aubert:2001pd} Collaborations.}
\label{tab:prod-eetab}
\begin{center}
\begin{tabular}{|ccc|ccc|}
\hline   
\hline   
\multicolumn{3}{|c|}{Belle} &
\multicolumn{3}{c|}{BaBar} \\
\hline
Range of $p^*$ (GeV) & $A$ & $\alpha$ & 
Range of $p^*$ (GeV) & $A$ & $\alpha$ \\
\hline
$2.0 < p^* < 2.6$ & $0.82^{+0.95}_{-0.63}$ & $-0.62^{+0.30}_{-0.24}$ &
        $p^* < 3.5$ & $0.05\pm 0.22$ & $-0.46\pm 0.21$ \\
$2.6 < p^* < 3.4$ & $1.44^{+0.42}_{-0.38}$ & $-0.34^{+0.18}_{-0.16}$ & 
                  &                       & \\
$3.4 < p^* < 5.0$ & $1.08^{+0.44}_{-0.33}$ & $-0.32^{+0.20}_{-0.18}$ &
        $3.5 < p^*$ & $1.5 \pm 0.6$ & $-0.80\pm0.09$ \\
\hline
\hline   
\end{tabular}
\renewcommand{\arraystretch}{1}
\end{center}
\end{table}
%%%%%%%%%%%%%%%%%%%%%%%%%%%%%%%%%%%%%%%%%%%%%%%%%%%%%%%%%%%%%%%%%%%%%%%%%%%%%%%%%%%%%%

The other variable that characterizes the momentum of the $J/\psi$ is
its angle $\theta^*$ with respect to the beam direction in the CM
frame.  The angular distribution $d\sigma/d(\cos \theta^*)$ is
proportional to $1+A\cos^2 \theta^*$, which defines an angular
asymmetry variable $A$.  The Belle \cite{Abe:2001za} and BaBar
\cite{Aubert:2001pd} Collaborations have measured $A$ in several bins
of $p^*$. Their results are shown in \Table~\ref{tab:prod-eetab}. The
NRQCD factorization approach predicts that $A\approx 0$ at small $p^*$
and $0.6<A<1.0$ at large $p^*$ \cite{Braaten:1995ez} . The
colour-singlet model predicts that $A\approx 0$ at small $p_T$ and
$A\approx -0.8$ at large $p^*$ \cite{Braaten:1995ez}. The Belle and
BaBar results favor the NRQCD factorization prediction, but the
uncertainties are large.

The polarization of the $J/\psi$ provides further information about
the production mechanism.  The polarization variable $\alpha$ for
$J/\psi$ production is defined by the angular distribution in
\Eq~(\ref{eq:prod-alphadef}).  In $e^+ e^-$ annihilation, the most
convenient choice for the polarization axis is the boost vector from
the quarkonium rest frame to the $e^+ e^-$ centre-of-momentum frame.
The Belle and BaBar Collaborations have measured the polarization
variable $\alpha$ in several bins of $p^*$. Their results are shown in
\Table~\ref{tab:prod-eetab}. The polarization of $J/\psi$'s from $e^+
e^-$ annihilation at the $B$ factories has not yet been calculated
within the NRQCD factorization approach. In contrast to the production
of $J/\psi$'s with large $p_T$ at the Tevatron, where the dominance of
gluon fragmentation into colour-octet ${}^3S_1$ $c \bar c$ states
implies a large transverse polarization, production of $J/\psi$'s at
the $B$ factories occurs at values of $p^*$ for which there are no
simple qualitative expectations for the polarization.  A comparison
between theory and experiment must await an actual calculation of the
$J/\psi$ polarization, including the effects of feeddown from higher
charmonium states. It may be necessary to include in such a
calculation resummations of the $v$ expansion and logarithms of $1-z$
in order to make precise quantitative statements. However, the effects
of these resummations is mainly to re-distribute the $J/\psi$'s that
are produced via the colour-octet mechanism over a range in $p^*$
without affecting the total number of such $J/\psi$'s.

A surprising result from the Belle Collaboration is that most of the
$J/\psi$'s that are produced in $e^+e^-$ annihilation at $10.6$~GeV are
accompanied by charmed hadrons. The presence of a charmed hadron
indicates the creation of a second $c \bar c$ pair in addition to the
pair that forms the $J/\psi$. A convenient measure of the probability
for creating the second $c \bar c$ pair is the ratio
\begin{eqnarray}
R_{\rm double} =
  \frac{\sigma[e^+e^-\to J/\psi+X_{c\bar c}]}
       {\sigma[e^+e^-\to J/\psi+X]} \;. 
\label{eq:prodsec-Rdouble}
\end{eqnarray}
The Belle Collaboration finds that $R_{\rm double} =0.82\pm 0.15\pm
0.14$ with $R_{\rm double} > 0.48$ at the 90\% confidence level
\cite{belle-eps2003}. The NRQCD factorization approach leads to the
prediction $R_{\rm double}\approx 0.1$
\cite{Cho:1996cg,Baek:1996kq,Yuan:1996ep}, which disagrees with the
Belle result by almost an order of magnitude. The discrepancy seems to
arise primarily from the cross-section in the numerator of
(\ref{eq:prodsec-Rdouble}). The Belle result for this cross-section is
about 0.6--1.1~pb \cite{Abe:2002rb}, while the prediction is about
0.10--0.15 pb \cite{Cho:1996cg,Baek:1996kq,Yuan:1996ep,Liu:2003jj}. At
leading order in $\alpha_s$, which is $\alpha_s^2$, the process of
$e^+e^-$ annihilation into $J/\psi+X_{c\bar c}$ proceeds through $(c
\bar c)+c\bar c$. The contributions to this cross-section in which the
$J/\psi$ is formed from a colour-octet $c\bar c$ pair are suppressed by
a factor $v^4\approx 0.1$, and they have been found to yield only
about 7\% of the total cross-section \cite{Liu:2003jj}. Corrections of
order $\alpha_s^3$ and higher are also not expected to be particularly
large. Thus, the source of the discrepancy between the Belle result
for $R_{\rm double}$ and theory remains a mystery.

%%%%%%%%%%%%%%%%%%%%%%%%%%%%%%%%%%%%%%%%%%%%%%%%%%%%%%%%%%%%%%%%%%%%%%%%%%%%%%%%%%%%%%%%
\begin{figure}[t]
\begin{center}
\includegraphics[width=\linewidth]{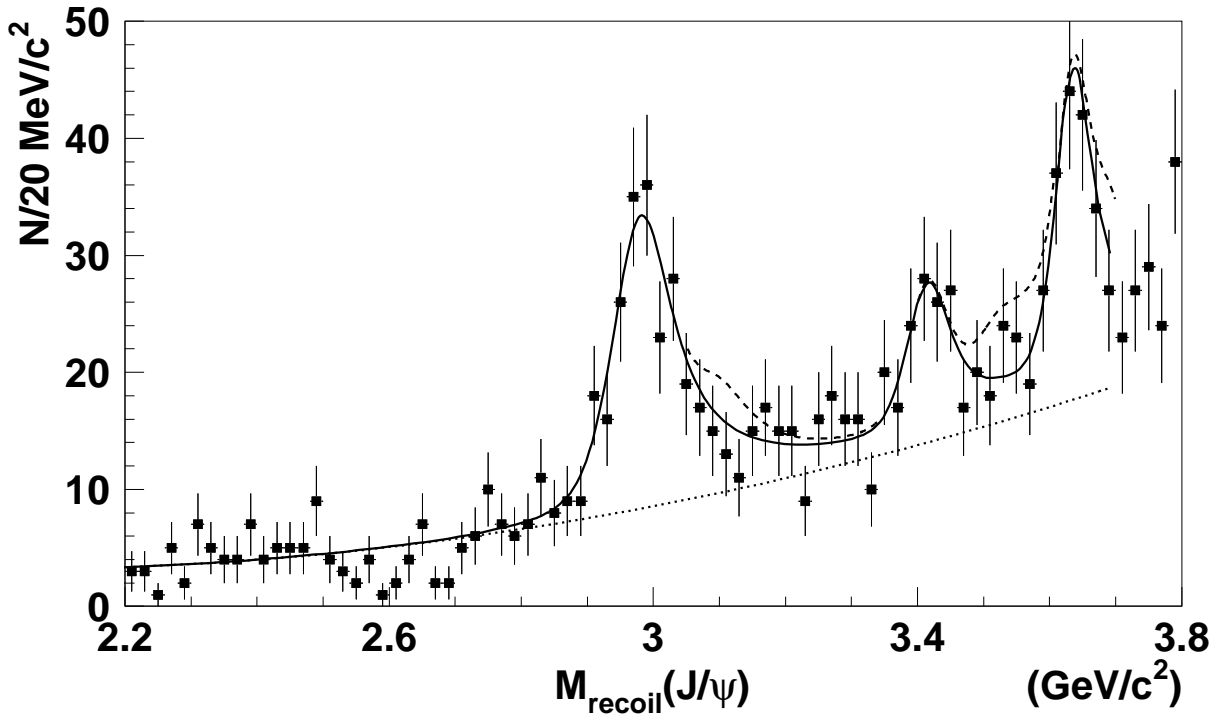}
\caption{Distribution of masses recoiling against the reconstructed
$J/\psi$ in inclusive $e^+e^- \rightarrow J/\psi X$ events at Belle
\cite{Abe:2004ww}. The solid line is the best fit, including
contributions from the $\eta_c$, $\chi_{c0}(1P)$, and $\eta_c(2S)$.
The dotted line is a fit in which additional contributions from the
$J/\psi$, $\chi_{c1}(1P)$, $\chi_{c2}(1P)$, and $\psi(2S)$ have been set
at their largest possible values within the 90\%-confidence-level
limits.
}
\label{fig:belledoublecharm}
\end{center}
\end{figure}
%%%%%%%%%%%%%%%%%%%%%%%%%%%%%%%%%%%%%%%%%%%%%%%%%%%%%%%%%%%%%%%%%%%%%%%%%%%%%%%%%%%%%%%%

There is also a large discrepancy between theory and experiment in an
exclusive double-$c \bar c$ cross-section. For the double-charmonium
process $e^+e^-\to J/\psi+\eta_c$, the Belle Collaboration measures the
product of the cross-section and the branching fraction for the $\eta_c$ to
decay into at least two charged tracks to be $25.6\pm
2.8\pm 3.4~\hbox{fb}$ \cite{Abe:2004ww}. In contrast, leading-order
calculations predict a cross-section of $2.31\pm 1.09~\hbox{fb}$
\cite{Braaten:2002fi,Liu:2002wq, brodsky-ji-lee}. There are some
uncertainties from uncalculated corrections of higher-order in
$\alpha_s$ and $v$ and from NRQCD matrix elements. However, because this
is an exclusive process, only colour-singlet matrix elements enter, and
these are fairly well determined from the decays $J/\psi\to e^+e^-$ and
$\eta_c\to \gamma\gamma$.

Since the Belle mass resolution is 110~MeV but the $J/\psi$--$\eta_c$
mass difference is only 120~MeV, it has been suggested that some of the
$J/\psi+\eta_c$ data sample may consist of $J/\psi+J/\psi$ events
\cite{Bodwin:2002fk,Bodwin:2002kk}. The state $J/\psi+J/\psi$ has
charge-parity $C=+1$, and consequently, is produced in a two-photon
process, whose rate is suppressed by a factor $(\alpha/\alpha_s)^2$
relative to the rate for $J/\psi+\eta_c$. However, as was pointed out in
Refs.~\cite{Bodwin:2002fk,Bodwin:2002kk}, the two-photon process
contains photon-fragmentation contributions that are enhanced by factors
$(E_{\rm beam}/2m_c)^4$ from photon propagators and $\log[8(E_{\rm
beam}/2m_c)^4]$ from a would-be collinear divergence. As a result, the
predicted cross-section  $\sigma[e^+e^-\to J/\psi+J/\psi]=8.70\pm
2.94$~fb is larger than the predicted cross-section $\sigma[e^+e^-\to
J/\psi+\eta_c]=2.31\pm 1.09$~fb \cite{Bodwin:2002fk,Bodwin:2002kk}.
Corrections of higher order in $\alpha_s$ and $v$ are likely to
reduce the prediction for the $J/\psi+J/\psi$ cross-section by about a
factor of three \cite{Bodwin:2002kk,Luchinsky:2003yh}. These
predictions spurred a re-analysis of the Belle data \cite{Abe:2003ja}.
The invariant mass distribution of $X$ in $e^+e^-\to J/\psi+X$ is shown
in \Figure~\ref{fig:belledoublecharm}. No significant $J/\psi+J/\psi$
signal was observed. The upper limit on the cross-section times the
branching fraction into at least two charged tracks \cite{Abe:2004ww} is
$\sigma[e^+e^-\to J/\psi+J/\psi]<9.1~\hbox{fb}$, which is consistent
with the prediction of Refs.~\cite{Bodwin:2002fk,Bodwin:2002kk}.

\subsection{Prospects at BaBar and Belle}

The BaBar and Belle detectors are accumulating ever larger data 
samples of charmonium that is produced directly in $e^+ e^-$ annihilation.
The simplicity of the initial state makes the theoretical analysis 
of this process particularly clean.  These two factors
make charmonium production in continuum $e^+ e^-$ annihilation a
particularly attractive process in which to compare theoretical
predictions with experiment.  The experimental results that have already 
emerged from these detectors provide further motivation for 
understanding this process.  There are 
significant discrepancies between previous measurements 
by BaBar and Belle.  There are also surprising results from Belle 
on double $c \bar c$ production that differ dramatically from theoretical
expectations.  The resolution of these problems will inevitably 
lead to progress in our understanding of charmonium production.

The surprising double-$c \bar c$ results from Belle include 
an inclusive measurement, the ratio $R_{\rm double}$ defined in 
\Eq~(\ref{eq:prodsec-Rdouble}), and exclusive double-charmonium 
cross-sections, such as $\sigma[e^+e^-\to J/\psi+\eta_c]$.
The discrepancies between theory and experiment in these measurements
are among the largest in the standard model. Theory and
experiment differ by about an order of magnitude\,---\,a discrepancy which
is larger than any known QCD $K$-factor.  It is important to
recognize that these discrepancies are problems not just for NRQCD
factorization, but for perturbative QCD in general. 
It is difficult to imagine how any perturbative calculation 
of $R_{\rm double}$ could give a value as large as 80\%.
With regard to the cross-section for $e^+e^-\to J/\psi+\eta_c$,
the perturbative QCD formalism for exclusive processes 
\cite{brodsky-ji-lee} gives a result that reduces to that of
NRQCD factorization \cite{Braaten:2002fi,Liu:2002wq} in the
nonrelativistic limit and is well-approximated by it if one uses
realistic light-cone wave functions for $J/\psi$ and $\eta_c$.\footnote{
The Belle result can be accommodated by using asymptotic light-cone
wave functions that are appropriate for light hadrons
\cite{Ma:2004qf}, but there is no justification for using such
wave functions for charmonium.} 
Clearly, it is very important to have independent checks of the Belle
inclusive and exclusive double-$c \bar c$ results. If the Belle
results are confirmed, then we would be forced to entertain some
unorthodox possibilities: the inapplicability of perturbative QCD to
double-$c \bar c$ production, new charmonium production mechanisms
within the standard model, or perhaps even physics beyond the standard
model. 

The larger data samples that are now available should allow much more 
accurate measurements of the inclusive process $e^+ e^- \to J/\psi X$,
including the momentum distribution of the $J/\psi$ and its polarization.
The measurements of the $J/\psi$ momentum distribution may allow the
determination of all the relevant NRQCD matrix elements.  A comparison
with the NRQCD matrix elements measured at the Tevatron would then provide 
a test of their universality.  Once the NRQCD matrix elements 
are determined, they can be used to predict the polarization of the 
$J/\psi$ as a function of its momentum, which would provide a stringent 
test of the NRQCD predictions for spin.
Instead of imposing cuts to suppress contributions from radiative return,
virtual photon radiation, and two-photon collisions, it might be better 
to choose cuts in order to maximize the precision of the measurements,
without any regard to the production mechanism.  The contributions 
from other mechanisms could instead be taken into account in the 
theoretical analyses.  

The large data samples of BaBar and Belle should also allow 
measurements of the inclusive production of other charmonium states, 
such as the $\psi(2S)$ and the $\chi_c(1P)$.  Such measurements could be used 
to determine the NRQCD matrix elements for these charmonium states.  
They are also important because they provide constraints on the 
contributions to inclusive $J/\psi$ production from decays of higher 
charmonium states.

There are some straightforward improvements that could be made
in the theoretical predictions for inclusive charmonium production 
in $e^+ e^-$ annihilation.  For example, there are only a few 
components missing from a complete calculation of all 
contributions through second order in $\alpha_s$.
In the contribution from the 
colour-octet ${}^3S_1$ channel, the virtual corrections at order 
$\alpha_s^2$ have not been calculated.  There are also colour-octet 
contributions to $e^+ e^- \to c \bar c c \bar c$ at order $\alpha_s^2$ 
that have not been calculated.  The theoretical predictions 
for inclusive charmonium production could 
also be improved by treating more systematically the contributions 
from the feeddown from decays of higher charmonium states and from 
other mechanisms, such as radiative return, virtual photon radiation, 
and two-photon collisions.

\section{Charmonium production in $B$-meson decays}
\label{sec:prodsec-bdecays}

$B$-meson decays are an excellent laboratory for studying 
charmonium production because $B$ mesons decay into 
charmonium with branching fractions greater than a percent.  
At a $B$ factory operating near the peak of the $\Upsilon(4S)$ 
resonance, about 25\% of the events consist of
a $B^+ B^-$ pair or a $B^0 \bar B^0$ pair.  The large sample 
of $B$ mesons accumulated by the CLEO experiment allowed 
the measurements of many exclusive and inclusive branching 
fractions into final states that include charmonium.
The Belle and BaBar experiments are accumulating even larger
samples of $B$ mesons, providing a new source of 
precise data on charmonium production in $B$ decays.

The inclusive branching fractions of $B$ mesons into charmonium states
can be measured most accurately for the mixture of $B^+$, $B^0$, and
their antiparticles that are produced in the decay of the
$\Upsilon(4S)$ resonance \cite{Balest:1994jf,Chen:2000ri}.  Those that
have been measured are listed in
\Table~\ref{tab:prodsec-Brbcharm}. The fraction of $J/\psi$'s that
come from decay of $\chi_c$'s, which is defined in
\Eq~(\ref{eq:prod-chifrac}), is $F_{\chi_c} = (11 \pm 2)$\%. This is
significantly smaller than the value that is measured at the Tevatron,
which is given in \Table~\ref{tab:prodsec-Jpsifractions}.  The
$\chi_{c1}$ to $\chi_{c2}$ ratio, which is defined in
\Eq~(\ref{eq:prod-chirat}), is $R_{\chi_c} = 5.1\pm 3.0$.  Although the
error bar is large, this ratio seems to be substantially larger than
the value that is measured at the Tevatron, which is given in
\Eq~(\ref{eq:Rchi12Tev}), and the values measured in fixed-target
experiments, which are given in \Table~\ref{tab:chi}.  Such
differences in $R_{\chi_c}$ and $F_{\chi_c}$ are contrary to the
predictions of the colour-evaporation model.

Inclusive branching fractions into charmonium states have also been
measured at LEP for the mixtures of $B^+$, $B^0$, $B^0_s$, $b$
baryons, and their antiparticles that are produced in $Z^0$ decay
\cite{Buskulic:1992wp,Adriani:1993ta,Abreu:1994rk}.  This mixture of
$b$ hadrons can be interpreted as the one that arises from the
fragmentation of a $b$ quark that is produced with a momentum of 45
GeV.  The branching fractions that have been measured are listed in
\Table~\ref{tab:prodsec-Brbcharm}. The branching fraction into
$\chi_{c1}(1P)$ seems to be significantly larger than for the mixture
from $\Upsilon(4S)$ decay. The difference could be due to the
contribution from $b$ baryons.  It is often assumed that the mixture
of $b$ hadrons that is produced at high-energy hadron colliders, such
as the Tevatron, is similar to that produced in $Z^0$ decay.  This
assumption could be tested by measuring ratios of inclusive
cross-sections for charmonium states that come from the decays of $b$
hadrons at the Tevatron.

%%%%%%%%%%%%%%%%%%%%%%%%%%%%%%%%%%%%%%%%%%%%%%%%%%%%%%%%%%%%%%%%%%%%%%%%%%%%%%%%%%
\begin{table}[ht]
\caption[Inclusive branching fractions]
        {Inclusive branching fractions (in units of $10^{-3}$) for
         mixtures of $b$ hadrons to decay into charmonium states.}
\label{tab:prodsec-Brbcharm}
\begin{center}
\begin{tabular}{|c|cccc|} 
\hline
mixture & $J/\psi$ & $\psi(2S)$ & $\chi_{c1}(1P)$ & $\chi_{c2}(1P)$ \\ 
\hline
from $\Upsilon(4S)$ decay      
& $11.5 \pm 0.6$ &  $3.5 \pm 0.5$ &  $3.6 \pm 0.5$ &  $0.7 \pm 0.4$ \\ 
from $Z^0$ decay   
& $11.6 \pm 1.0$ &  $4.8 \pm 2.4$ & $11.5 \pm 4.0$ &              \\ 
\hline
\end{tabular}
\end{center}
\end{table}
%%%%%%%%%%%%%%%%%%%%%%%%%%%%%%%%%%%%%%%%%%%%%%%%%%%%%%%%%%%%%%%%%%%%%%%%%%%%%%%%%%

The observed inclusive branching fractions of $B$ mesons into $J/\psi$ and
$\psi(2S)$ are larger than the predictions of the colour-singlet model by
about a factor of three. Ko, Lee, and Song applied the NRQCD
factorization approach to the production of $J/\psi$ and $\psi(2S)$ 
in $B$ decays~\cite{Ko:1995iv}. The colour-octet ${}^3S_1$ term in the
production rate is suppressed by a factor of $v^4$ that comes from
the NRQCD matrix element. However, the production rate also involves
Wilson coefficients that arise from evolving the effective weak
Hamiltonian from the scale $M_W$ to the scale $m_b$. The Wilson
coefficient for the colour-octet ${}^3S_1$ term is significantly larger
than that for the colour-singlet term, although the smallness of the
colour-singlet term may be the result of an accidental cancellation that
occurs in the leading-order treatment of the evolution of the
coefficients.  Moreover, the colour-singlet contribution is
decreased by the relativistic correction of order $v^2$. 
The inclusion of the colour-octet
${}^3S_1$ term allows one to explain the factor of three discrepancy
between the data and the colour-singlet-model prediction.

The observed branching fraction for decays of $B$ directly into
$J/\psi$, which excludes the feeddown from decays of $\psi(2S)$ and
$\chi_c$, is much larger than the prediction of the colour-evaporation
model.  The CEM prediction for the direct branching fraction is
$0.24-0.66$~\cite{Ko:1999zx}, where the range comes from the uncertainty
in the CEM parameters. The CLEO collaboration has made a precise
measurement of the direct branching fraction of $B$ into
$J/\psi$~\cite{Anderson:2002md}: 
${\rm Br}_{\rm dir}[B \to J/\psi+X] = (0.813\pm 0.041)\%$.  
The CEM prediction is significantly 
smaller than the data. As we have already mentioned, the data
can be accommodated within the 
NRQCD factorization approach by including colour-octet terms. 

Beneke, Maltoni, and Rothstein \cite{Beneke:1998ks} have calculated the
inclusive decay rates of $B$ mesons into $J/\psi$ and $\psi(2S)$ to
next-to-leading order in $\alpha_s$. They used their results to
extract NRQCD matrix elements from the data. Their results for the
linear combinations of NRQCD matrix elements defined in
\Eq~(\ref{eq:prod-lincomb}) are $M_{3.1}^{J/\psi}=(1.5_{-1.1}^{+0.8})\times
10^{-2}\hbox{~GeV}^3$ and $M_{3.1}^{\psi(2S)}=(0.5\pm 0.5)\times
10^{-2}\hbox{~GeV}^3$. The uncertainties arise from experiment and from
imprecise of knowledge of the matrix elements $\langle {\cal
O}^H_8({}^3S_1)\rangle$ and $\langle {\cal O}^H_1({}^3S_1)\rangle$. Ma,
taking into account initial-state hadronic corrections, has extracted
slightly different linear combinations of matrix elements
\cite{Ma:2000bz}: $M_{3.4}^{J/\psi}=2.4\times 10^{-2}\hbox{~GeV}^3$ and
$M_{3.4}^{\psi(2S)}=1.0\times 10^{-2}\hbox{~GeV}^3$. In both
extractions, the colour-octet matrix elements are considerably smaller
than those from the Tevatron fits, but the uncertainties are large.

The effects of colour-octet terms on the polarization of $J/\psi$ in $B$
decays were considered by Fleming, Hernandez, Maksymyk, and
Nadeau~\cite{Fleming:1996pt} and by Ko, Lee, and Song~\cite{Ko:1999zx}.
The polarization variable $\alpha$ for $J/\psi$ production is defined 
by the angular distribution in \Eq~(\ref{eq:prod-alphadef}).
In $B$ meson decays, the most convenient choice of the polarization axis 
is the direction of the boost vector from the $J/\psi$ rest frame 
to the rest frame of the $B$ meson.
The colour-evaporation model predicts no polarization.  The predictions of
NRQCD factorization and of the colour-singlet model depend on the mass
of the $b$ quark.  For $m_b = 4.7 \pm 0.3$~GeV, the prediction of
NRQCD factorization is $\alpha = -0.33 \pm 0.08$~\cite{Fleming:1996pt}
and the prediction of the colour-singlet model is $\alpha = -0.40 \pm
0.07$~\cite{Fleming:1996pt}. The uncertainties that arise from $m_b$
have been added in quadrature with other uncertainties. We note that the 
uncertainty in $m_b$ that was used in this calculation is 
considerably larger than the uncertainty of 2.4\% that is given in 
\Chapter~\ref{chapter:precisiondeterminations}. Measurements of
the polarization by the CLEO Collaboration have given the results
$\alpha=-0.30\pm 0.08$ for $J/\psi$ and $\alpha=-0.45\pm 0.30$ for
$\psi(2S)$~\cite{Anderson:2002md}. The result for $J/\psi$ strongly
disfavors the colour-evaporation model and is consistent with the
predictions of the NRQCD factorization approach and the colour-singlet
model.

Bodwin, Braaten, Yuan, and Lepage have applied the NRQCD factorization
approach to the production of the P-wave charmonium states
$\chi_{cJ}$ in $B$ decays~\cite{Bodwin:1992qr}. For P-wave
quarkonium production, there is a colour-octet NRQCD matrix element
that is of the same order in $v$ as the leading colour-singlet matrix
element. Therefore, the factorization formula must include both the
colour-singlet P-wave and the colour-octet S-wave contributions. The
short-distance coefficient in the colour-singlet ${}^3P_J$ term for
$\chi_{cJ}$ production vanishes at leading order in $\alpha_s$ for
$J=0,2$~\cite{Kuhn:1979zb,Kuhn:1983ar}. The colour-octet ${}^3S_1$ term
for $\chi_{cJ}$ production is proportional to the number of spin
states $2J+1$. Thus, the relative importance of the colour-singlet and
colour-octet terms can vary dramatically among the three $\chi_{cJ}$
states.  The prediction of the colour-singlet model at leading order in
$\alpha_s$ that the direct production rate of $\chi_{c2}$ should
vanish can be tested.  The feeddown from $\psi(2S)$ decay contributes
$(0.24 \pm 0.04) \times 10^{-3}$ to the inclusive branching fraction
into $\chi_{c2}$ given in \Table~\ref{tab:prodsec-Brbcharm}.  The
remainder $(0.5 \pm 0.4) \times 10^{-3}$ is consistent with zero, and
hence it is compatible with the prediction of the colour-singlet model,
but it is also compatible with a small colour-octet contribution.

\section{$B_c$ production}
\label{sec:prodsec-Bc}

The $B_c$ and $B_c^*$ are the ground state and the first excited state
of the $\bar b c$ quarkonium system. Their total angular momentum and
parity quantum numbers are $J^P$ = $0^-$ and $1^+$, and their dominant
Fock states have the angular momentum quantum numbers ${}^1S_0$ and
${}^3S_1$, respectively. In the following discussion, we will refer to
general $\bar b c$ quarkonium states as $B_c$ mesons and use the terms
$B_c$ and $B_c^*$ specifically for the ground state and the first 
excited state.

In contrast to charmonium
and bottomonium states, which have ``hidden flavour,'' $B_c$ mesons
contain two explicit flavours. As a consequence, the $B_c$ decays only
through the weak interactions, and the $B_c^*$ decays through an
electromagnetic transition into the $B_c$ with almost $100\%$
probability. The higher-mass $B_c$ mesons below the $B D$ threshold
decay through hadronic and electromagnetic transitions into lower-mass
$B_c$ mesons with almost $100\%$ probability. They cascade down through
the $\bar b c$ spectrum, eventually producing a $B_c$ or a $B_c^*$.
Another consequence of the explicit flavours is that the most important
production mechanisms for $B_c$ mesons are completely different from
those for hidden-flavour quarkonia. In the production of $B_c$ mesons by
strong or electromagnetic interactions, two additional heavy quarks
$\bar{c}$ and $b$ are always produced. The production cross-sections for
$B_c$ mesons are suppressed compared with the production cross-sections
for hidden-flavour quarkonia because the leading-order diagrams are of
higher order in the coupling constants and also because the phase-space
is suppressed, owing to the presence of the additional heavy quarks.

The small cross-sections for producing $B_c$ mesons make the
prospects for observing the $B_c$ at $e^+e^-$ and $ep$ colliders rather
bleak. A possible exception to this assessment exists for the case of
production at an $e^+e^-$ collider with energy at the $Z^0$ peak,
for which the production rate of the $B_c$ is predicted to be
marginal for observation \cite{Chang:bb}. The $B_c$ was not
discovered at LEP, despite careful searches
\cite{Abreu:1996nz,Barate:1997kk,Ackerstaff:1998zf}. It was finally
discovered at the Tevatron by the CDF collaboration in 1998
\cite{Abe:1998wi,Abe:1998fb}. We restrict our attention in the remainder
of this subsection to the production of $B_c$ mesons at hadron colliders.

The production of $B_c$ mesons was studied before the discovery of the
$B_c$
\cite{Chang:bb,Braaten:1993jn,Chang:1992jb,Chang:1994aw,Chang:1996jt,
Kolodziej:1995nv,Berezhnoy:an,Baranov:wy,Berezhnoi:1997fp}. If one
assumes that all nonperturbative effects in the production of the
$B_c$ in hadron--hadron collisions can be absorbed into the hadrons'
parton distribution functions (PDF's), then the inclusive production
cross-section can be written in the factorized form
\begin{eqnarray}
d\sigma[h_A h_B \rightarrow B_c + X] &=& \sum_{ij} 
\int dx_{1} dx_{2} \; f^{h_A}_i(x_{1},\mu) f^{h_B}_j(x_{2},\mu)\; 
d\hat{\sigma}[ij \rightarrow B_c + X] \;.
\label{eq:prodsec-Bccross}
\end{eqnarray}
The NRQCD factorization formula for the parton--parton cross-section is
\begin{eqnarray}
d\hat{\sigma}[ij \rightarrow B_{c} + X] =
\sum_n d\hat{\sigma}[ij \rightarrow (\bar b c)_n + X] \; 
\langle {\cal O}_n^{B_c} \rangle \;,
\label{eq:prodsec-Bcsubcross}
\end{eqnarray}
where the sum is over 4-fermion operators that create and annihilate
$\bar b c$. At the leading order in $\alpha_s$, which is $\alpha_s^4$,
the parton--parton process is $i j \to (\bar b c) + b \bar{c}$, where $ij
= gg$ (gluon fusion) or $q \bar q$ (quark--antiquark annihilation). Since
$m_{B_c}> m_b > m_c\gg \Lambda_{QCD}$, the leading-order parton--parton
process involves only hard propagators, even at small $p_T$.
Nevertheless, because of soft-gluon interactions, for example between
the $B_c$ and the recoiling $b$ and $\bar c$ quarks, it is not clear
that hard-scattering factorization holds at small $p_T$.

According to the velocity-scaling rules of NRQCD, the matrix element for
$B_c$ production that is of leading order in $v$ is 
$\langle {\cal O}^{B_c}_1({}^1S_0) \rangle$. 
The vacuum-saturation approximation
can be used to show that it is proportional to $F_{B_c}^2$, where
$F_{B_c}$ is the decay constant of the $B_c$, up to corrections of order
$v^4$. The leading matrix element for $B_c^*$ production is 
$\langle {\cal O}^{B_c^*}_1({}^3S_1) \rangle$. The vacuum-saturation
approximation and heavy-quark spin symmetry can be used to show that
this matrix element is also proportional to $F_{B_c}^2$, up to
corrections of order $v^3$.  The leading colour-octet matrix elements
are suppressed as $v^3$ or $v^4$. The colour-octet terms in
(\ref{eq:prodsec-Bcsubcross}) are probably not 
as important for $B_c$ mesons as
they are for hidden-flavour quarkonia. In the case of $J/\psi$
production, the short-distance coefficients of the colour-octet matrix
elements are enhanced relative to those for the colour-singlet matrix
element by an inverse power of the QCD-coupling $\alpha_s$, by factors
of $p_T/m_c$ at large $p_T$, by factors of $m_c/p_T$ at small $p_T$, and
by colour factors. The only one of these enhancement factors that may
apply to the $B_c$ is the colour factor. Because there are many Feynman
diagrams that contribute to the parton process $i j \to (\bar b c) + b
\bar c$ at order $\alpha_s^4$, the colour correlations implied by
individual Feynman diagrams tend to average out. We therefore expect a
$\bar b c$ pair to be created in a colour-octet state roughly eight times
as often as in a colour-singlet state. We will assume that, in spite of
the enhancement from this colour factor,  the colour-octet contributions
to the production cross-sections for $B_c$ and $B_c^*$ are small
compared with the leading colour-singlet contributions. This assumption
is equivalent to using the colour-singlet model to calculate the
production rate.
 
Two approaches have been used to compute the cross-sections for $B_c$
mesons: the complete order-$\alpha_s^4$ approach
\cite{Chang:1992jb,Chang:1994aw,Chang:1996jt,%
Kolodziej:1995nv,Berezhnoy:an,Baranov:wy} and the fragmentation
approach \cite{Braaten:1993jn,Cheung:1999ir}.  In the complete
order-$\alpha_s^4$ approach, the parton cross-section in
\Eq~(\ref{eq:prodsec-Bcsubcross}) is computed at leading order in
$\alpha_s$, where the only subprocesses are $i j \to (\bar b c) + b
\bar c $:
\begin{eqnarray}
d\hat{\sigma}[ij \rightarrow B_{c} + X] =
d\hat{\sigma}[ij \rightarrow \bar b c_1({}^1S_0) + b \bar c] \;
\langle  {\cal O}^{B_c}_1(^1S_0) \rangle \;.
\label{eq:prodsec-BcsubcrossLO}
\end{eqnarray}
The fragmentation approach is based on the fact that, for
asymptotically large $p_T \gg M_{B_c}$, the cross-section
(\ref{eq:prodsec-Bcsubcross}) can be further factored into a
cross-section for producing $\bar b$ and a fragmentation function
$D_{\bar b \to B_c}(z,\mu)$ that gives the probability for the $\bar
b$ to fragment into a $B_c$ carrying a fraction $z$ of the $\bar b$
momentum:
\begin{eqnarray}
d\hat{\sigma}[ij \rightarrow B_{c} + X] \approx
\int dz \; d\hat{\sigma}[ij \rightarrow \bar b  + b ] \;
D_{\bar b \to B_c}(z,\mu) \;.
\label{eq:prodsec-BcsubcrossFRAG}
\end{eqnarray}
If both factors are calculated only at leading order in $\alpha_s$,
this is just an approximation to the complete order-$\alpha_s^4$
cross-section in \Eq~(\ref{eq:prodsec-Bcsubcross}). One advantage of
the fragmentation approach is that the expressions for the $\bar b$
production cross-section $ d\hat{\sigma}$ and the fragmentation
function $D_{\bar b \to B_c}$ in \Eq~(\ref{eq:prodsec-BcsubcrossFRAG})
can be written down in a few lines. For $p_T \gg m_{B_c}$, the
fragmentation approach has another advantage in that the
Altarelli--Parisi evolution equations can be used to sum the leading
logarithms of $p_T/m_c$ to all orders. Unfortunately, as was pointed
out in Ref.~\cite{Chang:1994aw,Chang:1996jt,Kolodziej:1995nv}, the
fragmentation cross-section does not become an accurate approximation
to the complete order-$\alpha_s^4$ cross-section until surprisingly
large values of $p_T$. For example, if the parton centre-of-mass
energy is $\sqrt{\hat{s}}=200$~GeV, the fragmentation cross-section is
a good approximation only for $p_T \geq 40$~GeV.  We will therefore
not consider the fragmentation approach further.

The authors of Ref.~\cite{Chang:2003cq} recently developed a Monte
Carlo event generator for hadronic $B_c$ and $B^*_c$ production,
using the complete order-$\alpha_s^4$ approach and taking advantage
of helicity amplitude techniques \cite{Mangano:1990by}. The generator
is a Fortran package, and  it is implemented in PYTHIA
\cite{Sjostrand:1993yb}, which allows one to generate complete events.
The complete order-$\alpha_s^4$ cross-section includes contributions
from gluon--gluon fusion and  quark--antiquark annihilation. At the
Tevatron, the gluon--gluon fusion mechanism is dominant over 
quark--antiquark annihilation, except in certain kinematics regions
\cite{Chang:1992jb,Chang:2003cr}. At the LHC, the gluon--gluon fusion
mechanism is always dominant. All the results below are obtained
from the gluon--gluon fusion mechanism only.

The inputs that are required to calculate the complete
order-$\alpha_s^4$ cross-sections are the masses $m_b, m_c$, and
$m_{B_c}$, the decay constant $F_{B_c}$, the PDF's, the QCD coupling
constant $\alpha_s$, and the factorization scale $Q$. The masses $m_b$
and $m_c$ are known with uncertainties of about 2.4\% and 8\%,
respectively.  In the NRQCD factorization approach, one sets $m_{B_c} =
m_c+m_b$ and $m_{B_c^*} = m_c+m_b$ in the short-distance coefficients.
Contributions from operators of higher order in $v$ then account for the
binding energy in $m_{B_c}$ and $m_{B_c^*}$. This procedure is also
required in order to preserve gauge invariance if one makes use of
on-shell spin-projection operators for the $B_c$ and $B_c^*$ states
\cite{Chang:1979nn,Guberina:1980fn}. Since an experimental measurement
of the decay constant $F_{B_c}$ is not available, one has to use a value
that is obtained from potential models
\cite{Chen:fq,Eichten:1994gt,Kiselev:1994rc,AbdEl-Hady:1999xh} or from
lattice gauge theory \cite{Davies:1996gi}. The uncertainty in the factor
$F_{B_c}^2$ is about 6\%. Since the order-$\alpha_s^4$ cross-section is
at leading order in perturbation theory, the running of $\alpha_s$ can
be taken at leading order, and LO versions of the PDF's can be used.

The running coupling constant and the PDF's depend on the
renormalization/factorization scale $\mu$, and, so, a
prescription for the scale $\mu$ is also required. There is no general
rule for choosing the scale in an LO calculation, especially in the
case of a $2\to 3$ subprocess, such as $ij \to B_c + b \bar c$. The
factorization formula (\ref{eq:prodsec-BcsubcrossFRAG}) for asymptotically
large $p_T$ suggests that an appropriate choice for the scales in the
fragmentation contribution to the cross-section might be to set $\mu=
m_{bT}\equiv (m_b^2 + p_T^2)^{1/2}$ in the PDF's and $\alpha_s^4
=\alpha_s^2(m_{bT}) \alpha_s^2(m_c)$ in the parton cross-section.
However, the fragmentation term does not dominate until very large
$p_T$, and there are important contributions to the cross-section
that have nothing to do with fragmentation
\cite{Chang:1994aw,Chang:1996jt,Kolodziej:1995nv}. For example, there
are contributions that involve the splitting of one of the colliding
gluons into a $c \bar c$ pair, followed by the creation of a $b \bar b$
pair in the hard scattering of the $c$ from the other gluon and
then by the recombination of the $\bar b$ and $c$ into a $B_c$. 
The sensitivity to the choice of $\mu$ could be decreased by 
carrying out a complete
calculation of the production cross-section for the $B_c$ at
next-to-leading order in $\alpha_s$, but this is, at present,
prohibitively difficult. In the absence of such a calculation, we
can use the variation in the complete order-$\alpha_s^4$ cross-section 
for several reasonable choices for the scale as an estimate of the
uncertainty that arises from the choice of scale.

%%%%%%%%%%%%%%%%%%%%%%%%%%%%%%%%%%%%%%%%%%%%%%%%%%%%%%%%%%%%%%%%%%%%%%%%%%%%%%%%%%%%%%
\begin{table}
\caption[cross-sections for direct production of $B_c$ and $B_c^*$]
        {The cross-sections (in nb) for direct production of $B_c$ and
        $B_c^*$ at the Tevatron and at the LHC for various values of
        the charm-quark mass $m_c$. The gluon distribution function is
        CTEQ5L, the running of $\alpha_s$ is leading order, the scale
        is $\mu^{2}=\hat{s}/4$, and the other parameters are
        $F_{B_c}=480$~MeV, and $m_b=4.9$~GeV.}
\label{tab:prod-Bctab}
\begin{center}
\begin{tabular}{|c|ccccc|ccccc|}
\hline\hline & \multicolumn{5}{|c}{~~~Tevatron~($\sqrt s=2$
TeV)~~~}& \multicolumn{5}{|c}{~~~LHC~($\sqrt s=14$~TeV)~~~}\\
\hline $m_c$ (GeV) & ~~$1.4$~~ & ~~$1.5$~~
&~~$1.6$~~&~~$1.7$~~ & ~~$1.8$~~ & ~~$1.4$~~
 & ~~$1.5$~~ &~~$1.6$~~ &~~$1.7$~~ &~~$1.8$~~\\
$\sigma[B_{c}]$ (nb) & 3.87 & 3.12 &
2.56 & 2.12 & 1.76 & 61.0 & 49.8 & 41.4 & 34.7 & 28.9 \\
$\sigma[B^{*}_c]$ (nb) & 9.53 & 7.39 & 5.92 &
4.77 & 3.87 & 153. & 121.& 97.5 & 80.0 & 66.2 \\
\hline\hline
\end{tabular}
\end{center}
\end{table}
%%%%%%%%%%%%%%%%%%%%%%%%%%%%%%%%%%%%%%%%%%%%%%%%%%%%%%%%%%%%%%%%%%%%%%%%%%%%%%%%%%%%%%

\begin{figure}
\centering
\includegraphics[width=0.49\textwidth]{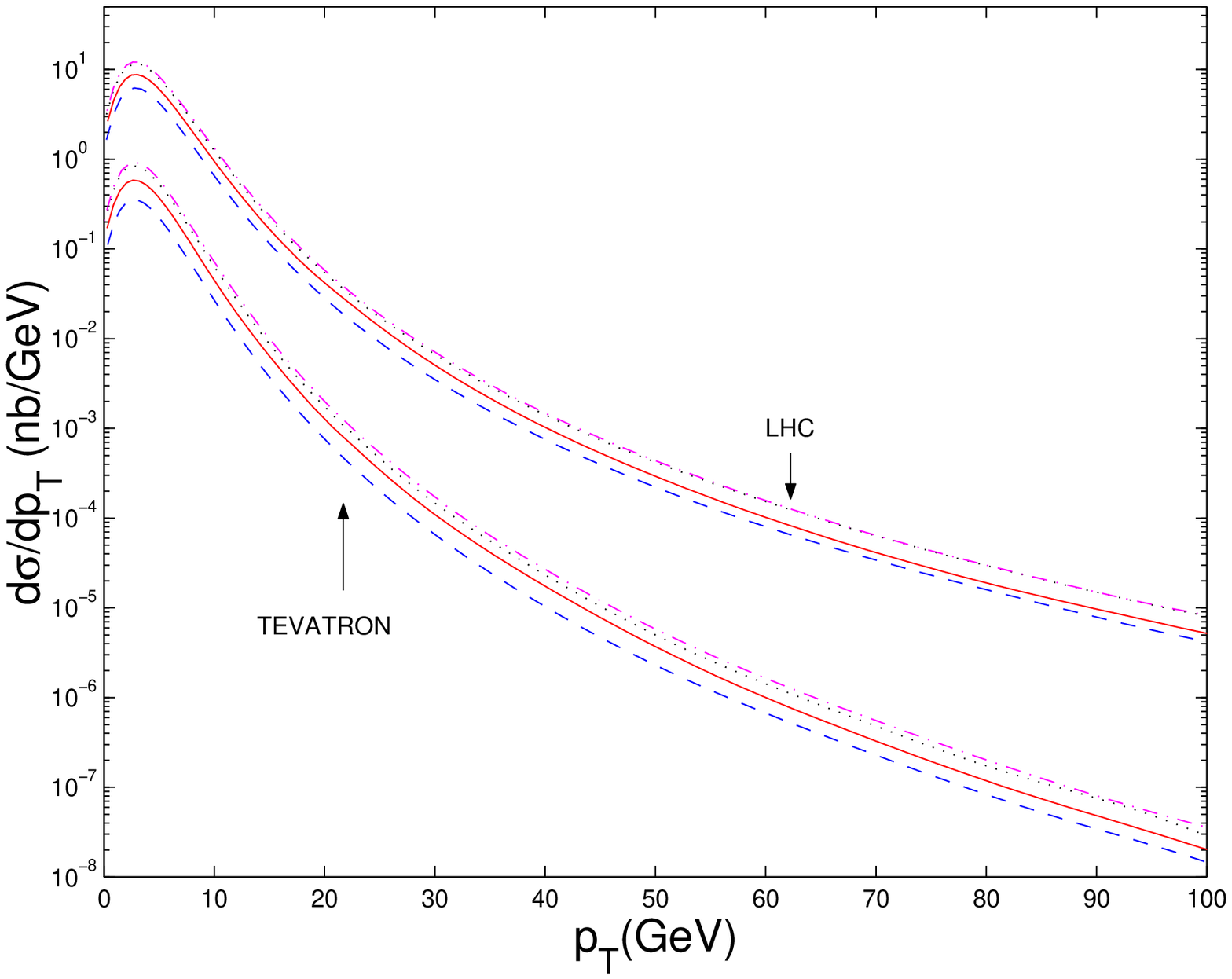}
\hfill
\includegraphics[width=0.49\textwidth]{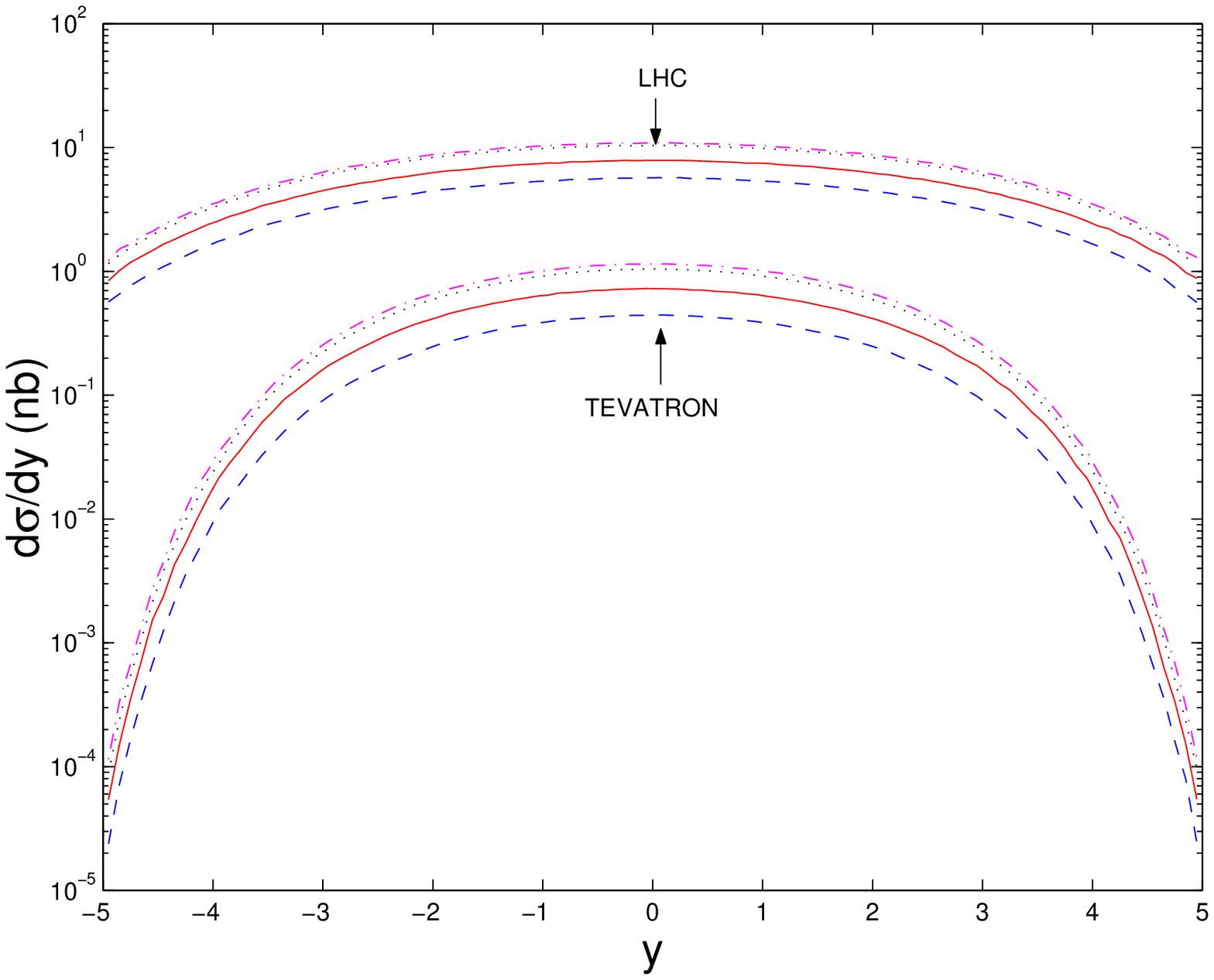}
\caption[Differential cross-sections for direct production of $B_c$]
        {The differential cross-sections for the direct production of
        the $B_c$ as a function of its transverse momentum $p_T$ and
        its rapidity $y$ at the Tevatron ($\sqrt{s}=2$~TeV) and at the
        LHC ($\sqrt{s}=14$~TeV) for four choices of the scale:
        $\mu^2=\hat{s}/4$ (solid line), $\mu^2=p_{T}^2+m_{B_c}^2$
        (dotted line), $\mu^2=\hat{s}$ (dashed line), and
        $\mu^2=p_{Tb}^2+m_b^2$ (dash--dot line).  The gluon
        distribution is CTEQ5L, the running of $\alpha_s$ is leading
        order, and the other parameters are $F_{B_c}=480$~MeV,
        $m_c=1.5$~GeV, and $m_b=4.9$~GeV.}
\label{fig:prod-Bcfig}
\end{figure}

The hadronic production cross-section for $B_c$ mesons depends
strongly on the collision energy. In \Table~\ref{tab:prod-Bctab}, we
give the direct cross-sections for $B_c$ and $B_c^*$ production at the
Tevatron and the LHC for several values of the charm quark mass $m_c$
and for typical values for the other parameters. The cross-section for
$B_c$ production at the LHC is larger than at the Tevatron by a factor
of about 16. The cross-sections for $B_c^*$ production are larger than
those for $B_c$ production by a factor of about 2.4. The
cross-sections are fairly sensitive to the charm-quark mass, varying
by more than a factor of two as $m_c$ is varied from 1.4 to 1.8
GeV. In \Figure~\ref{fig:prod-Bcfig}, we show the differential
cross-sections for $B_c$ production as a function of the $B_c$
transverse momentum $p_T$ and $B_c$ rapidity $y$ at the Tevatron and
the LHC, using four different prescriptions for the scale $\mu$. At
central rapidity, the variations among the four choices of scale is
about a factor of three at the Tevatron and a factor of two at the
LHC. The differential cross-sections decrease more slowly with $p_T$
and $|y|$ at the LHC than at the Tevatron. The total uncertainty from
combining all of the uncertainties in the direct cross-section for
$B_c$ production is less than an order of magnitude. The uncertainty
in the ratio of the direct-production cross-sections for the $B_c^*$
and the $B_c$ is much smaller because many of the uncertainties cancel
in the ratio.

%%%%%%%%%%%%%%%%%%%%%%%%%%%%%%%%%%%%%%%%%%%%%%%%%%%%%%%%%%%%%%%%%%%%%%%%%%%%%%%%%%%%%%
%%%%%%%%%%%%%%%%%%%%%%%%%%%%%%%%%%%%%%%%%%%%%%%%%%%%%%%%%%%%%%%%%%%%%%%%%%%%%%%%%%%%%%

The results presented above are for the direct production of the $B_c$
and the $B_c^*$. Experiments at the Tevatron and the LHC will measure
the inclusive cross-sections, including the feeddown from all of the
higher states of the $\bar b c$ system. The $\bar b c$ system has a
rich spectrum of excited states below the $B D$ threshold.  They
include an additional S-wave multiplet, one or two P-wave
multiplets, and a D-wave multiplet. After being produced, these
excited $B_c$ mesons all cascade eventually down to the ground state
$B_c$. Since the $B_c^*$ decays into the $B_c$ with a probability of
almost 100\%, the feeddown from directly-produced $B_c^*$'s increases
the cross-section for the $B_c$ by about a factor of 3.4. The complete
order-$\alpha_s^4$ cross-sections for $B_c$ and $B_c^*$ production can
be applied equally well to the $2S$ multiplet. The direct-production
cross-sections for these states are smaller than those for the $1S$
states by the ratio of the squares of the wave functions at the
origin, which is about 0.6. Thus, the inclusive cross-section for
$B_c$ production, including the effect of feeddown from the direct
production of all of the S-wave $B_c$ states, is larger than the
cross-section for direct $B_c$ production, which is given in
\Table~\ref{tab:prod-Bctab} and shown in \Figure~\ref{fig:prod-Bcfig},
by a factor of about 5.4.

%%%%%%%%%%%%%%%%%%%%%%%%%%%%%%%%%%%%%%%%%%%%%%%%%%%%%%%%%%%%%%%%%%%%%%%%%%%%%%%%%%%%%%
\begin{figure}[t]
\begin{center}
\includegraphics[width=10cm]{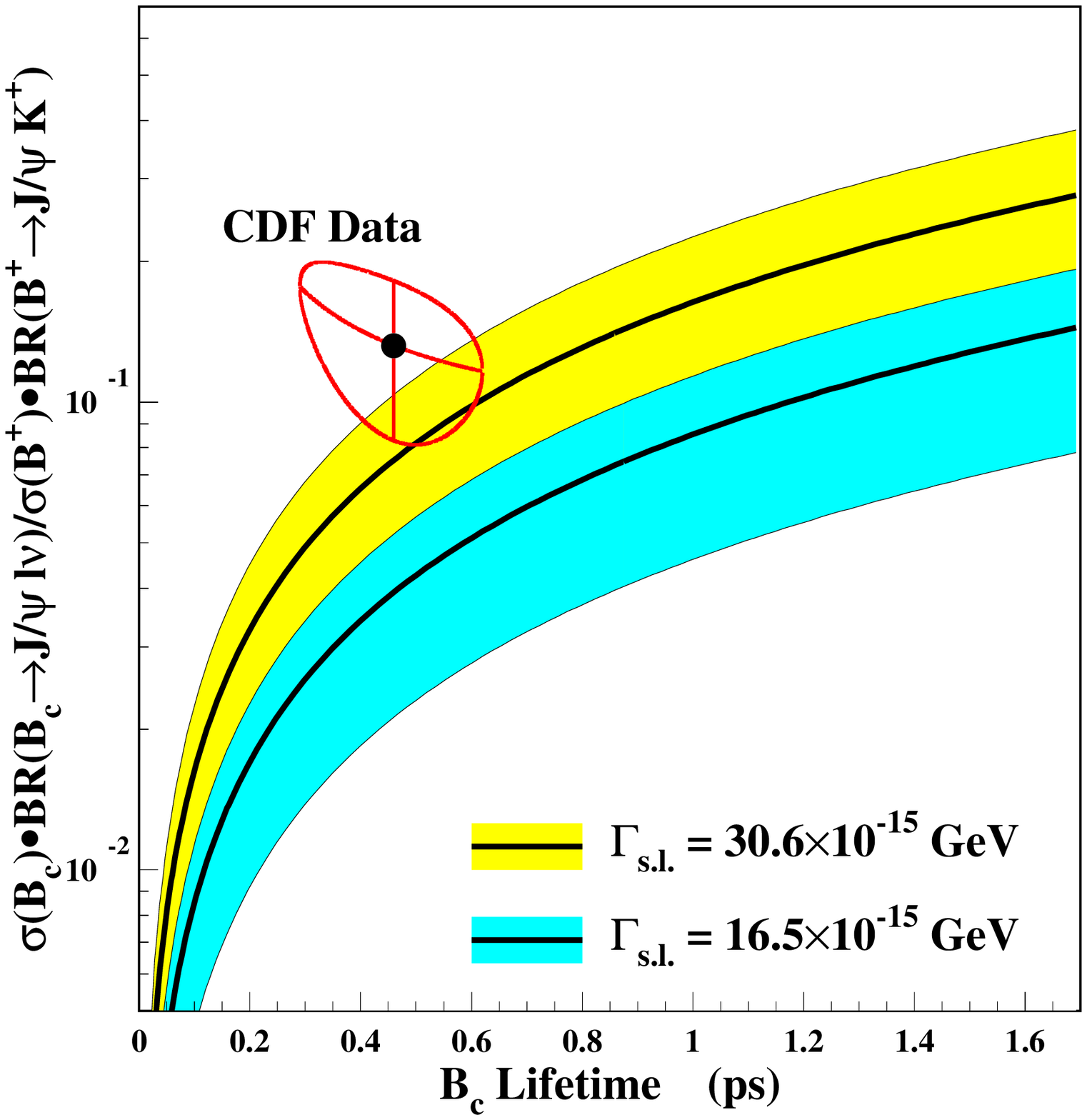}
\end{center}
\caption[The ratio \ensuremath{R[J/\psi l\nu]}]
        {The ratio $R[J/\psi l\nu]$, which is defined in
         \Eq~(\ref{eq:prodsec-Bcratio}), versus the $B_c$ lifetime.  The
         point, surrounded by a one-standard-deviation contour, shows
         the values of $R[J/\psi l\nu]$ and the $B_c$ lifetime that
         were measured by CDF \cite{Abe:1998wi,Abe:1998fb}.  The
         shaded region represents the theoretical predictions and
         their uncertainty bands from Refs.~\cite{Lusignoli1,Scora}
         for two different values of the semileptonic width
         $\Gamma_{s.l.} = \Gamma[B_c \rightarrow J/\psi l \nu]$. }
\label{fig:bcxsec_cdf}
\end{figure}
%%%%%%%%%%%%%%%%%%%%%%%%%%%%%%%%%%%%%%%%%%%%%%%%%%%%%%%%%%%%%%%%%%%%%%%%%%%%%%%%%%%%%%

The production of $B_c$ in $p \bar p$ collisons at $\sqrt{s} = 1.8$~TeV 
has been measured at the Tevatron by the CDF collaboration 
\cite{Abe:1998wi,Abe:1998fb}. 
CDF has measured the ratio
\begin{eqnarray}
R[J/\psi l\nu] = 
\frac{\sigma[B_c] {\rm Br}[B_c^+ \rightarrow J/\psi l^+\nu]}
     {\sigma[B^+] {\rm Br}[B^+ \rightarrow J/\psi K^+]}
\label{eq:prodsec-Bcratio}
\end{eqnarray}
for $B_c^+$ and $B^+$ with transverse momenta $p_T >$ 6.0~GeV 
and with rapidities $|y| <$ 1.0. 
Their result is $R[J/\psi l\nu] = 0.132^{+0.041}_{-0.037}(\rm stat.)
\pm 0.031 (\rm syst.) ^{+0.032}_{-0.020}(\rm lifetime)$. 
This result is consistent with results from previous 
searches \cite{Abreu:1996nz,Barate:1997kk,Ackerstaff:1998zf}. 
\Figure[b]~\ref{fig:bcxsec_cdf} compares the CDF
measurements of $R[J/\psi l\nu]$ and the $B_c$ lifetime
with theoretical predictions from Refs.~\cite{Lusignoli1,Scora}
for two different values of the semileptonic width 
$\Gamma_{s.l.} = \Gamma[B_c \rightarrow J/\psi l \nu]$. 
The theoretical predictions use the values 
$|V_{cb}| = 0.041 \pm 0.005$~\cite{PDG96}, 
$\sigma[B^+_c]/\sigma[\bar{b}] = 1.3 \times 10^{-3}$~\cite{Lusignoli2}, 
$\sigma[B^+]/\sigma[\bar{b}] = 0.378 \pm 0.022$~\cite{PDG96}, 
and ${\rm Br}[B^+ \rightarrow J/\psi K^+] 
        = (1.01 \pm 0.14) \times 10^{-3}$~\cite{PDG96}.
The predictions and the measurement are consistent within 
experimental and theoretical uncertainties.  
\shortpage

Quantitative predictions for the contribution to the inclusive $B_c$
production cross-section from the feeddown from P-wave states would
require complete knowledge of the order-$\alpha_s^4$ cross-sections for
the production of P-wave states. It is theoretically inconsistent to
use the colour-singlet model to calculate these cross-sections for the
P-wave states. There are colour-octet terms in the P-wave production
cross-sections that are of the same order in both $v$ and $\alpha_s$ as
the colour-singlet terms, and they must be included. The colour-singlet
production matrix elements for the P-wave states can be estimated from
potential models or determined from lattice gauge theory.  The
colour-octet production matrix elements for the P-wave states can
perhaps be estimated by interpolating between the corresponding matrix
elements for charmonium and bottomonium states.

In summary, the order-$\alpha_s^4$ colour-singlet production
cross-section for S-wave $\bar b c$ mesons can be used to predict the
$B_c$ production cross-section, including feeddown from excited S-wave
states.  The uncertainty in the normalization of that prediction is
less than an order of magnitude. If the inclusive cross-section for
$B_c$ production that is measured at the Tevatron or the LHC is much
larger than that prediction, it could indicate that there is a large
contribution from the feeddown from P-wave or
higher-orbital-angular-momentum states. It could also indicate that
the colour-octet contributions to the direct production of the $B_c$
and the $B_c^*$ are important.

\section{Summary and outlook}
\label{sec:prodsec-summary}
\shortpage

NRQCD factorization, together with hard-scattering factorization,
provides a systematic formalism for computing inclusive quarkonium
production rates in QCD.  Nonperturbative effects associated with the
binding of a $Q \bar Q$ pair into a quarkonium are factored into NRQCD
matrix elements that scale in a definite manner with the typical
relative velocity $v$ of the heavy quark in the quarkonium.  The NRQCD
matrix elements are predicted to be universal, \ie independent of the
process that creates the $Q \bar Q$ pair. The NRQCD factorization
formula for inclusive cross-sections is believed to hold when $p_T \gg
\Lambda_{\rm QCD}$, where $p_T$ is the transverse momentum of the
quarkonium with respect to the colliding particles.  It is
well-motivated by the effective field theory NRQCD and by
factorization theorems that have been proven for simpler
hard-scattering processes.  Explicit proofs of factorization for
quarkonium production would be welcome, because they would help
quantify the sizes of corrections to the factorization formula. It is
important to bear in mind that conventional proofs of hard-scattering
factorization fail at small $p_T$. Consequently, NRQCD factorization
formulas, even those that include soft-gluon resummation, may be
unreliable in this region. It also follows that the NRQCD
factorization approach may not be applicable to total cross-sections
if they are dominated by contributions at small $p_T$.

The NRQCD factorization approach incorporates elements of both the
colour-singlet model and the colour-evaporation model.  It includes
the colour-singlet model terms, for which the NRQCD matrix elements
can be determined from annihilation decays.  It also includes
colour-octet production mechanisms, as in the colour-evaporation
model.  The NRQCD factorization approach extends these models into a
systematically improvable framework.  The colour-singlet model is
emphatically ruled out by the observation of prompt $J/\psi$ and
$\psi(2S)$ production at the Tevatron at rates that are more than an
order of magnitude larger than the colour-singlet-model predictions.
The colour-evaporation model is ruled out by the observations of
nonzero polarization of $J/\psi$'s in $B$ meson decays and in $e^+e^-$
annihilation at 10.6~GeV and by the observation of nonzero
polarization of $\Upsilon(2S)$'s and $\Upsilon(3S)$'s in fixed-target
experiments.  It is also ruled out by the fact that different values
of the fraction of $J/\psi$'s that come from $\chi_c$ decays are
measured at the Tevatron and in $B$-meson decays.  Despite having been
ruled out, the colour-singlet model and the colour-evaporation model
can still play useful roles as ``straw men" with which to compare the
predictions of NRQCD factorization. The colour-evaporation model has
not yet been ruled out, for example, as a description of differential
cross-sections at the Tevatron and in fixed-target experiments.

The NRQCD factorization approach provides a general phenomenological 
framework that cannot be ruled out easily.  The factorization 
formula involves infinitely many NRQCD matrix elements, most of which are 
adjustable parameters.  It is only the truncation in $v$ that reduces 
those parameters to a finite set. The standard truncation has 
four independent NRQCD matrix 
elements for each S-wave multiplet and two independent NRQCD matrix 
elements 
for each P-wave multiplet.  NRQCD factorization with the standard 
truncation in $v$ remains a phenomenologically viable description of 
inclusive quarkonium production.  As one tests NRQCD factorization at
higher levels of precision, the standard truncation must ultimately
fail. The NRQCD factorization approach itself may remain viable if one
truncates at a higher order in $v$, but only at the cost of
introducing many new adjustable parameters.

In the effort to make the predictions of the NRQCD factorization 
approach more quantitative, the most 
urgent need is to extend all calculations to next-to-leading order (NLO) in 
$\alpha_s$.  For hadron collisions at small $p_T$ ($p_T \ll m$), the 
leading-order parton process is $ij \to Q \bar Q$. NLO calculations of 
that process are 
already available, but a resummation of multiple gluon emissions is required 
in order to tame large logarithms of $m^2/p_T^2$ and 
to turn the singular $p_T$ distribution into a smooth distribution.  For 
very large $p_T$ ($p_T \gg m$), the production of quarkonium is dominated 
by gluon fragmentation.  The leading-order fragmentation process is 
$g \to Q \bar Q_8 ( ^3S_1 )$, and the NLO calculation of the gluon 
fragmentation function into $Q \bar Q$ is available.  What is still lacking 
is the NLO calculation at intermediate $p_T$, for which the 
leading-order parton process is $ij \to  Q \bar Q + k$.  By taking 
into account the NLO 
corrections in $\alpha_s$, one should significantly decrease some of the 
uncertainties in the NRQCD factorization predictions.

One problematic source of uncertainties in the NRQCD factorization 
predictions is relativistic corrections.  The first relativistic 
corrections of order $v^2$ in the channel that corresponds to the 
colour-singlet model have been calculated for many processes.  In 
many cases, they have large coefficients that cast doubt on 
the validity of the 
expansion in powers of $v$ for charmonium, and even for bottomonium.  
The success of lattice NRQCD in describing bottomonium spectroscopy 
ensures the 
applicability of the velocity expansion for this system in some form.  It 
is possible that some reorganization or resummation of the velocity 
expansion might be necessary in order to make precise quantitative 
calculations of charmonium production.

The best individual experiments for determining the NRQCD production 
matrix elements for both charmonium and bottomonium are probably 
those at the Tevatron, because of the large range of $p_T$
that is accessible. It will be important to take advantage of the
measurements down to small $p_T$ that were achieved at the CDF
detector for bottomonium in Run I and for charmonium in Run II.  This
will require taking
into account the effects of multiple gluon emission in the theoretical
analysis.  
Measurements of charmonium production in other experiments
are also important because they provide tests of the universality of the
production matrix elements. These experiments include those that measure
charmonium production in $ep$ collisions at HERA,
in $e^+ e^-$ annihilation at the $B$ factories, and in
$B$ meson decays at the $B$ factories.  
One can use these experiments to extract values of
the NRQCD matrix elements or, as has typically been the practice to
date, one can use the matrix elements that have been extracted from the
Tevatron data to make predictions for charmonium production in other  
experiments.

The ratios of the production cross-sections for different quarkonium
states may also provide important tests of NRQCD factorization.
(Here, particularly, one must keep in mind the {\it caveats} about the
applicability of the NRQCD factorization approach to total
cross-sections.) The uncertainties in the predictions for ratios of
cross-sections are much smaller than those in the individual
cross-sections because many of the uncertainties cancel in the
ratio. The variations of the ratios from process to process and as
functions of kinematic variables provide important information about
the production mechanisms.  The ratios of production rates of
spin-triplet S-wave states, such as the $\psi(2S)$ to $J/\psi$ ratio,
do not seem to vary much.  However, a significant variation has been
observed in a ratio of the production rates of P-wave and S-wave
states, namely the fraction of $J/\psi$'s that come from decays of
$\chi_c$'s.  A substantial variation has also been observed in a ratio
of production rates of P-wave states, namely the $\chi_{c1}$ to
$\chi_{c2}$ ratio.  More precise measurements of these and other
ratios would be valuable.  Of particular importance would be
measurements of ratios of production rates of spin-singlet and
spin-triplet quarkonium states, such as the $\eta_c$ to $J/\psi$
ratio.  The absence of clean signatures for spin-singlet quarkonium
states makes such measurements difficult.

The polarization of quarkonium is another important test of NRQCD
factorization.  The standard truncation in $v$ leads to unambiguous
predictions for the ratios of production rates of different
spin states,  without introducing any new parameters.  The
predictions are most easily tested for the quarkonium
states with $J^{PC} = 1^{--}$, but they can also be tested for other 
states.  It is extremely important to test the simple qualitative 
predictions that in hadron collisions the $1^{--}$ states should become 
transversely polarized at sufficiently large $p_T$.  More careful 
quantitative estimates of the polarization of the $J/\psi$, the 
$\psi(2S)$, and the
$\Upsilon (nS)$ as functions of $p_T$ at the Tevatron and the LHC
would be useful.  More precise measurements of the polarization of
the $J/\psi$ and the $\psi(2S)$ in other production processes,
such as $ep$ collisions, $e^+e^-$ annihilations, and $B$ decay, would
also be valuable.

The most puzzling experimental results in quarkonium production in recent 
years have been the double-$c \bar c$ results from $e^+e^-$ annihilation at 
the $B$ factories.  The measurements by the Belle collaboration of the 
fraction of $J/\psi$'s that are accompanied by charmed hadrons and 
of the 
exclusive cross-section for $J/\psi + \eta_c$ production are both much 
larger than 
expected.  No satisfactory theoretical explanation of these results has 
emerged.  It would be worthwhile to measure the fraction of $J/\psi$'s 
accompanied by charm hadrons in other processes, such as $p \bar p$ 
annihilation at the Tevatron and $ep$ collisions at HERA, to see if there 
are any surprises.

The outlook for progress in understanding quarkonium production is very 
bright.  The NRQCD factorization approach provides a general framework for 
describing inclusive quarkonium production.  Current experiments will 
provide severe tests of NRQCD factorization with the standard truncation of 
the velocity expansion.  These tests will either provide a firm 
foundation for predictions of quarkonium production in future experiments
or lead us to new insights into the physics of quarkonium production.

\BLKP
%13/12/2004

%\documentclass[11pt,twoside]{cernrep}
%\usepackage{epsfig,graphicx}
%\usepackage{here,cite}

% Counter commands
%\setcounter{page}{1}
%\setcounter{chapter}{5}
%\setcounter{secnumdepth}{3}
%\setcounter{tocdepth}{3}

%\input{newcommand.tex}

%\begin{document}
\bibliographystyle{ckm}

\chapter{PRECISION DETERMINATIONS OF QCD PARAMETERS FROM QUARKONIA}
\label{chapter:precisiondeterminations}
{{\it Conveners:} S.~Eidelman, A.H.~Hoang, M.~Jamin}
\par\noindent
{{\it Authors:} S.~Boogert, C.T.H.~Davies, S.~Eidelman, A.H.~Hoang,
  M.~Jamin, A.S.~Kronfeld, P.B.~Mackenzie, A.~Penin, A.~Pineda,
  I.W.~Stewart, T.~Teubner}

\section{Introduction} 
\shortpage

The accurate knowledge of the parameters of the Standard Model (SM) is
an important requirement in the indirect search for new physics based
on observables that can be predicted with small theoretical
uncertainties and that are measurable experimentally with high
precision. Among the parameters of QCD, for example, the precise
knowledge of the top quark mass plays a crucial role in the relation
of the electroweak precision observables $M_Z$, $M_W$ and the weak
mixing angle, which is sensitive to the vacuum structure and to non-SM
virtual particles. On the other hand, for the analysis of inclusive
B-decay rates the bottom and charm quark masses are needed as an
input. 

Heavy quarkonium systems provide an ideal instrument to extract the
heavy quark masses and to get constraints on the strong coupling using
perturbative as well as non-perturbative methods. The perturbative
methods rely on the fact that the heavy quark masses are larger than
the hadronization scale $\Lambda_{\rm QCD}$ and that non-perturbative
effects affecting the dynamics can be suppressed. Based on the concept
of effective theories, on new techniques to compute higher order
perturbative corrections and on an improved understanding of the
higher order behaviour of perturbation theory, a number of powerful
methods were developed in recent years that led to an improved
understanding of the perturbative structure of heavy quarkonium
systems and to more realistic estimates of the uncertainties. For the
determination of the masses of the bottom and the charm quarks sum
rules based on moments of the hadronic cross-section in
$e^+e^-$-annihilation are the most reliable tool. While theoretically
one needs to predict the moments with high precision, dedicated
experiments are needed to provide measurements of the hadronic
cross-section with small uncertainties. A different method to
determine the masses of the bottom and charm quarks employs the masses
of the low lying bottomonium and charmonium resonances with the
assumption that the perturbative contributions dominate.

An alternative method relies on lattice simulations of heavy
quarkonium systems where theoretical predictions are made
non-perturbatively. Here the main issues are the control of lattice
artifacts, unquenching and the proper extrapolation to physical quark
masses, and the matching to the continuum theory. Continuous
improvements on this approach have been observed in recent years.

A heavy quarkonium system that can be studied at a future $e^+e^-$
Linear Collider is the top--antitop threshold at a centre-of-mass
energy of approximately twice the top quark mass. Although the top
quark lifetime in the SM is predicted to be too small to allow the
production of separated resonances, the top--antitop system is
governed by bound-state-type non-relativistic dynamics. Moreover, the
large top quark width provides a very effective protection against the
influence of non-perturbative effects making the non-relativistic
top--antitop systems almost entirely perturbative for predictions of
inclusive quantities. A number of precise measurements of top quark
properties can be carried out at the top--antitop threshold. Among
them the measurement of the top quark mass with an uncertainty at the
level of 100~MeV is the most prominent one, exceeding the capabilities
of hadron colliders by an order of magnitude.

In this chapter an overview is given on the current status of precision
determinations of QCD parameters from quarkonium systems. In
\Section~\ref{sec:R_measurements} experimental aspects of measurements of the
total cross-section $\sigma(e^+e^-\to \mbox{hadrons})$ in the charmonium and
bottomonium energy regions are reviewed. These measurements are the basis for
charm and bottom quark mass extractions using QCD sum rules. In
\Section~\ref{sec:mcmb} the theoretical aspects of charm and bottom mass
determinations from QCD sum rules and from the quarkonium ground state
masses are reviewed. The emphasis is placed on perturbative methods,
but also the status of lattice simulations is
summarised. \Section~\ref{sec:alphas} contains a brief review of
determinations of the strong coupling from quarkonium properties using
perturbative methods as well as lattice simulations of the quarkonium
spectrum. Some conceptual aspects of Nonrelativistic QCD and, in
particular, the issue of summing logarithms of the velocity in the
theoretical description of the quarkonium dynamics in the framework of
vNRQCD (``velocity NRQCD'') are summarised in
\Section~\ref{sec:vnrqcd}. Finally, in \Section~\ref{sec:topthreshold} the
experimental and theoretical aspects of top quark pair production
close to threshold at a future Linear Collider are reviewed. In
particular, the prospects of measurements of the top quark mass, its
width and its couplings to the Higgs boson and to gluons are
discussed.

\section{\boldmath{$R$}-Measurements in Heavy Quarkonium Regions} 
\label{sec:R_measurements}

The so-called $R$-ratio is the total cross-section of producing
hadrons in $e^+e^-$ collisions corrected for initial state
radiation and normalised to the lowest order QED
cross-section of the reaction $e^+e^- \to \mu^+\mu^-$,
\begin{equation}
R(s)= \frac{\sigma(e^+e^- \to \mbox{\rm hadrons})(s)}
         {\sigma^{(0)}(e^+e^- \to \mu^+\mu^-)(s)}
\,,
\end{equation}
where $\sqrt{s}$ is the c.m.\ energy.  Measurements of $R$ or of the
cross-section for hadrons containing a specific quark flavour such as
charm or bottom can be used for a variety of fundamental tests of
perturbative QCD using the assumption of duality and the operator
product expansion (OPE)~\cite{Shifman:1979bx,Reinders:1985sr}.  For
example, using QCD sum rules, which deal with moments of the hadronic
cross-section, one can extract values for the quark masses, and for
the quark and gluon condensates that parametrise non-perturbative
effects in the OPE~\cite{Shifman:1979bx,Reinders:1985sr}.  Through
dispersion relations $R$ measurements give an important input to the
calculations of the hadronic corrections to various fundamental
quantities that are influenced by the photonic vacuum polarisation at
low energies: the anomalous magnetic moment of the
muon~\cite{Davier:2003pw}, the running fine structure constant
$\alpha(s)$~\cite{Burkhardt:2001xp}, hyperfine splitting in
muonium~\cite{Czarnecki:2001yx} \etc  From the size of higher order
QCD corrections it is also possible to get constraints on the strong
coupling $\alpha_{\rm s}$~\cite{Kuhn:2001dm}.  Depending on the
problem, different energy ranges are of importance.

\begin{figure}
\centering\includegraphics[width=.75\textwidth]{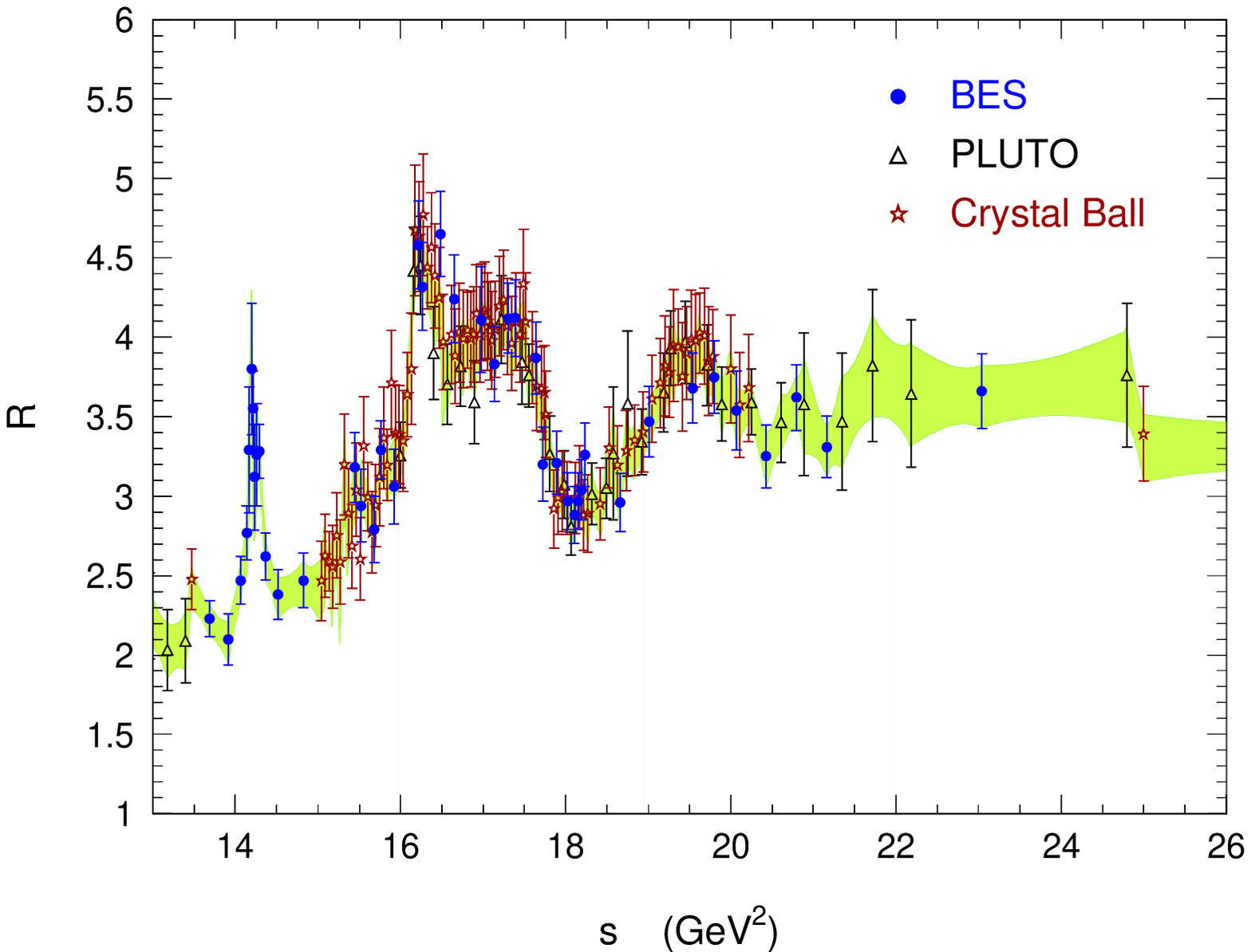}
\caption{Results of $R$ Measurements between 3.6 and 5~GeV}
\label{fig:figint}

\medskip

\begin{center}
\begin{minipage}[b]{.51\linewidth}
\centering\includegraphics[height=105mm]{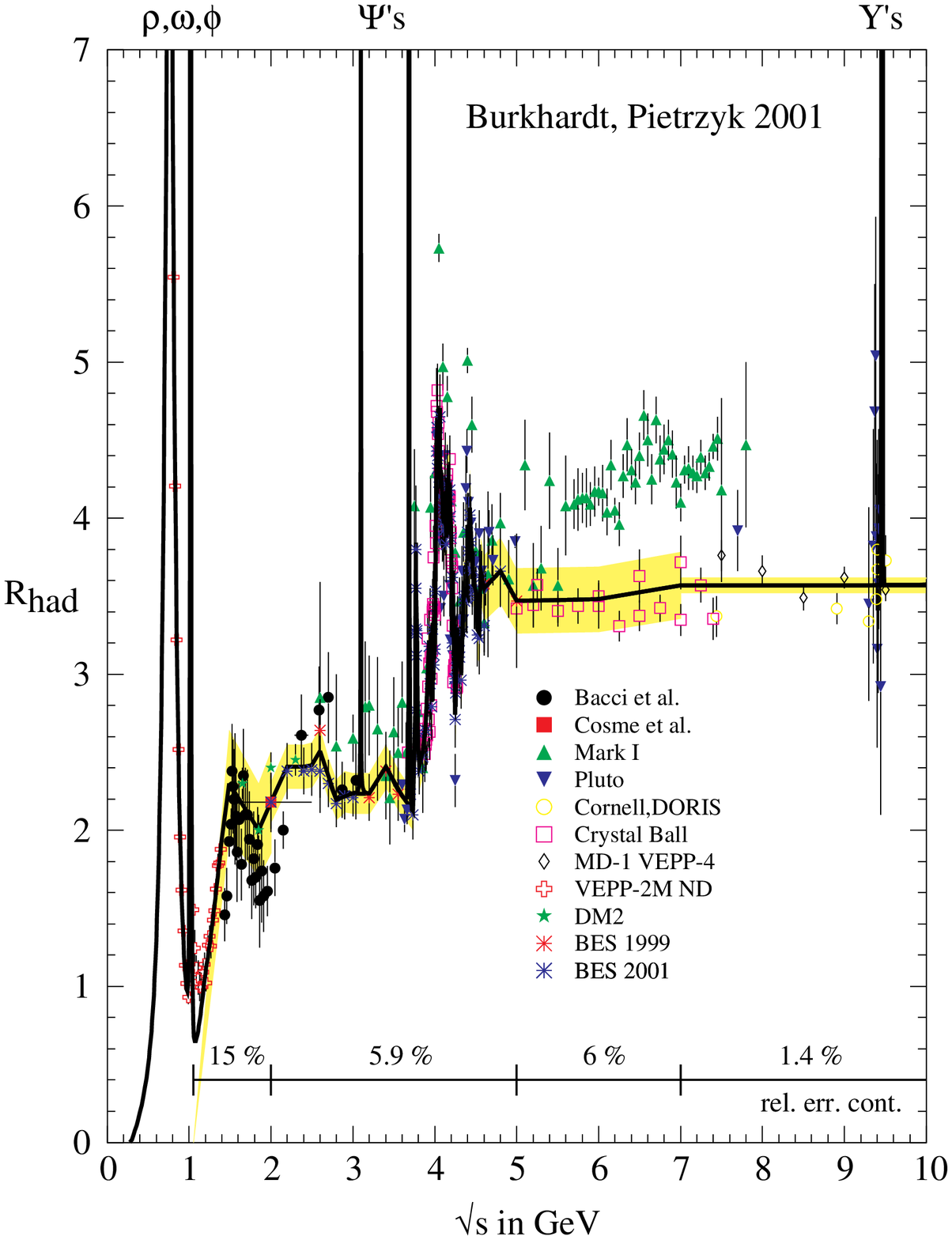}
\end{minipage}
\hfill
\begin{minipage}[b]{.47\linewidth}
\centering\includegraphics[height=105mm]{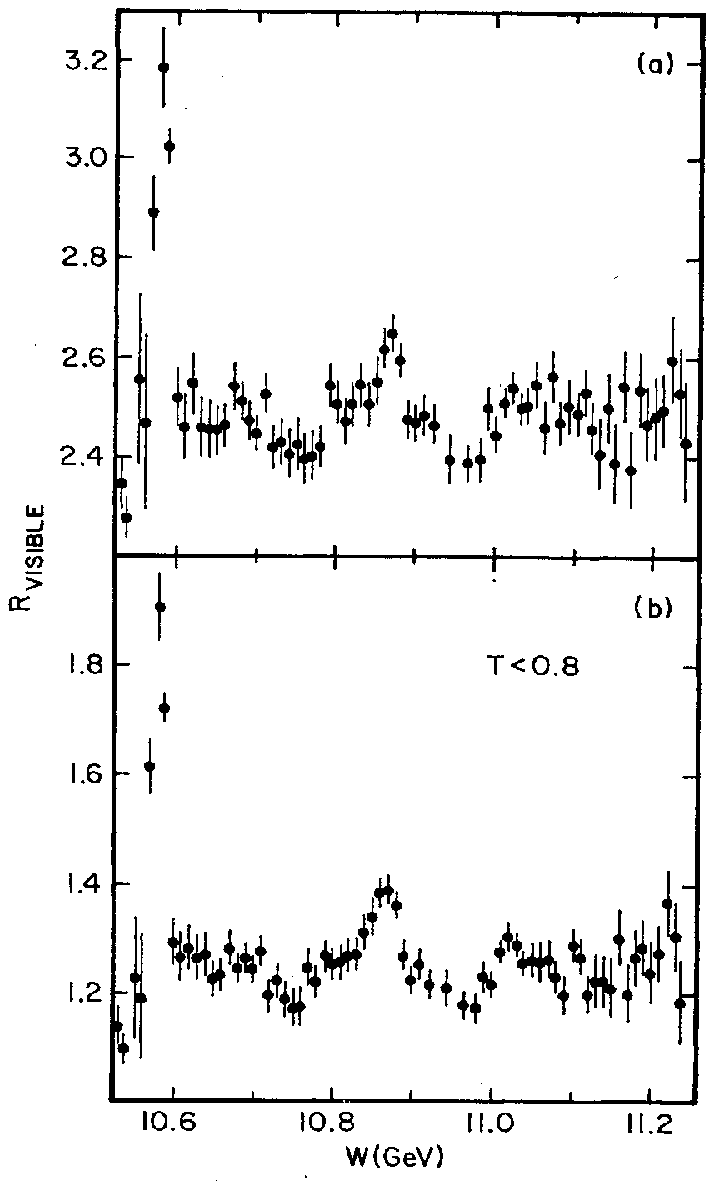}
\end{minipage}\\
\begin{minipage}[t]{.52\linewidth}
\caption{Results of $R$ Measurements below 10~GeV}
\label{fig:figall}
\end{minipage}
\hfill
\begin{minipage}[t]{.47\linewidth}
\caption{(a) $R_{\rm vis}$ vs. total energy W measured by CUSB.
(b) Same as (a) with a thrust cut to suppress  continuum.}
\label{fig:figcusb}
\end{minipage}
\end{center}
\end{figure}

In the c.m. energy range between 3 and 5~GeV, measurements of $R$ were
carried out by many experimental groups studying various states just
above the charmonium threshold: PLUTO~\cite{Burmester:1977mn},
DASP~\cite{Brandelik:1978ei},
MARK~I~\cite{Siegrist:1976br,Siegrist:1982zp}, Crystal
Ball~\cite{Osterheld:1986hw} and BES~\cite{Bai:1999pk,Bai:2001ct}.  In
general, these measurements are consistent.  Of all the analyses the
results by BES are the most precise. In the first measurement $R$ was
measured at 6 energy points from 2.6 to 5.0~GeV with a systematic
uncertainty between 5.9\% and 8.4\%~\cite{Bai:1999pk} while in the
second one the energy range from 2 to 5~GeV was scanned at 85 energy
points with an average systematic uncertainty of about
7\%~\cite{Bai:2001ct}.  Despite the rather detailed information on $R$
collected by BES, no data on the cross-sections of exclusive channels
or on the parameters of broad charmonia in this energy range are
available yet from this experiment. As a result, the resonance
properties of $\psi(4040), \psi(4160)$ and $\psi(4415)$ are still
determined by the older DASP~\cite{Brandelik:1978ei} and
MARK~I~\cite{Siegrist:1976br} measurements. A comparison of these
experiments is presented in \Table~\ref{tab:inter} whereas
\Figure~\ref{fig:figint} shows the energy dependence of $R$ between 3.6 and
5~GeV measured by PLUTO, Crystal Ball and BES.

\begin{table}
\caption{$R$ Measurement at charmonium threshold}
\label{tab:inter}
\begin{center}
\begin{tabular} {|c|c|c|c|c|c|}
\hline
Group & PLUTO  \cite{Burmester:1977mn}   & 
  DASP \cite{Brandelik:1978ei} &
MARK~I \cite{Siegrist:1982zp} &
Cr. Ball \cite{Osterheld:1986hw} &
BES \cite{Bai:1999pk,Bai:2001ct} \\
\hline
Solid angle, $\Omega/4\pi$ & 0.86 & 0.6 & 0.65 & 0.93 & 0.80 \\
\hline
Energy $\sqrt{s}$,~GeV &
 3.1 -- 4.8 & 3.1 -- 5.2 & 2.6 -- 5.0 & 
3.87 -- 4.50 & 3.0 -- 5.0 \\
\hline
 $\int{Ldt}$, nb$^{-1}$ &  $\sim$ 3000 & 7500 &  $\sim$ 1500 
&  $\sim$ 2000  &  $\sim$ 2500 \\
\hline
Events &  $\sim$ 20000 & $\sim$ 40000 & $\sim$ 7000 
& $\sim 22000$ & $\sim 35000$  \\
\hline
Av. efficiency, \% & 70 -- 80 & 35 -- 40 & 30 -- 60 & 85 
&  70 -- 80 \\
\hline
Syst. error, \% & 12 & 15 & 10 -- 20 & 10 & 7 \\
\hline
\end{tabular}
\end{center}
\end{table}

A controversial situation exists between 5 and 7~GeV where $R$ values
measured by MARK~I~\cite{Siegrist:1982zp} are substantially higher
than both those of Crystal Ball~\cite{Edwards:1990pc} and the
prediction of perturbative QCD, see \Figure~\ref{fig:figall}. The
result of Crystal Ball is in fair agreement with the QCD
prediction. Two groups, LENA~\cite{Niczyporuk:1982ya} and
MD-1~\cite{Blinov:1991dj,Blinov:1996fw}, performed measurements in the
broad energy range from 7.4 to 9.4~GeV and 7.2 to 10.34~GeV,
respectively. Information on these experiments is summarised in
\Table~\ref{tab:high}.

Various groups have measured the value of $R$ in the narrow energy range 
in the vicinity of the $\Upsilon$-family 
resonances~\cite{Berger:1979bp,Bock:1980ag,Albrecht:1982bq,Rice:1982br,Giles:1984vq,Jakubowski:1988cd,Albrecht:1992vp,Ammar:1998sk}. The
highest systematic accuracy of 1.9\% was reached by CLEO~\cite{Ammar:1998sk}. 
We summarise the obtained values of $R$ in Table~\ref{tab:r}. No energy 
dependence is observed within the experimental accuracy, which is 
not surprising taking into account that most of 
the measurements were made below the open beauty threshold. 
 
\begin{table}
\caption{$R$ Measurements from 5 to 10~GeV }
\label{tab:high}
\begin{center}
\begin{tabular} {|c|c|c|c|c|}
\hline
Group & MARK~I \cite{Siegrist:1982zp} &
Crystal Ball \cite{Edwards:1990pc} &
LENA \cite{Niczyporuk:1982ya} & MD-1 \cite{Blinov:1996fw} \\
\hline
Solid angle, $\Omega/4\pi$ & 0.65 & 0.93 & 0.75 & 0.7  \\
\hline
Energy $\sqrt{s}$,~GeV &
 5.0 -- 7.8 & 5.0 -- 7.4 & 7.4 -- 9.4 & 
7.80 -- 10.45 \\
\hline
 $\int{{Ldt}}$, nb$^{-1}$ & $\sim$ 3300 & $\sim$ 1500 &1140 
&  16000 \\
\hline
Events & $\sim 20000$ & $\sim$ 11000 & 4050 & $ 48000$  \\
\hline
Av. efficiency, \% & 60  & 85 & 82 -- 90 &  50 \\
\hline
Syst. error, \% & 10 & 10 & 7 & 3.9 \\
\hline
\end{tabular}
\end{center}
\end{table}

In \Figure~\ref{fig:figall}, we present the results of $R$
Measurements below 10~GeV~\cite{Burkhardt:2001xp}. Only statistical
errors are shown. The relative uncertainty assigned by the authors of
Ref.~\cite{Burkhardt:2001xp} to their parameterisation, displayed as
the solid line, is shown as a band and given with the numbers at the
bottom.

The hadronic cross-section above the $B\bar{B}$ threshold (a
centre-of-mass energy range from 10.60 to 11.25~GeV) was measured by
the CUSB~\cite{Lovelock:1985nb} and the CLEO~\cite{Besson:1985bd}
collaborations with an integrated luminosity of 123 pb$^{-1}$ and 70
pb$^{-1}$, respectively.  This energy range is of substantial interest
since in quarkonium potential models two excited states are expected
there~\cite{Eichten:1978tg}. Moreover the coupled-channel models also
predict a rich structure in $R$ due to the turn-on of various
exclusive states~\cite{Tornqvist:1984fx}. Both groups observe similar
structures and provide compatible parameters for the two highest
states at 10.865~GeV and 11.019~GeV, tentatively referred to as
$\Upsilon(5S)$ and $\Upsilon(6S)$. However, the values of these
parameters are obtained under different assumptions, thus, their
formal averaging presently applied by the PDG~\cite{Eidelman:2004wy}
hardly makes sense, as noted in Ref.~\cite{Corcella:2002uu}. There is
also a visible step between the continuum points below the
$\Upsilon(4S)$ and the average level above it. In
\Figure~\ref{fig:figcusb}(a) we show the results for the visible $R$ ratio,
$R_{\rm vis}$, in this energy range obtained by
CUSB~\cite{Lovelock:1985nb}. Their results with an additional thrust
cut to suppress the continuum are shown in
\Figure~\ref{fig:figcusb}(b).

\begin{table}
\caption{Values of $R$ in the $\Upsilon$-family range}
\label{tab:r}
\begin{center}
\begin{tabular} {|l|c|c|}
\hline
Group & Energy,  $\sqrt{s}$,~GeV & $R$ \\
\hline
PLUTO~\cite{Berger:1979bp} & 9.4 & $3.70 \pm 0.30 \pm 0.56$ \\
\hline
DHHM~\cite{Bock:1980ag} & 9.4--9.5 & $3.80 \pm 0.27 \pm 0.42$ \\
  & 9.98--10.1 & $3.60 \pm 0.36 \pm 0.54$ \\
\hline
LENA~\cite{Niczyporuk:1982ya} & 7.4--9.4 & $3.37 \pm 0.06 \pm 0.23$ \\
\hline
ARGUS~\cite{Albrecht:1982bq} & 9.40--9.44, & $3.73 \pm 0.16 \pm 0.28$ \\
             & 9.49--9.60  &   \\
\hline
CUSB~\cite{Rice:1982br} &  10.34--10.52 &  $3.54 \pm 0.05 \pm 0.39$ \\
\hline
CLEO~\cite{Giles:1984vq} & 10.49 & $3.77 \pm 0.06 \pm 0.24$ \\
\hline
Crystal Ball~\cite{Jakubowski:1988cd} & 9.39 & $3.48 \pm 0.04 \pm 0.16$ \\
\hline
ARGUS~\cite{Albrecht:1992vp} & 9.36 & $3.46 \pm 0.03 \pm 0.13$ \\
\hline
MD-1~\cite{Blinov:1996fw} & 7.34--10.24 & $3.58 \pm 0.02 \pm 0.14$ \\
\hline
CLEO~\cite{Ammar:1998sk} & 10.52  & $3.56 \pm 0.01 \pm 0.07$ \\
\hline
\end{tabular}
\end{center}
\end{table}

It is important to note that for various applications, \eg  for
extracting quark masses from spectral moments, it is necessary to know
the component of $R$ coming from a specific quark flavour,
particularly in the threshold energy range. Experimentally, this is a
rather complicated problem. One of the theory-driven possibilities can
be illustrated by a method used in a recent charm mass
determination~\cite{Hoang:2004xm}.  For the energy range from 3.73 to
4.8~GeV the authors employ the data for the total $R$ obtained by the
BES collaboration~\cite{Bai:2001ct}.  To obtain the charm component of
$R$, $R_{\rm cc}$, they first fit the non-charm $R$ ratio, $R_{\rm
nc}$, assuming its energy independence and using the last four data
points below 3.73~GeV. Finally, $R_{\rm cc}$ is obtained by
subtracting the fitted values of $R_{\rm nc}$ from $R_{\rm tot}$. Note
that to estimate the final error of $R_{\rm cc}$ in this method a
sophisticated analysis of various experimental uncertainties is
needed. Another possibility is to reconstruct all exclusive final
states containing particles with a corresponding quark flavour. For
example, in the vicinity of the charm threshold one can assume that
the corresponding $R$ component, $R_{\rm cc}$, is saturated by the
contributions from the $D\bar{D}, D\bar{D}^*, D^*\bar{D}^*$ final
states. Clearly, this method crucially depends on the assumptions made
and requires reliable reconstruction of various final states.  One
should hope that an analysis of the already collected data samples as
well as that of the new energy scan of the relevant energy range
planned by BES will clarify the situation. In the future, substantial
progress in the charmonium energy range can be expected from CLEOc and
the $c-\tau$ factory.  Prospects for the bottomonium energy range are
less clear since both B-factories are running at the peak of the
$\Upsilon(4S)$ only.

One should also take into account new possibilities suggested by the
method of radiative return or initial state radiation
(ISR)~\cite{Binner:1999bt,Benayoun:1999hm}. Recently BaBar
successfully applied this method to the detection of the $J/\psi \to
\mu^+\mu^-$ decay~\cite{Aubert:2003sv}.  The peak cross-section of
this process was measured with a 2.2\% systematic error and, using the
world average values for the leptonic branching ratios, the total and
leptonic width was obtained with an accuracy better than the world
average. Also studied were various exclusive final states with pions
and kaons. For example, the $2\pi^+2\pi^-$ final state with a hadronic
mass from 900 to 3000~MeV was successfully studied and the
corresponding cross-section already has better accuracy than all other
experiments at fixed energy~\cite{Davier:2003gp}. Also an inclusive
approach was applied, where the hadronic mass is extracted from the
ISR photon energy. However, in this method the resolution deteriorates
rapidly for low recoil masses.

 Now let us briefly discuss which accuracy seems ultimately feasible
in future experiments under ``ideal'' detection and analysis
conditions. The consideration of the experimental papers on $R$
measurements, which claimed the smallest systematic uncertainty in the
charmonium region (those of BES~\cite{Bai:1999pk,Bai:2001ct}) and in
the bottomonium region (those of MD-1~\cite{Blinov:1996fw} and
CLEO~\cite{Ammar:1998sk}) allows the following conclusions.
\begin{itemize}
\item The four main sources of systematic uncertainties are selection
      criteria, luminosity determination, detection efficiency and
      radiative corrections.
\item The uncertainty caused by selection criteria is dominated by 
      the knowledge of background. To large extent its level depends
      on the detector performance and ranges from 0.5\% for CLEO to
      2.0\% for MD-1.  We consider the value of 0.5\% as a very
      aggressive one, which can hardly be improved and will further
      assume the value (0.5--1.0)\%.
\item The uncertainty due to the luminosity determination was 1\% for CLEO  
      after a very thorough analysis of the three normalisation
      processes Bhabha scattering, $e^+e^- \to \mu^+\mu^-$ and $e^+e^-
      \to \gamma\gamma$~\cite{Crawford:1994gq}. One of the factors
      restricting the accuracy at that time was the knowledge of
      radiative corrections for the normalisation process, which was
      of the order of 1\%~\cite{Berends:1981jk}.  Today the
      cross-section of the main QED processes is known to better than
      0.2\%~\cite{Arbuzov:1997pj}. Taking also into account the
      experience gained at LEP, it doesn't look impossible to achieve
      the level of 0.5 to 0.7\%.
\item The detection efficiency is usually obtained by Monte Carlo 
      simulations of the experiment using the package
      JETSET~\cite{Sjostrand:1994yb} or its modification adapted to
      the charmonium energy range for the BES
      experiment~\cite{Andersson:1999yy}.  The corresponding
      contribution to the uncertainty was 1\% for CLEO and 2 to 3\%
      for BES. Note that this contribution is very sensitive to the
      solid angle coverage. Therefore, a value of 1\% for the future
      4$\pi$ detectors seems accessible.
\item The modern approach to the calculation of the radiative corrections 
      for the final hadronic state based on structure functions yields
      formulae with an intrinsic accuracy of about
      0.5\%~\cite{Kuraev:1985hb}. Some uncertainty is induced by the
      calculational procedure itself, \eg  by the choice of the
      maximum allowed energy. This leads to an error of about 0.7 to
      0.8\%.
\end{itemize}    

Finally, assuming that all four sources of the uncertainty are
independent and adding the corresponding estimates in quadrature, one
obtains that the ultimate systematic uncertainty could be as low as
1.4 to 1.8\%.  This is of course a very optimistic estimate. However,
it should be possible to reach 3\% in a dedicated $R$ measurement.
 
\section{Bottom and charm quark mass determinations}
\label{sec:mcmb}

At the ongoing and future B-physics experiments the values of the
bottom and charm quark masses and realistic estimates of their
uncertainties will become increasingly important for the measurements
of the CKM parameters and the search for new physics. However, due to
confinement and the non-perturbative aspects of the strong
interaction, the concept of quark masses cannot be tied to an
intuitive picture of the weight or the rest mass of a particle, such
as for leptons, which are to very good approximation insensitive to
the strong interactions. Rather, analogous to the strong coupling
$\alpha_{\rm s}$, quark masses should be considered as couplings of
the Standard Model Lagrangian that have to be determined from
processes that depend on them. As such, quark masses are quantities
that depend on the renormalisation scheme that is used and in general
also on the renormalisation scale.  Although physical quantities do
not depend on the renormalisation scheme, the question of which quark
mass scheme should be employed can be relevant in order to avoid
correlations to other unknown parameters or a badly converging
perturbative expansion.

\subsection{Quark mass definitions in perturbation theory}
\label{sec:massdefs}

In principle, one is free to employ any renormalisation scheme, or
definition for the quark masses. In the framework of QCD perturbation
theory, the difference between two mass schemes can be expressed as a
series in powers of the strong coupling $\alpha_{\rm s}$. Therefore,
higher-order terms in the perturbative expansion of a quantity that
depends on quark masses are affected by the renormalisation scheme
that is employed. Moreover, certain schemes turn out to be more
appropriate or more convenient for some purposes than others. In the
following, we review some of the most common quark mass definitions,
focusing for definiteness on the case of the bottom quark.

\subsubsection*{Pole mass}

The bottom quark pole mass $m_{b,{\rm pole}}$ is defined as the solution to
the full inverse quark propagator,
\begin{equation}
\Slash{p}-m_{b,{\rm pole}}-\Sigma(p,m_{b,{\rm pole}})
\Big|_{p^2=m_{b,{\rm pole}}^2} \,=\, 0 \,,
\end{equation}
where $\Sigma(p,m_b)$ is the bottom quark self energy. The pole mass
definition is gauge-invariant and infrared-safe to all orders in
perturbation theory \cite{Tarrach:1981up,Kronfeld:1998di} and has been
used as the standard mass definition of many perturbative computations
in the past. By construction, the pole mass is directly related to the
concept of the mass of a free quark, which is, however, problematic
because free quarks do not appear in nature. In practical applications
the pole mass has the additional disadvantage that the perturbative
series relating it to physical quantities are in general not well
convergent, due to a strong sensitivity of the pole mass definition to
small scales\cite{Bigi:1994em,Beneke:1994sw}.  This property of the
pole mass, often called the ``pole-mass renormalon problem'', is not
related to any physical effect.  The bad convergence of the
perturbative expansion of quantities in the pole mass scheme indicates
that the concept of the pole mass is ambiguous to an amount of order
$\Lambda_{\rm QCD}$. This issue is reviewed in more detail in
\Section~\ref{sec:mbmc1S}.

Nevertheless, in low orders of perturbation theory, there is in
principle nothing wrong to employ the pole mass as an intermediate
quantity, as long as it is used in a consistent way. In particular,
the presence of a renormalon
ambiguity~\cite{Bigi:1994em,Beneke:1994sw} requires considering the
numerical value of the pole mass as an order-dependent
quantity. Because this makes estimates of uncertainties difficult, the
pole mass definition should be avoided for analyses where quark mass
uncertainties smaller than $\Lambda_{\rm QCD}$ are desired. The
problems of the pole mass definition can be easily avoided if one uses
quark mass definitions that are less sensitive to small momenta and do
not have an ambiguity of order $\Lambda_{\rm QCD}$. Such quark mass
definitions are generically called ``short-distance'' masses. In
contrast to the pole mass, short-distance masses have a parametric
ambiguity of order $\Lambda_{\rm QCD}^2/m_b$ or smaller.

\subsubsection*{\MSB mass}

The most common short-distance quark mass parameter is the \MSB mass
$\overline m_b(\mu)$, which is defined by regulating QCD with
dimensional regularisation and subtracting only the
$1/\epsilon$-divergences in the \MSB scheme \cite{Bardeen:1978yd}.
Besides the renormalisation scheme, the \MSB mass also depends on the
renormalisation scale $\mu$. Since the subtractions do not contain any
infrared sensitive terms, the bottom \MSB mass is only sensitive to
scales of order or larger than $m_b$. The relation between the pole
mass and the \MSB mass is known to $\cO(\alpha_{\rm s}^3)$
\cite{Gray:1990yh,Chetyrkin:1999qi,Melnikov:2000qh} and reads for
massless light quarks $(\bar\alpha_{\rm s}\equiv\alpha_{\rm
s}^{(n_l=4)}(\overline m_b(\overline m_b)))$:
\begin{equation}
\label{polemsbar}
\frac{m_{b,{\rm pole}}}{\overline m_b(\overline m_b)} \,=\,
1 + \frac{4}{3}\,\frac{\bar\alpha_{\rm s}}{\pi} + \Big( 13.44 - 1.04\, n_l\Big)
\biggl(\frac{\bar\alpha_{\rm s}}{\pi}\biggr)^2 \! + \Big( 190.6 - 26.7\, n_l +
0.65\,n_l^2\Big)\biggl(\frac{\bar\alpha_{\rm s}}{\pi}\biggr)^3 \! + \ldots \,.
\end{equation}
The corrections coming from light quark masses at order $\alpha_{\rm
s}^2$ are fully known~\cite{Gray:1990yh}, while at order $\alpha_{\rm
s}^3$ only the dominant light quark mass corrections have been
determined~\cite{Hoang:2000fm}.  The bottom quark \MSB mass arises
naturally in processes where the bottom quark is far off-shell. The
scale $\mu$ in the \MSB mass is typically chosen of the order of the
characteristic energy scale of the process under consideration since
perturbation theory contains logarithmic terms $\sim\alpha_{\rm
s}(\mu)^n\ln(Q^2/\mu^2)$ that would be large otherwise. Using the
renormalisation group equation for $\overline m_b(\mu)$, the values of
the \MSB mass for different $\mu$ can be related to each other. The
\MSB mass definition is less useful for processes in which the bottom
quark is close to its mass-shell, which also includes when the bottom
quark has non-relativistic energies.

\subsubsection*{Threshold masses}

The shortcomings of the pole and the \MSB masses in describing
non-relativistic bottom quarks can be resolved by so-called threshold
masses~\cite{Hoang:2000yr}. The threshold masses do not possess the
ambiguity of order $\Lambda_{\rm QCD}$ and, at the same time, are
defined through subtractions that contain contributions that are
universal for the dynamics of non-relativistic quarks. Since the
subtractions are not unique, an arbitrary number of threshold masses
can be constructed. In the following some threshold mass definitions
that appear in the literature are briefly reviewed.

\vspace{0.2cm}\noindent
{\em Kinetic mass}\\
The kinetic mass is defined as~\cite{Bigi:1995ga,Bigi:1997si}
\begin{equation}
\label{kindef}
m_{b,{\rm kin}}^{}(\mu_{\rm kin}^{}) \,=\, m_{b,{\rm pole}} -
\Big[\bar\Lambda(\mu_{\rm kin}^{})\Big]_{\rm pert} -
\Biggl[\frac{\mu^2_\pi(\mu_{\rm kin}^{})}
{2 m_{b,{\rm kin}}(\mu_{\rm kin}^{})}\Biggr]_{\rm pert} + \ldots \,,
\end{equation}
where $[\bar\Lambda(\mu^{}_{\rm kin})]_{\rm pert}$ and
$[\mu_\pi^2(\mu^{}_{\rm kin})]_{\rm pert}$ are perturbative
evaluations of HQET matrix elements that describe the difference
between the pole mass and the B meson mass. The ellipses indicate
matrix elements of operators with higher dimension, which have not
been included in any analysis so far.

The relation between the kinetic mass and the \MSB mass is known to
$\cO(\alpha_{\rm s}^2)$ and $\beta_0 \cO(\alpha_{\rm s}^3)$
\cite{Czarnecki:1998sz,Melnikov:1998ug}.  The formulae for
$[\bar\Lambda(\mu_{\rm kin})]_{\rm pert}$ and $[\mu^2_\pi(\mu_{\rm
kin})]_{\rm pert}$ at $\cO(\alpha_{\rm s}^2)$ read
\cite{Melnikov:1998ug}
\begin{eqnarray}
\label{eq:Lambda_pert}
\Big[\bar\Lambda (\mu_{\rm kin})\Big]_{\rm pert} &\!\!\!=\!\!&
\frac{4}{3}\,C_F\mu_{\rm kin} \frac{\alpha_{\rm s}(\overline m)}{\pi} \Biggl\{
1 + \frac{\alpha_{\rm s}}{\pi} \biggl[
\biggl(\frac{4}{3} - \frac{1}{2}\ln\frac{2\mu_{\rm kin}}{\overline m}\biggr)
\beta_0 - C_A \biggl(\frac {\pi^2}{6} - \frac {13}{12}\biggr)\biggr]\Biggr \}
\,, \\
\smvs
\label{eq:mu2pert}
\Big[\mu_{\pi}^2 (\overline m)\Big]_{\rm pert} &\!\!\!=\!\!&
C_F\mu_{\rm kin}^2 \frac{\alpha_{\rm s}(\overline m)}{\pi} \Biggl\{
1 + \frac {\alpha_{\rm s}}{\pi} \biggl[ \biggl(\frac{13}{12} - \frac{1}{2} \ln
\frac{2\mu_{\rm kin}}{\overline m} \biggr) \beta_0 - C_A \biggl(
\frac{\pi^2}{6} - \frac{13}{12}\biggr)\biggr] \Biggr\} \,,
\end{eqnarray}
where $\overline m\equiv\overline m_b(\overline m_b)$, $C_F=4/3$, and
$\beta_0=11-2 n_f/3$ is the one-loop beta function. For $\mu_{\rm
kin}\,\to\;0$ the kinetic mass reduces to the pole mass.

\subsubsection*{Potential-subtracted mass}

The potential-subtracted (PS) mass is similar to the kinetic mass, but
arises considering the static energy of a bottom--antibottom quark pair
in NRQCD \cite{Beneke:1998rk}. The PS mass is known to
$\cO(\alpha_{\rm s}^3)$ and its relation to the pole mass reads
\begin{eqnarray}
m_{b,{\rm PS}}(\mu^{}_{\rm PS}) &\!\!=\!\!& m_{b,{\rm pole}} - 
C_F\mu_{\rm PS}\frac{\alpha_{\rm s}(\mu)}{\pi}\Biggl\{ 1+\frac{\alpha_{\rm s}(\mu)}{4\pi}
\biggl[a_1-\beta_0\biggl(\ln\frac{\mu_{\rm PS}^2}{\mu^2}-2\biggr)\biggr]
\\ \smvs
&&\hspace*{-1.5cm}+\biggl(\frac{\alpha_{\rm s}(\mu)}{4\pi}\biggr)^2 \biggl[a_2-
(2 a_1 \beta_0+\beta_1)\biggl(\ln\frac{\mu_{\rm PS}^2}{\mu^2}-2\biggr)
+\beta_0^2\biggl(\ln^2\frac{\mu_{\rm PS}^2}{\mu^2}-4 \ln\frac{\mu_{\rm PS}^2}
{\mu^2}+8\biggr)\biggr]\Biggr\} \,, \nonumber 
\end{eqnarray}
where $\beta_1=102-38 n_f/3$ is the coefficient of the two-loop beta
function, and $a_1=31/3-10 n_f/9$ as well as $a_2=456.749 - 66.354
\,n_f + 1.235\, n_f^2$ \cite{Schroder:1998vy,Peter:1997ig} are 
perturbative coefficients appearing in the static heavy $q\bar q$
potential. For $\mu_{\rm PS}\,\to\;0$ the PS mass also reduces to the
pole mass.

\subsubsection*{1S mass}

The kinetic and the potential-subtracted mass depend on an explicit
subtraction scale to remove the universal infrared sensitive
contributions associated with the non-relativistic heavy quark
dynamics. The 1S mass \cite{Hoang:1998ng,Hoang:1998hm,Hoang:1999zc}
achieves the same task without a factorisation scale, since it is
directly related to a physical quantity. The bottom 1S mass is defined
as one half of the perturbative contribution to the mass of the $n=1$,
${}^{2s+1}L_j={}^3S_1$ quarkonium bound state in the formal limit
$m_b\gg m_b v\gg m_b v^2\gg\Lambda_{\rm QCD}$. To three loop order (or
NNLO in the non-relativistic expansion) the 1S-pole mass relation
reads
\begin{eqnarray}
\label{eq:M1Sdef}
\frac{m_{b,{\rm 1S}}^{}}{m_{b,{\rm pole}}} &\!\!=\!\!& 1 - \frac{1}{8}\,
\Big(C_F \alpha_{\rm s}(\mu)\Big)^2\,\Biggl\{\,1 + \frac{\alpha_{\rm s}(\mu)}{\pi}\,
\biggl[\,\beta_0\,\Big(L + 1\Big) + \frac{a_1}{2}\,\biggr] \nonumber \\
\smvs
&& \hspace{3mm} +\,\biggl(\frac{\alpha_{\rm s}(\mu)}{\pi}\biggr)^2\,\biggl[\,
\beta_0^2\,\bigg(\,\frac{3}{4}\,L^2 + L + \frac{\zeta_3}{2} + \frac{\pi^2}{24}
+ \frac{1}{4} \,\bigg) + \beta_0\,\frac{a_1}{2}\,\bigg(\, \frac{3}{2}\,L + 1
\,\bigg) \\
\smvs
&& \hspace{28mm} +\,\frac{\beta_1}{4}\,\Big(L + 1\Big) + \frac{a_1^2}{16} +
\frac{a_2}{8} + \bigg(\,C_A - \frac{C_F}{48} \,\bigg)\, C_F \pi^2 \,\bigg]
\,\Biggr\} \nonumber \,,
\end{eqnarray}
where $L\equiv\ln(\mu/(C_F\alpha_{\rm s}(\mu)\,m_{b,{\rm pole}}))$ and
$\zeta_3=1.20206$.  The expression for the 1S mass is derived in the
framework of the non-relativistic expansion, where powers of the
bottom quark velocity arise as powers of $\alpha_{\rm s}$ in the 1S
mass definition. Thus, to achieve the renormalon cancellation in the
1S mass scheme for quantities which are not defined in the
non-relativistic power counting, such as for B decays, it is mandatory
to treat terms of order $\alpha_{\rm s}^{n+1}$ in \Eq~\ref{eq:M1Sdef} as
being of order $\alpha_{\rm s}^n$.  This prescription is called
``Upsilon expansion''~\cite{Hoang:1998ng,Hoang:1998hm} (see also
Ref.~\cite{Kiyo:2002rr}) and arises because of the difference between
the non-relativistic power counting and the usual counting in the
numbers of loops (or in powers of $\alpha_{\rm s}$).  In the upsilon
expansion it is crucial that the renormalisation scale of all
$\alpha_{\rm s}$ terms is chosen equal.

\subsubsection*{Renormalon-subtracted mass}

The renormalon-subtracted mass~\cite{Pineda:2001zq} is formally
defined as the perturbative series that results from subtracting all
non-analytic pole terms from the Borel transform of the pole-\MSB mass
relation at $u=1/2$ with a fixed choice for the renormalisation scale
$\mu=\mu_{\rm RS}\,$. The scale $\mu_{\rm RS}$ is then kept
independent from the renormalisation scale used for the computation of
the quantities of interest.  The expression for the relation between
the RS and pole mass reads
\begin{equation}
\label{RSdef}
m_{{\rm RS}({\rm RS}')}(\mu_{{\rm RS}})=m_{\rm pole}-\sum_{n=0(1)}^\infty
N_m\,\mu_{{\rm RS}}\,
\left(\frac{\beta_0}{2\pi}\right
)^n \als^{n+1}(\mu_{{\rm RS}})\,\sum_{k=0}^\infty c_k
  \frac{\Gamma(n+1+b-k)}{\Gamma(1+b-k)}
\,,
\end{equation}
where the coefficients $c_1$ and $c_2$ are known from
Refs.~\cite{Beneke:1998ui,Pineda:2001zq}. An approximation to the term
$N_m$ has been determined in
Refs.~\cite{Pineda:2001zq,Lee:2003hh}. The values given in
\Table~\ref{tab:tabmasses} refer to the $RS^\prime$ scheme.

In \Table~\ref{tab:tabmasses} $b$-quark threshold mass parameters are
compared numerically taking the \MSB mass $\overline m_b(\overline
m_b)$ as a reference value for different values of the strong
coupling. Each entry corresponds to the mass using the respective
1-loop/2-loop/3-loop relations.
\begin{table}[t] % tabmasses
\caption[Numerical values for $b$-quark masses for a given
         \MSB mass $\overline m_b(\overline m_b)$, $n_l=4$ and three
         values of $\alpha_{\rm s}^{(5)}(m_Z)$]
        {Numerical values for $b$-quark masses in units of {\rm GeV}
         for a given \MSB mass $\overline m_b(\overline m_b)$, $n_l=4$
         and three values of $\alpha_{\rm s}^{(5)}(m_Z)$. Flavour
         matching was carried out at $\mu=\overline m_b(\overline
         m_b)$. Numbers with a star are given in the large-$\beta_0$
         approximation. The 1S and kinetic masses are frequently used
         in the theoretical description of inclusive B meson
         decays~\cite{Battaglia:2003in}.}
\label{tab:tabmasses}    
\begin{center}
\begin{tabular}{|c|c|c|c|c|c|} \hline
$\overline m_b(\overline m_b)$ & 
$m_{b,\rm pole}$ &
$m_{b,\rm kin}(1\,\mbox{GeV})$ &
$m_{b,\rm PS}(2\,\mbox{GeV})$ & 
$m_{b,{\rm 1S}}$ & 
$m_{b,{\rm RS^\prime}}(2\,\mbox{GeV})$
\\[0.05cm]
\hline
\multicolumn{6}{|c|}{{\small $\alpha_{\rm s}^{(5)}(m_Z)=0.116$ 
}}\\
\hline
{\small 4.10} & 
{\small 4.48/4.66/4.80} &
{\small 4.36/4.42/4.45$^*$} & 
{\small 4.29/4.37/4.40} & 
{\small 4.44/4.56/4.60} & 
{\small 4.48/4.58/4.62}\\
\hline
{\small 4.15} & 
{\small 4.53/4.72/4.85} &
{\small 4.41/4.48/4.50$^*$} & 
{\small 4.35/4.42/4.45} & 
{\small 4.49/4.61/4.65} &
{\small 4.53/4.64/4.67}\\
\hline
{\small 4.20} & 
{\small 4.59/4.77/4.90} &
{\small 4.46/4.53/4.56$^*$} & 
{\small 4.40/4.48/4.51} & 
{\small 4.54/4.66/4.71} &
{\small 4.59/4.69/4.73}\\
\hline
{\small 4.25} & 
{\small 4.64/4.83/4.96} & 
{\small 4.52/4.59/4.61$^*$} & 
{\small 4.46/4.53/4.56} & 
{\small 4.60/4.72/4.76} &
{\small 4.64/4.75/4.78}\\
\hline
{\small 4.30} & 
{\small 4.69/4.88/5.01} &
{\small 4.57/4.64/4.67$^*$} & 
{\small 4.51/4.59/4.62} & 
{\small 4.65/4.77/4.81} &
{\small 4.69/4.80/4.84}\\
\hline
\multicolumn{6}{|c|}{{\small $\alpha_{\rm s}^{(5)}(m_Z)=0.118$ 
}}\\
\hline
{\small 4.10} & 
{\small 4.49/4.69/4.84} & 
{\small 4.37/4.44/4.46$^*$} & 
{\small 4.30/4.38/4.41} & 
{\small 4.45/4.57/4.62} &
{\small 4.49/4.60/4.64}\\
\hline
{\small 4.15} & 
{\small 4.55/4.74/4.89} & 
{\small 4.42/4.49/4.52$^*$} & 
{\small 4.36/4.43/4.47} & 
{\small 4.50/4.63/4.67} &
{\small 4.55/4.66/4.70}\\
\hline
{\small 4.20} & 
{\small 4.60/4.80/4.94} & 
{\small 4.47/4.55/4.57$^*$} & 
{\small 4.41/4.49/4.52} & 
{\small 4.55/4.68/4.73} &
{\small 4.60/4.71/4.75}\\
\hline
{\small 4.25} & 
{\small 4.65/4.85/5.00} & 
{\small 4.52/4.60/4.63$^*$} & 
{\small 4.46/4.54/4.58} & 
{\small 4.61/4.73/4.78} &
{\small 4.65/4.77/4.81}\\
\hline
{\small 4.30} & 
{\small 4.71/4.91/5.05} & 
{\small 4.58/4.66/4.69$^*$} & 
{\small 4.52/4.60/4.63} & 
{\small 4.66/4.79/4.84} &
{\small 4.71/4.82/4.86}\\
\hline
\multicolumn{6}{|c|}{{\small $\alpha_{\rm s}^{(5)}(m_Z)=0.120$ 
}}\\
\hline
{\small 4.10} & 
{\small 4.51/4.72/4.88} & 
{\small 4.37/4.45/4.48$^*$} & 
{\small 4.31/4.39/4.43} & 
{\small 4.46/4.59/4.64} &
{\small 4.51/4.63/4.67}\\
\hline
{\small 4.15} & 
{\small 4.56/4.77/4.93} & 
{\small 4.43/4.51/4.54$^*$} & 
{\small 4.36/4.45/4.48} & 
{\small 4.51/4.64/4.70} &
{\small 4.56/4.68/4.72}\\
\hline
{\small 4.20} & 
{\small 4.61/4.83/4.99} & 
{\small 4.48/4.56/4.59$^*$} & 
{\small 4.42/4.50/4.54} & 
{\small 4.56/4.70/4.75} &
{\small 4.61/4.74/4.78}\\
\hline
{\small 4.25} & 
{\small 4.67/4.88/5.04} & 
{\small 4.54/4.62/4.65$^*$} & 
{\small 4.47/4.56/4.59} & 
{\small 4.62/4.75/4.80} &
{\small 4.67/4.79/4.83}\\
\hline
{\small 4.30} & 
{\small 4.72/4.94/5.10} & 
{\small 4.59/4.67/4.71$^*$} & 
{\small 4.53/4.61/4.65} & 
{\small 4.67/4.81/4.86} &
{\small 4.72/4.85/4.89}\\
\hline
\end{tabular}
\end{center}
\end{table}

\subsection{Bottom quark mass from Upsilon sum rules}
\label{sec:mbsumrules}

The spectral sum rules for $\sigma(e^+e^-\to\,b\,\bar b)$ start from
the correlator of two electromagnetic bottom quark currents
\begin{equation}
(q_\mu\,q_\nu-g_{\mu\nu}\,q^2)\,\Pi(q^2) \,=\, \int dx\,e^{i\,qx}\,
\langle\, 0\,|\,T\,j^b_\mu(x)\,j^b_\nu(0)\,|\,0\, \rangle \,,
\end{equation}
where $j^b_\mu(x)\equiv\bar b(x)\gamma_\mu b(x)$. Using causality and
the optical theorem one can relate theoretically calculable
derivatives of $\Pi(q^2)$ at $q^2=0$ to moments of the total
cross-section $\sigma(e^+e^-\to\,b\,\bar b)$,
\begin{equation}
\label{eq:Mdef}
{\cal M}_n \,=\, \frac{12\,\pi^2\,Q_b^2}{n!}\,
\bigg(\frac{d}{d q^2}\bigg)^n\,\Pi(q^2)\bigg|_{q^2=0} \,=\,
\int \frac{d s}{s^{n+1}}\,R_{bb}(s) \,,
\end{equation}
where $R_{bb}\equiv\sigma(e^+e^-\to\,b\,\bar
b)/\sigma(e^+e^-\to\,\mu^+\mu^-)$.  From the comparison of the
theoretical moments and those based on experimental data for $R_{bb}$,
it is possible to determine the bottom quark mass
\cite{Novikov:1977tn}. Since the sum rules in \Eq~(\ref{eq:Mdef})
involve inclusive quantities referring only to global duality, they
are believed to be one of the most reliable tools to extract QCD
parameters. However, it is necessary to restrict the values of $n$ to
ensure that the moments are indeed sufficiently inclusive.

In general, one can distinguish between two regions in $n$, which
require a different theoretical treatment. For low values of $n$ the
moments are dominated by relativistic dynamics and scales of order of
the heavy quark mass $m_b$. This allows to apply the usual expansion
in the number of loops for the theoretical computations, and the \MSB
scheme is an appropriate choice for the heavy quark mass
parameter. However, the lack of data for $R_{bb}$ in the continuum
regions above the quarkonium resonances introduces model-dependent
errors. On the other hand, for large values of $n$ the continuum
regions are suppressed and the moments become dominated by the
quarkonium resonance region where good sets of data have been obtained
in the past. However, the theoretical predictions of moments for large
values of $n$ is more complicated since the usual loop expansion
breaks down and the size of non-perturbative effects increases. Here,
summations of higher order contributions proportional to powers of
$(\alpha_{\rm s}\sqrt{n})$ need to be carried out in order to capture
the relevant non-relativistic perturbative
information~\cite{Voloshin:1995sf,Hoang:1998uv}, and the threshold
masses discussed above are appropriate choices for the heavy quark
mass parameter.
\longpage

\begin{figure}[t]
\begin{center}
\includegraphics[angle=270, width=.8\linewidth]{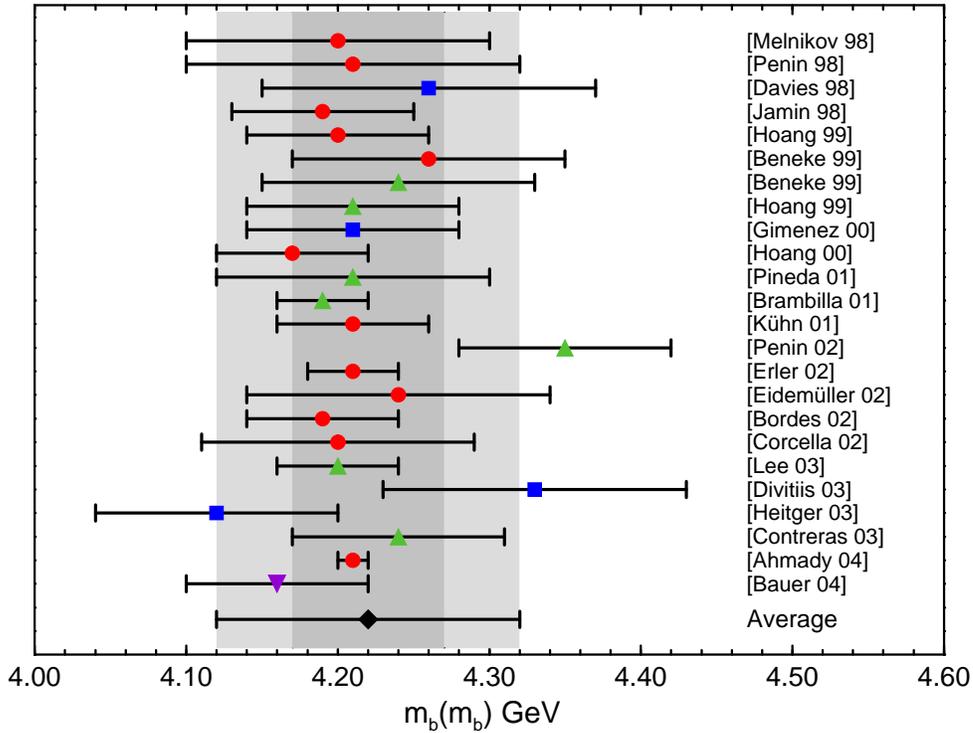}
\end{center}
\caption[Collection in historical order of recent bottom quark mass
         determinations]
        {Collection in historical order of recent bottom quark mass
         determinations. The red circles represent sum rule results,
         the green triangles Upsilon 1S determinations, the blue
         squares lattice QCD results and the purple upside down
         triangle a determination from semileptonic $B$ decays. The
         full diamond gives our global average for $\overline
         m_b(\overline m_b)$. The darker and lighter shaded areas
         represent our subjective error estimates corresponding to a
         $1\sigma$ error and a range respectively. Our average and the
         related error estimates are further discussed in 
         \Section~\ref{sec:mbmcaverage}.}
\label{fig:figmb}
\end{figure}

Moreover, there is an upper ``duality'' bound for the possible choices
of $n$ since the energy range contributing to the moments, which is of
order $m_q/n$, needs to be larger than the typical hadronization scale
$\Lambda_{\rm QCD}$ \cite{Poggio:1976af,Hoang:1998uv}. In the case of
$R_{bb}$ and the determination of the bottom mass, this bound is
around $n=10$. The low-$n$ and the large-$n$ ranges are believed to be
well separated with their boundary being approximately at $n=4$.  A
good number of analyses exists for small and large values of $n$ and
respecting the cancellation of the ${\cal O}(\lQ)$ renormalon
contributions associated to the choice of the quark mass
definition. For low as well as for large values of $n$ it is presently
believed that non-perturbative effects are negligibly small, based on
the size of the contributions from the non-perturbative gluon and
quark condensates in the OPE~\cite{Shifman:1979bx}. Alternative views
about the validity of the standard OPE (see
\eg Ref.~\cite{Chernodub:2000bk}) have not been accounted for in any
analysis so far.  In the following the advantages and disadvantages of
the two types of sum rules are reviewed. Results for bottom quark
masses obtained in recent sum rule analyses have been collected in
\Table~\ref{tab:tabmbsr} and a graphical summary is presented as the red
circles in \Figure~\ref{fig:figmb}.
\begin{table}[t!]  % tabmbsr
\caption[Collection of recent bottom quark mass determinations 
         from spectral sum rules]
        {Collection in historical order of recent bottom quark mass
         determinations in units of GeV from spectral sum rules. The
         uncertainties quoted in the respective references have been
         added quadratically. All numbers have been taken from the
         respective publications.}
\label{tab:tabmbsr}
\begin{center}
\setlength{\extrarowheight}{1mm}\small
\begin{tabular}{|l|l|c|l|} \hline
 Author & $\overline m_b(\overline m_b)$ & other mass  & comments, Ref.
\\ \hline\hline
\multicolumn{4}{|c|}{ spectral sum rules }\\
\hline\hline
  Voloshin  \hfill 95
    &  
    & $m^{}_{\rm pole}=4.83\pm 0.01$
    & {$8<n<20$, NLO; no theo. uncert.}~\cite{Voloshin:1995sf}
\\ \hline
  K\"uhn  \hfill 98 
    & 
    & $m^{}_{\rm pole}=4.78\pm 0.04$
    & $10<n<20$, NLO~\cite{Kuhn:1998uy}
\\ \hline
  Hoang  \hfill 98
    & %$4.26\pm 0.09$~* 
    & $m^{}_{\rm pole}=4.88\pm 0.09$
    & $4<n<10$, NNLO~\cite{Hoang:1998uv}
\\ \hline
  Melnikov \hfill  98
    & $4.20\pm 0.10$
    & $M^{1\mbox{\tiny GeV}}_{\rm kin}=4.56\pm 0.06$
    & $x<n<x$, NNLO~\cite{Melnikov:1998ug}
\\ \hline
  Penin  \hfill  98 
    & $4.21\pm 0.11$
    & $m^{}_{\rm pole}=4.80\pm 0.06$
    & $8<n<12$, NNLO~\cite{Penin:1998kx}
\\ \hline
  Jamin  \hfill 98
    & $4.19\pm 0.06$
    & 
    & $7<n<15$~\cite{Jamin:1997rt,Jamin:1998ra}
\\ \hline
  Hoang  \hfill 99
    & $4.20\pm 0.06$ 
    & $M^{}_{\rm 1S}=4.71\pm 0.03$
    & $4<n<10$, NNLO~\cite{Hoang:1999ye}
\\ \hline
  Beneke  \hfill 99
    & $4.26\pm 0.09$ 
    & $M^{2\mbox{\tiny GeV}}_{\rm PS}=4.60\pm 0.11$
    & $6<n<10$, NNLO~\cite{Beneke:1999fe}
\\ \hline
  Hoang \hfill  00
    & $4.17\pm 0.05$ 
    & $M^{}_{\rm 1S}=4.69\pm 0.03$
    & {$4<n<10$, NNLO, $m_c\neq 0$}~\cite{Hoang:2000fm}
\\ \hline
  K\"uhn  \hfill 01  
    & $4.21\pm 0.05$ 
    & 
    & $1<n<4$, $\cO(\alpha_{\rm s}^2)$~\cite{Kuhn:2001dm}
\\ \hline
  Erler  \hfill 02  
    & $4.21\pm 0.03$ 
    & & ${\cal O}(\alpha_{\rm s}^2)$~\cite{Erler:2002bu}
\\ \hline
  Eidem\"uller  \hfill 02  
    & $4.24\pm 0.10$ 
    & $M^{2\mbox{\tiny GeV}}_{\rm PS}=4.56\pm 0.11$
    & $3<n<12$~\cite{Eidemuller:2002wk}
\\ \hline
  Bordes  \hfill 02  
    & $4.19\pm 0.05$ 
    & 
    & ${\cal O}(\alpha_{\rm s}^2)$~\cite{Bordes:2002ng}
\\ \hline
  Corcella  \hfill 02  
    & $4.20\pm 0.09$ 
    & 
    & $1<n<3$, ${\cal O}(\alpha_{\rm s}^2)$~\cite{Corcella:2002uu}
\\ \hline
  Ahmady  \hfill 04
    & $4.21\pm 0.01$
    &
    & $1<n<4$, $\cO(\alpha_{\rm s}^2)$ only scale $+$ mom. uncert.~\cite{Ahmady:2004er}
\\ \hline
\end{tabular}
\end{center}
\end{table}

\subsubsection*{Non-relativistic sum rules}
\longpage

The large-$n$ sum rules have the advantage that the experimentally
unknown parts of the $b\bar b$ continuum cross-section above the
$\Upsilon$ resonance region are suppressed. A crude model for the
continuum cross-section is sufficient and causes an uncertainty in the
$b$-quark mass below the $10$~MeV level. Depending on which moment is
used the overall experimental uncertainties in the $b$-quark mass are
between $15$ and $20$~MeV.  Theoretically, large-$n$ sum rules are
characterised by the fact that the dynamics is non-relativistic. It
can be shown that the average three-momentum and the average kinetic
energies of the quarks scale like $m_b/\sqrt{n}$ and $m_b/n$,
respectively. Thus for $n\leq 10$ the moments can be considered as
being dominated by perturbation theory up to non-perturbative effects
that can be described by the standard OPE~\cite{Shifman:1979bx} in
terms of local condensates.  The leading-order gluon condensate
contributions to the large-$n$ sum rules were determined in
Refs.~\cite{Voloshin:1995sf,Onishchenko:2000yy} and shown to
contribute at the level of permille for $n\leq
10$\,\cite{Voloshin:1995sf}.  None of the analyses discussed below
therefore included any non-perturbative effects. It should be noted,
however, that in practice a strong hierarchy of all relevant dynamical
scales, $m_b\gg m_b/\sqrt{n}\gg m_b/n\gg \Lambda_{\rm QCD}$, is
difficult to achieve numerically. So a successful application of the
large-$n$ sum rules is based on a balance between a good
non-relativistic expansion and a good convergence of the OPE series.
Over the past years there has been a revived interest in
non-relativistic sum rules because new theoretical developments
allowed for the systematic determination of ${\cal O}(v^2)\sim {\cal
O}(1/n)$ (NNLO) corrections to the spectral moments
\cite{Penin:1998kx,Hoang:1998uv,Melnikov:1998ug,Hoang:1999ye,Beneke:1999fe,Hoang:2000fm}.
\clearpage

All analyses found that the NNLO corrections were as large or even
larger than the NLO corrections and various different methods were
devised to extract numerical values for the bottom quark mass. In
Refs.~\cite{Melnikov:1998ug,Hoang:1999ye,Beneke:1999fe,Hoang:2000fm}
threshold masses were implemented accounting for the renormalon
problem. This removed one source of the bad perturbative behaviour,
but it was found that a considerable theoretical uncertainty remained,
coming from the theoretical description of the production and
annihilation probability of the $b\bar b$ pair. In
Refs.~\cite{Melnikov:1998ug} and \cite{Beneke:1999fe} the kinetic and
the PS mass were determined from fits of individual moments. It was
found that the NLO and NNLO results for the bottom mass differ by
about $200$~MeV. In Ref.~\cite{Melnikov:1998ug} it was argued that the
results form an alternating series and a value of $m_{b,{\rm
kin}}(1\,{\rm GeV})=4.56\pm 0.06(\mbox{ex,th})$~GeV was determined. In
Ref.~\cite{Beneke:1999fe} only the NNLO results were accounted based
on consistency arguments with computations of the $\Upsilon({\rm 1S})$
mass and the result $m_{b,{\rm PS}}(2~\mbox{GeV})= 4.60\pm
0.02(\mbox{ex})\pm 0.10(\mbox{th})$~GeV was obtained. In
Ref.~\cite{Hoang:1999ye} the 1S mass was employed and a $\chi^2$-fit
based on four different moments was carried out. It was found that the
large normalisation uncertainties drop out at NLO and NNLO and that
the results for the mass at NLO and NNLO showed good convergence. The
result was $m_{b,{\rm 1S}}=4.71\pm 0.02(\mbox{ex})\pm
0.02(\mbox{th})$~GeV. It was also shown that the sum rule analysis
gives only very weak constraints on the value of the strong coupling
and that sum rules are in fact not a very competitive tool to
determine $\alpha_{\rm s}$ with high precision. A subsequent
analysis~\cite{Hoang:2000fm} which included the effects of the nonzero
charm mass yielded $m_{b,{\rm 1S}}= 4.69\pm 0.02(\mbox{ex})\pm
0.02(\mbox{th})$~GeV.

\subsubsection*{Relativistic sum rules}

The small-$n$ sum rules have the disadvantage that the unknown parts
of the $b\bar b$ continuum cross-section above the $\Upsilon$
resonance region constitute a substantial contribution to the spectral
moments. The advantage is that the computation of the theoretical
moments is less involved since usual perturbation theory in powers of
$\alpha_{\rm s}$ can be employed. In Ref.~\cite{Kuhn:2001dm} the
theoretical moments at order ${\cal O}(\alpha_{\rm s}^2)$ were used
and it was found that the perturbative behaviour of the theoretical
moments is good if the \MSB mass scheme is used such that the pole
mass ambiguity of order $\Lambda_{\rm QCD}$ is properly cancelled. For
the bottom quark mass determination it was assumed that the unknown
experimental continuum cross-section agrees with the perturbation
theory prediction and subsequently the result $\overline m_b(\overline
m_b)=4.21\pm 0.05$~GeV was determined.  Compatible results were also
obtained in an earlier analysis at larger $n$
\cite{Jamin:1997rt,Jamin:1998ra} which employed the \MSB mass at a lower scale
$\mu\approx 3$~GeV, in order to improve the stability of the sum
rule. A more conservative ${\cal O}(\alpha_{\rm s}^2)$ analysis in
Ref.~\cite{Corcella:2002uu} using the same approach as employed in
Ref.~\cite{Kuhn:2001dm}, but accounting also for finite charm mass
effects, uncertainties coming from the experimental continuum region
and for inconsistencies in the averages for the $\Upsilon(4S)$ and
$\Upsilon(5S)$ data obtained the result $\overline m_b(\overline
m_b)=4.20\pm 0.09$~GeV. In the same analysis non-perturbative effects
in terms of the gluon condensate based on two-loop matching
coefficients~\cite{Broadhurst:1994qj} were analysed and found to be
negligible.

\subsubsection*{Alternative approaches}

Besides the relativistic and non-relativistic sum rules discussed
above, also alternative approaches employing other types of QCD sum
rules have been investigated in the literature. These shall be briefly
discussed in what follows. In Ref.~\cite{Erler:2002bu}, Erler and Luo
performed an analysis based on the dispersion relation
\begin{equation}
12\pi^2\,\Big[ \Pi(0) - \Pi(-t) \Big] \,=\,
t\!\int \frac{ds}{s}\,\frac{R_{bb}}{s+t} \,,
\end{equation}
in the limit $t\to\infty$, which is rather sensitive to the continuum
contribution, together with low-$n$ moments. The parameters of a
certain ansatz for the shape of the continuum were constrained from
the resulting sum rule, and a combined analysis led to $\overline
m_b(\overline m_b)=4.21\pm0.03\,$GeV. In
Ref.~\cite{Eidemuller:2002wk}, on the other hand, besides the
conventional moments of \Eq~\ref{eq:Mdef} which are evaluated at
$q^2=0$, Eidem\"uller also studied moments being evaluated at $q^2 =
-\,4m_b^2\xi$ rather than at $q^2=0$. A variation of the parameter
$\xi$ then allows to modify the relative size of the various
theoretical as well as phenomenological contributions, thus gaining
further information on the system under investigation. Furthermore,
for the theoretical spectral function at small velocities, a
non-relativistic description was employed, whereas the relativistic
description was chosen at large velocities. In addition, in the
intermediate region, different choices for the matching of the two
regions were studied. The resulting bottom quark mass then turned out
to be $\overline m_b(\overline m_b)=4.24\pm0.10\,$GeV. Finally, in
Ref.~\cite{Bordes:2002ng}, Bordes, Penarrocha and Schilcher
investigate finite energy sum rules similar to the analysis of the
hadronic $\tau$ decay width. In the dispersion integral, a third
degree polynomial was added, and the parameters were chosen such as to
minimise the effect of the continuum contribution and the
corresponding uncertainties. The final result for the bottom quark
mass obtained from this analysis was $\overline m_b(\overline
m_b)=4.19\pm0.05\,$GeV.

Recently, the analysis of inclusive semileptonic B decays has emerged
as a new precise tool to determine the bottom mass taking advantage of
the large amount of statistics accumulated at B-factories and the
theoretical developments in heavy quark physics. In a global fit of
various shape variables in semileptonic B$\to$D decays the bottom 1S
mass was determined as $m_{b,{\rm 1S}}=4.68\pm 0.04$~GeV
\cite{Bauer:2004ve} which is in very good agreement with the large-$n$
sum rule analysis of Hoang~\cite{Hoang:2000fm} which also included the
effects of a non-zero charm quark mass.

\subsection{Charm quark mass from Charmonium sum rules}

The determination of the charm quark mass from charmonium sum rules
proceeds in principle in analogy to the bottom case.  However, one
needs to account for the fact that for $R_{cc}$ and the determination
of the charm mass, the distinction between large- and low-$n$ moments
is much more delicate because $m_c$ is much smaller than $m_b$ and in
fact not much larger than $\Lambda_{\rm QCD}$. Here, the upper duality
bound for $n$ is already around $3$ or $4$ and non-perturbative
contributions need to be included numerically. The small range of
allowed values of $n$ leaves basically no space at all to carry out
the non-relativistic summations that can be applied in the bottom
quark case because the corresponding techniques are only valid for
large $n$ and as long as $m_q/n$ is larger than $\Lambda_{\rm
QCD}$. On the other hand, even for $n\leq 4$ the non-relativistic
region close to the $c\bar c$ threshold can have a considerable
contribution to the moments, while the model-dependences from the
experimentally unknown continuum region can still be significant.  Due
to the recent BES data~\cite{Bai:2001ct} for the $e^+e^-$ total
cross-section in the range up to $5$~GeV a good part of the charm
continuum region can be deduced using reasonable assumptions for the
non-charm cross-section.  For a reliable (error) analysis these issues
need to be taken into account.

\begin{figure}
\begin{center}
\includegraphics[angle=270, width=.8\linewidth]{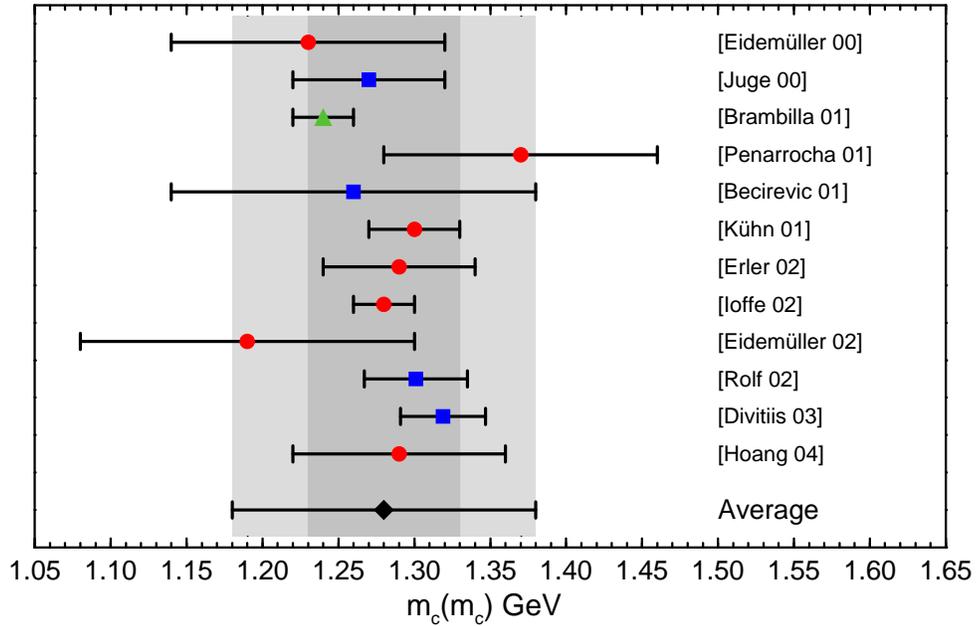}
\end{center}
\caption[Collection in historical order of recent charm quark mass
         determinations]
        {Collection in historical order of recent charm quark mass
         determinations. The red circles represent sum rule results,
         the green triangles J/$\Psi$ 1S determinations and the blue
         squares quenched lattice QCD results.  The full diamond gives
         our global average for $\overline m_c(\overline m_c)$.  The
         darker and lighter shaded areas represent our subjective
         error estimates corresponding to a $1\sigma$ error and a
         range respectively. Our average and the related error
         estimates are further discussed in
         \Section~\ref{sec:mbmcaverage}.}
\label{fig:figmc}
\end{figure}

In \Table~\ref{tab:tabmcsr}, determinations of $m_c$ from Charmonium
sum rules within the last years have been summarised. A graphical
representation of these results is also shown as the full circles in
\Figure~\ref{fig:figmc}. A conservative investigation of 
the issues discussed above in the case of the relativistic sum rule
was carried out in the very recent work \cite{Hoang:2004xm}, where
also a discussion of the previous works
\cite{Penarrocha:2001ig,Kuhn:2001dm,Erler:2002bu,Ioffe:2002be} was
given. In particular, it was found in Ref.~\cite{Hoang:2004xm} that
different ways to compute the perturbative series for the moments
appear to converge to different predictions if the standard methods to
estimate theoretical errors are employed, a situation known also from
predictions for hadronic $\tau$ decays~\cite{LeDiberder:1992te}.
non-relativistic sum rules for charmonium have been investigated in
Refs.~\cite{Eidemuller:2000rc,Eidemuller:2002wk}. Generally, $m_c$
turns out to be somewhat lower in the non-relativistic case, but due
to the low scale the resummation also introduces large
uncertainties. To conclude, for the charm quark mass determinations
relativistic sum rules appear to be more reliable than
non-relativistic ones.

\begin{table}
\caption[Collection in historical order of recent charm quark 
         mass determinations from spectral sum rules]
        {Collection in historical order of recent charm quark mass
         determinations in units of GeV from spectral sum rules. The
         uncertainties quoted in the respective references have been
         added quadratically. All numbers have been taken from the
         respective publications.}
\label{tab:tabmcsr}
\begin{center}
\begin{small}
\begin{tabular}{|l|l|c|l|} \hline
 Author & $\overline m_c(\overline m_c)$ & other mass & comments, Ref.
\\ \hline\hline
\multicolumn{4}{|c|}{ spectral sum rules }\\
\hline\hline
  Eidem\"uller  \hfill 00  
    & $1.23\pm 0.09$ 
    & $M^{1\mbox{\tiny GeV}}_{\rm PS}=1.35\pm 0.10$
    & $3<n<7$~\cite{Eidemuller:2000rc}
\\ \hline
  Penarrocha \hfill  01
    & $1.37\pm 0.09$ 
    &
    & FESR, $\cO(\alpha_{\rm s}^2)$~\cite{Penarrocha:2001ig}
\\ \hline
  K\"uhn  \hfill 01  
    & $1.30\pm 0.03$ 
    & 
    & $1<n<4$, $\cO(\alpha_{\rm s}^2)$~\cite{Kuhn:2001dm}
\\ \hline
  Erler  \hfill 02  
    & $1.29\pm 0.05$ 
    & 
    & ${\cal O}(\alpha_{\rm s}^2)$~\cite{Erler:2002bu}
\\ \hline
  Ioffe  \hfill 02  
    & $1.28\pm 0.02$ 
    & 
    & ${\cal O}(\alpha_{\rm s}^2)$~\cite{Ioffe:2002be}
\\ \hline
  Eidem\"uller  \hfill 02  
    & $1.19\pm 0.11$ 
    & $M^{1\mbox{\tiny GeV}}_{\rm PS}=1.30\pm 0.12$
    & $4<n<7$~\cite{Eidemuller:2002wk}
\\ \hline
  Hoang  \hfill 04  
    & $1.29\pm 0.07$ 
    & 
    & $n=2,3$, ${\cal O}(\alpha_{\rm s}^2)$~\cite{Hoang:2004xm}
\\ \hline
\end{tabular}
\end{small}
\end{center}
\end{table}

\subsection{Bottom and charm quark mass from the 1S resonances}
\label{sec:mbmc1S}

Compared to the sum rule methods described in the previous sections
the heavy quarkonium masses are more exclusive quantities. However,
since heavy quarkonia are colour singlet objects and their
interactions with the QCD vacuum is suppressed at least by $p^2/m^2$,
$p$ being the average quark three-momentum, it is worth to consider
also the heavy quarkonium masses as an alternative method to extract
QCD parameters, and in particular the heavy quark masses.  At this
point it is instructive to mention that the momentum transferred
between the heavy quark and antiquark is Euclidean. Therefore,
computations of the spectrum do, at least to low orders in the
perturbative expansion, not rely on local-duality and the crucial
issue is whether perturbative calculations are applicable for the
range of the relevant dynamical scales and whether the influence of
the long-range parts of the potential is significant.  It is obvious
that for precise determinations of the heavy quark masses one has to
be in a situation where the dynamics can be described to good
approximation by a weak coupling analysis and that non-perturbative
effects are subleading.  While in the sum rules the average
three-momentum and kinetic energy of the quarks can be influenced by
adjusting the value of $n$, for the quarkonium states these scales are
fixed by internal QCD dynamics. Thus some care has to be applied in
the determination and the interpretation of the results obtained for
the quarkonium masses. The average three-momentum transfer in the
bottomonium is of order of or below $2$~GeV whereas in the charmonium
case is of order of or below $1$~GeV. The corresponding average values
of the quark kinetic energies are even lower.  Thus it appears clear
from the very beginning that only the ground states should be
considered as tools for extractions of QCD parameters, cf. however
\Chapter~\ref{chapter:spectroscopy}, \Section~\ref{sec:spnrqcdwc}
for a perturbative calculation of the levels. In the following, recent
work is reviewed where for the bottomonium and charmonium ground
states is was assumed that a weak coupling analysis is possible.

\subsubsection*{Heavy Quarkonium Mass}

It is convenient to work within an effective field theory
framework. Since in the cases of bottomonium and charmonium the
assumption $mv^2\gg\lQ$ appears unrealistic, one might start with
``optimistic'' counting that $mv^2 \sim \lQ$. The effective theory
pNRQCD has been applied to describe this
situation~\cite{Pineda:1998bj,Pineda:1998ie,Pineda:1998kn,Brambilla:1999xf}.
(The effective theory vNRQCD has been constructed for the case
$mv^2\gg\lQ$~\cite{Luke:1999kz,Manohar:1999xd,Hoang:2002yy}.)  In the
pole mass (OS) scheme the heavy quarkonium mass has the following
structure,
\begin{equation}
\label{eq:Mnlj}
M_{nlj} \,=\, 2m_{\rm pole}+\sum_{m=2}^{\infty}A_{nlj}^{m,\OS}(\nu_{us})\als^m +
\delta M_{nlj}^{\rm US}(\nu_{us}) \,,
\end{equation}
where $\nu_{us}$ is a cutoff scale of order $mv^2$ that cancels in the
sum (for the perturbative sum this dependence first appears in
$A_{nlj}^{5,\OS}$) and
\begin{equation}
\delta M_{nlj}^{\rm US}(\nu_{us})
\,=\,
\frac{T_F}{3 N_c}  \int_0^\infty \!\! dt \langle n,l |{\bf r} e^{-t(H_o^{\OS}-
E_n^{\OS})} {\bf r}| n,l \rangle \langle g{\bf E}^a(t) \phi(t,0)^{\rm adj}_{ab}
g{\bf E}^b(0) \rangle(\nu_{us}) \, 
\label{eq:energyUS}
\end{equation}
 
The present status of the perturbative computations is as follows.
The complete NNLO result has been computed in
Refs.~\cite{Pineda:1998hz,Pineda:1998ja,Melnikov:1998pr,Penin:1998kx},
the $\cO(m\als^5\log\als)$ NNNLO terms in
Refs.~\cite{Brambilla:1999qa,Brambilla:1999xj,Kniehl:1999ud,Kniehl:1999mx,Hoang:2001rr},
the complete resummation of logs at NNLL in
Refs.~\cite{Pineda:2001ra,Hoang:2002yy}, the NNNLO large-$\beta_0$
result in Refs.~\cite{Kiyo:2000fr,Hoang:2000fm}, and some computations
that complete the NNNLO result (up to the still missing three loop
corrections to the static potential) in
Refs.~\cite{Kniehl:2001ju,Kniehl:2002br,Penin:2002zv}. It should be
noted that these NNNLO results were obtained assuming that
$mv^2\gg\lQ$, \ie that the ultrasoft scale is perturbative.
Nevertheless these computations are useful if one can identify the
coefficient $A_{nlj}^{5,\OS}(\nu_{us})$ by separating, in a specific
scheme, the contributions coming from the soft scale and those coming
from the ultrasoft scale.  The result obtained in
Ref.~\cite{Penin:2002zv} represents, up to the three-loop static
potential coefficient, the sum
\begin{equation}
A_{nlj}^{5,\OS}(\nu_{us})\als^5 +
\delta M_{nlj}^{\rm US}(\nu_{us})|_{\rm \cO(\als^5)\; pert.} \,.
\end{equation}
In principle, for the ground state of bottomonium, also finite mass charm
effects have to be taken into account, since the soft scale is of the order
of the charm mass. Estimates of finite charm mass effects are known up to
NNLO~\cite{Hoang:1999us,Melles:1998dj,Eiras:2000rh,Hoang:2000fm}. 
 
For the non-perturbative piece in \Eq~\ref{eq:energyUS}, assuming $mv^2
\gg \lQ$, one can perform an expansion in local condensates.  The
leading and subleading terms are known
\cite{Voloshin:1979hc,Leutwyler:1981tn,Voloshin:1982,Pineda:1997uk}. An
approach how to estimate the effect of even higher order condensates
based on a delocalised expansion of short-distance effects was
proposed in Ref.~\cite{Hoang:2002nt,Hoang:2002ax}.  However, as
discussed above, in the more realistic situation where $mv^2 \sim \lQ$
the explicit functional form of the chromoelectric correlator is
needed.  On the other hand, in the situation $\lQ \gg mv^2$ the
non-perturbative corrections to the potential scale as $\sim r^2$ (see
\cite{Balitsky:1985iw,Brambilla:1999xf}) and new non-perturbative
effects could exist~\cite{Brambilla:2003mu}.  Another point of concern
relevant for the case $mv^2 \sim \lQ$ was pointed out in
Ref.~\cite{Hoang:2002yy} using the vNRQCD framework. It was pointed
out that there could be more non-perturbative effects than those
encoded in \Eq~(\ref{eq:energyUS}) based on pNRQCD, since the soft and
ultrasoft renormalisation scales are correlated and the running of the
potential coefficients is affected by ultrasoft mixing effects for all
scales below $m$. Therefore, once the ultrasoft scale approaches $\lQ$
also the coefficients of the potentials can become affected by
non-perturbative effects.

\subsubsection*{Renormalons}

If one applies the results above expressed in the pole mass scheme to heavy
quarkonium ground states a quite bad convergence of the perturbative
series, $A_{nlj}^{m,\OS} \sim m!$, is found. This situation is symptomatic for
any quantity that has a strong dependence on the heavy quark mass and also
exists for the sum rules reviewed in \Section~\ref{sec:mbsumrules}.
Here, the bad convergence is
coming from the perturbative corrections to the static 
potential~\cite{Aglietti:1995tg}, which renders the prediction of the binding
energy ambiguous to an 
amount of order $\lQ$ although the left-hand side of \Eq~\ref{eq:Mnlj}
is an observable and ambiguity-free. This problem is directly related
to the existence of the pole mass renormalon~\cite{Bigi:1994em,Beneke:1994sw}
mentioned before in \Section~\ref{sec:massdefs} and represents a general
feature of the pole mass scheme. It is related to an artificially strong
sensitivity to small momenta in the pole mass scheme that renders the pole
mass definition ambiguous to an amount of order $\Lambda_{\rm QCD}$, but it
is, to the present knowledge, not related to any physical effect.
In fact, it can be shown in the Schr\"odinger equation that 
the ambiguities (and the dominant large corrections) cancel in the sum of
twice the pole mass and the static 
potential \cite{Pin98,Hoang:1998nz,Beneke:1998rk}. Thus the resolution to the
problem comes by obviating the pole mass and expressing it in terms of
so-called threshold masses
\cite{Bigi:1994em,Beneke:1998rk,Hoang:1998ng,Pineda:2001zq} (see
\Section~\ref{sec:massdefs}) so that the ambiguities cancel explicitly
within the coefficients of the perturbative series; cf. also
\Chapter~\ref{chapter:spectroscopy}, \Section~\ref{sec:spnrqcdwc} 
for a discussion about the renormalon subtraction in the spectrum.

One may also employ the \MSB mass scheme using the upsilon
expansion~\cite{Hoang:1998ng,Hoang:1998hm}. Numerically the resulting
series have been shown to converge well~\cite{Kiyo:2002rr,Kiyo:2000fr}
leading to reliable predictions. However, there are a few conceptual
issues that should be mentioned. Employing the \MSB mass scheme in the
effective theory framework introduces a bilinear quark mass term of
$\cO(m\als)$ into the action, which formally breaks the
non-relativistic power-counting since all leading order terms in
threshold mass schemes are of order $m
v^2$~\cite{Beneke:1998rk,Hoang:1999cj}. (In fact, the same happens in
HQET.) Numerically this means the first order corrections in the \MSB
scheme are larger than in threshold mass schemes.  The corresponding
effect in $t\bar t$ physics has been demonstrated to be
substantial~\cite{Hoang:1999cj} while for the bottom, the $\cO(m\als)$
term does not seem to be that large numerically, being much smaller
than the typical values of the soft scale in the $\Upsilon(1S)$.
Conceptually this means the \MSB mass extracted from the quarkonium
mass has a smaller parametric precision than threshold masses.
Another issue related to the use of the \MSB mass is that it is
impossible to avoid parametrically large logarithmic terms. If one
uses a low scale $\sim m\als$ one obtains logarithms of $m\als/m$ from
the series of the pole-\MSB mass relation; if one uses the high scale
$\sim m$, one obtains logarithms of $m\als/m$ and $m\als^2/m$ from the
series in \Eq~(\ref{eq:Mnlj}). For the bottomonium case these logarithms
are, however, not numerically large and do not seem to spoil the
perturbative expansion in practice.  On the other hand, if one uses a
prescription different from the upsilon expansion to avoid large
logarithmic terms the cancellation of the renormalon ambiguity is
incomplete.

\subsubsection*{Determination of the bottom and charm mass}

\begin{table}
\caption[Collection in historical order of recent bottom and charm quark mass
         determinations from the $\Upsilon$(1S) and J/$\Psi$(1S)
         resonances]
        {Collection in historical order of recent bottom and charm
         quark mass determinations in units of~GeV from the
         $\Upsilon$(1S) and J/$\Psi$(1S) resonances. The uncertainties
         quoted in the respective references have been added
         quadratically. All numbers have been taken from the
         respective publications.}
\label{tab:mb1S}
\begin{center}
\begin{small}
\begin{tabular}{|l|l|c|l|} \hline
 Author & $\overline m_b(\overline m_b)$ & other mass  & comments, Ref.
\\ \hline\hline
\multicolumn{4}{|c|}{ $\Upsilon({\rm 1S})$ mass }\\
\hline\hline
  Beneke \hfill  99
    & $4.24\pm 0.09$ 
    & $M^{2\mbox{\tiny GeV}}_{\rm PS}=4.58\pm 0.08$
    & NNLO~\cite{Beneke:1999fe}
\\ \hline
  Hoang  \hfill 99
    & $4.21\pm 0.07$ 
    & \mbox{}\hspace{3.5mm}$M^{}_{\rm 1S}=4.73\pm 0.05$
    & {NNLO}~\cite{Hoang:1999cj}
\\ \hline
  Pineda \hfill 01
    & $4.21\pm 0.09$ 
    & $M_{\rm RS}^{2\mbox{\tiny GeV}} = 4.39\pm 0.11$
    & {NNLO}~\cite{Pineda:2001zq}
\\ \hline
  Brambilla \hfill 01
  & $4.19\pm 0.03$ & 
  & NNLO, pert. th. only~\cite{Brambilla:2001qk}
\\ \hline
  Penin \hfill 02
  & $4.35\pm 0.07$ & 
  & {NNNLO}~\cite{Penin:2002zv}
\\ \hline
  Lee \hfill 03
  & $4.20\pm 0.04$ & 
  & {NNNLO}~\cite{Lee:2003hh}
\\ \hline
  Contreras \hfill 03
  & $4.24\pm 0.07$ & 
  & {NNNLO}~\cite{Contreras:2003zb}
\\ \hline\hline
  & $\overline m_c(\overline m_c)$ & J/$\Psi$(1S) & \\ \hline\hline
  Brambilla \hfill 01
  & $1.24\pm 0.02$ & 
  & {NNLO}~\cite{Brambilla:2001fw}
\\ \hline
\end{tabular}
\end{small}
\end{center}
\end{table}

In the following the different determinations of the bottom quark mass
available in the literature are reviewed.  The results are collected
in \Table~\ref{tab:mb1S}. They have also been included as the full
triangles in \Figures~\ref{fig:figmb} and \ref{fig:figmc}. In the
following the main features of these analysis are summarised. In the
first three references as well as in Ref.~\cite{Penin:2002zv} no
finite charm mass effects were included. In Ref.~\cite{Beneke:1999fe}
a NNLO analysis was made in the PS mass scheme. The result obtained
for the \MSB mass was less precise as at the time of the analysis the
conversion from the pole to the $\MS$ scheme was not yet known with
the required accuracy. Reference~\cite{Hoang:1999cj} contained a NNLO
analysis in the 1S scheme, while in Ref.~\cite{Pineda:2001zq} the RS
scheme was used at NNLO. The latter reference also contained an
analysis at NNNLO including the logs at this order as well as the
large $\beta_0$ result. In Ref.~\cite{Brambilla:2001qk} a NNLO
analysis was made including charm mass effects in the 1S
scheme.\footnote{The importance of the charm mass effects were first
pointed out in Ref.~\cite{Hoang:1999us}.}  non-perturbative effects
were not taken into account. In Ref.~\cite{Penin:2002zv} a computation
at NNNLO in the pole mass scheme was achieved, up to the still missing
three-loop corrections to the static potential, which were taken from
the Pade-estimates of Ref.~\cite{Chishtie:2001mf}.\footnote{All NNNLO
analyses mentioned below used the estimates of
Ref.~\cite{Chishtie:2001mf}.}  The difference with the other results
could be due to the presence of the renormalon as well as the fact
that the ultrasoft contribution was computed within perturbation
theory. Moreover, specific choices for renormalisation scale have been
employed. In fact, using the upsilon expansion the authors of
Ref.~\cite{Penin:2002zv} also obtained $\overline m_b(\overline
m_b)=4.274$ as the central value, which is consistent with the other
results.  In Ref.~\cite{Lee:2003hh} a NNNLO analysis was made in a
scheme similar to the RS one. In this reference the ultrasoft
contribution was treated
perturbatively. Reference~\cite{Contreras:2003zb} contains a very
similar analysis but the ultrasoft contribution was treated in a
different way than the soft contribution. Because the situation
$mv^2\gg\lQ$ is not quite realistic, it is illustrative reanalysing
these NNNLO results without the US contribution. In the results of
Ref.~\cite{Contreras:2003zb} it is possible to eliminate the ultrasoft
contribution in a scheme-dependent way, which shifts the \MSB bottom
mass by around $-50$~MeV.  There has also been a determination of the
charm mass from the $J/\Psi$(1S) mass \cite{Brambilla:2001fw}. The
authors performed a NNLO analysis in the 1S scheme but did not account
for non-perturbative contributions.

In the analyses discussed above, only the cancellation of the leading
IR renormalon of the pole mass and the singlet static potential have
been taken into account. But there are also subdominant renormalons
that eventually could play a role and the following parametric
consideration is useful (see also the discussion in
\Chapter~\ref{chapter:spectroscopy}, \Section~\ref{sec:spnrqcdwc}). On
the singlet static potential side, one expects the first subleading
ambiguities from a $\cO(\lQ^3r^2)$ IR renormalon. From the pole mass
in \Eq~(\ref{eq:Mnlj}) there is a $\cO(\lQ^2/m)$ renormalon that is
not cancelled in general in threshold mass schemes.  It depends on the
relative size of $\lQ$ and $m\als^2$ which of the ambiguities is
parametrically larger.  In the case $mv^2\sim m\als^2
\gg \lQ$, where a description in terms of local condensates is
appropriate, the leading {\it genuine} non-perturbative corrections to
the quarkonium mass scale like $m(\lQ/m\als)^4$. However, this
quantity is parametrically much smaller than the non-perturbative
effects associated to the subleading renormalons, either from the pole
mass or from the singlet static potential. In this case the leading
remaining ambiguity comes from the subleading pole mass renormalon of
$\cO(\lQ^2/m)$ and the actual accuracy of the result is set by the
perturbative calculation and not by non-perturbative effects.  In the
more realistic situation where $m\als^2 \sim \lQ$, on the other hand,
the subleading renormalon ambiguities from the singlet static
potential and the pole mass are of the same order as the {\it genuine}
non-perturbative corrections.

\subsection{Bottom and charm quark masses from lattice gauge theory}
\label{sec:mbmcLat}

Lattice determinations of parameters of the QCD Lagrangian have two components.
First, long-distance Monte Carlo calculations are used to fix the bare lattice
quark masses $m_{u0}$, $m_{d0}$, \etc and the bare lattice coupling constant
$\alpha_0(a)$ to make hadron masses and decays to match experiment. ($a$ is the
lattice spacing.) Second, short-distance calculations are used to ensure that
short-distance physics with the lattice regulator is the same as short-distance
physics with dimensional regularisation.  The lattice parameters are converted
to the \MSB parameters by calculating short distance quantities in
both regulators and making them agree. For example, a perturbative relation
between the lattice bare quark mass and the \MSB mass can be
obtained by calculating the on-shell quark propagator with both regulators and
requiring that they agree. The relation between QCD parameters in the two
regulators may also be obtained from less usual quantities like the PCAC
related quantity formed from the pseudoscalar density and the four-divergence
of the axial vector current
\begin{equation}
m_{\rm PCAC} =   \frac{\langle S_1|\partial \cdot A|S_2\rangle}
                      {\langle S_1|P|S_2\rangle}.
\end{equation}
($S_1$ and $S_2$ may be any states.) The lattice short-distance quantities may
be calculated with perturbation theory as usual. They may also be calculated 
non-perturbatively, which makes possible  robust lattice short distance
calculations.

Through the 1990s, most lattice phenomenology had some degree of quantitative
control over all sources of systematic error except one:  the quenched
approximation.  In the last five years,  unquenched calculations have 
have become more and more common, allowing more serious comparison of
different lattice calculations. There are three main families of methods for
including sea quarks  in unquenched calculations (for more details see 
the lattice introduction in \Chapter~\ref{chapter:commontheoreticaltools}).
Each has distinct advantages and complications.
Wilson and clover
 fermions break chiral symmetry strongly at the lattice spacing,
and have practical complications in recovering it.  In unquenched
calculations, they have trouble reaching light quark masses
much below $m_s/2.$
Staggered and naive fermions can reach lightest quark masses with
greatest precision and least 
computer time, but have theoretical complications due to fermion doubling.
Domain wall and overlap fermions are theoretically the cleanest, but 
appear to be much more expensive in computer time.

To fix the parameters of the lattice QCD Lagrangian,
including the heavy quark masses,
we are free to use the hadronic quantities that are
simplest for the lattice to calculate accurately.
 Stable particles, particularly mesons, require simpler lattice methods
than unstable particles.  Processes with a single hadron present at
a time are simpler than multihadron processes.
The numerous masses and mass splittings in the
charmonium and bottomonium systems are especially suitable.
Since the quarks are heavy, the difficult extrapolation of the valence
quarks to the chiral limit is unnecessary.
More importantly, the fact that the valence quarks are nonrelativistic
means that one can apply nonrelativistic arguments, EFT and phenomenological 
potential 
models to gain a more solid understanding of systematic uncertainties
than is possible with light hadrons.
It implies that we have better than usual understanding 
of the importance of the various higher dimension correction operators
that are used to improve the lattice Lagrangian.
We expect in advance, for example, that some of the uncertainty arising from
 imprecise settings of
 correction operators like $\overline{\psi}\Sigma \cdot B \psi$
is cancelled in spin-averaged masses like $(3M_{\psi}+M_{\eta_c})/4.$
On the other hand, the fact that quark momenta in quarkonia are larger than
they are in light hadrons makes some discretisation errors larger.
 A meson like the $D_s$, which also has a tame chiral extrapolation,
is also particularly simple to calculate on the lattice.

A variety of methods has been proposed to obtain \MSB masses from bare lattice
quark masses (see \cite{Lubicz:2000ch} for a review of some of the methods).
The standard perturbative way is to calculate the on-shell quark propagator
on the lattice and in the \MSB scheme to a given order in perturbation theory
and define renormalisation constants so that they are equal. Vector and axial
vector Ward identities may be used to define a renormalised mass
\cite{Bochicchio:1985xa} which has been used in several ways to obtain the
\MSB quark mass. Methods using Schr\"odinger functional and step-scaling
functions have been developed \cite{Rolf:2002gu}.

Almost all existing unquenched lattice determinations of $m_b$ use perturbation
theory to relate the lattice quark mass to the \MSB mass. The best
determination of the $b$ quark mass from Bottomonium is the relatively old
calculation of Davies et~al.~\cite{Davies:1994pz,Hornbostel:1998ki}, 
using NRQCD and two flavours of Wilson sea quarks. They used the mass of the
$\Upsilon$ to set $m_b$. The \MSB mass was obtained with first-order
perturbation theory from the bare lattice mass (or more precisely, from the
energy shift between the meson mass and the lattice mass). They  obtained for 
the \MSB mass $\overline{m}_b(\overline{m}_b)=4.26(4)(3)(10)$~GeV.
This agrees well with lattice determinations of the quark mass from the
$B$ meson mass.
For example, Gimenez et~al.~\cite{Gim04,Gimenez:2000cj}
 use the static approximation for the $b$ quark, 
and use two flavours of unquenched Wilson fermions.
Incorporating a  stochastic estimate of the third-order
perturbative correction \cite{Ren04} they
obtain $\overline{m}_b(\overline{m}_b)=4.21(3)(5)(4)$~GeV. 
In the quenched approximation,
Heitger and Sommer \cite{Heitger:2003nj} use step-scaling methods with
 the static approximation for the $b$ quark.
They use non-perturbative Ward identity based methods to calculate 
the renormalisation constants, and
 obtain $\overline{m}_b(\overline{m}_b)=4.12(7)(4)$~GeV.
Divitiis et~al.~\cite{deDivitiis:2003iy}
use step-scaling methods with a relativistic formulation of both the heavy
and light quarks, and non-perturbative Ward identity based determinations
of the renormalisation constants.
They obtain $\overline{m}_b(\overline{m}_b)=4.33(10)$~GeV.

Unquenched determinations of $m_c$ are just beginning to appear.
A consistent lattice picture of the charm quark mass
 exists from quenched calculations from a few years ago.
Using the charmonium spin-averaged 1S mass to set $m_c$ and 
the 1P--1S splitting to fix the lattice spacing,
Juge \cite{Juge:2000ev}, following Kronfeld \cite{Kronfeld:1997zc}, obtained
$\overline{m}_c(\overline{m}_c)=1.27 (5)$~GeV.
For the charm quark action, he used the clover action with the Fermilab
heavy quark interpretation.
He used a second order perturbative expansion to relate the lattice 
and \MSB masses.
The one-loop coefficient was obtained in the usual way.
The two-loop coefficient was calculated by calculating the charm quark
propagator at several very small values of $\alpha_{\rm s}$ and fitting the results
to a perturbative expansion.

This result is likewise compatible with results using the $D_s$
meson to set the quark mass.  For example,
Becirevic, Lubicz, and Martinelli \cite{Becirevic:2001yh}
related the quark masses of \MSB and the lattice 
using the vector and axial vector Ward identities.
The lattice part of the calculation was done non-perturbatively.
They obtained $\overline{m}_c(\overline{m}_c)=1.26 (3) (12)$~GeV.
Rolf and Sint \cite{Rolf:2002gu} have used the Schr\"odinger functional 
approach to calculate the renormalisation factors for the quark mass.  
They obtained $\overline{m}_c(\overline{m}_c)=1.301 (34)$~GeV
when  $F_K$ was used to set the lattice spacing.  
Divitiis et~al.~\cite{deDivitiis:2003iy},
using the methods described previously,
 obtain $\overline{m}_c(\overline{m}_c)=1.319(28)$~GeV.
All of these quenched calculations contain an additional
uncertainty comparable to their stated uncertainties
 from the quantity used to set the lattice spacing, arising
from the quenched approximation.
Preliminary unquenched results by Dougall, 
Maynard, and McNeile at Lattice 2004
\cite{Dou04,Dou02} are consistent with this picture.
Other unquenched $m_c$ determinations are in progress and will
appear this year.

\subsection{Final averages for \boldmath{$m_b$} and \boldmath{$m_c$}}
\label{sec:mbmcaverage}

In the previous sections we have presented the various bottom and
charm quark mass determinations available in the literature and gave
detailed discussions on the methods that were used to obtain the
respective central values and uncertainties. A compilation of the
individual numerical results from QCD sum rules, the 1S resonance and
lattice QCD is displayed in \Figure~\ref{fig:figmb} for the bottom
quark mass and in \Figure~\ref{fig:figmc} for the charm quark mass.

One of the general features of the analyses is that the theoretical component
of the uncertainty is substantial, sometimes the dominant component. Some of
the more precise results relied on additional assumptions and on specific
prescriptions, and some of the more recent analyses obtained larger
uncertainties than older ones because they considered new theoretical aspects.
Therefore, the quoted errors are subjective and do not have any statistical
meaning. Procedures such as taking a weighted average are meaningless a priori.

For this reason, first of all, the central values for bottom and charm
quark masses are obtained by simply taking the mean value of all
presented determinations with equal weights. Concerning our current
knowledge of the corresponding uncertainties, we have decided to
present two different approaches. For the first approach, we try to
infer from \Figures~\ref{fig:figmb} and \ref{fig:figmc} what a
sensible $1\sigma$ deviation should be if the distribution of the
various determinations could be interpreted statistically.  In
\Figures~\ref{fig:figmb} and \ref{fig:figmc} this 
``one standard deviation'' uncertainty has been displayed by the
darker gray area which numerically corresponds to
\begin{equation}
\overline m_b(\overline m_b) \,=\, 4.22 \,\pm\, 0.05 \;\gev
\qquad \mbox{and} \qquad
\overline m_c(\overline m_c) \,=\, 1.28 \,\pm\, 0.05 \;\gev \,.
\end{equation}
Our second approach consists in presenting ranges for the quark mass values
in which the respective \MSB masses are located to some high
degree of probability. These ``ranges'' then read
\begin{equation}
\overline m_b(\overline m_b) \,=\, 4.12 - 4.32 \;\gev
\qquad \mbox{and} \qquad
\overline m_c(\overline m_c) \,=\, 1.18 - 1.38 \;\gev \,.
\end{equation}
and correspond to the lighter gray area in \Figures~\ref{fig:figmb}
and \ref{fig:figmc}.

\subsection{Future Opportunities}
\label{Outlookmass}

\vspace{3mm}\noindent
The determination of the bottom and charm quark masses from heavy quarkonium
data represents a by now classic problem initiated by the early works on the
QCD operator product expansion~\cite{Shifman:1979bx}. In recent years an
impressive array of developments has led to a more refined understanding of
the uncertainties inherent to the methods that can be applied and to a number
of new higher order perturbative computations. Nevertheless, there are a
number of issues that are still open. 

So far all numerical analyses based on non-relativistic quantities relied on
fixed-order perturbation theory. The renormalisation group improved
computations already applied in the context of top pair production at
threshold could be applied here as well. In contrast to the top quark case,
however, where the hierarchy of relevant scales is large and a renormalisation
group improved treatment appears indispensable, the summation of logarithms
for bottomonium and charmonium quantities will mainly serve as a cross-check
for the fixed-order methods. Moreover, the proper renormalisation group
treatment of the charmonium and the higher excited bottomonium states is not
fully understood yet because for these systems the ultrasoft scale appears to
be below the hadronic scale $\Lambda_{\rm QCD}$. In any case, analyses based
on renormalisation group improved perturbation theory would represent a
valuable achievement toward a better understanding of the behaviour of
perturbation theory to bottomonium and maybe also charmonium states.

A number of perturbative results at the NNNLO level exists for the quarkonium
energy levels. As mentioned in the review, all these analyses determined the
contributions coming from the ultrasoft scale perturbatively, which is,
however, not realistic, particularly for higher radial bottomonium states and
the charmonium states in general. Here, the systematic factorisation of the
ultrasoft effects and a treatment based on non-perturbative methods could lead
to a better understanding and more realistic estimate of the theoretical
errors. In particular, the non-perturbative treatment of ultrasoft effects
might also shed more light on the validity of the assumption that the leading
order solution corresponds to a Coulomb-type bound state for quarkonium
systems for which the ultrasoft scale is non-perturbative.
Such an analysis might be even useful for quantities like the large-$n$
moments used in bottom quark mass determinations and for the bottomonium
ground state where there are good reasons for the assumption that the
ultrasoft scale can be treated perturbatively.   

A complementary approach toward a better understanding of non-perturbative
effects, particularly for quarkonium systems where the ultrasoft scale is
believed to be perturbative, would be the determination of quark and gluon
condensate contributions beyond leading order. Here, first principles
determinations of the quark and gluon condensates from lattice simulations
would be quite important. At present the matching coefficients of the
condensates are only known at leading-order in the
non-relativistic expansion. Results at the next-to-leading order level
would provide further tests for the approximation in terms of a series of local
condensates. Likewise, effects of subleading condensates in the non-relativistic
framework could be analysed more systematically.

Much programmatic work remains to be done combining the various ingredients
that have been applied to heavy quark mass determinations on the lattice. All
calculations need to be repeated with three flavours of unquenched light sea
quarks. It should be possible to extend all the perturbative calculations to
two-loop order at least. When this has been accomplished, an agreement between
perturbative and non-perturbative determinations of $m_b$ and $m_c$ to around
30--50~MeV should be possible.

\section{Strong coupling constant from Quarkonia}
\label{sec:alphas}

\subsection{Strong coupling from Upsilon decays and sum rules}

Heavy quarkonia leptonic and non-leptonic inclusive decay rates, in
principle, provide means to determine the strong coupling $\alpha_{\rm s}$
using perturbative QCD. Precise
data are available from the decay widths of the $1^{--}$ $J/\psi(1S)$ and
$\psi$ states and the $\Upsilon$ resonances. Assuming 
that the hadronic and leptonic decay widths of the heavy quarkonium states can
be factorised into a non-perturbative part,
 and a calculable perturbative part, ratios of partial decay widths
can be predicted. The ratio of the total hadronic decay width of
the $\Upsilon(1S)$ and its leptonic partial width is then given by 
\cite{Barbieri:1980yp,Mackenzie:1981sf,Bodwin:1995jh}
\begin{eqnarray}
\label{eq:Rmu}
R_\mu(\Upsilon) \!\!\!&=&\!\!\! \frac{\Gamma(\Upsilon\,\to\,{\rm hadrons})}
{\Gamma(\Upsilon\,\to\,\mu^+\mu^-)} \nonumber \\
\smvs
&=&\!\!\! \frac{10(\pi^2-9)\alpha_{\rm s}^3(M_b)}{9\pi\alpha^2_{{\rm em}}}
\Biggl[\,1+\frac{\alpha_{\rm s}}{\pi}\biggl(-19.36+\frac{3\beta_0}{2}\biggl(
1.161+\ln\frac{2M_b}{M_\Upsilon}\biggr)\biggr)\Biggr] \,. 
\end{eqnarray}
Theoretical corrections to \Eq~\ref{eq:Rmu} arise from two
sources. Corrections of order $v^2$, due to the relativistic nature of
the $Q\bar Q$ system have been analysed in
\cite{Bodwin:1995jh}. Further corrections of non-perturbative nature,
due to the annihilation from higher Fock states (``colour-octet''
contribution), can only be estimated and have been discussed in
\cite{Gremm:1997dq,Brambilla:2002nu}. Both types of corrections are more severe
for the charmonium and the higher $\Upsilon$ states. Thus for a determination
of $\alpha_{\rm s}$ the $\Upsilon(1S)$ state should be used. 

Employing the experimental value $R_\mu(\Upsilon)\,=\,39.11\pm 0.4$
\cite{Eidelman:2004wy}, the section on Quantum Chromodynamics in the Review
of Particle Physics quotes $\alpha_{\rm s}(M_b)=0.177\pm 0.010$
without reference, whereas the original work \cite{Gremm:1997dq}
obtains $\alpha_{\rm s}(M_b)=0.186\pm 0.032$. The uncertainty is fully
dominated by theory, and mainly originates from the above mentioned
colour-octet contributions as well as the residual scale
dependencies. Varying the renormalisation scale as well as the pole
quark mass $M_b=4.6-4.9$~GeV, the next-to-leading order $\alpha_{\rm
s}^4$ correction in \Eq~\ref{eq:Rmu} is around 30--40\%.  Such a large NLO
corrections entails that also the still unknown NNLO term could be
sizeable.  Altogether, the error estimate $\Delta\alpha_{\rm
s}(M_b)\approx 0.03$ appears realistic at the present stage, and it is
unclear, how the result $\Delta\alpha_{\rm s}(M_b)= 0.01$, quoted in
the PDG, can be justified.\footnote{The small uncertainty presented
for $\alpha_{\rm s}$ from quarkonia in the PDG~\cite{Eidelman:2004wy},
pulls down the global $\alpha_{\rm s}$ average quite noticeably, as
can also be seen from figure~9.1 in Ref.~\cite{Eidelman:2004wy}.}

In the past there have also been a number of analysis
\cite{Jamin:1997rt,Kuhn:1998uy,Penin:1998zh,Hoang:1998uv,Hoang:1999ye}
attempting the determination of $\alpha_{\rm s}$ from the large-$n$
Upsilon sum rules described in \Section~\ref{sec:mbsumrules} with
realistic error estimates. The analyses were based on simultaneous
fits for the bottom quark mass and $\alpha_{\rm s}$ using several
moments. In these analyses uncertainties $\Delta\alpha_{\rm
s}(M_b)\approx 0.03-0.05$ were found. In view of these remarks, the
determination of $\alpha_{\rm s}$ from heavy quarkonium properties
does not appear to be a method that can compete with other more
precise methods based on perturbation theory at higher scales or from
the hadronic $\tau$-decay rate
\cite{Bethke:2002rv,Eidelman:2004wy}. However, heavy quarkonium analyses may
provide a useful cross-check for other methods.

\subsection{Strong coupling constant from lattice QCD}
 
Lattice determinations of the strong coupling constant have the same
two components as lattice determinations of other standard model
parameters: fixing bare lattice parameters from hadronic data, and
conversion of lattice parameters into \MSB parameters.  Spin-averaged
splittings in charmonium and bottomonium are excellent quantities for
determining the lattice spacing in GeV.  Davies
et~al.~\cite{Davies:2003ik} recently reported unquenched lattice
calculations of the simplest heavy and light quark quantities with 2+1
(two light and one strange) flavours of unquenched staggered fermions.
Unlike the previous unquenched calculation mentioned, these used the
physical number of light quark flavours.  In addition, the use of
staggered fermions allowed the use of much lighter light quark masses,
as low as $m_s/6.$ The results, which were dominated by quarkonium
splittings, showed good, few per cent agreement among lattice spacings
obtained from the various quantities.  In particular, the 1P--1S
splitting in the $\psi$ system and the 1P--1S, 2S--1S,
\etc splittings in the $\Upsilon$ system all yielded the same lattice
spacing to high accuracy.

The factor relating the bare lattice and \MSB couplings, however,
is large, roughly a factor of two at typical lattice spacings,
making the conversion demanding.
Early lattice determinations of the physical coupling
used a mean-field improved coupling constant, $\alpha_{\rm s}=\alpha_0/{\rm Tr} U_P$,
to reduce the uncertainty due to the large tadpole contributions to the conversion which 
are responsible for much of the poor perturbative behaviour.  
(${\rm Tr} U_P$
is the trace of the plaquette operator on the lattice.)
 \cite{El-Khadra:vn}.
Subsequent work on perturbative methods proposed obtaining
improved coupling constants from Monte Carlo measurements of small
Wilson loops \cite{Lepage:1992xa}.
L\"uscher and collaborators proposed
non-perturbative methods for obtaining physical couplings via
the Schr\"odinger functional \cite{Luscher:1992an}.

Very recently, Mason et~al.~\cite{Mason:2004} have reported third-order
perturbative results relating the couplings via many small Wilson loops and
Creutz ratios. The many determinations agree with each other and agree with
asymptotic freedom over a large range of $q^2$.
Combined with the lattice spacings obtained in \cite{Davies:2003ik}, this
work yields $\alpha_{\overline{\rm MS}}(M_Z)=0.117(1)$, in good agreement with
the world average.  Quark masses and other standard model parameters based
on the calculations of Ref.~\cite{Davies:2003ik} will soon appear.

Booth et~al.~\cite{Booth:2001qp} have used two-flavour simulations of the less
physically transparent $r_0$ (defined from the heavy quark potential) to set
the lattice spacing, and perturbation theory via the plaquette to obtain
physical coupling constant. They obtain a result significantly lower than the
above and than the world average. However, the perturbation theory resulting
from Wilson sea quarks contains large, poorly convergent contributions from the
fermionic graphs, and the calculation employs two rather than three light sea
quarks.
 
The third-order perturbative determination of Mason et~al.~\cite{Mason:2004}
is unlikely to be extended with perturbative methods in the near future.
Future progress in $\alpha_{\rm s}$ determinations from the lattice may await the
fruition of unquenched non-perturbative methods for $\alpha_{\rm s}$ determination.

\section{NRQCD and the Velocity Renormalisation Group}
\label{sec:vnrqcd}

In this section we discuss some of the more conceptual aspects
involved in using effective theory methods for non-relativistic
QCD. In particular we review issues regarding the power counting and
renormalisation of these theories.  For motivation it is useful to ask
what properties we desire from an effective field theory of
non-relativistic particles. The EFT should
\begin{enumerate}
\item have degrees of freedom that reproduce the IR divergences of the full
      theory in its entire region of validity and thus have no large
      logs in matching calculations,

\item have a well defined power counting in $v$ so that the expansion is
      systematic,

\item exhibit all the expected symmetries of the physical problem, spin
      symmetries, gauge symmetries \etc,

\item start with a regulator independent Lagrangian, so that different choices
      of regulators and renormalisation schemes can be systematically
      implemented,

\item have a consistent renormalisation procedure so that UV divergences leave
      the theory well defined, and anomalous dimensions and
      renormalisation group evolution can be computed.
\end{enumerate}

The original NRQCD~\cite{Caswell:1985ui,Thacker:1990bm,Bodwin:1994jh}
(see also 
%\marginpar{Chapter 1 has no sections 2.2 and 2.3, but ch. 3 has.}
%%%%%\Chapter~\ref{chapter:commontheoreticaltools}
\Chapter~\ref{chapter:spectroscopy}
\Sections~\ref{sec:nrqcd} and \ref{sec:spnrqcd}) 
was formulated with the aim of separating the short distance physics
at the scale $m$ from the long distance physics at the
non-relativistic momentum and energy scales, $m v$ and $m v^2$.  This
effective theory has one distinct quantum field for each of the low
energy quarks, antiquarks, and gluons. It succeeds at the majority of
the above criteria but has well studied issues regarding item 2. In
particular, the EFT matrix elements and diagrams do not have a unique
scaling with $v$ as emphasised in
Refs.~\cite{Luke:1996hj,Manohar:1997qy}.\footnote{This version of
NRQCD still satisfies the power counting in a weaker form since the
subleading terms in the $v$ expansion can in principle be obtained
from higher order terms of the leading diagrams. This subtlety is
therefore not important until calculations are done beyond leading
order in $v$ where both soft and ultrasoft effects become important.}
From a modern viewpoint this is due to the fact that one gluon field
is used to describe both soft and ultrasoft gluon effects, and the
power counting for these two gluons
differs~\cite{Luke:1997ys,Beneke:1997zp,Griesshammer:1997wz}.  In
particular the soft gluons are responsible for binding while the
ultrasoft gluons give real radiation and have couplings which need to
be multipole expanded~\cite{Labelle:1996en,Grinstein:1997gv}.  A
second complication is that, although a small $v\ll 1$ guarantees that
the scales $m\gg mv\gg mv^2$ are well separated, the dispersion
relation, $E={\bf p}^2/(2m)$, couples the $mv^2$ energy and $mv$
momentum scales together and affects the renormalisation group
evolution~\cite{Luke:1996hj}.

One approach to resolving the power counting issue is to consider in
sequence the $mv$ and $mv^2$ scales and the soft and ultrasoft gluons.
One first considers an NRQCD theory with soft gluons, and then
``integrates out'' the $mv$ scale into nonlocal potential operators
for the effective theory with ultrasoft gluons.  This EFT is known as
pNRQCD~\cite{Pineda:1998bj,Brambilla:1999xf}.  While the soft NRQCD
theory does not have a $v$ power counting, the $v$ power counting is
obtained in the final pNRQCD theory. This construction has the
advantage of avoiding the double counting gluon effects in a simply
way, and also corresponds to the Wilsonian picture of integrating out
scales in step-wise fashion, $m\to mv\to mv^2$.  Further details about
the pNRQCD approach can be found in
\Chapter~\ref{chapter:spectroscopy}.
 
A second approach to resolving the power counting issue accounts for
the correlation in energy and momentum scales from the start. In this
case there is only one EFT below the scale $m$ and it simultaneously
contains soft and ultrasoft gluons. This EFT is known as
vNRQCD~\cite{Luke:1996hj,Manohar:1999xd,Manohar:2000hj,Hoang:2002yy}.
It has the advantage that power counting automatically induces the
correct correlation between the ultrasoft $\mu_U$ and soft $\mu_S$
renormalisation scales, $\mu_U=\mu_S^2/m\equiv m\nu^2$, that preserves
the dispersion relation.  Renormalisation group evolution in $\nu$ is
known as the velocity renormalisation group (vRGE), and incorporates
the correspondence between $\mu_U$ and $\mu_S$. In this section we
give a brief review of the vNRQCD approach for the case $m\gg mv \gg
mv^2 \gg \Lambda_{\rm QCD}$ relevant for QED and non-relativistic
$t\bar t$ systems.\footnote{Up to NNLO this situation is basically
equivalent to the case where $mv^2\sim \Lambda_{\rm QCD}$, since the
only obvious difference is that we must be careful to stop
renormalisation group running before any coupling $\alpha_{\rm
s}(\mu)$ becomes non-perturbative, so $\mu\ge \rm{few}\times
\Lambda_{\rm QCD}$. The main difference at higher order is that in the
latter scenario matrix elements involving ultrasoft gluons which are
N${}^3$LO become non-perturbative. }

The effective vNRQCD Lagrangian can be separated into ultrasoft, soft,
and potential components, ${\cal L}={\cal L}_u + {\cal L}_s + {\cal
L}_p$.  The presence of both soft and ultrasoft gluons immediately
brings up the issue of double counting.  To avoid double counting the
effective theory is constructed such that the ultrasoft gluons
reproduce only the physical gluon poles where $k^0\sim {\bf k}\sim
mv^2$, while soft gluons give only those with $k^0\sim {\bf k}\sim
mv$. The scales for the gluon momenta are influenced by the quark
propagators, so the quark-gluon interactions must be constructed in
such a way that we will not upset this scaling.  In ${\cal L}_u$ this
is achieved by the multipole expansion of interactions, which is
enforced by a phase redefinition that separates label ${\bf p}\sim mv$
and residual $\sim mv^2$ momenta. This ensures that ultrasoft gluon
momenta are always much smaller than the quark three-momenta. In
${\cal L}_s$ this is achieved by integrating out the intermediate
static HQET like fermion propagators into the effective soft
vertices~\cite{Luke:1996hj} and by the pull-up
mechanism~\cite{Hoang:2001rr}.  The pull-up mechanism refers to the
manner by which all soft loops in vNRQCD are made infrared finite
while at the same time the ultraviolet divergences in ultrasoft loops
are made to correspond directly to the hard scale $m$.

For example, the first few terms in the vNRQCD  ultrasoft Lagrangian are
\begin{eqnarray} \label{eq:Lus}
{\cal L}_u  &=& 
\sum_{\mathbf p}\,\bigg\{
   \psi_{\bmp}^\dagger   \bigg[ i D^0 - \frac {\left({\bf p}-i{\bf D}\right)^2}
   {2 m} +\frac{{\mathbf p}^4}{8m^3} + \ldots \bigg] \psi_{\bmp}
 + (\psi \to \chi)\,\bigg\}
 -\frac{1}{4}G_u^{\mu\nu}G^u_{\mu \nu} +\ldots \,,
\end{eqnarray}
where $G_u^{\mu\nu}$ is the ultrasoft field strength and $\psi_{\bf
p}$ is the quark field with label momentum ${\bf p}$, while $\chi$ is
the field for antiquarks. In dimensional regularisation the covariant
derivative has the form $D^\mu=\partial^\mu+i \mu_U^\epsilon\, g_u
A^\mu$, where $\mu_U=m\nu^2$ and $g_u=g_u(\mu_U)$ is the renormalised
ultrasoft QCD coupling.  The soft Lagrangian ${\cal L}_s$ can be found
in~\cite{Luke:1996hj,Manohar:1999xd,Manohar:2000hj} and couples soft
fields to the potential quarks with the soft QCD coupling
$g_s=g_s(\mu_S)$ ($\mu_S=m\nu$).  Finally the potential Lagrangian has
terms like
\begin{eqnarray} \label{eq:Lp}
{\cal L}_p &=& -\mu_S^{2\epsilon}\, V({\bmp,\bmp^\prime})\,
   \psi_{\bmp^\prime}^\dagger \psi_{\bmp}
   \chi_{-\bmp^\prime}^\dagger \chi_{-\bmp}
  + \ldots \,,
\label{vNRQCDLagrangian}
\end{eqnarray}  
where the coefficient function $V({\bmp},{\bmp^\prime})$ acts like a potential
(as do time ordered products of ${\cal L}_s$ vertices).  For equal mass fermions
in QCD perturbatively matching onto the first three orders in $v$ gives terms
\begin{eqnarray}
 && V({\bmp},{\bmp^\prime}) =   (T^A \otimes \bar T^A) \bigg[
 \frac{{\cal V}_c^{(T)}}{\bmk^2}
 + \frac{{\cal V}_k^{(T)}\pi^2}{m|{\bmk}|}
 + \frac{{\cal V}_r^{(T)}({\bmp^2 + \bmp^{\prime 2}})}{2 m^2 \bmk^2}
 + \frac{{\cal V}_2^{(T)}}{m^2}
 + \frac{{\cal V}_s^{(T)}}{m^2}{\bmS^2} \nn \\[2mm] 
&&\quad + \frac{{\cal V}_\Lambda^{(T)}}{m^2}\Lambda({\bmp^\prime ,\bmp}) 
 + \frac{{\cal V}_t^{(T)}}{m^2}\,
 T({\bmk}) + \ldots \bigg] 
 + (1\otimes 1)\bigg[
 \frac{{\cal V}_c^{(1)}}{\bmk^2}
 + \frac{{\cal V}_k^{(1)}\pi^2}{m|{\bmk}|}
 + \frac{{\cal V}_2^{(1)}}{m^2}
 + \frac{{\cal V}_s^{(1)}}{m^2}{\bmS^2} +\ldots \bigg]
\,, \nonumber \\[2mm]
 &&\quad {\bmS} \, = \, \frac{ {{\bmsigma}_1 + {\bmsigma}_2} }{2}\,,
 \quad 
 \Lambda({\bmp^\prime},{\bmp}) \, = \, -i \frac{{\bmS} \cdot ( {\bmp^\prime}
 \times {\bmp}) }{  {\bmk}^2 }\,,\quad
 T({\bmk}) \, = \, {{\bmsigma}_1 \cdot {\bmsigma}_2} - \frac{3\, {{\bmk}
 \cdot {\bmsigma}_1}\,  {{\bmk} \cdot {\bmsigma}_2} }{ {\bmk}^2} \,,
\label{vNRQCDpotential}
\end{eqnarray}
where $\bmk=\bmp'-\bmp$ is the momentum transfer and the colour and
spin factors have the appropriate index contractions with the fields
in \Eq~(\ref{eq:Lp}).  In \Eqs~(\ref{eq:Lus}) and (\ref{eq:Lp}) the
couplings are defined with dimensional regularisation and it is worth
noting that the factors of $\mu_U^\epsilon$ and $\mu_S^\epsilon$ in
${\cal L}_{u,s,p}$ are uniquely determined by mass dimension and $v$
power counting in $d=4-2\epsilon$ dimensions~\cite{Manohar:2000hj}
(see~\cite{Hoang:2002yy} for further detail). The renormalised
couplings depend on the parameter $\nu$, \ie ${\cal V}_i = {\cal
V}_i(\nu)$. In \Section~\ref{sec:topthresholdtheory} additional Wilson
coefficients are defined for the dominant current relevant for heavy
quark pair production in $e^+e^-$ annihilation, $c_1(\nu)$, as well as
coefficients for subleading production currents, $c_{2,3}(\nu)$. Apart
from the renormalisation scale for the couplings, the $v$ power
counting of an arbitrary diagram is universally determined by the
powers of $v$ assigned to operators, and therefore have no explicit
dependence on the choice of regulator.  Ambiguities regarding how
multipole expanded interactions renormalise operator products were
resolved in Ref.~\cite{Hoang:2002yy}.

Since we are considering the case $mv^2\gg \Lambda_{\rm QCD}$ the majority of
computations are purely perturbative. In this case they can be performed in
three stages, 
\begin{itemize}
 \item[i)] matching QCD at $\mu_U=\mu_S=m$ ($\nu=1$) onto vNRQCD,
 \item[ii)] running with the velocity renormalisation group from $\nu=1$
  to $\nu=v_0$ where $v_0$ is a number of order the typical velocity in 
  the non-relativistic bound state ($v_0\simeq 0.15$ for $t\bar t$), and 
 \item[iii)] computing the EFT matrix
elements at $\nu=v_0$.
\end{itemize}  
From an EFT point of view the most novel aspect of these
steps is the resummation of logarithms by running in $\nu$. The correlation of
energy and momentum cutoff scales in the evolution turns out to be crucial to
the proper resummation of logarithms in many cases. In
Ref.~\cite{Manohar:2000rz} it was shown for the first time that anomalous
dimensions and the vRGE could be used to predict $\ln\alpha$ contributions in
QED bound states like Hydrogen, positronium, and muonium.  For positronium the
$\alpha^7\ln^2\alpha$ hyperfine splitting corrections and $\alpha^3\ln^2\alpha$
corrections to decay rates were correctly reproduced, and the
$\alpha^8\ln^3\alpha$ Lamb shift was predicted.\footnote{For Hydrogen the
  $\alpha^8\ln^3\alpha$ terms resolved a controversy in favour of Karshemboim's
  original result~\cite{Karshenboim:1993xx}.} These results give a non-trivial
consistency check on the vNRQCD approach to renormalisation. Further consistency
checks from subdivergences are discussed in
Refs.~\cite{Hoang:2001mm,Hoang:2003ns}. In Ref.~\cite{Manohar:2000mx} it was
shown that the correlation of energy and momentum scales is necessary to compute
these QED corrections involving $\ln^k\alpha$ with $k\ge 2$. In QCD the
correlation in energy and momentum cutoffs is also known to be crucial to
resumming logs in the production
current~\cite{Luke:1996hj,Manohar:2000kr,Pineda:2001et,Hoang:2003ns}.

The running of the nonrelativistic QCD potentials ${\cal V}_i$ in vNRQCD was
worked out in~\cite{Manohar:1999xd,Manohar:2000kr,Hoang:2001rr} and for the
$c_i$ current coefficients in
Refs.~\cite{Luke:1999kz,Manohar:2000kr,Hoang:2001mm}. The running of the static
potential due to ultrasoft effects in the pNRQCD formalism was determined in
Ref.~\cite{Pineda:2000gz}. In Ref.~\cite{Pineda:2001ra} and also
~\cite{Pineda:2001et,Hoang:2002yy} an error was corrected in some of the
previous vNRQCD publications from a missing set of spin-independent
operators.\footnote{Numerically, the change turned out to be insignificant for
phenomenological $t\bar t$ applications as discussed in
\Section~\ref{sec:topthresholdrgi}.}  
All results for the vector current coefficient $c_1$ at NLL order and
the running of the potentials ${\cal V}_{c,k,2,r,s,\Lambda,t}$ have
now been checked by two independent groups. The renormalisation group
improved results were applied to $t\bar t$ production near threshold
in Refs.~\cite{Hoang:2000ib,Hoang:2001mm}, and significantly reduced
the uncertainty in previous fixed order NNLO
computations~\cite{Hoang:2000yr}. For renormalisation group improved
$t\bar t$ production near threshold all the necessary components in
steps i), ii), and iii) are known at NNLL order except for the full
NNLL order computation for $c_1$, and are included in the state of the
art $t\bar t$ analyses.

The necessary additional computations for $c_1$ at NNLL were discussed
in Refs.~\cite{Hoang:2001mm,Hoang:2003ns}.  A partial result was
obtained for the NNLL running of $c_1$ which fully incorporates the
non-mixing part of the anomalous dimension~\cite{Hoang:2003ns}. This
calculation required correctly associating the divergences in
three-loop vertex graphs with renormalisation of operators in the
effective theory.\footnote{The two-loop soft corrections to the
$1/m|{\bf k}|$ potentials obtained in Ref.~\cite{Kniehl:2001ju} were
used as an input.} At N${}^3$LO the results are consistent with the
fixed order computation in Ref.~\cite{Kniehl:2002yv}. Updates for the
changes due to a corrected operator basis and coming from the new
results for the NNLL order running of the dominant vector current were
made in Ref.~\cite{Hoang:2002yy} and \cite{Hoang:2003xg}
respectively. More details on the $t\bar t$ analyses are given in
\Section~\ref{sec:topthresholdrgi}. Finally, computations of the running of
the spin-dependent potentials beyond the leading logarithmic
approximation order and some applications for $b\bar b$, $b\bar c$ and
$c\bar c$ systems can be found in
Refs.~\cite{Kniehl:2003ap,Penin:2004xi,Penin:2004ay} for pNRQCD.  In
Ref.~\cite{Penin:2004ay} also the spin-dependent part of the mixing
contributions to $c_1$ at NNLL was determined. In these references the
renormalisation group running is called the nonrelativistic
renormalisation group.

\section{Top pair production at threshold in $e^+e^-$ collisions} 
\label{sec:topthreshold}

\subsection{Physics of the top threshold}
\label{sec:topthresholdintroduction}

With a weight of around $175$~GeV, the top quark is the heaviest known
quark flavour and plays an important role in the understanding of the
mass generation in the Standard Model and of electroweak symmetry
breaking. Top--antitop quark pair production close to the threshold,
\ie for centre of mass (c.m.) energies $\sqrt{s}\approx 2 m_t\approx 350$~GeV,
will provide an integral part of the top quark physics program at the
next $e^+e^-$ Linear Collider, which is supposed to be the next major
accelerator project after the
LHC~\cite{Aguilar-Saavedra:2001rg,Abe:2001nq}. In this kinetic regime
the $t\bar t$ cross-section rises as soon as sufficient amount of energy is
available to produce the top--antitop quark pair. Since almost all
energy is spent on the top quark masses, the top quarks are
non-relativistic and their velocities are small, $v\ll 1$. Due to the
large top quark width, 
\begin{equation}
\Gamma_t(t\to b W)\,\approx\,\frac{G_F}{\sqrt{2}}\frac{m_t^3}{8\pi}
\,\approx\, 1.5\,\,\mbox{GeV}\,\,\gg\Lambda_{\rm QCD}
\,,
\end{equation}
the would-be toponium resonances overlap and the total $t\bar t$
production cross-section line-shape is a smooth function of the
energy. Theoretically this means that non-perturbative effects are
strongly suppressed and can be neglected for studies of the total
cross-section~\cite{Kuhn:1981gw, Bigi:1986jk, Fadin:1987wz}.

The total cross-section for $e^+e^-\to \gamma^*, Z^*\to t\bar t$ has the form 
\begin{eqnarray}
  \sigma_{\rm tot}^{\gamma,Z}(s) = \sigma_{\rm pt}\,
  \Big[\, F^v(s)\,R^v(s) +  F^a(s) R^a(s) \Big] \,,
\label{eq:totalcross}
\end{eqnarray}
where $\sigma_{\rm pt}=4\pi\alpha^2/(3 s)$ is the muon pair
cross-section. The vector and axial-vector $R$-ratios are
\begin{eqnarray} \label{eq:fullR}
 R^v(s) \, =  \,\frac{4 \pi }{s}\,\mbox{Im}\,\left[-i\int d^4x\: e^{i q\cdot x}
  \left\langle\,0\,\left|\, T\, j^v_{\mu}(x) \,
  {j^v}^{\mu} (0)\, \right|\,0\,\right\rangle\,\right] \,, \nn\\[2pt]
 R^a(s) \, =  \,\frac{4 \pi }{s}\,\mbox{Im}\,\left[-i\int d^4x\: e^{i q\cdot x}
  \left\langle\,0\,\left|\, T\, j^a_{\mu}(x) \,
  {j^a}^{\mu} (0)\, \right|\,0\,\right\rangle\,\right] \,, 
\end{eqnarray}
where $q=(\sqrt{s},0)$ and $j^{v}_\mu$ ($j^{a}_\mu$) is the vector
(axial-vector) current in the Standard Model that produces a
quark--antiquark pair.  With both $\gamma$
and $Z$ exchange the prefactors in \Eq~(\ref{eq:totalcross}) are
\begin{eqnarray}
  F^v(s) &=& \bigg[\, Q_q^2 - 
   \frac{2 s\, v_e v_q Q_q}{s-m_Z^2} + 
   \frac{s^2 (v_e^2+a_e^2)v_q^2}{(s-m_Z^2)^2}\, \bigg]\,,
\qquad %\nn\\[2mm]
  F^a(s) = \frac{s^2\, (v_e^2+a_e^2)a_q^2}{ (s-m_Z^2)^2 } \,,
\end{eqnarray}
where 
\begin{eqnarray}
  v_f = \frac{T_3^f-2 Q_f \sin^2\theta_W}{2\sin\theta_W \cos\theta_W}\,,
  \qquad\qquad
  a_f = \frac{T_3^f}{2\sin\theta_W \cos\theta_W} \,.
\end{eqnarray}
Here, $Q_f$ is the charge of fermion $f$, $T_3^f$ is the third
component of weak 
isospin, and $\theta_W$ is the weak mixing angle. In the threshold
region the axial-vector contribution coming from the Z exchange is
suppressed by order $v^2$ and, numerically, at the level of a few
percent~\cite{Kuhn:1999hw,Hoang:1999zc}.

\begin{figure}[t] % fig:outlook
\begin{center}
%\epsffile[220 580 420 710]{asvar}
\includegraphics[width=.48\linewidth]{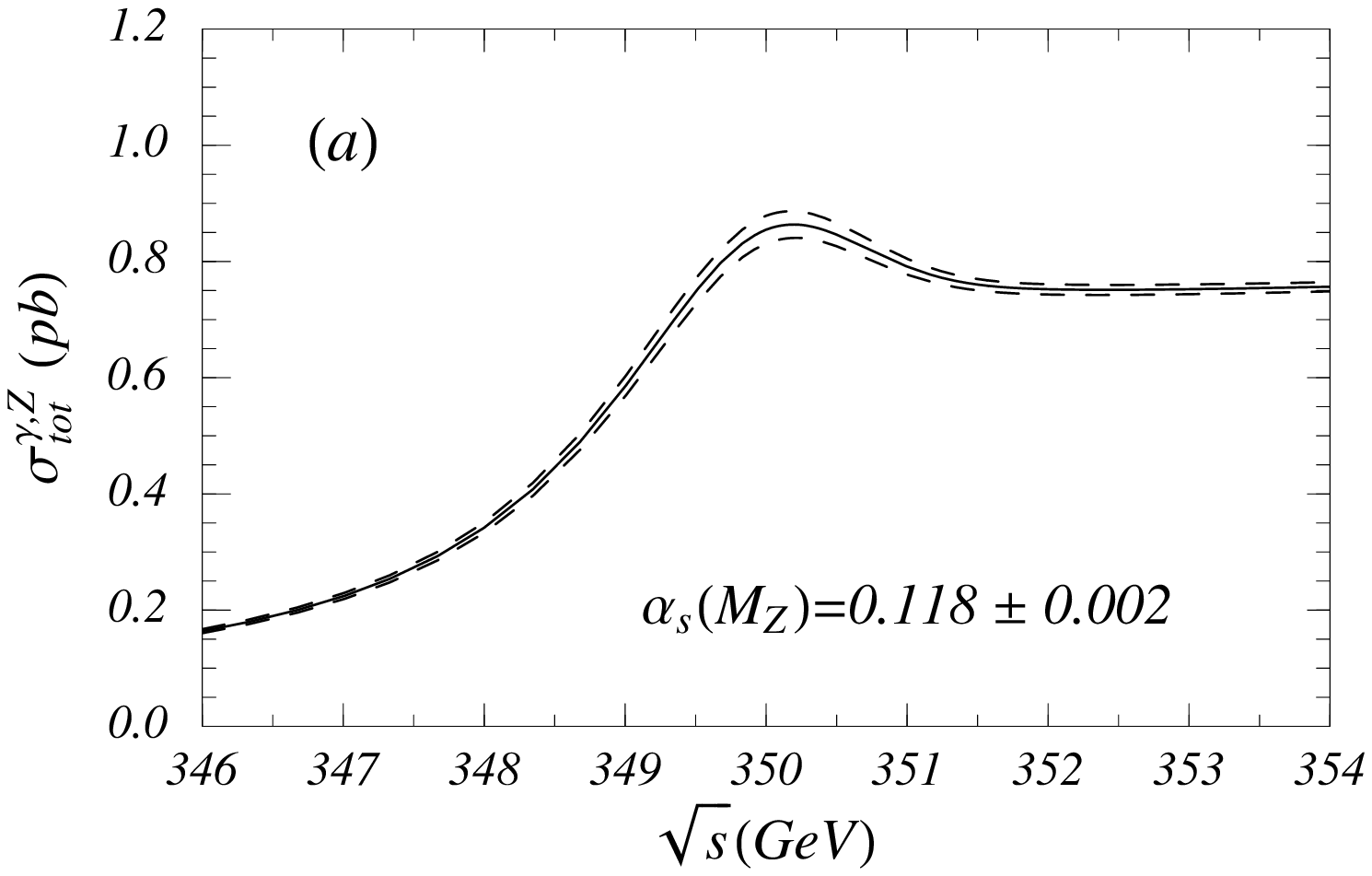}
\hfill
\includegraphics[width=.48\linewidth]{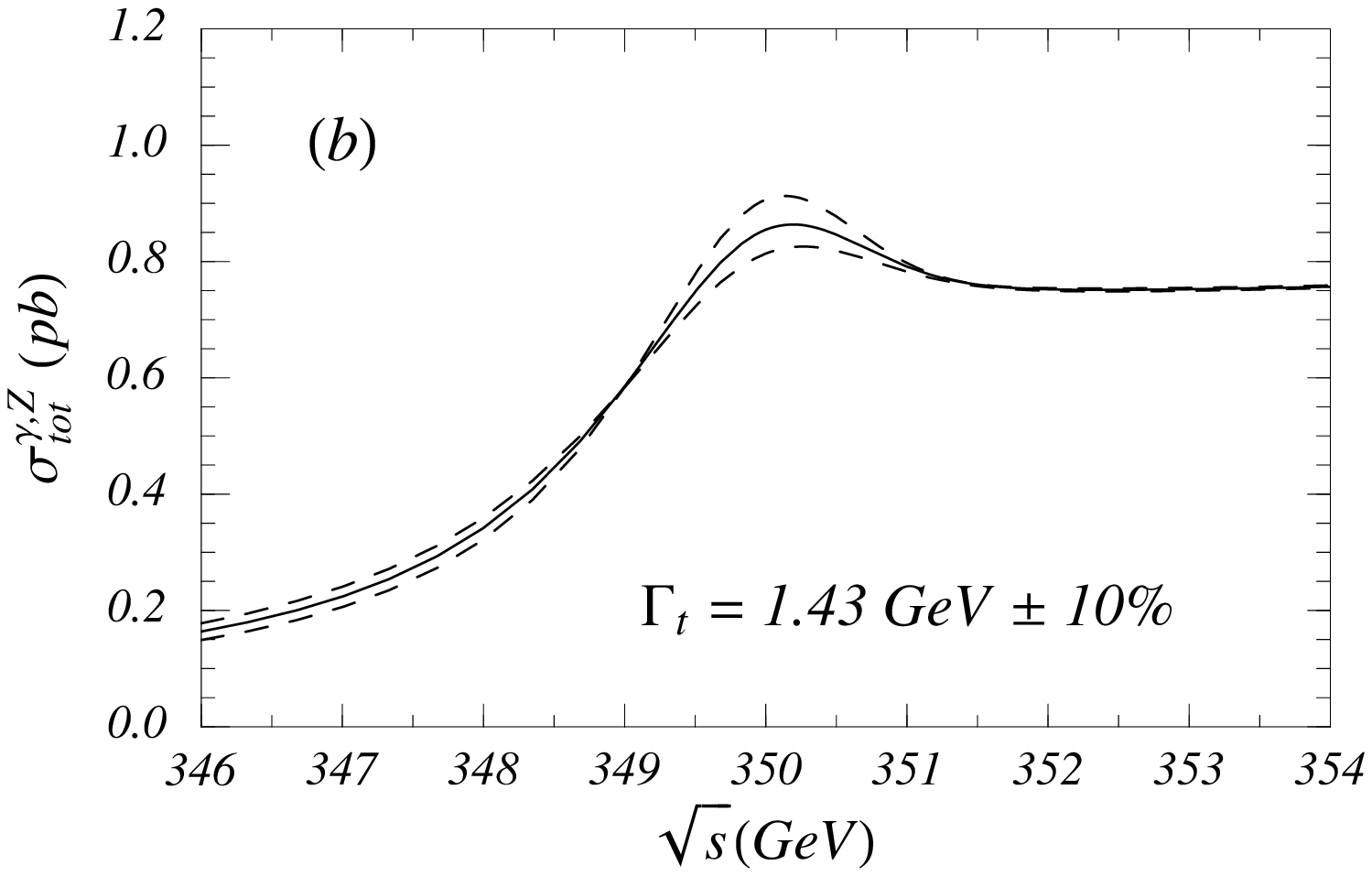}\\[4mm]
%\epsffile[220 580 420 710]{gamvar}
\includegraphics[width=.48\linewidth]{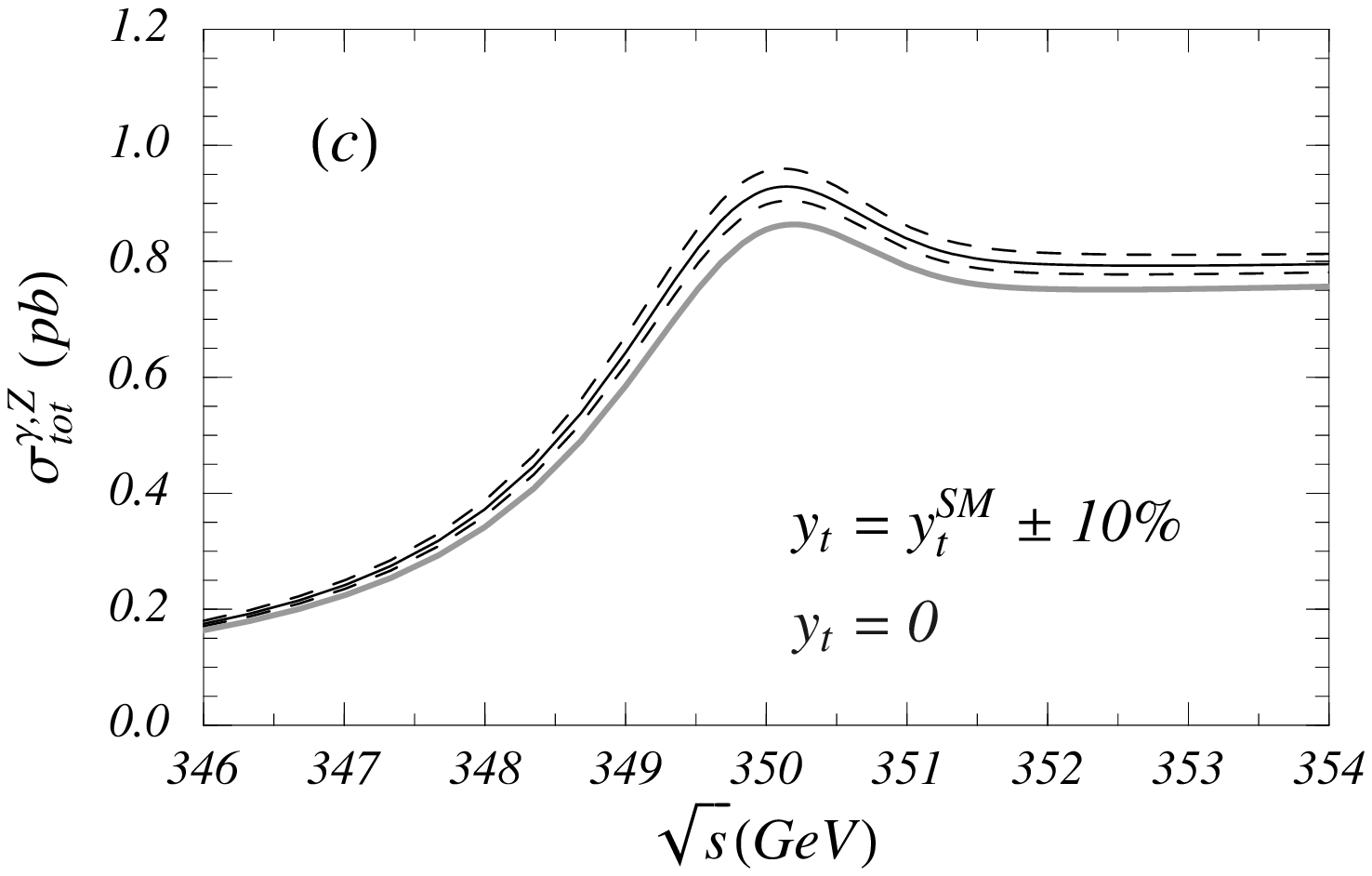}
%\epsffile[220 580 420 710]{gtthvar}
\end{center}
 \caption[Theoretical prediction for the top pair production
         cross-section] {Theoretical prediction for the top pair
         production cross-section without beam effects and its
         dependence on changes of a) the value of the strong coupling,
         b) the top quark width, and c) the inclusion of a Standard
         Model (SM) Higgs boson. Changes relative to the central value
         (solid lines) are shown by dashed red lines. In c) there are
         two solid lines, the lower black line is the decoupling limit
         for the Higgs boson ($m_H\to\infty$), and the upper blue line
         is for a SM Higgs with mass $m_H=115\,{\rm GeV}$.  The
         parameters $M_{\rm 1S}=175$~GeV, $\Gamma_t=1.43$~GeV,
         $\alpha_{\rm s}(M_Z)=0.118$, $y_t=0$ and $\nu=0.15$ have been
         used unless stated otherwise. The figures have been obtained
         in Ref.~\cite{Hoang:2000ib,Hoang:2001mm} using vNRQCD at NNLL
         order.}
\label{fig:outlook} 
\end{figure}

The cross-section rises rapidly at the point where the remnant 
of a toponium 1S resonance can be formed. From the energy where this
increase occurs, the top quark mass (in threshold mass schemes)
can be determined with uncertainties at the level of $100$~MeV,
whereas shape and height of 
the cross-section near  threshold  can be used to determine
$\Gamma_t$, the coupling strength of top quarks to gluons, $\alpha_{\rm s}$,
and, if the Higgs boson is not heavy, the top Yukawa
coupling $y_t=(2\sqrt{2}G_F
m_t^2)^{1/2}$ \cite{Strassler:1991nw,Harlander:1996dp}. For the
determination of the top quark width only very few other methods
are known. In \Figure~\ref{fig:outlook} the dependences of the
prediction for the total $t\bar t$ cross-section at threshold on
changes of $\alpha_{\rm s}$, $\Gamma_t$ and $y_t$ are displayed. 
The dependence of the cross-section on $\alpha_{\rm s}$ and $y_t$ comes to a
good approximation from the potential between the top quark pair.
The QCD Coulomb potential, which is responsible for the
binding-effects at the $t\bar t$ threshold, has the form
\begin{eqnarray}
V_{\rm QCD}(r) & = & -C_F\,\frac{\alpha_{\rm s}(\mu)}{r}
\end{eqnarray}
at leading order 
and deepens when $\alpha_{\rm s}$ is increased. This leads to an increase of
the cross-section as the $t\bar t$ pair is bound together more
strongly. Similarly, in the Standard Model the dominant effect of a
light Higgs exchange  can be understood from a Yukawa-type potential
of the form 
\begin{eqnarray}
 V_{tth}(r) 
 & = & - \frac{y_t^2}{4\pi}\ \frac{e^{-m_H r}}{r}
\,,
\end{eqnarray}
which also deepens for a stronger Yukawa coupling. In scenarios beyond
the Standard Model such as Supersymmetry, the effects of Higgs
exchanges can be larger than in the Standard Model due to modified
couplings and due to the fact that Higgs mass limits are lower,
Finally, the dependence on $\Gamma_t$ can be easily understood from
the fact that for smaller $\Gamma_t$ the resonance structure becomes
more pronounced.  The prediction in \Figure~\ref{fig:outlook} have
been made, exemplarily, in the 1S mass scheme.  In threshold mass
schemes the position of the 1S peak has, in general, only little
dependence on $\alpha_{\rm s}$, $y_t$, $\Gamma_t$ and other parameters
such as the renormalisation scale. Therefore in threshold mass schemes
the top mass measurement has only little correlation with other
theoretical parameters.  From differential quantities, such as the top
momentum distribution,\cite{Jezabek:1992np,Sumino:1993ai} the
forward--backward asymmetry or certain leptonic
distributions,\cite{Murayama:1993mg,Harlander:1995ac,Harlander:1997vg,Peter:1998rk}
one can obtain measurements of $\Gamma_t$, the top quark spin and
possible anomalous couplings \cite{Jezabek:2000gr}.

The top mass determination
represents the most important task of a threshold line-shape
measurement. In addition to the quite small expected error on the top
quark mass, the $t\bar t$ line-shape measurement has the feature that
the corresponding mass scheme is unambiguously defined since the
location where the cross-section rises is a stable (and perturbatively
calculable) function if threshold masses are employed. With
respect to both aspects the top mass measurement from a threshold scan
is superior to the reconstruction method as applied \eg at hadron
colliders. This is because the line-shape measurement relies on
counting the number of colour singlet $t\bar t$ pairs, while
reconstruction is based on the determination of a top four-momentum
from the top decay products, which unavoidably leads to larger
ambiguities since the top quark is coloured. 

\subsection{Experimental simulations for $e^+e^-$ collisions}
\label{sec:topthresholdexperiment}

A considerable number of experimental studies were carried out in the
past to assess the feasibility of top threshold measurements of the
total cross-section and certain distributions
\cite{Bagliesi:1992yf,Fujii:1994mk,Comas:1996kt,Cinabro:1999tq,Peralta:1999,Martinez:2002st}.
Apart from the standard experimental issues related to the event
selection and the identification of $t\bar t$ pairs, the treatment of
background, efficiencies and detector effects, which need to be
accounted for to measure the cross-section line-shape, a crucial role
is played by the luminosity spectrum and the absolute energy scale of
the $e^+e^-$ beam. (For a recent experimental review see
Ref.~\cite{Frey:1997sg}.) The luminosity spectrum arises from the
effects of initial state radiation, beamstrahlung and the beam energy
spread and leads to a loss of luminosity and a redistribution of
collision energy down to lower energies. The latter effect is
particularly important since it effectively smears out the resonance
structure visible in \Figure~\ref{fig:outlook}. Likewise a precise
knowledge of the absolute energy scale is crucial.

\begin{figure}[t] % fig:lumieffect
\begin{center}
\includegraphics[width=6cm]{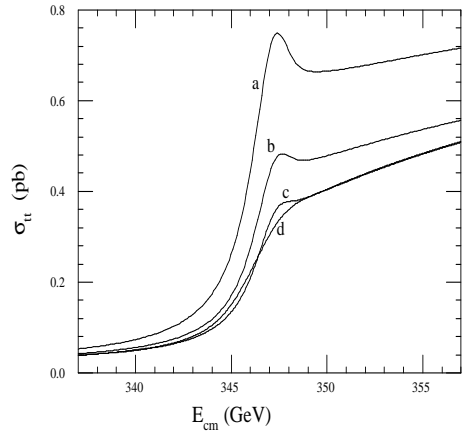}
\end{center}
\caption[Production cross-section for $t\bar t$ pairs at threshold for
         $m_t=175$~GeV]
        {Production cross-section for $t\bar t$ pairs at threshold for
         $m_t=175$~GeV. The theoretical cross-section without beam
         effects is given by curve (a). The following energy
         redistribution and luminosity losing effects have been
         applied to the theory for the remaining curves: (b)
         initial-state radiation (ISR); (c) ISR and beamstrahlung; (d)
         ISR, beamstrahlung and single-beam energy spread. The figure
         has been taken from Ref.~\protect\cite{Frey:1997sg}.}
\label{fig:lumieffect} 
\end{figure}
In \Figure~\ref{fig:lumieffect} (taken from Ref.~\cite{Frey:1997sg})
the effects of the three sources of the luminosity spectrum 
on the cross-section obtained for a fixed nominal c.m. energy are shown. The
effects are substantial. Thus the precise knowledge of the luminosity spectrum
for any nominal c.m. energy of the $e^+e^-$ beam is crucial for the
measurement of the threshold line-shape. Since the luminosity spectrum is
partially machine-dependent it has become standard for top threshold
studies to account for the complete luminosity spectrum (including
initial state radiation) during the experimental simulation. 
Thus it is the convention not to include the luminosity spectrum for
the theoretical predictions. Since the luminosity spectrum cannot be
fully predicted a priori in a machine-independent way it needs to be 
constantly measured while the top threshold scan is being performed. 
The luminosity spectrum can be deduced by using the acollinearity
distribution of $e^+e^-\to e^+e^-$ Bhabha scattering which depends on
the energy difference of the initial $e^+$ and $e^-$
beams\,\cite{Frary:1991aa}, while the absolute energy scale can be
determined from spectrometers.

\begin{figure}[t] % fig:expscan
\begin{center}
%\epsffile[10 10 545 560]{Experiment}
\includegraphics[width=10cm]{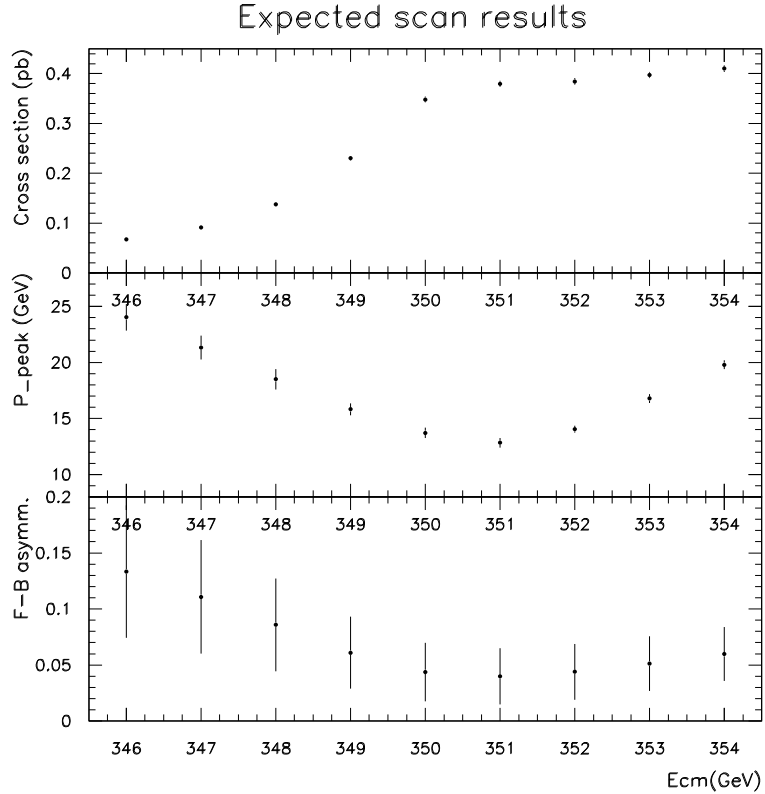}
\end{center}
 \caption[Energy dependence of three observables]
         {The energy dependence of the three observables $\sigma_{\rm
           tot}$, the peak position of the momentum distribution and
           $A_{\rm FB}$ as expected from a simulation of the threshold
           scan.  For the top quark mass $M_{\rm 1S}=175$~GeV is
           assumed (figure taken from Ref.~\cite{Martinez:2002st}.)}
\label{fig:expscan} 
\end{figure}

In a more recent simulation study by Martinez and Miquel \cite{Martinez:2002st}
the size of the experimental uncertainties for simultaneous
measurements of the 1S mass, $\alpha_{\rm s}(M_Z)$, $\Gamma_t$ and $y_t$ in a
top threshold run at an $e^+e^-$ Linear Collider was examined, based on
measurements of the total cross-section, the top three-momentum
distribution and the forward--backward asymmetry for a 9+1 point threshold
scan using a total integrated luminosity of $300$~fb$^{-1}$, see
\Figure~\ref{fig:expscan}. 
The analysis assumed perfect knowledge of the
luminosity spectrum and the absolute energy scale and thus reflects
the experimental uncertainties from other sources.
Fixing $y_t$ to the Standard Model (SM) value Martinez et al. obtained 
$\Delta M_{\rm 1S}=19$~MeV, $\Delta\alpha_{\rm s}(M_Z)=0.0012$ and
$\Delta\Gamma_t=32$~MeV from a three-parameter fit.
In a one parameter fit, fixing all other parameters, they obtained
$\delta y_t/y_t = ^{+0.18}_{-0.25}$. In a fit where $\Gamma_t$ is
fixed to the SM value and $\alpha_{\rm s}$ is constrained to
$\alpha_{\rm s}(M_Z)=0.120\pm 0.001$ they obtained $\delta M_{\rm
  1S}^{}=27$~MeV and $\delta y_t/y_t = ^{+0.33}_{-0.54}$.  
In a fit where only $\alpha_{\rm s}$ is constrained to
$\alpha_{\rm s}(M_Z)=0.120\pm 0.001$ they obtained $\delta M_{\rm
  1S}^{}=31$~MeV, $\delta\Gamma_t=34$~MeV
and $\delta y_t/y_t = ^{+0.35}_{-0.65}$ for a Higgs mass of $m_h=120$~GeV.
Note that the sensitivity on the top Yukawa coupling strongly decreases
for increasing Higgs mass. It is therefore likely that only weak 
constraints on the Yukawa coupling can be obtained if the Higgs is
considerably larger than the current LEP limit. 

Recently, a first study of the uncertainties in the measurements of
the luminosity spectrum (excluding the effect of linac energy spread)
has been carried out by Boogert \cite{Boogert_talk:2004}.
The luminosity spectrum is typically expressed in terms of the scaled
c.m. energy $x = \sqrt{s^{\prime}}/\sqrt{s}$, where
$\sqrt{s^{\prime}}$ is the event c.m. energy after beamstrahlung,
  ISR and beam spread and $\sqrt{s}$ is the nominal pre-collision
  energy. The cross-section after the inclusion of the luminosity
  spectrum is
\begin{equation}
\sigma^{\prime}_{tot}(s) = \int_{0}^{1} dx \: L(x) \sigma_{tot}(x^2 s)
\end{equation}
where $L(x)$ is a probability distribution representing the luminosity
spectrum. Unlike the analysis performed by Martinez and Miquel, a near perfect
detector was considered with a constant detector efficiency for
all scan points and only the cross-section information was used. The
theoretical cross-section smeared with the luminosity spectrum was
fitted with the same theoretical cross-section smeared by the
corresponding reconstructed luminosity spectrum. This fit was
performed for an 11 point threshold scan with a integrated luminosity
per scan point of $30$~fb$^{-1}$, with $M_{\rm 1S}$ and
$\alpha_{\rm s}(M_Z)$ as free parameters, while $\Gamma_{t}$ is
fixed. Systematic shifts of $\delta M_{\rm 1S}=M^{\rm fit}_{\rm
 1S}-M_{\rm 1S}=-44$~MeV and $\delta(\alpha_{\rm s})=\alpha^{\rm
 fit}_s-\alpha_{s}=-0.0016$, were observed, significantly larger than
the statistical errors obtained in the three parameter fit of Ref.
\cite{Martinez:2002st}.

The results of Boogert and similar studies must be taken with caution
as the luminosity spectrum and its reconstruction varies significantly
between different linear collider designs and the detail in which the
particle acceleration, focusing and collision dynamics are
simulated. The development of linked accelerator, collision dynamics,
hard scattering and detector reconstruction simulations will enable
more realistic determination of the expected systematic errors on the
top mass.

Finally, other observables such as the forward--backward asymmetry and top
three-momentum distribution,  discussed in
\Section~\ref{sec:topthresholddistributions}, must also be modified due to the
effects of the luminosity spectrum and included into a complete
analysis of the top threshold with realistic luminosity spectra.

\subsection{Theoretical status for $e^+e^-$ collisions}
\label{sec:topthresholdtheory}

With the excellent experimental prospect in view it is obvious that a
careful analysis and assessment of theoretical uncertainties in the
prediction of the total cross-section and various distributions 
is mandatory. Initially, a number of leading order\,\cite{Fadin:1988fn} and
next-order computations,\,
\cite{Strassler:1991nw,Jezabek:1992np,Sumino:1993ai,Murayama:1993mg,Harlander:1995ac,Harlander:1997vg,Peter:1998rk,Kwong:1991iy}   
were carried out. The latter relied basically on QCD-inspired
potential models that used phenomenological input from $\Upsilon$ and
charmonium data. As such they did not represent true first-principles
QCD calculations and there was no systematic way how the computations
could be consistently improved to include higher order radiative or
relativistic corrections. Moreover it was not clear, at a level of
precision of order 100~MeV, how the top quark mass appearing in 
these computations relates to a Lagrangian mass in QCD. 

\subsubsection{Fixed order approach}
\label{sec:topthresholdfixedorder}

The fixed order expansion of the non-relativistic heavy quark pair production
cross-section has the schematic form 
\begin{eqnarray}
 R \, = \, \frac{\sigma_{t\bar t}}{\sigma_{\mu^+\mu^-}}
 \, = \,
 v\,\sum\limits_k \left(\frac{\alpha_{\rm s}}{v}\right)^k
 \bigg\{1\,\mbox{(LO)}; \alpha_{\rm s}, v\,\mbox{(NLO)}; 
 \alpha_{\rm s}^2, \alpha_{\rm s} v, v^2\,\mbox{(NNLO)}\bigg\}
 \,,
 \label{eq:RNNLOorders}
\end{eqnarray}
where $v$ is the top velocity and
where the indicated terms are of leading (LO), next-to-leading
(NLO), and next-to-next-to-leading order (NNLO). 
The LO terms proportional to $(\alpha_{\rm s}/v)^n$ are the well-known Coulomb
singularities and arise from the iteration of the interaction between
the $t\bar t$ pair created by the QCD Coulomb potential.
Parametrically, one counts $\alpha_{\rm s}/v$ of order $1$ for the entire
threshold regime.
The systematic computation of the expansion in this scheme became possible
after adopting the concepts of effective theories in the framework
of NRQCD\,\cite{Caswell:1986ui,Bodwin:1995jh}. Consistent fixed order computations
up to NNLO were then worked out in 
Refs.~\cite{Hoang:1997sj,Hoang:1997ui,Czarnecki:1998vz,Beneke:1998jm,Hoang:1998vs,Hoang:1998xf,Melnikov:1998pr,Yakovlev:1998ke,Beneke:1999qg,Nagano:1999nw,Hoang:1999zc,Penin:1998ik,Penin:1998mx,Yakovlev:2000pv}.
The expansion in the fixed order scheme is obtained by identifying the  
corresponding contributions in the cross-section from the various momentum
regions. Since the contribution from each region has an unambiguous
scaling in the top velocity the expansion can be carried out
systematically. The contributions of the same order can then be summed
by using the Schr\"odinger equation and time-independent perturbation
theory. 

The results obtained in this scheme were not just some new higher
order corrections, but led to a number of surprising and important
insights. Although the methods and techniques used to perform the
computations differed among the groups, it was generally found that
the NNLO corrections to the location where the cross-section
rises and to the height of the cross-section were found to be much larger
than expected from the results at NLO. The large corrections to  
the location of the rise obtained
initially\,\cite{Hoang:1998xf,Melnikov:1998pr,Yakovlev:1998ke} were found 
to be an artifact of the use of the on-shell pole mass definition and
it was realized that  the top pole mass cannot be extracted with an
uncertainty smaller than ${\cal{O}}(\Lambda_{\rm QCD})$ from
non-relativistic heavy quark--antiquark
systems\,\cite{Aglietti:1995tg,Hoang:1998nz,Beneke:1998rk}. (See
\Section~\ref{sec:mbmc1S}.) 
Subsequently, carefully designed threshold mass definitions
(\Section~\ref{sec:massdefs}) were proposed to allow for a stable
extraction of the top quark mass parameter
\cite{Beneke:1998rk,Hoang:1998ng,Hoang:1998hm,Hoang:1999zc,Yakovlev:2000pv}.

\begin{figure}[t] % figthresholdmasses
\begin{center}
\includegraphics[width=.43\linewidth]{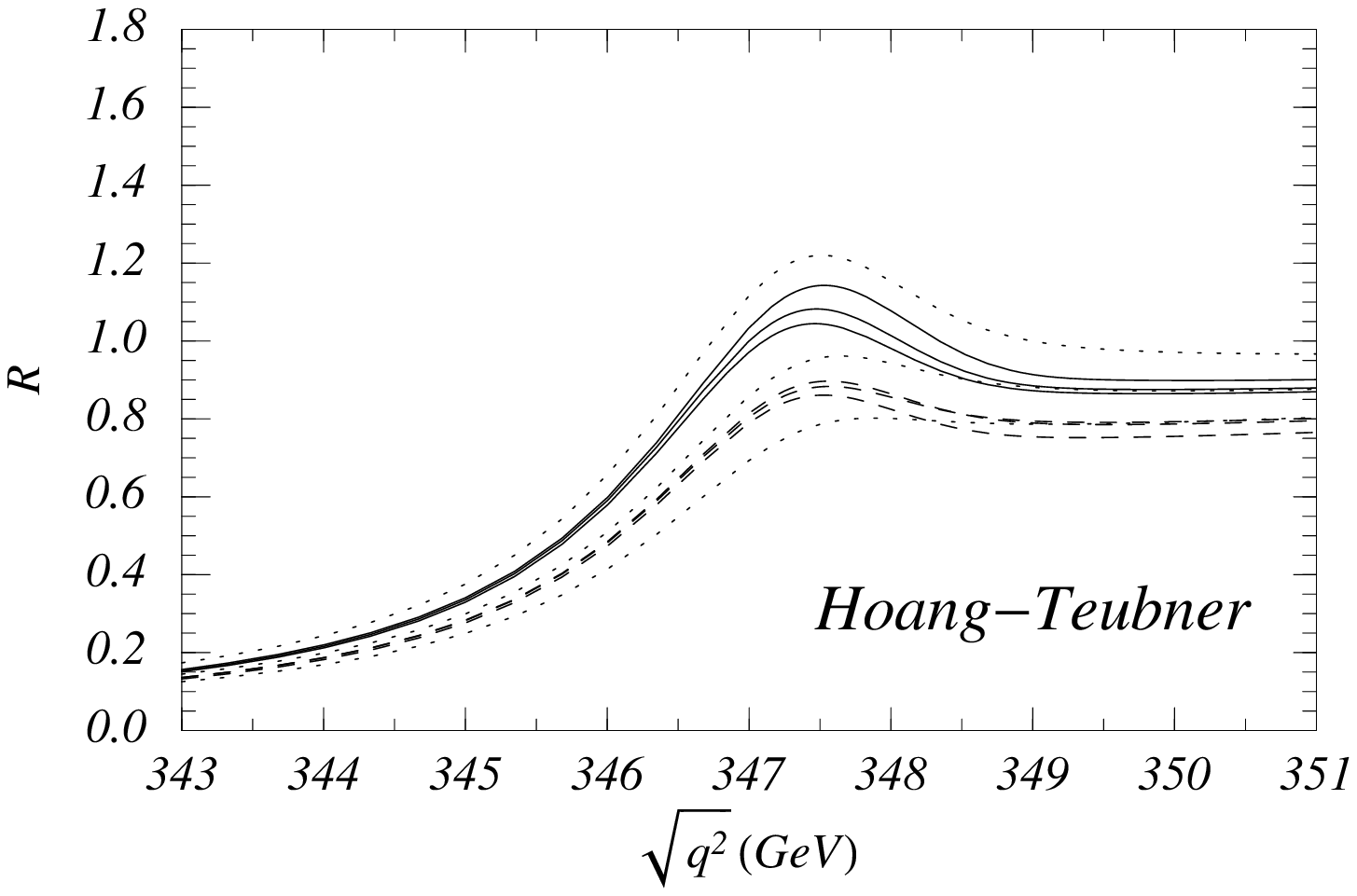}
\qquad
\includegraphics[width=.43\linewidth]{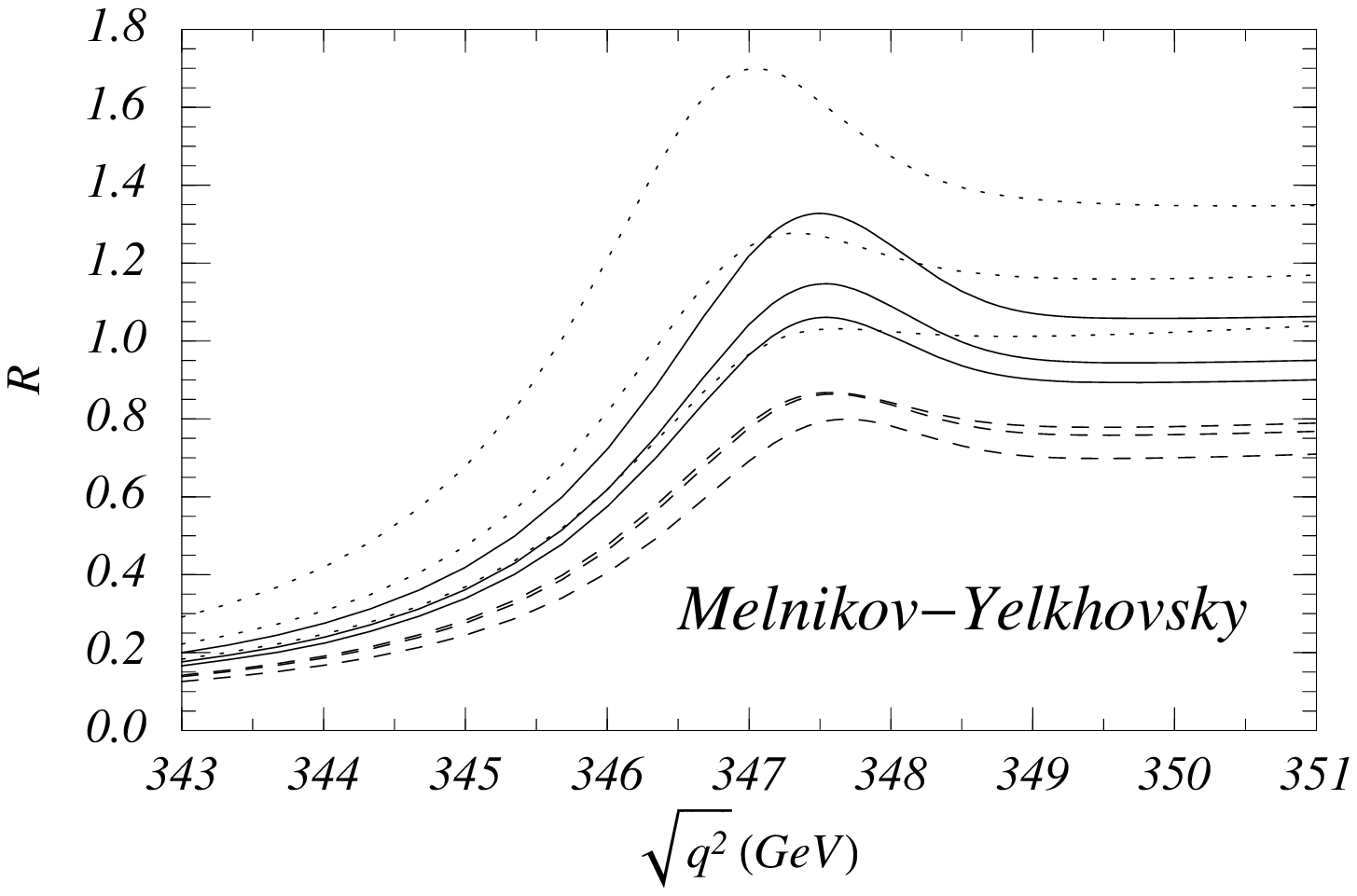}\\[4mm]
\includegraphics[width=.43\linewidth]{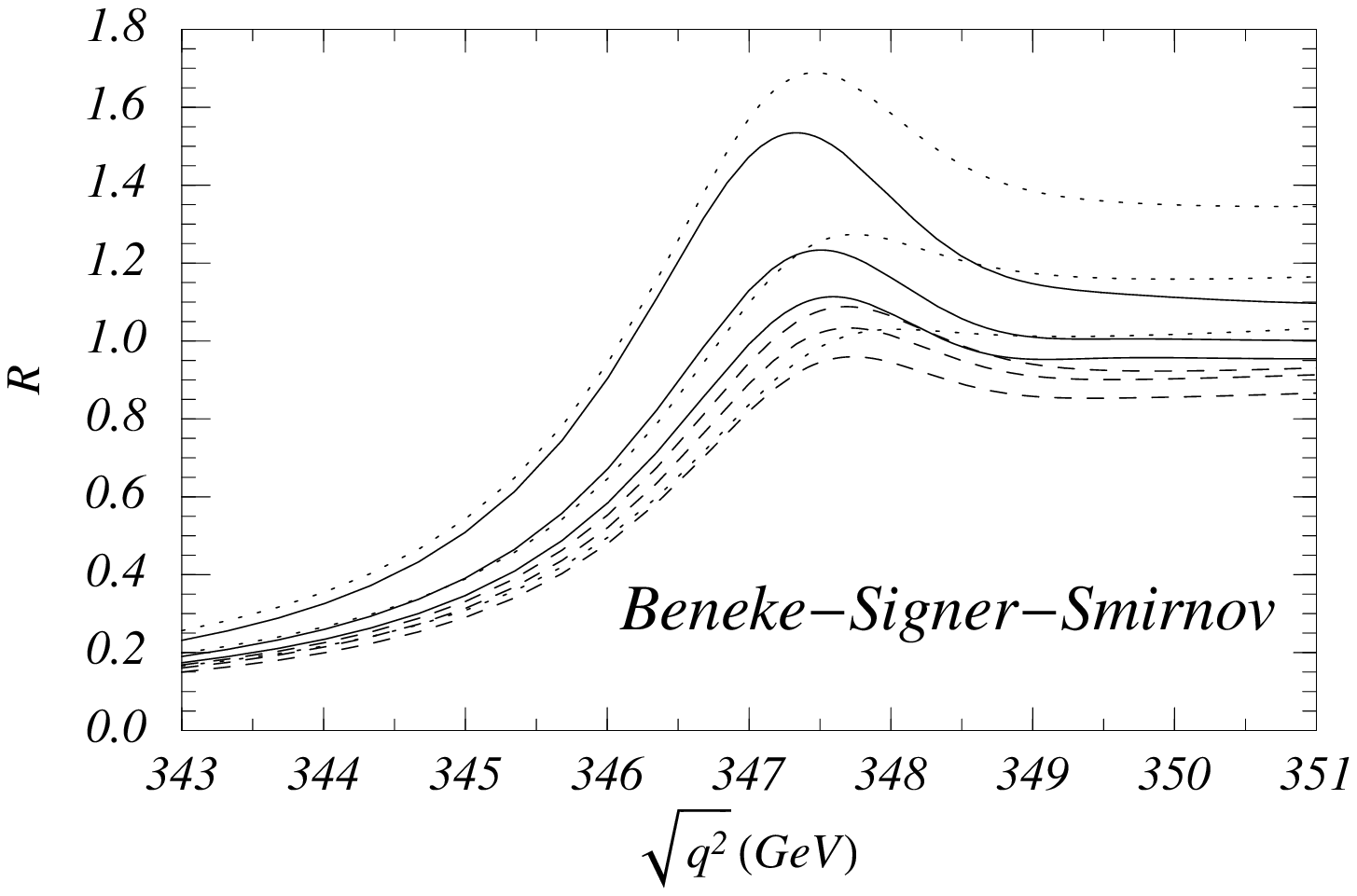}
\qquad
\includegraphics[width=.43\linewidth]{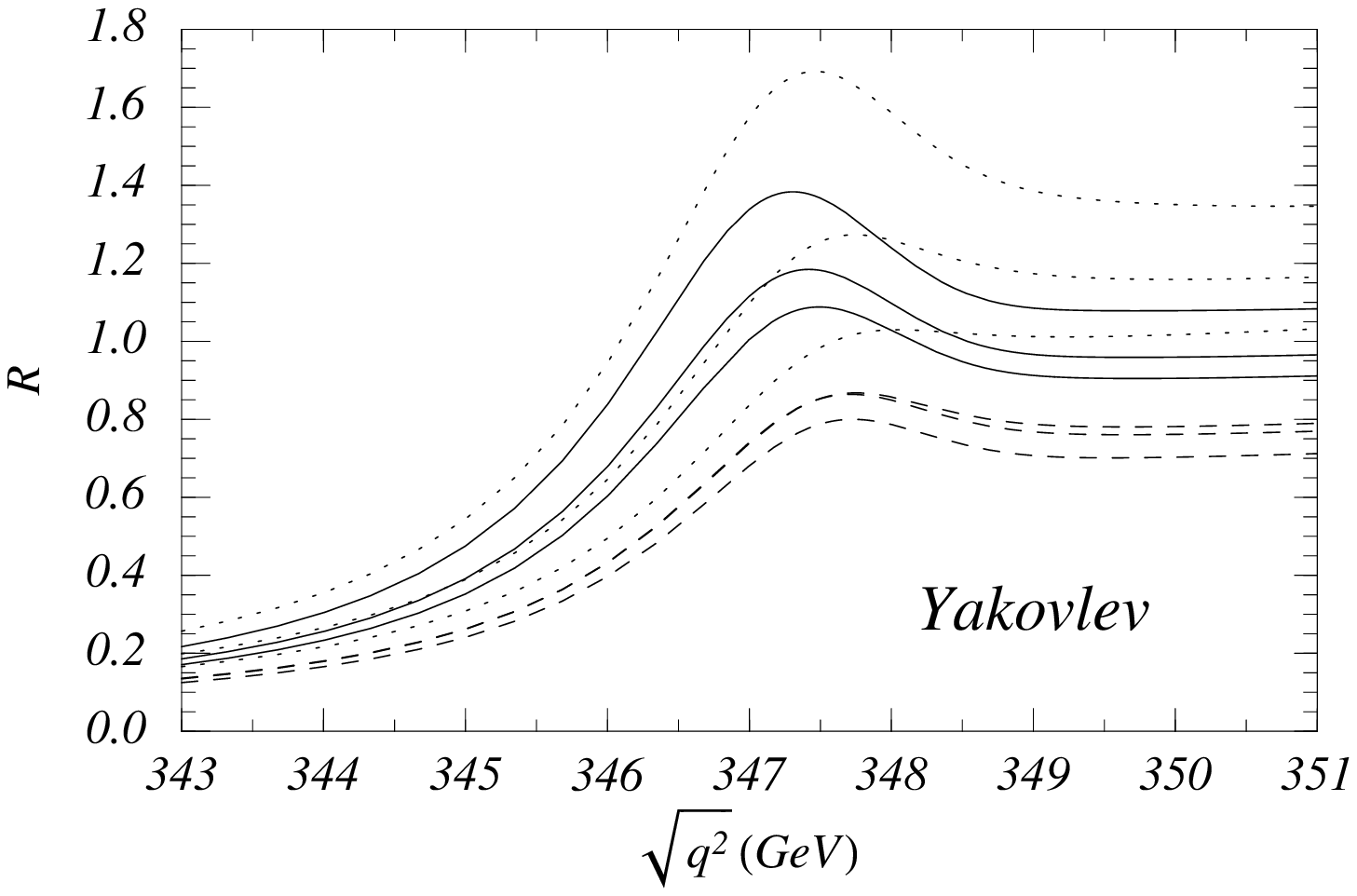}
\end{center}
\caption[Total normalised photon-induced $t\bar t$ cross-section  $R$ at
         the Linear Collider]
        {The total normalised photon-induced $t\bar t$ cross-section
         $R$ at the Linear Collider versus the c.m. energy in the
         threshold regime at LO (dotted curves), NLO (dashed) and NNLO
         (solid) for $\alpha_{\rm s}(M_Z)=0.119$, $\Gamma_t=1.43$~GeV
         and $\mu=15$, $30$, $60$~GeV. Hoang--Teubner used the 1S mass
         $M_{1S}^{}=173.68$~GeV, Melnikov--Yelkhovsky the kinetic
         $M_{\rm kin}(15~\mbox{GeV})=173.10$~GeV, and
         Beneke--Signer--Smirnov and Yakovlev the PS mass $M_{\rm
         PS}(20~\mbox{GeV})=173.30$~GeV.  The plots were given in
         Ref.~\cite{Hoang:2000yr} from results provided by
         Hoang--Teubner, Melnikov--Yelkhovsky, Beneke--Signer--Smirnov
         and Yakovlev. The effects of the luminosity spectrum are not
         taken into account.}
\label{fig:thresholdmasses}   
\end{figure}

In Ref.~\cite{Hoang:2000yr} the results for the normalised total
photon-induced cross-section $Q_t^2 R^v$ of a number of groups were
compiled and compared numerically in detail using an equivalent set of
parameters to analyse the scheme-dependence of the fixed order
approach. Since the axial-vector current contributions are only a
small correction at the percent level it is justified to consider only
the photon-induced cross-section. \Figure[b]~\ref{fig:thresholdmasses}
shows the results obtained in Ref.~\cite{Hoang:2000yr} in different
threshold mass schemes where the respective values of the threshold
masses were obtained from the top \ms mass value $\overline
m_t(\overline m_t)=165$~GeV as a reference point.
It was concluded that the perturbative 
uncertainty in the determination of threshold masses from the peak
position (i..e\ when beam smearing effects are neglected), is between
$50$ and $80$~MeV. It was also pointed out that the \ms mass
can be determined with a comparable perturbative precision, only if
$\alpha_{\rm s}(M_Z)$ is known with an uncertainty of around $0.001$. This
restriction arises from the relatively large order $\alpha_{\rm s}$ correction
in the relation between threshold masses and \ms mass.
For example, for a given measurement of the 1S mass, let's say
$M_{\rm 1S}^{}=175$~GeV$\,\pm\,\delta M_{\rm 1S}^{}$, and
$\alpha_{\rm s}(M_Z)=0.118\pm x\,0.001$ the result for $\overline
m_t(\overline m_t)$ reads\,\cite{Hoang:1999zc}
\begin{eqnarray}
\overline m_t(\overline m_t)  & = & 
\bigg[\,
175 - 7.58\,  - \, 
0.96\,  - \, 
0.23\,
%\nonumber
%\\[3mm] & & \hspace{3cm}
\, \pm \, \delta M_{1S}^{}
\, \pm \,  x\,0.07 
\,\bigg]~\mbox{GeV}
\,\qquad
\label{MSmassestimate}
\end{eqnarray} 
where the first four numbers represent the perturbative series up to
NNLO (three-loops). For a discussion of the behaviour of perturbative
corrections at the next order see
Refs.~\cite{Penin:2002zv,Kiyo:2002rr}.  The results in
\Figure~\ref{fig:thresholdmasses} also show a large uncertainty in the
normalisation of the cross-section. This uncertainty is particularly
puzzling, because there is no obvious physical reason for its
existence. At present it does not seem to be related to
renormalon-type higher order corrections, although it has also been
speculated that the large size of the corrections could have some
infrared origin that might be cured by accounting for off-shell
effects of the toponium dynamics~\cite{Kiyo:2001zm}. On the other
hand, it has been shown in Ref.~\cite{Kniehl:1999mx,Kniehl:2002yv}
that the dominant N$^3$LO logarithmic corrections to the cross-section
of (relative) order $\alpha_{\rm s}^3 \ln^2v$ and $\alpha_{\rm s}^3
\ln v$ are not small.  It was estimated in Ref.~\cite{Hoang:2000yr}
that the present theoretical relative uncertainty in the normalisation
of the cross-section in the fixed order approach is at least at the
level of $20$\%.  This seems to jeopardise precise measurements of the
top width, the top quark coupling to gluons and the Higgs boson.  At
present the only way to possibly improve the situation in the fixed
order approach seems to be the determination of N$^3$LO or even higher
order corrections in order to learn more about the structure of the
perturbative series.

\subsubsection{Renormalisation group improved approach}
\label{sec:topthresholdrgi}

Apart from the Coulomb singularities proportional to powers of
$\alpha_{\rm s}/v$, which have to be summed according to the scheme shown in
\Eq~(\ref{eq:RNNLOorders}) there is one additional source of potentially
large higher corrections coming from logarithms of the top mass
$m_t=175$~GeV, the average top three-momentum ${\bmp}_t\sim 25$~GeV and
the average kinetic energy $E_t\sim 4$~GeV. Terms such as
$\alpha_{\rm s}(m_t)\ln(m_t^2/E_t^2)\sim 0.8$ can spoil the fixed order
expansion, as discussed before, and should to be summed to all
orders. The scheme where all 
such logarithmic terms are summed consistently has the form  
\begin{eqnarray}
 R & = & \frac{\sigma_{t\bar t}}{\sigma_{\mu^+\mu^-}}
 \, = \,
 v\,\sum\limits_k \left(\frac{\alpha_{\rm s}}{v}\right)^k
 \sum\limits_i \left(\alpha_{\rm s}\ln v \right)^i \times
\nonumber \\[2mm]
& &
 \times \bigg\{1\,\mbox{(LL)}; \alpha_{\rm s}, v\,\mbox{(NLL)}; 
 \alpha_{\rm s}^2, \alpha_{\rm s} v, v^2\,\mbox{(NNLL)}\bigg\}
 \,,
 \label{eq:RNNLLorders}
\end{eqnarray}
where the indicated terms are of
leading-logarithmic (LL), next-to-leading-logarithmic (NLL), and
next-to-next-to-leading-logarithmic order (NNLL). The expansion in
\Eq~(\ref{eq:RNNLLorders}) is called ``renormalisation group improved''
perturbation theory. 
Since the fixed-order approach relies on the identification of the 
contributions from the hard, soft, potential and ultra soft momentum
regions in full QCD diagrams, it only accounts for the anomalous
dimension associated with the running of the strong coupling and cannot 
be used to carry out renormalisation group improved computations for
the top threshold cross-section. Moreover, in fixed order expansions
it is not clear a priori which scale to use for the couplings in the 
highest computed order.

To achieve the expansion scheme in \Eq~(\ref{eq:RNNLLorders}) one
needs to apply a full effective theory description where all
resonating degrees of freedom responsible for the $t\bar t$ low energy
dynamics are implemented as fields of an effective theory and where
all contributions from off-shell effects are integrated out.  The
logarithmic terms mentioned above are associated with UV-divergences
in the effective theory and can be summed up after renormalisation of
the effective theory operators and solution of the resulting
renormalisation group equations. In this formulation all logarithmic
terms are contained in the Wilson coefficients of the effective theory
operators after they have been run to the low energy scale. On the
other hand, all matrix elements are free of any large logarithmic
contributions.  Since the logarithmic terms mentioned above involve
only perturbative scales much larger than $\Lambda_{\rm QCD}$, the
effective field theory description for the $t\bar t$ total
cross-section at threshold allows for purely perturbative computations
of the Wilson coefficients as well as of the low-energy matrix
elements.

Renormalisation group improved computations that can be applied to the total
top threshold cross-section have been carried out in pNRQCD and
vNRQCD. In addition to the operators in the respective effective
Lagrangians also external currents need to be defined that describe the 
production of the top quark pair for non-relativistic momenta through the
electroweak interactions. 
The vector (${}^3S_1$) currents relevant
at NNLL order have the form ${\bf J}^v_{\sitbf{p}}= c_1 \O{p}{1} + c_2
\O{p}{2}$, where~\cite{Hoang:2001mm}
\begin{eqnarray}\label{Ov}
 \O{p}{1} & = & {\psixp{p}}^\dagger\, \bsigma(i\sigma_2)\, {\chip{-p}^*} \,, 
   \\[2mm]
 \O{p}{2} & = & \frac{1}{m^2}\, {\psixp{p}}^\dagger\, 
    \bmp^2\bsigma (i\sigma_2)\, {\chip{-p}^*} \,, \nn
\end{eqnarray} 
and the relevant axial-vector (${}^3P_1$) current is ${\bf J}^a_{\sitbf{p}}=
c_3 \O{p}{3} $, where
\begin{eqnarray}\label{Oa}
 \O{p}{3} & = & \frac{-i}{2m}\, {\psixp{p}}^\dagger\, 
      [\,\bsigma,\bsigma\cdot\bmp\,]\,(i\sigma_2)\,
   {\chip{-p}^*} \,. 
\end{eqnarray}  
The corresponding annihilation currents ${\O{p}{i}^\dagger}$ are obtained by
complex conjugation. The currents $\O{p}{2,3}$ contribute only at the NNLL
level.  The NNLL total cross-section is then written in the form
\begin{eqnarray}
\sigma_{t\bar t} & = &
c_1^2 \,\mbox{Im}[\,{\cal A}_{11}(\sqrt{s})\,]
+ c_1 c_2 \,\mbox{Im}[\,{\cal A}_{12}(\sqrt{s})+\mbox{h.c.}\,]
+c_3^2\,\mbox{Im}[\,{\cal A}_{33}(\sqrt{s})\,]
\,,
\end{eqnarray}
where the $c_i$ are the Wilson coefficients of the currents and the
${\cal A}_{ij}$ are the Fourier transforms of time ordered products of
the currents ${\O{p}{i}}$ and ${\O{p}{i}^\dagger}$.  The first term
contributes at the LL level, while the second and third terms
contribute at NNLL order only.  The ${\cal A}_{ij}$ are obtained from
the zero-distance Green function
$G(\bmr=0,\bmr^\prime=0,\sqrt{s}-2m_t)$ of a two-body Schr\"odinger
equation. At present, analyses of the total top pair cross-section at
threshold at NNLL order within renormalisation group improved
perturbation theory have only been presented in the framework of
vNRQCD~\cite{Hoang:2000ib,Hoang:2001mm} (see
\Section~\ref{sec:vnrqcd}). The Wilson coefficients of the potentials
in the Schr\"odinger equation and of the $v^2$-suppressed currents are
known at NNLL order for the
cross-section~\cite{Manohar:1999xd,Manohar:2000hj,Manohar:2000kr,Pineda:2000gz,Hoang:2001rr,Pineda:2001ra,Pineda:2001et,Hoang:2002yy},
whereas the Wilson coefficient $c_1$ of the leading order vector
current is only fully known at NLL
order~\cite{Luke:1999kz,Manohar:2000kr,Pineda:2001et,Hoang:2002yy}.
Recently, the NNLL order non-mixing contributions to the anomalous
dimension of $c_1$ coming from genuine three-loop vertex diagrams in
vNRQCD were computed in Ref.~\cite{Hoang:2003ns}. These corrections
were found to be comparable to the NLL contributions. The NNLL order
corrections associated to the higher order running of the couplings
that mix into the NLL order anomalous dimension of $c_1$ are still
unknown.
\begin{figure}[t] % figtopplots
\begin{center}
\iffalse
\leavevmode
\epsfxsize=3.8cm
\leavevmode
\epsffile[230 585 428 710]{figb}
\vskip  2.6cm
\fi
\includegraphics[width=85mm]{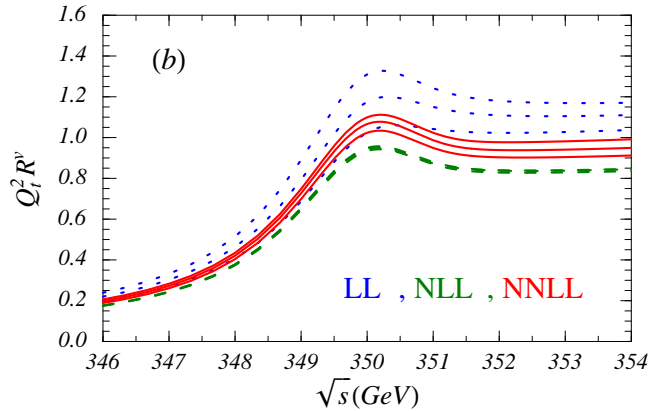}
\end{center}
\caption[The photon-induced top pair threshold cross-section for  
         $M^{\rm 1S}=175$\,GeV, $\alpha_{\rm s}(M_z)=0.118$ and
         $\Gamma_t=1.43$\,GeV]
        {The photon-induced top pair threshold cross-section for
         $M^{\rm 1S}=175$\,GeV, $\alpha_{\rm s}(M_z)=0.118$ and
         $\Gamma_t=1.43$\,GeV in renormalisation group improved
         perturbation theory in vNRQCD at LL (dotted lines), NLL
         (dashed lines) and NNLL (solid lines) order.  For each order
         curves are plotted for the renormalisation parameter
         $\nu=0.15$, $0.20$, and $0.3$.  The effects of the luminosity
         spectrum are not included.  }
\label{fig:topplots} 
\end{figure}
\Figure[b]~\ref{fig:topplots} shows the photon-induced total cross-section in 
renormalisation group improved perturbation theory obtained in
Ref.~\cite{Hoang:2003xg}. 
The curves show the LL (dotted blue lines), NLL (dashed green lines) and NNLL
(solid red lines) cross-section for the vNRQCD scaling parameter $\nu=0.15,
0.2$ and $0.3$, $\alpha_{\rm s}(M_z)=0.118$, $\Gamma_t=1.43$ and 
$m_{\rm 1S}=175$~GeV. Compared to the fixed order results discussed
above there is an improvement in the convergence and the remaining
scale variation, but the NNLL order curves are still shifted by a positive
correction similar to the fixed order predictions (see
\Figure~\ref{fig:thresholdmasses}). In Ref.~\cite{Hoang:2003xg} the
present theoretical uncertainty in the normalisation of the NNLL order
cross-section was estimated as $\delta\sigma_{t\bar t}/\sigma_{t\bar
t}=6\%$. Here, the determination of the still missing mixing
contributions to the NNLL anomalous dimension of $c_1$ is needed to
get a more complete picture on the theoretical uncertainties. The
spin-dependent mixing contributions have been determined in
Ref.~\cite{Penin:2004ay}

\subsubsection{Treatment of unstable particles}
\label{sec:topthresholdunstable}

For the theoretical description of the top threshold dynamics it is
not possible to treat the top quark as a stable particle because
within the Standard Model the top quark is expected to have a width of
approximately $\Gamma_t\approx 1.5$~GeV, which is comparable to the
average kinetic energy of the produced top quarks. For the total
cross-section, as long as one is not interested in any differential
information of the decay process, the finite width effects can be
implemented as modifications of the effective theory Wilson
coefficients that arise when the effective theory is matched to the
Standard Model. These modifications lead to additional real and
imaginary contributions. Such imaginary contributions are a well known
concept in quantum mechanics of inelastic processes where particle
decay and absorption are described by potentials and coefficients
carrying complex coefficients.  Due to the unitarity of the Standard
Model the optical theorem used in
\Eq~(\ref{eq:fullR}) still holds in the effective theory.

At leading order in the non-relativistic expansion, due to gauge
invariance, the only
modification in the effective theory matching conditions caused 
by electroweak effects are related to an additional imaginary mass
term proportional to the total on-shell top width\,\cite{Fadin:1987wz},
\begin{eqnarray}
\delta{\cal L} & = & 
\sum_{\bmp} \psixp{p}^\dagger i\,\frac{\Gamma_t}{2}\psixp{p} +
\sum_{\bmp} \chip{p}^\dagger i\,\frac{\Gamma_t}{2} \chip{p}
\,.
\end{eqnarray} 
Since one has to apply the counting rule $\Gamma_t/m_t\sim\alpha_{\rm s}^2\sim
v^2$~\cite{Hoang:1999zc} this leads to 
\begin{equation}
\frac{i}{k^0-\frac{\bmp^2}{2m_t}+\delta m_t+\frac{i}{2}\Gamma_t}
\end{equation}
for the form of the top quark propagator, where $\delta m_t$ is related to
the top mass scheme that is used for the computations. For the total
cross-section this is equivalent to a shift of the c.m.\ energy into
the positive complex plane, $\sqrt{s}\, \to \, \sqrt{s} +
i\Gamma_t$.~\cite{Fadin:1987wz} 
For the total cross-section the results of
Refs.~\cite{Fadin:1987wz,Melnikov:1994np,Fadin:1994dz,Beenakker:1999ya}
show that this prescription remains valid even at
next-to-leading order due to cancellation of QCD non-factorisable
corrections connecting the top quark production and decay.
The full set of electroweak corrections to the effective theory
matching conditions at next-to-next-to-leading
order is currently unknown, although some contributions have been
identified at this
order.~\cite{Guth:1992ab,Modritsch:1994hv,Kummer:1995id,Hoang:1999zc,Hollik:1998md,Kolodziej:2001xe,Biernacik:2002tt} 
These results indicate that these corrections are at the level of a
few percent. 

For differential observables a number of leading order analyses have
been made in the past in order to assess their physics impact (see
\Section~\ref{sec:topthresholddistributions}).  
However, a systematic description of electroweak effects in
differential observables, which would require a consistent theory
describing unstable top and antitop quarks, does not exist at this time.
First steps towards  such a theory have been
undertaken in Ref.~\cite{Beneke:2003xh,Beneke:2004km}.

\subsubsection{Non-perturbative effects}
\label{sec:topthresholdnon-perturbative}

As discussed above, all scales in the effective field theory
calculation are large compared to $\Lambda_{\rm QCD}$, and $t\bar
t$ production can be calculated in perturbative QCD.  Nevertheless, as
we are aiming at a very high accuracy, the question about the size of
possible residual non-perturbative effects is legitimate.  Such effects can
be estimated through the interaction of the non-relativistic $t\bar t$ system
with long-range fluctuations of the gluon field, see 
\Chapter~\ref{chapter:spectroscopy}, \Section~\ref{sec:spnrqcd}.
In the approximation of a constant (w.r.t. the size and
lifetime of the $t\bar t$ state) chromoelectric field these
corrections can be parametrised in form of the gluon condensate
$ \langle 0 | \alpha_{\rm s} G_{\mu\nu} G^{\mu\nu} | 0 \rangle $.  In
Ref.~\cite{Fadin:1991jh} an explicit formula was derived, and it was found that
the corrections due to the gluon-condensate are strongly energy
dependent but small and decreasing with increasing $m_t$, $\Gamma_t$.
For a realistic top mass and width they are completely negligible (of
the order $10^{-4}$ for the cross-section) compared to the uncertainties
from the perturbative treatment, see also \cite{Hoang:2002ae}.  This
result also agrees with studies of the influence of the long-distance
part of phenomenological QCD potentials.~\cite{Strassler:1991nw,
  ttdipl}  There, the effect from a linearly rising potential (or
even more drastic deviations from the coulombic form) was shown to be
irrelevant, a reflection of the extremely short lifetime of the
$t\bar t$ system.

\subsection{Studies of distributions and polarised beams}
\label{sec:topthresholddistributions}

With the expected high luminosity at TESLA it is clear that, apart
from the scan of the total cross-section through the threshold regime,
detailed measurements of distributions will become feasible.  Such
distributions will help to disentangle correlations between $m_t$ and
$\alpha_{\rm s}$ and improve the accuracy of the determination of the
top quark mass.  Even more important, the measurement of distributions
will allow for detailed studies of the top quark's couplings (and
possible deviations from SM expectations) in production and decay of
$t\bar t$.  Detailed analyses have been carried out by the European
and Asian Linear Collider study groups in the past years, partly
before the NNLO (and NNLL) improved calculations for the total
cross-section became available.  However, important information may be
gained from those leading order (partly higher order improved)
analyses, and, eventually, higher order corrections should be
calculated also for observables beyond the total inclusive
cross-section.

In the following we will discuss the momentum distribution of top
quarks, measurements of the forward--backward asymmetry $A_{\rm FB}$
and the issue of top quark polarisation and polarised $e^-$ (and
possibly $e^+$) beams.

\subsubsection{Momentum distribution of top quarks}

\begin{figure}[t] % fig:momdis
\centering\includegraphics[width=.6\linewidth]{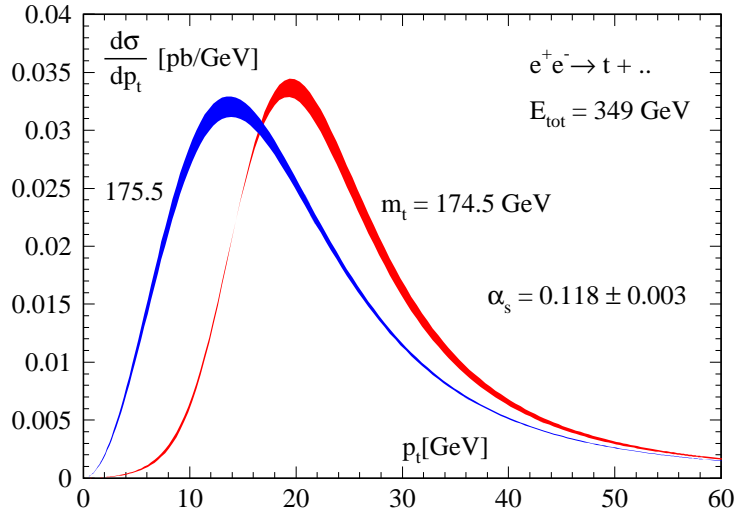}
\caption[Momentum distribution of the top quark for two different
         values of $m_t$]
        {Momentum distribution of the top quark for two different
         values of $m_t$ at the same fixed c.m. energy $E_{\rm tot} =
         349$~GeV.  The bands arise from variation of $\alpha_{\rm s}$
         as indicated (figure taken from Ref.~\cite{Accomando:1998wt}.)}
\label{fig:momdis} 
\end{figure}

As is clear from the picture of non-relativistic, quasi-bound top
quarks having a sizeable decay width, even at fixed c.m. 
energy $\sqrt{s}$, $t$ and $\bar 
t$ are not produced with a well defined momentum but with a broad
distribution.  In the picture of non-relativistic Quantum Mechanics
this momentum distribution is proportional to the square of the
(momentum-space) wave function and can be measured by reconstruction
of the top decays.
\Figure[b]~\ref{fig:momdis} shows theoretical
predictions for typical momentum distributions for two different
values of $m_t$ but the same fixed c.m. energy.  The shaded
bands come from the variation of $\alpha_{\rm s} = 0.118 \pm 0.003$.
It is clear from \Figure~\ref{fig:momdis} that the peak position of the
momentum distribution is much more sensitive to $m_t$ than to
$\alpha_{\rm s}$ which mainly influences the normalisation.

\subsubsection{The forward--backward asymmetry $A_{\rm FB}$}

Top pair production through a virtual $Z$ boson leads to a small $P$
wave in addition to the dominant S-wave contribution.  Interference
of S and P-wave results in a forward--backward asymmetry defined as
\begin{equation}
A_{\rm FB} = \frac{1}{\sigma_{\rm tot}} 
\left( \int_0^1 {\rm d}\,\cos\Theta - \int_{-1}^0 {\rm d}\,\cos\Theta
\right) \frac{{\rm d}\sigma}{{\rm d}\,\cos\Theta}
\end{equation}
which is of order 10\% but energy dependent with a minimum of about
5\% close to the energy of the $1S$ peak (see
\Figure~\ref{fig:expscan} in
\Section~\ref{sec:topthresholdexperiment}).  As $A_{\rm FB}$ comes
from the {\em overlap} of the (smeared out) $1S$ and $1P$ would-be
resonances, it is sensitive to the top quark width $\Gamma_t$ and, to
a lesser extent, to $\alpha_{\rm s}$.

In experimental simulations, the momentum distribution and the
forward--backward asymmetry have been used in addition to the total
cross-section in order to add valuable information and to 
increase the accuracy of the determination of $m_t$, $\alpha_{\rm s}$,
$\Gamma_t$ and possibly even the top Yukawa coupling $y_t$, see the
discussion in \Section~\ref{sec:topthresholdexperiment} above.

\subsubsection{Polarisation}

Near threshold, S-wave production dominates, and the spins of $t$,
$\bar t$ are aligned with the electron beam direction.  Therefore,
even for unpolarised incoming $e^+$, $e^-$ beams the longitudinal top
quark polarisation is still $\sim 40\%$.  With polarised beams one can 
obtain a sample of highly polarised top quarks near
threshold, and with a tunable polarisation of the $e^-$ beam the top
quark polarisation could be selected.  Quantitatively, for a realistic
longitudinal $e^-$ polarisation, $P_{e^-} = +80\%$ ($-80\%$), and
unpolarised positrons, $P_{e^+} = 0$, one would arrive at $+60\%$
($-90\%$) polarised top quarks!  This level would be further enhanced
if the $e^+$ beam would be polarised as well.

Because of the large top quark width
this polarisation is transferred to the top decay products practically
without hadronization effects.  As in addition the top spin can be
well measured through the angular distribution of $W$ decay leptons,
$t\bar t$ production near threshold is also a very interesting field
for spin physics.

While the top quarks are mainly polarised longitudinally, polarisation
transverse and normal to the production plane is induced through S--P
wave interference effects and the QCD threshold dynamics, see
\cite{Harlander:1995ac,Harlander:1997vg,Peter:1998rk} for details.
It is interesting to note that normal polarisation is sensitive to the
(electric, weak and strong) EDM of the top quark and could, if measured to be
larger than predicted, signal CP violation beyond the SM.
With tunable $e^{\pm}$ polarisations and measurements at several 
c.m. energies, the different effects from the top quarks
couplings to $\gamma$, $Z$ and $g$ could be
disentangled~\cite{Jezabek:2000gr}.  Such analyses in the threshold region
would complement top coupling measurements in the continuum at larger
c.m. energies and can have an unexpectedly high sensitivity due to
the tunable polarisation.

\subsubsection{Rescattering corrections}

Since the typical energy transfer at threshold is comparable to the
top quark width there is in general no factorisation between
production and decay of $t$ and $\bar t$. In principle one would
have to calculate $e^+ e^- \to 
W^+\,b\,W^-\,\bar b$, including non-resonant backgrounds. However,
the non-factorisable corrections due to gluonic cross-talk between $t
\leftrightarrow \bar b$, $\bar t \leftrightarrow b$ and $b
\leftrightarrow \bar b$ are suppressed in the
total cross-section
as mentioned
above \cite{Fadin:1987wz,Melnikov:1994np,Fadin:1994dz,Beenakker:1999ya,Harlander:1997vg,Peter:1998rk}.
The situation is different for exclusive observables which are in general
affected by final state corrections at NLO.  For the momentum distribution,
$A_{\rm FB}$ and the top quark polarisation there exist results for
these rescattering corrections in the non-relativistic approximation,
see \cite{Harlander:1997vg,Peter:1998rk}.  They are of the expected parametric size
(order $\alpha_{\rm s}$) and slightly reduce the momentum of the top
quarks.  Depending on the c.m. energy and the $e^{\pm}$
polarisation, they also lead to a sizeable change of $A_{\rm FB}$.
Similarly, the different components of the top quark
polarisation can be strongly affected.  Therefore, realistic
experimental studies should take the rescattering corrections into account.  
Nevertheless, observables have been constructed which are independent
of the $t\bar t$ production dynamics and rescattering corrections
and only probe the decay of the polarised $t$, $\bar
t$, even in the presence of anomalous decay
vertices~\cite{Harlander:1997vg,Peter:1998rk,Sumino:1997ve,Sumino:1998zt}.

\subsection{Future Opportunities}
\label{sec:top_future}

The study of the problem of top quark pair production at threshold
has, starting with the first pioneering works about 15 years ago, 
led to impressive results, including the recent development of
effective field theories for heavy quarkonia.  Nevertheless, many
problems remain to be solved.  Among them are, as discussed above, the
complete computation of the NNLL corrections (renormalisation group
improvement) for the total cross-section.  Complementary
information about the behaviour of the perturbative expansion and
independent cross-checks will also be gained through further advances in
fixed-order calculations beyond NNLO. Such results would also contribute
to the matching conditions of a NNNLL renormalisation group improved
computation of the cross-section. Some of the N$^3$LO
contributions are already known~\cite{Kniehl:1999mx,Kniehl:2002yv}, and more
is to be expected as the technology to carry out such computations is in
principle available.  

On the conceptual side, a consistent treatment of
electroweak corrections including the instability of the top quark and
interferences with non-resonant final states is
still missing.  These effects should be more important for differential
cross-sections, but a relevant impact on the total cross-section
cannot be excluded.  Closely connected to this issue is the problem of
rescattering corrections.  NLO calculations have shown their relevance
for distributions, but a consistent treatment within the effective
field theory context will require further breakthroughs in the treatment of
unstable particles.

With the Tevatron Run II and at the future LHC, hadroproduction of $t\bar t$
is growing out of the discovery era and entering the phase of precision
measurements. Consequently top threshold dynamics may also play a role there. 
Similarly, the production process $\gamma\gamma \to t\bar t$
(\eg through the option of laser backscattering at a future $e^+e^-$
collider) is not explored with the same accuracy as the production in
$e^+e^-$ annihilation~\cite{Penin:1998mx,Czarnecki:2001gi}.
The formulation of the threshold dynamics in the framework of
effective field theories may also be suitable to calculate the
threshold production of coloured particles in extended models, like
\eg squark pairs in supersymmetry, systematically and with higher precision.

Hopefully, in a not too far future from now, a future $e^+ e^-$ Linear
Collider will be in operation.  To make best use of the anticipated
$t\bar t$ threshold data, fully realistic simulations including
experimental cuts are needed. This requires the theoretical treatment
of differential distributions, width and rescattering effects as
discussed above, but also the simulation of detector and beam effects
as discussed in \Section~\ref{sec:topthresholdexperiment}.  In this
respect further collaboration between experiment and theory will be
crucial to extend the existing simulations which were performed within
the regional study groups preparing for the next $e^+ e^-$ Linear
Collider.

%\bibliography{chapter6}
\hyphenation{Post-Script Sprin-ger}

%\end{document}

\BLKP

%10/12/2004

%\documentclass[11pt]{cernrep} 
%\usepackage{graphicx} 
%\usepackage{here} 
%\usepackage{epsfig} 
%\usepackage{rotating}
%\include{rlfig} 

%\begin{document} 

%\centerline{\bf CHARM AND BEAUTY IN MEDIA }

\chapter{CHARM AND BEAUTY IN MEDIA }
\label{chapter:charm-beauty-in-media}

\noindent
{\it Conveners:} D.~Kharzeev, M.~P.~Lombardo, C.~Louren\c{c}o, M.~Rosati, H.~Satz \par\noindent
{{\it Authors:} S.~Datta, O.~Kaczmarek, F.~Karsch, D.~Kharzeev, S.~R.~Klein, V.~Laporta, 
M.~P.~Lombardo, C.~Louren\c{c}o, L.~Maiani, P.~Petreczky, 
F.~Piccinini, A.~D.~Polosa, L.~Ramello, R.~Rapp, V.~Riquer, M.~Rosati, 
H.~Satz, E.~Scomparin, R.~Vogt, F.~Zantow } 
\par\noindent 

\section[Introduction]{Introduction 
                       $\!$\footnote{Authors: D.~Kharzeev, 
                       M.~P.~Lombardo, C.~Louren\c{c}o, M.~Rosati, H.~Satz}}

Quarkonium in media is a topic which is central to the
ultrarelativistic heavy ion program.  In recent years this subject has
been among the focal points of discussion at meetings such as ``Quark
Matter''.  (``The Quark Matter'' series is traditionally the main
forum of the high energy heavy ion community).  Indeed, in these
collisions\,---\,``little bangs''\,---\,one hopes to recreate matter
as it was at the very beginning of the universe: a hot system with
deconfined quarks and gluons and no chiral symmetry breaking.
\shortpage

This fascinating possibility calls for a number of theoretical
and phenomenological studies. The QCD phase
diagram and the mechanisms of chiral symmetry restoration,
screening, and deconfinement at high temperature and baryon density
need to be understood.  A theory of the initial conditions must be developed 
and the equilibration of the plasma, if any, must be assessed
in real experiments.  It is also necessary to
identify the thermodynamical region which is being explored and to 
study nonequilibrium effects.  Finally, observables must be defined 
which provide physical signatures in real experiments. 

Quarkonium plays a very
important role in these phenomena.  Indeed,
quarkonium suppression was long ago suggested as a signal of
deconfinement \cite{Matsui:1986dk}. Due to
their small size, quarkonia can, in principle, survive the deconfinement phase 
transition. However, because of colour screening, no bound state can exist at 
temperatures $T>T_D$ when the screening radius, $1/\mu_D(T)$, becomes 
smaller than the typical bound-state size \cite{Matsui:1986dk}.
Later it was realized that dramatic changes in the gluon momentum 
distributions at the deconfinement phase transition result in a sharp 
increase in the quarkonium dissociation 
rates \cite{Kharzeev:1994pz,Kharzeev:1995kz,Xu:1995eb}. Both the 
magnitude \cite{Shuryak:1978ij} and the energy dependence 
\cite{Bhanot:1979vb} of charmonium 
dissociation by gluons result in significant suppression   
of the $c \overline c$ states even for $T<T_D$ but higher than the
deconfinement transition temperature, $T_c$. Moreover, close to 
$T_D$ the thermal activation mechanism is expected to dominate
\cite{Kharzeev:1995ju,Wong:2001uu}. The relative importance of 
gluon dissociation and thermal activation is governed by the ratio of the 
quarkonium binding energy $\epsilon(T)$ and the temperature $T$, 
$X(T) \equiv \epsilon(T)/T$ 
\cite{Kharzeev:1996se}.  At $X(T) \ll 1$ thermal activation 
dominates while for $X(T) \gg 1$ the dominant mechanism is ``ionization'' 
by gluons. 

Dissociation due to colour screening was studied using potential models
with different parameterizations of the heavy quark potential
\cite{Digal:2001ue,Ropke:zz,Karsch:1987pv,Hashimoto:1987hf} to
determine $T_D$.  All these studies predicted that excited charmonium
states ($\chi_c$, $\psi'$) will essentially dissolve at $T_c$ while
the ground state $J/\psi$ will dissociate at $1.1$--$1.3~T_c$.  Some
potential models also predicted a strong change in the binding energy,
see \eg Ref.~\cite{Karsch:1987pv}. Recently, charmonium properties
were investigated using lattice calculations
\cite{Umeda:2002vr,Datta:2002ck} which indicate that the ground states
exist with essentially unchanged properties at temperatures around
$1.5 T_c$.

Lattice investigations suggest that at low temperatures, $T<1.5 T_c$, 
screening is not efficient and therefore gluon dissociation may be the 
appropriate source of quarkonium suppression. 
One should also keep in mind that non-equilibrium effects in the very
early stages of a heavy-ion collision, when the energy density is highest,
should be considered for quarkonium suppression. 
Not much is known about these effects.  However,
they may be an even more important source of quarkonium suppression than
a thermalized system, see \eg Ref.~\cite{Digal:2002bm}.

To use heavy quarkonium as an effective probe of the state of matter
in QCD, we should also have good theoretical control over the
scattering amplitudes in a hadronic gas.

While addressing the issues outlined above, we will face questions
familiar to the heavy quark and the thermodynamics communities.  To
fully understand the behaviour of quarkonium in media, these two
communities should communicate and interact.  For further background
material and a review of the results as of Summer 2003, see
Ref.\cite{cern1}.

The behavior of quarkonium in cold nuclear matter can be used to
better understand the nuclear medium.  For example, quarkonium can
also be used to study nuclear parton distributions, particularly that
of the gluon.  In addition to hadroproduction studies, quarkonium
photoproduction is directly sensitive to the nuclear gluon
distribution.

\section[QCD in media, and the lattice approach]
{QCD in media, and the lattice approach
$\!$\footnote{Author: M.~P.~Lombardo}}

This section introduces strong interactions `in media' using the field
theory of QCD as the basic theoretical framework.  We build on work
presented in the `Introduction to QCD' and `General Tools' sections of
this Yellow Report and briefly discuss basic aspects which are not
covered there but are used in the rest of this Chapter.
\shortpage

We first discuss the phases of QCD, \ie the fate of chiral symmetry
and confinement at high temperature and density.  We then introduce
lattice thermodynamics, the main theoretical tool for studying
equilibrium characteristics of the phase diagram.  In doing so,
deconfinement, screening, and spectral modifications, which are
discussed at length later, will be briefly touched upon.

Our main interest is in heavy ion colliders experiments where the
baryon density is relatively small. Lattice high density calculations
are far less mature then those at high temperature. In particular, no
results for heavy quarks have been obtained in the high density
regime.  Hence we will focus on the physics of QCD at high
temperature.

Let us first consider the fate of confinement at high temperature.
A sketchy view of the screening mechanism, 
already at work at $T=0$, is the recombination
of a (heavy) quark and antiquark with pairs generated by the
vacuum, $Q \overline Q \rightarrow \overline q Q + q \overline Q$. 
At high temperature it becomes easier to produce light
$q \overline q$ pairs from the vacuum.
Hence it is easier to `break' the colour string  
between a (heavy) quark, $Q$,  and antiquark $\overline Q$. 
In other words, we expect colour screening to increase (sharply) at a 
phase boundary,
eventually leading to quark and gluon liberation, the quark gluon
plasma. Lattice
simulations indicate that enhanced screening occurs at about
$T\sim 200$~MeV, in a range accessible to collider experiments
(see \eg \cite{Karsch:2003jg} for a recent review). 

Consider now the fate of chiral symmetry at high temperature obtained
by following the behaviour of the
chiral condensate $\langle \overline \psi \psi \rangle$.
One picture of the high $T$
QCD transition can be drawn by using the ferromagnetic analogy of
the chiral transition where
$\overline \psi \psi$ can be thought of as a spin
field taking values in real space, 
but oriented in the chiral sphere. 
Chiral symmetry breaking occurs
when $\langle \overline \psi \psi \rangle \ne 0 $, corresponding 
to the ordered phase.
By increasing $T$, disorder increases, and 
$\langle \overline \psi \psi\rangle \to 0$,
restoring chiral symmetry. Lattice
simulations suggest that chiral symmetry restoration and enhanced
screening happen at the same temperature within the numerical
accuracy.

These seemingly simple pictures of chiral symmetry restoration and
deconfinement at high temperature are complicated by a
number of considerations.  A finite quark mass breaks the
chiral symmetry of the Lagrangian while screening can be 
rigorously related to deconfinement only at infinite quark 
mass. Rigorously speaking, the two mechanisms
we are concerned with are defined in two opposite limits:
zero mass for chiral symmetry and infinite mass for confinement.
For two (massless) flavours the transition seems to be of second order while 
the transition with three (massless) flavour  turns out to be first
order. We have to find out how the two flavour picture morphs with
the three flavour one. In addition, it may well be that
the $U_A(1)$ symmetry, broken at zero temperature, is effectively
restored at high temperature, further complicating
the patterns of chiral symmetry. 
It is also  worth mentioning in this brief introduction
that, even if the string `breaks', 
bound states might well survive, giving rise to complicated, 
nonperturbative dynamics above the critical temperature,
characterising the strongly-interacting quark gluon plasma.
In addition to these conceptual difficulties, there are 
calculational problems since the phenomena we are concerned with 
are well outside the reach of perturbative calculations.

\begin{figure}
\begin{center}
\includegraphics[width=.44\textwidth]{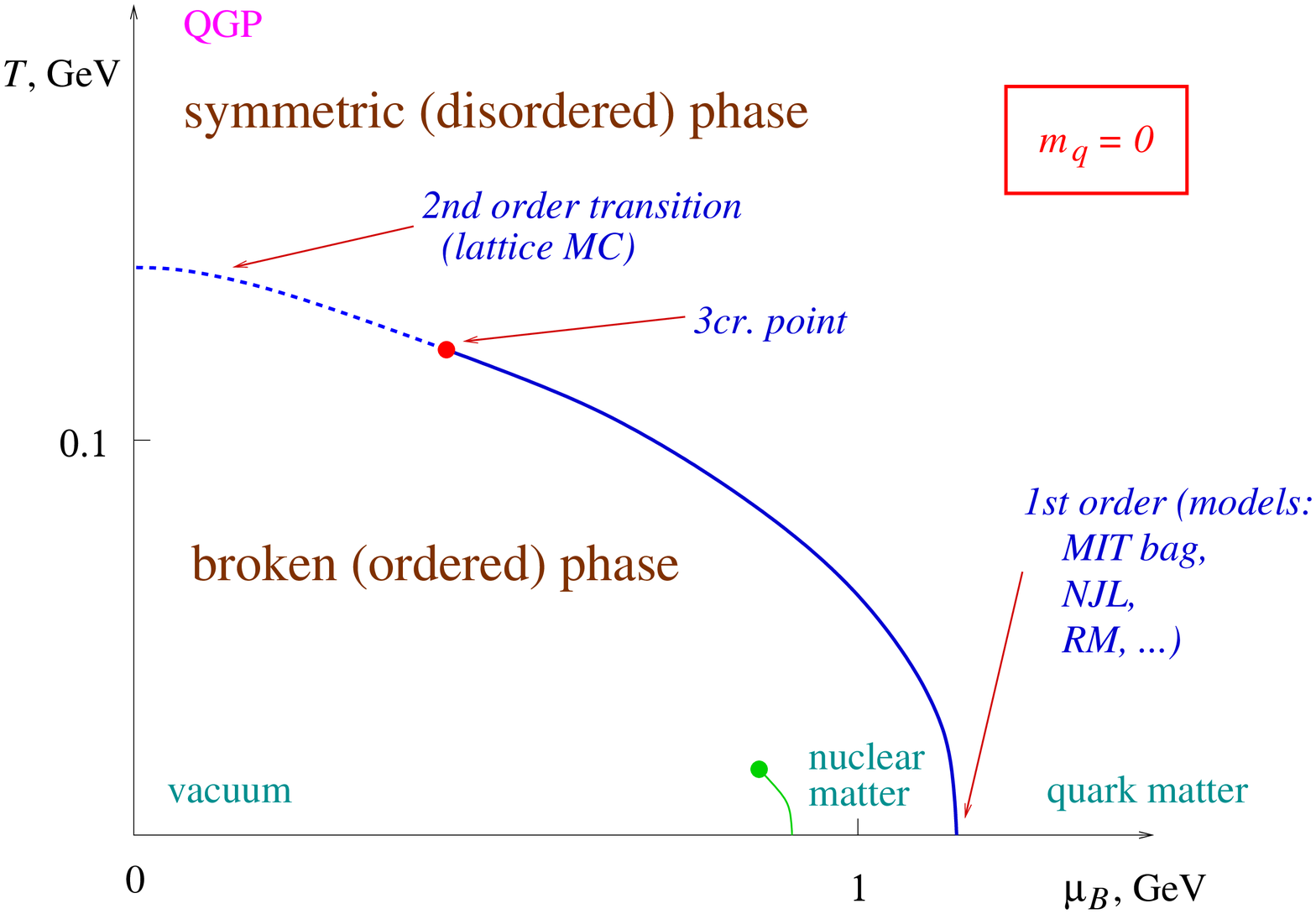}
\hfill
\includegraphics[width=.55\textwidth]{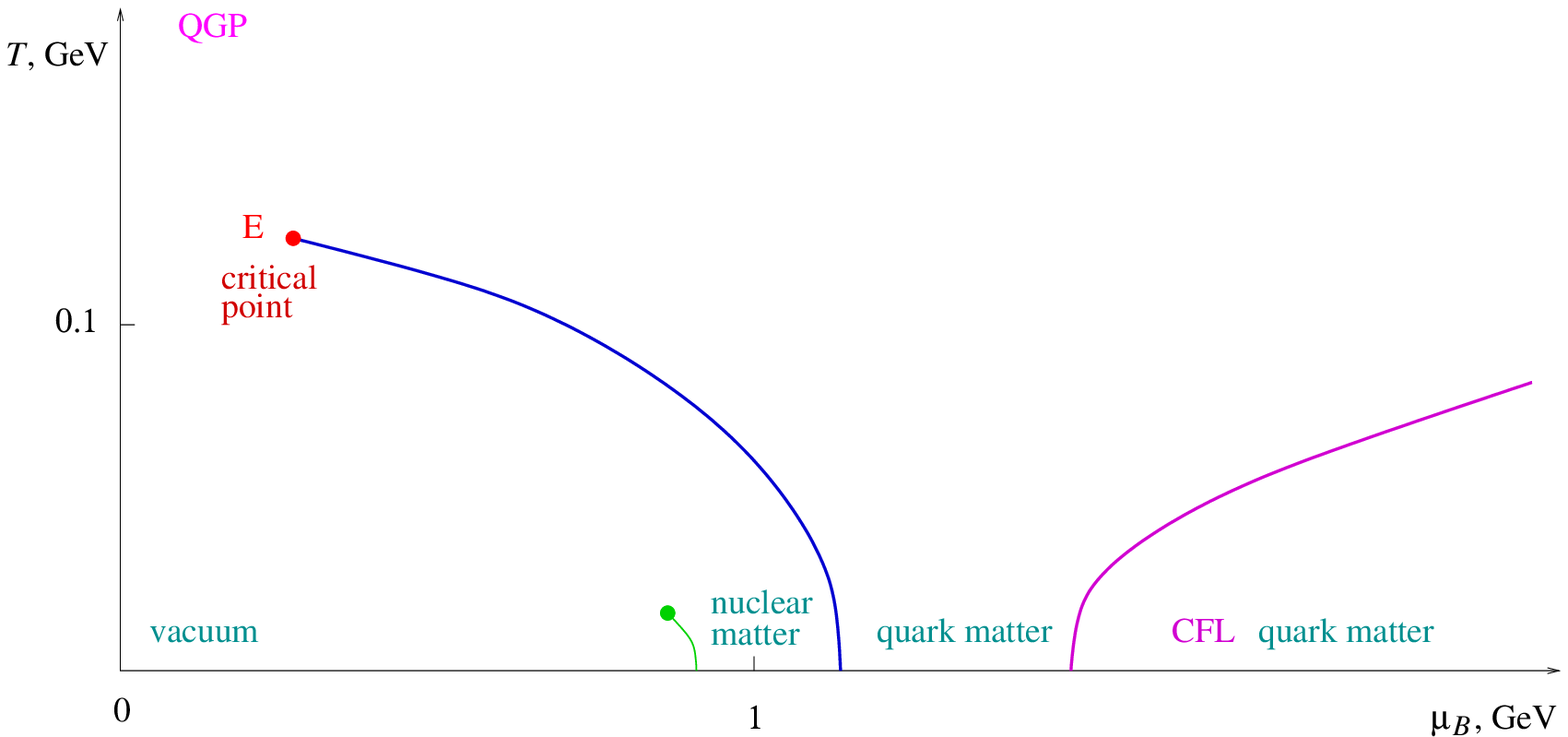}
\end{center}
\caption[The phase diagram of QCD]
        {The phase diagram of QCD for two massless flavours (left) and
         what is expected for realistic values of the quark masses
         (right).  From M. Stephanov \cite{Stephanov:2004wx}.}
\label{fig:phasediagram}
\end{figure}

One goal of our subgroup is to understand how all of these phenomena
affect the heavy quark spectrum when the phase boundary in
\Figure~\ref{fig:phasediagram} is crossed. Here, we show the phase
diagram of QCD in the temperature and chemical potential plane. The
diagrams for `light QCD' (two massless quarks), and for realistic
values of the $u$, $d$ and $s$ masses, were built using symmetry
arguments and model calculations.  The informal discussion presented
here should already suggest that the physics of heavy quark bound
states in media is driven by a rich and complex admixture of chiral
symmetry restoration and enhanced screening, requiring both lattice
and analytic studies.  Recent advances will be presented in the next
sections.

As a last introductory remark, we review a few basic facts
about finite temperature  field theory \cite{Kapusta}
and its formulation on the lattice \cite{Karsch:2001cy}.  In
equilibrium field theory, the grand canonical partition function,
$Z (V, T, \mu)$ completely determines
the thermodynamic state of a system  according to:
\begin{eqnarray}
P &=& T \frac {\partial \ln Z}{\partial V} \\
N &=& T \frac {\partial \ln Z}{\partial \mu} \\
S &=& \frac {\partial [T \ln Z]}{\partial T} \\
E &=& -PV + TS +\mu N 
\end{eqnarray}
while the physical observables $\langle O \rangle$ can be computed as
\begin{equation}
\langle O \rangle = \frac {\tr (O \hat \rho)} Z \, \, .
\end{equation}
In short, 
the  problem is to represent  $Z$ for QCD at finite temperature
and design a calculational scheme to describe it.  

The partition function, 
$Z$,  is the trace of the density matrix of the system, $\hat \rho$, so that
\begin{eqnarray}
Z &=& \tr \, \hat \rho \\ 
\hat \rho &=& \exp[(-H - \mu \hat N)/T]
\label{eq:stati}
\end{eqnarray}
where $H$ is the Hamiltonian, $T$ is the temperature and $\hat N$ is any
conserved number operator. 
Introducing the integral,
$S(\phi, \psi)$, of the Lagrangian density where 
$Te$ is the temporal extent in Euclidean space,
\begin{equation}
S(\phi, \psi) = \int_0^{Te} dt \int d^3 x {\cal L}(\phi, \psi) \, \, ,
\end{equation}
and $Z$ is defined as
\begin{equation} 
Z = \int d \phi \, d \psi \, \exp[-S(\phi, \psi)] \, \, .
\label{eq:path}
\end{equation}

Comparing the  path integral representation of the partition function,
\Eq~(\ref{eq:path}), 
with the statistical mechanics representation, \Eq~(\ref{eq:stati}),
we can identify the finite temporal extent of the
Euclidean space, $Te$, with the reciprocal of the system temperature
\cite{Kapusta}.
The only missing ingredients are the boundary conditions for
the fields in \Eq~(\ref{eq:path}) which follow from the (anti)commuting
properties of the (fermi)bose fields, implying
\begin{equation}
\hat \phi(\vec x, 0) = \hat \phi (\vec x, \beta)
\end{equation}
for bosons and
\begin{equation}
\hat \psi(\vec x, 0) = -\hat \psi (\vec x, \beta)
\end{equation}
for fermions.
Fermions and bosons obey antiperiodic and periodic 
boundary conditions, respectively,  in the
time direction.

Finite temperature lattice field theory is then straightforward since 
temperature comes for free. Because the lattice has a 
finite extent, $N_t a$, temperature is given by
$T = 1 / N_t a $. The discretization procedure is the same as
at zero temperature and most of the standard
lattice techniques, reviewed in the introductory Sections
above, carry over to finite temperature.
Such a field theoretic approach
to finite temperature QCD allows us to put 
thermodynamics  and spectral calculations
on the same footing.

%%%%%%%%%%%%%%%%%%%%%%%%%%%%%%%%%%%%%%%%%%%%%%%%%%%

\section[QCD at finite temperature: colour screening and quarkonium
         suppression]
        {QCD at finite temperature: colour screening
         and quarkonium suppression
           $\!$\footnote{Section coordinator: P.~Petreczky;
               Authors:  S.~Datta, O.~Kaczmarek, D.~Kharzeev, F.~Karsch, 
               M.~P.~Lombardo, P.~Petreczky, H.~Satz, F.~Zantow }} 
\label{sec:QCDfinitetemperature}

On quite general grounds, it is expected and, in fact,
confirmed by lattice simulations that strongly
interacting matter undergoes a transition to quark gluon plasma
at high temperature and density.
One of the most prominent properties of this state of matter
is the screening of colour forces between static quarks. The 
associated screening length (often referred to as the
chromoelectric or non-Abelian Debye length) is inversely
proportional to the temperature. 

Heavy quarkonia, unlike usual (light) hadrons, may exist
in the quark gluon plasma due to their relative small
sizes. However, above some temperature the screening radius
eventually becomes smaller than the typical quarkonia radii,
leading to their dissolution. This physical picture was 
used by Matsui and Satz to propose quarkonium suppression
as a signal for deconfinement in heavy ion collisions \cite{Matsui:1986dk}. 
In fact, it was found that the $\jpsi$ dissolves above but close to the deconfinement temperature $T_c$. Estimates of the dissociation temperature based on the
screening picture have some shortcomings. It is not clear
to what extent many body effects present in the strongly
coupled quark gluon plasma can be approximated by modification of the
interaction between the two heavy quarks. Unfortunately, it is also not trivial to define the screening radius.

Although a detailed understanding of screening phenomena at 
large distances is still missing, it is evident that in this regime 
the temperature is the dominant scale and consequently will control the 
running  of the QCD coupling, \ie $g\simeq g(T)$ for 
($rT \gg 1$, $T\gg T_c$)\footnote{We use the deconfinement temperature 
$T_c$ as a characteristic energy scale rather than a more conventionally
used $\Lambda$-parameter.}. However, at short distances, 
$r\cdot \max (T,T_c)\ll 1$, hard processes dominate the physics 
of the quark gluon plasma even at high temperature and it is expected that 
a scale appropriate for this short distance regime will control
the running of the QCD coupling, \ie $g\simeq g(r)$. 
The interplay between short and large distance length scales plays a 
crucial role for a quantitative understanding of hard as well as soft 
processes in dense matter. It will, for instance, determine the range
of applicability of perturbative calculations for thermal dilepton rates 
or the production of jets as well as  the analysis of processes that can 
lead to thermalization of the dense matter produced in heavy ion collisions. 
Moreover, the short and intermediate distance regime also is most
relevant for the discussion of in-medium modifications of heavy quark bound 
states which are sensitive to thermal modifications of the heavy quark
potential as well as the role of quasi-particle excitations in the 
quark--gluon plasma.

Another powerful tool to study the dissolution of 
quarkonia states in the plasma is the
corresponding meson spectral functions.

The first subsection will be devoted to a general discussion of
screening in hot matter and the running coupling and its observable
implications.  Next, we discuss direct lattice calculations of the
spectral functions. These simulations grow extremely expensive at
larger quark mass, which require a fine spacing, and $b$ quark physics
seems to be out of reach within this approach.  It is then natural to
consider NRQCD thermodynamics, a new theoretical tool.  We will
comment on this new possibility at the end.

\subsection{Colour screening and running coupling}

The simplest way to understand the screening phenomenon is
to consider the potential between an arbitrarily heavy (but not static)
quark and antiquark in perturbation theory. At zero 
temperature the potential can be calculated from the Born
heavy quark--antiquark scattering amplitude in the
non-relativistic limit. In the Born approximation, the potential
in momentum space is just the scattering amplitude which, at 
lowest order in the non-relativistic expansion, is
\begin{equation}
-\frac{4}{3} g^2 D_{00}(k)
\end{equation}
where $D_{00}(k)$ is the propagator in the Coulomb gauge \cite{febr}.
Using the leading order perturbative form, $D_{00}(k)=1/k^2$, we
recover the Coulomb potential in coordinate space.

Now consider the high temperature plasma phase. Assuming that
the heavy quark and antiquark are well defined quasi-particles,
the scattering amplitude is
\begin{equation}
-\frac{4}{3} g^2 \frac{1}{k^2 + \Pi_{00}(k)} \, \, ,
\end{equation}
where $\Pi_{00}(k)$ is the medium induced gluon self-energy.
At leading order and small momenta, $\Pi_{00}(k)$ is gauge independent and leads to a non-zero mass term in the gluon propagator,
$\displaystyle m_D^2\equiv\lim_{k\to0}\Pi_{00}(k)=\frac{1}{3} g^2 T^2 (N +\frac{1}{3} N_f)$. 
Thus we obtain the screened Coulomb potential
\begin{equation}
-\frac{4}{3}\frac{g^2}{4 \pi r}\exp(-m_D r) 
\label{eq:yuk}
\end{equation}
as a function of distance $r$.

Another way to discuss colour screening which
is also suitable for nonperturbative (lattice) study, 
is to consider the partition function in the presence of a 
static quark--antiquark pair normalized by the partition function $Z(T)$
of the system without static charges
which can be written as \cite{McLerran}
\begin{eqnarray}
\frac{Z_{q \overline q}(r,T)}{Z(T)}&=&\frac{1}{Z(T)}
\int D A_{\mu} D \overline \psi D \psi
e^{-\int_0^{1/T} d \tau \int d^3 x {\cal L}_{QCD}(\tau,\vx)}
W(\vr) W^{\dagger}(0)\nonumber \\
&=&\langle W(\vr) W^{\dagger}(0) \rangle \nonumber
\end{eqnarray}
where the Wilson line or Polyakov loop is defined as
$$
\displaystyle
W(\vx)=P e^{ig\int_0^{1/T} d\tau A_0(\tau,\vx)} \, \, .
$$
The above partition function contains all colour
orientations of the static $Q \overline Q$ pair. Using
projection operators one can formally define the partition function
for colour singlet (1), colour octet (8), and colour average (av) channels as
\cite{Brown,Nadkarni}
\begin{equation}
\frac{Z^{(1)}_{q \overline q}(r,T)}{Z(T)}=
\frac{1}{3} \tr \langle W(\vr) W^{\dagger}(0) \rangle
\label{eq:ass1}
\end{equation}

\begin{equation}
\frac{Z^{(8)}_{q \overline q}(r,T)}{Z(T)}=
\frac{1}{8} \langle \tr W(r) \tr W^{\dagger}(0) \rangle-\frac{1}{24}
\tr \langle  W(r) W^{\dagger}(0) \rangle
\label{eq:ass8}
\end{equation}

\begin{equation}
\frac{Z^{({\rm av})}_{q \overline q}(r,T)}{Z(T)}=
\frac{1}{9} \langle \tr W(r) \tr W^{\dagger}(0) \rangle \, \, .
\label{eq:assa}
\end{equation}

Only the colour average partition function is manifestly gauge
invariant. To define the singlet and octet partition functions one may
replace the Wilson line by dressed gauge invariant Wilson lines
\cite{Philipsen:2002az} or one may fix the Coulomb gauge in
\Eq~(\ref{eq:ass1}). These two definitions were shown to be equivalent
\cite{Philipsen:2002az} (see also \cite{jahn04}).  Having defined the
partition function in the presence of a static quark--antiquark pair,
the change in the free energy, $F_i$, the internal energy, $V_i$, and
the entropy, $S_i$, of the static quark--antiquark pair relative to a
system with no static charges are calculated as
\begin{eqnarray}
F_{i}(r,T) & = & -T \ln \biggr( \frac{Z^{(i)}_{q\overline q}(r,T)}{Z(T)} 
\biggl) \;=\; V_i(r,T) - T S_i(r,T)\\
V_i(r,T) & = & T^2 \frac{\partial}{\partial T}
\ln \biggr( \frac{Z_{q \overline q}^{(i)}(r,T)}{Z(T)}\biggl)
\,\, = \,\, -T^2 \frac{\partial [F_i(r,T)/T]}{\partial T}\\
S_i(r,T)& = & -\frac{\partial F_i(r,T)}{\partial T} \,\,\,\, i=1,8,{\rm av} \,
\, .
\end{eqnarray}
We concentrate on the colour singlet case in the following.
In leading order perturbation theory,
$F_1$ is dominated by 1-gluon exchange and is therefore also
given by \Eq~(\ref{eq:yuk}) \cite{McLerran}. For this reason
the free energies were (mis)interpreted as potentials.
However, $F_1$ generally contains an $r$-dependent 
entropy contribution which starts at order $g^3$ in
perturbation theory so that $V_1=F_1$ only at leading order.

\begin{figure}
\begin{center}
\includegraphics[width=.49\textwidth]{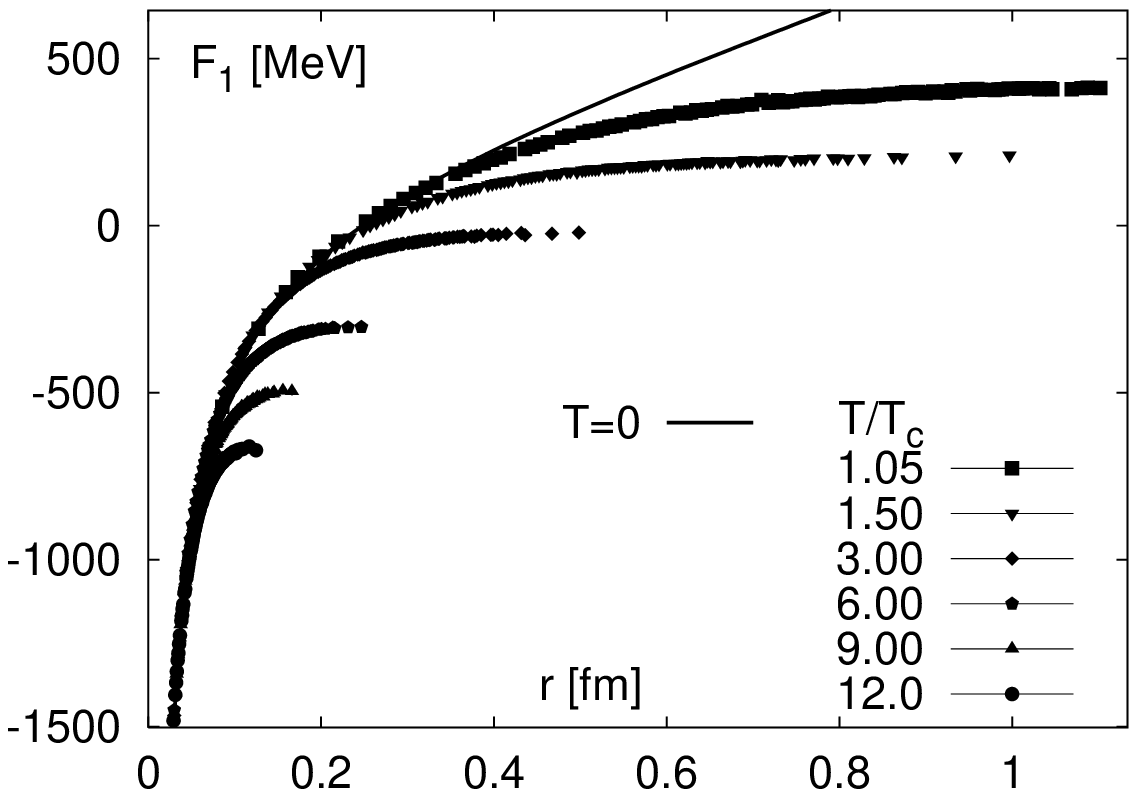}
\hfill
\includegraphics[width=.49\textwidth]{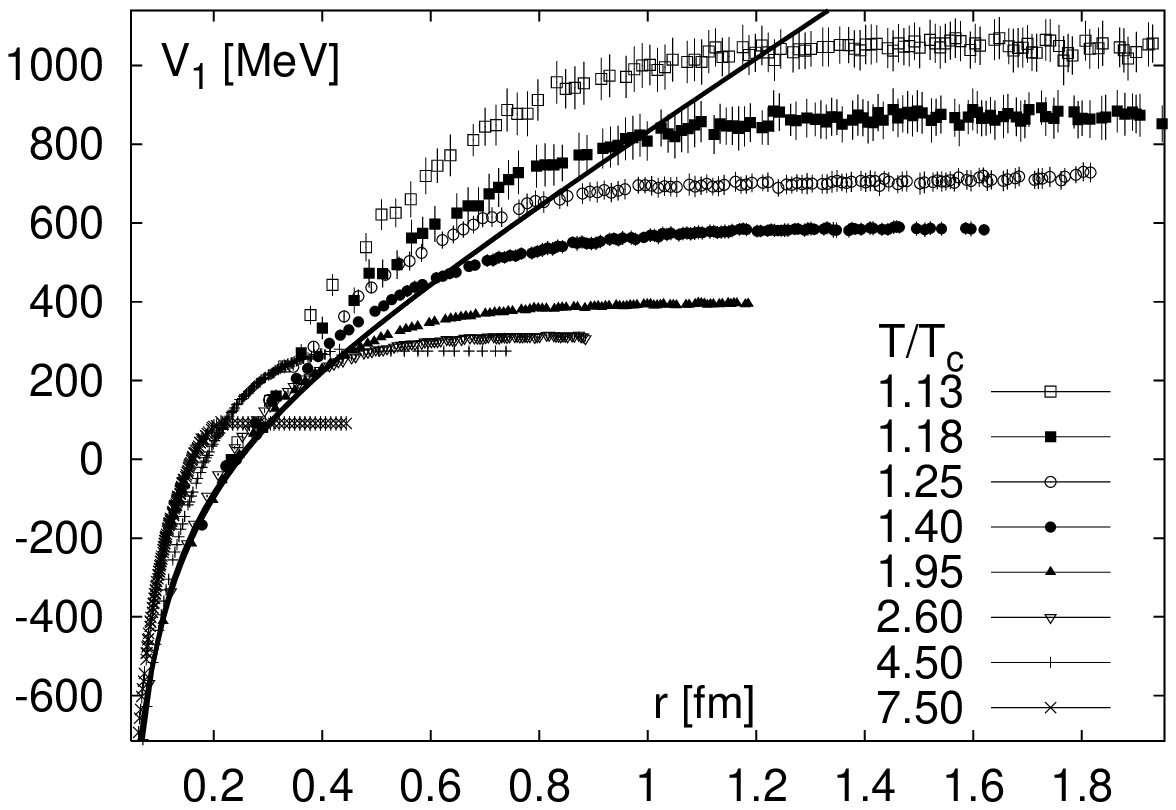}
\end{center}
\caption[The free energy and the internal energy of static $Q
         \overline Q$ pair in the plasma phase]
        {The free energy (left) and the internal energy (right) of
         static $Q \overline Q$ pair in the plasma phase.}
\label{fig:f1v1}
\end{figure}

Recently the colour singlet free and internal energies of static quark
antiquark pair have been studied in lattice simulations of SU(3) gauge
theory \cite{kaczmarek02,kaczmarek04}.  Below the deconfinement
transition temperature, $T<T_c$, $F_1(r,T)$ shows a linear rise with
$r$, as expected in the confined phase.  In the plasma phase ($T>T_c$)
both $F_1(r,T)$ and $V_1(r,T)$ approach a finite value for $r
\rightarrow \infty$, indicating the presence of colour screening. The
numerical results for $F_1$ and $V_1$ in the plasma phase are shown in
\Figure~\ref{fig:f1v1}.
For small distances, $r< 0.2$~fm, and temperatures close to $T_c$ both
$F_1$ and $V_1$ coincide with the $T=0$ potential, as expected since,
at small distances, medium effects are negligible and the free energy
of the static $Q \overline Q$ pair is simply the interaction energy,
\ie the heavy quark potential at zero temperature. In general,
however, the free energy and the internal energy show quite different
$T$- and $r$-dependences.

The perturbative short and large distance relations for the singlet
free energy have recently been used to define a running coupling at
finite temperature \cite{kaczmarek2004},
\begin{equation}
\alpha_{\rm qq} (r,T) = \frac{3r^2}{4} \frac{{\rm d} F_1(r,T)}{{\rm d}r}
\quad .
\label{eq:alpha}
\end{equation}
In general, however, the definition of a running coupling in QCD is
not unique beyond the validity range of 2-loop perturbation theory;
aside from the scheme dependence of higher order coefficients in the
QCD $\beta$-functions it will strongly depend on non-perturbative
contributions to the observable used for its definition.

\begin{figure}
\epsfxsize=9cm
\centerline{\epsffile{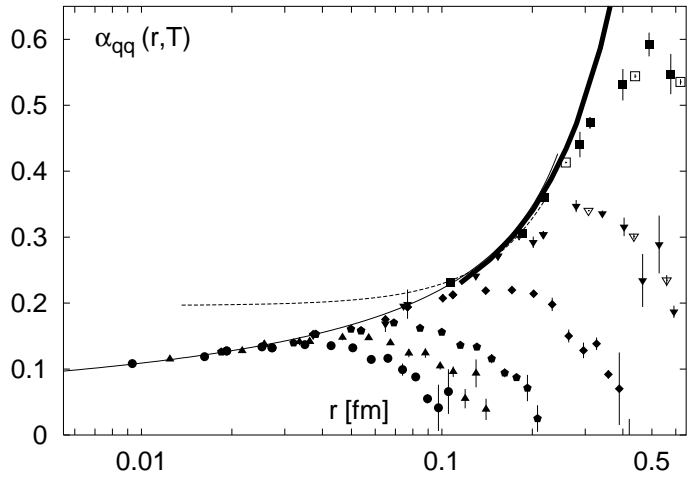}}
\caption[The running coupling in the $qq$-scheme]
        {The running coupling in the $qq$-scheme determined on
         lattices of size $32^3\times N_\tau$ with $N_\tau = 4$ (open
         symbols) and 8 (filled symbols) from derivatives the short
         distance part of the singlet free energy ($T=0$: from the
         force) at different temperatures. The relation of different
         symbols to the values of the temperature are as in the
         previous figure (left). The various lines correspond to the
         string picture (dashed), perturbative (thin) and numerical
         (thick) studies at zero temperature.}
\label{fig:alphaeff}
\end{figure}

We compare the finite temperature results (symbols) to calculations
performed at zero temperature (lines)
\cite{necco02,necco01,Peter,Schroder} in
\Figure~\ref{fig:alphaeff}. These numerical results on $\alpha_{\rm
qq}$ at distances smaller than $0.1$~fm cover also distances
substantially smaller than those analyzed so far at $T=0$. They
clearly show the running of the coupling with the dominant length
scale $r$ also in the QCD plasma phase. For temperatures below $3 T_c$
one finds that $\alpha_{\rm qq}$ agrees with the zero temperature
perturbative result in its entire regime of validity, \ie for
$r\; \lsim\; 0.1$~fm. At these temperatures thermal effects only
become visible at larger distances and lead, as expected, to a
decrease of the coupling relative to its zero temperature value; at
distances larger than $r\; \simeq\; 0.1$~fm non-perturbative effects
clearly dominate the properties of $\alpha_{\rm qq}$. It thus is to be
expected (and found) that the properties of a running coupling will
strongly depend on the physical observable used to define it
\cite{kaczmarek2004}.

\subsection{Real time properties of finite temperature QCD, spectral functions}

Most of the dynamic properties of 
the finite temperature system are incorporated 
in the spectral functions. The spectral function, $\sgh$, for a given 
mesonic channel $H$ in a system at temperature $T$ can be defined 
through the Fourier transform of the real time two point functions
$D^{>}$ and $D^{<}$ or equivalently as the imaginary part of 
the Fourier-transformed retarded 
correlation function \cite{lebellac},
\begin{eqnarray}
\sgh &=& \frac{1}{2 \pi} (D^{>}_H(p_0, \vec{p})-D^{>}(p_0, \vec{p}))
\nonumber\\
& = &
\frac{1}{\pi} {\rm Im} D^R_H(p_0, \vec{p}) \nonumber \\
 D^{>(<)}_H(p_0, \vec{p}) &=& \int\frac{d^4 p}{(2\pi)^4} 
 e^{i p.x} D^{>(<)}_H(x_0,\vec{x}) \label{eq:defspect} \\
D^{>}_H(x_0,\vec{x}) &=& \langle
J_H(x_0, \vec{x}), J_H(0, \vec{0}) \rangle \nonumber\\
D^{<}_H(x_0,\vec{x}) &=& \langle
\langle J_H(0, \vec{0}), J_H(x_0,\vec{x}) \rangle \, \, , x_0>0 \, \, . \
\end{eqnarray} 
The correlators $D^{>(<)}_H(x_0,\vec{x})$ satisfy the 
well-known Kubo--Martin--Schwinger
(KMS) condition \cite{lebellac}
\begin{equation}
D^{>}_H({x_0},\vec{x})= D^{<}({x_0}+i/T,\vec{x}) \, \, .
\label{eq:kms}
\end{equation}
Inserting a complete set of states and using \Eq~(\ref{eq:kms}), one
gets the expansion 
\begin{eqnarray} 
\sgh = \frac{(2 \pi)^2}{Z} \sum_{m,n} &{}&
(e^{-E_n / T} \pm e^{-E_m / T}) |\langle n | J_H(0) | m \rangle|^2
\nonumber \\ & \times & \delta^4(p_\mu - k^n_\mu + k^m_\mu)
\label{eq:specdef}
\end{eqnarray} 
where \eg $k^n$ refers to the four-momenta of the state $| n \rangle $.

A stable mesonic state contributes a $\delta$ function-like
peak to the spectral function,
\begin{equation}
\sgh = | \langle 0 | J_H | H \rangle |^2 \epsilon(p_0)
\delta(p^2 - m_H^2) \, \, ,
\label{eq:stable}
\end{equation}
where $m_H$ is the mass of the state. For an unstable
particle, a smoother peak is obtained with a width
related to the decay width. For sufficiently small
decay widths, a Breit--Wigner form is commonly used.
As the temperature increases, the contributions from
states in the spectral function changes due to collision broadening, and, 
at sufficiently high temperatures, these states may 
be too broad to contribute to the resonance any longer. Such a 
change in the contributions to the
resonance states and eventual `disappearance of resonances'
in the thermal spectral function has been studied
analytically, for example, in the Nambu--Jona--Lasinio 
model in Ref.~\cite{hatsuda}. The spectral function as defined in
\Eq~(\ref{eq:specdef}) is directly accessible in high energy
heavy ion experiments. For example, the spectral function for the vector 
current is directly related to the differential thermal cross-section 
for the production of lepton pairs \cite{Braaten:1990wp}
\begin{equation}
\frac{dW}{dp_0 d^3p}  = \frac{5 \al^2}{27 \pi^2} 
\frac{1}{p_0^2 (e^{p_0/T}-1)} \sigma(p_0, \vec{p}) \, \, .
\label{eq:dilepton} 
\end{equation}
Then presence or absence of a bound state in the spectral function
will manifest itself in the peak structure of the differential 
dilepton rate.

In finite temperature lattice calculations, one calculates
Euclidean time propagators, usually
projected to a given spatial momentum
\begin{equation}
G_H(\tau, \vec{p}) = \int d^3x e^{i \vec{p}.\vec{x}} 
\langle T_{\tau} J_H(\tau, \vec{x}) J_H(0,
\vec{0}) \rangle_T
\end{equation}
where $\langle ... \rangle_T$ indicates a thermal trace, as in
\Eq~(\ref{eq:defspect}), and $T_\tau$ refers to ordering in
Euclidean time $\tau$. This quantity is the analytical continuation
of $D^{>}(x_0,\vec{p})$
\begin{equation}
G_H(\tau,\vec{p})=D^{>}(-i\tau,\vec{p}) \, \, .
\end{equation}
Using this equation and the KMS condition           one can
easily show that $G_H(\tau,\vec{p})$ is related to the 
spectral
function, \Eq~(\ref{eq:defspect}), by an integral equation
\begin{eqnarray} 
G_H(\tau, \vec{p}) &=& \int_0^{\infty} d \omega
\sg K(\omega, \tau) \label{eq:spect} \\
K(\omega, \tau) &=& \frac{\cosh(\omega(\tau-1/2
T))}{\sinh(\omega/2 T)} \, \, .
\label{eq:kernel}
\end{eqnarray} 
\Eq[b]~(\ref{eq:spect}) lies at the heart of attempts to extract
spectral functions and properties of hadrons from
correlators calculated in lattice QCD. In what follows, we
use \Eq~(\ref{eq:spect}) to extract the behavior of degenerate
heavy meson systems in a thermal medium from finite
temperature mesonic correlators. \Eq[b]~(\ref{eq:kernel})
is valid only in the continuum. It is not clear in general 
whether the $G_H(\tau, \vec{p})$ measured on the lattice will satisfy
the same spectral representation but it was shown in Ref.~\cite{Karsch:2003wy}
that this is the case for the free theory.

\subsection{Charmonium at finite temperature: recent results
on correlators and spectral functions}

Direct investigations of the charmonia temperature modifications,
using the Matsubara correlators of suitable operators, have been
available over the past 3--4 years. All such studies available at
present are based on quenched lattices, that is, they do not include
any quark loops, even thermal quark loops (which indicates scattering
of thermal quarks off the medium). Excluding such loops may result in
exclusion of important physics below $\tc$ since the thermal
excitation of pions and meson resonances is considered to be one of
the main driving mechanisms of the QCD phase transition. Above the
transition, at least for \jpsi\ dissociation, the existing wisdom is
that the thermal gluons are the more relevant degrees of freedom.
Therefore, the description of quenched lattices should be at least
qualitatively correct.

In the following, the current status of the direct lattice studies on
charmonia are summarized. Only the main results are described here.
For detailed analysis of the systematics and other effects, please
refer to the original papers
\cite{Datta:2003ww,Asakawa:2003re,umeda02}.  As already mentioned in
the previous section, detailed information about the hadronic
properties are contained in the spectral function.  In order to study
thermal modifications of hadrons, ideally \Eq~(\ref{eq:spect}) should
be inverted to extract \sgt\ from the thermal correlators. (In the
following, we will only be interested in mesonic states at rest in the
heat-bath frame, \ie $\vec{p}=0$ in \Eq~(\ref{eq:spect}), and will
omit the $\vec p$ index altogether.) However, such an inversion is a
notoriously ill-defined problem, since the continuous function, \sgt\,
must be extracted from the correlators at a discrete number of
points. Progress in solving this problem became possible with the
introduction of Bayesian techniques in lattice data analysis
\cite{Asakawa:2000tr}. In the Maximum Entropy Method (MEM), one can
determine the most probable spectral function which describes the
data, subject to known constraints like positivity, asymptotic
behavior, \etc \cite{jarrel}.  At temperatures a few times $\tc$ or
higher, however, as we will see below, it is tricky to extract the
spectral function even with these techniques.

\begin{figure}[t]
\begin{center}
\includegraphics[width=.49\textwidth]{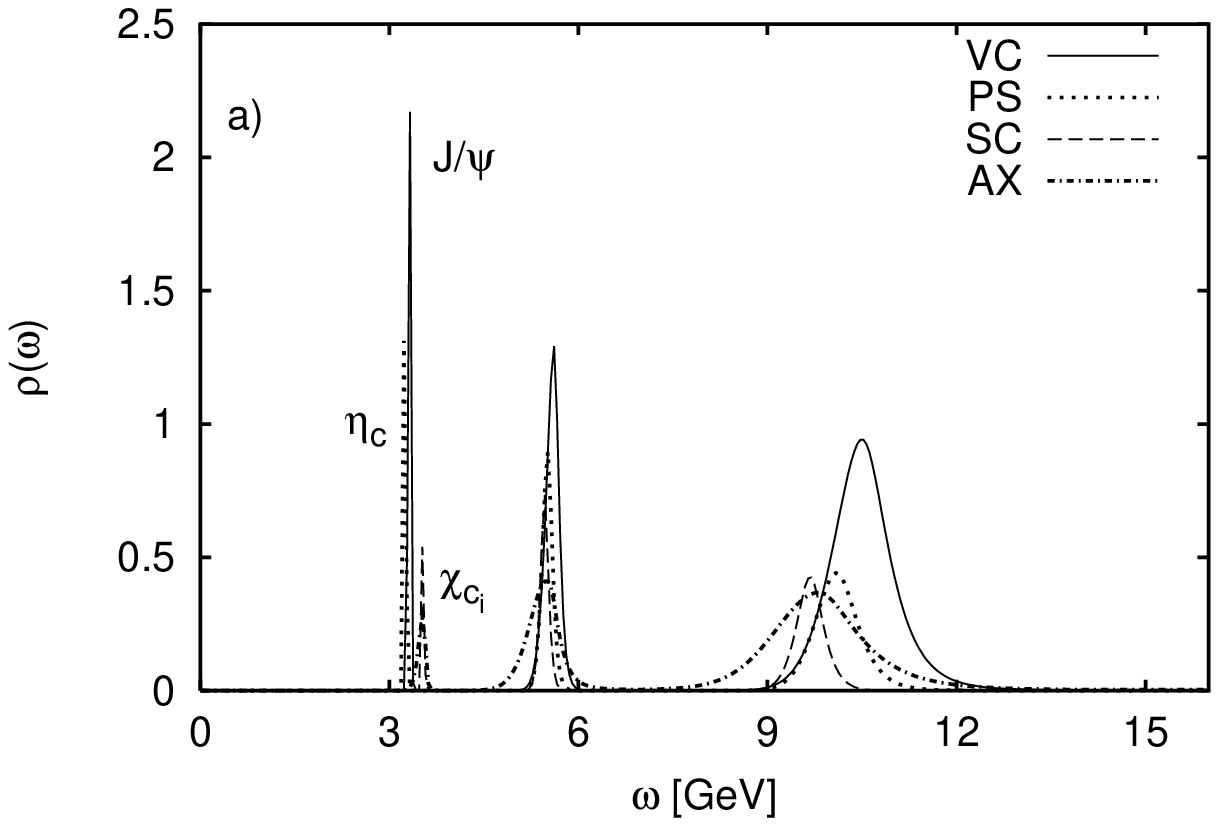}
\hfill
\includegraphics[width=.49\textwidth]{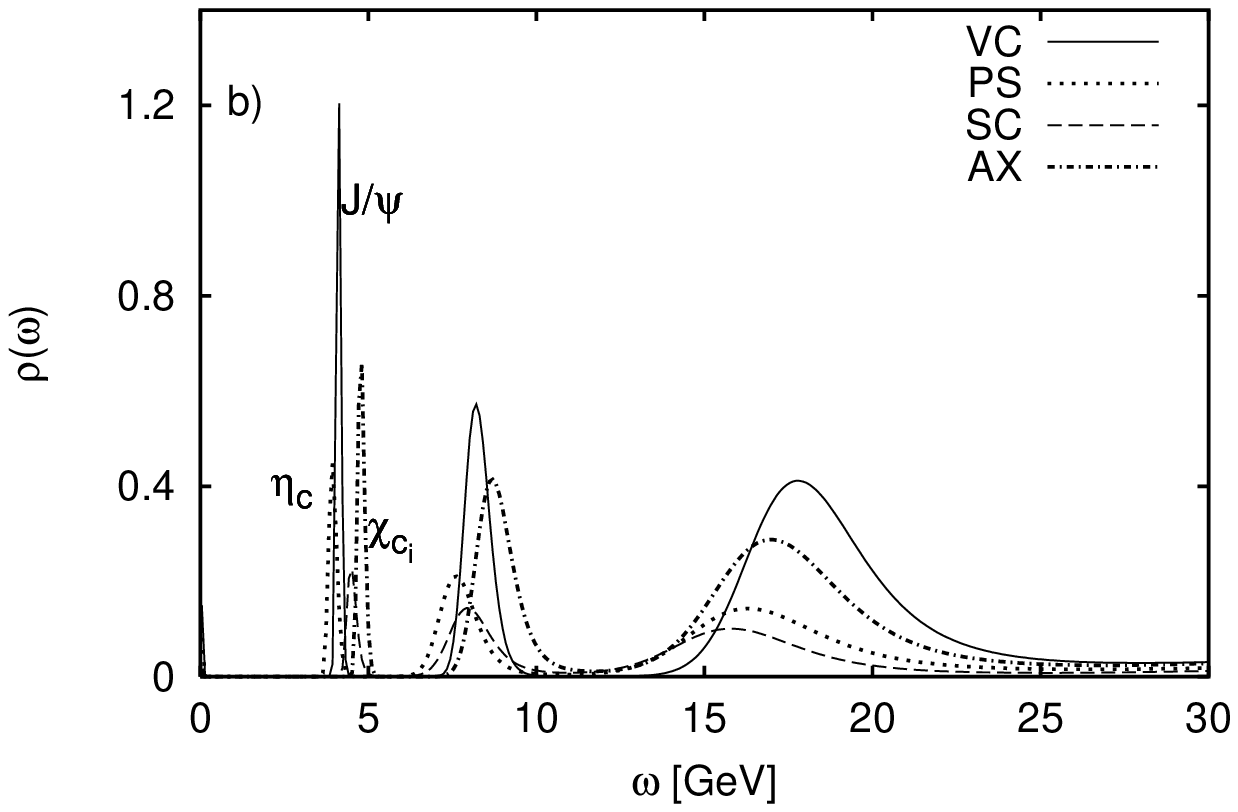}
\end{center}
\caption[Spectral functions for the vector, pseudo-scalar, 
         scalar and axial vector operators]
        {Spectral functions for the vector, pseudos-calar, scalar and
         axial vector operators \cite{Datta:2003ww}. (a) Lattices at
         0.75 $\tc$ with $a = 0.04$~fm spacing and (b) at 0.9 $\tc$
         and $a =0.02$~fm.  The ground state peaks correspond to
         $\jpsi$, $\ec$, $\ass$ and $\asx$, respectively.  The label
         $\chi_{c_i}$ is used because the $\ass$ and $\asx$ are
         difficult to resolve on the figure.}
\label{fig:su719}
\end{figure}

We begin with the low temperature phase where the larger extent of the
temporal direction makes it easier to extract the spectral
functions. In \Figure~\ref{fig:su719} we show the spectral functions
in the hadronic phase for the currents $J_H$ in different
representations.  \Figure[b]~\ref{fig:su719}(a) shows the spectral
functions obtained at 0.75 \tc\ on lattices with 0.04~fm spacing. The
peaks at low \om\ correspond to the ground state $\jpsi$ (VC), $\ec$
(PS), $\ass$ (SC) and $\asx$ (AX) respectively.  Note that we do not
discuss the $2S$ and $2P$ states because these are indistinguishable
from lattice artifacts.  The properties of the lowest states are
reproduced quite well by \Figure~\ref{fig:su719} since the peak
position and the integrated width of the peak are in reasonable
agreement with the mass and residue obtained from a
fit. \Figure[b]~\ref{fig:su719}(b) shows spectral functions obtained from
lattices with 0.02~fm spacing at $T=0.9 T_c$. A comparison of the two
figures helps explain the nature of the peaks at higher $\om$, since
they can be seen to scale approximately as the inverse lattice spacing
and therefore are probably dominated by lattice artifacts
\cite{Datta:2003ww,yamazaki}. It is interesting to note that even at
very high \om\ the structure of the spectral function is quite
different from the free theory.

\begin{figure}[t]
\begin{center}
\includegraphics[width=.49\textwidth]{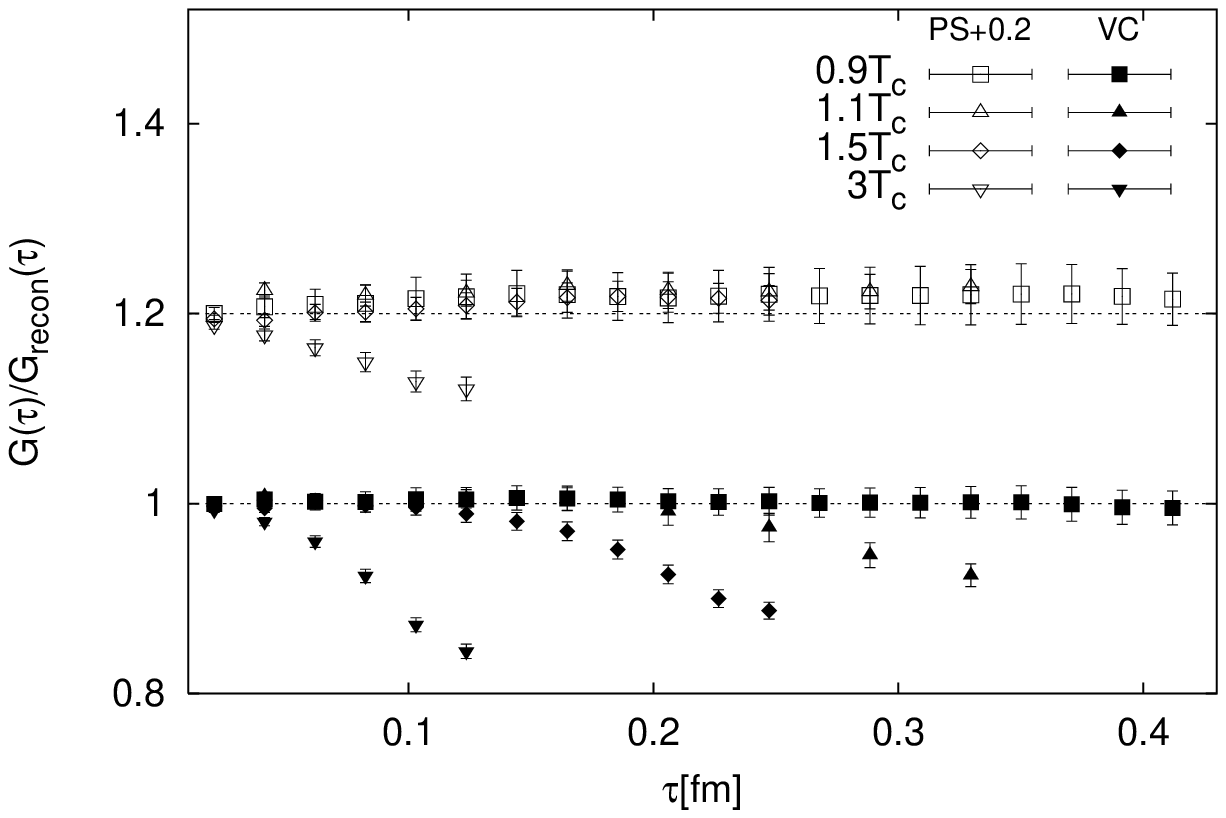}
\hfill
\includegraphics[width=.49\textwidth]{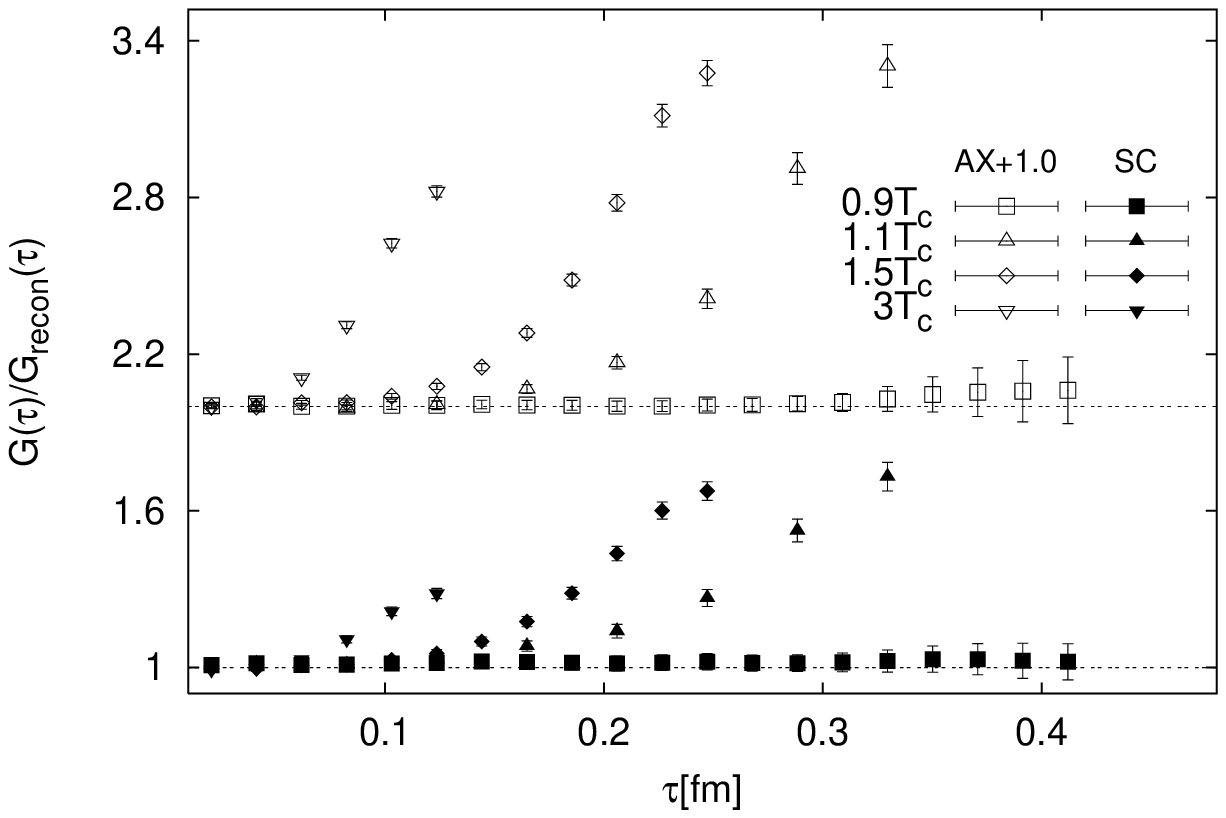}
\end{center}
\caption[$G/G_{\rm recon}$ as a function of Euclidean time]
        {$G/G_{\rm recon}$ as a function of Euclidean time for
         different temperatures for vector and pseudo-scalar channels
         (top) and in the scalar and axial vector channels (bottom).
         The figures above mix two sets of data in
         Ref. \cite{Datta:2003ww} with different lattice spacings and
         slightly different quark masses, with the data at 1.1 $\tc$
         using $T^*$ = 0.75 \tc\ and the others using $T^*$ = 0.9
         $\tc$. The data for the pseudo-scalar and axial vector
         channels have been shifted by a constant for better
         visualization.}
\label{fig:recon}
\end{figure}

It is possible to get a first idea of the temperature modification of
the charmonia above \tc\ by looking at the Matsubara correlators
measured at these temperatures. To factor out the trivial temperature
dependence of the kernel (see \Eq~(\ref{eq:kernel})), one can
construct `model correlators' by using the spectral function from the
hadron phase (\Figure~\ref{fig:su719}). The measured correlators,
$G(\tau, T)$, are then compared with these reconstructed correlators
using \Eq~(\ref{eq:kernel}):
\begin{equation}
\grecon = \int_0^{\infty} d \omega
\sigma(\om, T^*) \frac{\cosh(\omega(\tau-1/2
T))}{\sinh(\omega/2 T)}.
\label{eq:recon}
\end{equation}
Here $T^*$ refers to a temperature below
$\tc$. \Figure[b]~\ref{fig:recon} shows such a comparison for the $1S$
and $1P$ channels in two figures from Ref.~\cite{Datta:2003ww}. While
the comparison at 1.1 \tc\ uses the spectral function constructed at
$T^*=0.75 \tc$, \Figure~\ref{fig:su719}(a), the other temperatures use
the spectral functions shown in \Figure~\ref{fig:su719}(b), \ie
$T^*=0.9\tc$. The figure clearly shows that the $1S$ states are not
strongly affected by the deconfinement transition. For the \ec\ we see
no statistically significant change up to temperatures of 1.5 \tc\ and
only very modest changes at 2.25 $\tc$. For the $\jpsi$, the ratio
$G(\tau,T)/\grecon$ shows no significant deviations from unity at
short Euclidean time, $\tau$, a little above $\tc$, while some small
but significant deviations are seen as one goes to higher
temperatures. For the $1P$ channels, large temperature modifications
of the correlator relative to the reconstructed correlator are seen
when crossing $T_c$, indicating that the $1S$ states undergo only very
modest modifications up to 1.5--2 \tc\ while the $1P$ states suffer
more serious modifications.

In order to further understand the nature of thermal modifications of
the states, it is necessary to extract the spectral function directly
from the correlators.  Three groups have presented results for
spectral functions for the pseudo-scalar and vector charmonium states
using the MEM. Since the extraction becomes tricky at high
temperatures and there are some differences in the results, it is
useful to keep in mind the initial differences between them. While all
three work within the quenched approximation and use Wilson-type
valence fermions, some differences exist in their approaches. The
Bielefeld group uses very fine isotropic lattices with the
nonperturbatively improved clover action for the valence
fermions. Asakawa and Hatsuda use space--time anisotropic lattices, to
allow more data points in the temporal direction and use the
unimproved Wilson action for the fermions. Umeda \etal ~also use
anisotropic lattices but use the tadpole-improved Fermilab
action. They also use smeared operators while the others use point
operators.

\begin{figure}[t]
\begin{center}
\includegraphics[width=.49\textwidth]{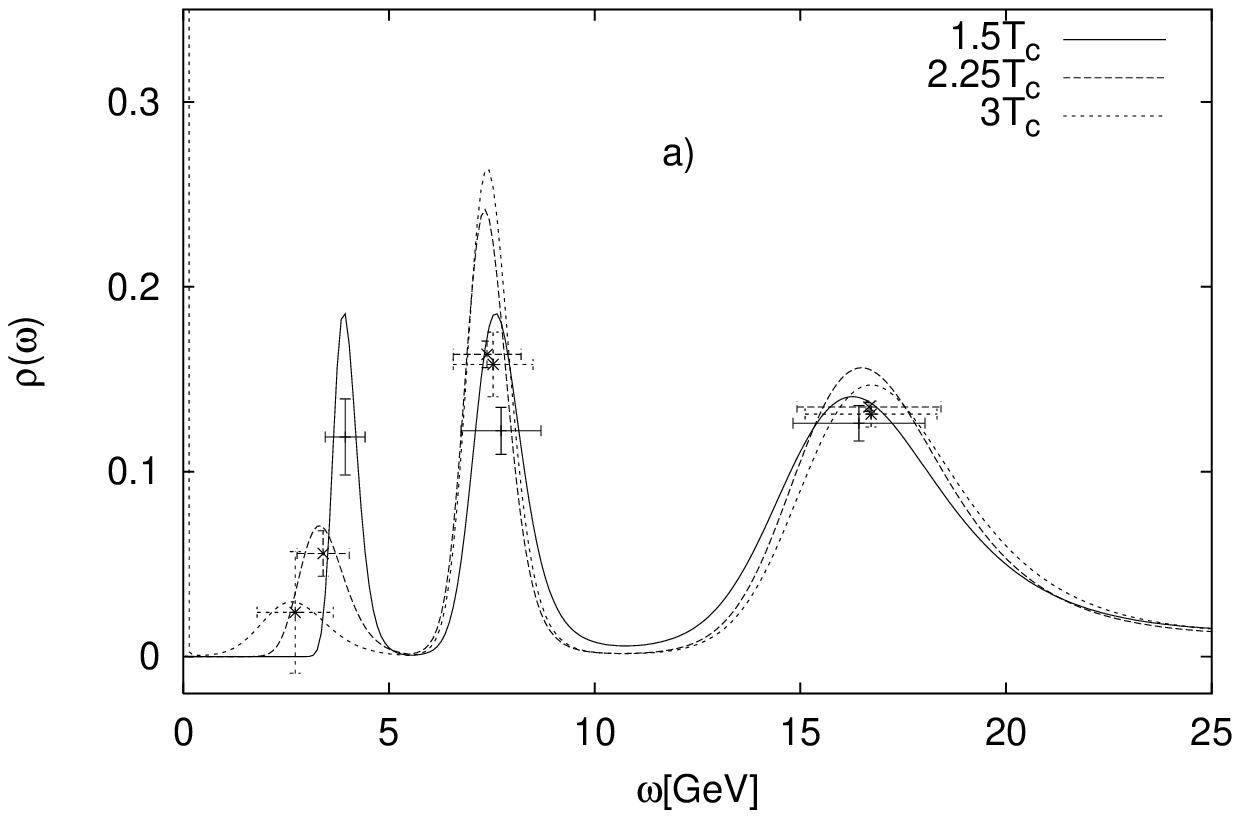}
\hfill
\includegraphics[width=.49\textwidth]{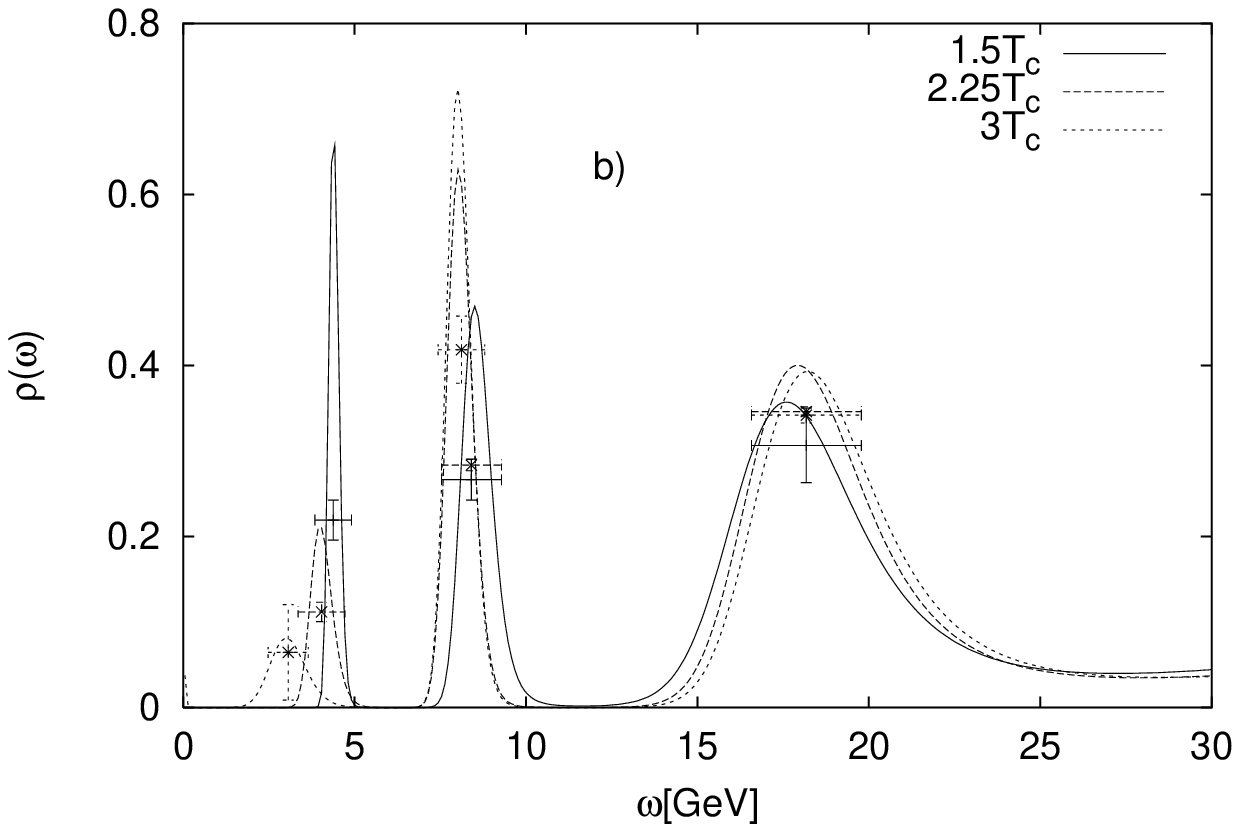}
\end{center}
\caption[Spectral functions above deconfinement 
         for pseudo-scalar and vector channels]
        {Spectral functions above deconfinement for (a) pseudo-scalar
         and (b) vector channels \cite{Datta:2003ww}. Here lattice
         spacing $a \approx 0.02$~fm and $m_\jpsi \approx 3.6$~GeV.}
\label{fig:psvc719}
\end{figure}

As \Figure~\ref{fig:su719} reveals, the structure of the spectral
function at high $\om$, for the interacting theory on lattice, is
considerably different from the free spectral function
\cite{yamazaki}.  The Bielefeld group uses this high energy structure
as part of the prior information when extracting the spectral
function. The default spectral function above $\tc$, in their
analysis, uses the high energy part of \Figure~\ref{fig:su719},
continuously matched to $m_1 \om^2$ at lower $\om$ where $m_1$ is
defined to match the spectral functions at temperatures below
$T_c$. The spectral functions for the $1S$ states, obtained with this
default model and the MEM analysis of Bryan \cite{bryan}, are shown in
\Figure~\ref{fig:psvc719}. The error bars shown in the figure are
standard deviations of the spectral functions averaged over the
$\omega$ interval indicated by the horizontal error band (see
\cite{jarrel}, \Eq~(5.13)). We see that, up to 1.5 \tc\, the $1S$
states persist as bound states with no significant weakening. There is
also no significant change in mass on crossing $\tc$.  At 2.25 $\tc$,
while the peak position is still not significantly changed, a
depletion of the peak strength, \ie the area under the peak, is
seen. Finally, at 3 \tc\ no statistically significant peak is seen.

\begin{figure}[t]
\includegraphics[width=.49\textwidth]{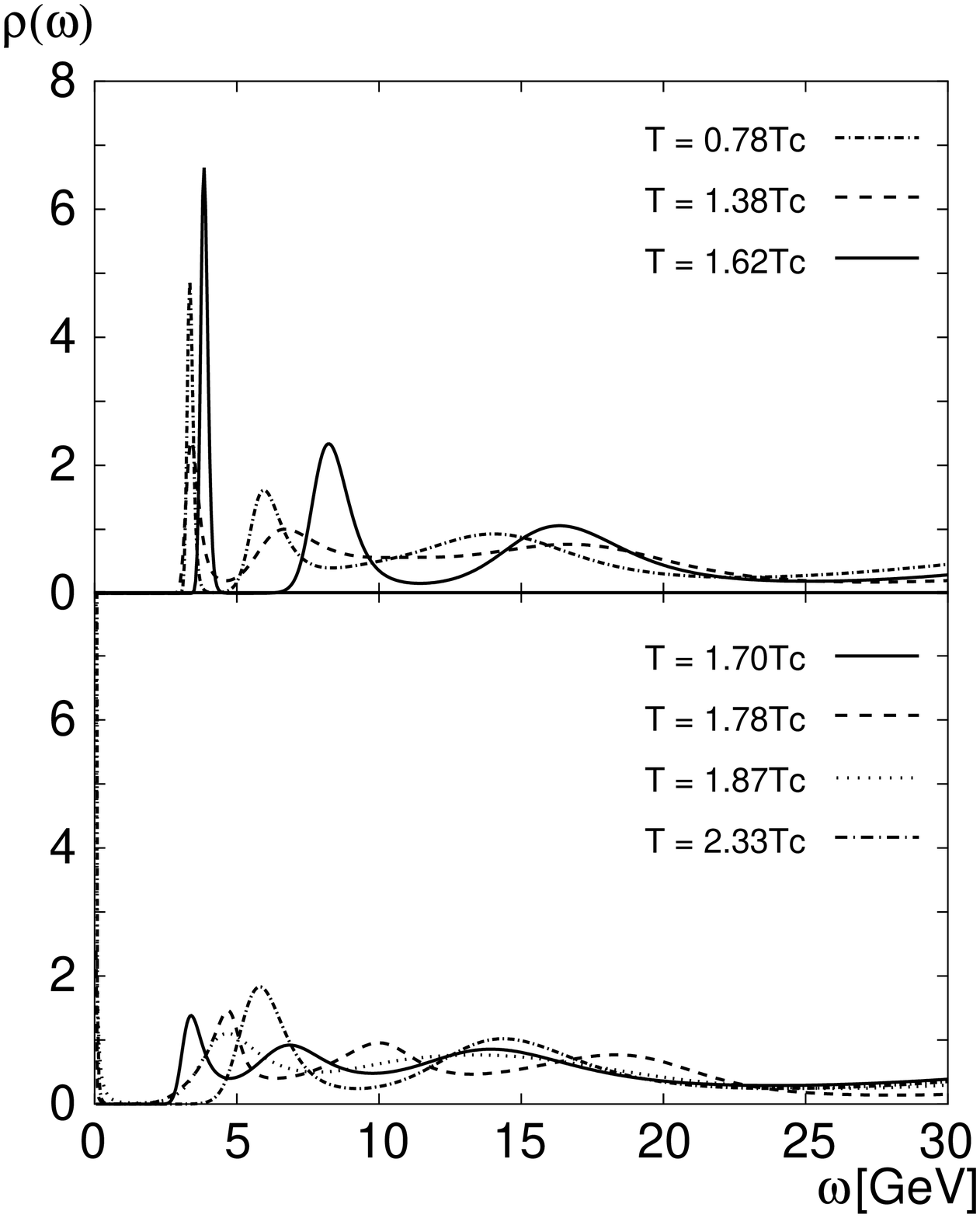}
\includegraphics[width=.49\textwidth]{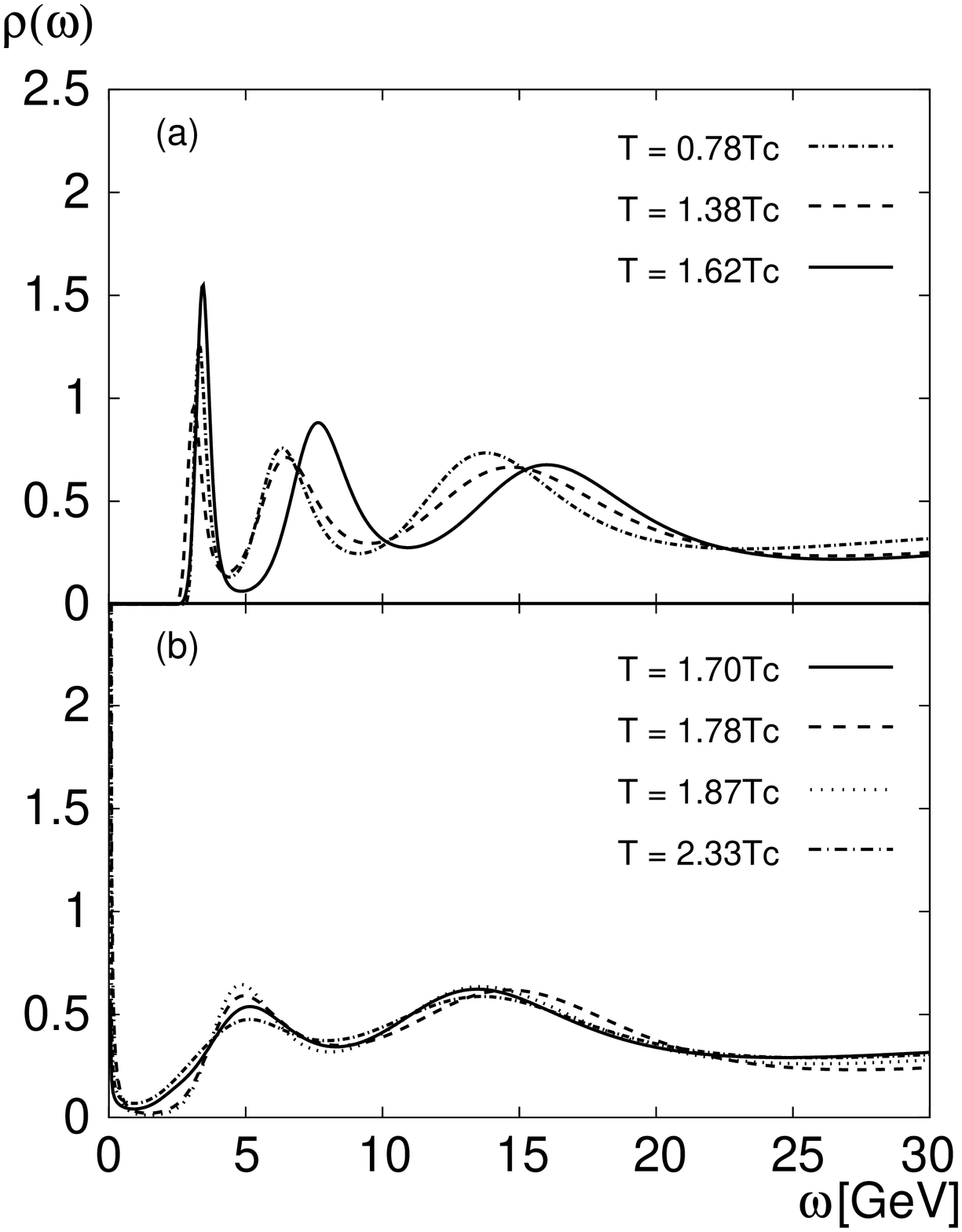}
\caption[Spectral functions above and below deconfinement 
         for pseudo-scalar and vector channels]
        {Spectral functions above and below deconfinement for
         pseudo-scalar (left) and vector (right) channels, for
         anisotropic lattices with $a_\tau \approx$ 0.0125~fm, as
         found by Asakawa \& Hatsuda \cite{Asakawa:2003re,bnl}. (a)
         Temperatures $\le$ 1.62 $\tc$, where a statistically
         significant ground state is seen. (b) Temperatures $\ge$ 1.70
         $\tc$ where the ground state of (a) is not seen any more. The
         peak at a slightly higher temperature is not statistically
         significant.}
\label{fig:asakawa}
\end{figure}

\Figure[b]~\ref{fig:psvc719} suggests that in a gluonic plasma the $1S$
charmonia survive as bound states till $\ge 2 \tc$, with no
significant weakening up to $\approx 1.5 \, \tc$ and then a gradual
weakening, perhaps due to collision broadening.  Since these results
are for point operators, the results for the vector current,
\Figure~\ref{fig:psvc719}(b), will be directly connected to the
thermal dilepton rate. Further results for point operators come from
Asakawa and Hatsuda who use the free continuum asymptotic form of the
spectral function, $\approx m_1 \om^2$, as their default model. Their
latest results \cite{bnl} are shown in \Figure~\ref{fig:asakawa}. They
find a sharp bound state, with little significant thermal
modification, up to 1.62 $\tc$, in complete agreement with
\Figure~\ref{fig:su719} up to 1.5 $\tc$. On going to higher temperatures,
however, their results seem to suggest a sharp disappearance of the
bound states at some temperature, with no statistically significant
peak being seen at 1.9 \tc\ \cite{Asakawa:2003re} and, in preliminary
results, already at 1.7 \tc\ \cite{bnl}. While the exact dissolution
temperature is probably not expected to match the Bielefeld group,
since the Bielefeld group uses a quark mass somewhat heavier than the
charm, if further analysis supports the dissolution at 1.7 \tc\ then
it will clearly suggest a qualitatively different picture from the
gradual dissolution shown in \Figure~\ref{fig:psvc719}, and possibly
suggest a drastic change in the properties of the plasma between 1.6
and 1.7 $\tc$. 
%%Although further analysis is needed to resolve this
%%issue, it is worth mentioning that no such sharp change is seen in the
%%behavior of the correlators just above 1.5 $\tc$, either by the
%%Bielefeld group (see \Figure~\ref{fig:recon}) or by Umeda \etal
%%~\cite{bnl}.

Unlike the two previous groups, Umeda \etal ~use smeared operators
with a smearing function of the form $w({\vec x}) = \exp (- a |{\vec
x}|^b)$ where the parameters $a$ and $b$ are chosen to optimize the
overlap with the ground state.  While use of a smeared operator has
the advantage of a good overlap with the ground state, so that it may
be possible to extract the properties of \jpsi\ and \ec\ more
reliably, it has two disadvantages.  First, bound state dissolution
must be carefully handled since smearing always mimics a bound state.
Second, the direct connection of the vector current correlator to the
dilepton rate, \Eq~(\ref{eq:dilepton}, is lost. The first problem can
be dealt with by comparing results using a different level of smearing
\cite{umeda02}.

\begin{figure}[t]
\begin{center}
\includegraphics[width=.49\textwidth]{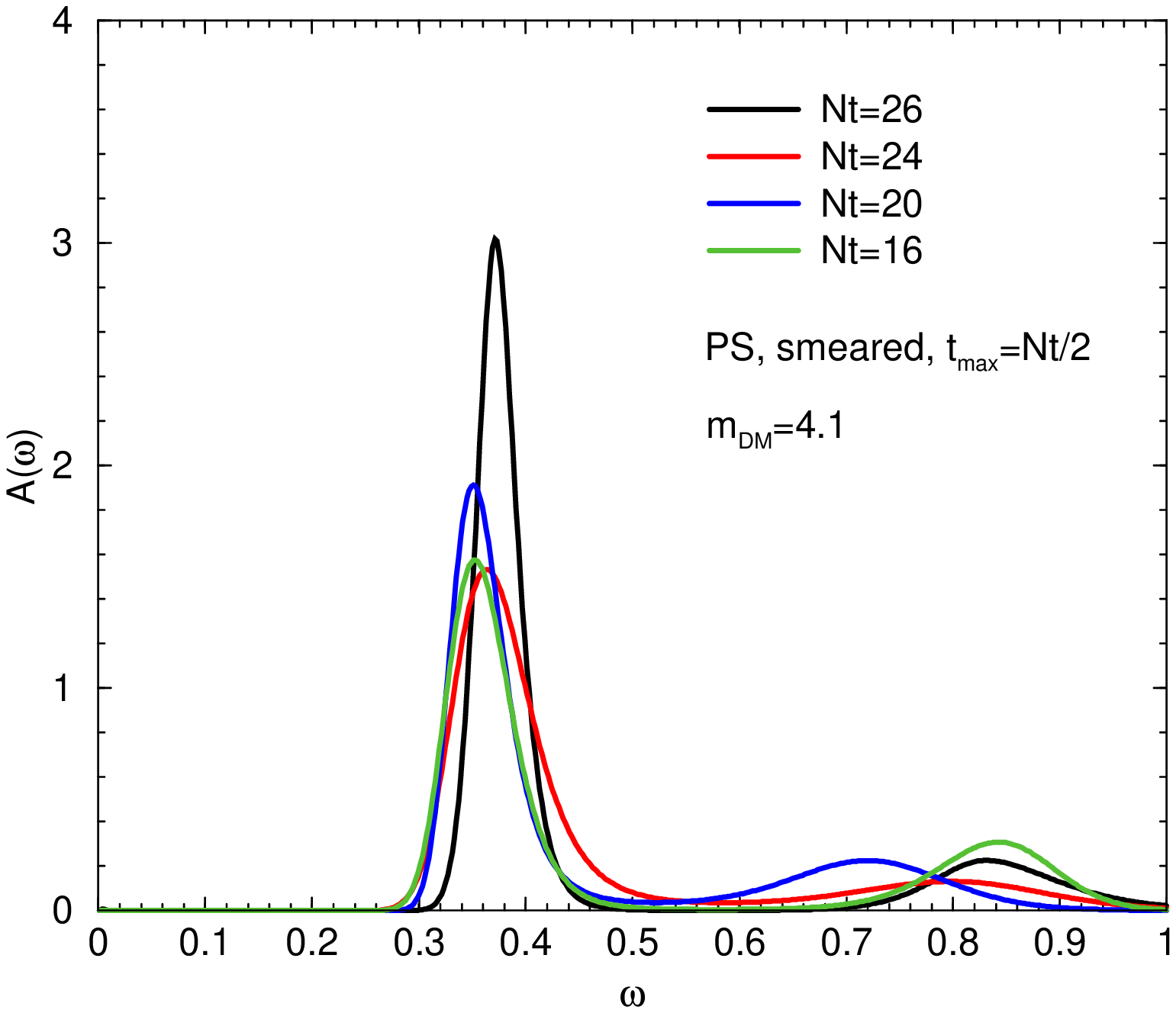}
\hfill
\includegraphics[width=.49\textwidth]{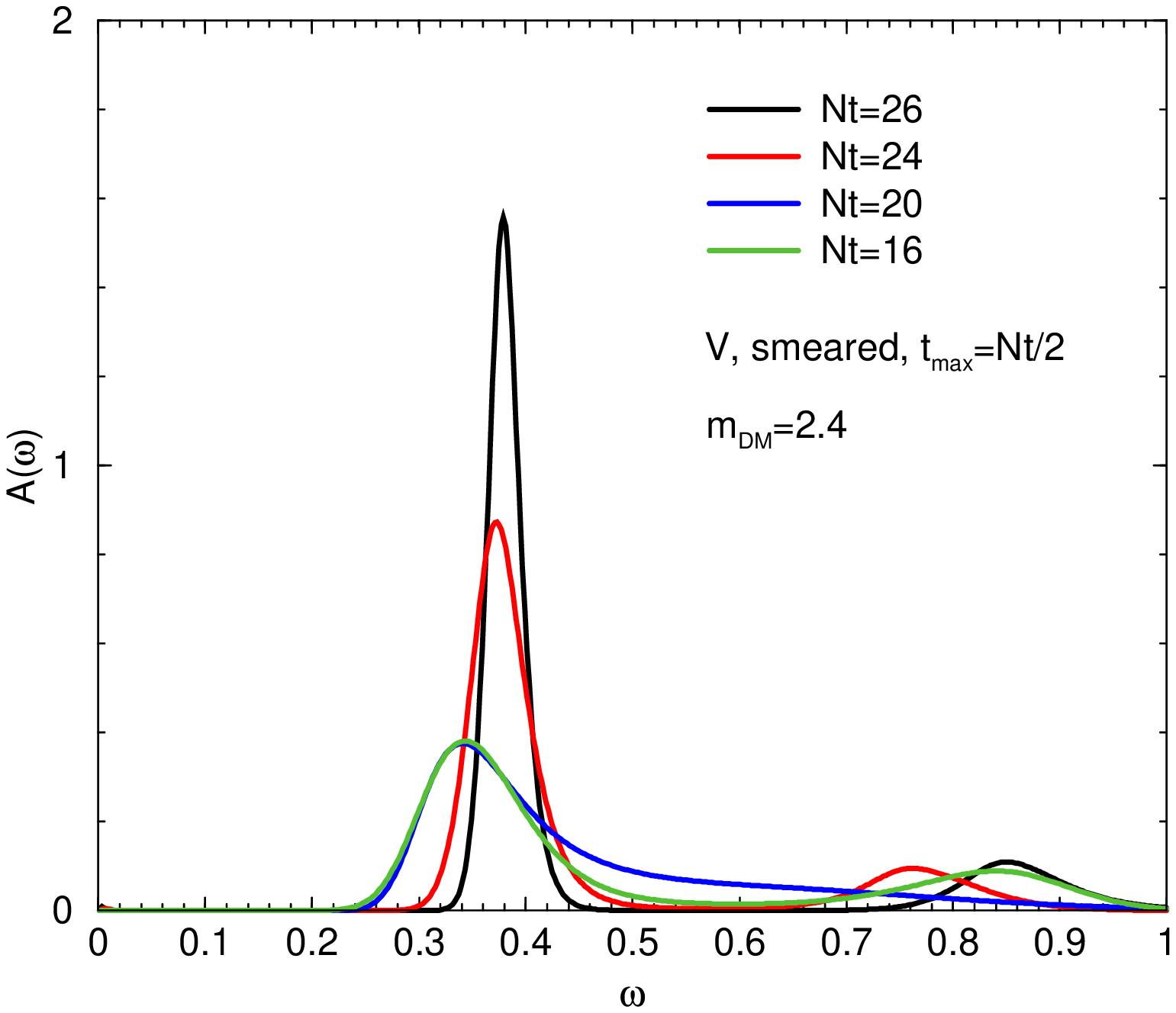}
\end{center}
\caption[Spectral functions above deconfinement 
         for pseudo-scalar and vector channels]
        {Spectral functions above deconfinement for pseudo-scalar
         (left) and vector (right) channels, for anisotropic lattices
         with $a_\tau \approx$ 0.025~fm and smeared operators (Umeda
         \etal ~\cite{umeda02,bnl}). The decreasing $Nt$ correspond to
         $T/T_c \approx$ 1.08, 1.17, 1.4 and 1.75, respectively.}
\label{fig:umeda}
\end{figure}

\Figure[b]~\ref{fig:umeda} shows their results for the $1S$ spectral
functions.  Evidence for the existence of a bound state above \tc\ is
found for the $1S$ states, in agreement with results of other groups
using point operators.  No dramatic change between 1.4 \tc\ and 1.75
\tc\ is indicated in the properties of the peak. In addition, Umeda
\etal ~\cite{umeda02,bnl} attempted a constrained fit of the
correlators to obtain more accurate information on the ground state
properties such as thermal modifications of the mass and width. This
analysis with extended operators indicate that the masses of ground
state charmonia do not change significantly up to temperatures 1.7
\tc. These studies also indicate a non-zero thermal width which
increases with temperature. The precise determination of the width,
however, appears to be difficult.

We now summarize what direct lattice studies have told us about the
properties of the $1S$ states in equilibrium with a plasma. The
studies all agree that the \jpsi\ and the \ec\ survive the
deconfinement transition with little significant change in their
properties. At least up to temperatures $\approx$ 1.5 $\tc$ such
states exist as bound states in the equilibrated plasma without
significant weakening of the state. Here, it may be worthwhile to also
mention earlier results from Umeda \etal ~for these states
\cite{umeda00}.  In this study, where they looked at the fall-off of
the $c \overline c$ spatial correlation, Umeda \etal ~previously
concluded that the $1S$ states survive as bound states, up to
temperatures of 1.5 $\tc$, in a gluonic plasma. Another important fact
revealed by the recent studies is that no significant $1S$ state mass
reduction is seen above $\tc$. If, as was the prevailing wisdom, these
states become Coulombic above $\tc$, one may have expected a
significant mass reduction which is not seen by any of the
groups. Beyond 1.5 \tc\ there seems to be some disagreement between
the different groups: while Ref. \cite{Datta:2003ww} and preliminary
results from Umeda \etal ~suggest a gradual disappearance of the
state, perhaps due to collision broadening, preliminary results from
Asakawa and Hatsuda suggest a sharp cutoff temperature below 1.7
$\tc$, beyond which the gluonic plasma cannot support these bound
states.

\begin{figure}[t]
\begin{center}
\includegraphics[width=.49\textwidth]{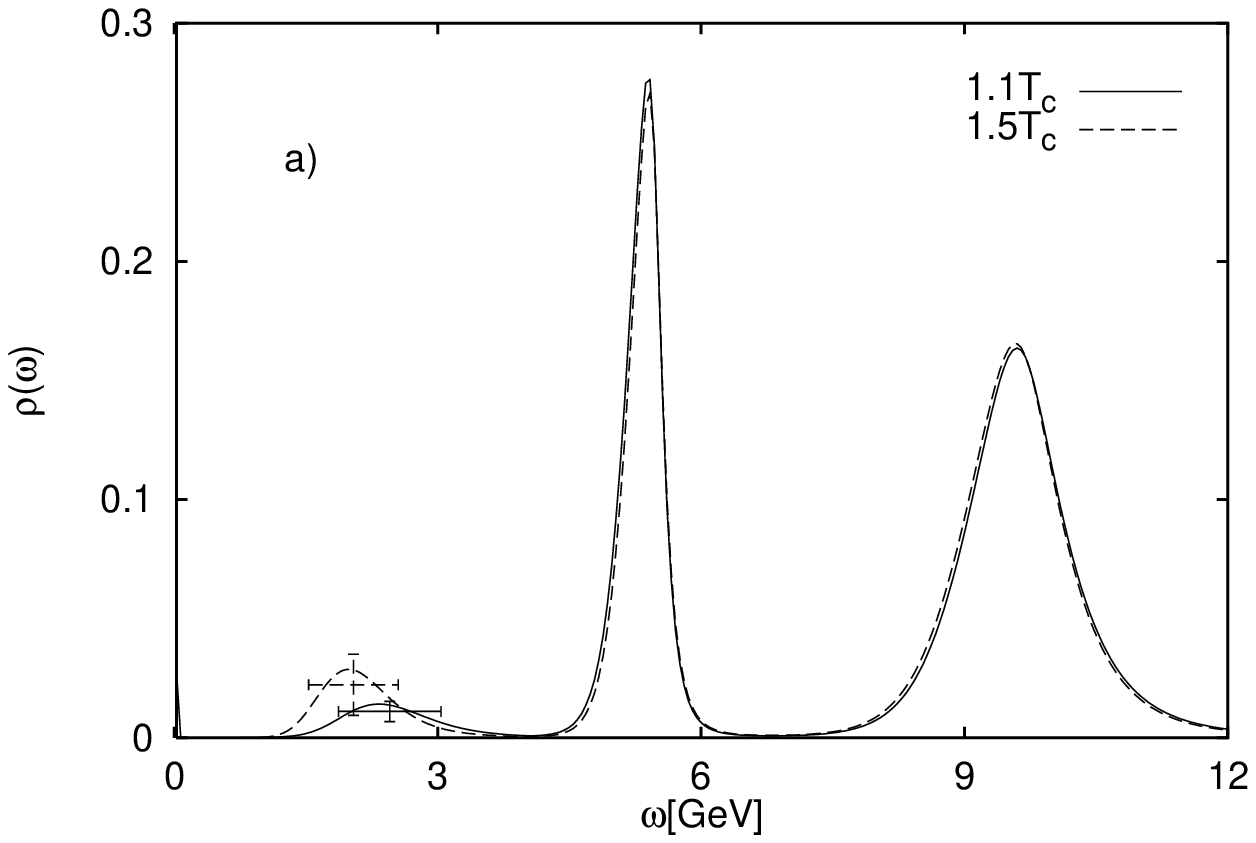}
\hfill
\includegraphics[width=.49\textwidth]{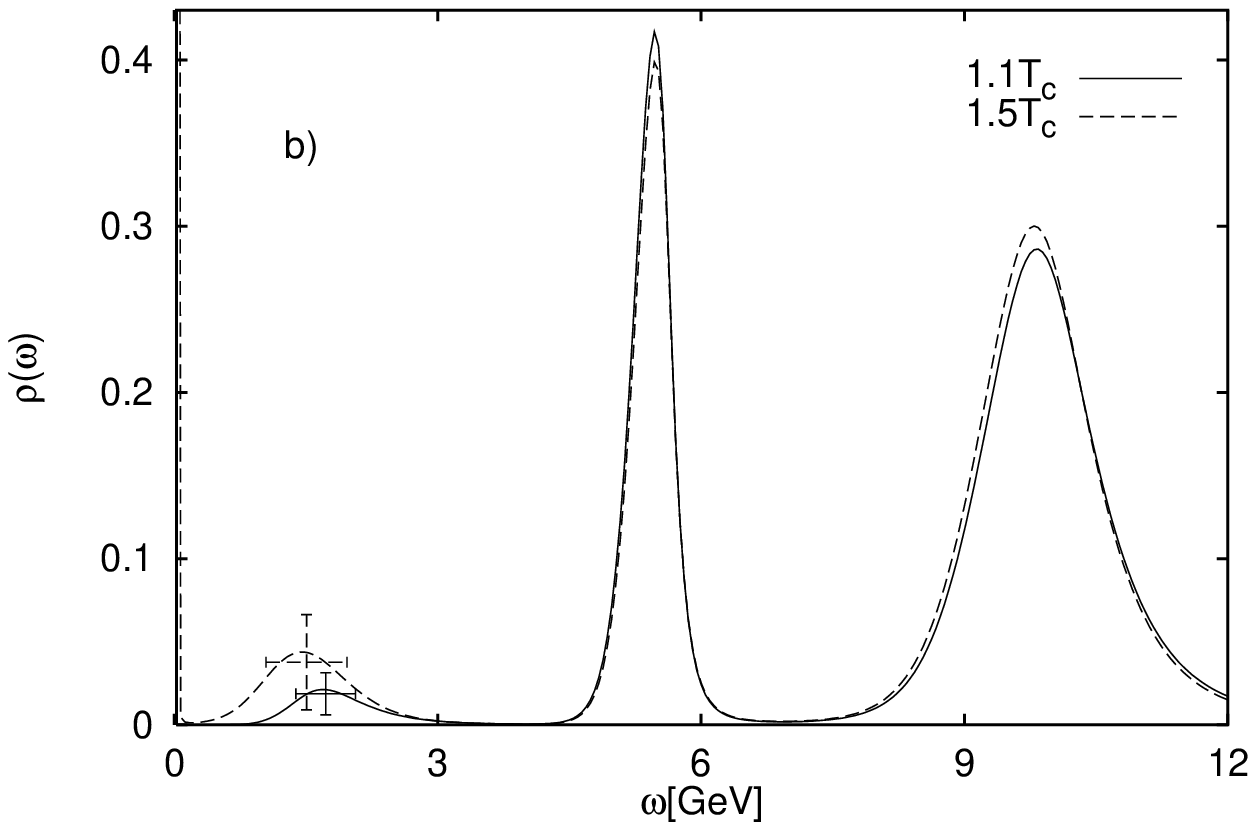}
\end{center}
\caption[Spectral functions for scalar and axial vector states above
         deconfinement]
        {Spectral functions for (a) scalar and (b) axial vector states
         above deconfinement, reconstructed from Matsubara correlators
         measured on lattices of $\approx$ 0.04~fm spacing
         \cite{Datta:2003ww}.}
\label{fig:scax664}
\end{figure}

Results for the orbitally excited $\chi_c$ states are available so far
only from the Bielefeld group. As was shown in
\Figure~\ref{fig:recon}, the behavior of these states is considerably
different from that of the $1S$ states since the correlators above
\tc\ differ substantially from the ones reconstructed from the
spectral function below $\tc$, suggesting serious modification of
these states due to deconfinement. The spectral functions can be
extracted from the correlators even though they are more noisy, making
such an extraction somewhat more difficult
\cite{Datta:2003ww}. \Figure[b]~\ref{fig:scax664}(a) shows the spectral
functions for the scalar channel. As before, as part of the prior
guess we provided the high \om\ structure of the lattice spectral
function in the interacting theory, as obtained below $\tc$. The
figure shows that the \ass\ peak below \tc\ is not present already at
1.1 $\tc$. \Figure[b]~\ref{fig:scax664}(b) shows a similar result for the
axial vector channel, indicating these states suffer serious system
modification, possibly dissolution, already just above $\tc$.

\subsection{Theory perspectives: NRQCD at $T>0$} 

All studies of the charmonium spectral functions mentioned above were
done either using relativistic Wilson fermions or Fermilab
fermions. In both formulations the so-called Wilson term is introduced
to remove the doublers. This, however, strongly distorts the quark
dispersion relation on the lattice leading to the artifact peaks in
the lattice spectral function. For heavy quarks an additional large
discretization error of order $m_q a$ appears where $m_q$ is the heavy
quark mass, making bottomonium studies with relativistic actions very
difficult although the Fermilab formalism can help to a certain
extent.  A useful alternative, at least close to $T_c$, where the
condition $m_q \gg T$ necessary for the approach to be valid is
satisfied, could be to employ NRQCD at $T > 0$.  In this formulation,
the scale related to $m_q$ is integrated out so that this formulation
has the advantage of being doubler free.  Furthermore, since the
spectral functions scale as $\sqrt{\omega}$ instead of the $\omega^2$
behavior of the relativistic formulation, the high energy part of the
spectral functions contribute less to the meson correlator. This can
simplify the reconstruction of meson properties, related to the low
energy part of the vector spectral function. Some preliminary studies
of the charmonium spectral function in NRQCD at zero temperature were
presented in Ref.~\cite{oevers}.

\section[\jpsi ~absorption in heavy-ion collisions]
        {\jpsi\ absorption in heavy-ion collisions at finite temperature
         $\!$\footnote{Section coordinator: A.~D.~Polosa;
         Authors: V.~Laporta, L.~Maiani, F.~Piccinini, 
         A.~D.~Polosa, V.~Riquer}} 

\subsection{Cross-section calculations}

Since the original paper by Matsui and Satz~\cite{Matsui:1986dk} on 
$\jpsi$ suppression in heavy-ion collisions, 
a number of studies on $\jpsi$ absorption in nuclear 
matter have been proposed to suggest other mechanisms than the quark 
gluon plasma screening of the $c\overline c$ potential.
We will use the words ``suppression'' and ``absorption'' to distinguish
between the plasma and non-plasma mechanisms responsible for the observed
reduced yield of $\jpsi$ particles in heavy ions collisions.
 
According to Ref.~\cite{Matsui:1986dk}, one should observe very few $\jpsi$ 
in heavy-ion collisions because 
plasma formation
could weaken the effectiveness of the quarkonium potential and prevent 
$\jpsi$ formation. But what if no plasma phase is generated?
The $\jpsi$ can be produced in the early stages of the 
collision since the energy is high enough to produce them. Their destiny
is then related to the interactions they will experience with the nuclear 
medium (the nuclear thickness the $\jpsi$ must traverse
during the intepenetration of the two colliding nuclei) and with the 
hadron gas (a possible state excited from vacuum left 
behind in the collision region by the two receding nuclei, assuming 
zero baryon number).

The $\jpsi N$ cross-section, $\sigma_{\rm abs}$,  
extracted from data is
$\sigma_{\rm abs} = 4.3\pm 0.6$~mb~\cite{Alessandro}.
The exponential absorption factor for the $\jpsi$ in 
nuclear matter is 
$\propto\mathrm{exp}(-x/\lambda)$,
where $x$ is the distance traveled and $\lambda$ is the mean free path, 
given by
\begin{equation} 
\lambda \approx 1/(\rho_A \sigma_{\rm abs}),
\end{equation}
where $\rho_A$ is the nuclear density.  
Nuclear absorption alone cannot explain the anomalous 
$\jpsi$ suppression described as a function of length
in Ref.~\cite{Abr99}. 
This argument has been used in favor of a plasma interpretation.
On the other hand, nuclear absorption does not take into account
interactions with 
all hadrons ($\pi,\eta,K,\rho,\omega,K^*,\phi...$) 
that can be excited from vacuum when very high 
energy densities are reached, as in heavy-ion collisions.
Each of these particles is presumably able to interact 
with the $\jpsi$, reshuffling its
$c\overline c$ content into an open charm final state such as
$D^{(*)}_{(s)}D^{(*)}_{(s)}$.  Can such interactions provide an 
explanation of the anomalous suppression?

The description of these interactions is difficult because 
they cannot be derived from first principles or 
extracted from independent experimental information. One 
has to resort to models of their dynamics
\cite{tutti,pQCDorig,pQCDimp,quarkintorig,quarkint,sumrules,memorig,mem}.

Dissociation of the $J/\psi$ by hadrons has been considered 
in several approaches, with rather different predictions 
for energy dependence and magnitude of the cross-sections near threshold. 
Basically, earlier calculations in the literature can be 
grouped into four classes: 
\begin{itemize}
\item perturbative QCD based calculations (pQCD); 
\item quark interchange models; 
\item QCD sum rule calculations; 
\item meson exchange models. 
\end{itemize}
The idea behind the pQCD method is that the interaction between heavy quark 
bound states and light hadrons can be described perturbatively when the 
heavy quark mass is sufficiently large. The small size of the heavy quark 
bound state allows for a multipole expansion of its interaction with external 
gluons where the colour-dipole interaction dominates at long range. 
Using pQCD, Peskin and Bhanot and later 
Kharzeev and Satz~\cite{pQCDorig} estimated the scattering cross-sections 
of $J/\psi$ with light hadrons, finding very low values, 
less than $0.1$~mb at about $\sqrt{s} = 5$~GeV. 
The pQCD result was further improved 
by including finite target mass corrections and relativistic phase space
corrections~\cite{pQCDimp}. 
However, the collisions between mesons and the $J/\psi$ in 
heavy-ion reactions occur at low energy where the application of pQCD 
is questionable. 

In the case of quark interchange models, the $J/\psi$ dissociation
cross-section is calculated in terms of non-perturbative quark
exchanges between the $J/\psi$ and light hadrons using explicit
non-relativistic quark model wave functions for the initial and final
hadrons. The largest contributions to the cross-section come from the
energy region just a few hundred MeV above threshold since the overlap
integrals are damped by the external meson wave functions at higher
momenta. The first calculation in this framework was performed by
Martins, Blaschke and Quack~\cite{quarkintorig}, finding a $J/\psi$
dissociation cross-section by pions of $\approx 7$~mb at about 1~GeV
above threshold. Subsequently, similar calculations have been carried
out with different treatments of the confining interaction, obtaining
lower cross-sections (of the order of 1.5~mb at $\sqrt{s}\approx
4$~GeV) and have been extended to $\rho J/\psi$ and $N J/\psi$
interactions~\cite{quarkint}.

Another independent approach is given by the QCD 
sum rules~\cite{sumrules}. 
This method relates the scattering amplitude to a sum of 
operator vacuum-expectation-values (VEVs). 
It gives a model independent result 
provided that the set of selected operators dominates the 
scattering amplitude in the chosen kinematic regime and 
that their VEVs are known experimentally. QCD sum rule
calculations for the reaction $\pi + J/\psi \to D^{(*)} 
\overline D^{(*)}$ give cross-sections at the mb level 
in the near-threshold region. 

\begin{figure}[t]
\begin{center}
\includegraphics[width=6cm]{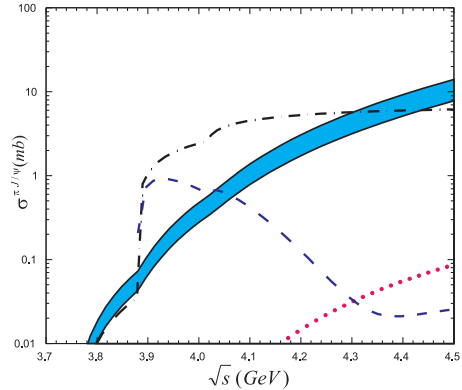}
\caption[Calculated $J/\psi$-hadron cross-sections]
        {The $J/\psi$-hadron cross-sections calculated with QCD sum
         rules (band), short-distance QCD (dotted line),
         meson-exchange models (dot--dashed lines) and the
         non-relativistic constituent quark model (dashed
         line)~\cite{figxsect}.}
\label{fig:xsect}
\end{center}
\end{figure}

Finally, the meson-exchange 
model~\cite{memorig}, 
is based on hadronic effective Lagrangians. Vector mesons are 
treated as gauge bosons 
mediating the interactions between pseudoscalar mesons.
An SU(4)-invariant effective meson Lagrangian is assumed. 
One starts from a free Lagrangian of the form 
${\cal L}_0 = {\rm Tr}(\partial_\mu P^\dagger \partial^\mu P) 
- 1/2 {\rm Tr}(F_{\mu \nu}^\dagger F^{\mu \nu})$ 
and derives the couplings between pseudoscalar and 
vector mesons with the minimal substitution 
\begin{eqnarray}
&& \partial_\mu P \to \partial_\mu P - i/2 g [V_\mu, P] \nonumber \\
&& F_{\mu \nu} \to \partial_\mu V_\nu - \partial_\nu V_\mu - i/2 g 
[V_\mu, V_\nu] \, \, ,  
\end{eqnarray}
giving rise to three-vector meson couplings and four-point couplings. 
The values of the 
couplings are obtained from experimental results via Vector Meson 
Dominance and
SU(4) relations, although the assumption of SU(4) symmetry is rather questionable.
The most studied channels are: 
$\pi J/\psi$, $\rho J/\psi$, $K J/\psi$ and $K^* J/\psi$ where 
quite large cross-sections, of the order of $1 - 10$~mb, 
have been obtained~\cite{mem}.
Since the exchanged mesons are not pointlike, several studies 
have introduced form factors at the interaction vertices at the 
price of introducing additional unknown or poorly known parameters. 
A strong dependence on the shape and cutoff values of the 
form factors is found. Some authors 
calculate the interaction vertices and form factors with the help of sum 
rules~\cite{sumrulesvert}. 
The state of the art of these calculations, including an uncertainty band, 
is summarized in 
\Figure~\ref{fig:xsect}~\cite{figxsect}. 
For a nice review of these topics see Ref.~\cite{barnes}.

\begin{figure}[t]
\begin{center}
\centering\includegraphics[width=10cm]{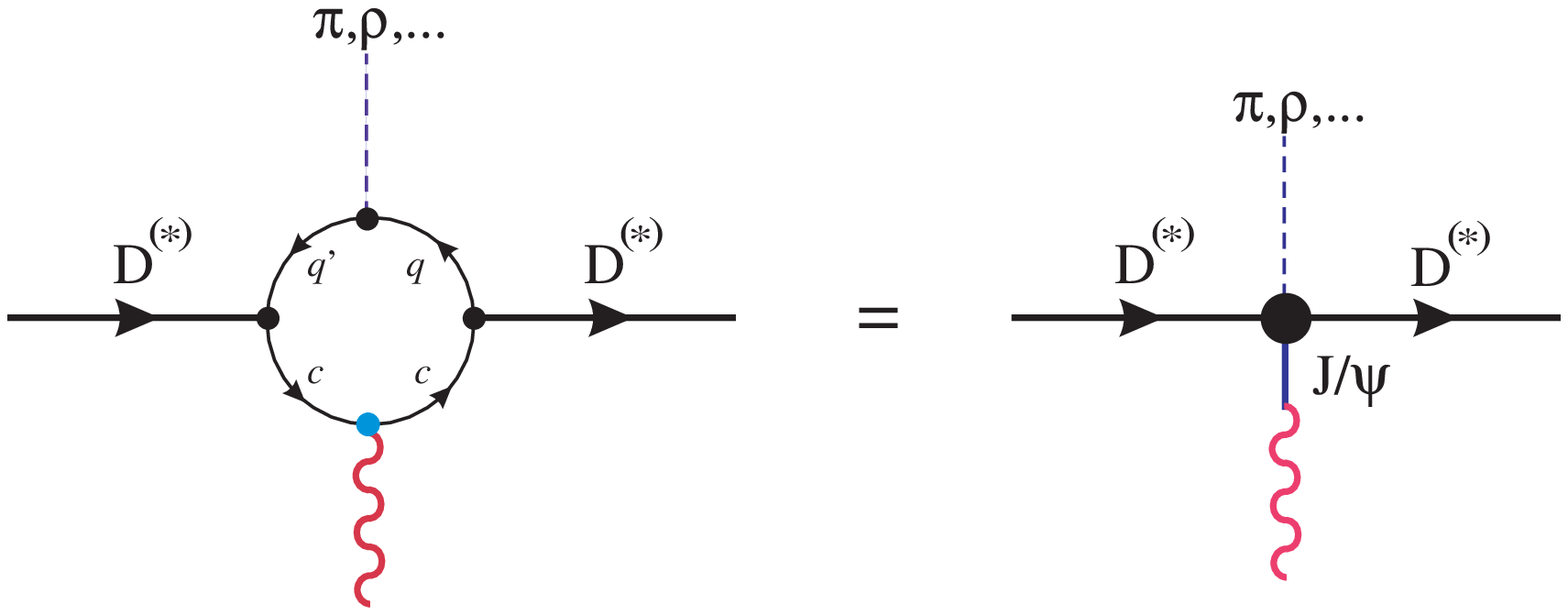}
\caption{Basic diagrammatic equation to compute the couplings $g_3$ and $g_4$.}
\label{fig:loop}

\medskip
\includegraphics[width=145mm]{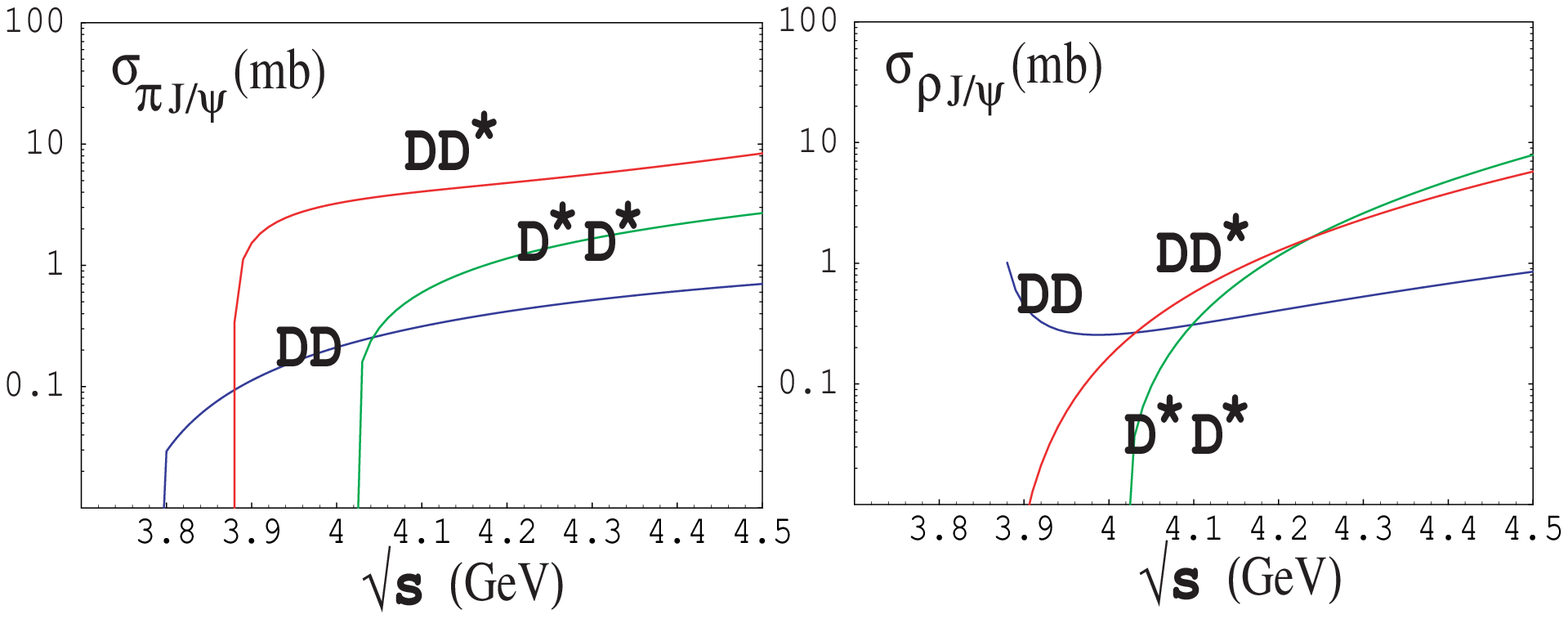}
\caption[Cross-sections in CQM model]
        {Cross-sections in CQM model. On the left-hand side, the
         cross-sections $\pi\jpsi\to D^{(*)} D^{(*)}$ modulated by a
         form factor (ff) directly derived from the model~\cite{pioni}
         are shown.  The case of $\rho\jpsi$ interactions is
         technically more complicated and it does not seem feasible to
         extract the dependency on $E_\rho$ of these cross-sections in
         the form of a polar ff. If we compute physical quantities
         such as the mean free paths determined by the inverse of
         thermal averages, $\langle \rho^{(\rho)}\sigma\rangle_T$,
         where $\rho^{(\rho)}$ is the $\rho$ number density, the
         Boltzmann factor will serve as an exponential ff cutting high
         energy tails faster than any polar ff.}
\label{fig:xsects}
\end{center}
\end{figure}

With the exception of pQCD, these approaches, each subject to
different limitations, give cross-sections on the order of mb, in
particular for initial $\pi$'s and $\rho$'s.  This is a clear
indication that the picture of $\jpsi$ absorption by nuclear matter as
the only alternative mechanism to plasma suppression is incomplete and
that interactions with {\it comoving particles} in the hadron gas have
to be taken into account.  We will not attempt to further review
existing approaches. We instead focus on some more recent calculations
which do not fall in the four classes described above.
%@
The $(\pi,\eta,K,\rho,\omega,K^{*},\phi)\jpsi\to
D^{(*)}_{(s)}D^{(*)}_{(s)}$ cross-sections have recently been
evaluated~\cite{ioniII}, based on the Constituent Quark Model (CQM).
The CQM was originally devised to compute exclusive heavy-light meson
decays and was tested on a large number of such processes~\cite{cqm}.
The CQM is based on an effective Lagrangian which incorporates the
heavy quark spin-flavour symmetries and chiral symmetry in the light
sector. In particular, it contains effective vertices between a heavy
meson and its constituent quarks, as shown on the left-hand side of
\Figure~\ref{fig:loop}, which occur when applying bosonization
techniques to the Nambu--Jona--Lasinio interaction terms of heavy and
light quark fields~\cite{ebert}. On this basis, we believe that the
CQM is a more solid approach than effective Lagrangian methods, often
based on SU(4) symmetry.

We compute the effective trilinear,
$g_3=(\pi,\eta,K,\rho,...)D^{(*)}_{(s)}D^{(*)}_{(s)}$, or $g_3=\jpsi
D^{(*)}_{(s)}D^{(*)}_{(s)}$ and quadrilinear,
$g_4=(\pi,\eta,K,\rho,...)\jpsi D^{(*)}_{(s)}D^{(*)}_{(s)}$,
couplings.
In \Figure~\ref{fig:loop} we show the diagrammatic equation which has
to be solved in order to obtain $g_4(g_3)$ in the various cases.  The
right-hand side represents the effective four-linear coupling to be
used in the cross-section calculation. To obtain the tri-linear
couplings we suppress either the $\jpsi$ or one of the dashed lines
representing the light resonances.  The effective interaction at the
meson level (right-hand side) is modeled as an interaction at the
quark-meson level (left-hand side of \Figure~\ref{fig:loop}).

The $\jpsi$ is introduced using a Vector Meson Dominance~(VMD) ansatz.
In the effective loop on the left-hand side of \Figure~\ref{fig:loop}
we have a vector current insertion on the heavy quark line~$c$ while
on the right-hand side the $\jpsi$ is assumed to dominate the tower of
$J^{PC}=1^{- -}$ $c\overline c$ states mixing with the vector current,
for more details see Refs.~\cite{cern1,pioni}.  Similarly, vector
particles coupled to the light quark component of heavy mesons such as
$\rho$ and $\omega$ when $q=(u,d)$ and $K^*$, $\phi$, when one or both
light quarks are strange, are also taken into account using VMD
arguments.  The pion and other pseudoscalar fields have a derivative
coupling to the light quarks of the Georgi--Manohar
kind~\cite{manogeo}.  The final results for $\sigma_{\pi\jpsi}$ and
$\sigma_{\rho\jpsi}$ are displayed in \Figure~\ref{fig:xsects}.

The authors of Ref.~\cite{Pietroetal} employ a relativistic quark
model \cite{model-1} to calculate amplitudes and cross-sections for
the same processes discussed above with only pions in the initial
state.  Their model is based on an effective Lagrangian which
describes the coupling of hadrons $H$ to their constituent quarks,
$q_1$ and $\overline q_2 $, given by:
\begin{equation}
{\cal L}_{\rm int}(x) = g_H H(x)\int\!\! dx_1
\!\!\int\!\! dx_2 F_H (x,x_1,x_2)\overline q_2(x_2)\Gamma_H\lambda_H
q_1(x_1) \, + {\rm h.c.}\, .
\end{equation}
Here, $\lambda_H$ and $\Gamma_H$ are the Gell-Mann and Dirac matrices
which describe the flavour and spin quantum numbers of the meson
field $H(x)$. The vertex function $F_H$ is given by:
\begin{equation}
F_H(x,x_1,x_2)=\delta\left (x - \frac{m_1}{m_1+m_2} x_1 - \frac{m_2}{m_1+m_2} x_2 \right )
\Phi_H\left((x_1-x_2)^2\right) ,
\end{equation}
where $\Phi_H$ is the correlation function of two constituent
quarks of mass $m_1$ and $m_2$. Moreover, for $\Phi_H$,
in momentum space, they chose the form
$\tilde\Phi_H(k^2_E) \doteq \exp(- k^2_E/\Lambda^2_H)$, where
$k_E$ is a Euclidean momentum \cite{model-2}.
The coupling $g_H$ is determined by the compositeness condition 
discussed in Ref.~\cite{Pietroetal}. By using  the corresponding
Feynman rules, the S-matrix elements describing hadronic interactions
are obtained in terms of a set of quark diagrams. 

In this approach, the dissociation processes 
are described by box and resonance diagrams. 
The details of the cross-section calculations can be found 
in Ref.~\cite{Pietroetal};
here we show the numerical results for the cross-sections
as a function of $\sqrt{s}$.

\begin{figure}[t]
\begin{center}
\includegraphics[width=.6\textwidth]{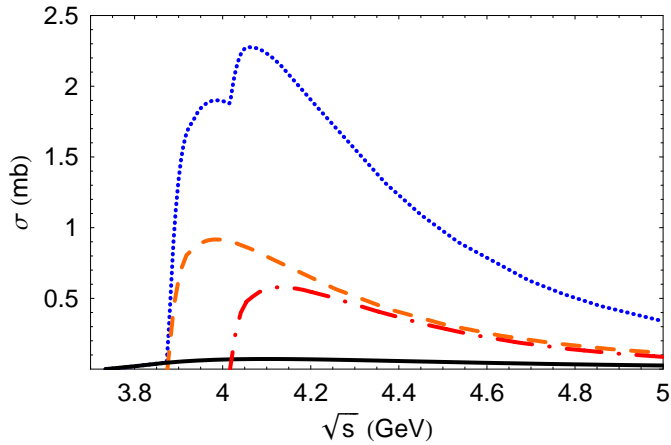}
\end{center}
\caption[The total cross-section together with various contributions]
        {The total cross-section (dotted line) together with the
         contributions coming from the $J/\psi+\pi\to D +
         \overline{D}$ (solid line), $J/\psi+\pi\to D^* +
         \overline{D}$ (dashed line) and $J/\psi+\pi\to D^* +
         \overline{D^*}$ (dot--dashed line) processes are shown.}
\label{fig:sigmatot}
\end{figure}

In \Figure~\ref{fig:sigmatot} the continuous line represents the
$J/\psi+\pi\to D + \overline{D}$ cross-section while the dashed and
the dot--dashed lines are for $J/\psi+\pi\to D^\ast + \overline{D}$ and
$J/\psi+\pi\to D^\ast + \overline{D^\ast}$, respectively.  The dotted
line shows the total cross-section as a function of $\sqrt{s}$.  One
can see that the maximum is about 2.3 mb at $\sqrt{s}\approx 4.1$~GeV.
This value turns out to be smaller than the previous one, but still in
the millibarn range.

To make a realistic computation of the effect of the interactions with
the hadronic gas, several things have to be considered.

\begin{enumerate}
\item There is a temperature dependence of absorption in the hadron
      gas due to the energy dependence of the cross-sections and to
      the fact that, as the temperature varies, the particle content
      and the characteristics of the hadron gas change. One should
      calculate thermal averages $\langle \rho \sigma\rangle_T\approx
      1/\lambda$.  This can be done, \eg by simply using the Bose
      distribution with zero chemical potential in an ideal gas
      approximation~\cite{ioni,ioniII}.

\item There is a problem of convergence. Including heavier resonances,
      on one hand, is disfavored because of the Boltzmann factor
      $\mathrm{exp}(-M/T)$. On the other hand, heavier resonances,
      $h$, integrate more cross-section at low momentum in the
      $h\jpsi\to D\overline{D}$ channel and have large spin
      multiplicities. The lowest lying vector meson nonet, for
      instance, has a charge-spin multiplicity 9 times larger than
      that of pions. It is difficult to asses a priori the relative
      weight of these two effects.
 
\item Some arguments indicate that there may be a limiting temperature
      of hadronic matter: the Hagedorn temperature, $T_H\simeq
      177$~MeV, \cite{hagedorn}. To be credible, a purely hadronic
      interpretation of $\jpsi$~absorption should be for temperatures
      $T\leq T_H$.

\item The SPS collaboration NA50 studies the $\jpsi$ yield as a
      function of the collision centrality.  At a certain centrality
      there is some evidence of a discontinuous breakdown in the
      $\jpsi$ yield ~\cite{na50web,Bor04,San04}.  Can such a
      discontinuity be explained by some hadronic absorption mechanism
      where cross-sections are expected to be mainly smooth
      polynomials?

\end{enumerate}

Studying absorption versus suppression is like estimating the
background to a weak signal. The shape and size of the background can
be crucial in assessing the reliability of the signal.

%@@@
%%%%%%%%%%%%%%%%%%%%%%%%%%%%%%%%%%%%%%%%%%%
\subsection{Comparison with data}

As stated earlier, nuclear absorption can be taken into account by the
factor $\exp(-\rho_A \sigma_{\rm abs} L)$, where $\rho_A$ is the
nuclear density and $L$ the path length the $\jpsi$'s traverse during
the intepenetration of the two colliding nuclei.

To compute the effect of the hadron gas also requires modeling the
fireball produced by the two receding nuclei.  We will briefly
describe the working hypotheses adopted in two recent
papers~\cite{ioni,ioniII}.

\begin{enumerate}
\item The fireball is a zero baryon density region of approximately
      spherical shape which the $\jpsi$ has to escape to be detected.

\item The fireball thermalizes as a hadron gas at temperature $T$ soon
      after its formation. Primary collisions give rise mostly to
      pions with an average energy of about $300$~MeV and a density of
      few/(fm$^3$) and interaction cross-sections of about
      $10$~mb. These parameters lead~\cite{bjorken} to mean free paths
      of a fraction of a fermi, much shorter than the linear
      dimensions of the fireball, $\approx 5-10$~fm.

\item The hadron gas is at zero chemical potential. This is an
      especially reasonable hypothesis for pions.

\item We calculate the thermal averages $\langle\rho\sigma\rangle_T$
      using an ideal gas at temperature $T$. Interactions can be taken
      into account by allowing higher and higher resonances in the
      gas.

\item The NA50 data~\cite{na50web,Bor04,San04} can be plotted as a
      function of collision centrality. The more central the
      collision, the larger the average size of the fireball and the
      higher the energy density.

\end{enumerate}

We take both nuclear and hadronic absorption into account by the
convolution of two exponentials, $\exp[-L(l)/\lambda]
\exp[-l/\lambda(T)]$ where $\lambda(T)$ is the mean free path of the
$J/\psi$ through the hadron gas, $l=2R_A-b$, $R_A$ is the nuclear
radius and $b$ is the impact parameter.  The first exponential is for
nuclear absorption while the second takes the thermal mean free path
through a hadron gas into account.  One can introduce such a
dependence on $l$ by assuming that the energy density, $\epsilon$,
depends on the number of nucleons divided by the effective surface
area of the nuclear interaction as a function of
$b$~\cite{bjorken}. Using the energy--temperature relation appropriate
to the hadron gas, one can determine the temperature profile as
function of centrality and the corresponding absorption profile.

\begin{figure}[t]
\centering\includegraphics[width=.8\linewidth]{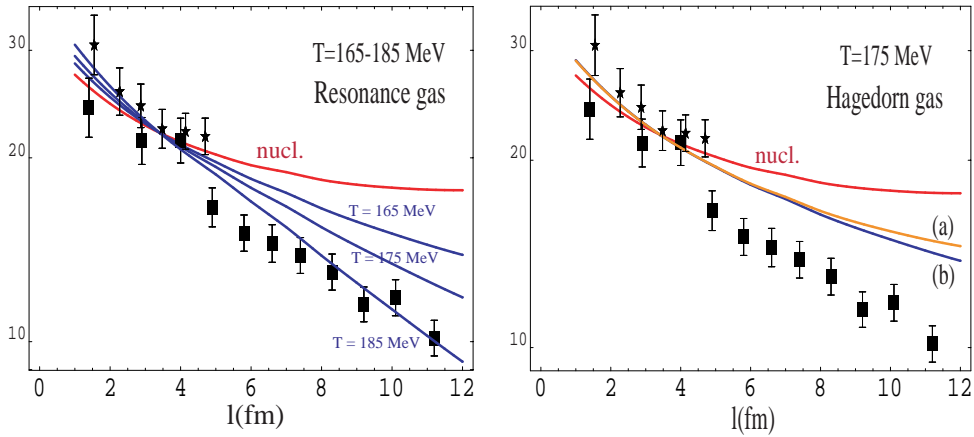}
\caption[The exponential attenuation model compared to Pb+Pb 
         and S+U $J/\psi$ normalized to Drell--Yan]
        {Left-hand side: The exponential attenuation model compared to
         Pb+Pb (boxes) and S+U (stars) $J/\psi$ normalized to
         Drell--Yan for three different values of the initial
         temperature in the hadron gas. The curve labeled as ``nucl.''
         includes only nuclear absorption effects. The other curves
         include both hadron gas and nuclear attenuation. Right-hand
         side: Same but for a Hagedorn gas. The curves labeled (a) and
         (b) show that there is no significant centrality dependence
         of the temperature since (one of the two curves shown has the
         geometrical effect switched off)}
\label{fig:resonancegas}
\end{figure}

In \Figure~\ref{fig:resonancegas}~(a) and (b) we show the calculated
absorption (nuclear + hadron gas) superimposed on the NA50 data
~\cite{na50web,Bor04,San04}. The boxes show the Pb+Pb data while the
stars give the S+U data.  The temperatures indicated on the curves are
the temperatures at low centrality (the first three points from the
left).  The hadron gas picture is more reliable in more peripheral
collisions.

It is very interesting to note that the estimated temperatures are in
the same region as the temperature expected for the phase transition
and close to the temperatures measured at freeze-out from various
hadron abundances. The curve labeled with $T=175$~MeV, where $T$
varies from $175$ to about $195$~MeV, fits the data for low centrality
but still falls short of reproducing the observed drop in $\jpsi$
production above $l=5$~fm.  The curve with $T=185$~MeV fits the low
centrality data and agrees relatively well with the data in central
collisions. However the temperature rises to $200$~MeV at $l\simeq
11$~fm, likely too high for a hadron gas (see below).

The increase in temperature due to the increase in energy density that
we find for the resonance gas is less pronounced than in the case of a
pure pion gas in Ref.~\cite{ioni} because the number of degrees of
freedom in the resonance gas increases appreciably with
temperature. The extra energy density has to be shared among more and
more degrees of freedom and the temperature increases less than with a
fixed $\epsilon= CT^4$ power law. This behaviour begins to reproduce
that expected from a Hagedorn gas with an exponentially increasing
resonance density per unit mass interval~\cite{hagedorn, rafelski,
Cab&Par}.

The extrapolation to increasing centrality using the
energy--temperature relation of the Hagedorn gas is shown in
\Figure~\ref{fig:resonancegas}~(b) with $T=175$~MeV.  The result is
quite spectacular. The sharp rise of the degrees of freedom due to the
vicinity of the Hagedorn temperature makes the temperature of the gas
nearly constant so that the dissociation curve cannot become harder
and the prediction falls far short of explaining the drop observed by
NA50. The simplest interpretation of \Figure~\ref{fig:resonancegas}~(b) is
that with increasing centrality, more energy goes into the excitation
of more and more thermodynamical degrees of freedom, leading to the
final transition to the quark--gluon plasma. The curve shown represents
the limiting absorption from the Hadron gas so that anything harder
would be due to the dissociation of the $\jpsi$ in the quark--gluon
plasma.

Some words of caution are in order. In the framework of our
calculation, it is certainly reasonable to expect the relevant
insertions in the quark loop of \Figure~\ref{fig:loop} to correspond
to the Dirac matrices $S$, $P$, $A$, $V$, and $T$ where the latter are
dominated by the lowest $q {\overline q}$, S-wave states we have
been considering. On the other hand, we cannot exclude that decreasing
couplings of the higher resonances may eventually resum up to a
significant effect which would change the picture obtained by
truncating the cross-section to include only the lowest states.

However, in all cases where this happens, such as in deep inelastic
lepton--hadron scattering, the final result reproduces what happens for
free quarks and gluons. In our case, this would mean going above the
Hagedorn temperature into the quark and gluon gas, precisely what
\Figure~\ref{fig:resonancegas}~(b) seems to tell us.
 
\section[Shadowing and absorption effects on $J/\psi$ production in d+Au
collisions]{Shadowing and absorption effects on $J/\psi$ production in d+Au
collisions 
$\!$\footnote{Author: R.~Vogt}}

The nuclear quark and antiquark distributions have been probed through
deep inelastic scattering (DIS) of leptons and neutrinos from nuclei.
These experiments showed that parton densities in free protons are
modified when bound in the nucleus \cite{Arn}.  This modification,
referred to collectively as shadowing, depends on the parton momentum
fraction $x$ and the square of the momentum transfer, $Q^2$.  Thus, in
addition to the already aforementioned nuclear absorption and
secondary hadronic scattering effects, initial state shadowing can
also play a role, especially as the energy increases, simultaneously
decreasing the values of parton momentum fraction, $x$, probed in the
collision.

Most models of shadowing predict that the modification should vary
depending on position within the nucleus\cite{ekkv4} but DIS
experiments are typically insensitive to this position dependence.
However, some spatial inhomogeneity has been observed in $\nu N$
scattering \cite{E745}.  Here we discuss the combined effects of
shadowing and absorption both in minimum bias d+Au collisions at RHIC
and as a function of centrality.

Our calculations employ the colour evaporation model (CEM) which treats
all charmonium production identically to $c \overline c$ production
below the $D \overline D$ threshold, neglecting colour and spin.  The
leading order (LO) rapidity distributions of $J/\psi$'s produced in
$dA$ collisions at impact parameter $b$ is
\begin{eqnarray} \label{eq:sigmajpsi} 
\frac{d\sigma_{{\rm d}A}}{dy d^2b d^2r}\!\!\! & = & \!\!\! 
2 F_{J/\psi} K_{\rm th} \int \,dz \, dz'
\int_{2m_c}^{2m_D} \!\!\!\!\!\!\!
M dM \left\{ F_g^d(x_1,Q^2,\vec{r},z) 
F_g^A(x_2,Q^2,\vec{b} - \vec{r},z')
\frac{\sigma_{gg}(Q^2)}{M^2} \right. \\
& & 
\mbox{} \left. \!\!\!\!\!\!\!\!\!\! \!\!\!\!\!\!\!\!\!\! \!\!\!\!\!\!\!\!\!\! 
\!\!\!\!\!\!\!\!\!\! + \sum_{q=u,d,s} [F_q^d(x_1,Q^2,\vec{r},z) 
F_{\overline q}^A(x_2,Q^2,\vec{b} - \vec{r},z') +
F_{\overline q}^d(x_1,Q^2,\vec{r},z) F_q^A(x_2,Q^2,\vec{b} - \vec{r},z')] 
\frac{\sigma_{q \overline 
q}(Q^2)}{M^2} \right\}  \, \,  .
\nonumber
\end{eqnarray}
The partonic cross-sections are given in Ref.~\cite{combridge}, $M^2 =
x_1x_2S_{NN}$ and $x_{1,2} = (M/\sqrt{S_{NN}}) \exp(\pm y) \approx
(m_{J/\psi}/\sqrt{S_{NN}})\exp(\pm y)$ where $m_{J/\psi}$ is the
$J/\psi$ mass.  The fraction of $c \overline c$ pairs below the $D
\overline D$ threshold that become $J/\psi$'s, $F_{J/\psi}$, is fixed
at next-to-leading order (NLO) \cite{HPC}.  Both this fraction and the
theoretical $K$ factor, $K_{\rm th}$, drop out of the ratios.  We use
$m_c=1.2$~GeV and $Q = 2m_c$ \cite{HPC}.

We assume that the nuclear parton densities, $F_i^A$, are the product of
the nucleon density in the nucleus, $\rho_A(s)$, the nucleon parton density,
$f_i^N(x,Q^2)$, and a shadowing ratio, $S^j_{{\rm P},{\rm S}}
(A,x,Q^2,\vec{r},z)$,
where $\vec{r}$ and $z$ are the 
transverse and longitudinal location of the parton in position space. 
The first subscript, P, refers to the choice of shadowing parameterization,
while the second,  S, refers to the spatial dependence.
Most available shadowing parameterizations ignore
effects in deuterium.  However, we take the proton and neutron
numbers of both nuclei into account.  Thus,
\begin{eqnarray}
F_i^d(x,Q^2,\vec{r},z) & = & \rho_d(s) f_i^N(x,Q^2) \\
F_j^A(x,Q^2,\vec{b} - \vec{r},z') & = & \rho_A(s') 
S^j_{{\rm P},{\rm S}}(A,x,Q^2,\vec{b} - \vec{r},z') 
f_j^N(x,Q^2) \, \, , \label{eq:fanuc}
\end{eqnarray}
where $s=\sqrt{r^2+z^2}$ and $s' = \sqrt{|\vec{b} - \vec{r}|^2 + z'^2}$.  
With no nuclear modifications, $S^j_{{\rm P},{\rm
S}}(A,x,Q^2,\vec{r},z)\equiv 1$.  The nucleon densities of the heavy
nucleus are assumed to be Woods--Saxon distributions with $R_{\rm Au} =
6.38$~fm and $R_{\rm Pb} = 6.62$~fm \cite{Vvv}.  We use the Hulthen
wave function\cite{hulthen} to calculate the deuteron density
distribution.  The densities are normalized so that $\int d^2r dz
\rho_A(s) = A$. We employ the MRST LO parton densities \cite{mrstlo}
for the free nucleon.

We have chosen shadowing parameterizations developed by two groups
which cover extremes of gluon shadowing at low $x$.  The Eskola {\it
et al.} parameterization, EKS98, is based on the GRV LO \cite{GRV}
parton densities.  Valence quark shadowing is identical for $u$ and
$d$ quarks.  Likewise, the shadowing of $\overline u$, $\overline d$
and $\overline s$ quarks are identical at $Q_0^2$.  Shadowing of the
heavier flavour sea, $\overline s$ and higher, is, however, calculated
and evolved separately at $Q^2 > Q_0^2$.  The shadowing ratios for
each parton type are evolved to LO for $1.5 < Q < 100$~GeV and are
valid for $x \geq 10^{-6}$ \cite{EKRS3,EKRparam}.  Interpolation in
nuclear mass number allows results to be obtained for any input $A$.
The parameterizations by Frankfurt, Guzey and Strikman (FGSo, the
original parameterization, along with FGSh and FGSl for high and low
gluon shadowing) combine Gribov theory with hard diffraction
\cite{FGS}.  They are based on the CTEQ5M \cite{cteq5} parton
densities and evolve each parton species separately to NLO for $4 <
Q^2 < 10^4$~GeV.  Although the $x$ range is $10^{-5} < x < 0.95$, the
sea quark and gluon ratios are unity for $x > 0.2$.  The EKS98 valence
quark shadowing ratios are used as input since Gribov theory does not
predict valence shadowing.  The FGSo parameterization is available for
four different values of $A$: 16, 40, 110 and 206 while FGSh and FGSl
also include $A = 197$.  We use $A = 206$ for the gold nucleus with
FGSo and $A=197$ for the other parameterizations.

\begin{figure}[t]
\begin{center}
\includegraphics[width=.8\textwidth]{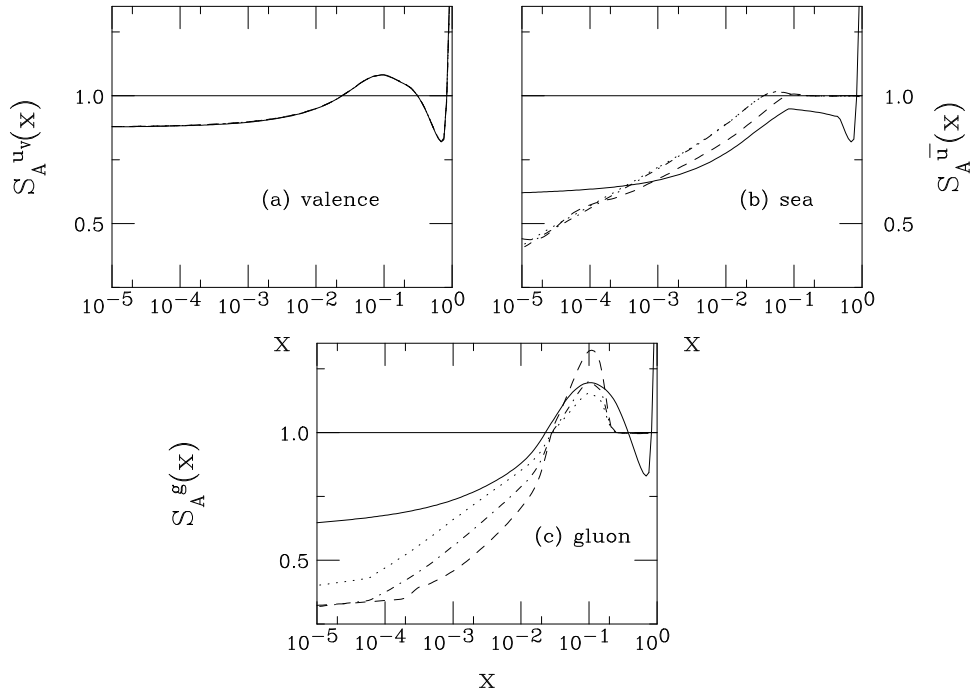}
\end{center}
\caption[Shadowing parameterizations compared at
         the scale $\mu = 2m_c = 2.4$~GeV]
        {The shadowing parameterizations are compared at the scale
         $\mu = 2m_c = 2.4$~GeV.  The solid curves are EKS98, the
         dashed, FGSo, dot--dashed, FGSh, and dotted, FGSl.}
\label{fig:fshadow}
\end{figure}

\Figure[b]~\ref{fig:fshadow} compares the four homogeneous ratios,
$S_{\rm EKS98}$ and $S_{\rm FGS}$ for $Q=2m_c$.  The FGSo calculation
predicts far more shadowing at small $x$ and larger antishadowing at
$x\sim 0.1$.  The difference is especially large for gluons.  At very
low $x$, the gluon ratios for FGSo and FGSh are quite similar but, in
the intermediate $x$ regime, the FGSh parameterization drops off more
smoothly.  On the other hand, the FGSl parameterization levels off at
a higher value of $S_P^i$ than the other two FGS parameterizations.
In the antishadowing regime, FGSh and FGSl are rather similar to the
EKS98 result.

We now turn to the spatial dependence of the shadowing.  
We show results for a parameterization proportional to the parton
path length through the nucleus \cite{psidaprl},
\begin{eqnarray}
S^j_{\rho}(A,x,Q^2,\vec{r},z) = 1 + N_\rho (S^j(A,x,Q^2) - 1)
\frac{\int dz \rho_A(\vec{r},z)}{\int dz \rho_A(0,z)} \, \, .
\end{eqnarray}
where $N_\rho$ is chosen so that $(1/A) \int d^2r dz \rho_A(s)
S^j_{\rho}(A,x,\mu^2,\vec b,z) = S^j(A,x,\mu^2)$. When $s \gg R_A$,
the nucleons behave as free particles while in the centre of the
nucleus, the modifications are larger than the average value $S^j$.
The normalization requires $(1/A) \int d^2r dz \rho_A(s) S^j_{{\rm
P},\rho} = S^j_{\rm P}$.  While there are three homogeneous FGS
parameterizations, only two inhomogeneous parameterizations are
provided.  No spatial dependence is given for FGSo, the case with the
strongest gluon shadowing.  We have checked the available dependencies
against those calculated using $S^j_{\rm FGSo,WS}$ and $S^j_{{\rm
FGSo},\rho}$ and found that, at similar values of the homogeneous
shadowing ratios, $S^j_{{\rm FGSo},\rho}$ is quite compatible with the
available FGS inhomogeneous parameterizations.  Therefore, to
characterize the spatial dependence of FGSo and EKS98, we use $S_{{\rm
P},\rho}^j$ while the given inhomogeneous parameterizations are used
for FGSh and FGSl.
 
\begin{figure}[t]
\begin{center}
\includegraphics[width=\textwidth]{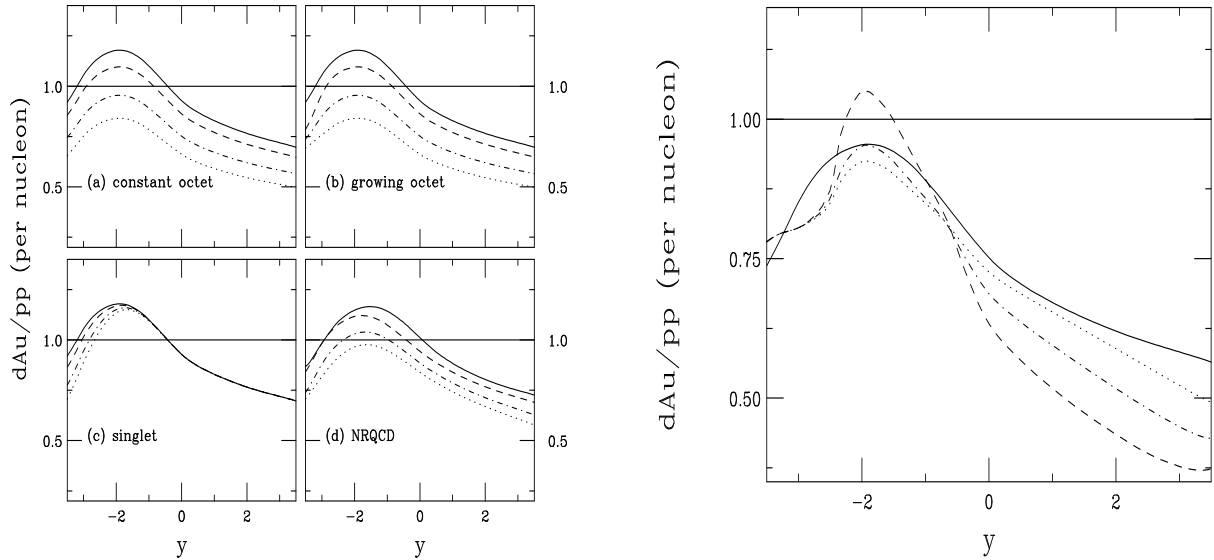}
\end{center}
\caption[The $J/\psi$ dAu/$pp$ ratio at 200~GeV]
        {Left-hand side: The $J/\psi$ dAu/$pp$ ratio with EKS98 at 200
         GeV as a function of rapidity for (a) constant octet, (b)
         growing octet, (c) singlet, all calculated in the CEM and (d)
         NRQCD.  For (a)--(c), the curves are no absorption (solid),
         $\sigma_{\rm abs} = 1$ (dashed), 3 (dot--dashed) and 5 mb
         (dotted).  For (d), we show no absorption (solid), 1 mb
         octet/1 mb singlet (dashed), 3 mb octet/3 mb singlet
         (dot--dashed), and 5 mb octet/3 mb singlet (dotted).
         Right-hand side: The $J/\psi$ dAu/$pp$ ratio at 200~GeV for a
         growing octet with $\sigma_{\rm abs} = 3$ mb is compared for
         four shadowing parameterizations.  We show the EKS98 (solid),
         FGSo (dashed), FGSh (dot--dashed) and FGSl (dotted) results as
         a function of rapidity.}
\label{fig:abs}
\end{figure}

To implement nuclear absorption on $J/\psi$ production in d$A$
collisions, the production cross-section in \Eq~(\ref{eq:sigmajpsi})
is weighted by the survival probability, $S^{\rm abs}$, so that
\begin{eqnarray} 
S^{\rm abs}(\vec b - \vec s,z^\prime) = \exp \left\{
-\int_{z^\prime}^{\infty} dz^{\prime \prime} 
\rho_A (\vec b - \vec s,z^{\prime \prime}) 
\sigma_{\rm abs}(z^{\prime \prime} - z^\prime)\right\} \, \, . 
\label{eq:nsurv} 
\end{eqnarray}
where $z^\prime$ is the longitudinal production point, as in
\Eq~(\ref{eq:fanuc}), and $z^{\prime \prime}$ is the point at which
the state is absorbed.  If shadowing is not considered and $S^{\rm
abs} = 1$, $\sigma_{{\rm d}A} = 2A \sigma_{pN}$.  For $S^{\rm abs}
\neq 1$, $\sigma_{{\rm d}A} = 2A^\alpha \sigma_{pN}$.  The nucleon
absorption cross-section, $\sigma_{\rm abs}$, depends on where the
state is produced and how far it travels through nuclear matter.  The
effective $A$ dependence is obtained from \Eqs~(\ref{eq:sigmajpsi})
and (\ref{eq:nsurv}) by integrating over $z'$, $z$, and $b$.  The
contribution to the full $A$ dependence in $\alpha(x_F)$ from
absorption alone is only constant if $\sigma_{\rm abs}$ is constant
and independent of the production mechanism \cite{rvherab}.  The
observed $J/\psi$ yield includes feed down from $\chi_{cJ}$ and
$\psi'$ decays, giving
\begin{eqnarray}
S_{J/\psi}^{\rm abs}(\vec b - \vec s,z') = 0.58 S_{J/\psi, \, {\rm dir}}^{\rm 
abs}(\vec b - \vec s,z')
+ 0.3 S_{\chi_{cJ}}^{\rm abs}(\vec b - \vec s,z') + 0.12 S_{\psi'}^{\rm 
abs}(\vec b - \vec s,z') \, \, . 
\label{eq:psisurv}
\end{eqnarray}
In colour singlet production, the final state absorption cross-section
depends on the size of the $c \overline c$ pair as it traverses the
nucleus, allowing absorption to be effective only while the
cross-section is growing toward its asymptotic size inside the target.
On the other hand, if the $c \overline c$ is only produced as a colour
octet, hadronization will occur only after the pair has traversed the
target except at very backward rapidity.  We have considered a
constant octet cross-section, as well as one that reverts to a colour
singlet at backward rapidities.  For singlets, $S_{J/\psi, \, {\rm
dir}}^{\rm abs} \neq S_{\chi_{cJ}}^{\rm abs} \neq S_{\psi'}^{\rm abs}$
but, with octets, one assumes that $S_{J/\psi, \, {\rm dir}}^{\rm abs}
= S_{\chi_{cJ}}^{\rm abs} = S_{\psi'}^{\rm abs}$.  As can be seen in
\Figure~\ref{fig:abs}, the difference between the constant and growing
octet assumptions is quite small at large $\sqrt{S}$ with only a small
singlet effect at $y< -2$.  Singlet absorption is also important only
at similar rapidities and is otherwise not different from shadowing
alone.  Finally, we have also considered a combination of octet and
singlet absorption in the context of the NRQCD model, see
Ref.~\cite{rvherab} for more details.  The combination of
nonperturbative singlet and octet parameters changes the shape of the
shadowing ratio slightly.  The results are shown integrated over
impact parameter for the EKS98 shadowing parameterization since it
gives good agreement with the trend of the PHENIX data shown later in
this chapter.
 
Several values of the asymptotic absorption cross-section,
$\sigma_{\rm abs} = 1$, 3 and 5 mb, corresponding to $\alpha = 0.98$,
0.95 and 0.92 respectively for absorption alone in \eg beryllium and
tungsten targets are shown in \Figure~\ref{fig:abs}.  These values of
$\sigma_{\rm abs}$ are somewhat smaller than those obtained for the
sharp sphere approximation where the relation between $\sigma_{\rm
abs}$ and $\alpha$ can be calculated analytically: $\sigma_{\rm abs} =
16\pi r_0^2 (1 - \alpha)/9$.  The diffuse surface of a real nucleus
and the longer range of the density distribution results in a smaller
value of $\sigma_{\rm abs}$ than that found for a sharp sphere
nucleus.

The right-hand side of \Figure~\ref{fig:abs} compares the EKS98
parameterization and $\sigma_{\rm abs} = 3$ mb with the FGS
parameterizations at the same value of $\sigma_{\rm abs}$.  In the
region that PHENIX can measure, the EKS98 and FGSl results are
essentially indistinguishable.  The FGSh result lies between the FGSo
and EKS98 results at forward rapidity but is also quite similar to
FGSh at negative rapidity.

\begin{figure}[t]
\begin{center}
\includegraphics[width=.95\textwidth]{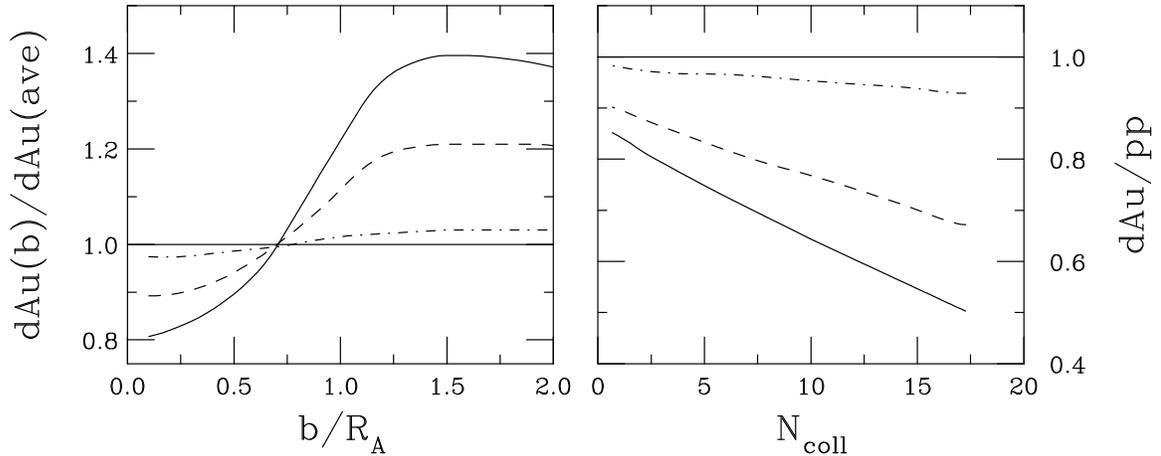}
\end{center}
\caption[The $J/\psi$ (dAu(b)/$pp$)/(dAu(ave)/$pp$) and dAu/$pp$ ratios]
        {Left-hand side: The $J/\psi$ (dAu(b)/$pp$)/(dAu(ave)/$pp$)
         ratio as a function of $b/R_A$.  Right-hand side: The ratio
         dAu/$pp$ as a function of $N_{\rm coll}$.  Results are shown
         for $y=-2$ (dot--dashed), $y=0$ (dashed) and $y=2$ (solid) at
         200~GeV for a growing octet with $\sigma_{\rm abs} = 3$ mb
         and the EKS98 parameterization.}
\label{fig:bdep}
\end{figure}

In central collisions, the inhomogeneous shadowing is stronger than
the homogeneous result.  The stronger the homogeneous shadowing, the
larger the inhomogeneity.  In peripheral collisions, the inhomogenous
effects are somewhat weaker than the homogenous results but some
shadowing is still present.  Shadowing persists in part because the
density in a heavy nucleus is large and approximately constant except
close to the surface and partly because the deuteron wave function has
a long tail.  We also expect absorption to be a stronger effect in
central collisions.  In \Figure~\ref{fig:bdep}, we show the
inhomogeneous shadowing and absorption results for EKS98 and
$\sigma_{\rm abs} = 3$ mb as a function of $b/R_A$ for the dAu/$pp$
ratio as a function of $b$ relative to the minimum bias ratio on the
left-hand side and the ratio dAu/$pp$ as a function of the number of
binary collisions, $N_{\rm coll}$, on the right-hand side.  The ratios
are shown for several values of rapidity to represent the behavior in
the anti-shadowing (large negative $y$), shadowing (large positive
$y$) and transition regions (midrapidity).

The (dAu$(b)$/$pp$)/(dAu/$pp$) ratios, denoted dAu$(b)$/dAu(ave) on
the $y$-axis of the left-hand figure, are all less than unity for
$b/R_A < 0.7$, with stronger than average shadowing and absorption,
and rise above unity for large $b/R_A$, weaker than average shadowing
and absorption.  The right-hand side shows the dAu/$pp$ ratios for the
same rapidity values as a function of the number of collisions,
$N_{\rm coll}$.  The dependence of the ratios on $N_{\rm coll}$ is
almost linear.  We do not show results for $N_{\rm coll} < 1$,
corresponding to $b/R_A > 1.3$ on the left-hand side, the point where
those ratios begin to flatten out.  The weakest $N_{\rm coll}$
dependence occurs where the shadowing effect itself is weakest, at $y
= -2$ at RHIC, in the antishadowing region, as expected.  The trends
of the ratios as a function of $N_{\rm coll}$ are consistent with the
PHENIX data from the north muon arm ($y = 2$) and the electron arms
($y=0$) but the PHENIX results from the south arm ($y=-2$) are much
stronger than our predictions.
\longpage

Thus the combination of shadowing and absorption seems to be in good
agreement with the PHENIX data from RHIC.  It is more difficult to
make predictions of shadowing for the CERN SPS since the average
values of $x$ at which the $J/\psi$ is produced are much higher, $x
\sim 0.16$ at $\sqrt{S_{NN}} = 20$~GeV.  In this $x$ region, the
nuclear gluon shadowing ratio is either nearly crossing unity into the
EMC region (EKS98) or assumed to be unity (FGS), see
\Figure~\ref{fig:fshadow}.  Therefore we have only shown results for
RHIC.  At the LHC, the $x$ values probed are significantly lower,
leading to a stronger shadowing effect over all rapidity, see
Ref.~\cite{psidaprl}.  Combining $J/\psi$ and $\Upsilon$ production
results in d$A$ collisions at RHIC and the LHC could help map out the
nuclear gluon distribution in $x$ and $Q^2$, both in minimum bias
collisions and as a function of centrality.

\section[Quarkonium dissociation in hot QCD matter]
        {Quarkonium dissociation in hot QCD matter
         $\!$\footnote{Author: D.~Kharzeev}}

The use of heavy quarkonium suppression as a signature of
deconfinement \cite{Matsui:1986dk} requires understanding dissociation
mechanisms in the confined and deconfined phases of QCD matter.  In
this section, we comment on the current calculations of quarkonium
dissociation rates and possible ways to improve them.
\longpage

In the confined hadron gas, the interactions of quarkonium are
traditionally treated in a low-density approximation to the kinetic
approach, where the survival probability is expressed through the
quarkonium--hadron dissociation cross-sections. An overview of
different approaches to calculating these cross-sections has been
given previously.  As discussed there, all of these approaches rely on
certain assumptions and approximations, the accuracy of which is often
difficult to assess {\it a priori}. Nonperturbative interactions of
light hadrons are still beyond the reach of reliable theroretical
calculations, so the hope is that the large heavy quark mass forming
the quarkonium bound state can lead to simplifications.

Among the approaches discussed before, the short-distance QCD approach
\cite{Bhanot:1979vb,Kharzeev:1994pz} is based on the assumption that
the heavy quark mass is sufficiently large, $m_q \gg \Lambda_{\rm
QCD}$, and the corresponding bound state size, $R \sim 1/(m_q
\alpha_s)$, is sufficiently small for interactions with light hadrons
of size $\sim 1/ \Lambda_{\rm QCD}$ can be treated by using the
multipole, or operator product, expansion.  In leading twist (or, in
the more intuitive language of the QCD multipole expansion, in the
leading electric dipole approximation), the cross-section is expressed
through Wilson coefficients (for the leading electric dipole operator,
they describe the electric polarizability of heavy quarkonium), and
the gluon structure functions of light hadrons at the scale determined
by the size of the $Q \overline{Q}$ state.

There are two main caveats to this approach. An analysis of the
applicability of the multipole expansion shows that not only the
quarkonium size should be small, $R \sim 1/(m_q \alpha_s) \ll
\Lambda_{\rm QCD}$, but also the binding energy (which determines, by
the uncertainty principle, the characteristic interaction time) has to
be large, $\epsilon \sim m_q \alpha_s^2 \gg \Lambda_{\rm QCD}$. The
latter inequality is only marginally justified for charmonium, at the
borderline between the perturbative and nonperturbative regimes. The
second caveat is the knowledge of the gluon structure function at
large Bjorken $x$, corresponding to low energies and relatively small
virtuality, $\sim \epsilon$.  For the $J/\psi$, the binding energy is
$\epsilon = 2M_D - M_{J/\psi} \simeq 0.64$~GeV. The gluon structure
function is not well determined at such low scales. Nevertheless, at
large $x$ the gluon structure functions have to be relatively
suppressed by quark counting rules which dictate at least a $\sim
(1-x)^4$ suppression at $x \to 1$ since the valence quarks dominate
the light hadron structure functions at large $x$.

This naturally brings us to the possibility that the light quark
exchanges, despite being suppressed in the heavy quark limit, become
important in the dissociation of charmonium states\,---\,in fact, most
of the approaches reviewed previously model such exchanges in
different ways.  At least in principle the picture can be clarified by
a lattice calculation of the quarkonium matrix elements involving
quark and gluon operators of different dimension. Such a calculation
could establish a hierarchy of different mechanisms of quarkonium
dissociation. It would also be important for understanding the
dynamics of quarkonium production and decay.  On the phenomenological
side, many of the approaches can be tested in quarkonium decays, \eg
$\psi' \to \psi X$ (see \eg Ref.~\cite{Fujii:1999xn} and references
therein).

Apart from the magnitude of the quarkonium dissociation cross-section,
one should also examine the validity of the low-density approximation
in the kinetic approach. This approach leads to the quarkonium
dissociation rate, $R = \sum_i v_i \sigma_i \rho_i$, where the sum
runs over different hadron species with densities $\rho_i$ and
dissociation cross-sections $\sigma_i$ and $v_i$ are the corresponding
relative quarkonium--hadron velocities. The survival probability of
heavy quarkonium is then obtained by integrating over the time history
of the hadron gas. For example, an isentropic longitudinal expansion
yields (see \eg Ref.~\cite{Kharzeev:1996yx})
\begin{equation}
S \sim \exp\left( - \sum_i v_i \sigma_i \rho_i 
\ln\left(\frac{\rho_i}{\rho_f}\right)\right),
\end{equation}        
where $\rho^i_f$ is a "freeze-out" density at which the system falls
apart and $\rho_i$ corresponds to the initial densities of the
different hadron species.

Such treatment assumes the dominance of two-body interactions of the
quarkonium and thus applies only at sufficiently low temperature. The
dissociation process, by the uncertainty principle, takes place over a
time inversely proportional to the binding energy, $\sim
1/\epsilon$. The typical time between subsequent thermal interactions
at temperature $T$ is $\sim 1/T$. The condition for the applicability
of the low-density approximation is thus $\epsilon(T)/T \gg 1$.  The
binding energy can be modified in a thermal system and is thus a
function of temperature.  Lattice results presented in the previous
sections, especially Section 3.3, indicate no substantial modification
of the binding energy below $T_c$ so that up to $T_c \sim 200$~MeV the
ratio $\epsilon(T)/T \simeq 3$ is likely to be large enough to justify
the kinetic approach.

In the deconfined phase, the original screening scenario can be seen
to correspond to the opposite ``weak coupling" limit of $
\epsilon(T)/T \ll 1$. Indeed, the heavy quarkonium state binding
energy vanishes when it is screened out of existence.  The lattice
results presented in Section 3 indicate that this does not happen for
the $J/\psi$ until $T \approx 1.5 \, T_c$. Moreover, the previous
lattice results indicate no significant change in the $J/\psi$ mass up
to these temperatures, suggesting that the weak coupling approach is
not appropriate even for $T_c \le T \le 1.5 \, T_c$.  However, this
does not mean that the quarkonia are not dissociated at temperatures
below $1.5 \, T_c$ since ``ionization" of heavy quarkonia by gluons
\cite{Kharzeev:1994pz,Shuryak:1978ij} is still possible.  An estimate
of the dissociation rate in this regime was given in
Ref. \cite{Kharzeev:1995ju},
\begin{equation}
R_{\rm act}= \frac{(LT)^2}{6\pi}\ m_q\ \exp(-\epsilon/T) \, \, .
\end{equation}
Here $L$ is the size of the $q \overline{q}$ system which is generally
temperature dependent and can, in principle, exceed the typical
hadronic size, $\sim 1/\Lambda_{\rm QCD}$, in the deconfined
phase. While this rate is moderate (at $T= 300$~MeV and $L = 1$~fm, we
find $R_{\rm act} \simeq 0.05$~fm$^{-1}$), a medium with a lifetime of
$\approx 10$~fm can reduce the survival probability by factor of
two. Thus, one cannot presently conclude from the lattice results that
there is no $J/\psi$ suppression in the deconfined phase below \, $1.5
T_c$.  A conclusive answer can be given by lattice calculations once a
reliable extraction of the $J/\psi$ width becomes possible.
 
Once the weak coupling limit of $ \epsilon(T)/T \ll 1$ is reached,
dissociation occurs very rapidly with a rate \cite{Kharzeev:1995ju}
\begin{equation}
R_{\rm act}= \frac{4}{L} \sqrt{\frac{T}{\pi m_q}}  \, \, . \label{eq:weakrate}
\end{equation}
This expression is easy to understand once we recall that the thermal
velocity of a free particle in three dimensions is $v_{\rm th}(T)=4
\sqrt{ T / \pi m_q}$ where we recover the classical high-temperature
limit for the thermal activation rate
\begin{equation}
R_{\rm act}= \frac{v_{\mathrm{th}}(T)}{L} \, \, .
\end{equation}
An estimate of \Eq~(\ref{eq:weakrate}) gives a large rate, $R_{\rm
act} \sim 1$/fm suggesting that once the binding energy of quarkonium
becomes small compared to the temperature, dissociation occurs very
rapidly and should lead to strong suppression, as envisioned in the
original scenario \cite{Matsui:1986dk}.

\section[Secondary charmonium production and charm--quark coalescence]
        {Secondary charmonium production and charm--quark coalescence
         $\!$\footnote{Author: R.~Rapp}}

\begin{figure}[!t]
\begin{center}
\includegraphics[width=.48\textwidth]{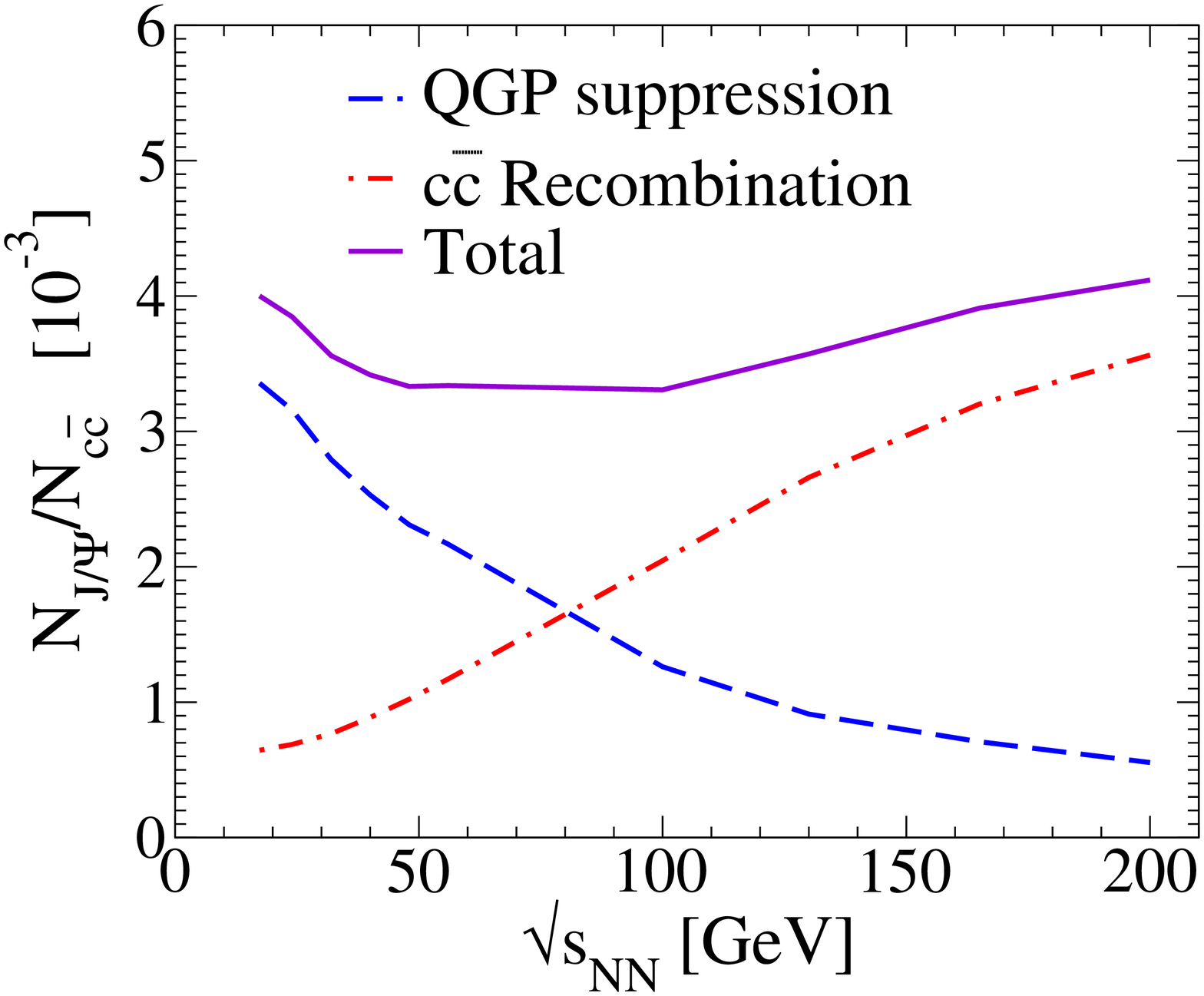}
\hfill
\includegraphics[width=.48\textwidth]{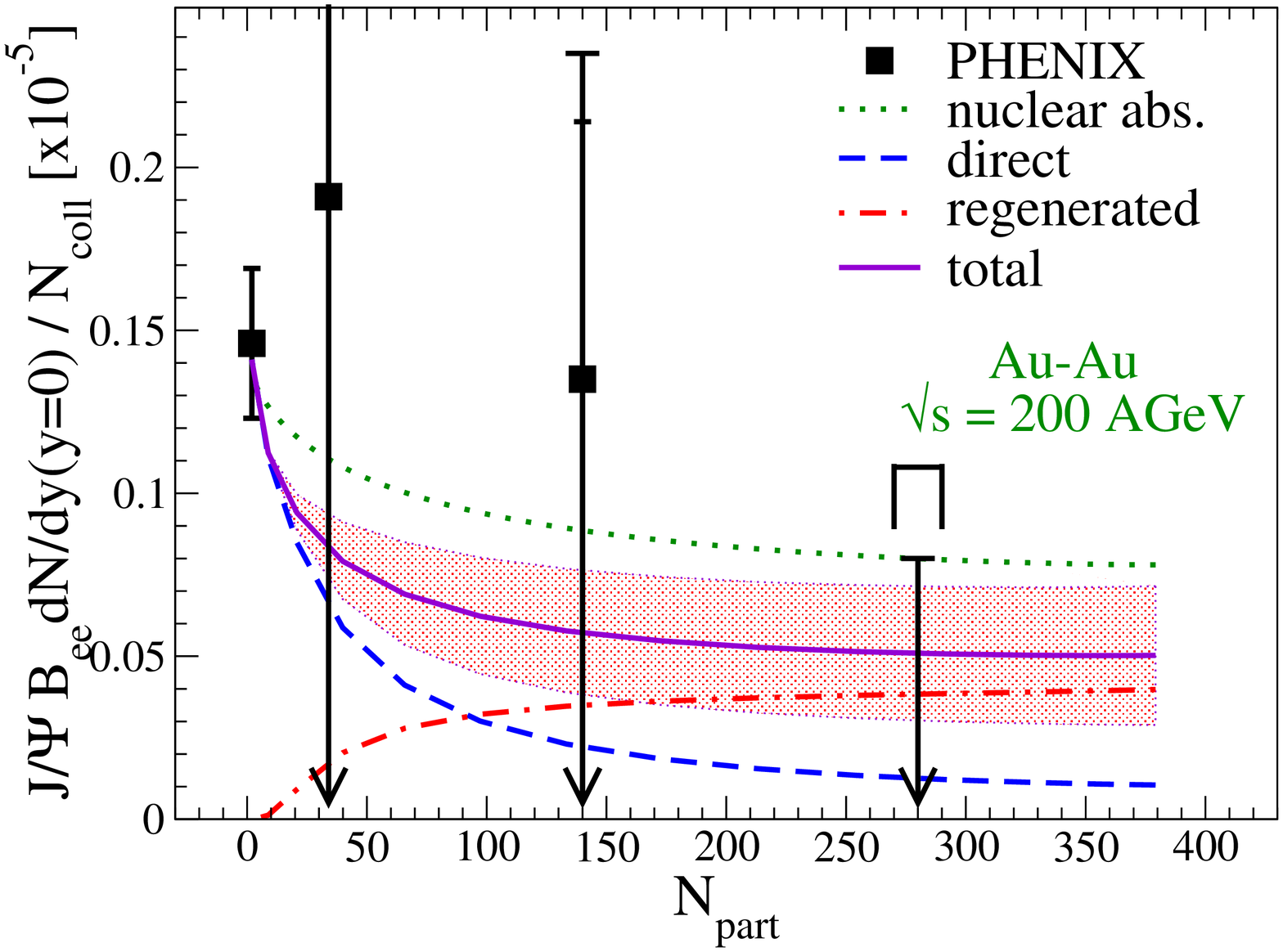}
\end{center}
\caption[Excitation function of the ratio of $J/\psi$'s over
         $c\overline c$ pairs in central heavy-ion collisions ($N_{\rm
         part}=360$) and centrality dependence of the $J/\psi$ yield]
        {Left-hand side: excitation function of the ratio of
         $J/\psi$'s over $c\overline c$ pairs in central heavy-ion
         collisions ($N_{\rm part}=360$) from SPS to RHIC
         energies~\cite{Grandchamp:2001pf,Grandchamp:2002wp} for the
         $J/\psi$ yield from quark gluon plasma suppressed primordial
         production (dashed), secondary $J/\psi$ production from
         $c\overline c$ coalescence (dot--dashed) and the sum (solid).
         Right-hand side: Centrality dependence of the $J/\psi$ yield,
         normalized to the number of primordial $NN$ collisions, in
         Au+Au collisions at RHIC within a kinetic theory framework
         including in-medium effects on both open and hidden charm
         states~\cite{Grandchamp:2003uw}. We show $J/\psi$ suppression
         with only the loss term in \Eq~(\ref{eq:rate2}) (dashed),
         secondary $J/\psi$ production (dot--dashed), primordial
         $J/\psi$'s subject to nuclear absorption only (dotted) while
         the band around the solid line, representing the full
         solution of the rate equation, indicates the uncertainty
         induced by varying the in-medium masses of open-charm
         hadrons.}
\label{fig:coal}
\end{figure}

In recent years, a new element has been added to charmonium production
in heavy-ion collisions with the realization that $c \overline c$
bound states might be recreated in later stages of the reaction. In
Refs.~\cite{Gazdzicki:1999rk,Braun-Munzinger:2000px,Gorenstein:2000ck,
Kostyuk:2001zd,Andronic:2003zv} secondary charmonium production was
evaluated in the statistical model for hadron production, motivated by
the success of this framework in the description of light hadron
species~\cite{Braun-Munzinger:2003zd}. Whereas in
Ref.~\cite{Gazdzicki:1999rk} the $J/\psi$ abundance was calculated
simply in terms of its thermal density at the hadronization
temperature, $T_c$, yielding fair agreement with NA50
data~\cite{Abreu:1999qw,Abreu:2000ni} at the SPS,
Refs.~\cite{Braun-Munzinger:2000px,Gorenstein:2000ck,Kostyuk:2001zd,
Andronic:2003zv} included the notion that the (rather heavy) charm
quarks are exclusively produced in primordial (hard) nucleon--nucleon
($NN$) collisions. This constraint is implemented by introducing a
charm--quark fugacity, $\gamma_c\equiv\gamma_{\overline c}$, to match
the primordial number of $c\overline c$ pairs, $N_{c\overline c}$, to
the total (hadronic) charm content in the fireball at hadronization,
\begin{equation}      
N_{c\overline c}= \frac{1}{2} N_{\rm op} \ \frac{I_1(N_{\rm op})}{I_0(N_{\rm 
op})}  \
+ \ V_{FB}(T) \ \gamma_c(T)^2 \sum\limits_{\Psi}  
n_\psi(T) \, \, .
\label{eq:Ncc}
\end{equation}
Here $N_{\rm op} = V_{FB}(T)~\gamma_c~\sum_C n_C(T,\mu_B)$ and $C =
D$, $\overline D$, $D^*$, $\overline D^*$, $\Lambda_c \dots$, runs
over all known open charm hadrons, $V_{FB}$ denotes the
(centrality-dependent) hadronic fireball volume covering an
appropriate range in rapidity and $I_{0,1}$ are modified Bessel
functions. The number of charmonium states ($\Psi = \eta_c$, $J/\psi$,
$\psi' \dots$) produced by ``statistical coalescence" then follows as
\begin{equation}
N^{\rm eq}_\Psi(T,\gamma_c) =  V_{FB}(T) \  
  d_\Psi \ \gamma_c^2 \ \int \frac{d^3q}{(2\pi)^3} f^\psi(m_\psi;T)
\label{eq:Npsi}
\end{equation}     
where $d_\Psi$ is the spin-degeneracy.  The statistical approach
correctly reproduces the $\psi'$ to $J/\psi$ ratio~\cite{Sitta:2004hj}
for sufficiently central Pb+Pb collisions ($N_{\rm part} \ge$ 200) at
the SPS.  However, to describe the absolute $J/\psi$ and $\psi'$
numbers in terms of statistical coalescence alone (implying that all
primordially produced charmonium states are suppressed), an
enhancement of total charm production over the standard expectation
from $NN$ collision-scaled $pp$ cross-sections by a factor of $\sim 3$
is necessary. This requirement, as well as the assumption of complete
suppression of primordial charmonia, has been relaxed in the
``two-component model" of
Refs.~\cite{Grandchamp:2001pf,Grandchamp:2002wp} where statistical
charmonium production has been combined with a primordial component
subject to suppression in both Quark--Gluon Plasma (QGP) and hadron gas
phases. As a result, it has been found that the (moderately)
suppressed primordial $J/\psi$ component prevails as the major yield
at SPS energies, whereas statistical recombination is the dominant
source in central Au+Au collisions at RHIC, with interesting
consequences for the excitation function, cf.\ the left-hand side of
\Figure~\ref{fig:coal}.

As discussed in \Section~\ref{sec:QCDfinitetemperature}, an important
new insight from QCD lattice calculations is that low-lying
charmonium states survive as resonance/bound states with finite width
in the QGP up to temperatures of about $1.5-2 \, T_c$. This implies
that these charmonium states can be formed not only at the
hadronization transition, but also in the QGP. Of course, at the same
time, dissociation reactions are operative, and the evolution of the
$J/\psi$ number should be described by kinetic theory within a
Boltzmann equation. In simplified form, the latter can be written as
\begin{equation}
\frac{dN_{\Psi}}{d\tau} = 
-\hat\Gamma_{D}  \ N_{\Psi} + \hat\Gamma_{F} \ N_c \ N_{\overline c} \  
\label{eq:rate1}
\end{equation}
where $\hat\Gamma_D$ and $\hat \Gamma_F$ are the charmonium
dissociation and formation rates respectively.  The inclusion of the
backward reaction in inelastic charmonium interactions such as,
$J/\psi + X_1 \leftrightharpoons X_2 + c + \overline c \, (D+\overline
D)$, is, in fact, mandated by the principle of detailed balance.  A
key question is then under which conditions regeneration becomes
quantitatively relevant.  First, the equilibrium number of $J/\psi$
mesons should be comparable to initial production (after nuclear
absorption).  While at the SPS, where $N_{c\overline c}\simeq 0.2$ in
central Pb+Pb collisions (based on binary $NN$ collision scaling),
this is not the case, the situation is more favorable at RHIC, where,
according to current measurements~\cite{Adams:2004fc,Adler:2004ta},
$N_{c\overline c}\simeq 15-30$ in central Au+Au collisions.  Also note
that the higher QGP temperatures at RHIC presumably lead to stronger
suppression of the primordial component.  Second, the amount of
regeneration depends on the momentum distributions of charm
quarks. Thus, if the latter are in thermal equilibrium (along with
gluons and light quarks) and as long as a well-defined $J/\psi$
(resonance) state persists, \Eq~(\ref{eq:rate1}) takes a particularly
simple and instructive form,
\begin{equation}
\frac{dN_{\Psi}}{d\tau} =
-\Gamma_{\Psi}(T) \left[ N_{\Psi} - N_{\Psi}^{\rm eq}(T,\gamma_c) \right] 
 \ . 
\label{eq:rate2}
\end{equation}
An important point here is that the key ingredients to
\Eq~(\ref{eq:rate2}) are rather directly related to equilibrium
in-medium properties of charm(onium) states, \ie quantities
which can be extracted from lattice QCD: (i) the reaction rate
$\Gamma_{\Psi}(T)$ is the (inelastic) width of the $\Psi$ spectral
function; (ii) the equilibrium charmonium abundance, $N_{\Psi}^{\rm
eq}$, given by \Eq~(\ref{eq:Npsi}), depends on the in-medium
charmonium mass and, via $\gamma_c$ in \Eq~(\ref{eq:Ncc}), on the
spectrum of open-charm states (including their in-medium masses).  A
recent calculation including current knowledge from lattice QCD has
been performed in Ref.~\cite{Grandchamp:2003uw}; results for RHIC are
shown on the right-hand side of \Figure~\ref{fig:coal}, indicating
appreciable sensitivity to the in-medium open-charm threshold, mostly
due to its impact on $N_{\Psi}^{\rm eq}$. This investigation has also
shown that most of the secondary $J/\psi$ production occurs through
resonance formation in the QGP. At full RHIC energy, relative chemical
equilibrium is reached close to the hadronization temperature and
frozen thereafter due to small reaction rates in hadronic matter,
implying approximate agreement with the limiting case of statistical
coalescence models applied at $T_c$, as discussed above.

The sensitivity of secondary production to the charm quark momentum
distributions (\ie deviations from thermal equilibrium) has been
addressed in Refs.~\cite{Thews:2002jg,Greco:2003vf}, as well as in
recent transport simulations~\cite{Zhang:2002ug,Bratkovskaya:2003ux}:
although the use of, \eg primordial (perturbative QCD) charm quark
distributions reduces the regenerated yield appreciably, it still
remains significantly larger than expectations based on suppression
scenarios alone.  Measurements of elliptic flow of $D$-mesons at RHIC,
testing their (early) thermalization, will therefore impose important
constraints on models for charmonium regeneration. Similarly, since in
coalescence approaches $N_{\Psi}^{\rm eq}\propto N_{c\overline c}^2$
(cf. \Eqs~(\ref{eq:Ncc}) and (\ref{eq:Npsi})), an accurate measurement of
open-charm production will be essential for reliable predictions of
charmonium yields and spectra.  Finally, it will be of great interest
to extend both experimental and theoretical investigations to the
bottomonium sector.

%-----------EXPERIMENTAL PART----------------------
\section[Quarkonium production in nuclear collisions]
        {Quarkonium production in nuclear collisions
         $\!$\footnote{Section coordinator: M. Rosati;
         Authors: C.~Louren\c{c}o, L.~Ramello, M.~Rosati, E.~Scomparin}}

\subsection{Charmonium suppression at the CERN SPS}
\def \ef    {\mbox{$E_{\scriptscriptstyle F}$}}
\def \nch   {\mbox{$N_{\scriptscriptstyle ch}$}}

The experimental study of charmonium production in collisions of light and 
heavy ions at ultrarelativistic energies was carried out at the CERN SPS 
over 15 years (1986--2000) by experiments NA38 (see Refs.~\cite{Bag89,Bag91} 
for initial results and \cite{Abr99a,Abr99b} for the most recent ones) 
and NA50 (see Refs.~\cite{Abr97a,Abr97b,Abr99,Abr00,Abr01}).
It is continuing with experiment NA60, taking data in 2003 and 2004.

In addition, NA51 provided a reference measurement of charmonium
production in $pp$ and $p$d collisions~\cite{Abr98a} while
proton--nucleus, $pA$, reference data were collected by
NA38~\cite{Abr98b,Abr99b} and, with higher statistics, by
NA50~\cite{Abr03,Ale04}.

The SPS energies, between 158 and 450~GeV per nucleon in a fixed
target configuration, are well suited for charmonium studies since the
\jpsi\ production cross-section in the forward hemisphere ($x_F > 0$)
in proton--proton interactions is between 50 and 100~nb/nucleon (see
\eg Ref.~\cite{Vog99}), well above threshold.  The high intensity
experimental area at the SPS provides beam rates of about
$10^7$~Pb~ions/s and $10^9$~protons/s which, with a branching ratio
$\jpsi\ \rightarrow \mu^+\mu^-$ of about 6\% and a typical acceptance
of 15\%, resulted in samples of several tens of thousand of \jpsi\
collected for a given system. The statistics collected (see
\Tables~\ref{tab:pastat77} and~\ref{tab:supbstat77}) made it possible
to perform detailed studies of \jpsi\ and $\psi'$ production as a
function of centrality in ion--ion collisions. The centrality was
evaluated via the transverse energy, \et, by NA38, as well as via the
forward energy in the zero degree calorimeter, $E_{\rm ZDC}$, and the
charged multiplicity, $N_{\rm ch}$ by NA50.

\begin{table}
\caption{Summary of proton--nucleus data collected by NA50.}
\label{tab:pastat77}
\begin{center}
\begin{tabular}{|c|c|c|c|} 
\hline
Years                     & 1996--1998 & 1998--2000 & 2000 \\ 
\hline
Energy (GeV/nucleon)      &   450     &    450    & 400  \\   
\hline
Target thickness $L/\lambda_I$ &  26--39\%  &  26--39\%  & 26--39\% \\
\hline
Targets                   & Be, Al, Cu, Ag, W & Be, Al, Cu, Ag, W & Be, Al,
Cu, Ag, W, Pb \\
\hline
Beam intensity ($p$/s) & $(4-13)\times 10^8$  &  $(0.8-2.5)\times 10^8$ &  
$(9-13)\times 10^8$ \\
\hline
$J/\psi$ ($\times 10^3$)  & 350--800  & 80--180  & 38--68 \\
\hline
\end{tabular}
\end{center}
\end{table}

\begin{table}
\caption{Summary of S+U data and Pb+Pb data collected by NA38 and NA50.}
\label{tab:supbstat77}
\begin{center}
\begin{tabular}{|c|c|c|c|c|c|} 
\hline
Year                    & 1992 & 1995 & 1996 & 1998 & 2000 \\ 
\hline
Energy (GeV/nucleon)     &  200 &  158 &  158 &  158 &  158 \\   
\hline
Target thickness $L/\lambda_I$ &  20.5\% &  17\%  & 30\%   &  7\%  & 9.3\%  \\
\hline
Beam--Target              &  S+U &  Pb+Pb &  Pb+Pb &  Pb+Pb &  Pb+Pb  \\
\hline
Beam intensity (ions/5 s) & $8\times 10^7$ & $3\times 10^7$ & $5\times 10^7$ 
& $5.5\times 10^7$ & $7\times 10^7$ \\
\hline
$J/\psi$ ($\times 10^3$)  & 113  & 50  & 190 & 49 & 129 \\
\hline
\end{tabular}
\end{center}
\end{table}

The experimental program at the CERN SPS was developed in successive
phases.  At first, the pioneering experiment NA38~\cite{Bag89}
collected data with light ions (oxygen and sulphur) at 200~GeV/nucleon
and with proton beams at 450~GeV, using the NA10 dimuon spectrometer
and an electromagnetic calorimeter as a centrality detector.  Then,
the second generation experiment NA50~\cite{Abr97a} studied \jpsi\
production in lead--lead collisions at 158~GeV/nucleon with new or
improved centrality detectors, and in proton--nucleus collisions with
much higher statistics relative to NA38.  Finally, experiment NA60
collected data with indium--indium collisions at 158~GeV/nucleon and
proton--nucleus collisions at 400~GeV with a new silicon vertex
spectrometer and a beam tracker (see next section).

In order to compare the hard production of charmonium states in
different collisions ranging from $pp$ to Pb+Pb, it is appropriate to
define the cross-section per nucleon--nucleon collision, obtained by
dividing the measured cross-section by the product of the mass
numbers, $AB$, of the colliding nuclei.  In the study of centrality,
the measured Drell--Yan cross-section can replace $AB$ since it has
been verified experimentally (see \eg Refs.~\cite{Abr97b,Bor04}) that
the Drell--Yan cross-section is proportional to the number of
nucleon--nucleon collisions.

The nuclear dependence of the charmonium cross-section is often
parametrized as $\sigma_{pA} = \sigma_{pp} A^\alpha$, where $\alpha =
1$ is expected for a hard process in the absence of nuclear absorption
effects. A more accurate description, valid also for light targets, is
provided by the Glauber formalism~\cite{Gla59}. A detailed description
of such formalism applied to both $pA$ and nucleus--nucleus collisions
is given in Ref.~\cite{Kharzeev:1996yx}. The distribution of matter
inside nuclei is described by 2-parameter or 3-parameter Woods--Saxon
distributions from a compilation of electron scattering
measurements~\cite{deJ74}.  When comparing the centrality evolution of
different systems, a useful variable is the average path of the
\ccbar\ pair through nuclear matter, denoted by $L$.

The first NA38 results, obtained with 200~GeV/nucleon oxygen and
sulphur beams, revealed~\cite{Bag89,Bag91} that \jpsi\ production is
suppressed in ion collisions, both relative to $p$U and as a function
of the transverse energy, \et.  However, it was later found that the
suppression pattern observed in S+U collisions was compatible with the
extrapolation of the trend observed in $pA$ collisions. NA38 then
collected a significantly larger sample of S+U events (see
\Table~\ref{tab:supbstat77}), obtaining~\cite{Abr99a} absolute
cross-sections for \jpsi, $B\sigma_{\psi} = 7.78 \pm 0.04 \pm
0.62$~$\mu$b, $\psi^\prime$, $B\sigma_{\psi^{\prime}} = 59.1 \pm 6.2
\pm 4.7$~nb, and Drell--Yan in the mass window $2.9<M<4.5$~GeV,
$\sigma_{\rm DY} = 310 \pm 10 \pm 25$~nb.  Comparing the suppression
pattern of the two resonances it was found that the $\psi'$ is more
suppressed than the \jpsi\ by at least a factor of two, even more so
for central collisions.  A global study of this result together with
$pp$ and $p$d results from NA51~\cite{Abr98a} and $pA$ results from
NA38~\cite{Abr98b} revealed (see Ref.~\cite{Abr99b}, in particular
\Figure~5) that \jpsi\ production exhibits a continuous decreasing
pattern from $pp$ to S+U reactions (including the centrality
dependence observed in S+U interactions) which can be accounted for by
normal nuclear absorption.  On the other hand, the $\psi'$ showed
extra suppression in S+U interactions.  Since the $\psi'$ state is
very loosely bound, it can be broken into a pair of open charm mesons
by purely hadronic interactions, independent of whether the produced
matter is confined or deconfined.

\begin{figure}[t]
\begin{center}
\begin{tabular}{cc}
\includegraphics[width=7cm]{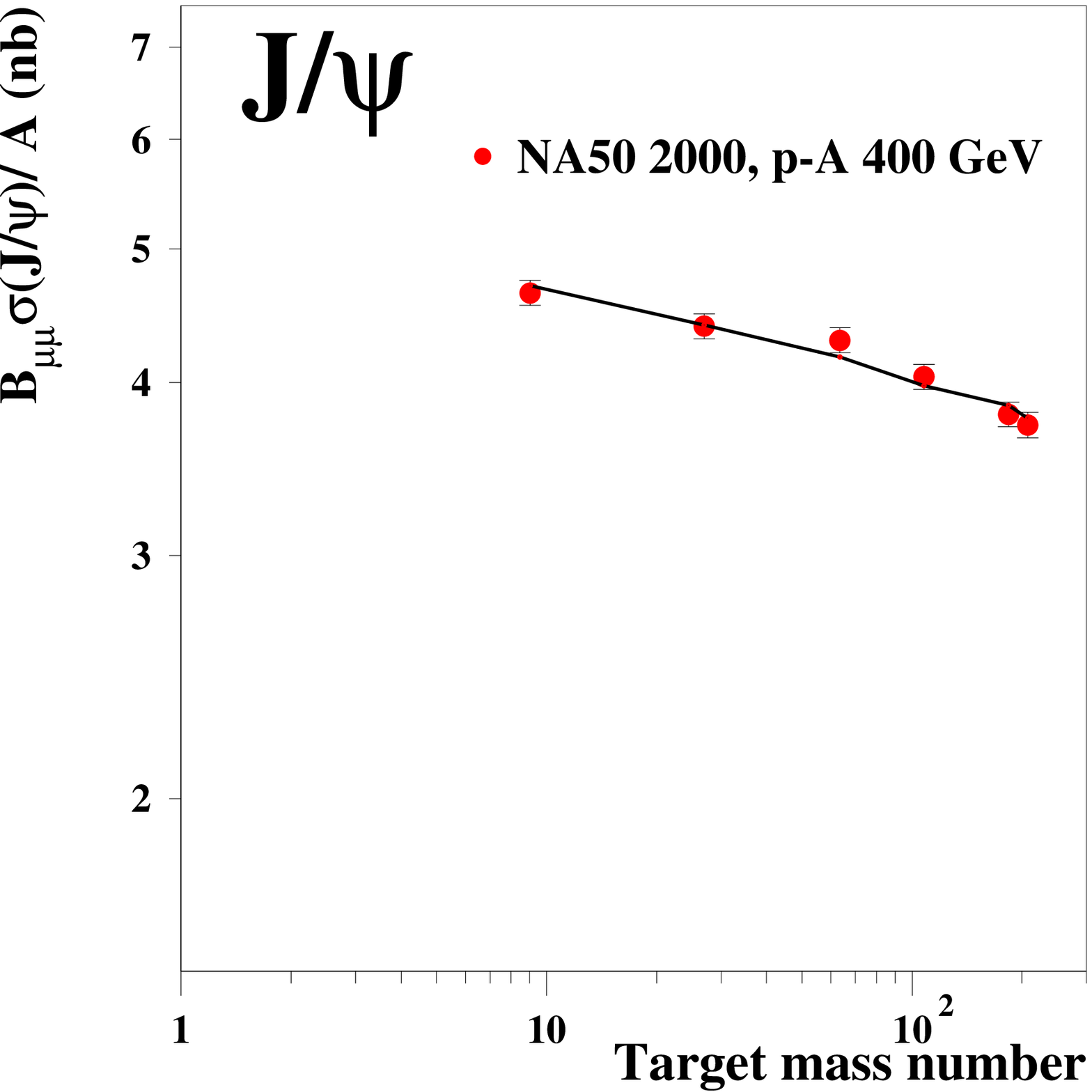} &
\includegraphics[width=7cm]{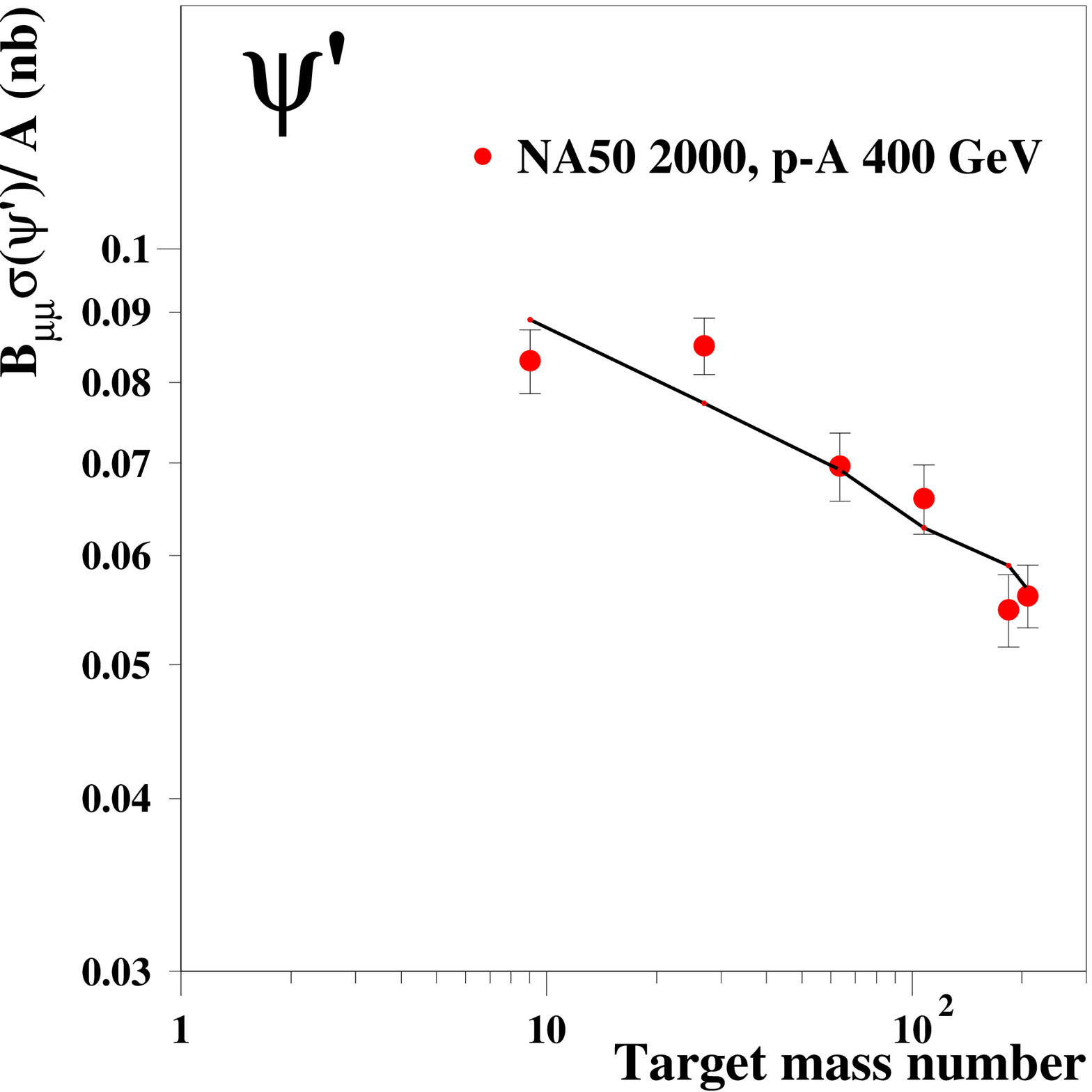} \\
\end{tabular}
\caption[The \jpsi\ and $\psi'$ cross-sections in 
         proton--nucleus collisions]
        {The \jpsi\ (left) and $\psi'$ (right) cross-sections in
         proton--nucleus collisions with six different targets at 
         400~GeV, measured by NA50, together with the result of the
         Glauber fits.}
\label{fig:pa77}
\end{center}
\end{figure}

The understanding of the reference proton--nucleus data improved
dramatically thanks to recent high statistics measurements by NA50,
see \Table~\ref{tab:pastat77}.  \Figure[b]~\ref{fig:pa77} (see
Ref.~\cite{Bor04}) shows the absolute \jpsi\ and $\psi'$
cross-sections, multiplied by the branching ratio to dimuons and
divided by the target mass number $A$ for the most recent $pA$ data
collected at 400~GeV.  A fit using the Glauber
formalism~\cite{Gla59,Kharzeev:1996yx}, more accurate than the usual
$A^{\alpha}$ parametrization, leads to absorption cross-sections
$\sigma_{\rm abs}^{\psi} = 4.2 \pm 0.5$~mb and $\sigma_{\rm
abs}^{\psi^{\prime}} = 9.6 \pm 1.6$~mb.  A difference between the
\jpsi\ and $\psi'$ absorption cross-sections is observed already in
proton--nucleus collisions, thanks to higher statistics and improved
systematics relative to NA38.

\begin{figure}[t]
\begin{center}
\begin{tabular}{cc}
\includegraphics[width=6cm]{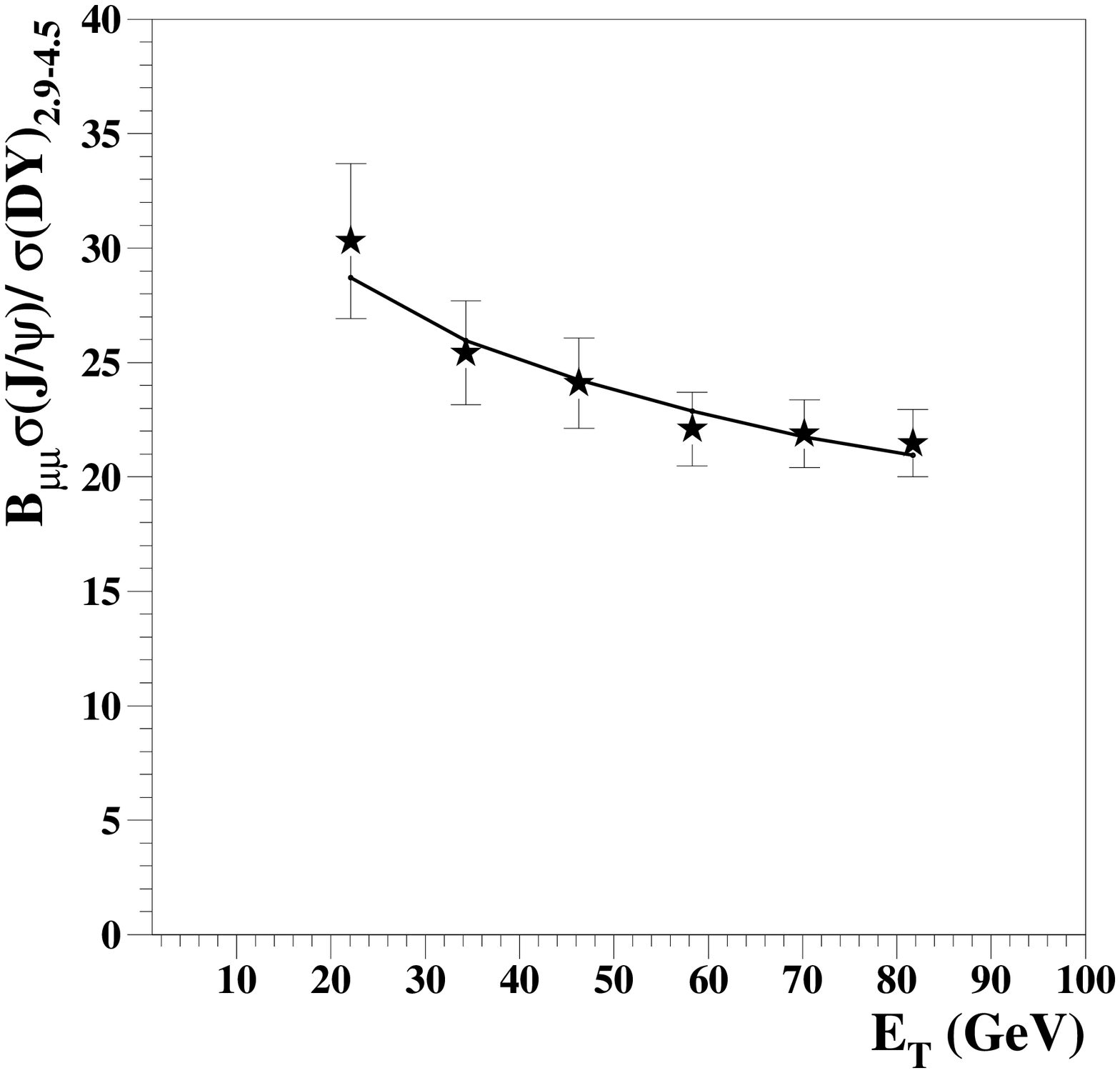} &
\includegraphics[width=6cm]{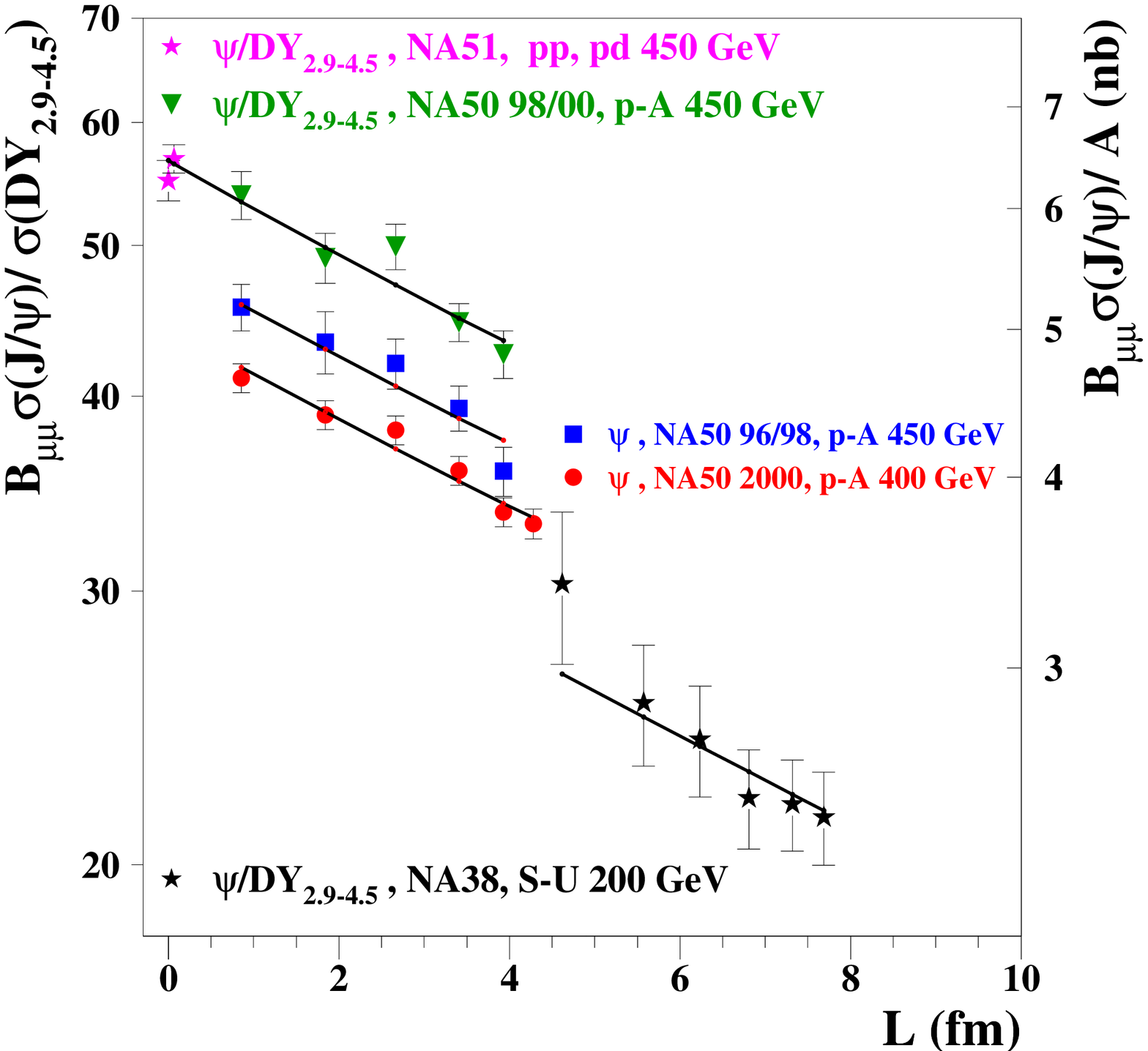} \\
\end{tabular}
\end{center}
\caption[The (\jpsi)/Drell--Yan ratio in S+U collisions as a 
         function of \et{} and of $L$]
        {Left-hand side: the (\jpsi)/Drell--Yan ratio as a function of
         centrality (determined by \et) in S+U collisions.  Right-hand
         side: the (\jpsi)/Drell--Yan as a function of $L$ for several
         systems.}
\label{fig:supa77}
\end{figure}

A more precise picture of normal nuclear absorption is obtained by
combining the absolute cross-sections with the (\jpsi)/Drell--Yan
ratios at all available beam energies.  Using, in addition to the
400~GeV p--A data, also the NA50 data collected with a 450~GeV proton
beam~\cite{Abr03,Ale04} and the NA51 $pp$ and pd
results~\cite{Abr98a}, a simultaneous Glauber fit gives~\cite{Bor04}
$\sigma_{\rm abs}^{\psi} = 4.3 \pm 0.3$~mb.  The NA38 S+U data at
200~GeV/nucleon have been reanalysed with the most recent techniques.
By fitting the reanalyzed data to a centrality-dependent Glauber
calculation for six different centrality regions (see
\Figure~\ref{fig:supa77} left), $\sigma_{\rm abs}^{\psi} = 7.3 \pm
3.3$~mb is obtained, statistically compatible with the $pA$
cross-section.  A global fit to $pp$, $p$d, $pA$ and S+U data, with
separate normalizations for the three different (energy and rapidity)
kinematical conditions, leads to $\sigma_{\rm abs}^{\psi} = 4.3 \pm
0.3$~mb (see \Figure~\ref{fig:supa77} right).  An extrapolation from
the 200~GeV/nucleon S+U to the 158~GeV/nucleon Pb+Pb kinematical
conditions is then made in order to obtain the normal absorption curve
against which the Pb+Pb results are compared.

The analysis of the Pb+Pb data collected in 1995, 1996 and 1998 (see
\Table~\ref{tab:supbstat77}) showed~\cite{Abr97a,Abr97b} that \jpsi\
production, relative to Drell--Yan, is anomalously suppressed with
respect to the normal nuclear absorption pattern. This integrated
result was complemented by detailed studies of the \jpsi\ suppression
pattern as a function of collision
centrality~\cite{Abr99,Abr00,Abr01}, determined from \et\ and $E_{\rm
ZDC}$, which suggests that this extra suppression sets in for
semi-central collisions, with the transition occurring over a rather
small range of centrality values.  The suppression pattern, showing a
departure from normal absorption and then no saturation at high \et,
is currently interpreted in the quark--gluon plasma scenario as the
sequential suppression of two \ccbar\ states, first the $\chi_c$ and
then the \jpsi.

\begin{figure}
\begin{center}
\begin{tabular}{cc}
\includegraphics[width=6cm]{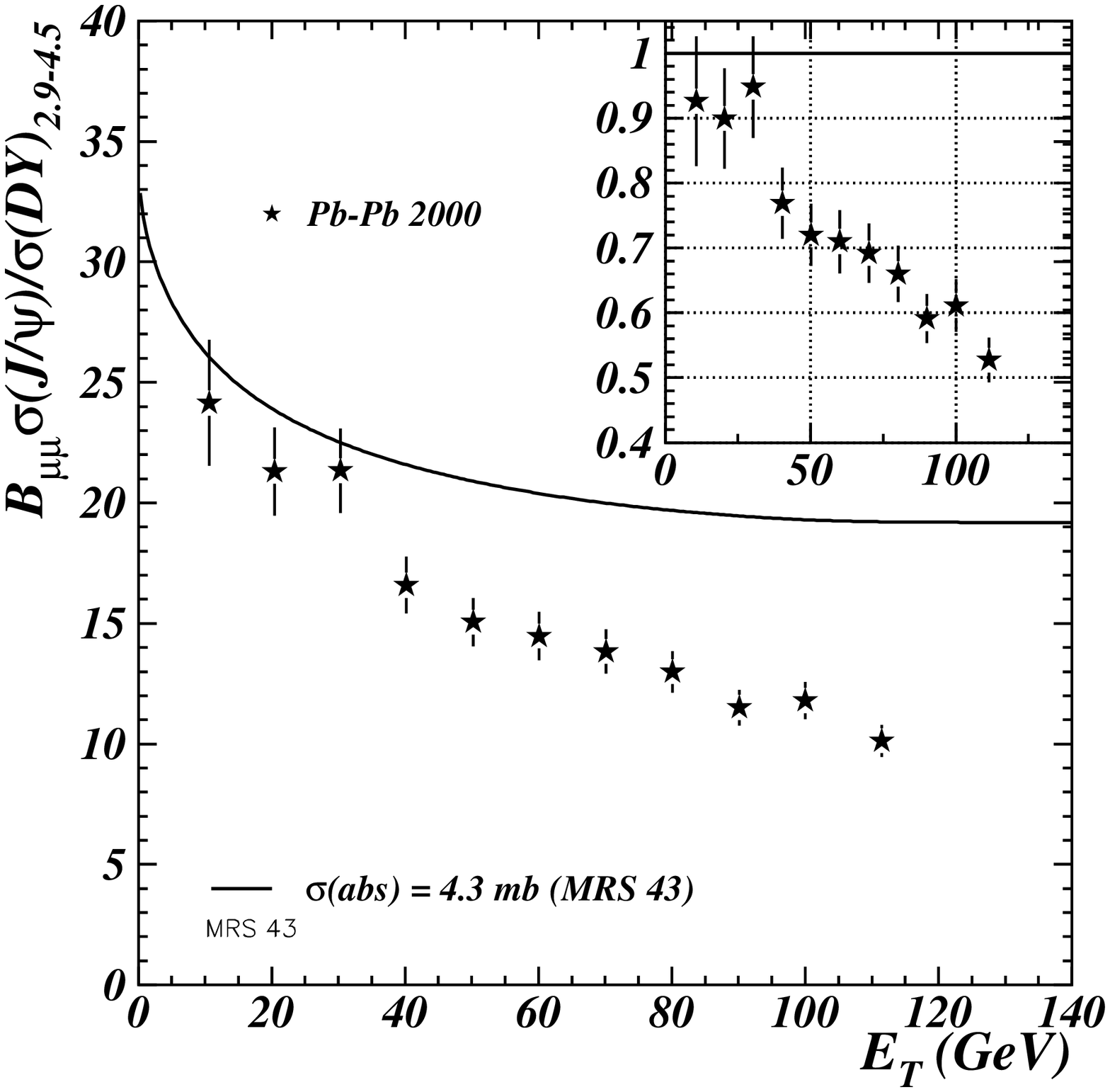} &
\includegraphics[width=6cm]{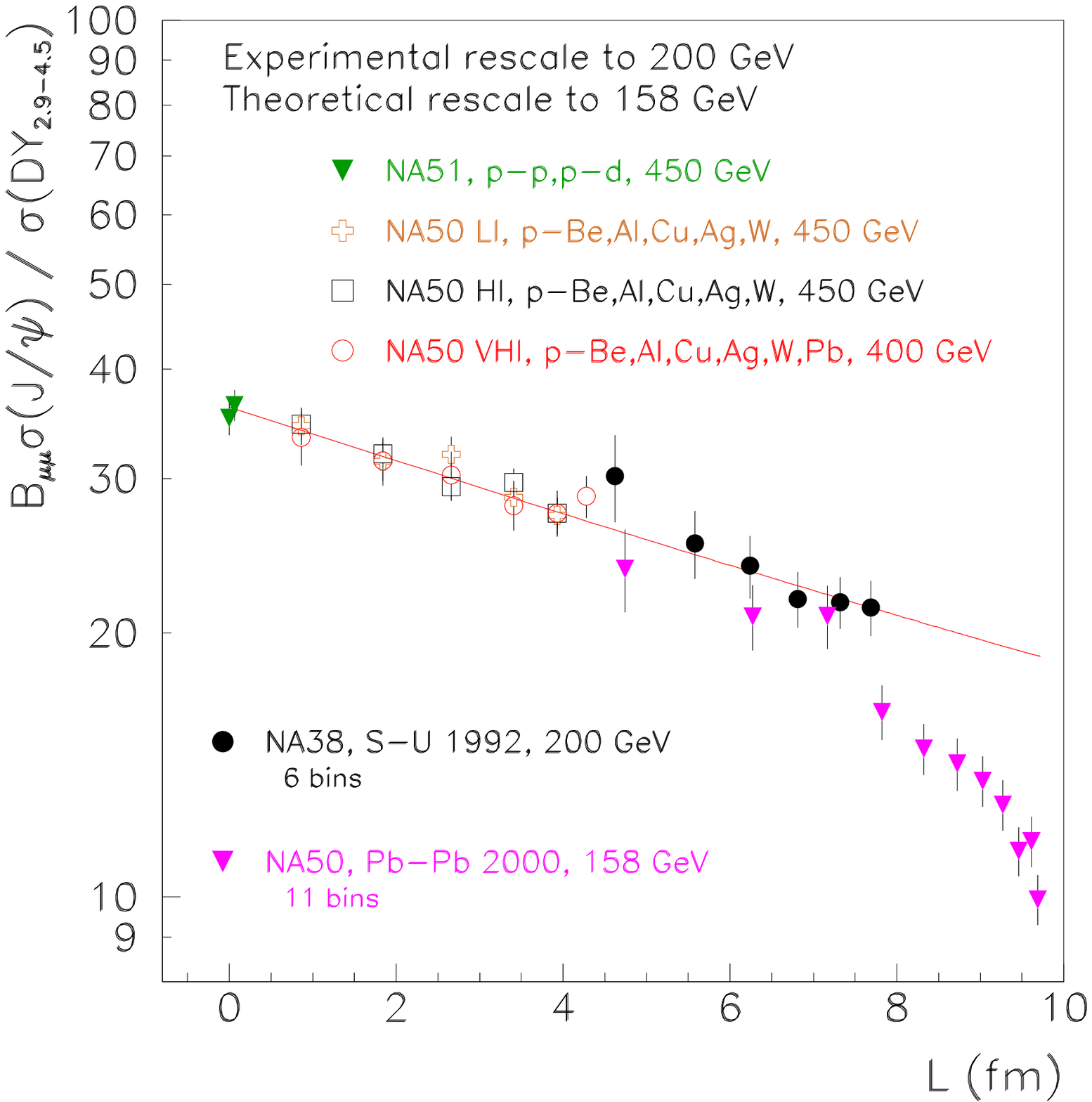} \\
\end{tabular}
\end{center}
\caption[The (\jpsi)/Drell--Yan ratio in Pb+Pb collisions as a 
         function of \et{} and of $L$]
        {Left-hand side: the (\jpsi)/Drell--Yan ratio as a function of
         centrality in Pb+Pb collisions at 158~GeV/nucleon.
         Right-hand side: the (\jpsi)/Drell--Yan ratio for several
         systems, from $pp$ to Pb+Pb, as a function of L.}
\label{fig:jpsi77}
\end{figure}

A more detailed analysis of Pb+Pb data revealed that the peripheral
centrality data was affected by a considerable admixture of Pb--air
interactions, especially in the multi-target configuration used in
1996. Therefore, more data were collected by NA50 in 2000 with a
single target under vacuum.  The 2000 Pb+Pb \jpsi\ result is shown on
the left-hand side of \Figure~\ref{fig:jpsi77} together with the
absorption curve derived from the analysis presented in
\Figure~\ref{fig:supa77}.  The departure from the ordinary nuclear
absorption is still, in the new data set, a striking feature.  The
Pb+Pb data are compared to other systems on the right-hand side of
\Figure~\ref{fig:jpsi77}.

\begin{figure}[t]
\begin{center}
\includegraphics[width=.48\textwidth]{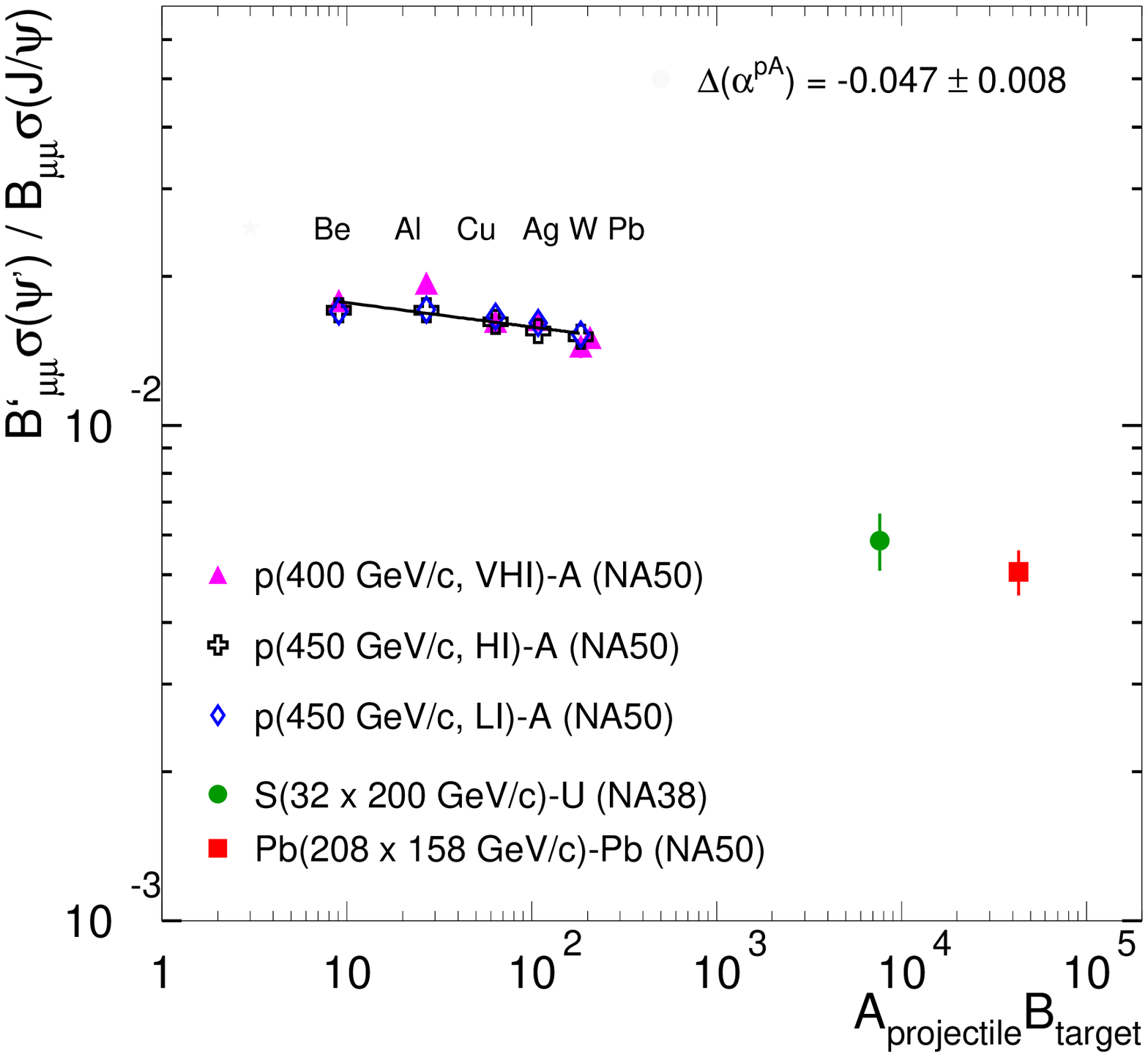} 
\hfill
\includegraphics[width=.48\textwidth]{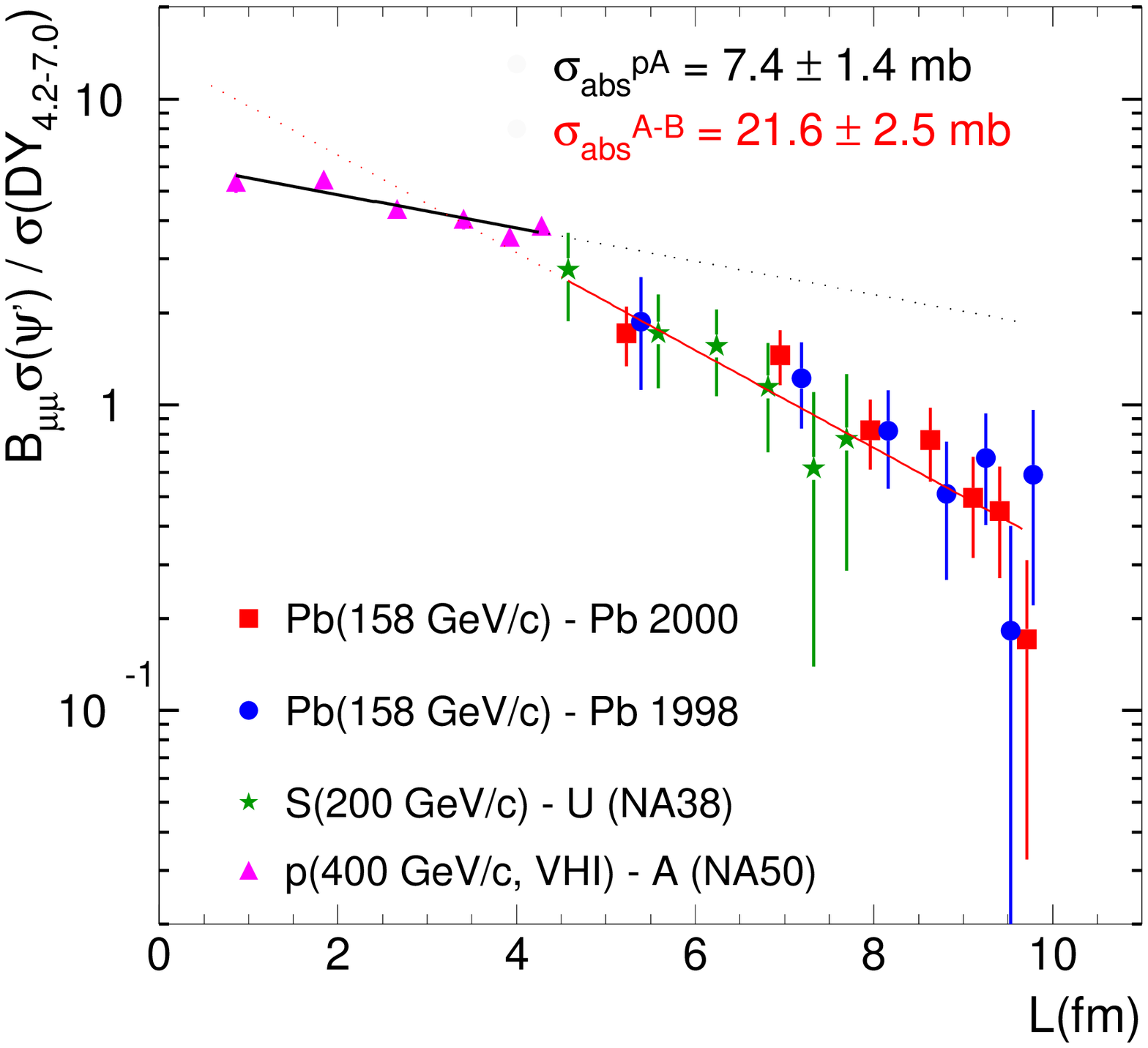} 
\end{center}
\caption[The ratio of $\psi'$ to \jpsi\ in different 
         collision systems]
        {Left-hand side: the ratio of $\psi'$ to \jpsi\ in different
         collision systems as a function of $AB$. Right-hand side: the
         $(\psi^\prime$)/Drell--Yan ratio as a function of $L$.}
\label{fig:psiprime77}
\end{figure}

A new result on $\psi'$ production, recently obtained~\cite{San04}
from the 1998 and 2000 Pb+Pb data samples, analysed with the most
recent procedures, is presented in \Figure~\ref{fig:psiprime77}. The
left-hand side shows the relative suppression of the two \ccbar\ bound
states for several systems ranging from $p$Be to Pb+Pb, as a function
of the product $AB$. As indicated above, the $\psi'$ is more absorbed
than the \jpsi\ already in $pA$ collisions.  Furthermore, a stronger
$\psi'$ suppression relative to the \jpsi\ is observed for the heavier
S+U and Pb+Pb systems.  On the right-hand side, $\psi'$ suppression
relative to Drell--Yan is presented as a function of centrality,
expressed by the path length $L$.  The $\psi'$ suppression is the same
in S+U and Pb+Pb interactions as a function of centrality and about
three times stronger than in $pA$ interactions.

\begin{figure}[t]
\begin{center}
\includegraphics[width=.49\textwidth]{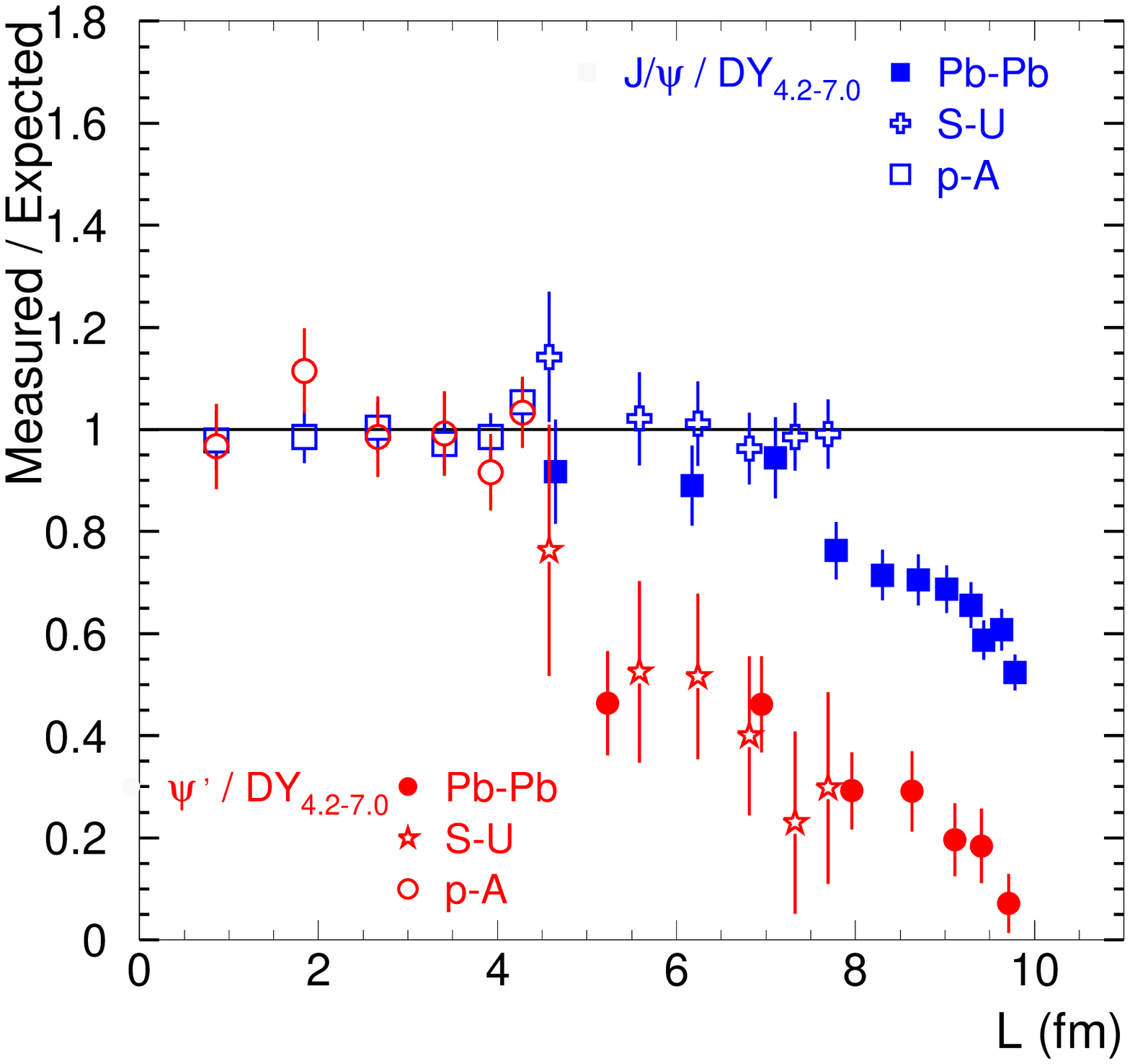} 
\end{center}
\caption{The Measured/Expected ratio for \jpsi\ and $\psi'$ in $pA$, 
S+U and Pb+Pb collisions at the CERN SPS.}
\label{fig:summary77}
\end{figure}

In conclusion, experiments NA38, NA50 and NA51 provided valuable
information on \jpsi\ and $\psi'$ production with proton and ion beams
at the SPS fixed target energies.  A synthesis of the main result, the
different suppression patterns of the two \ccbar\ states, is presented
in \Figure~\ref{fig:summary77}.

\subsection{Recent studies of charmonium production in heavy-ion 
collisions at the SPS}

Charmonium physics in ultrarelativistic nucleus--nucleus collisions is
now being studied at the SPS by NA60. The experimental apparatus
includes the muon spectrometer already used by NA50 and a vertex
spectrometer, based on silicon pixel detectors, that allows an
accurate determination of primary and secondary interaction vertices
with a resolution better than 50~$\mu$m. By matching the muons
measured in the muon spectrometer to tracks in the pixel telescope,
simultaneously using coordinate and momentum information, it is
possible to overcome the uncertainties introduced by the multiple
scattering and energy loss fluctuations induced by the 5.5~m long
hadron absorber positioned in front of the muon spectrometer.  The
consequent improvement in dimuon mass resolution is extremely
impressive, and reaches a factor $\sim$~4 for low mass dimuons ($M <
1$~GeV), where multiple scattering is quantitatively more important.
In the charmonia mass region, NA60 reaches a mass resolution of
$\approx 70$~MeV as opposed to $\approx 105$~MeV in NA50, particularly
important for $\psi'$ studies.

In 2003, NA60 collected more than 230 million dimuon triggers in In+In
collisions at 158~GeV/nu\-cle\-on.  The choice of Indium, an
intermediate mass nucleus, is particularly interesting for the study
of the anomalous $J/\psi$ suppression observed by NA50 in Pb+Pb
collisions.  The onset of the anomalous $J/\psi$ suppression occurs in
semiperipheral Pb+Pb events, a region with potentially important
systematic uncertainties due to the presence of out-of-target
events. For the lighter In+In system, the anomaly should show up for
comparatively more central events, thereby confirming this effect in a
region where systematics are easier to control.

Furthermore, other useful insights into the suppression mechanisms can
be obtained by studying the $J/\psi$ yield as a function of centrality
for both In+In and Pb+Pb, plotting the results as a function of
centrality using several variables such as $L$, the thickness of
nuclear matter traversed by the charmonium state, the number of
participant nucleons, $N_{\rm part}$, and the energy density,
$\epsilon$.  In this way it may become possible to precisely identify
the centrality variable that governs the anomalous suppression and,
therefore, its origin.  In particular, one could distinguish between
the thermal (QGP) and geometrical (percolation) phase transitions,
both resulting in the suppression of $J/\psi$ production but as a
function of different variables and with different thresholds in
collision centrality.

\iffalse
Currently, preliminary results on $J/\psi$ suppression have been
obtained by integrating the events over the centrality range $E_{\rm
ZDC} <15$~TeV.  In \Figure~\ref{fig:na60-psi} we show the
$J/\psi$\,/\,DY ratio for In+In collisions, corresponding to about 50
\% of the available statistics, compared with previous results from
NA38, NA50 and NA51.  The comparison has been carried out both as a
function of $L$ (left plot) and $N_{\rm part}$ (right plot).  In the
latter figure the data points are plotted with respect to the normal
nuclear absorption curve, determined from the $pA$ measurements, shown
by the line on the left-hand side.  The In+In measurement, when
normalized to the absorption curve, gives the value $0.87\pm0.07$.
While the In--In point sits (at 7.0~fm) to the left of the most central
S+U value as a function of $L$, as a function of $N_{\rm part}$ we see
the opposite.  Once all the Indium data have been analyzed, we should
be able to probe the $J/\psi$ suppression pattern as a function of
centrality, for $5.5<L<7.8$~fm and $50<N_{\rm part}<200$,
corresponding to the onset of anomalous $J/\psi$ suppression.

\begin{figure}[t]
\begin{center}
\includegraphics[width=.4\textwidth]{All-psi-vs-L}
\qquad
\includegraphics[width=.4\textwidth]{All-Npart}
\end{center}
\caption[The $J/\psi$ suppression pattern versus $L$ and $N_{\rm
         part}$]
        {The $J/\psi$ suppression pattern versus $L$ (left) and $N_{\rm
         part}$ (right), including the NA60 indium--indium measurement.}
\label{fig:na60-psi}
\end{figure}
\fi

\begin{figure}[t]
\begin{center}
\includegraphics[width=.43\textwidth]{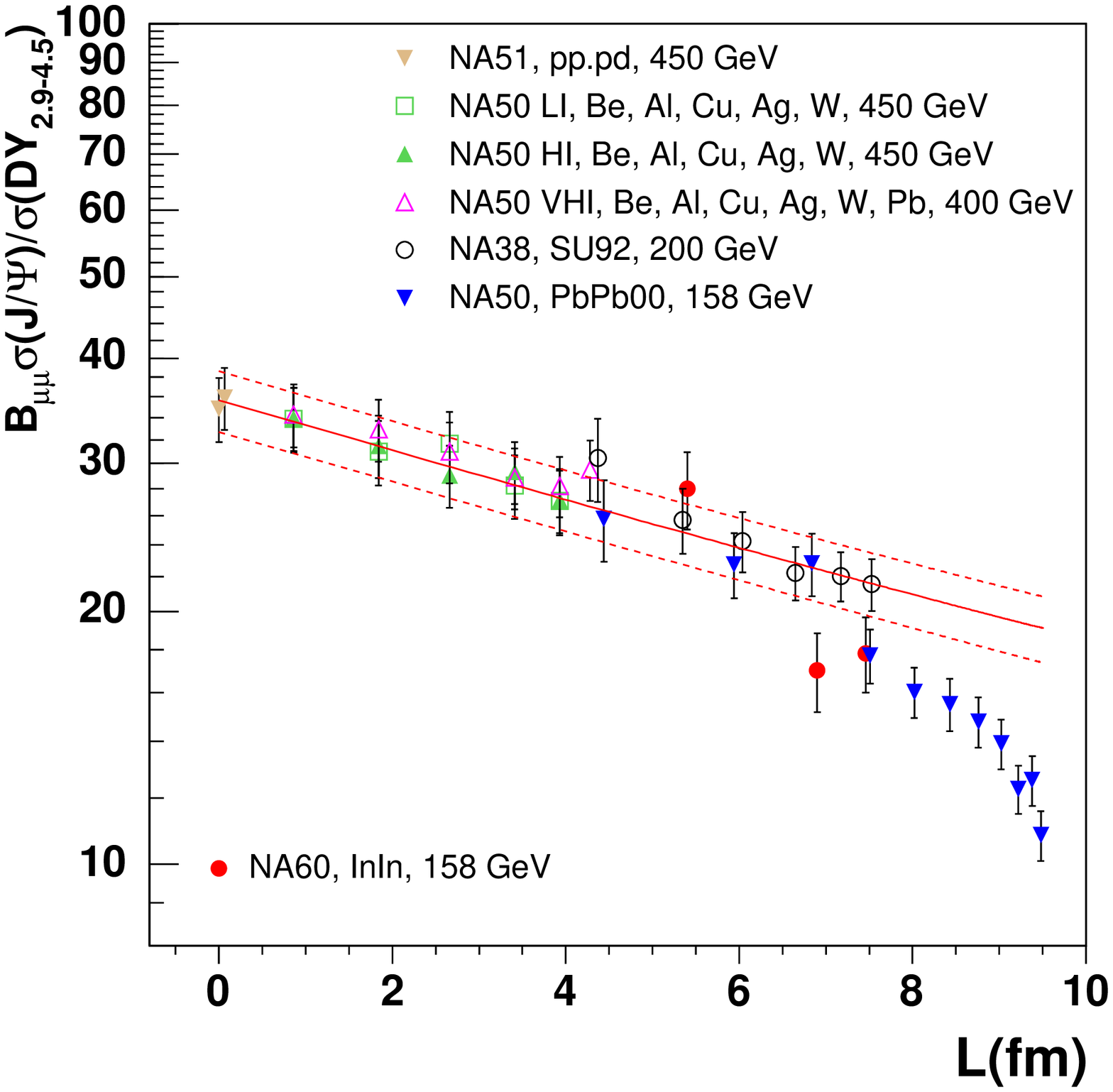}
\qquad
\includegraphics[width=.4\textwidth]{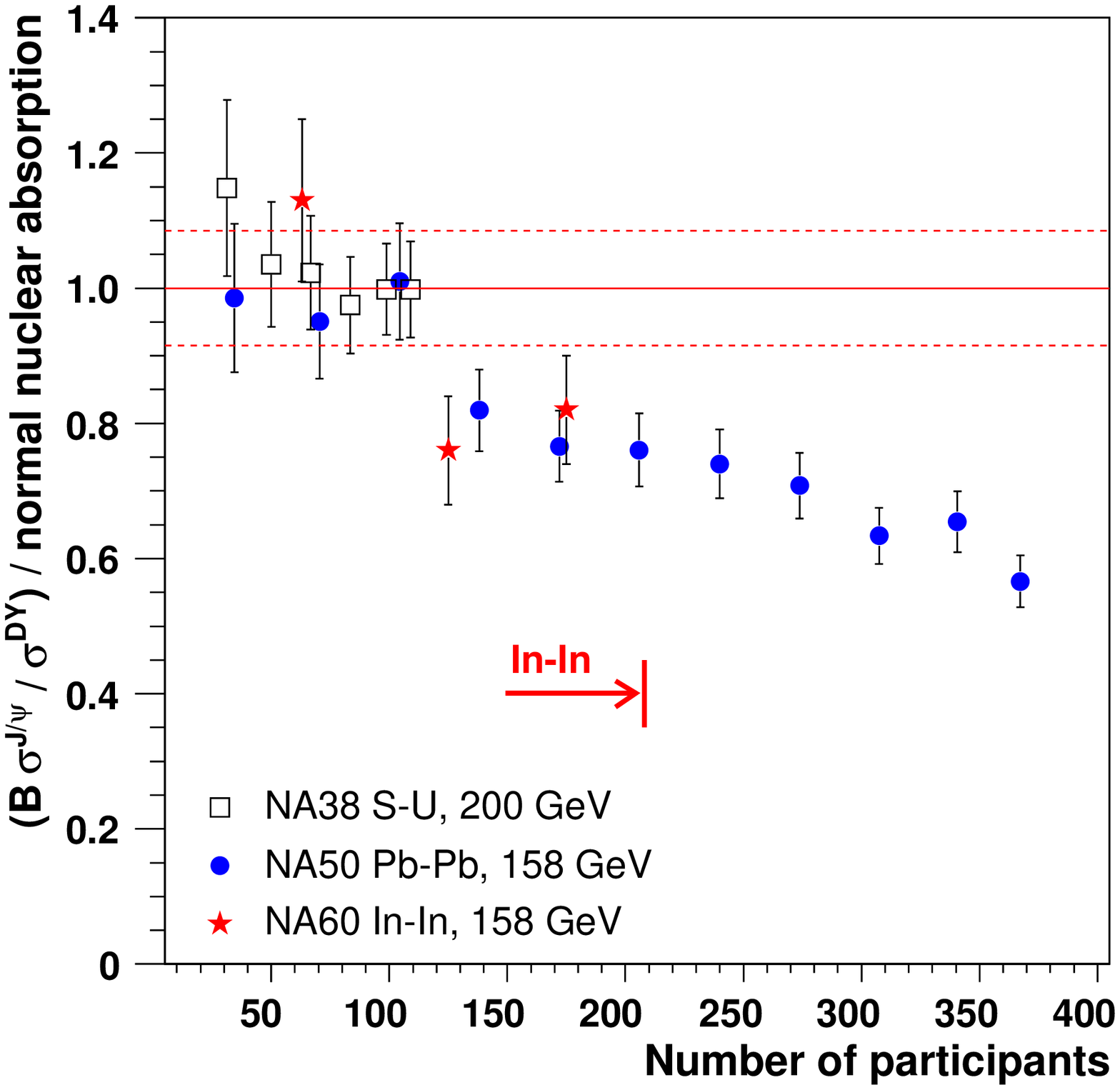}
\end{center}
\caption[The $J/\psi$ suppression pattern versus $L$ and $N_{\rm
         part}$]
        {The $J/\psi$ suppression pattern versus $L$ (left) and, after
         dividing by the normal nuclear absorption curve, versus
         $N_{\rm part}$ (right), including the NA60 indium--indium
         measurements.}
\label{na60-psi}
\end{figure}

\Figure[b]~\ref{na60-psi} shows the ratio between the $J/\psi$ and the
Drell--Yan production cross-sections measured in In--In collisions, in
three centrality bins, either as a function of $L$ (the distance of
nuclear matter crossed by the $J/\psi$ mesons after production) or
$N_{\rm part}$.  On the right panel the $J/\psi$ suppression pattern is
divided by the normal nuclear absorption curve, defined by p--nucleus
data.  The $J/\psi$ and Drell--Yan cross-sections are evaluated in the
phase space window $2.92 < y_{\rm lab} < 3.92$ and
$-0.5 < \cos\theta_{\rm CS} <0.5$, where $\theta_{\rm CS}$ is the polar
decay angle of the muons in the Collins--Soper reference system.
The Drell--Yan value is given in the 2.9--4.5~GeV/$c^2$ mass window.
We see that, unlike what happens in the S--U collisions studied by NA38,
the $J/\psi$ production is suppressed in indium--indium collisions beyond
the normal nuclear absorption.  When the $J/\psi$ over Drell--Yan ratio
is plotted as a function of $N_{\rm part}$ the indium data points seem
to agree with the suppression pattern measured in Pb--Pb.  The two sets
of data points \emph{do not} overlap as a function of $L$.  To clarify
the origin of the anomalous $J/\psi$ suppression, the statistical
significance of the results must be increased and reference processes
alternative to Drell--Yan are presently under study in NA60.  Also the
study as a function of the energy density is in progress.

\subsection{Charmonium production in proton--nucleus collisions at 158~GeV}

\begin{figure}[p]
\mbox{}
\vfill
\begin{center}
\begin{tabular}{@{}cc@{}}
\includegraphics[width=.44\textwidth]{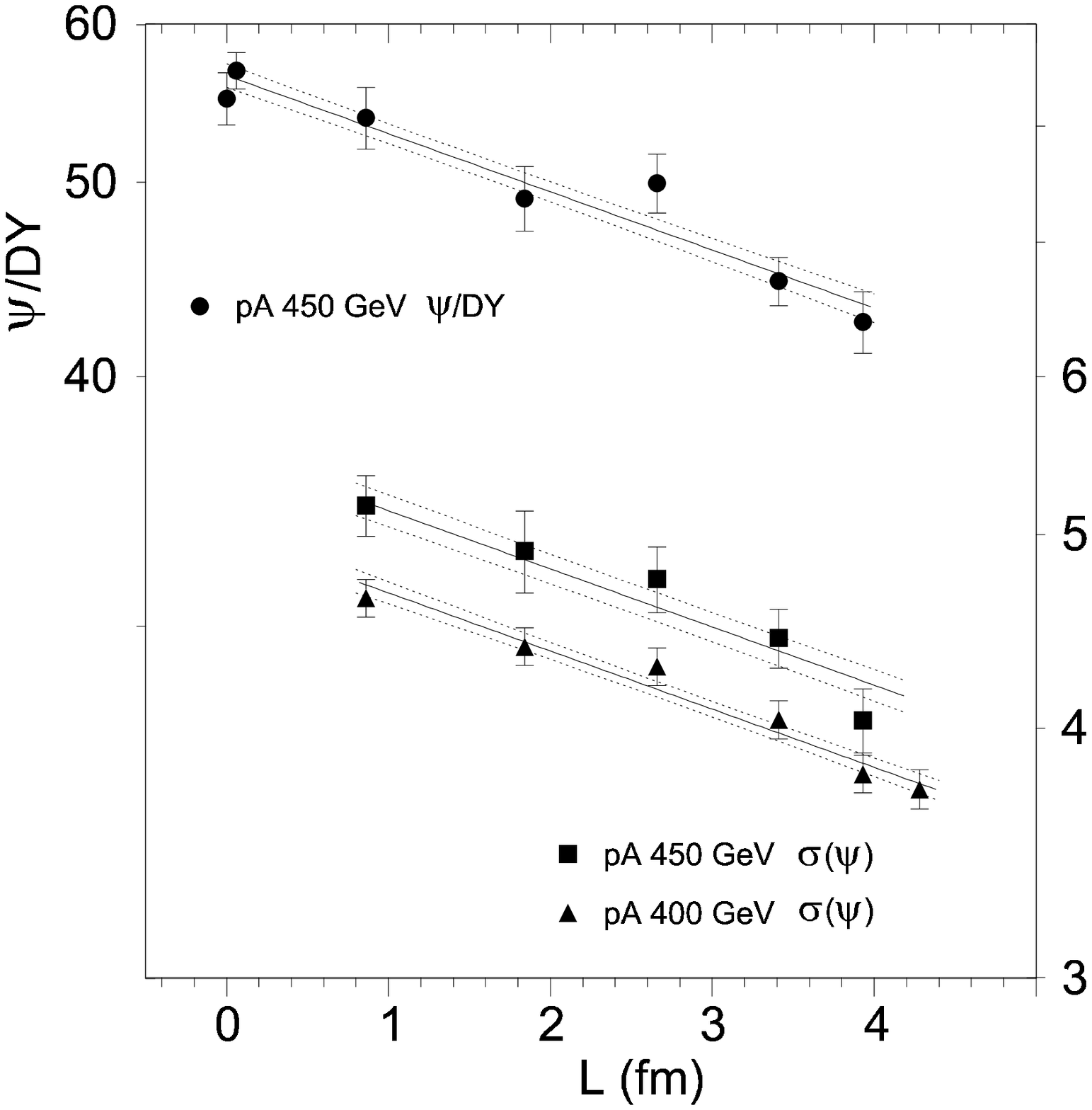}
&
\includegraphics[width=.48\textwidth]{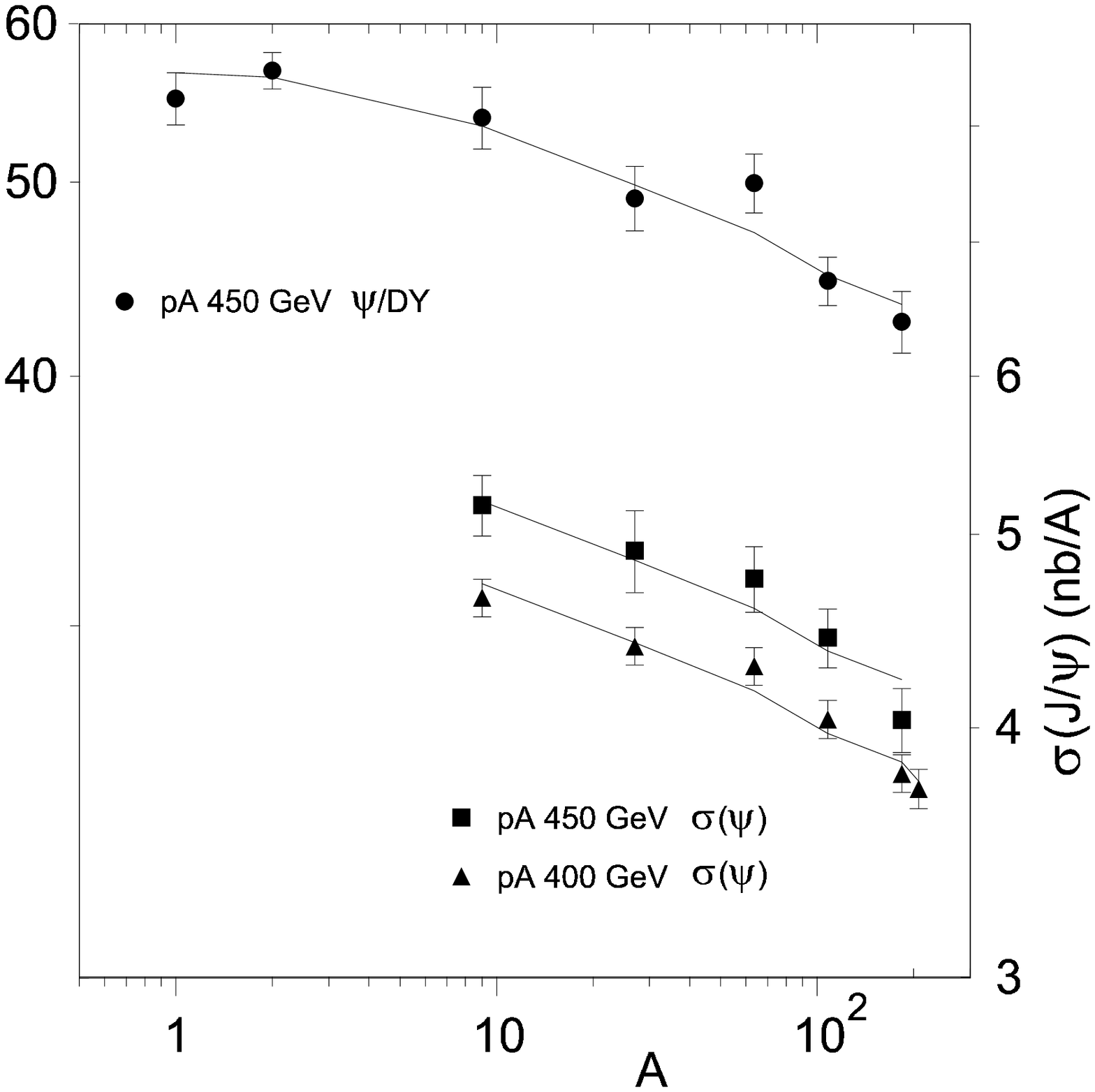}
\end{tabular}
\end{center}
\caption[The \jpsi\ production cross-sections in $pA$ 
         collisions at 400 and 450~GeV]
        {The \jpsi\ production cross-sections in $pA$ collisions at
         400 and 450~GeV times the branching ratio to dimuons, in the
         phase space window of the measurements (right axis), or with
         respect to the yield of Drell--Yan dimuons in the mass range
         $2.9<M<4.5$~GeV (left axis).  The data are plotted as a
         function of $L$ (left-hand side) and of the mass number of
         the nuclei (right-hand side).  The lines represent a fit in
         the framework of the Glauber nuclear absorption model, and
         result in a common absorption cross-section $\sigma_{\rm
         abs}=4.3\pm0.3$~mb.  The uncertainties on the result of the
         fit are represented by the dotted lines on the left-hand
         side.}
\label{fig:na60-pA}

\bigskip

\begin{center}
\includegraphics[width=.48\textwidth]{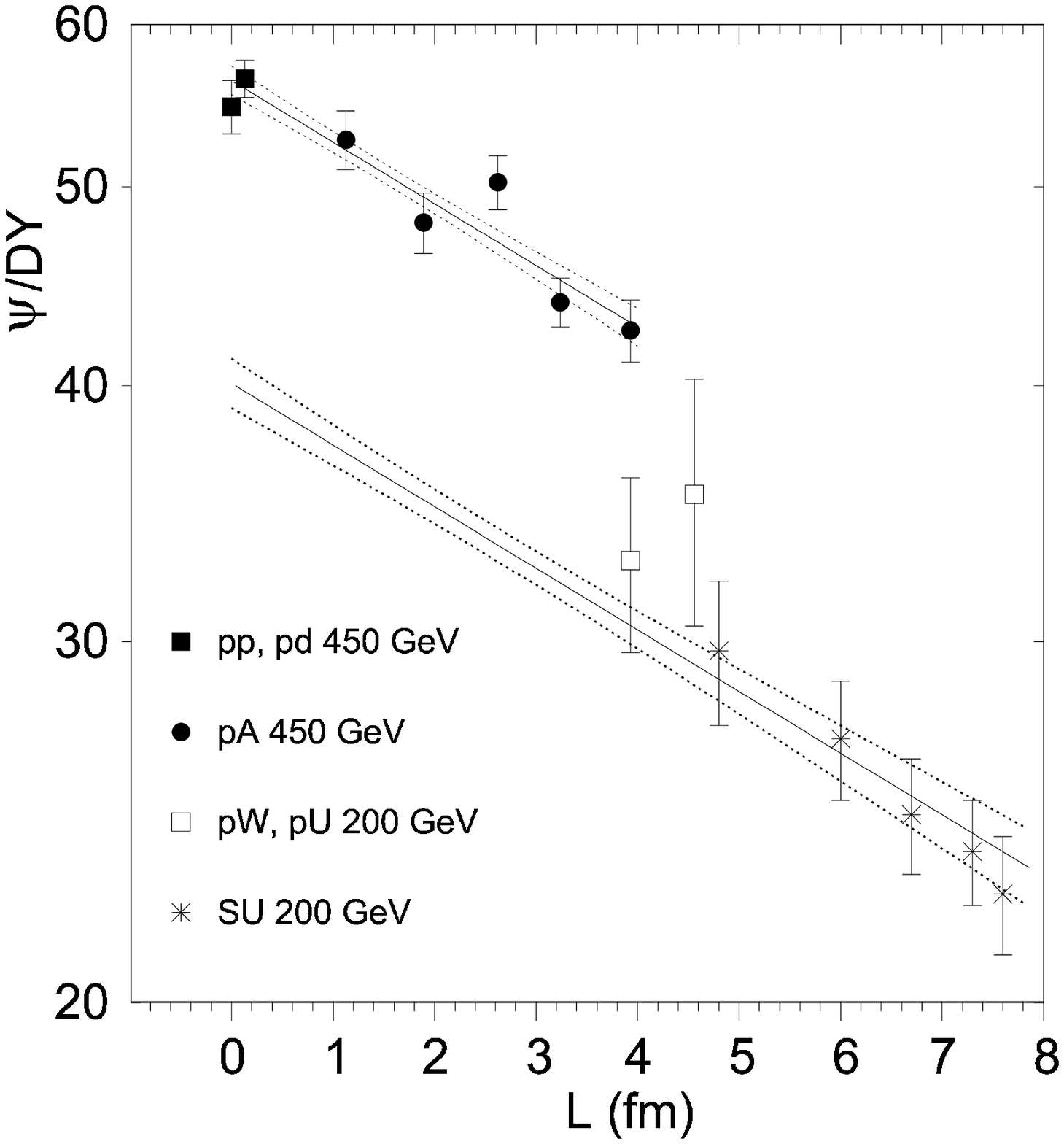}
\hfill
\includegraphics[width=.48\textwidth]{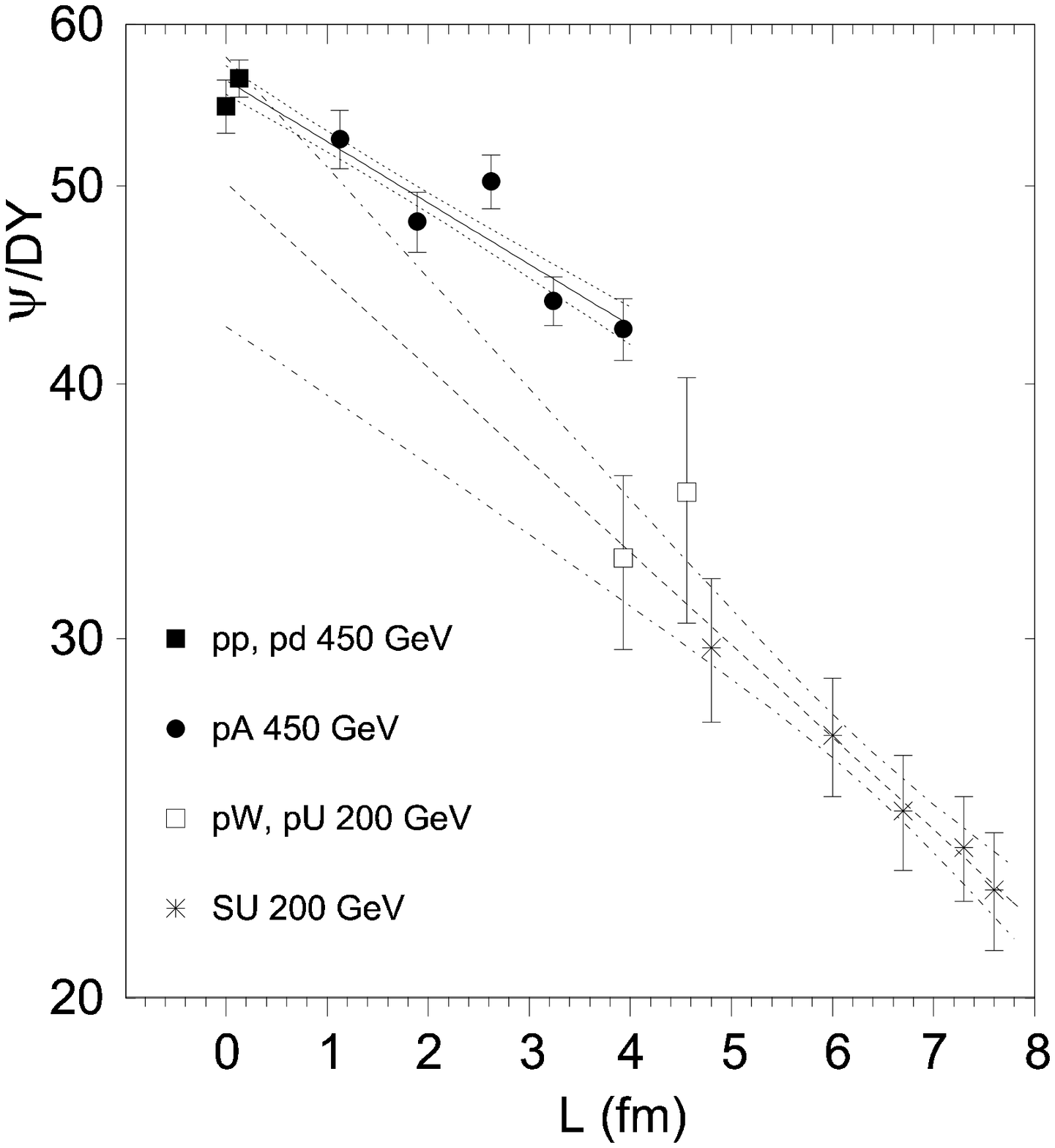}
\end{center}
\caption[Ratio of the \jpsi\ and Drell--Yan yields]
        {Ratio of the \jpsi\ and Drell--Yan yields measured in $pA$ 
         and S+U collisions, as a function of $L$.}
\label{fig:na60-pasu}
\vfill
\end{figure}

The NA60 experiment was proposed to clarify several physics questions
resulting from specific experimental measurements made by previous SPS
experiments, including the observation that \jpsi\ production is
suppressed in heavy-ion collisions with respect to the yields
extrapolated from proton--nucleus data.  The NA60 In+In measurements
have been made at the same energy as Pb+Pb to minimize systematic
uncertainties in their comparison.  However, to fully interpret the
\jpsi\ production patterns observed in nuclear collisions as a
function centrality, it is crucial to have a proper ``reference
baseline'', to extract any ``anomalous behaviour'' specific to heavy
ion collisions.  Only with such a ``normal nuclear absorption'' curve
we can look for signatures of quark--gluon plasma formation in the
heavy-ion data.  However, this reference has so far been based on
proton--nucleus data collected at a rather different beam energies,
450~GeV (and a small data sample at 400~GeV).
\Figure[b]~\ref{fig:na60-pA} summarizes those results.

The NA50 collaboration has also made use of the S+U data set collected
by NA38 at 200~GeV.  While it is certainly true that the rescaling
from 200 to 158~GeV is much more robust than from 450~GeV, this data
set has been used making the extra assumption that there is nothing
new happening between the proton--nucleus reference and the S+U
collision system.  We know, however, that the $\psi'$ resonance is
considerably suppressed in S+U collisions with respect to its ``normal
nuclear absorption'' pattern, established by exclusively studying
proton--nucleus interactions.  Therefore, even though this assumption
is based on the compatibility of the results obtained from $p$, O and
S induced reactions, it remains nevertheless a questionable assumption
which must be verified with a precise measurement.  This problem does
not prevent us from directly comparing the In and Pb data since both
sets were taken at exactly the same energy, 158~GeV.  However, the
interpretation of the measured pattern in terms of ``new physics''
requires the comparison to an ``expected'' pattern, based on a purely
conventional ``normal nuclear absorption''.  Presently, this
comparison is mostly limited by the accuracy with which we can rescale
the measured proton--nucleus points to the energy and kinematical
domain of the heavy-ion measurements.

In 2004 the NA60 experiment has collected three days of data to study
\jpsi\ production in proton--nucleus collisions with a high intensity
158~GeV primary proton beam.  This data sample will allow us to
directly establish a normal nuclear absorption reference based on
proton induced interactions, minimizing systematic uncertainties and
model-dependent assumptions.

The normal nuclear absorption pattern can be determined by comparing
the measured \jpsi\ production cross-sections (or production yield
with respect to high mass Drell--Yan) in proton--nucleus collisions for
several target nuclei with a calculation based on the Glauber
scattering formalism.  It can be approximately expressed as an
exponential function of the average length of nuclear matter the
produced charmonium state needs to traverse to get out of the nucleus,
$\sigma_{pA}(L) = \sigma_{pp} \exp(-\rho_A L \sigma_{\rm abs})$.  This
calculation uses the Woods--Saxon nuclear density profiles.  We can
describe the measured data points by adjusting a normalization
coefficient and the absorption cross-section, \sabs, to get the
absorption rate.

A priori, it may very well happen that the absorption cross-section
depends on the energy of the interactions.  In fact, it is well known
that the NA50 experiment measured stronger \jpsi\ absorption than
E866, for the same \xf\ range (close to 0). The main difference is the
proton beam energy: 800~GeV in E866 and 450~GeV in NA50.  Expressed in
terms of the simple $A^\alpha$ parameterization, E866 gives values of
$\alpha$ around 0.95 while NA50 gives values closer to 0.92.  If the
difference is due to the change in energy, we can easily imagine that
at 158~GeV, the energy of the In and Pb beams, the value of $\alpha$
would be even smaller, equivalent to having a higher normal nuclear
absorption cross-section.  Unfortunately, the energy is not the only
difference between NA50 and E866 and the change of $\alpha$ is not
understood well enough to extrapolate to lower energy.  For instance,
the \jpsi\ mesons produced at 800~GeV have, on average, higher values
of \pt\ and the value of $\alpha$ increases with \pt\ (Cronin effect).

\Figure[b]~\ref{fig:na60-pasu} shows the ratio between the \jpsi\ and
Drell--Yan (in the mass range $2.9<M<4.5$~GeV) production
cross-sections, as a function of $L$, for the proton--nucleus and S+U
data, collected either at 450 or 200~GeV.  The fit of the 400 and
450~GeV data points leads to $\sigma_{\rm abs}=4.3\pm0.5$~mb.  On the
left-hand side, we show what happens if we impose this \sabs\ when
fitting the 200~GeV data points, leaving the normalization of the
curve free.  On the right-hand side the 200~GeV data were
independently fitted, resulting in $\sigma_{\rm abs}=6.8\pm1.8$~mb.
The dotted lines indicate the uncertainty band, including both the
errors on \sabs\ and on the normalization.  These values indicate that
the absorption cross-section seems to increase when the collision
energy decreases, a tendency that would match the 800~GeV data
collected by E866.

Besides the possible change of \sabs\ from 450 to 158~GeV, another
very important unknown is the normalization of the absorption curve at
158~GeV, needed to compare the In and Pb data.  In principle, the
energy dependence of the \jpsi\ production cross-section should be
calculable so that it would be possible to estimate the normalization
at 158~GeV from the 450~GeV data.  In practice, however, such
calculations are not unique and are severely limited by
nonperturbative contributions.  The accuracy we need can only be
obtained from a measurement made at the same energy and in the same
experiment that measured the nuclear data.

Even if we assume that \sabs\ is not energy dependent and use the
value determined at 400 and 450~GeV, $\sigma_{\rm abs}=4.3\pm0.5$~mb,
to build the nuclear absorption curve at 158~GeV, we still need to
determine its normalization.  The NA50 collaboration has taken the
normalization value determined at 200~GeV, assuming that the S+U and
$pA$ data share the same \sabs\ value, obtain the normalization at
158~GeV using the so-called ``Schuler parameterization''.  This
procedure has the big advantage that the energy difference is very
small, thereby reducing the importance of the uncertainties on the
energy dependence.  However, it has the disadvantage of imposing the
extra assumption that the \sabs\ value is the same in proton--nucleus
and S+U collisions (besides ignoring any energy dependence of \sabs).
If we accept that the S+U data does not need to be described by the
\sabs\ value derived from the $pA$ data, we must start from the
450~GeV normalization and scale it down to 158~GeV, requiring accurate
knowledge of the energy dependence of \jpsi\ production.

Since we are comparing data expressed as the ratio between the \jpsi\
and Drell--Yan cross-sections, the energy dependence of the Drell--Yan
process also needs to be accurately known in the same phase space
window to evaluate the scaling factor needed to normalize the 158~GeV
data from 450~GeV.  Different parton distribution functions may give
somewhat different energy dependences although the calculations are
more robust.  Unfortunately, the statistics collected in a few days
will not allow us to verify the energy dependence of the Drell--Yan
production cross-section.

It should be clear by now that it is crucial to measure the \jpsi\
absolute production cross-section in proton--nucleus collisions at
158~GeV, if we want to fully understand \jpsi\ suppression.  Such
measurements started in 2004 but should be repeated in the near
future, with much higher statistics, to ensure a proper baseline for
$\psi'$ and Drell--Yan production in heavy-ion collisions.  We
conclude by summarizing the issue:

\begin{itemize}
\item
The 450~GeV data points alone are not enough to determine the normal
nuclear absorption at 158~GeV since the energy rescaling factors are
too uncertain.

\item
The existing 200~GeV data points are also not enough because
of their poor precision.

\item
The 450 and 200~GeV data sets, used together, would solve the problem
if the absorption cross-section is the same for the two sets, an
assumption presently without solid experimental evidence.
\end{itemize}

Thus proton data at 158~GeV must be collected in order to establish a
robust reference baseline with respect to which the In+In and Pb+Pb
\jpsi\ suppression patterns can be directly compared to place the
existence of ``anomalous'' effects in the heavy-ion data on more solid
ground.  The present systematic errors due to the energy (and phase
space) corrections and to the absence of solid evidence that the
absorption cross-section remains the same from 450 to 158~GeV, are the
main sources of uncertainty in the interpretation of the data
collected in nuclear collisions.

\subsection{Charmonium production at RHIC}

We review here the first results on charmonia produced in nuclear
collisions at the Relativistic Heavy Ion Collider (RHIC).  PHENIX is
specifically designed to make use of high luminosity ion--ion,
proton--ion, and proton--proton collisions at the RHIC to sample rare
physics probes including the $J/\psi$ and other heavy quarkonium
states.  The PHENIX experiment reported on $J/\psi$ production in p--p,
d--Au and Au--Au reactions at
$\sqrt{s_{NN}}$~=~200~GeV~\cite{Adler:2003qs,deCassagnac:2004kb,Adler:2003rc}.

The PHENIX experiment is able to measure $J/\psi$'s through their
dilepton decay in four spectrometers: two central arms covering the
mid-rapidity region of $|\eta|<0.35$ and twice $\pi/2$ in azimuth and
two forward muon arms covering the full azimuth and $1.2 < |\eta| <
2.4$ in pseudorapidity.  The central spectrometers are comprised, from
the inner radius outward, of a Multiplicity and Vertex Detector (MVD),
Drift Chambers (DC), Pixel Pad Chambers (PC), Ring Imaging Cerenkov
Counters (RICH), a Time-of-Flight Scintillator Wall (TOF), Time
Expansion Chambers (TEC), and two types of Electromagnetic
Calorimeters (EMC).  This combination of detectors allows for the
clean electron identification over a broad range of transverse
momentum.  Each forward spectrometer consists of a precision muon
tracker (MuTr) comprised of three stations of cathode-strip readout
chambers followed by a muon identifier (MuID) comprised of multiple
layers of steel absorbers instrumented with low resolution planar
drift tubes.  Muons at the vertex must have a mean energy of at least
1.9~GeV to reach the MuID system.  Further details of the detector
design and performance are given in Ref.~\cite{Adcox:2003zm}.

\begin{figure}[t]
 \begin{minipage}[b]{.46\linewidth}
  \centering\includegraphics[width=\linewidth]{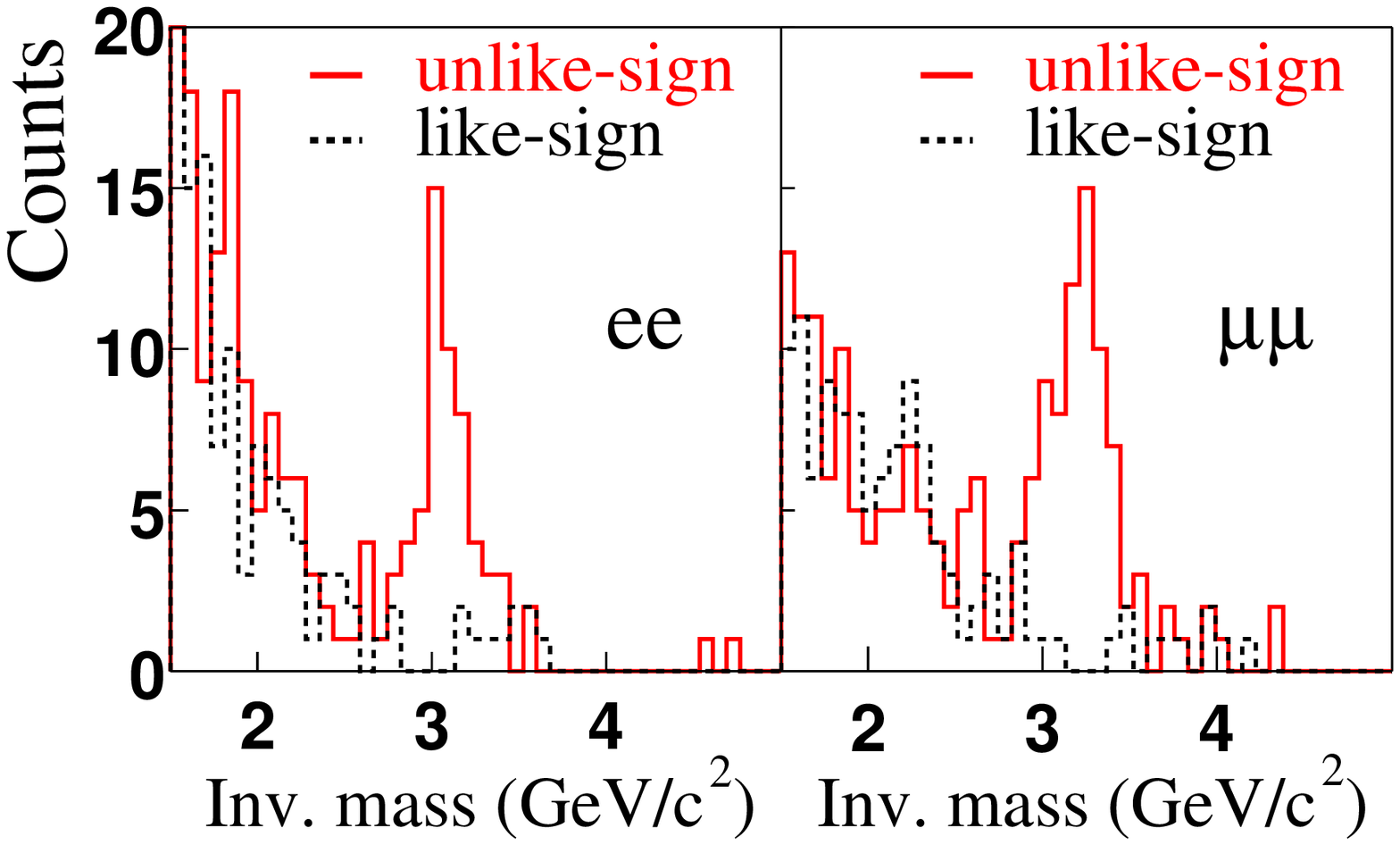}
%  \vspace*{2ex}
  \caption[The invariant mass spectra for dielectron and dimuon pairs in 
           the Run 2 data sample]
          {The invariant mass spectra for dielectron and dimuon pairs in 
           the Run 2 data sample. 
           Unlike-sign pairs are shown as solid lines, like-sign pairs as
           dashed lines.}
   \label{fig:phenix1}
 \end{minipage} \hfill 
 \begin{minipage}[b]{.46\linewidth}
  \centering\includegraphics[width=.60\linewidth]{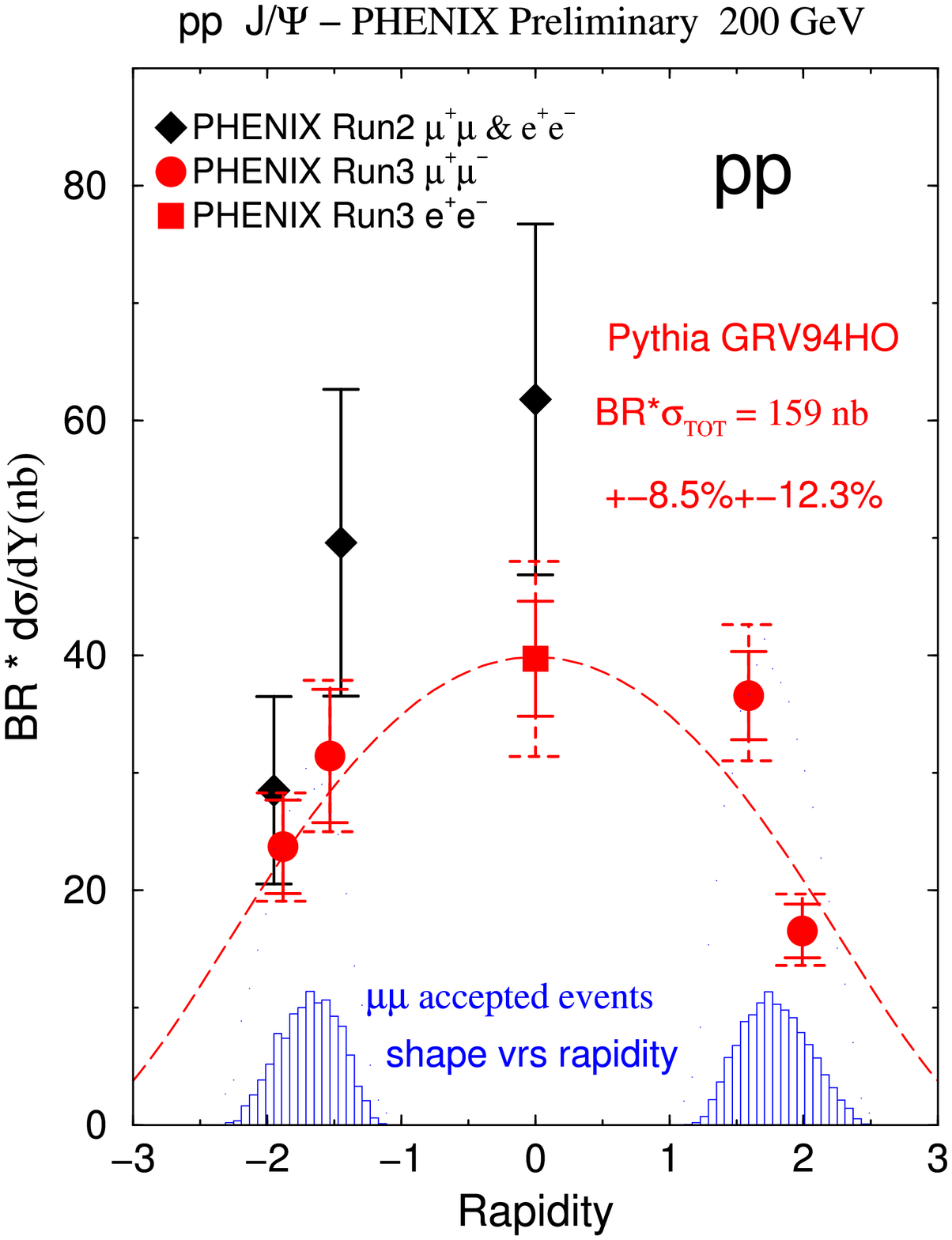}
%  \vspace*{2ex}
  \caption[$J/\psi$ differential cross-section multiplied by the
           dilepton branching ratio versus rapidity]
          {$J/\psi$ differential cross-section, multiplied by the
           dilepton branching ratio, versus rapidity as measured by
           the central and muon spectrometers.}
  \label{fig:phenix2}
 \end{minipage} 
\end{figure}

The data were recorded during the 2001/2002 and 2003 runs at
$\sqrt{s}$ = 200~GeV with 150~nb$^{-1}$ and 350~nb$^{-1}$ p--p
collisions.  Event samples were selected using online triggers and
offline reconstruction criteria as described in
Ref.~\cite{Adler:2003qs}.  Unlike-sign pairs and, for background
estimation, like-sign pairs were combined to form invariant mass
spectra.  In \Figure~\ref{fig:phenix1}, unlike-sign and like-sign
invariant mass spectra from the entire Run2 $pp$ data set are shown
together.  For electrons, the net yield in the mass region 2.8--3.4
GeV is 46, for muons in the range 2.71--3.67~GeV is 65.

The $J/\psi$ cross-sections were determined from the measured yields
using
\begin{equation}
B_{ll}
\frac{ d^2\sigma_{J/\psi} }{ dy dp_T} \,=\,
\frac{ N_{J/\psi} }{ (\int{{\cal L}dt}) \, {\Delta}y \, {\Delta}p_T} \,
\frac{ 1 } { A \, \epsilon }
\label{eq:dsigdpt_alt}
\end{equation}
where $N_{J/\psi}$ is the measured $J/\psi$ yield, $\int{{\cal L}dt}$
is the integrated luminosity, $B_{ll}$ is the branching fraction for
the $J/\psi$ to either $e^+e^-$ or $\mu^+\mu^-$ pairs (PDG average
value 5.9\%\cite{pdg}), and $A \, \epsilon$ is the acceptance times
efficiency for detecting a $J/\psi$.  The $J/\psi$ rapidity
distribution obtained by combining the dielectron and dimuon
measurements is shown in \Figure~\ref{fig:phenix2} with the muon arm data
divided into two rapidity bins.  A fit to a shape generated with
PYTHIA using the GRV94HO parton distribution functions is performed
and gives a total cross-section, multiplied by the dilepton branching
ratio, equal to:
\begin{equation}
{\rm BR} \times \sigma_{pp}^{J/\psi} = 159 \; {\rm nb} \pm 8.5\% \,({\rm fit}) \pm
 12.3\% \,({\rm abs})
\end{equation}
where the first uncertainty comes from the fit and thus includes both
the statistical and point-to-point systematics. The second uncertainty
accounts for normalization systematic errors.

\begin{figure}[t]
\begin{center}
\includegraphics[width=55mm]{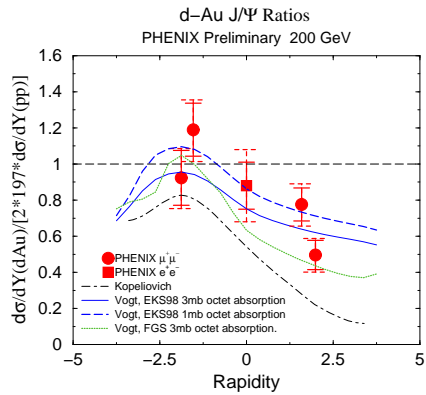}
\end{center}
\caption[Ratio between d--Au and p--p $J/\psi$ differential cross-sections
         versus rapidity]
        {Ratio between d--Au and p--p $J/\psi$ differential cross-sections, 
         divided by $2\times 197$, versus rapidity.}
\label{fig:phenix3}
\end{figure}

Preliminary analysis is now available of the data recorded during the
2003 run at $\sqrt{s}$ = 200~GeV with 2.74~nb$^{-1}$ d--Au collisions.
In d--Au collisions, PHENIX is able to measure $J/\psi$ production at
forward, backward and central rapidity probing moderate to low x
regions of the Au nucleus. The covered rapidity region spans the
expected shadowing, antishadowing and no shadowing regions. The ratio
between the $J/\psi$ yields observed in d--Au and p--p collisions
divided by $2\times197$ is shown in \Figure~\ref{fig:phenix3}.  Solid
error bars represent statistical and point to point systematic
uncertainties.  The dashed error bars stand for the systematic
uncertainties common to one spectrometer.  An additional 13.4\% global
error bar is not displayed.
 
While this ratio is close to unity at backward rapidity, it is
significantly lower at forward rapidity, where parton distributions
are expected to be shadowed in a heavy nucleus. Theoretical
predictions ~\cite{Klein:2003dj,Kopeliovich:2001ee} are displayed on
the figure for comparison.  The shape is consistent with shadowing at
low $x$ and less shadowing at larger $x$.  Unfortunately, the
statistical and systematic error bars make it difficult to distinguish
among various shadowing models and models with various amounts of
nuclear absorption.

The Au--Au data at $\sqrt{s_{NN}}=200$~GeV used in this analysis were
recorded during Run 2 at RHIC in the fall of 2001.  For three
exclusive centrality bins, 0--20\%, 20--40\%, and 40--90\% of the total
Au--Au cross-section, we determined the branching fraction of $J/\psi
\rightarrow e^{+}e^{-}$ (B=$5.93 \pm 0.10 \times 10^{-2}$~\cite{pdg})
times the invariant yield at mid-rapidity $dN/dy|_{y=0}$.

\begin{figure}[p]
\centering\includegraphics[width=.65\linewidth]{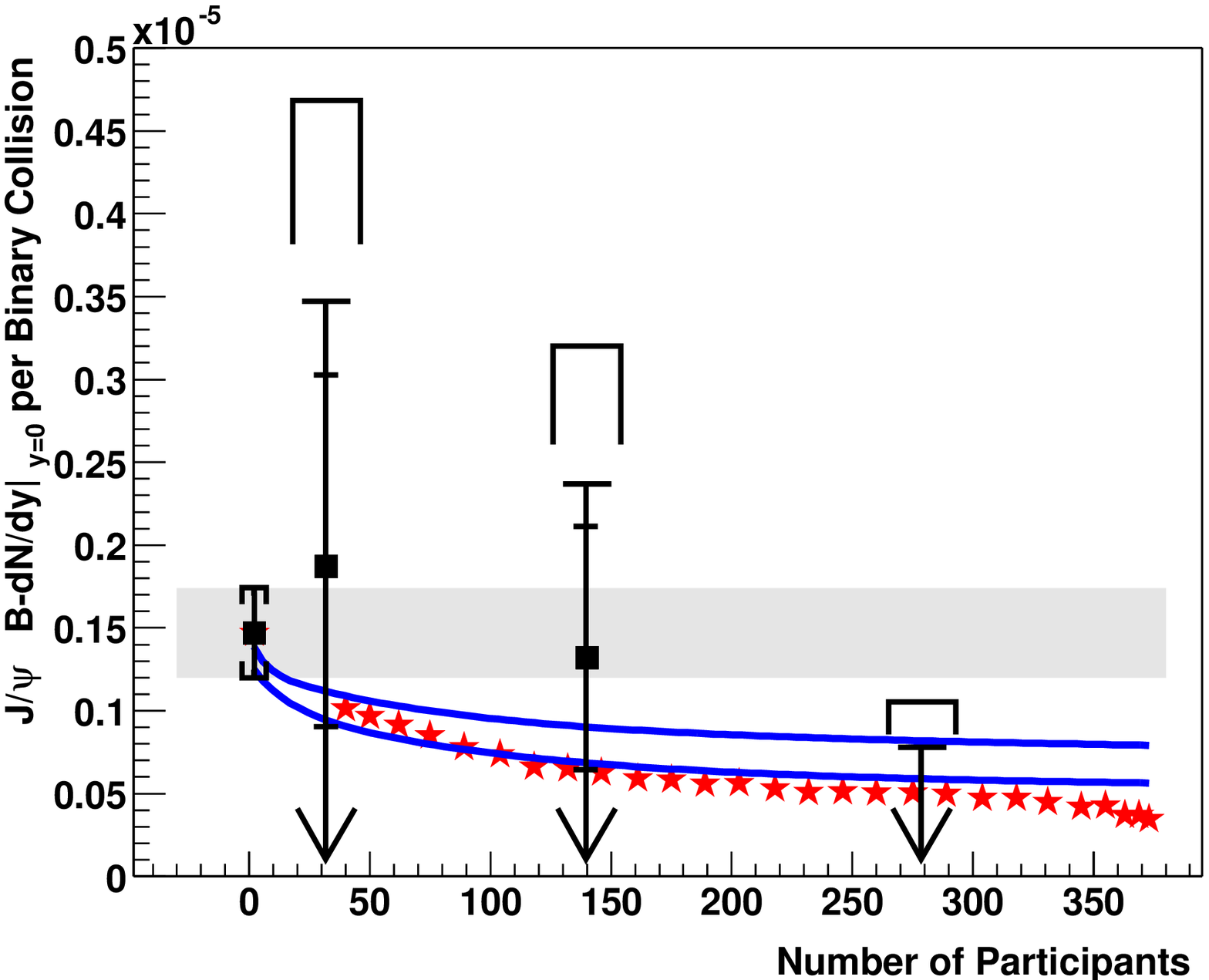}
\caption[The $J/\psi$ yield per binary collision]
        {The $J/\psi$ yield per binary collision is shown from
         proton--proton reactions and three exclusive centrality
         ranges of Au--Au reactions all at $\sqrt{s_{NN}}$ = 200~GeV.
         The solid line is the theoretical expectation from ``normal''
         nuclear absorption with $\sigma_{c\overline{c}-N}$ = 4.4 mb
         (upper curve) and 7.1 mb (lower curve) cross-section.  The
         stars are the $J/\psi$ per binary collision measured by the
         NA50 experiment at lower collision energy.  In order to
         compare the shapes of the distribution, we have normalized
         the NA50 data to match the central value for our
         proton--proton results.}
\label{fig:phenix-auau1}

\medskip

\centering\includegraphics[width=.65\linewidth]{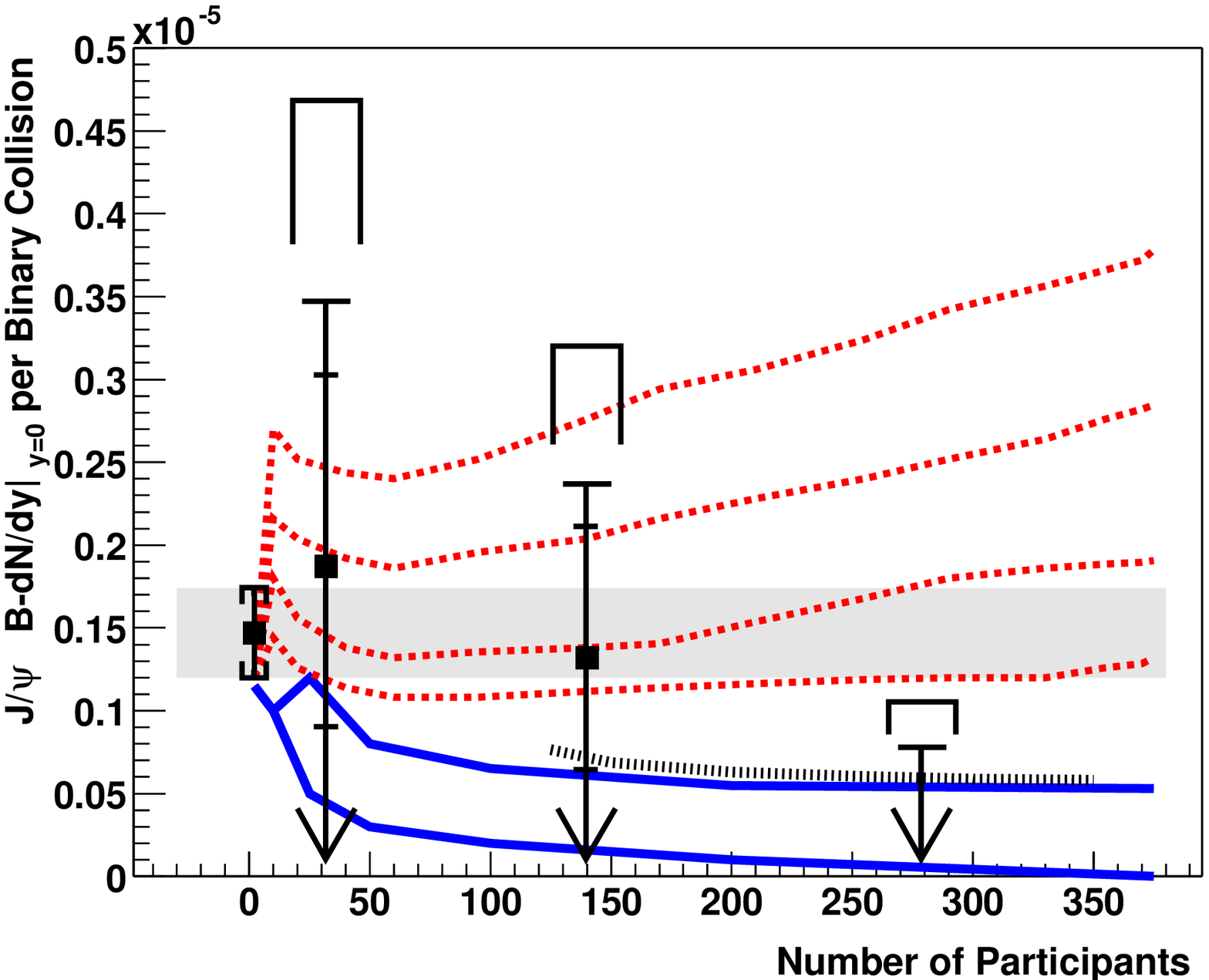}

\caption[More $J/\psi$ yields per binary collision]
        {The $J/\psi$ yield per binary collision is shown from
         proton--proton reactions and three exclusive centrality
         ranges of Au--Au reactions all at $\sqrt{s_{NN}}$ = 200~GeV.
         The lowest curve is a calculation including ``normal''
         nuclear absorption in addition to substantial absorption in a
         high temperature quark--gluon
         plasma~\cite{Grandchamp:2001pf,Grandchamp:2002wp}.  The curve
         above this is including backward reactions that recreate
         $J/\psi$.  The statistical model~\cite{Andronic:2003zv}
         result is shown as a dotted curve for mid-central to central
         collisions just above that.  The four highest dashed curves
         are from the plasma coalescence model~\cite{Thews:2000rj} for
         a temperature parameter of $T$ = 400~MeV and different charm
         rapidity widths.}
\label{fig:phenix-auau2}
\end{figure}

In \Figure~\ref{fig:phenix-auau1} we show the results from the three
Au--Au centrality bins and the proton--proton data normalized per binary
nucleon--nucleon collision as a function of the number of participating
nucleons.  Note that for proton--proton reactions, there are two
participating nucleons and one binary collision.  Despite the limited
statistical significance of these first $J/\psi$ results, we can
address some important physics questions raised by the numerous
theoretical frameworks in which $J/\psi$ rates are calculated.  The
binary scaling expectations are also shown as a gray band.  We also
show a calculation of the suppression expected from ``normal'' nuclear
absorption using a $\sigma_{\mathrm{abs}} = 4.4$ mb~\cite{Abr03} and $7.1$
mb~\cite{Kharzeev:1996yx}.  We show the NA50 suppression pattern
relative to binary scaling~\cite{Abr00,Abr01}, normalized to match our
proton--proton data point at 200~GeV.  The data disfavor binary scaling
while they are consistent with ``normal'' nuclear absorption alone and
also the NA50 suppression pattern measured at lower energies, within
the large statistical errors.
 
One model calculation~\cite{Grandchamp:2001pf,Grandchamp:2002wp}
including just the ``normal'' nuclear and plasma absorption components
at RHIC energies is shown in \Figure~\ref{fig:phenix-auau2}.  The
higher temperature ($T$) and longer time duration of the system at
RHIC lead to a predicted larger suppression of $J/\psi$ relative to
binary collision scaling.

Many recent theoretical calculations also include the possibility for
additional late stage re-creation or coalescence of $J/\psi$ states.
In Ref.~\cite{Grandchamp:2001pf,Grandchamp:2002wp}, both break-up and
creation reactions $D + \overline{D} \leftrightarrow J/\psi + X$ are
included.  At the lower fixed target CERN energies, this represents a
very small contribution due to the small charm production
cross-section.  However, at RHIC energies, where in central Au--Au
collisions around 10 $c\overline{c}$ pairs are produced, the
contribution is significant.

The sum of the initial production, absorption, and re-creation is
shown in \Figure~\ref{fig:phenix-auau2}.
 
A different calculation~\cite{Thews:2000rj} assumes the formation of a
quark--gluon plasma in which the mobility of heavy quarks in the
deconfined region leads to increased $c\overline{c}$ coalescence.
This leads to a very large enhancement of $J/\psi$ production at RHIC
energies for the most central reactions.  The model considers the
plasma temperature ($T$) and the rapidity width ($\Delta y$) of charm
quark production as input parameters.  Shown in
\Figure~\ref{fig:phenix-auau2} are the calculation results for $T$ =
400~MeV and $\Delta y$ = 1, 2, 3, and 4.  The narrower the rapidity
window in which all charm quarks reside, the larger the probability
for $J/\psi$ formation.  All of these parameters within this model
predict a $J/\psi$ enhancement relative to binary collisions scaling,
disfavored by our data.
 
Another framework for determining quarkonia yields is to assume a
statistical distribution of charm quarks that may then form quarkonia.
A calculation assuming thermal, but not chemical
equilibration~\cite{Andronic:2003zv}, is shown in
\Figure~\ref{fig:phenix-auau2}.
 
Significantly larger data sets are required to address the various
models that are still consistent with our first measurement.  Key
tests will be the $p_{T}$ and $x_{F}$ dependence of the $J/\psi$
yield, and how these compare with other quarkonium states such as the
$\psi'$.

\section[Quarkonium photoproduction at hadron colliders]
        {Quarkonium production in nuclear collisions
         $\!$\footnote{Author: S.~R.~Klein}}
 
Quarkonium may be produced via photoproduction at hadron colliders.
The electromagnetic field of a proton or ion projectile acts as a
field of virtual photons.  These photons may collide with an oncoming
target nucleus to produce quarkonium.  The photon flux is high enough
to allow detailed studies of charmonium photoproduction. The
cross-section is sensitive to the gluon content of the target.
Photoproduction at the LHC reaches an order of magnitude higher
energies than are possible at HERA.  For ions, the advance beyond
existing data is even larger which may allow the first real low-$x$
measurements of gluon shadowing.

%begin section

Photoproduction has traditionally been studied with fixed target
photon beams and at electron--proton colliders. However, energetic
hadrons also have significant electromagnetic fields and high energy
$pp$, $\overline{p}p$ and $AA$ colliders can be used to study
photoproduction at photon energies higher than those currently
accessible.  These photoproduction reactions are of interest as a way
to measure the gluon distribution in protons at low Feynman~$x$.  The
Fermilab Tevatron, RHIC and the LHC (with both proton and ion beams)
all produce significant quantities of heavy quarkonium.  Indeed,
$J/\psi$ photoproduction may have already been seen by the CDF
collaboration.

One unique aspect of photoproduction at hadron colliders is that the
initial system is symmetric so that photoproduction can occur from
either ion.  Since these two possibilities are indistinguishable, the
amplitudes must be added.  This interference significantly affects the
$p_T$ spectrum of the produced quarkonium.  The relative sign of the
two amplitudes depends on the symmetry of the system.  The symmetry is
different for $p\overline p$ than $pp$ and $AA$ colliders, leading to
significantly different $p_T$ spectra for particle--particle and
particle--antiparticle colliders.

\subsection{Cross-section calculation}

The cross-section for vector meson production is the convolution of
the photon flux $dn/dk$ for photon energy $k$ with the photon--proton
or photon--nucleus cross-sections $\sigma_{\gamma A}$
\cite{ppupsilon,usrate}:
\begin{equation}
   \sigma(A+A \rightarrow A+A+V) = 2 \int_{0}^{\infty} \frac{dn}{dk}
   \, \sigma_{\gamma A} (k) \, d k \; .
\label{eq:sigma}
\end{equation}
Here, $A$ refers to any ion, including protons.  The '2' is because
either nucleus can emit the photon or be the target.  Interference
between the two possibilities alters the $p_T$ distribution 
but does not significantly affect the total cross-section.

The rapidity, $y$, of a produced state with mass $M_V$ is related to
the photon energy through $y = \ln(2k/M_V)$. Using this relation in
\Eq~(\ref{eq:sigma}) and differentiating gives
\begin{equation}
\frac{d \sigma}{dy} = k \frac{dn}{dk} \sigma_{\gamma 
A \rightarrow V A} (k) \; .
\end{equation}
At mid-rapidity, $y=0$, the photon energies for the two possibilies
(as to which nucleus emitted the photon) are equal.  However, away
from mid-rapidity, the photon energies are different,
\begin{equation}
k_{1,2} = \frac{M_V}{2} \exp(\pm y)
\end{equation}
so that the amplitudes for the two possibilities are
also different.

We now consider the ingredients in the cross-section in turn.  The
photon flux from a relativistic hadron is given by the
Weizs\"acker--Williams formalism.  One important detail is the form
factor of the emitter.

For protons, the form factor was considered by Drees and Zeppenfield.
They use a dipole form factor $F(Q^2) = 1/(1 +Q^2/(0.71 \, {\rm
GeV}^2) )^2$ for the proton, and found \cite{Drees:1988pp}:
\begin{eqnarray}
\frac{dn}{dk} = \frac{\alpha}{2 \pi k} \left[ 1 + (1 - z)^2 \right] 
\left( \ln{I} - \frac{11}{6} + \frac{3}{I}  - \frac{3}{2 I^2} + \frac{1}{3 I^3} \right) 
\label{eq:pdndk}
\end{eqnarray}
where $z=W_{\gamma p}^2/s$, $A = 1 + (0.71 \, {\rm GeV}^2)/Q_{\rm
min}^2$ and $Q_{\rm min}^2 \approx (k/ \gamma)^2$.  Here, $W_{\gamma
p}$ is the $\gamma p$ centre of mass energy, and $\sqrt{s}$ is the
proton--proton centre of mass energy.  This photon spectrum is similar
to that of a point charge with a minimum impact parameter of $b_{\rm
min} = 0.7$~fm.

Drees and Zeppenfeld neglected the magnetic form factor of the proton which
 is important only at very high energies\cite{Kniehl:1990iv}.
They also required that the proton remain intact.  If the proton is
allowed to be excited, the effective flux increases considerably
\cite{Ohnemus:1993qw}.  However, in this case, one or both protons
dissociate, producing hadronic debris.  We will ignore this
possibility here since the debris considerably complicates event
selection.

For ion--ion collisions, the cutoff conditions are somewhat different.
For a vector meson to be observable, the two nuclei must miss each
other, with $b_{\rm min} = 2R_A$, where $R_A$ is the nuclear radius.
The effective photon flux from a charge $Z$ nucleus is the flux
striking an incoming nucleus subject to that constraint.  This is
within 15\% of the flux integrated over the requirement $r>2R_A$,
given analytically by \cite{baurreview}
\begin{equation}
\frac{dn}{dk} = \frac{2Z^2\alpha}{\pi k} 
\big(XK_0(X)K_1(X) - \frac{X^2}{2}\big[K_1^2(X)-K_0^2(X)\big]\big)
\label{eq:ffions}
\end{equation}
where $X=kr/\gamma\hbar$.  More detailed calculations find the photon
flux numerically by determing the photon flux from one nucleus that
interacts in another, subject to the criteria that the two nuclei do
not interact hadronically.

As a check of the photon flux rates, we consider an alternate
calculation that replaces the proton form factor with a hard cutoff,
$b_{\rm min} = 1.0$~fm.  This stricter requirement slightly decreases
the effective photon flux.

The leading-order vector meson photoproduction cross-section for a
vector meson with mass $M_V$ is \cite{Ryskin:1992ui}
\begin{equation}
\left. \frac{d \sigma (\gamma p \rightarrow Vp)}{dt} \right|_{t=0} =
\frac{\alpha_s ^2 \Gamma_{ee}}{3 \alpha M_V ^5} 16 \pi^3 \left[ x
g(x,M_V^2/4) \right]^2.
\end{equation}
More recent and more sophisticated calculations have considered the
use of relativistic wave functions, off-diagonal parton distributions,
and NLO contributions \cite{Frankfurt:1998yf,Martin:1999rn}.
Parton--hadron duality has also been used to study quarkonium
production.  Calculations give cross-sections $\sim$30--50\% larger
than the pQCD results, depending on $W_{\gamma p}$
\cite{Martin:1999rn}.  The QCD calculations are in reasonable
agreement with data on $J/\psi$ \cite{HERAjpsi} and $\Upsilon$
\cite{Adloff:2000vm}\cite{Breitweg:1998ki} production at HERA.  The
$\Upsilon$ data has limited statistics, and, consequently, significant
uncertainties.

To assess the viability of photoproduction at hadron colliders, we use
simple cross-section parameterizations to estimate the rates:
\begin{equation}
\sigma_{\gamma p}(W_{\gamma p}) = 1.5 W_{\gamma p}^{0.8} \, {\rm pb}
\end{equation}
for $J/\psi$, and
\begin{equation}
\sigma_{\gamma p}(W_{\gamma p}) = 0.06 W_{\gamma p}^{1.7} \, {\rm pb}
\end{equation}
for $\Upsilon(1S)$ where $W_{\gamma p}$ is in GeV.  Both H1 and
ZEUS estimate that roughly 70\% of the signal is from the
$\Upsilon(1S)$ state.  The cross-section for the $\psi'$ is expected
to be about 15\% of that for the $J/\psi$.

One drawback for this parameterization is that there is a significant
discontinuity at threshold, $W_{\gamma p}=m_p + m_{J/ \Psi}$.  In this
region, either better data or a more sophisticated calculation is
needed.

For heavy mesons, the cross-section $\gamma A\rightarrow VA$ is not
well measured.  There is little data for the $J/\psi$ and none on the
$\Upsilon$.  The ion-target cross-sections depends on the square of
the gluon density.  Thus vector meson photoproduction can provide a
sensitive measurement of gluon shadowing.  At LHC, mid-rapidity
production of the $J/\psi$ and $\Upsilon$ corresponds to $x\approx
5\times 10^{-4}$ and $1.7\times 10^{-3}$ respectively.  In this
region, shadowing will likely reduce the cross-sections by 30--50\% for
Pb+Pb collisions at the LHC.

Neglecting shadowing, the cross-section for vector meson
photoproduction in a nucleus, $\gamma A\rightarrow VA$, may be
determined using data on photoproduction off a proton target as input
to a Glauber calculaton.  However, because the cross-section for a
$c\overline c$ or $b\overline b$ pair to interact in a nucleus is
relatively small, multiple interactions are unlikely and the
calculation simplifies so that the forward scattering cross-section
scales with $A^2$ \cite{usrate,strikmanjpsi}
\begin{equation}
\frac{d \sigma (\gamma A \rightarrow \Upsilon A)}{dt}  = A^2 \, 
\left. \frac{d \sigma (\gamma p \rightarrow \Upsilon p)}{dt} 
\right|_{t=0} \, | F(t) |^2 \; .
\label{eq:A2_scaling}
\end{equation}
A Woods--Saxon distribution can be used for the nuclear form factor
$F(t)$.  The total photonuclear cross-section is the integral of
\Eq~(\ref{eq:A2_scaling}) over all momentum transfers, $t > t_{\rm
min} = [M_{\Upsilon}^2/4 k \gamma]^2$. For protons,
$d\sigma/dt|_{t=0}$ is determined from HERA data.  For the $\Upsilon$,
an exponential $t$-dependence is assumed with the same slope, 4
GeV$^{-2}$, as for $J/ \psi$ production, leading to a forward
scattering amplitude about 5\% lower than if the proton form factor
had been used.  With these ingredients, the ion--ion photoproduction
cross-section may be calculated from \Eq~(\ref{eq:sigma}).

This $A^2$ scaling applies for coherent production with $p_T <
\hbar/R_A$.  At significantly larger $p_T$, the scattering is
incoherent and the cross-section should scale as $A$.  The coherence
leads to a large peak in the production at small $p_T$, providing a
clean experimental signature which greatly simplifies the data
analysis.

For lighter mesons, the interaction cross-section is larger and the
Glauber calculation is required to determine the $A$ scaling.  The
STAR collaboration has studied $\rho$ photoproduction in Au+Au
collisions \cite{Adler:2002sc} and has found that the cross-sections
agree with a calculation based on \Eq~(\ref{eq:sigma}) and a Glauber
calculation.

Exclusive $J/\psi$ photoproduction in $pp$ interactions was also
considered by Khoze {\it et al.}\cite{Khoze:2002dc}. They use a very
different approach, based on the proton energy lost.

\subsection{Experimental prospects}

We consider $pp$ and $AA$ collisions at RHIC, $p\overline p$ at the
Tevatron, and $pp$ and Pb+Pb the LHC.  The cross-sections and rates,
as well as the assumed energies and luminosities, are shown in
\Table~\ref{tab:photosigma}.  Since the LHC is primarily a $pp$
collider, assume $10^7$ s of $pp$ and $10^6$s of $AA$ running per
year.  Because RHIC is primarily an ion collider, we reverse the two
durations.  We assume $10^7$ s of Tevatron operation per year.
Although it is interesting, we do not consider d$A$ collisions at RHIC
or $pA$ or d$A$ collisions at the LHC.

\begin{table}[t]
\caption[Cross-sections and rates for production of the $J/\psi$ and
         the $\Upsilon$] 
        {Cross-sections and rates for production of the $J/\psi$ and
         the $\Upsilon$.  The rates are for $10^7$ s of running at the
         Tevatron, $pp$ at the LHC and $AA$ at RHIC, and $10^6$ s of
         running with $AA$ at the LHC and $pp$ at RHIC.}
\label{tab:photosigma}
\begin{tabular}{l|ccc|rr|rr}
Collider & Species & $\sqrt{s_{NN}}$ & Luminosity &
\multicolumn{2}{c|}{$J/\psi$}  & \multicolumn{2}{c}{$\Upsilon$}\\
& & (GeV) & (cm$^{-2}$s$^{-1}$)
& \multicolumn{1}{c}{$\sigma$ ($\mu$b)} 
& \multicolumn{1}{c|}{Rate} 
& \multicolumn{1}{c}{$\sigma$ (nb)} 
& \multicolumn{1}{c}{Rate} \\
\hline
RHIC     & $pp$ & 500 & $10^{31}$ &
0.007 & $7.0\!\times\!10^{4}$    & 0.012   & 120 \\

RHIC     & Au+Au & 200 & $2\times10^{26}$ &
290 & $5.8\times10^5$  & --  & -- \\

Tevatron & $p \overline p$ & 1960 & $2\times10^{32}$ & 
0.023 & $4.6\!\times\!10^{7}$ & 0.12 & $2.4\!\times\!10^5$ \\

LHC      & $pp$ & 14000 &  $10^{34}$ &
0.120 & $1.2\!\times\!10^{10}$ & 3.5 & $3.5\!\times\!10^8$ \\

LHC & Pb+Pb & 5500 & $10^{27}$ & 32000 &
$3.2\!\times\!10^{6}$ & 170000 & $1.7\!\times\!10^5$ \\

\hline
\end{tabular}
\end{table}

\begin{figure}[t]
\begin{center}
\includegraphics[width=.75\linewidth]{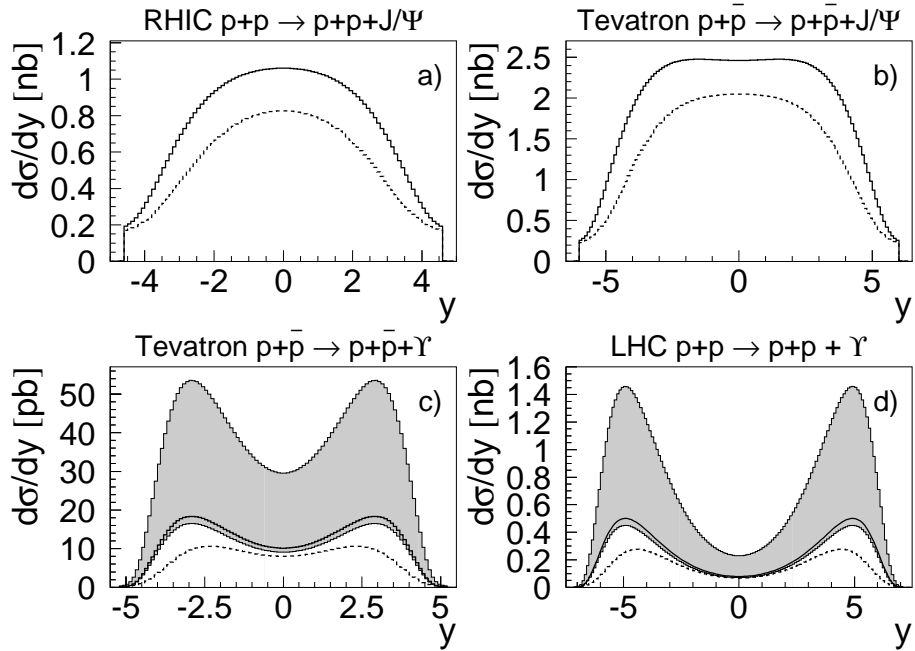} 
\caption[Rapidity distributions for photoproduction of
         $J/\psi$ and $\Upsilon$($1S$) mesons in $p p$ and $p \overline{p}$
         interactions]
        {Rapidity distributions for photoproduction of $J/\psi$ and
         $\Upsilon$($1S$) mesons in $p p$ and $p \overline{p}$
         interactions at RHIC, the Tevatron, and the LHC
         \cite{ppupsilon}. The curves are explained in the text.}
\label{fig:photorap}
\end{center}
\end{figure}

\Figure[b]~\ref{fig:photorap} shows the rapidity distribution for vector
meson production in $pp$ and $\overline pp$ collisions.  The solid
histogram is with the dipole form factor discussed above.  For the
$\Upsilon$, the shaded bands show the uncertainty in the production
rates, based on the uncertainties in the HERA cross-sections but
neglecting extrapolation uncertainties.  The dashed line is an
alternative calculation, with $b_{\rm min}=1.0$~fm replacing the form
factor.  For the $J/\psi$, the abrupt dropoff at large $|y|$ is due to
the discontinuity at threshold.

\begin{figure}
\centering\includegraphics[width=.85\linewidth]{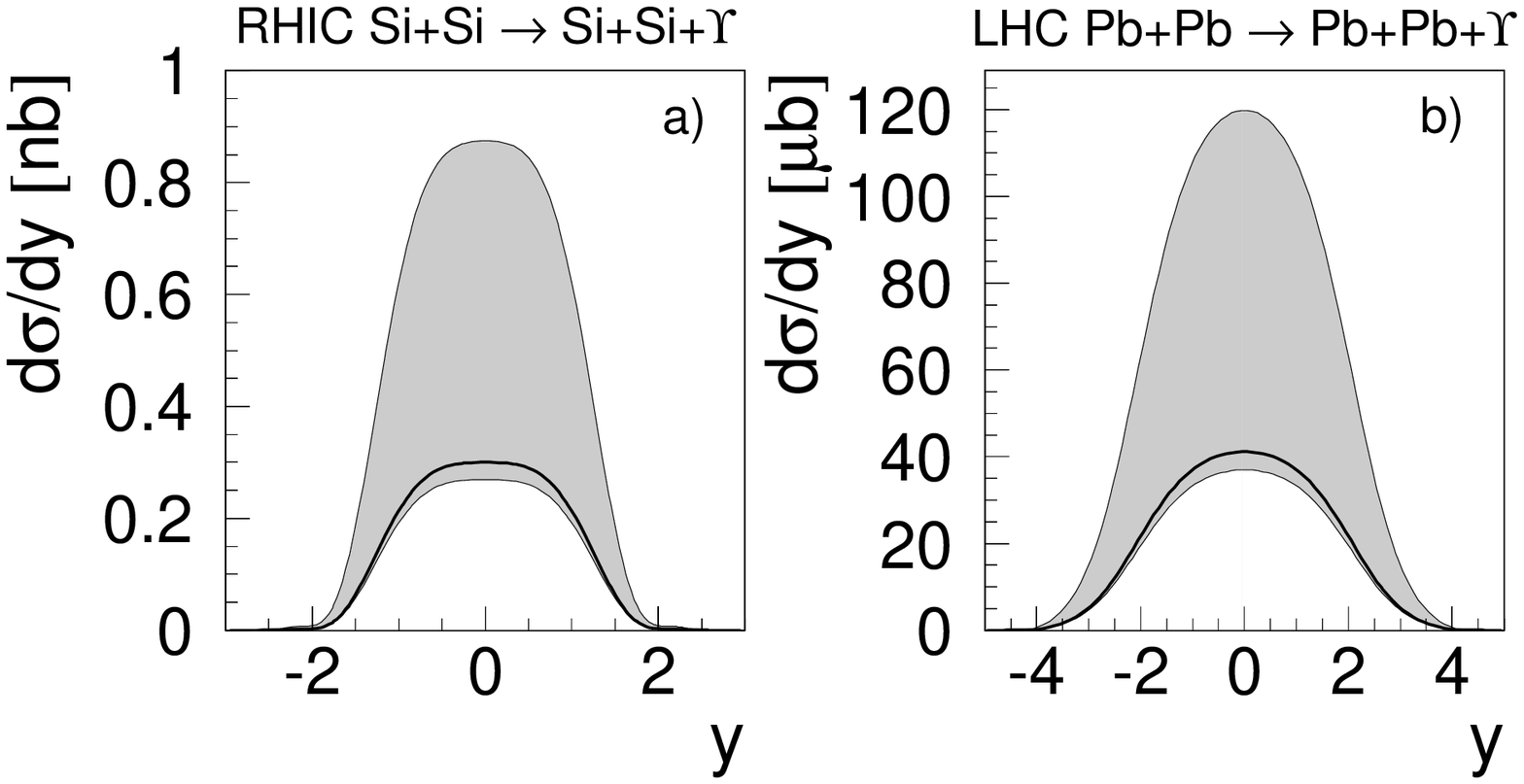} 
\caption[Rapidity distributions of $\Upsilon$ mesons
         produced in coherent photonuclear interactions]
        {Rapidity distributions of $\Upsilon$ mesons produced in
         coherent photonuclear interactions at RHIC and the LHC
         \cite{ppupsilon}. The solid curves correspond to the
         parameterization $\sigma_{\gamma p} (W_{\gamma p}) = 0.06
         W_{\gamma p}^{1.7}$ (pb) and the grey bands show the
         uncertainty in $\sigma_{\gamma p}$}
\label{fig:AA} 
\end{figure}

\Figure[b]~\ref{fig:AA} shows the rapidity distribution for $\Upsilon$
production in Si+Si collisions at RHIC (the $\Upsilon$ is below the
effective threshold in Au+Au), and Pb+Pb collisions at the LHC.  The
solid histogram is for the dipole form factor, and the shaded bands
show the uncertainty due to the uncertainties in the HERA
cross-sections, but neglecting uncertainties due to the extrapolation.

Coherent $\Upsilon$ production at the LHC was also studied by
Frankfurt {\it et al.} \cite{Frankfurt:2003qy}. The result (solid
curve in \Figure~\ref{fig:AA}) is about 10\% higher than their result
for the impulse approximation (no shadowing).  The difference may be
due to the slightly different photon spectrum and slope of $d
\sigma/dt$ in photon--proton interactions.

At the LHC, the cross-sections are very large for both $pp$ and $AA$
collisions, and obtaining good event samples should be relatively
straightforward.  For the $\Upsilon$ in $pp$ collisions, production
tends to be at large $|y|$.  The rapid rise in cross-section,
$\sigma\propto W^{1.7}$ outweighs the $dn/dk\propto 1/k$ photon
spectrum.  For the $J/\psi$, the $W_{\gamma p}$ dependence is smaller
and nearly 'cancels' the photon spectrum, leading to a rather flat
$d\sigma/dy$.  In both cases, it should be easy to obtain good
statistics, even with only a central detector.

At the Tevatron, the cross-sections are smaller.  However, the CDF
collaboration has already observed an apparent exclusive $J/\psi$
signal which appears consistent with photoproduction \cite{angela}.
With an appropriate trigger and more data, the $\Upsilon$ should be
within reach.

Very roughly, the photoproduction cross-section is about 1/1000 of the
equivalent hadroproduction cross-sections \cite{lowxupsilon} with
moderate variation depending on species and beam energy.  Despite the
small fraction, because of the very different event characteristics,
clean separation of photoproduction seems possible.

Hadronically produced vector mesons have $p_T\sim M_V$.  In contrast,
almost all photoproduced mesons have $p_T < 1$~GeV
(cf. \Figure~\ref{fig:interference}).  Such a cut eliminates about
94\% of the hadroproduced $\Upsilon$ at the Tevatron \cite{Abe:1995an}
while retaining almost all of the photoproduction.

As long as both protons remain intact, the vector meson will not be
accompanied by any other particles in the same event. In contrast, in
hadronic events, additional particles are distributed over the
available phase space.  A moderate requirement for two rapidity gaps
around the vector meson should remove most of the remaining hadronic
background \cite{lowxupsilon}, even at RHIC energies.  The Fermilab
results on exclusive $J/\psi$ production appear to bear this out.

For heavy-ion collisions, the situation is even simpler since most of
the coherent vector meson production is at $p_T < 100$~MeV/c.  In
addition, the ions can be required to remain intact and/or rapidity
gaps can be required.  These techniques were effective in isolating
the $\rho^0$ in STAR \cite{Adler:2002sc}.  In 2004, the RHIC
experiments collected a large data sample of Au+Au collisions at
$\sqrt{s_{NN}}=200$~GeV.  Thus large experiments should be able to
observe $J/\psi$ photoproduction and measure gluon shadowing in the
region $x\approx0.015$.

\subsection{Interference and the $p_T$ spectrum}

Photoproduction in $p p$ and $p \overline{p}$ collisions differs from
production in $ep$ or $eA$ collisions in that either projectile can
act as photon source or target.  For small meson $p_T$, the two
possibilities are indistinguishable so that the amplitudes add
\cite{usinterfere}.  The vector meson production is well localized to
in or near (within 1~fm of) the two sources, giving a cross-section of
\begin{equation}
\sigma = \big|A_1(y) \mp A_2(y)\exp{i(\vec{p_T}\cdot\vec{b})}\big|^2
\end{equation}
where $A_1(y)$ and $A_2(y)$ are the amplitudes for photoproduction at
the two sources and the propagator, $\exp{i(\vec{p_T}\cdot\vec{b})}$,
accounts for the ion--ion separation.  The relative sign of the two
amplitudes depends on the symmetry of the system.  For $pp$ and $AA$
collisions, transforming from nucleus 1 emitting a photon which
interacts with nucleus 2, to nucleus 2 emitting a photon which
interacts with nucleus 1 is a parity transformation.  Vector mesons
are negative parity, so the sign is negative.  For $p\overline p$
collisions, the transformation is $CP$, positive for vector mesons,
giving a positive sign.

At mid-rapidity, $A_1=A_2$ and
\begin{equation}
\sigma = \sigma_0 \bigg(1 \mp \cos{(\vec{p_T}\cdot\vec{b})}\bigg)
\label{eq:costerm}
\end{equation}
where $\sigma_0$ is the cross-section without interference.  The
interference depends on the unknown impact parameter which must be 
integrated over.

\begin{figure}
\centering\includegraphics[width=.85\linewidth]{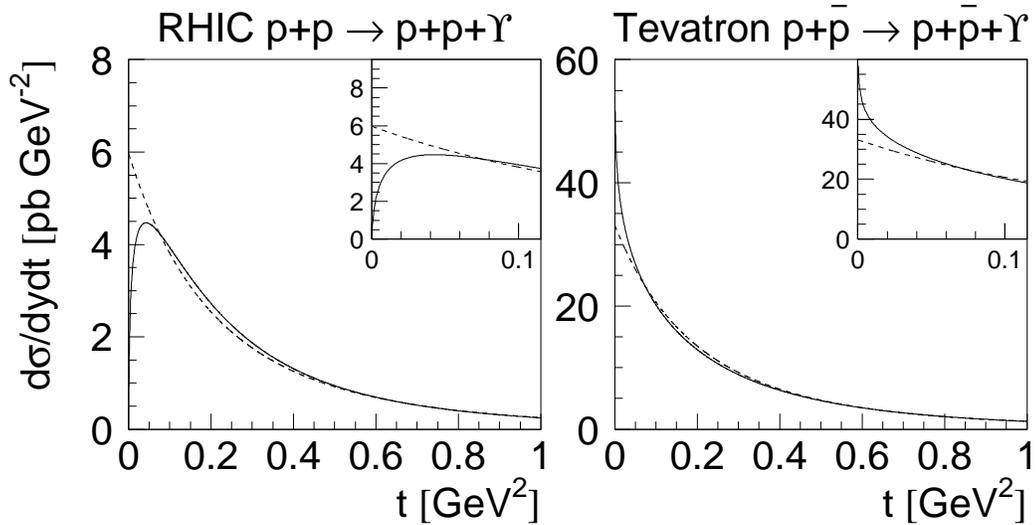}
\caption[The $\Upsilon(1S)$ photoproduction cross-section at
         mid-rapidity in $p p$ and $p \overline{p}$ collisions]
        {The $\Upsilon(1S)$ photoproduction cross-section, $d \sigma /
         dy dt$, at mid-rapidity in $p p$ and $p \overline{p}$
         collisions at RHIC and the Tevatron \cite{ppupsilon}. The
         inset has an expanded $t$ scale.}
\label{fig:interference}
\end{figure}

Without interference, the $p_T$ spectrum is that for production off a
single (anti)proton. This spectrum is the convolution of the photon
transverse momentum spectrum with the spectrum of transverse momentum
transfers from the target \cite{usinterfere}.  For $p_T >
\hbar/\langle b\rangle$, $\cos(\vec{p}_T \cdot \vec{b} )$ oscillates
rapidly as $b$ varies, giving a zero net contribution to the integral.
For small transverse momenta, however, $\vec{p} \cdot \vec{b} < \hbar$
for all relevant impact parameters and interference alters the
spectrum. This alteration has been observed in $\rho^0$ production at
RHIC \cite{STARinterfere}.

\Figure[b]~\ref{fig:interference} compares $d^2\sigma / dy dt$ with and
without interference at RHIC and the Tevatron.
\Figure[b]~\ref{fig:interference} includes a  $b_{min}=1.0$~fm cut which
has a small effect on the spectrum.  The interference is large for $t
< 0.05$~GeV$^{2}$/$c^{2}$. The different sign of the interference in
$p p$ and $p \overline{p}$ is clearly visible.

Photoproduction of other final states should also be accessible at
existing and future $p \overline{p}$ and $pp$ colliders.  Open charm,
bottom and even top quark production should be accessible
\cite{opencb} in ion--ion and deuterium--ion collisions and could be
used to measure the charge of the top quark as well as determine the
nuclear gluon distributions.  These events would have only a single
rapidity gap but the experimental techniques should be similar.
Because of the large rates, at least for ion--ion collisions, the
signal to noise ratios should be high.

To summarize, we have calculated the heavy vector meson
photoproduction cross-sections in $p p$ and $p \overline{p}$
collisions. The cross-sections are large enough for this reaction
channel to be observed experimentally. The $d \sigma / dt$
distribution is distinctly different in $p p$ and $p \overline{p}$
collisions because of the interference between the production
sources. The cross-section for producing $\Upsilon$ mesons in coherent
photonuclear Pb+Pb interactions at the LHC is large.  Because of the
distinctive experimental signature, these reactions should be easy to
detect.

\section[Outlook]{Outlook
          $\!$\footnote{Authors: D.~Kharzeev, 
              M.~P.~Lombardo, C.~Louren\c{c}o, M.~Rosati, H.~Satz}}

Recent advances in lattice and analytical calculations described in
this Chapter have significantly improved the theoretical understanding
of heavy quarkonium dynamics in hot QCD matter.  In fact, since many
of the recent results came as a surprise, they still need to be
analyzed, improved upon, and clarified.

The list of open issues which can be addressed by theory and
experiment in the near future includes:

\begin{itemize}
 \item Lattice QCD at finite temperature: \\ 
       What is the critical temperature at which different quarkonium
       states dissociate?  What is the influence of temperature on the
       masses, dispersion relations, and widths of quarkonia?  How big
       is the influence of light quarks?  What are the properties of
       strongly interacting matter in the vicinity of the deconfining
       transition and how do they compare with phenomenological
       models?

 \item Lattice QCD at zero temperature: \\
       What are the matrix elements of different gluon and quark
       operators of various dimension which are related to quarkonium
       dissociation?
 
 \item Analytic theoretical calculations at zero temperature: \\
       Analyze the expressions for quarkonium dissociation amplitudes,
       and relate them to the matrix elements which can be computed on
       the lattice.  Apply and test theoretical approaches developed
       for the studies of quarkonium dissociation to exclusive
       quarkonium decays and quarkonium production; understand the
       range of validity of various approaches.
 
 \item Analytic theoretical calculations at finite temperature: \\
       Understand the properties of Quark--Gluon Plasma in the vicinity
       of the deconfinement phase transition; identify the dynamical
       degrees of freedom and apply this knowledge to understanding
       the lattice results at finite $T$. Try to develop a
       quantitative theory of quarkonium interactions with hot QCD
       matter.
 
 \item Experiment: \\
       Collect data on the nuclear dependence of the production of
       different quarkonium states, including the $\chi$ family, in
       $pA$ collisions in a broad range of $x_F$ (rapidity), and
       different energies.  Refine the data on quarkonium suppression
       in nucleus--nucleus collisions at SPS energy; improve
       statistics on quarkonium suppression in Au+Au (and some other,
       lighter, $AA$ system) collisions at RHIC; prepare for the
       LHC. Measure photoproduction cross-sections to determine the
       nuclear gluon density.
 \end{itemize}  

To summarize, impressive progress in the physics of quarkonium
interactions with QCD media has been made in recent years.  It has
already become clear that the physics of heavy quark bound states in
QCD media is a rich and promising field of theoretical and
experimental studies. Nevertheless, even a brief examination of the
list above shows that a lot more has to be done before one can claim
an understanding of quarkonium interactions in media.

%\end{document}

\BLKP
%10/12/04

\chapter{BEYOND THE STANDARD MODEL}
\label{chapter:beyondstandardmodel}

\noindent
{\it Convener:} M.~Sanchis-Lozano \\
{\it Authors:} J.~P.~Ma, M.~Sanchis-Lozano

\section[General overview]
{General overview}

During the last years, a large amount of new data on charmonium and
bottomonium production and decays have been collected in
$B$-factories, Tevatron, HERA and BEPC, greatly improving the accuracy
of the measured widths and branching fractions. Such measurements,
together with the soundness of the theoretical background based on
effective field theories, could show up possible deviations from SM
expectations, thereby pointing out the existence of NP. Lepton flavour
and CP violation in heavy quarkonia decays are good examples of such
precision physics. Moreover, in the past radiative decays of heavy
quarkonium were employed in the search for axions and Higgs particles
according to the Wilczek mechanism \cite{Wilczek:1977zn}. Recently,
the possibility of relatively light non-standard Higgs bosons (which
might have evaded LEP searches) has been pointed out in different
scenarios beyond the SM
\cite{Hiller:2004ii,Dobrescu:2000yn,Carena:2002bb}.  Therefore,
discovery strategies should be conducted to detect possible signals of
new physics from heavy quarkonia decays.

\section[Prospects to detect new physics]
{Prospects to detect new physics}

Heavy quarkonium offers an interesting place where probing NP which
would manifest experimentally in different ways: a) slight but
observable modifications of decay rates and branching fractions; b)
unexpected topologies in decays; c) CP and lepton flavour violation,
\etc  Along this chapter we will discuss in some detail three
proposals to search for new physics and the prospects to detect
non-standard light particles based on decays of heavy quarkonium:
\begin{itemize}
\item CP test with $J/\psi$ decays, probing the electric and
      chromo-dipole moments of charm quarks
\item Lepton flavour violation in $J/\psi$'s two-body decays
\item non-standard Higgs-mediated leptonic decays of Upsilon resonances
\end{itemize}
Moreover, let us mention other possibilities 
(not developed further in this chapter)
to seek NP:
\begin{itemize}
\item Inspired by string-like scenarios, field theories formulated in
      noncommutative spaces should be explored. In particular,
      noncommutative QCD corrections to the gluonic decays of heavy
      quarkonia have been analyzed in \cite{Devoto:2004qv}.  Despite
      proving the consistency of perturbative calculations in this
      model, the inclusion of such corrections does not change
      substantially the magnitudes of the hadronic widths, thereby
      making difficult the experimental test.

\item A relatively light bottom squark and gluino sector in
      supersymmetry was put forward some time ago \cite{Berger:2000mp}
      to explain the longstanding discrepancy on the bottom
      hadroproduction cross-section between theory and experiment
      found at Tevatron \cite{Abbott:1999se,Abe:1996zt}.  Under this
      hypothesis, interesting consequences could show up in
      bottomonium phenomenology
      \cite{Berger:2001jb,Berger:2002gu,Berger:2002bz}, \eg the decay
      modes
\[ 
\Upsilon\ \to\  \tilde{b}\tilde{b^*}\ \ \ ;\ \ \  
\chi_b\ \to\ \tilde{b}\tilde{b}^*,\ \ \ 
\Upsilon\ \to\ \gamma\ \tilde{S}\ \ \ \ ;\ \ \ 
\Upsilon\ \to\ \gamma\ \eta_{\tilde{g}} 
\]     
      If the bottom squark was relatively $\lq\lq$stable'' it might
      yield a $\tilde{B}^-$ or a $\tilde{B}^0$ $\lq\lq$mesino'' (the
      superpartner of the B meson) by picking up a $\bar{u}$ or a
      $\bar{d}$ quark, respectively. Such a meson has baryon number
      zero but would act like a heavy $\bar{p}$ (of mass $\sim 3 - 7$~GeV).  
      In fact at LHC experiments it could fake a heavy muon in
      muon chambers but leaving some activity in the hadron
      calorimeter; ionization, time-of-flight and Cherenkov
      measurements would be consistent with a particle whose mass is
      heavier than a proton.  However, recently a more accurate
      description of the $b$-quark fragmentation function has
      substantially reduced the difference between theoretical
      expectations and experimental results in bottom hadroproduction
      \cite{Cacciari:2002pa,Cacciari:2003uh}.  Although the situation
      is not definitely settled, now the claim for a new physics
      contribution in bottom production is not compelling at
      all. Besides, a throughout analysis of the $e^+e^- \to hadrons$
      cross-section from PEP, PETRA, TRISTAN, SLC and LEP allows the
      $95\%$ C.L. exclusion of sbottom with mass below $7.5$~GeV
      \cite{Janot:2004cy}. Also a light gluino mass less than $6.3$~GeV 
      has been excluded \cite{Janot:2003cr}.
\end{itemize} 

\section[Precision tests using $J/\psi$ decays]
{Precision tests using $J/\psi$ decays}

Huge amount of data (to be) collected in $e^+e^-$ factories like BEPC
(and the upgraded BEPCII) and CLEO should allow to test some aspects
of the SM to an unprecedented accurracy. In the following sections we
describe two research lines based on $J/\psi$ rare decays.

\subsection[CP test with $J/\psi$ decays]
           {CP test with $J/\psi$ decays $\!$\footnote{Author:
            Jian-Ping Ma}}

We open this review on searches for new physics by remarking that a
nonzero electrical dipole moment (EDM) of a quark or a lepton implies
that CP symmetry is violated. Actually, EDM's of quarks and leptons
are very small from the SM (see
\cite{He:1989xj,Bernreuther:1990jx,Korner:1990zk} and references
therein). If the EDM of a quark is found to be nonzero, it is likely
an indication of new physics.

Since the operator for EDM does not converse helicities of quarks, its
effect is suppressed in a high energy process by a factor $m_q/E$,
where $m_q$ is the quark mass and $E$ is a large energy scale. For
light quarks, useful information can be obtained through measurement
of the EDM of the neutron \cite{He:1989xj}.  So far there is no
experimental information about EDM's of heavy quarks, like charm- and
bottom-quark.  $J/\psi$ decays can provide information of EDM of
charm quark and has the advantage that the effect of EDM will be not
suppressed, because the large energy scale is around $m_c$. Since in
radiative decays a $c\bar c$ pair is annihilated into a photon and
gluons, it also provides a way to detecting the chromodipole moment of
the charm quark.  These moments are defined by the effective
Lagrangian:
\begin{eqnarray}
L_{CP} = -i\frac{d_c}{2}\bar c\gamma_5\sigma_{\mu\nu}F^{\mu\nu}c
          -i\frac{\tilde{d}_c}{2} \bar
c\gamma_5\sigma_{\mu\nu}G^{\mu\nu}c,
\end{eqnarray}
where $d_c$ is the electric dipole moment, $\tilde d_c$ is the chromodipole
moment.
\par

In general a CP symmetry test requires a large data sample because the
effect of its possible violation is expected to be very small.  In the
following we focus on $J/\psi$ decays \cite{Ma:2003yf} as large data
samples already exist or will be collected at BEPC and CLEO-c. Indeed,
such huge data samples (with $10^7 \sim 10^{10}$ $J/\psi$'s) are very
suited for CP tests. However, not every decay mode of $J/\psi$ can be
used for this purpose. For a $J/\psi$ decay into a particle and its
antiparticle, a CP test is not possible if these particles are
spinless or their polarizations are not observed
\cite{Bernreuther:jr,Korner:1990zk}. It is only possible if
polarizations of decay products are measured.  The decay $J/\psi\to
\Lambda\bar\Lambda$ is an example, where the polarizations can be
determined through subsequential decays of $\Lambda$ and $\bar\Lambda$
\cite{He:1992ng}.

On the other hand, a CP test can be carried out for three-body decays,
even without knowing the polarizations of the decay products. This is
the case of the $J/\psi\to\gamma\phi\phi$ decay mode, which can
provide useful information about the charm quark EDM.  The reason for
choosing this channel is because $\phi$ is a very narrow resonance,
just above $K\bar K$ threshold, and can be clearly identified by its
$K^+K^-$ decay mode in experiment. In principle
$J/\psi\to\gamma\rho\rho$ could also serve for the purpose, but
experimentally the broad width of $\rho$ meson makes it impossible to
get a clean sample from this channel.  Therefore, let us consider the
decay in the rest-frame of $J/\psi$ produced at a $e^+e^-$ collider:
\begin{equation}
e^+( k_+) + e ^- (k_-) \to J/\psi(P) \to \gamma(k) +\phi(p_1) +\phi(p_2),
\end{equation}
where momenta are given in brackets. Because the two $\phi$ mesons are
identical particles, we require $p_1^0 > p_2^0$ to distinguish them in
experiment.  In our case two CP-odd observables can be constructed:
\begin{equation}
O_1={\bf \hat k_+}\cdot {\bf \hat p_1} {\bf \hat k_+}\cdot
  ({\bf\hat p_1}\times {\bf\hat p_2}),
~~O_2={\bf \hat k_+}\cdot {\bf \hat p_2} {\bf \hat k_+}\cdot
  ({\bf\hat p_1}\times {\bf\hat p_2}),
\end{equation}
where momenta with a hat  denote their directions.
From these oberservables, one can define the CP-asymmetry as
\begin{equation}
B_i=\langle \theta(O_i)-\theta(-O_i)\rangle~~(i=1,2),
\end{equation}
where $\theta (x)=1$ if $x>0$ and is zero if $x<0$.  If these
asymmetries are not zero, CP symmetry is violated.

In calculating these asymmetries, we will use nonrelativistic
wave-functions for $J/\psi$ and also for $\phi$ mesons.  It should be
noted that reliable predictions for various distributions can not be
obtained with this approximation.  Neverthesless, one may expect that
for integrated asymmetries it could become a good approximation,
especially because the integrated asymmetries will not depend on the
wave functions at the origin.

The following CP asymmetries are obtained:
\begin{eqnarray}
B_1 &=& 4.2\left [ \frac { d_c}{10^{-10} e~{\rm cm}}\right ] -1.2
  \left [ \frac{ \tilde d_c}{10^{-10} e~{\rm cm}}\right ] ,
\nonumber\\
B_2 &=& -3.9 \left [ \frac { d_c}{10^{-10} e~{\rm cm}}\right ]
    +1.3 \left [ \frac{ \tilde d_c}{10^{-10} e~{\rm cm}}\right ].
\end{eqnarray}
A statistic sensitivity to $d_c$ and $\tilde d_c$ can be determined
from these results by requiring that the asymmetry generated by these
dipole moments should be larger than the statistical error.
With the $5.8\times 10^7~J/\psi$ data sample at BES, the sensitivities of
such CP asymmetries to these dipole moments are
\begin{equation}
d_c \sim 1.4\times 10^{-13} e~{\rm cm}, \ \ \ \
\tilde d_c \sim 4.5\times 10^{-13} e~{\rm cm}.
\end{equation}
With a $10^{10}$ data sample which will be collected in the near future,
the sensitivities are:
\begin{equation}
d_c \sim 1.2\times 10^{-14} e~{\rm cm}, \ \ \ \
\tilde d_c \sim 3.6 \times 10^{-14} e~{\rm cm}.
\end{equation}

To conclude this section: With large date samples of $J/\psi$, which
are collected at BES and will be collected with BES~III and CLEO-c
program, a CP test is possible with $J/\psi$ decays. By using the
decay mode $J/\psi\to\gamma\phi\phi$, the electric- and chromo-dipole
moment can be probed at order of $10^{-13}e~{\rm cm} \sim
10^{-14}e~{\rm cm}$.

\subsection[Lepton flavour violation]
{Lepton flavour violation}

In the SM, lepton flavour number is independently conserved provided
that neutrinos are massless, although (being a global symmetry) there
is no fundamental dynamical principle requiring its conservation.
Actually, lepton flavour is violated in many extensions of the SM,
such as grand unified theories \cite{Pati:1974yy}, supersymmetric
models \cite{Haber:1984rc}, left-right symmetric models
\cite{Mohapatra:1974hk} and models where the electroweak symmetry is
dynamically broken \cite{Hill:2002ap}.  Recent results
\cite{Fukuda:1998mi,Eguchi:2002dm} indicate that neutrinos indeed have
nonzero masses and can mix with each other; therefore, lepton flavour
is a broken symmetry in nature.  Here, we focus on lepton flavour
violation (LFV) via the two-body $J/\psi$ decay (which conserves total
lepton number):
\[ J/\psi \to \ell\ell' \] 
with $\ell$ and $\ell'$ denoting charged leptons of different species.
This process could occur at tree-level induced by leptoquarks,
sleptons (both in the $t$-channel) or mediated by $Z'$ bosons (in the
$s$-channel) \cite{Zhang:2000ri,Huo:2002mz} in correspondence with the
aforementioned scenarios.

The large sample ($5.8 \times 10^7$ events) collected in leptonic
decays of $J/\psi$ resonances at BEPC and analized by BES up to now
makes this search especially interesting; in fact, upper limits for
different lepton combinations have already been set at 90$\%$
C.L.\cite{Bai:2003sv,unknown:2004nn}:
\begin{eqnarray}
{\cal B}(J/\psi \to \mu\tau) & < & 2.0\ \times\ 10^{-6} \nonumber \\
{\cal B}(J/\psi \to e\tau) & < & 8.3\ \times\ 10^{-6} \nonumber \\
{\cal B}(J/\psi \to e\mu) & < & 1.1\ \times\ 10^{-6} \nonumber
\end{eqnarray}

In the future, larger samples collected at BEPC(II) should allow to
test LFV at a higher precision, severely constraing new physics
models.  Similarly, estimates of the LFV Upsilon decay $\Upsilon \to
\ell\ell'$ can be found in \cite{Huo:2002mz}.

\section[Searches for light pseudoscalars in $\Upsilon$ decays]
        {Searches for light pseudoscalars in $\Upsilon$ decays}

In many extensions of the SM, new scalar and pseudoscalar states
appear in the physical spectrum. Admittedly, the masses of these
particles are typically of the same order as the weak scale and, in
principle, a fine-tuning is required to make them much
lighter. Nevertheless, if the theory possesses a global symmetry, its
spontaneous breakdown gives rise to a massless Goldstone boson, the
$\lq\lq$axion''.  The original axion \cite{Peccei:1977ur} was
introduced in the framework of a two-Higgs doublet model (2HDM) to
solve the strong CP problem. However, such an axial U(1) symmetry is
anomalous and the pseudoscalar acquires a (quite low) mass which has
been ruled out experimentally.  Thus, theorists have looked for other
models (by relaxing model parameter constraints) and axion-like
particles, not running into conflict with present terrestrial
experiments and astrophysical limits (see \cite{Masso:2002ip} and
references therein).

On the other hand, if the global symmetry is explicitly (but slightly)
broken, one expects a pseudo-Nambu--Goldstone boson in the theory
which, for a range of model parameters, still can be significantly
lighter than the other scalars. A good example is the so-called next
to minimal supersymmetric standard model (NMSSM) where a new
gauge-singlet superfield is added to the Higgs sector \cite{gunion}.
The mass of the lightest CP-odd Higgs can be naturally small due to a
global symmetry of the Higgs potential only softly broken by
trillinear terms \cite{Hiller:2004ii}. Moreover, the smallness of the
mass is protected from renormalization group effects in the large
$\tan{\beta}$ limit.  Actually, there are other scenarios containing a
light\footnote{By $\lq\lq$light'' we consider here a broad interval
               which might reach a $\simeq 10$~GeV mass value} 
pseudoscalar Higgs boson which could have escaped detection in the
searches at LEP-II, \eg a MSSM Higgs sector with explicit CP
violation \cite{Carena:2002bb}. Another example is a minimal composite
Higgs scenario \cite{Dobrescu:2000yn} where the lower bound on the
CP-odd scalar mass is quite loose, as low as $\sim 100$~MeV (from
astrophysical constaints).

Thus we conclude that the existence of a relatively light pseudoscalar
Higgs (to be denoted as $A^0$ hereafter) is not in contradiction with
current experimental data and could be accomodated within well
motivated extensions of the SM.  Therefore, it is worth to revisit
some of the $\lq\lq$old'' techniques to search for non-standard
particles in quarkonia decays, also exploring new possibilities like a
possible breakdown of lepton universality in $\Upsilon$ decays.

\subsection[$\Upsilon(J/\psi) \to \gamma + X^0$]
{$\Upsilon(J/\psi) \to \gamma + X^0$}

Heavy resonances have been helpful so far putting limits in the
searches for extensions of the SM through the radiative decay channel
\[ \Upsilon(J/\psi)\ \to\ \gamma\ +\ X^0 \]
where $X^0$ stands for a weakly interacting (experimentally unseen)
particle. This decay mode represents, in essence, the Wilczek
mechanism \cite{Wilczek:1977zn} for the real emission of either a
Higgs boson or an axion from quarkonium.  The experimental signature
would be very clean: the observation of a single photon with a
considerable missing energy in the event. Let us observe that this
would be so if the $X^0$ is sufficiently stable, \ie the probability
to decay inside the detector (of typical size $r \sim 10$ m) is quite
small, $\Gamma_{X^0}<<E_{X^0}/m_{X^0}r$, where $E_{X^0}$ and $m_{X^0}$
denote the (laboratory) energy and mass of the unseen particle,
respectively.  Notice, however, that the chances to leave unseen the
detector decrease for values of $m_{X^0}$ close to $E_{X^0}$ as the
Lorentz dilation factor approaches unity. To date, no evidence has
been found and limits have been set as a function of the mass of the
$X^0$ particle \cite{Balest:1994ch}. Note that such limits in
$\Upsilon$ decays only exclude particles below $5-7$~GeV!
\cite{Hagiwara:fs} Thus, in view of the renewed interest in
pseudoscalars whose mass may lie around 10~GeV, an open mind should be
kept in those and related searches.

\subsection[Non-standard Higgs-mediated leptonic decays of $\Upsilon$ 
resonances]
{Non-standard Higgs-mediated leptonic decays of $\Upsilon$ resonances}

\begin{figure}
\centering\includegraphics[width=.6\linewidth]{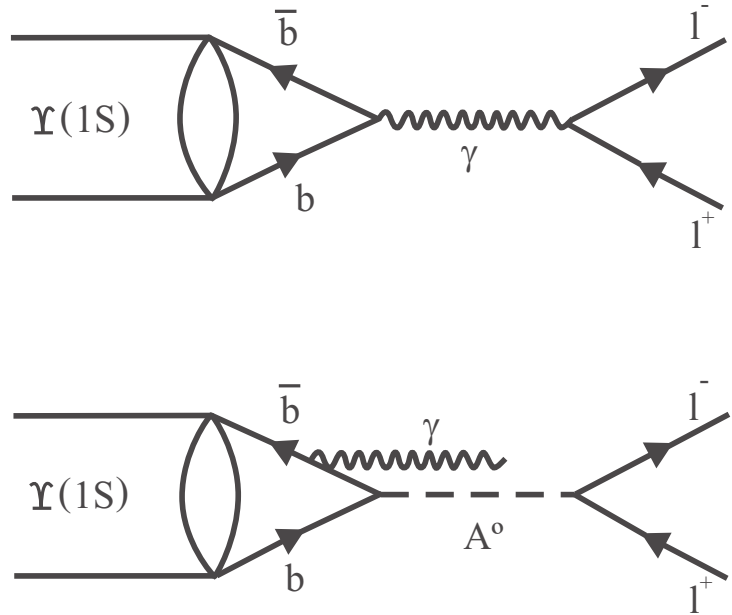}
\caption[Electromagnetic annihilation of a
         $\Upsilon(1S)$ resonance into a charged lepton pair]
        {$\Upsilon(1S)$ resonance into a charged lepton pair through a
         virtual photon; (b)[lower panel]: Hypothetical annihilation
         of an intermediate $\eta_b^*$ state (subsequent to a M1
         structural transition yielding a final-state soft photon)
         into a charged lepton pair through a CP-odd Higgs-like
         particle denoted by $A^0$.}
\label{fig:1BSM} 
\end{figure}

In the previous section we considered the possibility of emission by
heavy quarkonium of a real, long-lived but unseen particle. However,
if the emitted particle width is large enough, the particle would
promptly decay and its products could make possible its observation by
the detector.  On the other hand, virtual poduction of (off-shell)
particles should be also analized. In this section we examine the
possible existence of a CP-odd Higgs mediating the annihilation of the
$b\bar{b}$ pair (subsequent to a magnetic dipole transition of the
Upsilon resonance) into a final-state dilepton (see
\Figure~\ref{fig:1BSM}). This channel would constitute a rare decay
mode of the $\Upsilon$ resonance, observable however if the Higgs mass
is not too far from the $\Upsilon$ mass and the couplings are not
small.  In fact, rare decays have been traditionally employed for
seeking new physics, in particular looking for extensions of the Higgs
sector of the SM. Let us mention, as a significant example, the
(flavor-changing neutral current) decays of B mesons into lepton pairs
(\eg $B_{s,d}^0 \to \mu^+\mu-$), where a non-standard Higgs-mediated
contribution could modify (enhancing) the SM decay rates
\cite{Dedes:2002rh}.

As pointed out in a series of recent papers
\cite{Sanchis-Lozano:2002pm,Sanchis-Lozano:2003ha,Sanchis-Lozano:2004zd},
bottomonium also offers the possibility of testing extensions of the
SM by looking at possible non-standard Higgs-mediated leptonic decay
channels of Upsilon resonances below the $B\bar{B}$ threshold, in
addition to the dominating electromagnetic mode
\[ 
\Upsilon(nS) \to \gamma^* \to \ell^+\ell^-\ \ \ \ (\ell=e,\mu,\tau,\ \
n=1,2,3) 
\]

We shall focus as a theoretical background on a general 2DHM of the
type II \cite{gunion} where down fermions couple to the Higgs boson
proportionally to the ratio ($\tan{\beta}$) of the two Higgs vacuum
expectation values. Nevertheless, the main conclusions can be extended
to different scenarios predicting other Higgs-like particles with
analogous phenomenological features.

\begin{figure}
\begin{center}
\includegraphics[width=.48\linewidth]{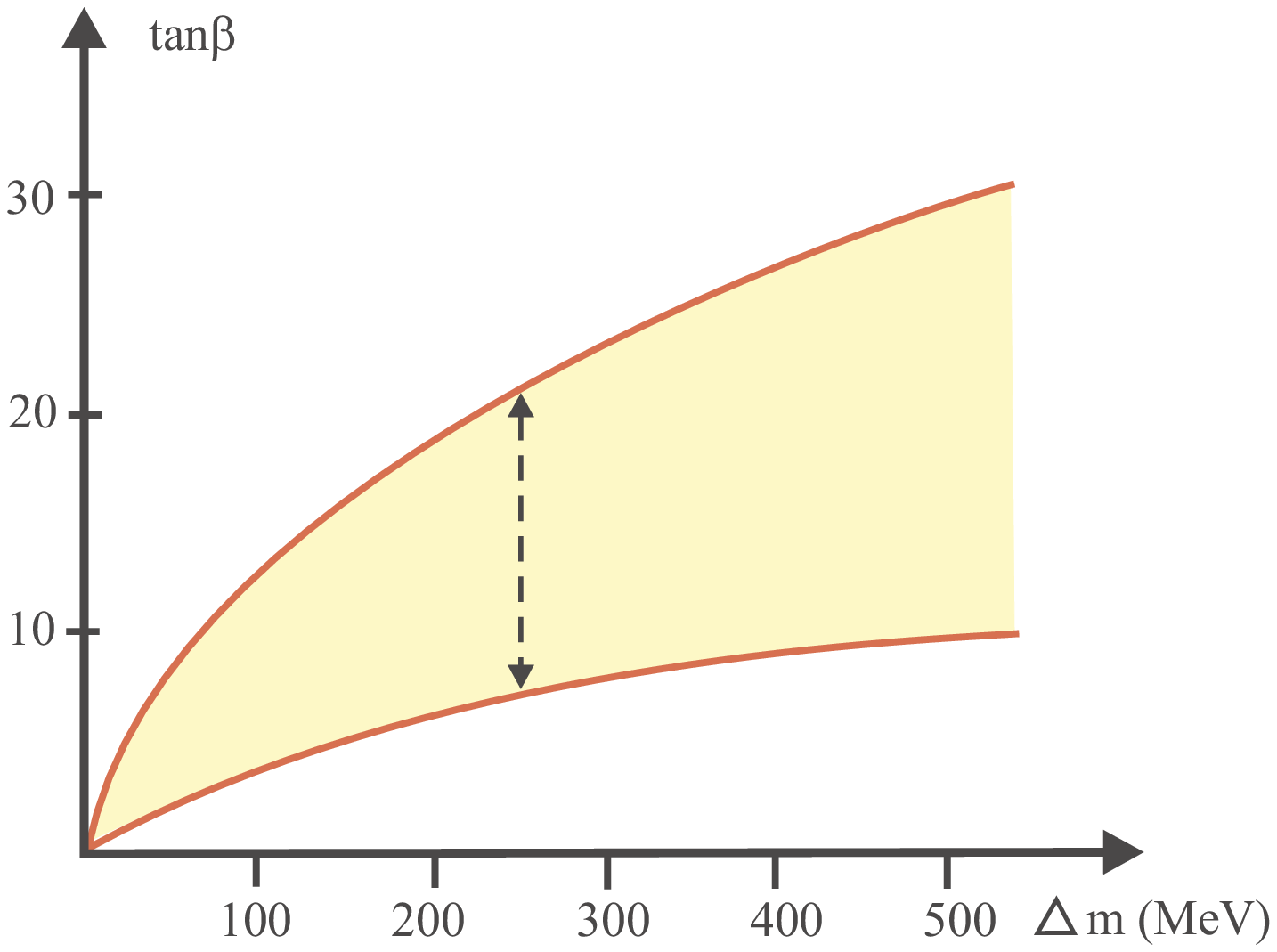}
\hfill
\includegraphics[width=.48\linewidth]{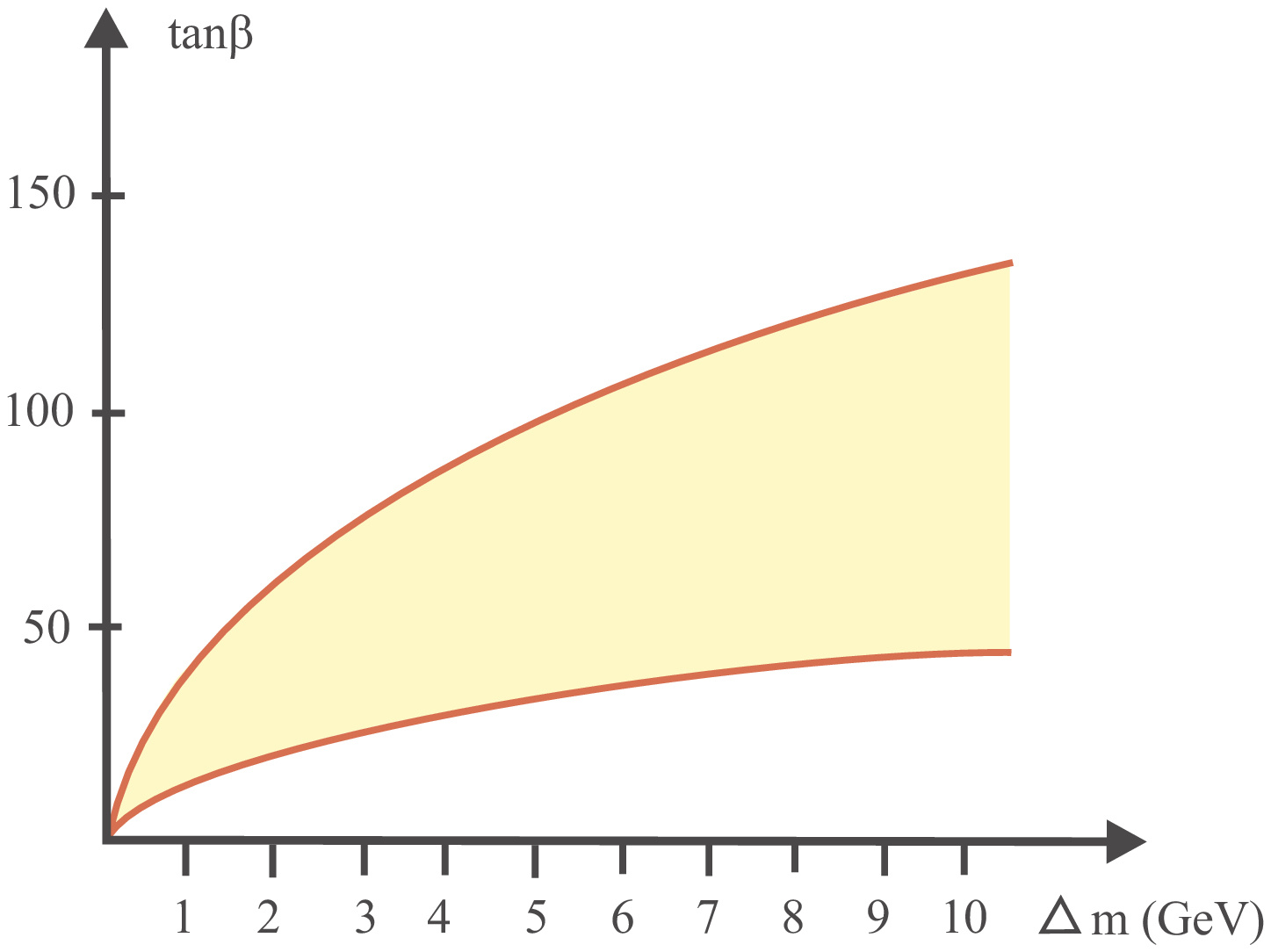}
\end{center}
\caption[Required $\tan{\beta}$ values as a function of
         $\Delta m$]
        {Required $\tan{\beta}$ values (shaded area) as a function of
         $\Delta m$ needed to account for a $\sim 10\%$ breakdown of
         lepton universality in $\Upsilon$ decays according to a
         2HDM(II). The vertical dotted line shows the range of
         $\tan{\beta}$ for $\Delta m = 250$~MeV used in
         \cite{Sanchis-Lozano:2003ha} as a reference value.}
\label{fig:2BSM} 
\end{figure}

Let us assume that a prior magnetic dipole (M1) direct transition from
the initial-state $\Upsilon$ can take place yielding a pseudoscalar
$b\bar{b}$ intermediate state as shown in \Figure~\ref{fig:1BSM},
subsequently annihilating into a lepton pair via a non-standard
$CP$-odd Higgs boson $A^0$:
\[
\Upsilon(nS)\ \rightarrow\ \gamma_s\ \eta_b^*(\rightarrow 
A^0{\rightarrow}\ \ell^+\ell^-)\ \ \ \ \ \ (\ell=e,\mu,\tau,\ \ 
n=1,2,3)
\]
where $\gamma_s$ stands for an undetected soft photon with energy in
the range 35--150~MeV, depending on the still unknown $\Upsilon-\eta_b$
hyperfine splitting.  As the photon is quite soft, the M1-transition
probability ${\cal P}^{\Upsilon}(\eta_b^*\gamma_s)$ was roughly
obtained in \cite{Sanchis-Lozano:2003ha,Sanchis-Lozano:2004zd} from a
textbook expression relating on-shell states.  A consequence of the
existence of this kind of NP would be the 
$\lq\lq$apparent''\footnote{In the sense that once the Higgs contribution 
                            were taken into account, lepton universality 
                            would be restored} 
breaking of lepton universality based on the two following keypoints:
\begin{itemize}
\item In the experimental determinations of the leptonic BF of the
      Upsilon resonances, the Higgs contribution would be unwittingly
      ascribed to the leptonic decay mode as the radiated photon would
      remain undetected. This would be especially the case for the
      $\tau^{\pm}$ channel\footnote{The leptonic mass squared with a
            final-state photon is given by
            $m_{\ell\ell}^2=m_{\Upsilon}^2(1-2E_{\gamma}/m_{\Upsilon})$.
            Hence $E_{\gamma}$ is much more limited by invariant mass
            reconstruction of either final-state electrons or muons than for
            tau's where such constraint is not applicable.}
\item The leptonic (squared) mass dependence in the width from the
      Higgs contribution would introduce a dependence on the leptonic
      species in the leptonic BF. The effect would only be noticeable
      in the tauonic decay mode as the electron and muon masses are
      much smaller than the tau mass.
\end{itemize}
\shortpage
 
\begin{table} [htb]
\caption[Measured leptonic BF's in $\Upsilon(1S)$ and $\Upsilon(2S)$]
        {Measured leptonic BF's and error bars in $\%$ of
         $\Upsilon(1S)$ and $\Upsilon(2S)$ (from \cite{Hagiwara:fs}).}
\label{tab:1BSM}
\begin{center}
\begin{tabular}{cccc}
\hline
channel: & $e^+e^-$ & $\mu^+\mu^-$ & $\tau^+\tau^-$  \\
\hline
$\Upsilon(1S)$ & $2.38 \pm 0.11$ & $2.48 \pm 0.06$ & $2.67 \pm 0.16$ \\
\hline
$\Upsilon(2S)$ & $1.34 \pm 0.20$ & $1.31 \pm 0.21$ & $1.7 \pm 1.6$ \\
\hline
\end{tabular}
\end{center}
\end{table}

Current experimental data (see \Table~\ref{tab:1BSM}) may indeed hint
that there is a difference of order $10\%$ in the BFs between the
tauonic channel on the one side, and the electronic and muonic modes
on the other side \cite{Sanchis-Lozano:2003ha}. The range of the
$\tan{\beta}$ needed to account for such an effect is shown in
\Figure~\ref{fig:2BSM} as a function of the mass difference ($\Delta
m$) between the non-standard Higgs and the $\eta_b(1S)$ resonance,
applying the factorization of the decay width used in
\cite{Sanchis-Lozano:2003ha}. The upper and lower curves correspond to
the maximal and mimimal estimates of the M1-transition probability
${\cal P}^{\Upsilon}(\eta_b^*\gamma_s)$, respectively.  For the large
values of $\Delta m$, only the lower values of the shaded region would
be acceptable, corresponding to the higher estimates of ${\cal
P}^{\Upsilon}(\eta_b^*\gamma_s)$.

In addition to the postulated breaking of lepton universality, other
experimental signatures which would eventually support the conjecture
on a CP-odd Higgs boson showing up in bottomonium spectroscopy and
decays are:
\begin{itemize}
\item A $\Upsilon-\eta_b$ hyperfine splitting larger than expected
      from quark potential models, caused by $A^0-\eta_b$ mixing. A
      mass splitting significantly larger than 100~MeV could be hardly
      accomodated within the SM
\item A rather large full width of the $\eta_b$ resonances due to the
      NP channel (especially for high values of $\tan{\beta}$)
\item If, instead, the $\eta_b$ state is not too broad (as this would
      be the case for the lowest values of $\tan{\beta}$ in
      \Figure~\ref{fig:2BSM}), one could look for monoenergetic photons
      with energy of order 100~MeV (hence above detection threshold)
      in those events mediated by the CP-odd Higgs boson (estimated to
      be about 10$\%$ of all $\Upsilon$ tauonic decays)
\end{itemize}
\shortpage

\subsubsection[Spectroscopic consequences for bottomonium states]
{Spectroscopic consequences for bottomonium states}

In view of our previous considerations, one can speculate about a
quite broad $\eta_b$ resonance (\eg $\Gamma_{\eta_b}\ {\gtrsim}\ 30$
MeV)\footnote{One expects $\Gamma_{\eta_b(1S)}\simeq 4\,$MeV using the
asymptotic expression $\Gamma_{\eta_b} \simeq m_b/m_c \times
[\alpha_s(m_b)/\alpha_s(m_c)]^5 \times \Gamma_{\eta_c}$ and setting
the measured $\Gamma_{\eta_c(1S)}=16\pm 3\,$MeV \cite{Hagiwara:fs}}
which might partially explain why there was no observed signal from
the hindered radiative decays of higher Upsilon resonances in the
search performed by CLEO \cite{Mahmood:2002jd,Coan:2003fe}.  Indeed
the signal peak (which should appear in the photon energy spectrum)
could be considerably smoothed\,---\,in addition to the spreading by
the experimental measurement\,---\,and thereby might not be
significantly distinguished from the background (arising primarly from
$\pi^0$'s decays).  Of course, the matrix elements for the hindered M1
transitions are expected to be small and difficult to predict as they
are generated by relativistic and finite size
corrections. Nevertheless, most of the theoretical calculations are
ruled out by CLEO results (at least) at a $90\%$ CL (see a compilation
in \cite{godfrey}), though substancially lower rates are obtained in
\cite{Lahde:1998ee} where exchange currents play an essential role and
currently cannot be excluded.  Notice finally that a large full width
of the $\eta_b$ resonance would bring negative effects on the
prospects for its detection at the Tevatron through the
double-$J/\psi$ decay: $\eta_b \to J/\psi+J/\psi$. Indeed, the
expected BF would drop by about one order of magnitude with respect to
the range between $7 \times 10^{-5}$ and $7 \times 10^{-3}$ assumed in
\cite{Braaten:2000cm}.

Furthermore, another interesting possibility is linked to a
$A^0-\eta_b$ mixing \cite{Drees:1989du} which could sizeably lower the
mass of the mixed (physical) $\eta_b$ state, especially for high
$\tan{\beta}$ values starting from similar masses of the unmixed
states \cite{Sanchis-Lozano:2003ha}. Then the signal peak in the
photon energy plot could be (partially) shifted off the search window
used by CLEO \cite{Mahmood:2002jd,Coan:2003fe} towards higher $\gamma$
energies (corresponding to a smaller 
$\eta_b$ mass\footnote{This would be the case if the (unmixed)  
   CP-odd Higgs boson had a mass greater than the (unmixed) 
   $\eta_b$ resonance \cite{Drees:1989du}} 
perhaps contributing additionally to the failure to find evidence
about the existence of the $\eta_b$ resonances to date.

The mass formula for the physical $A^0$ and $\eta_b$ states in terms
of the unmixed states (denoted as $A_0^0$ and $\eta_{b0}$
respectively), and the off-diagonal mass matrix element $\delta m^2
\simeq 0.146 \times \tan{\beta}$~GeV$^2$, for quite narrow resonances
(\ie $\Gamma_{\eta_{b0}},\ \Gamma_{A_0^0}\ \ll m_{\eta_{b0}},\
m_{A_0^0}$) reads \cite{Sanchis-Lozano:2003ha}:
\[
m_{\eta_b,A^0}^2\ \simeq\ \frac{1}{2}(m_{A_0^0}^2
+m_{\eta_{b0}}^2)\ \mp\ \ \frac{1}{2}\biggl[(m_{A_0^0}^2
-m_{\eta_{b0}}^2)^2+4(\delta m^2)^2\biggr]^{1/2} 
\]
which yields in the case of the physical $\eta_b$ and $A^0$ particles
for different mass intervals:
\begin{eqnarray}
m_{\eta_b,A^0} &  \simeq & m_{\eta_{b0}}\ \mp\ \frac{\delta
m^2}{2m_{\eta_{b0}}} 
\ \ ;\ \ \ 0 < m_{A_0^0}^2-m_{\eta_{b0}}^2 << 2\ \delta m^2 \nonumber \\
m_{\eta_b,A^0} & \simeq  & m_{\eta_{b0},A_0^0}\ \mp\ \frac{(\delta m^2)^2}
{2m_{\eta_{b0}}(m_{A_0^0}^2-m_{\eta_{b0}}^2)} 
\ \ ;\ \ m_{A_0^0}^2-m_{\eta_{b0}}^2 >> 2\ \delta m^2 \nonumber
\end{eqnarray}
As a particular but significant example, assuming for the masses of
the unmixed states $m_{\eta_{b0}} \simeq m_{A_0^0} =9.4$~GeV and the
moderate $\tan{\beta}=12$ value, one gets for the physical states
$m_{A^0} \simeq 9.5$~GeV and $m_{\eta_b} \simeq 9.3$~GeV respectively,
which corresponds to a mass difference
$m_{\Upsilon(1S)}-m_{\eta_b(1S)} \simeq 160$~MeV. Higher $\tan{\beta}$
values would, in principle, lead to larger mass shifts. However a
caveat is in order: the hyperfine splitting (enhanced by the mixing)
cannot raise unlimitedly, since the dependence on the third power of
the photon energy in ${\cal P}^{\Upsilon}(\eta_b^*\gamma_s)$
(corresponding to a magnetic dipole transition) would eventually push
up the new physics contribution for the tauonic BF beyond the
postulated ${\cal O}(10\%)$ effect.

To end this section, let us point out that CLEO has completed detailed
scans of the $\Upsilon(nS)$ ($n=1,2,3$) resonances and we want to
stress the relevance of these measurements (aside other applications)
for testing more accurately the possible existence of NP by a more
precise determination of the electronic, muonic and tauonic BFs of
{\em all three} resonances below open bottom threshold. In case no
lepton universality breaking is definitely found, some windows in the
$\tan{\beta}$--$m_{A^0}$ plane for such a non-standard CP-odd light
Higgs boson would be closed.

\section[Summary]
{Summary}

Quarkonium phenomenology should play an important role to explore new
physics as it did in the past to develop the SM.  Annihilation and
radiative decays of resonances are well suited for testing symmetry
conservation laws, as well as searching for (relatively) light
particles arising in diverse scenarios beyond the SM, in addition to a
much heavier sector.

The expected large statistics of $J/\psi$ and $\Upsilon$ resonances,
to be collected at $e^+e^-$ and hadronic colliders along the next few
years, makes heavy quarkonium physics especially convenient to conduct
high precision studies and the quest for new particles and new
phenomena. In this chapter, we have particularly developed three
issues concerning CP and lepton-flavour violation in $J/\psi$ decays,
and a possible lepton universality breaking in $\Upsilon$ decays
indicating the existence of a non-standard light Higgs boson. An open
mind should be kept regarding those and other possible phenomena
beyond the SM in heavy quarkonium physics.

\BLKP
%10/12/2004
\chapter{FUTURE EXPERIMENTAL FACILITIES}
\label{chapter:futureexpfacilities}
{{\it Conveners:}   S.~Godfrey, M.~A.~Sanchis-Lozano}\par\noindent
{{\it Authors:} D.~Bettoni, P.~Crochet, S.~Godfrey, F.~A.~Harris, A.~Hoang,
O.~Iouchtchenko, A.~Nairz, J.~Napolitano, S.~Olsen, P.~Petreczki, 
M.~A.~Sanchis-Lozano, O.~Schneider, A.~Zieminski}
\par\noindent

\bigskip

\noindent
Opportunities for quarkonium physics abound from a broad range of
complementary facilities; CESR-c/CLEO-c, BECP~II/BES~III, B-factories,
CDF and D0 (and BTEV) at the Tevatron, RHIC, GSI, and the LHC. In this
chapter we look at these future facilities reviewing and suggesting
experimental measurements that can be used as a roadmap for future
directions in quarkonium physics.

\section{Tevatron}

The Tevatron is an existing facility offering an exciting programme on
future opportunities for heavy quarkonium physics in the short, medium
and long terms (\eg BTeV, with data taking foreseen in 2009).  By
August 2004 shutdown, each experiment (CDF and D0) had collected
approximately 500 pb$^{-1}$ of data on tape.  The number of
reconstructed quarkonium states is quite impressive: 2.5 million and 4
million $J/\psi$ candidates collected by D0 and CDF, respectively. D0
reported over 50~000 $\Upsilon(1S)$ events in the data sample
corresponding to 200 pb$^{-1}$.

Run~II at the Tevatron will provide a substantial increase in
luminosity (about 1.4 fb$^{-1}$ delivered to CDF and D0 by the end of
2005 and 8.5 fb$^{-1}$ by 2009) and will allow the collider
experiments to determine the $J/\psi$, $\psi(2S)$ and $\chi_c$
cross-sections more precisely and at a larger $p_{\mathrm{T}}$
range. An accurate measurement of the $J/\psi$ and $\psi(2S)$
polarization at large transverse momentum will be the most crucial
test of NRQCD factorization. In addition, improved data on the
$J/\psi$ and $\psi(2S)$ cross-sections will help to reduce some of the
ambiguities in extracting the colour-octet matrix elements.
 
With increased statistics it might be possible to access 
charmonium states such as the $\eta_c(nS)$ or the $h_c(nP)$.  Heavy-quark
spin symmetry provides approximate relations between the
non-perturbative matrix elements that describe spin-singlet and
spin-triplet states. The matrix elements for $\eta_c(nS)$ are related to
those for $\psi(nS)$, while the leading matrix elements for $h_c(nP)$
can be obtained from those for $\chi_c(nP)$. Within NRQCD, the rates for
$\eta_c(nS)$ and $h_c(nP)$ production can thus be predicted unambiguously in
terms of the non-perturbative matrix elements that describe the $J/\psi$,
$\psi(2S)$ and $\chi_c$ production cross-sections. A comparison of the
various charmonium production rates would therefore provide a stringent
test of NRQCD factorization and the heavy-quark spin symmetry.  The
cross-sections for producing the $\eta_c$ and the $h_c$ at Run~II of the
Tevatron are large~\cite{Mathews:1998nk,Sridhar:1996vd}, but the
acceptances and efficiencies for observing the decay modes on which one
can trigger are, in general, small, and detailed experimental studies
are needed to quantify the prospects.  Other charmonium processes that
have been studied in the literature include the production of D-wave
states~\cite{Qiao:1997wb}, $J/\psi$ production in association with
photons~\cite{Kim:1997bb,Mathews:1999ye}, and double gluon fragmentation
to $J/\psi$ pairs~\cite{Barger:1996vx}.

On the other hand, Run~II of the Tevatron will also allow the collider
experiments to improve the measurement of the bottomonium
cross-sections. As-yet-undiscovered states, such as the $\eta_b(1S)$,
could be detected, for example, in the decay $\eta_b \to J/\psi +
J/\psi$~\cite{Braaten:2000cm} or in the decay $\eta_b \to D^* +
D^{(*)}$~\cite{Maltoni:2004hv}, and the associated production of
$\Upsilon$ and electroweak bosons might be
accessible~\cite{Braaten:1999th}. If sufficient statistics can be
accumulated, the onset of transverse $\Upsilon(nS)$ polarization may
be visible at $p_{\mathrm{T}}(\Upsilon)> 15$~GeV.

\newpage

In sum, the future large statistics data will be used for:

\begin{description}

\item[Production studies]\mbox{}
\vspace*{0.2cm}
\begin{itemize}
\item[--] 
      Detailed differential cross-section measurements covering the
      transverse momentum range up to at least 30~GeV and rapidity
      range up to probably 1.1 (CDF) and 2 (D0), respectively.
\item[--]
      Determination of fractions of quarkonium states produced through
      the radiative $\chi$ decays. The $\gamma\to ee$ conversions
      provide a mass resolution sufficient to separate contributions
      from individual $\chi$ states (both for the $\chi_c$ and
      $\chi_b$ case).  Both experiments have already demonstrated
      their potential to do such studies.
\item[--]
      Establishment of cross-sections for direct production of
      quarkonium states ($J/\psi$, $\psi(2S)$, and $\Upsilon(1S)$).
\item[--]
      Measurement of polarization of $J/\psi$, $\psi(2S)$ and $\Upsilon$ 
      states.
\item [--]
      Associated production of quarkonium states, \eg double $J/\psi$
      production.
\item [--] 
      Associated production of quarkonium states and heavy quarks, \eg
      $J/\psi$ production in association with $c\bar{c}$ pairs.
\end{itemize}
\vspace*{0.2cm}
\item [Quarkonium decays]\mbox{}
\vspace*{0.2cm}\\
  A large sample of $\psi(2S) \to J/\psi \pi^+ \pi^-$
  decays can be used for a better determination of the dependence of
  the decay matrix element on the invariant di-pion mass. The observed
  event accumulation rate is over 5000/fb$^{-1}$ (D0) and
  ~25~000/fb$^{-1}$ (CDF). However, BES accumulated statistics for
  this channel will be difficult to beat.
\vspace*{0.2cm}
\item[{\boldmath $X(3872)$} state properties]\mbox{}
\vspace*{0.2cm}\\
  The two experiments collected approximately 500 (D0) and 750 (CDF)\\
  $X(3872) \to J/\psi \pi \pi$ decays per 230 pb$^{-1}$ of
  data.  Studies of the $X(3872)$ properties will continue as the
  statistics increase. The quantities to measure include the fractions
  of $X(3872)$ states produced via $b$-quark decays as a function of
  the production transverse momentum, the decay matrix element
  dependence on the invariant mass of two final-state pions,
  $m(\pi\pi)$, and the di-pion resonance contribution to the decay
  process.
\vspace*{0.2cm}
\item[Searches for exotics decaying into final states involving
  quarkonia]\mbox{}
\vspace*{0.2cm}\\
  CDF searches for strange and char\-med pentaquarks have been widely
  publicized. CDF is also looking for the ($u$ $d$ $u$ $s$ $\bar{b}$) 
  pentaquarks decaying into the $J/\psi p$ final state.
\vspace*{0.2cm}
\item[Hadron decays into charmonia]\mbox{}
\vspace*{0.2cm}\\
  Exclusive B hadron decays into final states involving a $J/\psi$
  have been used for the world's best determination of the $\rm B_s$
  and $\Lambda_b$ masses and their lifetimes. The $\rm B \to VV$
  decays (where one of the vector decay products is a $J/\psi$) are
  being used for time-dependent polarization amplitude studies. These
  studies have led to the determination of lifetime differences of the
  CP-odd and CP-even $\rm B_s$ mass eigenstates. They also provide
  tests of factorization, \ie representing the weak decay matrix
  element as a product of two independent hadronic currents.
\vspace*{0.2cm}
\item[{\boldmath $\rm B_c$} studies]\mbox{}
\vspace*{0.2cm}\\
  Both collaborations are advancing their analyses of the $\rm B_c$
  mass and lifetime with Run II data. The yield of observed events in
  the $\rm B_c \to J/\psi \mu + X$ semileptonic decay channel is
  approximately 250 events/250 pb$^{-1}$ (D0). These studies require
  large statistics and a good understanding of the fake muon
  background in order to reduce systematic uncertainties. A search for
  exclusive $\rm B_c \to J/\psi \pi$ decays is also under way in both
  experiments.
\end{description}

\section{CLEO-c}
CLEO-c is an experiment that makes use of the upgraded CLEO~III detector, 
at the upgraded CESR storage ring. The storage ring (dubbed CESR-c) will use 
12 wiggler magnet systems to give increased luminosity at low energies. 
This section reviews the CLEO-c experimental programme
with particular emphasis on 
charmonium physics. The full details of CLEO-c and CESR-c are available 
in Ref.~\cite{Briere:2001rn}.

\subsection{Charmonium physics with the $\psi(2S)$}
\label{sec:psip}

\begin{figure}[t]
\begin{center}
\includegraphics[width=3.5in]{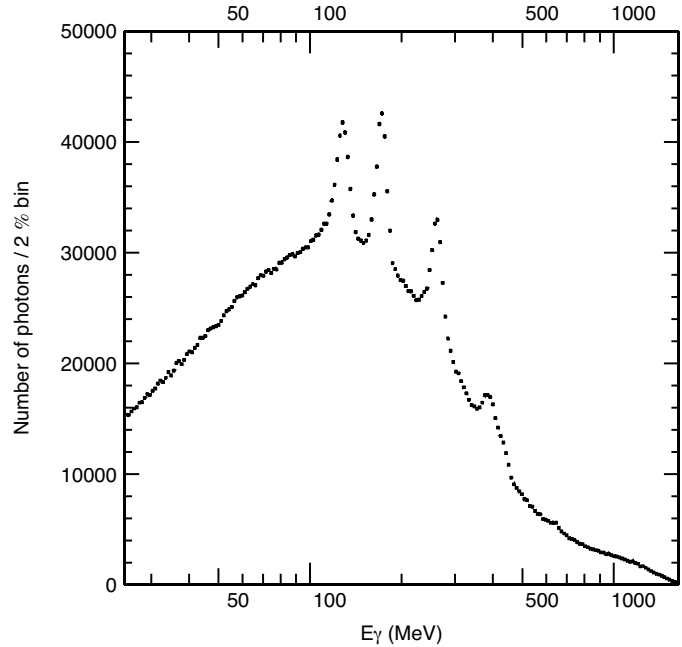}
\end{center}
\caption[The inclusive photon spectrum from $\psi(2S)$ decay]
        {The inclusive photon spectrum from $\psi(2S)$ decay. The data
         were taken with the CLEO~III detector, prior to the CLEO-c
         upgrade.  The peaks from single photon transitions between
         charmonium states are evident. This already exceeds the data
         volume acquired by the Crystal Ball at SLAC.}
\label{fig:psipgams}
\end{figure}

CLEO-c has already begun taking data on the $\psi(2S)$.  The inclusive
photon spectrum from the first data set is shown in
\Figure~\ref{fig:psipgams}.

More running, with an upgraded CESR and higher instantaneous
luminosity, is in the planning stages for the $\psi(2S)$. (In fact, at
the time of writing, more data has already been taken.) Depending on
running conditions and the will of the collaboration (see
\Section~\ref{sec:Summary}), a very large $\psi(2S)$ data set may
eventually be accumulated. It is possible that this running may be
traded for integrated luminosity on the \Jpsi~(see
\Section~\ref{sec:Jpsi}).

Potential physics opportunities with the $\psi(2S)$ include the following:
\begin{itemize}
\item Inclusive photons. Absolute branching fractions for
      $\psi(2S)\to\gamma\eta_c$ and $\psi(2S)\to\gamma\eta_c(2S)$ will
      be measured, although no signal for the former is evident (yet).
      Note that the mass of the $\eta_c(2S)$ is now accurately known
      from B-meson decay at Belle~\cite{Choi:2002na} and from
      $\gamma\gamma$ fusion at CLEO~\cite{Asner:2003wv}.
\item Detailed studies of $\chi_{c0}$, $\chi_{c1}$, and $\chi_{c2}$.
      High-statistics single-photon tags of the intermediate $\chi_c$
      states will allow various measurements of their decay branching
      fractions.
\item Hadronic decays of $\psi(2S)$. A prime goal is to search for the
      $h_c$ in $\psi(2S)\to\pi^0h_c$. The decay $\psi(2S)\to\rho\pi$
      will also be searched for, a branching ratio which is
      anomalously small. 
\item Exotica from $\psi(2S)$ decay. Radiative decay of high mass
      vectors is expected to be a prime source for glue-rich final
      states.  (See the discussion in Section~\ref{sec:Jpsi}.)
      Although one expects the majority of this data to come from
      \Jpsi\ running, $\psi(2S)$ decay would also allow flavour
      tagging through the hadronic decays where a low-mass vector
      meson (\ie $\rho$, $\omega$, $\phi$) replaces the radiative
      photon.  The possibility of studying \Jpsi\ decay using
      $\psi(2S)$ running and tagging the \Jpsi\ from
      $\psi(2S)\to\pi\pi J/\psi$ is also being investigated.
\end{itemize}

\subsection{Physics at the $\psi(3770)$}
\label{sec:psipp}

CLEO-c has already begun taking a large data sample (eventually
3~fb$^{-1}$) at the $\psi(3770)$. The main goal of this running is to
acquire a large sample of tagged $\rm D\bar{D}$ events, but the
opportunity presents itself for charmonium studies as well.

Measurements include searches for (presumably) rare decays of the
$\psi(3770)$.  Examples are $\psi(3770)\to\pi^+\pi^-J/\psi$; inclusive
photons from $\psi(3770)\to\gamma X$, where one would expect to detect
transitions to $X=\chi_{cJ}$ if the branching ratio is greater than
about $10^{-3}$; and the double cascade decay
$\psi(3770)\to\gamma\chi_{c1,c2}\to\gamma\gamma
J/\psi\to\gamma\gamma\ell^+\ell^-$ in which one might detect some ten
events or so if there is no background and the branching ratio for
$\psi(3770) \to \gamma \chi_{c1,c2}$ is more than
$\approx3\times10^{-4}$.  These results would provide information on
$1D_1/2S_1$ mixing.

\subsection{Decays of the \Jpsi}
\label{sec:Jpsi}

\begin{figure}[t]
\begin{center}
\includegraphics[width=3.5in]{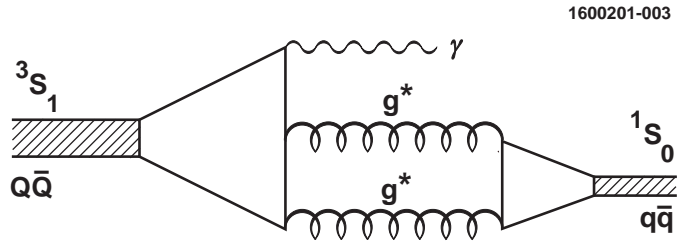}
\end{center}
\caption[Radiative decay of vector mesons to a photon and two gluons]
        {Radiative decay of vector mesons, such as \Jpsi, to a photon
         and two gluons. This process is expected to be a copious
         source of glueballs $X$ via $J/\psi\to\gamma X$.
\label{fig:psiggg}}
\end{figure}

An important goal for CLEO-c and CESR-c is to acquire $\approx10^9$
events at the \Jpsi\ peak. In addition to various rare decay
processes, a prime focus will be to study gluonic excitations through
radiative decay, \ie $J/\psi\to\gamma X$. The basic idea is shown in
\Figure~\ref{fig:psiggg}. A vector resonance can decay to three (but not
two) vector particles. If one of these decay products is a photon,
then there is a fair probability that the remaining two are
gluons. Hence, this process is expected to give rise to final-state
glueballs $X$~\cite{Close:1996yc}.

\subsection{Summary: the CLEO-c schedule}
\label{sec:Summary}

CLEO-c began taking data in October 2003, and has been running smoothly. 
An upgrade of CESR-c, adding more wigglers for higher luminosity, is on 
track for spring and summer 2004. The facility will run for approximately 
three years.

The values of the actual beam energies are decided upon dynamically by
the collaboration, and will depend on technical issues as well as
emerging physics cases. The original plan~\cite{Briere:2001rn} is as
follows:
\begin{itemize}
\item Act I: Obtain 3~fb$^{-1}$ at the $\psi(3770)$, yielding
      $\sim1.5\times10^6$ $\rm D\bar{D}$ events.
\item Act II: Obtain 3~fb$^{-1}$ at $\sqrt{s}\approx4.1$~GeV, yielding
      $\sim3\times10^5$ tagged $\rm D_s$ decays.
\item Act III: Obtain $10^9$ \Jpsi.
\end{itemize}
CLEO-c was encouraged by the QWG to consider running at the $\psi(2S)$
for a long period of time, making use of $\psi(2S)\to\pi\pi J/\psi$ to
tag \Jpsi. This prospect is being studied, including determination of
the final-state polarization of the \Jpsi\ and detector acceptance
issues, as they relate to the ability to carry out a partial wave
analysis.

\section{The BEPCII/BES~III project}

The Beijing Electron--Positron Collider (BEPC) is going to have a major
upgrade, called the BEPCII project. The BEPCII feasibility report has been
officially approved by the Chinese government.

The main physics goals of BEPCII are precision measurements and
searches for new particles and new phenomena, mainly in the energy
region from the $J/\psi$ to the $\psi(3770)$.  For example, precision
measurements of $\rm D$ and $\rm D_s$ meson decays will be essential to allow
the CKM matrix parameters, $V_{cs}$ and $V_{cd}$, to be determined
with a precision of a few per cent.  Studies of light hadron
spectroscopy and glueball candidates with very high statistics will be
necessary to test QCD, in particular lattice QCD calculations, which
should reach precisions of a few per cent in the coming years.
Searches for $\rm D_0 \bar{D}_0$ mixing are important to look for physics
beyond the Standard Model.  The number of important physics topics in the
$\tau$-charm energy region is very large.

Our physics goals require major upgrades of the BEPC to increase its
luminosity by two orders of magnitude and the BES detector to reduce
its systematic errors, as well as to adapt to the small bunch spacing
and high event rates. The large-scale upgrade will enable BEPC to
approach the specifications of a factory-type machine, whose main
parameters are listed in \Table~\ref{tab:bepcii}, along with a comparison
to those of the current BEPC.

\begin{table}[t]
\caption{Main parameters of BEPCII in comparison with BEPC} 
\label{tab:bepcii}
\begin{center}
\small
\begin{tabular}{|l|c|c|c|}
\hline 
 {\bf{Parameters}}               & {\bf{Unit}} & {\bf{BEPCII}}     & {\bf{BEPC}} \\ \hline \hline
Operation energy ($E$)      & GeV  & 1.0--2.1    & 1.0--2.5 \\ \hline
Injection energy ($E_{inj}$)   & GeV  & Up to 1.89 & 1.3   \\ \hline
Circumference ($C$)         & m    & 237.5      & 240.4 \\ \hline
Revolution frequency ($f_r$) & MHz  & 1.262      & 1.247  \\ \hline
Lattice type              &      &  FODO + micro-$\beta$ 
& FODO + low--$\beta$ \\ \hline
$\beta^*$--function at IP ($\beta^*_x/\beta^*_y$) & 
cm & 100/1.5 & 120/5 \\ \hline
Natural energy spread ($\sigma_e$) & & 2. 73E$\times 10^{-4}$ &
2.64E$\times 10^{-4}$ \\ \hline
Damping time  ($\tau_x$/$\tau_y$/$\tau_e$) & & 25/25/12.5 at 1.89 GeV & 
28/28/14 at 1.89 GeV \\ \hline
RF frequency ($f_{rf}$)   &  MHz  & 499.8     & 199.533  \\ \hline 
Harmonic number ($h$)       &       & 396    &  160  \\ \hline
RF voltage per ring ($V_{rf}$)  & MV & 1.5 & 0.6--1.6 \\ \hline
Bunch number ($N_b$)      &     &  93      &  2$\times$1 \\ \hline
Bunch spacing             & m   &  2.4 & 240.4 \\ \hline
Bunch current ($I_b$)     & mA  &  9.8 at 1.89 GeV & 35 at 1.89 GeV \\ \hline 
Beam current (colliding)  & mA  & 910 at 1.89 GeV & 2$\times$35 at 1.89
 GeV \\ \hline
Bunch length ($\sigma_l$)   & cm  & $\sim$1.5 &  $\sim$5 \\ \hline 
Impedance ($|Z/n|_0$)       & $\Omega$ & $\sim$0.2  & $\sim$4 \\ \hline
Crossing angle            & mrad & $\pm11$  &  0  \\ \hline
Vert. beam--beam param. ($\xi_y$) & &  0.04   &  0.04 \\ \hline
Beam lifetime             & hrs &  $\sim$2.7  & 6--8 \\ \hline
Luminosity at 1.89 GeV       & $10^{31}$ cm$^{-2}$s$^{-1}$  & 100 &1\\ \hline 
\end{tabular}
\end{center}
\end{table}

BEPCII will be a double-ring collider with superconducting
micro-$\beta$ magnets, a 500 MHz RF system with superconducting
cavities, and a low-impedance antechamber beam pipe. The second ring
can be accommodated in the existing BEPC tunnel. BEPCII will have a
large horizontal crossing angle of 11 mrad at the southern interaction
region. There will be 93 bunches per ring with a total current of
910~mA in each ring. The peak luminosity of BEPCII will be
10$^{33}$cm$^{-2}$ s$^{-1}$ at the beam energy of 1.89~GeV, which is
about 100 times higher than that of the BEPC. The peak luminosity at
the $J/\psi$ and at 4.1~GeV c.m. energy will be about
$0.6\times10^{33}$cm$^{-2}$ s$^{-1}$. The upgrade of the linac will
provide full energy injection up to 1.89~GeV for `topping off' the
beam.  The positron injection rate will reach 50~mA/min compared to
the present rate of about 5~mA/min.  The number of events expected for
one year of running for various physics topics is given in
\Table~\ref{tab:events}.

\begin{table}[htb]
\caption{Number of events expected in one year of running} 
\label{tab:events}
\begin{center}
\begin{tabular}{|l|c|c|c|c|}
\hline
Physics   & CM      & Peak       & Physics     & Number of  \\
channel   &  energy & luminosity & cross-section   & events per year  \\
          & (GeV)   & ($10^{33}$cm$^{-2}$s$^{-1}$) & (nb)  & \\ \hline
$J/\psi$  & 3.097   &  0.6       & $\sim$3400  &  $10\times10^9$
\\ \hline
$\tau$    & 3.67    &  1.0       & $\sim$2.4   &  $12\times10^6$
\\ \hline
$\psi(2S)$ & 3.686   &  1.0       & $\sim$640   &  $3.0\times10^9$
\\ \hline
$D$       & 3.770   &  1.0       & $\sim$5     &  $25\times10^6$
\\ \hline
$D_s$     & 4.030   &  0.6       & $\sim$0.32  &  $1.0\times10^6$
\\ \hline
$D_s$     & 4.140   &  0.6       & $\sim$0.67  &  $2.0\times10^6$
\\ \hline
\end{tabular}
\end{center}
\end{table}

Most of the existing utility facilities of the BEPC, after some
upgrading, will be used for BEPCII. A cryogenics system of 1000 W at
4.2 K will be installed for the three different superconducting
devices. The design of BEPCII will keep the electron beam in the outer
ring during the dedicated synchrotron radiation running, and all
synchrotron radiation beam lines and the experimental stations will be
unchanged, but the beam current will be increased from 140~mA at
2.2~GeV to 250~mA at 2.5~GeV.

\begin{figure}[t]
\begin{center}
\hspace*{1.2cm}
\includegraphics[width=12cm]{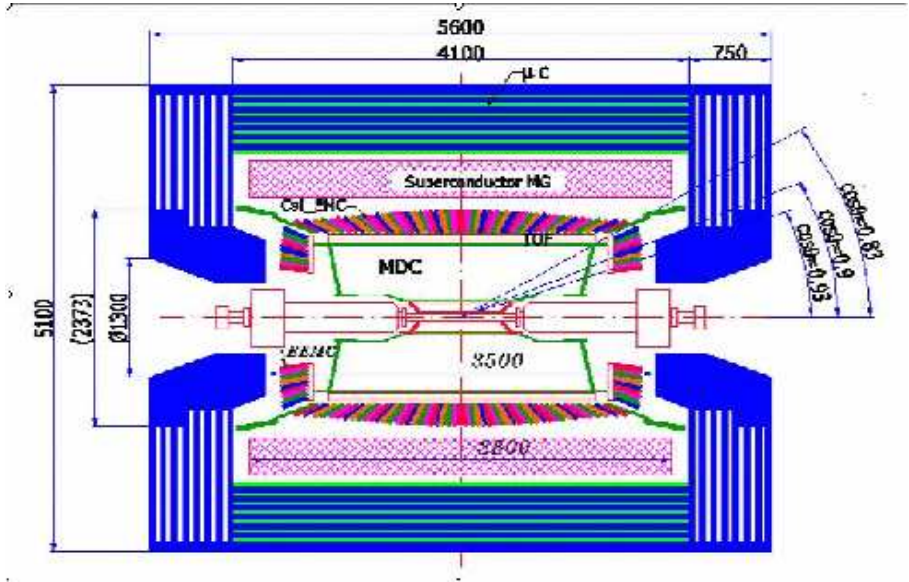}
\end{center}
\caption{Schematic view of the BES~III detector}
\label{fig:besiii}
\end{figure}

BEPCII is a high luminosity, multi-bunch collider, which requires a
comparable high quality detector with modern technology.  The main
features of the detector are as follows (a schematic view of the
BES~III detector is shown in \Figure~\ref{fig:besiii}):

\begin{itemize}
\item Main draft chamber (MDC): the design features small cell
      structure, aluminium filled wires, and He-based gas, with
      expected performances of $\sigma_{xy}$ = 130 $\mu$m, $\Delta
      p/p$ = 0.5\% at 1~GeV, and $dE/dx$ = 6--7\%.  The stepped end
      plates provide space for the superconducting micro-$\beta$
      magnets;
\item Electromagnetic calorimeter (EMCAL): CsI crystals of 15
      radiation length (28~cm), with expected performances of $\Delta
      E/E$ = 2.5 \% at 1~GeV and $\sigma_{pos}$ = 0.5~cm/$\sqrt{E}$;
\item TOF: plastic scintillators with $\Delta T$ = 90 ps for the barrel
      part and $\Delta T$ = 100 ps for the end caps;
\item 1 tesla superconducting solenoidal magnet;
\item Resistive Plate Chambers (RPC) for muon identification: nine layers
      interleaved with the iron plates of the return yoke;
\item Trigger: largely based on FPGA technology and using information
      of the MDC tracks and EMCAL showers, pipelined with a time
      latency of 6.4 $\mu$s.
\end{itemize}

The total estimated budget for the BEPCII will be around 640 million
Chinese Yuan (about \$77 million). The Chinese government will provide
funding to cover the costs of the machine and the major part of the
detector.  Part of the detector cost is expected to be provided by
international contributions. International cooperation is already
helping IHEP with the design and R\&D of BEPCII, as well as the
production of some key devices, \eg Brookhaven National laboratory is
helping with the superconducting micro-$\beta$ magnets; KEK is helping
with the superconducting RF cavities and the superconducting solenoid
magnet.

The preliminary design of BEPCII has been finished. The engineering
design is under way and most parts are finalized. Contracts are
already signed for many important items. The project is expected to be
completed by the end of 2006, and physics running is scheduled in
2007.

The great physics potential of BEPCII/BES~III will attract the interest
of many physicists who are warmly welcomed to join the BEPC/BES
upgrade and its physics programme. The completion of the BEPC II will
add a new machine of `factory class' to the fabric of
high-energy physics, thus adding new momentum to the research efforts
in $\tau$-charm physics.

\section{B-factories}

One of the goals of a Super B-factory is the search for new physics in
B-meson decays. In addition, a very interesting programme on
quarkonium physics can be undertaken since $b$ decays are good sources
of $c\bar{c}$ states. In this section we shall focus on the KEKB
facility, although similar expectations and physical potential apply
equally to the SLAC PEPIII project.  In particular, there is a
proposal to upgrade KEKB to a Super KEKB with a design luminosity of
$5\times 10^{35}$~cm$^{-2}$s$^{-1}$ which is 50 times the peak
luminosity achieved by KEKB. The target annual integrated luminosity
is 5~ab$^{-1}$.

Along with the luminosity upgrade the Belle detector would be
upgraded.  The largest challenge will be the very harsh background
environment due to the high beam current.  Another challenge is the
background level in the end-cap and an upgrade to another advanced
technology is necessary.  Among the candidates, pure CsI crystals with
photo tetrode readout is the most promising for the end-cap
electromagnetic calorimeter (EECL) and tiles of plastic scintillator
with silicon photomultipliers (SiPM) is a good candidate for the
end-cap $K_L^0$ muon system (EKLM).  The trigger and data acquisition
systems should be upgraded to handle the 20 times higher occupancy
level from the higher beam current.  Computing is another
technological challenge with online data having to be recorded at a
speed of 250 MB/s after online reconstruction and reduction amounting
to the data size of 5 PB/year.  Including Monte Carlo simulations, a
storage system holding 10--20 PB is needed at the beginning which
should be expandable to several tens of PB.

In B-meson decays the $b\to c\bar{c}s$ subprocess is CKM-favoured so 
that final states containing charmonium particles are common. A 
super-B factory would provide superb opportunities for high-sensitivity 
measurements of the charmonium system and the 
discovery of missing charmonium states such as the $h_c$, or the 
D-wave states:
\begin{itemize}
\item in the continuum: $e^+e^- \to \chi_{c1}\ (c\bar{c}); \eta_c\
      (c\bar{c})$; search for $C=-1$ states ($h_c$, \etc) in the
      $\chi_{c1},\ \eta_c$ recoil spectra;
\item in B decays, either with exclusive channels like $\rm B \to K\
      1^3D_2$ (how big is the suppression factor?) or with inclusive
      channels like $\rm B \to 1^3D_2 \to \pi^+\pi^-J/\psi$ at the
      $\Upsilon(4S)$.
\end{itemize}

Another exciting possibility is the discovery of one or more
charmonium hybrid states. B-factories can also shed light on the
possibility of $\rm D\bar{D}$ molecules. The discovery of the
$X(3872)$ just above $\rm D^0\bar{D}^{*0}$ threshold by the Belle
Collaboration has led to speculation that the $X(3872)$ is a $\rm
D\bar{D}^*$ molecule or some other 4-quark object. The high statistics
available at a super B-factory would allow detailed studies of the
$X(3872)$ and other new states, including $\rm D\bar{D}$ molecules if
they exist.

\section{GSI}

\subsection{Introduction}

The charmonium spectroscopy physics programme of the PANDA
($\overline{\mathrm{P}}$ ANnihilations at DArmstadt) experiment using
$\overline{\mathrm{p}}\mathrm{p}$ annihilations at GSI~\cite{gsiref1}
is an extension of successful experiments performed recently at the
Fermilab antiproton accumulator. Advanced $\overline{p}$ cooling
techniques and a more versatile detector setup will be employed,
allowing for the first time the measurement of both electromagnetic
and hadronic decays. The goal is to make comprehensive measurements of
the spectroscopy of the charmonium system and hence provide a detailed
experimental study of the QCD confining forces in the charm region to
complement theoretical investigation.

Unlike $e^+e^-$, where only states with the quantum numbers of the
photon ($J^{PC}=1^{--}$) can be formed directly, all quantum numbers
are directly accessible in $\overline{\mathrm{p}}\mathrm{p}$
annihilation.  Charmonium states are studied by accelerating the
$\overline{\mathrm{p}}$ beam to the energy of the resonance, which is
then scanned by changing the beam momentum in small steps.

The experimental programme of PANDA also includes the study of gluonic
excitations (glueballs and hybrids) in the charmonium sector, as well
as the study of charmonium in nuclei.

\subsection{Experimental apparatus}

The PANDA experiment will be installed at the High Energy Storage Ring
(HESR), a major component of the recently approved new accelerator
facility at GSI in Darmstadt, Germany \cite{gsiref1}. The antiproton
beam will be produced by a primary proton beam from the planned fast
cycling, superconducting 100~T$\cdot$m synchrotron ring. The
antiprotons will be produced with a rate of approximately $2\times
10^7/$s and then stochastically cooled; after $5\times 10^{10}$
$\overline{\mathrm{p}}$ have been stored, they will be transferred to
the HESR where internal experiments in the momentum range from 1 to 15
GeV can be performed.  Two modes of operation are foreseen: a
high-luminosity mode, where peak luminosities of $2\times 10^{32}$
cm$^{-2}$s$^{-1}$ will be reached with a beam momentum spread $\delta
p/p \approx 10^{-4}$, achieved by means of stochastic cooling in the
HESR ring, and a high-resolution mode, where for beam momenta below 8
GeV electron cooling will yield a smaller beam momentum spread $\delta
p/p \approx 10^{-5}$ at a reduced luminosity of
$10^{31}$cm$^{-2}$s$^{-1}$.

\begin{figure}[t]
\begin{center}
\includegraphics[width=\textwidth]{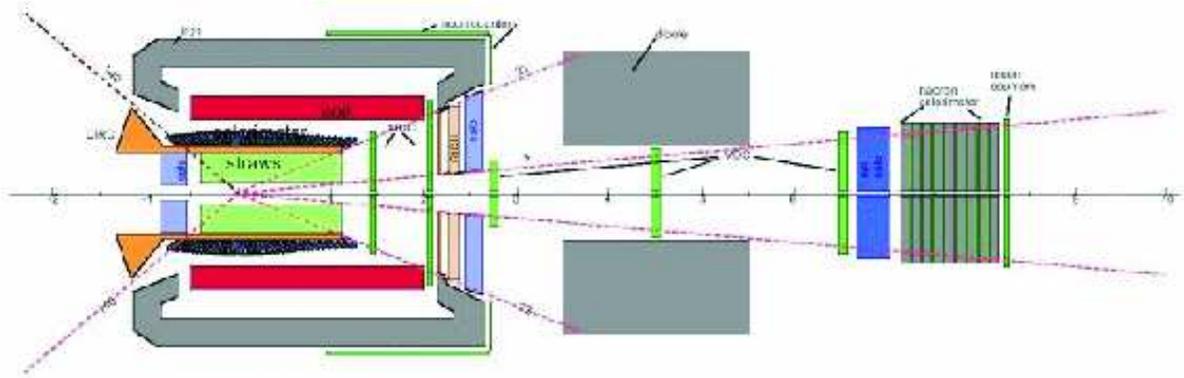}
\end{center}
\caption{Schematic view of the PANDA detector (side view)
\label{fig:pandadet}}
\end{figure}

The proposed PANDA detector is being designed to study the structure
of hadrons in the charmonium mass range as well as the spectroscopy of
double hypernuclei. The detector must provide (nearly) full
solid-angle coverage, it must be able to handle high rates ($2\times
10^7$ annihilations/s) with good particle identification and momentum
resolution for $\gamma$, e, $\mu$, $\pi$, K and p.  Additional
requirements include vertex reconstruction capability and, for
charmonium, a pointlike interaction region, efficient lepton
identification, and excellent calorimetry (both resolution and
sensitivity to low-energy photons).

A schematic view of the present detector concept is shown in
\Figure~\ref{fig:pandadet}.  The antiprotons circulating in the HESR
hit an internal hydrogen pellet (or jet) target, while for the nuclear
target part of the experimental programme wire or fibre targets will
be used.  The detector consists of a Target Spectrometer (TS) and a
Forward Spectrometer (FS).

The TS, for the measurement of particles emitted at laboratory polar
angles larger than 5$^{\circ}$, is located inside a solenoidal magnet,
2.5~m in length and 0.8~m in inner radius.  Its main components are
four diamond or silicon start detectors surrounding the interaction
region followed by a 5-layer silicon microvertex detector; 15 layers
of crossed straw tubes, for the measurement of charged-particle
trajectories; a cylindrical DIRC and a forward aerogel \v{C}erenkov
detector for particle identification; an electromagnetic calorimeter
consisting of PbWO$_4$ crystals with Avalanche Photo Diode (APD)
readout. The region between the calorimeter and the end-cap will be
instrumented with two sets of mini drift chambers; scintillator strips
used for muon identification will be located behind the return yoke of
the magnet.

The FS will measure particles emitted at polar angles below
10$^{\circ}$ in the horizontal and 5$^{\circ}$ in the vertical
direction. It will consist of a dipole magnet with a 1~m gap, with
MDCs before and after for charged-particle tracking. Particle
identification will be achieved by means of a TOF stop and a
dual-radiator RICH detector.  Other components of the FS are an
electromagnetic and a hadronic calorimeter followed by a set of muon
chambers.

Detailed simulations of the detector concept presented here show its
ability to measure electrons, muons, pions, kaons and photons over a
large phase space region.  Combining a momentum resolution of 1--2\%
with a high discriminating power for particle identification and a
nearly 4$\pi$ solid angle coverage allows the application of strong
kinematical constraints, which will serve to achieve an excellent
level of final-state identification and background suppression.

The PANDA project is part of the recently approved new accelerator
facility at GSI. An international collaboration is currently forming
to develop a detailed technical proposal for the design and
construction of the detector system.
\longpage

\subsection{Physics programme}

\subsubsection{The ground state of charmonium, $\eta_c(1^1S_0)$}

Despite the abundance of experimental measurements, it is
disappointing how little is known about the ground state of
charmonium, $\eta_c(1^1S_0)$.  Five new measurements of its mass were
reported in 2002 and 2003, disagreeing by more than 5
MeV~\cite{gsiref2}. The fit to all existing measurements of the
$\eta_c$ mass yields an error of 1~MeV (adequate to the accuracy of
present theoretical model calculations), but the consistency between
the various measurements is fairly poor (CL = 0.5\%). In addition to
that, as the accuracy of theoretical calculation increases it will be
mandatory to measure its mass with a precision better than 1~MeV.  The
width of the $\eta_c$ is even more uncertain. Four new measurements
were reported in 2002 and 2003, and the fit to all data yields a width
value of ($25.0 \pm 3.3$)~MeV, with a CL of 0.05\%.  It is important
to know the width of the $\eta_c$ because a width value as large as 25
MeV is difficult to reconcile with simple quark models, and it has
been suggested that instanton effects may be responsible
\cite{gsiref3}.

It must be stressed that unlike the E760/E835 experiments
\cite{gsiref4a,gsiref4b}, which were obliged to identify $\eta_c$
formation in the extremely weak two-photon decay channel (BR($\eta_c
\rightarrow \gamma \gamma) \simeq 3\times10^{-4}$), the PANDA detector
at the new facility at GSI is being designed to detect both
electromagnetic and hadronic final states.  This will make it possible
to study the $\eta_c$ in several decay channels, which have hundred
times larger branching ratios: $\eta_c \rightarrow 2(K^+K^-),
K\overline{K}\pi, 2(\pi^+\pi^-), \eta\pi\pi,$ \etc

\subsubsection{The radial excitation of charmonium, $\eta_c(2S)$}

The $\eta_c(2S)$ was discovered by the Belle experiment in the
hadronic decays of the B meson \cite{gsiref6}, with a mass of ($3654
\pm 6 \pm 8$)~MeV, incompatible with the Crystall Ball observation
\cite{gsiref7}.  The $\eta_c(2S)$ was then seen also by BaBar
\cite{gsiref8} and Cleo \cite{gsiref9} in $\gamma \gamma$ collisions.
The mass measurements of the three experiments are consistent
\cite{gsiref2} and yield a value of ($3637.7 \pm 4.4$)~MeV (CL =
14\%); this value is only marginally consistent with most model
calculations and it has been suggested that coupled channels effects
may shift the $\eta_c(2S)$ mass \cite{gsiref10}.  The present accuracy
on the $\eta_c(2S)$ width is only 50\%, the measured value being ($19
\pm 10$)~MeV. More precise measurements of the mass and width are
clearly needed.

A search for the $\eta_c(2S)$ was performed by the experiments E760
and E835 at Fermilab in the process $\overline{\mathrm{p}}\mathrm{p}
\rightarrow \eta_c(2S) \rightarrow \gamma\gamma$
\cite{gsiref4a,gsiref11}.  No signal was observed by either
experiment. The technique employed by E760/E835 suffered from the
severe limitations due to the relatively high background from
$\pi^0\pi^0$ and $\pi^0\gamma$ compared to the small $\gamma \gamma$
signal \cite{gsiref12}.  Further measurements using this channel will
require increased statistics and a substantial reduction of the
background.  The real significant improvement of PANDA with respect to
the Fermilab experiments will be the ability to detect the hadronic
decay modes, such as $\eta_c(2S) \rightarrow K^*\overline{K}^*$ or
$\eta_c(2S) \rightarrow \phi\phi$, which will allow a clean
identification of this state.

\subsubsection{The $h_c(^1P_1)$ resonance of charmonium}
\longpage

The singlet P resonance of charmonium $h_c(^1P_1)$ is of extreme
importance in determining the spin-dependent component of the
$q\overline{q}$ confinement potential.

If the recent observation of $h_c$, described in
\Chapter~\ref{chapter:spectroscopy}, is confirmed during this decade,
the precise measurement of its width will have to wait for the high
statistics to be accumulated by the PANDA experiment.  By comparing
the total width with the probably dominant radiative width to
$\eta_c\gamma$, it will be possible to measure its partial width to
light hadrons, relevant for NRQCD calculations.  It must be pointed
out that owing to its very narrow width ($\leq$ 1~MeV) and expected
low yields, only a $\overline{\mathrm{p}}\mathrm{p}$ formation
experiment like PANDA will be able to perform this measurement and to
carry out a systematic study of its decay modes.  The study of the
$h_c$ constitutes a central part of the PANDA charmonium physics
programme.

\subsubsection{Radiative transitions of the $\chi_J(^3P_{0,1,2})$ 
               charmonium states} 

The measurement of the angular distributions in the radiative decay of
the $\chi_1$ and $\chi_2$ states formed in
$\overline{\mathrm{p}}\mathrm{p}$ annihilations provides insight into
the dynamics of the formation process, the multipole structure of the
radiative decay, and the properties of the $c\overline{c}$ bound
state.  A comparison of the E760 result at the $\chi_{c2}$
\cite{gsiref15} with the Crystal Ball result at the $\chi_{c1}$
\cite{gsiref16} is not consistent with theory, and may suggest the
existence of additional contributions to the theoretical predictions
for the M2 amplitudes.  The simultaneous measurement of both angular
distributions has been recently performed by E835
\cite{gsiref17}. They too observed a discrepancy with respect to
theoretical predictions, which could indicate the presence of
competing mechanisms, leading to the cancellation of the M2 amplitude
at the $\chi_{c1}$.  The effect seen by E835 is at the 2.5$\sigma$
level, therefore further high-statistics measurements are clearly
needed to increase the significance of this result.

\subsubsection{Charmonium states above the $\rm D\overline{D}$ threshold}

The energy region above the $\rm D\overline{D}$ threshold at 3.73~GeV
will be object of many studies during this decade. This is the region
in which narrow $\rm ^1D_2$, $\rm ^3D_2$ states (which are narrow
because they cannot decay to $\rm D\overline{D}$) and the first radial
excitations of the singlet and triplet P states are expected to exist,
as shown in \Chapter~\ref{chapter:spectroscopy}. The discovery of
X(3872) has raised further interest in this energy region: the nature
of this new, narrow state is not yet clear, and speculation ranges
from a $\rm D^0\overline{D^{0*}}$ molecule to a $\rm ^3D_2$
state. There are theoretical problems with all these interpretations,
and further, more accurate measurements of its width and particularly
of its decay modes are needed to shed light on this state
\cite{gsiref10}.  This kind of study is ideally suited for a
$\overline{\mathrm{p}}\mathrm{p}$ formation experiment.  The study of
the energy region above the $\rm D\overline{D}$ threshold is a central
part of the charmonium physics programme of PANDA. It will require
high-statistics, small-step scans of the entire energy region
accessible at GSI.

\subsubsection{Charmonium hybrids}
 
Predictions for hybrids come mainly from calculations based on the bag
model, flux tube model, constituent gluon model and recently, with
increasing precision, from lattice QCD (LQCD) \cite{gsiref20}.  For
these calculations the parameters are fixed according to the
properties of the known $Q\overline{Q}$ states.  All model predictions
and LQCD calculations agree that the masses of the lowest-lying
charmonium hybrids are between 3.9 and 4.5~GeV and that the state with
the lowest mass has $J^{PC}=1^{-+}$ \cite{gsiref21}. Some of the
charmonium hybrids have spin exotic quantum numbers, so mixing effects
with nearby $c\overline{c}$ states are excluded for them, thus making
their experimental identification easier. Predictions for the widths
of these states range from a few MeV to several tens of MeV.
Cross-sections for the formation and production of charmonium hybrids
are estimated to be similar to those of normal charmonium states.

In PANDA two kinds of experiments can be done: formation and
production. Formation experiments would generate non-exotic charmonium
hybrids, while production experiments would yield a charmonium hybrid
together with another particle, such as a $\pi$ or an $\eta$. In
$\overline{\mathrm{p}}\mathrm{p}$ annihilation, production experiments
are the only way to obtain exotic quantum numbers.  This distinction
is a very powerful tool from the experimental point of view: the
detection of a state in production and its non-detection in formation
is a clear, unique signature for exotic behaviour.

\subsubsection{Charmonium in nuclei}

The proposed experimental programme of PANDA will address open
problems of in-medium modifications of hadrons with charmed quarks in
nuclei and the interaction of these hadrons with nuclei. This is, on
the one hand, an extension of the present chiral dynamics studies with
partial restoration of chiral symmetry in the hadronic environment,
from the light quark to the open charm quark sector. On the other
hand, this programme is focused on the first experimental studies of
the charmonium--nucleon and charmonium--nucleus interaction, which is
also of basic importance for ultra-relativistic heavy-ion collisions.

\section{Jefferson Lab 12 GeV upgrade}

The Jefferson Laboratory has plans to upgrade the Continuous Electron
Beam Accelerator Facility (CEBAF) to 12~GeV \cite{jlab}.  The 12~GeV
electron beam will be used to produce 9~GeV photons in the new Hall D.
Photon fluxes of up to $10^8$ photons/s with 50\% linear polarization
are achievable.  In Hall D, a tagged coherent bremsstrahlung beam and
solenoidal detector will be constructed in support of a programme of
gluonic spectroscopy.  The detector has been optimized to provide
nearly hermetic acceptance for both charged particles and photons.  In
addition, a combination of particle identification systems will allow
very good $K$--$\pi$ separation.  Optimization will allow the detector
to fully reconstruct exclusive many-body final states.  In conjunction
with high statistics, this will allow excellent partial wave analyses
of many final states.  The $4\pi$ acceptance of the Hall D detector
and the energy resolution of its tagged beam could help to reduce the
background considerably.

The threshold production of charmonium and open charm production open 
up a new window into QCD dynamics; in particular, these reactions are 
sensitive to multiquark, gluonic, and so-called `hidden colour' 
correlations in nucleons and nuclei.  In contrast to diffractive charm 
production at high energy, which tests the behaviour of the gluon 
structure functions at small $x$, charm production near threshold 
tests the structure of the target near $x=1$ and its short-range 
behaviour. This difference results from the kinematics of the reaction 
products.  For $J/\psi$ production off the nucleon, the threshold 
energy is $E_\gamma=8.2$~GeV and because of the large mass of the 
charmed quark the $c\bar{c}$ fluctuation of the photon travels over a 
short coherence length.  Charm production near threshold implies a 
small impact parameter so that all five valence quarks must be in the 
same small interaction volume and all the quarks must be involved in 
the reaction mechanism.  For nucleon targets this implies that three-gluon 
exchange may dominate two-gluon and one-gluon exchange.  

Even though the $c\bar{c}$ pair is created with rather high momentum at 
threshold, it may be possible to observe reactions where the pair is 
captured by the target nucleus forming `nuclear-bound quarkonium'.   
The discovery of such qualitatively new states of matter would be 
significant.  

\section{LHC (ATLAS/CMS)}

\newcommand{\llumi} {\ensuremath{10^{33}\;{\mathrm{cm}}^{-2}{\mathrm{s}}^{-1}}}
\newcommand{\hlumi} {\ensuremath{10^{34}\;{\mathrm{cm}}^{-2}{\mathrm{s}}^{-1}}}
\newcommand{\ifb}   {\ensuremath{{\mathrm{fb}}^{-1}}}

The Large Hadron Collider (LHC) is a proton--proton collider currently
being built at CERN and scheduled to start in the second half of
2007. It will provide many opportunities for studying heavy quarkonia,
which will be produced at unprecedented rates and energies.
Significant contributions may be expected in the fields of heavy
quarkonia production and decays, whereas high background rates will
make dedicated studies on heavy quarkonia spectroscopy difficult (as
at other hadron machines).
   
This section concentrates on future opportunities at the two
multipurpose LHC experiments ATLAS and CMS. Aspects related to the
dedicated B-physics experiment LHCb and to studies `in media' are
covered in separate sections.

Heavy quarkonia production issues at the LHC in general are discussed
in \Section~\ref{sec:HQprodLHC}, and therefore apply to both ATLAS
and CMS. \Section~\ref{sec:HQATLAS} presents selected topics from
current, heavy-quarkonia-related ATLAS studies. Results from CMS were
not available at the time of writing. The selected topics are by no
means comprehensive and are intended to serve as illustrative examples
only.

\subsection{Heavy quarkonia production at the LHC}
\label{sec:HQprodLHC}

Thanks to its high collision energy (design centre-of-mass energy
14~TeV) and high luminosity (design luminosity ${\cal L}=\hlumi$), the
LHC will be able to explore a new high-energy frontier at the TeV
scale. It is expected, however, that the LHC will not operate at its
design luminosity from the beginning, but rather at an initial
luminosity of ${\cal L}=2\times\llumi$. This initial period will be
best suited for dedicated studies on heavy quarkonia at the LHC, both
in view of affordable trigger rates, modest pile-up (\ie minimum-bias
events superimposed on interesting signal), event reconstruction, \etc

The production rates for heavy quark flavours at the LHC will be
huge. The total cross-section at the LHC is about 100~mb; the expected
total cross-section for charm production is 7.8~mb, for bottom
production 0.5~mb, and for top production 0.8~nb, respectively
\cite{PhTDR}. Thus, for an integrated luminosity of only 1~\ifb
(\ie about one week of running at initial luminosity), as many as
$7.8\times 10^{12}$ charm events, $0.5\times 10^{12}$ bottom events,
and $0.8\times 10^{6}$ top events will be produced.

\begin{figure}[t]
\begin{center}
  \includegraphics[width=85mm]{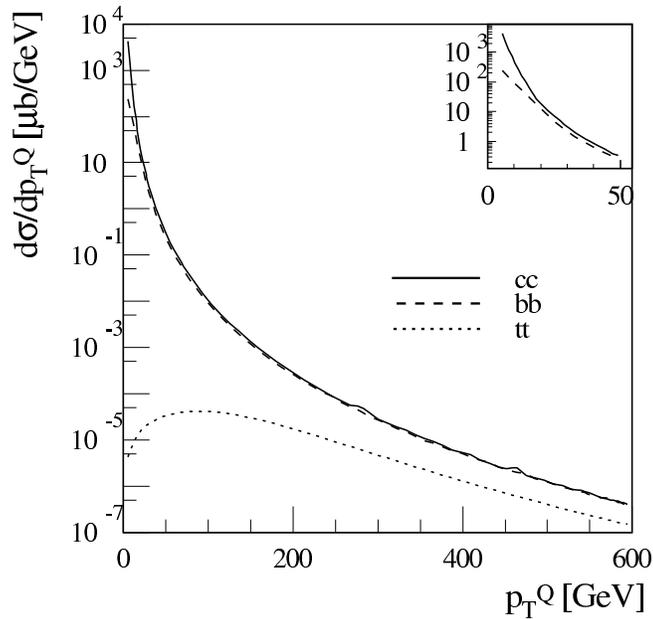}
\end{center}
\caption[Differential cross-section for heavy-quark-pair production]
        {Differential cross-section for heavy-quark-pair production as
         a function of the transverse momentum $p_{\mathrm{T}}^{\, Q}$
         of the heavy quark. The smaller figure shows the region
         $p_{\mathrm{T}}^{\, Q} < 50$~GeV for charm and bottom
         production \cite{PhTDR}.}
\label{fig:XS_heavyQ}
\end{figure} 

\begin{figure}[t] 
\begin{center}
  \includegraphics[width=.49\textwidth]{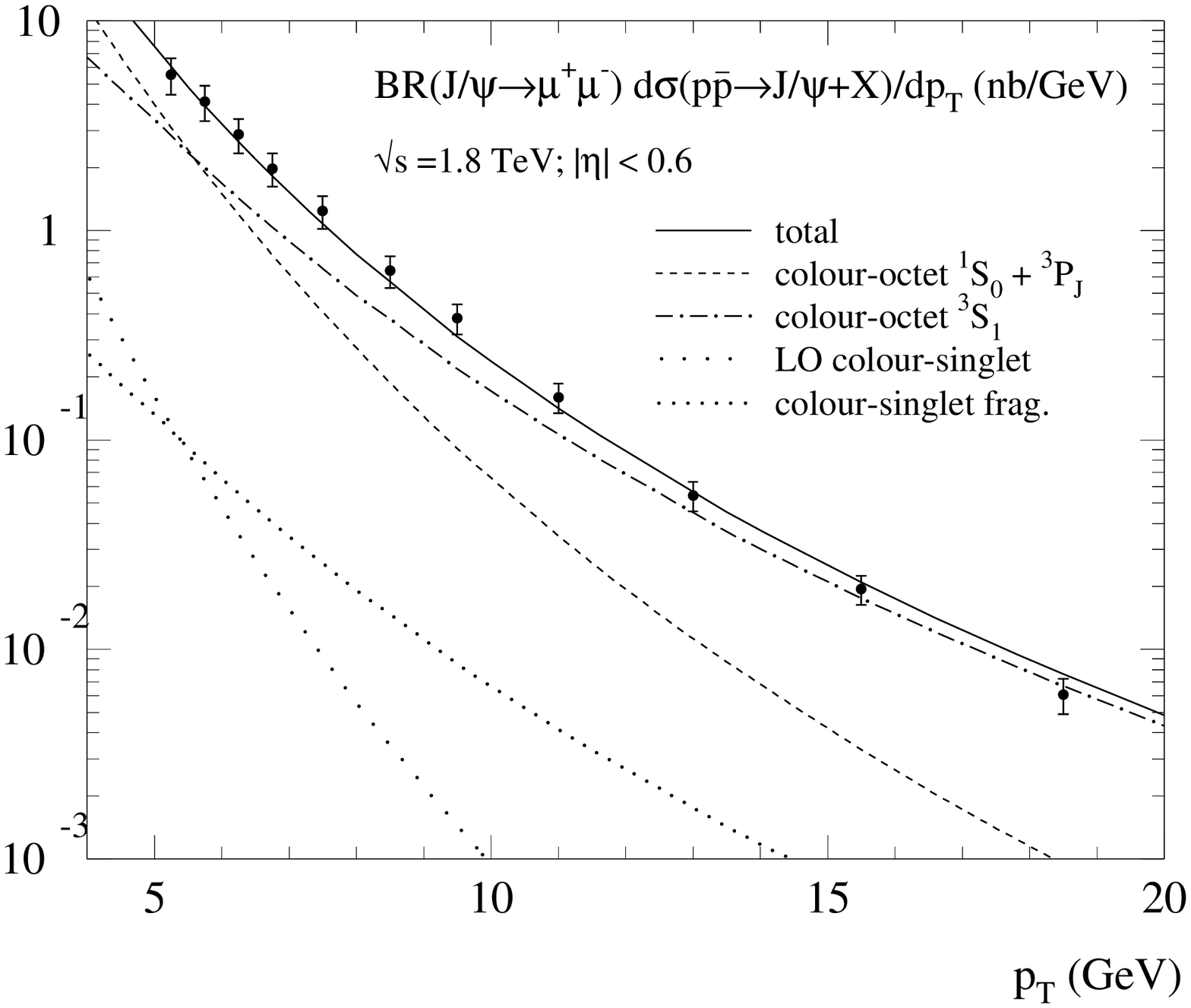}
  \hfill
  \includegraphics[width=.49\textwidth]{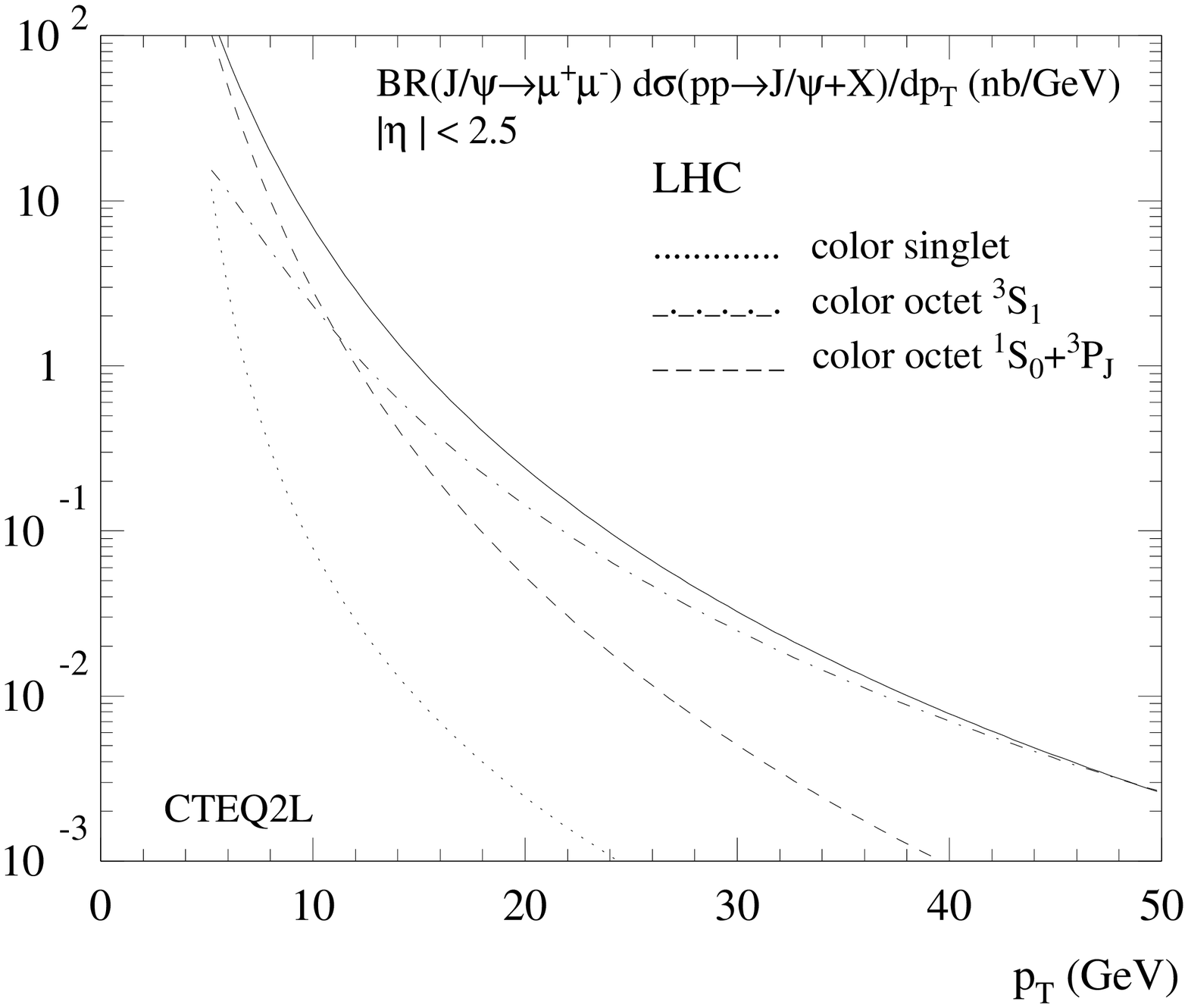}
\end{center}
\caption[Colour-singlet and colour-octet contributions
         to direct $J/\psi$ production in $p\bar{p}\to J/\psi + X$]
        {Left: Colour-singlet and colour-octet contributions to direct
         $J/\psi$ production in $p\bar{p}\to J/\psi + X$ at the
         Tevatron, together with experimental data from
         CDF~\cite{CDF1997}. Right: cross-sections for $J/\psi$
         production in $pp\to J/\psi + X$ at the LHC, as obtained with
         the Monte arlo generator PYTHIA supplemented by leading order
         colour-octet processes~\cite{MASL1999,MASL2000}. Plots are
         taken from Refs.~\cite{Kramer2001,SM1999}.}
\label{fig:C1_C8_Jpsi}
\end{figure} 

\begin{figure}[t]
\begin{center}
\includegraphics[width=85mm]{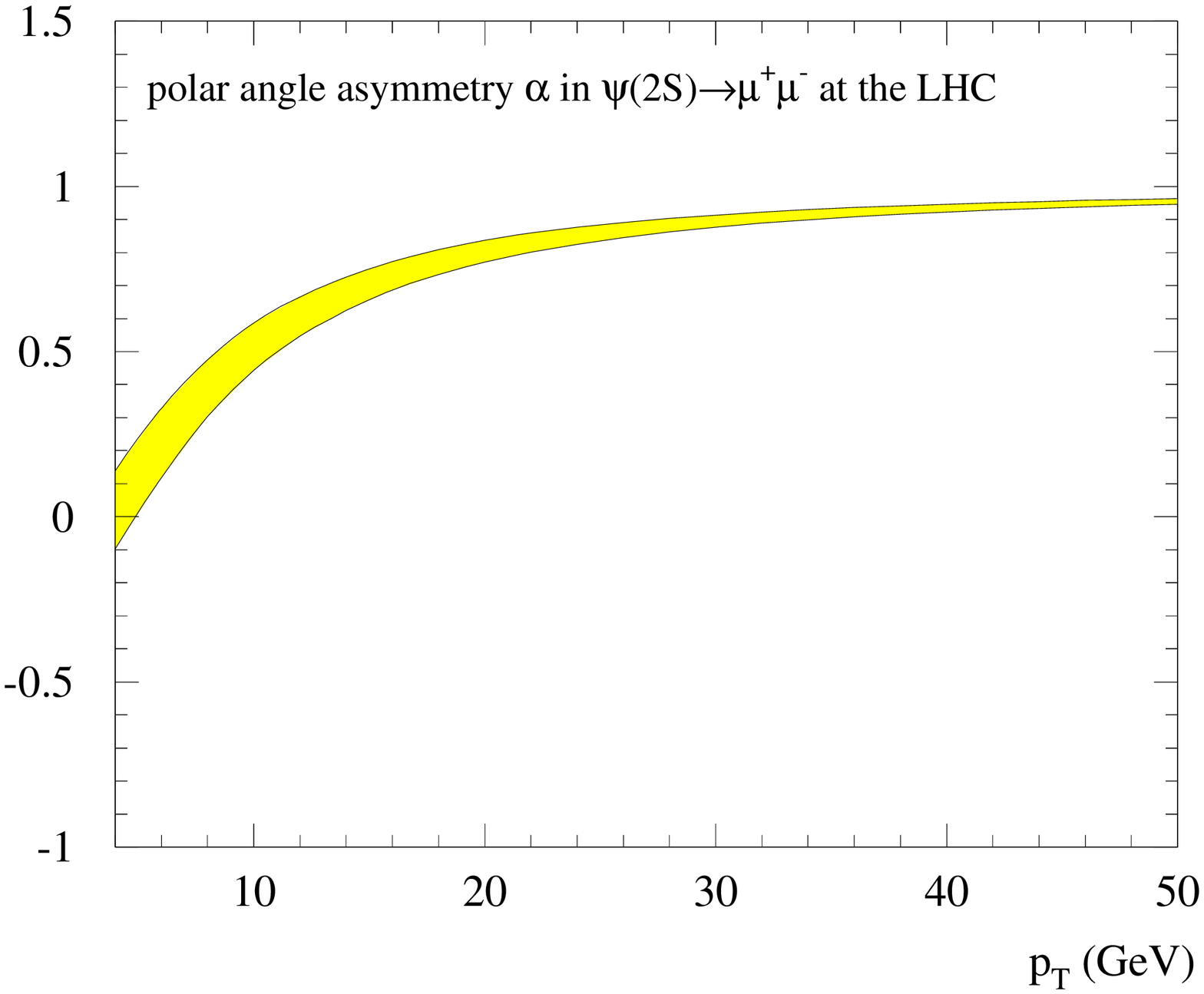}
\end{center}
\caption[Polar angle asymmetry $\alpha =
         (\sigma_{T}-2\sigma_{L})/(\sigma_{T}+2\sigma_{L})$ for
         $\psi(2S)$ production]
        {Polar angle asymmetry $\alpha =
         (\sigma_{T}-2\sigma_{L})/(\sigma_{T}+2\sigma_{L})$ for
         $\psi(2S)$ production in $pp\to \psi(2S)(\to\mu^{+}\mu^{-}) +
         X$ at the LHC as a function of $p_{\mathrm{T}}$
         \cite{Kramer2001}.}
\label{fig:asymm_psi2S}
\end{figure} 

\Figure[b]~\ref{fig:XS_heavyQ} shows the differential cross-section for
heavy-quark pair production at the LHC as a function of the transverse
momentum of the heavy quark~\cite{PhTDR}.  Up to next-to-leading-order
(NLO) perturbative QCD, the $c$ and $b$ cross-sections are identical
for high $p_{\mathrm{T}}$; differences due to mass effects show up
only for very small $p_{\mathrm{T}}$ values
($p_{\mathrm{T}}<20$~GeV). For orders higher than NLO, the spectrum
for $c$ quarks is expected to become softer, and differences might
become visible even for high $p_{\mathrm{T}}$.

In order to predict the production rates for heavy quarkonia at the
LHC, the available models (including Monte Carlo generators) are tuned
with Tevatron data and extrapolated to LHC energies; see~\cite{SM1999}
for a detailed description. \Figure~\ref{fig:C1_C8_Jpsi} illustrates
the results of this procedure applied to the prediction of the
$J/\psi$ production cross-section. An eventual measurement of heavy
quarkonia production rates at the LHC will help in understanding, for
high energies and $p_{\mathrm{T}}$, the roles and importance of
individual production mechanisms (\eg colour-singlet vs.\
colour-octet) and the applicability of concepts used so far in the
calculations (\eg factorisation in NRQCD).  It might also be possible
to probe the gluon density in the proton \cite{MASL2002}.
   
The fact that the LHC will produce heavy quarkonia with high
transverse momentum in large numbers will also allow for a better
discrimination between different models of heavy quarkonia
polarization, like NRQCD and the colour-evaporation model; see
Ref.~\cite{SM1999}. For example, NRQCD predicts transversely polarized
$J/\psi$ and $\psi(2S)$ (see \Figure~\ref{fig:asymm_psi2S}) at high
$p_{\mathrm{T}}$.  This seems not to be supported by CDF
data~\cite{CDF2000}, although the statistics is too low to draw
definitive conclusions. Measurements at the LHC will help in resolving
the issue of quarkonium polarization.

\subsection{Heavy quarkonia studies with ATLAS: selected topics}
\label{sec:HQATLAS}

The ATLAS experiment has been designed both to maximize the discovery
potential for new physics and to enable high-accuracy
measurements. ATLAS also accommodates features which make it possible
to incorporate an ambitious B-physics programme, in particular in the
first years of running at low luminosity. Most of the foreseen studies
on heavy quarkonia will be performed in the context of the B-physics
programme. For a full review of the ATLAS detector and physics
performance, see Ref.~\cite{PhTDR}.

\subsubsection{ATLAS B-Physics Trigger Issues}
\label{sec:ATLASBtrig}

ATLAS will have a flexible and efficient multi-level trigger
system. The ATLAS trigger will consist of three levels, reducing the
trigger rates from 40~MHz to ${\cal O}$(20)~kHz at level 1, to ${\cal
O}$(1--5)~kHz at level 2, and to ${\cal O}$(200)~Hz at level 3 (`Event
Filter', EF).

The classical B-physics trigger scenario~\cite{HLTTP} foresees to
trigger for a muon with $p_{\mathrm{T}}>6$~GeV and pseudo-rapidity
$|\eta|<2.4$ at level 1; to confirm this muon with better resolution
and efficiency at level 2, together with performing a full scan of the
Inner Detector (ID) to search for interesting signatures; and to
refine the search at the EF level, where offline algorithms will be
used and calibration and alignment data will be available.
\longpage

In view of tight funding constraints, changes in detector geometry,
the possible usage of a reduced detector at start-up, and a changed
luminosity target at start-up ($1\to 2\times\llumi$), the classical
scenario had to be revised; see Ref.~\cite{HLTTDR} for details. In
addition to more flexibility with respect to varying luminosity
conditions, the revised scenario foresees additional trigger objects
at level 1 (\eg muon, `Regions-of-Interest'/RoIs from calorimeter jet
or electromagnetic triggers), and the RoI-guided search for tracks in
the ID, in order to avoid the resource-intensive ID full scan. Studies
are still ongoing, but first results look promising.

In the context of heavy quarkonia studies, the di-muon trigger will be
the most important one. \Figure~\ref{fig:dimuon_rates} shows the
expected rates at a luminosity of ${\cal L} =
\llumi$~\cite{HLTTDR}. The di-muon trigger will allow for an effective
selection of channels with $J/\psi(\mu^{+}\mu^{-})$, rare decays like
$\rm B\to\mu^{+}\mu^{-}(X)$, \etc\ Minimum possible thresholds in the
level 1 muon trigger are $p_{\mathrm{T}}>5$~GeV (barrel) and
$p_{\mathrm{T}}>3$~GeV (end-cap), but the actual thresholds will be
determined by the (yet incompletely known) level 1 muon rate. At
higher trigger levels (level 2, EF), the muons from level 1 will be
confirmed using the ID and Muon Detector precision
chambers. Preliminary studies yield modest di-muon trigger rates of
$\sim$200~Hz after level 2, and of $\sim$10~Hz after the EF, for the
initial luminosity scenario of ${\cal L} = 2\times\llumi$.

\begin{figure}[t] 
 \begin{center}
  \includegraphics[width=85mm]{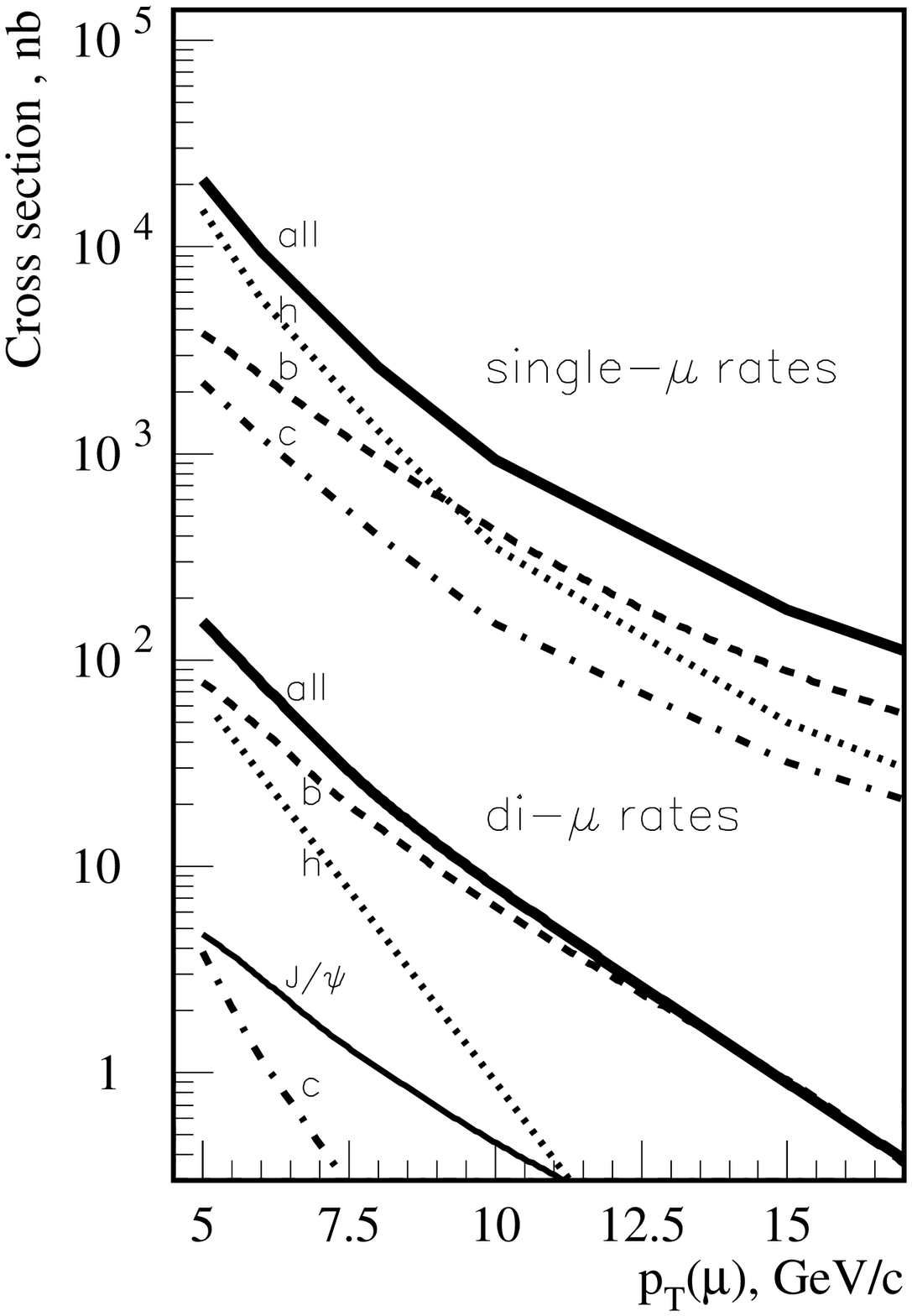}
 \end{center}
\caption[Single-muon and di-muon cross-sections]
        {Single-muon and di-muon cross-sections. Curves are shown for
         muons from $K$ and $\pi$ in-flight decays (labelled `h'), $b$
         and $c$ decays, and for the sum of all sources (`all'). Muons
         are considered within $|\eta|<2.5$. For di-muons, the
         horizontal axis shows the $p_{\mathrm{T}}$ of the
         lower-$p_{\mathrm{T}}$ muon. At least one muon must have
         $p_{\mathrm{T}}>6$~GeV and $|\eta|<2.4$ according to level 1
         trigger conditions~\cite{HLTTDR}.}
\label{fig:dimuon_rates}
\end{figure} 
    
\subsubsection{Recent ATLAS Studies on \boldmath{$J/\psi$}}
\label{sec:ATLASJpsi}

The main emphasis in ongoing ATLAS physics-related studies lies on
technical issues like validation and optimization of the architecture
of trigger and offline software, performance, \etc, not on doing
full-fledged, detailed physics analyses. The results presented here
are taken from a study on measuring the direct $J/\psi$ production
cross-section, carried out in a wider context of studies on the
performance of a staged (\ie incomplete) detector in an initial
commissioning period of 1~\ifb\ (corresponding to one year at 5\% of
the planned start-up luminosity).
\longpage

The determination of the direct $J/\psi$ production cross-section will
be one of the first B-physics measurements in ATLAS. There will be a
large $J/\psi$ rate after the level 1 trigger, whose direct $J/\psi$
contribution will not be known.  The measurement of the direct
$J/\psi$ production cross-section is, among other things, important in
order to find the best strategy for selecting $b$-events, \eg  to
optimize the interplay between $p_{\mathrm{T}}$ and vertexing cuts.

Events of the type $pp\to J/\psi(\mu^{+}\mu^{-}) + X$ were generated
with a version of PYTHIA which includes colour-octet
processes~\cite{MASL2000,MASL1999,MASL1997}.  One of the muons coming
from $J/\psi$ was required to have $p_{\mathrm{T}}>6$~GeV, the second
to have $p_{\mathrm{T}}>3$~GeV. For this purpose, functionality to
enable filtering at generation time~\cite{Natalia_note} was
implemented into PYTHIA.  Taking muons with $p_{\mathrm{T}}$ as low as
3~GeV is only possible when information from the hadronic calorimeter
(Tile Calorimeter) is additionally taken into account, to allow for
muon/hadron separation~\cite{PhTDR}.

As a result, one obtains a cross-section for direct $J/\psi$
production of about 5~nb~\cite{Natalia_note}. Typical values for
relevant resolutions are primary vertex resolution $\sigma_{PV} <
15$~$\mu$m (given by the LHC beam cross-section); secondary vertex
resolution $\sigma_{xy}(\mathrm{core})\simeq 70$~$\mu$m and
$\sigma_{xy}(\mathrm{tail})\simeq 150$~$\mu$m; mass resolution
$\sigma_{J/\psi}\simeq 40$~MeV. Preliminary studies suggest that,
based on those performance parameters, a good separation of direct
$J/\psi$'s and $J/\psi$'s from B-decays will be feasible. For a
qualitative illustration, see \Figure~\ref{fig:JpsiSeparation}.

\begin{figure}[t] 
\begin{center}
  \includegraphics[width=100mm]{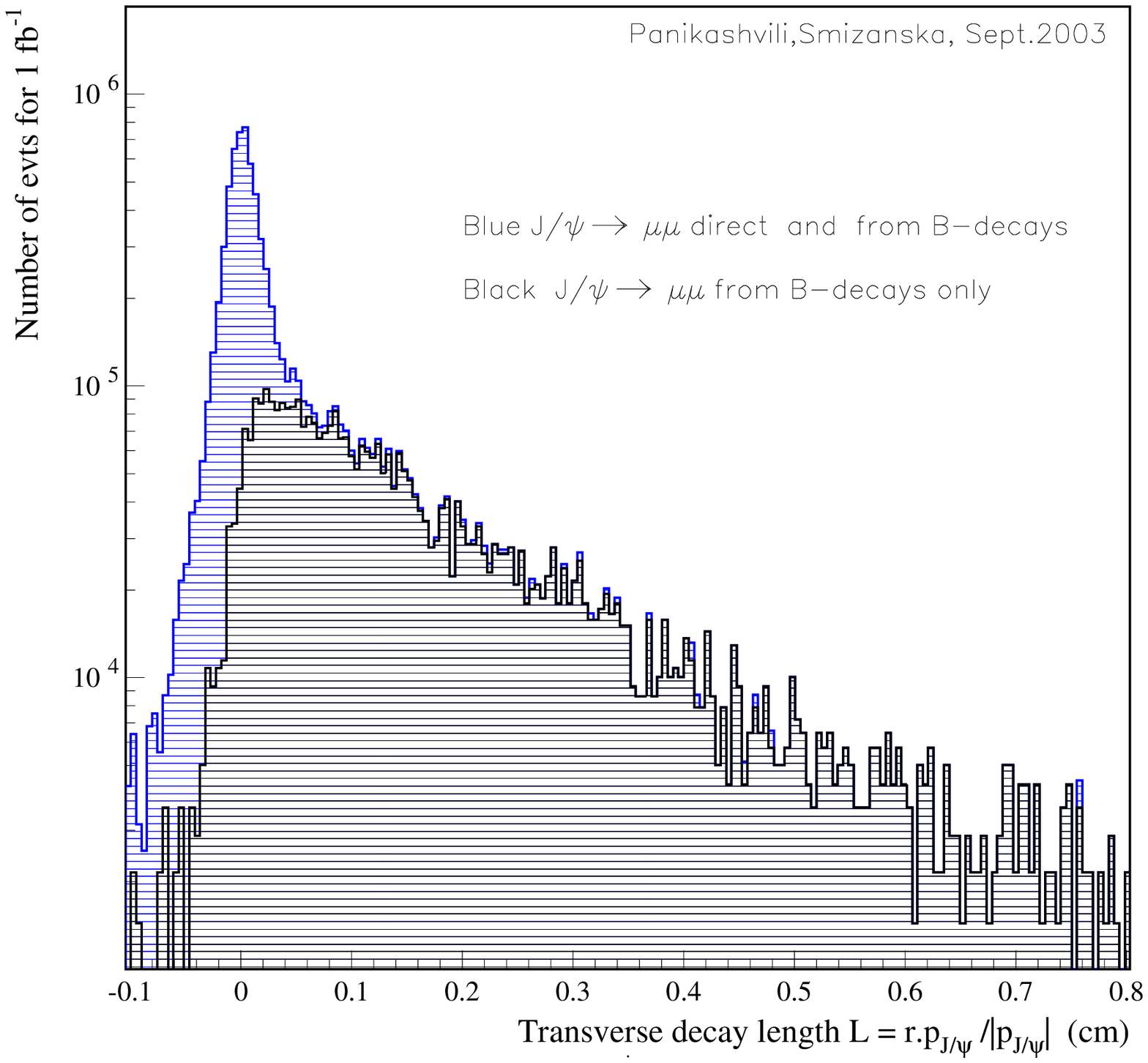}
\end{center}
\caption[Expected ATLAS potential for separating direct
         $J/\psi$'s and $J/\psi$'s from B-decays]
        {The expected ATLAS potential for separating direct $J/\psi$'s
         and $J/\psi$'s from B-decays~\cite{Natalia_note}. As
         discriminating variable, $L =
         \vec{r}\cdot\vec{p}_{J/\psi}/|\vec{p}_{J/\psi}|$ is used,
         where $\vec{r}$ and $\vec{p}_{J/\psi}$ denote the vectors of
         transverse $J/\psi$ decay radius and transverse $J/\psi$
         momentum, respectively.}
\label{fig:JpsiSeparation}
\end{figure} 

\subsubsection{Recent ATLAS Studies on $B_{c}$}
\label{sec:ATLASBc}

The expected large production rates at the LHC will also allow for
precision measurements of $\rm B_{c}$ properties. Assuming a branching
ratio $f(b\to {\rm B_{c}})\simeq 10^{-3}$, an integrated luminosity of
20~\ifb\ (\ie one year at initial luminosity), and requiring a level
1 muon with $p_{\mathrm{T}}>6$~GeV and $|\eta|<2.4$, recent estimates
show that ATLAS will be able to record about 5600 events of the type
$\rm B_{c}\to J/\psi\,\pi$, and about 100 events of the type $\rm
B_{c}\to \rm B_{s}\,\pi$.

The channels studied so far in ATLAS are $\rm B_{c}\to J/\psi\,\pi$
for $\rm B_{c}$ mass measurement, and $\rm B_{c}\to J/\psi\,\mu\nu$,
since it provides a clean signature and can be used as an ingredient
for determination of the CKM matrix element
$|V_{cb}|$~\cite{PhTDR}. Examples of older studies can be found in
Ref.~\cite{Bc_1995}.

Since the production of $\rm B_{c}$ is suppressed by the hard
production of an additional $c\bar{c}$ pair, Monte Carlo generation of
$\rm B_{c}$ events using standard tools (\eg PYTHIA) is CPU
intensive. As an example, out of 100~000 PYTHIA pp events, one obtains
about one $\rm B_{c}$ event, which in turn does not necessarily
survive the ATLAS level 1 trigger selection. Recent developments in
ATLAS have therefore concentrated on implementing dedicated $\rm
B_{c}$ generators into PYTHIA. One approach is via the so-called
`Fragmentation Approximation Model', the other the so-called `Full
Matrix Element' (FME) approach~\cite{Bc_2003}.

The FME approach is based on the concept of extended helicity,
\ie the grouping of Feynman diagrams into gauge-invariant sub-groups
to simplify the calculations, an approach never before followed in
$gg\to QQ$ processes. It takes into account matrix elements from PQCD
up to ${\cal O}(\alpha_{s}^{4})$ (36 diagrams). Results obtained with
the FME generator (subroutine BCVEGPY) in PYTHIA are shown in
\Figures~\ref{fig:BCVEGPY_1} (total cross-sections of $\rm B_{c}$ and
$\rm B_{c}^{\,\ast}$ productions) and \ref{fig:BCVEGPY_2} ($\rm B_{c}$
pseudo-rapidity distribution). In terms of CPU performance, BCVEGPY is
six times faster than other Monte Carlo generators available.

\begin{figure}[t]
\begin{minipage}[t]{.48\linewidth}
\raisebox{3mm}{\includegraphics[width=\linewidth]{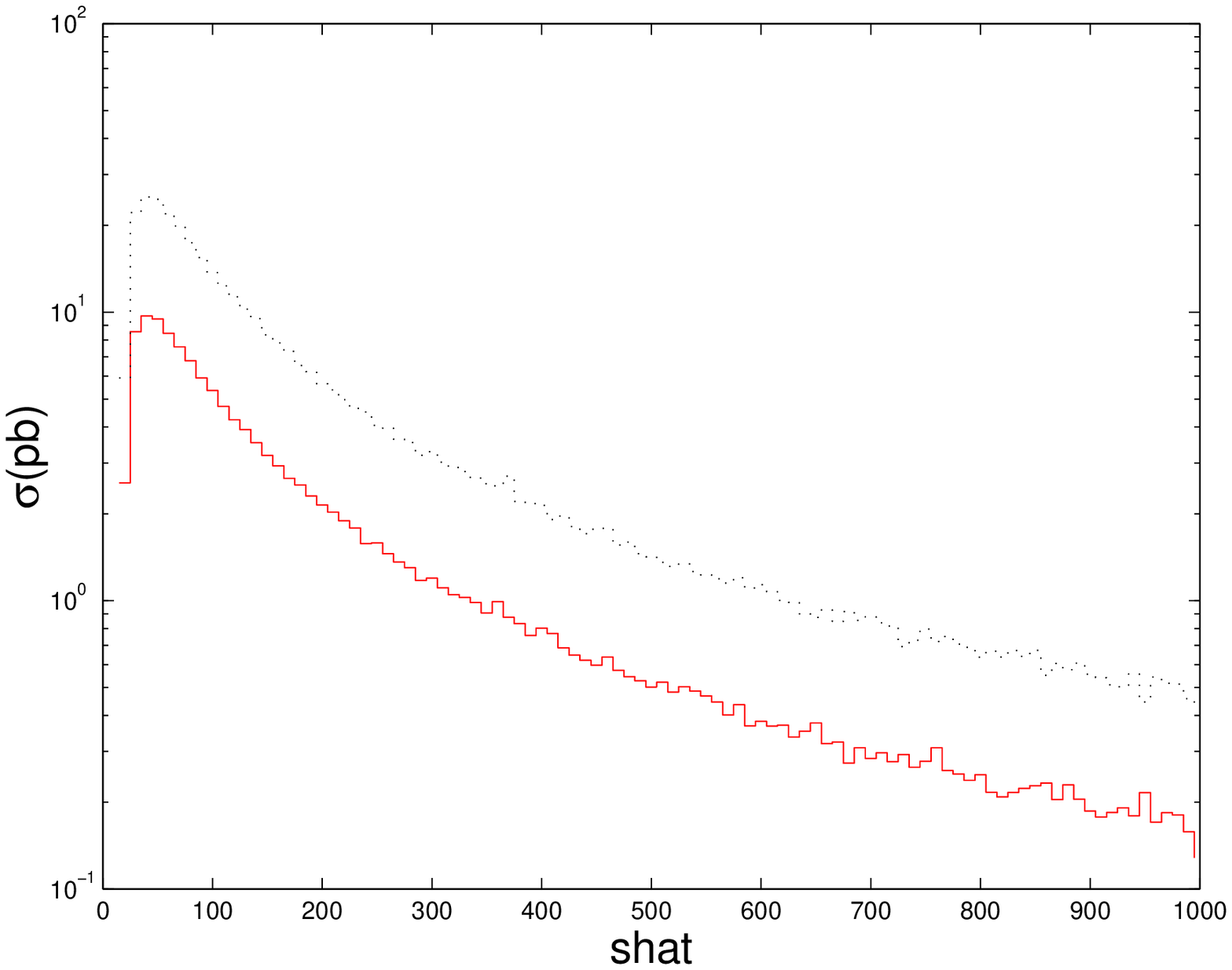}}
\caption[Integrated cross-sections for the sub-process
         $gg\to {\rm B_{c}}(\rm B_{c}^{\,\ast})+b+\bar{c}$]
        {Integrated cross-sections for the sub-process $gg\to {\rm
         B_{c}}(\rm B_{c}^{\,\ast})+b+\bar{c}$ as obtained with the
         BCVEGPY Monte Carlo generator~\cite{Bc_2003}. The solid
         (dotted) line corresponds to $\rm B_{c}$ ($\rm
         B_{c}^{\,\ast}$) production. `shat' denotes $\sqrt{s}$, the
         centre-of-mass energy of the
         sub-process.}
\label{fig:BCVEGPY_1}
\end{minipage}
\hfill
\begin{minipage}[t]{.48\linewidth}
\includegraphics[width=\linewidth]{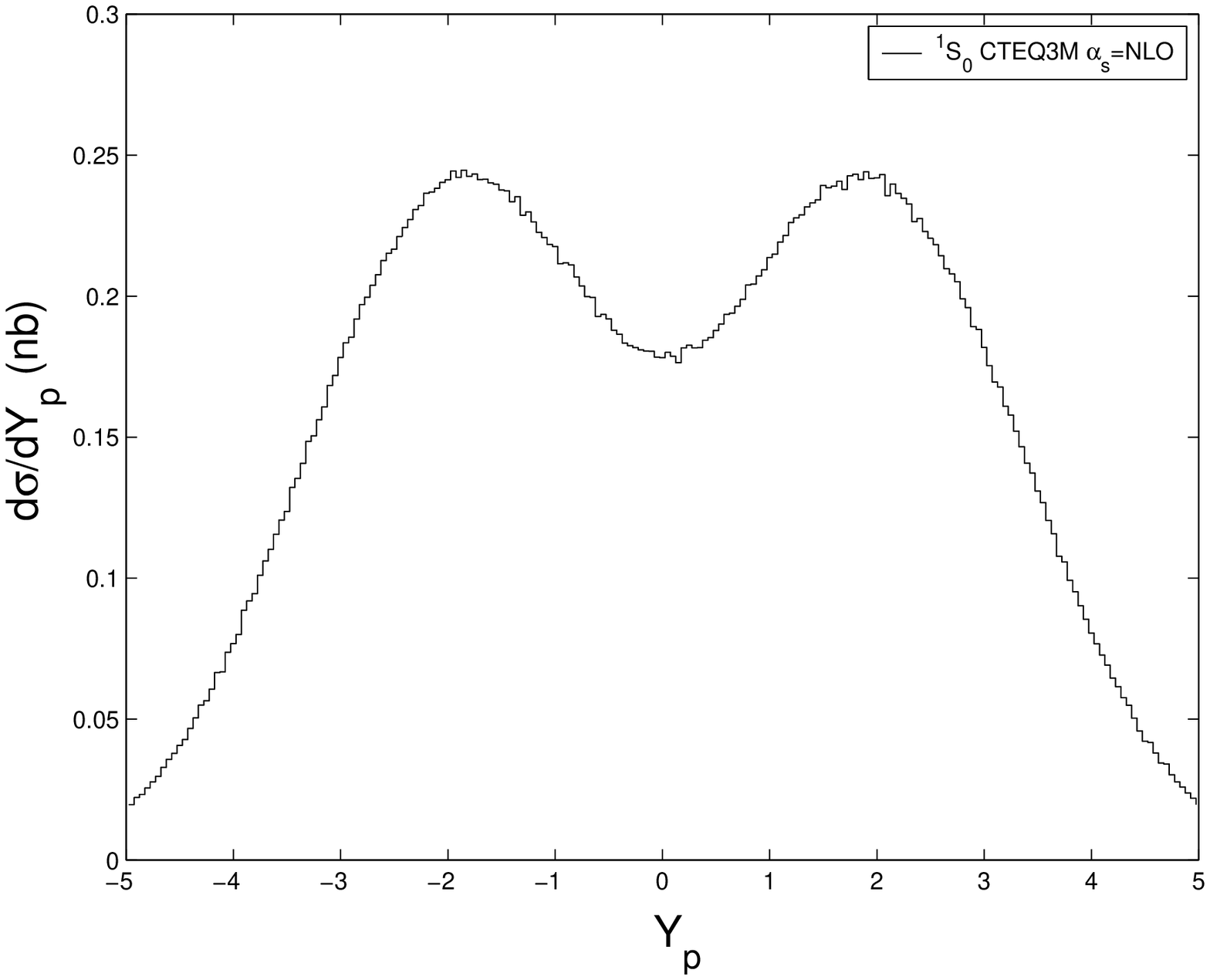}
\caption[The $\rm B_{c}$ pseudo-rapidity distribution]
        {The $\rm B_{c}$ pseudo-rapidity distribution obtained with
         the parton distribution function CTEQ3M. For details see
         Ref.~\cite{Bc_2003}.}
\label{fig:BCVEGPY_2}
\end{minipage}
\end{figure}

Studies on $\rm B_{c}$ physics performance in ATLAS are ongoing, where
events generated by BCVEGPY /PYTHIA are passed through a full GEANT3
detector simulation and are subsequently reconstructed. As a
preliminary result, a mass resolution of $\sigma_{B_{c}}=74$~MeV was
obtained in the channel $\rm B_{c}\to J/\psi\,\pi$ for a
staged-detector scenario.

\subsection{Conclusions and outlook}
\label{sec:HQLHCConcl}

The LHC will be a heavy quarkonia factory, producing them in high
statistics up to large transverse momenta. Measurements of heavy
quarkonia production rates at the LHC will be valuable for a deepened
understanding of the production mechanisms involved and the
applicability of the present theoretical approaches.

The multi-purpose experiments ATLAS (and CMS) at the LHC have the
potential to play important roles in exploring the properties of heavy
quarkonia. A deeper knowledge of heavy quarkonia properties is not
only interesting as such, but also of vital interest for other physics
fields, \eg to understand the backgrounds occurring there.

ATLAS is planning to continue the studies on heavy quarkonia in the
future, including topics not covered until now like cross-section
measurements for $\psi(2S)$ or for the process $\chi_{c}\to
J/\psi\,\gamma$, polarization measurements, and studies on $\Upsilon$
states.

\section{LHCb}

\subsection{The LHCb detector}

The LHCb detector is designed to exploit the large number of b-hadrons
produced at the LHC in order to make precision studies of CP
asymmetries and of rare decays in the B-meson systems. It has a
high-performance trigger which is robust and optimized to collect B
mesons efficiently, based on particles with large transverse momentum
and displaced decay vertices.

The detector can reconstruct a B-decay vertex with very good
resolution and provide excellent particle identification for charged
particles. Excellent vertex resolution is essential for studying the
rapidly oscillating $\rm B_s$ mesons and in particular their CP
asymmetries. It also helps to reduce combinatoric background when
reconstructing rare decays.

The LHCb experiment plans to operate with an average luminosity of
$2\times 10^{32}$~$\rm cm^{-2}$~$\rm s^{-1}$, which should be obtained
from the beginning of the LHC operation.  Running at this luminosity
has further advantages. The detector occupancy remains low, and
radiation damage is reduced.  Events are dominated by single pp
interactions that are easy to analyse. The luminosity at the LHCb
interaction point can be kept at its nominal value while the
luminosities at the other interaction points are being progressively
increased to their design values. This will allow the experiment to
collect data for many years under constant conditions.  About
$10^{12}$ $\rm b \overline{b}$ pairs are expected to be produced in
one year of data taking.

In addition to investigating CP violation in B-meson decays, the
physics programme of the LHCb experiment will include studies of rare
B and $\tau$ decays, D--$\mathrm{\overline{D}}$ oscillations and 
$\mathrm{B_c}$-meson decays.

\Figure~\ref{fig:LHCb-det} shows the layout of the LHCb detector. It
consists of the beam pipe, VELO (VErtex LOcator), dipole magnet,
tracking system, two Ring Imaging Cherenkov detectors with three
radiators (RICH1 and RICH2), calorimeter system and muon system.

\begin{figure}[t]
\centerline{\epsfig{file=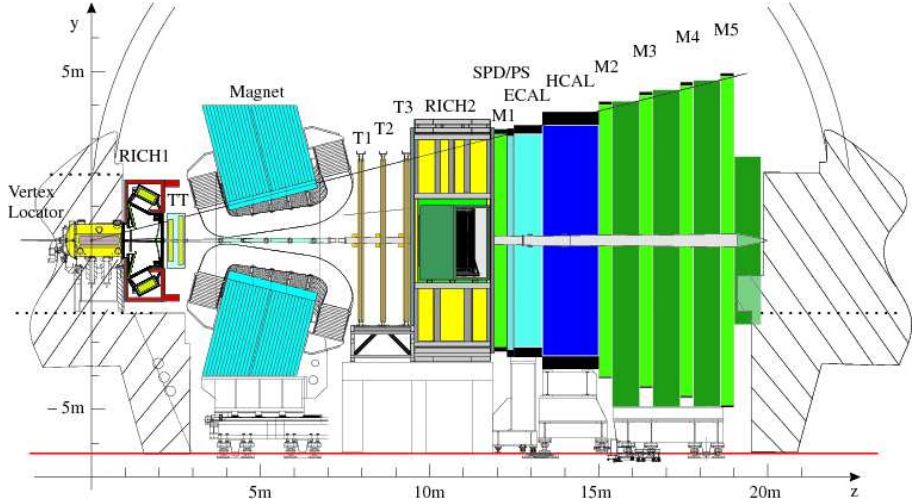,width=12cm}}
\caption{LHCb detector layout}
\label{fig:LHCb-det}
\end{figure}
 
The trigger system is designed to suppress the initial rate of about
40~MHz to approximately 200~Hz by selecting the events with
high-$p_{\mathrm{T}}$ hadrons, leptons and photons, requiring
secondary vertices, and performing partial event online reconstruction
to select the desired b-hadron decays. The trigger system is designed
to be flexible, robust and efficient. Events are selected by various
criteria that can be easily adjusted according to the experimental
conditions.  For a full review of the LHCb detector and physics
performance, see Refs.~\cite{LHCB1,LHCB2}.

\subsection{Recent LHCb studies on $\rm B_c$}

The $\rm B_c$ meson is the ground state of the $\bar{b}c$ system which
in many respects is an intermediate between charmonium and bottomonium
systems. However, since the $\rm B_c$ mesons carry flavour, they
provide a window for studying heavy-quark dynamics very different from
that of by $c\bar{c}$- and $b\bar{b}$-quarkonia.

The $\bar{b}c$ system exhibits a rich spectroscopy of orbital and
angular-momentum excitations, radiative and weak decays. In addition,
the $\rm B_c$ also provides a good place for extracting the
Cabibbo--Kobayashi--Maskawa (CKM) matrix elements $V_{cb}$ and
$V_{ub}$. Two-body hadronic decays of $\rm B_c$ mesons can play an
important role for the exploration of CP violation. For a recent
review of the $\rm B_c$ physics issues, see Ref.~\cite{LHCB4}.

The production of $\rm B_c$ states at high $p_{\mathrm{T}}$ is well
described by b-quark fragmentation, while the complete $O(\alpha_s^4)$
calculations show the dominance of the recombination mechanism in the
low-$p_{\mathrm{T}}$ region (see \Figure~\ref{fig:bcpt} where the
$p_{\mathrm{T}}$ dependence of $\rm B_c$ and $\rm B^*_c$ production is
shown).

The low-$p_{\mathrm{T}}$ region is dominated by high-momentum $\rm
B_c$ (see \Figure~\ref{fig:bcang}) that gives additional advantages
for the LHCb detector where $\rm B_c$ meson decays will produce
secondary vertices well separated from the primary one. This fact
ensures that $\rm B_c$ decays will satisfy the LHCb trigger conditions
and provides the possibility of strong background suppression.

\begin{figure}[t]
\begin{minipage}[t]{.48\linewidth}
\includegraphics[width=\linewidth]{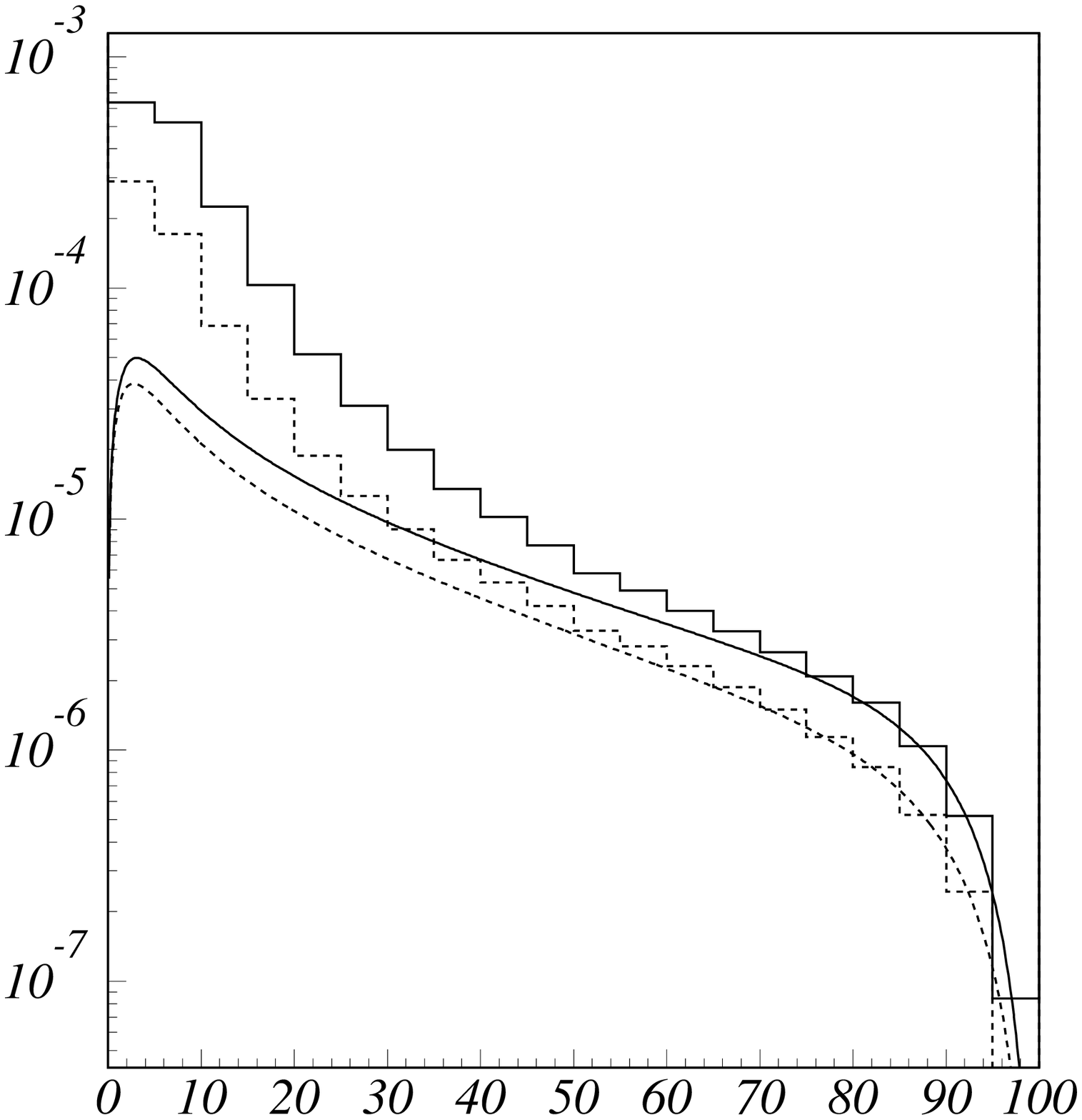}
\caption[$O(\alpha_s^4)$ calculations and fragmentation model 
         for vector and pseudo-scalar $\rm B_c$ mesons production]
        {$O(\alpha_s^4)$ calculations (histograms) and fragmentation
         model (curves) for vector (solid) and pseudo-scalar (dashed)
         $\rm B_c$ mesons production at $\hat{s} = 200$~GeV. The
         horizontal axis denotes the $\rm B_c$ transverse
         momentum.}
\label{fig:bcpt}
\end{minipage}
\hfill
\begin{minipage}[t]{.48\linewidth}
\includegraphics[width=\linewidth]{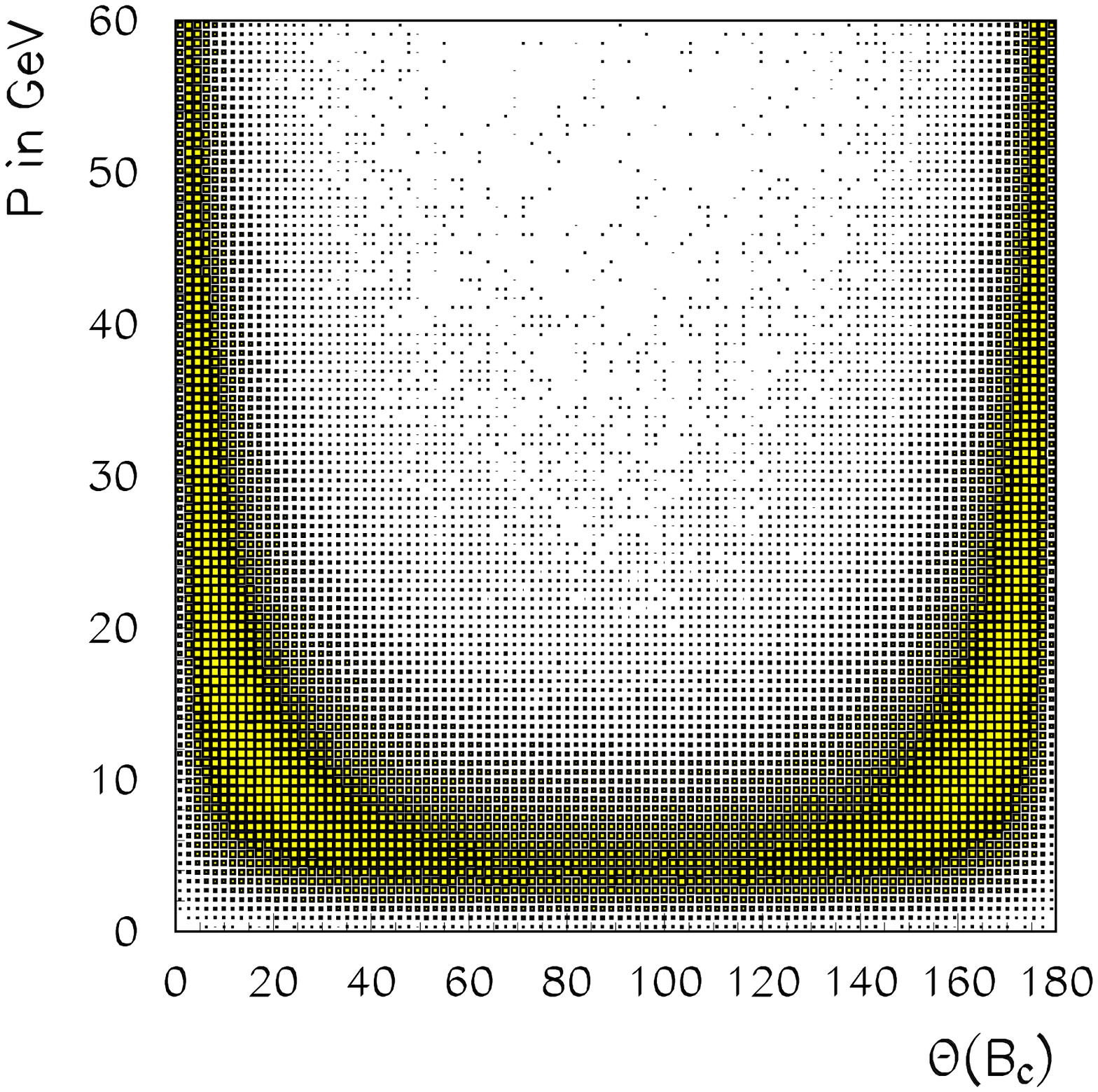}
\caption[The momentum $P(\rm B_c)$ vs. the angle between the 
         momentum of the $\rm B_c$ and the $z$-axis]
        {The momentum $P(\rm B_c)$ vs. the angle between the momentum
         of the $\rm B_c$ and the $z$-axis in the laboratory system at
         LHC energies.}
\label{fig:bcang}
\end{minipage}
\end{figure}

The exclusive decay channel $\rm B_c^{\pm} \to J/\psi \pi^{\pm},\;\;
J/\psi \to \mu^+\mu^-$ has been studied in Ref.~\cite{LHCB3}.  $\rm
B_c$ mesons with $m=6.4$~GeV and $\tau=0.47$~ps were used for the
signal Monte Carlo.  The main background comes from the prompt
$J/\psi$ production ($\sim 0.8$ mb) and $\rm B \to J/\psi X$ decays.
The complete GEANT simulation of the signal and background events
(including minimum-bias background) has been performed with the
detailed description of the detector response.  Trigger algorithms
have been applied, $\rm B_c \to J/\psi \pi$ candidates have been
reconstructed using full pattern recognition, and specific offline
cuts have been applied to reject background.

The mass resolution is shown in \Figure~\ref{fig:bcmass} with the
shaded area representing the surviving background. A clean and narrow
signal is observable, with an expectation of about 14~k $\rm B_c \to
J/\psi \pi$ decays reconstructed per year with a $B/S$ ratio estimated
to be $< 0.8$.

\begin{figure}[t]
\begin{minipage}[t]{.48\linewidth}
\includegraphics[width=\linewidth]{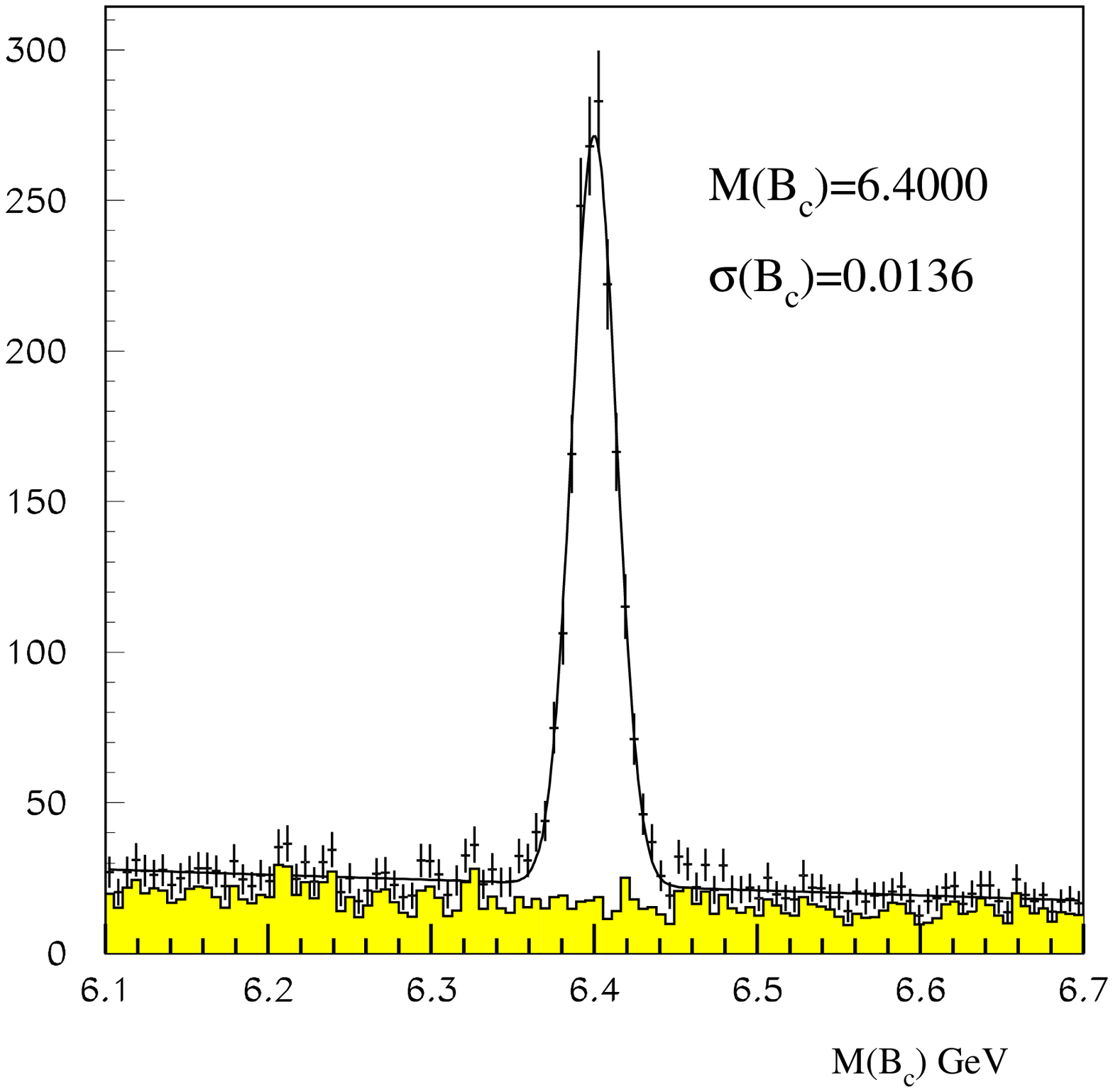}
\caption[The reconstructed $\rm B_c$ mass with background]
        {The reconstructed $\rm B_c$ mass with background.
         The curve is a Gaussian plus polynomial fit.}
\label{fig:bcmass}
\end{minipage}
\hfill
\begin{minipage}[t]{.48\linewidth}
\includegraphics[width=\linewidth]{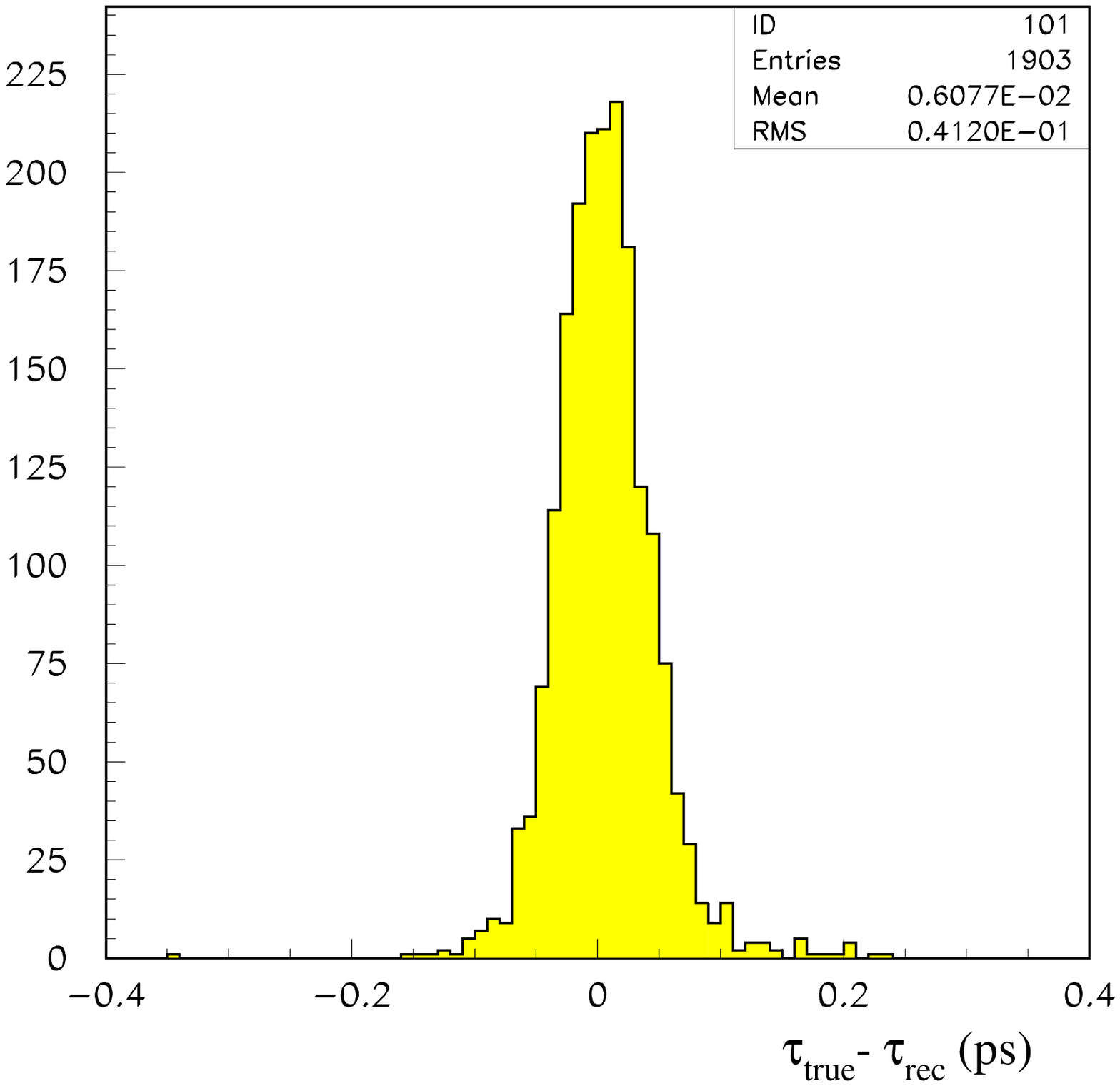}
\caption[Comparison of the reconstructed and generated $\rm B_c$ lifetime]
        {Comparison of the reconstructed and generated $\rm B_c$ lifetime}
\label{fig:bctof}
\end{minipage}
\end{figure}

The reconstructed events were used to determine the $\rm B_c$ lifetime
as well.  The difference between true and reconstructed lifetime is
shown in \Figure~\ref{fig:bctof}.  The proper time resolution is about
0.04 ps and can be improved using high-momentum $\rm B_c$ mesons.

LHCb is planning to study $\rm B_c$ mesons production including
radiative $\rm B_c^*$ decays as well as other ground-state decay
channels.

\section{ALICE}

ALICE is the dedicated heavy-ion experiment at the LHC.  The apparatus
will investigate the properties of strongly interacting matter at
extreme energy density where the formation of quark--gluon plasma (QGP)
is expected~\cite{ALICEPPR}.  For this purpose, heavy quarkonium
states are especially relevant since they provide, via their leptonic
decays, an essential probe of the medium produced in the collision. A
lot of effort has been devoted to the subject (for reviews see
Refs.~\cite{Satz:2000bn,Vogt:cu}) since the early predictions of
charmonium suppression by Debye screening in a deconfined
medium~\cite{Matsui:1986dk}.  The LHC energy is unique for such
studies since it allows, for the first time, the spectroscopy of
charmonium and bottomonium resonances in heavy-ion collisions.  In
particular, because the $\Upsilon$(1$S$) is expected to dissolve only
significantly above the critical
temperature~\cite{Digal:2001ue,Wong:2001uu}, the spectroscopy of the
$\Upsilon$ family at LHC energies should reveal unique information on
the characteristics of the QGP~\cite{Gunion:1996qc}.  On the other
hand, the study of heavy-quark resonances in heavy-ion collisions at
the LHC is subject to significant differences with respect to lower
energies.  First, in addition to prompt charmonia produced directly
via hard scattering, secondary charmonia can be produced from bottom
decay~\cite{Eidelman:pdg2004}, $\rm D\bar{D}$
annihilation~\cite{Braun-Munzinger:2000dv}, and by coalescence
mechanisms which could result in enhancement rather than
suppression~\cite{thews}.  Then, in the environment of a heavy-ion
reaction, in-medium effects such as shadowing and heavy-quark energy
loss may substantially modify the final yields and
spectra~\cite{thews}.  Open charm and open beauty production is
another important issue addressed by ALICE since they provide the most
natural normalization to quarkonia suppression/enhancement.  On the
other hand, in a QGP, an additional source of charm quarks may arise
from secondary parton scattering~\cite{muller, geiger} and information
on the properties of the deconfined medium could be revealed by the
kinematical properties of heavy mesons.

ALICE will allow measurements of heavy flavours in both the muon and
the electron channels as well as full reconstruction of $\rm D$ mesons
in the hadronic channel.  Although the apparatus is dedicated to
heavy-ion collisions, proton--proton and proton--nucleus collisions
are also an important part of the ALICE physics programme in order to
unravel initial and final-state medium effects.

\subsection{ALICE detector}

The ALICE detector~\cite{ALICEWEB} is designed to cope with large
particle multiplicities which, in central Pb--Pb collisions, are
expected to be between 2000 and 8000 per unit rapidity at mid
rapidity.  It consists of a central part, a forward muon spectrometer,
and forward/backward small acceptance detectors.  The central part of
ALICE consists of four layers of detectors placed in the solenoidal
field ($B < 0.5~{\rm T}$) provided by the LEP L3 magnet.  From the
inner side to the outer side, these detectors are the Inner Tracker
System (ITS), the large Time Projection Chamber (TPC), the Transition
Radiation Detector (TRD) and the Time of Flight system (TOF).  They
provide charged-particle reconstruction and identification in the
pseudorapidity range $|\eta|<0.9$, with full azimuthal coverage and a
broad $p_{\mathrm{T}}$ acceptance.  These large-area devices are
complemented by smaller acceptance detectors: the High-Momentum
Particle Identification (HMPID), the PHOton Spectrometer (PHOS) and
the Photon Multiplicity Detector (PMD).  In the forward/backward
region, the charged multiplicity and the zero degree energy will be
measured by additional detectors (T0, V0, FMD, ZDC) which will allow
fast characterization and selection of the events.  Finally, a forward
muon spectrometer covering the pseudo rapidity range $2.5 < \eta <4$
is placed on the right side of the central part.  It consists of a
front absorber, a dipole magnet, ten high-granularity tracking
chambers, a muon filter, and four large-area trigger chambers.

\subsection{Muons}

The goal of the forward muon spectrometer~\cite{MUONTDR} is to measure
the full set of onium resonances from the $\phi$ to the $\Upsilon$,
with high statistics, a low background, and a high resolution.  The
spectrometer is equipped with a dimuon trigger based on the selection
of pairs of muons with large transverse momentum.  An important
specification of the spectrometer is its mass resolution which has to
be about $100~{\rm MeV}$ at $10~{\rm GeV}$ to allow the separation of
the $\Upsilon$ substates.  Detailed simulations have shown that this
goal should be reached, even in the worst scenario of background
environment that could be foreseen.  The acceptance for $J/\psi$ and
$\Upsilon$ is fairly uniform in $p_{\mathrm{T}}$ and allows both
$J/\psi$ and $\Upsilon$ to be detected down to $p_{\mathrm{T}}=0$.
The statistics expected in a $10^6$~s run, roughly corresponding to
one month of data taking, is of $\sim 500 \times 10^3$ $J/\psi$ and
$\sim 10 000$ $\Upsilon$ in minimum-bias Pb--Pb collisions.  For
$J/\psi$, the rate and signal-over-background (S/B) are very good and
permit a high-precision measurement of the differential cross-section.
The $\psi(2S)$ can be measured at best with an accuracy of the order
of $10\%$ because of less favourable S/B.  For the $\Upsilon$ family,
the S/B is larger than unity and the significance is 70, 31 and 22 for
$\Upsilon(1S)$, $\Upsilon(2S)$ and $\Upsilon(3S)$.  In addition to
quarkonia measurements, the spectrometer will allow measurements of
the differential open bottom cross-section.  This will be achieved
both in the single muon and the dimuon channels.

\subsection{Electrons}

The measurement of dielectrons in the central barrel of ALICE is
complementary to the dimuon channel.  It extends quarkonia
measurements from the forward rapidity region to mid-rapidity and
allows one to measure secondary $J/\psi$ from bottom decay thanks to
the vertex capabilities of the ITS.  This will permit the distinction
between primary and secondary $J/\psi$ and also lead to a direct
measurement of the B-meson production cross-section.  Furthermore,
single high-$p_{\mathrm{T}}$ electrons with displaced vertex give
access to the inclusive open charm and open bottom cross-sections.
The centrepiece for dielectron physics is the TRD which provides an
electron trigger and identification~\cite{TRDTDR}.  Its acceptance is
identical to that of the ITS/TPC ($|\eta| < 0.9$ with full azimuthal
coverage).  Its expected pion rejection factor in a high-multiplicity
environment has been investigated by means of detailed simulations.
These simulations were adjusted to test beam data for well isolated
tracks and then performed for various track multiplicities.  Going
from well isolated tracks to a full multiplicity event (8000 charged
particles per unit of rapidity at mid-rapidity), a worsening of the
pion rejection by a factor 6--7 is observed.  For an electron
efficiency of $90\%$ the pion rejection factor is still better than
50.  On the other hand, the track reconstruction using information
from the ITS, TPC and TRD leads to a $\Upsilon$ mass resolution good
enough for the separation of the $\Upsilon$ substates.

\subsection{Hadrons}

In the central part of ALICE, heavy mesons can be fully reconstructed
from their charged-particle decay products in the ITS, TPC and TOF.
Thus, not only their integrated yields, but also their
$p_{\mathrm{T}}$ distributions can be measured.  The most promising
decay channel for open charm detection is the $\rm D^0 \rightarrow
K^-\pi^+$ decay (and its charge conjugate) which has a branching ratio
of about 3.8$\%$ and $c\tau=124~\mu{\rm m}$.  The expected rates (per
unit of rapidity at mid rapidity) for $\rm D^0$ (and $\rm \bar{D}^0$)
mesons, decaying in a $K^\mp\pi^\pm$ pair, in central ($5\%$) Pb--Pb
at $\sqrt{s}=5.5~{\rm TeV}$ and in pp collisions at $\sqrt{s}=14~{\rm
TeV}$ are $5.3\times 10^{-1}$ and $7.5\times 10^{-4}$ per event,
respectively.  The selection of this decay channel allows the direct
identification of the $\rm D^0$ particles by computing the invariant
mass of fully-reconstructed topologies originating from displaced
secondary vertices.  The expected statistics are $\simeq 13~000$
reconstructed $\rm D^0$ in $10^7$ central Pb--Pb events and $\simeq
20~000$ in $10^9$ pp events.  The significance is larger than 10 for
up to about $p_{\mathrm{T}}=10~{\rm GeV}$ both in Pb--Pb and in pp.
The cross-section can be measured down to $p_{\mathrm{T}} \simeq
1~{\rm GeV}$ in Pb--Pb collisions and down to almost
$p_{\mathrm{T}}=0$ in pp collisions.

\BLKP
%10/12/2004 
%\documentclass[11pt]{cernrep}
%\usepackage{epsfig}
%\usepackage{graphicx}
%\input{newcommand.tex}

%\begin{document}
\chapter{OUTLOOK}
\label{chapter:outlook}
\noindent
{{\it Conveners:} S.~Godfrey, M.~A.~Sanchis-Lozano}\par\noindent
{{\it Authors:} G.~Bodwin, E.~Braaten, N.~Brambilla, A.~Deandrea,
E.~Eichten, S.~Godfrey, A.~Hoang, M.~Kr\"a\-mer, R.~Mussa,
P.~Petreczki, M.~A.~Sanchis-Lozano, A.~Vairo
} 

\section{The renaissance of heavy quarkonium physics}
\label{sec:Xrenaissance}

Quarkonium physics has played a fundamental role in the development of
quantum chromodynamics. It may play an even more relevant role now for
QCD, the Standard Model, and physics beyond the Standard Model.  Heavy
quarkonium, being a multiscale system, offers a precious window into
the transition region between high-energy and low-energy QCD and
thus a way to study the behaviour of the perturbative series and the
nontrivial vacuum structure. The existence of energy levels below,
close to and above threshold, as well as the several production
mechanisms, allows one to test the population of the QCD Fock space in
different regimes and eventually to search for novel states with
nontrivial glue content (hybrids, glueballs).  Precise determinations
of Standard Model parameters from quarkonium systems have become
possible because of the level of precision reached by the experimental
data and by the most recent developments in effective field theories
and lattice QCD. Moreover, the clean signature of heavy quarkonium in
heavy-ion collisions provides a perfect probe of in-media phenomena,
and eventually of quark--gluon plasma formation and the
confinement--deconfinement transition in QCD. The expected large
statistics of $\psi$ and $\Upsilon$ resonances to be collected in the
next few years at $e^+e^-$ and hadronic colliders makes heavy
quarkonium physics also suitable for searches for new particles and
new phenomena. A number of new physics scenarios can be constrained or
discovered in the near future, ranging from the contribution of
supersymmetric particles or extended Higgs sectors in quarkonium
decay, to lepton flavour violation tests, CP tests and chromo-dipole
moments of quarks.  All these studies will play a major role in the
test of extensions of the Standard Model, and will be complementary to
direct searches at colliders like the LHC or a future linear collider.

\section{Opportunities in theory and experiment}
\label{sec:Xopportunities}

The future relevance of quarkonium physics will be proportional to the
number of observables that can be rigorously described in terms of the
Standard Model and its parameters and well measured by experiments.
The enormous progress made in this direction in recent years is the
reason for the quarkonium renaissance that we are witnessing today and
that has been documented in this report. It comes mainly from QCD
effective field theories in either their continuum or in their lattice
versions, while phenomenological models have played (and will still
play in the future) a crucial role in suggesting experimental search
strategies and interpreting new results. In order to achieve further
progress it will be important to complete the following general
programme.
\begin{itemize}
\item[(1)]{Adopt a common, model-independent, EFT-based, language to
     describe all aspects of heavy quarkonium physics. This has not
     been achieved yet for all observables, and, noteworthily, not for
     observables sensitive to threshold effects, where
     phenomenological models still provide the only available
     theoretical tool.}
\item[(2)]{Improve the determination of the nonperturbative parameters
     that describe the low-energy dynamics either by experimental data
     or by lattice calculations.  In an EFT context the number of
     these parameters is finite.  Therefore, precise quarkonium data
     are important today more than ever.  They may check
     factorization, allow for precise extractions of the Standard
     Model parameters, and severely constrain theoretical
     determinations and predictions.}
\end{itemize}
We note that the more progress there is in (1), the more importance that
experimental data will acquire for (2). 

In the following we discuss progress expected or invoked for some
specific set of observables.

\subsection{Quarkonium ground and   lower states}
\label{sec:Xquarkonium}

Ground-state observables and to a lesser extent low lying
quarkonium-state observables may be studied in the framework of
perturbative QCD.  These studies are relevant because they may allow,
in principle, the precise extraction of some of the fundamental
parameters of the Standard Model, such as the heavy quark masses and
the strong coupling constant (see
\Chapter~\ref{chapter:precisiondeterminations}).  $B_c$ will be
copiously produced at future hadron colliders and will allow the
determination of the electroweak parameters of the Standard Model,
such as the CKM matrix elements and CP violating parameters (see
\Chapter~\ref{chapter:decay}).  
However, the accuracy with which the fundamental parameters can be
measured is at present limited by nonperturbative contributions whose
form is in many cases known, but whose size is not known with
sufficient precision.  Therefore the main theoretical challenge is the
precise determination of these nonperturbative contributions (see
\Chapters~\ref{chapter:spectroscopy} and \ref{chapter:decay}).  On the
other hand we could take the opposite approach and use the lower
quarkonium states as a theoretically clean environment to study the
interplay of perturbative and nonperturbative effects in QCD and
extract nonperturbative contributions by comparison with data.  A few
examples are:

\begin{itemize}
  
\item The $\eta_b$ has been intensively searched for at the Tevatron
      and CLEO. Theoretically several observables related to the
      production mechanism (\Chapter~\ref{chapter:production}),
      spectroscopy (\Chapter~\ref{chapter:spectroscopy}), and decay
      (\Chapter~\ref{chapter:decay}) have been studied.  Most likely,
      the $\eta_b$ discovery will come from the Tevatron experiments
      CDF and D0.  NRQCD predictions suggest
      ($\sigma_{\eta_b+X}\approx 2.5~{\mu}$b at 1.96~TeV), so that
      $\eta_b$ should be found during Run II.  However, as the decay
      rates are expected to be very low, reliable theory estimates for
      the decays into $\jpsi\jpsi$, $D\bar{D}\pi$, $K\bar{K}\pi$ are
      important.  Indications can also come from the efforts made to
      detect its charmonium analogue, the $\eta_c$, in hadronic
      collisions.  An eventual discovery will put severe constraints
      on the size of the nonperturbative corrections and confirm or
      disprove our current understanding of the bottomonium ground
      state. In case this system, as expected, turns out to be mainly
      perturbative, it will provide, combined with the $\Upsilon(1S)$,
      a very precise measurement of $\alpha_{\rm s}$.
  
\item The perturbative $\Upsilon(1S)$ mass is used for a competitive
      determination of $m_b$.  However, at present accuracy,
      perturbation theory has difficulty reproducing the measured
      width $\Upsilon(1S)~ \to ~ e^+e^-$.  Given the importance of
      this quantity, the origin of these difficulties should be
      clarified.  Furthermore, the experimental determination of the
      $\Upsilon(1S)$ polarization at the Tevatron is roughly
      consistent with the NRQCD prediction, and fixed-target
      experiments find an almost transverse polarization for the
      $\Upsilon(2S)$ and $\Upsilon(3S)$ (although the experimental
      result disagrees with NRQCD for the $\Upsilon(1S)$). This
      provides a strong motivation for measuring the polarization of
      all three resonances at the Tevatron. However, because of the
      large bottom mass, the fragmentation mechanism does not dominate
      until relatively high values of $p_{\mathrm{T}}$ are reached
      ($p_{\mathrm{T}}>$ 10~GeV). LHC experiments will likely play a
      decisive role in settling this issue because of the broader
      $p_{\mathrm{T}}$ range.
  
\item The $B_c$ mass determined by experiments is affected at present
      by about 400~MeV uncertainty, while theoretical calculations
      based on perturbative QCD are affected by errors which are not
      larger than 30~MeV (see
      \Chapter~\ref{chapter:spectroscopy}). Again a precise
      determination of the $B_c$ mass will strongly constrain the size
      of the nonperturbative contributions and confirm or disprove our
      understanding of this system in terms of perturbative QCD. An
      analogous argument holds for the yet undiscovered $B_c^*$.
  
\item Baryons with two or three heavy quarks and in particular the yet
      undiscovered baryons with two bottom quarks will offer a
      completely new system to test our understanding of low-lying
      heavy quarkonium resonances (see
      \Chapter~\ref{chapter:spectroscopy}). The study of these systems
      from QCD is just beginning with lattice simulations just
      starting to analyse these systems. Further progress is expected
      in the future, in particular if driven by new experimental
      findings.
  
\item As the $\eta_c(2S)$ and $h_c$ complete the low-mass charmonium
      multiplets, the theoretical understanding of fine and hyperfine
      splittings is far from the precision reached by experiments (see
      \Chapter~\ref{chapter:spectroscopy}).  
      Further progress in unquenched lattice calculations is needed.
      The plans to produce very large ($>10^9$) samples of $\jpsi$'s
      and $\psi(2S)$'s can open the era of 1--2\% precision
      measurements on many radiative transitions, allowing access to
      the suppressed (M1, M2 and E2) amplitudes, which are mostly
      dependent on higher-order corrections and better test different
      theoretical approaches.  Runs at the $\psi(2S)$ energy will also
      provide a very large sample of tagged $\jpsi$ decays (as more
      than half of these mesons decay to $\jpsi$), but are also an
      excellent source of $\chi_c$'s, and, as recently shown, of
      $h_c$'s.
  
\item At last, B factories will allow us to reach accuracies better
      than 10\% on the $\gamma\gamma$ widths of the $\etac(1,2S)$ and
      $\chi_{c0,2}$, by the proper combination of their data with
      measurements from $\mathrm{p}\bar{\mathrm{p}}$ and $\tau$-charm
      factories.  Electromagnetic and hadronic decay widths, whose
      experimental accuracy is already sensitive to NLO corrections
      (see \Chapter~\ref{chapter:decay}), may in perspective provide a
      competitive measurement of $\als$ at charmonium energies.
  
\item In the LHC era, very large samples ($>$10$^{10}$ events) of
      $\jpsi$ and $\psi(2S)$ mesons will allow the high-precision test
      of lepton flavour violation, severely constraining new physics
      models.  Lepton flavour violation can be tested via two-body
      decay, $J/\psi \to \ell\ell'$ (which conserves total lepton
      number), with $\ell$ and $\ell'$ denoting charged leptons of
      different species.  This process (discussed in
      \Chapter~\ref{chapter:beyondstandardmodel}) could occur at
      tree-level induced by leptoquarks, sleptons (both in the $t$
      channel) or mediated by $Z'$ bosons (in the $s$ channel).

\end{itemize}

\subsection{Higher quarkonium states}
\label{sec:Xhigherquarkomium}

The rigorous study of higher quarkonium states, including exotic
states like hybrids, will mostly rely on lattice calculations.
However, at the moment, phenomenological models still play a major
role in describing states above the open flavour threshold.  In the
framework of nonrelativistic EFTs on the lattice, further progress
will need:
\begin{itemize}
  
\item[--] the calculation in lattice perturbation theory of the Wilson
     coefficient of the EFT at least at NLO (see
     \Chapters~\ref{chapter:commontheoreticaltools} and
     \ref{chapter:spectroscopy});
  
\item[--] the lattice implementation of lower-energy EFTs like pNRQCD.
     In this framework the lattice data would provide the form of the
     potentials and the states would be determined by solving the
     corresponding Schr\"odinger-like equation.

\end{itemize}

The observation of the $X(3872)$ is the start of challenging searches
for non-vector states across the open flavour threshold. This is
probably the richest experimental field of research on heavy quarkonia
at present.  As mentioned above, phenomenological models have played a
particularly important role in predicting which states are likely to
be narrow enough to be observed and suggesting the most promising
channels for their observation.

\begin{itemize}
  
\item Studies on the nature of the $X(3872)$, described in
      \Chapters~\ref{chapter:spectroscopy}, \ref{chapter:decay}, and
      \ref {chapter:production} can benefit from data taking at B
      factories, Tevatron, and even $\tau$-charm factories: these
      should have high priority, as emphasized throughout the report.
  
\item Given the excellent momentum resolution of the B factories and
      the unexpected double charm process, $\jpsi$ recoil techniques
      also have good discovery potential.  More conventional methods,
      like the study of the production of pairs of open charm mesons
      near threshold (in B decays and hadronic collisions) are now
      reaching the statistics necessary to allow the discovery of new
      resonances in the $c \bar c$ system with quantum numbers other
      than $1^{--}$.  In particular, the remaining $1D$ states, some
      of the $2P$ states and the $3 ^1S_0$ are likely to be observable
      in this way.  Observation of new states with different quantum
      numbers is also beneficial for the understanding of the
      mechanism of charmonium production.
  
\item The current CLEO-c run at $\psi(3770)$ energy, presently
      measuring $f_D$ from $\bar DD$ decays, can also look for rare
      radiative and hadronic decays to lower $c \bar c$ states. This
      study can give a unique insight into the S--D mixing and
      coupling to decay channel effects. \ It may also give clues to
      the understanding of the $\rho$--$\pi$ puzzle.

\end{itemize}

\subsection{Production}
\label{sec:Xproduction}

If measurements of quarkonium production are to be exploited fully to
test theoretical models, then the precision of the theoretical
predictions should be improved. Several theoretical tools that are, by
now, standard could be applied to increase the precision of
theoretical predictions for charmonium and bottomonium production
rates. These tools include calculations at NLO in $\alpha_{\rm s}$ and
$v$, resummation of logarithms of $m^2/p_{\mathrm{T}}^2$ or
$m^2/(p^*)^2$, resummation of logarithms of $p_{\mathrm{T}}^2/m^2$ or
$(p*)^2/m^2$, resummation of logarithms of $s/m^2$, resummation of
logarithms of $1-z$, and lattice calculations of quarkonium matrix
elements.

\begin{itemize}
  
\item Calculations of cross-sections at NLO in $\alpha_{\rm s}$
      already exist for total cross-sections and for some quarkonium
      fragmentation functions. NLO calculations of quarkonium
      differential cross-sections in the colour-singlet model also
      exist. However, full NLO calculations in the NRQCD factorization
      approach are lacking, in general, for quarkonium differential
      cross-sections and, in particular, for the important cases of
      quarkonium cross-sections that are differential in
      $p_{\mathrm{T}}$.
  
\item Some calculations of corrections of higher order in the
      heavy-quark velocity $v$ have already been carried out and have
      yielded large corrections. It is important to investigate such
      higher-order corrections for all quarkonium production processes
      and to develop a phenomenology of the higher-order NRQCD matrix
      elements. It is also important to understand the origins of
      large corrections of higher order in $v$, with the aim of
      controlling them to all orders in the $v$ expansion.
  
\item Logarithms of $m^2/p_{\mathrm{T}}^2$ are important at small
      $p_{\mathrm{T}}$. Their resummation involves the introduction of
      non-perturbative $k_{\mathrm{T}}$-dependent parton
      distributions. The effects of these distributions are small for
      bottomonium, but are important for charmonium. It may be
      possible to work out a phenomenology of such distributions by
      exploiting their universality properties to extract them from
      processes other than quarkonium production, such as Drell--Yan
      production of lepton pairs.
  
\item Logarithms of $p_{\mathrm{T}}^2/m^2$ are important at large
      $p_{\mathrm{T}}$. They may have a large effect on, for example,
      extractions of the ${}^3S_1$ colour-octet matrix elements that
      dominate $J/\psi$ and $\Upsilon$ production at large
      $p_{\mathrm{T}}$.
  
\item Logarithms of $s/m^2$ may play an important r\^ole in
      diffractive quarkonium production and quarkonium production in
      which sub-processes involve small momentum transfer. They are
      often resummed in existing calculations in the
      $k_{\mathrm{T}}$-factorization approach by making use of the
      BFKL equation. Large corrections that occur at NLO in the BFKL
      equation cast some doubt on the accuracy of such resummed
      calculations.
  
\item The quantity $1-z$ generally measures the departure of a
      quarkonium production process from a kinematic
      endpoint. Examples of $z$ are the quarkonium energy fraction and
      the quarkonium longitudinal-momentum fraction. It follows that
      logarithms of $1-z$ are important near the kinematic limits of
      cross-sections.  Their resummation involves nonperturbative
      shape functions. It would be useful to develop a phenomenology
      of these shape functions so that information from, say,
      quarkonium production in $e^+e^-$ collisions could be used to
      make predictions for other quarkonium production processes, such
      as photoproduction at HERA.
  
\item Lattice techniques can be used to compute colour-singlet NRQCD
      production matrix elements in the vacuum-saturation
      approximation, and, hence, to supplement information on the
      values of these matrix elements that can be obtained from the
      phenomenology of quarkonium decay and production. Unfortunately,
      it is not yet known, except within the vacuum-saturation
      approximation, how to formulate the problem of the calculation
      of production matrix elements in lattice field theory. In
      particular, no lattice method exists for the computation of
      colour-octet production matrix elements.
\end{itemize}

In addition to the logarithmic contributions that we have already
mentioned, large non-logarithmic contributions appear in some
calculations of production cross-sections at NLO in $\alpha_{\rm s}$.
Examples of large corrections also exist in NLO calculations of
quarkonium decay rates. It is important to understand the origins of
such large corrections and to bring them under control to all orders
in $\alpha_{\rm s}$.

A significant theoretical issue is the correctness of the NRQCD
factorization formula for production. An all-orders perturbative proof
of the factorization formula would be an important step forward. Such
a proof might establish that there is a range of validity of the
factorization formula. For example, the formula might hold at large
$p_{\mathrm{T}}$, but not at small $p_{\mathrm{T}}$ or for total
cross-sections. Most of the experimental data are at small
$p_{\mathrm{T}}$. Experiments at small $p_{\mathrm{T}}$ are important
to fix the values of certain NRQCD matrix elements and, hence, to test
matrix-element universality. However, theoretical confidence in NRQCD
factorization is highest at values of $p_{\mathrm{T}}$ that are
significantly larger than the heavy-quark mass. Therefore, it is also
important for experiments to obtain data points with the highest
possible statistics at the largest accessible values of
$p_{\mathrm{T}}$. Such high-$p_{\mathrm{T}}$ measurements are
particularly important in testing the key prediction that, owing to
the colour-octet mechanism, there should be a large transverse
polarization in spin-triplet quarkonium produced at large
$p_{\mathrm{T}}$.

The results from the Belle Collaboration on inclusive and exclusive
double $c\bar c$ production in $e^+e^-$ collisions are strongly at
odds with current theoretical calculations. These calculations are
carried out, essentially, within the colour-singlet model. However,
colour-octet corrections are absent at leading twist in the exclusive
case and are expected to be small in the inclusive case. An
independent check of the Belle Collaboration results would be welcome.
If these results are confirmed, they would pose a severe challenge to
the current theoretical thinking about double $c\bar c$ production in
$e^+e^-$ collisions. A measurement of the double $c\bar c$ production
cross-section in ${\mathrm{p}}\bar{\mathrm{p}}$ collisions at the
Tevatron might give some additional clues as to the nature of the
production mechanisms.

Polarized beams are available at RHIC and may be available at the LHC.
Polarized-beam measurements of quarkonium production rates might be
useful in discriminating between various models for quarkonium
production and between various production mechanisms within the NRQCD
factorization approach. Exploratory theoretical work on quarkonium
production cross-sections for polarized beams is needed in order to
understand the potential of such measurements.

\subsection{In media}
\label{sec:Xinmedia}

The study of quarkonia in media is relevant because one may use the
media as a filter to study the time and length scale associated with
quarkonia production and thus learn more about the production
mechanism.  Moreover, one may use quarkonia as a test of the medium to
find out its properties (\eg whether it is hadronic or deconfined).
It has been suggested that production and suppression of quarkonia in
heavy-ion collisions could signal deconfinement and quark--gluon plasma
formation and eventually estimate its temperature (see
\Chapter~\ref{chapter:charm-beauty-in-media}).  To use quarkonium to
study the quark--gluon plasma one needs to understand its properties
inside the plasma using lattice QCD and understand its formation in a
high density environment.  A list of future challenges related to
these issues may consist of the following.

\begin{itemize}
\item Several lattice QCD studies show that ground-state charmonia
      survive in the plasma at temperatures as high as
      $1.5\,T_{\mathrm{c}}$. At the same time all these studies show
      strong lattice artefacts in the charmonium spectral
      functions. Better lattice actions are needed to reduce these
      artefacts (\eg lattice NRQCD, perfect actions).
  
\item It would be important to understand what governs the quarkonium
      suppression in the plasma. In this respect it would be extremely
      interesting to study the bottomonium system and see whether or
      not the $1P$ bottomonium states ($\chi_b$), which are expected
      to have the same size as the $J/\psi$, melt at the same
      temperature.
  
\item It is expected that the medium created in heavy-ion collisions
      is not fully equilibrated. Thus it would be very important to
      determine the thermal width of different quarkonium states on
      the lattice to make contact with experiments.
  
\item In heavy-ion collisions at very high energy, the initial state
      is given by coherent gluon fields, the so-called colour glass
      condensate with typical momentum scale (saturation scale ) $Q_s
      \simeq 1$--$2$~GeV. In the presence of colour glass condensate,
      heavy-quark and quarkonium production may be quite different
      from superposition of proton--proton production at the same
      energy. A detailed proton--nucleus study is needed together with
      nucleus--nucleus experiments to study this issue.
  
\item Hydrodynamic models suggest a fast equilibrization in heavy-ion
      collisions at RHIC which will be even faster at LHC. The time
      scale of quark--gluon plasma formation $\tau \simeq 0.6$~fm is
      comparable to the time scale of the quarkonium formation.  In
      order to disentangle between possible effects in quarkonium
      yield due to in-medium production from true medium effects,
      \eg screening, a detailed analysis of quarkonium yields versus
      rapidity, energy, and $p_{\mathrm{T}}$ is required.
\end{itemize}

\subsection{Top--antitop production}
\label{sec:Xtopantitop}
\shortpage

Top--antitop quark-pair production will provide an integral part of
the top-quark physics programme at the International Linear Collider
(ILC), which after the decision for cold, superconducting technology,
is now becoming an inter-regional endeavour. A precise measurement of
the threshold cross-section can provide a top-quark mass determination
with uncertainties at the level of 100~MeV, a measurement of the total
top decay rate $\Gamma_t$, the coupling strength of top quarks to
gluons, $\alpha_{\mathrm{s}}$, and, if the Higgs boson is light, the
top Yukawa coupling.  For more details see
\Chapter~\ref{chapter:precisiondeterminations}.

On the theoretical side there are a number of problems that need to be
solved. Among them are the still missing NNLL corrections to the
top-pair production current, which affect the theoretical
uncertainties of the cross-section normalization.  Also, a complete
fixed-order prediction at NNNLO of the total cross-section would be
desirable as a cross-check of the importance of the summation of
logarithms and as a starting point to renormalization group improved
calculations beyond the NNLL level. An important conceptual issue to
be solved is a consistent treatment of the interplay of electroweak
and QCD effects. Since the top quark decay is a leading order effect
in the threshold region, this problem includes the consistent
treatment of interference effects of top quark production and decay,
of factorizable and non-factorizable corrections and also of
non-resonant final states. Closely related is also the problem of
rescattering corrections.

Top quark threshold dynamics may also play a role in other processes
at high energy. For example, the process $\gamma\gamma\to t\bar t$,
accessible through the option of laser backscattering at the ILC,
should be explored with the same accuracy as the production in
$e^+e^-$ annihilation. Likewise, top threshold dynamics also governs
associated top-pair production processes in specific kinematic
end-point regions, such as $e^+e^-\to t\bar t H$ in the limit of large
Higgs energy. The importance of the top threshold region might also be
systematically explored in the framework of hadron collisions. These
threshold effects are not important at the present level of accuracy,
but they could become relevant at the LHC where top pairs will be
produced with very high statistics.

From the beginning of Linear Collider studies more than 10 years ago,
a number of experimental studies analysing the principle feasibility
of top threshold measurements have been carried out. But only recently
were machine-specific and more subtle effects such as the influence of
uncertainties in measurements of the luminosity spectrum considered
and found to be very important. After the recent technology decision
for the ILC, such design-dependent studies need to be driven further
to fully explore the potential of threshold measurements and to derive
the optimal scan strategy. The most important missing tool, however,
is a fully implemented top threshold event generator. This endeavour
will be enormous, since such a generator will have to describe
production, propagation, and the decay of quasi-bound top quarks in
one single consistent framework.  A close collaboration between
experiment and theory will be crucial.

\section{The Superlab}
\label{sec:Xsuperlab}
\shortpage

In recent years, the field of experimental high-energy physics has
experienced a substantial reduction in the number of running
experiments, together with the growth of large, expensive facilities
used by an increasing number of physicists. In this respect,
quarkonium physics is only one of the many physics studies competing
for manpower and resources.  In any of the experiments: CLEO-c,
BES~III, PANDA, Belle and BaBar at the B factories, and CDF and D0 at
the Tevatron, the physics potential for quarkonia is many times larger
than the analyses that are likely to actually be made.  Because of
this, many physics opportunities are lost.  Heavy quarkonium studies
will probably be a high-priority topic for BES~III and possibly for
PANDA.  The currently running experiments, together with the BES~III
and LHC experiments, are likely to yield in the next five years an
increase by one to two orders of magnitude of the number of
reconstructed heavy quarkonia.

{\itshape 
Initiatives involving transversally the different experiments
are needed, to sort priorities between the different analyses to be
done on the datasets, and to provide, whenever needed, a guideline for
the best possible usage of available luminosity. By thinking of future
studies on heavy quarkonia as an ideal `Quarkonium Superlab', the
QWG aims at merging theoretical and experimental efforts in this subject
in a more effective and constructive way.}

Recently, heavy quarkonium discoveries and confirmations in different
production mechanisms have been blossoming in a concerted way at all
the accelerator facilities. The observation of the $\eta_c(2S)$ and
the $X(3872)$ are cases in point.  While the $\eta_c(2S)$ was first
observed by Belle in B decay and double charm production, it was
subsequently confirmed by CLEO and BaBar in $\gamma\gamma$ collisions.
Similarly, the $X(3872)$ was discovered by Belle in B decay and
subsequently confirmed by both CDF and D0 in high-energy
$\mathrm{p}\bar{\mathrm{p}}$ collisions at the Tevatron. Discoveries
of the same state with different production mechanisms are very
important in order to gain information on the nature of the state.
The study of quarkonia production in high-energy, heavy-ion
collisions, as a signature of the production of a deconfined state of
QCD matter, started at CERN in 1986, and gave rise to one of the most
exciting observations made in the field of experimental QCD
thermodynamics.  Here again, the comparison of heavy quarkonium
production results in $\mathrm{p}\mathrm{p}$,
${\mathrm{p}}\bar{\mathrm{p}}$, $\mathrm{p}\mathrm{A}$ and
$\mathrm{A}\mathrm{A}$ collisions is crucial to have full control of
the underlying physics mechanisms.

Future expectations on heavy quarkonium experiments may be grouped as
follows:

\begin{itemize}
  
\item In the short term (2004--2006), dedicated measurements on
      charmonium systems, mainly above the open charm threshold will
      be performed by CLEO-c, by running at $\psi(3770)$ between 4 and
      4.4~GeV. At the same time, the two asymmetric B factories are
      poised to reach a total $\sim 1000$ fb$^{-1}$ of data at the
      $\Upsilon(4S)$ resonance, yielding record samples of
      $\eta_c(1,2S)$, $X(3872)$ mesons with probably a few new
      surprises. The collider experiments CDF and D0 will likely
      accumulate about 20 times the data samples taken during Run I,
      and will hopefully shed new light on the issue of heavy
      quarkonium production and polarization.  A precise determination
      of the mass of the $B_c$ meson, done using exclusive
      $\jpsi\,\pi$ decays, will be available soon (see {\it Fermilab
      Today}, December 3, 2004).  Results from the NA60 experiment at
      CERN and the STAR and PHENIX experiments at RHIC are expected to
      provide further evidence of QGP.  With the expected significant
      increase in statistics of the upgraded HERA collider, it might
      be possible to study inelastic photoproduction of bottomonium
      states for the first time, with the caveat that production rates
      of $\Upsilon$ resonances are suppressed by more than two orders
      of magnitude compared with the $J/\psi$.
  
\item In the mid term (2007--2009), high-precision results are
      expected from the CLEO-c run at the $\jpsi$, and hopefully at
      the $\psi(2S)$, as recommended by the QWG.  In the meantime, the
      fully upgraded BES~III detector will take over the precision
      studies of charmonium started by CLEO-c.  BES~III could run on
      higher $\psi$ resonances, high above the $D\bar{D}$ threshold
      and with high enough statistics to produce the higher $P$, $D$
      states via transitions from the $\psi$'s.  While the two running
      Tevatron experiments will complete their period of data taking,
      reaching total final samples of about 16 fb$^{-1}$, CERN will
      return to the scene in 2008, when the first beams are expected
      to circulate in the LHC accelerator.  The first period of data
      taking, at low luminosity, will be very important to establish
      the physics potential of the ATLAS, CMS and LHCb experiments on
      heavy quarkonium production and $B_c$ studies. Later, three
      experiments will be able to study quarkonia production in
      heavy-ion collisions at the LHC: ALICE, CMS and ATLAS.  ALICE
      will measure the charmonia states through their decay into
      electron and muon pairs.  Before the LHC start-up, the CERN
      experiment COMPASS also has the possiblility of obtaining new
      information on the doubly charmed baryons discovered by SELEX at
      FNAL.  At CEBAF, high-intensity 9~GeV photon beams will allow
      the study of the $\jpsi$ photoproduction mechanism at low
      energy.
  
      The future of the asymmetric B factories is unclear at present.
      If at least one of the two facilities increases its luminosity
      to become a super B factory, a large number of charmonium
      measurements will be within reach, and even the simple ISR
      production mechanism will yield record samples of
      $\Upsilon(1,2,3S)$ decays. A promising opportunity, recommended
      by the QWG, is to turn one of the two asymmetric B factories
      into an $\Upsilon(1,2,3S)$ factory during the last fraction of
      its running period.  There is considerable $b\bar{b}$ physics
      that remains to be done.  A few examples are: $\eta_b(n^1S_0)$
      searches using M1 radiative transitions from the
      $\Upsilon(n^3S_1)$ states, $h_b$ searches via hadronic
      transitions $\Upsilon(3S)\to h_b + \pi^0 \to \eta_b +\gamma +
      \pi^0$ or $\Upsilon(3S)\to h_b + \pi\pi \to \eta_b +\gamma +
      \pi\pi$, and measuring radiative and hadronic transitions in the
      $b\bar{b}$ system.
  
\item In the long term (2010 and beyond), PANDA at GSI offers the
      opportunity to study many charmonium states with non $1^{--}$
      quantum numbers and can challenge BES~III on high-statistic
      studies of the narrow charmonium states. Given the huge yields
      of heavy quarkonia expected from LHC experiments when running at
      full luminosity, the main problem will be the choice of
      high-priority physics cases and the main challenge will come
      from the need to keep trigger rates under control.  At the same
      time the BTeV experiment at Tevatron will challenge LHCb on
      $B_c$ physics in the forward region.  Finally, the ILC, in the
      next decade, will test NRQCD at $\ttbar$ threshold, as well as a
      number of still unpredicted future hot issues in heavy
      quarkonium physics.
\end{itemize}

%\end{document}

\BLKP
\fancyhead{}
%9/12/2004 
%%%%%%%%%%%%%%%%%%%%%%%%%%%%%%%%%%%%%%%%%%%%%%%%%%%%%%%%%%%%
{\Huge\bf  Acknowledgements}\\
\vspace{3mm}
%%%%%%%%%%%%%%%%%%%%%%%%%%%%%%%%%%%%%%%%%%%%%%%%%%%%%%%%%%%%

\noindent
We acknowledge the following contracts/funds agencies:
%Bali/Davies
EC Hadron Physics I3 Contract No. RII3-CT-2004-506078
PPARC grants PPA/A/S/2000/00271, PPA/G/0/2002/0463;
%Bodwin
%Work by G.T.~Bodwin in the High Energy Physics Division at Argonne National
%Laboratory is supported by 
the U.S.~Department of Energy, Division of
High Energy Physics, under Contract No.~W-31-109-ENG-38;
%Braaten
%This research was supported in part by 
the U.S.~Department of Energy under
grant DE-FG02-91-ER4069;
%Brambilla-Vairo
Iniziativa Specifica INFN MI23; Azioni Integrate Italy-Spain IT1824;
Azione Esplorativa U. Milano;
%Carnegie Mellon/Tom Ferguson
US Department of Energy under contract number DE-FG02-91ER40682;
%Godfrey
The Natural Sciences and Engineering Research Council of Canada;
%Faustov
The DFG grant No.Eb 139/2-2 and
The RFBR grant No.03-02-27251z;
%Harris
The US Department of Energy under contract number DE-FG02-04ER41291;
%Matthias Jamin, supported in part by 
the German BMBF under contract 05HT4WOA/3;
%Kiselev:
Grant of the president of Russian Federation for scientific schools
NSc-1303.2003.2 and the Russian Foundation for 
Basic Research, grant 04-02-17530;
%Kronfeld
%Fermilab is operated by Universities Research Association, Inc.,
%under contract with the U.S. Department of Energy.
%Kuang
The National Natural Science Foundation of China under the grant number 90103008;
%Morningstar
the NSF under the grant PHY-0354982;
%Vaia Papadimitriou
%(DOE grants)
the US Department of Energy under grants
DE-FG03-95ER40938,
DE-AC02-76CH03000;
%Sanchis-Lozano
the grants FPA2002-00612 and GV-GRUPOS03/094;
%Soto-Pineda
the CICYT-INFN (Spain-Italy) 2003 and 2004 collaboration contracts;
MCyT and Feder (Spain) grant FPA2001-3598;
CIRIT (Catalonia) grant 2001SGR-00065;
network EURIDICE (EU) HPRN-CT2002-00311;
Acciones Integradas (Spain-Italy) HI2003-0362;
Acciones Integradas (Spain-Germany) HA2003-0141;
the Italian MIUR program ``incentivazione alla mobilit\`a di studiosi 
stranieri e italiani residenti all'estero (rientro dei cervelli)''.

\end{document}